Ioan Doré Landau • Rogelio Lozano •
Mohammed M'Saad • Alireza Karimi

# Adaptive Control

## Algorithms, Analysis and Applications

Second Edition Third edition in progress

Prof. Ioan Doré Landau
Département d'Automatique
GIPSA-LAB (CNRS/INPG/UJF)
PO Box 46
38402 St. Martin d'Heres
France
ioan-dore.landau@gipsa-lab.grenoble-inp.fr

Prof. Rogelio Lozano
UMR-CNRS 6599
Centre de Recherche de Royalieu
Heuristique et Diagnostic des Systèmes Complexes
Université de Technologie de Compiègne
PO Box 20529
60205 Compiègne
France
Rogelio.Lozano@hds.utc.fr

Prof. Mohammed M'Saad
Centre de Recherche (ENSICAEN)
Laboratoire GREYC
École Nationale Supérieure d'Ingénieurs de Caen
Campus Côte de Nacre
bd. Maréchal Juin 6
14032 Caen Cedex
France
msaad@greyc.ensicaen.fr

Prof. Alireza Karimi
Laboratoire d'Automatique
École Polytechnique Fédérale de Lausanne
1015 Laussanne
Switzerland
alireza.karimi@epfl.ch

*To Lina, Leticia, Salima, Noushin*
*Vlad, Jessica, Rogelio, Azadeh and Omid*

*Ce qui est simple est toujours faux*
*Ce qui ne l'est pas est inutilisable*

*Paul Valéry*
*Mauvaises Pensées*

# Preface

Adaptive control provides techniques for the automatic adjustment of control parameters in real time either to achieve or to maintain a desired level of control system performance when the dynamic parameters of the process to be controlled are unknown and/or time-varying. The main characteristic of these techniques is the ability to extract significant information from real data in order to tune the controller and they feature a mechanism for adjusting the parameters of either the plant model or the controller. The history of adaptive control is long, significant progress in understanding and applying its ideas having begun in the early nineteen-seventies. The growing availability of digital computers has also contributed to the progression of the field. The early applications provided important feedback for the development of the field and theoretical innovations allowed a number of basic problems to be solved. The aim of this book is to provide a coherent and comprehensive treatment of the field of adaptive control. The presentation takes the reader from basic problem formulation to analytical solutions the practical significance of which is illustrated by applications. A unified presentation of adaptive control is not obvious. One reason for this is that several design steps are involved and this increases the number of degrees of freedom. Another is that methods have been proposed having different applications in mind but without a clear motivation for the intermediate design steps. It is our belief, however, that a coherent presentation of the basic techniques of adaptive control is now possible. We have adopted a discrete-time formulation for the problems and solutions described to reflect the importance of digital computers in the application of adaptive control techniques and we share our understanding and practical experience of the soundness of various control designs with the reader. Throughout the book, the mathematical aspects of the synthesis and analysis of various algorithms are emphasized; however, this does not mean that they are sufficient in themselves for solving practical problems or that ad hoc modifications of the algorithms for specific applications are not possible. To guide readers, the book contains various applications of control techniques but it is our belief that without a solid mathematical understanding of the adaptation techniques available, they will not be able to apply them creatively to new and difficult situations. The book has grown out of several survey papers, tutorial and courses delivered to various audiences (graduate students, practicing engineers, etc.) in various countries, of the research in the field done by the authors (mostly at Laboratoire d'Automatique de Grenoble, now the Control Department of GIPSA-LAB (Institut National Polytechnique de Grenoble/CNRS), HEUDYASIC (Université Technologique de Compiègne/CNRS), CINVESTAV (Mexico), GREYC (Caen) and the Laboratoire d'Automatique of EPFL (Lausanne)), and of the long and rich practical experience of the authors. On the one hand, this new edition reflects new developments in the field both in terms of techniques and applications and, on the other, it puts a number of techniques into proper perspective as a result of feedback from applications.

**Expected Audience** The book is intended as a textbook for graduate students as well as a basic reference for practicing engineers facing the problem of designing adaptive control systems. Control researchers from other areas will find a comprehensive presentation of the field with bridges to various other control design techniques.

**About the Content** It is widely accepted that stability analysis in a deterministic environment and convergence analysis in a stochastic environment constitute a basic grounding for analysis and design of adaptive control systems and so these form the core of the theoretical aspects of the book. Parametric adaptation algorithms (PAAs) which are present in all adaptive control techniques are considered in greater depth.

Our practical experience has shown that in the past the basic linear controller designs which make up the background for various adaptive control strategies have often not taken robustness issues into account. It is both possible and necessary to accommodate these issues by improving the robustness of the linear control designs prior to coupling them with one of the adaptation algorithms so the book covers this.

In the context of adaptive control, robustness also concerns the parameter adaptation algorithms and this issue is addressed in detail. Furthermore, multiple-model adaptive control with switching is an illustration of the combination of robust and adaptive control and is covered in depth in the new edition. In recent years, plant model identification in closed-loop operation has become more and more popular as a way of improving the performance of an existing controller. The methods that have arisen as a result are directly relevant to adaptive control and will also be thoroughly treated. Adaptive regulation and adaptive feedforward disturbance compensation have emerged as new adaptive control problems with immediate application in active vibration control and active noise control. These aspects are now covered in this second edition.

The book is organized as follows:

- Chapter 1 provides an introduction to adaptive control and a tutorial presentation of the various techniques involved.
- Chapter 2 presents a brief review of discrete-time linear models for control with emphasis on optimal predictors which are often used throughout the book.
- Chapter 3 is a thorough coverage of parameter adaptation algorithms (PAA) operating in a deterministic environment. Various approaches are presented and then the stability point of view for analysis and design is discussed in detail.

- Chapter 4 is devoted to the analysis of parameter adaptation algorithms in a stochastic environment.
- Chapter 5 discusses recursive plant model identification in open loop which is an immediate application of PAAs on the one hand and an unavoidable step in starting an adaptive controller on the other.
- Chapter 6 is devoted to the synthesis of adaptive predictors.
- Chapter 7 covers digital control strategies which are used in adaptive control. One step ahead predictive control and long-range predictive control are presented in a unified manner.
- Chapter 8 discusses the robust digital control design problem and provides techniques for achieving required robustness by shaping the sensitivity functions.
- Digital control techniques can be combined with the recursive plant model identification in closed loop to obtain an adaptive controller. These recursive identification techniques are discussed in Chap. 9.
- The issue of robustification of parameter adaptation algorithm in the context of adaptive control is addressed in Chap. 10.
- For special types of plant model structures and control strategies, appropriate parametrization of the plant model allows direct adjustment of the parameters of the controllers yielding so called *direct adaptive control schemes*. Direct adaptive control is the subject of Chap. 11.
- Indirect adaptive control which combines in real-time plant model parameter estimation in closed loop with the redesign of the controller is discussed in Chap. 12.
- Multimodel adaptive control with switching, which combines robust control and adaptive control, is discussed in Chap. 13 (new in the second edition).
- Rejection of unknown disturbances is the objective of adaptive regulation which is the subject of Chap. 14 (new in the second edition).
- Adaptive feedforward compensation of disturbances is discussed in Chap. 15 (new in the second edition).
- Chapter 16 is devoted to the practical aspects of implementing adaptive controllers.

Chapters 5, 9, 12, 13, 14 and 15 include applications using the techniques presented in these chapters. A number of appendices which summarize important background topics are included.

Problems and simulation exercises are included in most of the chapters.

**Pathways Through the Book** The book was written with the objective of presenting comprehensive coverage of the field of adaptive control and of making the subject accessible to a large audience with different backgrounds and interests. Thus the book can be read and used in different ways.

For those only interested in applications we recommend the following sequence: Chaps.: 1, 2, 3 (Sects. 3.1 and 3.2), 5 (Sects. 5.1, 5.2, 5.7 through 5.9), 7 (Sects. 7.1, 7.2, 7.3.1 and 7.3.2), 8 (Sects. 8.1, 8.2 and 8.3.1), 9 (Sects. 9.1 and 9.6), 10 (Sect. 10.1), 11 (Sects. 11.1 and 11.2), 12 (Sects. 12.1 and 12.2.1), 13 (Sects. 13.1, 13.2 and 13.4), 14 (Sects. 14.1, 14.2, 14.4 and 14.7), 15 (Sects. 15.1, 15.2 and 15.5) and Chap.16. Most of the content of Chaps. 14 and 15 can also be

The subjects of chapters 14 and 15 have been expanded in the book "Adaptive and Robust Active Vibration Control" (Landau & al.) Springer 2017. See also :
http:// http://www.gipsa-lab.grenoble-inp.fr/~ioandore.landau/benchmark_adaptive_regulation/

**Fig. 1** Logical dependence of the chapters

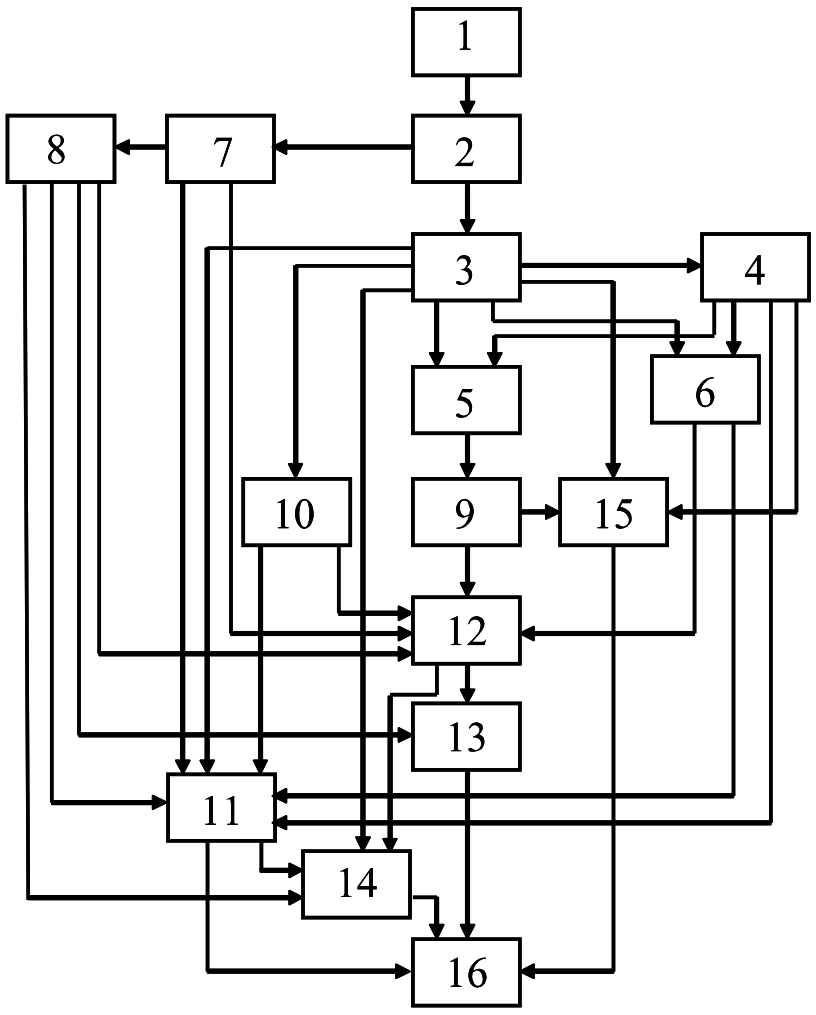

read just after Chap. 3. The sequence above (till Chap. 15) can also serve as an introductory course in adaptive control.

For a more in-depth study of the field a course should include in addition the following Sects.: 3.3, 3.4, 4.1, 4.2, 5.3 through 5.6, 6.1, 6.2, 7.3.3 through 7.7, 8.3 through 8.6, 9.2 through 9.6, 10.2, 10.3, 10.4, 10.6, 11.4.1, 11.4.2 and 11.6, 12.2.2 through 12.3.1, 12.4 and 12.7, 13.3, 14.3, 14.5, 15.3 and 15.4. A graduate course in adaptive control might include all chapters of the book.

The material has been organized so that readers can easily see how the more technical parts of the book can be bypassed. Figure 1 shows the logical progression of the chapters.

**The Website** Complementary information and material for teaching and applications can be found on the book website: ~~http://www.landau-adaptivecontrol.org.~~
http://http://www.gipsa-lab.grenoble-inp.fr/~ioandore.landau/adaptivecontrol/

**Acknowledgments** We wish to acknowledge the large number of contributors on whose work our presentation is partly based. In particular, we wish to mention: G. Zames, V.M. Popov, L. Ljung, G. Goodwin, D. Clarke, K.J. Aström, B.D.O. Anderson, A.S. Morse, P. Kokotovic from whom we learned many things.

In our research activity we had the privilege of interacting with a number of colleagues among whom we would like to mention: M. Tomizuka, R. Bitmead, M. Gevers, C.R. Johnson, H.M. Silveira, C. Samson, L. Praly, R. Ortega, Ph. de Larminat, K. Najim, E. Irving, F. Giri, B. Egardt, L. Dugard, J.M. Dion, B. Brogliato, G. Béthoux, B. Courtiol, H. Duong, A. Besançon and H. Prochazka. We would like to express our appreciation for their contributions.

The long term support of the Centre National de la Recherche Scientifique (CNRS) and of the Institut National Polytechnique de Grenoble is gratefully acknowledged.

We would like to thank J. Langer, A. Constantinescu, J. Chebassier and M. Alma for their effective contribution to this project.

This second edition has also been made possible by Stéphane Mocanu who succeed in finding the “lost” electronic files of the first edition.

We would also like to thank Oliver Jackson from Springer whose enthusiasm and professionalism has helped us to finalize this new edition of the book.

Writing takes a lot of time and most of the writing has been done on overtime. We would like to thank our families for their patience.

Grenoble, France

Ioan Doré Landau
Rogelio Lozano
Mohammed M'Saad
Alireza Karimi

# Contents

# Abbreviations

*Acronyms*

| | |
|---|---|
| a.s. | Almost sure convergence |
| ANC | Active noise control |
| ARMA | Auto regressive moving average |
| ARMAX | Auto regressive moving average with exogenous input |
| AVC | Active vibration control |
| FOE | Filtered output error algorithm |
| GPC | Generalized predictive control |
| CLOE | Closed loop output error recursive algorithm |
| EFR | Equivalent feedback representation |
| ELS | Extended least squares algorithm |
| FOL | Filtered open loop identification algorithm |
| G-CLOE | Generalized closed loop output error algorithm |
| GLS | Generalized least squares algorithm |
| IVAM | Instrumental variable with auxiliary model |
| LHS | Left hand side |
| MRAS | Model reference adaptive system |
| OE | Recursive output error algorithm |
| OEAC | Output error with adjustable compensator |
| OEEPM | Output error with extended prediction model |
| OEFC | Output error with fixed compensator |
| PAA | Parameter adaptation algorithm |
| PRBS | Pseudo random binary sequence |
| PSMR | Partial state model reference control |
| RHS | Right hand side |
| RLS | Recursive least squares algorithm |
| RML | Recursive maximum likelihood algorithm |
| SPR | Strictly positive real |
| X-CLOE | Extended closed loop output error algorithm |

*Notation*

| | |
|---|---|
| $f_s$ | Sampling frequency |
| $T_s$ | Sampling period |
| $t$ | Continuous time or normalized discrete time (with respect to the sampling period) |
| $u(t)$, $y(t)$ | Plant input and output |
| $e(t)$ | Discrete-time Gaussian white noise |
| $\hat{y}(t+j/t)$ | $j$-steps ahead prediction of $y(t)$ |
| $q^{-1}$ | Backward shift operator ($q^{-1}y(t+1)=y(t)$) |
| $\tau$ | Time delay (continuous time systems) |
| $s, z$ | Complex variables ($z=e^{sT_s}$) |
| $d$ | Delay of the discrete-time system (integer number of sampling periods) |
| $A(q^{-1})$ | Polynomial in the variable $q^{-1}$ |
| $\hat{A}(t,q^{-1})$ | Estimation of the polynomial $A(q^{-1})$ at instant $t$ |
| $\hat{a}_i(t)$ | Estimation of the coefficients of the polynomials $A(q^{-1})$ (they are the coefficients of the polynomial $A(t,q^{-1})$) |
| $\theta$ | Parameter vector |
| $\hat{\theta}(t)$ | Estimated parameter vector |
| $\tilde{\theta}(t)$ | Parameter error vector |
| $\phi(t)$, $\Phi(t)$ | Measurement or observation vector |
| $F$, $F(t)$ | Adaptation gain |
| $\hat{y}^0(t)$ | A priori output of an adjustable predictor |
| $\hat{y}(t)$ | A posteriori output of an adjustable predictor |
| $\varepsilon^0(t)$ | A priori prediction error |
| $\varepsilon(t)$ | A posteriori prediction error |
| $\nu^0(t)$ | A priori adaptation error |
| $\nu(t)$ | A posteriori adaptation error |
| $P(z^{-1})$ | Polynomial defining the closed loop poles |
| $P_D(z^{-1})$ | Polynomial defining the dominant closed loop poles |
| $P_F(z^{-1})$ | Polynomial defining the auxiliary closed loop poles |
| $A, M, F$ | Matrices |
| $F>0$ | Positive definite matrix |
| $\omega_0$ | Natural frequency of a 2nd order system |
| $\zeta$ | Damping coefficient of a 2nd order system |
| $\mathbf{E}\{\cdot\}$ | Expectation |
| $R(i)$ | Autocorrelation or cross-correlation |
| $RN(i)$ | Normalized autocorrelation or cross-correlation |

# Chapter 1
# Introduction to Adaptive Control

## 1.1 Adaptive Control—Why?

*Adaptive Control* covers a set of techniques which provide a systematic approach for automatic adjustment of controllers in *real time*, in order to achieve or to maintain a desired level of control system performance when the parameters of the plant dynamic model are unknown and/or change in time.

Consider first the case when the parameters of the dynamic model of the plant to be controlled are unknown but constant (at least in a certain region of operation). In such cases, although the structure of the controller will not depend in general upon the particular values of the plant model parameters, the correct tuning of the controller parameters cannot be done without knowledge of their values. Adaptive control techniques can provide an automatic tuning procedure in closed loop for the controller parameters. In such cases, the effect of the adaptation vanishes as time increases. Changes in the operation conditions may require a restart of the adaptation procedure.

Now consider the case when the parameters of the dynamic model of the plant change unpredictably in time. These situations occur either because the environmental conditions change (ex: the dynamical characteristics of a robot arm or of a mechanical transmission depend upon the load; in a DC-DC converter the dynamic characteristics depend upon the load) or because we have considered simplified linear models for nonlinear systems (a change in operation condition will lead to a different linearized model). These situations may also occur simply because the parameters of the system are slowly time-varying (in a wiring machine the inertia of the spool is time-varying). In order to achieve and to maintain an acceptable level of control system performance when large and unknown changes in model parameters occur, an *adaptive control* approach has to be considered. In such cases, the adaptation will operate most of the time and the term *non-vanishing adaptation* fully characterizes this type of operation (also called *continuous adaptation*).

Further insight into the operation of an adaptive control system can be gained if one considers the design and tuning procedure of the "good" controller illustrated in Fig. 1.1. In order to design and tune a good controller, one needs to:

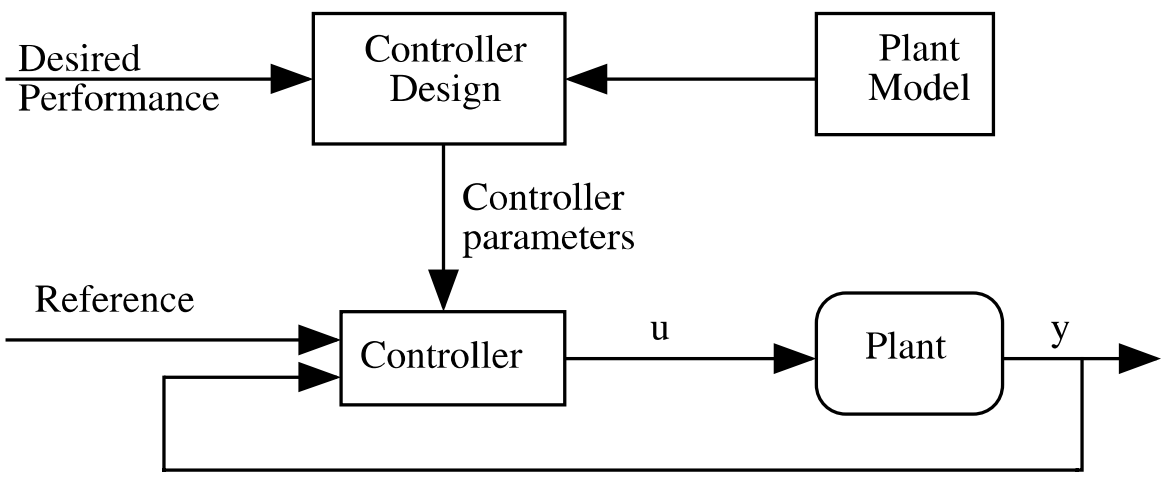

**Fig. 1.1** Principles of controller design

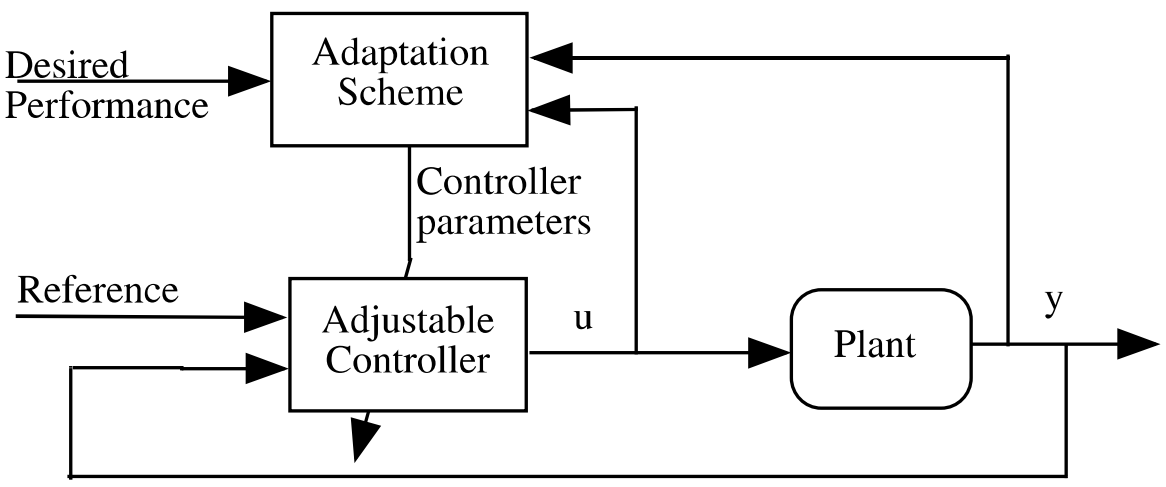

**Fig. 1.2** An adaptive control system

(1) Specify the desired control loop performances.
(2) Know the dynamic model of the plant to be controlled.
(3) Possess a suitable controller design method making it possible to achieve the desired performance for the corresponding plant model.

The dynamic model of the plant can be identified from input/output plant measurements obtained under an experimental protocol in open or in closed loop. One can say that the design and tuning of the controller is done from data collected on the system. An adaptive control system can be viewed as an implementation of the above design and tuning procedure in real time. The tuning of the controller will be done in real time from data collected in real time on the system. The corresponding adaptive control scheme is shown in Fig. 1.2.

The way in which information is processed in real time in order to tune the controller for achieving the desired performances will characterize the various adaptation techniques. From Fig. 1.2, one clearly sees that an adaptive control system is nonlinear since the parameters of the controller will depend upon measurements of system variables through the adaptation loop.

The above problem can be reformulated as nonlinear stochastic control with incomplete information. The unknown parameters are considered as auxiliary states (therefore the linear models become nonlinear: $\dot{x} = ax \Longrightarrow \dot{x}_1 = x_1 x_2$, $\dot{x}_2 = v$ where $v$ is a stochastic process driving the parameter variations). Unfortunately, the resulting solutions (*dual control*) are extremely complicated and cannot be implemented in practice (except for very simple cases). Adaptive control techniques can be viewed as approximation for certain classes of nonlinear stochastic control problems associated with the control of processes with unknown and time-varying parameters.

## 1.2 Adaptive Control Versus Conventional Feedback Control

The unknown and unmeasurable variations of the process parameters degrade the performances of the control systems. Similarly to the disturbances acting upon the controlled variables, one can consider that the variations of the process parameters are caused by disturbances acting upon the parameters (called parameter disturbances). These parameter disturbances will affect the performance of the control systems. Therefore the disturbances acting upon a control system can be classified as follows:

(a) disturbances acting upon the controlled variables;
(b) (parameter) disturbances acting upon the performance of the control system.

Feedback is basically used in conventional control systems to reject the effect of disturbances upon the controlled variables and to bring them back to their desired values according to a certain performance index. To achieve this, one first measures the controlled variables, then the measurements are compared with the desired values and the difference is fed into the controller which will generate the appropriate control.

A similar conceptual approach can be considered for the problem of achieving and maintaining the desired performance of a control system in the presence of parameter disturbances. We will have to define first a *performance index* (IP) for the control system which is a measure of the performance of the system (ex: the damping factor for a closed-loop system characterized by a second-order transfer function is an IP which allows to quantify a desired performance expressed in terms of "damping"). Then we will have to measure this IP. The *measured* IP will be compared to the *desired* IP and their difference (if the measured IP is not acceptable) will be fed into an *adaptation mechanism*. The output of the *adaptation mechanism* will act upon the parameters of the controller and/or upon the control signal in order to modify the system performance accordingly. A block diagram illustrating a basic configuration of an adaptive control system is given in Fig. 1.3.

Associated with Fig. 1.3, one can consider the following definition for an adaptive control system.

**Definition 1.1** *An adaptive control system* measures a certain performance index (IP) of the control system using the inputs, the states, the outputs and the known disturbances. From the comparison of the measured performance index and a set of given ones, the adaptation mechanism modifies the parameters of the adjustable controller and/or generates an auxiliary control in order to maintain the performance index of the control system close to the set of given ones (i.e., within the set of acceptable ones).

Note that the control system under consideration is an *adjustable dynamic system* in the sense that its performance can be adjusted by modifying the parameters of the controller or the control signal. The above definition can be extended straightforwardly for "adaptive systems" in general (Landau 1979).

A conventional feedback control system will monitor the controlled variables under the effect of disturbances acting on them, but its performance will vary (it is not monitored) under the effect of parameter disturbances (the design is done assuming known and constant process parameters).

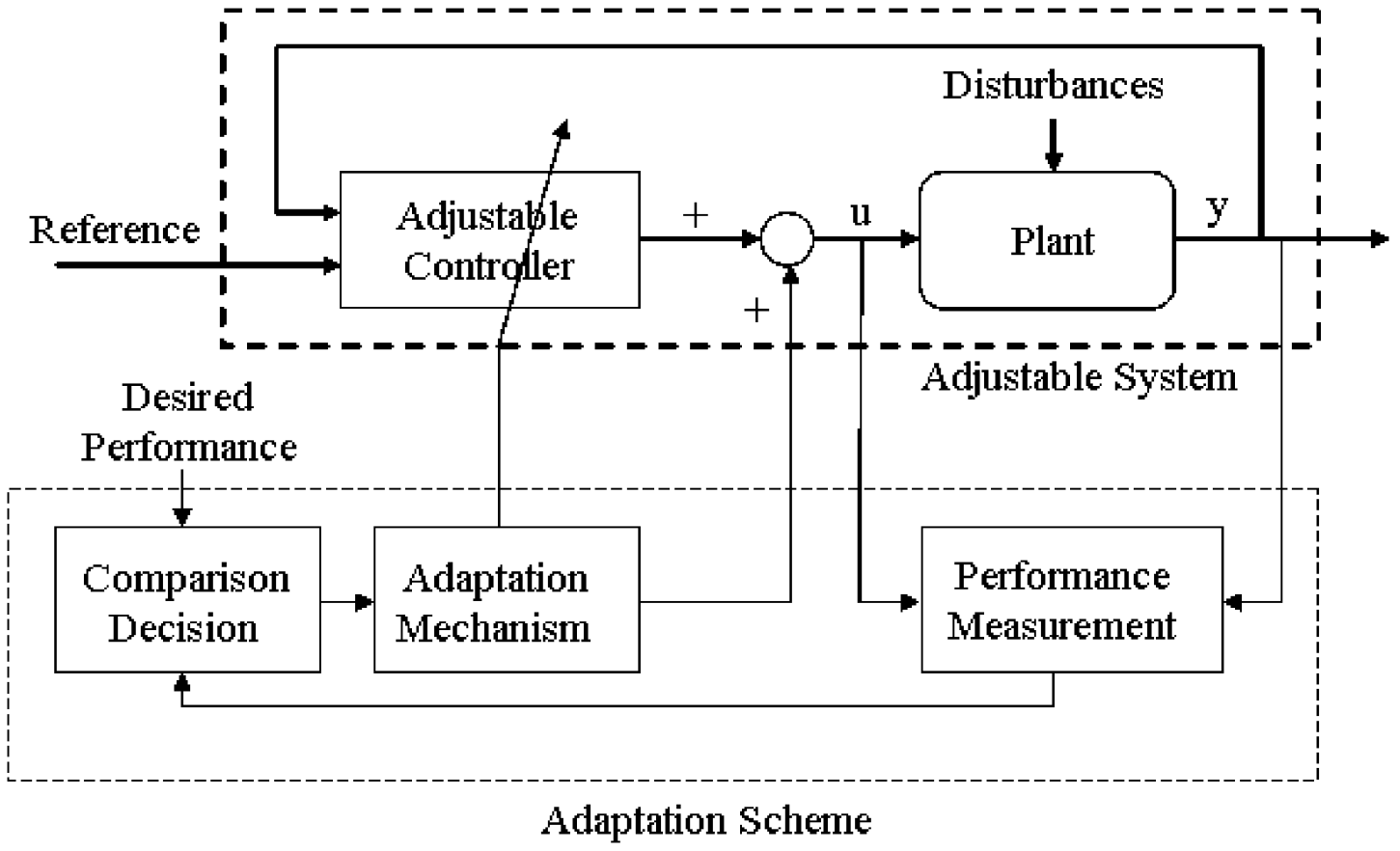

**Fig. 1.3** Basic configuration for an adaptive control system

An adaptive control system, which contains in addition to a feedback control with adjustable parameters a supplementary loop acting upon the adjustable parameters of the controller, will monitor the performance of the system in the presence of parameter disturbances.

Consider as an example the case of a conventional feedback control loop designed to have a given damping. When a disturbance acts upon the controlled variable, the return of the controlled variable towards its nominal value will be characterized by the desired damping if the plant parameters have their known nominal values. If the plant parameters change upon the effect of the parameter disturbances, the damping of the system response will vary. When an adaptation loop is added, the damping of the system response will be maintained when changes in parameters occur.

Comparing the block diagram of Fig. 1.3 with a conventional feedback control system, one can establish the correspondences which are summarized in Table 1.1.

*While the design of a conventional feedback control system is oriented firstly toward the elimination of the effect of disturbances upon the controlled variables, the design of adaptive control systems is oriented firstly toward the elimination of the effect of parameter disturbances upon the performance of the control system.* An adaptive control system can be interpreted as a feedback system where the controlled variable is the *performance index* (IP).

One can view an adaptive control system as a hierarchical system:

- Level 1: conventional feedback control;
- Level 2: adaptation loop.

In practice often an additional "monitoring" level is present (Level 3) which decides whether or not the conditions are fulfilled for a correct operation of the adaptation loop.

**Table 1.1** Adaptive control versus conventional feedback control

| Conventional feedback control system | Adaptive control system |
|---|---|
| Objective: monitoring of the "controlled" variables according to a certain IP for the case of known parameters | Objective: monitoring of the performance (IP) of the control system for unknown and varying parameters |
| Controlled variable | Performance index (IP) |
| Transducer | IP measurement |
| Reference input | Desired IP |
| Comparison block | Comparison decision block |
| Controller | Adaptation mechanism |

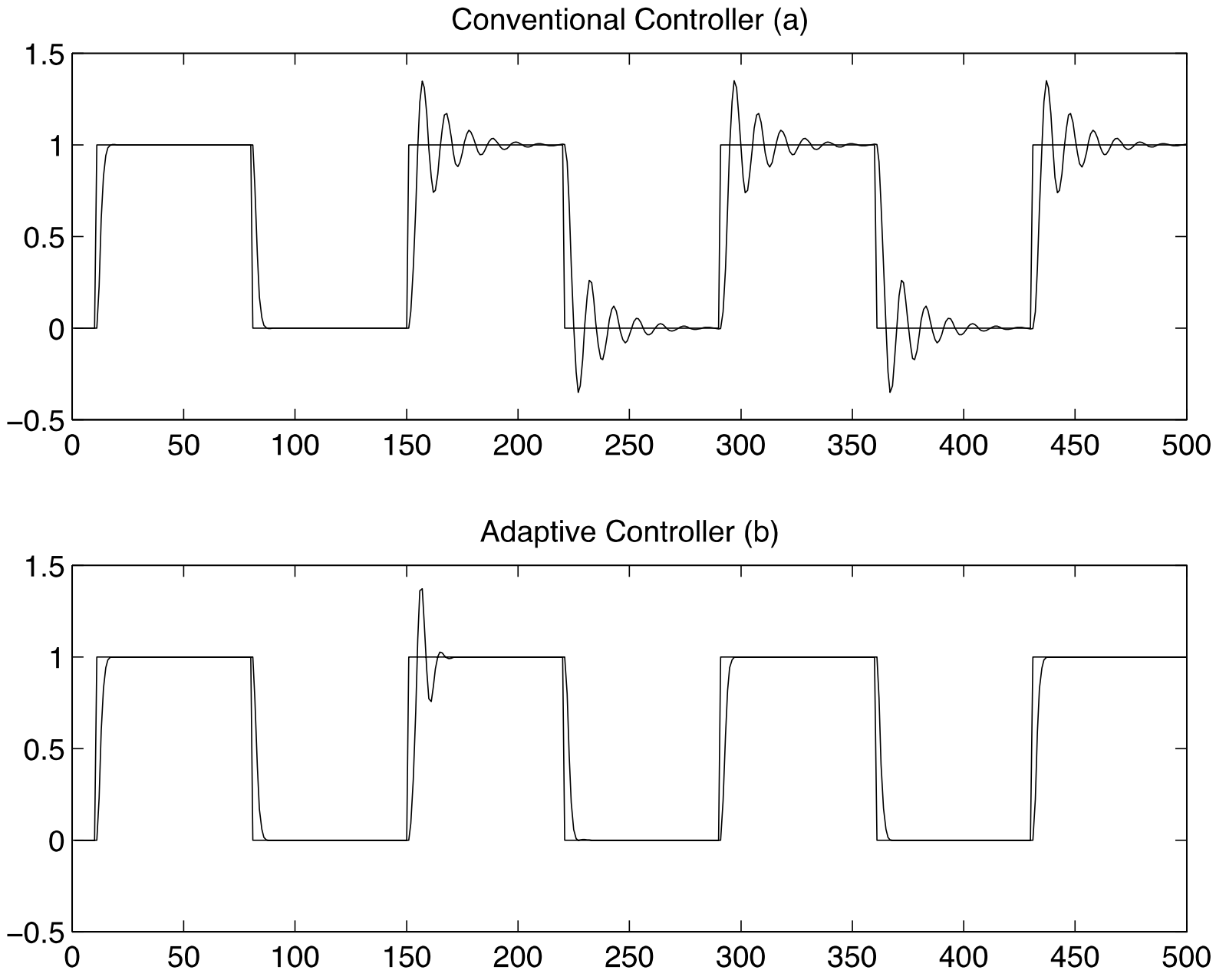

**Fig. 1.4** Comparison of an adaptive controller with a conventional controller (fixed parameters), (**a**) fixed parameters controller, (**b**) adaptive controller

Figure 1.4 illustrates the operation of an adaptive controller. In Fig. 1.4a, a change of the plant model parameters occurs at $t = 150$ and the controller used has constant parameters. One can see that poor performance results from this parameter change. In Fig. 1.4b, an adaptive controller is used. As one can see, after an adaptation transient the nominal performance is recovered.

### 1.2.1 Fundamental Hypothesis in Adaptive Control

The operation of the adaptation loop and its design relies upon the following fundamental hypothesis: *For any possible values of plant model parameters there is a controller with a fixed structure and complexity such that the specified performances can be achieved with appropriate values of the controller parameters.*

In the context of this book, the plant models are assumed to be linear and the controllers which are considered are also linear.

Therefore, *the task of the adaptation loop is solely to search for the "good" values of the controller parameters.*

This emphasizes the importance of the control design for the known parameter case (the *underlying control design problem*), as well as the necessity of a priori information about the structure of the plant model and its characteristics which can be obtained by *identification* of a model for a given set of operational conditions.

In other words, an adaptive controller is not a "black box" which can solve a control problem in real time without an initial knowledge about the plant to be controlled. This a priori knowledge is needed for specifying achievable performances, the structure and complexity of the controller and the choice of an appropriate design method.

### 1.2.2 Adaptive Control Versus Robust Control

In the presence of model parameter variations or more generally in the presence of variations of the dynamic characteristics of a plant to be controlled, *robust control design* of the conventional feedback control system is a powerful tool for achieving a satisfactory level of performance for a family of plant models. This family is often defined by means of a *nominal model* and a size of the uncertainty specified in the parameter domain or in the frequency domain.

The range of uncertainty domain for which satisfactory performances can be achieved depends upon the problem. Sometimes, a large domain of uncertainty can be tolerated, while in other cases, the uncertainty tolerance range may be very small. If the desired performances cannot be achieved for the full range of possible parameter variations, adaptive control has to be considered in addition to a robust control design. Furthermore, the tuning of a robust design for the true nominal model using an adaptive control technique will improve the achieved performance of the robust controller design. Therefore, robust control design will benefit from the use of adaptive control in terms of performance improvements and extension of the range of operation. On the other hand, using an underlying robust controller design for building an adaptive control system may drastically improve the performance of the adaptive controller. This is illustrated in Figs. 1.5, 1.6 and 1.7, where a comparison between conventional feedback control designed for the nominal model, robust control design and adaptive control is presented. To make a fair comparison the presence of unmodeled dynamics has been considered in addition to the parameter variations.

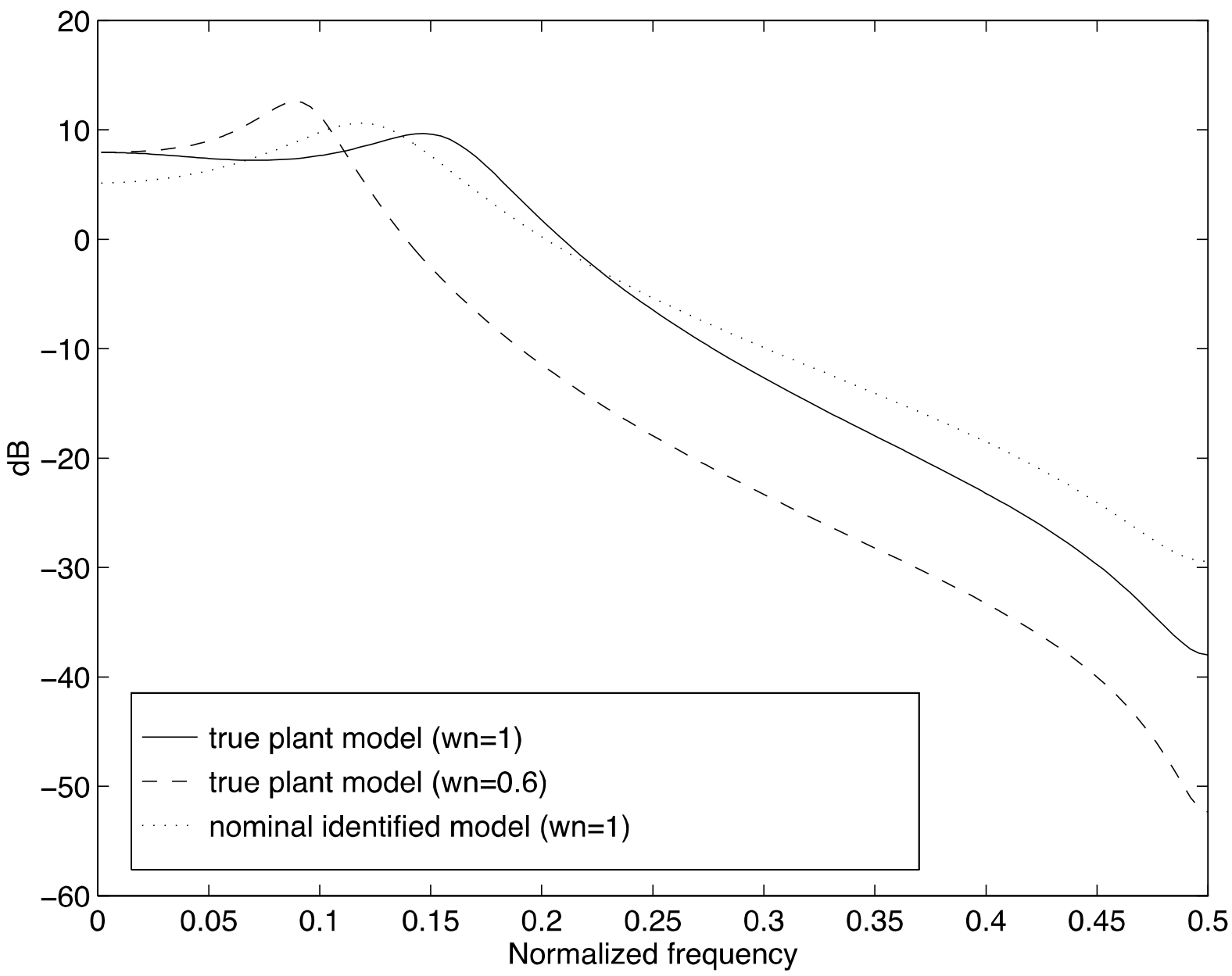

**Fig. 1.5** Frequency characteristics of the true plant model for $\omega_0 = 1$ and $\omega_0 = 0.6$ and of the identified model for $\omega_0 = 1$

For each experiment, a nominal plant model is used in the first part of the record and a model with different parameters is used in the second part.

The plant considered for this example is characterized by a third order model formed by a second-order system with a damping factor of 0.2 and a natural frequency varying from $\omega_0 = 1$ rad/sec to $\omega_0 = 0.6$ rad/sec and a first order system. The first order system corresponds to a high-frequency dynamics with respect to the second order. The change of the damping factor occurs at $t = 150$.

The nominal system (with $\omega_0 = 1$) has been identified using a second-order model (lower order modeling). The frequency characteristics of the true model for $\omega_0 = 1$, $\omega_0 = 0.6$ and of the identified model for $\omega_0 = 1$ are shown in Fig. 1.5.

Based on the second-order model identified for $\omega_0 = 1$ a conventional fixed controller is designed (using pole placement—see Chap. 7 for details). The performance of this controller is illustrated in Fig. 1.6a. One can see that the performance of the closed-loop system is seriously affected by the change of the natural frequency. Figure 1.6b shows the performance of a robust controller designed on the basis of the same identified model obtained for $\omega_0 = 1$ (for this design pole placement is combined with the shaping of the sensitivity functions—see Chap. 8 for details). One can observe that the nominal performance is slightly lower (slower step response) than for the previous controller but the performance remains acceptable when the characteristics of the plant change.

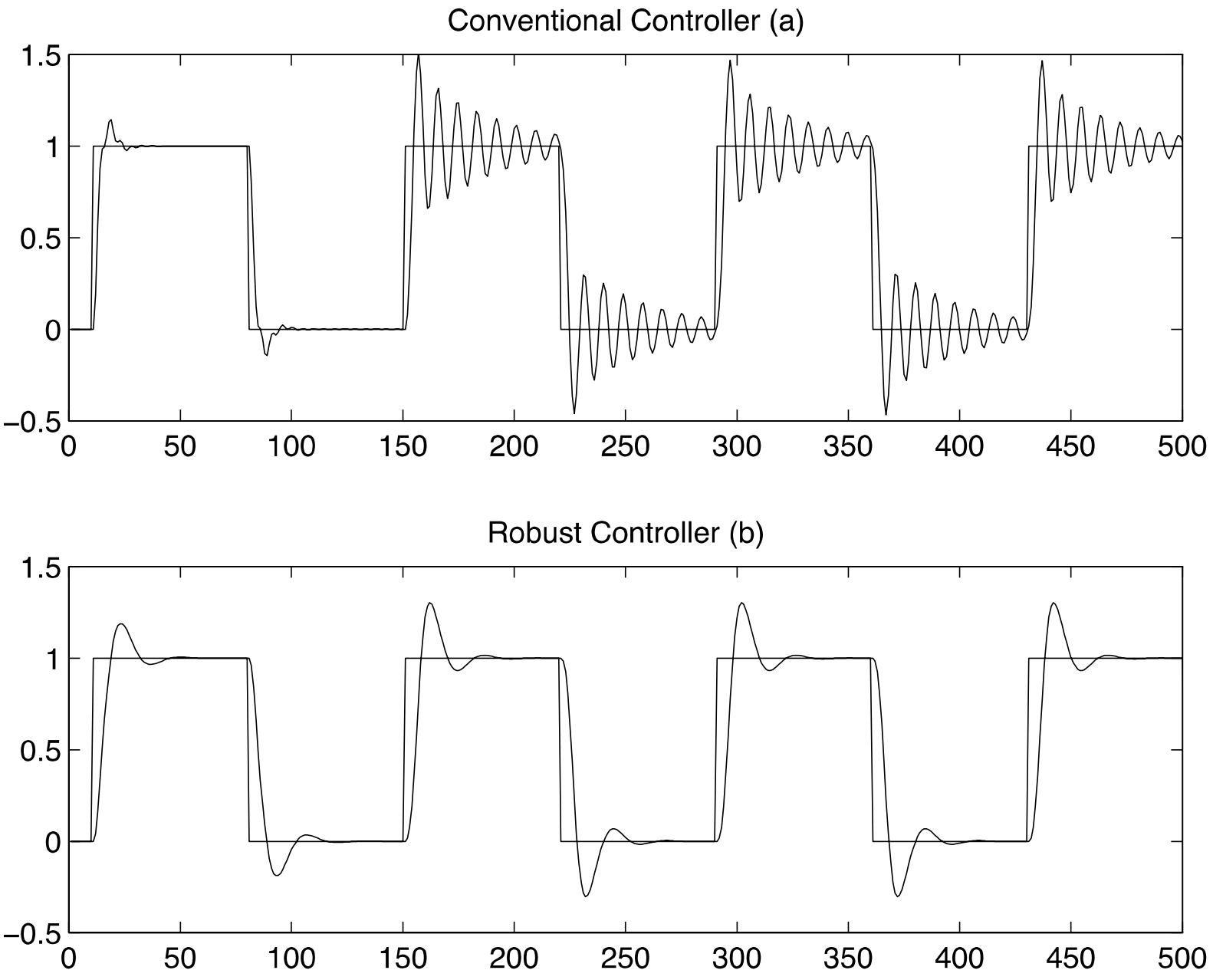

**Fig. 1.6** Comparison of conventional feedback control and robust control, (**a**) conventional design for the nominal model, (**b**) robust control design

Figure 1.7a shows the response of the control system when the parameters of the conventional controller used in Fig. 1.6a are adapted, based on the estimation in real time of a second-order model for the plant. A standard parameter adaptation algorithm is used to update the model parameters. One observes that after a transient, the nominal performances are recovered except that a residual high-frequency oscillation is observed. This is caused by the fact that one estimates a lower order model than the true one (but this is often the situation in practice). To obtain a satisfactory operation in such a situation, one has to "robustify" the adaptation algorithm (in this example, the "filtering" technique has been used—see Chap. 10 for details) and the results are shown in Fig. 1.7b. One can see that the residual oscillation has disappeared but the adaptation is slightly slower.

Figure 1.7c shows the response of the control system when the parameters of the robust controller used in Fig. 1.6b are adapted using exactly the same algorithm as for the case of Fig. 1.7a. In this case, even with a standard adaptation algorithm, residual oscillations do not occur and the transient peak at the beginning of the adaptation is lower than in Fig. 1.7a. However, the final performance will not be better than that of the robust controller for the nominal model.

After examining the time responses, one can come to the following conclusions:

1. Before using adaptive control, it is important to do a robust control design.
2. Robust control design improves in general the adaptation transients.

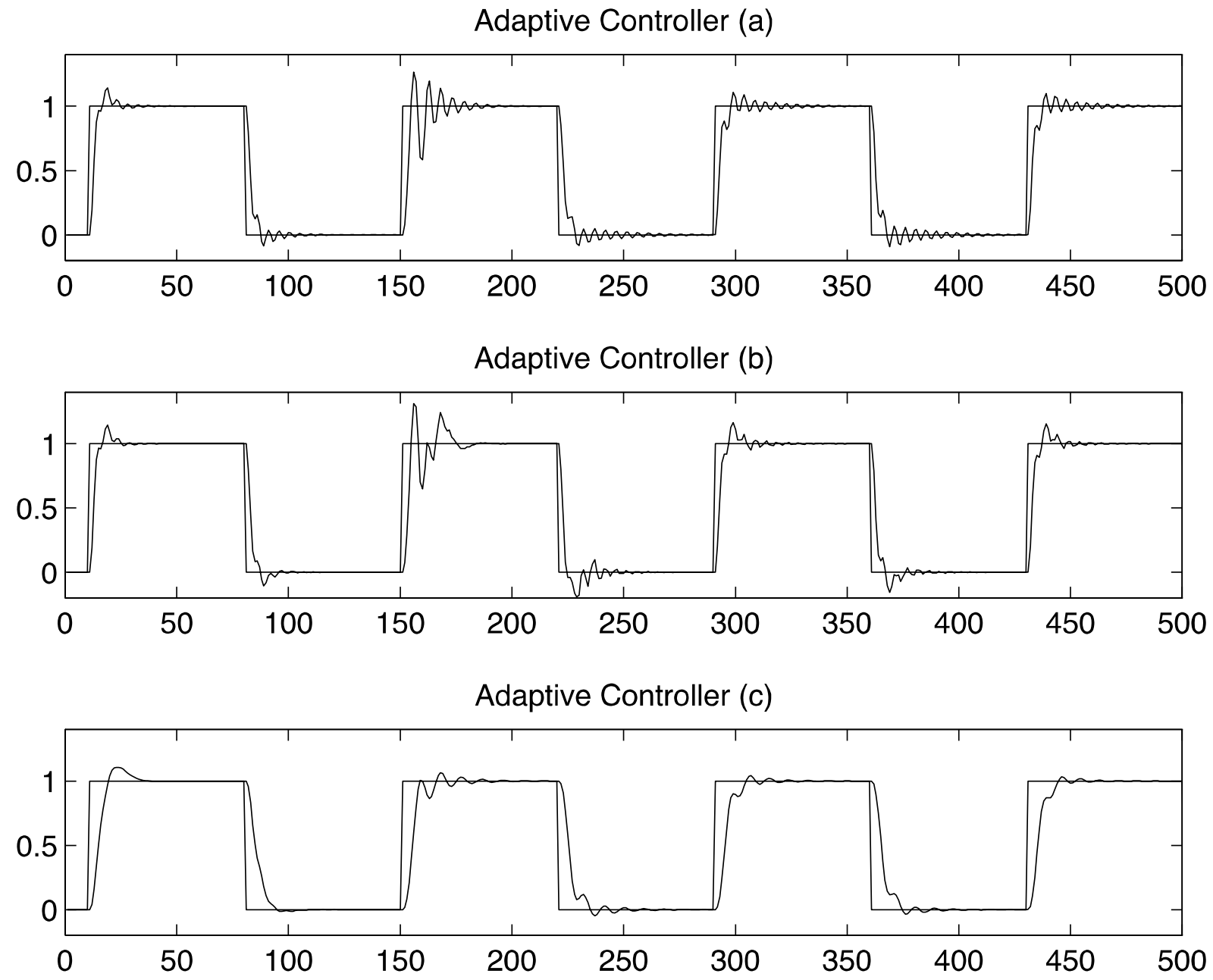

**Fig. 1.7** Comparison of adaptive controller, (**a**) adaptation added to the conventional controller (Fig. 1.6a), (**b**) robust adaptation added to the conventional controller (Fig. 1.6a), (**c**) adaptation added to the robust controller (Fig. 1.6b)

3. A robust controller is a "fixed parameter" controller which instantaneously provides its designed characteristics.
4. The improvement of performance via adaptive control requires the introduction of additional algorithms in the loop and an "adaptation transient" is present (the time necessary to reach the desired performance from a degraded situation).
5. A trade-off should be considered in the design between robust control and robust adaptation.

## 1.3 Basic Adaptive Control Schemes

In the context of various adaptive control schemes, the implementation of the three fundamental blocks of Fig. 1.3 (performance measurement, comparison-decision, adaptation mechanism) may be very intricate. Indeed, it may not be easy to decompose the adaptive control scheme in accordance with the basic diagram of Fig. 1.3. Despite this, the basic characteristic which allows to decide whether or not a system is truly "adaptive" is the presence or the absence of the closed-loop control of a certain performance index. More specifically, an *adaptive control* system will use information collected in real time to improve the tuning of the controller in order to achieve or to maintain a level of desired performance. There are many control systems which are designed to achieve acceptable performance in the presence of parameter variations, but they do not assure a closed-loop control of the performance and, as such, they are not "adaptive". The typical example is the *robust control design* which, in many cases, can achieve acceptable performances in the presence of parameter variations using a fixed controller.

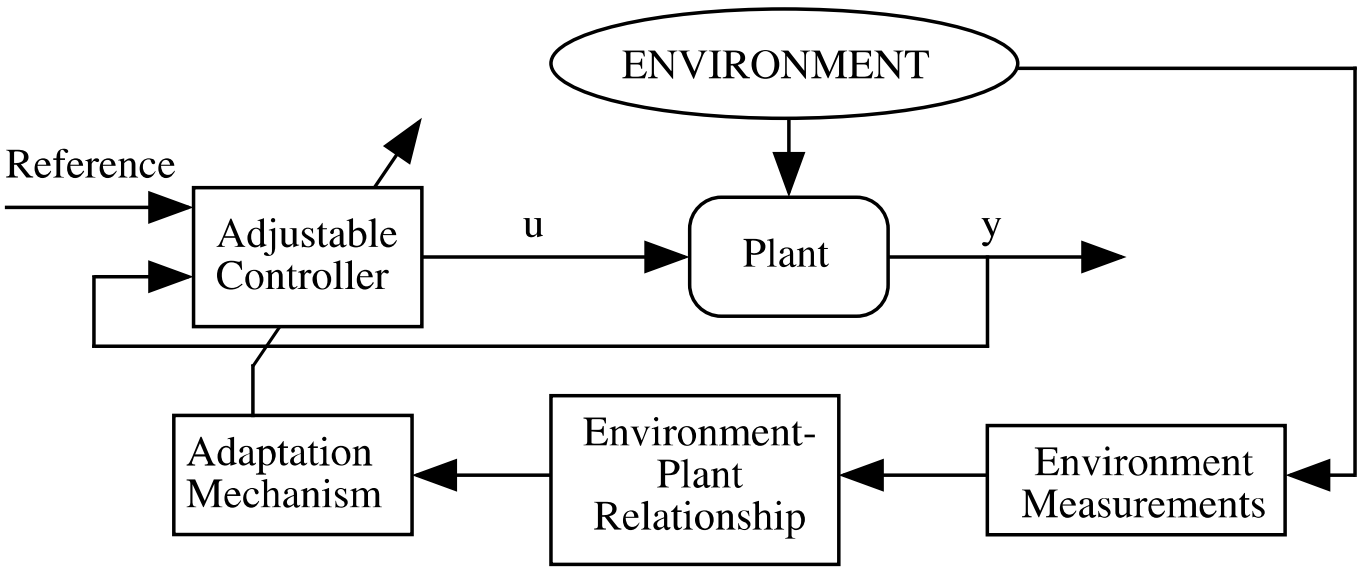

**Fig. 1.8** Open-loop adaptive control

We will now go on to present some basic schemes used in adaptive control.

### *1.3.1 Open-Loop Adaptive Control*

We shall consider next as an example the "gain-scheduling" scheme which is an *open-loop adaptive control system*. A block diagram of such a system is shown in Fig. 1.8. The adaptation mechanism in this case is a simple look-up table stored in the computer which gives the controller parameters for a given set of environment measurements. This technique assumes the existence of a rigid relationship between some measurable variables characterizing the environment (the operating conditions) and the parameters of the plant model. Using this relationship, it is then possible to reduce (or to eliminate) the effect of parameter variations upon the performance of the system by changing the parameters of the controller accordingly.

This is an *open-loop adaptive* control system because the modifications of the system performance resulting from the change in controller parameters are not measured and feedback to a comparison-decision block in order to check the efficiency of the parameter adaptation. This system can fail if for some reason or another the rigid relationship between the environment measurements and plant model parameters changes.

Although such *gain-scheduling systems* are not fully adaptive in the sense of Definition 1.1, they are widely used in a variety of situations with satisfactory results. Typical applications of such principles are:

1. adjustments of autopilots for commercial jet aircrafts using speed and altitude measurements,

2. adjustment of the controller in hot dip galvanizing using the speed of the steel strip and position of the actuator (Fenot et al. 1993),

and many others.

Gain-scheduling schemes are also used in connection with adaptive control schemes where the gain-scheduling takes care of rough changes of parameters when the conditions of operation change and the adaptive control takes care of the fine tuning of the controller.

Note however that in certain cases, the use of this simple principle can be very costly because:

1. It may require additional expensive transducers.
2. It may take a long time and numerous experiments in order to establish the desired relationship between environment measurements and controller parameters.

In such situations, an adaptive control scheme can be cheaper to implement since it will not use additional measurements and requires only additional computer power.

### 1.3.2 Direct Adaptive Control

Consider the basic philosophy for designing a controller discussed in Sect. 1.1 and which was illustrated in Fig. 1.1.

One of the key points is the specification of the desired control loop performance. In many cases, the desired performance of the feedback control system can be specified in terms of the characteristics of a dynamic system which is a *realization* of the desired behavior of the closed-loop system. For example, a tracking objective specified in terms of rise time, and overshoot, for a step change command can be alternatively expressed as the input-output behavior of a transfer function (for example a second-order with a certain resonance frequency and a certain damping). A regulation objective in a deterministic environment can be specified in terms of the evolution of the output starting from an initial disturbed value by specifying the desired location of the closed-loop poles. In these cases, the controller is designed such that for a given plant model, the closed-loop system has the characteristics of the *desired* dynamic system.

The design problem can in fact be equivalently reformulated as in Fig. 1.9. The *reference model* in Fig. 1.9 is a realization of the system with *desired* performances. The design of the controller is done now in order that:

(1) the error between the output of the plant and the output of the reference model is identically zero for identical initial conditions;
(2) an initial error will vanish with a certain dynamic.

When the plant parameters are unknown or change in time, in order to achieve and to maintain the desired performance, an adaptive control approach has to be considered and such a scheme known as *Model Reference Adaptive Control* (*MRAC*) is shown in Fig. 1.10.

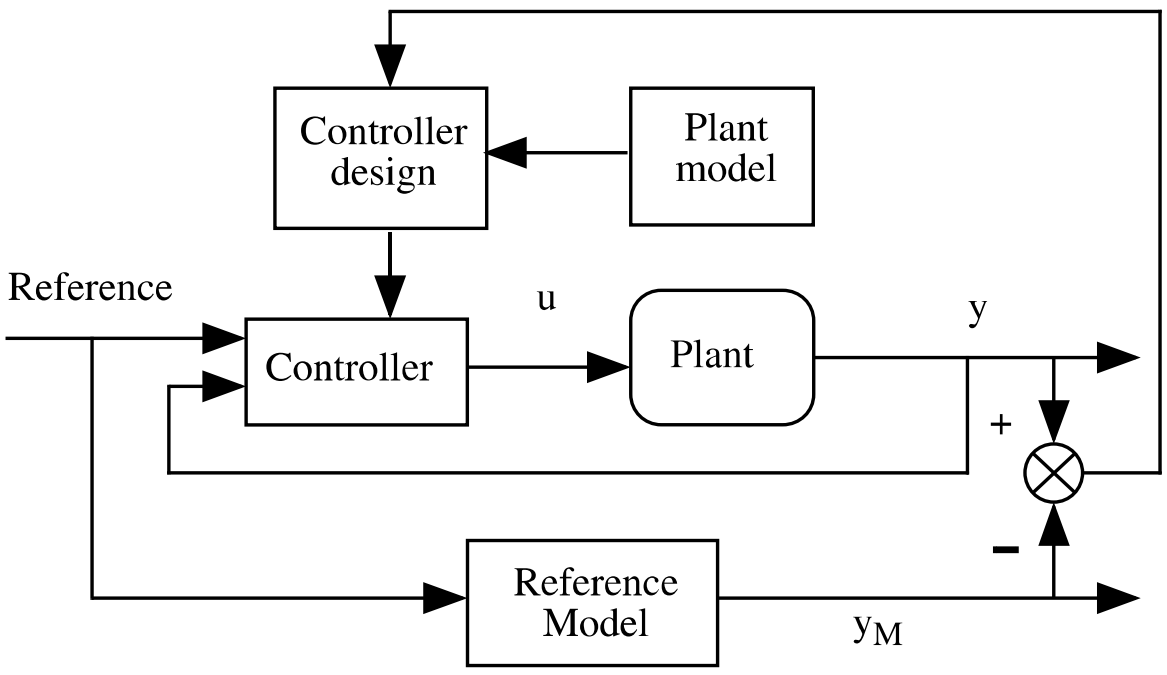

**Fig. 1.9** Design of a linear controller in deterministic environment using an explicit reference model for performance specifications

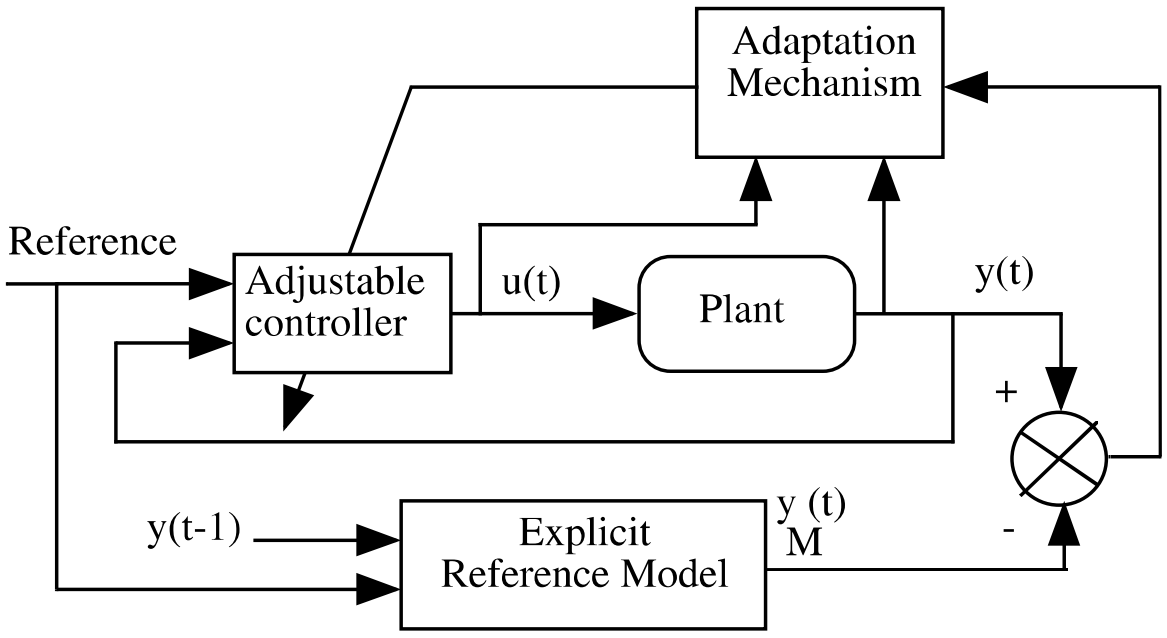

**Fig. 1.10** Model Reference Adaptive Control scheme

This scheme is based on the observation that the difference between the output of the plant and the output of the reference model (called subsequently plant-model error) is a measure of the difference between the real and the desired performance. This information (together with other information) is used by the *adaptation mechanism* (subsequently called *parameter adaptation algorithm*) to directly adjust the parameters of the controller in real time in order to force asymptotically the plant-model error to zero. This scheme corresponds to the use of a more general concept called *Model Reference Adaptive Systems* (*MRAS*) for the purpose of control. See Landau (1979). Note that in some cases, the reference model may receive measurements from the plant in order to predict future desired values of the plant output.

The model reference adaptive control scheme was originally proposed by Whitaker et al. (1958) and constitutes the basic prototype for direct adaptive control.

The concept of model reference control, and subsequently the concept of direct adaptive control, can be extended for the case of operation in a stochastic environment. In this case, the disturbance affecting the plant output can be modeled as an ARMA process, and no matter what kind of linear controller with fixed parameter will be used, the output of the plant operating in closed loop will be an ARMA model. Therefore the control objective can be specified in terms of a desired ARMA model for the plant output with desired properties. This will lead to the concept of *stochastic reference model* which is in fact a *prediction reference model*. See Landau (1981). The prediction reference model will specify the desired behavior of the predicted output. The plant-model error in this case is the *prediction error* which is used to directly adapt the parameters of the controller in order to force asymptotically the plant-model stochastic error to become an *innovation* process. The *self tuning minimum variance controller* (Åström and Wittenmark 1973) is the basic example of direct adaptive control in a stochastic environment. More details can be found in Chaps. 7 and 11.

**Fig. 1.11** Indirect adaptive control (principle)

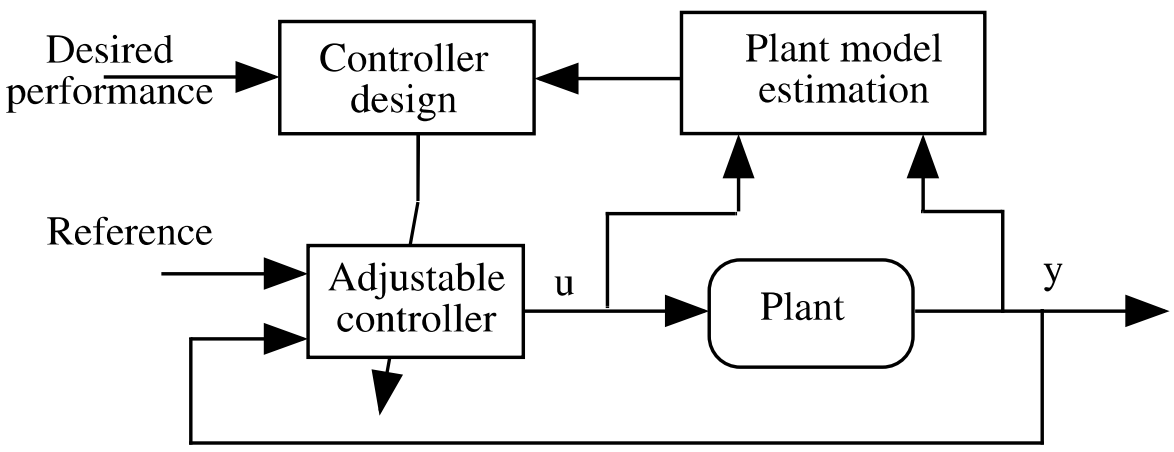

Despite its elegance, the use of *direct adaptive control* schemes is limited by the hypotheses related to the underlying linear design in the case of known parameters. While the performance can in many cases be specified in terms of a reference model, the conditions for the existence of a feasible controller allowing for the closed loop to match the reference model are restrictive. One of the basic limitations is that one has to assume that the plant model has in all the situations stable zeros, which in the discrete-time case is quite restrictive.[1] The problem becomes even more difficult in the multi-input multi-output case. While different solutions have been proposed to overcome some of the limitations of this approach (see for example M'Saad et al. 1985; Landau 1993a), direct adaptive control cannot always be used.

### *1.3.3 Indirect Adaptive Control*

Figure 1.11 shows an indirect adaptive control scheme which can be viewed as a real-time extension of the controller design procedure represented in Fig. 1.1. The basic idea is that a suitable controller can be designed on line if a model of the plant is estimated on line from the available input-output measurements. The scheme is termed *indirect* because the adaptation of the controller parameters is done in two stages:

(1) on-line estimation of the plant parameters;
(2) on-line computation of the controller parameters based on the current estimated plant model.

---

[1]Fractional delay larger than half sampling periods leads to unstable zeros. See Landau (1990a). High-frequency sampling of continuous-time systems with difference of degree between denominator and numerator larger or equal to two leads to unstable zeros. See Åström et al. (1984).

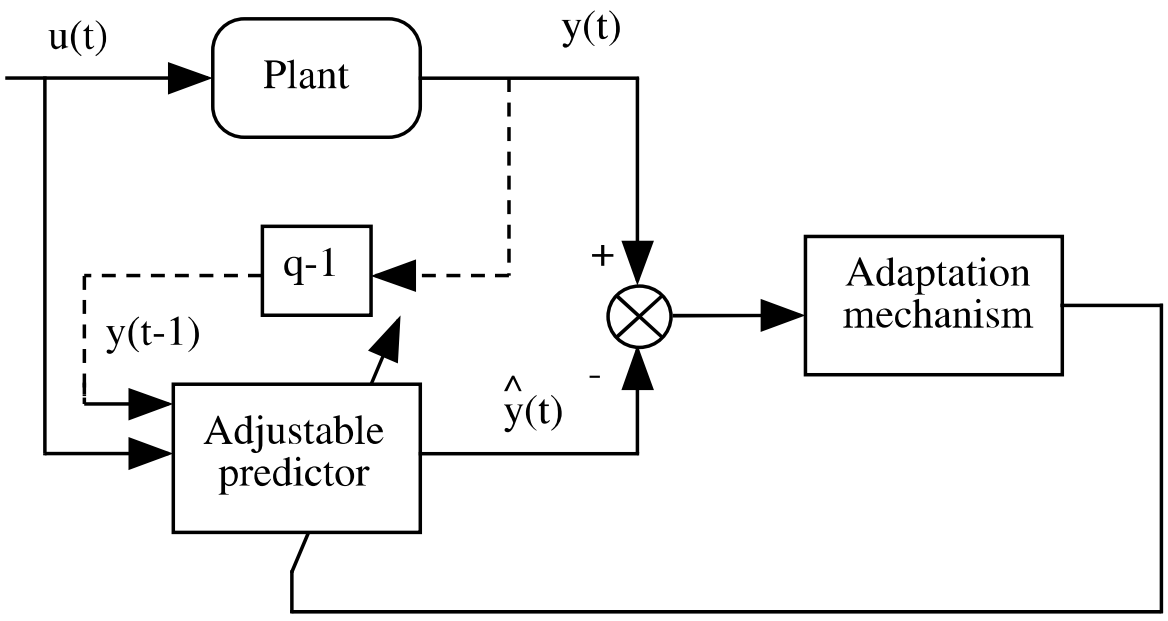

**Fig. 1.12** Basic scheme for on-line parameter estimation

This scheme uses current plant model parameter estimates as if they are equal to the true ones in order to compute the controller parameters. This is called the *ad-hoc certainty equivalence* principle.[2]

The indirect adaptive control scheme offers a large variety of combinations of control laws and parameter estimation techniques. To better understand how these indirect adaptive control schemes work, it is useful to consider in more detail the on-line estimation of the plant model.

The basic scheme for the on-line estimation of plant model parameters is shown in Fig. 1.12. The basic idea is to build an *adjustable predictor* for the plant output which may or may not use previous plant output measurements and to compare the predicted output with the measured output. The error between the plant output and the predicted output (subsequently called *prediction error* or *plant-model error*) is used by a *parameter adaptation algorithm* which at each sampling instant will adjust the parameters of the adjustable predictor in order to minimize the prediction error in the sense of a certain criterion. This type of scheme is primarily an *adaptive predictor* which will allow an estimated model to be obtained asymptotically giving thereby a correct input-output description of the plant for the given sequence of inputs.

This technique is successfully used for the plant model identification in open-loop (see Chap. 5). However, in this case special input sequences with a rich frequency content will be used in order to obtain a model giving a correct input-output description for a large variety of possible inputs.

The situation in indirect adaptive control is that in the absence of external rich excitations one cannot guarantee that the excitation will have a sufficiently rich spectrum and one has to analyze when the computation of the controller parameters based on the parameters of an adaptive predictor will allow acceptable performance to be obtained asymptotically.

Note that on-line estimation of plant model parameters is itself an adaptive system which can be interpreted as a *Model Reference Adaptive System* (MRAS). The plant to be identified represents the reference model. The parameters of the adjustable predictor (the adjustable system) will be driven by the PAA (parameter adaptation algorithm) in order to minimize a criterion in terms of the adaptation error (prediction error).

[2]For some designs a more appropriate term will be *ad-hoc separation theorem*.

**Table 1.2** Duality of model reference adaptive control and adaptive prediction

| Model reference adaptive control | Adaptive predictor |
|---|---|
| Reference model | Plant |
| Adjustable system (plant + controller) | Adjustable predictor |

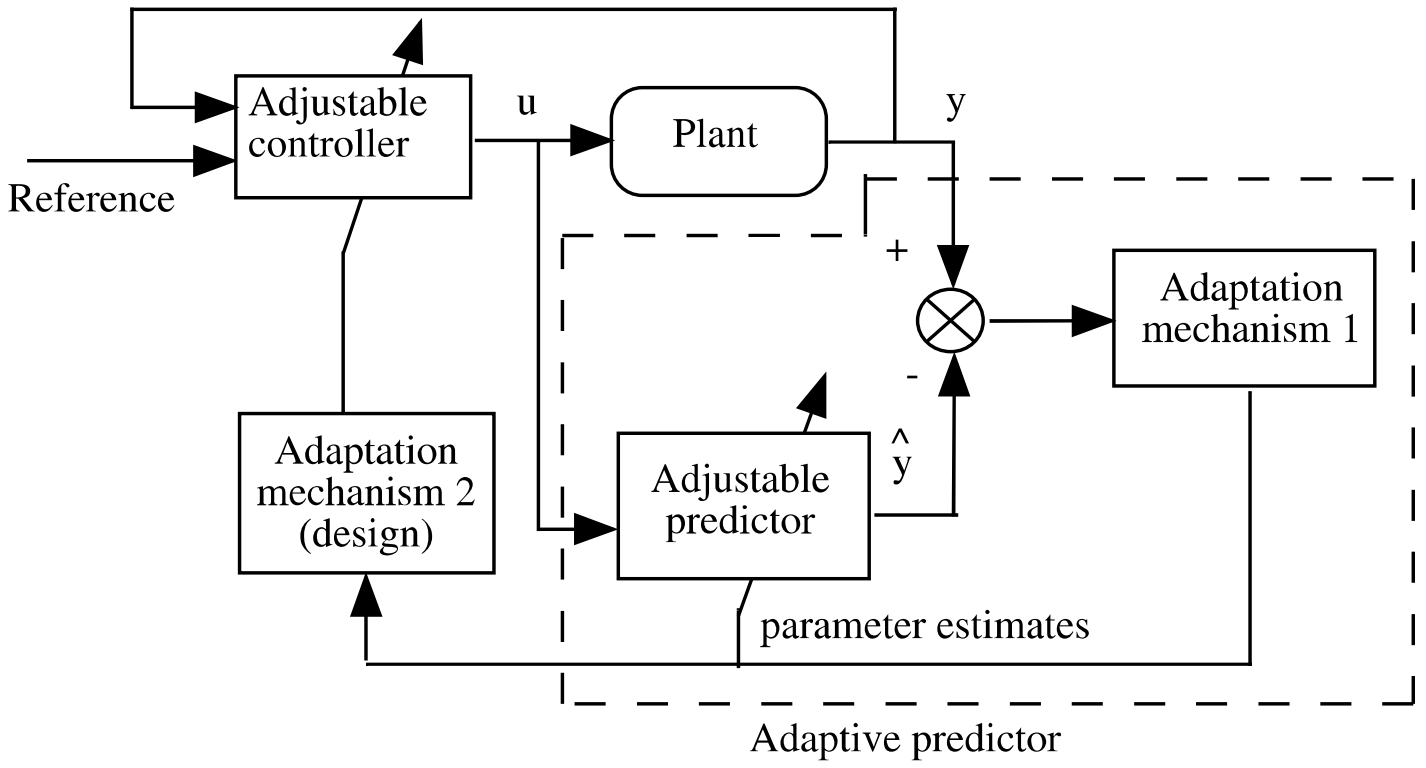

**Fig. 1.13** Indirect adaptive control (detailed scheme)

The scheme of Fig. 1.12 is the *dual* of Model Reference Adaptive Control because they have a similar structure but they achieve different objectives. Note that one can pass from one configuration to the other by making the following substitutions (Landau 1979) (see Table 1.2).

Introducing the block diagram for the plant model parameter estimation given in Fig. 1.12 into the scheme of Fig. 1.11, one obtains the general configuration of an *indirect adaptive control* shown in Fig. 1.13. Using the indirect adaptive control schemes shown in Fig. 1.13, one can further elaborate on the ad-hoc use of the "certainty equivalence" or "separation theorem" which hold for the linear case with known parameters.

In terms of *separation* it is assumed that the adaptive predictor gives a good prediction (or estimation) of the plant output (or states) when the plant parameters are unknown, and that the prediction error is independent of the input to the plant (this is false however during adaptation transients). The adjustable predictor is a system for which full information is available (parameters and states). An appropriate control for the predictor is computed and this control is also applied to the plant. In terms of *certainty equivalence*, one considers the unknown parameters of the plant model as additional states. The control applied to the plant is the same as the one applied when all the "states" (i.e., parameters and states) are known exactly, except that the "states" are replaced by their estimates. The indirect adaptive control was originally introduced by Kalman (1958).

However, as mentioned earlier, the parameters of the controller are calculated using plant parameter estimates and there is no evidence, therefore, that such schemes will work (they are not the exact ones, neither during adaptation, nor in general, even asymptotically). A careful analysis of the behavior of these schemes should be done. In some cases, external excitation signals may be necessary to ensure the convergence of the scheme toward desired performances. As a counterpart adaptation has to be stopped if the input of the plant whose model has to be estimated is not *rich* enough (meaning a sufficiently large frequency spectrum).

Contributions by Gevers (1993), Van den Hof and Schrama (1995) have led to the observation that in indirect adaptive control the objective of the plant parameter estimation is to provide the best prediction for the behavior of the closed loop system, for given values of the controller parameters (in other words this allows to assess the performances of the controlled system). This can be achieved by either using appropriate data filters on plant input-output data or by using adaptive predictors for the closed-loop system parameterized in terms of the controller parameters and plant parameters. See Landau and Karimi (1997b), Chaps. 9 and 16.

### *1.3.4 Direct and Indirect Adaptive Control: Some Connections*

Comparing the direct adaptive control scheme shown in Fig. 1.10 with the indirect adaptive control scheme shown in Fig. 1.13, one observes an important difference. In the scheme of Fig. 1.10, the parameters of the controller are directly estimated (adapted) by the adaptation mechanism. In the scheme of Fig. 1.13, the adaptation mechanism 1 tunes the parameters of an adjustable predictor and these parameters are then used to compute the controller parameters.

However, in a number of cases, related to the desired control objectives and structure of the plant model, by an appropriate parameterization of the adjustable predictor (reparameterization), the parameter adaptation algorithm of Fig. 1.13 will directly estimate the parameter of the controller yielding to a direct adaptive control scheme. In such cases the adaptation mechanism 2 (the design block) disappears and one gets a *direct adaptive control* scheme. In these schemes, the output of the adjustable predictor (whose parameters are known at each sampling) will behave as the output of a reference model. For this reason, such schemes are also called "implicit model reference adaptive control" (Landau 1981; Landau and Lozano 1981; Egardt 1979). This is illustrated in Fig. 1.14.

To illustrate the idea of "reparameterization" of the plant model, consider the following example. Let the discrete-time plant model be:

$$y(t+1) = -a_1 y(t) + u(t) \tag{1.1}$$

where $y$ is the plant output, $u$ is the plant input and $a$ is an unknown parameter. Assume that the desired objective is to find $u(t)$ such that:

$$y(t+1) = -c_1 y(t) \tag{1.2}$$

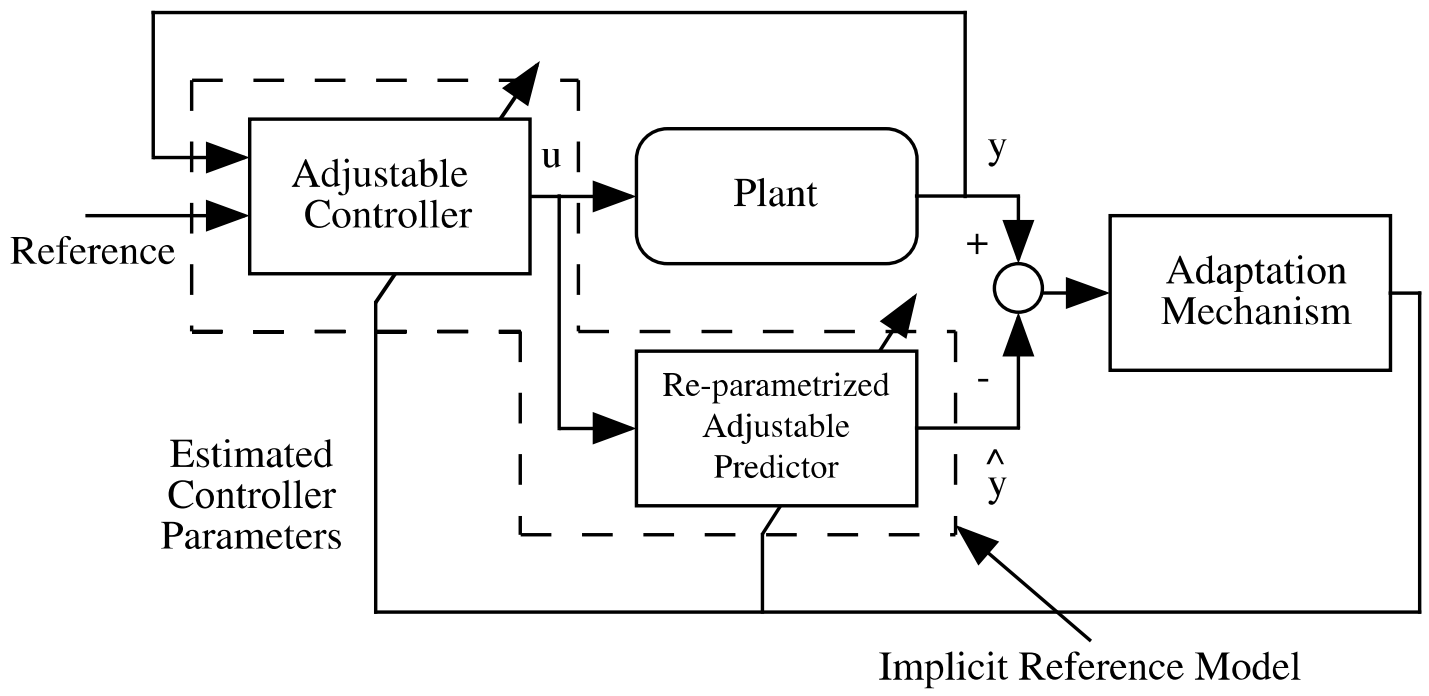

**Fig. 1.14** Implicit model reference adaptive control

(The desired closed-loop pole is defined by $c_1$). The appropriate control law when $a_1$ is known has the form:

$$u(t) = -r_0 y(t); \quad r_0 = c_1 - a_1 \tag{1.3}$$

However, (1.1) can be rewritten as:

$$y(t+1) = -c_1 y(t) + r_0 y(t) + u(t) \tag{1.4}$$

and the estimation of the unknown parameter $r_0$ will directly give the parameter of the controller. Using an adjustable predictor of the form:

$$\hat{y}(t+1) = -c_1 y(t) + \hat{r}_0(t) y(t) + u(t) \tag{1.5}$$

and a control law derived from (1.3) in which $r_0$ is replaced by its estimates:

$$u(t) = -\hat{r}_0(t) y(t) \tag{1.6}$$

one gets:

$$\hat{y}(t+1) = -c_1 y(t) \tag{1.7}$$

which is effectively the desired output at $(t+1)$ (i.e., the output of the implicit reference model made from the combination of the predictor and the controller).

A number of well known adaptive control schemes (minimum variance self-tuning control—Åström and Wittenmark 1973, generalized minimum variance self-tuning control—Clarke and Gawthrop 1975) have been presented as indirect adaptive control schemes, however in these schemes one directly estimates the controller parameters and therefore they fall in the class of direct adaptive control schemes.

### 1.3.5 Iterative Identification in Closed Loop and Controller Redesign

In indirect adaptive control, the parameters of the controller are generally updated at each sampling instant based on the current estimates of the plant model parameters.

However, nothing forbids us to update the estimates of the plant model parameters at each sampling instant, and to update the controller parameters only every $N$ sampling instants. Arguments for choosing this procedure are related to:

- the possibility of getting better parameter estimates for control design;
- the eventual reinitialization of the plant parameters estimation algorithm after each controller updating;
- the possibility of using a more sophisticated control design procedure (in particular robust control design) requiring a large amount of computation.

If the plant to be controlled has constant parameters over a large time horizon, one can consider a large horizon $N$ for plant parameters estimation, followed by the redesign of the controller based on the results of the identification in closed loop. Of course, this procedure can be repeated. The important feature of this approach is that identification in closed loop is done in the presence of a linear fixed controller (which is not the case in indirect adaptive control where plant parameter estimates and controller parameters are updated at each sampling instant).

This approach to indirect adaptive control is called *iterative identification in closed loop and controller redesign*. See Gevers (1993), Bitmead (1993), Van den Hof and Schrama (1995).

This technique can be used for:

- retuning and redesign of an existing controller without opening the loop;
- retuning of a controller from time to time in order to take into account possible change in the model parameters.

It has been noticed in practice that this technique often allows to improve the performances of a controller designed on the basis of a model identified in open loop. See Bitmead (1993), Van den Hof and Schrama (1995), Langer and Landau (1996) and Chap. 9.

The explanation is that identification in closed loop is made with an effective plant input which corresponds to the external excitation filtered by a sensitivity function. This sensitivity function will enhance the signal energy in the frequency range around the band pass of the closed loop and therefore will allow a more accurate model to be obtained in this region, which is critical for the design.

This technique emphasizes the role of identification in closed loop as a basic step for controller tuning based on data obtained in closed-loop operation.

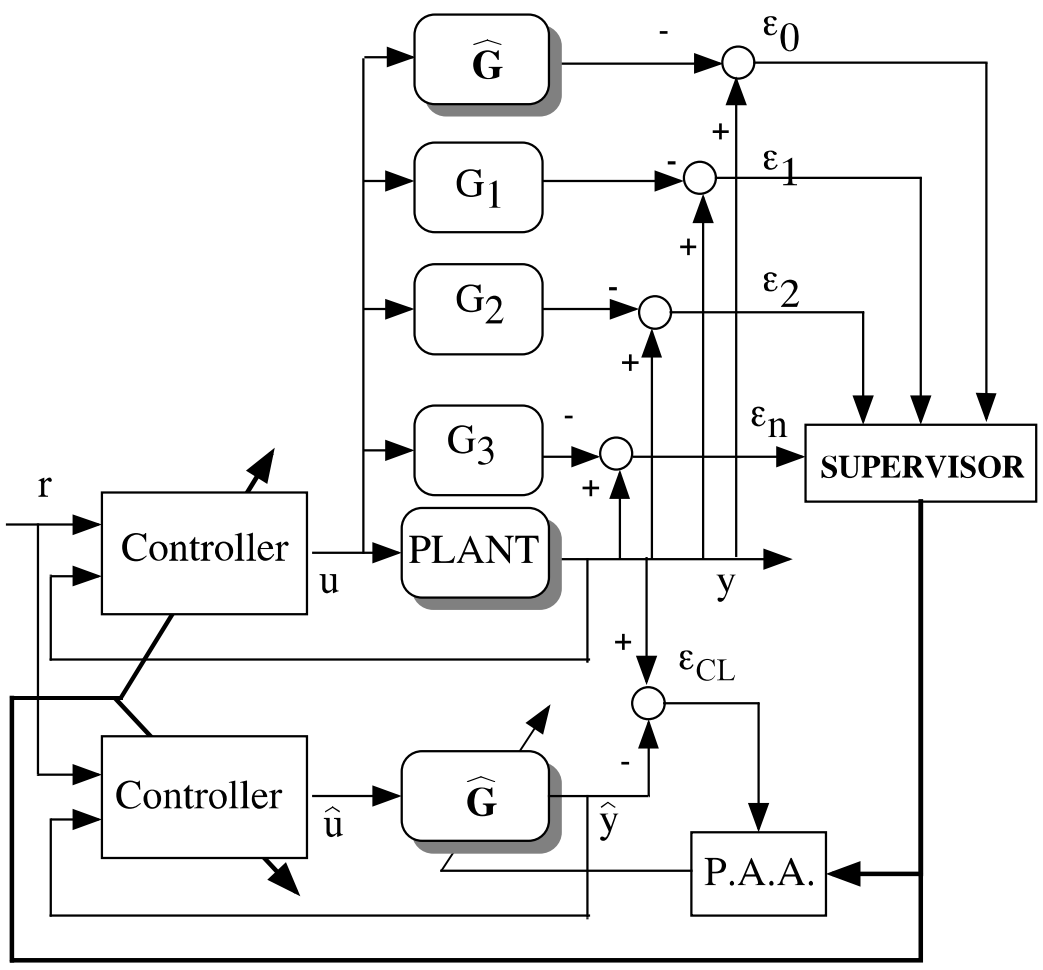

**Fig. 1.15** Schematic diagram of the multiple-model adaptive control approach

## 1.3.6 Multiple Model Adaptive Control with Switching

When large and rapid variations of the plant model parameters occur, the adaptation transients in classical indirect adaptive control schemes are often unsatisfactory. To improve the adaption transients the use of the so called "multiple model adaptive control" offers a very appealing solution. The basic idea of this approach is to select in real time the best model of the plant from an a priori known set of models and apply the output of the corresponding predesigned controller to the plant. The performance can be improved by on-line adaptation of an adjustable model in order to find a more accurate plant model. The block diagram of such a system is presented in Fig. 1.15. The system contains a bank of fixed models ($G_1$, $G_2$, $G_3$), an adaptive model estimator $\hat{G}$ (using a closed-loop type parameter estimation scheme) and an adjustable controller.

The system operates in two steps:

Step 1: The best available fixed model (with respect to an error criterion) is selected by a switching procedure (implemented in the supervisor).

Step 2: The parameters of an adjustable plant model are updated. When its performance in term of the error criterion is better than the best fixed model, one switches to this model and one computes a corresponding controller.

This approach has been developed in Morse (1995), Narendra and Balakrishnan (1997), Karimi and Landau (2000) among other references.

## 1.3.7 Adaptive Regulation

Up to now we have considered that the plant model parameters are unknown and time varying and implicitly it was assumed the disturbance (and its model) is known.

However, there are classes of applications (active vibration control, active noise control, batch reactors) where the plant model can be supposed to be known (obtained by system identification) and time invariant and where the objective is to reject the effect of disturbances with unknown and time varying characteristics (for example: multiple vibrations with unknown and time varying frequencies). To reject disturbances (asymptotically), the controller should incorporate the model of the disturbance (the internal model principle). Therefore in adaptive regulation, the internal model in the controller should be adapted in relation with the disturbance model. Direct and indirect adaptive regulation solutions have been proposed. For the indirect approach the disturbance model is estimated and one computes the controller on the basis of the plant and disturbance model. In the direct approach, through an appropriate parametrization of the controller, one adapts directly the internal model. These techniques are discussed in Chap. 14. Among the first references on this approach (which include applications) see Amara et al. (1999a, 1999b), Valentinotti (2001), Landau et al. (2005).

### *1.3.8 Adaptive Feedforward Compensation of Disturbances*

In a number of applications (including active vibration control, active noise control) it is possible to get a measurement highly correlated with the disturbance (an image of the disturbance). Therefore one can use an (adaptive) feedforward filter for compensation of the disturbance (eventually on top of a feedback system). This is particularly interesting for the case of wide band disturbances where the performance achievable by feed back only may be limited (limitations introduced by the Bode "integral" of the output sensitivity function).

The feedforward filter should be adapted with respect to the characteristics of the disturbance. It is important to mention that despite its "open-loop character", there is an inherent positive feedback in the physical system, between the actuator and the measurement of the image of the disturbance. Therefore the adaptive feedforward filter operates in closed loop with positive feedback. The adaptive feedforward filter should stabilize this loop while simultaneously compensating the effect of the disturbance. The corresponding block diagram is shown in Fig. 1.16. These techniques are discussed in Chap. 15.

### *1.3.9 Parameter Adaptation Algorithm*

The *parameter adaptation algorithm* (PAA) forms the essence of the adaptation mechanism used to adapt either the parameter of the controller directly (in direct adaptive control), or the parameters of the adjustable predictor of the plant output.

The development of the PAA which will be considered in this book and which is used in the majority of adaptive control schemes assumes that the "models are linear

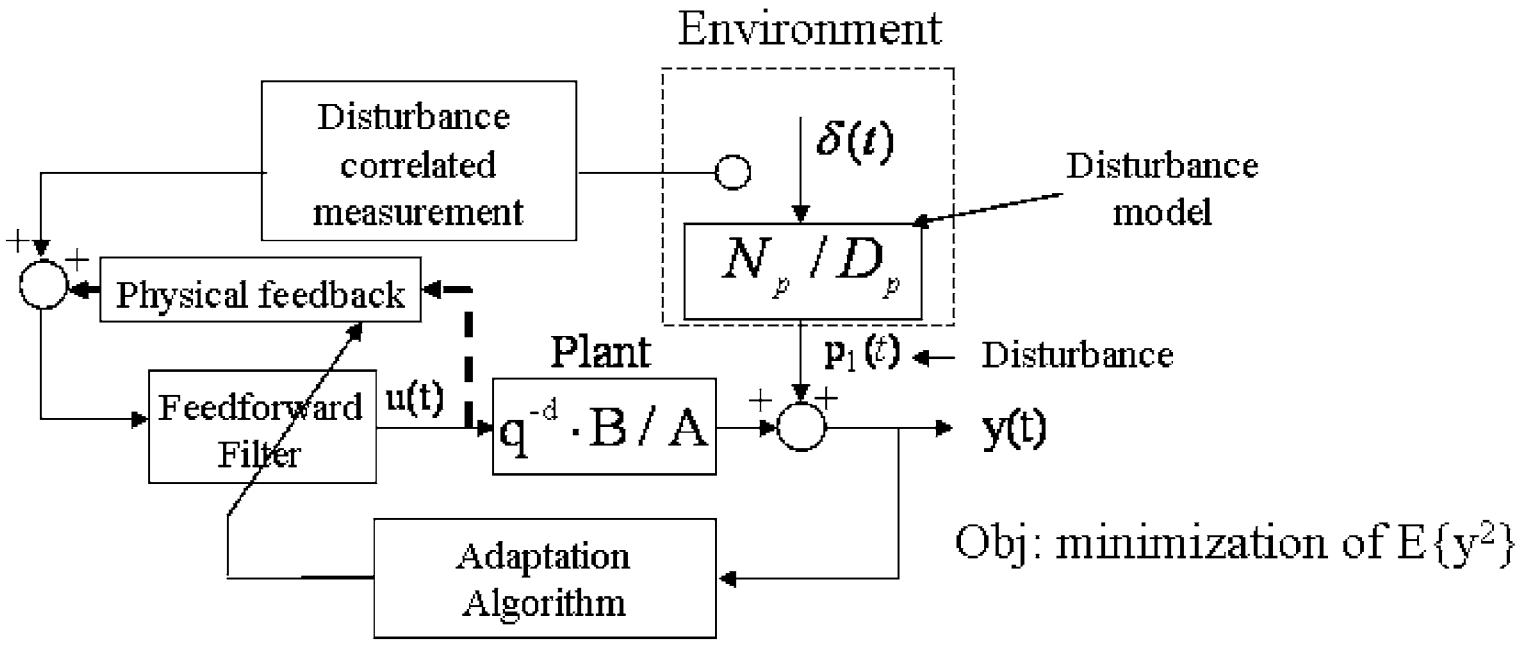

**Fig. 1.16** Schematic diagram of the adaptive feedforward disturbance compensation

in parameters",[3] i.e., one assumes that the plant model admits a representation of the form:

$$y(t+1) = \theta^T \phi(t) \tag{1.8}$$

where $\theta$ denotes the vector of (unknown) parameters and $\phi(t)$ is the vector of measurements. This form is also known as a "linear regression". The objective will be to estimate the unknown parameter vector $\theta$ given in real time $y$ and $\phi$. Then, the estimated parameter vector denoted $\hat{\theta}$ will be used for controller redesign in indirect adaptive control. Similarly, for direct adaptive control it is assumed that the controller admits a representation of the form:

$$y^*(t+1) = -\theta_c^T \phi(t) \tag{1.9}$$

where $y^*(t+1)$ is a desired output (or filtered desired output), $\theta_c$ is the vector of the unknown parameters of the controller and $\phi(t)$ is a vector of measurements and the objective will be to estimate $\theta_c$ given in real time $y^*$ and $\phi$.

The parameter adaptation algorithms will be derived with the objective of minimizing a criterion on the error between the plant and the model, or between the desired output and the true output of the closed-loop system.

The parameter adaptation algorithms have a recursive structure, i.e., the new value of the estimated parameters is equal to the previous value plus a correcting term which will depend on the most recent measurements.

The general structure of the parameter adaptation algorithm is as follows:

$$\begin{bmatrix}\text{New estimated}\\ \text{parameters}\\ \text{(vector)}\end{bmatrix} = \begin{bmatrix}\text{Previous estimated}\\ \text{parameters}\\ \text{(vector)}\end{bmatrix} + \begin{bmatrix}\text{Adaptation}\\ \text{gain}\\ \text{(matrix)}\end{bmatrix}$$

$$\times \begin{bmatrix}\text{Measurement}\\ \text{function}\\ \text{(vector)}\end{bmatrix} \times \begin{bmatrix}\text{Prediction error}\\ \text{function}\\ \text{(scalar)}\end{bmatrix}$$

[3]The models may be linear or nonlinear but linear in parameters.

which translates to:

$$\hat{\theta}(t+1) = \hat{\theta}(t) + F(t)\phi(t)\nu(t+1) \tag{1.10}$$

where $\hat{\theta}$ denotes the estimated parameter vector, $F(t)$ denotes the adaptation gain, $\phi(t)$ is the observation (regressor) vector which is a function of the measurements and $\nu(t+1)$ denotes the adaptation error which is the function of the plant-model error.

Note that the adaptation starts once the latest measurements on the plant output $y(t+1)$ is acquired (which allows to generate the plant-model error at $t+1$). Therefore $\hat{\theta}(t+1)$ will be only available after a certain time $\delta$ within $t+1$ and $t+2$, where $\delta$ is the computation time associated with (1.10).

## 1.4 Examples of Applications

### *1.4.1 Open-Loop Adaptive Control of Deposited Zinc in Hot-Dip Galvanizing*

Hot-dip galvanizing is an important technology for producing galvanized steel strips. However, the demand, particularly from automotive manufacturers, became much sharper in terms of the coating uniformity required, both for use in exposed skin panels and for better weldability. Furthermore, the price of zinc rose drastically since the eighties and a tight control of the deposited zinc was viewed as a means of reducing the zinc consumption (whilst still guaranteeing the minimum zinc deposit). Open-loop adaptive control is one of the key elements in the Sollac hot-dip galvanizing line at Florange, France (Fenot et al. 1993).

The objective of the galvanizing line is to obtain galvanized steel with formability, surface quality and weldability equivalent to uncoated cold rolled steel. The variety of products is very large in terms of deposited zinc thickness and steel strip thickness. The deposited zinc may vary between 50 to 350 $\text{g/m}^2$ (each side) and the strip speed may vary from 30 to 180 m/mn.

The most important part of the process is the hot-dip galvanizing. The principle of the hot-dip galvanizing is illustrated in Fig. 1.17. Preheated steel strip is passed through a bath of liquid zinc and then rises vertically out of the bath through the stripping “air knives” which remove the excess zinc. The remaining zinc on the strip surface solidifies before it reaches the rollers, which guide the finished product. The measurement of the deposited zinc can be made only on the cooled finished strip and this introduces a very large and time-varying pure time delay. The effect of air knives depends on the air pressure, the distance between the air knives and the strip, and the speed of the strip. Nonlinear static models have been developed for computing the appropriate pressure, distance and speed for a given value of the desired deposited zinc.

The objective of the control is to assure a good uniformity of the deposited zinc whilst guaranteeing a minimum value of the deposited zinc per unit area. Tight control (i.e., small variance of the controlled variable) will allow a more uniform coating and will reduce the average quantity of deposited zinc per unit area. As a consequence, in addition to quality improvement, a tight control on the deposited zinc per unit area has an important commercial impact since the average consumption for a modern galvanizing line is of the order of 40 tons per day.

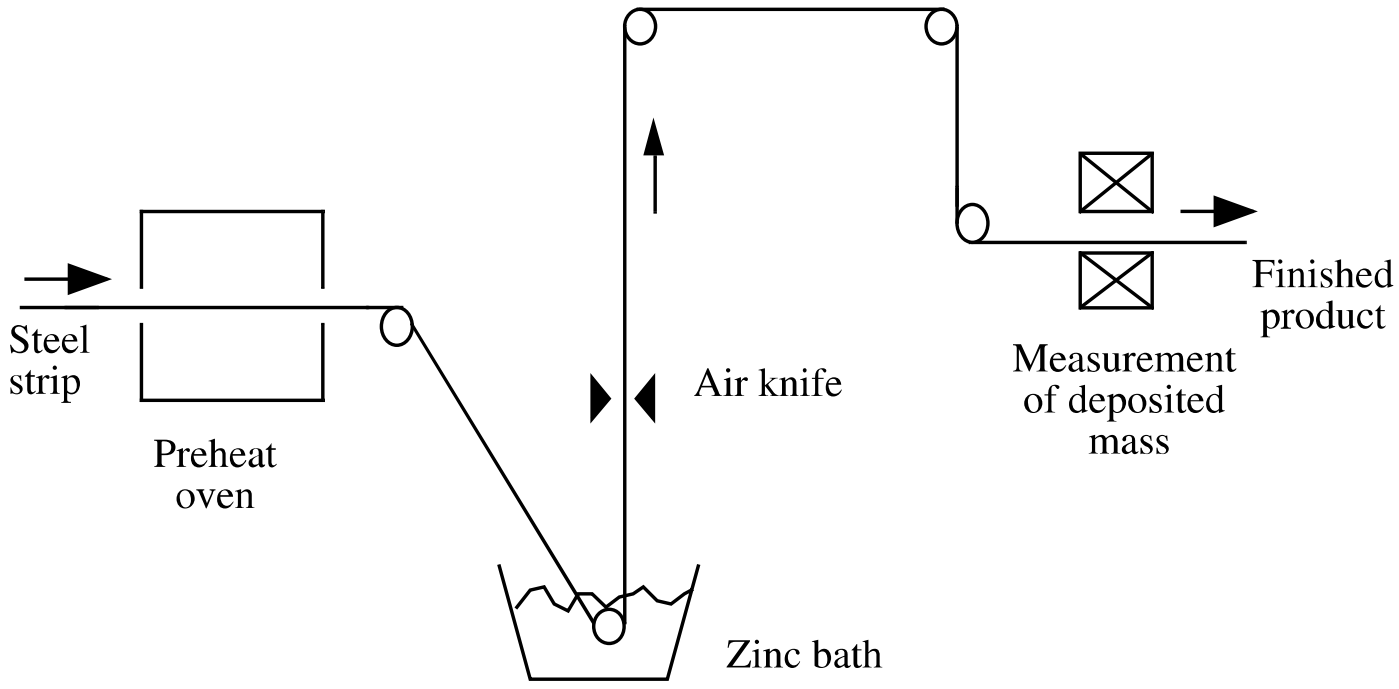

**Fig. 1.17** Hot-dip galvanizing process

The pressure in the air knives, which is the control variable, is itself regulated through a pressure loop, which can be approximated by a first order system. The delay of the process will depend linearly on the speed. Therefore a continuous-time linear dynamic model relating variations of the pressure to variations of the deposited mass, of the form:

$$H(s) = \frac{Ge^{-s\tau}}{1+sT}; \quad \tau = \frac{L}{V}$$

can be considered, where $L$ is the distance between the air knives and the transducers and $V$ is the strip speed. When discretizing this model, the major difficulty comes from the variable time-delay. In order to obtain a controller with a fixed number of parameters, the delay of the discrete-time model should remain constant. Therefore, the sampling period $T_S$ is tied to the strip speed using the formula:

$$T_s = \frac{\frac{L}{V} + \delta}{d}; \quad (d = \text{integer})$$

where $\delta$ is an additional small time-delay corresponding to the equivalent time-delay of the industrial network and of the programmable controller used for pressure regulation and d is the discrete-time delay (integer). A linearized discrete-time model can be identified.

However, the parameters of the model will depend on the distance between the air knives and the steel strip and on the speed $V$.

In order to assure satisfactory performances for all regions of operation an "open-loop adaptation" technique has been considered. The open-loop adaptation is made with respect to:

- steel strip speed;
- distance between the air knives and the steel strip.

The speed range and the distance range have been split into three regions giving a total of nine operating regions. For each of these operating regions, an identification has been performed and robust controllers based on the identified models have been designed for all the regions and stored in a table.

A reduction of the dispersion of coating is noticed when closed-loop digital control is used. This provides a better quality finished product (extremely important in the automotive industry, for example). The average quantity of deposited zinc is also reduced by 3% when open-loop adaptive digital control is used, still guaranteeing the specifications for minimum zinc deposit and this corresponds to a very significant economic gain.

### *1.4.2 Direct Adaptive Control of a Phosphate Drying Furnace*

This application has been done at the O.C.P., Beni-Idir Factory, Morocco (Dahhou et al. 1983). The phosphate, independently of the extraction method, has about 15% humidity. Before being sold its humidity should be reduced to about 1.5% using a rotary drying furnace. The drying process requires a great consumption of energy. The objective is to keep the humidity of the dried phosphate close to the desired value (1.5%) independently of the raw material humidity variations (between 7 and 20%), feedflow variations (100 to 240 t/h) and other perturbations that may affect the drying process.

The dynamic characteristics of the process vary as a consequence of the variable moisture and the nature of the damp product. A direct adaptive control approach has been used to achieve the desired performances over the range of possible changes in the process characteristics.

A block diagram of the system is shown in Fig. 1.18. The drying furnace consists of a:

- feeding system,
- combustion chamber,
- rotary drying tube,
- dust chamber,
- ventilator and chimney.

The combustion chamber produces the hot gas needed for the drying process. The hot gas and the phosphate mix in the rotary drying tube. In the dust chamber, one recaptures the phosphate fine particles which represent approximately 30% of the dried phosphate. The final product is obtained on the bottom of the dust chamber and recovered on a conveyor. The temperature of the final product is used as an indirect measure of its humidity. The control action is the fuel flow while the other variable for the burner (steam, primary air) are related through conventional loops to the fuel flow.

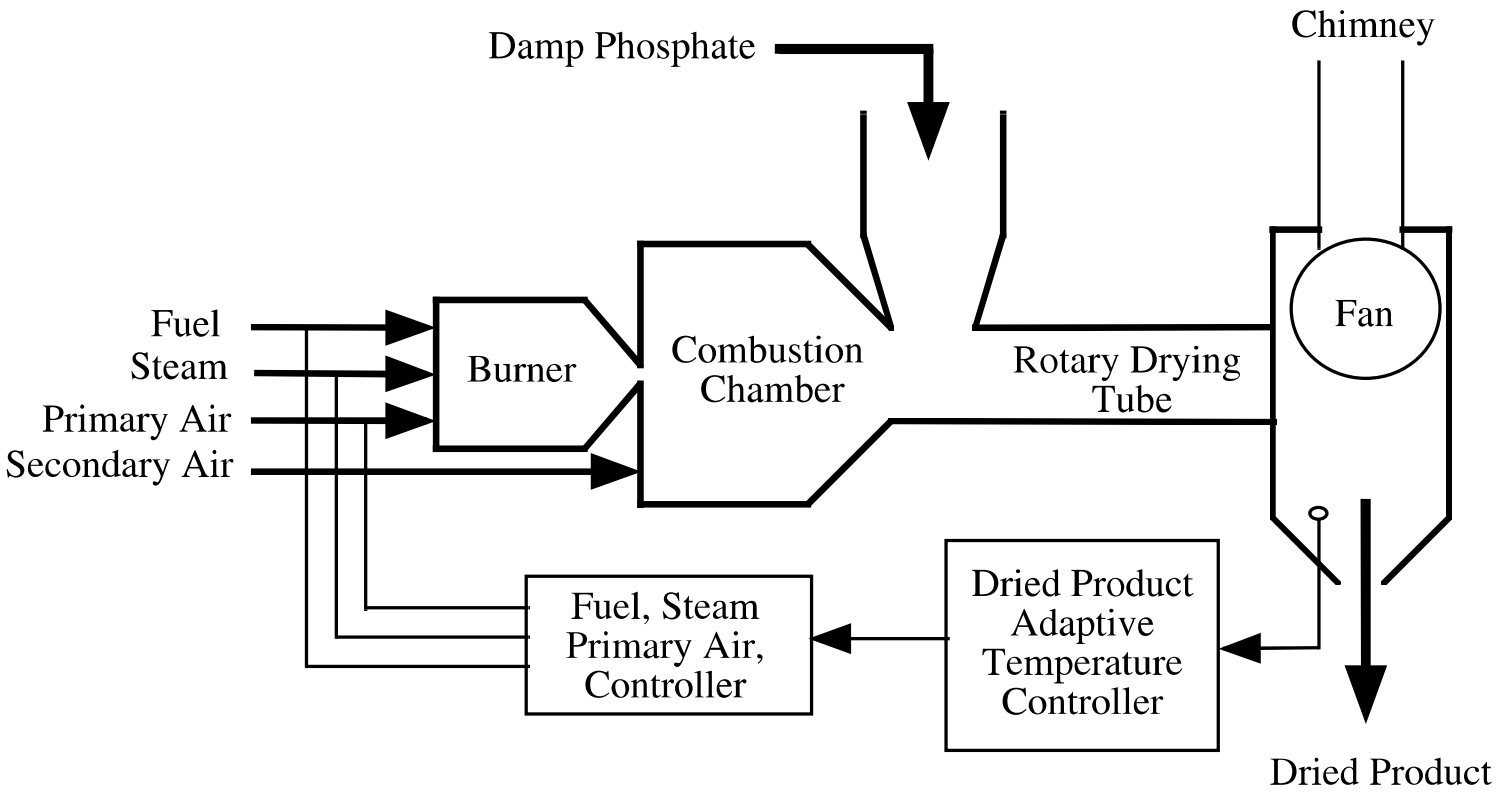

**Fig. 1.18** Phosphate drying furnace

Significant improvement in performance was obtained with respect to a standard PID controller (the system has a delay of about 90 s). The improved regulation has as a side effect an average reduction of the fuel consumption and a reduction of the thermal stress on the combustion chamber walls allowing to increase the average time between two maintenance operations.

### *1.4.3 Indirect and Multimodel Adaptive Control of a Flexible Transmission*

The flexible transmission built at GIPSA-LAB, Control Dept. (CNRS-INPG-UJF), Grenoble, France, consists of three horizontal pulleys connected by two elastic belts (Fig. 1.19). The first pulley is driven by a D.C. motor whose position is controlled by local feedback. The third pulley may be loaded with disks of different weight. The objective is to control the position of the third pulley measured by a position sensor. The system input is the reference for the axis position of the first pulley. A PC is used to control the system. The sampling frequency is 20 Hz.

The system is characterized by two low-damped vibration modes subject to a large variation in the presence of load. Fig. 1.20 gives the frequency characteristics of the identified discrete-time models for the case without load, half load (1.8 kg) and full load (3.6 kg). A variation of 100% of the first vibration mode occurs when passing from the full loaded case to the case without load. In addition, the system features a delay and unstable zeros. The system was used as a benchmark for robust digital control (Landau et al. 1995a), as well as a test bed for indirect adaptive control, multiple model adaptive control, identification in open-loop and closed-loop operation, iterative identification in closed loop and control redesign. The use of various algorithms for real-time identification and adaptive control which will be discussed throughout the book will be illustrated on this real system (see Chaps. 5, 9, 12, 13 and 16).

Fig. 1.19 The flexible transmission (GIPSA-LAB, Grenoble), (**a**) block diagram, (**b**) view of the system

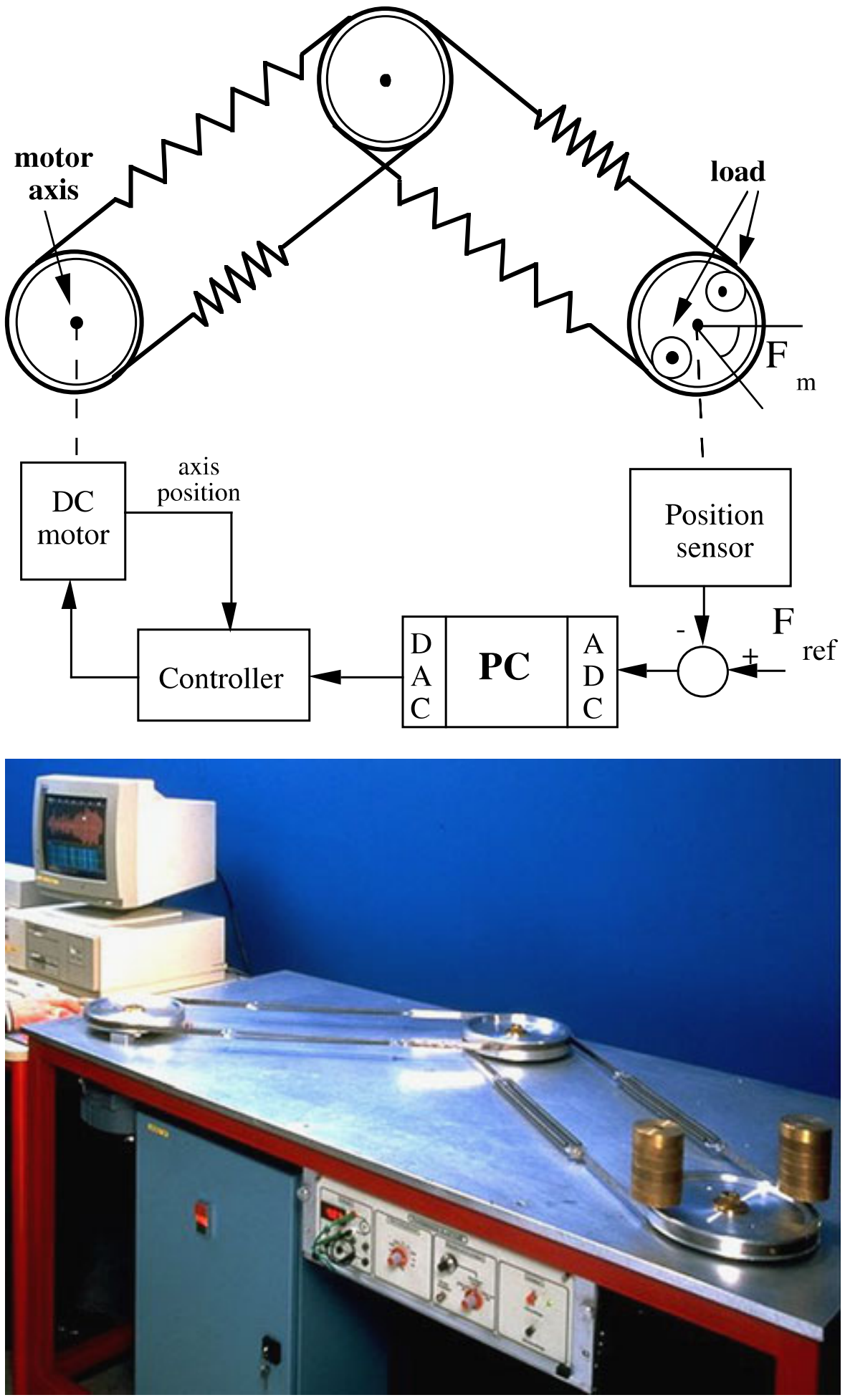

## *1.4.4 Adaptive Regulation in an Active Vibration Control System*

The active vibration control system built at GIPSA-LAB, Control Dept. (CNRS-INPG-UJF), Grenoble, France for benchmarking of control strategies is shown in Fig. 1.21 and details are given in Fig. 1.22. For suppressing the effect of vibrational disturbances one uses an inertial actuator which will create vibrational forces to counteract the effect of vibrational disturbances (inertial actuators use a similar principle as loudspeakers). The load is posed on a passive damper and the inertial actuator is fixed to the chassis where the vibrations should be attenuated. A shaker posed on the ground is used to generate the vibration. The mechanical construction of the load is such that the vibrations produced by the shaker, are transmitted to the upper side of the system. The controller will act (through a power amplifier) on the position of the mobile part of the inertial actuator in order to reduce the residual force. The sampling frequency is 800 Hz.

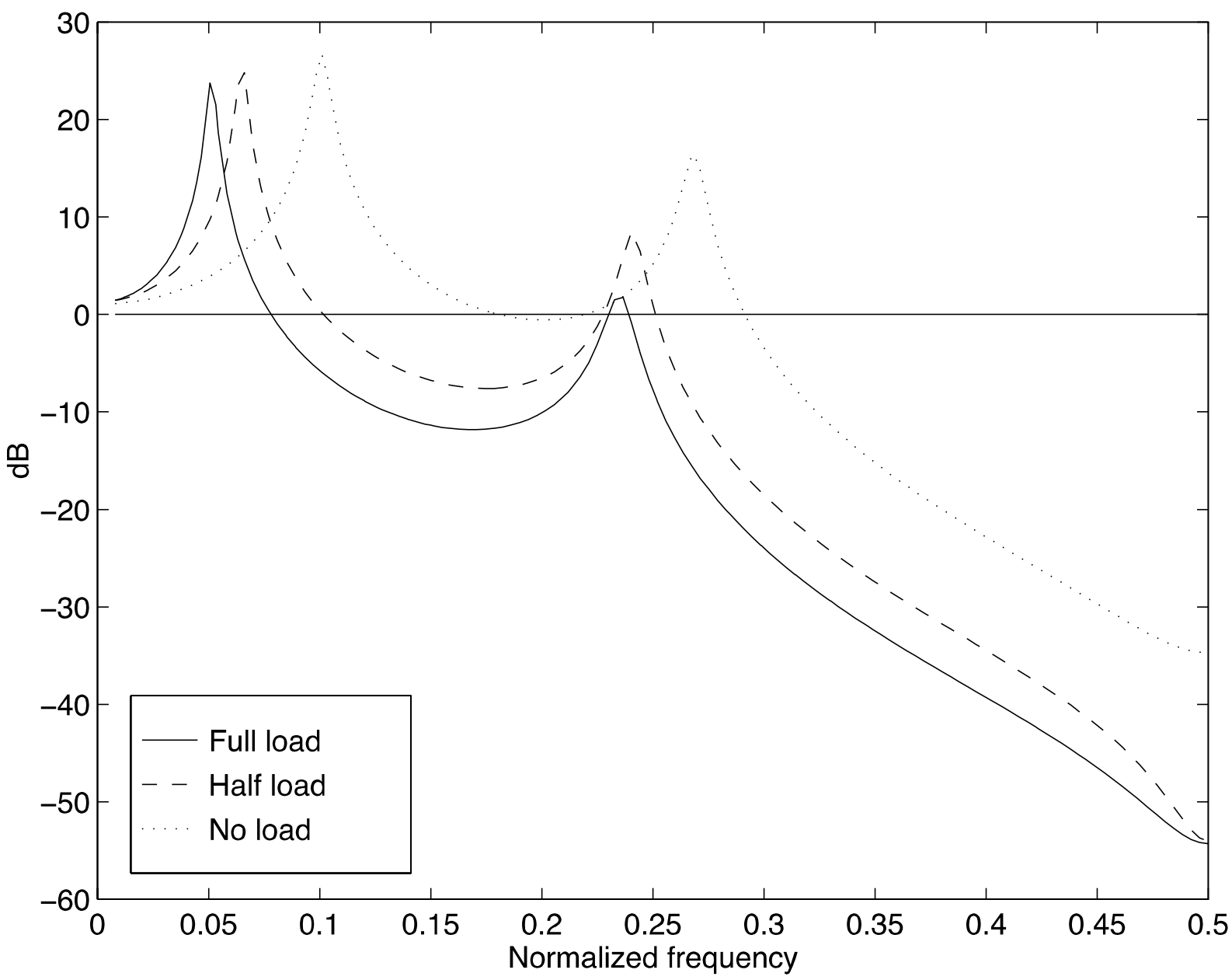

**Fig. 1.20** Frequency characteristics of the flexible transmission for various loads

The system was used as a benchmark for direct and indirect adaptive regulation strategies. The performance of the algorithms which will be presented in Chap. 14 will be evaluated on this system.

### *1.4.5 Adaptive Feedforward Disturbance Compensation in an Active Vibration Control System*

The system shown in Fig. 1.23 is representative of distributed mechanical structures encountered in practice where a correlated measurement with the disturbance (an image of the disturbance) is made available and used for feedforward disturbance compensation (GIPSA-LAB, Control Dept., Grenoble, France). A detailed scheme of the system is shown in Fig. 1.24. It consists on five metal plates connected by springs. The second plate from the top and the second plate from the bottom are equipped with an inertial actuator. The first inertial actuator will excite the structure (disturbances) and the second will create vibrational forces which can counteract the effect of these vibrational disturbances. Two accelerometers are used to measure the displacement of vibrating plates. The one posed on the second plate on the bottom measures the residual acceleration which has to be reduced. The one posed above gives an image of the disturbance to be used for feedforward compensation. As it results clearly from Figs. 1.23 and 1.24, the actuator located down side will compensate vibrations at the level of the lowest plate but will induce forces upstream beyond this plate an therefore a positive feedback is present in the system which modifies the effective measurement of the image of the disturbance. The algorithms which will be presented in Chap. 15 will be evaluated on this system.

**Fig. 1.21** Active vibration control system using an inertial actuator (photo)

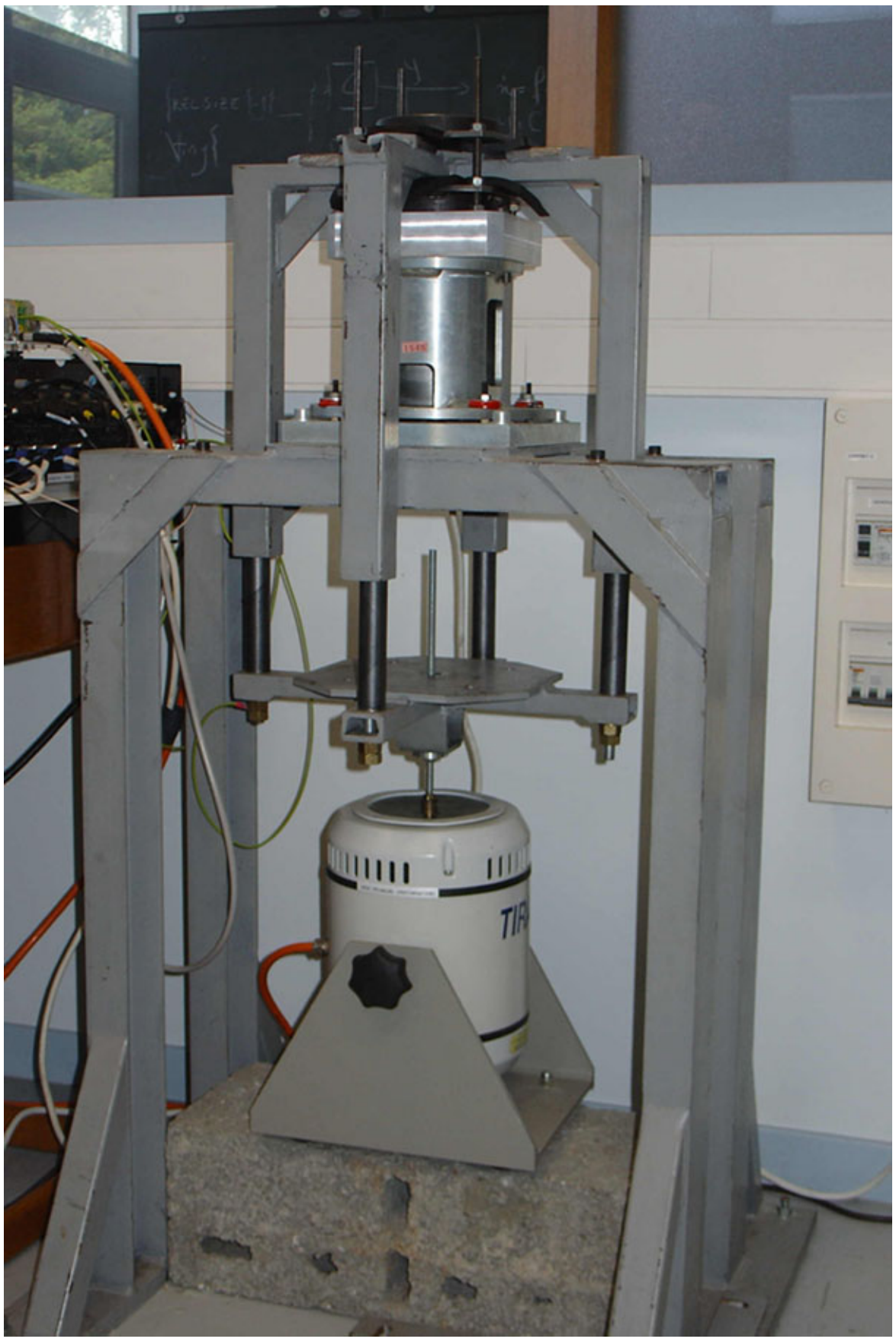

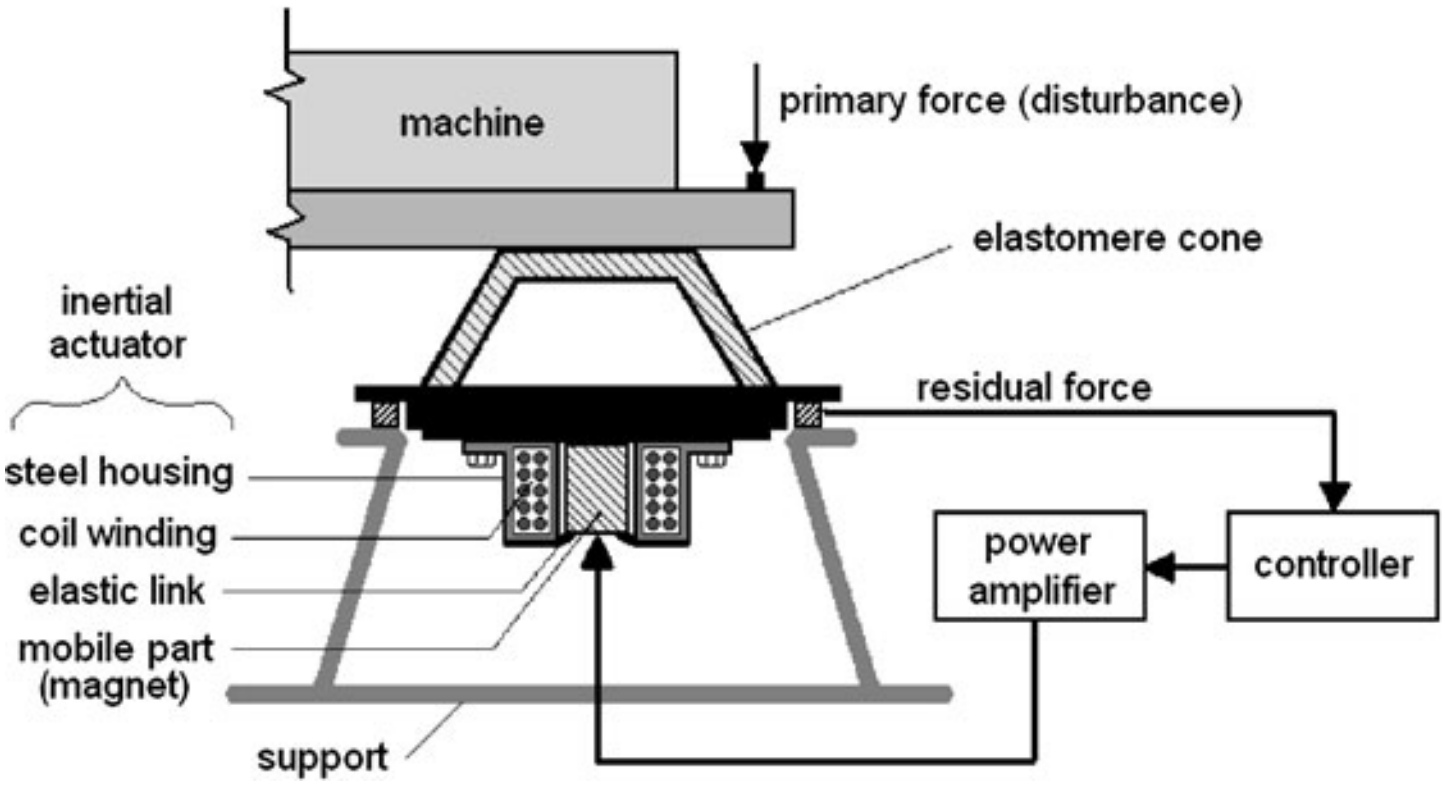

**Fig. 1.22** Active vibration control using an inertial actuator (scheme)

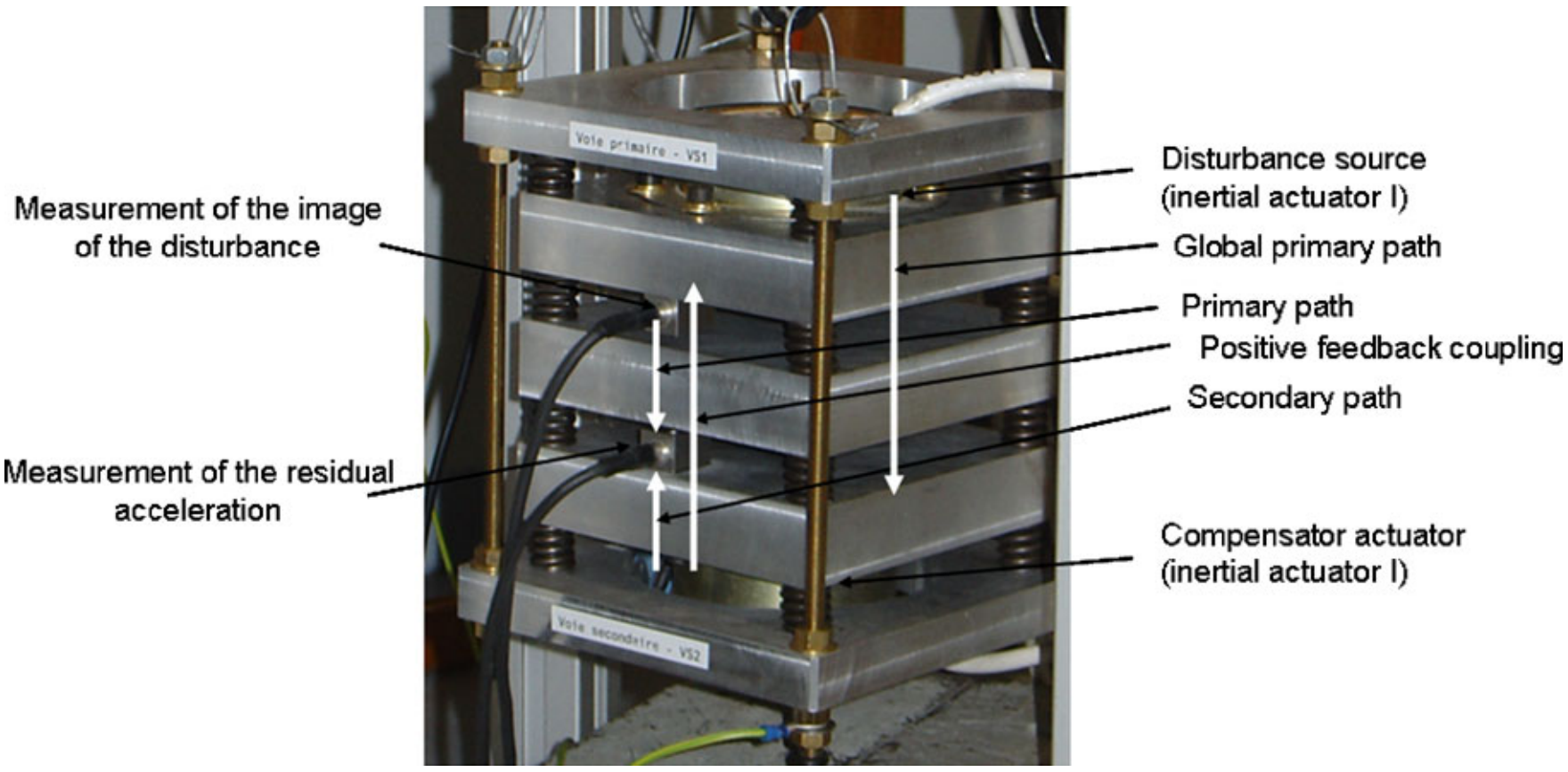

**Fig. 1.23** An active vibration control using feedforward compensation (photo)

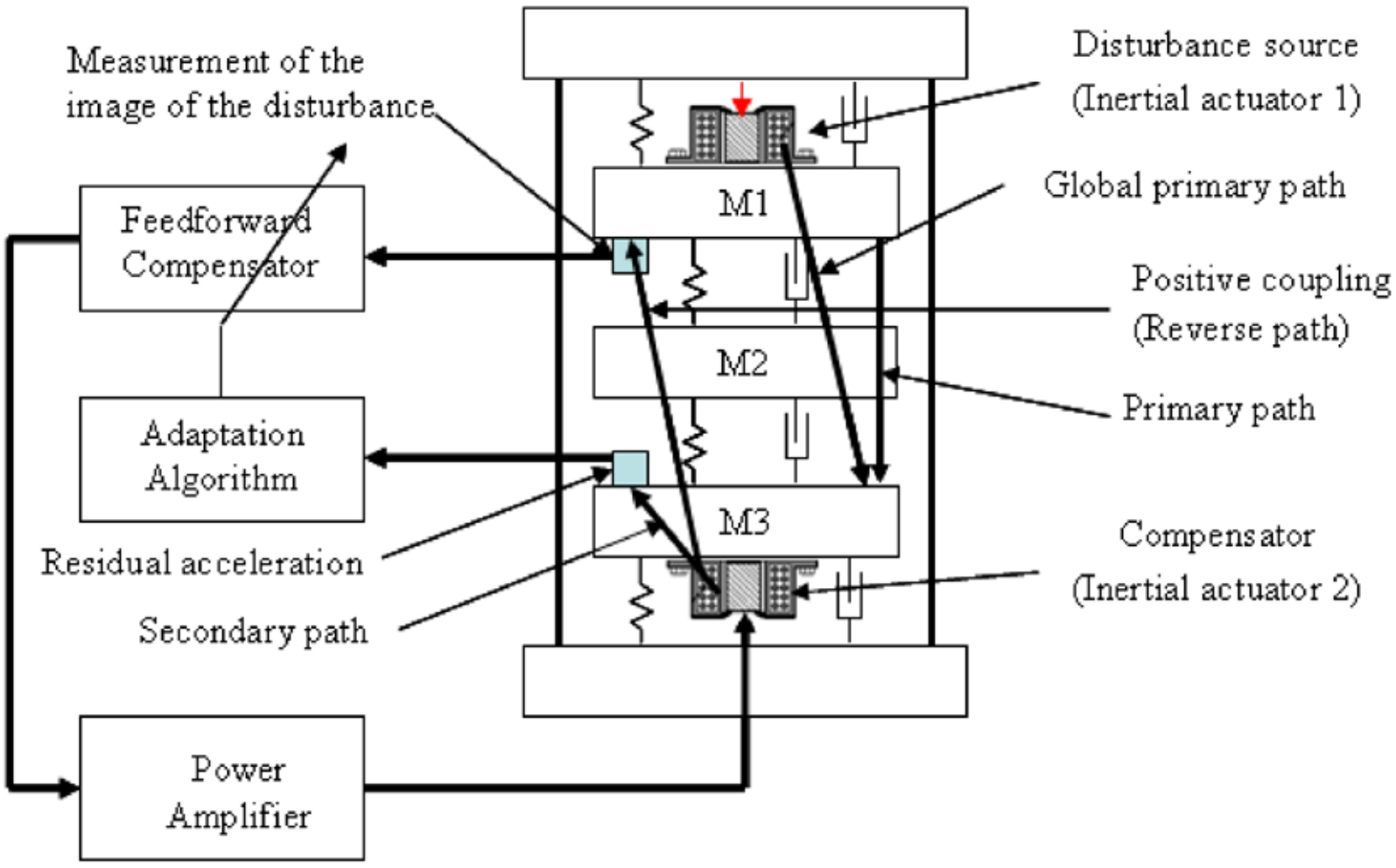

**Fig. 1.24** An active vibration control using feedforward compensation (scheme)

## 1.5 A Brief Historical Note

This note is not at all a comprehensive account of the evolution of the adaptive control field which is already fifty years old. Its objective is to point out some of the moments in the evolution of the field which we believe were important. In particular, the evolution of discrete-time adaptive control of SISO systems (which is the main subject of this book) will be emphasized (a number of basic references to continuous-time adaptive control are missed).

The formulation of adaptive control as a stochastic control problem (dual control) was given in Feldbaum (1965). However, independently more ad-hoc adaptive control approaches have been developed.

*Direct adaptive control* appeared first in relation to the use of a model reference adaptive system for aircraft control. See Whitaker et al. (1958). The *indirect adaptive control* was probably introduced by Kalman (1958) in connection with digital process control.

Earlier work in model reference adaptive systems has emphasized the importance of *stability* problems for these schemes. The first approach for synthesizing stable model reference adaptive systems using Lyapunov functions was proposed in Butchart and Shakcloth (1966) and further generalized in Parks (1966). The importance of positive realness of some transfer functions for the stability of MRAS was pointed out for the first time in Parks (1966). The fact that model reference adaptive systems can be represented as an equivalent feedback system with a linear time invariant feedforward block and a nonlinear time-varying feedback block was pointed out in Landau (1969a, 1969b) where an input-output approach based on hyperstability (passivity) concepts was proposed for the design. For detailed results along this line of research see Landau (1974, 1979). While initially direct adaptive control schemes have only been considered in continuous time, synthesis of discrete-time direct adaptive schemes and applications appeared in the seventies. See Landau (1971, 1973), Bethoux and Courtiol (1973), Ionescu and Monopoli (1977). For an account of these earlier developments see Landau (1979).

The *indirect adaptive control* approach was significantly developed starting with Åström and Wittenmark (1973) where the term "self-tuning" was coined. The resulting scheme corresponded to an adaptive version of the minimum variance discrete-time control. A further development appeared in Clarke and Gawthrop (1975). In fact, the *self-tuning minimum variance controller* and its extensions are a *direct adaptive control* scheme since one estimates directly the parameters of the controller. It took a number of years to understand that discrete-time model reference adaptive control systems and stochastic self-tuning regulators based on minimization of the error variance belong to the same family. See Egardt (1979), Landau (1982a).

Tsypkin (1971) also made a very important contribution to the development and analysis of discrete-time parameter adaptation algorithms.

Despite the continuous research efforts, it was only at the end of the 1970's that full proofs for the stability of discrete-time model reference adaptive control and stochastic self-tuning controllers (under some ideal conditions) became available.

See Goodwin et al. (1980b, 1980a) (for continuous-time adaptive control see Morse 1980; Narendra et al. 1980).

The progress of the adaptive control theory on the one hand and the availability of microcomputers on the other hand led to a series of successful applications of direct adaptive controllers (either deterministic or stochastic) in the late 1970's and early 1980's. However, despite a number of remarkable successes in the same period a number of simple counter-examples have since shown the limitation of these approaches.

On the one hand, the lack of robustness of the original approaches with respect to noise, unmodeled dynamics and disturbances has been emphasized. On the other hand, the experience with the various applications has shown that one of the major assumptions in these schemes (i.e., the plant model is a discrete-time model with stable zeros and a fixed delay) is not a very realistic one (except in special applications). Even successful applications have required a careful selection of the sampling frequency (since fractional delay larger than half of the sampling period generates a discrete-time unstable zero). The industrial experiences and various counter-examples were extremely beneficial for the evolution of the field. A significant research effort has been dedicated to robustness issues and development of robust adaptation algorithms. Egardt (1979), Praly (1983c), Ortega et al. (1985), Ioannou and Kokotovic (1983) are among the basic references. A deeper analysis of the adaptive schemes has also been done in Anderson et al. (1986). An account of the work on robustness of adaptive control covering both continuous and discrete-time adaptive control can be found in Ortega and Tang (1989). For continuous-time only, see also Ioannou and Datta (1989, 1996).

The other important research direction was aimed towards adaptive control of discrete-time models with unstable zeros which led to the development of indirect adaptive control schemes and their analysis. The problems of removing the need of persistence of excitation and of the eventual singularities which may occur when computing a controller based on plant model parameter estimates have been addressed, as well as the robustness issues (see Lozano and Zhao 1994 for details and a list of references). Various underlying linear control strategies have been considered: pole placement (de Larminat 1980 is one of the first references), linear quadratic control (Samson 1982 is the first reference) and generalized predictive control (Clarke et al. 1987). Adaptive versions of pole placement and generalized predictive control are the most popular ones.

The second half of the nineties has seen the emergence of two new approaches to adaptive control. On one hand there is the development of plant model identification in closed loop (Gevers 1993; Van den Hof and Schrama 1995; Landau and Karimi 1997a, 1997b) leading to the strategy called "iterative identification in closed loop and controller redesign" (see Chap. 9). On the other hand the "multiple model adaptive control" emerged as a solution for improving the transients in indirect adaptive control. See Morse (1995), Narendra and Balakrishnan (1997), Karimi and Landau (2000) and Chap. 13.

End of the nineties and beginning of the new century have seen the emergence of a new paradigm: *adaptive regulation*. In this context the plant model is assumed to be known and invariant and adaptation is considered with respect to the disturbance model which is unknown and time varying (Amara et al. 1999a; Valentinotti 2001; Landau et al. 2005 and Chap. 14). In the mean time it was pointed out that adaptive feedforward compensation of disturbances which for a long time has been considered as an "open-loop" problem has in fact a *hidden feedback* structure bringing this subject in the context of adaptive feedback control. New solutions are emerging, see Jacobson et al. (2001), Zeng and de Callafon (2006), Landau and Alma (2010) and Chap. 15.

## 1.6 Further Reading

It is not possible in a limited number of pages to cover all the aspects of adaptive control. In what follows we will mention some references on a number of issues not covered by this book.

- Continuous Time Adaptive Control (Ioannou and Sun 1996; Sastry and Bodson 1989; Anderson et al. 1986; Åström and Wittenmark 1995; Landau 1979; Datta 1998)
- Multivariable Systems (Dugard and Dion 1985; Goodwin and Sin 1984; Dion et al. 1988; Garrido-Moctezuma et al. 1993; Mutoh and Ortega 1993; de Mathelin and Bodson 1995)
- Systems with Constrained Inputs (Zhang and Evans 1994; Feng et al. 1994; Chaoui et al. 1996a, 1996b; Sussmann et al. 1994; Suarez et al. 1996; Åström and Wittenmark 1995)
- Input and Output Nonlinearities (Tao and Kokotovic 1996; Pajunen 1992)
- Adaptive Control of Nonlinear Systems (Sastry and Isidori 1989; Marino and Tomei 1995; Krstic et al. 1995; Praly et al. 1991; Lozano and Brogliato 1992a; Brogliato and Lozano 1994; Landau et al. 1987)
- Adaptive Control of Robot Manipulators (Landau 1985; Landau and Horowitz 1988; Slotine and Li 1991; Arimoto and Miyazaki 1984; Ortega and Spong 1989; Nicosia and Tomei 1990; Lozano 1992; Lozano and Brogliato 1992b, 1992c)
- Adaptive Friction Compensation (Gilbart and Winston 1974; Canudas et al. 1995; Armstrong and Amin 1996; Besançon 1997)
- Adaptive Control of Asynchronous Electric Motors (Raumer et al. 1993; Marino et al. 1996; Marino and Tomei 1995; Espinoza-Perez and Ortega 1995)

## 1.7 Concluding Remarks

In this chapter we have presented a number of concepts and basic adaptive control structures. We wish to emphasize the following basic ideas:

1. Adaptive control provides a set of techniques for automatic adjustment of the controllers in real time in order to achieve or to maintain a desired level of control system performance, when the parameters of the plant model are unknown and/or change in time.

2. While a *conventional feedback control* is primarily oriented toward the elimination of the effect of disturbances acting upon the controlled variables, an *adaptive control system* is mainly oriented toward the elimination of the effect of parameter disturbances upon the performances of the control system.
3. A control system is truly *adaptive* if, in addition to a conventional feedback, it contains a closed-loop control of a certain *performance index*.
4. *Robust control design* is an efficient way to handle known parameter uncertainty in a certain region around a nominal model and it constitutes a good underlying design method for adaptive control, but it is not an *adaptive control system*.
5. *Adaptive control* can improve the performance of a robust control design by providing better information about the nominal model and expanding the uncertainty region for which the desired performances can be guaranteed.
6. One distinguishes adaptive control schemes with *direct* adaptation of the parameters of the controller or with *indirect* adaptation of the parameters of the controller (as a function of the estimates of plant parameters).
7. One distinguishes between adaptive control schemes with non-vanishing adaptation (also called continuous adaptation) and adaptive control schemes with vanishing adaptation. Although in the former, adaptation operates most of the time, in the latter, its effect vanishes in time.
8. The use of adaptive control is based on the assumption that for any possible values of the plant parameters there is a controller with a fixed structure and complexity such that the desired performances can be achieved with appropriate values of the controller parameters. The task of the adaptation loop is to search for the *good* values of the controller parameters.
9. There are two control paradigms: (1) *adaptive control* where the plant model parameters are unknown an time varying while the disturbance model is assumed to be known; and (2) *adaptive regulation* where the plant model is assumed to be known and the model of the disturbance is unknown and time varying.
10. Adaptive control systems are nonlinear time-varying systems and specific tools for analyzing their properties are necessary.

# Chapter 2
# Discrete-Time System Models for Control

## 2.1 Deterministic Environment

### *2.1.1 Input-Output Difference Operator Models*

We will consider single-input single-output time invariant systems described by input-output discrete-time models of the form:

$$y(t) = -\sum_{i=1}^{n_A} a_i y(t-i) + \sum_{i=1}^{n_B} b_i u(t-d-i) \tag{2.1}$$

where $t$ denotes the normalized sampling time (i.e., $t = \frac{t}{T_S}$, $T_S =$ sampling period), $u(t)$ is the input, $y(t)$ is the output, $d$ is the integer number of sampling periods contained in the time delay of the systems, $a_i$ and $b_i$ are the parameters (coefficients) of the models. As such the output of the system at instant $t$ is a weighted average of the past output over an horizon of $n_A$ samples plus a weighted average of past inputs over an horizon of $n_B$ samples (delayed by $d$ samples). This input-output model (2.1) can be more conveniently represented using a coding in terms of forward or backward shift operators defined as:

$$q \stackrel{\Delta}{=} \text{forward shift operator } (qy(t) = y(t+1)) \tag{2.2}$$

$$q^{-1} \stackrel{\Delta}{=} \text{backward shift operator } (q^{-1}y(t) = y(t-1)) \tag{2.3}$$

Using the notation:

$$1 + \sum_{i=1}^{n_A} a_i q^{-i} = A(q^{-1}) = 1 + q^{-1}A^*(q^{-1}) \tag{2.4}$$

where:

$$A(q^{-1}) = 1 + a_1 q^{-1} + \cdots + a_{n_A} q^{-n_A} \tag{2.5}$$

$$A^*(q^{-1}) = a_1 + a_2 q^{-1} + \cdots + a_{n_A} q^{-n_A+1} \tag{2.6}$$

and:

$$\sum_{i=1}^{n_B} b_i q^{-i} = B(q^{-1}) = q^{-1} B^*(q^{-1}) \tag{2.7}$$

where:

$$B(q^{-1}) = b_1 q^{-1} + b_2 q^{-2} + \cdots + b_{n_B} q^{-n_B} \tag{2.8}$$

$$B^*(q^{-1}) = b_1 + b_2 q^{-1} + \cdots + b_{n_B} q^{-n_B+1} \tag{2.9}$$

Equation (2.1) can be rewritten as:

$$A(q^{-1})y(t) = q^{-d} B(q^{-1})u(t) = q^{-d-1} B^*(q^{-1})u(t) \tag{2.10}$$

or forward in time:

$$A(q^{-1})y(t+d) = B(q^{-1})u(t) \tag{2.11}$$

as well as:[1]

$$y(t+1) = -A^* y(t) + q^{-d} B^* u(t) = -A^* y(t) + B^* u(t-d) \tag{2.12}$$

Observe that (2.12) can also be expressed as (the *regressor form*):

$$y(t+1) = \theta^T \varphi(t) \tag{2.13}$$

where $\theta$ defines the vector of parameters

$$\theta^T = [a_1, \ldots, a_{n_A}, b_1, \ldots, b_{n_B}] \tag{2.14}$$

and $\varphi(t)$ defines the vector of measurements (or the regressor)

$$\varphi^T(t) = [-y(t), \ldots, -y(t-n_A+1), u(t-d), \ldots, u(t-d-n_B+1)] \tag{2.15}$$

The form of (2.13) will be used in order to estimate the parameters of a system model from input-output data. Consider (2.10). Passing the quantities in the left and in the right through a filter $\frac{1}{A(q^{-1})}$ one gets:

$$y(t) = G(q^{-1})u(t) \tag{2.16}$$

where:

$$G(q^{-1}) = \frac{q^{-d} B(q^{-1})}{A(q^{-1})} \tag{2.17}$$

is termed the *transfer operator*.

Computing the $z$-transform of (2.1), one gets the pulse transfer function characterizing the input-output model of (2.1):[2]

$$G(z^{-1}) = \frac{z^{-d} B(z^{-1})}{A(z^{-1})} \tag{2.18}$$

[1] In many cases, the argument $q^{-1}$ will be dropped out, to simplify the notation.

[2] A number of authors prefer to use the notation $G(z)$ for this quantity, instead of $G(z^{-1})$, in order to be coherent with the definition of the $z$-transform.

Observe that the transfer function of the input-output model of (2.1) can be formally obtained from the *transfer operator* by replacing the time operator $q$ by the complex variable $z$. However, one should be careful since the domain of these variables is different. Nevertheless in the linear case with constant parameters one can use either one and their appropriate signification will result from the context.

Note also that the transfer operator $G(q^{-1})$ can be defined even if the parameters of the model (2.1) are time varying, while the concept of pulse transfer function does simply not exist in this case.

While in most of the developments throughout the book we will not need to associate a state-space form to the input-output model of (2.1), this indeed clarifies a number of properties of these models, in particular the definition of the "order" of the system.

**Theorem 2.1** *The order $r$ of the system model* (2.1), *is the dimension of the minimal state space representation associated to the input-output model* (2.1) *and in the case of irreducible transfer function it is equal to*:

$$r = \max(n_A, n_B + d) \tag{2.19}$$

*which corresponds also to the number of the poles of the irreducible transfer function of the system.*

The order of the system is immediately obtained by expressing the transfer operator (2.17) or the transfer function (2.18) in the forward operator $q$ and respectively the complex variable $z$. The passage from $G(z^{-1})$ to $G(z)$ is obtained multiplying by $z^r$:

$$G(z) = \frac{\bar{B}(z)}{\bar{A}(z)} = \frac{z^{r-d}B(z^{-1})}{z^r A(z^{-1})} \tag{2.20}$$

*Example*

$$G(z^{-1}) = \frac{z^{-3}(b_1 z^{-1} + b_2 z^{-2})}{1 + a_1 z^{-1}} \quad \Longrightarrow \quad r = \max(1, 5) = 5$$

$$G(z) = \frac{b_1 z + b_2}{z^5 + a_1 z^4}$$

To see that $r$ effectively corresponds to the dimension of the minimal state space representation let us consider the following observable canonical form:

$$x(t+1) = A_0 x(t) + B_0 u(t) \tag{2.21}$$

$$y(t) = C_0 x(t) \tag{2.22}$$

where $x(t)$ is the state vector and the matrices (or vectors) $A_0$, $B_0$, $C_0$ are given by:

(a) $n_B + d > n_A$

$$
A_0 = \begin{bmatrix}
-a_1 & 1 & & & & & \\
\vdots & & \ddots & & & 0 & \\
\vdots & & & \ddots & & & \\
-a_{n_A} & \cdots & \cdots & \cdots & 1 & \cdots & \cdots \\
0 & & & & \ddots & & \\
\vdots & & 0 & & & \ddots & \\
0 & & & & & & 1
\end{bmatrix}
\begin{array}{l}
\left.\vphantom{\begin{array}{c}1\\1\\1\\1\end{array}}\right\} n_A \\
\left.\vphantom{\begin{array}{c}1\\1\\1\end{array}}\right\} n_B + d - n_A
\end{array}
$$

$$
B_0 = \begin{bmatrix} 0 \\ \vdots \\ 0 \\ b_1 \\ \vdots \\ \vdots \\ b_{n_B} \end{bmatrix}
\begin{array}{l}
\left.\vphantom{\begin{array}{c}1\\1\\1\end{array}}\right\} d \\
\left.\vphantom{\begin{array}{c}1\\1\\1\\1\end{array}}\right\} n_B
\end{array}
$$

$$
C_0 = [1, 0, \ldots, 0]
$$

(b) $n_A \geq n_B + d$

$$
A_0 = \begin{bmatrix}
-a_1 & 1 & \cdots & \cdots & \cdots & 0 \\
\vdots & & \ddots & & & \vdots \\
\vdots & & & \ddots & & \vdots \\
\vdots & & & & \ddots & 0 \\
-a_{n_A} & & & & 0 & 1
\end{bmatrix}, \qquad
B_0 = \begin{bmatrix} 0 \\ \vdots \\ 0 \\ b_1 \\ \vdots \\ b_{n_B} \\ 0 \\ \vdots \\ 0 \end{bmatrix}
\begin{array}{l}
\left.\vphantom{\begin{array}{c}1\\1\\1\end{array}}\right\} d \\
\left.\vphantom{\begin{array}{c}1\\1\\1\end{array}}\right\} n_B \\
\left.\vphantom{\begin{array}{c}1\\1\\1\end{array}}\right\} n_A - (n_B + d)
\end{array}
$$

$$
C_0 = [1, 0, \ldots, 0]
$$

The input-output transfer function is given by:

$$
G(z) = C_0(zI - A_0)^{-1} B_0 = \frac{\bar{B}(z)}{\bar{A}(z)} = \frac{z^{r-d} B(z^{-1})}{z^r A(z^{-1})} \tag{2.23}
$$

*Remarks*

- If $n_B + d > n_A$, the system will have $n_B + d - n_A$ poles at the origin ($z = 0$).
- One has assumed that $A_0$, $B_0$, $C_0$ is a minimal state space realization of the system model (2.1), i.e., that the eventual common factors of $A(z^{-1})$ and $B(z^{-1})$ have been canceled.

In general we will assume that the model of (2.1) and the corresponding transfer function (2.18) is irreducible. However, situations may occur where this is indeed not the case (an estimated model may feature an almost pole zeros cancellation). The properties of the system model in such cases are summarized below:

- The presence of pole zeros cancellations correspond to the existence of unobservable or uncontrollable modes.
- If the common poles and zeros are stable (they are inside the unit circle) the system is termed *stabilizable*, i.e., there is a feedback law stabilizing the system.
- If the common poles and zeros are unstable, the system is *not stabilizable* (a feedback law stabilizing the system does not exist).

The co-primeness of $A(z^{-1})$ and $B(z^{-1})$ is an important property of the model. A characterization of the co-primeness of $A(z^{-1})$ and $B(z^{-1})$ without searching the roots of $A(z^{-1})$ and $B(z^{-1})$ is given by the Sylvester Theorem.

**Theorem 2.2** (Wolowich 1974; Kailath 1980; Goodwin and Sin 1984) *The polynomials $A(q^{-1})$, $q^{-d}B(q^{-1})$ are relatively prime if and only if their eliminant matrix $M$ (known also as the Sylvester Matrix) is nonsingular, where $M$ is a square matrix $r \times r$ with $r = \max(n_A, n_B + d)$ given by*:

$$\left.\underbrace{\begin{bmatrix} \overbrace{\begin{matrix} 1 & 0 & \dots & 0 \\ a_1 & 1 & & \vdots \\ a_2 & & & 0 \\ & & & 1 \\ & & & a_1 \\ a_{n_A} & & & a_2 \\ 0 & & & \vdots \\ 0 & \dots & 0 & a_{n_A} \end{matrix}}^{n_B+d} & \overbrace{\begin{matrix} 0 & \dots & \dots & 0 \\ b'_1 & & & \\ b'_2 & & & b'_1 \\ \vdots & & & b'_2 \\ \vdots & & & \vdots \\ b'_{n_B} & & & \vdots \\ 0 & \dots & \dots & \ddots \\ 0 & 0 & 0 & b'_{n_{B'}} \end{matrix}}^{n_A} \end{bmatrix}}_{n_A+n_B+d}\right\} n_A + n_B + d \tag{2.24}$$

*where*: $b'_i = 0$ *for* $i = 0, 1, \dots, d$; $b'_i = b_{i-d}$ *for* $i \geq d + 1$.

*Remarks*

- The nonsingularity of the matrix $M$ implies the controllability and the observability of the associated state space representation.
- The condition number of $M$ allows to evaluate the ill conditioning of the matrix $M$, i.e., the closeness of some poles and zeros.
- The matrix $M$ is also used for solving the diophantine equation (Bezout identity):

$$A(z^{-1})S(z^{-1}) + z^{-d}B(z^{-1})R(z^{-1}) = P(z^{-1})$$

for $S$ and $R$, given $P$ (see Sect. 7.3 Pole Placement).

### 2.1.2 Predictor Form (Prediction for Deterministic SISO Models)

The model of (2.1) or of (2.12) is a one step ahead predictive form. A problem of interest is to predict from the model (2.1) future values of the output beyond $(t+1)$ based on the information available up to the instant $t$, assuming in the deterministic case that disturbances will not affect the system in the future. A typical example is the control of a system with delay $d$ for which we would like either to predict its output $d+1$ step ahead based on the measurements and the controls applied up to and including the instant $t$ or to compute $u(t)$ in order to reach a certain value of the output at $t+d+1$.

Therefore the objective is in general to compute (for $1 \le j \le d+1$):

$$\hat{y}(t+j/t) = f[y(t), y(t-1), \ldots, u(t), u(t-1), \ldots] \tag{2.25}$$

or a filtered predicted value:

$$P(q^{-1})\hat{y}(t+j/t) = f_P[y(t), y(t-1), \ldots, u(t), u(t-1), \ldots] \tag{2.26}$$

where:

$$\begin{aligned} &P(q^{-1}) = 1 + p_1 q^{-1} + \cdots + p_{n_p} q^{-n_p} = 1 + q^{-1} P^*(q^{-1}) \\ &n_p \le n_A + j - 1 \end{aligned} \tag{2.27}$$

Note that in deterministic case, due to the hypothesis of the absence of disturbances in the future, the future values of the output can be exactly evaluated.

To simplify the notation, $\hat{y}(t+j/t)$ will be replaced by $\hat{y}(t+j)$. We will start with the case $j=1$, in order to emphasize some of the properties of the predictor. For the case $j=1$, taking account of (2.12) and (2.27), one has:

$$P(q^{-1})y(t+1) = (P^* - A^*)y(t) + q^{-d}B^*u(t) \tag{2.28}$$

and, therefore, this suggests to consider as predictor the form:

$$P(q^{-1})\hat{y}(t+1) = (P^* - A^*)y(t) + q^{-d}B^*u(t) \tag{2.29}$$

The prediction error:

$$\varepsilon(t+1) = y(t+1) - \hat{y}(t+1) \tag{2.30}$$

will be governed by:

$$P(q^{-1})\varepsilon(t+1) = 0 \tag{2.31}$$

which is obtained by subtracting (2.29) from (2.28). Observe also that the prediction equation has a "feedback" form since it can be rewritten as:

$$\hat{y}(t+1) = (P^* - A^*)[y(t) - \hat{y}(t)] - A^*\hat{y}(t) + q^{-d}Bu(t) \tag{2.32}$$

(the prediction error drives the prediction). Note also that for $P = A$, (2.32) leads to an "open-loop predictor" or "output error" predictor governed by:

$$\hat{y}(t+1) = -A^*(q^{-1})\hat{y}(t) + q^{-d}B^*(q^{-1})u(t) \tag{2.33}$$

i.e., the predicted output will depend only on the input and previous predictions.

A state space observer can be immediately associated with the predictor of (2.29). Associating to (2.1), the observable canonical state space representation (to simplify the presentation, it is assumed that $d = 0$ and $n_A = n_B = r$):

$$x(t+1) = \begin{bmatrix} -a_1 & \vdots & & \\ \vdots & \vdots & I_{r-1} & \\ \vdots & \ddots & \cdots & \cdots \\ -a_r & & & \end{bmatrix} x(t) + \begin{bmatrix} b_1 \\ \vdots \\ \vdots \\ b_r \end{bmatrix} u(t)$$

$$y(t) = [1, 0, \ldots, 0]x(t) \tag{2.34}$$

the observer will have the form:

$$\hat{x}(t+1) = \begin{bmatrix} -a_1 & \vdots & & \\ \vdots & \vdots & I_{r-1} & \\ \vdots & \ddots & \cdots & \cdots \\ -a_r & 0 & & 0 \end{bmatrix} \hat{x}(t) + \begin{bmatrix} b_1 \\ \vdots \\ \vdots \\ b_r \end{bmatrix} u(t) + \begin{bmatrix} p_1 - a_1 \\ \vdots \\ \vdots \\ p_n - a_n \end{bmatrix} [y(t) - \hat{y}(t)]$$

$$\hat{y}(t) = [1, \ldots, 0]\hat{x}(t) \tag{2.35}$$

and the state error $\tilde{x}(t) = x(t) - \hat{x}(t)$ is governed by:

$$\tilde{x}(t+1) = \begin{bmatrix} -p_1 & \vdots & & \\ \vdots & \vdots & I_{r-1} & \\ \vdots & \ddots & \cdots & \cdots \\ -p_r & 0 & 0 & \end{bmatrix} \tilde{x}(t) \tag{2.36}$$

(i.e., the polynomial $P(q^{-1})$ defines the dynamics of the observer).

For the case $1 \le j \le d+1$, one has the following result:

**Theorem 2.3** *The filtered predicted value* $P(q^{-1})\hat{y}(t+j)$ *can be expressed as*:

$$P(q^{-1})\hat{y}(t+j) = F_j(q^{-1})y(t) + E_j(q^{-1})B^*(q^{-1})u(t+j-d-1) \tag{2.37}$$

*with*:

$$\deg P(q^{-1}) \le n_A + j - 1 \tag{2.38}$$

*where* $E(q^{-1})$ *and* $F(q^{-1})$ *are solutions of the polynomial equation.*

$$AE_j(q^{-1}) + q^{-j}F_j(q^{-1}) = P(q^{-1}) \tag{2.39}$$

*with*:

$$\begin{aligned} E_j(q^{-1}) &= 1 + e_1 q^{-1} + \cdots + e_{j-1}q^{-j+1} \\ F_j(q^{-1}) &= f_0^j + f_1^j q^{-1} + \cdots + f_{n_F}^j q^{-n_f}; \quad n_F \le \max(n_A - 1, n_p - j) \end{aligned} \tag{2.40}$$

*and the prediction error is governed by*:

$$P(q^{-1})[y(t+j) - \hat{y}(t+j)] = 0 \tag{2.41}$$

*Proof* Using (2.39) one can write:

$$P(q^{-1})y(t+j) = A(q^{-1})E_j(q^{-1})y(t+j) + F_j(q^{-1})y(t) \tag{2.42}$$

But taking into account (2.10), one finally gets:

$$P(q^{-1})y(t+j) = F_j(q^{-1})y(t) + E_j(q^{-1})B^*(q^{-1})u(t+j-d-1) \tag{2.43}$$

from which (2.37) is obtained. Subtracting (2.37) from (2.43), the prediction error equation (2.41) is obtained. □

Since one takes advantage of (2.10), it is clear that if $j > d+1$ (long range prediction), the predicted values $\hat{y}(t+j)$ will depend also on future input values (beyond $t$) and a "scenario" for these future values is necessary for computing the prediction (see Sect. 7.7).

The orders of polynomials $E_j(q^{-1})$ and $F_j(q^{-1})$ assure the unicity of the solutions of (2.39).

Equation (2.39) can be expressed in matrix form:

$$Mx^T = p \tag{2.44}$$

where:

$$\begin{aligned} x^T &= [1, e_1, \ldots, e_{j-1}, f_o^j, \ldots, f_{n_F}^j] \\ p &= [1, p_1, \ldots, p_{n_A}, p_{n_{A+1}}, \ldots, p_{n_A+j-1}] \end{aligned} \tag{2.45}$$

and $M$ is a lower-triangular matrix of dimension $(n_A + j) \times (n_A + j)$.

$$M = \left.\left[\begin{array}{ccccccccc}
1 & 0 & \cdots & \cdots & \cdots & \cdots & \cdots & \cdots & 0 \\
a_1 & 1 & 0 & & & & & & \vdots \\
a_2 & a_1 & 1 & & & & & & \vdots \\
\vdots & & & 0 & & & & & \vdots \\
a_{j-1} & a_{j-2} & & 1 & 0 & & & & \vdots \\
a_j & a_{j-1} & & a_1 & 1 & 0 & & & \vdots \\
a_{j+1} & a_j & & \vdots & a_1 & 1 & 0 & & \vdots \\
\vdots & \vdots & & \vdots & \vdots & 0 & 1 & & \vdots \\
a_{n_A} & \vdots & & \vdots & \vdots & \vdots & 0 & & \vdots \\
0 & \vdots & & \vdots & \vdots & \vdots & \vdots & 0 & \vdots \\
\vdots & \vdots & \cdots & \cdots & \vdots & \vdots & \vdots & 1 & 0 \\
0 & \cdots & \cdots & 0 & a_{n_A} & 0 & \vdots & 0 & 1 \\
\multicolumn{5}{c}{\underbrace{\hphantom{a_{j-1}\quad a_{j-2}\quad 1\quad 0\quad 0}}_{j}} & \multicolumn{4}{c}{\underbrace{\hphantom{0\quad 0\quad 0\quad 0}}_{n_A}}
\end{array}\right]\right\} n_A + j \tag{2.46}$$

Because of the structure of the matrix $M$ (lower-triangular), there is always an inverse and one has:

$$x^T = M^{-1} p \tag{2.47}$$

*Remark* The $j$ step ahead predictor can be obtained by successive use of the one step ahead predictor (see Problem 2.1). One replaces: $y(t+j-1)$ by $\hat{y}(t+j-1) = f[y(t+j-2)\ \ldots]$, then $y(t+j-2)$ by $\hat{y}(t+j-2) = f[y(t+j-3)\ \ldots]$ and so on.

### Regressor Form

Taking into account the form of (2.37), the $j$ step ahead predictor can be expressed also in a regressor form as:

$$P(q^{-1})\hat{y}(t+j) = \theta^T \phi(t) \tag{2.48}$$

where:

$$\theta^T = [f_0^j, \ldots, f_{n_F}^j, g_0, \ldots, g_{n_B+j-2}] \tag{2.49}$$

$$\phi^T(t) = [y(t), \ldots, y(t-n_F), u(t+j-d-1), \ldots, u(t-n_B-d+1)] \tag{2.50}$$

$$G_j(q^{-1}) = E_j B^* = g_0 + g_1 q^{-1} + \cdots + g_{j+n_B-2} q^{-j-n_B+2} \tag{2.51}$$

## 2.2 Stochastic Environment

### *2.2.1 Input-Output Models*

In many practical situations, the deterministic input-output model given in (2.1) cannot take into account the presence of stochastic disturbances. A model is therefore needed which accommodates the presence of such disturbances.

An immediate extension of the deterministic input-output model is:

$$y(t+1) = -A^*(q^{-1})y(t) + q^{-d} B^*(q^{-1})u(t) + v(t+1) \tag{2.52}$$

where $v(t)$ is a stochastic process which describes the effect upon the output of the various stochastic disturbances. However, we need to further characterize this disturbance in order to predict the behavior of the system, and to control it.

Stationary stochastic disturbances having a rational spectrum can be modeled as the output of a dynamic system driven by a Gaussian white noise sequence (Factorization Theorem—Åström 1970; Faurre et al. 1979; Åström et al. 1984).

The Gaussian discrete-time white noise is a sequence of independent, equally distributed (Gaussian) random variables of zero mean value and variance $\sigma^2$. This sequence will be denoted $\{e(t)\}$ and characterized by $(0, \sigma)$, where the first number indicates the mean value and second number indicates the standard deviation.

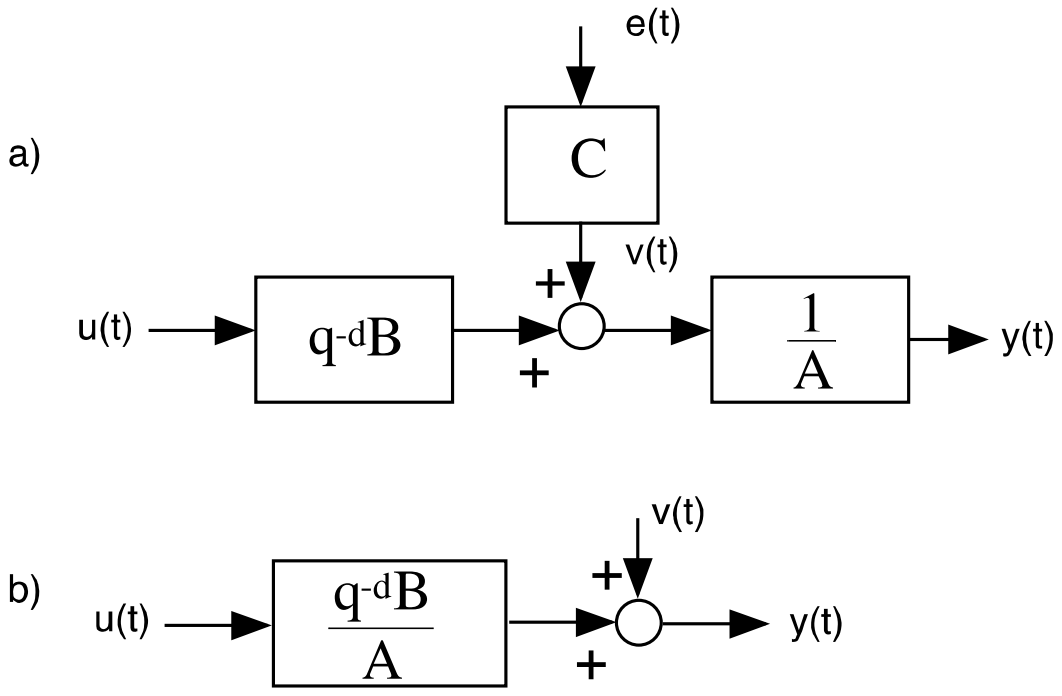

**Fig. 2.1** Configuration for additive stochastic disturbances, (**a**) equation error model structure, (**b**) output error model structure

A large class of stochastic processes of interest for applications will be described by the output of a poles and zeros system driven by white noise called the Auto Regressive Moving Average (ARMA) process.

$$v(t) = \frac{C(q^{-1})}{D(q^{-1})} e(t) \tag{2.53}$$

where:

$$C(q^{-1}) = 1 + c_1 q^{-1} + \cdots + c_{n_C} q^{-n_C} \tag{2.54}$$

$$D(q^{-1}) = 1 + d_1 q^{-1} + \cdots + d_{n_D} q^{-n_D} \tag{2.55}$$

$D(z^{-1})$ has all its roots inside the unit circle and it will be assumed that $C(z^{-1})$ also has all its roots inside the unit circle.[3]

For the case $C(q^{-1}) = 1$, one has an autoregressive (AR) model and for the case $D(q^{-1}) = 1$ one has a moving average model (MA). However, $v(t)$ may act in different part of the system. Equation (2.52) corresponds to the block diagram shown in Fig. 2.1a (known also as the *equation error model*). If in addition $v(t)$ is modeled by (2.53) with $D(q^{-1}) = 1$, one obtains the Autoregressive Moving Average with Exogenous Input (ARMAX) model (Fig. 2.1a):

$$y(t+1) = -A^*(q^{-1})y(t) + q^{-d}B^*(q^{-1})u(t) + C(q^{-1})e(t+1) \tag{2.56}$$

or:

$$y(t) = \frac{q^{-d}B(q^{-1})}{A(q^{-1})} u(t) + \frac{C(q^{-1})}{A(q^{-1})} e(t) \tag{2.57}$$

In general: $n_c \leq n_A$.

Equation (2.57) gives an equivalent form for (2.56) which corresponds to the disturbance added to the output, but filtered by $1/A(q^{-1})$. However, the disturbance may be represented as directly added to the output as shown in Fig. 2.1b.

$$y(t) = \frac{q^{-d}B(q^{-1})}{A(q^{-1})} u(t) + v(t) \tag{2.58}$$

[3]Factorization theorem for rational spectrum allows for zeros of $C(z^{-1})$ on the unit circle (Åström et al. 1984).

This configuration is known as the *output error model* which can further be written as:

$$y(t+1) = -A^*(q^{-1})y(t) + q^{-d}B^*(q^{-1})u(t) + A(q^{-1})v(t) \tag{2.59}$$

A model of the form (2.53) can be associated to $v(t)$ in (2.58) (this is the Box-Jenkins model). However, in many cases, one simply makes the hypothesis that $\{u(t)\}$ and $\{v(t)\}$ are independent without considering a special model for $v(t)$ (except that it is assumed to be of finite power).

All these models allow a representation of the form:

$$y(t+1) = \theta^T \varphi(t) + w(t+1) \tag{2.60}$$

where $w(t+1)$ is different depending on the context, and $\theta$ and $\varphi$ are given by (2.14) and (2.15), respectively.

### 2.2.2 Predictors for ARMAX Input-Output Models

In the stochastic context, the prediction takes its full significance since the future values of the output will be affected by the disturbances for $j \geq 1$. However, the information upon the disturbance model will be taken into account for constructing a predictor. We will consider the ARMAX model:

$$y(t+1) = -A^*(q^{-1})y(t) + q^{-d}B^*(q^{-1})u(t) + C(q^{-1})e(t+1) \tag{2.61}$$

The objective is to find a linear predictor as a function of the available data up to and including time $t$:

$$\hat{y}(t+j/t) = \hat{y}(t+j) = f[y(t), y(t-1), \ldots, u(t), u(t-1), \ldots] \tag{2.62}$$

such that:

$$\mathbf{E}\{[y(t+j) - \hat{y}(t+j)]^2\} = \min \tag{2.63}$$

One has the following results.

**Theorem 2.4** (Åström 1970) *For the system of* (2.61) *provided that* $C(q^{-1})$ *is asymptotically stable and* $e(t)$ *is a discrete-time white noise, the optimal* $j$ *step ahead predictor minimizing* (2.63) *is given by*:

$$\hat{y}(t+j) = \frac{F_j(q^{-1})}{C(q^{-1})}y(t) + \frac{E_j(q^{-1})B^*(q^{-1})}{C(q^{-1})}u(t+j-d-1) \tag{2.64}$$

*where* $F_j(q^{-1})$ *and* $E_j(q^{-1})$ *are solutions of the polynomial equation*:

$$C(q^{-1}) = A(q^{-1})E_j(q^{-1}) + q^{-j}F_j(q^{-1}) \tag{2.65}$$

*where*:

$$E_j(q^{-1}) = 1 + e_1 q^{-1} + \cdots + e_{j-1} q^{-j+1} \tag{2.66}$$

$$F_j(q^{-1}) = f_0^j + f_1^j q^{-1} + \cdots + f_{n_F} q^{-n_F}$$

$$n_F = n_A - 1;\ n_C \le n_A \tag{2.67}$$

*and the optimal prediction error is a moving average given by*:

$$\varepsilon(t+j)|_{opt} = y(t+j) - \hat{y}(t+j)\,|_{opt} = E_j e(t+j) \tag{2.68}$$

*Note that for* $j = 1$, *the optimal prediction error is a white noise.*

*Proof* From (2.65) one has:

$$C(q^{-1})y(t+j) = E_j A y(t+j) + F_j y(t) \tag{2.69}$$

and using (2.61) one has:

$$Ay(t+j) = B^* u(t+j-d-1) + C(q^{-1})e(t+j) \tag{2.70}$$

Introducing (2.70) in (2.69) one gets:

$$C(q^{-1})y(t+j) = F_j y(t) + E_j B^* u(t+j-d-1) + E_j C e(t+j) \tag{2.71}$$

and respectively:

$$y(t+j) = \frac{F_j}{C} y(t) + \frac{E_j B^*}{C} u(t+j-d-1) + E_j e(t+j) \tag{2.72}$$

The expression of $y(t+j)$ given by (2.72) has what is called an "innovation" form, i.e.:

$$\begin{aligned} y(t+j) = {} & f[y(t), y(t-1), \ldots, u(t+j-d-1), \ldots] \\ & + g[e(t+1), e(t+2), \ldots, e(t+j)] \end{aligned} \tag{2.73}$$

where the first term depends on the data available up to and including time $t$, and the second term is the "true" stochastic term describing the future behavior which by no means can be predicted at time $t$. Introducing (2.72) in the criterion (2.63) leads to:

$$\begin{aligned} &\mathbf{E}\{[y(t+j) - \hat{y}(t+j)]^2\} \\ &\quad = \mathbf{E}\left\{\left[\frac{F_j}{C} y(t) + \frac{E_j B^*}{C} u(t+j-d-1) - \hat{y}(t+j)\right]^2\right\} \\ &\qquad + 2\mathbf{E}\left\{[E_j e(t+j)]\left[\frac{F_j}{c} y(t) + \frac{E_j B^*}{c} u(t+j-d-1)\right]\right\} \\ &\qquad + \mathbf{E}\{[E_j e(t+j)]^2\} \end{aligned} \tag{2.74}$$

The second term of the right hand side is null because $E_j e(t+j)$ contains $e(t+1)$, $\ldots, e(t+j)$, which are all independent of $y(t), y(t-1), \ldots, u(t+j-d-1), \ldots$. The third term does not depend on the choice of $\hat{y}(t+j)$ and therefore the minimum is obtained by choosing $\hat{y}(t+j)$ given in (2.64) such that the first term will become null.

Subtracting $\hat{y}(t+j)$ given by (2.64) from (2.72) gives the expression (2.68) of the prediction error. □

### Rapprochement with the Deterministic Case

Comparing Theorem 2.4 with Theorem 2.3, one can see immediately that the predictors for deterministic case and the stochastic case are the same if $P(q^{-1}) = C(q^{-1})$.

In other words, a deterministic predictor is an optimal predictor in the stochastic environment for an ARMAX type model with $C(q^{-1}) = P(q^{-1})$. Conversely, the polynomial $C(q^{-1})$ will define the decaying dynamics of the prediction error in the deterministic case (and of the initial conditions in the stochastic case).

### Rapprochement with the Kalman Predictor

Consider the ARMAX model of (2.61) with $j = 1, d = 0, n_B < n_A$, $n_C = n_A$ to simplify the discussion. This model admits a state space "innovation" representation in observable canonical form given by:

$$x(t+1) = \begin{bmatrix} -a_1 & \vdots & & \\ \vdots & \vdots & I_{n_A-1} & \\ \vdots & \vdots & \cdots & \cdots \\ -a_{n_A} & 0 & \cdots & 0 \end{bmatrix} x(t) + \begin{bmatrix} b_1 \\ \vdots \\ b_{n_B} \\ \vdots \\ 0 \end{bmatrix} u(t) + \begin{bmatrix} c_1 - a_1 \\ \vdots \\ \vdots \\ c_{n_C} - a_{n_A} \end{bmatrix} e(t) \tag{2.75}$$

$$y(t) = [1, 0, \ldots, 0]x(t) + e(t) \tag{2.76}$$

The optimal predictor of (2.64) for $d = 0, j = 1$ can be rewritten as ($E_1 = 1$, $F_1 = C^* - A^*$):

$$C\hat{y}(t+1) = (C^* - A^*)y(t) + B^*u(t) \pm A^*\hat{y}(t) \tag{2.77}$$

yielding:

$$\hat{y}(t+1) = -A^*\hat{y} + B^*u(t) + (C^* - A^*)[y(t) - \hat{y}(t)] \tag{2.78}$$

i.e., the predictor is driven by the prediction error which is an innovation process.

The associated state space representation of the predictor takes the form:

$$\hat{x}(t+1) = \begin{bmatrix} -a_1 & \vdots & & \\ \vdots & \vdots & I_{n_A-1} & \\ \vdots & \ddots & \cdots & \cdots \\ a_{n_A} & 0 & \cdots & 0 \end{bmatrix} \hat{x}(t) + \begin{bmatrix} b_1 \\ \vdots \\ b_{n_B} \\ \vdots \\ 0 \end{bmatrix} u(t) + \begin{bmatrix} c_1 - a_1 \\ \vdots \\ \vdots \\ c_{n_A} - a_{n_A} \end{bmatrix} [y(t) - \hat{y}(t)] \tag{2.79}$$

$$\hat{y}(t) = [1, 0, \ldots, 0]\hat{x}(t) \tag{2.80}$$

which is nothing else than the steady state Kalman predictor for the system (2.75)–(2.76). The steady state Kalman predictor can be directly obtained by inspection, since the model (2.75)–(2.76) is in *innovation* form. The objective is to obtain asymptotically an output prediction error which is an innovation sequence, i.e.

$$\lim_{t\to\infty}[y(t)-\hat{y}(t)]=e(t) \tag{2.81}$$

Subtracting (2.79) from (2.75), one obtains for the dynamics of the state estimation error $\tilde{x}(t)=x(t)-\hat{x}(t)$:

$$\tilde{x}(t+1)=\begin{bmatrix} -c_1 & \vdots & & \\ \vdots & \vdots & I_{n_C-1} & \\ \vdots & \ddots & \dots & \dots \\ -c_{n_C} & 0 & \dots & 0 \end{bmatrix}\tilde{x}(t) \tag{2.82}$$

Therefore the poles associated with the state estimation error are defined by $C(z^{-1})$ which is assumed to be asymptotically stable.

Taking this into account, it results that:

$$\lim_{t\to\infty}[y(t)-\hat{y}(t)]=[1,0,\dots,0]\lim_{t\to\infty}\tilde{x}(t)+e(t)=e(t) \tag{2.83}$$

## Regressor Form

Taking into account (2.64), the $j$ step ahead predictor can be expressed as:

$$C\hat{y}(t+j)=F_j y(t)+G_j u(t+j-d-1) \tag{2.84}$$

where:

$$G_j=E_jB^*=g_0+g_1q^{-1}+\dots+g_{j+n_B-2}q^{-j-n_B+2} \tag{2.85}$$

from which one obtains:

$$C\hat{y}(t+j)=\theta^T\phi(t) \tag{2.86}$$

where:

$$\theta^T=[f_0^j,\dots,f_{n_F}^j,g_0,\dots,g_{j+n_B-2}] \tag{2.87}$$

$$\phi^T(t)=[y(t),\dots,y(t-n_F),u(t+j-d-1),\dots,u(t-n_B-d+1)] \tag{2.88}$$

An alternative regressor form is:

$$\hat{y}(t+j)=\theta_e^T\phi_e(t) \tag{2.89}$$

where:

$$\theta_e^T=[\theta^T,c_1,\dots,c_{n_C}] \tag{2.90}$$

$$\phi_e(t)=[\phi^T(t),-\hat{y}(t+j-1),\dots,-\hat{y}(t+j-n_C)] \tag{2.91}$$

For the case $j = 1$, the predictor equation takes the form:

$$\hat{y}(t+1) = (C^* - A^*)y(t) + q^{-d}B^*u(t) - C^*\hat{y}(t) \tag{2.92}$$

allowing the representation shown above. However, this predictor can also be rewritten as:

$$\hat{y}(t+1) = -A^*y(t) + q^{-d}B^*u(t) + C^*[y(t) - \hat{y}(t)] \tag{2.93}$$

leading to the regressor form representation:

$$\hat{y}(t+1) = \theta_e^T \phi_e(t) \tag{2.94}$$

where:

$$\theta_e^T = [a_1, \ldots, a_{n_A}, b_1, \ldots, b_{n_B}, c_1, \ldots, c_{n_C}] \tag{2.95}$$

$$\begin{aligned}\phi_e^T(t) = [&-y(t), \ldots, -y(t - n_A + 1), u(t-d), \ldots, u(t - d - n_B + 1),\\ &\varepsilon(t), \ldots, \varepsilon(t - n_C + 1)]\end{aligned} \tag{2.96}$$

$$\varepsilon(t) = y(t) - \hat{y}(t) \tag{2.97}$$

Equation (2.93) can also be equivalently rewritten as (by adding $\pm A^*\hat{y}(t)$):

$$\hat{y}(t+1) = -A^*\hat{y}(t) + q^{-d}B^*u(t) + [C^* - A^*][y(t) - \hat{y}(t)] \tag{2.98}$$

leading to the regression form representation (2.94) where now:

$$\theta_e^T = [a_1, \ldots, a_{n_A}, b_1, \ldots, b_{n_B}, h_1, \ldots, h_{n_C}] \tag{2.99}$$

$$\begin{aligned}\phi_e^T(t) = [&-\hat{y}(t), \ldots, -\hat{y}(t - n_A + 1), u(t-d), \ldots, u(t - d - n_B + 1),\\ &\varepsilon(t), \ldots, \varepsilon(t - n_C + 1)]\end{aligned} \tag{2.100}$$

$$h_i = c_i - a_i; \quad i = 1, \ldots, n_C \tag{2.101}$$

These different forms of one step ahead predictor will be used for the estimation of the parameters of a plant model in a stochastic environment.

An important characteristic of the predictors for ARMAX models is that the predicted values depend upon (i) current and previous values of the input and output and (ii) current and previous values of the prediction.

### *2.2.3 Predictors for Output Error Model Structure*

Let us consider now the case where the stochastic disturbance $v(t)$ affects directly the output:

$$y(t+1) = \frac{q^{-d}B^*(q^{-1})}{A(q^{-1})}u(t) + v(t+1) \tag{2.102}$$

under the hypothesis that $\{v(t)\}$ and $\{u(t)\}$ are independent and that $\{v(t)\}$ has finite variance.

The objective is to find a linear predictor depending on the information available up to and including $t$ which minimizes the criterion:

$$\mathbf{E}\{[y(t+1)-\hat{y}(t+1)]^2\} \tag{2.103}$$

Introducing (2.102) in (2.103), one gets:

$$\begin{aligned}
&\mathbf{E}\{[y(t+1)-\hat{y}(t+1)]^2\} \\
&\quad = \mathbf{E}\left\{\left[\frac{q^{-d}B^*}{A^*}u(t)-\hat{y}(t+1)\right]^2\right\}+\mathbf{E}\{v(t)^2\} \\
&\quad + 2\mathbf{E}\left\{v(t+1)\left[\frac{q^{-d}B^*}{A^*}u(t)-\hat{y}(t+1)\right]\right\}
\end{aligned} \tag{2.104}$$

The third term of this expression is null, since $v(t)$ is independent with respect to $u(t), u(t-1), \ldots$ and linear combinations of these variables which will serve to generate $\hat{y}(t+1)$. The second term does not depend upon the choice of $\hat{y}(t+1)$ and the criterion will be minimized if the first term becomes null. This leads to:

$$\hat{y}(t+1)=\frac{q^{-d}B^*}{A}u(t) \tag{2.105}$$

or:

$$\hat{y}(t+1)=-A^*\hat{y}(t)+q^{-d}B^*u(t) \tag{2.106}$$

known as the *output error predictor*.

Its main characteristic is that the predicted output will depend only upon (i) the current and previous inputs and (ii) the current and previous predicted outputs (in the ARMAX case the predicted output depends also upon current and past measurements).

The output error predictor can be expressed also in the regressor form:

$$\hat{y}(t+1)=\theta^T\phi(t) \tag{2.107}$$

where:

$$\theta^T=[a_1,\ldots,a_{n_A},b_1,\ldots,b_{n_B}] \tag{2.108}$$

$$\phi^T(t)=[-\hat{y}(t),\ldots,-\hat{y}(t-n_A+1),u(t-d),\ldots,u(t-d-n_B+1)] \tag{2.109}$$

## 2.3 Concluding Remarks

1. In a deterministic environment the discrete-time single input, single output models describing the systems to be controlled are of the form:

$$y(t)=-\sum_{i=1}^{n_A}a_i y(t-i)+\sum_{i=1}^{n_B}b_i u(t-d-i) \tag{$*$}$$

where $y(t)$ is the output and $u(t)$ is the input. Using the delay operator $q^{-1}(y(t-1) = q^{-1}y(t))$, the model $(*)$ takes the form:

$$y(t+1) = -A^*(q^{-1})y(t) + q^{-d}B^*(q^{-1})u(t)$$

where:

$$A(q^{-1}) = 1 + \sum_{i=1}^{n_A} a_i q^{-i} = 1 + q^{-1}A^*(q^{-1})$$

$$B(q^{-1}) = \sum_{i=1}^{n_B} b_i q^{-i} = q^{-1}B^*(q^{-1})$$

2. The filtered predicted values of $y(t)$, $j$ steps ahead (for $j$ up to $d+1$) can be expressed as a function of current and previous input-output measurements:

$$P(q^{-1})\hat{y}(t+j) = F_j(q^{-1})y(t+j) + E_j(q^{-1})B^*(q^{-1})u(t+j-d-1)$$

with:

$$\deg P(q^{-1}) \le n_A + j - 1$$

where $E_j(q^{-1})$ and $F_j(q^{-1})$ are solutions of the polynomial equation:

$$P(q^{-1}) = A(q^{-1})E_j(q^{-1}) + q^{-j}F_j(q^{-1})$$

3. In a stochastic environment, the input-output model takes the form:

$$y(t+1) = -A^*y(t) + q^{-d}B^*u(t) + v(t+1)$$

or:

$$y(t+1) = -A^*y(t) + q^{-d}B^*u(t) + Av(t+1)$$

where $v(t+1)$ represents the effect of a stochastic disturbance. The first model corresponds to the *equation error model* and the second model corresponds to the *output error model*.

4. Typical forms for the stochastic disturbances are:
   (i) $v(t) = \frac{C(^{-1})}{D(q^{-1})}e(t)$ where $e(t)$ is a zero mean white noise
   (ii) $\{v(t)\}$ is a zero mean stochastic process of finite power and independent of $u(t)$. For the case $v(t) = C(q^{-1})e(t)$, the *equation error model* is an ARMAX model:

$$y(t+1) = -A^*y(t) + q^{-d}B^*u(t) + Ce(t+1)$$

5. The optimal $j$ step ahead predictor for ARMAX models has the form:

$$\hat{y}(t+j) = \frac{F_j}{C}y(t) + \frac{E_jB^*}{C}u(t+j-d-1)$$

where $E_j(q^{-1})$ and $F_j(q^{-1})$ are solutions of the polynomial equation:

$$C(q^{-1}) = A(q^{-1})E_j(q^{-1}) + q^{-j}E_j(q^{-1})$$

6. Predictors for the deterministic models and stochastic ARMA models are the same if $P(q^{-1}) = C(q^{-1})$.
7. The one step ahead optimal predictor for the *output error model*, when $v(t)$ is independent with respect to the input, is:

$$\hat{y}(t+1) = \frac{q^{-d}B^*}{A}u(t)$$

(it depends only upon the input).

## 2.4 Problems

**2.1** Show that the $j$-step ahead predictor (2.37) for $y(t)$ given by (2.1) can be obtained by successive use of one step ahead predictors.

**2.2** From (2.64) it results that the $j$-step ahead predictor of $y(t+1)$ given by (2.60) can be expressed as:

$$\hat{y}(t+j) = f[\hat{y}(t+j-1), \hat{y}(t+j-2), \ldots, y(t), y(t-1), \ldots, \\ u(t+j-d-1), u(t+j-d-2), \ldots]$$

Using a second polynomial division show that one can express $\hat{y}(t+j)$ as:

$$\hat{y}(t+j) = f[y(t), y(t-1), \ldots, u(t+j-d-1), u(t+j-d-2), \ldots]$$

**2.3** Give a recursive formula for the computation of the polynomials $E_{j+1}(q^{-1})$ and $F_{j+1}(q^{-1})$ in (2.39), knowing the solutions $E_j(q^{-1})$ and $F_j(q^{-1})$.

**2.4** Construct the one step ahead optimal predictor for:

$$y(t) = \frac{q^{-d}B(q^{-1})}{A(q^{-1})}u(t) + \frac{1}{C(q^{-1})A(q^{-1})}e(t)$$

where $e(t)$ is a white noise.

**2.5** Construct the one step ahead optimal predictor for:

$$y(t) = \frac{q^{-d}B(q^{-1})}{A(q^{-1})}u(t) + \frac{C(q^{-1})}{D(q^{-1})}e(t)$$

where $e(t)$ is a white noise.

**2.6** Consider the ARMAX model:

$$y(t) = \frac{q^{-d}B(q^{-1})}{A(q^{-1})}u(t) + \frac{C(q^{-1})}{A(q^{-1})}e(t)$$

where:

$$u(t) = -\frac{R(q^{-1})}{S(q^{-1})}y(t) + r(t)$$

Construct the one step ahead optimal predictor for $y(t)$.

**2.7** Consider the continuous-time transfer function:

$$G(s) = \frac{Ge^{-s\tau}}{1+sT}$$

Compute the zero-hold, discrete-time equivalent model. The relation between the continuous-time delay and the discrete-time delay is:

$$\tau = dT_S + L; \quad 0 < L < T_S$$

where $T_S$ is the sampling period and $L$ is the fractional delay. Examine the position of the discrete-time zeros for various values of $L$. Does the system has always stable discrete-time zeros?

**2.8** Give examples of discrete-time models featuring:

(a) unstable discrete-time zeros but "minimum phase" behavior
(b) unstable discrete-time zeros and "non-minimum phase" behavior
(c) stable discrete-time zeros but "non-minimum phase" behavior

(non-minimum phase behavior = the time-response to a positive step starts with negative values).

**2.9** Consider the discrete-time model:

$$\frac{b_1q^{-1} + b_2q^{-2}}{(1+f_1q^{-1})(1+f_2q^{-2})}$$

Verify that for $f_2 = -\frac{b_2}{b_1}$, the determinant of the Sylvester matrix (2.24) is null.

# Chapter 3
# Parameter Adaptation Algorithms—Deterministic Environment

## 3.1 The Problem

On-line estimation of the parameters of a plant model or of a controller is one of the key steps in building an adaptive control system. As shown in Chap. 1 the direct estimation of the controller parameters (when possible) can also be interpreted as a plant model estimation in a reparameterized form. Therefore the problem of on-line estimation of plant model parameters is a generic problem in adaptive control. Such systems will feature a parametric adaptation algorithm which will up-date the estimated parameters at each sampling instant.

On-line plant parameter estimation can also be viewed as a technique for achieving system identification in general by using recursive parametric estimation methods which process a pair of input-output measurements sequentially, as opposed to non recursive (or off-line) identification methods which process an input-output data file acquired over a certain time horizon at once. The problem of recursive plant model identification which is broader than just plant parameter estimation will be discussed in Chap. 5.

The on-line parameter estimation principle for sampled models is illustrated in Fig. 3.1.

A discrete-time model with adjustable parameters is implemented on the computer. The error between the system output at instant $t$, $y(t)$ and the output predicted by the model $\hat{y}(t)$ (called plant-model error or prediction error) is used by the *parameter adaptation algorithm*, which, at each sampling instant, will modify the model parameters in order to minimize this error (in the sense of a certain criterion).

The input to the system is either a low level and frequency rich signal generated by the computer itself for the case of plant model identification, or the signal generated by the controller in the case of an adaptive control system. It can also be the combination of both (e.g., Identification in closed loop).

The key element for implementing the on-line estimation of the plant model parameters is the *parameter adaptation algorithm* (PAA) which drives the parameters of the adjustable prediction model from the data acquired on the system at each sampling instant. This algorithm has a *recursive* structure, i.e., the new value of the estimated parameters is equal to the previous value plus a correcting term which will depend on the most recent measurements.

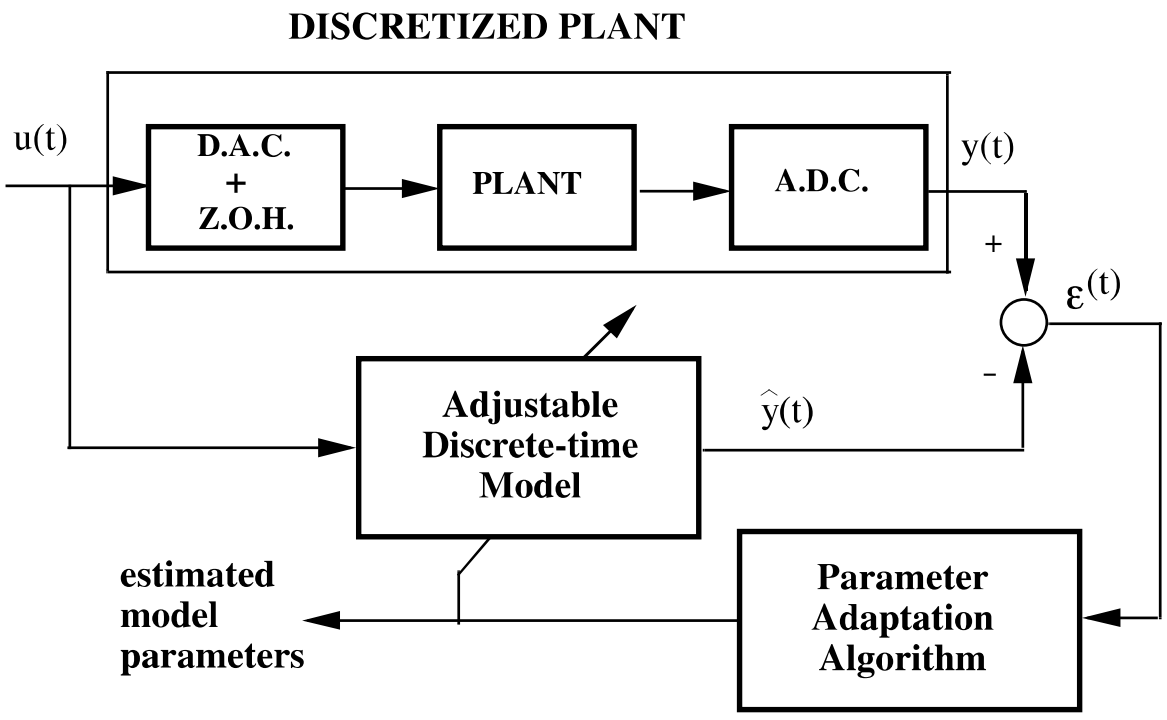

**Fig. 3.1** Parameter estimation principle

In general a *parameter vector* is defined. Its components are the different parameters that should be estimated.

Parameter adaptation algorithms generally have the following structure:

$$\begin{bmatrix} \text{New estimated} \\ \text{parameters} \\ \text{(vector)} \end{bmatrix} = \begin{bmatrix} \text{Previous estimated} \\ \text{parameters} \\ \text{(vector)} \end{bmatrix} + \begin{bmatrix} \text{Adaptation} \\ \text{gain} \\ \text{(matrix)} \end{bmatrix}$$
$$\times \begin{bmatrix} \text{Measurement} \\ \text{function} \\ \text{(vector)} \end{bmatrix} \times \begin{bmatrix} \text{Prediction error} \\ \text{function} \\ \text{(scalar)} \end{bmatrix}$$

This structure corresponds to the so-called *integral type adaptation algorithms* (the algorithm has memory and therefore maintains the estimated value of the parameters when the correcting terms become null). The algorithm can be viewed as a discrete-time integrator fed at each instant by the correcting term. The measurement function vector is generally called the *observation vector*. The prediction error function is generally called the *adaptation error*.

As will be shown, there are more general structures where the integrator is replaced by other types of dynamics, i.e., the new value of the estimated parameters will be equal to a function of the previous parameter estimates (eventually over a certain horizon) plus the correcting term.

The adaptation gain plays an important role in the performances of the parameter adaptation algorithm and it may be constant or time-varying.

The problem addressed in this chapter is the synthesis and analysis of parameter adaptation algorithms in a deterministic environment.

## 3.2 Parameter Adaptation Algorithms (PAA)—Examples

Several approaches can be considered for deriving parameter adaptation algorithms:

- heuristic approach;
- gradient technique;
- least squares minimization;
- stability;
- rapprochement with Kalman filter.

We will consider first for pedagogical reasons the gradient technique followed by the least squares approach and we will use the Kalman filter approach for interpreting the parametric adaptation algorithms. The stability approach will be used in Sect. 3.3 for both synthesis and analysis of PAA in a deterministic environment. For the heuristic approach, see Landau (1990b).

### 3.2.1 Gradient Algorithm

The aim of the gradient parameter adaptation algorithm is to minimize a quadratic criterion in terms of the prediction error.

Consider the discrete-time model of a plant described by:

$$y(t+1) = -a_1 y(t) + b_1 u(t) = \theta^T \phi(t) \tag{3.1}$$

where the unknown parameters $a_1$ and $b_1$ form the components of the *parameter vector* $\theta$:

$$\theta^T = [a_1, b_1] \tag{3.2}$$

and

$$\phi^T(t) = [-y(t), u(t)] \tag{3.3}$$

is the *measurement vector*.

The adjustable prediction model will be described in this case by:

$$\hat{y}^0(t+1) = \hat{y}[(t+1)/\hat{\theta}(t)] = -\hat{a}_1(t)y(t) + \hat{b}(t)u(t) = \hat{\theta}^T(t)\phi(t) \tag{3.4}$$

where $\hat{y}^0(t+1)$ is termed the a priori predicted output depending on the values of the estimated parameters at instant $t$.

$$\hat{\theta}^T(t) = [\hat{a}_1(t), \hat{b}_1(t)] \tag{3.5}$$

is the *estimated parameter vector* at instant $t$.

As it will be shown later, it is very useful to consider also the a posteriori predicted output computed on the basis of the new estimated parameter vector at $t+1$, $\hat{\theta}(t+1)$, which will be available somewhere between $t+1$ and $t+2$. The a posteriori predicted output will be given by:

$$\begin{aligned}\hat{y}(t+1) &= \hat{y}[(t+1)|\hat{\theta}(t+1)] \\ &= -\hat{a}_1(t+1)y(t) + \hat{b}_1(t+1)u(t) = \hat{\theta}^T(t+1)\phi(t)\end{aligned} \tag{3.6}$$

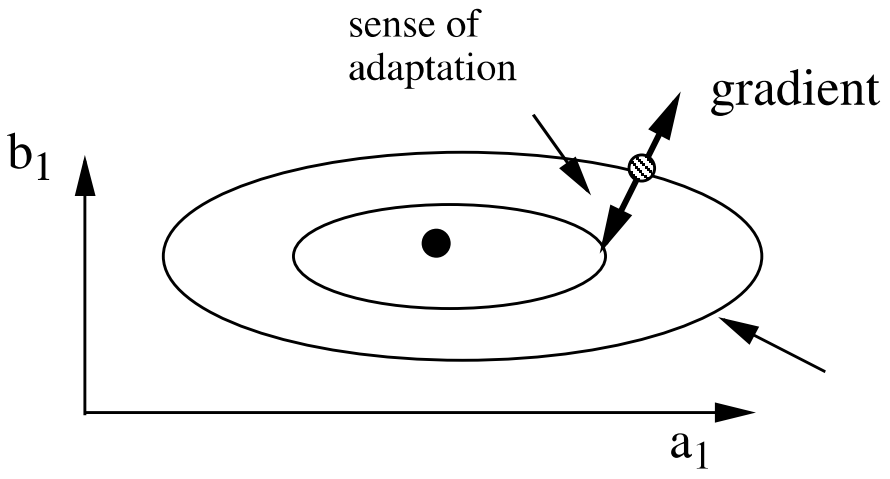

**Fig. 3.2** Principle of the gradient method

One defines an a priori prediction error as:

$$\varepsilon^0(t+1) = y(t+1) - \hat{y}^0(t+1) \tag{3.7}$$

and an a posteriori prediction error as:

$$\varepsilon(t+1) = y(t+1) - \hat{y}(t+1) \tag{3.8}$$

The objective is to find a recursive parameter adaptation algorithm with memory. The structure of such an algorithm is:

$$\hat{\theta}(t+1) = \hat{\theta}(t) + \Delta\hat{\theta}(t+1) = \hat{\theta}(t) + f[\hat{\theta}(t), \phi(t), \varepsilon^0(t+1)] \tag{3.9}$$

The correction term $f[\hat{\theta}(t), \phi(t), \varepsilon^0(t+1)]$ must depend solely on the information available at the instant $(t+1)$ when $y(t+1)$ is acquired (last measurement $y(t+1)$, $\hat{\theta}(t)$, and a finite amount of information at times $t, t-1, t-2, \ldots, t-n$). The correction term must enable the following criterion to be minimized at each step:

$$\min_{\hat{\theta}(t)} J(t+1) = [\varepsilon^0(t+1)]^2 \tag{3.10}$$

A solution can be provided by the gradient technique.

If the iso-criterion curves ($J = \text{const}$) are represented in the plane of the parameters $a_1$, $b_1$, concentric closed curves are obtained around the minimum value of the criterion, which is reduced to the point $(a_1, b_1)$ corresponding to the parameters of the plant model. As the value of $J = \text{const}$ increases, the iso-criterion curves move further and further away from the minimum. This is illustrated in Fig. 3.2.

In order to minimize the value of the criterion, one moves in the opposite direction of the gradient to the corresponding iso-criterion curve. This will lead to a curve corresponding to $J = \text{const}$, of a lesser value, as is shown in Fig. 3.2. The

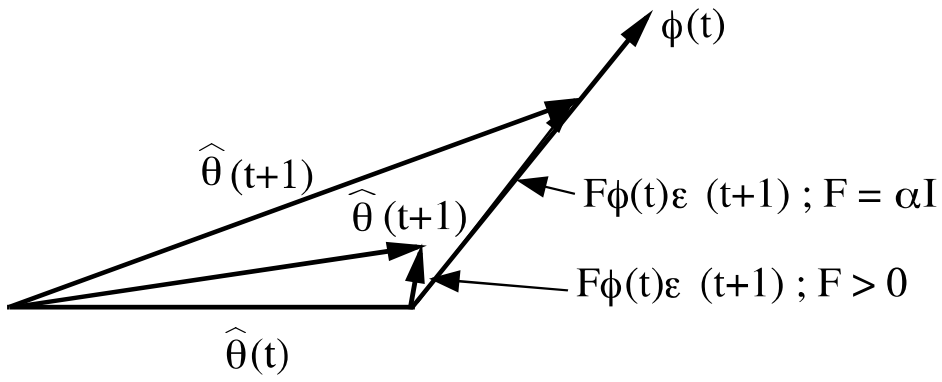

**Fig. 3.3** Geometrical interpretation of the gradient adaptation algorithm

corresponding parametric adaptation algorithm will have the form:

$$\hat{\theta}(t+1) = \hat{\theta}(t) - F\frac{\delta J(t+1)}{\delta\hat{\theta}(t)} \tag{3.11}$$

where $F = \alpha I(\alpha > 0)$ is the matrix adaptation gain (I—unitary diagonal matrix) and $\delta J(t+1)/\delta\hat{\theta}(t)$ is the gradient of the criterion given in (3.10) with respect to $\hat{\theta}(t)$. From (3.10) one obtains:

$$\frac{1}{2}\frac{\delta J(t+1)}{\delta\hat{\theta}(t)} = \frac{\delta\varepsilon^0(t+1)}{\delta\hat{\theta}(t)}\varepsilon^0(t+1) \tag{3.12}$$

But:

$$\varepsilon^0(t+1) = y(t+1) - \hat{y}^0(t+1) = y(t+1) - \hat{\theta}^T(t)\phi(t) \tag{3.13}$$

and

$$\frac{\delta\varepsilon^0(t+1)}{\delta\hat{\theta}(t)} = -\phi(t) \tag{3.14}$$

Introducing (3.14) in (3.12), the parameter adaptation algorithm of (3.11) becomes:

$$\boxed{\hat{\theta}(t+1) = \hat{\theta}(t) + F\phi(t)\varepsilon^0(t+1)} \tag{3.15}$$

where $F$ is the matrix adaptation gain. There are two possible choices:

1. $F = \alpha I; \alpha > 0$
2. $F > 0$ (positive definite matrix)[1]

The resulting algorithm has an integral structure. Therefore it has memory (for $\varepsilon^0(t+1) = 0$, $\hat{\theta}(t+1) = \hat{\theta}(t)$).

The geometrical interpretation of the PAA of (3.15) is given in Fig. 3.3.

The correction is done in the direction of the observation vector (which in this case is the measurement vector) in the case $F = \alpha I, \alpha > 0$ or within ±90 deg around this direction when $F > 0$ (a positive definite matrix may cause a rotation of a vector for less than 90 deg).

[1]A symmetric square matrix $F$ is termed positive definite if $x^T F x > 0$ for all $x \neq 0$, $x \in \mathbb{R}^n$. In addition: (i) all the terms of the main diagonal are positive, (ii) the determinants of all the principals minors are positive.

The parameter adaptation algorithm given in (3.15) presents instability risks if a large adaptation gain (respectively a large $\alpha$) is used. This can be understood by referring to Fig. 3.2. If the adaptation gain is large near the optimum, one can move away from this minimum instead of getting closer.

The following analysis will allow to establish necessary conditions upon the adaptation gain in order to avoid instability.

Consider the parameter error defined as:

$$\tilde{\theta}(t) = \hat{\theta}(t) - \theta \tag{3.16}$$

From (3.1) and (3.4) it results:

$$\varepsilon^0(t+1) = y(t+1) - \hat{y}^0(t+1) = \theta^T\phi(t) - \hat{\theta}^T(t)\phi(t) = -\phi^T(t)\tilde{\theta}(t) \tag{3.17}$$

Subtracting $\theta$ in the two terms of (3.15) and using (3.17) one gets:

$$\begin{aligned}\tilde{\theta}(t+1) &= \tilde{\theta}(t) - F(t)\phi(t)\phi^T(t)\tilde{\theta}(t) = [I - F\phi(t)\phi^T(t)]\tilde{\theta}(t)\\ &= A(t)\tilde{\theta}(t)\end{aligned} \tag{3.18}$$

Equation (3.18) corresponds to a time-varying dynamical system. A necessary stability condition (but not sufficient) is that the eigen values of $A(t)$ be inside the unit circle at each instant $t$. This leads to the following condition for the choice of the adaptation gain as $F = \alpha I$:

$$\alpha < \frac{1}{\phi^T(t)\phi(t)} \tag{3.19}$$

## Improved Gradient Algorithm

In order to assure the stability of the PAA for any value of the adaptation gain $\alpha$ (or of the eigenvalues of the gain matrix $F$) the same gradient approach is used but a different criterion is considered:

$$\min_{\hat{\theta}(t+1)} J(t+1) = [\varepsilon(t+1)]^2 \tag{3.20}$$

The equation:

$$\frac{1}{2}\frac{\delta J(t+1)}{\delta\hat{\theta}(t+1)} = \frac{\delta\varepsilon(t+1)}{\delta\hat{\theta}(t+1)}\varepsilon(t+1) \tag{3.21}$$

is then obtained. From (3.6) and (3.8) it results that:

$$\varepsilon(t+1) = y(t+1) - \hat{y}(t+1) = y(t+1) - \hat{\theta}^T(t+1)\phi(t) \tag{3.22}$$

and, respectively:

$$\frac{\delta\varepsilon(t+1)}{\delta\hat{\theta}(t+1)} = -\phi(t) \tag{3.23}$$

Introducing (3.23) in (3.21), the parameter adaptation algorithm of (3.11) becomes:

$$\boxed{\hat{\theta}(t+1) = \hat{\theta}(t) + F\phi(t)\varepsilon(t+1)} \tag{3.24}$$

This algorithm depends on $\varepsilon(t+1)$, which is a function of $\hat{\theta}(t+1)$. For implementing this algorithm, $\varepsilon(t+1)$ must be expressed as a function of $\varepsilon^0(t+1)$, i.e. $\varepsilon(t+1) = f[\hat{\theta}(t), \phi(t), \varepsilon^0(t+1)]$.

Equation (3.22) can be rewritten as:

$$\varepsilon(t+1) = y(t+1) - \hat{\theta}^T(t)\phi(t) - [(\hat{\theta}(t+1) - \hat{\theta}(t)]^T\phi(t) \tag{3.25}$$

The first two terms of the right hand side correspond to $\varepsilon^0(t+1)$, and from (3.24), one obtains:

$$\hat{\theta}(t+1) - \hat{\theta}(t) = F\phi(t)\varepsilon(t+1) \tag{3.26}$$

which enables to rewrite (3.25) as:

$$\varepsilon(t+1) = \varepsilon^0(t+1) - \phi^T(t)F\phi(t)\varepsilon(t+1) \tag{3.27}$$

from which the desired relation between $\varepsilon(t+1)$ and $\varepsilon^0(t+1)$ is obtained:

$$\boxed{\varepsilon(t+1) = \frac{\varepsilon^0(t+1)}{1+\phi^T(t)F\phi(t)}} \tag{3.28}$$

and the algorithm of (3.24) becomes:

$$\boxed{\hat{\theta}(t+1) = \hat{\theta}(t) + \frac{F\phi(t)\varepsilon^0(t+1)}{1+\phi^T(t)F\phi(t)}} \tag{3.29}$$

which is a *stable algorithm* irrespective of the value of the gain matrix $F$ (positive definite). For a stability analysis see Sect. 3.3.1.

The division by $1 + \phi^T(t)F\phi(t)$ introduces a normalization with respect to $F$ and $\phi(t)$ which reduces the sensitivity of the algorithm with respect to $F$ and $\phi(t)$. In this case, the equation for the evolution of the parametric error $\hat{\theta}(t)$ is given by:

$$\tilde{\theta}(t+1) = \left[I - \frac{F\phi(t)\phi^T(t)}{1+\phi^T(t)F\phi(t)}\right]\tilde{\theta}(t) = A(t)\tilde{\theta}^T(t) \tag{3.30}$$

and the eigenvalues of $A(t)$ will always be inside or on the unit circle, but this is not enough to conclude upon the stability of the algorithm.

### *3.2.2 Recursive Least Squares Algorithm*

When using the Gradient Algorithm, $\varepsilon^2(t+1)$ is minimized at each step or, to be more precise, one moves in the quickest decreasing direction of the criterion, with a step depending on $F$. The minimization of $\varepsilon^2(t+1)$ at each step does not necessarily lead to the minimization of:

$$\sum_{i=1}^{t} \varepsilon^2(i+1)$$

on a time horizon, as is illustrated in Fig. 3.4. In fact, in the vicinity of the optimum, if the gain is not low enough, oscillations may occur around the minimum. On the other hand, in order to obtain a satisfactory convergence speed at the beginning when the optimum is far away, a high adaptation gain is preferable. In fact, the least squares algorithm offers such a variation profile for the adaptation gain. The same equations as in the gradient algorithm are considered for the plant, the prediction model, and the prediction errors, namely (3.1) through (3.8).

**Fig. 3.4** Evolution of an adaptation algorithm of the gradient type

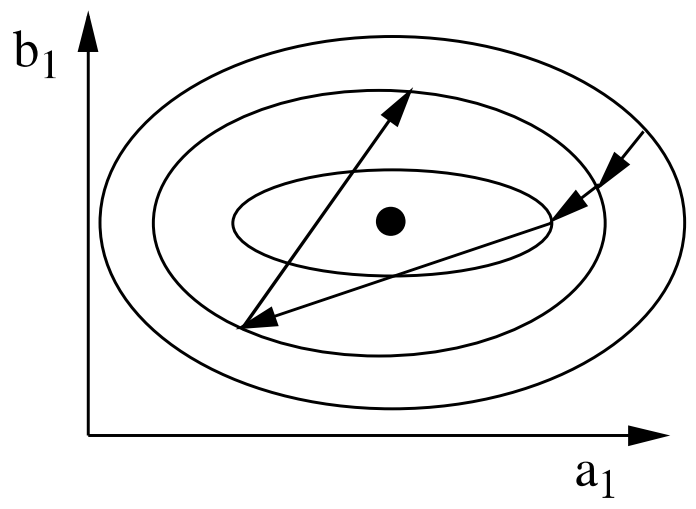

The aim is to find a recursive algorithm of the form of (3.9) which minimizes the *least squares* criterion:

$$\min_{\hat{\theta}(t)} J(t) = \sum_{i=1}^{t} [y(i) - \hat{\theta}^T(t)\phi(i-1)]^2 \tag{3.31}$$

The term $\hat{\theta}(t)^T\phi(i-1)$ corresponds to:

$$\hat{\theta}^T(t)\phi(i-1) = -\hat{a}_1(t)y(i-1) + \hat{b}_1(t)u(i-1) = \hat{y}[i \mid \hat{\theta}(t)] \tag{3.32}$$

Therefore, this is the prediction of the output at instant $i (i \leq t)$ based on the parameter estimate at instant $t$ obtained using $t$ measurements.

First, a parameter $\theta$ must be estimated at instant $t$ so that it minimizes the sum of the squares of the differences between the output of the plant and the output of the prediction model on a horizon of $t$ measurements. The value of $\hat{\theta}(t)$, which minimizes the criterion (3.31), is obtained by seeking the value that cancels $\delta J(t)/\delta\hat{\theta}(t)$:

$$\frac{\delta J(t)}{\delta \hat{\theta}(t)} = -2\sum_{i=1}^{t} [y(i) - \hat{\theta}^T(t)\phi(i-1)]\phi(i-1) = 0 \tag{3.33}$$

From (3.33), taking into account that:

$$[\hat{\theta}^T(t)\phi(i-1)]\phi(i-1) = \phi(i-1)\phi^T(i-1)\hat{\theta}(t)$$

one obtains:

$$\left[\sum_{i=1}^{t}\phi(i-1)]\phi^T(i-1)\right]\hat{\theta}(t)=\sum_{i=1}^{t}y(i)\phi(i-1)$$

and left multiplying by:[2]

$$\left[\sum_{i=1}^{t}\phi(i-1)\phi^T(i-1)\right]^{-1}$$

one obtains:

$$\hat{\theta}(t)=\left[\sum_{i=1}^{t}\phi(i-1)\phi^T(i-1)\right]^{-1}\sum_{i=1}^{t}y(i)\phi(i-1)=F(t)\sum_{i=1}^{t}y(i)\phi(i-1) \tag{3.34}$$

in which:

$$F(t)^{-1}=\sum_{i=1}^{t}\phi(i-1)\phi^T(i-1) \tag{3.35}$$

This estimation algorithm is not recursive. In order to obtain a recursive algorithm, the estimation of $\hat{\theta}(t+1)$ is considered:

$$\hat{\theta}(t+1)=F(t+1)\sum_{i=1}^{t+1}y(i)\phi(i-1) \tag{3.36}$$

$$F(t+1)^{-1}=\sum_{i=1}^{t+1}\phi(i-1)\phi^T(i-1)=F(t)^{-1}+\phi(t)\phi^T(t) \tag{3.37}$$

We can now express $\hat{\theta}(t+1)$ as a function of $\hat{\theta}(t)$:

$$\hat{\theta}(t+1)=\hat{\theta}(t)+\Delta\hat{\theta}(t+1) \tag{3.38}$$

From (3.36) one has:

$$\hat{\theta}(t+1)=F(t+1)\left[\sum_{i=1}^{t}y(i)\phi(i-1)+y(t+1)\phi(t)\right] \tag{3.39}$$

Taking into account (3.34), (3.39) can be rewritten as:

$$\hat{\theta}(t+1)=F(t+1)[F(t)^{-1}\hat{\theta}(t)+y(t+1)\phi(t)] \tag{3.40}$$

[2] It is assumed that the matrix $\sum_{i=1}^{t}\phi(i-1)\phi^T(i-1)$ is invertible. As it will be shown later this corresponds to an *excitation* condition.

From (3.37) after post-multiplying both sides by $\hat{\theta}(t)$ one gets:

$$F(t)^{-1}\hat{\theta}(t) = F(t+1)^{-1}\hat{\theta}(t) - \phi(t)\phi^T(t)\hat{\theta}(t) \tag{3.41}$$

and (3.40), becomes:

$$\hat{\theta}(t+1) = F(t+1)\left\{F(t+1)^{-1}\hat{\theta}(t) + \phi(t)[y(t+1) - \hat{\theta}^T(t)\phi(t)]\right\} \tag{3.42}$$

Taking into account the expression of $\varepsilon^0(t+1)$ given by (3.13), the result is:

$$\boxed{\hat{\theta}(t+1) = \hat{\theta}(t) + F(t+1)\phi(t)\varepsilon^0(t+1)} \tag{3.43}$$

The adaptation algorithm of (3.43) has a recursive form similar to the gradient algorithm given in (3.15) except that the gain matrix $F(t+1)$ is now time-varying since it depends on the measurements (it automatically corrects the gradient direction and the step length). A recursive formula for $F(t+1)$ remains to be given from the recursive formula $F(t+1)^{-1}$ given in (3.37). This is obtained by using the *matrix inversion* lemma.

**Lemma 3.1** (Matrix Inversion Lemma) *Let* $F$ *be a* $(n \times n)$ *dimensional nonsingular matrix*, $R$ *a* $(m \times m)$ *dimensional nonsingular matrix and* $H$ *a* $(n \times m)$ *dimensional matrix of maximum rank*, *then the following identity holds*:

$$(F^{-1} + HR^{-1}H^T)^{-1} = F - FH(R + H^T FH)^{-1}H^T F \tag{3.44}$$

*Proof* By direct multiplication one finds that:

$$[F - FH(R + H^T FH)^{-1}H^T F][F^{-1} + HR^{-1}H^T] = I$$

□

For the case of (3.37), one chooses $H = \phi(t)$, $R = 1$ and one obtains from (3.37) and (3.44):

$$F(t+1) = F(t) - \frac{F(t)\phi(t)\phi^T(t)F(t)}{1 + \phi^T(t)F(t)\phi(t)} \tag{3.45}$$

and, putting together the different equations, a first formulation of the recursive least squares (RLS) parameter adaptation algorithm (PAA) is given below:

$$\boxed{\hat{\theta}(t+1) = \hat{\theta}(t) + F(t+1)\phi(t)\varepsilon^0(t+1)} \tag{3.46}$$

$$\boxed{F(t+1) = F(t) - \frac{F(t)\phi(t)\phi^T(t)F(t)}{1 + \phi^T(t)F(t)\phi(t)}} \tag{3.47}$$

$$\boxed{\varepsilon^0(t+1) = y(t+1) - \hat{\theta}^T(t)\phi(t)} \tag{3.48}$$

An equivalent form of this algorithm is obtained by introducing the expression of $F(t+1)$ given by (3.47) in (3.46), where:

$$\begin{aligned}[\hat{\theta}(t+1)-\hat{\theta}(t)] &= F(t+1)\phi(t)\varepsilon^0(t+1) \\ &= F(t)\phi(t)\frac{\varepsilon^0(t+1)}{1+\phi^T(t)F(t)\phi(t)}\end{aligned} \tag{3.49}$$

However, from (3.7), (3.8) and (3.49), one obtains:

$$\begin{aligned}\varepsilon(t+1) &= y(t+1)-\hat{\theta}^T(t+1)\phi(t) \\ &= y(t+1)-\hat{\theta}(t)\phi(t)-[\hat{\theta}(t+1)-\hat{\theta}(t)]^T\phi(t) \\ &= \varepsilon^0(t+1)-\phi^T(t)F(t)\phi(t)\frac{\varepsilon^0(t+1)}{1+\phi^T(t)F(t)\phi(t)} \\ &= \frac{\varepsilon^0(t+1)}{1+\phi^T(t)F(t)\phi(t)}\end{aligned} \tag{3.50}$$

which expresses the relation between the a posteriori prediction error and the a priori prediction error. Using this relation in (3.49), an equivalent form of the parameter adaptation algorithm for the recursive least squares is obtained:

$$\boxed{\hat{\theta}(t+1)=\hat{\theta}(t)+F(t)\phi(t)\varepsilon(t+1)} \tag{3.51}$$

$$\boxed{F(t+1)^{-1}=F(t)^{-1}+\phi(t)\phi^T(t)} \tag{3.52}$$

$$\boxed{F(t+1)=F(t)-\frac{F(t)\phi(t)\phi^T(t)F(t)}{1+\phi^T(t)F(t)\phi(t)}} \tag{3.53}$$

$$\boxed{\varepsilon(t+1)=\frac{y(t+1)-\hat{\theta}^T(t)\phi(t)}{1+\phi^T(t)F(t)\phi(t)}} \tag{3.54}$$

For the recursive least squares algorithm to be exactly equivalent to the non-recursive least squares algorithm, it must be started from a first estimation obtained at instant $t_0=\dim\phi(t)$, since normally $F(t)^{-1}$ given by (3.35) becomes nonsingular for $t<t_0$. In practice, the algorithm is started up at $t=0$ by choosing:

$$F(0)=\frac{1}{\delta}I=(GI)I; \quad 0<\delta\ll 1 \tag{3.55}$$

a typical value being $\delta=0.001$ ($GI=1000$). It can be observed in the expression of $F(t+1)^{-1}$ given by (3.37) that the influence of this initial error decreases with the time. In this case one minimizes the following criterion:

$$\min_{\hat{\theta}(t)} J(t)=\sum_{i=1}^{t}[y(i)-\hat{\theta}^T(t)\phi(i-1)]^2+[\theta-\hat{\theta}^T(0)]^T F(0)^{-1}[\theta-\hat{\theta}(0)]^T \tag{3.56}$$

A rigorous analysis (based on the stability theory—see Sect. 3.3) shows nevertheless that for any positive definite matrix $F(0)[F(0) > 0]$,

$$\lim_{t \to \infty} \varepsilon(t+1) = 0$$

The recursive least squares algorithm is an algorithm with a decreasing adaptation gain. This is clearly seen if the estimation of a single parameter is considered. In this case, $F(t)$ and $\phi(t)$ are scalars, and (3.53) becomes:

$$F(t+1) = \frac{F(t)}{1 + \phi(t)^2 F(t)} \leq F(t); \quad \phi(t), F(t) \in R^1$$

The same conclusion is obtained observing that $F(t+1)^{-1}$ is the output of an integrator which has as input $\phi(t)\phi^T(t)$. Since $\phi(t)\phi^T(t) \geq 0$, one conclude that if $\phi(t)\phi^T(t) > 0$ in the average, then $F(t)^{-1}$ will tends towards infinity, i.e., $F(t)$ will tends towards zero.

The recursive least squares algorithm in fact gives less and less weight to the new prediction errors and thus to the new measurements. Consequently, this type of variation of the adaptation gain is not suitable for the estimation of time-varying parameters, and other variation profiles for the adaptation gain must therefore be considered. Under certain conditions (see Sect. 3.2.4), the adaptation gain is a measure of the evolution of the covariance of the parameter estimation error.

The least squares algorithm presented up to now for $\hat{\theta}(t)$ and $\phi(t)$ of dimension 2 may be generalized for any dimensions resulting from the description of discrete-time systems of the form:

$$y(t) = \frac{q^{-d} B(q^{-1})}{A(q^{-1})} u(t) \tag{3.57}$$

where:

$$A(q^{-1}) = 1 + a_1 q^{-1} + \cdots + a_{n_A} q^{-n_A} \tag{3.58}$$

$$B(q^{-1}) = b_1 q^{-1} + \cdots + b_{n_B} q^{-n_B} \tag{3.59}$$

Equation (3.57) can be written in the form:

$$y(t+1) = -\sum_{i=1}^{n_A} a_i y(t+1-i) + \sum_{i=1}^{n_B} b_i u(t-d-i+1) = \theta^T \phi(t) \tag{3.60}$$

in which:

$$\theta^T = [a_1, \ldots, a_{n_A}, b_1, \ldots, b_{n_B}] \tag{3.61}$$

$$\phi^T(t) = [-y(t), \ldots, -y(t-n_A+1), u(t-d), \ldots, u(t-d-n_B+1)] \tag{3.62}$$

The a priori adjustable predictor is given in the general case by:

$$\hat{y}^0(t+1) = -\sum_{i=1}^{n_A} \hat{a}_i(t)y(t+1-i) + \sum_{i=1}^{n_B} \hat{b}_1(t)u(t-d-i+1)$$
$$= \hat{\theta}^T(t)\phi(t) \quad (3.63)$$

in which:

$$\hat{\theta}^T(t) = [\hat{a}_1(t), \ldots, \hat{a}_{n_A}(t), \hat{b}_1(t), \ldots, \hat{b}_{n_B}(t)] \quad (3.64)$$

and for the estimation of $\hat{\theta}(t)$, the algorithm given in (3.51) through (3.54) (or in (3.46) through (3.48)) is used, with the appropriate dimension for $\hat{\theta}(t)$, $\phi(t)$ and $F(t)$.

### 3.2.3 Choice of the Adaptation Gain

The recursive formula for the inverse of the adaptation gain $F(t+1)^{-1}$ given by (3.52) is generalized by introducing two weighting sequences $\lambda_1(t)$ and $\lambda_2(t)$, as indicated below:

$$F(t+1)^{-1} = \lambda_1(t)F(t)^{-1} + \lambda_2(t)\phi(t)\phi^T(t)$$
$$0 < \lambda_1(t) \leq 1; \; 0 \leq \lambda_2(t) < 2; \; F(0) > 0 \quad (3.65)$$

Note that $\lambda_1(t)$ and $\lambda_2(t)$ in (3.65) have the opposite effect. $\lambda_1(t) < 1$ tends to increase the adaptation gain (the gain inverse decreases); $\lambda_2(t) > 0$ tends to decrease the adaptation gain (the gain inverse increases). For each choice of sequences, $\lambda_1(t)$ and $\lambda_2(t)$ corresponds a *variation profile* of the adaptation gain and an interpretation in terms of the error criterion, which is minimized by the PAA. Equation (3.65) allows to interpret the inverse of the adaptation gain (for constant weighting sequences) as the output of a filter $\lambda_2/(1-\lambda_1 q^{-1})$ having as an input $\phi(t)\phi^T(t)$ and as an initial condition $F(0)^{-1}$.

Using the *matrix inversion lemma* given by (3.44), one obtains from (3.65):

$$F(t+1) = \frac{1}{\lambda_1(t)}\left[F(t) - \frac{F(t)\phi(t)\phi^T(t)F(t)}{\frac{\lambda_1(t)}{\lambda_2(t)} + \phi^T(t)F(t)\phi(t)}\right] \quad (3.66)$$

Next, a certain number of choices for $\lambda_1(t)$ and $\lambda_2(t)$ and their interpretations will be given.

**A.1: Decreasing (Vanishing) Gain (RLS)** In this case:

$$\lambda_1(t) = \lambda_1 = 1; \qquad \lambda_2(t) = 1 \quad (3.67)$$

and $F(t+1)^{-1}$ is given by (3.52), which leads to a decreasing adaptation gain. The minimized criterion is that of (3.31). This type of profile is suited to the estimation of the parameters of stationary systems.

**A.2: Constant Forgetting Factor** In this case:

$$\lambda_1(t) = \lambda_1; \quad 0 < \lambda_1 < 1; \qquad \lambda_2(t) = \lambda_2 = 1 \tag{3.68}$$

The typical values for $\lambda_1$ are:

$$\lambda_1 = 0.95 \text{ to } 0.99$$

The criterion to be minimized will be:

$$J(t) = \sum_{i=1}^{t} \lambda_1^{(t-i)} [y(i) - \hat{\theta}^T(t)\phi(i-1)]^2 \tag{3.69}$$

The effect of $\lambda_1(t) < 1$ is to introduce increasingly weaker weighting on the old data $(i < t)$. This is why $\lambda_1$ is known as the *forgetting factor*. The maximum weight is given to the most recent error.

This type of profile is suited to the estimation of the parameters of slowly time-varying systems.

The use of a constant forgetting factor without the monitoring of the maximum value of $F(t)$ causes problems in adaptive regulation if the $\{\phi(t)\phi^T(t)\}$ sequence becomes null in the average (steady state case) because the adaptation gain will tend towards infinity. In this case:

$$F(t+i)^{-1} = (\lambda_1)^i F(t)^{-1}$$

and:

$$F(t+i) = (\lambda_1)^{-i} F(t)$$

For: $\lambda_1 < 1$, $\lim_{i \to \infty} (\lambda_1)^{-i} = \infty$ and $F(t+i)$ will become asymptotically unbounded.

**A.3: Variable Forgetting Factor** In this case:

$$\lambda_2(t) = \lambda_2 = 1 \tag{3.70}$$

and the forgetting factor $\lambda_1(t)$ is given by:

$$\lambda_1(t) = \lambda_0 \lambda_1(t-1) + 1 - \lambda_0; \quad 0 < \lambda_0 < 1 \tag{3.71}$$

The typical values being:

$$\lambda_1(0) = 0.95 \text{ to } 0.99; \qquad \lambda_0 = 0.5 \text{ to } 0.99$$

($\lambda_1(t)$ can be interpreted as the output of a first order filter $(1-\lambda_0)/(1-\lambda_0 q^{-1})$ with a unitary steady state gain and an initial condition $\lambda_1(o)$.)

Relation (3.71) leads to a forgetting factor that asymptotically tends towards 1. The criterion minimized will be:

$$J(t)=\sum_{i=1}^{t}\left[\prod_{j=1}^{t}\lambda_1(j-i)\right][y(i)-\hat{\theta}^T(t)\phi(i-1)]^2 \tag{3.72}$$

As $\lambda_1$ tends towards 1 for large $i$, only the initial data are forgotten (the adaptation gain tends towards a decreasing gain).

This type of profile is highly recommended for the model identification of stationary systems, since it avoids a too rapid decrease of the adaptation gain, thus generally resulting in an acceleration of the convergence (by maintaining a high gain at the beginning when the estimates are at a great distance from the optimum).

Other types of evolution for $\lambda_1(t)$ can be considered. For example:

$$\lambda_1(t)=1-\frac{\phi^T(t)F(t)\phi(t)}{1+\phi^T(t)F(t)\phi(t)}$$

This forgetting factor depends upon the input/output signals via $\phi(t)$. It automatically takes the value 1 if the norm of $\phi(t)\phi^T(t)$ becomes null. In the cases where the $\phi(t)$ sequence is such that the term $\phi^T(t)F(t)\phi(t)$ is significative with respect to one, the forgetting factor takes a lower value assuring good adaptation capabilities (this is related to the concept of "persistently exciting" signal—see Sect. 3.4).

Another possible choice is:

$$\lambda_1(t)=1-\alpha\frac{[\varepsilon^0(t)]^2}{1+\phi^T(t)F(t)\phi(t)};\quad \alpha>0$$

The forgetting factor tends towards 1, when the prediction error tends towards zero. Conversely, when a change occurs in the system parameters, the prediction error increases leading to a forgetting factor less than 1 in order to assure a good adaptation capability.

**A4: Constant Trace** In this case, $\lambda_1(t)$ and $\lambda_2(t)$ are automatically chosen at each step in order to ensure a constant trace of the gain matrix (constant sum of the diagonal terms):

$$\operatorname{tr} F(t+1)=\operatorname{tr} F(t)=\operatorname{tr} F(0)=nGI \tag{3.73}$$

in which $n$ is the number of parameters and $GI$ the initial gain (typical values: $GI=0.1$ to 4), the matrix $F(0)$ having the form:

$$F(0)=\begin{bmatrix} GI & & 0\\ & \ddots & \\ 0 & & GI\end{bmatrix} \tag{3.74}$$

The minimized criterion is of the form:

$$J(t) = \sum_{i=1}^{t} \left[ \prod_{j=1}^{t} \lambda_1(j-i) \right] \mu(i-1) \left[ y(i) - \hat{\theta}^T(t)\phi(i-1) \right]^2 \tag{3.75}$$

with:

$$\mu(t) = \frac{1 + \lambda_2(t)\phi^T(t)F(t)\phi(t)}{1 + \phi^T(t)F(t)\phi(t)} \tag{3.76}$$

Using this technique, at each step there is a movement in the optimal direction of the RLS, but the gain is maintained approximately constant. The value of $\lambda_1(t)$ and $\lambda_2(t)$ are determined from the equation:

$$\operatorname{tr} F(t+1) = \frac{1}{\lambda_1(t)} \operatorname{tr} \left[ F(t) - \frac{F(t)\phi(t)\phi^T(t)F(t)}{\alpha(t) + \phi^T(t)F(t)\phi(t)} \right] \tag{3.77}$$

fixing the ratio $\alpha(t) = \lambda_1(t)/\lambda_2(t)$ ((3.77) is obtained from (3.66)).

This type of profile is suited to the model identification of systems with time-varying parameters and for adaptive control with non-vanishing adaptation.

**A.5: Decreasing Gain + Constant Trace** In this case, A1 is switched to A4 when:

$$\operatorname{tr} F(t) \le nG; \quad G = 0.01 \text{ to } 4 \tag{3.78}$$

in which $G$ is chosen in advance. This profile is suited to the model identification of time-varying systems and for adaptive control in the absence of initial information on the parameters.

**A.6: Variable Forgetting Factor + Constant Trace** In this case A3 is switched to A4 when:

$$\operatorname{tr} F(t) \le nG \tag{3.79}$$

The use is the same as for A5.

**A.7: Constant Gain (Gradient Algorithm)** In this case:

$$\lambda_1(t) = \lambda_1 = 1; \qquad \lambda_2(t) = \lambda_2 = 0 \tag{3.80}$$

and thus from (3.66), one obtains:

$$F(t+1) = F(t) = F(0) \tag{3.81}$$

The improved gradient adaptation algorithm given by (3.24) or (3.29) is then obtained.

This algorithm can be used for the identification and adaptive control of stationary or time-varying systems with few parameters ($\le 3$) and in the presence of a reduced noise level. This type of adaptation gain results in performances that are inferior to those provided by the A1, A2, A3, and A4 profiles, but it is simpler to implement.

**Choice of the Initial Gain $F(0)$** The initial gain $F(0)$ is usually chosen as a diagonal matrix of the form given by (3.55) and, respectively, (3.74).

In the absence of initial information upon the parameters to be estimated (typical value of initial estimates = 0), a high initial gain ($GI$) is chosen for reasons that have been explained in Sect. 3.2.3 (3.56). A typical value is $GI = 1000$ (but higher values can be chosen).

If an initial parameter estimation is available (resulting for example from a previous identification), a low initial gain is chosen. In general, in this case $GI < 1$.

Since in standard RLS the adaptation gain decreases as the true model parameters are approached (a significant measurement is its trace), the adaptation gain may be interpreted as a measurement of the accuracy of the parameter estimation (or prediction). This explains the choices of $F(0)$ proposed above. Note that under certain hypotheses, $F(t)$ is effectively a measurement of the quality of the estimation (see Sect. 3.2.4).

This property can give indications upon the evolution of a parameter estimation procedure. If the trace of $F(t)$ did not decrease significantly, the parameter estimation is in general poor. This may occur in system identification when the level and type of input used are not appropriate. The importance of the input choice for parameter convergence will be discussed in Sect. 3.4.

### Parameter Adaptation Algorithms with Scalar Adaptation Gain

This concerns an extension of PAA with constant adaptation gain of the form $F = \alpha I$ for $\alpha > 1$ (see the improved gradient algorithm) for the cases where $\alpha(t) = \frac{1}{p(t)}$, i.e.,

$$F(t) = \alpha(t) I = \frac{1}{p(t)} I \tag{3.82}$$

Some PAA's of this type are mentioned next:

(1) Improved Gradient

$$p(t) = \text{const} = 1/\alpha > 0 \quad \Longrightarrow \quad F(t) = F(0) = \alpha I \tag{3.83}$$

(2) Stochastic Approximation

$$p(t) = t \quad \Longrightarrow \quad F(t) = \frac{1}{t} I \tag{3.84}$$

This is the simplest PAA with time decreasing adaptation gain (very useful for the analysis of PAA in the presence of stochastic disturbances).

(3) Controlled Adaptation Gain

$$p(t+1) = \lambda_1(t)p(t) + \lambda_2(t)\phi^T(t)\phi(t)$$
$$p(0) > 0;\ 0 < \lambda_1(t) \le 1;\ 0 \le \lambda_2(t) < 2 \tag{3.85}$$

There is a connection between the matrix adaptation gain updated using (3.65) and the adaptation gain using a $p(t)$ updated by (3.85). If:

$$p(0) = \operatorname{tr} F(0)^{-1} \quad \Longrightarrow \quad p(t) = \operatorname{tr} F(t)^{-1}$$

The principal interests in using these algorithms are:

- simpler implementation than those using a matrix adaptation gain updating;
- simplified analysis both in deterministic and stochastic environments.

Their disadvantage is that their performances are lower than those of the PAA using a matrix adaptation gain.

### *3.2.4 Recursive Least Squares and Kalman Filter*

Consider a dynamic system:

$$x(t+1) = Ax(t) + v(t) \tag{3.86}$$

$$y(t) = Cx(t) + e(t) \tag{3.87}$$

where $x(t)$ and $v(t)$ are $n$-dimensional vectors, $y$ and $e$ are scalars. $v(t)$ and $e(t)$ are independent Gaussian white noise processes characterized by:

$$\mathbf{E}\{v(t)\} = 0; \qquad \mathbf{E}\{v(t)v^T(t)\} = Q > 0 \tag{3.88}$$

$$\mathbf{E}\{e(t)\} = 0; \qquad \mathbf{E}\{e^2(t)\} = \eta^2 \tag{3.89}$$

For estimating the state $x(t)$ from the measurement of the output $y(t)$, one uses the Kalman filter (predictor):

$$\hat{x}(t+1) = A\hat{x}(t) + K(t+1)[y(t) - \hat{y}(t)] \tag{3.90}$$

$$\hat{y}(t) = C\hat{x}(t) \tag{3.91}$$

Let us define the state estimation error:

$$\tilde{x}(t) = \hat{x}(t) - x(t) \tag{3.92}$$

and its covariance:

$$\mathbf{E}\{\tilde{x}(t)\tilde{x}^T(t)\} = F(t) \tag{3.93}$$

The gain sequence $K(t+1)$ is computed in order to minimize the estimation error:

$$\min_{K(t+1)} l^T \mathbf{E}\{\tilde{x}(t+1)\tilde{x}^T(t+1)\}l; \quad l \in \mathcal{R}^n \tag{3.94}$$

The corresponding solution is (Åström 1970):

$$K(t+1)\,|_{opt} = AF(t)C^T[\eta^2 I + CF(t)C^T]^{-1} \tag{3.95}$$

Using the optimal gain sequence, the evolution of the covariance of the state estimation error is given:

$$F(t+1) = AF(t)A^T + Q - AF(t)C^T[\eta^2 I + CF(t)C^T]^{-1}CF(t)A \tag{3.96}$$

Consider now the problem of estimating a constant vector of parameters. This vector will be characterized by the following state equation:

$$\theta(t+1) = \theta(t) = \theta \tag{3.97}$$

Associated to this state equation consider an observation equation:

$$y(t+1) = \phi^T(t)\theta(t) + e(t+1) \tag{3.98}$$

where $y(t)$ corresponds to the output of the plant to be identified. The only difference with respect to (3.60) is the presence of the noise term $e$.

Let us particularize (3.86) and (3.87) for the case of the system defined by (3.97) and (3.98) i.e.:

$$\begin{aligned}
&x(t) = \theta(t); \qquad A = I; \qquad v(t) = 0; \qquad Q = 0; \qquad \hat{x}(t) = \hat{\theta}(t) \\
&y(t) = y(t+1); \qquad C = \phi^T(t); \qquad e(t) = e(t+1); \qquad \eta^2 = 1 \\
&\hat{y}(t) = \hat{y}^0(t+1)
\end{aligned}$$

(Note however that in this case $C$ is time-varying.)

The corresponding Kalman predictor associated to the (3.97) and (3.98) results from (3.90) and (3.91):

$$\hat{\theta}(t+1) = \hat{\theta}(t) + K(t+1)[y(t+1) - \hat{y}^0(t+1)] \tag{3.99}$$

$$\hat{y}^0(t+1) = \phi^T(t)\hat{\theta}(t) \tag{3.100}$$

The parameter estimation error will be given by:

$$\tilde{\theta}(t) = \hat{\theta}(t) - \theta(t) = \hat{\theta}(t) - \theta \tag{3.101}$$

and its covariance by:

$$\mathbf{E}\{\tilde{\theta}(t)\tilde{\theta}^T(t)\} = F(t) \tag{3.102}$$

The objective is to find $K(t+1)$ which minimizes:

$$\min_{K(t+1)} l^T \mathbf{E}\{\tilde{\theta}(t)\tilde{\theta}^T(t)\}l; \quad l \in \mathcal{R}^n \tag{3.103}$$

From (3.95) one gets:

$$K(t+1) = \frac{F(t)\phi(t)}{1 + \phi^T(t)F(t)\phi(t)} \tag{3.104}$$

and (3.99) will corresponds to the (3.51) of the recursive least squares ($y(t+1) - \hat{y}^0(t+1) = \varepsilon^0(t+1)$.

On the other hand, the evolution of the covariance of the parameter error will be given by:

$$\mathbf{E}\{\tilde{\theta}(t+1)\tilde{\theta}^T(t+1)\} = F(t+1) = F(t) - \frac{F(t)\phi(t)\phi^T(t)F(t)}{1 + \phi^T(t)F(t)\phi(t)} \tag{3.105}$$

which is nothing else than the updating equation for the adaptation gain in the case of the recursive least squares algorithm.

It results that *the adaptation gain in the case of recursive least squares is an image of the evolution of the covariance of the parameter error*, if $e(t+1)$ in (3.98) is a white noise with $\mathbf{E}\{e^2(t)\} = 1$.

Several policies for updating the adaptation gain when using the least squares predictor of (3.100) have interpretations in terms of a specific Kalman Predictor.

Consider for example the estimation of a time-varying parameter vector, with the associated equations:

$$\theta(t+1) = \alpha(t)\theta(t) + v(t) \tag{3.106}$$

$$y(t+1) = \phi^T(t)\theta(t) + e(t+1) \tag{3.107}$$

where $\alpha(t) \neq 0$ and $Q = \mathbf{E}\{v(t)v^T(t)\} \neq 0$ ($\alpha(t)$ and $v(t)$ causes variations of $\theta(t)$).

Provided that the system (3.106), (3.107) is uniformly observable,[3] the corresponding Kalman predictor takes the form:

$$\hat{\theta}(t+1) = \hat{\theta}(t) + \frac{\alpha(t)F(t)\phi(t)\varepsilon^0(t+1)}{\eta^2 + \phi^T(t)F(t)\phi(t)} \tag{3.108}$$

$$F(t+1) = \alpha^2(t)\left[F(t) - \frac{F(t)\phi(t)\phi^T(t)F(t)}{\eta^2 + \phi^T(t)F(t)\phi(t)} + Q\right] \tag{3.109}$$

Several interpretations are possible:

Case 1 $\alpha^2(t) = \frac{1}{\lambda_1(t)}$; $\eta^2(t) = \frac{\lambda_1(t)}{\lambda_2(t)}$; $Q = 0$

[3] This condition corresponds to "persistent excitation" condition on the observation vector $\phi(t)$. See Sect. 3.4.

One obtains the updating formula for $F(t+1)$ corresponding to the evolution of the adaptation gain given by (3.66):

$$F(t+1)^{-1} = \lambda_1(t)F(t)^{-1} + \lambda_2(t)\phi(t)\phi^T(t)$$

Since in general $\lambda_1(t)/\lambda_2(t)$ is chosen as a constant (see Sect. 3.2.3, the constant trace algorithm), then if $\lambda_1(t)/\lambda_2(t) = 1$, (3.108) corresponds to the standard PAA for $\hat{\theta}(t+1)$ with the approximation of $\alpha(t)$ term. However, in general $\lambda_1(t)$ and $\lambda_2(t)$ are close to one and $\alpha(t) = \frac{1}{\sqrt{\lambda_1(t)}}$ will also be close to one and the difference with respect to the formula issued from the Kalman Predictor is small. This rapprochement is valid also for the fixed or time-varying forgetting factor (in this case $\lambda_2(t) \equiv 1$).

Case 2 $Q \neq 0$; $\eta^2 = 1$; $\alpha(t) \equiv 1$

This corresponds to the problem of estimating a time-varying parameter with a constant average value. The magnitude of this variation will depend upon $Q = \mathbf{E}\{v(t)v^T(t)\}$. In order to be able to follow such parameter variations, it is necessary to add a constant term (in the expression of the adaptation gain given by (3.45)). If such a term is introduced only at certain instants, for reinitializing the value of the adaptation gain, this can be interpreted as a jump of $v(t)$ and therefore as a jump of the values of the parameters to be estimated.

This technique is currently used in practice for the reinitialization of the adaptation gain.

### *3.2.5 Some Remarks on the Parameter Adaptation Algorithms*

The parameter adaptation algorithms (PAA) presented up to now (integral type) have the form ($\lambda_1 = 1$; $\lambda_2 = 1$):

$$\hat{\theta}(t+1) = \hat{\theta}(t) + F(t+1)\phi(t)\varepsilon^0(t+1) = \hat{\theta}(t) + \frac{F(t)\phi(t)\varepsilon^0(t+1)}{1+\phi^T(t)F(t)\phi(t)}$$

where $\hat{\theta}(t)$ is the vector of estimated parameters and $F(t+1)\phi(t)\varepsilon^0(t+1)$ represents the correcting term at each sample.

$F(t)$ is the adaptation gain (constant or time-varying), $\phi(t)$ is the observation vector and $\varepsilon^0(t+1)$ is the a priori prediction error (or in general the adaptation error), i.e., the difference between the measured output at the instant $t+1$ and the predicted output at $t+1$ based on the knowledge of $\hat{\theta}(t)$.

There is always a relationship between the a priori prediction (adaptation) error $\varepsilon^0(t+1)$ and the a posteriori prediction (adaptation) error $\varepsilon(t+1)$ defined on the basis of the knowledge of $\hat{\theta}(t+1)$. This relation is:

$$\varepsilon(t+1) = \frac{\varepsilon^0(t+1)}{1+\phi^T(t)F(t)\phi(t)}$$

Several types of updating formulas for the adaptation gain can be used in connection with the type of the parameter estimation problem to be solved (systems with fixed or time-varying parameters), with or without availability of initial information for the parameter to be estimated.

The PAA examined up to now are based on:

- minimization of a one step ahead criterion;
- off-line minimization of a criterion over a finite horizon followed by a sequential minimization;
- sequential minimization of a criterion over a finite horizon starting at $t = 0$;
- relationship with the Kalman predictor.

However, from the point of view of real-time identification and adaptive control, the parameter adaptation algorithms are supposed to operate on a very large number of measurements ($t \to \infty$). Therefore, it is necessary to examine the properties of parameter adaptation algorithms as $t \to \infty$. Specifically, one should study the conditions which guarantee:

$$\lim_{t\to\infty} \varepsilon(t+1) = 0$$

This corresponds to the study of the stability of parameter adaptation algorithms. Conversely, other parameter adaptation algorithms will be derived from the stability condition given above.

## 3.3 Stability of Parameter Adaptation Algorithms

### *3.3.1 Equivalent Feedback Representation of the Parameter Adaptation Algorithms and the Stability Problem*

In the case of recursive least squares or of the improved gradient algorithm, the following a posteriori predictor has been used:

$$\hat{y}(t+1) = \hat{\theta}^T(t+1)\phi(t) \tag{3.110}$$

where

$$\hat{\theta}^T(t) = [-\hat{a}_1(t), \ldots, -\hat{a}_{n_A}(t), \hat{b}_1(t), \ldots, \hat{b}_{n_B}(t)]$$
$$\phi^T(t) = [-y(t), \ldots, -y(t-n_A+1), u(t-d), \ldots, u(t-d-n_B+1]$$

The PAA has the following form:

$$\hat{\theta}(t+1) = \hat{\theta}(t) + F(t)\phi(t)\varepsilon(t+1) \tag{3.111}$$

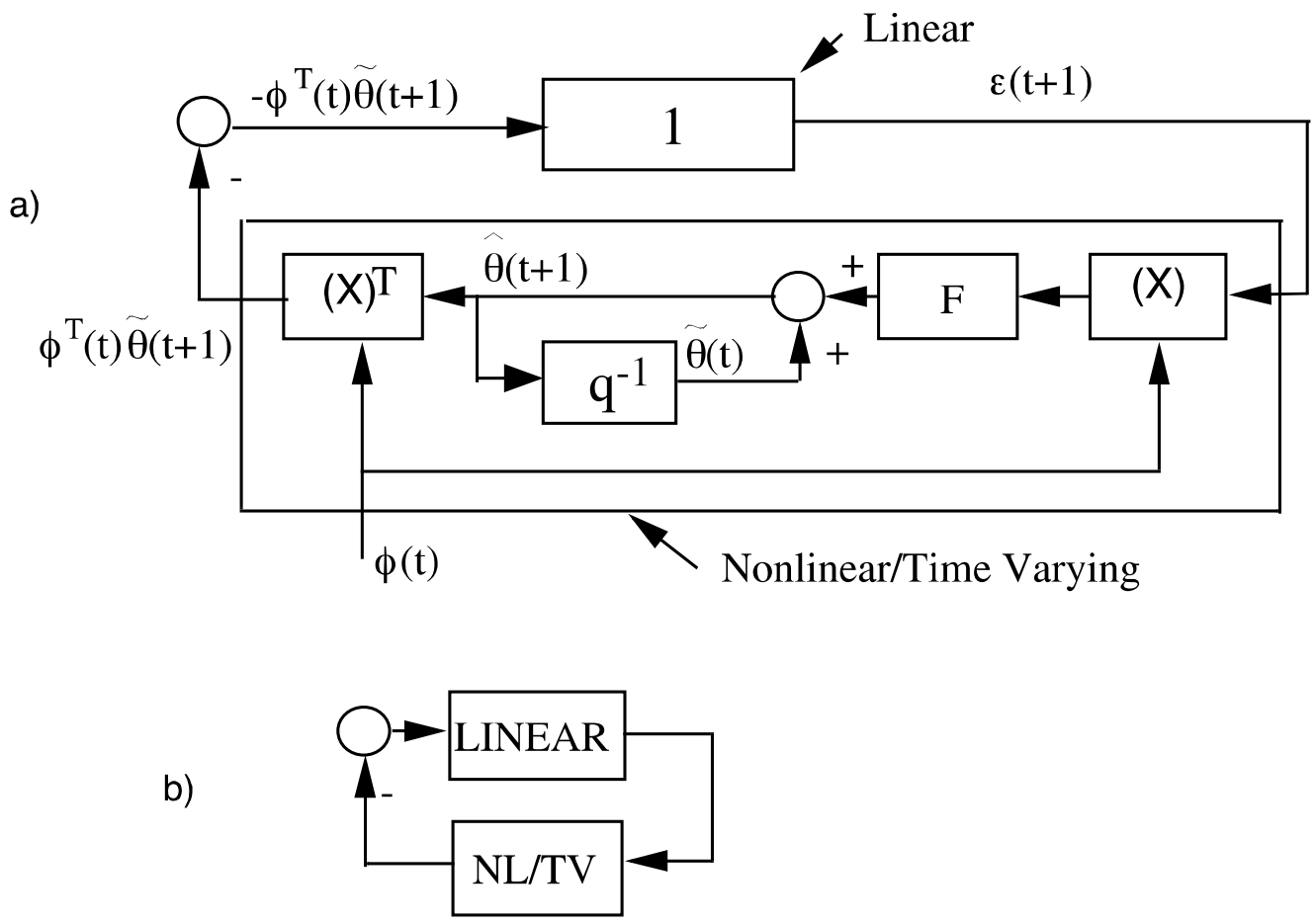

**Fig. 3.5** Equivalent feedback representation of PAA, (**a**) the case of RLS, (**b**) generic equivalent representation

The parameter error is defined as:

$$\tilde{\theta}(t) = \hat{\theta}(t) - \theta \tag{3.112}$$

Subtracting $\theta$ in both sides of (3.111) and, taking into account (3.112), one obtains:

$$\tilde{\theta}(t+1) = \tilde{\theta}(t) + F(t)\phi(t)\varepsilon(t+1) \tag{3.113}$$

From the definition of the a posteriori prediction error $\varepsilon(t+1)$ given by (3.8) and taking into account (3.60) and (3.112), one gets:

$$\varepsilon(t+1) = y(t+1) - \hat{y}(t+1) = \phi^T(t)\theta - \phi^T(t)\hat{\theta}(t+1) = -\phi^T(t)\tilde{\theta}(t+1) \tag{3.114}$$

and using (3.113), one can write:

$$\phi(t)\tilde{\theta}(t+1) = \phi^T(t)\tilde{\theta}(t) + \phi^T(t)F(t)\phi(t)\varepsilon(t+1) \tag{3.115}$$

Equations (3.113) through (3.115) defines an equivalent feedback system represented in Fig. 3.5. Equation (3.114) defines a linear block with constant parameters on the feedforward path, whose input is $-\phi^T(t)\tilde{\theta}(t+1)$. In the case of least squares predictor (known also as equation error predictor or series-parallel predictor) this block is characterized by a unitary gain. Equations (3.113) and (3.115) define a nonlinear time-varying block in the feedback path. For the particular case of estimation in open loop operation using recursive least squares this block is only time-varying, since $\phi(t)$ does not depend neither upon $\varepsilon$, nor upon $\hat{\theta}$.

The objective of a PAA in a deterministic environment (absence of noise), is to drive the estimated parameter vector $\hat{\theta}$ towards a value such that the a posteriori prediction error vanishes asymptotically, i.e., $\lim_{t\to\infty} \varepsilon(t+1) = 0$. This objective can be expressed as a condition for the asymptotic stability of the equivalent feedback system associated to the PAA.

It is assumed of course, that the structure chosen for the adjustable model is such that there is a value of the parameter vector which gives a perfect input-output description of the plant (i.e., the plant model and the adjustable predictor have the same structure).

Similar *equivalent feedback representation* (EFR) can be obtained for other PAA and adjustable predictors. In general, the linear feedforward block will be characterized by a transfer function and the feedback block will be time-varying and nonlinear.

Two types of problems can be considered:

*Analysis*: Given an adjustable predictor and a PAA, find the most general stability conditions.

*Synthesis*: For a given adjustable predictor find the largest family of PAA for which stability conditions can be established.

The equivalent feedback representation associated to PAA is always formed by two blocks (see Fig. 3.5b): a linear time-invariant block (LINEAR) and a nonlinear time-varying (NL/TV). It is therefore sensible to use specific stability analysis tools dedicated to this type of system.

Passivity (hyperstability) properties or Lyapunov functions with two (or several) terms are well suited for the stability analysis of feedback systems. The passivity (hyperstability) approach is more natural and systematic, but the same results can be obtained by using Lyapunov functions of particular form (for a discussion of the relationship between these two approaches see Brogliato et al. 1993).

The passivity approach concerns input-output properties of systems and the implications of these properties for the case of feedback interconnection.

We shall next present a pragmatic approach without formal generalized and proven results concerning the use of the passivity approach for the analysis and the synthesis of PAA. Detailed results can be found in Sects. 3.3.3 and 3.3.4 and the Appendix C.

The norm $L_2$ is defined as:

$$\|x(t)\|_2 = \left(\sum_0^{\infty} x^2(t)\right)^{1/2}$$

(it is assumed that all signals are 0 for $t < 0$).

To avoid the assumption that all signals go to zero as $t \to \infty$, one uses the so-called extended $L_2$ space denoted $L_2e$ which contains the *truncated* sequences:

$$x_T(t) = \begin{cases} x(t) & 0 \le t \le T \\ 0 & t > T \end{cases}$$

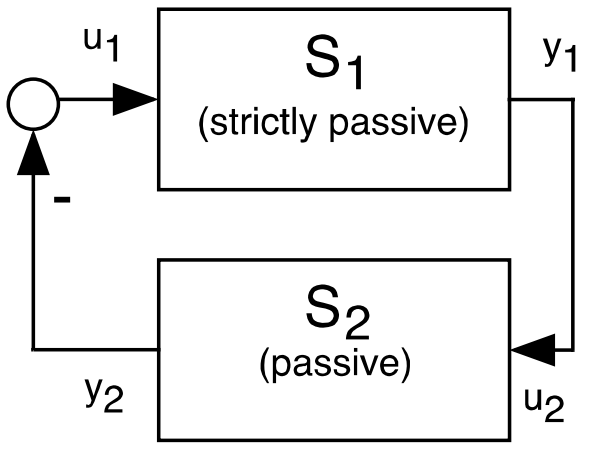

**Fig. 3.6** Feedback interconnection of two passive blocks

Consider a SISO system $S$ with input $u$ and output $y$. Let us define the input-output product:

$$\eta(0,t_1) = \sum_{t=0}^{t_1} u(t)y(t)$$

A system $S$ is termed *passive* if:

$$\eta(0,t_1) \geq -\gamma^2; \quad \gamma^2 < \infty; \ \forall t_1 \geq 0$$

A system $S$ is termed (*input*) *strictly passive* if:

$$\eta(0,t_1) \geq -\gamma^2 + \kappa\|u\|_2^2; \quad \gamma^2 < \infty; \ \kappa > 0; \ \forall t_1 \geq 0$$

and (*output*) *strictly passive* if:

$$\eta(0,t_1) \geq -\gamma^2 + \delta\|y\|_2^2; \quad \gamma^2 < \infty; \ \delta > 0; \ \forall t_1 \geq 0$$

Consider now the feedback interconnection between a block $S$ which is strictly passive (for example strictly input passive) and a block $S$ which is passive as illustrated in Fig. 3.6.

One has:

$$\eta_1(0,t_1) = \sum_0^{t_1} u_1(t)y_1(t) \geq -\gamma_1^2 + \kappa\|u\|_2^2; \quad \gamma_1^2 < \infty; \ \kappa > 0; \ \forall t_1 \geq 0$$

$$\eta_2(0,t_1) = \sum_0^{t_1} u_2(t)y_2(t) \geq -\gamma_2^2; \quad \gamma_2^2 < \infty; \ \forall t_1 \geq 0$$

The feedback connection corresponds to:

$$u_1 = -y_2; \qquad u_2 = y_1$$

Taking into account the feedback connection, it results that:

$$\eta_1(0,t_1) = \sum_0^{t_1} u_1(t)y_1(t) = -\sum_0^{t_1} u_2(t)y_2(t) = -\eta_2(0,t_1)$$

and therefore:

$$-\gamma_1^2 + \kappa \|u\|_2^2 \leq \sum_0^{t_1} u_1(t) y_1(t) \leq \gamma_2^2$$

from which one concludes:

$$\kappa \|u\|_2^2 \leq \gamma_1^2 + \gamma_2^2$$

i.e.,

$$\lim_{t_1 \to \infty} \sum_{t=0}^{t_1} u_1^2(t) \leq \gamma_1^2 + \gamma_2^2 < \infty$$

which implies that $u(t)$ is bounded and that:

$$\lim_{t \to \infty} u_1(t) = 0$$

If, in addition the block $S_1$ is a linear asymptotically stable system, then:

$$\lim_{t \to \infty} u_1(t) \to \lim_{t \to \infty} y_1(t) = 0$$

Let us see now how these results can be used for the stability analysis of the *improved gradient algorithm* (which corresponds to recursive least squares using a constant adaptation gain $F(t+1) = F(t) = F > 0$).

The equivalent feedback block is characterized by (3.113) and (3.115) with $F(t) = F$:

$$\tilde{\theta}(t+1) = \tilde{\theta}(t) + F\phi(t)\varepsilon(t+1) \tag{3.116}$$

$$\begin{aligned} y_2(t) &= \phi^T(t)\tilde{\theta}(t+1) = \phi^T(t)\tilde{\theta}(t) + \phi^T(t)F\phi(t)\varepsilon(t+1) \\ &= \phi(t)\tilde{\theta}(t) + \phi^T(t)F\phi(t)u_2(t) \end{aligned} \tag{3.117}$$

The input is the a posteriori prediction error $\varepsilon(t+1)$ and the output is $\phi^T(t)\tilde{\theta}(t+1)$.

If we want to apply the passivity approach, we should verify first that the feedback path is passive. Taking advantage of (3.116), one gets:

$$\begin{aligned} \sum_{t=0}^{t_1} y_2(t)u_2(t) &= \sum_{t=0}^{t_1} \tilde{\theta}^T(t+1)\phi(t)\varepsilon(t+1) \\ &= \sum_{t=0}^{t_1} \tilde{\theta}^T(t+1)F^{-1}[\tilde{\theta}(t+1) - \tilde{\theta}(t)] \\ &= \sum_{t=0}^{t_1} [\tilde{\theta}^T(t+1)F^{-1}\tilde{\theta}(t+1) - \tilde{\theta}^T(t+1)F^{-1}\tilde{\theta}(t)] \end{aligned} \tag{3.118}$$

Observe that:

$$\begin{aligned}
&[\tilde{\theta}(t+1)-\tilde{\theta}(t)]^T F^{-1}[\tilde{\theta}(t+1)-\tilde{\theta}(t)] \\
&\quad = \tilde{\theta}^T(t+1)F^{-1}\tilde{\theta}(t+1)+\tilde{\theta}^T(t)F^{-1}\tilde{\theta}(t)-2\tilde{\theta}^T(t+1)F^{-1}\tilde{\theta}(t) \\
&\quad \geq 0
\end{aligned} \tag{3.119}$$

and therefore:

$$-\tilde{\theta}^T(t+1)F^{-1}\tilde{\theta}(t) \geq -\frac{1}{2}[\tilde{\theta}^T(t+1)F^{-1}\tilde{\theta}(t+1)+\tilde{\theta}^T(t)F^{-1}\tilde{\theta}(t)] \tag{3.120}$$

Introducing this expression in (3.118), one obtains:

$$\begin{aligned}
\sum_{t=0}^{t_1} y_2(t)u_2(t) &\geq \frac{1}{2}\sum_{t=0}^{t_1}\tilde{\theta}^T(t+1)F^{-1}\tilde{\theta}(t+1)-\frac{1}{2}\sum_{t=0}^{t_1}\tilde{\theta}^T(t)F^{-1}\tilde{\theta}(t) \\
&= \frac{1}{2}\tilde{\theta}^T(t+1)F^{-1}\tilde{\theta}(t+1)-\frac{1}{2}\tilde{\theta}^T(0)F^{-1}\tilde{\theta}(0) \\
&\geq -\frac{1}{2}\tilde{\theta}^T(0)F^{-1}\tilde{\theta}(0) = -\gamma_2^2; \quad \gamma_2^2 < \infty
\end{aligned} \tag{3.121}$$

One concludes therefore that the equivalent feedback path is passive. Taking now into account the feedback connection and the corresponding linear path specified in (3.114), one gets:

$$\begin{aligned}
\sum_{t=0}^{t_1} y_1 u_1 &= \sum_{t=0}^{t_1}\varepsilon^2(t+1) \\
&= -\sum_{t=0}^{t_1} y_2 u_2 = \sum_{t=0}^{t_1} -\tilde{\theta}^T(t+1)\phi(t)\varepsilon(t+1) \\
&\leq \frac{1}{2}\tilde{\theta}^T(0)F^{-1}\tilde{\theta}(0)
\end{aligned} \tag{3.122}$$

and, therefore:

$$\lim_{t_1\to\infty}\sum_{t=0}^{t_1}\varepsilon^2(t+1) \leq \frac{1}{2}\tilde{\theta}^T(0)F^{-1}\tilde{\theta}(0) = \gamma_2^2 \tag{3.123}$$

from which one concludes that $\varepsilon(t+1)$ is bounded and:

$$\lim_{t_1\to\infty}\varepsilon(t+1) = 0 \tag{3.124}$$

i.e., the global asymptotic convergence to zero of the a posteriori prediction (adaptation) error for any finite initial condition on the parameter error and any adaptation gain $F > 0$.

Taking into account the relationship between the a priori and the a posteriori prediction error:

$$\varepsilon(t+1) = \frac{\varepsilon^0(t+1)}{1+\phi^T(t)F\phi(t)} \tag{3.125}$$

one also concludes that $\lim_{t\to\infty} \varepsilon(t+1) = 0$ implies $\lim_{t\to\infty} \varepsilon^0(t+1) = 0$ when $\phi(t)$ is bounded. In this example, $\phi(t)$ contains the inputs and the outputs of a system assumed to be stable and excited by an input assumed to be bounded and then, it will be bounded.

Therefore, the conclusion of this analysis is that the *improved gradient algorithm* is asymptotically stable for any finite value of the adaptation gain $F$, which is positive definite.

Similar equivalent feedback systems can be associated with various PAA and various types of adjustable predictors as it will be shown in the subsequent sections, where more general results will be given.

### 3.3.2 Stability Approach for the Synthesis of PAA Using the Equivalent Feedback Representation

In Sect. 3.2, we derived parameter adaptation algorithms either by the minimization at each step of an error criterion (for a given structure of the adjustable predictor) or by transforming an off-line parameter estimation procedure in a recursive parameter estimation algorithm. This second approach gave us the structure of the adjustable predictor and of the PAA. To analyze the behavior of the algorithms for $t \to \infty$, as well as the effect of various modifications (discussed in Sect. 3.2.3), a stability analysis has to be considered. This has been done for the particular case of a constant adaptation gain in Sect. 3.3.1.

In this section, we will take a different approach for obtaining parameter adaptation algorithms. We will first define the structure of the adjustable predictor and then we will synthesize a PAA such that the prediction (adaptation) error be globally asymptotically stable for any initial values of the parameter error and adaptation error, and for any finite values of the adaptation gain. The synthesis of an asymptotically stable PAA (in the sense of the adaptation error) will take advantage of the equivalent feedback structure (EFR), which can be associated with the PAA, and of the passivity concepts which are useful in this context.

As an example for illustrating this approach for the synthesis of PAA, we will discuss the synthesis of a so-called *output error adaptive predictor* (known also as *parallel model reference adaptive system*—Landau 1979).

#### Output Error Adaptive Predictor

As in Sect. 3.2, one considers that the process model is described by:

$$y(t+1) = -a_1 y(t) + b_1 u(t) = \theta^T(t)\varphi(t) \tag{3.126}$$

where:

$$\theta^T = [a_1, b_1]; \qquad \varphi^T(t) = [-y(t), u(t)] \tag{3.127}$$

The *output error adjustable predictor* is described by:

$$\hat{y}^0(t+1) = -\hat{a}_1(t)\hat{y}(t) + \hat{b}_1(t)u(t) = \hat{\theta}^T(t)\phi(t) \tag{3.128}$$

where $\hat{y}^0(t+1)$ is the a priori output of the predictor, and:

$$\hat{y}(t+1) = -\hat{a}_1(t+1)\hat{y}(t) + \hat{b}_1(t+1)u(t) = \hat{\theta}^T(t+1)\phi(t) \tag{3.129}$$

is the a posteriori output of the predictor.

$$\hat{\theta}^T(t) = [\hat{a}_1(t), \hat{b}_1(t)]; \qquad \phi^T(t) = [-\hat{y}(t), u(t)] \tag{3.130}$$

The difference with respect to the adjustable predictor used in RLS or in the "gradient algorithm" can be found in the replacement of the output measurement $y(t)$ at instant $t$ by $\hat{y}(t)$ which represents the a posteriori predicted output at instant $t$. Since $\hat{y}(t)$ should converge asymptotically to $y(t)$, $\hat{y}(t)$ is an approximation of the output $y(t)$ which will improve as the time passes.

The reason for using this type of adjustable predictor is that it can provide better performance when the plant output measurement is disturbed by noise. This can be explained by the fact that the output of this predictor does not directly depend upon the measured variables disturbed by noise. The output of the predictor depends indirectly upon the measurements through the adaptation algorithm but this dependence can decrease in time by using a decreasing adaptation gain. The fact that only the inputs and the previous outputs of the predictor are used, motivates the term *parallel* MRAS or *output error*. The predictor used in RLS is termed *series-parallel* (known also as *equation error*). The difference between the two types of predictors is illustrated in Fig. 3.7.

The a priori prediction error will have the expression:

$$\varepsilon^0(t+1) = y(t+1) - \hat{y}^0(t+1) = y(t+1) - \hat{\theta}^T(t)\phi(t) \tag{3.131}$$

and the a posteriori prediction error will be given by:

$$\varepsilon(t+1) = y(t+1) - \hat{y}(t+1) \tag{3.132}$$

The objective is now to write an equation for the a posteriori prediction error as a function of the parameter error. Observe first that by adding and subtracting the term $\pm a_1\hat{y}(t)$ to (3.126), the output of the plant can be expressed as:

$$y(t+1) = -a_1\hat{y}(t) + b_1u(t) - a_1\varepsilon(t) = -a_1\varepsilon(t) + \theta^T\phi(t) \tag{3.133}$$

Taking into account (3.129) and (3.133), (3.132) becomes:

$$\begin{aligned} \varepsilon(t+1) &= -a_1\varepsilon(t) + \phi^T(t)[\theta - \hat{\theta}(t+1)] \\ &= -a_1\varepsilon(t) - \phi^T(t)\tilde{\theta}(t+1) \end{aligned} \tag{3.134}$$

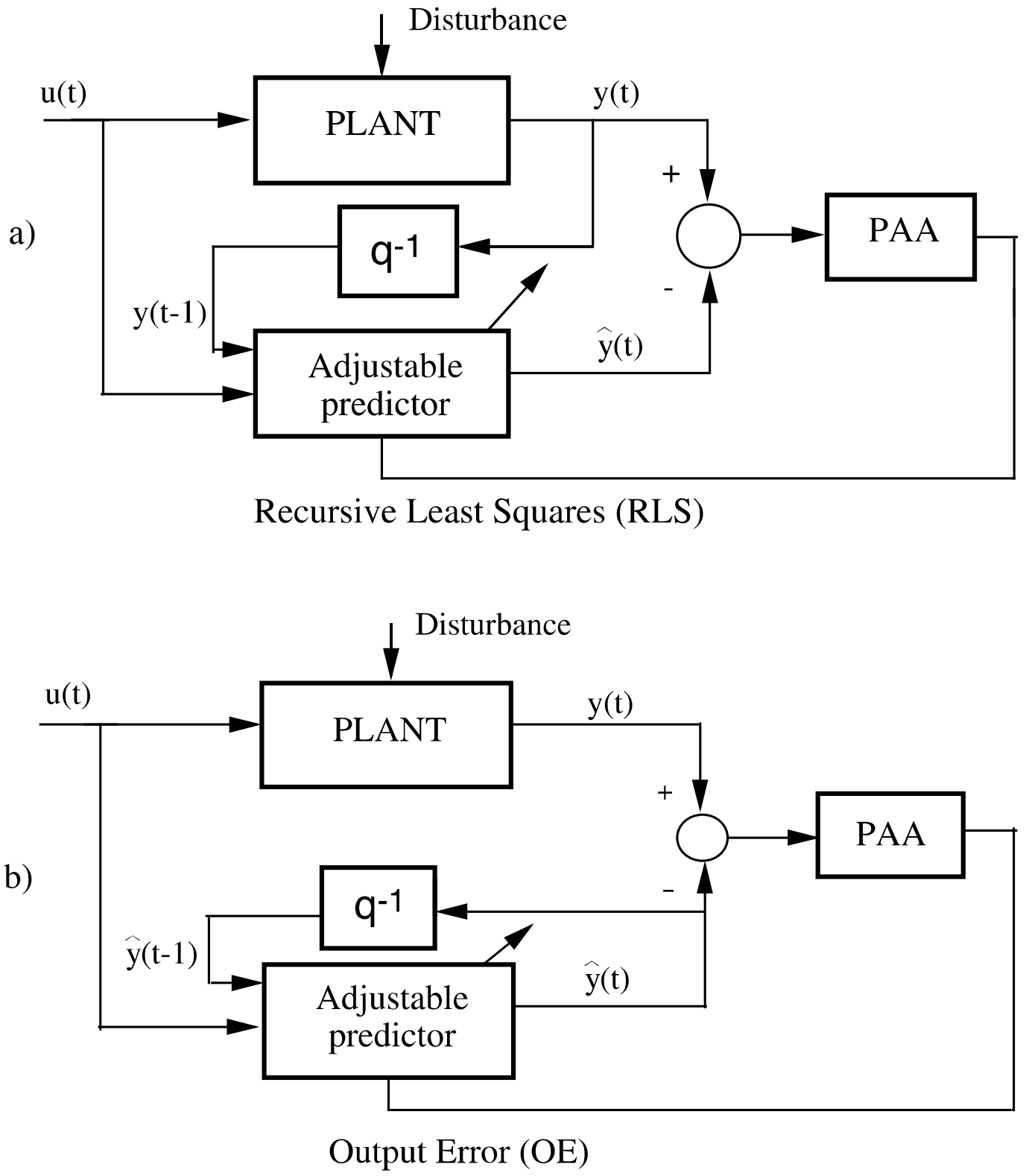

**Fig. 3.7** Comparison between two adjustable predictor structures, (**a**) recursive least squares (equation error), (**b**) output error

which can be rewritten as:

$$A(q^{-1})\varepsilon(t+1) = -\phi^T(t)\tilde{\theta}(t+1) \tag{3.135}$$

where:

$$A(q^{-1}) = 1 + a_1 q^{-1} = 1 + q^{-1}A^*(q^{-1}) \tag{3.136}$$

from which one obtains:

$$\varepsilon(t+1) = \frac{1}{A(q^{-1})}[-\phi^T(t)\tilde{\theta}(t+1)] \tag{3.137}$$

This result remains valid even for higher order predictors where $a_1$ is replaced by $A^*(q-1) = a_1 + a_2 q^{-1} + \cdots + a_{n_A} q^{-n_A}$. In other words, the a posteriori prediction error is the output of a linear block characterized by a transfer function $1/A(z^{-1})$, whose input is $-\phi^T(t)\tilde{\theta}(t+1)$.

Once the equation for the a posteriori prediction error has been derived, the PAA synthesis problem can be formulated as:

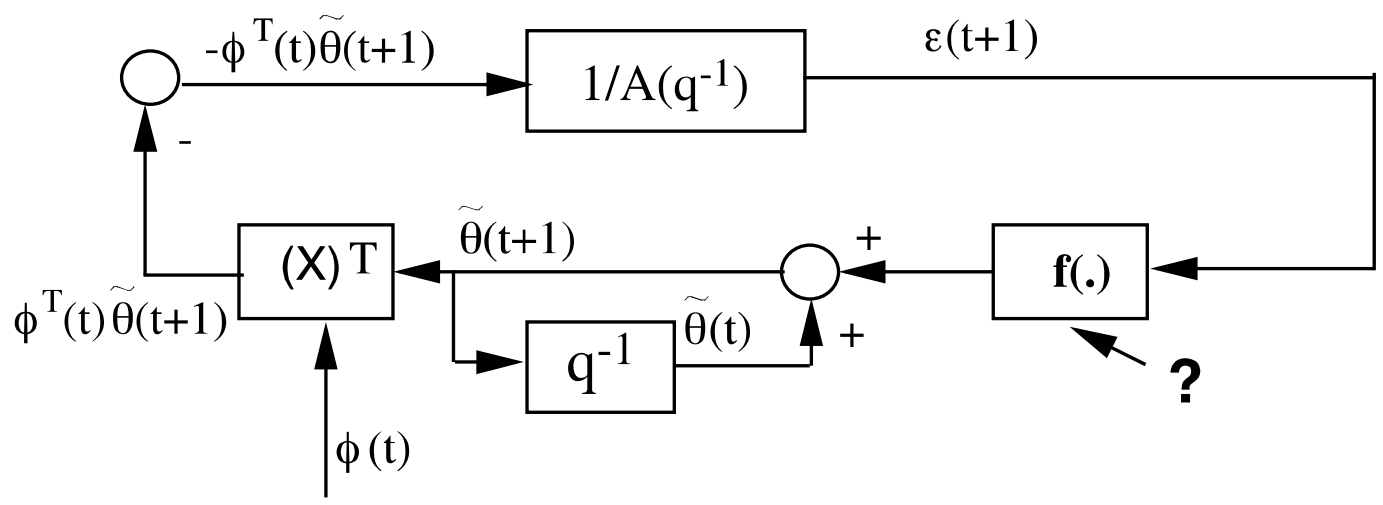

**Fig. 3.8** Equivalent feedback system associated to the output error predictor

*Find a PAA of the form*:

$$\hat{\theta}(t+1) = \hat{\theta}(t) + f_\theta[\phi(t), \hat{\theta}(t), \varepsilon(t+1)] \tag{3.138}$$

$$\varepsilon(t+1) = f_\varepsilon[\phi(t), \hat{\theta}(t), \varepsilon^0(t+1)] \tag{3.139}$$

*such that* $\lim_{t\to\infty} \varepsilon(t+1) = 0$ *for all initial conditions* $\varepsilon(0)$, $\hat{\theta}(0)$ (*or* $\tilde{\theta}(0)$).

Note that the structure of (3.138) assures the memory of the PAA (integral form), but other structures can be considered. The structure of (3.139) assures the causality of the algorithm.

From (3.138) subtracting $\theta$ from both sides, one gets:

$$\tilde{\theta}(t+1) = \tilde{\theta}(t) + f_\theta[\phi(t), \hat{\theta}(t), \varepsilon(t+1)] \tag{3.140}$$

and multiplying both sides by $\phi^T(t)$, yields:

$$\phi^T(t)\tilde{\theta}(t+1) = \phi^T(t)\tilde{\theta}(t) + \phi^T(t) f_\theta[\phi(t), \hat{\theta}(t), \varepsilon(t+1)] \tag{3.141}$$

Equations (3.137), (3.140) and (3.141) define an equivalent feedback system associated to the *output error* predictor as shown in Fig. 3.8.

Based on the passivity approach, our objective will be to first find $f_\theta(.)$ such that the equivalent feedback path be passive and then we will see under what conditions the feedforward path is strictly passive.

The passivity of the equivalent feedback path requires that:

$$\sum_{t=0}^{t_1} \tilde{\theta}^T(t+1)\phi(t)\varepsilon(t+1) \geq -\gamma_2^2; \quad \gamma_2^2 < \infty; \ \forall t_1 \geq 0 \tag{3.142}$$

On the other hand, from (3.118) through (3.121), one has the following basic results:

$$\sum_{t=0}^{t_1} \tilde{\theta}^T(t+1) F^{-1}[\tilde{\theta}(t+1) - \tilde{\theta}(t)] \geq -\frac{1}{2}\tilde{\theta}^T(0)F^{-1}\tilde{\theta}(0); \quad F > 0; \ \forall t_1 \geq 0 \tag{3.143}$$

The comparison of (3.142) and (3.143) suggests we take:

$$f_\theta[\phi(t), \hat{\theta}(t), \varepsilon(t+1)] = \tilde{\theta}(t+1) - \tilde{\theta}(t) = F\phi(t)\varepsilon(t+1) \tag{3.144}$$

and respectively:

$$\hat{\theta}(t+1) = \hat{\theta}(t) + F\phi(t)\varepsilon(t+1) \tag{3.145}$$

Note that in this PAA one uses in the observation vector $\phi(t)$, the previous a posteriori predicted output $\hat{y}(t), \hat{y}(t-1), \ldots$ instead of the previous output measurements $y(t), y(t-1), \ldots$ used in recursive least squares or improved gradient algorithm.

There are many other possibilities for the structure of (3.144) and (3.145) which lead to an equivalent passive feedback path (see subsequent sections).

To make the PAA implementable it is necessary to establish a relationship between $\varepsilon(t+1)$ and $\varepsilon^0(t+1)$, the last one being a measurable quantity at $t+1$. Subtracting (3.128) from (3.133), one gets:

$$\varepsilon^0(t+1) = [\theta - \hat{\theta}(t)]^T \phi(t) - a_1 \varepsilon(t) \tag{3.146}$$

and using also (3.134), one obtains:

$$\varepsilon(t+1) - \varepsilon^0(t+1) = -[\hat{\theta}(t+1) - \hat{\theta}(t)]^T \phi(t) \tag{3.147}$$

But, from (3.145) one has:

$$\hat{\theta}(t+1) - \hat{\theta}(t) = F\phi(t)\varepsilon(t+1) \tag{3.148}$$

and (3.147) becomes:

$$\varepsilon(t+1) + \phi^T(t)F\phi(t)\varepsilon(t+1) = \varepsilon^0(t+1) \tag{3.149}$$

which leads to the desired relationship between the a posteriori prediction error and the a priori prediction error:

$$\varepsilon(t+1) = \frac{\varepsilon^0(t+1)}{1 + \phi^T(t)F\phi(t)} \tag{3.150}$$

It now remains to be shown under what conditions the equivalent feedforward block characterized by the discrete-time transfer function:

$$H(z^{-1}) = \frac{1}{A(z^{-1})} \tag{3.151}$$

is strictly passive. To do this we will use the Parseval theorem which connects the time domain with the frequency domain in order to evaluate in the frequency domain the input-output properties of the equivalent feedforward path.

We will consider a truncated input sequence:

$$\bar{u}_1(t) = \begin{cases} u_1(t) & 0 \le t \le t_1 \\ 0 & t > t_1 \end{cases} \tag{3.152}$$

With this assumption, the $z$ transform of $\bar{u}_1(t)$:

$$\bar{U}(z^{-1}) = \sum_{t=0}^{\infty} \bar{u}_1(t) z^{-t} \tag{3.153}$$

always exists and:

$$\eta_1(0, t_1) = \sum_{t=0}^{t_1} u_1(t) y_1(t) = \sum_{t=0}^{\infty} \bar{u}_1(t) \bar{y}_1(t) \tag{3.154}$$

But since $\bar{y}_1(t)$ is the output of a linear system characterized by a transfer function $H(z^{-1})$, one has:

$$y_1(t) = \frac{1}{2\pi} \int_{-\pi}^{\pi} H(e^{j\omega}) \bar{U}(e^{j\omega}) e^{jt\omega} d\omega \tag{3.155}$$

This integral exists under the assumption that $H(z^{-1})$ is asymptotically stable (i.e., all its poles lie in $|z| < 1$).

Introducing (3.155) in (3.154) one obtains:

$$\eta_1(0, t_1) = \frac{1}{2\pi} \sum_{t=0}^{\infty} \int_{-\pi}^{\pi} u(t) e^{jt\omega} H(e^{j\omega}) \bar{U}(e^{j\omega}) d\omega \tag{3.156}$$

Interchanging the sum and the integral in (3.156), one obtains:

$$\begin{aligned} \eta_1(0, t_1) &= \frac{1}{2\pi} \int_{-\pi}^{\pi} \left( \sum_{t=0}^{\infty} \bar{u}_1(t) e^{jt\omega} \right) H(e^{j\omega}) \bar{U}(e^{j\omega}) d\omega \\ &= \frac{1}{2\pi} \int_{-\pi}^{\pi} \bar{U}(e^{-j\omega}) H(e^{j\omega}) \bar{U}(e^{j\omega}) d\omega \\ &= \frac{1}{4\pi} \int_{-\pi}^{\pi} \bar{U}(e^{-j\omega}) [H(e^{j\omega}) + H(e^{-j\omega})] \bar{U}(e^{j\omega}) d\omega \end{aligned} \tag{3.157}$$

In order to satisfy the strict passivity condition it is necessary and sufficient that:

$$\frac{1}{2}[H(e^{j\omega}) + H(e^{-j\omega})] = \operatorname{Re} H(e^{j\omega}) > \delta > 0; \quad \forall -\pi < \omega < \pi \tag{3.158}$$

which implies that $H(z^{-1})$ should be a *strictly positive real transfer function* formally characterized by:

- $H(z^{-1})$ is real for real $z$,
- All the poles of $H(z^{-1})$ lie in $| z |< 1$,
- $H(z^{-1}) > 0$ for all $| z |= 1$.

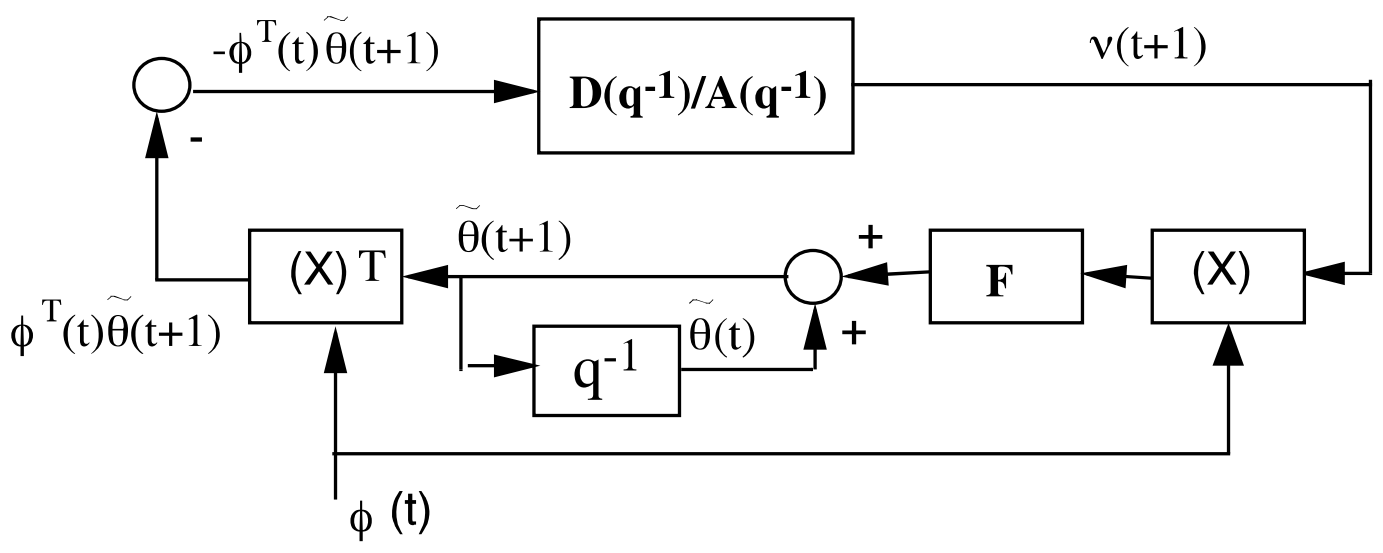

**Fig. 3.9** Equivalent feedback representation for the output error with fixed compensator

Applying this result, one concludes that in order to have $\lim_{t\to\infty} \varepsilon(t+1) = 0$, it is sufficient that $H(z^{-1})$ given by (3.149) which characterizes the equivalent linear feedforward path, *be a strictly positive real discrete transfer function.*

This condition may be restrictive in certain situations. In order to overcome this difficulty, one may consider filtering the a posteriori prediction error before its use in the PAA. One defines the a posteriori adaptation error as:

$$\nu(t+1) = D(q^{-1})\varepsilon(t+1) = \varepsilon(t+1) + \sum_{i=1}^{n_D} d_i \varepsilon(t+1-i) \tag{3.159}$$

where:

$$D(q^{-1}) = 1 + \sum_{i=1}^{n_D} d_i q^{-i} \tag{3.160}$$

is an asymptotically stable polynomial with $n_D \le n_A$ ($n_A$—degree of the polynomial $A$). Using (3.137), $\nu(t+1)$ can be expressed as:

$$\nu(t+1) = \frac{D(q^{-1})}{A(q^{-1})}[-\phi^T(t)\tilde{\theta}(t+1)] \tag{3.161}$$

and the parameter adaptation algorithm satisfying the passivity condition will take the form:

$$\hat{\theta}(t+1) = \hat{\theta}(t) + F\phi(t)\nu(t+1) \tag{3.162}$$

The equivalent feedback representation associated with the output error predictor using the above PAA is shown in Fig. 3.9. The equivalent feedback path with input $\nu(t+1)$ and output $\tilde{\theta}^T(t+1)\phi(t)$ is passive and $\lim_{t\to\infty} \nu(t+1) = 0$ will be assured if:

$$H(z^{-1}) = \frac{D(z^{-1})}{A(z^{-1})} \tag{3.163}$$

is strictly positive real.

To make the algorithm of (3.162) implementable, one has to give a relation between the a posteriori adaptation error given in (3.159) and the a priori adaptation

error defined as:

$$\nu^0(t+1) = \varepsilon^0(t+1) + \sum_{i=1}^{n_D} d_i \varepsilon(t+1-i) \tag{3.164}$$

Note that the a posteriori prediction errors $\varepsilon(t), \varepsilon(t-1), \ldots$ are available at $t+1$. Subtracting (3.164) from (3.159), one gets:

$$\nu(t+1) - \nu^0(t+1) = \varepsilon(t+1) - \varepsilon^0(t+1) = -[\hat{\theta}(t+1) - \hat{\theta}(t)]^T \phi(t) \tag{3.165}$$

But, from (3.162), one obtains:

$$\hat{\theta}(t+1) - \hat{\theta}(t) = F\phi(t)\nu(t+1) \tag{3.166}$$

and (3.165) becomes:

$$\nu(t+1) + F\phi(t)\nu(t+1) = \nu^0(t+1) \tag{3.167}$$

from which one obtains:

$$\nu(t+1) = \frac{\nu^0(t+1)}{1 + \phi^T(t)F\phi(t)} \tag{3.168}$$

In fact not only $\nu(t+1)$ is bounded and $\lim_{t\to\infty} \nu(t+1) = 0$, but also $\varepsilon(t+1)$ is bounded and $\lim_{t\to\infty} \varepsilon(t+1) = 0$ since:

$$\varepsilon(t+1) = \frac{A(q^{-1})}{D(q^{-1})}\nu(t+1) \tag{3.169}$$

with $D(z^{-1})$ being an asymptotically stable polynomial (consequence of the strict positive realness of $D(z^{-1})/A(z^{-1})$ and of the design itself). This fact allows to show that $\hat{y}(t), \hat{y}(t-1), \ldots$ are bounded (and respectively $\phi(t)$).

Assuming that $u(t)$ is a bounded sequence and $A(z^{-1})$ being an asymptotically stable polynomial, the sequence $y(t)$ will also be a bounded sequence. But from (3.132) one has:

$$\hat{y}(t) = y(t) - \varepsilon(t) \tag{3.170}$$

Since $\varepsilon(t)$ is bounded and $\lim_{t\to\infty} \varepsilon(t+1) = 0$, it results that $\hat{y}(t)$ is bounded. This implies that $\phi(t)$ is bounded and from (3.168) one can conclude also that $\lim_{t\to\infty} \nu(t+1) = 0$ implies $\lim_{t\to\infty} \nu^0(t+1) = 0$, and from (3.164) one concludes also that $\lim_{t\to\infty} \varepsilon^0(t+1) = 0$.

For the case of a model of the form:

$$y(t+1) = -A^*(q^{-1})y(t) + B^*(q^{-1})u(t-d) = \theta^T \varphi(t) \tag{3.171}$$

$$\theta^T = [a_1, \ldots, a_{n_A}, b_1, \ldots, b_{n_B}] \tag{3.172}$$

$$\varphi^T(t) = [-y(t), \ldots, -y(t-n_A+1), u(t-d), \ldots, u(t-d-n_B+1)] \tag{3.173}$$

The *output error* algorithm with constant adaptation gain can therefore summarized as follows:

$$\hat{\theta}(t+1) = \hat{\theta}(t) + F\phi(t)\nu(t+1) \tag{3.174}$$

$$\nu(t+1) = \frac{\nu^0(t+1)}{1+\phi^T(t)F\phi(t)} \tag{3.175}$$

$$\nu^0(t+1) = \varepsilon^0(t+1) + D^*(q^{-1})\varepsilon(t)$$

$$D^*(q^{-1}) = d_1 + d_2 q^{-1} + d_{n_D} q^{-n_D+1} \tag{3.176}$$

$$\varepsilon^0(t+1) = y(t+1) - \hat{\theta}^T(t)\phi(t) \tag{3.177}$$

$$\varepsilon(t-i) = y(t-i) - \hat{y}(t-i); \quad i = 0, \dots, n_A - 1 \tag{3.178}$$

$$\hat{y}(t-i) = \hat{\theta}^T(t-i)\phi(t-i-1); \quad i = 0, \dots, (n_A - 1) \tag{3.179}$$

$$\begin{aligned}\phi^T(t) = [&-\hat{y}(t), \dots, -\hat{y}(t-n_A+1), u(t-d), \dots, \\ &u(t-d-n_B+1)]\end{aligned} \tag{3.180}$$

### 3.3.3 Positive Real PAA Structures

Examining the various equivalent feedback representations obtained up to now, one observes that the equivalent feedback path contains an integrator whose input is the product between the a posteriori adaptation error and a time-varying vector which is the one appearing in the equation expressing the prediction or adaptation error as a function of the parameter error. In the equivalent feedback representation, the same vector multiplied by the output of the integrator (which is $\tilde{\theta}(t+1)$) generates the output of the equivalent feedback path.

On the other hand, the integrator (in fact, a multi-input multi-output integrator) is a passive linear block characterized by a positive real transfer matrix. A positive real transfer matrix is characterized by the following properties (see Appendix C):

1. All elements of $H(z)$ are analytic outside the unit circle (i.e. they do not have poles in $|z| > 1$).
2. The eventual poles of any element of $H(z)$ on $|z| = 1$ are simple and the associated residue matrix is a positive semidefinite Hermitian.
3. The matrix $H(z) + H^T(z^{-1})$ is a positive semidefinite Hermitian for all $|z| = 1$ which are not a pole of $H(z)$.

Therefore, one can consider to replace the integrator by a more general passive linear system leading to a PAA of the form:

$$x(t+1) = Ax(t) + B\phi(t)\nu(t+1) \tag{3.181}$$

$$\hat{\theta}(t+1) = Cx(t) + D\phi(t)\nu(t+1) \tag{3.182}$$

where $x(t)$ is the state of the passive linear system.

The case of integral adaptation corresponds to: $A = I$, $B = D = F$, $C = I$. In other terms, the integrator is replaced by a dynamic system:

$$x(t+1) = Ax(t) + Bu(t) \tag{3.183}$$

$$y(t) = Cx(t) + Du(t) \tag{3.184}$$

characterized also by the matrix transfer function:

$$H_{PAA}(z) = C(zI - A)^{-1}B + D \tag{3.185}$$

with a pole at $z = 1$ if we want that the PAA has memory.

One has the following general result:

**Theorem 3.1** *For a PAA having the form of* (3.181) *and* (3.182), *the equivalent feedback path characterized by*:

$$x(t+1) = Ax(t) + B\phi(t)\nu(t+1) \tag{3.186}$$

$$\phi^T(t)\tilde{\theta}(t+1) = \phi^T(t)Cx(t) + \phi^T(t)D\phi(t)\nu(t+1) \tag{3.187}$$

*is passive*, *i.e.*,

$$\eta(0, t_1) = \sum_{t=0}^{t_1} \nu(t+1)\phi^T(t)\tilde{\theta}(t+1) \geq -\gamma^2; \quad \gamma^2 < \infty; \ \forall t \geq 0 \tag{3.188}$$

*if the associated linear system described by* (3.183) *and* (3.184) *is passive*, *or alternatively*, *if* $H_{PAA}(z)$ *given in* (3.185) *is a positive real transfer matrix.*

*Proof* If the associated linear system (3.183) and (3.184) is passive, $H(z)$ is a positive real transfer matrix and the matrices $A$, $B$, $C$, $D$ verify the Positive Real Lemma (Appendix C, Lemma C.3). As a consequence, there is a positive definite matrix $P$, positive semidefinite matrices $Q$ and $R$ and a matrix $S$ such that:

$$A^TPA - P = -Q \tag{3.189}$$

$$C - B^TPA = S^T \tag{3.190}$$

$$D + D^T - B^TPB = R \tag{3.191}$$

$$\begin{bmatrix} Q & S \\ S^T & R \end{bmatrix} \geq 0 \tag{3.192}$$

The passivity inequality of (3.188) can be interpreted as the input-output product for a system with the input $\nu(t+1)$ and the output $\phi^T(t)\tilde{\theta}(t+1)$ and described by (3.186) and (3.187), which represent a time-varying system characterized by the time-varying matrices:

$$A(t) = A; \qquad B(t) = B\phi(t); \qquad C(t) = \phi^T(t)C; \qquad D(t) = \phi^T(t)D\phi(t) \tag{3.193}$$

From the properties of time-varying passive systems (see Appendix C, Lemma C.6), one concludes that the inequality of (3.188) will be satisfied if there are sequences of positive semidefinite time-varying matrices $P(t)$, $Q(t)$ and $R(t)$, and a sequence of time-varying matrices $S(t)$ such that:

$$A^T(t)P(t)A(t) - P(t) = -Q(t) \tag{3.194}$$

$$C(t) - B^T(t)P(t)A(t) = S^T(t) \tag{3.195}$$

$$D^T(t) + D^T(t) - B^T(t)P(t+1)B(t) = R(t) \tag{3.196}$$

$$\begin{bmatrix} Q(t) & S(t) \\ S^T(t) & R(t) \end{bmatrix} \geq 0, \quad \forall t \geq 0 \tag{3.197}$$

Replacing $A(t)$, $B(t)$, $C(t)$, $D(t)$ by the values given in (3.193), (3.194) through (3.197) become (taking also into account (3.189) through (3.192)):

$$A^T PA - P = -Q = -Q(t) \tag{3.198}$$

$$\phi^T(t)C - \phi^T(t)B^T PA = \phi^T(t)S^T = S^T(t) \tag{3.199}$$

$$\phi^T(t)(D + D^T)\phi(t) - \phi^T(t)B^T PB\phi(t) = \phi^T(t)R\phi(t) = R(t) \tag{3.200}$$

$$\begin{aligned} \begin{bmatrix} Q(t) & S(t) \\ S^T(t) & R(t) \end{bmatrix} &= \begin{bmatrix} Q & S\phi(t) \\ \phi^T(t)S^T & \phi^T(t)R\phi(t) \end{bmatrix} \\ &= \begin{bmatrix} 1 & 0 \\ 0 & \phi^T(t) \end{bmatrix} \begin{bmatrix} Q & S \\ S^T & R \end{bmatrix} \begin{bmatrix} 1 & 0 \\ 0 & \phi(t) \end{bmatrix} \geq 0, \quad \forall t \geq 0 \end{aligned} \tag{3.201}$$

and therefore (3.186) and (3.187) define a passive block satisfying the inequality (3.188). □

If we would like the PAA to have memory, $H(z)$ should have a pole at $z = 1$.

## "Integral + Proportional" Parameter Adaptation Algorithm

A first particularization of the above result is obtained for:

$$A = I; \qquad B = F_I; \qquad C = I; \qquad D = F_I + F_P \tag{3.202}$$

which corresponds to an associated transfer matrix:

$$H_{PAA}(z^{-1}) = \frac{1}{1 - z^{-1}} F_I + F_P; \quad F_I > 0; \ F_P = \alpha F_i; \ \alpha > -0.5 \tag{3.203}$$

where $F_I$ is called the *integral adaptation gain* and $F_p$ *the proportional adaptation gain*.

The algorithm is in general written under the equivalent form:

$$\hat{\theta}_I(t+1) = \hat{\theta}_I(t) + F_I\phi(t)\nu(t+1); \quad F_I > 0 \tag{3.204}$$

$$\hat{\theta}_P(t+1) = F_p\phi(t)\nu(t+1); \quad F_p = \alpha F_I; \ \alpha \geq -0.5 \tag{3.205}$$

$$\hat{\theta}(t+1) = \hat{\theta}_I(t+1) + \hat{\theta}_P(t+1) \tag{3.206}$$

The adjustable predictor has the form:

$$\hat{y}^0(t+1) = \hat{\theta}_I^T(t)\phi(t)$$
$$y(t+1) = \hat{\theta}^T(t+1)\phi(t)$$

The associated passivity condition on the matrices $F_I$ and $F_p$ takes the form:

$$P - P = -Q \tag{3.207}$$

$$I - F_I P = S^T \tag{3.208}$$

$$2(F_I + F_P) - F_I P F_I = R \tag{3.209}$$

Taking $P = F_I^{-1}$, which implies to take $F_I > 0$, the remaining condition is:

$$F_I + 2F_P \geq 0 \tag{3.210}$$

which leads to the "surprising" condition:

$$F_P \geq -0.5F_I \qquad (\text{or } F_P = \alpha F_I; \quad \alpha \geq -0.5) \tag{3.211}$$

i.e., *negative* proportional adaptation gain can be used provided that the above condition is satisfied.

One can ask what is the influence of the proportional term upon the adaptation algorithm.

Proportional + Integral PAA with *positive* proportional gain leads to the improvement of the convergence of the adaptation error. However, high value of the proportional gain will slow down the convergence of the parameters.

Proportional + Integral PAA with small *negative* proportional adaptation gain improves in general the convergence of the parameters, but large negative values (below the limit) will slow down the convergence of the adaptation error.

These phenomena are illustrated in the following simulation example.

*Simulation example* Figure 3.10a illustrates the evolution of $J(N) = \sum_0^N \nu^2(t+1)$ and Fig. 3.10b illustrates the evolution of the parametric distance defined as

$$D(N) = \{[\theta - \hat{\theta}(t)]^T[\theta - \hat{\theta}(t)]\}^{1/2}$$

for the case of integral + proportional adaptation.

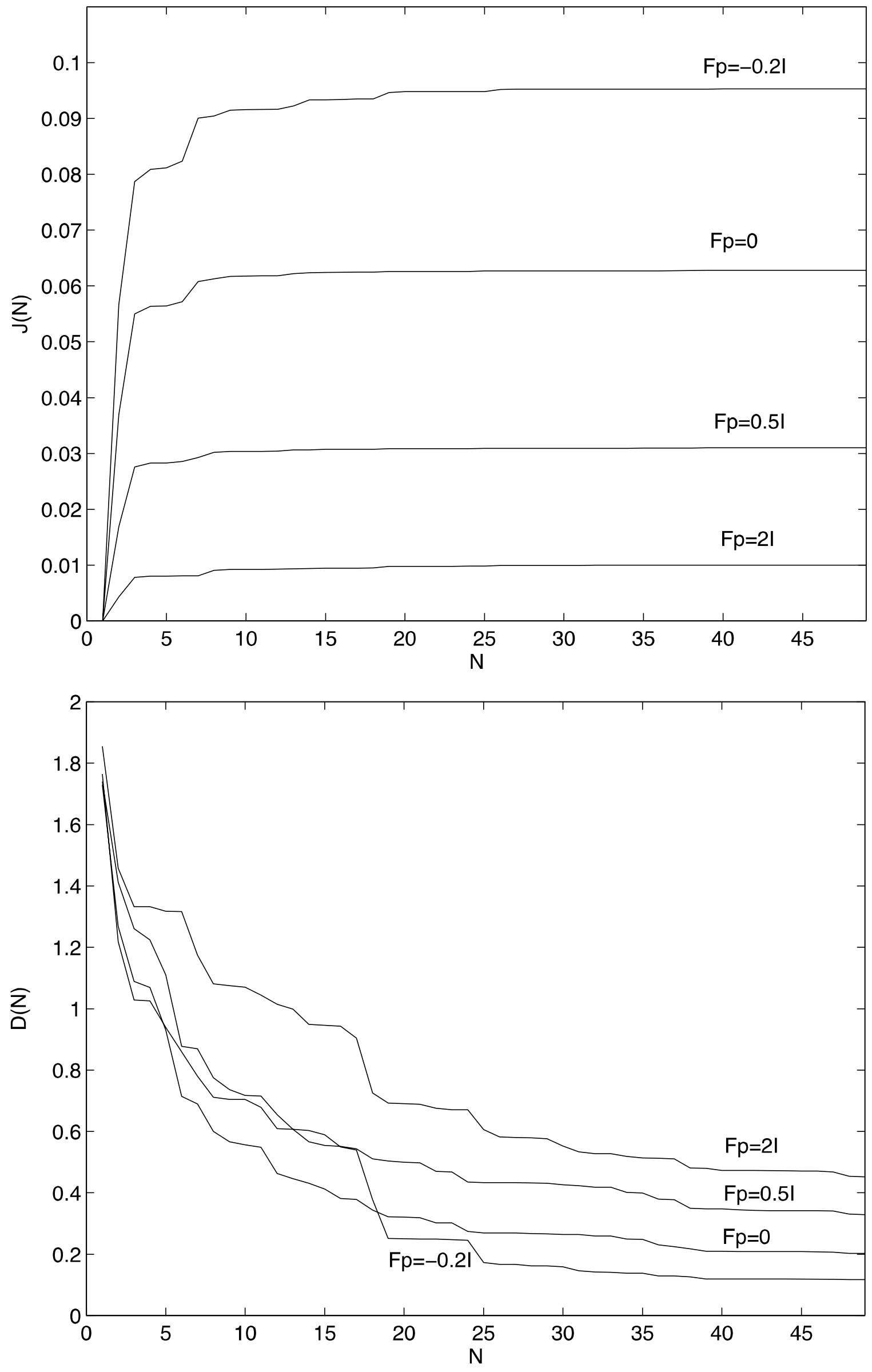

**Fig. 3.10** (**a**) Evolution of $J(N)=\sum_0^N \nu^2(t+1)$, (**b**) evolution of the parametric distance $D(N)=\{[\theta-\hat{\theta}(t)]^T[\theta-\hat{\theta}(t)]\}^{1/2}$

The nominal system is:

$$G(q^{-1})=\frac{q^{-1}(q^{-1}+0.5q^{-2})}{1-1.5q^{-1}+0.7q^{-2}}$$

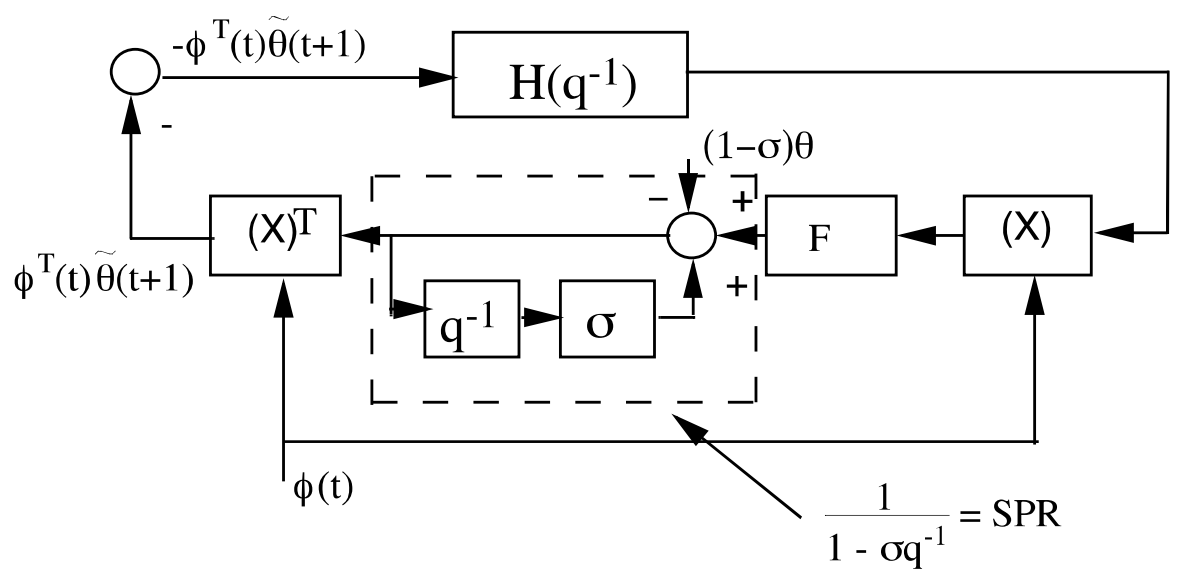

**Fig. 3.11** Equivalent feedback representation for the output error with leakage

All the estimated parameters have been initialized at zero ($\hat{\theta}(0) = 0$) and the input is a PRBS. A least squares type adaptive predictor with a parameter adaptation gain using a constant integral matrix adaptation gain $I$ and a proportional matrix adaptation gain which takes the values: $-0.2I, 0, 0.5I, 2I$ has been used.

One can effectively see that positive proportional adaptation gain speed up the convergence of the a posteriori adaptation error (in this case the a posteriori prediction error) but it slows down the convergence of the estimated parameters. A low negative proportional adaptation has the opposite effect. Similar results are obtained for output error adaptive predictors.

### Parameter Adaptation Algorithm with Leakage

For the case of tracking slowly time-varying parameters where there is not a steady state parameter to be reached, the integrator may not be justified. In this case, one can replace the integrator by a first order system i.e. the PAA of (3.181) and (3.182) is particularized for:

$$A = \sigma I; \quad 0 < \sigma < 1; \qquad B = F; \qquad C = I; \qquad D = F \tag{3.212}$$

which corresponds to an associated transfer matrix:

$$H_{PAA}(z^{-1}) = \frac{1}{1 - \sigma z^{-1}} F; \quad 0 < \sigma < 1 \tag{3.213}$$

The PAA takes the form:

$$\hat{\theta}(t+1) = \sigma \hat{\theta}(t) + F\phi(t)\nu(t+1); \quad 0 < \sigma < 1 \tag{3.214}$$

and the parameter error is driven by:

$$\tilde{\theta}(t+1) = \sigma \tilde{\theta}(t) + F\phi(t)\nu(t+1) - (1-\sigma)\theta \tag{3.215}$$

The term $(1-\sigma)\theta$ corresponds to an exogenous bounded input to the EFR as indicated in Fig. 3.11.

If the linear feedforward path is also passive, the EFR has a BIBO property, and this exogenous input will generate a bounded adaptation error $\nu(t+1) \neq 0$ even for the case $\theta = \text{const}$. (The algorithm does not have a memory.)

Assume that $\theta$ is time-varying as:

$$\theta(t+1) = \sigma\theta(t); \quad 0 < \sigma < 1$$

then the exogenous input in (3.215) disappears and $\lim_{t\to\infty} \nu(t+1) = 0$ if $H(z^{-1})$ characterizing the linear feedforward path is strictly positive real.

This remark suggests that it may be possible to perfectly track asymptotically the output of a system with time-varying parameters if the law of variation is known and it satisfies some conditions.

### PAA for Systems with Time-Varying Parameters

The last remark immediately suggests the following solution for the case of estimating time-varying parameters governed by:

$$\theta(t+1) = A\theta(t) \quad \text{with } \theta(0) = \theta_0 \tag{3.216}$$

where $\det(zI - A)$ has all its roots inside or on the unit circle, and the roots on the unit circle are simple and their residues are positive (for example: periodic time-varying parameters).

Choose the PAA:

$$\hat{x}(t+1) = A\hat{x}(t) + B\phi(t)\nu(t+1) \tag{3.217}$$

$$\hat{\theta}(t+1) = C\hat{x}(t) + D\phi(t)\nu(t+1) \tag{3.218}$$

Define:

$$\left.\begin{aligned} \tilde{x}(t+1) &= \hat{x}(t+1) - \theta(t+1) \\ \tilde{\theta}(t+1) &= \hat{\theta}(t+1) - C\theta(t+1) \end{aligned}\right\} \tag{3.219}$$

one gets:

$$\left.\begin{aligned} \tilde{x}(t+1) &= A\tilde{x}(t) + B\phi(t)\nu(t+1) \\ \tilde{\theta}(t+1) &= C\tilde{x}(t) + D\phi(t)\nu(t+1) \end{aligned}\right\} \tag{3.220}$$

If:

$$H_{PAA}(z^{-1}) = C(zI - A)^{-1}B + D \tag{3.221}$$

is *positive real*, the equivalent feedback path is passive and $\lim_{t\to\infty} \nu(t+1) = 0$ provided that $H_{PAA}(z^{-1})$ characterizing the equivalent feedforward path is *strictly positive real*. For a detailed discussion and evaluation see Glower (1996).

### 3.3.4 Parameter Adaptation Algorithms with Time-Varying Adaptation Gain

In the case of time-varying adaptation gain of the form:

$$F(t+1)^{-1} = \lambda_1(t)F(t)^{-1} + \lambda_2(t)\phi(t)\phi^T(t)$$
$$0 < \lambda_1(t) \le 1;\ 0 \le \lambda_2(t) \le 2;\ F(0) > 0 \tag{3.222}$$

one cannot show that the equivalent feedback block will be passive for $\lambda_2(t) > 0$ (and in particular for the case $\lambda_1(t) \equiv 1$ and $\lambda_2(t) > 0$, corresponding to a decreasing adaptation gain). This will require to take advantage of an excess of passivity of the equivalent feedforward block in order to compensate for the lack of passivity of the equivalent feedback block. Another interpretation is that an additional equivalent loop transformation allows to get a transformed equivalent feedback system featuring an equivalent feedback path with the desired passivity properties, but the passivity condition on the new equivalent linear feedforward path will be more restrictive.

Let us consider as an example the case of recursive least squares, to be analyzed using the passivity approach as discussed in Sect. 3.3.1. One obtains an equivalent feedback representation defined by (3.113) through (3.115). The equivalent feedback path will be characterized by ($\nu(t+1) = \varepsilon(t+1)$ in this case):

$$\tilde{\theta}(t+1) = \tilde{\theta}(t) + F(t)\phi(t)\nu(t+1) \tag{3.223}$$

$$\begin{aligned} y_2(t) &= \phi^T(t)\tilde{\theta}(t+1) \\ &= \phi^T(t)\tilde{\theta}(t) + \phi^T(t)F(t)\phi(t)\nu(t+1) \\ &= \phi^T(t)\tilde{\theta}(t) + \phi^T(t)F(t)\phi(t)u_2(t) \end{aligned} \tag{3.224}$$

where:

$$F(t+1)^{-1} = F(t)^{-1} + \lambda_2(t)\phi(t)\phi^T(t); \quad 0 < \lambda_2(t) < 2 \tag{3.225}$$

Equations (3.225) which is a particular form of (3.222), corresponds to a time decreasing adaptation gain. (The standard recursive least squares uses $\lambda_2(t) \equiv 1$.)

In order to apply the passivity approach, one should evaluate first the sum of the input-output product of the feedback path. For this type of adaptation gain, one has the following property:

**Lemma 3.2** *For the PAA of* (3.222) *and* (3.223), *with* $\lambda_1(t) \equiv 1$ *one has* $\forall t_1 \ge 0$:

$$\sum_{t=0}^{t_1} \tilde{\theta}^T(t+1)\phi(t)\nu(t+1)$$
$$= \frac{1}{2}\tilde{\theta}^T(t+1)F(t+1)^{-1}\tilde{\theta}(t+1) - \frac{1}{2}\tilde{\theta}^T(0)F(0)^{-1}\tilde{\theta}(0)$$

$$+\frac{1}{2}\sum_{t=0}^{t_1}\phi^T(t)F(t)\phi(t)\nu^2(t+1)-\frac{1}{2}\sum_{t=0}^{t_1}\lambda_2(t)[\tilde{\theta}^T(t+1)\phi(t)]^2$$

$$\geq -\frac{1}{2}\tilde{\theta}^T(0)F(0)^{-1}\tilde{\theta}(0)-\frac{1}{2}\sum_{t=0}^{t_1}\lambda_2(t)[\tilde{\theta}^T(t+1)\phi(t)]^2 \tag{3.226}$$

*Proof* Taking into account (3.223), one has:

$$\begin{aligned}
\tilde{\theta}^T(t+1)\phi(t)\nu(t+1) &= \tilde{\theta}^T(t+1)F(t)^{-1}[\tilde{\theta}(t+1)-\tilde{\theta}(t)] \\
&= \tilde{\theta}^T(t+1)[F(t+1)^{-1}-\lambda_2(t)\phi(t)\phi^T(t)] \\
&\quad \times [\tilde{\theta}(t+1)-\tilde{\theta}(t)] \\
&= \tilde{\theta}^T(t+1)F(t+1)^{-1}\tilde{\theta}(t+1)-\lambda_2(t)[\phi^T(t)\tilde{\theta}(t+1)]^2 \\
&\quad - \tilde{\theta}^T(t+1)F(t)^{-1}\tilde{\theta}(t)
\end{aligned}$$

On the other hand:

$$\begin{aligned}
&[\tilde{\theta}(t+1)-\tilde{\theta}(t)]^T F(t)^{-1}[\tilde{\theta}(t+1)-\tilde{\theta}(t)] \\
&\quad = \tilde{\theta}^T(t+1)F(t)^{-1}\tilde{\theta}(t+1)+\tilde{\theta}^T(t)F(t)^{-1}\tilde{\theta}(t)-2\tilde{\theta}^T(t+1)F(t)^{-1}\tilde{\theta}(t) \\
&\quad \geq 0
\end{aligned}$$

from which it results:

$$\begin{aligned}
-\tilde{\theta}^T(t+1)F(t)^{-1}\tilde{\theta}(t) &= -\frac{1}{2}[\tilde{\theta}^T(t+1)F(t)^{-1}\tilde{\theta}(t+1)+\tilde{\theta}^T(t)F(t)^{-1}\tilde{\theta}(t)] \\
&\quad + \frac{1}{2}\phi^T(t)F(t)\phi(t)\nu^2(t+1) \\
&= -\frac{1}{2}\tilde{\theta}^T(t+1)[F(t+1)^{-1}-\lambda_2(t)\phi(t)\phi^T(t)]\tilde{\theta}(t+1) \\
&\quad - \frac{1}{2}\tilde{\theta}^T(t)F(t)^{-1}\tilde{\theta}(t)+\frac{1}{2}\phi^T(t)F(t)\phi(t)\nu^2(t+1)
\end{aligned}$$

and, therefore:

$$\begin{aligned}
\tilde{\theta}^T(t+1)\phi(t)\nu(t+1) &= \tilde{\theta}^T(t+1)F(t+1)^{-1}\tilde{\theta}(t+1)-\lambda_2(t)[\phi^T(t)\tilde{\theta}(t+1)]^2 \\
&\quad - \frac{1}{2}\tilde{\theta}^T(t+1)[F(t+1)^{-1}-\lambda_2(t)\phi(t)\phi^T(t)]\tilde{\theta}(t+1) \\
&\quad - \frac{1}{2}\tilde{\theta}^T(t)F(t)^{-1}\tilde{\theta}(t)+\frac{1}{2}\phi^T(t)F(t)\phi(t)\nu^2(t+1) \\
&= \frac{1}{2}\tilde{\theta}^T(t+1)F(t+1)^{-1}\tilde{\theta}(t+1)-\frac{1}{2}\tilde{\theta}^T(t)F(t)^{-1}\tilde{\theta}(t) \\
&\quad - \frac{1}{2}\lambda_2(t)[\phi^T(t)\tilde{\theta}(t+1)]^2+\frac{1}{2}\phi^T(t)F(t)\phi(t)\nu^2(t+1)
\end{aligned}$$

and summing up from $t=0$ to $t_1$, one gets (3.226). □

Therefore, the equivalent feedback path satisfies:

$$\sum_{t=0}^{t_1} y_2(t)u_2(t) = \sum_{t=0}^{t_1} \tilde{\theta}^T(t+1)\phi(t)\nu(t+1)$$
$$\geq -\frac{1}{2}\tilde{\theta}^T(0)F(0)^{-1}\tilde{\theta}(0) - \frac{1}{2}\sum_{t=0}^{t_1}\lambda_2(t)[\tilde{\theta}^T(t+1)\phi(t)]^2$$

On the other hand, the equivalent linear block which has a unitary gain, satisfies the inequality:

$$\sum_{t=0}^{t_1} y_1(t)u_1(t) = \sum_{t=0}^{t_1} -\tilde{\theta}^T(t+1)\phi(t)\varepsilon(t+1)$$
$$= \sum_{t=0}^{t_1}[\phi^T(t)\tilde{\theta}(t+1)]^2 \geq 0 \tag{3.227}$$

The feedback connection is still valid and from (3.226) and (3.227), one obtains:

$$\sum_{t=0}^{t_1} y_1(t)u_1(t) = \sum_{t=0}^{t_1}[\phi^T(t)\tilde{\theta}(t+1)]^2$$
$$\leq \frac{1}{2}\tilde{\theta}^T(0)F(0)^{-1}\tilde{\theta}(0) + \frac{1}{2}\sum_{t=0}^{t_1}\lambda_2(t)[\tilde{\theta}(t+1)\phi^T(t)]^2 \tag{3.228}$$

and, this leads to:

$$\sum_{t=0}^{t_1}\left(1 - \frac{\lambda_2(t)}{2}\right)[\phi^T(t)\tilde{\theta}(t+1)]^2 = \sum_{t=0}^{t_1}\left(1 - \frac{\lambda_2(t)}{2}\right)\varepsilon^2(t+1)$$
$$\leq \frac{1}{2}\tilde{\theta}^T(0)F(0)^{-1}\tilde{\theta}(0) \tag{3.229}$$

But $(1 - \frac{\lambda_2(t)}{2}) > 0$ since $\lambda_2(t) < 2$ and one concludes from (3.229) that $\varepsilon(t+1)$ is bounded and:[4]

$$\lim_{t\to\infty} \varepsilon(t+1) = 0 \tag{3.230}$$

There are two possible interpretations of this result:

[4] In fact, taking advantage of the term $\sum_{t=0}^{t_1}\phi^T(t)F(t)\phi(t)\nu^2(t+1)$ in (3.226) and the fact that the equivalent linear block is characterized by a unitary gain, one can conclude that: $\lim_{t\to\infty}[1 + \phi^T(t)F(t)\phi(t)]\varepsilon^2(t+1) = 0$.

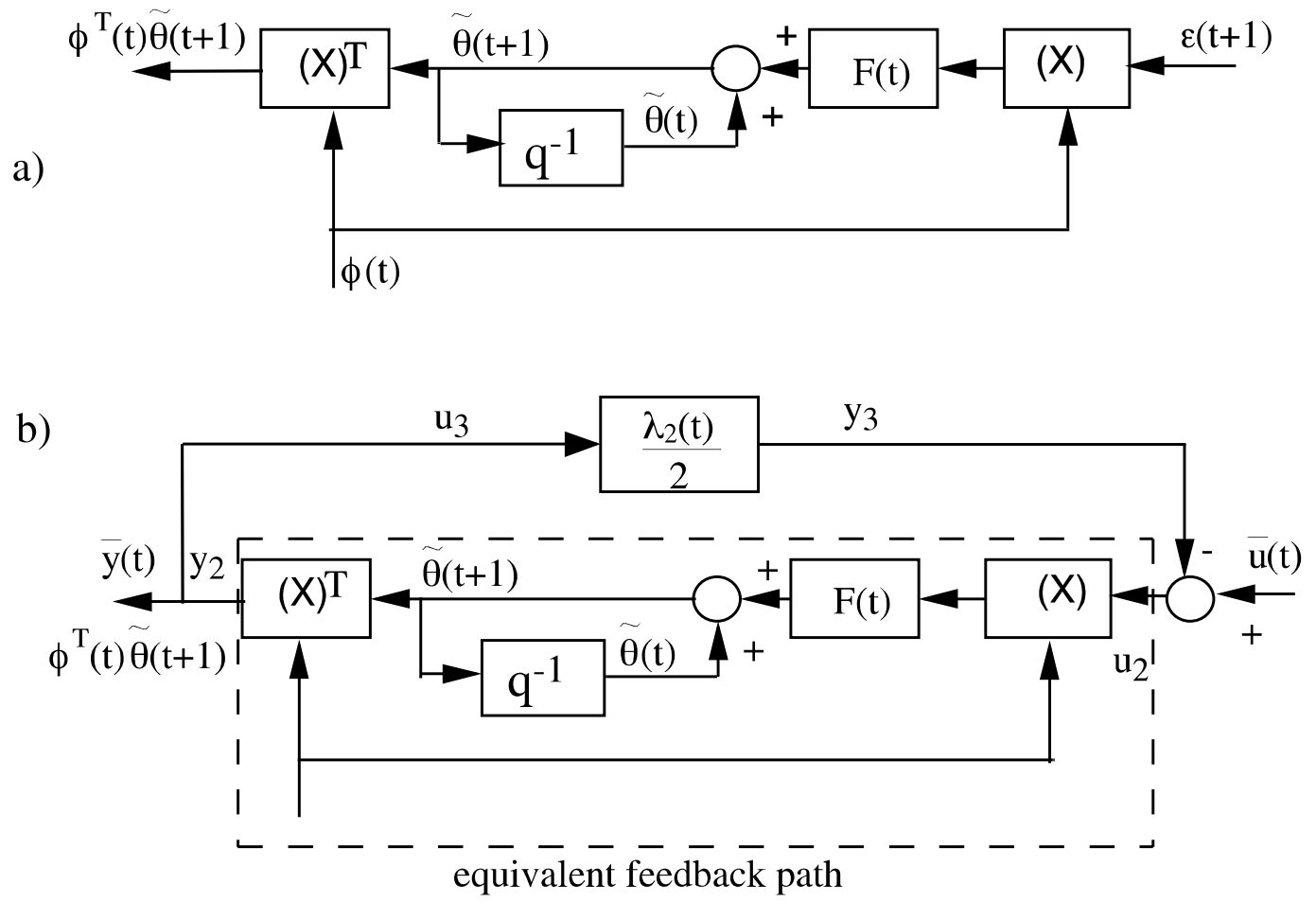

**Fig. 3.12** The feedback composite block, (**a**) original equivalent feedback block (RLS), (**b**) feedback composite block

- *First interpretation*

  The equivalent feedback path has a lack of passivity which is expressed in terms of the input-output product of a block with a gain $-\frac{\lambda_2(t)}{2}$ and having the output of the equivalent feedback block $\phi^T(t)\tilde{\theta}(t+1)$ as input. On the other hand, this is compensated by an excess of passivity of the equivalent feedforward block which is strictly input passive with an excess $\delta = 1$ which makes $(1 - \frac{\lambda_2(t)}{2})$ positive as long as $\lambda_2(t) < 2$ ($\lambda_2(t) \equiv 1$ in the case of recursive least squares).
- *Second interpretation*

  Let us consider a composite block obtained from the equivalent feedback path by adding a negative feedback through a time-varying gain $\frac{\lambda_2(t)}{2}$ as shown in Fig. 3.12.

The resulting input and output relationships are:

$$u_2 = \bar{u} - y_3$$
$$\bar{y} = y_2 = u_3$$

Therefore the input-output product of the composite block is expressed as:

$$\begin{aligned}
\sum \bar{u}(t)\bar{y}(t) &= \sum y(t)[u_2 + y_3] = \sum y_2(t)u_2(t) + \sum u_3(t)y_3(t) \\
&= \sum y_2(t)u_2(t) + \sum \frac{\lambda_2(t)}{2}(\tilde{\theta}^T(t+1)\phi(t))^2 \\
&\geq -\frac{1}{2}\tilde{\theta}(0)F(0)^{-1}\tilde{\theta}(0)
\end{aligned}$$

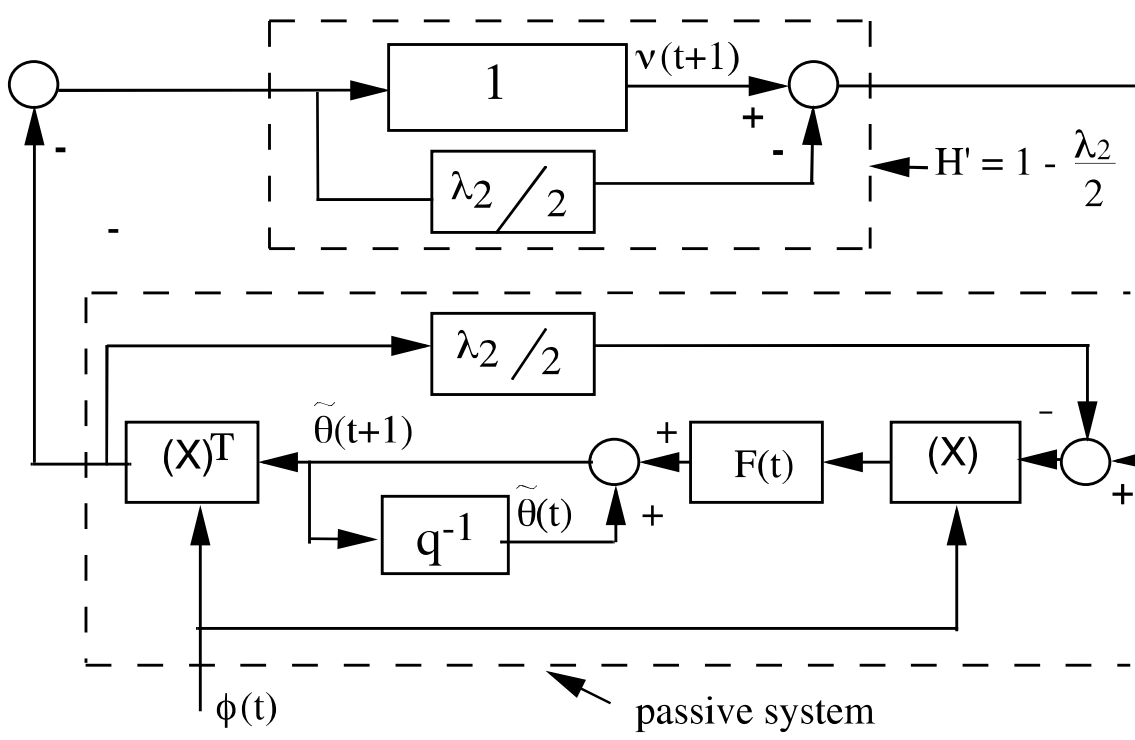

**Fig. 3.13** Transformed equivalent feedback system representation for the case of recursive least squares with $\lambda_2 = \text{const}$

The consequence of this relationship is that the equivalent feedback block in the case of a decreasing adaptation gain in feedback connection with a block having the gain $\frac{\lambda_2(t)}{2}$ is passive.

Let us consider next the case $\lambda_2(t) \equiv \lambda_2 = \text{const}$. In order that the equivalent feedback system remains unchanged, it will be necessary to add a block with a gain $\frac{\lambda_2}{2}$ in parallel to the feedforward block, as indicated in the transformed EFR shown in Fig. 3.13. The strict passivity condition on the transformed equivalent feedforward path becomes:

$$1 - \frac{\lambda_2}{2} > 0$$

This obviously generalizes for the cases where the equivalent feedforward path is a transfer function, leading to the condition:

$$H'(z^{-1}) = H(z^{-1}) - \frac{\lambda_2}{2}$$

be strictly positive real.

The case $\lambda_2(t)$ requires us to consider a feedback around the equivalent feedback path with a gain $\frac{\lambda_2}{2}$ where $\lambda_2 \geq \max_t \lambda_2(t)$. Both types of argument can be used for dealing with the general case of a time-varying adaptation gain given by (3.222).

### A General Structure and Stability of PAA

One can consider as a general structure for the PAA (integral type):

$$\boxed{\hat{\theta}(t+1) = \hat{\theta}(t) + F(t)\phi(t)\nu(t+1)} \tag{3.231}$$

$$\boxed{\nu(t+1) = \frac{\nu^0(t+1)}{1 + \phi^T(t)F(t)\phi(t)}} \tag{3.232}$$

$$\boxed{\begin{aligned}&F(t+1)^{-1}=\lambda_1(t)F(t)^{-1}+\lambda_2(t)\phi(t)\phi^T(t)\\&\quad 0<\lambda_1(t)\le 1;\ 0\le\lambda_2(t)<2\\&\quad F(0)>0;\ F(t)^{-1}>\alpha F(0)^{-1};\ \infty>\alpha>0\end{aligned}} \tag{3.233}$$

where $\hat{\theta}(t)$ is the adjustable parameter vector, $F(t)$ is the adaptation gain, $\phi(t)$ is the observation vector, $\nu^0(t+1)$ is the a priori adaptation error and $\nu(t+1)$ is the a posteriori adaptation error. The a priori adaptation error $\nu^0(t+1)$ depends only on the adjustable parameter vector $\hat{\theta}(i)$ up to and including $i=t$. $\nu^0(t+1)$ is in fact the prediction of $\nu(t+1)$ based on these $\hat{\theta}(i)$, i.e.:

$$\nu^0(t+1)=\nu(t+1/\hat{\theta}(t),\hat{\theta}(t-1),\ldots)$$

The adaptation gain matrix $F(t)$ is computed recursively using the *matrix inversion lemma*, and (3.3.124) becomes:

$$F(t+1)=\frac{1}{\lambda_1(t)}\left[F(t)-\frac{F(t)\phi(t)\phi^T(t)F(t)}{\frac{\lambda_1(t)}{\lambda_2(t)}+\phi^T(t)F(t)\phi(t)}\right] \tag{3.234}$$

Associated with the PAA of (3.231) through (3.233) one considers the class of adaptive systems for which the a posteriori adaptation error satisfies an equation of the form:

$$\boxed{\nu(t+1)=H(q^{-1})[\theta-\hat{\theta}(t+1)]^T\phi(t)} \tag{3.235}$$

where:

$$H(q^{-1})=\frac{H_1(q^{-1})}{H_2(q^{-1})} \tag{3.236}$$

with:

$$H_j(q^{-1})=1+q^{-1}H_j^*(q^{-1})=1+\sum_{i=1}^{n_j}h_i^jq^{-i};\quad j=1,2 \tag{3.237}$$

and $\theta$ is a fixed value of the unknown parameter vector.

The relationship between a priori and a posteriori adaptation errors given in (3.232), can be alternatively expressed using (3.231) as:

$$\nu(t+1)=[\hat{\theta}(t)-\hat{\theta}(t+1)]^T\phi(t)+\nu^0(t+1) \tag{3.238}$$

From (3.235) and (3.236) one gets:

$$\begin{aligned}\nu(t+1)&=[\theta-\hat{\theta}(t+1)]^T\phi(t)-H_2^*(q^{-1})\nu(t)\\&\quad+H_1^*(q^{-1})[\theta-\hat{\theta}(t)]^T\phi(t-1)\end{aligned} \tag{3.239}$$

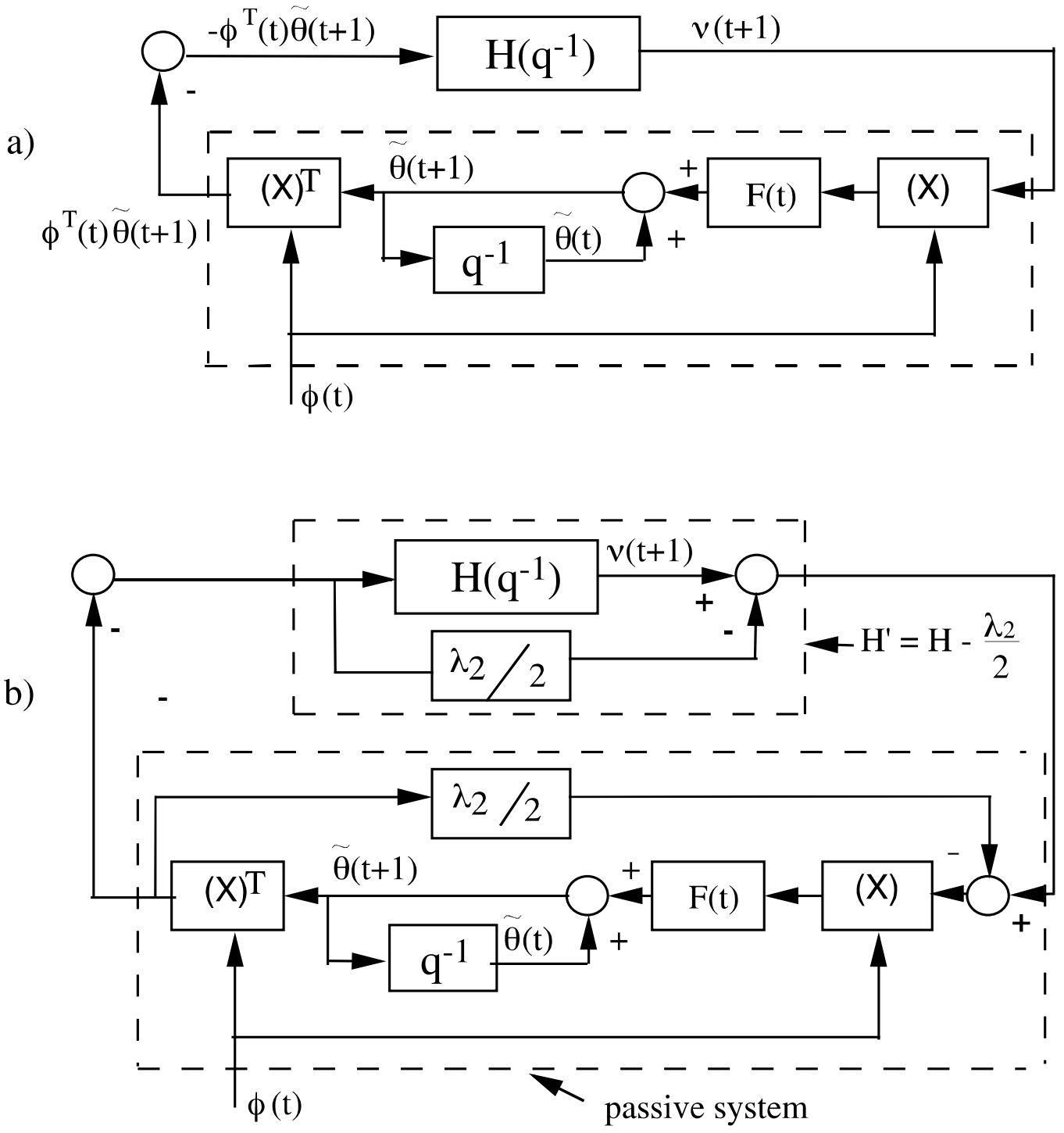

**Fig. 3.14** Transformed equivalent feedback systems associated to the PAA with time-varying gain

Adding and subtracting the term $\hat{\theta}^T(t)\phi(t)$ in the right hand side of (3.239), one gets:

$$\begin{aligned}\nu(t+1) &= [\hat{\theta}(t) - \hat{\theta}(t+1)]^T\phi(t) \\ &\quad + \{[\theta - \hat{\theta}(t)]^T\phi(t) + H_1^*(q^{-1})[\theta - \hat{\theta}(t)]^T\phi(t-1) \\ &\quad - H_2^*(q^{-1})\nu(t)\}\end{aligned} \tag{3.240}$$

Comparing (3.238) and (3.240), one observes that:

$$\begin{aligned}\nu^0(t+1) = [\theta - \hat{\theta}(t)]^T\phi(t) - H_1^*(q^{-1})[\theta - \hat{\theta}(t)]^T\phi(t-1) \\ - H_2^*(q^{-1})\nu(t)\end{aligned} \tag{3.241}$$

and one clearly sees that $\nu^0(t+1)$ depends upon $\hat{\theta}(i)$ for $i \leq t$.

The PAA of (3.231) through (3.233), together with (3.235), define an equivalent feedback system with a linear time-invariant feedforward block and a time-varying and/or nonlinear feedback block (see Fig. 3.14a).

Exploiting the input-output properties of the equivalent feedback and feedforward block, one has the following general result.

**Theorem 3.2** *Consider the parameter adaptation algorithm given by* (3.231) *through* (3.233). *Assume that the a posteriori adaptation error satisfies* (3.235) *where* $\phi(t)$ *is a bounded or unbounded vector sequence*, $H(z^{-1})$ *is a rational discrete transfer function* (*ratio of monic polynomials*) *and* $\theta$ *is a constant vector. Then, if*:

$$H'(z^{-1}) = H(z^{-1}) - \frac{\lambda_2}{2} \tag{3.242}$$

*is strictly positive real where*:

$$\max_t [\lambda_2(t)] \leq \lambda_2 < 2 \tag{3.243}$$

*one has for any bounded* $\nu(0), \hat{\theta}(0)$:

$$(1) \quad \lim_{t_1 \to \infty} \sum_{t=0}^{t_1} \nu^2(t+1) < C(\nu(0), \hat{\theta}(0)); \quad 0 < C < \infty \tag{3.244}$$

$$(2) \quad \lim_{t \to \infty} \nu(t+1) = 0 \tag{3.245}$$

$$(3) \quad \lim_{t \to \infty} [\theta - \hat{\theta}(t+1)]^T \phi(t) = 0 \tag{3.246}$$

$$(4) \quad \lim_{t \to \infty} [\hat{\theta}(t+1) - \hat{\theta}(t)]^T F(t)^{-1} [\hat{\theta}(t+1) - \hat{\theta}(t)] = 0 \tag{3.247}$$

$$(5) \quad [\hat{\theta}(t) - \theta]^T F(t)^{-1} [\hat{\theta}(t) - \theta] < M_1 < \infty \tag{3.248}$$

(6) *if* $H'(z^{-1})$ *is output strictly passive*:

$$\lim_{t \to \infty} [1 + \phi^T(t) F(t) \phi(t)] \nu^2(t+1) = \lim_{t \to \infty} \frac{[\nu^0(t+1)]^2}{1 + \phi^T(t) F(t) \phi(t)} = 0 \tag{3.249}$$

*If, in addition*:

$$F(t)^{-1} \geq \alpha F(0)^{-1}; \quad F(0) > 0; \ \alpha > 0; \ \forall t \geq 0 \tag{3.250}$$

*then*:

$$\lim_{t \to \infty} F(t)\phi(t)\nu(t+1) = \lim_{t \to \infty} [\hat{\theta}(t+1) - \hat{\theta}(t)] = 0 \tag{3.251}$$

$$\lim_{t \to \infty} \|\hat{\theta}(t+k) - \hat{\theta}(t)\| = 0; \quad k < \infty \tag{3.252}$$

$$\|\hat{\theta}(t)\| \leq M_2 < \infty; \quad \forall t \geq 0 \tag{3.253}$$

### Interpretation of the Results

1. $\lim_{t \to \infty} \nu(t+1) = 0$ can be interpreted as the output of the equivalent linear block which asymptotically tend to 0.

2. Since $H(z^{-1})$ is strictly positive real, it is asymptotically stable as well as its inverse. Therefore, its input will also tend to zero, which taking into account the notation, leads to (3.246).
3. Equation (3.247) indicates that the asymptotic variations of the adjustable parameters tend to zero if $F(t) > 0$.
4. Equation (3.248) indicates that the parameter errors remain bounded for all $t$ if $F(t) > 0$ for all $t$.
5. The result of (3.249) is particularly interesting. Its indicates that not only $\nu(t+1)$ goes to zero, but also $\nu(t+1)^2$ multiplied by a factor $\geq 1$. Taking into account the relationship between $\nu(t+1)$ and $\nu^0(t+1)$, one can express this result also in terms of $\nu^0(t+1)$. This result is useful for proving the boundedness of $\phi(t)$ in certain schemes and the convergence of the a priori errors. An alternative form is:
$$\lim_{t\to\infty} \frac{\nu^0(t+1)}{[1+\phi^T(t)F(t)\phi(t)]^{1/2}} = 0 \tag{3.254}$$
6. The condition (3.250) is automatically satisfied for $\lambda_1(t) \equiv 1$ or when $\lambda_1(t)$ and $\lambda_2(t)$ are chosen such that the trace of $F(t)$ has a specified value. For $\lambda_2(t) \equiv 0$, and $\lambda_1(t) \leq 1$, in order to satisfy (3.250), one should have $\lambda_1(t) = 1$ for $t > t_1$ where $t_1$ is determined by $\prod_{i=0}^{t_1} \lambda_1(i) \geq \varepsilon > 0$. When using a *variable forgetting factor*, the condition (3.250) is also satisfied. In this case, $\lim_{t_1\to\infty} \prod_{i=0}^{t_1} \lambda_1(i) \approx e^{-\lambda_0}(\frac{1-\lambda_1(0)}{1-\lambda_0})$ (Ljung and Söderström 1983).

*Remark* Note that $\lim_{t\to\infty} \nu(t+1) = 0$ does not imply $\lim_{t\to\infty} \nu^0(t+1) = 0$ since:
$$\nu(t+1) = \frac{\nu^0(t+1)}{1+\phi^T(t)F(t)\phi(t)}$$

If $\phi(t)$ is unbounded, then $\nu(t+1)$ can be zero with $\nu^0(t+1) \neq 0$. To conclude that $\lim_{t\to\infty} \nu^0(t+1) = 0$, one should show that $\phi(t)$ is bounded (assuming that $F(t)$ is bounded).

*Proof* The proof of Theorem 3.2 is a straightforward application of Theorem C.1, Appendix C, which gives the properties of the feedback connection of a linear system belonging to the class $L(\Lambda)$, i.e., which is strictly input passive when in parallel with a gain $-\frac{1}{2}\Lambda$, and a nonlinear time-varying block which belongs to the class $N(\Gamma)$ which is passive when in feedback connection with a time-varying gain $\frac{1}{2}\Gamma(t)$.

Equations (3.231) through (3.233) and (3.235) define an equivalent feedback system:
$$\nu(t+1) = H(q^{-1})v(t+1) \tag{3.255}$$
$$v(t+1) = -\omega(t+1) \tag{3.256}$$
$$\begin{aligned}\omega(t+1) &= \tilde{\theta}^T(t+1)\phi(t)\\ &= \tilde{\theta}^T(t)\phi(t) + \phi^T(t)F(t)\phi(t)\nu(t+1)\end{aligned} \tag{3.257}$$

The feedback block with input $\nu(t+1)$ and output $\omega(t+1)$, belongs to the class $N(\Gamma)$. Identifying:

$$A(t) = I; \qquad B(t) = F(t)\phi(t); \qquad C(t) = \phi^T(t); \qquad D(t) = \phi^T(t)F(t)\phi(t)$$
$$\bar{x}(t) = \tilde{\theta}(t); \qquad \bar{u}(t) = \nu(t+1); \qquad \bar{y}(t) = \omega(t+1) \tag{3.258}$$

Lemma C.7, Appendix C will be satisfied for:

$$\Gamma(t) = \lambda_2(t)I \tag{3.259}$$

$$P(t) = F(t)^{-1} \tag{3.260}$$

$$P(t+1) = F(t+1)^{-1} = \lambda_1(t)F(t)^{-1} + \lambda_2(t)\phi(t)\phi^T(t) \tag{3.261}$$

$$Q(t) = [1-\lambda_1(t)]\,F(t)^{-1} \tag{3.262}$$

$$S(t) = [1-\lambda_1(t)]\,\phi(t) \tag{3.263}$$

$$R(t) = [2-\lambda_1(t)]\,\phi^T(t)F(t)\phi(t) \tag{3.264}$$

The condition of (C.38) can be alternatively expressed by:

$$[\bar{x}^T(t)\ \bar{u}^T(t)]\begin{bmatrix} Q(t) & S(t) \\ S^T(t) & R(t) \end{bmatrix}\begin{bmatrix} \bar{x}(t) \\ \bar{u}(t) \end{bmatrix}$$
$$= [\bar{x}(t) + S(t)\bar{u}(t)]^T Q^{-1}(t)[\bar{x}(t) + S(t)\bar{u}(t)]$$
$$+ \bar{u}(t)[R(t) - S^{-1}(t)Q^{-1}(t)S(t)]\bar{u}(t) \geq 0 \tag{3.265}$$

which is equivalent to the condition:

$$R(t) - S^T(t)Q^{-1}(t)S(t) \geq 0 \tag{3.266}$$

Using (3.262) through (3.264), (3.266) becomes:

$$\phi^T(t)F(t)\phi(t) \geq 0 \tag{3.267}$$

which is always satisfied since $F(t) \geq 0$.

The linear block with input $\upsilon(t+1) = u(t)$ and output $\nu(t+1) = y(t)$ belongs to the class $L(\Lambda)$ for:

$$\Lambda = \lambda_2 \geq \max_t \lambda_2(t) \tag{3.268}$$

Therefore, the condition (C.39) of Theorem C.1:

$$\Lambda - \Gamma(t) \geq 0 \tag{3.269}$$

is satisfied and all the conclusions of Theorem C.1 translate in the results (3.244) through (3.248) of Theorem 3.2 taking into account the meaning of $\upsilon(t+1)$, $\nu(t+1)$, $\tilde{\theta}(t+1)$, $\omega(t+1)$.

Equation (3.249) is also obtained from (C.47) taking into account that a term of the form $\sum_{t=0}^{t_1} y^T(t)\Delta y(t)$ is added in the left hand side when $H'(z^{-1})$ is output strictly passive. This term which has the specific form $\delta \sum_{t=0}^{t_1} \nu^2(t+1), \delta > 0$ together with $\sum_{t=0}^{t_1} \phi^T(t)F(t)\phi(t)\nu^2(t+1)$ resulting from (3.266) yields the result.

The result of (3.252) is obtained from (3.251) by using the Schwartz inequality and the fact that $k$ is finite:

$$\|\hat{\theta}(t+k) - \hat{\theta}(t)\|^2 \leq k \sum_{i=1}^{k} \|\hat{\theta}(t+i) - \hat{\theta}(t+i-1)\|^2$$

□

The transformed equivalent feedback associated to Theorem 3.2 is shown in Fig. 3.14b. The equivalent feedback path has a local negative feedback with a gain $\frac{\lambda}{2} \geq \max_t \frac{\lambda_2(t)}{2}$, which makes the transformed feedback path passive. However, in order that the whole feedback system remains unchanged, a gain $-\frac{\lambda_2}{2}$ should be added in parallel to the feedforward path and, therefore, the equivalent transformed linear path will be characterized by the transfer function $[H(z^{-1}) - \frac{\lambda_2}{2}]$.

The same positive real condition of Theorem 3.2 holds for the case of "integral + proportional" with time-varying adaptation gain, i.e., for:

$$\hat{\theta}_I(t+1) = \hat{\theta}_I(t) + F_I(t)\phi(t)\nu(t+1) \tag{3.270}$$

$$\hat{\theta}_p(t+1) = F_p(t)\phi(t)\nu(t+1) \tag{3.271}$$

$$F_I(t+1)^{-1} = \lambda_1(t)F_I(t)^{-1} + \lambda_2(t)\phi(t)\phi^T(t) \tag{3.272}$$

$$F_p(t) = \alpha(t)F_I(t); \quad \alpha(t) > -0.5 \tag{3.273}$$

$$\hat{\theta}(t+1) = \hat{\theta}_I(t+1) + \hat{\theta}_p(t+1) \tag{3.274}$$

See Landau (1979).

Note that positive values of the proportional gain related to the magnitude of $\phi$ allows to relax the positive real condition. See Tomizuka (1982) and Sect. 3.3.5.

### *3.3.5 Removing the Positive Real Condition*

In a number of cases, it may be interesting to remove the positive real condition upon the equivalent feedforward path required for stability (for example in system identification, if, like in the output error algorithm, this condition depends upon an unknown transfer function).

Removing the positive real condition will require a modification of the adjustable predictor or use of adjustable compensators on the adaptation error and this will imply an augmentation of the number of adjustable parameters.

We will illustrate this approach for the case of the "output error" algorithm presented in the Sect. 3.3.2.

The model of the plant is described by:

$$y(t+1) = -A^* y(t) + B^* u(t) = \theta^T \varphi(t) \tag{3.275}$$

where:

$$\theta^T = [a_1, \ldots, a_{n_A}, b_1, \ldots, b_{n_B}] \tag{3.276}$$

$$\varphi^T(t) = [-y(t), \ldots, -y(t-n_A+1), u(t), \ldots, u(t-n_B+1)] \tag{3.277}$$

The standard "output error predictor" has the form:

$$\hat{y}^0(t+1) = \hat{\theta}^T(t)\phi(t) \tag{3.278}$$

$$\hat{y}(t+1) = \hat{\theta}^T(t+1)\phi(t) \tag{3.279}$$

where:

$$\hat{\theta}^T(t) = [\hat{a}_1(t), \ldots, \hat{a}_{n_A}(t), \hat{b}_1(t), \ldots, \hat{b}_{n_B}(t)] \tag{3.280}$$

$$\phi^T(t) = [-\hat{y}(t), \ldots, -\hat{y}(t-n_A+1), u(t), \ldots, u(t-n_B+1)] \tag{3.281}$$

Using (3.134) or (3.135), the a posteriori output prediction error can be expressed as:

$$\varepsilon(t+1) = -A^*\varepsilon(t) + \phi^T(t)[\theta - \hat{\theta}(t+1)] \tag{3.282}$$

and the positive real condition will come from the presence of the term $-A^*\varepsilon(t)$. The idea in the *output error with extended prediction model* is to add a term in (3.278) and (3.279) with adjustable parameters such that the a posteriori prediction error takes the form:

$$\varepsilon(t+1) = \phi_e^T(t)[\theta_e - \hat{\theta}_e(t+1)] \tag{3.283}$$

where $\phi_e(t)$ and $\hat{\theta}_e(t)$ are "extended" observation and parameters vectors.

#### Output Error with Extended Prediction Model

From (3.279) and (3.283), it results that the *extended output error* adjustable predictor should have the structure:

$$\hat{y}^0(t+1) = \hat{\theta}^T(t)\phi(t) + \sum_{i=1}^{n_A} \hat{c}_i(t)\varepsilon(t+1-i) = \hat{\theta}_e^T(t)\phi_e(t) \tag{3.284}$$

$$\hat{y}(t+1) = \hat{\theta}(t)\phi(t) + \sum_{i=1}^{n_A} \hat{c}_i(t+1)\varepsilon(t+1-i) = \hat{\theta}_e^T(t+1)\phi_e(t) \tag{3.285}$$

where:

$$\hat{\theta}_e^T(t) = [\hat{\theta}^T(t), \hat{c}_1(t), \dots, \hat{c}_{n_A}(t)] \tag{3.286}$$

$$\phi_e^T(t) = [\phi^T(t), \varepsilon(t), \dots, \varepsilon(t - n_A + 1)] \tag{3.287}$$

Subtracting now (3.285) from (3.275), one gets:

$$\begin{aligned}
\varepsilon(t+1) &= y(t+1) - \hat{y}(t+1) \\
&= \phi^T(t)[\theta - \hat{\theta}(t+1)] + \sum_{i=1}^{n_A}[-a_i - \hat{c}_1(t+1)]\varepsilon(t+1-i) \\
&= \phi_e^T(t)[\theta_e - \hat{\theta}_e(t+1)]
\end{aligned} \tag{3.288}$$

with:

$$\theta_e = [a_1, \dots, a_{n_A}, b_1, \dots, b_{n_B}, -a_1, \dots, -a_{n_A}] \tag{3.289}$$

By applying Theorem 3.1 directly, one concludes that $\lim_{t\to\infty} \varepsilon(t+1) = 0$ when using the adaptation algorithm of (3.231) through (3.233) in which $\hat{\theta}(t)$, $\phi(t)$, $\nu(t)$ are replaced by $\hat{\theta}_e(t)$, $\phi_e(t)$, $\varepsilon(t)$ given by (3.284), (3.285) and (3.286) and without any positive real condition to be satisfied. For similar solutions see also Sect. 5.5, Anderson and Landau (1994) and Shah and Franklin (1982).

### Signal Dependent Condition

A direction for relaxing the positive real condition is to use the "excess" of passivity of the equivalent feedback path which depends upon the magnitude of $\phi(t)$ (in particular, by adding a proportional adaptation gain) and transferring it to the feedforward path, taking into account the fact that for SISO linear discrete-time systems there is always a negative feedback gain which makes the system strictly positive real. One has the following result (Tomizuka 1982):

**Theorem 3.3** *Consider the "output error" algorithm where the adjustable predictor is given by* (3.278) *and* (3.279), *and the PAA by* (3.204) *through* (3.206) *with* $\nu(t+1) = \varepsilon(t+1) = y(t+1) - \hat{y}(t+1)$. *Then* $\lim_{t\to\infty} \varepsilon(t+1) = 0$ *provided that*:

(1) *The adaptation gains $F_I$ and $F_P$ and the observation vector $\phi(t)$ satisfy the condition*:

$$\phi^T(t)\left(F_P + \frac{1}{2}F_I\right)\phi(t) > K > 0; \quad \forall t \tag{3.290}$$

(2) *The gain $K$ is selected such that*:

$$\frac{1}{1+K}\sum_{i=1}^{n_A}|a_i| < 1 \tag{3.291}$$

*Remarks*

(1) Condition 2 implies that the transfer function $\frac{1}{A(z^{-1})}$ in feedback with the gain $K$ is strictly positive real.
(2) Theorem 3.2 can be generalized for the case of time varying adaptation gains.

*Proof* Using the following lemma (Tomizuka 1982):

**Lemma 3.3** *The discrete transfer function* $\frac{1}{A(z^{-1})}$ *is strictly positive real if*

$$\sum_{i=1}^{n_A} |a_i| < 1$$

It results that $\frac{1}{A(z^{-1})}$ in feedback with a gain $K$ will be SPR if:

$$\frac{1}{1+K}\sum_{i=1}^{n_A} |a_i| < 1 \tag{3.292}$$

This also implies that the equivalent feedforward path characterized by the transfer function $\frac{1}{A(z^{-1})}$ satisfies the property:

$$\sum_{t=0}^{t_1} u_1(t)y_1(t) \geq -\gamma_1^2 - \sum_{t=0}^{t_1} K\varepsilon^2(t+1) \tag{3.293}$$

The input-output product equivalent feedback path can be expressed as:

$$\begin{aligned}\sum_{t=0}^{t_1} u_2(t)y_2(t) &= \sum_{t=0}^{t_1} \tilde{\theta}^T(t+1)\phi(t)\varepsilon(t+1)\\ &= \frac{1}{2}\tilde{\theta}_I^T(t+1)F_I^{-1}\tilde{\theta}_I(t+1) - \frac{1}{2}\tilde{\theta}_I^T(0)F_I^{-1}\tilde{\theta}(0)\\ &\quad + \sum_{t=0}^{t_1}\phi^T(t)\left(F_P + \frac{1}{2}F_I\right)\phi(t)\varepsilon^2(t+1)\end{aligned} \tag{3.294}$$

Taking into account the feedback connection, one obtains from (3.293) and (3.294):

$$\lim_{t_1\to\infty}\sum_{t=0}^{t_1}\left[\phi^T(t)\left(F_P + \frac{1}{2}F_I\right)\phi(t) - K\right]\varepsilon^2(t+1) \leq \gamma_1^2 + \frac{1}{2}\tilde{\theta}_I^T(0)F_I^{-1}\tilde{\theta}_I(0) \tag{3.295}$$

and the result of Theorem 3.2 follows. □

Another "signal dependent" approach for relaxing the positive real condition is considered in Anderson et al. (1986) using "averaging" techniques of analysis (ap-plicable for slow adaptation i.e. small adaptation gains). This approach will be used in Chap. 15.

## 3.4 Parametric Convergence

### 3.4.1 The Problem

As will be shown, the convergence toward zero of the adaptation or prediction error does not imply in every case that the estimated parameters will converge toward the true parameters. The objective will be to determine under what conditions the convergence of the adaptation (prediction) error will imply the convergence toward the true parameters.

We will make the hypothesis that such a value of parameter vector exists, i.e.,

$$\exists\theta; \quad \nu(t+1)\mid_{\hat{\theta}=\theta} = 0$$

This is based on the assumption that the structure of the adjustable model is identical to that of the system and that the orders $n_A$, $n_B$ of the adjustable model are equal or higher than those of the system model.

A first indication concerning the parametric convergence is provided by the equivalent feedback representation of the PAA.

Let us consider as an example the EFR associated to the *improved gradient algorithm* (corresponding to the RLS with $\lambda_1(t) \equiv 1$ and $\lambda_2(t) \equiv 0$) shown in Fig. 3.15a.

This EFR can alternatively be represented as shown in Fig. 3.15b, where the feedforward block corresponds to an integrator (diagonal matrix) whose output is the parameter error vector, and the feedback path is a time-varying block characterized by the gain $F\phi(t)\phi^T(t)$.

Since the equivalent linear block in Fig. 3.15b is an integrator, in order to have $\lim_{t\to\infty}\tilde{\theta}(t+1) = 0$, the gain of the feedback path should be positive in the average (one transforms the integrator in a first order time-varying system). Taking into account that $F > 0$ the above requirement translates in:

$$\lim_{t_1\to\infty}\frac{1}{t_1}\sum_{t=1}^{t_1}\phi(t)\phi^T(t) > 0 \tag{3.296}$$

The observation vectors $\phi(t)$ satisfying condition (3.296) are called *persistently exciting signals of order $n$*, where $n$ is the dimension of $\phi(t)$.

In order to illustrate the influence of the excitation signal for the parametric convergence, let us consider the discrete-time system model described by:

$$y(t+1) = -a_1 y(t) + b_1 u(t) \tag{3.297}$$

and consider an estimated model described by:

$$\hat{y}(t+1) = -\hat{a}_1 y(t) + \hat{b}_1 u(t) \tag{3.298}$$

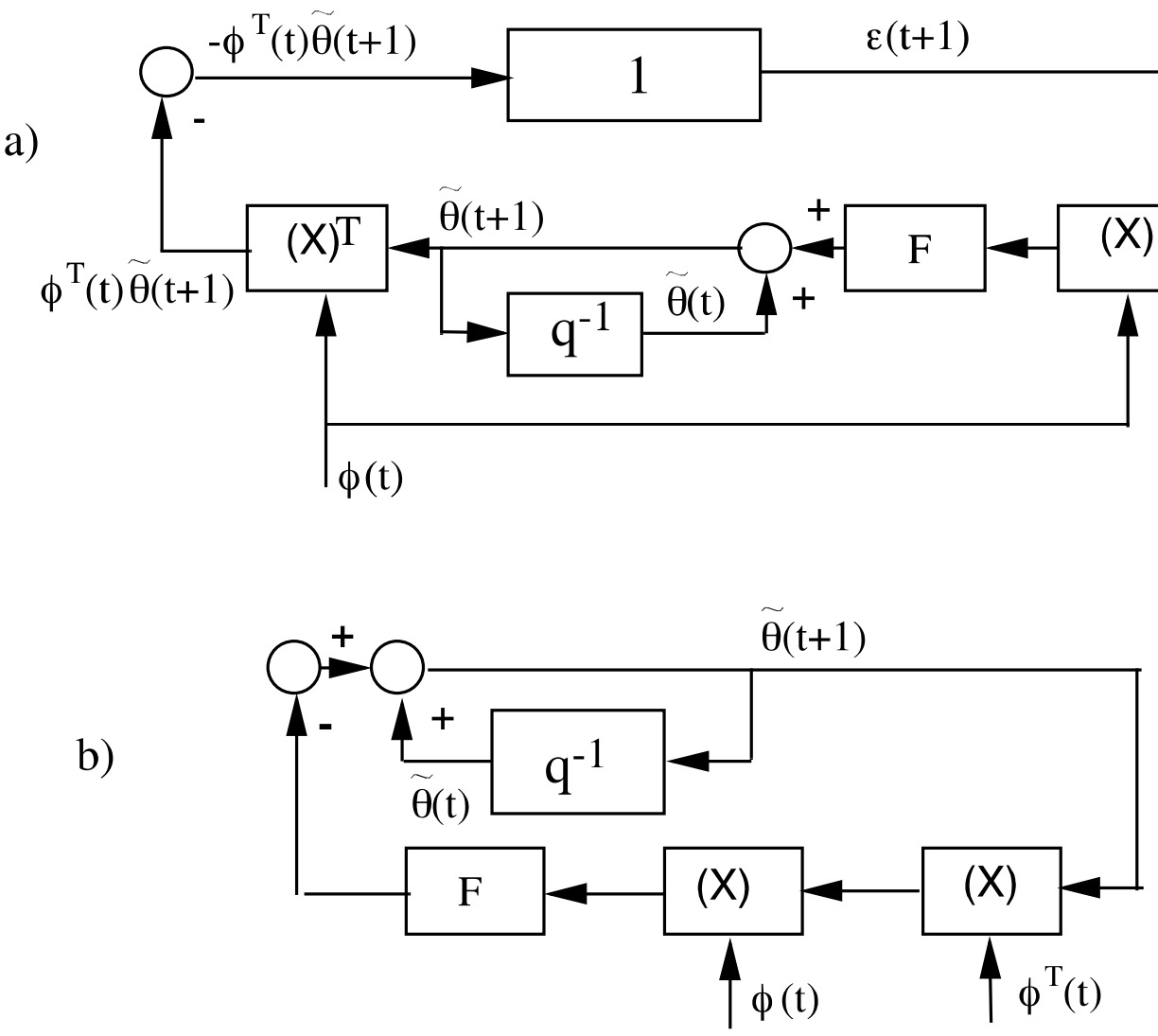

**Fig. 3.15** Equivalent feedback representation of the improved gradient PAA, (**a**) standard configuration, (**b**) equivalent transformation

in which $\hat{y}(t+1)$ is the output predicted by the estimation model with the constant parameters $\hat{a}_1, \hat{b}_1$.

Now assume that $u(t) = \text{constant}$ and that the parameters $a_1, b_1, \hat{a}_1, \hat{b}_1$ verify the following relation:

$$\frac{b_1}{1+a_1} = \frac{\hat{b}_1}{1+\hat{a}_1} \tag{3.299}$$

i.e., the steady state gains of the system and of the estimated model are equal even if $\hat{b}_1 \neq b_1$ and $\hat{a}_1 \neq a_1$. Under the effect of the constant input $u(t) = u$, the plant output will be given by:

$$y(t+1) = y(t) = \frac{b_1}{1+a_1} u \tag{3.300}$$

and the output of the estimated prediction model will be given by:

$$\hat{y}(t+1) = \hat{y}(t) = \frac{\hat{b}_1}{1+\hat{a}_1} u \tag{3.301}$$

However taking into account (3.299), it results that:

$$\begin{aligned} \varepsilon(t+1) &= y(t+1) - \hat{y}(t+1) = 0 \\ &\text{for } u(t) = \text{const};\ \hat{a}_1 \neq a_1;\ \hat{b}_1 \neq b_1 \end{aligned} \tag{3.302}$$

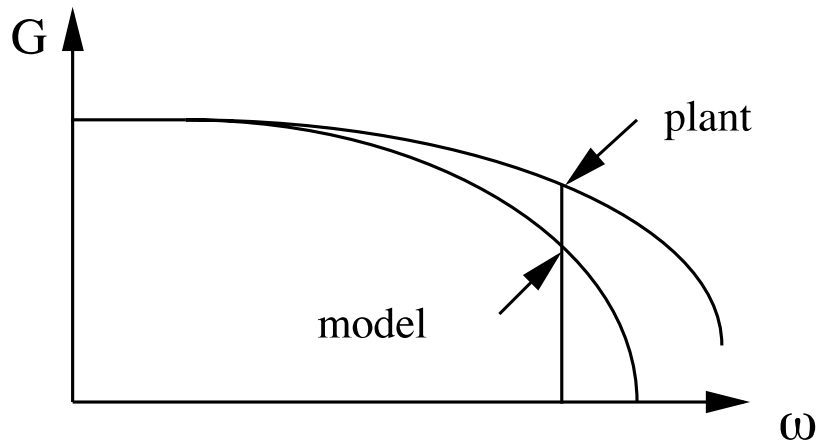

**Fig. 3.16** Gain frequency characteristics of two systems with the same steady state gain

It can thus be concluded from this example that the application of a constant input does not allow to distinguish the two models, since they both have the same steady state gain.

If the frequency characteristics of both systems are represented, they will superpose each other at zero frequency and the difference between them will appear for frequencies other than zero since the poles of the two systems are different. Such a situation is shown in Fig. 3.16. Figure 3.16 indicates that in order to highlight the difference between the two models (i.e., between the parameters) a signal $u(t) = \sin \omega t (\omega \neq 0)$ must be applied instead of signal $u(t) = \text{const}$.

Let us analyze the phenomenon in more detail. From (3.297) and (3.298), one obtains:

$$\begin{aligned}\varepsilon(t+1) &= y(t+1) - \hat{y}(t+1) \\ &= (a_1 - \hat{a}_1)y(t) + (b_1 - \hat{b}_1)u(t) = 0\end{aligned} \tag{3.303}$$

From (3.297), $y(t)$ can be expressed as a function of $u(t)$ using the system transfer operator:

$$y(t) = \frac{b_1 q^{-1}}{1 + a_1 q^{-1}} u(t) \tag{3.304}$$

Introducing the expression $y(t)$ given by (3.304) in (3.303) and after multiplying by $(1 + a_1 q^{-1})$, one obtains:

$$\begin{aligned}\varepsilon(t+1) &= [(a_1 - \hat{a}_1)b_1 q^{-1} + (b_1 - \hat{b}_1)(1 + a_1 q^{-1})]u(t) \\ &= [(b_1 - \hat{b}_1) + q^{-1}(b_1 \hat{a}_1 - a_1 \hat{b}_1)]u(t) = 0\end{aligned} \tag{3.305}$$

We are concerned with finding the characteristics of $u(t)$ so that (3.305) implies zero parametric errors. Denoting:

$$b_1 - \hat{b}_1 = \alpha_0; \qquad b_1 \hat{a}_1 - a_1 \hat{b}_1 = \alpha_1 \tag{3.306}$$

Equation (3.305) is thus written as:

$$(\alpha_0 + \alpha_1 q^{-1})u(t) = 0 \tag{3.307}$$

which is a difference equation having a solution of the discretized exponential type.

Let us take:

$$u(t) = z^t = e^{sT_s t} \tag{3.308}$$

where $T_s$ is the sampling period. Equation (3.307) is then written:

$$(\alpha_0 + z^{-1}\alpha_1)z^t = (z\alpha_0 + \alpha_1)z^{t-1} = 0 \tag{3.309}$$

and it will be verified for $z$, which is the solution of the characteristic equation:

$$z\alpha_0 + \alpha_1 = 0 \tag{3.310}$$

One obtains:

$$z = -\frac{\alpha_1}{\alpha_0} = e^{\sigma T_s}; \quad \sigma = \text{real}; \left(\frac{\alpha_1}{\alpha_0} < 0\right) \tag{3.311}$$

and the nonperiodic solution:

$$u(t) = e^{\sigma T_s t} \tag{3.312}$$

leads to the verification of (3.307) and (3.305) respectively without $\hat{b}_1 = b_1$ and $\hat{a}_1 = a_1$. Indeed the signal $u(t) = \text{const}$ previously considered, corresponds to $\sigma = 0$, i.e., $-\alpha_1 = \alpha_0$. However:

$$\begin{aligned} -\alpha_1 = \alpha_0 \quad &\Longrightarrow \quad b_1 - \hat{b}_1 = a_1\hat{b}_1 - b_1\hat{a}_1 \\ &\Longrightarrow \quad \frac{b_1}{1 + a_1} = \frac{\hat{b}_1}{1 + \hat{a}_1} \end{aligned} \tag{3.313}$$

In conclusion, if $u(t) = \text{const}$, only the steady state gain of the system is correctly estimated. In order to correctly estimate the system model parameters, $u(t)$ must thus be found such that $\varepsilon(t) = 0$ implies $\hat{b}_1 = b_1$ and $\hat{a}_1 = a_1$. This will be obtained if $u(t)$ is not a possible solution of (3.307).

Let us:

$$u(t) = e^{j\omega T_s t} \quad \text{or} \quad e^{-j\omega T_s t} \tag{3.314}$$

For $u(t) = e^{j\omega^{T_s} t}$, (3.307) becomes:

$$(e^{j\omega T_s}\alpha_0 + \alpha_1)e^{j\omega T_s(t-1)} = 0 \tag{3.315}$$

Since $\alpha_0$ and $\alpha_1$ are real, $e^{j\omega^{T_s} t}$ cannot be a root of the characteristic equation (3.315) and therefore $\varepsilon(t + 1) = 0$ will be obtained only if:

$$\alpha_0 = \alpha_1 = 0 \quad \Longrightarrow \quad \hat{b}_1 = b_1; \quad \hat{a}_1 = a_1 \tag{3.316}$$

It was this type of input that was previously proposed ($\sin \omega t = (e^{j\omega t} - e^{-j\omega t})/2j$) when the frequency characteristics of the two models were examined. A non zero frequency sinusoid is thus required in order to identify two parameters. The signal $u(t)$ which in this case is a sinusoid, is a *persistently exciting signal* of order 2 (allowing to estimate 2 parameters).

We are interested next in the characterization of the *persistently exciting signals*.

### 3.4.2 Persistently Exciting Signals

In the general case, one considers $u(t)$ signals bounded in average verifying the property:

$$\lim_{t_1 \to \infty} \frac{1}{t_1} \sum_{t=1}^{t_1} u^2(t) < \infty \tag{3.317}$$

Defining $\varphi^T(t) = [u(t), u(t-1), \ldots, u(t-n+1)]$, the signal $u(t)$ is said to be a *persistently exciting of order n* if:

$$\lim_{t_1 \to \infty} \frac{1}{t_1} \sum_{t=1}^{t_1} \varphi(t)\varphi^T(t) > 0$$

This corresponds in fact to the condition (3.296) if we are concerned with the estimation of the parameters of a FIR model.

One has the following result (Åström and Wittenmark 1995):

**Theorem 3.4** *$u(t)$ is a persistently exciting signal of order n if:*

$$\lim_{t_1 \to \infty} \frac{1}{t_1} \left[ \sum_{t=1}^{t_1} L(q^{-1})u(t) \right]^2 > 0 \tag{3.318}$$

*for all nonzero polynomials $L(q^{-1})$ of order $n-1$.*

This theorem is nothing else that a generalization of the example considered in Sect. 3.4.1. In this example, $L(q^{-1})$ was a polynomial of order 1 (formed by differences between true and estimated parameters) and we have searched a persistently exciting signal $u(t)$ of order 2. The resulting signal was a sinusoid of nonzero frequency. It results that a sum of $n/2$ sinusoids of distinct frequencies is a persistently exciting signal of order $n$. Effectively, in the case of the estimation of $n$ parameters, $\varepsilon(t+1) = 0$ leads to an equation $L(q^{-1})u(t) = 0$ where $L(q^{-1})$ is a polynomial of order $n-1$ whose coefficients depend upon the difference between the estimated parameters and the true parameters. Depending on whether $(n-1)$ is odd or even, the equation $L(q^{-1})u(t) = 0$ admits as solution a sum of $p$ sinusoids of distinct frequencies.

$$u(t) = \sum_{t=1}^{p} \sin \omega_i T_s t$$

with $p \leq (n-1)/2$. This leads to the conclusion that a sum of $p$ sinusoids of distinct frequencies with $p \geq n/2$ ($n =$ even) or $p \geq (n+1)/2$ ($n =$ odd) cannot be a solution of the equation $L(q^{-1})u(t) = 0$, and therefore:

$$p = \frac{n}{2}; \quad \forall n = \text{even}$$

$$p = \frac{n+1}{2}; \quad \forall n = \text{odd}$$

leads to a persistent excitation signal of order $n$, allowing to estimate $n$ parameters.

The persistently exciting signals have a frequency interpretation which is more general than a sum of sinusoids of distinct frequencies. Using the Parseval theorem, (3.318) becomes:

$$\lim_{t_1 \to \infty} \frac{1}{t_1} \sum_{t=1}^{t_1} [L(q^{-1})u(t)]^2 = \frac{1}{2\pi} \int_{-\pi}^{\pi} \mid L(e^{j\omega}) \mid^2 \phi_u(\omega) d\omega > 0$$

where $\phi_u(\omega)$ is the spectral density of the input signal. Since $L(q^{-1})$ is of order $n-1$, $L(e^{j\omega}) = 0$ at most $n-1$ points between $[-\pi, \pi]$. If $\phi_u(\omega) \neq 0$ for at least $n$ points in the interval $-\pi \leq \omega \leq \pi$, $u(t)$ is a persistently exciting signal of order $n$.

A signal whose spectrum is different from zero over all the interval $0 \leq f \leq 0.5 f_s$ is a persistently exciting signal of any order. Such an example is the discrete Gaussian white noise which has a constant spectral density between 0 and $0.5 f_s$.

In practice, one uses for system identification *pseudo-random binary sequences* (PRBS) which approach the characteristic of a white noise. Their main advantage is that they have a constant magnitude (the stochastic character coming from the pulse width variation) which permits the precise definition of the level of instantaneous stress on the process (or actuator). The PRBS will be examined in detail in Chap. 5.

### *3.4.3 Parametric Convergence Condition*

For the parameter estimation of a model of the form:

$$y(t+1) = -A^* y(t) + B^* u(t) = \theta^T \varphi(t) \tag{3.319}$$

where:

$$\theta^T = [a_1, \ldots, a_{n_A}, b_1, \ldots, b_{n-B}] \tag{3.320}$$

$$\varphi^T(t) = [-y(t), \ldots, y(t - n_A + 1), u(t), \ldots, u(t - n_B + 1)] \tag{3.321}$$

using an *equation error* type adjustable predictor:

$$\hat{y}(t+1) = \hat{\theta}^T(t+1)\phi(t) = \hat{\theta}^T(t+1)\varphi(t) \tag{3.322}$$

one has the following result:

**Theorem 3.5** *Given the system model described by* (3.319) *and using an adjustable predictor of the form of* (3.322), *the parameter convergence, i.e.*

$$\lim_{t \to \infty} \hat{a}_1(t) = a_i; \quad i = 1, \ldots, n_A$$

$$\lim_{t\to\infty} b_i(t) = b_i; \quad i = 1, \ldots, n_B$$

*is assured if*:

(1) *One uses a PAA which assures*

$$\lim_{t\to\infty} \varepsilon(t+1) = \lim_{t\to\infty} [y(t+1) - \hat{y}(t+1)] = 0$$

(2) *The orders* $n_A$ *and* $n_B$ *are known exactly.*
(3) *The plant model to be identified is characterized by an irreducible transfer function in* $z^{-1}$ (*i.e.*, $A(q^{-1})$ *and* $B(q^{-1})$ *are relatively prime*).
(4) *The input* $u(t)$ *is a persistently exciting signal of order* $n = n_A + n_B$.

*Proof* The proof is a generalization of the example considered in Sect. 3.4.1 combined with the definition and properties of persistently exciting signals.

Under the assumption that $\lim_{t\to\infty} \varepsilon(t+1) = 0$, the adjustable predictor becomes asymptotically a fixed predictor described by:

$$y(t+1) = \hat{\theta}^T \varphi(t) \tag{3.323}$$

where:

$$\hat{\theta}^T = [\hat{a}_1, \ldots, \hat{a}_{n_A}, \hat{b}_1, \ldots, \hat{b}_{n_B}] \tag{3.324}$$

contains the final values of the estimated parameters ($\hat{a}_i = \lim_{t\to\infty} \hat{a}_i(t)$, $\hat{b}_i = \lim_{t\to\infty} \hat{b}_i(t)$), and one has:

$$\lim_{t\to\infty} \varepsilon(t+1) = -\left(\sum_{i=1}^{n_A} (a_i - \hat{a}_i) q^{-i+1}\right) y(t) + \left(\sum_{i=1}^{n_B} (b_i - \hat{b}_i) q^{-i+1}\right) u(t) = 0 \tag{3.325}$$

Taking into account assumption (3) and replacing $y(t)$ by:

$$y(t) = \frac{B(q^{-1})}{A(q^{-1})} u(t) \tag{3.326}$$

Equation (3.325) becomes:

$$\left[-\left(\sum_{i=1}^{n_A} (a_i - \hat{a}_i) q^{-i+1}\right) B(q^{-1}) + \left(\sum_{i=1}^{n_B} (b_i - \hat{b}_i) q^{-i+1}\right) A(q^{-1})\right] u(t)$$
$$= \left(\sum_{i=0}^{n_A+n_B-1} \alpha_i q^{-i}\right) u(t) = 0 \tag{3.327}$$

with:

$$\alpha_0 = \Delta b_1 \tag{3.328}$$

$$[\alpha_1, \ldots, \alpha_{n_A+n_B-1}] = \Delta b_1[a_1, \ldots, a_{n_A}, 0, \ldots, 0] + [\Delta b_2, \ldots, \Delta b_{n_B}, -\Delta a_{n_A}, \ldots, -\Delta a_1]R(a_i, b_i) \tag{3.329}$$

and condition (4) assures the desired persistence of excitation. □

Further insight on the influence of the input characteristics upon the convergence of various adaptive schemes can be found in Anderson and Johnson (1982).

## 3.5 Concluding Remarks

In this chapter we have presented discrete-time *parameter adaptation algorithms* (PAA) and we have examined their properties in a deterministic environment.

We wish to emphasize the following basic ideas:

1. The PAA has in general the following recursive form (integral adaptation):

$$\hat{\theta}(t+1) = \hat{\theta}(t) + F(t)\phi(t)\nu(t+1)$$

   where $\hat{\theta}(t)$ is the adjustable parameter vector. At each step the correcting term is formed by the product of the adaptation error $\nu(t+1)$, the observation vector $\phi(t)$ and the adaptation gain matrix $F(t)$. The adaptation error $\nu(t+1)$ is computed from the measurements up to and including $t+1$ and the estimated parameters up to $t$. The above form of the PAA features memory and it can be interpreted as the state equation of an integrator with state $\hat{\theta}(t)$ and input $F(t)\phi(t)\nu(t+1)$. Other forms of PAA involving more general state equations have also been presented (integral + proportional adaptation, adaptation with leakage, adaptation for time-varying parameters).
2. Several approaches can be used for the derivation of PAA among which we have considered:

   - recursive minimization of a criterion in term of the adaptation error;
   - transformation of an off-line parameter estimation into a recursive parameter estimation;
   - rapprochement with the Kalman Filter;
   - stability considerations.

   However since the resulting system is nonlinear, a stability analysis is mandatory.
3. An equivalent feedback system (EFR) can be associated with the PAA in the cases where the adaptation error equation features the parameter error explicitly. The use of the EFR simplifies drastically the stability analysis (or synthesis) via the use of the properties of passive systems connected in feedback.
4. For general classes of adaptation error equations and PAA, stability conditions for the resulting adaptive systems have been established.
5. A variety of choices for the adaptation gain are possible. The choice depends upon the specific application.

## 3.6 Problems

**3.1** Show that least squares criterion (3.31) can be expressed as:

$$\min_{\hat{\theta}} J(t) = \frac{1}{N} \|Y(t) - R(t)\hat{\theta}\|^2$$

where:

$$Y^T(t) = [y(t), y(t-1), \ldots, y(1)]$$
$$R(t) = [Y(t-1), Y(t-2), \ldots, Y(t-n_A), U(t-1), U(t-2), \ldots, U(t-n_B)]$$
$$U^T(t) = [u(t), u(t-1), \ldots, u(1)]$$

Develop the least squares algorithm for this formulation of the criterion.

**3.2** Find the off-line and the recursive parameter adaptation algorithm which minimizes criterion (3.56).

**3.3** Show that the PAA with fixed and variable forgetting factor are obtained when minimizing (3.69) and (3.72) respectively.

**3.4** Show that the PAA with constant trace is obtained when minimizing (3.75).

**3.5** Consider the improved gradient algorithm in which (3.29) is replaced by:

$$\hat{\theta}(t+1) = \hat{\theta}(t) + \frac{\gamma \phi(t) \varepsilon^0(t+1)}{\alpha + \phi^T(t)\phi(t)}$$

Show that asymptotic stability of the adaptation error is assured for: $0 < \gamma < 2$, $\alpha > 0$.

**3.6** Establish the stability conditions when using an output error adjustable predictor ((3.128) and (3.129)) with the stochastic approximation scalar adaptation gain (3.84).

**3.7** Develop a parameter adaptation algorithm and establish the stability conditions for an output error adaptive predictor using filtered inputs and outputs, i.e.,

- Plant model: $y(t+1) = -A^*(q^{-1})y(t) + B^*(q^{-1})u(t)$
- Filtered inputs and outputs: $L(q^{-1})y_f(t) = y(t)$; $L(q^{-1})u_f(t) = u(t)$
- A priori adjustable predictor: $\hat{y}_f^0(t+1) = -\hat{A}^*(t, q^{-1})\hat{y}_f(t) + \hat{B}^*(t, q^{-1})u_f(t)$
- A posteriori adjustable predictor:

$$\hat{y}_f(t+1) = -\hat{A}^*(t+1, q^{-1})\hat{y}_f(t) + \hat{B}^*(t+1, q^{-1})u_f(t)$$

- Prediction errors:

$$\varepsilon_f^0(t+1) = y_f(t+1) - \hat{y}_f^0(t+1); \qquad \varepsilon_f(t+1) = y_f(t+1) - \hat{y}_f(t+1)$$

**3.8** What will be the parameter adaptation algorithm and the stability conditions if in the output error adaptive predictor one replaces (3.128) and (3.129) by:

$$\hat{y}^0(t+1) = \hat{\theta}^T(t)\phi(t) + \sum_{i=1}^{n_A} l_i \varepsilon(t+1-i)$$

$$\hat{y}(t+1) = \hat{\theta}^T(t+1)\phi(t) + \sum_{i=1}^{n_A} l_i \varepsilon(t+1-i)$$

**3.9** Give the details of the "Integral + Proportional" parameter adaptation algorithm when it is used with a least squares type adjustable predictor or with an output error adjustable predictor.

**3.10** Develop a PAA and establish the stability condition for an output error adaptive predictor where (3.159) is replaced by:

$$\nu(t+1) = \varepsilon(t+1) + \sum_{i=1}^{n_A} \hat{d}_i(t+1)\varepsilon(t+1-i)$$

**3.11** Find the positivity domain in the parameter plane $a_1 - a_2$ for the discrete-time transfer function:

$$H(z^{-1}) = \frac{1 + d_1 z^{-1} + d_2 z^{-2}}{1 + a_1 z^{-1} + a_2 z^{-2}}$$

for the following cases:

(a) $d_1 = d_2 = 0$

(b) $d_1 = 0.5, \qquad d_2 = 0$

(c) $d_1 = 0.5, \qquad d_2 = d_1 - 1$

**3.12** Give conditions for the selection of $d_i$:

(a) $d_1 \neq 0, \qquad d_2 = 0$

(b) $d_1 \neq 0, \qquad d_2 \neq 0$

such that the transfer function from Problem 3.1 be positive real.

**3.13** Use a state space realization of $H(z^{-1})$ in Problem 3.1 and the positive real lemma (Appendix C, Lemma C.3) in order to compute $d_i$ for obtaining a positive real transfer function.

Hint: Denoting the elements of $P = [P_{ij}]$ select $Q$ such that $P_{nn} \leq 2$.

# Chapter 4
# Parameter Adaptation Algorithms—Stochastic Environment

## 4.1 Effect of Stochastic Disturbances

Consider the estimation of a plant model disturbed by a stochastic process which can be represented by:

$$\begin{aligned} y(t+1) &= -A^*(q^{-1})y(t) + B^*(q^{-1})u(t) + w(t+1) \\ &= \theta^T \phi(t) + w(t+1) \end{aligned} \tag{4.1}$$

where $u$ is the input, $y$ is the output, $w$ is a zero mean stationary stochastic disturbance with finite moments and:

$$\theta^T = [a_1, \dots, a_{n_A}, b_1, \dots, b_{n_B}] \tag{4.2}$$

$$\phi^T(t) = [-y(t), \dots, -y(t-n_A+1), u(t), \dots, u(t-n_B+1)] \tag{4.3}$$

Let's consider an *equation error* type adjustable predictor (like in RLS). In this case, the a posteriori output of the adjustable predictor is given by:

$$\hat{y}(t+1) = \hat{\theta}^T(t+1)\phi(t) \tag{4.4}$$

where:

$$\hat{\theta}^T(t) = [\hat{a}_1(t), \dots, \hat{a}_{n_A}(t), \hat{b}_1(t), \dots, \hat{b}_{n_B}(t)] \tag{4.5}$$

The a posteriori prediction error is characterized by the following equation:

$$\varepsilon(t+1) = y(t+1) - \hat{y}(t+1) = [\theta - \hat{\theta}(t+1)]^T \phi(t) + w(t+1) \tag{4.6}$$

and the PAA will have the form:

$$\hat{\theta}(t+1) = \hat{\theta}(t) + F(t)\phi(t)\varepsilon(t+1) \tag{4.7}$$

*The first remark to be made is that in the presence of a zero mean stochastic disturbance, even if $\hat{\theta}(t) \equiv \theta$, the prediction error will not be null. Therefore, if we want that the estimated parameter vector $\hat{\theta}(t)$ tends toward a constant value, the adaptation gain $F(t)$ in (4.7) should tend toward zero when $t \to \infty$ (vanishing adaptation gain).*

This means that in the case of a time-varying matrix adaptation gain updated by:

$$F(t+1)^{-1} = \lambda_1(t)F(t)^{-1} + \lambda_2(t)\phi(t)\phi^T(t) \tag{4.8}$$

$\lambda_1(t)$ and $\lambda_2(t)$ should be chosen accordingly, i.e.:

$$0 < \lambda_2(t) < 2 \tag{4.9}$$

and the possible choices for $\lambda_1(t)$ are:

$$\left.\begin{array}{ll}\text{(a)} & \lambda_1(t) \equiv 1 \\ \text{(b)} & \lambda_1(t) = 1 \quad \text{for } t \geq t_0 \\ \text{(c)} & \lim_{t\to\infty} \lambda_1(t) = 1\end{array}\right\} \tag{4.10}$$

Similar reasoning applies when $F(t)$ is of the form:

$$F(t) = \frac{1}{p(t)} \tag{4.11}$$

where $p(t)$ should go to $\infty$ as $t \to \infty$. In particular, the *stochastic approximation algorithm* which uses:

$$F(t) = \frac{1}{t} \tag{4.12}$$

satisfies this property.[1]

The equivalent feedback system associated to (4.6) and (4.7) is shown in Fig. 4.1a. Figure 4.1b gives the equivalent feedback in the general case when the a posteriori adaptation error equation takes the form:

$$\nu(t+1) = H(q^{-1})(\theta - \hat{\theta}(t+1))^T\phi(t) + w(t+1) \tag{4.13}$$

where $w(t+1)$ is the image of the disturbance in the adaptation error equation. Therefore, one gets a feedback system with an external input. Several major questions require an answer:

1. Is $\hat{\theta}(t) = \theta$ a possible equilibrium point of the system?
2. If this is not the case, what are the possible equilibrium points?
3. Does the algorithm converge to the equilibrium points for any initial conditions?

The answer to these questions will depend upon the nature of the disturbance, as well as upon the structure of the adjustable predictor, the way in which the adaptation error is generated and the choice of the observation vector.

If the estimated parameters do not converge to the values corresponding to the deterministic case (under same input richness conditions), the resulting estimates will be called *biased* estimates, and the corresponding error will be termed *bias* (or more exactly asymptotic bias).

Looking to Fig. 4.1, one can ask under what conditions (using a decreasing adaptation gain) $\hat{\theta} = \theta$ is a possible equilibrium point. Alternatively, this can be formulated as follows: under what conditions the input into the integrator in the feedback path (which generates $\tilde{\theta}(t)$) will be null in the average. A first answer is given by the condition:

$$\mathbf{E}\{\phi(t)w(t+1)\} = 0 \tag{4.14}$$

i.e., the observation vector and the image of the stochastic disturbance in the adaptation error equation should be uncorrelated. This observation is used in an averaging method for the analysis of PAA in a stochastic environment.

[1] The vanishing adaptation gain should in addition satisfy the following property $\sum_{t=1}^{\infty} \frac{1}{p(t)} = \infty$.

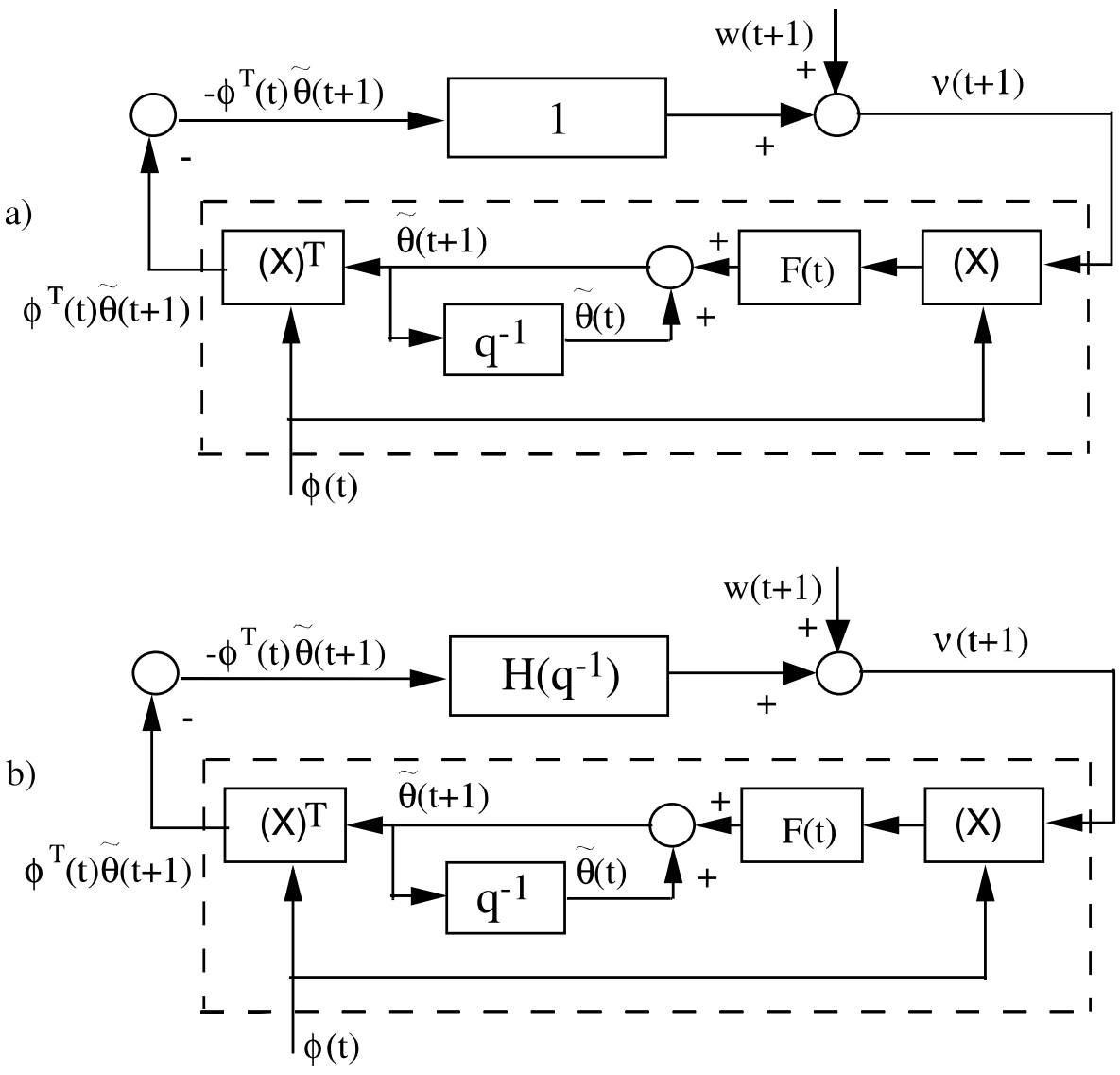

**Fig. 4.1** Equivalent feedback representation in the presence of stochastic disturbances, (**a**) recursive least squares, (**b**) general case

A second answer, using the conditional probabilities, takes the form:

$$\mathbf{E}\{F(t)\phi(t)w(t+1)\} = \mathbf{E}\{F(t)\phi(t)\}\mathbf{E}\{w(t+1)/\mathcal{F}_t\} = 0 \tag{4.15}$$

where $\mathcal{F}_t$ is the collections of all observations and their combinations up to and including $t$. Since $\mathbf{E}\{F(t)\phi(t)\} \neq 0$, one concludes that a condition for getting unbiased estimates is that:

$$\mathbf{E}\{w(t+1)/\mathcal{F}_t\} = 0 \tag{4.16}$$

which means that the image of the disturbance in the equation of the adaptation error should be uncorrelated with the past observations. This allows to characterize $w(t+1)$ as a *martingale difference sequence* leading to an analysis method exploiting the properties of martingale sequences.

We will try next to further explore the effect of the stochastic disturbances by examining the behavior of the PAA for large $t$, when the adaptation gain is very small by using an averaging approach. To simplify the analysis we will consider the PAA corresponding to (4.1) through (4.7) and the updating formula for the adapta-tion gain given by (4.12). The updating formula for the estimated parameter vector becomes:

$$\hat{\theta}(t+1) = \hat{\theta}(t) + \frac{1}{t}\phi(t)\varepsilon(t+1) \tag{4.17}$$

In the presence of the disturbance $w(t+1)$ the a posteriori prediction error can be split into the sum of two terms as it results from (4.6):

$$\varepsilon(t+1) = [\theta - \hat{\theta}(t+1)]^T\phi(t) + w(t+1) = \underline{\varepsilon}(t+1) + w(t+1) \tag{4.18}$$

where $\underline{\varepsilon}(t+1)$ is the prediction error produced mostly by the parameter error (similar to the deterministic case) and $w(t+1)$ is the disturbance itself (in general the image of the disturbance acting upon the plant). Therefore, the PAA of (4.17) can be written:

$$\hat{\theta}(t+1) = \hat{\theta}(t) + \frac{1}{t}\phi(t)\underline{\varepsilon}(t+1) + \frac{1}{t}\phi(t)w(t+1) \tag{4.19}$$

For high signal/noise ratio one can consider that $\underline{\varepsilon}(t+1)$ is a good approximation of the prediction error in the deterministic case and therefore the effect of the disturbance is reflected in the third term of the right hand side of (4.19). In order to obtain the same results as in the deterministic case, the cumulative effect of the third term upon $\hat{\theta}(t+1)$ when $t \to \infty$ should be null. If this is not the case, a bias error will appear leading to a biased estimate of $\theta$.

We will try next to see what will be the effect of the disturbance upon the PAA for large $t$, when the magnitude of the second and third terms are comparable (at the beginning of the adaptation process the magnitude of $\underline{\varepsilon}(t+1)$ will be much larger than that of the disturbance as a consequence of the large initial parameter error).

Let's consider the PAA of (4.19) for $t \gg 1$. Choosing a number of samples $N$ with the property $1 \ll N \ll t$ (for example $t = 10^5$; $N = 10^3$), the estimated parameter vector at the instant $t + N + 1$ is given by:

$$\begin{aligned}\hat{\theta}(t+N+1) = \hat{\theta}(t) &+ \sum_{i=0}^{N}\frac{1}{t+i}\phi(t+i)\underline{\varepsilon}(t+1+i)\\ &+ \sum_{i=0}^{N}\frac{1}{t+i}\phi(t+i)w(t+1+i)\end{aligned} \tag{4.20}$$

Since $t \gg N$ and $t \gg 1$, the various gains $\frac{1}{t+i}$ for $i = 0, 1, \dots, N$, can be approximated by their mean value:

$$\frac{1}{N+1}\left(\sum_{i=0}^{N}\frac{1}{t+i}\right)$$

and the above expression for $\hat{\theta}(t+N+1)$ can be approached by:

$$\hat{\theta}(t+N+1) \approx \hat{\theta}(t) + \left(\sum_{i=0}^{N} \frac{1}{t+i}\right) \frac{1}{N+1} \sum_{i=0}^{N} \phi(t+i)\underline{\varepsilon}(t+1+i) + \left(\sum_{i=0}^{N} \frac{1}{t+i}\right) \frac{1}{N+1} \sum_{i=0}^{N} \phi(t+i)w(t+1+i) \quad (4.21)$$

Now, since $N \gg 1$ and taking into account that in this example $\phi(t)$ and $w(t+1)$ do not depend upon $\hat{\theta}(t)$, one has:

$$\frac{1}{N+1} \sum_{i=0}^{N} \phi(t+i)w(t+1+i) \approx \mathbf{E}\{\phi(t+i)w(t+1+i)\} \quad (4.22)$$

(under the ergodicity assumption). From (4.22), one concludes that in order to obtain unbiased estimates, one should have $\mathbf{E}\{\phi(t)w(t+1)\} = 0$. Taking into account how $y(t)$ entering in $\phi(t)$ is generated through (4.1) it results that in the case of an *equation error* recursive identification scheme (like RLS) unbiased estimates will be obtained only if $w(t+1)$ is a stationary white noise sequence.

Let us now consider an *output error* recursive identification scheme using a *parallel* adjustable predictor and a parameter adaptation algorithm of the form of (4.17). The a posteriori output of the adjustable predictor will be given in this case by:

$$\begin{aligned} \hat{y}(t+1) &= -\sum_{i=1}^{n_A} \hat{a}_i(t+1)\hat{y}(t+1-i) + \sum_{i=1}^{n_B} \hat{b}_i(t+1)u(t+1-i) \\ &= \hat{\theta}^T(t+1)\phi(t) \end{aligned} \quad (4.23)$$

where:

$$\hat{\theta}(t)^T = [\hat{a}_1(t), \ldots, \hat{a}_{n_A}(t), \hat{b}_1(t), \ldots, \hat{b}_{n_B}(t)] \quad (4.24)$$

and:

$$\phi(t) = [-\hat{y}(t), \ldots, -\hat{y}(t-n_A+1), u(t), \ldots, u(t-n_B+1)] \quad (4.25)$$

(note that $\phi(t)$ in (4.4) and (4.23) are not the same). The a posteriori prediction error equation is given by (for details, see Sect. 3.3):

$$\begin{aligned} \varepsilon(t+1) &= y(t+1) - \hat{y}(t+1) \\ &= -\sum_{i=1}^{n_A} a_i \varepsilon(t+1-i) + [\theta - \hat{\theta}(t+1)]^T \phi(t) + w(t+1) \end{aligned} \quad (4.26)$$

which again can be written as:

$$\varepsilon(t+1) = \underline{\varepsilon}(t+1) + w(t+1) \quad (4.27)$$

where $\underline{\varepsilon}(t+1)$ is a good approximation of the prediction error in the absence of the disturbance.

Using the PAA algorithm of (4.17), for $t \gg 1$ and $1 \ll N \ll t$, one obtains (as in the previous example):

$$\hat{\theta}(t+N+1) \approx \hat{\theta}(t) + \left(\sum_{i=0}^{N} \frac{1}{t+i}\right) \frac{1}{N+1} \sum_{i=0}^{N} \phi(t+i)\underline{\varepsilon}(t+1+i)$$
$$+ \left(\sum_{i=0}^{N} \frac{1}{t+i}\right) \frac{1}{N+1} \sum_{i=0}^{N} \phi(t+i)w(t+1+i) \quad (4.28)$$

For very large $t$, the parameter $\hat{\theta}(t)$ is slowly varying since the adaptation gain become very small. Therefore as $t \to \infty$, $\phi(t)$ which contains $\hat{y}(t)$ generated through (4.23), becomes almost a stationary process which depends only upon $u(t)$, and one can write that:

$$\frac{1}{N+1} \sum_{i=0}^{N} \phi(t+i)w(t+1+i) \approx \mathbf{E}\{\phi(t+i, \hat{\theta}(t)), w(t+1+i)\} \quad (4.29)$$

where $\phi(t+1, \hat{\theta}(t))$ corresponds to the stationary process generated through (4.23) when $\hat{\theta}(t+i) = \hat{\theta}(t)$ for all $i \geq 0$. One concludes that (4.29) is equal to zero for any zero mean finite power stationary disturbance $w(t+1)$ which is independent with respect to the plant input $u(t)$.

Therefore, the use of a *parallel* adjustable predictor (output error) can lead to an unbiased estimate for various types of disturbances provided that they are independent with respect to the input, while the use of a *series-parallel* adjustable predictor (equation error) will lead to unbiased estimates only when $w(t+1)$ is a white noise. *Therefore, the choice of an appropriate structure for the adjustable predictor is crucial for obtaining unbiased parameter estimates in the presence of disturbances.*

## 4.2 The Averaging Method for the Analysis of Adaptation Algorithms in a Stochastic Environment

Consider again an algorithm with a scalar decreasing adaptation gain of the form:

$$\hat{\theta}(t+1) = \hat{\theta}(t) + \frac{1}{t}\phi(t)\nu(t+1) \quad (4.30)$$

where $\phi(t)$ is the observation (or measurement vector) and $\nu(t+1)$ is the a posteriori adaptation error.

We will try next to describe the behavior of the algorithm for large $t$ ($t \gg 1$). Taking a number of steps $N$ such that $1 \ll N \ll t$, one obtains from (4.30):

$$\hat{\theta}(t+N+1) = \hat{\theta}(t) + \sum_{i=0}^{N} \frac{1}{t+i}\phi(t+i)\nu(t+1+i) \quad (4.31)$$

Taking again into account the fact that $t \gg 1$ and $t \gg N$, the gains $\frac{1}{t+i}$, $i = 0, 1, \ldots, N$ can be approximated by their mean value and one can write:

$$\hat{\theta}(t+N+1) \approx \hat{\theta}(t) + \left(\sum_{i=0}^{N} \frac{1}{t+i}\right) \frac{1}{N+1} \sum_{i=0}^{N} \phi(t+i)\nu(t+1+i) \quad (4.32)$$

In general, $\phi(t+i)$ and $\nu(t+1+i)$ will depend on the values of $\hat{\theta}(t+i)$. However, for $t \gg 1$ and $1 \ll N \ll t$ since $\frac{1}{t+i}$ is very small, $\hat{\theta}(t+i)$ is slowly varying and one can write:

$$\begin{aligned}
&\frac{1}{N+1}\sum_{i=0}^{N}\phi(t+i)\nu(t+1+i) \\
&\quad\approx \frac{1}{N+1}\sum_{i=0}^{N}\phi(t+i,\hat{\theta}(t))\nu(t+1+i,\hat{\theta}(t))
\end{aligned} \tag{4.33}$$

where $\phi(t+i,\hat{\theta}(t))$ and $\nu(t+1+i,\hat{\theta}(t))$ are stationary quantities generated when for $i \geq 0$, one fixes $\hat{\theta}(t+i) = \hat{\theta}(t)$. Now if $N$ is sufficiently large (but still $N \ll t$):

$$\begin{aligned}
\frac{1}{N+1}\sum_{i=0}^{N}\phi(t+i,\hat{\theta}(t))\nu(t+1+i,\hat{\theta}(t)) &\approx \mathbf{E}\{\phi(t+i,\hat{\theta}(t))\nu(t+1+i,\hat{\theta}(t))\} \\
&= f(\hat{\theta}(t))
\end{aligned} \tag{4.34}$$

and (4.32) becomes:

$$\hat{\theta}(t+N+1) - \hat{\theta}(t) \approx \left(\sum_{i=0}^{N}\frac{1}{t+i}\right) f(\hat{\theta}(t)) \tag{4.35}$$

When $t \to \infty$, (4.35) can be considered as a discrete-time approximation of the deterministic ordinary differential equation (ODE):

$$\frac{d\hat{\theta}}{d\tau} = f(\hat{\theta}) \tag{4.36}$$

where:

$$\Delta\tau_t^{N+1} \triangleq \sum_{i=0}^{N}\frac{1}{t+i} \tag{4.37}$$

and:

$$f(\hat{\theta}) = \mathbf{E}\{\phi(t,\hat{\theta}),\, \nu(t+1,\hat{\theta})\} \tag{4.38}$$

The correspondence between the discrete-time $t$ and the continuous-time $\tau$ is given by:

$$\tau = \sum_{i=1}^{t-1}\frac{1}{i} \tag{4.39}$$

Equation (4.35) can be interpreted as a discretized version of (4.36) with a decreasing sampling time (as $t \to \infty$, the approximation will be more precise). Equation (4.36) is an *average type equation* since the right hand side corresponds to the averaging of the product of stationary processes $\phi(t,\hat{\theta})\nu(t+1,\hat{\theta})$ evaluated for a value $\hat{\theta}$ of the estimated parameters.

Since (4.35) which describes the behavior of the parameter adaptation algorithm of (4.30) for large $t$ is related to the ODE of (4.36), it is natural to expect that the asymptotic behavior of (4.36) (i.e. equilibrium points and stability) will give information upon the asymptotic behavior of the parameter adaptation algorithm in the stochastic environment. These connections have been examined in detail in Ljung (1977a), Ljung and Söderström (1983) and will be summarized next. The first assumptions which one should made, in order to use this analogy, is the existence of the stationary processes $\phi(t,\hat{\theta})$ and $\nu(t+1,\hat{\theta})$ for any possible value $\hat{\theta}(t)$. This means that the possible trajectories of the algorithm of (4.30) should lie infinitely often in a domain in the parameter space for which these stationary processes exist. In more concrete terms, this signifies that the system generating $\phi(t)$ and $\nu(t+1)$ are stable for any possible value $\hat{\theta}(t)$ generated by (4.30). Once this assumption is satisfied, the ODE of (4.36) can be used for the analysis of the asymptotic behavior of the adaptation algorithm of (4.30). The following correspondences between (4.30) and (4.36) have been established:

1. The equilibrium points of (4.36) correspond to the only possible convergence points of the algorithm of (4.30).
2. The global (or local) asymptotic stability of an equilibrium point $\theta^*$ of (4.36) corresponds to the global (local) convergence w.p.1 (with probability 1) of the algorithm of (4.30) toward $\theta^*$.

*Example* We will illustrate the application of this technique for the case of a recursive estimation scheme using a *parallel* adjustable predictor given by (4.23) (output error), and using the parameter adaptation algorithm of (4.30) where we will take $\nu(t+1)=\varepsilon(t+1)$.

In order to apply the ODE of (4.36) for convergence analysis, one needs to calculate $f(\hat{\theta})$ for this scheme as defined in (4.34).

From (4.26) one has:

$$\begin{aligned}\nu(t+1) &= \varepsilon(t+1)\\ &= -\sum_{i=1}^{n} a_i \nu(t+1-i) + [\theta - \hat{\theta}(t+1)]^T \phi(t) + w(t+1)\end{aligned} \tag{4.40}$$

For analysis purposes only, making $\hat{\theta}(t)=\hat{\theta}$, one gets (after passing the term $-\sum_{i=1}^{n} a_i \nu(t+1-i)$ in the right hand side):

$$A(q^{-1})\nu(t+1,\hat{\theta}) = (\theta-\hat{\theta})^T \phi(t,\hat{\theta}) + w(t+1) \tag{4.41}$$

where:

$$A(q^{-1}) = 1 + a_1 q^{-1} + \cdots + a_n q^{-n} = \frac{1}{H(q^{-1})} \tag{4.42}$$

Passing both sides of (4.41) through $H(q^{-1})$, one gets:

$$\nu(t+1,\hat{\theta}) = H(q^{-1})(\theta-\hat{\theta})^T \phi(t,\hat{\theta}) + w'(t+1) \tag{4.43}$$

where:

$$w'(t+1) = H(q^{-1})w(t+1) \tag{4.44}$$

Using (4.43), $f(\hat{\theta})$ will be given by:

$$\begin{aligned} f(\hat{\theta}) &= \mathbf{E}\{\phi(t,\hat{\theta})\nu(t+1,\hat{\theta})\} \\ &= \mathbf{E}\{\phi(t,\hat{\theta})H(q^{-1})\phi^T(t,\hat{\theta})\}(\theta - \hat{\theta}) + \mathbf{E}\{\phi(t,\hat{\theta})w'(t+1)\} \end{aligned} \tag{4.45}$$

and the associated ODE takes the form:

$$\frac{d\hat{\theta}}{d\tau} = -\mathbf{E}\{\phi(t,\hat{\theta})H(q^{-1})\phi^T(t,\hat{\theta})\}(\hat{\theta} - \theta) + \mathbf{E}\{\phi(t,\hat{\theta})w'(t+1)\} \tag{4.46}$$

Since $u(t)$ and $w(t+1)$ are assumed to be independent, one concludes that $\mathbf{E}\{\phi(t,\hat{\theta})w'(t+1)\} = 0$ ($\phi(t,\hat{\theta})$ depends only on $u(t)$) and the equilibrium points of (4.46) are given by:

$$\phi^T(t,\hat{\theta})[\hat{\theta} - \theta] = 0 \tag{4.47}$$

This corresponds to a set $D_c : \{\hat{\theta} \mid \phi^T(t,\hat{\theta})[\hat{\theta} - \theta] = 0\}$ of convergence points assuring a correct input-output behavior of the predictor (in the absence of the disturbance) and with $A(q^{-1})$ asymptotically stable, one sees from (4.41) that the prediction error will become asymptotically zero in the absence of the disturbance. If $u(t)$ is a persistently exciting signal of appropriate order, (4.47) will have only a unique solution $\hat{\theta} = \theta$, which will be the unique possible convergence point of the parameter adaptation algorithm.

We must now examine the stability of the (4.46) without the forcing term which is null:

$$\frac{d\hat{\theta}}{d\tau} = -\mathbf{E}\{\phi(t,\hat{\theta})H(q^{-1})\phi^T(t,\hat{\theta})\}[\hat{\theta} - \theta] \tag{4.48}$$

The global asymptotic stability of (4.48) is assured if $\mathbf{E}\{\phi(t,\hat{\theta})H(q^{-1})\phi^T(t,\hat{\theta})\}$ is a positive definite matrix. But this means that for any vector $l$ one should have:

$$\begin{aligned} l^T\mathbf{E}\{\phi(t,\hat{\theta})H(q^{-1})\phi^T(t,\hat{\theta})\}l &= \mathbf{E}\{l^T\phi(t,\hat{\theta})H(q^{-1})\phi^T(t,\theta)l\} \\ &= \mathbf{E}\{v(t,\hat{\theta})\tilde{v}(t,\hat{\theta})\} > 0 \end{aligned} \tag{4.49}$$

where:

$$\begin{aligned} v(t,\hat{\theta}) &= l^T\phi(t,\hat{\theta}) \\ \tilde{v}(t,\hat{\theta}) &= H(q^{-1})l^T\phi(t,\hat{\theta}) = H(q^{-1})v(t,\hat{\theta}) \end{aligned} \tag{4.50}$$

$v(t,\hat{\theta})$ and $\tilde{v}(t,\hat{\theta})$ can be interpreted as the input and the output respectively of a block with the transfer function $H(z^{-1})$. If $H(z^{-1})$ is a strictly positive real transfer function, and considering a state space realization of it (state vector: $x(t)$), one can write (using Lemma C.4, Appendix C):

$$
\begin{aligned}
\mathbf{E}\{v(t,\hat{\theta})\tilde{v}(t,\hat{\theta})\} &= \lim_{N\to\infty}\frac{1}{N}\sum_{t=1}^{N} v(t,\hat{\theta})\tilde{v}(t,\hat{\theta}) \\
&\geq \lim_{N\to\infty}\frac{1}{N}\left[\frac{1}{2}x(N+1)^T P x(N+1) - \frac{1}{2}x(1)^T P x(1)\right] \\
&\quad + \lim_{N\to\infty}\frac{1}{N}\sum_{t=1}^{N}[x(t)^T, v(t,\hat{\theta})] M_0 \begin{bmatrix} x(t) \\ v(t,\hat{\theta}) \end{bmatrix} > 0 \qquad (4.51)
\end{aligned}
$$

where $P > 0$ and $M_0 > 0$. Therefore, in the case of the *output error* scheme defined by the predictor of (4.23) and the adaptation algorithm of (4.30) if $H(z^{-1}) = \frac{1}{A(z^{-1})}$ is a strictly positive real transfer function, global convergence w.p.1 of the estimated parameter vector $\hat{\theta}(t)$ towards $D_c$ (or $\theta^*$ under input richness conditions) will be assured.

A similar approach can be used in order to associate an ordinary differential equation for PAA using matrix type adaptation gains of the form:

$$\hat{\theta}(t+1) = \hat{\theta}(t) + F(t)\phi(t)\nu(t+1) \qquad (4.52)$$

$$F(t+1)^{-1} = F(t)^{-1} + \lambda_2\phi(t)\phi(t)^T$$

$$0 < \lambda_2 < 2; \;\; F(0) > 0 \qquad (4.53)$$

In order to use the previous results for the *stochastic approximation* adaptation gain (4.17), it is convenient to replace $F(t)$ by:

$$R(t) = \frac{1}{t}F(t)^{-1} \qquad (4.54)$$

and therefore (4.52) and(4.53) will take the form:

$$\hat{\theta}(t+1) = \hat{\theta}(t) + \frac{1}{t}R^{-1}(t)\phi(t)\nu(t+1) \qquad (4.55)$$

$$R(t+1) = \hat{R}(t) + \frac{1}{t+1}[\lambda_2\phi(t)\phi(t)^T - R(t)] \qquad (4.56)$$

Following the same type of developments as in the case of stochastic approximation algorithm and using the relation:

$$\frac{1}{N+1}\sum_{i=0}^{N}\frac{1}{t+i}R^{-1}(t+i) \approx \frac{1}{N+1}\left(\sum_{i=0}^{N}\frac{1}{t+i}\right)R^{-1}(t)$$

for $t \gg 1$ , $1 \ll N \ll t$ one gets the following associated ODE (Ljung 1977a; Ljung and Söderström 1983; Dugard and Landau 1980):

$$\frac{d\hat{\theta}}{d\tau} = R^{-1}(\tau)f(\hat{\theta}) \qquad (4.57)$$

$$\frac{dR}{d\tau} = \lambda_2 G(\hat{\theta}) - R(\tau) \qquad (4.58)$$

where:

$$f(\hat{\theta}) = \mathbf{E}\{\phi(t,\hat{\theta})\nu(t+1,\hat{\theta})\} \tag{4.59}$$

$$G(\hat{\theta}) = \mathbf{E}\{\phi(t,\hat{\theta})\phi^T(t,\hat{\theta})\} \tag{4.60}$$

The first step for effectively applying the associated ODE for convergence analysis of a certain scheme is the evaluation of $f(\hat{\theta})$ (see also the example discussed earlier) which requires the knowledge of the equation governing $\nu(t+1,\hat{\theta})$. However, for many recursive schemes used for identification, adaptive state estimation and adaptive control, the equation governing $\nu(t+1,\hat{\theta})$ has a normalized form. The use of this normalized form will allow to simplify the analysis procedure through ODE.

We will next focus upon the averaging method for the class of recursive schemes for which the a posteriori adaptation error equation for $\hat{\theta}(t) = \hat{\theta}$ takes the form:

$$\nu(t+1,\hat{\theta}) = H(q^{-1})\phi^T(t,\hat{\theta})[\theta^* - \hat{\theta}] + \omega(t+1,\hat{\theta}) \tag{4.61}$$

where $\theta^*$ is a fixed parameter vector and $\omega(t+1,\hat{\theta})$ is the image of the disturbance in the adaptation error for $\hat{\theta}(t) = \hat{\theta}$.

Studying the properties of the associated ODE for the case of $\nu(t+1,\hat{\theta})$ generated by equations of the form of (4.61), one obtains the following stochastic convergence theorem which can then be straightforwardly used:

**Theorem 4.1** *Consider the parameter adaptation algorithm*:

$$\hat{\theta}(t+1) = \hat{\theta}(t) + F(t)\phi(t)\nu(t+1) \tag{4.62}$$

$$F(t+1)^{-1} = F(t)^{-1} + \lambda_2(t)\phi(t)\phi^T(t)$$

$$0 < \lambda_2(t) < 2; \; F(0) > 0 \tag{4.63}$$

- Assume that the stationary processes $\phi(t,\hat{\theta})$ and $\nu(t+1,\hat{\theta})$ can be defined for $\hat{\theta}(t) \equiv \hat{\theta}$.
- *Assume that $\hat{\theta}(t)$ generated by the algorithm belongs infinitely often to the domain $(D_s)$ for which the stationary processes $\phi(t,\hat{\theta})$ and $\nu(t+1,\hat{\theta})$ can be defined.*
- *Assume that $\nu(t+1,\hat{\theta})$ is given by an equation of the form*:

$$\nu(t+1,\hat{\theta}) = H(q^{-1})\phi^T(t,\hat{\theta})[\theta^* - \hat{\theta}] + \omega(t+1,\hat{\theta}) \tag{4.64}$$

*where $H(q^{-1})$ is a discrete transfer operator, ratio of monic polynomials.*
- *Assume that either*:

(a) $\omega(t+1,\hat{\theta})$ *is a sequence of independent equally distributed normal random variables* $(0,\sigma)$, *or*

(b)
$$\mathbf{E}\{\phi(t,\hat{\theta}),\omega(t+1,\hat{\theta})\} = 0 \tag{4.65}$$

*Define the convergence domain $D_c$*:

$$D_c : \{\hat{\theta} : \phi^T(t,\hat{\theta})[\theta^* - \hat{\theta}]\} = 0 \tag{4.66}$$

*Then if there is $\lambda_2 : \max_t \lambda_2(t) \le \lambda_2 < 2$ such that*:

$$H'(z^{-1}) = H(z^{-1}) - \frac{\lambda_2}{2} \tag{4.67}$$

*is a strictly positive real discrete transfer function, one has*:

$$\text{Prob}\left\{\lim_{t\to\infty}\hat{\theta}(t)\in D_c\right\}=1 \tag{4.68}$$

**Corollary 4.1** *If*:

$$\phi^T(t,\hat{\theta})[\theta^*-\theta]=0 \tag{4.69}$$

*has a unique solution* $\hat{\theta}=\theta^*$, *the condition that* $H'(z^{-1})$ *given by* (4.67) *be strictly positive real implies*:

$$\text{Prob}\left\{\lim_{t\to\infty}\hat{\theta}(t)=\theta^*\right\}=1 \tag{4.70}$$

**Corollary 4.2** *If the adaptation gain given by* (4.63) *is replaced by*:

$$F(t)=\frac{1}{t}F;\qquad F>0 \tag{4.71}$$

*The condition of* (4.67) *is replaced by the condition that* $H(z^{-1})$ *be a strictly positive real discrete transfer function.*

*Proof* Taking into account (4.64), one has:

$$f(\hat{\theta})=\mathbf{E}\{\phi(t,\hat{\theta})H(q^{-1})\phi^T(t,\hat{\theta})[\theta^*-\hat{\theta}]\}+\mathbf{E}\{\phi(t,\hat{\theta})\omega(t+1,\hat{\theta})\} \tag{4.72}$$

If conditions (a) or (b) are satisfied, one has:

$$\mathbf{E}\{\phi(t,\hat{\theta})\omega(t+1,\hat{\theta})\}=0 \tag{4.73}$$

and (4.72) becomes:

$$f(\hat{\theta})=\mathbf{E}\{\phi(t,\hat{\theta})H(q^{-1})\phi^T(t,\hat{\theta})\}[\theta^*-\hat{\theta}]=\tilde{G}(\hat{\theta})[\theta^*-\hat{\theta}] \tag{4.74}$$

where:

$$\tilde{G}(\hat{\theta})=\mathbf{E}\{\phi(t,\hat{\theta})H(q^{-1})\phi^T(t,\hat{\theta})\} \tag{4.75}$$

The stationary points of the ODE are given by:

$$f(\hat{\theta})=\tilde{G}(\theta)[\theta^*-\theta]=0 \tag{4.76}$$

Taking into account the structure of $\tilde{G}(\theta)$, this translates to the condition:

$$\phi^T(t,\hat{\theta})[\theta^*-\hat{\theta}]=0 \tag{4.77}$$

which defines the domain $D_c$.

It remains to show the global asymptotic stability of the associated ODE, which in this case is defined by (4.57) with $f(\hat{\theta})$ given by (4.74) and by:

$$\frac{dR}{d\tau}=G_\lambda(\hat{\theta})-R(\tau) \tag{4.78}$$

(which replaces (4.58)) where:

$$G_\lambda(\hat{\theta})=\mathbf{E}\{\lambda_2(t)\phi(t,\hat{\theta})\phi^T(t,\hat{\theta})\}\le\lambda_2 G(\hat{\theta}) \tag{4.79}$$

with $G(\hat{\theta})$ given by (4.60).

Let us consider the Lyapunov function candidate:

$$V(\hat{\theta}, R) = (\hat{\theta} - \theta^*)^T R(\hat{\theta} - \theta^*) \tag{4.80}$$

Along the trajectories of (4.57) and (4.78), one has:

$$\begin{aligned}\frac{d}{d\tau}V(\hat{\theta}(\tau), R(\tau)) &= -(\hat{\theta} - \theta^*)^T [\tilde{G}(\hat{\theta}) + \tilde{G}^T(\hat{\theta}) - \lambda_2 G(\hat{\theta})][\hat{\theta} - \theta^*] \\ &\quad - (\hat{\theta} - \theta^*)^T [R + \lambda_2 G(\hat{\theta}) - G_\lambda(\hat{\theta})](\hat{\theta} - \theta^*)\end{aligned} \tag{4.81}$$

Therefore, in order to have global convergence, it is sufficient to show that:

$$\bar{G}(\hat{\theta}) = \tilde{G}(\hat{\theta}) + \tilde{G}^T(\hat{\theta}) - \lambda_2 G(\hat{\theta}) \tag{4.82}$$

is a positive semi-definite matrix in order to assure that:

$$\frac{d}{d\tau}V(\hat{\theta}(\tau), R) < 0 \tag{4.83}$$

Equation (4.82) can be rewritten as:

$$\bar{G}(\hat{\theta}) = \mathbf{E}\{\phi(t, \hat{\theta})[H(q^{-1}) + H(q^{-1}) - \lambda_2]\phi^T(t, \hat{\theta})\} \tag{4.84}$$

taking into account that $H(q^{-1})$ is a scalar transfer operator and $\lambda_2$ is a scalar. Using the same developments as in (4.48) through (4.51), one concludes that $\bar{G}(\hat{\theta})$ is a positive definite matrix if the condition of (4.67) is satisfied. □

*Example 4.1* (Output Error) One can illustrate the use of this tool by applying it to the analysis of the convergence of the *output error* scheme which uses a *parallel* adjustable predictor and a filtered prediction error as adaptation error (for details see Sect. 3.3 and the previous example given in Sect. 4.1).

The plant model to be identified is given by (4.1) with the assumption that $u(t)$ and $w(t)$ are independent. The a posteriori output of the adjustable predictor is given by (4.23). The a posteriori prediction error is given by (4.26) and the a posteriori adaptation error is given by:

$$\nu(t + 1) = D(q^{-1})\varepsilon(t + 1) \tag{4.85}$$

where:

$$D(q^{-1}) = 1 + d_1 q^{-1} + \cdots + d_n q^{-n} \tag{4.86}$$

is an asymptotically stable polynomial.

In order to apply Theorem 4.1, one should first derive the equations for $\nu(t + 1)$. From (4.26), one obtains:

$$\varepsilon(t + 1) = \frac{1}{A(q^{-1})}[\theta - \hat{\theta}(t + 1)]^T \phi(t) + \frac{1}{A(q^{-1})} w(t + 1) \tag{4.87}$$

where:

$$A(q^{-1}) = 1 + a_1 q^{-1} + \cdots + a_{n_A} q^{-n_A} \tag{4.88}$$

From (4.85) and (4.87), one obtains:

$$\nu(t + 1) = H(q^{-1})[\theta - \hat{\theta}(t + 1)]^T \phi(t) + H(q^{-1}) w(t + 1) \tag{4.89}$$

where:

$$H(q^{-1}) = \frac{D(q^{-1})}{A(q^{-1})} \tag{4.90}$$

Making $\hat{\theta}(t) = \hat{\theta}$, one obtains:

$$\nu(t+1, \hat{\theta}) = H(q^{-1})[\theta - \hat{\theta}]^T \phi(t, \hat{\theta}) + H(q^{-1})w(t+1) \tag{4.91}$$

which has the form of (4.64) of Theorem 4.1. Since $w(t)$ and $u(t)$ are independent, one concludes from (4.23), that $\phi(t, \hat{\theta})$ and $w(t+1, \hat{\theta})$ are independent for $\hat{\theta}(t) \equiv \hat{\theta}$, and therefore:

$$\mathbf{E}\{\phi(t, \hat{\theta})\omega(t+1, \hat{\theta})\} = \mathbf{E}\{\phi(t, \hat{\theta}), H(q^{-1})w(t+1)\} = 0 \tag{4.92}$$

It results that global convergence toward unbiased estimates can be obtained w.p.1 if:

$$H'(z^{-1}) = \frac{D(z^{-1})}{A(z^{-1})} - \frac{\lambda_2}{2} \tag{4.93}$$

is strictly positive real (note that a similar convergence condition has been obtained in the deterministic case).

*Example 4.2* (Recursive Least Squares) Applying the same analysis to RLS ((4.1), (4.4), (4.5), (4.52) and (4.53)), one finds that:

$$\nu(t+1, \hat{\theta}) = \varepsilon(t+1, \hat{\theta}) = [\theta - \hat{\theta}]^T \phi(t, \hat{\theta}) + w(t+1) \tag{4.94}$$

where $\phi(t)$ is given by (4.3). In this case, $H(q^{-1}) = 1$, $H(q^{-1}) - \frac{\lambda_2}{2} > 0$, and convergence w.p.1 to the unbiased parameters will be obtained if $w(t+1)$ is a sequence of independent equally distributed normal random variables (for other types of disturbance $\phi(t)$ and $w(t+1)$ will be correlated).

An important remark concerning the use of the averaging method is the fact that for a complete analysis, one has to show by other means that the hypothesis $\hat{\theta}(t)$ belongs infinitely often to $D_s$ (in fact, this condition can be replaced by the condition that $\lim_{N\to\infty} \frac{1}{N}\phi^T(t)\phi(t) < \infty$). However, this aspect of the analysis is justified to be carried on only if the ODE gives satisfactory results. Another solution is to incorporate a stability test in the algorithm and to use a projection into $D_s$ each time $\hat{\theta}(t)$ leaves $D_s$ (see Sect. 10.5 and Ljung and Söderström 1983).

## 4.3 The Martingale Approach for the Analysis of PAA in a Stochastic Environment

Consider the PAA and the associated equation for the generation of the a posteriori adaptation error given in Sect. 3.4. for the case of a decreasing adaptation gain ($\lambda_1(t) \equiv 1$) and with an image of the disturbance acting on the system which reflects as an additive term in the equation of the a posteriori adaptation error, i.e.,

$$\hat{\theta}(t+1) = \hat{\theta}(t) + F(t)\phi(t)\nu(t+1) \tag{4.95}$$

$$F(t+1)^{-1} = F(t)^{-1} + \lambda_2(t)\phi(t)\phi^T(t)$$
$$F(0) > 0;\ 0 < \lambda_2(t) < 2 \tag{4.96}$$

$$\nu(t+1) = \frac{\nu^0(t+1)}{1+\phi^T(t)F(t)\phi(t)} \tag{4.97}$$

$$\nu(t+1) = H(q^{-1})[\theta - \hat{\theta}(t+1)]^T\phi(t) + \omega(t+1) \tag{4.98}$$

where $H(q^{-1}) = \frac{H_1(q^{-1})}{H_2(q^{-1})}$ is a ratio of monic polynomials.

The associated equivalent feedback representation is shown in Fig. 4.1b. From the analysis made in Sect. 4.1, it was concluded that a condition for $\hat{\theta} = \theta$, be an equilibrium point of the system (or in other term, a condition for an asymptotic unbiased estimate) is that:

$$\mathbf{E}\{\omega(t+1) \mid \mathcal{F}_t\} = 0$$

where $\mathcal{F}_t$ are all the observations generated up to and including $t$.

For convergence analysis, the sequence $\{\omega(t)\}$ will be taken to be a *martingale difference sequence* (see Appendix A) defined on a probability space $(\Omega, \mathcal{A}, \mathcal{P})$ adapted to the sequence of an increasing $\sigma$*-algebras* $\mathcal{F}_t$ generated by the observations up to and including time $t$. The sequence $\{\omega(t+1)\}$ is assumed to satisfy the following:

$$\mathbf{E}\{\omega(t+1) \mid \mathcal{F}_t\} = 0 \tag{4.99}$$

$$\mathbf{E}\{\omega^2(t+1) \mid \mathcal{F}_t\} = \sigma^2 \tag{4.100}$$

$$\lim_{N\to\infty} \sup \frac{1}{N}\sum_{t=1}^{N} \omega^2(t) < \infty \tag{4.101}$$

Note that the usual model for a white noise sequence as a sequence for equally distributed normal random variables (0, 1) satisfies also the above properties.

The observation vector $\phi(t)$ used in the PAA depends on the particular structure of the recursive estimation or adaptive control scheme. In general, it has the following form:

$$\phi(t) = f_\phi(\hat{\theta}(t), \hat{\theta}(t-1), \dots, \omega(t), \omega(t-1), \dots) \tag{4.102}$$

For example, in the *output error* scheme $\phi(t)$ has this general form. However, in other scheme like RLS, the components of $\phi(t)$ do not depend upon $\hat{\theta}(t)$, $\hat{\theta}(t-1), \dots$.

As in the deterministic case, (4.95) and (4.98) are related with the a priori adaptation error (or prediction error) denoted by $\nu^0(t+1)$ which is effectively used for implementing the adaptation algorithm of (4.95).

Taking into account the structure of the monic polynomials:

$$H_j(q^{-1}) = 1 + h_i^j q^{-1} + \cdots + h_n^j q^{-n} = 1 + q^{-1}H_j^*(q^{-1});\quad j = 1, 2 \tag{4.103}$$

Equation (4.97) can be written as:

$$\begin{aligned}\nu(t+1) &= [\hat{\theta}(t) - \hat{\theta}(t+1)]^T \phi(t) + [\theta - \hat{\theta}(t)]^T \phi(t) \\ &\quad + H_1^*(q^{-1})[\theta - \hat{\theta}(t)]^T \phi(t-1) \\ &\quad - H_2^*(q^{-1})\nu(t) + H_2^*(q^{-1})\omega(t) + \omega(t+1) \\ &= [\hat{\theta}(t) - \hat{\theta}(t+1)]^T \phi(t) + \nu^0(t+1) \end{aligned} \tag{4.104}$$

where:

$$\begin{aligned}\nu^0(t+1) &= [\theta - \hat{\theta}(t)]^T \phi(t) + H_1^*(q^{-1})[\theta - \hat{\theta}(t)]^T \phi(t-1) \\ &\quad - H_2^*(q^{-1})\nu(t) + H_2^*(q^{-1})\omega(t) + \omega(t+1) \end{aligned} \tag{4.105}$$

Replacing the term $[\hat{\theta}(t) - \hat{\theta}(t+1)]$ in (4.104) by the one obtained from (4.95), one gets (4.97) and the PAA can be rewritten as:

$$\begin{aligned}\hat{\theta}(t+1) &= \hat{\theta}(t) + \frac{F(t)\phi(t)\nu^0(t+1)}{1+\phi^T(t)F(t)\phi(t)} \\ &= \hat{\theta}(t) + \gamma(t)F(t+1)\phi(t)\nu^0(t+1) \end{aligned} \tag{4.106}$$

where:

$$\gamma(t) = \frac{1+\lambda_2(t)\phi^T(t)F(t)\phi(t)}{1+\phi^T(t)F(t)\phi(t)} \tag{4.107}$$

which as a consequence of the possible choices for $\lambda_2(t)$ in (4.96) satisfies the double inequality:

$$0 < \gamma(t) < 2 \tag{4.108}$$

Note from (4.104) and (4.105) that the quantity:

$$\hat{\nu}(t+1/t) = \nu^0(t+1) - \omega(t+1) = f_\nu[\hat{\theta}(t), \hat{\theta}(t-1), \ldots, \omega(t), \omega(t-1), \ldots] \tag{4.109}$$

is the one step ahead prediction of the a posteriori adaptation error $\nu(t+1)$.

An important remark which will be subsequently used in the analysis is that $[\nu^0(t+1) - \omega(t+1)]$, $\phi(t)$ and $F(t+1)$ are all $\mathcal{F}_t$ measurable as it results from (4.105), (4.102) and (4.96).

Equations (4.95) and (4.98) define an equivalent feedback stochastic system. As in the deterministic case, defining the parameter error as:

$$\tilde{\theta}(t) = \hat{\theta}(t) - \theta \tag{4.110}$$

one gets from (4.95) that:

$$\tilde{\theta}(t+1) = \tilde{\theta}(t) + F(t)\phi(t)\nu(t+1) \tag{4.111}$$

Introducing the notation:

$$\bar{y}(t+1) = -u(t+1) = \phi^T(t)\tilde{\theta}(t+1) \tag{4.112}$$

and associating to $H(z^{-1})$ a minimal state space realization $[A, b, c^T, \delta]$ (i.e., $H(z^{-1}) = c^T(I - z^{-1}A)^{-1}b + \delta$), (4.98) defines an equivalent linear feedforward block which can be written as:

$$x(t+1) = Ax(t) + bu(t+1) \tag{4.113}$$

$$\nu(t+1) = c^T x(t) + \delta u(t+1) + \omega(t+1) \tag{4.114}$$

Equations (4.111) and (4.112) define an equivalent feedback block with state $\tilde{\theta}(t)$ and input $\nu(t+1)$.

$$\tilde{\theta}(t+1) = \tilde{\theta}(t) + F(t)\phi(t)\nu(t+1) \tag{4.115}$$

$$\begin{aligned} \bar{y}(t+1) &= \tilde{\theta}^T(t+1)\phi(t) \\ &= \phi^T(t)\tilde{\theta}(t) + \phi^T(t)F(t)\phi(t)\nu(t+1) \end{aligned} \tag{4.116}$$

Equations (4.113) through (4.116) define an equivalent stochastic feedback system shown in Fig. 4.1b. The output of the linear feedforward block is disturbed by the martingale difference sequence $\{\omega(t+1)\}$. The feedback block defined by (4.115) and (4.116) can be considered in a first step as a linear time-varying block and, in a second step, the fact that $\phi(t)$ is generated by an equation of the form (4.102) is taken into account. The reason is (like in deterministic environment) that some properties of this block (related to passivity) do not depend on how $\phi(t)$ is generated. Clearly, from (4.98) it results that for $\hat{\theta}(t) \equiv \theta$ ($\tilde{\theta}(t) \equiv 0$) one has:

$$\nu(t+1)\,|_{\hat{\theta}(t)\equiv\theta} = \omega(t+1)$$

Therefore, the objective of the analysis will be to show that for any initial conditions, $\nu(t+1)$ will tend toward $\omega(t+1)$ in a certain sense to be defined. Furthermore, one will have to show that the input to the linear block which characterizes the error between the input-output behavior of the model and of the adjustable system goes to zero in a certain sense. We will also be interested in the convergence toward zero of the parameter error, and on the boundedness of $\phi(t)$. The results of this analysis are summarized in the following theorem.

**Theorem 4.2** *Consider the PAA given by* (4.95) *through* (4.97)

(i) *Assume that the a posteriori adaptation error satisfies* (4.98) *where* $\{\omega(t+1)\}$ *is a martingale difference sequence satisfying* (4.99) *through* (4.101), *and* $H(q^{-1})$ *is a rational discrete transfer operator* (*ratio of monic polynomials*).
(ii) *Assume that the a priori adaptation error satisfies* (4.105).
(iii) *Assume that* $\phi(t)$ *is* $\mathcal{F}_t$ *measurable.*

*If*:

1.
$$\lim_{N\to\infty} \frac{1}{N}\sum_{t=1}^{N} \phi^T(t)\phi(t) < \infty \tag{4.117}$$

2. *There is* $\lambda_2 : \max_t \lambda_2(t) \le \lambda_2 < 2$, *such that*:

$$H'(z^{-1}) = H(z^{-1}) - \frac{\lambda_2}{2} \tag{4.118}$$

*is a strictly positive real transfer function*, *then*:

$$\text{(a)} \quad \lim_{N\to\infty} \frac{1}{N}\sum_{t=1}^{N}[\nu(t)-\omega(t)]^2 = 0 \quad a.s. \tag{4.119}$$

$$\text{(b)} \quad \lim_{N\to\infty} \frac{1}{N}\sum_{t=1}^{N}\nu^2(t) = \lim_{N\to\infty} \frac{1}{N}\sum_{t=1}^{N}\omega^2(t) \quad a.s. \tag{4.120}$$

$$\text{(c)} \quad \lim_{N\to\infty} \frac{1}{N}\sum_{t=1}^{N}[[\hat{\theta}(t)-\theta]^T\phi(t-1)]^2 = 0 \quad a.s. \tag{4.121}$$

$$\text{(d)} \quad \lim_{N\to\infty} [\hat{\theta}(t)-\theta]^T\left(\frac{F(t)^{-1}}{t}\right)[\tilde{\theta}(t)-\theta] = 0 \quad a.s. \tag{4.122}$$

(e) *If, in addition*:

$$\lim_{t\to\infty} \frac{1}{t}F(t)^{-1} > 0 \quad a.s. \tag{4.123}$$

*then*:

$$\lim_{t\to\infty} \hat{\theta}(t) = \theta \quad a.s. \tag{4.124}$$

*Remarks*

- Equation (4.121) which characterizes the behavior of the input of the equivalent linear feedforward block, assures in the case of identification a correct input-output description and is one of the results desired for adaptive control and adaptive prediction.
- Equation (4.122) concerns the behavior of the parameter error and is of interest in identification schemes. The additional condition ($e$) is directly related to the richness conditions upon the plant input.
- From these results, it is possible to go further and to show that under certain conditions $\lim_{N\to\infty}\frac{1}{N}\sum_{t=1}^{N}[\nu^0(t)-\omega(t)]^2 = 0$ a.s. which is important in the context of adaptive control.
- The restrictive part of Theorem 4.2, is the boundedness condition (4.117). To evaluate (4.117), one should take into account how $\phi(t)$ is generated in each scheme. Note that this is not taken into account in the assumptions of Theorem 4.2 except the assumption that $\phi(t)$ is $\mathcal{F}_t$ measurable. A class of adaptive schemes which satisfy the condition of (4.117) will be presented in Lemma 4.1.

**Lemma 4.1** *For the PAA of* (4.95) *through* (4.97) *with* $\lambda_2(t)\equiv 1$, *if*:

1.

$$\frac{1}{N}\sum_{t=1}^{N}\phi_i^2(t) \le K_1 + \frac{K_2}{N}\sum_{t=1}^{N}\nu^2(t-j)$$
$$0\le K_1;\ K_2<\infty;\ 0\le j<\infty;\ i=1,2,\dots,n \tag{4.125}$$

*where*:

$$\phi^T(t) = [\phi_1(t),\dots,\phi_n(t)] \tag{4.126}$$

*and*:

2.
$$\lim_{N\to\infty} \frac{1}{N} \sum_{t=1}^{N} [\hat{\theta}^T(t)\phi(t) + \nu^0(t+1)]^2 < \infty \tag{4.127}$$

*then*:

$$\text{(a)} \quad \lim_{N\to\infty} \frac{1}{N} \sum_{t=1}^{N} \nu^2(t) < \infty \tag{4.128}$$

$$\text{(b)} \quad \lim_{N\to\infty} \frac{1}{N} \sum_{t=1}^{N} \phi^T(t)\phi(t) < \infty \tag{4.129}$$

Application of these results is basically done as follows:

1. Write the equation of the a posteriori adaptation error in the presence of the disturbance, searching for a value $\theta$ such that an expression of the form of (4.98) is obtained where $\{\omega(t+1)\}$ is a martingale difference sequence (i.e., the image of the disturbance is a white noise which is a particular case of a martingale difference sequence).
2. Check if the observation vector $\phi(t)$ is bounded in the sense of (4.117) using for example Lemma 4.1.
3. Apply Theorem 4.2 which means checking the strictly real positivity condition of (4.118).

*Example* (Output Error with Extended Prediction Model (OEEPM)) This algorithm has been presented in a deterministic context in Sect. 3.3.5. In a stochastic context, the model of the system to be estimated is considered to be an ARMAX model:

$$y(t+1) = -A^*(q^{-1})y(t) + B^*(q^{-1})u(t) + C(q^{-1})\omega(t+1) \tag{4.130}$$

where $\{\omega(t+1)\}$ is a martingale difference sequence (which can be viewed as a generalization of a zero mean white noise) and:

$$A(q^{-1}) = 1 + a_1 q^{-1} + \cdots + a_{n_A} q^{-n_A} = 1 + q^{-1}A^*(q^{-1}) \tag{4.131}$$

$$B(q^{-1}) = b_1 q^{-1} + \cdots + b_{n_B} q^{-n_B} = q^{-1}B^*(q^{-1}) \tag{4.132}$$

$$C(q^{-1}) = 1 + C_1 q^{-1} + \cdots + c_{n_C} q^{-n_C} = 1 + q^{-1}C^*(q^{-1}) \tag{4.133}$$

It is assumed that the polynomials $A(q^{-1})$ and $C(q^{-1})$ are asymptotically stable.

The adjustable predictor is given by:

$$\hat{y}^0(t+1) = \hat{\theta}^T(t)\phi(t) \tag{4.134}$$

$$\hat{y}(t+1) = \hat{\theta}^T(t+1)\phi(t) \tag{4.135}$$

where:

$$\hat{\theta}^T(t) = [\hat{a}_1(t), \ldots, -\hat{a}_{n_A}(t), \hat{b}_1(t), \ldots, \hat{b}_{n_B}(t), \hat{h}_1(t), \ldots, \hat{h}_{n_H}(t)] \tag{4.136}$$

$$\phi^T(t) = [-\hat{y}(t), \ldots, -\hat{y}(t-n_A+1), u(t), \ldots, u(t-n_B+1), \nu(t), \ldots, \nu(t-n_H+1)] \tag{4.137}$$

The a priori adaptation (prediction) error is given by:

$$\nu^0(t+1) = \varepsilon^0(t+1) = y(t+1) - \hat{\theta}^T(t)\phi(t) \tag{4.138}$$

The a posteriori adaptation (prediction) error is given by:

$$\nu(t+1) = \varepsilon(t+1) = y(t+1) - \hat{\theta}^T(t+1)\phi(t) \tag{4.139}$$

and one uses the PAA of (4.95) through (4.97).

*Step* 1: Derivation of the a posteriori adaptation error equation
Taking into account (4.130), (4.139) can be rewritten as (after adding and subtracting the terms $\pm A^*\hat{y}(t) \pm C^*\nu(t)$):

$$\begin{aligned}\nu(t+1) &= -A^*\hat{y}(t) + B^*u(t) - A^*[y(t) - \hat{y}(t)] + C^*\nu(t) \\ &\quad - C^*\nu(t) - \hat{\theta}^T(t+1)\phi(t) + C(q^{-1})\omega(t+1)\end{aligned} \tag{4.140}$$

defining $\theta$ as:

$$\theta = [a_1, \ldots, a_{n_A}, b_1, \ldots, b_{n_B}, h_1, \ldots, h_{n_H}] \tag{4.141}$$

where:

$$h_i = c_i - a_i \tag{4.142}$$

Equation (4.140) becomes:

$$\nu(t+1) = [\theta - \hat{\theta}(t+1)]^T\phi(t) - C^*(q^{-1})\nu(t) + C(q^{-1})\omega(t+1) \tag{4.143}$$

Passing the term $\pm C^*(q^{-1})\nu(t)$ in the left hand side and passing both sides through $1/C(q^{-1})$ one obtains:

$$\nu(t+1) = \frac{1}{C(q^{-1})}[\theta - \hat{\theta}(t+1)]^T\phi(t) + \omega(t+1) \tag{4.144}$$

which has the desired form given in (4.98). From (4.144) taking into account that $H_1(q^{-1}) = 1$ and $H_2(q^{-1}) = C(q^{-1})$ one deduces that $\nu^0(t+1)$ satisfies:

$$\nu^0(t+1) = [\theta - \hat{\theta}(t)]^T\phi(t) + C(q^{-1})\omega(t+1) - C^*(q^{-1})\nu(t) \tag{4.145}$$

and one concludes that $[\nu^0(t+1) - \omega(+1)]$ is $\mathcal{F}_t$ measurable.

*Step* 2: Boundedness of $\phi(t)$
From (4.138) one has:

$$\hat{\theta}^T(t)\phi(t) + \nu^0(t+1) = y(t+1) \tag{4.146}$$

and condition (2) of Lemma 4.1 is satisfied since $A(q^{-1})$ and $C(q^{-1})$ are asymptotically stable and $u(t)$ is bounded which implies that $y(t)$ is bounded (i.e., $\lim_{N\to\infty} \frac{1}{N}\sum_{t=1}^{N} y^2(t) < \infty$). On the other hand:

$$\frac{1}{N}\sum_{t=1}^{N}\hat{y}^2(t) = \frac{1}{N}\sum_{t=1}^{N}[y(t) - \nu(t)]^2 \le K_1 + \frac{K_2}{N}\sum_{t=1}^{N}\nu^2(t) \tag{4.147}$$

which implies that the first $n_A$ components of $\phi(t)$ satisfy the condition (1) of Lemma 4.1 Since $u(t)$ is bounded, the next $n_B$ components of $\phi(t)$ will also satisfy condition (1) of Lemma 4.1 and the last $n_H$ components of $\phi(t)$ will obviously satisfy this condition (with $K_1 = 0$ and $K_2 = 1$). This implies that condition (1) of Theorem 4.2 is satisfied.

*Step* 3: Application of Theorem 4.2.

Applying Theorem 4.2, one concludes that the convergence condition assuming the properties (4.119) through (4.122) is:

$$H'(z^{-1}) = \frac{1}{C(z^{-1})} - \frac{1}{2} \tag{4.148}$$

be a strictly positive real transfer function ($\lambda_2(t) \equiv 1$).

*Remarks*

- This convergence condition (4.148) is similar to that obtained using the ODE method (Dugard and Landau 1980 and Problem 4.1) but is more powerful, since through this approach, one shows that monitoring of the estimated parameters in order to force them in the stability domain $D_s$ for all $t$ is not necessary.
- This algorithm has been originally developed for removing the SPR condition in the deterministic context, but a SPR condition for convergence appears in a stochastic environment (it depends only upon the noise model).

*Proof* The proof of Theorem 4.2 uses essentially the basic result of Theorem D.2, Appendix D which gives the behavior of a general feedback system formed by a linear feedforward block, whose output is disturbed by a martingale difference sequence and a linear time-varying feedback block. Taking into account the equivalent system given by (4.113) through (4.116), one can apply Theorem D.2 by making the following substitutions:

$$\begin{aligned}
&A(t) = I; \qquad B(t) = F(t)\phi(t); \qquad C(t) = \phi^T(t) \\
&D(t) = \phi^T(t)F(t)\phi(t); \qquad A = A; \qquad B = b; \qquad C = c^T \\
&D = \delta; \qquad \bar{x}(t) = \tilde{\theta}(t) \\
&y(t+1) = \nu(t+1) - \omega(t+1); \qquad \tilde{y}(t+1) = \nu(t+1) = \bar{u}(t+1) \\
&\bar{y}(t+1) = -u(t+1) = \tilde{\theta}^T(t+1)\phi(t)
\end{aligned} \tag{4.149}$$

The feedback block defined by (4.115) and (4.116) belongs to the class $N(\Gamma)$ for $N(\Gamma) = \lambda_2(t)$ (see the proof of Theorem 3.1, Sect. 3.3.4).

The condition (3) of Theorem D.2 will be satisfied if there is $\lambda_2 \geq \max_t \lambda_2(t)$ such that $H(z^{-1}) - \frac{\lambda_2}{2}$ be a strictly positive real discrete transfer function (i.e., $H(z^{-1})$ has an excess of passivity to compensate for the lack of passivity of the feedback block). Since $H(z^{-1})$ is a ratio of monic polynomials, $\lambda_2 < 2$ is a necessary condition for strict positive realness of $H(z^{-1}) - \frac{\lambda_2}{2}$. This covers the possible domain allowed for the variation of $\lambda_2(t)$ as indicated in (4.96).

Condition (1) of Theorem D.2 requires to evaluate $\alpha(t)$ given by (D.12). In this specific case taking into account the equivalence of notation given in (4.149), one has to check that:

$$\alpha(t+1) = \frac{1}{t+1}\mathbf{E}\{\tilde{\theta}^T(t+1)\phi(t)\omega(t+1) \mid \mathcal{F}_t\} \geq 0; \quad \forall t \geq 0 \tag{4.150}$$

Taking into account (4.112) and (4.106), one has:

$$\begin{aligned}\bar{y}(t+1) &= \phi^T(t)\tilde{\theta}(t+1)\\ &= \phi^T(t)[\hat{\theta}(t)-\theta]^T + \gamma(t)\phi^T(t)F(t+1)\phi(t)[\nu^0(t+1)-\omega(t+1)]\\ &\quad + \gamma(t)\phi^T(t)F(t+1)\phi(t)\omega(t+1)\end{aligned} \tag{4.151}$$

The important remark to be made is that $\nu^0(t+1)$ satisfies (4.105) from which one concludes that $[\nu^0(t+1)-\omega(t+1)]$ is $\mathcal{F}_t$ measurable. On the other hand from (4.96), one concludes that $F(t+1)$ is $\mathcal{F}_t$ measurable and from (4.95) and assumption (iii) of Theorem 4.2 upon $\phi(t)$, one concludes that $\hat{\theta}(t)$ is $\mathcal{F}_t$ measurable. Multiplying both sides of (4.151) by $\omega(t+1)$ and taking the conditional expectation, one obtains:

$$\begin{aligned}\mathbf{E}\{\bar{y}(t+1)\omega(t+1)\mid\mathcal{F}_t\} &= \gamma(t)\phi^T(t)F(t+1)\phi(t)\mathbf{E}\{\omega^2(t+1)\mid\mathcal{F}_t\}\\ &= \gamma(t)\phi^T(t)F(t+1)\phi(t)\sigma^2\end{aligned} \tag{4.152}$$

because:

$$\mathbf{E}\{\phi^T(t)[\hat{\theta}(t)-\theta]\omega(t+1)\mid\mathcal{F}_t\}=0 \tag{4.153}$$

and:

$$\mathbf{E}\{\gamma(t)\phi^T(t)F(t+1)\phi(t)[\nu^0(t+1)-\omega(t+1)]\omega(t+1)\mid\mathcal{F}_t\}=0 \tag{4.154}$$

in view of the remarks above on $F(t+1), \phi(t), \hat{\theta}(t)$ and $[\nu^0(t+1)-\omega(t+1)]$. Introducing (4.152) in (4.150), one concludes that condition (1) of Theorem D.2 is always verified.

Condition (2) of Theorem D.2 becomes:

$$\begin{aligned}\sum_{t=1}^{\infty}\alpha(t) &= \sigma^2\sum_{t=1}^{\infty}\mu(t)\frac{\phi^T(t)F(t+1)\phi(t)}{t+1}\\ &< 2\sigma^2\sum_{t=1}^{\infty}\frac{\phi^T(t)F(t+1)\phi(t)}{t+1}<\infty\end{aligned} \tag{4.155}$$

To show that condition (1) of Theorem 4.2 implies that (4.155) is true, one needs some intermediate relationship.

**Lemma 4.2** *For the adaptation gain satisfying* (4.96), *one has*:

$$\sum_{t=0}^{\infty}\phi^T(t)F(t+1)^2\phi(t)<\infty \tag{4.156}$$

*Proof* From (4.96) (after using the matrix inversion lemma), one gets by left multiplication with $\phi^T(t)$ and right multiplication with $F(t+1)\phi(t)$:

$$\begin{aligned}\phi^T(t)F(t+1)^2\phi(t) &= \phi^T(t)F(t)F(t+1)\phi(t)\\ &\quad - \frac{\phi^T(t)F(t)\phi(t)\phi^T(t)F(t)F(t+1)\phi(t)}{\phi^T(t)F(t)\phi(t)}\end{aligned} \tag{4.157}$$

from which one concludes taking into account (4.96) that:

$$\begin{aligned}\phi^T(t)F(t+1)^2\phi(t) &\le \phi^T(t)F(t)F(t+1)\phi(t)\\ &= \mathrm{tr}[\phi(t)\phi^T(t)F(t+1)F(t)]\\ &\le \frac{1}{\lambda_{2\min}}\mathrm{tr}\{[F(t+1)^{-1}-F(t)^{-1}]F(t+1)F(t)\}\\ &= \frac{1}{\lambda_{2\min}}\mathrm{tr}\,[F(t)-F(t+1)]\end{aligned} \tag{4.158}$$

and:

$$\sum_{t=1}^{\infty}\phi^T(t)F(t+1)^2\phi(t) \le \frac{1}{\lambda_{2\min}}\mathrm{tr}\,[F(0)-F(\infty)] < \infty \tag{4.159}$$

□

On the other hand (Solo 1979):

$$\begin{aligned}&\phi^T(t)F(t+1)\phi(t)\\ &\quad = \phi^T(t)F(t+1)F(t+1)^{-1}F(t+1)\phi(t)\\ &\quad \le \mathrm{tr}[F(t+1)^{-1}]\phi^T(t)F(t+1)^2\phi(t)\\ &\quad = \left\{\sum_{i=0}^{t}\phi^T(i)\phi(i)+\mathrm{tr}[F(0)^{-1}]\right\}\phi^T(t)F(t+1)^2\phi(t)\end{aligned} \tag{4.160}$$

and, therefore:

$$\sum_{t=1}^{\infty}\frac{\phi^T(t)F(t+1)\phi(t)}{t+1} \le \sum_{t=1}^{\infty}\frac{\mathrm{tr}[F(t+1)^{-1}]}{t+1}\phi^T(t)F(t+1)^2\phi(t) < \infty \tag{4.161}$$

because of (4.156) and condition (1) of Theorem 4.2 which assures $\mathrm{tr}[F(t+1)^{-1}]/(t+1) < \infty$ and therefore (4.155) is true. Therefore, the conditions imposed by Theorem D.2 correspond to those of Theorem 4.2, (4.119), (4.121), (4.122) and (4.124) result from (D.15), (D.16) and (D.17) taking into account the corresponding substitutions given in (4.149) and the fact that $\bar{P}(t) = F(t)^{-1}$. Equation (4.120) results from (4.116) as it will be shown below (Solo 1979):

$$\begin{aligned}\lim_{N\to\infty}\frac{1}{N}\sum_{t=1}^{N}\nu^2(t) &= \lim_{N\to\infty}\frac{1}{N}\sum_{t=1}^{N}[\nu(t)-\omega(t)]^2 + \lim_{N\to\infty}\frac{1}{N}\sum_{t=1}^{N}2\omega(t)[\nu(t)-\omega(t)]\\ &\quad + \lim_{N\to\infty}\frac{1}{N}\sum_{t=1}^{N}\omega^2(t)\end{aligned} \tag{4.162}$$

But in the above equation:

$$\begin{aligned}&\lim_{N\to\infty}\frac{1}{N}\sum_{t=1}^{N}\omega(t)[\nu(t)-\omega(t)]\\ &\quad \le \lim_{N\to\infty}\left[\left\{\frac{1}{N}\sum_{t=1}^{N}\omega^2(t)\right\}\left\{\frac{1}{N}\sum_{t=1}^{N}[\nu(t)-\omega(t)]^2\right\}\right]^{1/2} = 0\end{aligned} \tag{4.163}$$

because of (4.101) and (4.119) and one concludes that (4.120) holds.

*Proof* Lemma 4.1 is proven as follows. From (4.106) and (4.96), one obtains for $\lambda_2(t) \equiv 1$:

$$\begin{aligned}\hat{\theta}(t+1)F(t+1)^{-1}\hat{\theta}(t+1) &= \hat{\theta}^T(t)F(t)^{-1}\hat{\theta}(t) + \nu^0(t+1)^2\phi^T(t)F(t+1)\phi(t) \\ &\quad + [\hat{\theta}^T(t)\phi(t)]^2 + 2[\hat{\theta}^T(t)\phi(t)]\nu^0(t+1) \\ &= \hat{\theta}^T(t)F(t)^{-1}\hat{\theta}(t) + [\hat{\theta}^T(t)\phi(t) + \nu^0(t+1)]^2 \\ &\quad - [\nu^0(t+1)]^2[1-\phi^T(t)F(t+1)\phi(t)]\end{aligned} \tag{4.164}$$

Summing up from 1 to $N$, one obtains from (4.164) after rearranging the various terms:

$$\begin{aligned}&\frac{1}{N}\hat{\theta}^T(N+1)F(N+1)^{-1}\hat{\theta}(N+1) - \frac{1}{N}\hat{\theta}^T(1)F(1)^{-1}\hat{\theta}(1) \\ &\quad + \frac{1}{N}\sum_{t=1}^{N}[\nu^0(t+1)]^2[1-\phi^T(t)F(t+1)\phi(t)] \\ &= \frac{1}{N}\sum_{t=1}^{N}[\hat{\theta}^T(t)\phi(t) + \nu^0(t+1)]^2\end{aligned} \tag{4.165}$$

Using condition (2) of Lemma 4.1, one concludes that:

$$\lim_{N\to\infty}\frac{1}{N}\sum_{t=1}^{N}[\nu^0(t+1)]^2[1-\phi^T(t)F(t+1)\phi(t)] < \infty \tag{4.166}$$

because the first term of the left hand side of (4.165) is positive or null as $N \to \infty$, and the second term is negative but tends to zero as $N \to \infty$.

Using the matrix inversion lemma, (4.97) can be rewritten as:

$$\nu(t+1) = \frac{\nu^0(t+1)}{1+\phi^T(t)F(t)\phi(t)} = \nu^0(t+1)[1-\phi^T(t)F(t+1)\phi(t)] \tag{4.167}$$

and this allows us to reformulate (4.166) as:

$$\lim_{N\to\infty}\frac{1}{N}\sum_{t=1}^{N}[\nu(t+1)]^2[1+\phi^T(t)F(t)\phi(t)] < \infty \tag{4.168}$$

which implies (4.128). (4.128) together with the condition (1) of the Lemma 4.1 implies that (4.129) holds. □

When the condition (1) of Theorem 4.2 cannot be directly verified, the following result is useful:

**Theorem 4.3** *Consider the PAA given by* (4.95) *through* (4.97) *and the assumption* (i) *through* (iii) *of Theorem* 4.2.

*If there is* $\lambda_2 : \max_t \lambda_2(t) \leq \lambda_2 < 2$ *such that*:

$$H'(z^{-1}) = H(z^{-1}) - \frac{\lambda_2}{2} \tag{4.169}$$

*is a strictly positive real transfer function*, *one has*:

(a) $$\lim_{N\to\infty} \sum_{t=0}^{N} \frac{[\hat{\theta}(t+1) - \theta]^T \phi(t)]^2}{r(t+1)} < \infty \quad a.s. \tag{4.170}$$

(b) $$\lim_{N\to\infty} \sum_{t=0}^{N} \frac{[\nu(t) - \omega(t)]^2}{r(t)} < \infty \quad a.s. \tag{4.171}$$

(c) $$\lim_{N\to\infty} \sum_{t=1}^{N} \frac{\phi^T(t)F(t)\phi(t)}{r(t)} \nu^2(t+1) < \infty \quad a.s. \tag{4.172}$$

*where*:

$$\begin{aligned} r(t+1) &= r(t) + \lambda_2(t)\phi^T(t)\phi(t) \\ r_0 &= \operatorname{tr} F_0^{-1}; \quad 0 < \lambda_2(t) < 2 \end{aligned} \tag{4.173}$$

*Remarks*

- This theorem shows that some normalized quantities are bounded.
- The results of Theorem 4.3 are weaker than those of Theorem 4.2, but from the specific properties of the sequence $r(t)$ it is possible in certain cases to conclude upon the convergence of the interesting variables (see Sect. 11.4 and Goodwin et al. 1980c; Goodwin and Sin 1984).
- If one can assume that in addition to condition (1), $\lim_{t\to\infty} r(t) = \infty$, using Kronecker lemma (Appendix D, Lemma D.1) one obtains from (4.170), (4.171) and (4.172) the results of Theorem 4.2
- The use of this theorem will be illustrated in Chap. 11 in the context of direct adaptive control in a stochastic environment.

*Proof* The proof is based on the use of Theorem D.3 (Appendix D) taking into account the substitutions considered for the proof of Theorem 4.2. $r(t+1)$ given in (4.173) satisfies condition (1) of Theorem D.3. Note that $r(t+1)$ is $\mathcal{F}_t$ measurable. One has to verify now conditions (2) and (3) of Theorem D.3.

Multiplying both sides of (4.151) by $\omega(t+1)/r(t+1)$ and then computing the conditional expectation by taking into account that $\hat{\theta}(t)$, $F(t+1)$, $[\nu^0(t+1) - \omega(t+1)]$ and $r(t+1)$ are $\mathcal{F}_t$ measurable, one obtains:

$$\mathbf{E}\left\{ \frac{\bar{y}(t+1)\omega(t+1)}{r(t+1)} \bigg| \mathcal{F}_t \right\} = \gamma(t) \frac{\phi^T(t)F(t+1)\phi(t)}{r(t+1)} \sigma^2 \geq 0 \tag{4.174}$$

Therefore condition (2) of Theorem D.3 is satisfied. Then:

$$\sum_{t=0}^{\infty} \alpha(t+1) \leq 2\sigma^2 \sum_{t=0}^{\infty} \frac{\phi^T(t)F(t+1)\phi(t)}{r(t+1)} \tag{4.175}$$

But:

$$\phi^T(t)F(t+1)\phi(t) \leq |F(t+1)^{-1}|\phi(t)F(t+1)^2\phi(t) \\ \leq \mathrm{tr}[F(t+1)^{-1}]\phi(t)F(t+1)^2\phi(t) \tag{4.176}$$

and therefore (4.174) becomes:

$$\sum_{t=0}^{\infty}\alpha(t+1) \leq 2\sigma^2\sum_{t=0}^{\infty}\frac{\mathrm{tr}[F(t+1)^{-1}]}{r(t+1)}\phi^T(t)F(t+1)^2\phi(t) \\ = 2\sigma^2\sum_{t=0}^{\infty}\phi^T(t)F(t+1)^2\phi(t) \leq \frac{2\sigma^2}{\lambda_{2\min}}\mathrm{tr}[F(0)-F(\infty)] < \infty \tag{4.177}$$

since by construction $r(t+1) = \mathrm{tr}[F(t+1)^{-1}]$. Therefore condition (3) of Theorem D.3 is also satisfied. Condition (4) of Theorem D.3 corresponds to condition (4) of Theorem 4.3. Then from (D.36) and (D.39) making the appropriate substitutions, one obtains (4.170) through (4.172). □

## 4.4 The Frequency Domain Approach

One can consider that the various recursive parameter adaptation algorithms try to minimize recursively a quadratic criterion in terms of the plant model prediction error or, in general, in terms of the adaptation error, and the optimal value of the estimated parameter vector is given by:

$$\hat{\theta}^* = \arg\min_{\hat{\theta}\in\mathcal{D}}\lim_{t\to\infty}\frac{1}{N}\sum_{t=1}^{N}\nu^2(t,\hat{\theta}) \approx \arg\min_{\hat{\theta}\in\mathcal{D}}\mathbf{E}\{\nu^2(t,\hat{\theta})\} \tag{4.178}$$

where $\mathcal{D}$ is the domain of admissible parameters related to the model set and $\nu(t)$ is the adaptation error. Under the basic assumption that:

$$\lim_{t\to\infty}\frac{1}{N}\sum_{t=1}^{N}\nu^2(t,\hat{\theta}) < \infty \tag{4.179}$$

(i.e., the adaptation error is mean square bounded or an extended $l_2$ space is used), one can use the Parseval Theorem to get a frequency interpretation of the criterion (4.178):

$$\hat{\theta}^* \approx \arg\min_{\hat{\theta}\in\mathcal{D}}\frac{1}{2\pi}\int_{-\pi}^{\pi}\phi_\nu(e^{-j\omega},\hat{\theta})d\omega \tag{4.180}$$

where $\phi_\nu(e^{-j\omega})$ is the spectral density of the adaptation error (the $\frac{1}{2\pi}$ can be dropped out). Equation (4.180) allows to assess the possible achievable performances of a given parameter adaptation algorithm. However, it does not provide any information if these performances will be obtained (i.e., a convergence analysis is still necessary). Nevertheless, the knowledge of the possible achievable performances are useful:

- for understanding the relative effect of the noise and input spectral characteristics;
- for understanding the effect of the various data and error filters which may be introduced;
- for assessing the frequency distribution of the (asymptotic) bias.

In addition, this approach also allows to assess qualitatively the performances of the adaptation algorithms when the plant model and the adjustable model do not have the same structure (i.e., when there is no value $\hat{\theta}$ for which the prediction error is zero in the absence of noise for all types of input signals). This method has been introduced by Ljung (1999). To illustrate this approach, we will consider the two basic models introduced in Sect. 2.2.

- *Equation error model*

$$y(t) = \frac{q^{-d}B(q^{-1})}{A(q^{-1})}u(t) + \frac{C(q^{-1})}{A(q^{-1})}e(t) \tag{4.181}$$

  where $e(t)$ is a zero mean white noise sequence.
- *Output error model*

$$y(t) = \frac{q^{-d}B(q^{-1})}{A(q^{-1})}u(t) + v(t) \tag{4.182}$$

  where $v(t)$ is a zero mean noise with finite power and independent with respect to the input $u(t)$.

Model (4.181) can be written under the form:

$$y(t) = G(q^{-1})u(t) + W(q^{-1})e(t) \tag{4.183}$$

and model (4.182) can be written under the form:

$$y(t) = G(q^{-1})u(t) + v(t) \tag{4.184}$$

For the model (4.183), the optimal predictor (in the sense that the prediction error tends asymptotically to a white noise when the estimated parameters are equal to the true ones) is given by:

$$\begin{aligned}\hat{y}(t) &= \hat{W}^{-1}(q^{-1})[\hat{G}(q^{-1})u(t) + (\hat{W}(q^{-1}) - 1)y(t)]\\ &= \hat{G}(q^{-1})u(t) + [\hat{W}(q^{-1}) - 1]\varepsilon(t)\end{aligned} \tag{4.185}$$

where $\hat{G}(q^{-1}, \hat{\theta})$ and $\hat{W}(q^{-1}, \hat{\theta})$ describing the estimated model, have been replaced by $\hat{G}(q^{-1})$ and $\hat{W}(q^{-1})$ in order to simplify the writing.

The prediction error takes the form:

$$\begin{aligned}\varepsilon(t) &= [G(q^{-1}) - \hat{G}(q^{-1})]u(t) + W(q^{-1})e(t) - [\hat{W}(q^{-1}) - 1]\varepsilon(t)\\ &= \hat{W}^{-1}(q^{-1})\{[G(q^{-1}) - \hat{G}(q^{-1})]u(t) + [W(q^{-1}) - \hat{W}]e(t)\} + e(t)\end{aligned} \tag{4.186}$$

which asymptotically tends to $e(t)$ when $\hat{G}(q^{-1}) = G(q^{-1})$, $\hat{W}(q^{-1}) = W(q^{-1})$ (under the assumption that $\hat{W}(q^{-1})$ is asymptotically stable).

For the model (4.184), the optimal predictor is given by:

$$\hat{y}(t) = \hat{G}(q^{-1})u(t) \tag{4.187}$$

and the output (prediction) error takes the form:

$$\varepsilon(t) = [G(q^{-1}) - \hat{G}(q^{-1})]u(t) + v(t) \tag{4.188}$$

To use (4.180), one proceeds as follows:

1. For a given recursive algorithm and a given assumption upon the noise, one writes the adaptation error expression for a constant $\hat{\theta}$.
2. One computes $\mathbf{E}\{\nu^2(t,\hat{\theta})\}$ taking into account the hypotheses upon the noise, the structure of the predictor and the principle used in the algorithm (asymptotic whitening of the adaptation error or asymptotic decorrelation between adaptation error and observation error).
3. One computes the Fourier transform of $\mathbf{E}\{\nu^2(t,\hat{\theta})\}$ and neglects the terms which do not depend upon the estimated parameters.

As an application of this approach, we will consider the recursive least squares and the output error algorithm. For recursive least squares $\hat{W}(q^{-1}) = \hat{A}^{-1}(q^{-1})$ and $\hat{y}(t)$ will have the form:

$$\begin{aligned}\hat{y}(t) &= q^{-d}\hat{B}(q^{-1})u(t) - q^{-1}\hat{A}^*(q^{-1})y(t)\\ &= \hat{A}(q^{-1})\{\hat{G}(q^{-1})u(t) + [\hat{A}^{-1}(q^{-1}) - 1]y(t)\}\end{aligned} \tag{4.189}$$

and the prediction error is given by:

$$\varepsilon(t) = \hat{A}(q^{-1})\{[G(q^{-1}) - \hat{G}(q^{-1})]u(t) + [W(q^{-1}) - \hat{A}^{-1}(q^{-1})]e(t)\} + e(t) \tag{4.190}$$

Therefore:

$$\begin{aligned}\mathbf{E}\{\varepsilon^2(t,\hat{\theta}\} &= |\hat{A}(q^{-1})|^2[|G(q^{-1}) - \hat{G}(q^{-1})|^2\mathbf{E}\{u^2(t)\}\\ &\quad + |W(q^{-1}) - \hat{A}^{-1}(q^{-1})|^2\mathbf{E}\{e^2(t)\}] + \mathbf{E}\{e^2(t)\}\end{aligned} \tag{4.191}$$

and taking the Fourier transform one gets:

$$\begin{aligned}\hat{\theta}^* = \arg\min_{\hat{\theta}\in\mathcal{D}} \int_{-\pi}^{\pi} &|\hat{A}(e^{-j\omega})|^2[|G(e^{-j\omega}) - \hat{G}(e^{-j\omega})|^2\phi_u(\omega)\\ &+ |W(e^{-j\omega}) - \hat{A}^{-1}(e^{-j\omega})|^2\phi_e(\omega)]d\omega\end{aligned} \tag{4.192}$$

and one clearly sees that:

1. Unbiased estimates can only be obtained for $W(q^{-1}) = A^{-1}(q^{-1})$ (assuming that $G$ and $\hat{G}$ have the same structure).
2. In the absence of noise, if $G$ and $\hat{G}$ do not have the same structure, the *bias* error (the difference between $G$ and $\hat{G}$) in the frequency domain will be weighted by $|\hat{A}(e^{-j\omega})|^2$ and $\phi_u(\omega)$, i.e., it will be small at the frequencies where $\phi(\omega)$ and $|\hat{A}(e^{-j\omega})|^2$ are large.

Similarly, for the output error recursive algorithm using (4.187) and (4.188), one gets for the case without compensator on the output error:

$$\hat{\theta}_{opt} = \arg\min_{\hat{\theta}\in\mathcal{D}} \int_{-\pi}^{\pi} |G(e^{-j\omega}) - \hat{G}(e^{-j\omega})|^2 \phi_u(\omega) d\omega \tag{4.193}$$

since $\phi_{uv}(\omega) = 0$.

One concludes that:

1. Unbiased parameters estimates are obtained when $G$ and $\hat{G}$ have the same structure.
2. When $G$ and $\hat{G}$ do not have the same structure, the frequency distribution of the bias is affected by the spectral distribution of $\phi_u(\omega)$.

When a linear compensator is used, i.e.

$$\nu(t) = D(q^{-1})\varepsilon(t)$$

one gets:

$$\hat{\theta}_{opt} = \arg\min_{\hat{\theta}\in\mathcal{D}} \frac{1}{2\pi} \int_{-\pi}^{\pi} |D(e^{-j\omega})|^2 |G(e^{-j\omega}) - \hat{G}(e^{-j\omega})|^2 \phi_u(\omega) d\omega \tag{4.194}$$

and $D(q^{-1})$ will introduce a frequency weighting on the bias when $G$ and $\hat{G}$ will not have the same structure.

The general formula for the bias distribution corresponding to the model (4.183), is obtained using (4.186):

$$\begin{aligned}\hat{\theta}_{opt} = \arg\min_{\hat{\theta}\in\mathcal{D}} \int_{-\pi}^{\pi} &|\hat{W}^{-1}(e^{-j\omega})|^2 [|G(e^{-j\omega}) - \hat{G}(e^{-j\omega})|^2 \phi_u^2(\omega) \\ &+ |W(e^{-j\omega}) - \hat{W}(e^{-j\omega})|^2 \phi_e(\omega)] d\omega\end{aligned} \tag{4.195}$$

For the case of the model (4.187), the bias distribution takes the form (4.193) (if the prediction error is not filtered), which results from (4.195) by taking $W(e^{-j\omega}) = \hat{W}(e^{-j\omega}) = 1$, $e(t) = v(t)$ and assuming $\phi_{uv}(\omega) = 0$ (since $u(t)$ and $v(t)$ are independent).

This approach is very useful to assess qualitatively

- the influence of the various filters which may be introduced on the data and prediction errors,
- the influence of input spectrum and noise spectrum,
- the effects of the operation in closed loop upon the parameter estimation (see Chap. 9),
- the behavior of the algorithms when the plant model and the adjustable model do not have the same structure.

## 4.5 Concluding Remarks

1. The behavior of the PAA in the presence of stochastic disturbances depends upon the structure of the adjustable predictor and the way in which the observation vector and the adaptation error is generated.

2. In order to obtain asymptotically constant parameter estimates in the presence of stochastic disturbances, one has to use PAA with decreasing adaptation gain.
3. The typical effect of the presence of stochastic disturbances is an error on the estimated parameters called *bias*. As a consequence, one of the major objectives is to develop adaptive schemes leading to unbiased parameter estimates in the presence of stochastic disturbances. Often such schemes assure unbiased parameter estimates only for specific types of stochastic disturbances.
4. Three methods for the analysis of adaptive schemes in a stochastic environment have been considered:
   - averaging method (known also as the associated ODE method);
   - martingale approach;
   - frequency domain approach.

   Other approaches for the analysis of PAA in a stochastic environment can be considered.

For further reading see Landau (1979), Kushner and Clark (1978), Guo (1996), Goodwin and Sin (1984), Ljung and Söderström (1983), Ljung (1999).

## 4.6 Problems

**4.1** Analyze the behavior of the output error with extended prediction model (Sect. 3.3.5) in a stochastic environment when the plant model is an ARMAX model using the averaging method.

**4.2** Analyze the behavior of the recursive least squares in a stochastic environment when the plant model is an ARX model using the martingale approach.

**4.3** Analyze the behavior of the output error algorithm in a stochastic environment using the martingale approach assuming that the plant model is described by:

$$y(t+1) = \frac{B(q^{-1})}{A(q^{-1})}u(t) + \frac{1}{D(q^{-1})}w(t+1)$$

where $w(t+1)$ is a martingale sequence and $D(q^{-1})$ is the linear compensator on the output prediction error.

**4.4** Give the expression for the asymptotic frequency distribution of the bias for the output error with extended prediction model (Sect. 3.3.5).

**4.5** Give the expression for the asymptotic frequency distribution of the bias for the recursive least squares algorithm operating on filtered inputs and outputs ($L(q^{-1})y_f(t) = y(t)$, $L(q^{-1})u_f(t) = u(t)$). Discuss the effect of the filter upon the asymptotic frequency distribution of the bias.

**4.6** Consider the linear time-varying stochastic system:

$$x(t+1) = A(t+1,t)x(t) + B(t+1,t)v(t) \qquad (*)$$

where $A(t+1,t)$, $B(t+1,t)$ are time-varying matrices and $v(t)$ is a zero mean random vector [$\mathbf{E}\{v(t)\} = 0$]. The system $(*)$ is termed *asymptotically stable in the mean* if and only if:

$$\lim_{t\to\infty} \mathbf{E}\{x(t)\} = 0$$

Using the following theorem (Mendel 1973):

**Theorem** *The system* $(*)$ *is* asymptotically stable in the mean *if*:

1. *The free dynamic system*:

$$x(t+1) = A(t+1,t)x(t)$$

   *is asymptotically stable.*
2. $\lim_{t\to\infty} \mathbf{E}\{B(t+1,t)v(t)\} = 0$.
3. $\lim_{t\to\infty} \mathbf{E}\{\prod_{i=j}^{t} A(i+1,i)B(j,j-1)v(j-1)\} = 0$ *for all* $j = 1, 2, \ldots, t$.

Find for the recursive least squares and output error with decreasing adaptation gain, the conditions for asymptotic unbiasedness in the mean ($\lim_{t\to\infty} \mathbf{E}\{\theta - \hat{\theta}(t)\} = 0$).

Hint: Define the vector $x^T(t) = [\tilde{\theta}^T(t), \varepsilon^T(t)]$ where $\tilde{\theta}^T(t) = \hat{\theta}(t) - \theta$, $\varepsilon^T = [\varepsilon(t), \varepsilon(t-1), \ldots]$ and write an equation of the form $(*)$.

# Chapter 5
# Recursive Plant Model Identification in Open Loop

## 5.1 Recursive Identification in the Context of System Identification

In this chapter, we will focus on the identification of plant models operating in open loop using recursive parameter estimation. The recursive parameter estimation algorithms play a crucial role in the identification of plant models in real time.

From a practical point of view, plant model identification is a key starting point for designing a high-performance linear controller, but it is also a major step toward the design of an appropriate adaptive control scheme. Furthermore, the use of recursive identification techniques in real time is in many cases a first step toward the implementation of an adaptive control. It also serves for the initialization of adaptive control systems. It is therefore logical to consider the use of recursive parameter estimation algorithms for doing system identification.

The background in recursive parameter estimation both in a deterministic and stochastic environment is presented in Chaps. 3 and 4, and can be used to analyze or to synthesize recursive parameter estimation algorithms. Nevertheless, recursive parameter estimation represents only one aspect of system identification and it is important to locate it in the general context of system identification.

Identification of dynamic systems is an experimental approach for determining a dynamic model of a system. It includes four steps:

1. input-output data acquisition under an experimental protocol;
2. selection (or estimation) of the model complexity (structure);
3. estimation of the model parameters;
4. validation of the identified model (structure of the model and values of the parameters).

A complete identification operation must comprise the four stages indicated above. The specific methods used at each step depend on the type of model desired (parametric or non parametric, continuous-time or discrete-time). In our case, we will focus on the identification of discrete-time input-output models described by difference equations using recursive parameter estimation methods. As will be seen, the algorithms which will be used for parameter estimation will depend on the assumptions made on the noise disturbing the measurements, assumptions which have to be confirmed by the model validation.

**Fig. 5.1** The identification methodology

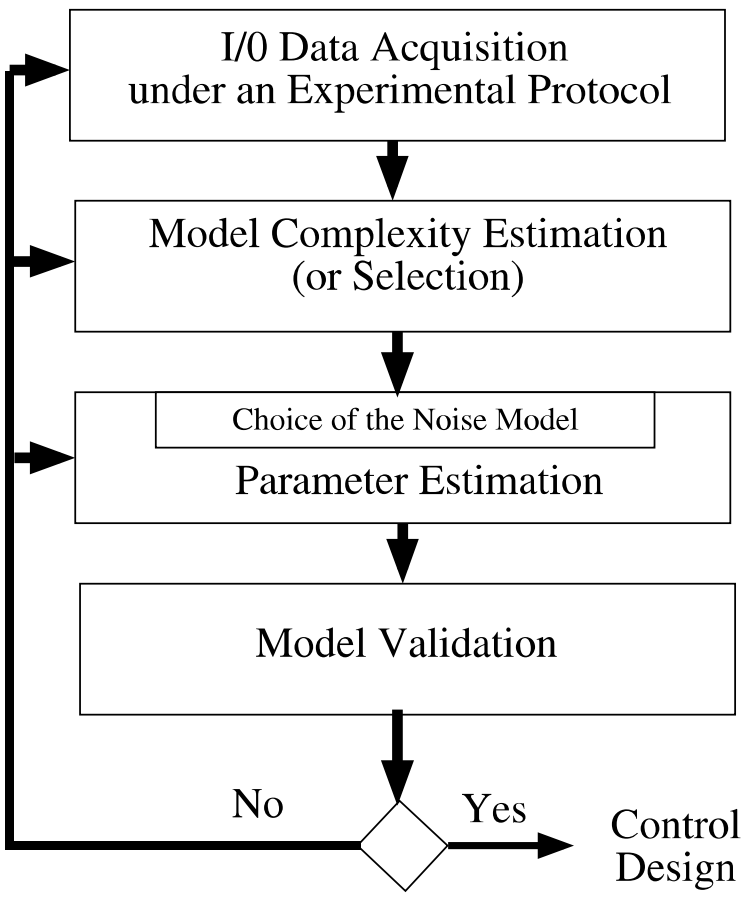

It is important to emphasize that no one single *plant* + *disturbance* structure exists that can describe all the situations encountered in practice. Furthermore, there is no parameter estimation algorithm that may be used with all possible plant + disturbance structures such that the estimated parameters are always unbiased. Furthermore, due to the lack of a priori information, the input-output data acquisition protocol may be initially inappropriate.

All these causes may lead to identified models which do not pass the validation test, and therefore the identification should be viewed as an iterative process as illustrated in Fig. 5.1. With respect to off-line parameter estimation techniques, the use of the recursive parameter estimation techniques offer three main advantages:

(a) An estimation of the model of the system can be obtained as the system evolves.
(b) The on-line estimation model contains all the information provided by the past input-output data. Therefore one achieves a significant data compression.
(c) Without major changes (just some choices in updating the adaptation gain ) one can follow the evolution of slowly time-varying systems.

It is also important to mention that the performance of these recursive algorithms makes them a competitive alternative to the off-line identification algorithm.

To keep the continuity of the text we will first discuss the recursive parameter estimation techniques and the model validation techniques. We will then go on to discuss the design of the input-output data acquisition protocol and the problem of model structure selection (estimation) which, in our case, concerns the selection (estimation) of the orders of the various polynomials and of the delay used in the discrete-time model of the system.

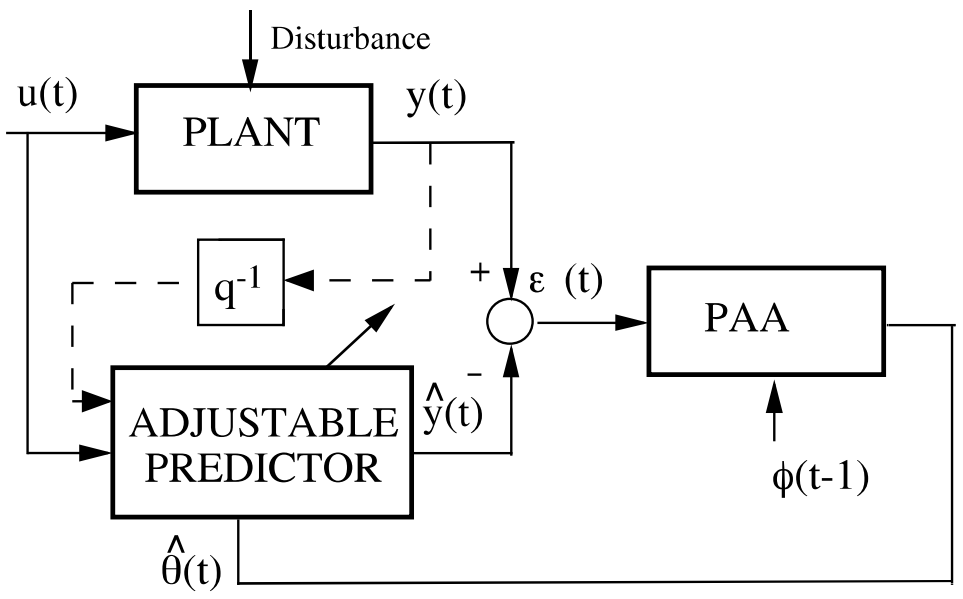

**Fig. 5.2** Structure of the recursive identification methods

## 5.2 Structure of Recursive Parameter Estimation Algorithms

All the recursive parameter estimation algorithms[1] fit within the schematic diagram shown in Fig. 5.2.

They all use the same structure for the parameter adaptation algorithm (PAA) with the different choices for the adaptation gain and this PAA has been extensively discussed in Sect. 3.3. All the methods use a one step ahead adaptive predictor which, under specific requirements upon the input, allows to obtain correct parameter estimates (under specific noises conditions). In other words, adaptive prediction is an intrinsic aspect of parameter estimation.

One of the major issues in parameter estimation for identification, is that for practical reasons, the excitation signal must be of very low magnitude. This constraint comes mainly from the fact that one often identifies systems which are in normal operation and the allowed additional excitation is very low. The fact that we would like to identify linear models around an operating point of the system, is also a reason for using low-excitation signal for certain types of systems.

Since, on the one hand, the allowed excitation is low, and on the other hand, the output measurements are always disturbed by noise, it results that the acquired data will be characterized in general by a significant noise/signal ratio. Therefore, the effect of the noise upon the parameter estimation algorithms cannot be neglected. As indicated in Chap. 4, the major effect is the bias error introduced on the estimated parameters. Removing the bias for different types of stochastic disturbances was one of the major driving factors in developing recursive parameter estimation algorithms. In fact, there is not a unique algorithm which can be used with all the possible disturbance structures such the estimated parameters are always unbiased.

The various algorithms are distinguished from each other by the following:

- predictor structure;
- nature of the components of the observation vector $\phi(t)$;
- dimension of the adjustable parameter vector $\hat{\theta}(t)$ and of the observation vector $\phi(t)$;
- way in which the prediction errors and respectively the adaptation errors are generated.

[1]We will often call them recursive identification methods (algorithms).

The convergence properties in the presence of random disturbances, will depend upon the different choices indicated above.

Classification of the various algorithms can be done using different criteria.

As indicated in Sect. 4.1, to avoid the bias error, it is necessary that:

$$E\{\phi(t)\varepsilon(t+1)\} = 0 \quad \text{for } \hat{\theta} = \theta$$

where $\phi(t)$ is the observation error and $\varepsilon(t+1)$ is the prediction (or adaptation) error. Therefore, the predictor structure, the observation vector and the adaptation error should be chosen accordingly. In fact the various algorithm will try to satisfy this condition by assuring asymptotically one of the two following criteria:

1. $\varepsilon(t+1)$ (or $\nu(t+1)$ = adaptation error) is a white noise for $\hat{\theta} = \theta$.
2. $\phi(t)$ and $\varepsilon(t+1)$ (or $\nu(t+1)$) are uncorrelated (or independent) for $\hat{\theta} = \theta$.

Therefore, the various recursive identification methods can be divided into two categories according to the principle used for obtaining unbiased estimated parameters:

(1) identification methods based on the whitening of the prediction error;
(2) identification methods based on the decorrelation of the observation vector ($\phi$) and the prediction error ($\varepsilon$).

Other classification of recursive identification methods are possible. One such classification is related to the structure of the predictor and of the observation vector and one can distinguish:

*Equation Error Methods*: which include recursive least squares and the different extensions. These methods are characterized by the fact that the measured output (and its delayed values) are used both in the predictor equation and in the observation vector (eventually filtered). Each method tends to obtain asymptotically a white prediction error (white noise) for a class of disturbance models by recursively modeling the disturbance.

*Instrumental Variable Methods*: Using these methods one tries to generate an observation vector which is not correlated with the prediction errors, by generating instrumental variables which are correlated with the noise free output of the system and uncorrelated with the noise.

*Output Error Methods*: These methods are characterized by the fact that the adjustable predictor uses the previous values of the predicted output instead of the previous measurements of the system output. Furthermore, in the observation vector the output measurements are replaced by the predicted outputs.

*Input Error Methods*: Using these methods, one identifies the inverse of the model. The adjustable predictor is in series with the plant and the output of the adjustable predictor is compared with a delayed input (Landau 1979).

Another possible classification is the way in which the algorithms are derived starting from appropriate off-line identification methods (Ljung and Söderström 1983). One can distinguish:

*Recursive Prediction Error Methods* (*RPEM*): Off-line identification procedures which minimize a criterion in terms of the prediction error (for a given disturbance) are transformed into recursive algorithms by sequential minimization of the criterion. The adjustable predictor takes the form of a predictor allowing to whiten asymptotically the prediction error. The observation vector multiplied by the prediction error is an estimate of the gradient of the criterion to be minimized. These algorithms involve on-line filtering of data through time-varying filters depending on the estimated parameters.

*Pseudo-Linear Regressions* (*PLR*): Include all the algorithms which uses as observation vector, the regressor vector used in the adjustable predictor. Most of them can be viewed as an approximation of the recursive prediction error methods (since they use the same type of predictor) or as a generalization of the recursive least squares in the sense that the observation vector depends itself on the previous values of the estimated parameters (which is not the case in RLS).

*Instrumental Variables* (*RIV*): See above.

However, independently of how these algorithms are derived, a stability analysis in deterministic environment and convergence analysis in stochastic environment is necessary.

One of the drawbacks of RPEM and RIV is that neither global stability nor global convergence can be insured and they require an initialization using other identification methods.

Derivation of pseudo-linear regression algorithms featuring global asymptotic stability property can be easily done using the method of equivalent feedback representation by (1) first writing the desired structure of the adjustable predictor (2) deriving an equation of the a posteriori prediction (adaptation) error of the form given in Theorem 3.2 and then using the corresponding PAA.

PLR methods are often subject to a positive real condition in a deterministic environment. Elimination of this positive real condition has motivated the development of a number of methods.

It is also interesting to remark that RPEM algorithms can be viewed as local solutions for relaxing the positive real conditions which occur in pseudo-linear regression algorithms (PLR).

From a practical point of view, the various classifications mentioned are not essential. What is important in practice is to be able to specify for a given algorithm:

(1) For what type of noise model, asymptotic unbiased estimates can be potentially obtained?
(2) What are the stability conditions in a deterministic environment and the convergence conditions in a stochastic environment?
(3) What are the initialization requirements (if any)?

Therefore, one has to consider the structure of the *plant + disturbance model* described by:

$$y(t) = \frac{q^{-d}B(q^{-1})}{A(q^{-1})}u(t) + v(t) \tag{5.1}$$

where $u(t)$ is the plant input, $y(t)$ is the measured output and $v(t)$ is the disturbance: Three models for the disturbance are considered in general.

(1)

$$v(t) = \text{zero mean stochastic process with finite moments independent of } u(t) \tag{5.2}$$

(2)

$$v(t) = \frac{C(q^{-1})}{A(q^{-1})} e(t) \tag{5.3}$$

where $e(t)$ is a zero mean white noise (or for technical reason defined as a martingale sequence). In this case, (5.1) is called an ARMAX model.

(3)

$$v(t) = \frac{C(q^{-1})}{A(q^{-1})D(q^{-1})} e(t) \tag{5.4}$$

The monic polynomials $C(q^{-1})$ and $D(q^{-1})$ are assumed to have their roots inside the unit circle (asymptotically stable polynomials) and they have the form:

$$C(q^{-1}) = 1 + c_1 q^{-1} + \cdots + c_{n_C} q^{-n_C} = 1 + q^{-1} C^*(q^{-1}) \tag{5.5}$$

$$D(q^{-1}) = 1 + d_1 q^{-1} + \cdots + d_{n_D} q^{-n_D} = 1 + q^{-1} D^*(q^{-1}) \tag{5.6}$$

Model (1) corresponds to output disturbances while models (2) and (3) correspond to the filtering of the disturbance through the poles of the system (termed also equation error disturbance).

We will also assume that the structure and the complexity (order the various polynomials) of the adjustable predictor is such that there is a parameter vector $\theta$ (or several) such that the prediction error $\varepsilon(t+1)$ in a deterministic environment be null and, in addition, in a stochastic environment either the prediction error is a white noise or $\mathbf{E}\{\varepsilon(t+1)\phi(t)\} = 0$, where $\phi(t)$ is the observation vector. In fact, one assumes that the true *plant + disturbance model* and the *adjustable predictor* have the same structure and, therefore, it is a value of the parameter vector $\theta$ for which the adjustable predictor will correspond to the true plant + disturbance model. One can say that the true plant + disturbance model is in the set of the possible estimated models parameterized by $\hat{\theta}$ or, in short, that "the true plant + disturbance model is in the *model set*" (Ljung 1999).

Table 5.1 summarizes a number of significant recursive parameter estimation techniques. They all use PAA of the form:

$$\hat{\theta}(t+1) = \hat{\theta}(t) + F(t)\Phi(t)\nu(t+1) \tag{5.7}$$

$$\begin{aligned} F(t+1)^{-1} &= \lambda_1(t)F(t)^{-1} + \lambda_2(t)\Phi(t)\Phi^T(t) \\ 0 &< \lambda_1(t) \le 1;\ 0 \le \lambda_2(t) < 2;\ F(0) > 0 \\ F^{-1}(t) &> \alpha F^{-1}(0);\ 0 < \alpha < \infty \end{aligned} \tag{5.8}$$

$$\nu(t+1) = \frac{\nu^0(t+1)}{1 + \Phi^T(t)F(t)\Phi(t)} \tag{5.9}$$

Integral + proportional PAA can also be used.

In the next section the various algorithms will be discussed and analyzed.

**Table 5.1** Recursive identification algorithms

(a)

| | | Recursive Least Squares (RLS) | Extended Least Squares (ELS) | Output Error with Extended Prediction Model (OEEPM) |
|---|---|---|---|---|
| Plant + noise model | | $y=\frac{q^{-d}B}{A}u+\frac{1}{A}e$ | $y=\frac{q^{-d}B}{A}u+\frac{C}{A}e$ | $y=\frac{q^{-d}B}{A}u+\frac{C}{A}e$ |
| Adjustable parameter vector | | $\hat{\theta}^T(t)=[\hat{a}^T(t),\hat{b}^T(t)]$<br>$\hat{a}^T(t)=[\hat{a}_1,\ldots,\hat{a}_{n_A}]$<br>$\hat{b}^T(t)=[\hat{b}_1,\ldots,\hat{b}_{n_B}]$ | $\hat{\theta}^T(t)=[\hat{a}^T(t),\hat{b}^T(t),\hat{c}^T(t)]$<br>$\hat{a}^T(t)=[\hat{a}_1,\ldots,\hat{a}_{n_A}]$<br>$\hat{b}^T(t)=[\hat{b}_1,\ldots,\hat{b}_{n_B}]$<br>$\hat{c}^T(t)=[\hat{c}_1\ldots\hat{c}_{n_C}]$ | $\hat{\theta}^T(t)=[\hat{a}^T(t),\hat{b}^T(t),\hat{h}^T(t)]$<br>$\hat{a}^T(t)=[\hat{a}_1,\ldots,\hat{a}_{n_A}]$<br>$\hat{b}^T(t)=[\hat{b}_1,\ldots,\hat{b}_{n_B}]$<br>$\hat{h}^T(t)=[\hat{h}_1\ldots\hat{h}_{n_H}]$<br>$n_H=\max(n_A,n_C)$ |
| Predictor regressor vector | | $\psi^T(t)=[-y(t),\ldots,-y(t-n_A+1),$<br>$u(t-d),\ldots,u(t-d-n_B+1)]$ | $\psi^T(t)=[-y(t),\ldots,-y(t-n_A+1),$<br>$u(t-d),\ldots,u(t-d-n_B+1),$<br>$\varepsilon(t),\ldots,\varepsilon(t-n_C+1)]$ | $\psi^T(t)=[-\hat{y}(t),\ldots,-\hat{y}(t-n_A+1),$<br>$u(t-d),\ldots,u(t-d-n_B+1),$<br>$\varepsilon(t),\ldots,\varepsilon(t-n_C+1)]$ |
| Predictor output | a priori | $\hat{y}^0(t+1)=\hat{\theta}^T(t)\psi(t)$ | | |
| | a posteriori | $\hat{y}(t+1)=\hat{\theta}^T(t+1)\psi(t)$ | | |
| Prediction error | a priori | $\varepsilon^0(t+1)=y(t+1)-\hat{y}^0(t+1)$ | | |
| | a posteriori | $\varepsilon(t+1)=y(t+1)-\hat{y}(t+1)$ | | |
| Adaptation error | | $\nu^\circ(t+1)=\varepsilon^\circ(t+1)$ | $\nu^\circ(t+1)=\varepsilon^\circ(t+1)$ | $\nu^\circ(t+1)=\varepsilon^\circ(t+1)$ |
| Observation vector | | $\Phi(t)=\psi(t)$ | $\Phi(t)=\psi(t)$ | $\Phi(t)=\psi(t)$ |
| Stability condition deterministic env. | | None | None | None |
| Convergence condition stochastic env. | | None | $\frac{1}{C}-\frac{\lambda_2}{2}=SPR$ | $\frac{1}{C}-\frac{\lambda_2}{2}=SPR$ |

**Table 5.1** (Continued)
(b)

| | Recursive Maximum Likelihood (RML) | Generalized Least Squares (GLS) | Output Error with Fixed Compensator (OEFC) |
|---|---|---|---|
| Plant + noise model | $y = \frac{q^{-d}B}{A}u + \frac{C}{A}e$ | $y = \frac{q^{-d}B}{A}u + \frac{C}{AD}e$ | $y = \frac{q^{-d}B}{A}u + v$ |
| Adjustable parameter vector | $\hat{\theta}^T(t) = [\hat{a}^T(t), \hat{b}^T(t), \hat{c}^T(t)]$<br>$\hat{a}^T(t) = [\hat{a}_1, \ldots, \hat{a}_{n_A}]$<br>$\hat{b}^T(t) = [\hat{b}_1, \ldots, \hat{b}_{n_B}]$<br>$\hat{c}^T(t) = [\hat{c}_1, \ldots, \hat{c}_{n_C}]$ | $\hat{\theta}^T(t) = [\hat{a}^T(t), \hat{b}^T(t), \hat{c}^T(t), \hat{d}^T(t)]$<br>$\hat{a}^T(t) = [\hat{a}_1, \ldots, \hat{a}_{n_A}]$<br>$\hat{b}^T(t) = [\hat{b}_1, \ldots, \hat{b}_{n_B}]$<br>$\hat{c}^T(t) = [\hat{c}_1, \ldots, \hat{c}_{n_C}]$<br>$\hat{d}^T(t) = [\hat{d}_1, \ldots, \hat{d}_{n_D}]$ | $\hat{\theta}^T(t) = [\hat{a}^T(t), \hat{b}^T(t)]$<br>$\hat{a}^T(t) = [\hat{a}_1, \ldots, \hat{a}_{n_A}]$<br>$\hat{b}^T(t) = [\hat{b}_1, \ldots, \hat{b}_{n_B}]$ |
| Predictor regressor vector | $\psi^T(t) = [-y(t), \ldots, -y(t-n_A+1),$<br>$u(t-d), \ldots, u(t-d-n_B+1),$<br>$\varepsilon(t), \ldots, \varepsilon(t-n_C+1)]$ | $\psi^T(t) = [-y(t), \ldots, -y(t-n_A+1),$<br>$u(t-d), \ldots, u(t-d-n_B+1),$<br>$\varepsilon(t), \ldots, \varepsilon(t-n_C+1),$<br>$-\alpha(t), \ldots, -\alpha(t-n_D+1)]$<br>$\alpha(t) = \hat{A}(t)y(t) - \hat{B}^*(t)u(t-d-1)$ | $\psi^T(t) = [-\hat{y}(t), \ldots, -\hat{y}(t-n_A+1),$<br>$u(t-d), \ldots, u(t-d-n_B+1)]$ |
| Predictor output: a priori<br>a posteriori | $\hat{y}^0(t+1) = \hat{\theta}^T(t)\psi(t)$<br>$\hat{y}(t+1) = \hat{\theta}^T(t+1)\psi(t)$ | | |
| Prediction error: a priori<br>a posteriori | $\varepsilon^0(t+1) = y(t+1) - \hat{y}^0(t+1)$<br>$\varepsilon(t+1) = y(t+1) - \hat{y}(t+1)$ | | |
| Adaptation error | $\nu^\circ(t+1) = \varepsilon^\circ(t+1)$ | $\nu^\circ(t+1) = \varepsilon^\circ(t+1)$ | $\nu^\circ(t+1) = \varepsilon^\circ(t+1) + \sum_{i=1}^{n_D} d_i \varepsilon(t+1-i)$ |
| Observation vector | $\Phi(t) = \frac{1}{\hat{C}(t,q^{-1})}\psi(t)$<br>$\hat{C}(t,q^{-1}) = 1 + \hat{c}_1(t)q^{-1} + \cdots$ | $\Phi(t) = \psi(t)$ | $\Phi(t) = \psi(t)$ |
| Stability condition deterministic env. | None | None | $\frac{D}{A} - \frac{\lambda_2}{2} = SPR$ |
| Convergence condition stochastic env. | Local convergence only<br>$\hat{C}(t,q^{-1}) =$ stable | (1) Unknown<br>(2) $D,\ C$ known $\frac{D}{A} - \frac{\lambda_2}{2} = SPR$ | $\frac{D}{C} - \frac{\lambda_2}{2} = SPR$<br>$w$ and $u$ are independent |

**Table 5.1** (Continued)
(c)

| | | Filtered Output Error | Output Error with Adjustable Compensator (AEAC) | Instrumental Variable with Auxiliary Model (IVAM) |
|---|---|---|---|---|
| Plant + noise model | | $y(t)=\frac{a^{-d}B}{A}u+v$ | $y=\frac{q^{-d}B}{A}u+v$ | $y=\frac{q^{-d}B}{A}u+v$ |
| Adjustable parameter vector | | $\hat{\theta}^T(t)=[\hat{a}^T(t),\hat{b}^T(t)]$<br>$\hat{a}^T(t)=[\hat{a}_1,\dots,\hat{a}_{n_A}]$<br>$\hat{b}^T(t)=[\hat{b}_1,\dots,\hat{b}_{n_B}]$ | $\hat{\theta}^T(t)=[\hat{a}^T(t),\hat{b}^T(t),\hat{c}^T(t)]$<br>$\hat{a}^T(t)=[\hat{a}_1,\dots,\hat{a}_{n_A}]$<br>$\hat{b}^T(t)=[\hat{b}_1,\dots,\hat{b}_{n_B}]$<br>$\hat{d}^T(t)=[\hat{d}_1,\dots,\hat{d}_{n_D}]$<br>$n_D=n_A$ | $\hat{\theta}^T(t)=[\hat{a}^T(t),\hat{b}^T(t)]$<br>$\hat{a}^T(t)=[\hat{a}_1,\dots,\hat{a}_{n_A}]$<br>$\hat{b}^T(t)=[\hat{b}_1,\dots,\hat{b}_{n_B}]$ |
| Predictor regressor vector | | $\psi^T(t)=[-\hat{y}(t),\dots,-\hat{y}(t-n_A+1),$<br>$u(t-d),\dots,u(t-d-n_B+1)]$ | $\psi^T(t)=[-\hat{y}(t),\dots,-\hat{y}(t-n_A+1),$<br>$u(t-d),\dots,u(t-d-n_B+1)]$ | $\psi^T(t)=[-y(t),\dots,-y(t-n_A+1),$<br>$u(t-d),\dots,u(t-d-n_B+1)]$ |
| Predictor output | a priori | $\hat{y}^0(t+1)=\hat{\theta}^T(t)\psi(t)$ | $\hat{y}^0(t+1)=[\hat{a}^T(t),\hat{b}^T(t)]\psi(t)$ | $\hat{y}^0(t+1)=\hat{\theta}^T(t)\psi(t)$ |
| | a posteriori | $\hat{y}(t+1)=\hat{\theta}^T(t+1)\psi(t)$ | $\hat{y}(t+1)=[\hat{a}^T(t+1),\hat{b}^T(t+1)]\psi(t)$ | $\hat{y}(t+1)=\hat{\theta}^T(t+1)\psi(t)$ |
| Prediction error | a priori | $\varepsilon^0(t+1)=y(t+1)-\hat{y}^0(t+1)$ | | |
| | a posteriori | $\varepsilon(t+1)=y(t+1)-\hat{y}(t+1)$ | | |
| Adaptation error | | $\nu^0(t+1)=\varepsilon^0(t+1)$ | $\nu^\circ(t+1)=\varepsilon^\circ(t+1)$<br>$+\sum_{i=1}^{n_D}\hat{d}_i(t)\varepsilon(t+1-i)$ | $\nu^\circ(t+1)=\varepsilon^\circ(t+1)$ |
| Observation vector | | $\Phi(t)=\frac{1}{L(t,q^{-1})}\psi(t)$ or $\frac{1}{\hat{A}(t,q^{-1})}\psi(t)$<br>$\hat{A}(t,q^{-1})=1+\hat{a}_1(t)q^{-1}+\cdots$ | $\Phi^T(t)=[\psi^T(t),-\varepsilon(t),\dots,-\varepsilon(t-n_d+1)]$ | $\Phi^T(t)=[-y_{IV}(t),\dots,-y_{IV}(t-n_A+1)$<br>$u(t-d),\dots,u(t-n_B-d+1)]$<br>$y_{IV}(t)=\hat{\theta}^T(t)\Phi(t-1)$ |
| Stability condition deterministic env. | | For $\hat{A}(t,q^{-1})\equiv L(q^{-1})$ None<br>$\frac{L}{A}-\frac{\lambda_2}{2}=SPR$ | Local | (Need initialization) |
| Convergence condition stochastic env. | | $\frac{L}{A}-\frac{\lambda_2}{2}=SPR$<br>$v$ and $u$ are independent | Local convergence<br>$\frac{D^*}{A}=SPR$<br>$D^*:E\{(D^*v)^2\}=\min$ | Local (need initialization) |

## 5.3 Recursive Identification Methods Based on the Whitening of the Prediction Error (Type I)

The following recursive identification methods belonging to this category will be presented:[2]

- Recursive Least Squares (RLS);
- Extended Least Squares (ELS);
- Output Error with Extended Prediction Model (OEEPM);
- Recursive Maximum Likelihood (RML) ;
- Generalized Least Squares (GLS) .

### *5.3.1 Recursive Least Squares (RLS)*

The recursive least squares algorithm has been presented in detail in Sect. 3.2. Global asymptotic stability in deterministic environment is assured (direct application of Theorem 3.2). Global convergence in the stochastic case toward unbiased parameter estimates is assured for a model generating the data of the form:

$$A(q^{-1})y(t) = q^{-d}B(q^{-1})u(t) + e(t) \tag{5.10}$$

where $e(t)$ is a zero mean white noise sequence. The analysis in the stochastic case can be done quite simply using the averaging method (Theorem 4.1), see Sect. 4.2, Example 4.2 or martingales (Theorem 4.2, Sect. 4.3).

The main disadvantage of this method is the fact that the noise model, for which unbiased estimates are obtained, is very unrealistic in practical situations, since in general the disturbance in (5.10) is not just a white noise. Therefore, despite its simplicity, in most applications it will provide erroneous parameter estimates.

### *5.3.2 Extended Least Squares (ELS)*

This method has been developed in order to identify without bias *plant + disturbance* models of the form (ARMAX model):

$$A(q^{-1})y(t) = q^{-d}B(q^{-1})u(t) + C(q^{-1})e(t) \tag{5.11}$$

The idea is to simultaneously identify the plant model and the disturbance model, in order to obtain a prediction (adaptation) error which is asymptotically white.

The model generating the data can be expressed as:

---

[2]Routines corresponding to these methods in Matlab and Scilab can be downloaded from the websites: http://www.landau-adaptivecontrol.org and http://landau-bookic.lag.ensieg.inpg.fr.

$$
\begin{aligned}
y(t+1) &= -A^*(q^{-1})y(t) + B^*(q^{-1})u(t-d) + C^*(q^{-1})e(t) + e(t+1) \\
&= \theta^T \phi(t) + e(t+1)
\end{aligned}
\tag{5.12}
$$

where:

$$
\theta^T = [a_1, \ldots, a_{n_A}, b_1, \ldots, b_{n_B}, c_1, \ldots, c_{n_C}] \tag{5.13}
$$

$$
\begin{aligned}
\phi_0^T(t) = [&-y(t), \ldots, -y(t-n_A+1), u(t-d), \ldots, u(t-d-n_B+1), \\
&e(t), \ldots, e(t-n_c+1)]
\end{aligned}
\tag{5.14}
$$

Assume that the parameters are known and construct a predictor that will give a white prediction error:

$$
\hat{y}(t+1) = -A^*(q^{-1})y(t) + B^*(q^{-1})u(t-d) + C^*(q^{-1})e(t) \tag{5.15}
$$

Furthermore, this predictor will minimize $\mathbf{E}\{[y(t+1) - \hat{y}(t+1)]^2\}$ (see Sect. 2.2). The prediction error, in the case of known parameters, is given by:

$$
\varepsilon(t+1) = y(t+1) - \hat{y}(t+1) = e(t+1) \tag{5.16}
$$

This enables (5.15) to be rewritten in the form:

$$
\hat{y}(t+1) = -A^*(q^{-1})y(t) + B^*(q^{-1})u(t-d) + C^*(q^{-1})\varepsilon(t) \tag{5.17}
$$

Subtracting now (5.15) from (5.12), one gets:

$$
\varepsilon(t+1) = -C^*(q^{-1})\,[\varepsilon(t) - e(t)] + e(t) \tag{5.18}
$$

i.e.,

$$
C(q^{-1})\,[\varepsilon(t+1) - e(t+1)] = 0 \tag{5.19}
$$

Since $C(q^{-1})$ is an asymptotically stable polynomial, it results that $\varepsilon(t+1)$ will become white asymptotically.

The adaptive version of this predictor is as follows. The a priori adjustable predictor will take the form:

$$
\begin{aligned}
\hat{y}^0(t+1) &= -\hat{A}^*(q^{-1},t)y(t) + \hat{B}^*(q^{-1},t)u(t) + \hat{C}^*(q^{-1},t)\varepsilon(t) \\
&= \hat{\theta}^T(t)\phi(t)
\end{aligned}
\tag{5.20}
$$

in which:

$$
\hat{\theta}^T(t) = [\hat{a}_1(t), \ldots, \hat{a}_{n_A}(t), \hat{b}_1(t), \ldots, \hat{b}_{n_A}(t), \hat{c}_1(t), \ldots, \hat{c}_{n_A}(t)] \tag{5.21}
$$

$$
\begin{aligned}
\phi^T(t) = [&-y(t), \ldots, -y(t-n_A+1), u(t-d), \ldots, u(t-d-n_B+1), \\
&\varepsilon(t), \ldots, \varepsilon(t-n_c+1)]
\end{aligned}
\tag{5.22}
$$

The a posteriori adjustable predictor will be given by:

$$
\hat{y}(t+1) = \hat{\theta}^T(t+1)\phi(t) \tag{5.23}
$$

The a posteriori prediction error $\varepsilon(t)$ which enters in the regressor vector of the predictor is given by:

$$
\varepsilon(t) = y(t) - \hat{y}(t) \tag{5.24}
$$

(where $\hat{y}(t)$ is now the a posteriori output of the adjustable predictor) and the a priori prediction error is given by:

$$\varepsilon^0(t+1) = y(t+1) - \hat{y}^0(t+1) \tag{5.25}$$

The a posteriori prediction equation is obtained subtracting (5.23) from (5.12) and observing that (5.11) can be alternatively expressed as:

$$y(t+1) = \theta^T \phi(t) - C^*(q^{-1})\varepsilon(t) + C(q^{-1})e(t) \tag{5.26}$$

(by adding and subtracting the term $\pm C^*(q^{-1})\varepsilon(t)$). One obtains:

$$\varepsilon(t+1) = -C^*(q^{-1})\varepsilon(t) + [\theta - \hat{\theta}(t+1)]^T \phi(t) + C(q^{-1})e(t) \tag{5.27}$$

from which it results that:

$$\varepsilon(t+1) = \frac{1}{C(q^{-1})}[\theta - \hat{\theta}(t+1)]^T \phi(t) + e(t) \tag{5.28}$$

In the deterministic case $C(q^{-1}) = 1$, $e(t) \equiv 0$. One sees that (5.28) has the appropriate format corresponding to Theorems 3.2, 4.1 and 4.2. One immediately concludes, using the PAA given in (5.7) through (5.9), with $\Phi(t) = \phi(t)$ that in the deterministic case, global asymptotic stability is assured without any positive real condition and in stochastic environment, either using ODE or martingales, the convergence is assured provided that:

$$H'(z^{-1}) = \frac{1}{C(z^{-1})} - \frac{\lambda_2}{2} \tag{5.29}$$

is a strictly positive real transfer matrix for $2 > \lambda_2 \geq \max \lambda_2(t)$.

Boundedness of $\varepsilon(t)$ in the stochastic case (which implies the boundedness of $\phi(t)$, since $u(t)$ and $y(t)$ are bounded) is a direct consequence of Lemma 4.1 since:

$$\varepsilon^0(t+1) + \theta^T(t)\phi(t) = y(t) \tag{5.30}$$

which leads to the satisfaction of the condition (2) of the lemma since $y(t)$ is bounded.

### 5.3.3 Output Error with Extended Prediction Model (OEEPM)

This algorithm is discussed in detail in Sects. 3.3.5 and 4.3. It can be used appropriately for stochastic models of the form of (5.11) (ARMAX model). It does not require any positive real condition in deterministic context, but in the stochastic context one has the convergence condition:

$$H'(z^{-1}) = \frac{1}{C(z^{-1})} - \frac{\lambda_2}{2} \tag{5.31}$$

be strictly positive real ($2 > \lambda_2 \geq \max \lambda_2(t)$) similar to that for ELS.

It turns out that the OEEPM can be interpreted as a variant of the ELS. To see this, consider the a priori output of the adjustable predictor for ELS (5.20), which can be rewritten as follows by adding and subtracting the term $\pm \hat{A}^*(q^{-1}, t)\hat{y}(t)$:

$$\hat{y}^0(t+1) = -\hat{A}^*(q^{-1},t)\hat{y}(t) + \hat{B}^*(q^{-1},t)u(t-d) + [\hat{C}^*(q^{-1},t)\varepsilon(t) - \hat{A}^*(q^{-1},t)[y(t)-\hat{y}(t)]] \quad (5.32)$$

Defining:

$$\hat{H}^*(q^{-1}) = \hat{C}^*(q^{-1},t) - \hat{A}^*(q^{-1},t) = \hat{h}_1(t) + q^{-1}\hat{h}_2(t) + \cdots$$

with:

$$\hat{h}_i(t) = \hat{c}_i(t) - \hat{a}_i(t); \quad i = 1,2,\ldots,\max(n_A, n_C)$$

Equation (5.32) can be rewritten as:

$$\hat{y}^0(t+1) = -\hat{A}^*(q^{-1},t)\hat{y}(t) + \hat{B}^*(q^{-1},t)u(t-d) + \hat{H}^*(q^{-1})\varepsilon(t) = \hat{\theta}^T(t)\phi(t) \quad (5.33)$$

where now:

$$\hat{\theta}^T(t) = [\hat{a}_1(t),\ldots,\hat{a}_{n_A},\hat{b}_1(t),\ldots,\hat{b}_{n_B}(t),\hat{h}_1(t),\ldots,\hat{h}_{n_H}(t)]$$

$$\phi^T(t) = [-\hat{y}(t),\ldots,\hat{y}(t-n_A+1),u(t-d),\ldots,u(t-d-n_B+1),\varepsilon(t),\ldots,\varepsilon(t-n_C+1)]$$

Equation (5.33) corresponds to the adjustable predictor for the output error with extended prediction model.

Despite their similar asymptotic properties, there is a difference in the first $n_A$ components of the vector $\phi(t)$. While ELS uses the measurements $y(t)$, $y(t-1),\ldots$ directly affected by the disturbance, the OEEPM uses the predicted a posteriori outputs $\hat{y}(t)$, $\hat{y}(t-1)$ which depend upon the disturbance only indirectly through the P.A.A. This explains why a better estimation is obtained with OEEPM than with ELS over short or medium time horizons (it removes the bias more quickly).

### 5.3.4 Recursive Maximum Likelihood (RML)

This algorithm is also dedicated to ARMAX type *plant + disturbance model*. It belongs to the class of RPEM. It can be interpreted as a modification of the ELS in order to eliminate (locally) the positive real condition for convergence.

Consider the equation of the a posteriori prediction error for the ELS but for a fixed value of the estimated parameter vector $\hat{\theta}$ (which is used in Theorem 4.1). From (5.28) one obtains:

$$\nu(t+1,\hat{\theta}) = \frac{1}{C(q^{-1})}[\theta - \hat{\theta}]^T\phi(t,\hat{\theta}) + e(t+1) \quad (5.34)$$

In the vector $\hat{\theta}$, an estimation of the coefficients of the polynomials $C(q^{-1})$ is available. This estimated polynomial will be denoted by $\hat{C}(q^{-1},\hat{\theta})$. Introduce the filtered regressor vector:

$$\phi_f(t,\hat{\theta}) = \frac{1}{\hat{C}(q^{-1},\hat{\theta})}\phi(t,\hat{\theta}) \quad (5.35)$$

Taking into account that $\hat{\theta}$ is fixed in (5.34), this equation can be rewritten as:

$$\nu(t+1,\hat{\theta}) = \frac{\hat{C}(q^{-1},\hat{\theta})}{C(q^{-1})}[\theta - \hat{\theta}]\phi_f(t,\hat{\theta}) + e(t+1) \tag{5.36}$$

In other words, the positive real condition will be relaxed in particular around the correct value $\hat{\theta} = \theta$. This suggests to use of $\phi_f(t)$ as observation vector. Therefore, in the RML, with respect to ELS, one replaces the observation vector $\phi(t)$ by its filtered version given in (5.35). One filters the regressor through the current estimate of the polynomial $C(q^{-1})$ to generate the observation vector (i.e. one filters the components of $\phi(t)$). More specifically, one has: (for $n_A = n_B = n_C$)

$$\phi_f(t) = \begin{bmatrix} \hat{c} & 0 & 0 \\ 0 & \hat{c} & 0 \\ 0 & 0 & \hat{c} \end{bmatrix} \phi_f(t-1) + \begin{bmatrix} 1 & 0 & 0 \\ 0 & 1 & 0 \\ 0 & 0 & 1 \end{bmatrix} \phi(t)$$

where:

$$\hat{c} = \begin{bmatrix} \hat{c}_1(t) & \dots & \dots & \hat{c}_n(t) \\ 0 & & & \\ 0 & & I_{n-1} & \\ 0 & & & \end{bmatrix}$$

However, no global convergence results are available for this algorithm. In fact, this algorithm cannot be implemented from $t = 0$, if a good estimation of $C(q^{-1})$ is not available. First, an *initialization horizon* must be considered during which the ELS is used in order to obtain a first estimation of $C(q^{-1})$ which of course should be at least stable (usual initialization horizon: 5 to 10 times the number of parameters).

Furthermore, filtering of the observations can only be carried out for stable estimation of $C(q^{-1})$. Therefore, a stability test should certainly be incorporated.

A smooth transition from the ELS toward RML may also be used by introducing a *contraction factor* in $\hat{C}(t, q^{-1})$. One filters the data by:

$$1 + \alpha(t)\hat{C}^*(q^{-1}, t) \quad \text{with } 0 \le \alpha(t) \le 1$$

This forces the polynomial roots inside the unit circle at the beginning of the estimation process and, in addition, one uses a contraction factor which asymptotically tends toward 1. This type of variation can be obtained by using the formula:

$$\alpha(t) = \alpha_0\alpha(t-1) + 1 - \alpha_0; \quad 0.5 \le \alpha_0 \le 0.99$$

The RML may be used to improve, if necessary, the results of ELS or OEEPM in the vicinity of the equilibrium point. However, if the above mentioned precautions (initialization, stability test) are not taken into account, the algorithm easily diverges.

### *5.3.5 Generalized Least Squares (GLS)*

This method is dedicated to the unbiased estimation of the plant + disturbance model of the form:

$$A(q^{-1})y(t) = q^{-d}B(q^{-1})u(t) + \frac{C(q^{-1})}{D(q^{-1})}e(t) \tag{5.37}$$

As in the case of ELS, the first step is to construct a predictor for the known parameter case which gives an asymptotically white prediction error and then replace the unknown parameters by their estimates. The model generating the data can be expressed as:

$$y(t+1) = -A^* y(t) + B^* u(t-d) + \frac{C(q^{-1})}{D(q^{-1})} e(t+1) \tag{5.38}$$

and after multiplying both sides by $D(q^{-1})$ and rearranging the various terms, one obtains:

$$\begin{aligned} y(t+1) = {} & -A^* y(t) + B^* u(t-d) - D^*[Ay(t) - B^* u(t-d-1)] \\ & + C(q^{-1}) e(t+1) \end{aligned} \tag{5.39}$$

This immediately suggests taking as an adjustable predictor:

$$\begin{aligned} \hat{y}^0(t+1) &= -\hat{A}^*(t) y(t) + \hat{B}^*(t) u(t-d) - \hat{D}^*(t)\alpha(t) + \hat{C}^*(t)\varepsilon(t) \\ &= \hat{\theta}^T(t)\phi(t) \end{aligned} \tag{5.40}$$

$$\hat{y}(t+1) = \hat{\theta}^T(t+1)\phi(t) \tag{5.41}$$

where:

$$\alpha(t) = \hat{A}(t) y(t) - \hat{B}^*(t) u(t-d-1) \tag{5.42}$$

$$\hat{\theta}^T(t) = [\hat{\theta}_1^T(t), \hat{\theta}_2^T(t)] \tag{5.43}$$

$$\hat{\theta}_1^T(t) = [\hat{a}_1(t), \ldots, \hat{a}_{n_A}(t), \hat{b}_1(t), \ldots, \hat{b}_{n_A}(t)] \tag{5.44}$$

$$\hat{\theta}_2^T(t) = [\hat{c}_1(t), \ldots, \hat{c}_{n_C}(t), \hat{d}_1(t), \ldots, \hat{d}_{n_D}(t)] \tag{5.45}$$

$$\phi^T(t) = [\phi_1^T(t), \phi_2^T(t)] \tag{5.46}$$

$$\phi_1^T(t) = [-y(t), \ldots, -y(t-n_A+1), u(t-d), \ldots, u(t-d-n_B+1)] \tag{5.47}$$

$$\phi_2^T(t) = [\varepsilon(t), \ldots, \varepsilon(t-n_C+1), -\alpha(t), \ldots, -\alpha(t-n_D+1)] \tag{5.48}$$

Equation (5.39) can be rewritten as:

$$\begin{aligned} y(t+1) = {} & -A^* y(t) + B^* u(t-d) + C^*\varepsilon(t) - D^*\alpha(t) - C^*\varepsilon(t) \\ & - D^*[(A^* - \hat{A}^*(t)) y(t-1) - (B - \hat{B}(t)) u(t-d-1)] \\ & + C(q^{-1}) e(t+1) \end{aligned} \tag{5.49}$$

and the a posteriori prediction error equation takes the form:

$$\begin{aligned} \varepsilon(t+1) = {} & [\theta - \hat{\theta}(t+1)]^T \phi(t) + D^*[\theta_1 - \hat{\theta}_1(t)]\phi_1(t-1) \\ & - C^*\varepsilon(t) + Ce(t+1) \end{aligned} \tag{5.50}$$

In the deterministic case, $e(t+1) = 0$, $D^*(q^{-1}) = C^*(q^{-1}) = 0$ and (5.45) reduces to:

$$\varepsilon(t+1) = [\theta_0 - \hat{\theta}(t+1)]^T \phi(t) \tag{5.51}$$

where:

$$\theta_0^T = [a_1, \ldots, a_{n_A}, b_1, \ldots, b_{n_B}, \ldots, 0, \ldots, \ldots, 0, \ldots]$$

According to Theorem 3.2, global asymptotic stability will be assured without any positive real condition. In the stochastic case, using the averaging method, one gets for a fixed value $\hat{\theta}$:

$$\begin{aligned}\varepsilon(t+1,\hat{\theta}) &= [\theta_1 - \hat{\theta}_1]^T \phi_1(t,\hat{\theta}) + D^*[\theta_1 - \hat{\theta}_1]^T \phi_1(t-1,\hat{\theta}) \\ &\quad + [\theta_2 - \hat{\theta}_2]\phi_2(t,\hat{\theta}) - C^*\varepsilon(t,\hat{\theta}) + Ce(t+1) \\ &= \frac{D(q^{-1})}{C(q^{-1})}[\theta_1 - \hat{\theta}_1]^T \phi_1(t,\hat{\theta}) + \frac{1}{C}[\theta_2 - \hat{\theta}_2]\phi_2(t,\hat{\theta}) + e(t+1)\end{aligned} \tag{5.52}$$

As such, Theorem 4.1 is not applicable despite that the noise term is white. However, for the case $C(q^{-1})$ and $D(q^{-1})$ known, (5.52) reduces to:

$$\varepsilon(t+1,\hat{\theta}) = \frac{D(q^{-1})}{C(q^{-1})}[\theta_1 - \hat{\theta}_1]^T \phi_1(t,\hat{\theta}) \tag{5.53}$$

and applying Theorem 4.1, one has the convergence condition:

$$H'(z^{-1}) = \frac{D(z^{-1})}{C(z^{-1})} - \frac{\lambda_2}{2} \tag{5.54}$$

is a strictly positive real transfer function.

As such, this algorithm is an extension of the GLS algorithm of Bethoux (1976) and Landau (1990a) which corresponds to the case $C(q^{-1}) = 1$.

This algorithm is more appropriate to be used than the ELS, for the case where the disturbance term in (5.37) has a narrow frequency spectrum which can usually be modeled as $\frac{1}{D(q^{-1})}e(t)$ where $D(q^{-1})$ is a low-damped second-order polynomial (while using a model of the form $C(q^{-1})e(t)$ will require a large number of parameters to be identified).

## 5.4 Validation of the Models Identified with Type I Methods

This section is concerned with the validation of models identified using identification methods based on the *whitening* of the prediction error. If the residual prediction error is a white noise sequence, in addition to obtaining unbiased parameter estimates, this also means that the identified model gives the best prediction for the plant output in the sense that it minimizes the variance of the prediction error. On the other hand, since the residual error is white and a white noise is not correlated with any other variable, then all the correlations between the input and the output of the plant are represented by the identified model and what remains unmodeled does not depend on the input.

The principle of the validation method is as follows:[3]

[3]Routines corresponding to this method in Matlab and Scilab can be downloaded from the websites: http://www.landau-adaptivecontrol.org and http://landau-bookic.lag.ensieg.inpg.fr.

- If the plant + disturbance structure chosen is correct, i.e., representative of reality.
- If an appropriate identification method for the structure chosen has been used.
- If the degrees of the polynomials $A(q^{-1}), B(q^{-1}), C(q^{-1})$ and the value of $d$ (delay) have been correctly chosen (the plant model is in the model set).

Then the prediction error $\varepsilon(t)$ asymptotically tends toward a white noise, which implies:

$$\lim_{t\to\infty} \mathbf{E}\{\varepsilon(t)\varepsilon(t-i)\} = 0; \quad i = 1, 2, 3, \ldots; \ -1, -2, -3, \ldots$$

The validation method implements this principle. It is made up of several steps:

1. Creation of an I/O file for the identified model (using the same input sequence as for the system).
2. Creation of a prediction error file for the identified model (minimum 100 data).
3. *Whiteness* (uncorrelatedness) test on the prediction errors sequence (also known as residual prediction errors).

### 5.4.1 Whiteness Test

Let $\{\varepsilon(t)\}$ be the centered sequence of the residual prediction errors (centered: measured value − mean value).

One computes:

$$R(0) = \frac{1}{N}\sum_{t=1}^{N} \varepsilon^2(t); \qquad RN(0) = \frac{R(0)}{R(0)} = 1 \tag{5.55}$$

$$R(i) = \frac{1}{N}\sum_{t=1}^{N} \varepsilon(t)\varepsilon(t-i); \qquad RN(i) = \frac{R(i)}{R(0)}; \quad i = 1, 2, 3, \ldots, i_{\max} \tag{5.56}$$

with $i_{\max} \geq n_A$ [degree of polynomial $A(q^{-1})$], which are estimations of the (normalized) autocorrelations. If the residual prediction error sequence is perfectly white (theoretical situation), and the number of samples is very large ($N \to \infty$), then $RN(0) = 1$; $RN(i) = 0, i \geq 1$.[4]

In real situations, however, this is never the case (i.e., $RN \neq 0; i \geq 1$), since on the one hand, $\varepsilon(t)$ contains residual structural errors (order errors, nonlinear effects, non-Gaussian noises), and on the other hand, the number of samples is in general relatively small (several hundreds). Also, it should be kept in mind that one always seeks to identify *good* simple models (with few parameters). One considers as a practical validation criterion (extensively tested on applications):

$$RN(0) = 1; \qquad |RN(i)| \leq \frac{2.17}{\sqrt{N}}; \quad i \geq 1 \tag{5.57}$$

[4]Conversely, for Gaussian data, uncorrelation implies independence. In this case, $RN(i) = 0, i \geq 1$ implies independence between $\varepsilon(t), \varepsilon(t-1), \ldots$, i.e., the sequence of residuals $\{\varepsilon(t)\}$ is a Gaussian white noise.

**Table 5.2** Confidence intervals for whiteness tests

| Level of significance | Validation criterion | $N=128$ | $N=256$ | $N=512$ |
|---|---|---|---|---|
| 3% | $\frac{2.17}{\sqrt{N}}$ | 0.192 | 0.136 | 0.096 |
| 5% | $\frac{1.96}{\sqrt{N}}$ | 0.173 | 0.122 | 0.087 |
| 7% | $\frac{1.808}{\sqrt{N}}$ | 0.16 | 0.113 | 0.08 |

where $N$ is the number of samples. This test has been defined taking into account the fact that for a white noise sequence $RN(i), i \neq 0$ has an asymptotically Gaussian (normal) distribution with zero mean and standard deviation:

$$\sigma = \frac{1}{\sqrt{N}}$$

The confidence interval considered in (5.57) corresponds to a 3% level of significance of the hypothesis test for Gaussian distribution. If $RN(i)$ obeys the Gaussian distribution $(o, 1/\sqrt{N})$, there is only a probability of 1.5% that $RN(i)$ is larger than $2.17/\sqrt{N}$, or that $RN(i)$ is smaller than $-2.17/\sqrt{N}$. Therefore, if a computed value $RN(i)$ falls outside the range of the confidence interval, the hypothesis $\varepsilon(t)$ and $\varepsilon(t-i)$ are independent should be rejected, i.e., $\{\varepsilon(t)\}$ is not a white noise sequence. Sharper confidence intervals can be defined. Table 5.2 gives the values of the validation criterion for various $N$ and various levels of significance.

The following remarks are important:

- An acceptable identified model has in general:

$$|RN(i)| \leq \frac{1.8}{\sqrt{N}} \cdots \frac{2.17}{\sqrt{N}}; \quad i \geq 1$$

- If several identified models have the same complexity (number of parameters), one chooses the model given by the methods that lead to the smallest $|RN(i)|$.
- A *too good* validation criterion indicates that model simplifications may be possible.
- To a certain extent, taking into account the relative weight of various non-Gaussian and modeling errors (which increases with the number of samples), the validation criterion may be slightly tightened for small $N$ and slightly relaxed for large $N$. Therefore, for simplicity's sake, one can consider as a basic practical numerical value for the validation criterion value:

$$|RN(i)| \leq 0.15; \quad i \geq 1$$

Note also that a complete model validation implies, after the validation using the input/output sequence employed for identification, a validation using a plant input/output sequence other than the one used for identification.

There is another important point to consider. If the level of the residual prediction errors is very low compared to the output level (let us say more than 60 dB), the whiteness tests lose their signification. This is because, on one hand, the noise level is so low that the bias on the RLS for example is negligible, and on the other hand, because the residual noise to a large extent may not be Gaussian (for example, noise caused by the propagation of round-off errors). This situation may occur when identifying simulated I/O data generated without disturbances.

## 5.5 Identification Methods Based on the Decorrelation of the Observation Vector and the Prediction Error (Type II)

The following recursive identification methods which fall in this category will be presented:[5]

- Output Error with Fixed Compensator (OEFC);
- Output Error with Adjustable Compensator (OEAC);
- Filtered Output Error (FOE);
- Instrumental Variable with Auxiliary Model (IVAM).

### *5.5.1 Output Error with Fixed Compensator*

The output error with fixed compensator is dedicated to the unbiased parameter estimation of *plant + disturbance* models of the form:

$$y(t) = \frac{q^{-d}B(q^{-1})}{A(q^{-1})}u(t) + v(t) \tag{5.58}$$

where $v(t+1)$ is a zero mean stochastic process with finite power and independent with respect to the input. The algorithm has been presented in Sect. 3.3.2 and analyzed in a stochastic environment in Sect. 4.2. The a posteriori adaptation error equation takes the form:

$$\nu(t+1) = \frac{D(q^{-1})}{A(q^{-1})}[\theta - \hat{\theta}(t+1)]^T \phi(t) + D(q^{-1})v(t+1) \tag{5.59}$$

In the deterministic context (using Theorem 3.2), as well as in a stochastic context using the averaging method (Theorem 4.1), one gets the stability and convergence condition:

$$H'(z^{-1}) = \frac{D(z^{-1})}{A(z^{-1})} - \frac{\lambda_2}{2} \tag{5.60}$$

be a strictly positive real discrete-time transfer function. If one makes a stronger assumption upon the disturbance, i.e.:

$$v(t+1) = \frac{1}{D(q^{-1})}e(t+1)$$

[5]Routines corresponding to these methods in Matlab and Scilab can be downloaded from the websites: http://www.landau-adaptivecontrol.org and http://landau-bookic.lag.ensieg.inpg.fr.

(one assumes in fact that the disturbance $w(t+1)$ is a known AR stochastic process) martingale approach (Theorem 4.2) can be applied yielding the same convergence condition of (5.60).

### 5.5.2 Output Error with Adjustable Compensator

The predictor and prediction error equations remain the same as for the output error with fixed compensator (see Table 5.1). The difference lies in the generation of the adaptation error where the fixed compensator is replaced by an adjustable one:

$$\nu(t+1) = \hat{D}(q^{-1}, t+1)\varepsilon(t+1) \tag{5.61}$$

where:

$$\hat{D}(q^{-1}, t) = 1 + \sum_{t=1}^{n_D} \hat{d}_i(t) q^{-i}; \quad n_D = n_A \tag{5.62}$$

The a priori adaptation error is given by:

$$\nu^0(t+1) = \varepsilon^0(t+1) + \sum_{i=1}^{n_D} \hat{d}_i(t)\varepsilon(t+1-i) \tag{5.63}$$

The equation of the a posteriori output error is:

$$\begin{aligned}\varepsilon(t+1) = &-\sum_{i=1}^{n_A} a_i \varepsilon(t+1-i) - \sum_{i=1}^{n_A} [a_i - \hat{a}_i(t+1)]\hat{y}(t+1-i) \\ &+ \sum_{i=1}^{n_B} [b_i - \hat{b}_i(t+1)] u(t+1-d-i)\end{aligned} \tag{5.64}$$

and the equation of the a posteriori adaptation error takes the form, using (5.61):

$$\nu(t+1) = [\theta - \hat{\theta}(t+1)]^T \phi(t) \tag{5.65}$$

where:

$$\begin{aligned}\theta^T &= [a_1, \ldots, a_{n_A}, b_1, \ldots, b_{n_B}, a_1, \ldots, a_{n_A}] \\ \hat{\theta}^T(t) &= [\hat{a}_1(t), \ldots, \hat{a}_{n_A}(t), \hat{b}_1(t), \ldots, \hat{b}_{n_B}(t), \hat{d}_1(t), \ldots, \hat{d}_{n_D}(t)] \\ \phi^T(t) &= [-\hat{y}(t), \ldots, -\hat{y}(t-n_A+1), u(t-d), \ldots, u(t-d-n_B+1), \\ &\quad -\varepsilon(t), \ldots, -\varepsilon(t-n_D+1)]\end{aligned}$$

and the positive real condition for convergence to zero of $\nu(t+1)$ is removed. However, strictly speaking this does not guarantee the convergence to zero of the a posteriori output prediction error $\varepsilon(t+1)$ if $\hat{D}(q^{-1}, t)$ is not asymptotically sta-ble. Therefore a stability test on $\hat{D}(q^{-1}, t)$ should be introduced. See Dugard and Goodwin (1985) for a detailed analysis.

In the stochastic case, using the averaging method for a certain value of $\hat{\theta}$, one gets:

$$\nu(t+1, \hat{\theta}) = [\theta - \hat{\theta}]^T \phi(t, \hat{\theta}) + \hat{D}(q^{-1}, \hat{\theta}) v(t+1) \tag{5.66}$$

Local convergence w.p.1 in a stochastic environment for a disturbance $w(t+1)$, which is independent with respect to $u(t)$ is guaranteed by $\frac{D^*(z^{-1})}{A(z^{-1})}$ strictly positive real, where $D^*(z^{-1})$ is given by Stoica and Söderström (1981):

$$\mathbf{E}\{[D^*(z^{-1})v(t)]^2\} = \min \tag{5.67}$$

Simulations have shown excellent robustness of this algorithm with respect to various types of disturbances.

### *5.5.3 Filtered Output Error*

To relax the positive real condition associated to the output error algorithm, one can use a filtered observation vector instead of the filtering of the output error prediction. In this case, one defines the filtered inputs $u_f(t)$ and outputs $y_f(t)$ as:

$$L(q^{-1}) y_f(t) = y(t) \tag{5.68}$$

$$L(q^{-1}) u_f(t) = u(t) \tag{5.69}$$

where $L(q^{-1})$ is an asymptotically stable monic polynomial:

$$L(q^{-1}) = 1 + q^{-1} L^*(q^{-1}) \tag{5.70}$$

The non-filtered adjustable predictor is defined by:

$$\hat{y}^0(t+1) = \hat{\theta}^T(t)\phi(t) \tag{5.71}$$

$$\hat{y}(t+1) = \hat{\theta}^T(t+1)\phi(t) \tag{5.72}$$

where:

$$\phi^T(t) = [\hat{y}(t), \ldots, \hat{y}(t - n_A + 1), u(t), \ldots, u(t - n_B + 1)] \tag{5.73}$$

The filtered adjustable predictor is defined by:

$$\hat{y}_f^0(t+1) = \hat{\theta}^T(t)\phi_f(t) \tag{5.74}$$

$$\hat{y}_f(t+1) = \hat{\theta}^T(t+1)\phi_f(t) \tag{5.75}$$

where:

$$L(q^{-1})\phi_f(t) = \phi(t) \tag{5.76}$$

The corresponding output prediction errors associated to the two adjustable predictors are given by:

$$\varepsilon^0(t+1) = y(t+1) - \hat{y}^0(t+1)$$
$$\varepsilon(t+1) = y(t+1) - \hat{y}(t+1) \tag{5.77}$$

$$\varepsilon_f^0(t+1) = y_f(t+1) - \hat{y}_f^0(t+1)$$
$$\varepsilon_f(t+1) = y_f(t+1) - \hat{y}_f(t+1) \tag{5.78}$$

Define the a posteriori adaptation error as:

$$\nu(t+1) = L(q^{-1})\varepsilon_f(t+1) \tag{5.79}$$

and the a priori adaptation error as:

$$\nu^0(t+1) = \varepsilon_f^0(t+1) + L^*(q^{-1})\varepsilon_f(t) \tag{5.80}$$

Taking into account that, one can write:

$$\begin{aligned} y_f(t+1) &= -A^*(q^{-1})y_f(t) + B^*(q^{-1})u_f(t) \\ &= -A^*(q^{-1})\varepsilon_f(t) + \theta^T\phi_f(t) \end{aligned} \tag{5.81}$$

where:

$$\theta^T = [a_1, \ldots, a_{n_A}, b_1, \ldots, b_{n_B}] \tag{5.82}$$

one gets from (5.78), using (5.81) and (5.75):

$$\varepsilon_f(t+1) = \frac{1}{A(q^{-1})}[\theta - \hat{\theta}(t+1)]^T\phi_f(t) \tag{5.83}$$

and from (5.79) it results that:

$$\nu(t+1) = \frac{L(q^{-1})}{A(q^{-1})}[\theta - \hat{\theta}(t+1)]^T\phi_f(t) \tag{5.84}$$

Global asymptotic stability in the deterministic environment and w.p.1 convergence in the stochastic environment for $v(t)$ independent with respect to $u(t)$ is obtained under the sufficient condition that:

$$H'(z^{-1}) = \frac{L(z^{-1})}{A(z^{-1})} - \frac{\lambda_2}{2} \tag{5.85}$$

be a strictly positive real transfer function.

Using the PAA (5.7) with $\Phi(t) = \phi_f(t)$ and (5.78), one can write:

$$\nu(t+1) = \varepsilon_f^0(t+1) + L^*(q^{-1})\varepsilon_f(t) - \phi_f^T(t)F(t)\phi_f(t)\nu(t+1) \tag{5.86}$$

from which (5.9) results.

Several approximations for this algorithm can be considered. Taking into account (5.68), (5.69) and (5.78) one can rewrite $\nu^0(t+1)$ as follows:

$$\begin{aligned} \nu^0(t+1) &= L(q^{-1})y_f(t+1) - L(q^{-1})\hat{y}_f^0(t+1) \\ &\quad + L^*(q^{-1})[\hat{y}_f^0(t+1) - \hat{y}_f(t+1)] \end{aligned} \tag{5.87}$$

But:

$$L(q^{-1})\hat{y}_f^0(t+1) = L(q^{-1})\left\{\hat{\theta}^T(t)\left[\frac{1}{L(q^{-1})}\phi(t)\right]\right\} \approx \hat{y}^0(t+1) \tag{5.88}$$

Neglecting the third term of (5.87) and taking in account (5.68) and (5.88) one gets:

$$\nu^0(t+1) \approx y(t+1) - \hat{y}^0(t+1) \tag{5.89}$$

This allows to implement the algorithm using the adjustable predictor (5.71), (5.72) and to replace the a priori adaptation error by the non-filtered output error. The only additional operations compared to the output error algorithm without fixed compensator will be the filtering of $\phi(t)$.

The second approximation is related to the on line updating of the filter $L(q^{-1})$ (like in recursive maximum likelihood). One replaces $L(q^{-1})$ by:

$$\hat{L}(q^{-1},t) = \hat{A}(q^{-1},t) \tag{5.90}$$

in order to remove at least around the equilibrium the positive real condition. Of course, a stability test has to be performed and initialization by non-filtered output error or recursive least squares may be necessary.

### 5.5.4 *Instrumental Variable with Auxiliary Model*

The general idea behind the instrumental variable methods consists in the construction of a new observation vector, that is highly correlated with the uncontaminated variables, but uncorrelated with the noise disturbance in order to obtain asymptotically $\mathbf{E}\{\phi(t)\varepsilon(t+1)\} = 0$.

One of such type of algorithm is the *instrumental variable with auxiliary model* (Young 1969). The *plant* + *disturbance* model considered in this case, is the same as for the output error (5.58). The method can be viewed either as an improvement of recursive least squares or as a midway between recursive least squares and output error algorithms. The adjustable predictor is the one used in recursive least squares:

$$\hat{y}^0(t+1) = \hat{\theta}^T(t)\phi(t) \tag{5.91}$$

where:

$$\hat{\theta}^T(t) = [\hat{a}_1(t), \ldots, \hat{a}_{n_A}(t), \hat{b}_1, \ldots, \hat{b}_{n_B}(t)] \tag{5.92}$$

$$\phi^T(t) = [-y(t), \ldots, -y(t-n_A+1), u(t-d), \ldots, u(t-d-n_B+1)] \tag{5.93}$$

and respectively:

$$\hat{y}(t+1) = \hat{\theta}^T(t+1)\phi(t) \tag{5.94}$$

The prediction errors are defined by:

$$\varepsilon^0(t+1) = y(t+1) - \hat{y}^0(t+1) \tag{5.95}$$

$$\varepsilon(t+1) = y(t+1) - \hat{y}(t+1) \tag{5.96}$$

An auxiliary prediction model is used to generate the *instrumental variable* $y_{IV}$:

$$\begin{aligned}y_{IV}(t) &= -\sum_{i=1}^{n_A} \hat{a}_i(t) y_{IV}(t-i) + \sum_{i=1}^{n_B} \hat{b}_i(t) u(t-d-i) \\ &= \hat{\theta}^T(t)\phi_{IV}^T(t-1) \end{aligned} \tag{5.97}$$

where:

$$\begin{aligned}\phi_{IV}^T(t) = \big[ &- y_{IV}(t), \ldots, -y_{IV}(t-n_A+1), \\ &u(t-d), \ldots, u(t-d-n_B+1)\big]\end{aligned} \tag{5.98}$$

The output of this auxiliary prediction model depends on the previous predicted outputs and not on the previous measurement (it is an output error type model). These new variables will be less affected by the disturbance that will act only through the PAA. If an asymptotically decreasing adaptation gain is used, asymptotically $y_{IV}(t)$ will solely depend upon the input which is not the case for $\hat{y}(t)$ given by the least squares predictor of (5.91). The new observation vector is defined by:

$$\Phi(t) = \phi_{IV}(t) \tag{5.99}$$

For $y_{IV}(t)$ to be representative of $y(t)$, an estimation of $a_i$ and $b_i$ must be available. This is why the method should be initialized by the recursive least squares (initialization horizon: 5 to 10 times the number of parameters). For a fixed value $\hat{\theta}$, the equation of the prediction error can be written as (see Problem 5.2):

$$\varepsilon(t+1, \hat{\theta}) = \frac{\hat{A}(q^{-1}, \hat{\theta})}{A(q^{-1})}[\theta - \hat{\theta}]\phi_{IV}(t, \hat{\theta}) + w(t+1) \tag{5.100}$$

$\phi_{IV}(t, \hat{\theta})$ and $w(t+1)$ are uncorrelated and the averaging method clearly indicates that $\hat{A}$ should be close enough to $A$, in order that the algorithm converge ($\hat{A}/A - \lambda_2/2$ should be strictly positive real). For a detailed coverage of instrumental variable methods see Ljung and Söderström (1983).

## 5.6 Validation of the Models Identified with Type II Methods

This section is concerned with the validation of models obtained using the identification methods based on the decorrelation between the observations and the prediction errors. The uncorrelation between the observations and the prediction error leads to an unbiased parameter estimation. However, since the observations include the predicted output which depends on the input, the uncorrelation between the residual prediction error and the observations implies also that the residual prediction error is uncorrelated with the input. The interpretation of this fact is that the residual prediction error does not contain any information depending upon the input, i.e., all the correlations between the input and the output of the plant are represented by the identified model.

The principle of the validation method is as follows:[6]

- If the disturbance is independent of the input ($\Longrightarrow \mathbf{E}\{w(t)u(t)\} = 0$).
- If the *model + disturbance* structure chosen is correct, i.e., representative of the reality.
- If an appropriate identification method has been used for the chosen structure.
- If the degrees of the polynomials $A(q^{-1})$, $B(q^{-1})$ and the value of $d$ (delay) have been correctly chosen (the plant model is in the model set).

then the predicted outputs $\hat{y}(t-1), \hat{y}(t-2), \ldots$ generated by a model of the form (output error type predictor):

$$\hat{A}(q^{-1})\hat{y}(t) = q^{-d}\hat{B}(q^{-1})u(t) \tag{5.101}$$

and the prediction error $\varepsilon(t)$ are asymptotically uncorrelated, which implies:

$$\mathbf{E}\{\varepsilon(t)\hat{y}(t-i)\} \approx \frac{1}{N}\sum_{t=1}^{N}\varepsilon(t)\hat{y}(t-i) = 0; \quad i = 1, 2, 3, \ldots$$

The validation method implements this principle. It is made up of several steps:

1. Creation of an I/O file for the identified model (using the same input sequence as for the system).
2. Creation of files for the sequences $\{y(t)\}$; $\{\hat{y}(t)\}$; $\{\varepsilon(t)\}$ (system output, model output, residual output prediction error). These files must contain at least 100 data in order that the test be significant.
3. *Uncorrelation* test between the residual output prediction error sequence and the delayed prediction model output sequences.

### *5.6.1 Uncorrelation Test*

Let $\{y(t)\}$ and $\{\hat{y}(t)\}$ be the centered output sequences of the plant and of the identified model, respectively. Let $\{\varepsilon(t)\}$ be the centered sequence of the residual output (prediction) errors (centered = measured values − mean values). One computes:

$$R(i) = \frac{1}{N}\sum_{t=1}^{N}\varepsilon(t)\hat{y}(t-i); \quad i = 0, 1, 2, \ldots, n_A, \ldots \tag{5.102}$$

$$RN(i) = \frac{R(i)}{[(\frac{1}{N}\sum_{t=1}^{N}\hat{y}^2(t))(\frac{1}{N}\sum_{t=1}^{N}\varepsilon^2(t))]^{1/2}}$$
$$i = 0, 1, 2, \ldots, n_A, \ldots \tag{5.103}$$

[6] Routines corresponding to this method in Matlab and Scilab can be downloaded from the websites: http://www.landau-adaptivecontrol.org and http://landau-bookic.lag.ensieg.inpg.fr.

which are estimations of the (normalized) cross-correlations (note that $RN(0) \neq 1$), and $n_A$ is the degree of polynomial $A(q^{-1})$. If $\varepsilon(t)$ and $\hat{y}(t-i), i \geq 1$ are perfectly uncorrelated (theoretical situation), then:

$$RN(i) = 0; \quad i = 1, 2, \ldots, n_A, \ldots$$

In practice, this is never the case, either because the duration of the identification was not long enough to approach the asymptotic character, or because the identification of *good* simple models is desired [with few parameters, i.e., under estimation of the orders of $A(q^{-1})$ and $B(q^{-1})$], which results in *structural* errors that are correlated with the input $u(t)$.

The validation criterion for Type II identification methods based on an uncorrelation test will be defined similarly to that for Type I identification methods, which also use an uncorrelation test but on different sequences of variables (see Sect. 5.4). One therefore considers as a practical validation criterion in this case:

$$|RN(i)| \leq \frac{2.17}{\sqrt{N}}; \quad i \geq 1$$

where $N$ is the number of samples. Table 5.2 can be used to find the numerical value of the validation criterion for various $N$ and the level of signification of the test.

All the comments made in Sect. 5.4 about the signification of the values of the validation criterion apply also in this case. In particular, the basic practical numerical value for the validation criterion, which is:

$$|RN(i)| < 0.15; \quad i \geq 1$$

is worth remembering. This test is also used when one would like to compare models identified with Type I method, with models identified with Type II method. In this case, only the set of estimated parameters of $A(q^{-1})$ and $B(q^{-1})$ are used on the output error predictor given in (5.101) and then the uncorrelation test is made.

## 5.7 Selection of the Pseudo Random Binary Sequence

In Sect. 3.4, it was shown that the correct parameter estimation requires the use of a rich signal (persistently exciting signal). It was pointed out that pseudo-random binary sequences (PRBS) offer on the one hand a signal with a large frequency spectrum approaching the white noise and, on the other hand, they have a constant magnitude. This allows to define precisely the level of the instantaneous stress on the process (or actuator).

In this section, we will discuss how to select the PRBS for a good identification taking into account the various constraints which are encountered in practice.

### *5.7.1 Pseudo Random Binary Sequences (PRBS)*

Pseudo random binary sequences are sequences of rectangular pulses, modulated in width, that approximate a discrete-time white noise and thus have a spectral content

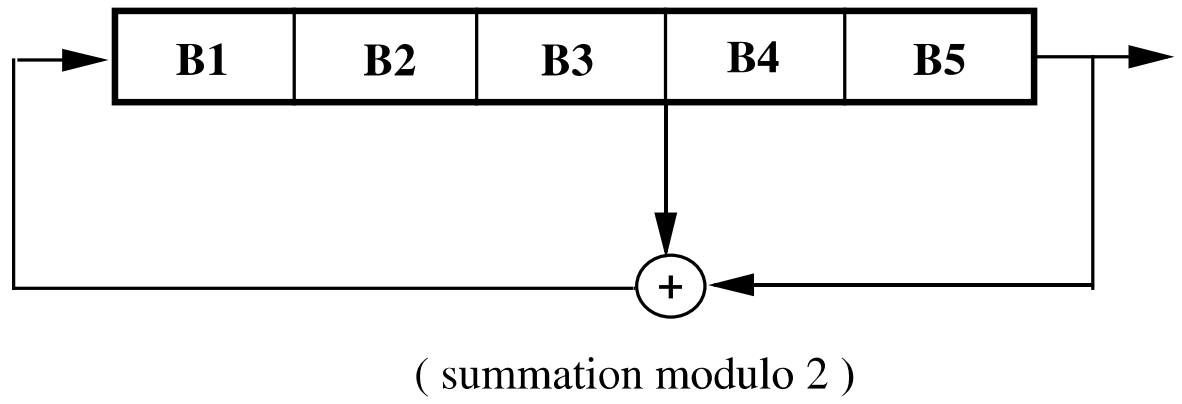

**Fig. 5.3** Generation of a PRBS of length $2^5 - 1 = 31$ sampling periods

**Table 5.3** Generation of maximum length PRBS

| Number of cells $N$ | Sequence length $L = 2^N - 1$ | Bits added $B_i$ and $B_j$ |
|---|---|---|
| 2 | 3 | 1 and 2 |
| 3 | 7 | 1 and 3 |
| 4 | 15 | 3 and 4 |
| 5 | 31 | 3 and 5 |
| 6 | 63 | 5 and 6 |
| 7 | 127 | 4 and 7 |
| 8 | 255 | 2, 3, 4 and 8 |
| 9 | 511 | 5 and 9 |
| 10 | 1023 | 7 and 10 |

*rich* in frequencies. They owe their name *pseudo random* to the fact that they are characterized by a *sequence length* within which the variations in pulse width vary randomly, but that over a large time horizon, they are periodic, the period being defined by the length of the sequence. In the practice of system identification, one generally uses just one complete sequence and we should examine the properties of such a sequence.

The PRBS are generated by means of shift registers with feedback (implemented in hardware or software).[7] The maximum length of a sequence is $2^N - 1$, in which $N$ is the number of cells of the shift register. Figure 5.3 presents the generation of a PRBS of length $31 = 2^5 - 1$ obtained by means of a 5-cells shift register.

Note that at least one of the $N$ cells of the shift register should have an initial logic value different from zero (one generally takes all the initial values of the $N$ cells equal to the logic value 1). Table 5.3 gives the structure enabling maximum length PRBS to be generated for different numbers of cells. Note also a very important characteristic element of the PRBS: *the maximum duration of a PRBS impulse is equal to N (number of cells)*. This property is to be considered when choosing a PRBS for system identification.

In order to correctly identify the steady-state gain of the plant dynamic model, the duration of at least one of the pulses (e.g., the maximum duration pulse) must be greater than the rise time $t_R$ of the plant. The maximum duration of a pulse being $NT_s$, the following condition results:

$$NT_s > t_R \tag{5.104}$$

which is illustrated in Fig. 5.4. From condition (5.104), one determines $N$ and therefore the length of the sequence, which is $2^N - 1$. Furthermore, in order to cover the entire frequency spectrum generated by a particular PRBS, the length of a test must be at least equal to the length of the sequence. In a large number of cases, the duration of the test $L$ is chosen equal to the length of the sequence. If the duration of the test is specified, it must therefore be ensured that:

$$(2^N - 1)T_s < L \quad (L = \text{test duration}) \tag{5.105}$$

[7]Routines for generating PRBS can be downloaded from the websites: http://www.landau-adaptivecontrol.org and http://landau-bookic.lag.ensieg.inpg.fr.

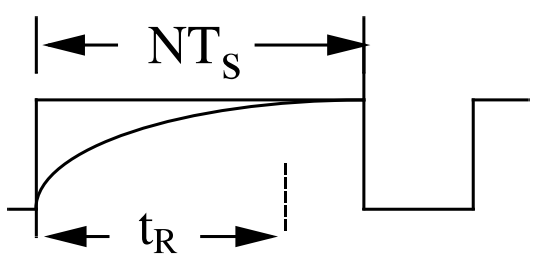

**Fig. 5.4** Choice of a maximum duration of a pulse in a PRBS

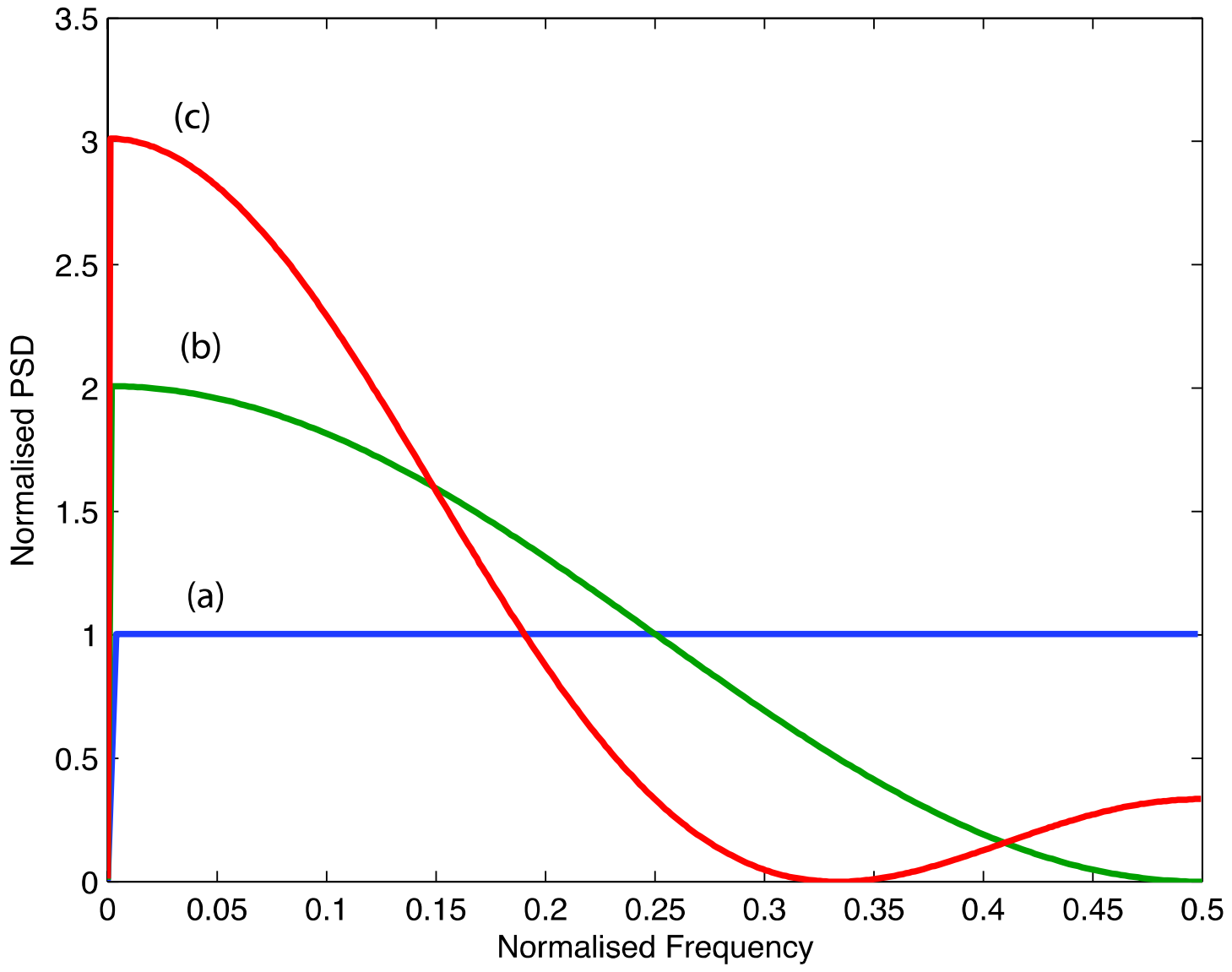

**Fig. 5.5** Power spectral density of a PRBS sequence, (**a**) $N = 8,\ p = 1$, (**b**) $N = 8,\ p = 2$, (**c**) $N = 8,\ p = 3$

Note that the condition (5.104) can result in fairly large values of $N$ corresponding to sequence lengths of prohibitive duration, either because $T_s$ is very large, or because the system to be identified may well evolve during the duration of the test. This is why, in a large number of practical situations, a submultiple of the sampling frequency is chosen as the clock frequency for the PRBS. If:

$$f_{PRBS} = \frac{f_s}{p}; \quad p = 1, 2, 3, \ldots \tag{5.106}$$

then condition (5.104) becomes:

$$pNT_s > t_R \tag{5.107}$$

This approach is more interesting than the increase of the sequence length by increasing $N$ in order to satisfy (5.104). Indeed, if one passes from $N$ to $N' = N + 1$, the maximum duration of a pulse passes from $NT_s$ to $(N + 1)T_s$ but the duration of the sequence is doubled $L' = 2L$. On the other hand, if $f_{PRBS} = f_s/2$ is chosen, the maximum duration of a pulse passes from $NT_s$ to $2NT_s$ for a doubled sequence duration $L' = 2L$. From a comparison of the two approaches, it results that the second approach (frequency division) enables a pulse of greater duration to be obtained for an identical duration of the sequence and thus of the test. If $p$ is the integer frequency divider, one has in the case of clock frequency division ($d_{\max}$ = maximum pulse duration)

$$d_{\max} = pNT_S; \qquad L' = pL; \quad p = 1, 2, 3, \ldots$$

By increasing $N$ by $(p - 1)$ and thus the length of the sequence (without changing the clock frequency), one obtains:

$$d_{\max} = (N + p - 1)T_S; \qquad L' = 2^{(p-1)}L; \quad p = 1, 2, 3, \ldots$$

Note that dividing the clock frequency of the PRBS will reduce the frequency range corresponding to a constant spectral density in the high frequencies while augmenting the spectral density in the low frequencies. In general, this will not affect the quality of identification, either because in many cases when this solution is considered, the plant to be identified has a low band pass or because the effect or the reduction of the signal/noise ratio at high frequencies can be compensated by the use of appropriate identification techniques. However, it is recommended to choose $p \le 4$.

Figure 5.5 shows the normalized power spectral density (PSD) of PRBS sequences generated with $N = 8$ for $p = 1, 2, 3$. As one can see, the energy of the spectrum is reduced in the high frequencies and augmented in the lower frequencies. Furthermore, for $p = 2$ and $p = 3$ we have no excitation at $f_s/p$.

Up to now, we have been concerned only with the choice of the length and clock frequency of the PRBS; however, the magnitude of the PRBS must also be considered. Although the magnitude of the PRBS may be very low, it should lead to output variations larger than the residual noise level. If the signal/noise ratio is too low, the length of the test must be augmented in order to obtain a satisfactory parameter estimation. Note that in a large number of applications, the significant increase in the PRBS level may be undesirable in view of the nonlinear character of the plants to be identified (we are concerned with the identification of a linear model around an operating point).

## 5.8 Model Order Selection

In the absence of a clear a priori information upon the plant model structure, two approaches can be used to select the appropriate orders for the values of $d, n_A, n_B$ characterizing the input-output plant model

(a) a practical approach based on trial and error,
(b) estimation of $d, n_A, n_B$ directly from data.

Even in the case of using an estimation of the orders directly from data it is useful to see to what extend the two approaches are coherent. We will discuss the two approaches next.

### 5.8.1 A Practical Approach for Model Order Selection

The *plant* + *disturbance* model to be identified is of the form:

$$A(q^{-1})y(t) = q^{-d}B(q^{-1})u(t) + w(t)$$

where, according to the structures chosen, one has:

$$\begin{aligned}
w(t) &= e(t) \\
w(t) &= A(q^{-1})v(t); \quad (v(t) \text{ and } u(t) \text{ are independent}) \\
w(t) &= C(q^{-1})e(t) \\
w(t) &= \frac{C(q^{-1})}{D(q^{-1})}e(t)
\end{aligned}$$

The degrees of the polynomials, $A(q^{-1})$, $B(q^{-1})$, $C(q^{-1})$ and $D(q^{-1})$ are respectively $n_A, n_B, n_C$ and $n_D$. In order to start the parameter estimation methods $n_A, n_B$ and $d$ must be specified. For the methods which estimate also the noise model, one needs in addition to specify $n_C$ and $n_D$.

#### A Priori Choice of $n_A$

Two cases can be distinguished:

1. Industrial plant (temperature, control, flow rate, concentration, and so on). For this type of plant in general:

   $$n_A \leq 3$$

   and the value $n_A = 2$, which is very typical, is a good starting value to choose.
2. Electromechanical systems. $n_A$ results from the structural analysis of the system.

*Example* Flexible Robot Arm with two vibration modes. In this case, $n_A = 4$ is chosen, since a second-order is required, to model a vibratory mode.

### Initial Choice of $d$ and $n_B$

If no knowledge of the time delay is available, $d = 0$ is chosen as an initial value. If a minimum value is known, an initial value $d = d_{\min}$ is chosen. If the time delay has been underestimated, during identification the first coefficients of $B(q^{-1})$ will be very low. $n_B$ must then be chosen so that it can both indicate the presence of the time delays and identify the transfer function numerator. $n_B = (d_{\max} - d_{\min}) + 2$ is then chosen as the initial value. At least two coefficients are required in $B(q^{-1})$ because of the *fractional* delay which is often present in applications. If the time delay is known, $n_B \geq 2$ is chosen, but 2 remains a good initial value.

### Determination of Time Delay $d$ (First Approximation)

**Method 5.1** One identifies using the RLS (Recursive Least Squares). The estimated numerator will be of the form:

$$\hat{B}(q^{-1}) = \hat{b}_1 q^{-1} + \hat{b}_2 q^{-2} + \hat{b}_3 q^{-3} + \cdots$$

If:

$$|\hat{b}_1| < 0.15|\hat{b}_2|$$

$b_1 \approx 0$ is considered and time delay $d$ is increased by $1 : d = d_{\min} + 1$ [since if $b_1 = 0$, $B(q^{-1}) = q^{-1}(b_2 q^{-1} + b_3 q^{-2})$]. If:

$$|\hat{b}_i| < 0.15|\hat{b}_{di+1}|; \quad i = 1, 2, \ldots, d_i$$

time delay $d$ is increased by $d_i : d = d_{in} + d_i$. After these modifications, identification is restarted.

**Method 5.2** An *impulse response* type model is identified with the RLS. The estimated model will have the form:

$$y(t) = \sum_{i=1}^{n_B} b_i u(t-i); \quad n_B = \text{ large (20 to 30)}$$

If there is a time delay, then:

$$|\hat{b}_i| < 0.15|\hat{b}_{d+1}|; \quad i = 1, 2, \ldots, d$$

After this delay estimation, identification is restarted with a poles-zeros model.

Both of these methods may, of course, be completed by a display of the step response of the estimated model. A more accurate estimation of time delay $d$ is carried out during the second passage in the identification algorithm. This is followed by a validation of the identified model. Note that if the system is contaminated by measurement system noise, an accurate estimation of the delay will only be carried out with the method enabling the identified model to be validated.

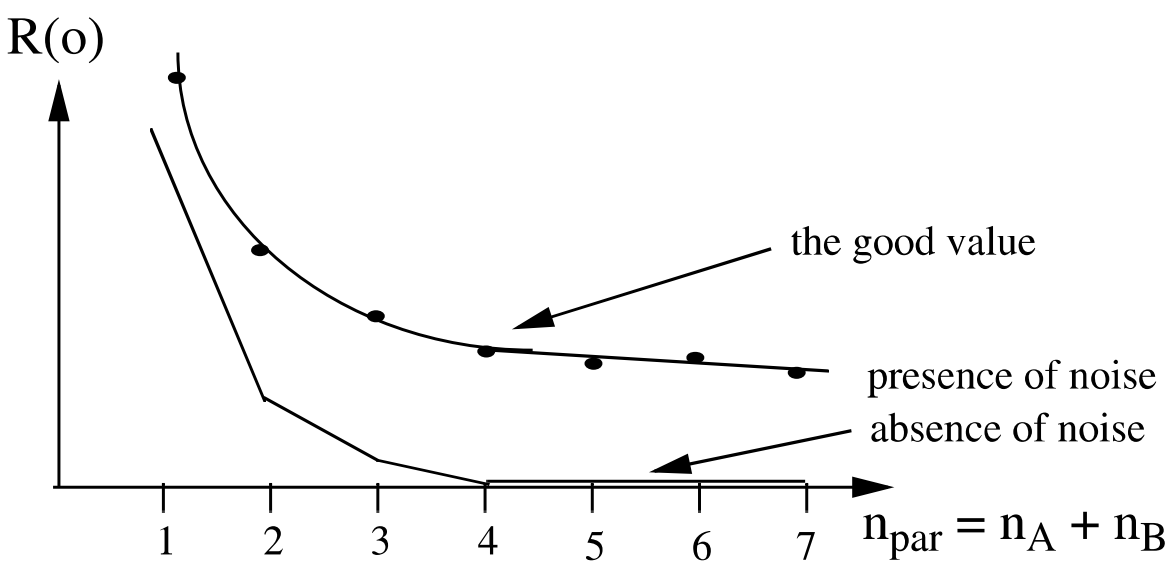

**Fig. 5.6** Evolution of the variance of residual errors as a function of the number of model parameters

## Determination of the $(n_A)_{\max}$ and $(n_B)_{\max}$

The aim is to obtain the simplest possible identified model that verifies the validation criteria. This is linked on the one hand to the complexity of the controller (which will depend on $n_A$ and $n_B$) but equally to the robustness of the identified model with respect to the operating conditions.

A first approach to estimate the values of $(n_A)_{\max}$ and $(n_B)_{\max}$ is to use the RLS and to study the evolution of the variance of the residual prediction errors, i.e., the evolution of:

$$R(0) = E\{\varepsilon^2(t)\} = \frac{1}{N}\sum_{t=1}^{N}\varepsilon(t)^2$$

as a function of the value of $n_A + n_B$. A typical curve is given in Fig. 5.6.

In theory, if the example considered is simulated and noise-free, the curve should present a neat elbow followed by a horizontal segment, which indicates that the increase in parameter number does not improve the performance. In practice, this elbow is not neat because measurement noise is present.

The practical test used for determining $n_A + n_B$ is the following: consider first $n_A, n_B$ and the corresponding variance of the residual errors $R(0)$. Consider now $n'_A = n_A + 1, n_B$ and the corresponding variance of the residual errors $R'(0)$. If:

$$R'(0) \geq 0.8R(0)$$

it is unwise to increase the degree of $n_A$ (same test with $n'_B = n_B + 1$). With the choice that results for $n_A$ and $n_B$, the model identified by the RLS does not necessarily verify the validation criterion. Therefore, while keeping the values of $n_A$ and $n_B$, other structures and methods must be tried out in order to obtain a *valid* model. If after all the methods have been tried, none is able to give a model that satisfies the validation criterion, then $n_A$ and $n_B$ must be increased.

For a more detailed discussion of various procedures for the estimation of $(n_A)$ max and $(n_B)$ max see Söderström and Stoica (1989).

## Initial Choice of $n_C$ and $n_D$ (Noise Model)

As a rule, $n_C = n_D = n_A$ is chosen.

### 5.8.2 Direct Order Estimation from Data

To introduce the problem of order estimation from data, we will start with an example: assume that the plant model can be described by:

$$y(t) = -a_1 y(t-1) + b_1 u(t-1) \tag{5.108}$$

and that the data are noise free. The order of this model is $n = n_A = n_B = 1$. Question: Is any way to test from data if the order assumption is correct? To do so, construct the following matrix:

$$\begin{bmatrix} y(t) & \vdots & y(t-1) & u(t-1) \\ y(t-1) & \vdots & y(t-2) & u(t-2) \\ y(t-2) & \vdots & y(t-3) & u(t-3) \end{bmatrix} = \begin{bmatrix} Y(t) & \vdots & R(1) \end{bmatrix} \tag{5.109}$$

Clearly, if the order of the model given in (5.108) is correct, the vector $Y(t)$ will be a linear combination of the columns of $R(1)$ ($Y(t) = R(1)\theta$ with $\theta^T = [-a_1, b_1]$) and the rank of the matrix will be 2 (instead of 3). If the plant model is of order 2 or higher, the matrix (5.109) will be full rank. Of course, this procedure can be extended for testing the order of a model by testing the rank of the matrix $[Y(t), R(\hat{n})]$ where:

$$R(\hat{n}) = [Y(t-1), U(t-1), Y(t-2), U(t-2), \ldots, Y(t-\hat{n}), U(t-\hat{n})] \tag{5.110}$$

$$Y^T(t) = [y(t), y(t-1), \ldots] \tag{5.111}$$

$$U^T(t) = [u(t), u(t-1), \ldots] \tag{5.112}$$

Unfortunately, as a consequence of the presence of noise, this procedure cannot directly be applied in practice.

A more practical approach results from the observation that the rank test problem can be approached by the searching of $\hat{\theta}$ which minimize the following criterion for an estimated value of the order $\hat{n}$

$$V_{LS}(\hat{n}, N) = \min_{\hat{\theta}} \frac{1}{N} \|Y(t) - R(\hat{n})\hat{\theta}\|^2 \tag{5.113}$$

where $N$ is the number of data. But this criterion is nothing else than an equivalent formulation of the least squares (Söderström and Stoica 1989). If the conditions for unbiased estimation using least squares are fulfilled, (5.113) is an efficient way for assessing the order of the model since $V_{LS}(\hat{n}) - V_{LS}(\hat{n}+1) \to 0$ when $\hat{n} \geq n$

In the mean time, the objective of the identification is to estimate lower order models (parsimony principle) and therefore, it is reasonable to add in the criterion (5.113) a term which penalizes the complexity of the model. Therefore, the criterion for order estimation will take the form:

$$CV_{LS}(\hat{n}, N) = V_{LS}(\hat{n}, N) + S(\hat{n}, N) \tag{5.114}$$

**Fig. 5.7** Evaluation of the criterion for order estimation

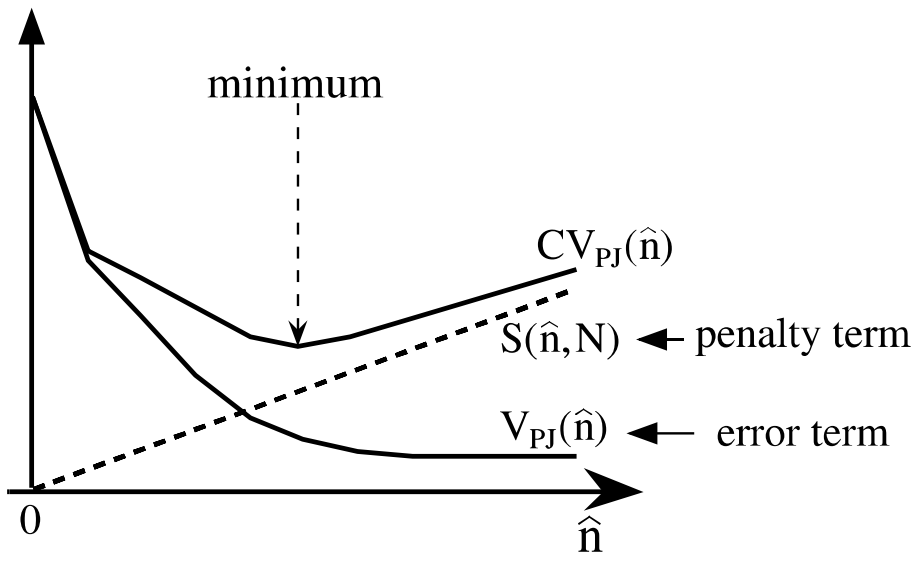

where typically:

$$S(\hat{n}, N) = 2\hat{n}X(N) \tag{5.115}$$

$X(N)$ in (5.115) is a function that decreases with $N$. For example, in the so called $BIC_{LS}(\hat{n}, N)$ criterion, $X(N) = \frac{\log N}{N}$ (other choices are possible—see Ljung 1999; Söderström and Stoica 1989; Duong and Landau 1996) and the order $\hat{n}$ is selected as the one which minimizes $CV_{LS}$ given by (5.114). Unfortunately, the results are unsatisfactory in practice because in the majority of situations, the conditions for unbiased parameter estimation using least squares are not fulfilled.

In Duong and Landau (1994) and Duong and Landau (1996), it is proposed to replace the matrix $R(\hat{n})$ by an instrumental variable matrix $Z(\hat{n})$ whose elements will not be correlated with the measurement noise. Such an instrumental matrix $Z(\hat{n})$ can be obtained by replacing in the matrix $R(\hat{n})$, the columns $Y(t-1)$, $Y(t-2)$, $Y(t-3)$ by delayed version of $U(t-L-i)$, i.e., where $L > n$:

$$Z(\hat{n}) = [U(t-L-1), U(t-1), U(t-L-2), U(t-2), \ldots] \tag{5.116}$$

and therefore, the following criterion is used for the order estimation:

$$CV_{PJ}(\hat{n}, N) = \min_{\hat{\theta}} \frac{1}{N} \|Y(t) - Z(\hat{n})\hat{\theta}\|^2 + \frac{2\hat{n}\log N}{N} \tag{5.117}$$

and:

$$\hat{n} = \min_{\hat{n}} CV_{PJ}(\hat{n}) \tag{5.118}$$

A typical curve of the evolution of the criterion (5.117) as a function of $\hat{n}$ is shown in Fig. 5.7. It is shown in Duong and Landau (1996) that using this criterion a consistent estimate of the order $\hat{n}$ is obtained under mild noise condition (i.e., $\lim_{N\to\infty} Pr(\hat{n} = n) = 1$). Comparisons with order estimation criterions are also provided in this reference.

Once an estimated order $\hat{n}$ is selected, one can apply a similar procedure to estimate $\hat{n}_A, \hat{n} - \hat{d}, \hat{n}_B + \hat{d}$, from which $\hat{n}_A, \hat{n}_B$ and $\hat{d}$ are obtained.[8]

[8]Routines corresponding to this method in Matlab and Scilab can be downloaded from the websites: http://www.landau-adaptivecontrol.org and http://landau-bookic.lag.ensieg.inpg.fr.

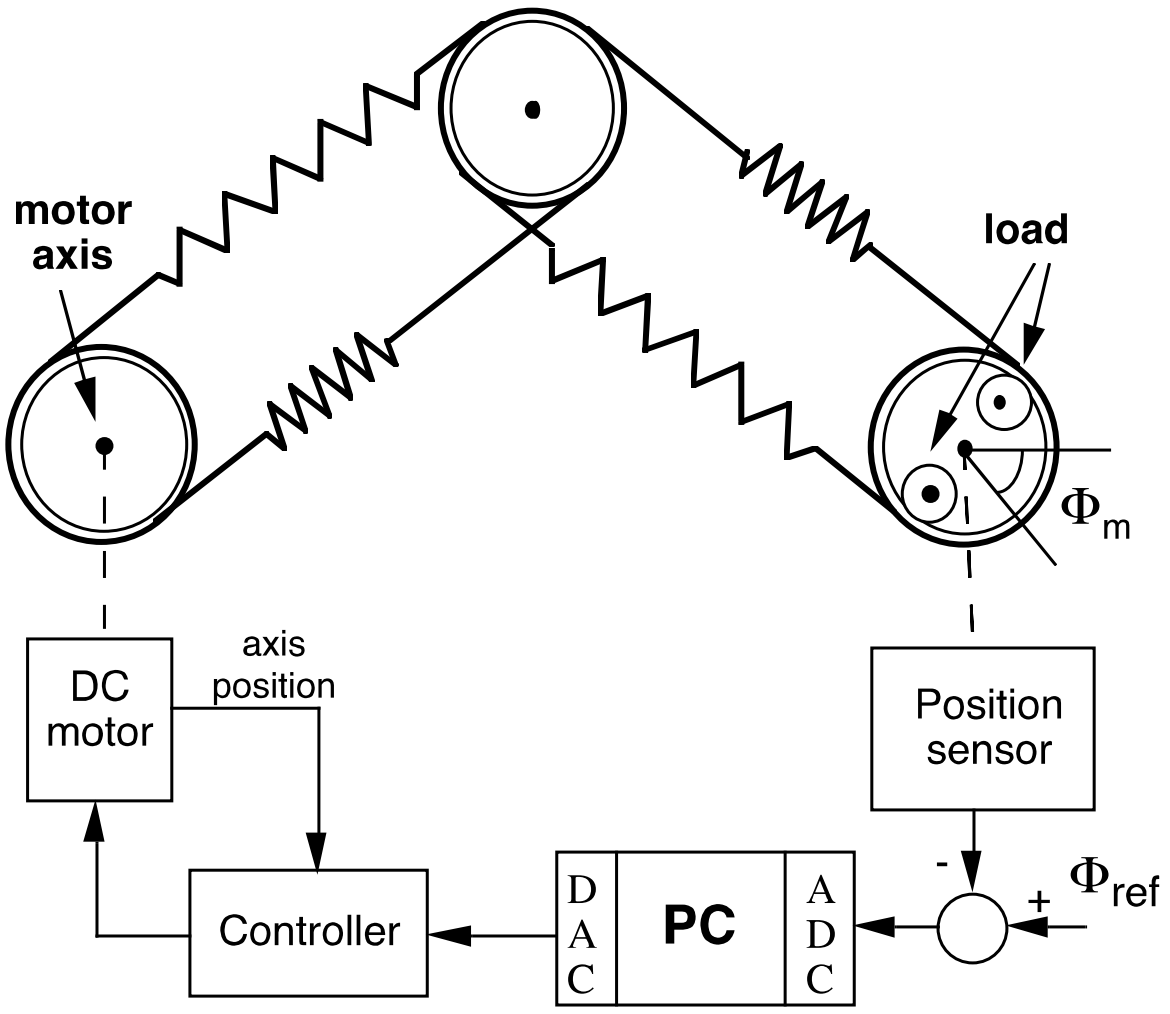

**Fig. 5.8** The flexible transmission

## 5.9 An Example: Identification of a Flexible Transmission

The flexible transmission is depicted in Fig. 5.8. and has been described in Sect. 1.4.3. It is formed by a set of three pulleys coupled with two very elastic belts. The system is controlled by a PC via an I/O board. The sampling frequency is 20 Hz (for more details see Sect. 1.4.3). The open-loop identification is carried out with a PC using WinTrac software for data acquisition and WinPim identification software (Adaptech 1988). The input sequence is a PRBS generated by a shift register with 7 cells and a frequency divider of 2 is used for the clock frequency of the PRBS generator. The length of the sequence is 256 samples. See Fig. 5.9.

After removing the d.c. component, order estimation from data has been performed using the criterion (5.117). The following results have been obtained: $n=4$, $n_A=4$, $n_B=2$, $d=2$. These values have been confirmed by using the trial and error method presented in Sect. 5.8.1.

The various methods presented in Table 5.1 have been used. It appears clearly that an ARMAX type model is representative for the system after comparing the validation results obtained with the models identified by different methods. Therefore, in the following, we will present the comparison of the models obtained with: ELS, OEEPM, RML for the case without load. In all cases, a vanishing adaptation gain has been used (without forgetting factor, i.e., $\lambda_1(t)\equiv 1$, $\lambda_2(t)\equiv 1$).

Table 5.4 summarizes both the results of the whiteness test on the residual prediction error and of the uncorrelation test when using an output error predictor.

The results obtained are very close, but priority has been given to the uncorrelation test using an output error predictor. ELS and OEEPM give extremely close results. The OEEPM model has been chosen since it gives a smaller value for $R_{e\hat{y}}(i)_{\max}$ (however, the frequency characteristics of the two models are almost identical).

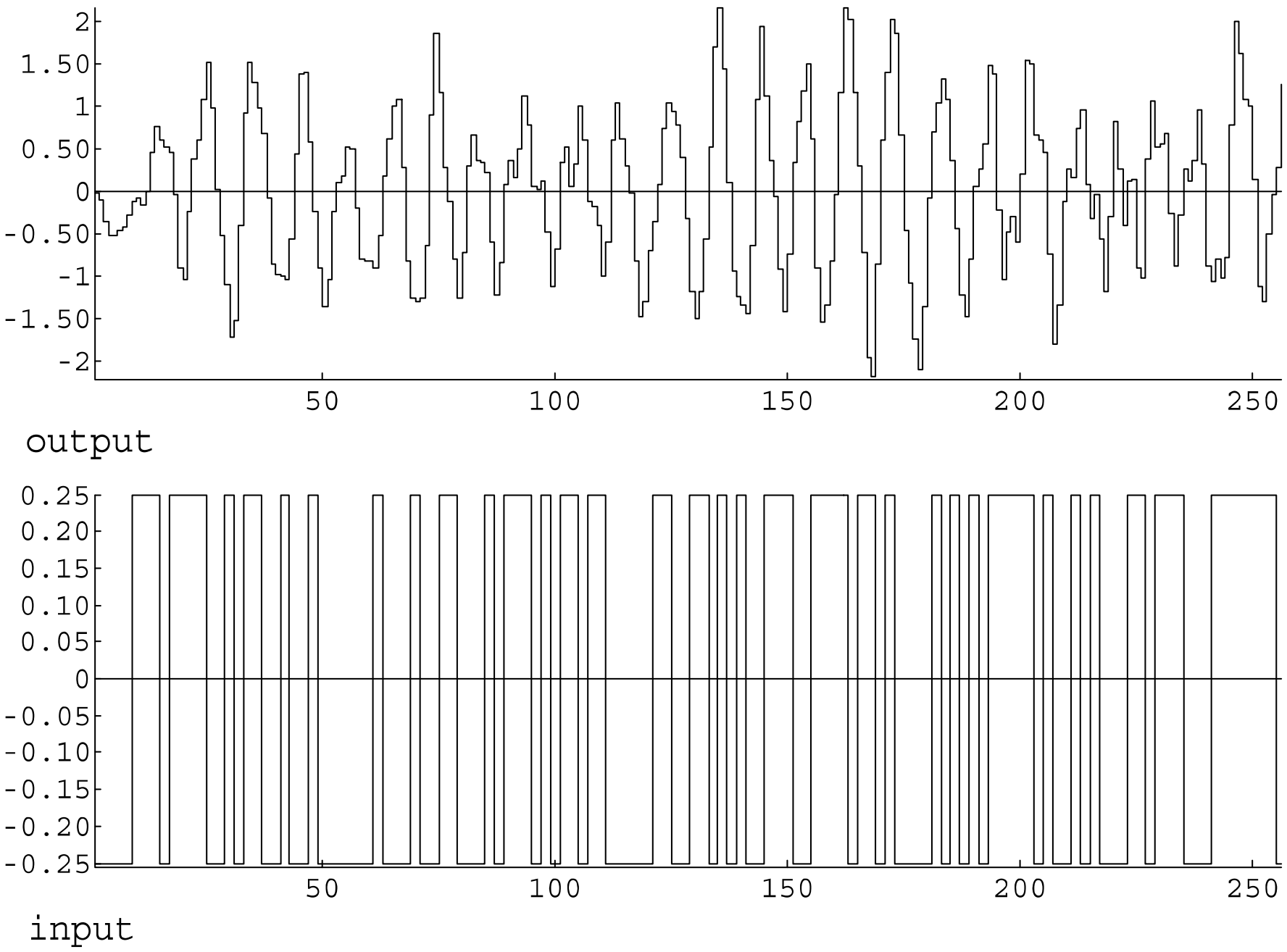

**Fig. 5.9** The input/output sequence used for identification

**Table 5.4** Comparative model validation

| | Whiteness test (prediction error) | | Uncorrelation test (output error) | |
|---|---|---|---|---|
| | $R(0)$ | $R_{\varepsilon\varepsilon}(i)_{\max}$ | $R(0)$ | $R_{\varepsilon\hat{y}}(i)_{\max}$ |
| ELS | 0.00239 | 0.1094 | 0.01037 | 0.12163 |
| OEEPM | 0.00238 | 0.1061 | 0.01051 | 0.11334 |
| RML | 0.0027 | 0.0644 | 0.01313 | 0.2440 |

Figure 5.10 shows the true output and the predicted output (ARMAX predictor) as well as the residual prediction error.

Figure 5.11 shows the true output and the predicted output for an output error type predictor, as well as the residual output error prediction. Similar results have been obtained for the case of 50% load (1.8 kg) and full load case (3.6 kg).

The resulting identified models are:

- No load ($L00$)

$$d = 2, \qquad B(q^{-1}) = 0.41156q^{-1} + 0.52397q^{-2}$$
$$A(q^{-1}) = 1 - 1.35277q^{-1} + 1.55021q^{-2} - 1.27978q^{-3} + 0.91147q^{-4}$$

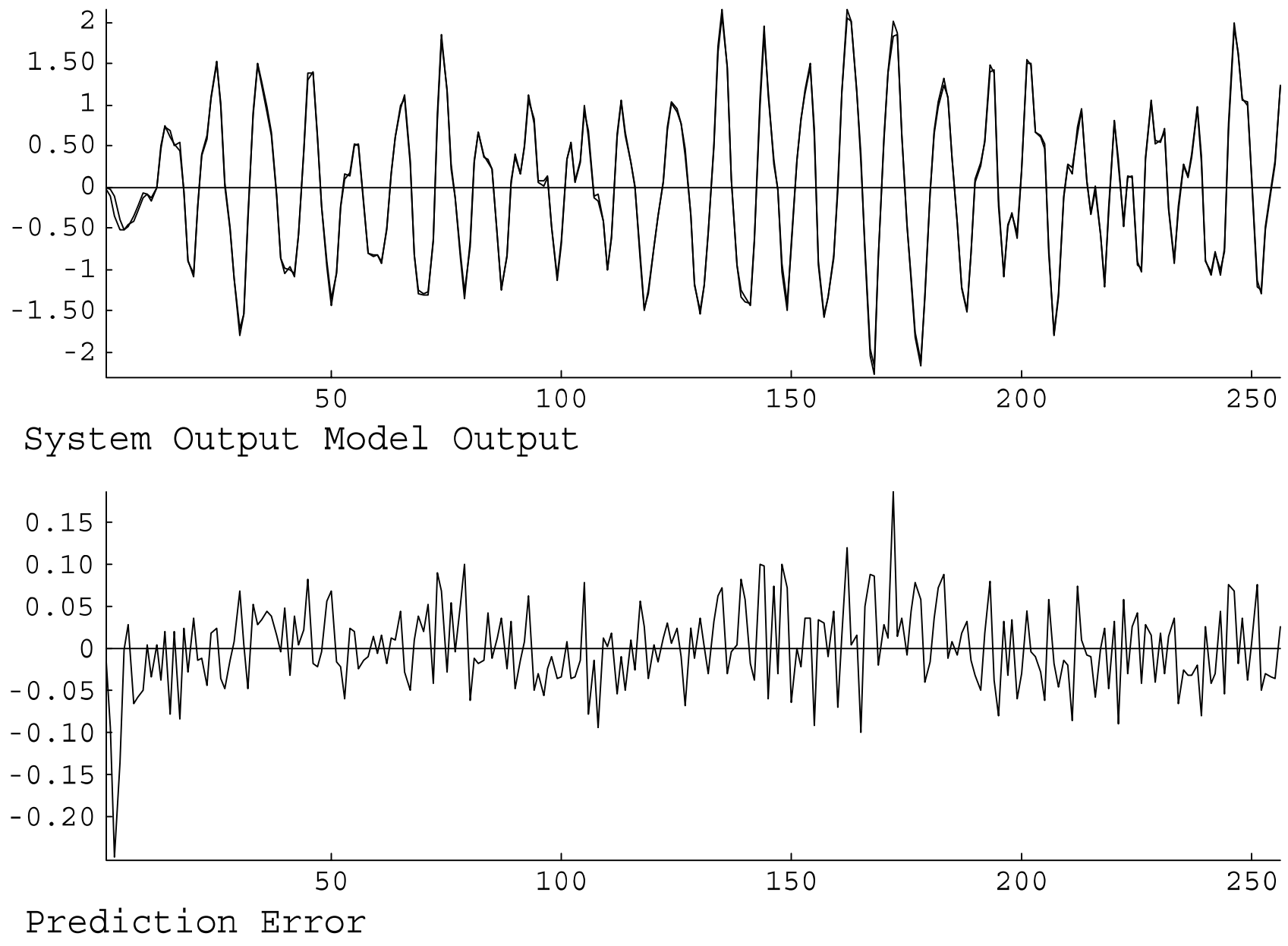

**Fig. 5.10** True output, predicted output (ARMAX predictor) and residual prediction error for the model identified using OEEPM algorithm

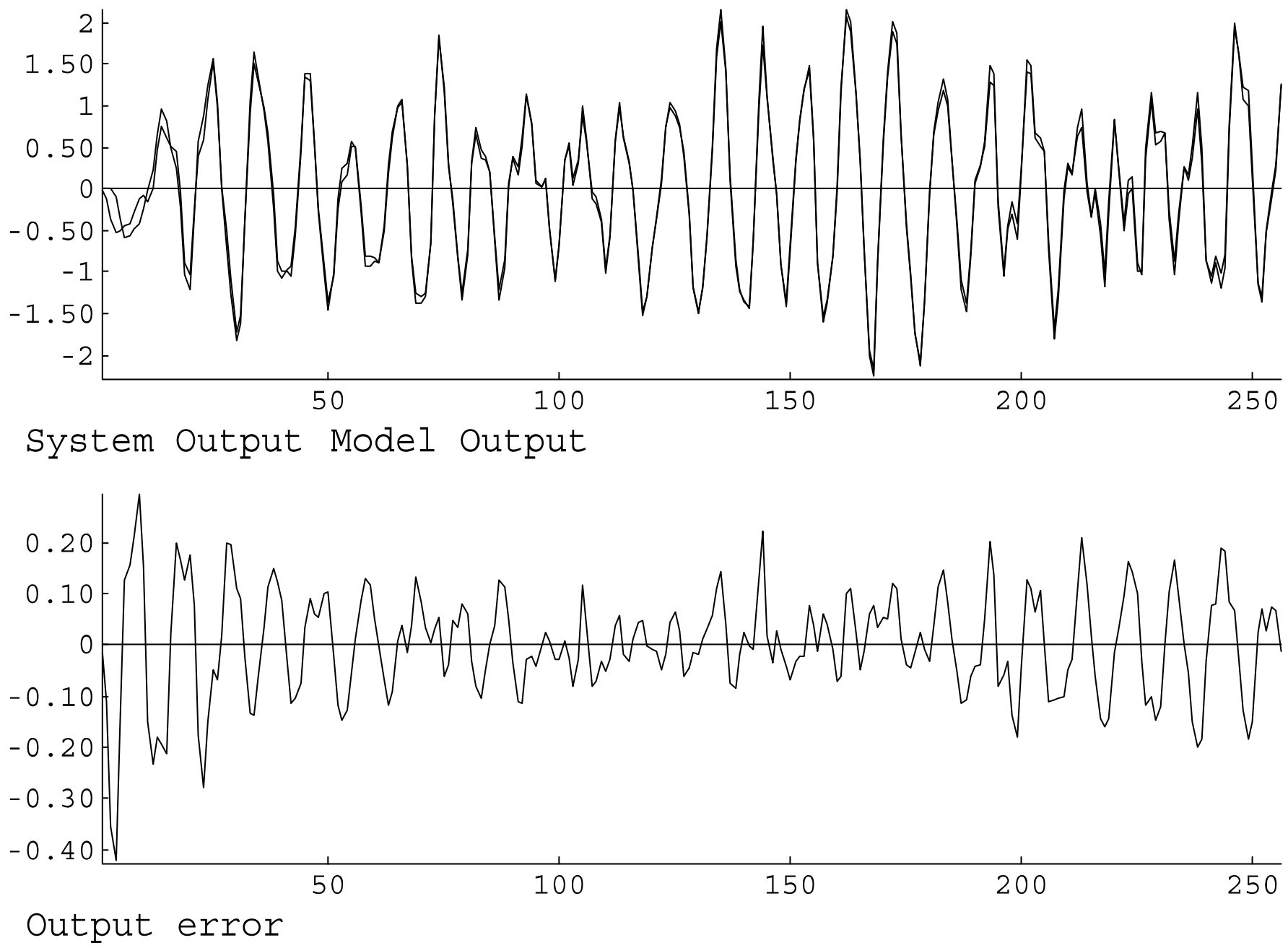

**Fig. 5.11** True output, predicted output (output error predictor) and residual output error prediction for the model identified using OEEPM algorithm

- 50% load ($L50$)

$$d = 2, \qquad B(q^{-1}) = 0.16586q^{-1} + 0.181329q^{-2}$$
$$A(q^{-1}) = 1 - 1.92005q^{-1} + 2.1151q^{-2} - 1.81622q^{-3} + 0.91956q^{-4}$$

- 100% load ($L100$)

$$d = 2, \qquad B(q^{-1}) = 0.09553q^{-1} + 0.11116q^{-2}$$
$$A(q^{-1}) = 1 - 2.05182q^{-1} + 2.23744q^{-2} - 1.91294q^{-3} + 0.90643q^{-4}$$

The models feature two very lightly damped vibration modes (see Fig. 1.20) and all have unstable zeros.

## 5.10 Concluding Remarks

Basic elements for the identification of discrete-time models for dynamical systems using recursive parameter estimation algorithms have been laid down in this chapter. The following facts have to be emphasized:

1. System identification includes four basic steps:
   - input/output data acquisition under an experimental protocol;
   - estimation or selection of the model complexity;
   - estimation of the model parameters;
   - validation of the identified model (structure of the model and values of parameters).

     This procedure has to be repeated (with appropriate changes at each step) if the validation of the model fails.
2. Recursive or off-line parameter estimation algorithms can be used for identification of the plant model parameters.
3. Recursive parameter estimation algorithms offer a number of advantages with respect to off-line algorithms:
   (a) estimation of the model parameters as the system evolves;
   (b) significant data compression of past data;
   (c) real-time identification capabilities.
   In addition, their performance makes them a competitive alternative to off-line parameter estimation algorithms.
4. The various recursive parameter estimation algorithms use the same structure for the PAA. They differ from each other in the following ways:
   - structure of the adjustable predictor;
   - nature of the components of the observation vector;
   - way in which the adaptation error is generated.
5. The stochastic disturbances, which contaminate the measured output, may cause errors in the parameter estimates (bias). For a specific type of disturbances, appropriate recursive identification algorithms providing asymptotically unbiased estimates are available.

6. A unique *plant* + *disturbance* model structure that describes all the situations encountered in practice does not exist, nor is there a unique identification method providing satisfactory parameter estimates (unbiased estimates) in all situations.
7. Selection of the complexity of an unknown plant model from data can be done either by trial and error approach or by using complexity estimation techniques.

## 5.11 Problems

**5.1** Analyze the behavior of the filtered output error (FOE) algorithm (Sect. 5.5.3) in a stochastic environment using the averaging method presented in Sect. 4.2.

**5.2** For the instrumental variable with auxiliary model (Sect. 5.5.4) derive the adaptation error for $\hat{\theta}(t) = \hat{\theta} =$ constant (5.100).

**5.3** Work out the details of convergence analysis of the extended least squares algorithm in a stochastic environment using the averaging method (Sect. 4.2) and the martingale approach (Sect. 4.3).

**5.4** For an ARMAX model of the form (5.11) develop an instrumental variable method using delayed observations $y_{VI}(t) = y(t-i)$ in order to obtain unbiased parameter estimates with a least squares type predictor (Banon and Aguilar-Martin 1972).

**5.5** Develop the expression of the asymptotic frequency distribution of the bias for the various identification algorithms given in Table 5.1.

**5.6** Generate an input file for a simulated model:

$$\begin{aligned} y(t+1) = 1.5y(t) - 0.7y(t-1) + u(t-1) + 0.5u(t-2) \\ + 0.5e(t-1) - 0.2e(t) + e(t+1) \end{aligned}$$

where $e(t)$ is a white noise, and $u(t)$ is a PRBS generated by a shift register with 8 cells (length: 255 samples).

(a) Examine the effect of the noise/signal ratio upon the bias of the estimated parameter when using recursive least squares.
(b) For a value of the noise/signal ratio which gives significant bias on the recursive least squares, try the other algorithms given in Table 5.1.

Use the parametric distance $[(\theta - \hat{\theta})^T (\theta - \hat{\theta})]^{1/2}$ to globally evaluate the bias. Discuss the validation results.

# Chapter 6
# Adaptive Prediction

## 6.1 The Problem

In Chap. 5, we saw how adjustable one-step-ahead predictors are used for recursive plant model parameter estimation. In Chap. 2, $j$-steps-ahead ($j > 1$) predictors were discussed for the case of known plant parameters both in deterministic and stochastic case. These predictors are computed from the knowledge of the plant model parameters in the deterministic case and from the knowledge of the plant and disturbance models parameters in the stochastic case.

When the plant model parameters are unknown (as well as those of the disturbance model), the idea is that, based on the available data, one will adapt the parameters of an adjustable $j$-steps ahead predictor such that in the deterministic case the prediction error goes asymptotically to zero and in the stochastic case it converges asymptotically (in a stochastic sense) toward the linear optimal prediction error.

This chapter will discuss the design of adaptive $j$-steps ahead predictors both in the deterministic and stochastic environment. Two approaches can be considered for the adaptation of $j$-steps ahead predictors.

(1) Direct adaptive prediction.
In this case the parameters of the $j$-steps ahead adjustable predictor are directly adapted from the measurement of the prediction error.
(2) Indirect adaptive prediction.
One estimates recursively the unknown parameters of the plant model (and disturbance model) and, at each sampling time, one computes the parameters of the $j$-steps ahead predictor based on the current estimates of the plant parameters.

The advantage of the direct approach is that no additional calculations will be necessary. A disadvantage of this approach may occur for long-range prediction (for example for systems with large delay $d$) since the predictor will have more parameters to estimate than there are in the plant model ($n_a + n_B + j$ instead of $n_A + n_B$). This will increase the complexity of the PAA and may slow down the parameters convergence rate.

The *indirect* approach has the advantage of leading to the estimation of fewer parameters for the case of long range predictors and to offer the possibility of generating a family of predictors for various prediction horizons (as are required for example in generalized predictive control—see Chap. 7). The disadvantage is that it requires the solution of a polynomial equation at each sampling time. However, recursive procedures also exist for this purpose (see Sect. 7.7).

## 6.2 Adaptive Prediction—Deterministic Case

### 6.2.1 Direct Adaptive Prediction

The plant model is given by:

$$A(q^{-1})y(t) = q^{-d-1}B^*(q^{-1})u(t) \tag{6.1}$$

Using the results of Theorem 2.3, the $d+1$ steps ahead filtered output for the plant model (6.1) has the form (to simplify the notations we have took $j = d+1$ and suppressed the index $j$):

$$P(q^{-1})y(t+d+1) = F(q^{-1})y(t) + G(q^{-1})u(t) = \theta^T\phi(t) \tag{6.2}$$

where:

$$G = EB^* \tag{6.3}$$

$F(q^{-1})$ and $E(q^{-1})$ are solutions of the polynomial equation:

$$AE + q^{-d-1}F = P \tag{6.4}$$

and $\theta$ and $\phi(t)$ are given by:

$$\theta^T = [f_0, \ldots, f_{n_F}, g_0, \ldots, g_{n_G}] \tag{6.5}$$

$$\phi^T(t) = [y(t), \ldots, y(t-n_F), u(t), \ldots, u(t-n_G)] \tag{6.6}$$

where:

$$n_F = \deg F = \max(n_A - 1, n_P - 1) \tag{6.7}$$

$$n_G = \deg B^* + \deg E = n_B + d + 1 \tag{6.8}$$

Using a similar procedure as for the one-step ahead predictor, one defines the a priori and the a posteriori outputs of the adjustable predictor as:

$$P\hat{y}^0(t+d+1) = \hat{\theta}^T(t)\phi(t) \tag{6.9}$$

$$P\hat{y}(t+d+1) = \hat{\theta}^T(t+d+1)\phi(t) \tag{6.10}$$

where:

$$\hat{\theta}^T(t) = [\hat{f}_0(t), \ldots, \hat{f}_{n_F}(t), \hat{g}_0(t), \ldots, \hat{g}_{n_G}(t)] \tag{6.11}$$

One defines the filtered a priori and a posteriori prediction error as:

$$\varepsilon^0(t+d+1) = Py(t+d+1) - P\hat{y}^0(t+d+1) \tag{6.12}$$

$$\varepsilon(t+d+1) = Py(t+d+1) - P\hat{y}(t+d+1) \tag{6.13}$$

This leads immediately to:

$$\varepsilon^0(t+d+1) = [\theta - \hat{\theta}(t)]^T\phi(t) \tag{6.14}$$

and:

$$\varepsilon(t+d+1) = [\theta - \hat{\theta}(t+d+1)]^T\phi(t) \tag{6.15}$$

Equation (6.15) has the standard form for the a posteriori adaptation error equation considered in Theorem 3.2. Applying Theorem 3.2, the PAA to be used for assuring $\lim_{t\to\infty} \varepsilon(t+d+1)$ is:

$$\hat{\theta}(t+d+1) = \hat{\theta}(t+d) + F(t)\phi(t)\varepsilon(t+d+1) \tag{6.16}$$

$$\begin{aligned} F(t+1)^{-1} &= \lambda_1(t)F(t)^{-1} + \lambda_2(t)\phi(t)\phi^T(t) \\ 0 < \lambda_1(t) &\le 1;\ 0 \le \lambda_2(t) < 2;\ F(0) > 0 \end{aligned} \tag{6.17}$$

However, to make this algorithm implementable we need to give an expression for $\varepsilon(t+d+1)$ which will depend on $\hat{\theta}(t+i)$ up to and including $i=d$. Taking into account (6.14) and (6.16), (6.15) can be written as:

$$\begin{aligned} \varepsilon(t+d+1) &= [\theta - \hat{\theta}(t)]^T\phi(t) + [\hat{\theta}(t+d) - \hat{\theta}(t+d+1)]^T\phi(t) \\ &\quad + [\hat{\theta}(t) - \hat{\theta}(t+d)]^T\phi(t) \\ &= \varepsilon^0(t+d+1) - \phi^T(t)F(t)\phi(t)\varepsilon(t+d+1) \\ &\quad + [\hat{\theta}(t) - \hat{\theta}(t+d)]^T\phi(t) \\ &= y(t+d+1) - \hat{\theta}^T(t+d)\phi(t) \\ &\quad - \phi^T(t)F(t)\phi(t)\varepsilon(t+d+1) \end{aligned} \tag{6.18}$$

from which one obtains:

$$\varepsilon(t+d+1) = \frac{y(t+d+1) - \hat{\theta}^T(t+d)\phi(t)}{1 + \phi^T(t)F(t)\phi(t)} \tag{6.19}$$

which can also be expressed as:

$$\varepsilon(t+d+1) = \frac{\varepsilon^0(t+d+1) - [\hat{\theta}(t+d) - \hat{\theta}(t)]^T\phi(t)}{1 + \phi^T(t)F(t)\phi(t)} \tag{6.20}$$

Using Theorem 3.2 and taking into account that $\phi(t)$ and $F(t)$ are bounded and that $\lim_{t\to\infty}[\hat{\theta}(t+d) - \hat{\theta}(t)]^T = 0$, one also concludes that $\lim_{t\to\infty} \varepsilon(t+d+1)$ will imply the asymptotic convergence to zero of the a priori prediction error ($\lim_{t\to\infty} \varepsilon^0(t+d+1) = 0$).

*Remark* Other adaptation algorithms discussed in Chap. 3 can be used. In particular, since we are more interested in a faster convergence of the prediction error towards zero than in the speed of convergence of the adjustable parameters, the use of *integral + proportional* PAA with a positive proportional adaptation gain ((3.270) through (3.274)) will be very beneficial.

### 6.2.2 Indirect Adaptive Prediction

The plant model is described by:

$$y(t+1) = -A^* y(t) + q^{-d} B^* u(t) = \theta^T \phi(t) \tag{6.21}$$

An indirect adaptive predictor can be obtained as follows:

*Step I*: Identify recursively the plant model parameters. The one step ahead adjustable predictor used for plant model parameter estimation has the form:

$$\hat{y}^0(t+1) = \hat{\theta}^T(t)\phi(t) = -\hat{A}^*(t)y(t) + q^{-d}\hat{B}^*(t)u(t) \tag{6.22}$$

$$\hat{y}(t+1) = \hat{\theta}^T(t+1)\phi(t) \tag{6.23}$$

where:

$$\hat{\theta}^T(t) = [\hat{a}_1(t), \ldots, \hat{a}_{n_A}(t), \hat{b}_1(t), \ldots, \hat{b}_{n_B}(t)] \tag{6.24}$$

$$\phi^T(t) = [-y(t), \ldots, -y(t-n_A+1), u(t-d), \ldots, u(t-d-n_B+1)] \tag{6.25}$$

The adaptation error is defined as:

$$\nu^0(t+1) = y(t+1) - \hat{y}^0(t+1) \tag{6.26}$$

$$\nu(t+1) = y(t+1) - \hat{y}(t+1) \tag{6.27}$$

and using the PAA

$$\hat{\theta}(t+1) = \hat{\theta}(t) + F(t)\phi(t)\nu(t+1) \tag{6.28}$$

$$F(t+1)^{-1} = \lambda_1(t)F(t)^{-1} + \lambda_2(t)\phi(t)\phi^T(t)$$

$$0 < \lambda_1(t) \le 1;\ 0 \le \lambda_2(t) < 2;\ F(0) > 0 \tag{6.29}$$

$$\nu(t+1) = \frac{\nu^0(t+1)}{1 + \phi^T(t)F(t)\phi(t)} \tag{6.30}$$

one assures $\lim_{t\to\infty} \nu(t+1) = \lim_{t\to\infty} \nu^0(t+1) = 0$. The information provided by the estimated plant model parameters is used for Step II.

*Step II*: Compute at each sampling an adjustable predictor of the form:

$$P(q^{-1})\hat{y}^0(t+d+1) = \hat{F}(t)y(t) + \hat{E}(t)\hat{B}^*(t)u(t) \tag{6.31}$$

where $\hat{F}(t)$ and $\hat{E}(t)$ are time-varying polynomials solutions of the polynomial equation:

$$\hat{E}(t)\hat{A}(t) + \hat{F}(t)q^{-d-1} = P \tag{6.32}$$

One has to establish that using indirect adaptive prediction the a priori prediction error defined as:

$$\varepsilon^0(t+d+1) = Py(t+d+1) - P\hat{y}^0(t+d+1) \tag{6.33}$$

goes asymptotically to zero. One has the following result.

**Lemma 6.1** *For the adaptive predictor* (6.31) *where* $\hat{F}(t)$ *and* $\hat{E}(t)$ *are solutions of* (6.32) *with* $\hat{A}(t)$ *and* $\hat{B}^*(t)$ *estimated using* (6.22) *through* (6.30), *provided that* $u(t)$ *and* $y(t)$ *are bounded, one has*:

$$\lim_{t\to\infty} \varepsilon^0(t+d+1) = 0 \tag{6.34}$$

*Proof* Observe first that from (6.23) and (6.27) one gets:

$$\begin{aligned}\nu(t+1) &= y(t+1) + \hat{A}^*(t+1)y(t) - \hat{B}^*(t+1)q^{-d}u(t)\\ &= \hat{A}(t+1)y(t+1) - \hat{B}^*(t+1)q^{-d}u(t)\end{aligned} \tag{6.35}$$

and respectively:

$$\nu(t+d+1) = \hat{A}(t+d+1)y(t+d+1) - \hat{B}^*(t+d+1)u(t) \tag{6.36}$$

Passing both terms of (6.36) through the transfer operator $\hat{E}(t)$, one gets:

$$\begin{aligned}\hat{E}(t)\nu(t+d+1) &= \hat{E}(t)\hat{A}(t+d+1)y(t+d+1) - \hat{E}(t)B^*(t+d+1)u(t)\\ &= \hat{E}(t)\hat{A}(t)y(t+d+1) - E(t)\hat{B}^*(t)u(t)\\ &\quad + \hat{E}(t)[\hat{A}(t+d+1) - \hat{A}(t)]y(t+d+1)\\ &\quad - \hat{E}(t)[\hat{B}^*(t+d+1) - \hat{B}^*(t)]u(t)\end{aligned} \tag{6.37}$$

But, from (6.32):

$$\hat{E}(t)\hat{A}(t) = P - \hat{F}(t)q^{-d-1} \tag{6.38}$$

and taking into account (6.38) and (6.31), (6.37) becomes:

$$\begin{aligned}\varepsilon^0(t+d+1) &= \hat{E}(t)\nu(t+d+1) - \hat{E}(t)[\hat{A}(t+d+1) - \hat{A}(t)]y(t+d+1)\\ &\quad + \hat{E}(t)[\hat{B}^*(t+d+1) - \hat{B}^*(t)]u(t)\end{aligned} \tag{6.39}$$

From Theorem 3.2, one has that:

- $\lim_{t\to\infty} \nu(t+d+1) = 0$;
- $\lim_{t\to\infty}[\hat{A}(t+d+1) - \hat{A}(t)] = \lim_{t\to\infty}[\hat{B}^*(t+d+1) - \hat{B}^*(t)] = 0$;
- $\hat{A}(t)$ and $\hat{B}^*(t)$ have bounded elements.

The latest result implies that the solutions $\hat{E}(t)$ and $\hat{F}(t)$ of (6.32) will have bounded elements. Taking also into account that $u(t)$ and $y(t)$ are bounded, one can conclude that the right term of the equality (6.39) goes asymptotically to zero, which implies (6.34). □

*Remark* The stability proof for the indirect adaptation of the predictor uses the following facts:

(1) The adaptation error in the parameter estimation schemes goes to zero.
(2) The variations of the estimated parameters go to zero.
(3) The parameter estimates are bounded for all $t$.
(4) The *design equation*, in this case (6.32), provides bounded solutions for bounded plant parameter estimates.
(5) $y(t)$ and $u(t)$ are bounded.

Therefore, a *design equation* can be combined with any parameter estimation scheme which assures properties 1 through 3 leading to an adaptive system for which a representative error (for performance evaluation) is asymptotically driven to zero provided that the computed parameters and the input and output of the plant are bounded. As will be shown for indirect adaptive control, the difficulty in some cases arises from the fact that properties (4) and (5) are not necessarily satisfied for all the possible values of the estimated plant parameters.

## 6.3 Adaptive Prediction—Stochastic Case

### *6.3.1 Direct Adaptive Prediction*

Consider the ARMAX plant + disturbance model:

$$y(t+1) = -A^*(q^{-1})y(t) + q^{-d}B^*(q^{-1})u(t) + C(q^{-1})e(t+1) \tag{6.40}$$

where $e(t)$ is a discrete-time Gaussian white noise $(0, \sigma)$. For the model (6.40), the expression of $y(t+d+1)$ takes the form (see Sect. 2.2):

$$Cy(t+d+1) = Fy(t) + Gu(t) + CEe(t+d+1) \tag{6.41}$$

where:

$$G = B^*E \tag{6.42}$$

and $E(q^{-1})$ and $F(q^{-1})$ are solutions of:

$$C = AE + q^{-d-1}F \tag{6.43}$$

The linear optimal predictor has the form:

$$\hat{y}(t+d+1) = -C^*\hat{y}(t+d) + Fy(t) + Gu(t) \tag{6.44}$$

While in the linear case with known parameters, the predictors for deterministic case and stochastic case look the same (except that $P$ is replaced by $C$ in the stochastic case), in the adaptive case there will be a significative difference since the polynomial $C$ (characterizing the disturbance model) is unknown. To accommodate this situation some modifications of the predictor (6.44) will be considered.

The natural extension for the adaptive prediction in stochastic case will be to define an adjustable predictor as:

$$\hat{y}^0(t+d+1) = -\hat{C}^*(t)\hat{y}(t+d) + \hat{F}(t)y(t) + \hat{G}(t)u(t) = \hat{\theta}^T(t)\phi(t)$$
$$\hat{y}(t+d+1) = \hat{\theta}^T(t+d+1)\phi(t)$$

where $\hat{y}^0(t+j)$ and $\hat{y}(t+j)$ will define the a priori and the a posteriori outputs of the adjustable predictor. Unfortunately, difficulties arise in the generation of the a posteriori predicted outputs for $j > 1$ (i.e., $d > 0$). An approximate algorithm can indeed be used by defining the adjustable predictor as:

$$\hat{y}^0(t+d+1) = -\hat{C}^*(t)\hat{y}^0(t+d) + \hat{F}(t)y(t) + \hat{G}(t)u(t) = \hat{\theta}^T(t)\phi(t) \tag{6.45}$$

where:

$$\hat{\theta}^T(t) = [\hat{c}_1(t), \ldots, \hat{c}_{n_C}(t), \hat{f}_0(t), \ldots, \hat{f}_{n_F}(t), \hat{g}_0(t), \ldots, \hat{g}_{n_G}(t)] \tag{6.46}$$

$$\phi^T(t) = [-\hat{y}^0(t+d), \ldots, -\hat{y}^0(t+d-n_C), y(t), \ldots, y(t-n_F), u(t), \ldots, u(t-n_G)] \tag{6.47}$$

and using the PAA:

$$\hat{\theta}(t+d+1) = \hat{\theta}(t) + F(t)\phi(t)\nu(t+d+1) \tag{6.48}$$

$$F^{-1}(t+1) = F^{-1}(t) + \lambda_2(t)\phi(t)\phi^T(t)$$
$$0 < \lambda_2(t) < 2; \; F(0) > 0 \tag{6.49}$$

$$\nu(t+d+1) = \frac{y(t+d+1) - \hat{\theta}^T(t+d)\phi(t)}{1 + \phi^T(t)F(t)\phi(t)} \tag{6.50}$$

An analysis of this algorithm using the averaging method (Theorem 4.1) can be carried on.

To overcome the difficulty of generating future a posteriori outputs, the idea is to further reparameterize the model (6.41) such that:

$$y(t+d+1) = f[y(t), y(t-1), \ldots, u(t), u(t-1), \ldots, e(t+d+1), e(t+d), \ldots,]$$

To achieve this, one considers a second polynomial division (Fuchs 1982; Goodwin and Sin 1984)

$$1 = C\tilde{E} + q^{-d-1}\tilde{H} \tag{6.51}$$

and one has the following result:

**Lemma 6.2** *The asymptotically optimal $d+1$ step ahead predictor for the ARMAX model* (6.40) *can be expressed as*:

$$\hat{y}(t+d+1) = \bar{F}y(t) + \bar{G}u(t) + \bar{H}\varepsilon(t) \tag{6.52}$$

*where*

$$\varepsilon(t) = y(t) - \hat{y}(t) \tag{6.53}$$

*and the prediction error $\varepsilon(t+d+1)$ obeys the equation*:

$$\bar{C}[\varepsilon(t+d+1) - Ee(t+d+1)] = 0 \tag{6.54}$$

*where*

$$\bar{F} = \tilde{H} + \tilde{E}F \tag{6.55}$$

$$\bar{G} = \tilde{E}B^*E \tag{6.56}$$

$$\bar{H} = -\tilde{H} \tag{6.57}$$

$$\bar{C} = C\tilde{E} \tag{6.58}$$

*$E$, $F$ and $\tilde{E}$, $\tilde{H}$ being the solutions of* (6.43) *and* (6.51) *respectively.*

*Proof* Using (6.51) and (6.41), one has:

$$
\begin{aligned}
C\tilde{E}y(t+d+1) &= (1-q^{-d-1}\tilde{H})y(t+d+1) \\
&= \tilde{E}Fy(t) + \tilde{E}B^*Eu(t) + C\tilde{E}Ee(t+d+1)
\end{aligned}
\tag{6.59}
$$

and respectively:

$$
y(t+d+1) = \bar{F}y(t) + \bar{G}u(t) + \bar{C}Ee(t+d+1) \tag{6.60}
$$

Subtracting (6.52) from (6.60), one obtains:

$$
\begin{aligned}
\varepsilon(t+d+1) &= -\bar{H}\varepsilon(t) + \bar{C}Ee(t+d+1) \\
&= (1-\bar{C})\varepsilon(t+d+1) + \bar{C}Ee(t+d+1)
\end{aligned}
\tag{6.61}
$$

from which (6.54) results, which implies that asymptotically the prediction error tends towards $Ee(t+d+1)$, the optimal prediction error. □

*Remark* The second polynomial division (6.51) can be used in various ways to generate $j$-step ahead adaptive predictors in the stochastic case. See Fuchs (1982), Goodwin and Sin (1984) for other forms of the $j$-steps ahead predictor.

Equation (6.52) suggests immediately to consider as an adjustable predictor:

$$
\hat{y}^0(t+d+1) = \hat{\bar{F}}(t)y(t) + \hat{\bar{G}}(t)u(t) + \hat{\bar{H}}(t)\varepsilon(t)\hat{\theta}^T(t)\phi(t) \tag{6.62}
$$

$$
\hat{y}(t+d+1) = \hat{\theta}^T(t+d+1)\phi(t) \tag{6.63}
$$

where:

$$
\varepsilon^0(t+d+1) = y(t+d+1) - \hat{y}^0(t+d+1) \tag{6.64}
$$

$$
\varepsilon(t+d+1) = y(t+d+1) - \hat{y}(t+d+1) \tag{6.65}
$$

$$
\hat{\theta}^T(t) = [\ldots, \hat{\bar{f}}_i(t), \ldots, \ldots, \hat{\bar{g}}_i(t), \ldots, \ldots, \hat{\bar{h}}_i(t), \ldots] \tag{6.66}
$$

$$
\phi^T(t) = [y(t), \ldots, u(t), \ldots, \varepsilon(t), \ldots] \tag{6.67}
$$

The equation of the a posteriori prediction error takes the form (by adding and subtracting $\bar{H}\varepsilon(t)$):

$$
\varepsilon(t+d+1) = [\theta - \hat{\theta}(t+d+1)]^T\phi(t) - \bar{H}\varepsilon(t) + \bar{C}Ee(t+d+1) \tag{6.68}
$$

where:

$$
\theta^T = [\ldots, \bar{f}_i, \ldots, \ldots, \bar{g}_i, \ldots, \ldots, \bar{h}_i, \ldots] \tag{6.69}
$$

and using the expression of $\bar{H}$ given by (6.57) and (6.51), one gets:

$$
\varepsilon(t+d+1) = \frac{1}{\bar{C}}[\theta - \hat{\theta}(t+d+1)]^T\phi(t) + Ee(t+d+1) \tag{6.70}
$$

This fits the format of the adaptation error equation used in Theorem 4.1 (averaging method) with the observation that for $\hat{\theta}(t+d+1) = \hat{\theta}$ one has:

$$
\mathbf{E}\{\phi(t,\hat{\theta})Ee(t+d+1)\} = 0 \tag{6.71}
$$

This suggests using the following PAA (similar to the deterministic case but with a decreasing adaptation gain):

$$\hat{\theta}(t+d+1) = \hat{\theta}(t+d) + F(t)\phi(t)\varepsilon(t+d+1) \tag{6.72}$$

$$F^{-1}(t+1) = F^{-1}(t) + \lambda_2(t)\phi(t)\phi^T(t); \quad 0 < \lambda_2(t) < 2 \tag{6.73}$$

$$\varepsilon(t+d+1) = \frac{\varepsilon^0(t+d+1) - [\hat{\theta}(t+d) - \hat{\theta}(t)]^T\phi(t)}{1+\phi^T(t)F(t)\phi(t)}$$

$$= \frac{y(t+d+1) - \hat{\theta}^T(t+d)\phi(t)}{1+\phi^T(t)F(t)\phi(t)} \tag{6.74}$$

and we will have:

$$\text{Prob}\left\{\lim_{t\to\infty}[\theta - \hat{\theta}(t)]^T\phi(t) = 0\right\} = 1 \tag{6.75}$$

provided that there is $\lambda_2 : \max_t \lambda_2(t) \le \lambda_2 < 2$, such that:

$$\frac{1}{\bar{C}(z^{-1})} - \frac{\lambda_2}{2} \tag{6.76}$$

is a strictly positive real transfer function. Equations (6.75) and (6.76) will also imply that:

$$\text{Prob}\left\{\lim_{t\to\infty}\varepsilon(t+d+1) = Ee(t+d+1)\right\} = 1 \tag{6.77}$$

Taking into account the form of (6.70), Theorem 4.2, together with Lemma 4.1 can also be used observing that the optimal linear prediction error $Ee(t+d+1)$ is an innovation process $\omega(t+d+1) = Ee(t+d+1)$ satisfying the martingale type properties:

$$\mathbf{E}\{\omega(t+d+1)|\mathcal{F}_t\} = 0$$

$$\mathbf{E}\{\omega^2(t+d+1)|\mathcal{F}_t\} = \mu^2 \le \left(\sum_{i=0}^{n_E} e_i^2\right)\pi^2$$

$$\lim_{N\to\infty}\sup\frac{1}{N}\sum_{t=1}^{N}\omega^2(t) < \infty$$

yielding to the same convergence condition (6.76) of the PAA.

### *6.3.2 Indirect Adaptive Prediction—Stochastic Case*

Consider again the ARMAX model given in (6.40). An indirect adaptive predictor can be obtained as follows:

*Step I*: Identify recursively the parameters of the ARMAX model (6.40) using OEEPM or ELS algorithm (see Chap. 5).

*Step II*: Compute at each sampling an adjustable predictor

$$\hat{y}^0(t+d+1) = -\hat{C}^*(t)\hat{y}^0(t+d) + \hat{F}(t)y(t) + \hat{E}(t)\hat{B}^*(t)u(t) \tag{6.78}$$

where $\hat{E}(t)$ and $\hat{F}(t)$ are solutions of the polynomial equation:

$$\hat{E}(t)\hat{A}(t) + \hat{F}(t)q^{-d-1} = \hat{C}(t) \tag{6.79}$$

computed only for the values of $\hat{C}(t)$ which are asymptotically stable (if $\hat{C}(t)$ is unstable, one replaces it by $\hat{C}(t-1)$).

One has the following result (Goodwin and Sin 1984):

**Lemma 6.3** *For the indirect adaptive predictor given in* (6.78) *and whose parameters are computed using* (6.79) *where the parameters of the ARMAX model are estimated using the OEEPM or ELS algorithms*, *the* $d+1$ *steps ahead prediction error*

$$\varepsilon^0(t+d+1) = y(t+d+1) - y^0(t+d+1) \tag{6.80}$$

*has the property that*:

$$\lim_{N\to\infty}\lim_{t_0\to\infty}\frac{1}{N}\sum_{t=t_0}^{N+t_0-1}[\varepsilon^0(t+d+1)]^2 = \left(\sum_{i=0}^{n_E} e_1^2\right)\sigma^2 \tag{6.81}$$

*where* $(\sum_{i=0}^{n_e} e_i^2)\sigma^2$ *represents the optimal* $d+1$ *steps ahead prediction error variance* ($e_i$ *are the coefficients of the polynomial* $E(q^{-1})$ *solution of* (6.43)) *provided that*:

$$\frac{1}{C(z^{-1})} - \frac{\lambda_2}{2} \tag{6.82}$$

*is strictly positive real and*:

$$\frac{F^{-1}(t)}{t} > 0; \quad \left(\textit{with} \lim_{t\to\infty}\frac{1}{t}F^{-1}(t) > 0,\ a.s\right) \tag{6.83}$$

For other indirect adaptive prediction schemes see for example (Fuchs 1982).

## 6.4 Concluding Remarks

1. For $j$-steps ahead adaptive predictors ($1 < j \leq d+1$) a direct or an indirect adaptive approach can be considered in both the deterministic and stochastic case.
2. Direct adaptive predictors are obtained by transforming the fixed linear predictor for the known parameter case into an adjustable predictor and using the prediction error as adaptation error (directly or after filtering it).

3. Indirect adaptive predictors are obtained through a two-step procedure. In the first step, one recursively identifies the parameters of the plant model and of the disturbance model (in the stochastic case), and in the second step, one computes at each sampling period an adjustable predictor from the estimated parameters (one uses the "ad hoc certainty principle"). Asymptotically, indirect adaptive predictors have the same properties as direct adaptive predictors.
4. The use of direct or indirect adaptive prediction is problem dependent. One of the decision criteria is the final numerical complexity of the algorithm. Direct adaptive prediction seems more appropriate for short-horizon predictors while indirect adaptive prediction seems more appropriate for long-range predictors or for the computation of a family of predictors with various prediction horizons.

## 6.5 Problems

**6.1** Derive an "integral + proportional" adaptation algorithm for direct deterministic adaptive predictors.

**6.2** Compare by simulation in the deterministic case for an example with $n_A = 2$, $n_B = 2$, $d = 3$ the performance of a 4 steps ahead adaptive predictor using:

- direct adaptive prediction with integral adaptation
- direct adaptive prediction with integral + proportional adaptation
- indirect adaptive prediction

(use the same integral adaptation gain in all cases).

**6.3** For $n_A = 2$, $n_B = 2$, $d = 3$ and $j = 4$, compare the complexity (in terms of number of additions and multiplications) of direct and indirect adaptive deterministic prediction (to simplify the problem consider first the case of constant diagonal adaptation gain).

**6.4** Use the averaging method (Sect. 4.2) for analyzing the adaptive stochastic predictor given by (6.45) through (6.50).

**6.5** Work out the details of the proof for Lemma 6.3.

# Chapter 7
# Digital Control Strategies

## 7.1 Introduction

We are concerned with the design of digital controllers for single-input/single-output systems described by discrete-time models in input-output form.

The control strategies and the corresponding design assume that the discrete-time plant model is known. In general, one has to solve a joint tracking and regulation problem. The design will also incorporate the a priori knowledge upon the disturbances. It is important to note that in many cases the tracking and regulation performances have to be decoupled (as much as possible). This leads us to consider a *two-degree of freedom digital controller*.

Since the plant model is given in the input/output form, it is also reasonable to search also for a controller structure which will be fed by the measurements of the output and the desired tracking trajectory (or reference signal) and will generate the control $u(t)$. This controller will also have an input/output form and will consist of three polynomials in the delay operator $q^{-1}$ related to the control $u(t)$, the output $y(t)$ and the desired tracking trajectory $y^{\star}(t)$.

In many cases the design can be done using a polynomial approach which in fact corresponds to a design in the frequency domain. This will allow the introduction of specifications in various frequency ranges both for assuring the nominal performances as well as robustness with respect to plant parameter variations, noise etc. The *pole placement* strategy is the basic prototype from which a number of designs result as particular cases. However, the same design can be obtained by a synthesis in the time domain starting from a control objective specification in the time domain. Other strategies are naturally introduced in the time domain by the specification of the control objective. However, the resulting design can always be interpreted in terms of a special pole placement.

Comparing the control objectives in the time domain associated with the various control strategies one can classify these strategies in two categories:

1. *One step ahead predictive control*. In these strategies one computes a prediction of the output at $t + d + 1$ ($d$ integer delay of the plant) namely $\hat{y}(t + d + 1)$ as a function of $u(t), u(t-1), \dots, y(t), y(t-1), \dots$ and one computes $u(t)$ such that a control objective in terms of $\hat{y}(t+d+1)$ be satisfied. This is illustrated in Fig. 7.1.

**Fig. 7.1** One step ahead predictive control

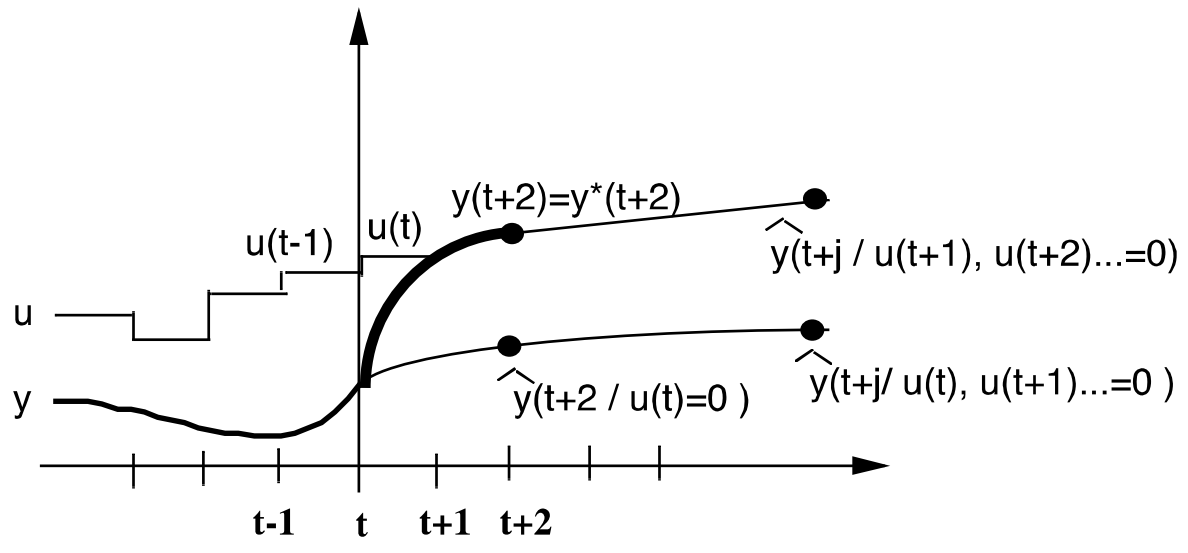

**Fig. 7.2** Long range predictive control

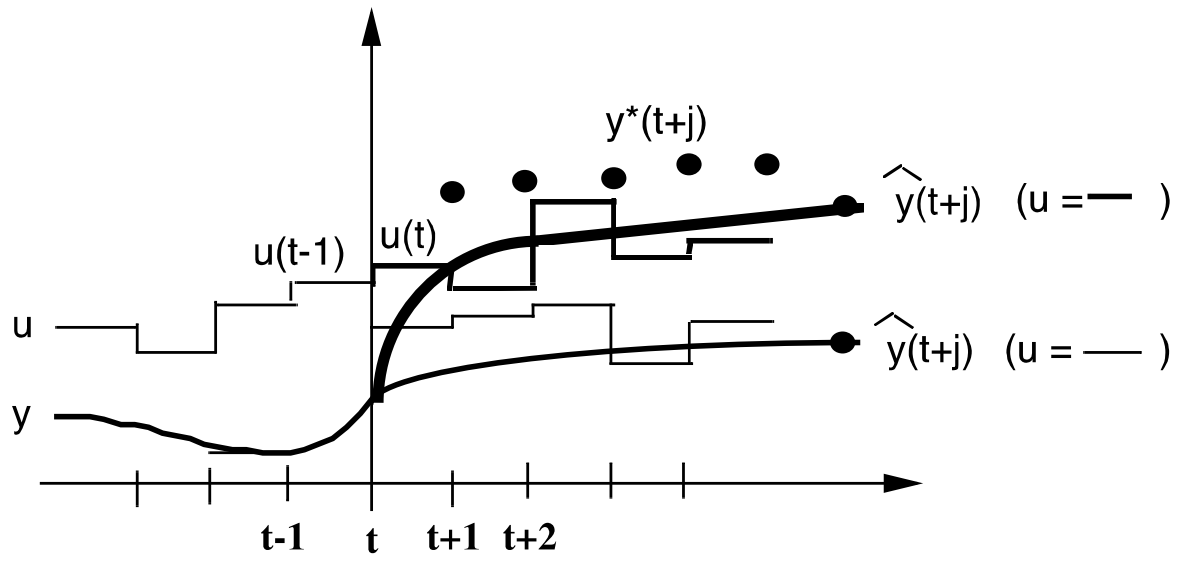

2. *Long range predictive control*. In these strategies the control objective is expressed in terms of the future values of the output over a certain horizon and of a sequence of future control values.

In order to solve the problem, one needs to compute: $\hat{y}(t+d+1)$, $\hat{y}(t+d+2)$, $\ldots$, $\hat{y}(t+d+j)$ which are expressed as:

$$\hat{y}(t+d+1) = f_1(y(t), y(t-1), \ldots, u(t), u(t-1), \ldots)$$

$$\vdots$$

$$\begin{aligned}\hat{y}(t+d+j) &= f_j(y(t), y(t-1), \ldots, u(t), u(t-1), \ldots)\\ &\quad + g_j(u(t+1), \ldots, u(t+j-1))\end{aligned}$$

To satisfy the control objective the sequence of present and future values of the control $u(t), u(t+1), \ldots, u(t+j-1)$ is computed but only the first one (i.e., $u(t)$) is applied to the plant and the same procedure is restarted at $t+1$. This is called the *receding horizon procedure*.

The principle of long range predictive control is illustrated in Fig. 7.2 where the sequence of desired values $y^\star$, of predicted values $\hat{y}$ and the future control sequence are represented (predicted values are represented for two different future control sequences).

All the control strategies concern linear design in the sense that constraints on the values of the admissible control applied to the plant are not considered. As a consequence all the control strategies will yield a linear controller of the same structure.

The use of one or another strategy corresponds finally to different values of the parameters of the controller for the same plant model used for design. Another important issue is that the control should be admissible (realizable) i.e., it should depend only on the information available up to and including time $t$ where the control $u(t)$ is computed.

As indicated at the beginning of this section the use of these control strategies requires the knowledge of the plant model. These plant models can be obtained by system identification in open loop or by identification in closed loop if a controller already exists.

If the parameters of the plant model vary during operation, a robust control design will be able to assure satisfactory performance in a region of parameter variations around the nominal model. The robustness issues will be discussed in Chap. 8. However, for large variations of the plant parameters or fine tuning of the controller an adaptive control solution has to be considered using in general the "ad hoc" certainty equivalence principle i.e., the plant models will be replaced by their estimates and the design will be done based on these estimates.

The canonical structure of the two-degree of freedom digital controllers which is common to all the control strategies will be presented in Sect. 7.2. Pole placement will be discussed in Sect. 7.3. Tracking and regulation with independent objectives which can be viewed either as a generalization of "model reference control" or as a particular case of pole placement will be examined in Sect. 7.4. The extension of the design for the case where a weighting on the control is considered will be presented in Sect. 7.5. Section 7.6 is dedicated to minimum variance tracking and regulation which is in fact the stochastic counterpart of tracking and regulation with independent objectives. Long range predictive control strategies will be presented in Sect. 7.7 (generalized predictive control) and in Sect. 7.8 (linear quadratic control).

## 7.2 Canonical Form for Digital Controllers

It is assumed that the discretized plant to be controlled is described by the input–output model:

$$A(q^{-1})y(t) = q^{-d}B(q^{-1})u(t) + A(q^{-1})v(t) \tag{7.1}$$

where $y(t)$ is the output, $u(t)$ is the input and $v(t)$ the disturbance added to the output (or more precisely an image of the disturbance).

A general (canonical) form of a two-degree of freedom digital controller is given by the equation:

$$S(q^{-1})u(t) + R(q^{-1})y(t) = T(q^{-1})y^{\star}(t+d+1) \tag{7.2}$$

where $y^{\star}(t+d+1)$ represents the *desired tracking trajectory* given with $d+1$ steps in advance which is either stored in the computer or generated from the reference signal $r(t)$ via a *tracking reference model*

$$y^{\star}(t+d+1) = \frac{B_m(q^{-1})}{A_m(q^{-1})}r(t) \tag{7.3}$$

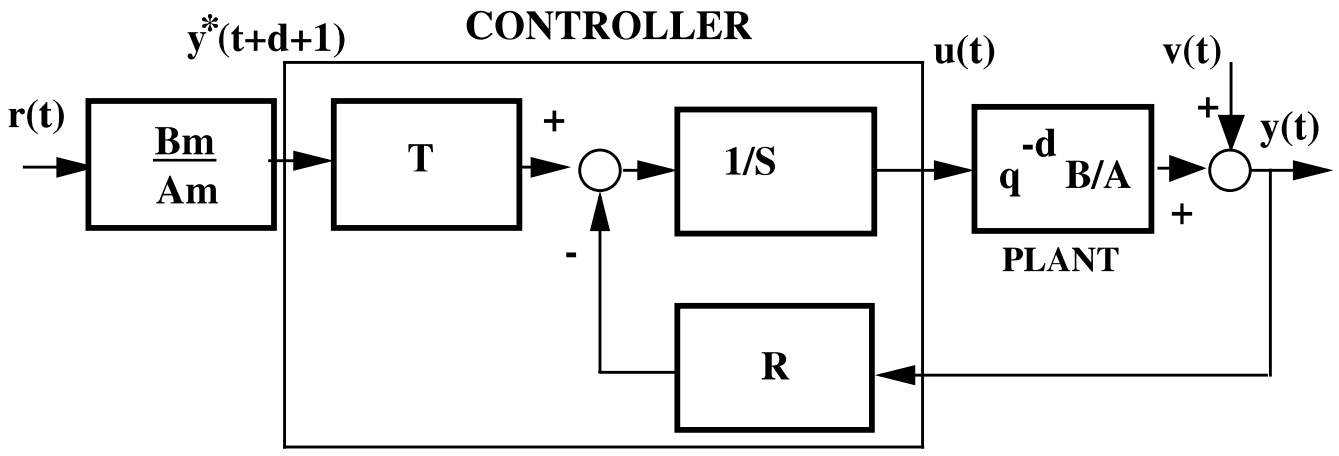

**Fig. 7.3** Control loop with RST digital controller

(with $B_m(q^{-1}) = b_{m0} + b_{m1}q^{-1} + \cdots$) and $A_m(q^{-1})$ a monic polynomial. The corresponding block diagram is shown in Fig. 7.3.

The controller of (7.2) is termed the RST controller and its *two-degree of freedom* capabilities come from the fact that the *regulation* objectives are assured by the $R$-$S$ part of the controller and the *tracking* objectives are assured by an appropriate design of the $T$ polynomial.

On the basis of (7.1) and (7.2) one can define several transfer operators defining the relationship between on the one hand the reference trajectory and the disturbance and on the other hand the output and the input of the plant. We first observe that the closed-loop poles are defined by the equation:

$$P(q^{-1}) = A(q^{-1})S(q^{-1}) + q^{-d-1}B^{\star}(q^{-1})R(q^{-1}) \tag{7.4}$$

and eliminating $u(t)$ between (7.1) and (7.2) one gets:

$$y(t) = H_{CL}(q^{-1})y^{\star}(t+d+1) + S_{yp}(q^{-1})v(t) \tag{7.5}$$

where:

$$H_{CL}(q^{-1}) = \frac{q^{-d-1}B^{\star}(q^{-1})T(q^{-1})}{P(q^{-1})} \tag{7.6}$$

and

$$S_{yp}(q^{-1}) = \frac{A(q^{-1})S(q^{-1})}{P(q^{-1})} \tag{7.7}$$

where $S_{yp}(q^{-1})$ is termed the *output sensitivity function*. Similarly one gets an expression for $u(t)$:

$$u(t) = H_U(q^{-1})y^{\star}(t+d+1) + S_{up}(q^{-1})v(t) \tag{7.8}$$

where

$$H_U(q^{-1}) = \frac{A(q^{-1})T(q^{-1})}{P(q^{-1})} \tag{7.9}$$

and

$$S_{up}(q^{-1}) = -\frac{A(q^{-1})R(q^{-1})}{P(q^{-1})} \tag{7.10}$$

where $S_{up}(q^{-1})$ is termed the *input sensitivity function*.

Reasoning in the frequency domain clearly indicates that the poles of the closed loop will have a crucial role in achieving the *regulation* specifications which in the time domain translates as the rejection of an initial output error in the absence of disturbances. This is seen from (7.5) for $v(t) \equiv 0$. The corresponding time-domain equation for $v(t) \equiv 0$ is:

$$P(q^{-1})y(t) - B^\star(q^{-1})T(q^{-1})y^\star(t) = 0 \tag{7.11}$$

and for $y^\star(t) \equiv 0$ (or constant), the equation governing the evolution of $y(t)$ is

$$P(q^{-1})y(t) = 0; \quad y(0) = y_0 \tag{7.12}$$

Looking to the transfer operator between the disturbance and the output and assuming that the disturbance $v(t)$ is generated by a model of the form

$$D(q^{-1})v(t) = e(t) \tag{7.13}$$

where $e(t)$ is a Dirac pulse in the deterministic case, then in order to make this disturbance unobservable at the output (perfect disturbance rejection) it results from (7.5) that the polynomial $S(q^{-1})$ should incorporate the disturbance model $D(q^{-1})$ (this is the *internal model principle*—Francis and Wonham 1976). The corresponding time-domain equation for the input in the absence of disturbances is:

$$P(q^{-1})u(t) - A(q^{-1})T(q^{-1})y^\star(t+d+1) = 0 \tag{7.14}$$

This is nothing other than an equivalent rewriting of (7.2) (it can be obtained in a straightforward way by passing this expression through the operator $A(q^{-1})$ and taking into account the expression of $P$ as well as (7.1)).

Looking to the transfer operator from disturbance to the plant input one sees that if for certain types of disturbances we would like to operate like in open loop, which translates in making unobservable the disturbances at the input, the polynomial $R(q^{-1})$ should include a model of the disturbance. Such situations occur in a number of industrial processes where periodic disturbances are voluntarily applied for technological reasons and the controller should not counteract this effect (e.g., continuous steel casting).

The perfect rejection of disturbances with known characteristics or conversely opening of the loop for certain disturbances lead to the conclusion that the general structure of $R$ and $S$ will be of the form:

$$S(q^{-1}) = S'(q^{-1})H_S(q^{-1}) \tag{7.15}$$

$$R(q^{-1}) = R'(q^{-1})H_R(q^{-1}) \tag{7.16}$$

where $H_S(q^{-1})$ and $H_R(q^{-1})$ are monic fixed polynomials which are introduced in the controller for achieving certain performances with respect to disturbances. Using this parameterization the closed-loop poles will be given by:

$$P(q^{-1}) = A(q^{-1})H_S(q^{-1})S'(q^{-1}) + q^{-d-1}B^\star(q^{-1})H_R(q^{-1})R'(q^{-1}) \tag{7.17}$$

Note that $H_S(q^{-1})$ and $H_R(q^{-1})$ can be interpreted as an "augmentation" of the plant model (for computation purposes).

The design of the RST controller can be done in the frequency domain using transfer functions (operators) or in the time domain. To use a design in the frequency domain (known also as the polynomial approach) the time-domain specifications should be translated in desired closed-loop poles and desired closed transfer functions. The frequency domain specifications have also to be taken into account.

The time-domain design is essentially based on forcing some performance indicators to zero (in finite time or asymptotically) or minimizing a quadratic criterion in terms of some performance indicators. However, these performance indicators in most practical cases will be weighted (or filtered) by filters which allow to take into account frequency domain specifications (robustness, disturbance rejection etc.).

For a number of control strategies the design can be done either in the frequency domain or in the time domain yielding the same results. For other strategies the design is done in the time domain but the results can be interpreted in the frequency domain. However, no matter what approach is used, the analysis of the resulting design in the frequency domain is absolutely necessary for assessing its robustness.

In the field of adaptive control preference has been given to the design of the underlying linear controller in the time domain, which allows a direct extension to the adaptive case. However, doing so, the robustness issues have been hidden.

In presenting the various control strategies we will consider both time domain and frequency domain. This will allow to enhance the rapprochements between the various control strategies and to put the adaptive versus robustness issues into real perspective (see Chap. 8).

## 7.3 Pole Placement

The pole placement strategy is applicable to plant models of the form of (7.1). We will make the following hypothesis upon the plant model of (7.1):

H1: No restrictions upon the orders of the polynomials $A(q^{-1})$, $B(q^{-1})$ and the value of the delay $d$.
H2: The orders $n_A, n_B$, the delay $d$ and the coefficients of $A(q^{-1})$ and $B(q^{-1})$ are known.
H3: The zeros of $B(q^{-1})$ can be inside or outside the unit circle.
H4: $A(q^{-1})$ and $B(q^{-1})$ do not have any common factors.
H5: The zeros of $A(q^{-1})$ can be inside or outside the unit circle.

The control law is of the form (7.2) and the polynomials $R(q^{-1})$ and $S(q^{-1})$ have the structure of (7.15) and (7.16).

### *7.3.1 Regulation*

The closed-loop behavior is defined by:

- the desired closed-loop poles;
- the choice of the fixed parts $H_R(q^{-1})$ and $H_S(q^{-1})$.

The desired closed-loop poles are chosen under the form:

$$P(q^{-1}) = P_D(q^{-1})P_F(q^{-1}) \tag{7.18}$$

where $P_D(q^{-1})$ defines the *dominant poles* and $P_F(q^{-1})$ defines the *auxiliary poles*. Often $P_D(q^{-1})$ is chosen as a second order polynomial resulting from the discretization of a continuous-time second-order with a given natural frequency $\omega_0$ and damping factor $\zeta$ (or a cascade of second order systems). The role of $P_F(q^{-1})$ is on the one hand to introduce a filtering effect at certain frequencies and on the other hand to improve the robustness of the controller (as it will be discussed in Chap. 8).

$H_R(q^{-1})$ and $H_S(q^{-1})$ are primarily chosen in relation with the desired nominal performances. $H_S(q^{-1})$ should incorporate the internal model of the disturbance to be rejected (for example $H_S(q^{-1}) = 1 - q^{-1}$, in order to have an integrator which will assure zero steady state error for a step disturbance). $H_R(q^{-1})$ will incorporate either the model of the disturbance which we would like not to be affected by the controller or a filter for reducing the actuator stress in a certain frequency range (by reducing the gain of the controller). As it will be shown in Chap. 8, robustness issues will lead also to the introduction of $H_R(q^{-1})$ and $H_S(q^{-1})$. Note that the introduction of $H_R$ and $H_S$ can be interpreted as the internal model associated with a deterministic disturbance given by the model

$$H_S(q^{-1})v(t) = H_R(q^{-1})\delta(t)$$

where $\delta(t)$ is the Dirac impulse. Therefore the controller polynomials $S(q^{-1})$ and $R(q^{-1})$ are given by:

$$S(q^{-1}) = S'(q^{-1})H_S(q^{-1}) \tag{7.19}$$

$$R(q^{-1}) = R'(q^{-1})H_R(q^{-1}) \tag{7.20}$$

where $S'$ and $R'$ are solutions of:

$$\begin{aligned} &A(q^{-1})H_S(q^{-1})S'(q^{-1}) + q^{-d-1}B^\star(q^{-1})H_R(q^{-1})R'(q^{-1}) \\ &\quad = P_D(q^{-1})P_F(q^{-1}) \end{aligned} \tag{7.21}$$

Defining

$$n_A = \deg A; \qquad n_B = \deg B$$

$$n_{HS} = \deg H_S; \qquad n_{HR} = \deg H_R$$

One has the following result:

**Theorem 7.1** *Under the hypotheses* H1 *through* H5 *and* H4′: $AH_S$ *and* $BH_R$ *do not have any common factors*, (7.21) *has a unique solution for* $S'$ *and* $R'$, *if*:

$$n_P = \deg P(q^{-1}) \leq n_A + n_{HS} + n_B + n_{HR} + d - 1 \tag{7.22}$$

$$n_{S'} = \deg S'(q^{-1}) = n_B + n_{HR} + d - 1 \tag{7.23}$$

$$n_{R'} = \deg R'(q^{-1}) = n_A + n_{HS} - 1 \tag{7.24}$$

*with*

$$S'(q^{-1}) = 1 + s'_1 q^{-1} + \cdots + s'_{n_S} q^{-n_S} \tag{7.25}$$

$$R'(q^{-1}) = r'_0 + r'_1 q^{-1} + \cdots + r'_{n_R} q^{-n_R} \tag{7.26}$$

*Proof* In order to effectively solve (7.21) one uses the fact that the polynomial equation (7.21) takes the matrix form

$$Mx = p \tag{7.27}$$

where:

$$x^T = \left[1, s'_1, \ldots, s'_{n_S}, r'_0, \ldots, r'_{n_R}\right] \tag{7.28}$$

$$p^T = \left[1, p_1, \ldots, p_{n_P}, 0, \ldots, 0\right] \tag{7.29}$$

and the matrix $M$ (known as Sylvester matrix or controllability matrix) has the form:

$$M = \underbrace{\left[\begin{array}{cccccccc}
\multicolumn{4}{c}{\overbrace{\hphantom{a'_1 \; 1 \; \ddots \; 0 \; 0}}^{n_{B'}}} & \multicolumn{4}{c}{\overbrace{\hphantom{b'_1 \; 0 \; \ddots \; 0 \; 0}}^{n_{A'}}} \\
1 & 0 & \cdots & 0 & 0 & \cdots & \cdots & 0 \\
a'_1 & 1 & \ddots & \vdots & b'_1 & 0 & \ddots & \vdots \\
\vdots & & \ddots & 0 & \vdots & & \ddots & \\
 & & & 1 & & & & 0 \\
\vdots & & & a'_1 & \vdots & & & b'_1 \\
a'_{n_{A'}} & & & & b'_{n_{B'}} & & & \\
0 & \ddots & & \vdots & 0 & \ddots & & \vdots \\
\vdots & \ddots & & & \vdots & \ddots & & \\
0 & \cdots & 0 & a'_{n_{A'}} & 0 & \cdots & 0 & b'_{n_{B'}}
\end{array}\right]}_{n_{A'}+n_{B'}} \left.\vphantom{\begin{array}{c}1\\1\\1\\1\\1\\1\\1\\1\\1\\1\end{array}}\right\} n_{A'} + n_{B'} \tag{7.30}$$

where $a'_i$ and $b'_i$ are the coefficients of

$$A'(q^{-1}) = A(q^{-1})H_S(q^{-1})$$
$$B'(q^{-1}) = q^{-d}B(q^{-1})H_R(q^{-1})$$

Note that $b'_i = 0$ for $i = 0, 1, \ldots, d$. The vector $x$ containing the coefficients of polynomials $R'(q^{-1})$ and $S'(q^{-1})$ is obtained by inverting the matrix $M$ which is non singular if $A'(q^{-1})$ and $B'(q^{-1})$ do not have any common factors (hypothesis H4 and H4′—see also Sect. 2.1)

$$x = M^{-1}p \tag{7.31}$$

If there are common factors between $A$ and $B$ the determinant of this matrix is null and (7.21) is not solvable (see Sect. 2.1). Various methods for solving this equation are available. □

### Choice of $H_R$ and $H_S$—Examples

1. Zero steady state error for a step disturbance.

   From (7.5) in the absence of a reference one has:

$$y(t) = \frac{A(q^{-1})H_S(q^{-1})S'(q^{-1})}{P(q^{-1})}v(t)$$

$$D(q^{-1})v(t) = \delta(t); \qquad D(q^{-1}) = 1 - q^{-1}$$

   The problem can be viewed as either imposing the cancellation of the disturbance model (in this case $(1 - q^{-1})$) or as choosing $H_S(q^{-1})$ such that the gain of the transfer function between $v(t)$ and $y(t)$ be zero for the zero frequency (i.e., for $q = z = 1$). Both points of view lead to:

$$H_S(q^{-1}) = (1 - q^{-1})$$

2. Perfect rejection of a harmonic disturbance.

   In this case the disturbance model is:

$$(1 + \alpha q^{-1} + q^{-2})v(t) = \delta(t)$$

   with

$$\alpha = -2\cos(\omega T_s) = -2\cos\left(2\pi \frac{f}{f_s}\right)$$

   and applying the same reasonment one finds:

$$H_S(q^{-1}) = (1 + \alpha q^{-1} + q^{-2})$$

   If we would like only a desired attenuation at the frequency $f$, one can use a pair of damped zeros in $H_S$ with a damping factor in relation with the desired attenuation.

3. Opening the loop.

   In a number of applications, the measured signal may contain specific frequencies which should not be attenuated by the regulation (they correspond in general to signals inherent to the technology of the process). In such cases the system should be in open loop at these frequencies. i.e., this disturbance is made unobservable at the input of the plant. From (7.8) in the absence of the reference, the input to the plant is given by:

$$u(t) = S_{up}(q^{-1})v(t) = \frac{A(q^{-1})H_R(q^{-1})R'(q^{-1})}{P(q^{-1})}v(t)$$

   and therefore in order to make the input sensitivity function zero at a given frequency $f$ one should introduce a pair of undamped zeros in $H_R(q^{-1})$ i.e.,

$$H_R(q^{-1}) = (1 + \beta q^{-1} + q^{-2})$$

where

$$\beta = -2\cos(\omega T_s) = -2\cos\left(2\pi \frac{f}{f_s}\right)$$

In many cases it is desired that the controller does not react to signals of frequencies close to $0.5 f_s$ (where the gain of the system is in general very low). In such cases one uses:

$$H_R(q^{-1}) = (1 + \beta q^{-1})$$

where

$$0 < \beta \leq 1$$

Note that $(1 + \beta q^{-1})^2$ corresponds to a second order with a damped resonance frequency equal to $\omega_s/2$:

$$\omega_0 \sqrt{1 - \zeta^2} = \frac{\omega_s}{2}$$

and the corresponding damping is related to $\beta$ by

$$\beta = e^{-\frac{\zeta}{\sqrt{1-\zeta^2}}\pi}$$

For $\beta = 1$, the system will operate in open loop at $f_s/2$.

## 7.3.2 Tracking

In the ideal case we would like to perfectly follow a desired trajectory $y^\star(t + d + 1)$, known $d + 1$ steps ahead which is stored in the computer or generated from the reference signal via a tracking reference model, i.e.,

$$H_m(q^{-1}) = \frac{B_m(q^{-1})}{A_m(q^{-1})} \tag{7.32}$$

with

$$B_m(q^{-1}) = b_{m0} + b_{m1} q^{-1} + \cdots + b_{mn_{Bm}} q^{-n_{Bm}} \tag{7.33}$$

$$A_m(q^{-1}) = 1 + a_{m1} q^{-1} + \cdots + a_{mn_{Am}} q^{-n_{Am}} \tag{7.34}$$

Often, this tracking reference model is determined from the desired tracking performances (rise time, settling time, overshoot) by selecting either an appropriate normalized second order system (defined by $\omega_0, \zeta$) or sometimes a cascade of two second orders. Once a continuous-time reference model is selected, one gets by discretization the discrete-time tracking reference model. Note, however, that the discrete-time tracking reference model can be directly defined in discrete time (see Landau 1990b).

The remaining design element of the controller is the polynomial $T(q^{-1})$. The transfer function from the desired reference trajectory to the output is:

$$H_{CL}(q^{-1}) = \frac{q^{-d-1}T(q^{-1})B^{\star}(q^{-1})}{P_D(q^{-1})P_F(q^{-1})} \tag{7.35}$$

Several situations may occur.

(a) The desired tracking dynamic $A_m(q^{-1})$ does not contain any of the poles of the closed loop.
(b) The desired tracking dynamic $A_m(q^{-1})$ contains some of the poles of $P(q^{-1})$ denoted by $P_0(q^{-1})$.
(c) The tracking and regulation dynamics are the same.

In the case (a) $H_{CL}(q^{-1})$ should have a steady stage gain of 1 and $T(q^{-1)}$ should compensate the closed-loop poles i.e.,

$$T(q^{-1}) = \beta P(q^{-1}) \tag{7.36}$$

where

$$\beta = \frac{1}{B^{\star}(1)} \tag{7.37}$$

The resulting transfer function from the reference to the output is:

$$H(q^{-1}) = \frac{q^{-d-1}B_m(q^{-1})B^{*}(q^{-1})}{A_m(q^{-1})B^{*}(1)} \tag{7.38}$$

In the case (b) assuming that

$$A_m(q^{-1}) = A'_m(q^{-1})P_0(q^{-1}) \tag{7.39}$$

and

$$P(q^{-1}) = P'(q^{-1})P_0(q^{-1}) \tag{7.40}$$

the tracking reference model will be

$$H_m(q^{-1}) = \frac{B'_m(q^{-1})}{A'_m(q^{-1})} \tag{7.41}$$

$$T(q^{-1}) = \beta P'(q^{-1}) \tag{7.42}$$

The resulting transfer function from the reference to the output is:

$$H(q^{-1}) = \frac{q^{-d-1}B_m(q^{-1})B^{*}(q^{-1})}{A'_m(q^{-1})P_0(q^{-1})B^{*}(1)} \tag{7.43}$$

In the case (c)

$$H_m(q^{-1}) = 1 \tag{7.44}$$

$$T(q^{-1}) = \beta P(1) \tag{7.45}$$

If $S$ contains an integrator then $P(1) = B(1)R(1)$ and therefore:

$$T(q^{-1}) = R(1) \tag{7.46}$$

**Fig. 7.4** Pole placement scheme for tracking and regulation

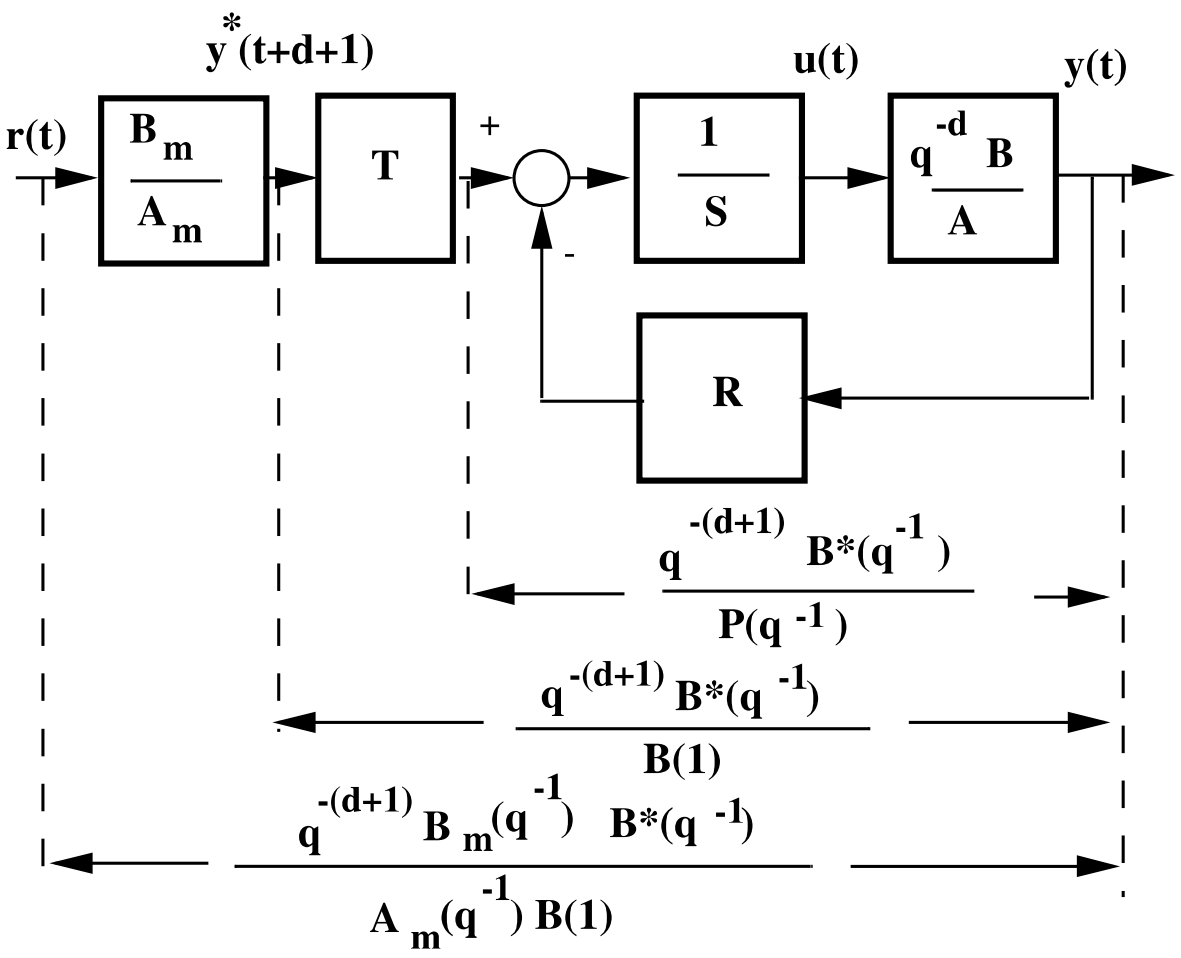

The resulting transfer function from the reference to the output is in this case:

$$H(q^{-1}) = \frac{q^{-d-1}B^{\star}(q^{-1})P(1)}{P(q^{-1})B^{\star}(1)} \tag{7.47}$$

Note that better tracking performances can be obtained if the stable zeros of $B^{\star}(q^{-1})$ are compensated and the unstable ones are partially compensated by using the stable reciprocal inverse (Tsypkin 1993). However, this requires a very good knowledge of the coefficient of $B^{\star}(q^{-1})$.

The complete block diagram of the pole placement for the case (a) (different dynamics for tracking and regulation) is shown in Fig. 7.4.

## *7.3.3 Some Properties of the Pole Placement*

### Time-Domain Design

The same results are obtained if the design starts in the time domain for the following objective: the control $u(t)$ given by (7.2) should be designed such that:

$$\begin{aligned} P(q^{-1})e_y(t+d+1) &= P(q^{-1})[y(t+d+1) - \beta B^{*}(q^{-1})y^{\star}(t+d+1)] \\ &= 0; \quad t \geq 0 \end{aligned} \tag{7.48}$$

for any initial error $e_y(0)$ where $\beta$ is given by (7.37). The objective of (7.48) can be interpreted as a regulation objective for $y^{\star}(t+d+1) \equiv 0$ or as a tracking objective for $y^{\star}(t+d+1) \neq 0$. It simply says that any initial difference between the output and the filtered desired trajectory through the process zeros, vanishes with the dynamics defined by $P(q^{-1})$. From (7.11) one sees immediately that taking $T(q^{-1})$ given by (7.36) and solving (7.4) produces the desired RST control law corresponding to the pole placement.

### Alternative Expression of $S$ and $R$ in the Case of Auxiliary Poles

Assume that $S_D$ and $R_D$ are solutions of the polynomial equation:

$$P_D(q^{-1}) = A(q^{-1})S_D(q^{-1}) + q^{-d-1}B^\star(q^{-1})R_D(q^{-1}) \tag{7.49}$$

Then the solutions for the case

$$P(q^{-1}) = P_D(q^{-1})P_F(q^{-1}) \tag{7.50}$$

can be expressed as:

$$S(q^{-1}) = P_F(q^{-1})S_D(q^{-1}); \qquad R(q^{-1}) = P_F(q^{-1})R_D(q^{-1}) \tag{7.51}$$

(multiply both sides of (7.49) by $P_F(q^{-1})$).

### Predictor Interpretation of the Pole Placement

Consider the plant model of (7.1) with a delay $d$ and $P(q^{-1}) = P_D(q^{-1})P_F(q^{-1})$. The controller will be denoted by:

$$S_d(q^{-1})u(t) = -R_d(q^{-1})y(t) + T_d(q^{-1})y^\star(t+d+1) \tag{7.52}$$

where $S_d$ and $R_d$ are solutions of:

$$P(q^{-1}) = P_D(q^{-1})P_F(q^{-1}) = A(q^{-1})S_d(q^{-1}) + q^{-d-1}B^\star(q^{-1})R_d(q^{-1}) \tag{7.53}$$

and

$$T_d(q^{-1}) = \frac{1}{B^\star(1)}P_D(q^{-1})P_F(q^{-1}) \tag{7.54}$$

**Theorem 7.2** *Controller* (7.52) *through* (7.54) *for the plant* (7.1) *is equivalent to the following one*:

$$\begin{aligned} P_F(q^{-1})S_0(q^{-1})u(t) = {} & -R_0(q^{-1})P_F(q^{-1})\hat{y}(t+d|t) \\ & + P_F(q^{-1})T_0(q^{-1})y^\star(t+d+1) \end{aligned} \tag{7.55}$$

$$P_F(q^{-1})\hat{y}(t+d|t) = F(q^{-1})y(t) + B(q^{-1})E(q^{-1})u(t) \tag{7.56}$$

*where $S_0$ and $R_0$ are solutions of*:

$$A(q^{-1})S_0(q^{-1}) + B(q^{-1})R_0(q^{-1}) = P_D(q^{-1}) \tag{7.57}$$

*$F(q^{-1})$ and $E(q^{-1})$ are solutions of*:

$$A(q^{-1})E(q^{-1}) + q^{-d}F(q^{-1}) = P_F(q^{-1}) \tag{7.58}$$

*and*

$$T_0(q^{-1}) = \frac{1}{B^\star(1)}P_D(q^{-1}) \tag{7.59}$$

Figure 7.5 gives the pole placement scheme associated with the controller of (7.55) and (7.56).

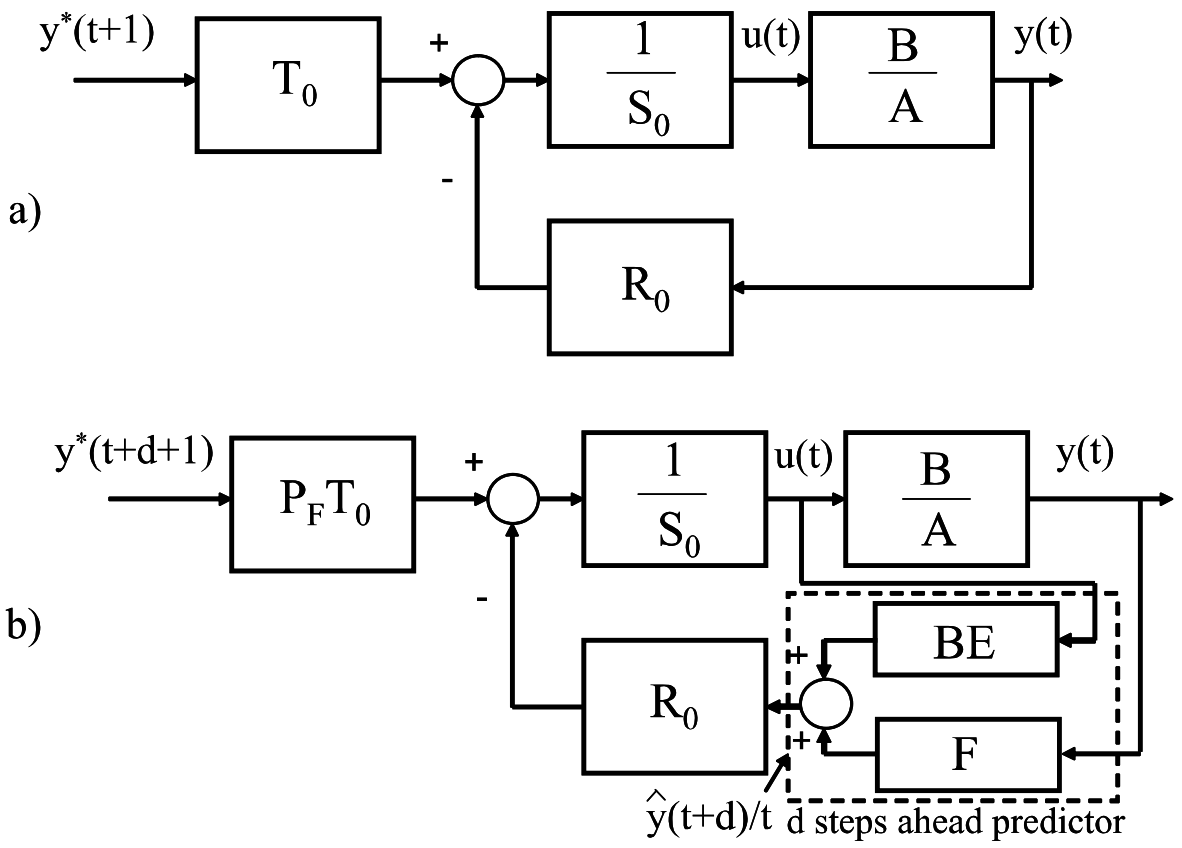

**Fig. 7.5** Equivalent pole placement scheme (predictor form), (**a**) the case $d = 0$, $P_F(q^{-1}) = 1$, (**b**) the general case

*Remarks*

- The controller $R_0$, $S_0$, $T_0$ correspond to the pole placement controller for the plant model (7.1) with $d = 0$ and with the desired closed-loop poles defined by $P_D(q^{-1})$.
- The controller of (7.55) and (7.56) shows for $P_F(q^{-1}) = 1$ that in order to control a system with delay using pole placement, one can use the controller for the case without delay by replacing the measured output $y(t)$ by its d-step ahead prediction $\hat{y}(t+d|t)$.
- The introduction of the auxiliary poles corresponds to replacing the measured output or its prediction by a filtered prediction.
- If (7.55) is replaced by:

$$S_0(q^{-1})u(t) = -R_0(q^{-1})\hat{y}(t+d|t) + T_0(q^{-1})y^\star(t+d+1) \tag{7.60}$$

  The auxiliary poles $P_F(q^{-1})$ can be interpreted as the poles of an observer since they are canceled in the input-output transfer function.
- Therefore implicitly an RST controller contains a predictor. Some of the closed-loop poles (or all) can be interpreted as the predictor poles. Conversely all the non-zero poles can be considered as design poles and the predictor is a dead-beat predictor (i.e., all the poles are at the origin).

*Proof* From (7.1) and (7.55) one has:

$$\begin{aligned} AP_F S_0 y(t+d+1) &= B^\star P_F S_0 u(t) \\ &= -B^\star R_0 P_F \hat{y}(t+d|t) + B^\star P_F T_0 y^\star(t+d+1) \end{aligned} \tag{7.61}$$

and taking also into account (7.56) one has:

$$\begin{aligned} &[AP_F S_0 + q^{-d-1}B^\star R_0 F + q^1 B^\star R_0 AE]y(t+d+1) \\ &\quad = B^\star P_F T_0 y^\star(t+d+1) \end{aligned} \tag{7.62}$$

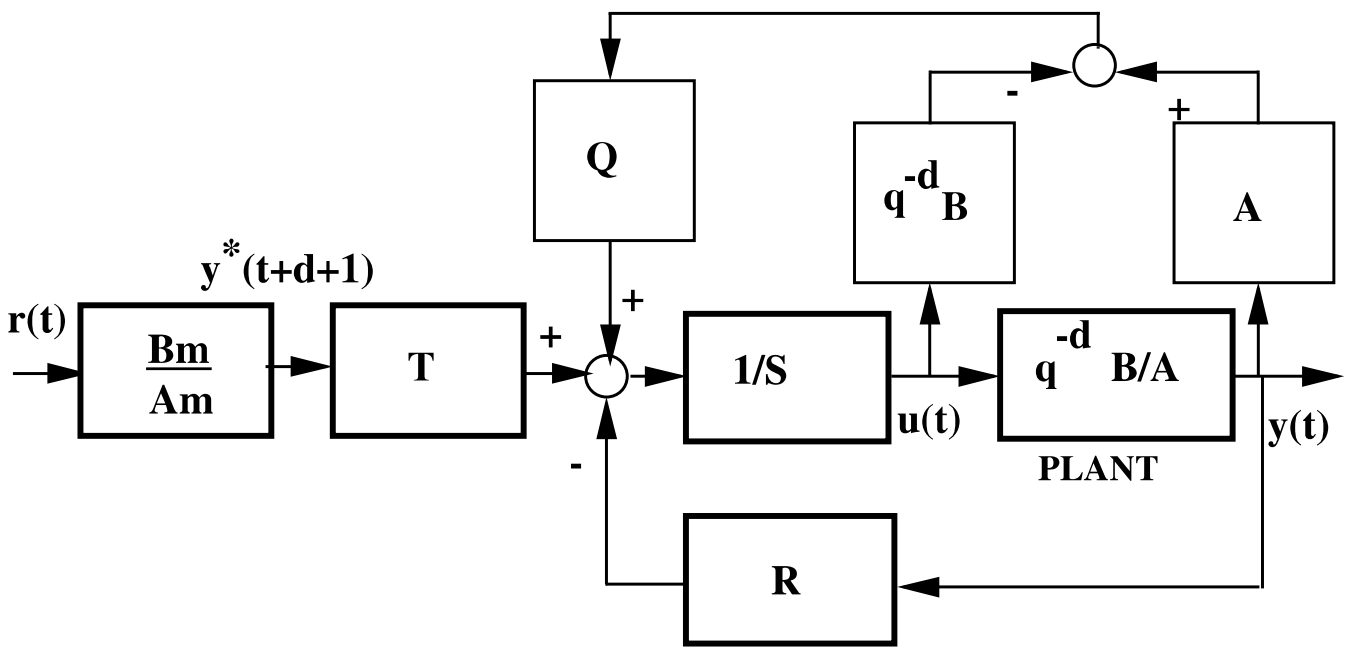

**Fig. 7.6** Pole placement scheme using Youla-Kucera parameterization

But:

$$AP_F S_0 + q^{-d-1}B^\star R_0 F + q^{-1}B^\star R_0 AE = AS_0 P_F + q^{-1}B^\star R_0[AE + q^{-d}F]$$
$$= P_D P_F \tag{7.63}$$

and one sees, taking into account the expression of $T_0$, that both closed-loop poles and the transfer operator from the reference to the output are the same. □

## Youla-Kucera Parameterization

The controllers computed till now will be termed "central" since using the Youla-Kucera parameterization:

$$S_Y = S - QBq^{-d} \tag{7.64}$$

$$R_Y = R + QA \tag{7.65}$$

where $Q$ is a stable polynomial, other controllers which are solution of (7.4) for a given $P(q^{-1})$ can be obtained. However, they are not necessarily of minimal order. While the desired closed-loop poles remain the same, the sensitivity functions may be different and this is exploited in some robustness design techniques (see Chap. 8). The structure of pole placement in this case is illustrated in Fig. 7.6.

## Another Time-Domain Interpretation

(M'Saad et al. 1986; Irving et al. 1986) We have seen that in the case of pole placement one has:

$$P(q^{-1})e_y(t+d+1) = 0; \quad t \geq 0 \tag{7.66}$$

where

$$e_y(t+d+1) = y(t+d+1) - \beta B^\star(q^{-1})y^\star(t+d+1) \tag{7.67}$$

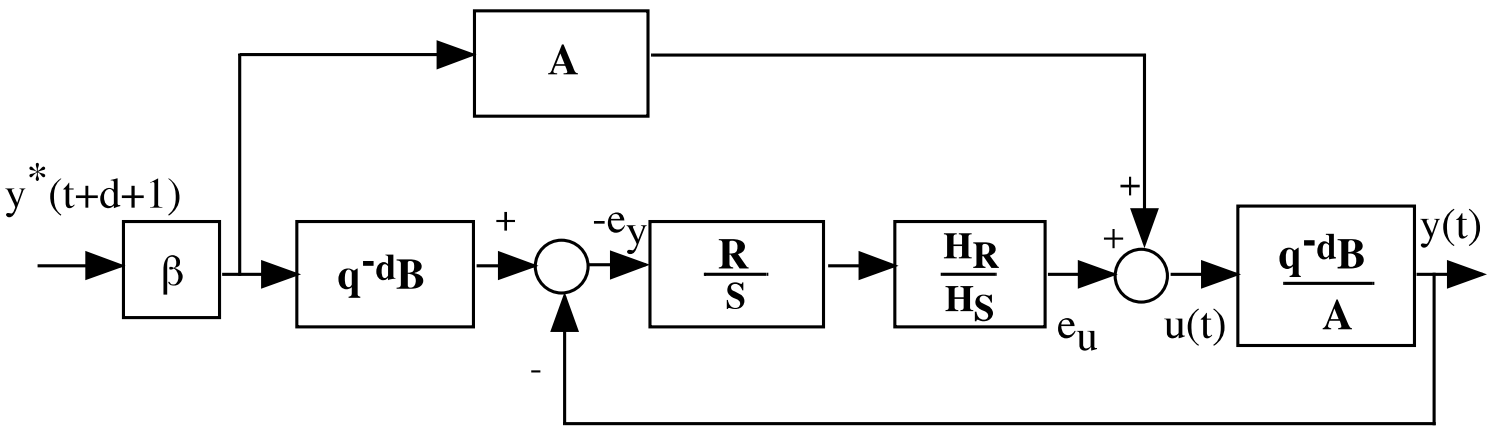

**Fig. 7.7** Equivalent pole placement scheme (partial state model reference control)

with

$$T(q^{-1}) = \beta P(q^{-1}) \tag{7.68}$$

Introducing this expression of $T(q^{-1})$ in (7.14) one gets

$$P(q^{-1})[u(t) - \beta A(q^{-1})y^\star(t+d+1)] = P(q^{-1})e_u(t) = 0 \tag{7.69}$$

where

$$e_u(t) = u(t) - \beta A(q^{-1})y^\star(t+d+1); \quad t > 0 \tag{7.70}$$

Therefore pole placement can be interpreted as forcing two performance indicators $e_y(t)$, $e_u(t)$ to zero:

$$P(q^{-1})e_y(t) = P(q^{-1})e_u(t) = 0; \quad t > 0 \tag{7.71}$$

On the other hand taking into account the expression of $T$ one gets:

$$\begin{aligned} S(q^{-1})u(t) = {} & (A(q^{-1})S(q^{-1}) + q^{-d-1}B^\star(q^{-1})R(q^{-1}))\beta y^\star(t+d+1) \\ & - R(q^{-1})y(t) \end{aligned} \tag{7.72}$$

or in other terms:

$$\begin{aligned} S(q^{-1})u(t) = {} & -R(q^{-1})[y(t) - B^\star(q^{-1})\beta y^\star(t)] \\ & + S(q^{-1})A(q^{-1})\beta y^\star(t+d+1) \end{aligned} \tag{7.73}$$

yielding:

$$u(t) = -\frac{R(q^{-1})}{S(q^{-1})}e_y(t) + A(q^{-1})\beta y^\star(t+d+1) \tag{7.74}$$

or

$$S(q^{-1})e_u(t) + R(q^{-1})e_y(t) = 0 \tag{7.75}$$

which leads to the equivalent pole placement scheme shown in Fig. 7.7 where instead of the polynomial $T$ acting on the desired trajectory, one applies it in two different points through two polynomials which are those of the plant model. This means that whatever values are given to $R$ and $S$ the poles of the closed loop will be compensated from the reference. This equivalent scheme is called *partial state model reference* (PSMR) control (M'Saad et al. 1986).

### Regressor Formulation

Taking into account the form of the polynomial $S$, $u(t)$ can be expressed as:

$$\begin{aligned} u(t) &= -S^\star(q^{-1})u(t-1) - R(q^{-1})y(t) + T(q^{-1})y^\star(t+d+1) \\ &= -\theta_C^T\phi_C^T(t) \end{aligned} \tag{7.76}$$

where:

$$\theta_C^T = [s_1, \dots, s_{n_S}, r_0, \dots, r_{n_R}, t_0, \dots, t_{n_\beta}] \tag{7.77}$$

$$\begin{aligned} \phi_C^T(t) = [&u(t-1), \dots, u(t-n_S), y(t), \dots, y(t-n_R), \\ &- y^\star(t+d+1), \dots, -y^\star(t+d+1-n_P)] \end{aligned} \tag{7.78}$$

## 7.3.4 *Some Particular Pole Choices*

### Internal Model Control (IMC)

IMC corresponds to a pole placement characterized by the fact that the desired closed loop contains the poles of the plant assumed to be asymptotically stable i.e., the RST controller is the solution of:

$$\begin{aligned} A(q^{-1})S(q^{-1}) + q^{-d}B(q^{-1})R(q^{-1}) &= A(q^{-1})P_D(q^{-1})P_F(q^{-1}) \\ &= A(q^{-1})P_0(q^{-1}) \end{aligned} \tag{7.79}$$

However, this implies that:

$$R(q^{-1}) = A(q^{-1})R'(q^{-1}) \tag{7.80}$$

i.e., (7.79) is replaced by

$$S(q^{-1}) + q^{-d}B(q^{-1})R'(q^{-1}) = P_0(q^{-1}) \tag{7.81}$$

and this allows the set of all controllers to be characterized as:

$$S(q^{-1}) = P_0(q^{-1}) - q^{-d}B(q^{-1})Q(q^{-1}) \tag{7.82}$$

$$R'(q^{-1}) = Q(q^{-1}) \tag{7.83}$$

where $Q$ is a stable polynomial in $q^{-1}$. The value of $Q$ is determined if a constraint on $S(q^{-1})$ is imposed. For example if $S(q^{-1})$ contains an integrator $S(1) = 0$ and therefore

$$P_0(1) - B(1)Q(1) = S(1) = 0 \tag{7.84}$$

One gets then:

$$Q = Q(1) = \frac{P_0(1)}{B(1)} \tag{7.85}$$

**Fig. 7.8** Equivalent pole placement scheme for $P(q^{-1}) = A(q^{-1})P_0(q^{-1})$ (Internal Model Control)

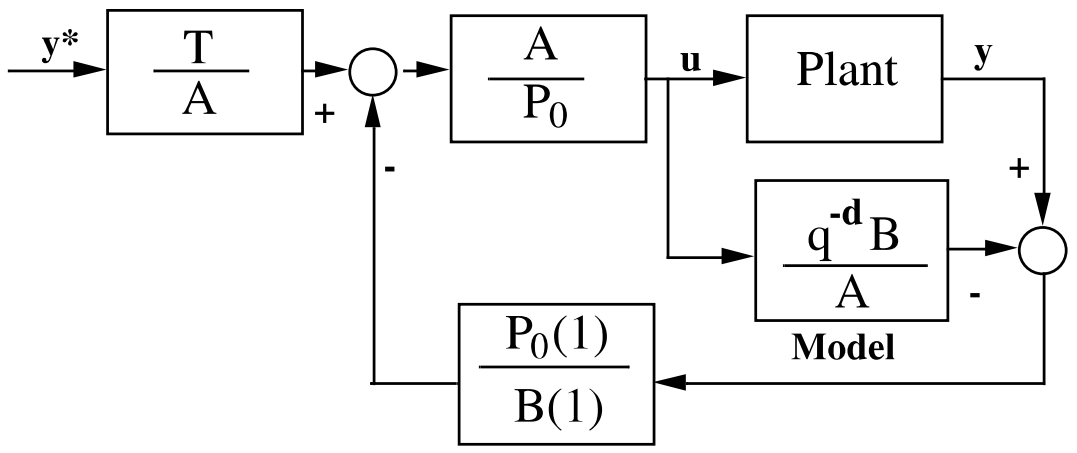

as the simplest solution yielding:

$$S(q^{-1}) = P_0(q^{-1}) - q^{-d}\frac{B(q^{-1})P_0(1)}{B(1)} \tag{7.86}$$

$$R(q^{-1}) = A(q^{-1})\frac{P_0(1)}{B(1)} \tag{7.87}$$

Note also that including $A(q^{-1})$ in the closed-loop poles the Diophantine equation is transformed in a polynomial division which has always a solution (see Sect. 7.4). Observes also that the controller equation takes the form:

$$S(q^{-1})u(t) = -R(q^{-1})y(t) + T(q^{-1})y^\star(t+d+1)$$

from which one obtains using (7.86) and (7.87):

$$P_0(q^{-1})u(t) = \frac{P_0(1)}{B(1)}[q^{-d}B(q^{-1})u(t) - A(q^{-1})y(t)] + T(q^{-1})y^\star(t+d+1)$$

and, respectively:

$$\frac{P_0(q^{-1})}{A(q^{-1})}u(t) = -\frac{P_0(1)}{B(1)}\left[y(t) - \frac{q^{-d}B(q^{-1})}{A(q^{-1})}u(t)\right] + \frac{T(q^{-1})}{A(q^{-1})}y^\star(t+d+1)$$

The term $\frac{q^{-d}B(q^{-1})}{A(q^{-1})}u(t)$ is interpreted as the output of the model of the plant and this leads to the typical IMC scheme (Morari and Zafiriou 1989) (see Fig. 7.8).

## Model Algorithmic Control

It corresponds to pole placement for the case where the plant model is represented by its truncated impulse response (Richalet et al. 1978), i.e.,

$$y(t+d+1) = B^\star(q^{-1})u(t)$$

Again in this case a simplification occurs and the Diophantine equation (7.4) becomes

$$S(q^{-1}) + q^{-d-1}B^\star(q^{-1})R(q^{-1}) = P(q^{-1})$$

which allows an easy computation of $S(q^{-1})$ and $R(q^{-1})$ (like for IMC).

## 7.4 Tracking and Regulation with Independent Objectives

This design method allows the desired tracking behavior independent of the desired regulation behavior to be obtained. Unlike the *pole placement*, this method results in the simplification of the zeros of the plant discrete-time model. This enables the desired tracking performances without approximations to be achieved (which is not the case for the pole placement).

This strategy can be considered as a particular case of the pole placement when hypothesis (H3) and (H4) for pole placement are replaced by:

(H3) $B^\star(q^{-1})$ has all its zeros inside the unit circle;
(H4) $B^\star(q^{-1})$ and $A(q^{-1})$ may have common factors;

and the closed-loop poles contain $B^\star(q^{-1})$.

This strategy can also be viewed as a generalization of the *model reference control* which was the underlying linear control design used in a large number of adaptive control schemes (Landau 1979, 1981; Goodwin and Sin 1984; Ionescu and Monopoli 1977). It is also related to the stochastic minimum variance control as it will be shown in Sect. 7.6. As for the pole placement, this control strategy can be applied for both stable and unstable systems

- without restriction on the degrees of the polynomials $A(q^{-1})$ and $B(q^{-1})$;
- without restriction on the integer delay $d$ of the system.

As a result of the simplification of the zeros (or equivalently as the result of including $B^\star(q^{-1})$ in the closed-loop poles), this strategy can be applied only to discrete-time models with stable zeros (from a practical point of view one should add that the zeros should be sufficiently damped). In particular this method is not applicable for the case of fractional time delay greater than $0.5T_s$ (this induces an unstable discrete-time zero) (Landau 1990b; Åström et al. 1984; Franklin et al. 1990). Therefore selection of the sampling interval to avoid this situation should be considered. The structure of the closed-loop system is represented in Fig. 7.9. The desired closed-loop poles are defined by a polynomial $P(q^{-1})$ which specifies the desired regulation behavior.

The output follows perfectly the desired tracking trajectory with a delay of $d+1$ samples. This is possible because $B^\star(q^{-1})$ is canceled in the closed-loop transfer function.

### *7.4.1 Polynomial Design*

#### Regulation (Computation of $R(q^{-1})$ and $S(q^{-1})$)

The transfer operator of the closed loop without precompensator is:

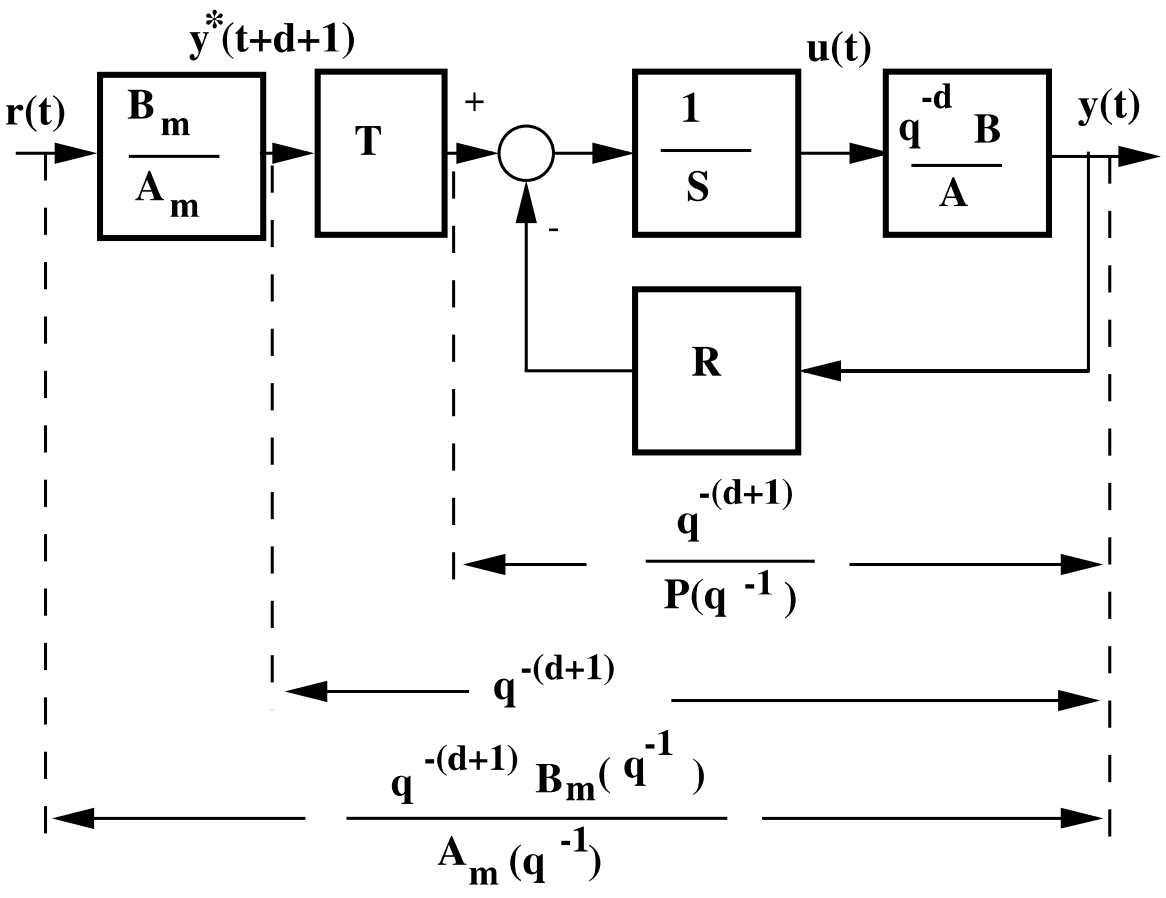

**Fig. 7.9** Tracking and regulation with independent objectives (generalized model reference control)

$$
\begin{aligned}
H'_{CL}(q^{-1}) &= \frac{q^{-d-1}B^{\star}(q^{-1})}{A(q^{-1})S(q^{-1}) + q^{-d-1}B^{\star}(q^{-1})R(q^{-1})} \\
&= \frac{q^{-d-1}}{P(q^{-1})} = \frac{q^{-d-1}B^{\star}(q^{-1})}{B^{\star}(q^{-1})P(q^{-1})}
\end{aligned} \tag{7.88}
$$

While $P(q^{-1})$ represents the desired closed-loop poles, the real closed-loop poles will include also $B^{\star}(q^{-1})$. Standard application of pole placement leads to:

$$A(q^{-1})S(q^{-1}) + q^{-d-1}B^{\star}(q^{-1})R(q^{-1}) = B^{\star}(q^{-1})P(q^{-1}) \tag{7.89}$$

The structure of this equation implies that $S(q^{-1})$ will be of the form:

$$S(q^{-1}) = s_0 + s_1 q^{-1} + \cdots + s_{n_S} q^{-ns} = B^{\star}(q^{-1})S'(q^{-1}) \tag{7.90}$$

In fact $S(q^{-1})$ will compensate the plant model zeros. Introducing the expression of $S(q^{-1})$ in (7.89) and after simplification by $B^{\star}(q^{-1})$, one obtains:

$$A(q^{-1})S'(q^{-1}) + q^{-d-1}R(q^{-1}) = P(q^{-1}) \tag{7.91}$$

**Theorem 7.3** (Landau and Lozano 1981) *The polynomial equation* (7.91) *has a unique solution for*:

$$n_P = \deg P \leq n_A + d \tag{7.92}$$

$$n_{S'} = \deg S' = d \tag{7.93}$$

$$n_R = \deg R(q^{-1}) = n_A - 1 \tag{7.94}$$

*and the polynomials* $R$ *and* $S'$ *have the form*:

$$R(q^{-1}) = r_0 + r_1 q^{-1} + \cdots + r_{n_A-1} q^{-n_A+1} \tag{7.95}$$

$$S'(q^{-1}) = 1 + s'_1 q^{-1} + \cdots + s'_d q^{-d} \tag{7.96}$$

*Proof* Equation (7.91) corresponds to the matrix equation

$$Mx = p \tag{7.97}$$

where $M$ is a lower triangular matrix of dimension $(n_A + d + 1) \times (n_A + d + 1)$ of the form:

$$M = \underbrace{\begin{bmatrix} \overbrace{\begin{matrix} 1 & 0 & \cdots & 0 \end{matrix}}^{d} & \overbrace{\begin{matrix} 0 & \cdots & \cdots & 0 \end{matrix}}^{n_A} \\ \begin{matrix} a_1 & 1 & \ddots & \vdots \\ \vdots & & \ddots & 0 \\ & & & 1 \\ \vdots & & & a_1 \\ a_{n_A} & & & \\ 0 & \ddots & & \vdots \\ \vdots & \ddots & & \\ 0 & \cdots & 0 & a_{n_A} \end{matrix} & \begin{matrix} \vdots & & & \vdots \\ \vdots & & & \vdots \\ 0 & \cdots & \cdots & 0 \\ 1 & \ddots & & \vdots \\ 0 & \ddots & \ddots & \vdots \\ \vdots & \ddots & \ddots & 0 \\ & & & \\ 0 & \cdots & 0 & 1 \end{matrix} \end{bmatrix}}_{n_A + d} \Bigg\} n_A + d \tag{7.98}$$

$$x^T = \left[1, s'_1, \ldots, s'_d, r_0, \ldots, r_{n_A - 1}\right] \tag{7.99}$$

$$p^T = \left[1, p_1, \ldots, p_{n_A}, p_{n_A+1}, \ldots, p_{n_A+d}\right] \tag{7.100}$$

Some of the coefficients $p_i$ can be zero. Since $M$ is a lower triangular matrix it is always non-singular which implies that (7.97) (and respectively (7.91)) always has a solution:

$$x = M^{-1} p$$

□

As in the case of the pole placement it is convenient to consider a parameterization of the controller polynomials $S(q^{-1})$ and $R(q^{-1})$ as:

$$S(q^{-1}) = S'(q^{-1}) H_S(q^{-1}) \tag{7.101}$$

$$R(q^{-1}) = R'(q^{-1}) H_R(q^{-1}) \tag{7.102}$$

where $H_R(q^{-1})$ and $H_S(q^{-1})$ represent the prespecified parts of $R(q^{-1})$ and $S(q^{-1})$. In this case $H_S(q^{-1})$ takes the specific form:

$$H_S(q^{-1}) = B^{\star}(q^{-1}) H'_S(q^{-1}) \tag{7.103}$$

and (7.91) becomes

$$A(q^{-1}) H'_S(q^{-1}) S'(q^{-1}) + q^{-d-1} H_R(q^{-1}) R'(q^{-1}) = P(q^{-1}) \tag{7.104}$$

All the consideration concerning $H_R(q^{-1})$ and $H'_S(q^{-1})$ discussed in Sect. 7.3 for the pole placement are applicable.

### Tracking (Computation of $T(q^{-1})$)

The precompensator $T(q^{-1})$ is computed in order to ensure (in accordance to Fig. 7.9) the following transfer operator between the reference $r(t)$ and $y(t)$:

$$H_{CL}(q^{-1}) = \frac{q^{-d-1}B_m(q^{-1})}{A_m(q^{-1})} = \frac{q^{-d-1}B_m(q^{-1})T(q^{-1})}{A_m(q^{-1})P(q^{-1})} \tag{7.105}$$

From (7.105) one obtains:

$$T(q^{-1}) = P(q^{-1}) \tag{7.106}$$

The input to $T(q^{-1})$ at sampling instant $t$ is the desired trajectory $y^\star$, $d+1$ steps ahead ($y^\star(t+d+1)$).

### Controller Equation

The controller equation is given by:

$$S(q^{-1})u(t) = -R(q^{-1})y(t) + P(q^{-1})y^\star(t+d+1) \tag{7.107}$$

Taking into account the form of $S(q^{-1})$

$$\begin{aligned} S(q^{-1}) &= s_0 + s_1 q^{-1} + \cdots + s_{n_S} q^{-ns} = s_0 + q^{-1}S^\star(q^{-1}) \\ &= B^\star(q^{-1})S'(q^{-1}) \end{aligned} \tag{7.108}$$

with

$$s_0 = b_1 \tag{7.109}$$

Equation (7.107) takes the form:

$$u(t) = \frac{1}{b_1}[P(q^{-1})y^\star(t+d+1) - S^\star(q^{-1})u(t-1) - R(q^{-1})y(t)] \tag{7.110}$$

or in a regressor form:

$$\theta_C^T \phi_C^T(t) = P(q^{-1})y^\star(t+d+1) \tag{7.111}$$

where:

$$\phi_C^T(t) = [u(t), \ldots, u(t-n_S), y(t), \ldots, y(t-n_R)] \tag{7.112}$$

$$\theta_C^T = [s_0, \ldots, s_{n_S}, r_0, \ldots, r_{n_R}] \tag{7.113}$$

Observe in (7.110) that low values of $|b_1|$ may lead to very large plant inputs. But this is exactly what happens when one has a significant fractional delay or an unstable zero.

### 7.4.2 Time Domain Design

The problem of independent tracking and regulation can be formulated as follows: Find the control

$$u(t) = f_u(y(t), y(t-1), \ldots, u(t-1), u(t-2), \ldots)$$

such that

$$P(q^{-1})[y(t+d+1) - y^{\star}(t+d+1)] = 0; \quad t > 0 \tag{7.114}$$

or equivalently which minimizes the criterion

$$J(t+d+1) = \{P(q^{-1})[y(t+d+1) - y^{\star}(t+d+1)]\}^2 \tag{7.115}$$

This means that in regulation ($y^{\star}(t) \equiv 0$) any initial disturbance $y(0)$ will be eliminated with the dynamics defined by

$$P(q^{-1})y(t+d+1) = 0 \tag{7.116}$$

and in tracking any initial error between the output and the desired trajectory is eliminated with the dynamics defined by $P(q^{-1})$.

**Step 1** One computes a $d+1$ step ahead filtered prediction of $y(t)$ i.e.:

$$\begin{aligned} P(q^{-1})y(t+d+1) &= f(y(t), y(t-1), \ldots, u(t), u(t-1), \ldots) \\ &= F(q^{-1})y(t) + B^{\star}(q^{-1})E(q^{-1})u(t) \end{aligned} \tag{7.117}$$

where $F(q^{-1})$ and $E(q^{-1})$ are solutions of the polynomial equation

$$P(q^{-1}) = A(q^{-1})E(q^{-1}) + q^{-d-1}F(q^{-1}) \tag{7.118}$$

**Step 2** The expression of $P(q^{-1})y(t+d+1)$ given by (7.117) is introduced in the criterion (7.114) yielding:

$$F(q^{-1})y(t) + B^{\star}(q^{-1})E(q^{-1})u(t) - P(q^{-1})y^{\star}(t+d+1) = 0 \tag{7.119}$$

from which one obtains the $R$-$S$-$T$ controller

$$S(q^{-1})u(t) = T(q^{-1})y^{\star}(t+d+1) - R(q^{-1})y(t) \tag{7.120}$$

where:

$$\begin{aligned} R(q^{-1}) &= F(q^{-1}); \qquad S(q^{-1}) = B^{\star}(q^{-1})E(q^{-1}) \\ T(q^{-1}) &= P(q^{-1}) \end{aligned} \tag{7.121}$$

which is exactly the same solution as in the polynomial approach. The design can be interpreted as an application of the *separation theorem*. Namely in the first step one designs the predictor and in the second step one forces the output of the predictor to be equal to the desired filtered trajectory.

### Predictor Interpretation of the Tracking and Regulation with Independent Objectives

We have a result similar to that for the pole placement. For the plant model of (7.1) with a delay $d$ and $P(q^{-1}) = P_D(q^{-1})P_F(q^{-1})$ the corresponding controller will be given by:

$$S_d(q^{-1})u(t) = -R_d(q^{-1})y(t) + P(q^{-1})y^\star(t+d+1) \tag{7.122}$$

where

$$S_d(q^{-1}) = B^\star(q^{-1})S'_d(q^{-1}) \tag{7.123}$$

and $S_D(q^{-1})$ and $R_D(q^{-1})$ are solutions of:

$$A(q^{-1})S'_d(q^{-1}) + q^{-d-1}R_d(q^{-1}) = P_D(q^{-1})P_F(q^{-1}) \tag{7.124}$$

**Theorem 7.4** (Landau 1993b) *Controller* (7.122) *through* (7.124) *for the plant* (7.1) *is equivalent to the following one*:

$$P_F(q^{-1})S_0(q^{-1})u(t) = -R_0(q^{-1})P_F(q^{-1})\hat{y}(t+d|t) + P_F(q^{-1})P_D(q^{-1})y^\star(t+d+1) \tag{7.125}$$

$$P_F(q^{-1})\hat{y}(t+d|t) = F(q^{-1})y(t) + B(q^{-1})E(q^{-1})u(t) \tag{7.126}$$

*where*

$$S_0(q^{-1}) = B^\star(q^{-1})S'_0(q^{-1}) \tag{7.127}$$

$S'_0(q^{-1})$ *and* $R_0(q^{-1})$ *are solutions of*

$$A(q^{-1})S'_0(q^{-1}) + q^{-1}R_0(q^{-1}) = P_D(q^{-1}) \tag{7.128}$$

*and* $F(q^{-1})$ *and* $E(q^{-1})$ *are solutions of*

$$A(q^{-1})E(q^{-1}) + q^{-d}F(q^{-1}) = P_F(q^{-1}) \tag{7.129}$$

*Proof* As for Theorem 7.2. □

*Remarks*

- The controller $R_0, S_0, T_0 = P_D(q^{-1})$ corresponds to the tracking and regulation with independent objectives for the plant model (7.1) with $d = 0$ and with the desired closed-loop poles defined by $P_D(q^{-1})$.
- The controller of (7.125) and (7.126) shows for $P_F(q^{-1}) = 1$ that in order to control a system with delay, the controller for the case without delay can be used by replacing the measured output $y(t)$ by its d-steps ahead prediction $\hat{y}(t+d|t)$.
- The other remarks from Sect. 7.3.3 for the pole placement are also applicable in this case.

### Taking into Account Measurable Disturbances

In a number of applications measurable disturbances act upon the output of the process through a certain transfer operator. The knowledge of this transfer would be useful for compensating the effect of the measurable disturbances. In this case, the plant output is described by:

$$A(q^{-1})y(t+d+1) = B^*(q^{-1})u(t) + C^*(q^{-1})v(t) \tag{7.130}$$

where $v(t)$ is the measurable disturbance and $C^*(q^{-1})$ is a polynomial of order $n_C$.

$$C^*(q^{-1}) = c_1 + c_2 q^{-1} + \cdots + c_{n_c} q^{-n_c+1} \tag{7.131}$$

Using the polynomial equation (7.118) the $d+1$ step ahead filtered prediction of $y(t)$ is:

$$\begin{aligned} P(q^{-1})y(t+d+1) &= F(q^{-1})y(t) + B^*(q^{-1})E(q^{-1})u(t) + C^*(q^{-1})E(q^{-1})v(t) \\ &= R(q^{-1})y(t) + S(q^{-1})u(t) + W(q^{-1})v(t) \end{aligned} \tag{7.132}$$

where:

$$\begin{aligned} &R(q^{-1}) = F(q^{-1}); \qquad S(q^{-1}) = B^*(q^{-1})E(q^{-1}) \\ &W(q^{-1}) = C^*(q^{-1})E(q^{-1}) \end{aligned} \tag{7.133}$$

Forcing the prediction to equal $P(q^{-1})y^*(t+d+1)$ yields the controller.

$$P(q^{-1})y^*(t+d+1) = R(q^{-1})y(t) + S(q^{-1})u(t) + W(q^{-1})v(t) = \theta_c^T \phi_c(t) \tag{7.134}$$

Here in this case:

$$\theta_c^T = [s_0, \ldots, s_{n_S}, r_0, \ldots, r_{n_R}, w_0, \ldots, w_{n_W}] \tag{7.135}$$

$$\phi_c^T(t) = [u(t), \ldots, u(t-n_S), y(t), \ldots, y(t-n_R), v(t), \ldots, v(t-n_W)] \tag{7.136}$$

which will assure (7.114) even in the presence of measurable disturbances.

## 7.5 Tracking and Regulation with Weighted Input

This control strategy can be considered as an extension of the tracking and regulation with independent objectives and was introduced by Clarke and Gawthrop (1975). This extension may be interpreted as weighting the control energy resulting from *tracking and regulation with independent objectives design method.*

This weighting results in the modification of the polynomial $S(q^{-1})$ of the controller. As a consequence the system zeros will not be simplified and the closed-loop poles will differ from the specified ones. Since the zeros of the plant model will not be simplified, this method may work for certain plant models with unstable zeros (in general for the plant models with stable poles).

In brief this design method makes it possible to

- reduce the control energy with a single tuning parameter;
- control certain types of plants having discrete-time models with unstable zeros but the resulting closed-loop poles will differ from the desired ones.

The hypothesis (H3) from tracking and regulation with independent objectives is replaced by:

(H3) $B^\star(q^{-1})$ may have stable or unstable zeros, but there is a polynomial $Q(q^{-1})$ and $\lambda > 0$ such that $\lambda Q(q^{-1})A(q^{-1}) + B^\star(q^{-1})P(q^{-1})$ has all its zeros inside the unit circle.

The design can be obtained by considering one of the following three control criteria in the time domain which lead to the same control law.

**Design problem** Find $u(t) = f_u(y(t), y(t-1), \ldots, u(t-1), u(t-2), \ldots)$ such that:

1.

$$\begin{aligned}\varepsilon_0(t+d+1) &= P(q^{-1})[y(t+d+1) - y^\star(t+d+1)] + \lambda Q(q^{-1})u(t)\\ &= 0\end{aligned} \tag{7.137}$$

where $\lambda > 0$ and $Q(q^{-1})$ is a monic polynomial,[1] or

2.

$$\begin{aligned}J_1(t+d+1) = \{&P(q^{-1})[y(t+d+1) - y^\star(t+d+1)]\\ &+ \lambda Q(q^{-1})u(t)\}^2\end{aligned} \tag{7.138}$$

is minimized, or

3.

$$\begin{aligned}J_2(t+d+1) = \{&P(q^{-1})[y(t+d+1) - y^\star(t+d+1)]\}^2\\ &+ b_1\lambda[Q(q^{-1})u(t)]^2\end{aligned} \tag{7.139}$$

is minimized.

A typical form for $Q(q^{-1})$ is

$$Q(q^{-1}) = 1 - q^{-1} \tag{7.140}$$

which means that the variations of $u(t)$ are weighted instead of $u(t)$. This allows to incorporate an integral action of the controller. Using the expression of the filtered prediction $P(q^{-1})y(t+d+1)$ given in (7.117) and taking into account equations (7.118) and (7.119) one has:

$$P(q^{-1})y(t+d+1) = S(q^{-1})u(t) + R(q^{-1})y(t) \tag{7.141}$$

and the criterion of (7.137) becomes:

[1]The quantity $y(t+d+1) + \lambda\frac{Q}{P}u(t)$ is often interpreted as a "generalized output".

$$\varepsilon_0(t+d+1) = [S(q^{-1}) + \lambda Q(q^{-1})]u(t) \\ + R(q^{-1})y(t) - P(q^{-1})y^\star(t+d+1) = 0 \quad (7.142)$$

yielding the controller

$$[S(q^{-1}) + \lambda Q(q^{-1})]u(t) = -R(q^{-1})y(t) + P(q^{-1})y^\star(t+d+1) \quad (7.143)$$

A similar result is obtained if one considers the criterion of (7.138) or (7.139). Using (7.141), (7.139) becomes:

$$J(t+d+1) = [b_1 u(t) + S^\star(q^{-1})u(t-1) + R(q^{-1})y(t) \\ - P(q^{-1})y^\star(t+d+1)]^2 + b_1\lambda[Q(q^{-1})u(t)]^2 \quad (7.144)$$

In order to minimize the quadratic criterion (7.144) one must find $u(t)$ such that

$$\frac{\nabla J(t+1)}{\nabla u(t)} = 0$$

From (7.144) one gets:

$$\frac{1}{2}\frac{\nabla J(t+1)}{\nabla u(t)} = b_1[S(q^{-1})u(t) + R(q^{-1})y(t) - P(q^{-1})y^\star(t+d+1)] \\ + b_1\lambda Q(q^{-1})u(t) = 0 \quad (7.145)$$

which is satisfied by the control law of (7.143). The closed-loop poles are now:

$$A(q^{-1})[S(q^{-1}) + \lambda Q(q^{-1})] + q^{-d-1}B^\star(q^{-1})R(q^{-1}) \\ = \lambda Q(q^{-1})A(q^{-1}) + B^\star(q^{-1})P(q^{-1}) \quad (7.146)$$

Therefore (7.146) indicates that the strategy can be used for plant models with unstable zeros provided that for a given weighting filter $Q(q^{-1})$ there is a value $\lambda > 0$ such that $\lambda Q(q^{-1})A(q^{-1}) + B^\star(q^{-1})P(q^{-1})$ is asymptotically stable. The transfer operator from the desired trajectory to the output is:

$$H_{CL}(q^{-1}) = \frac{q^{-d-1}B^\star(q^{-1})T(q^{-1})}{\lambda Q(q^{-1})A(q^{-1}) + B^\star(q^{-1})P(q^{-1})} \quad (7.147)$$

If $Q(1) = 0$ the steady state gain remains unchanged for $T(q^{-1}) = P(q^{-1})$ with respect to tracking and regulation with independent objectives. However, we do not have more perfect tracking of the desired trajectory. In order to preserve the tracking performances independently of the value of $\lambda$, one should choose:

$$T(q^{-1}) = \beta[\lambda Q(q^{-1})A(q^{-1}) + B^\star(q^{-1})P(q^{-1})] \\ \beta = \frac{1}{B^\star(1)} \quad (7.148)$$

The design is done in two steps:

1. One computes a RST controller using tracking and regulation with independent objective even if $B^\star(q^{-1})$ has unstable zeros.
2. One searches for a value $\lambda > 0$ such that $\lambda Q(q^{-1})A(q^{-1}) + B^\star(q^{-1})P(q^{-1})$ is asymptotically stable and that the input or its filtered value has a desired behavior.

*Remark* The input energy can be reduced also by changing the initial desired poles as well as by introducing filters in $H_R(q^{-1})$ in certain frequency ranges.

*Example* Consider the case: $d = 0$, $B^\star(q^{-1}) = b_1 + b_2 q^{-1}$ with $|\frac{b_2}{b_1}| > 1$, i.e. $B^\star(q^{-1})$ has an unstable zero. Then

$$S(q^{-1}) = B^\star(q^{-1}) = b_1 + b_2 q^{-1}$$

Choosing

$$\lambda Q(q^{-1}) = \lambda(1 - q^{-1})$$

one obtains

$$S(q^{-1}) + \lambda(1 - q^{-1}) = (b_1 + \lambda) + (b_2 - \lambda)q^{-1}$$

If we choose $\lambda$ such that:

$$\left|\frac{b_2 - \lambda}{b_1 + \lambda}\right| < 1$$

the controller will be stable, and it remains to check the stability of the resulting closed-loop poles.

**Regressor form** The control law can be expressed as:

$$\theta_C^T \phi_C(t) = T(q^{-1})y^\star(t + d + 1) - \lambda Q(q^{-1})u(t)$$

where $\theta_C$ and $\phi_C(t)$ are given by (7.112) and (7.113).

*Remark* Weighting factors in the form of the ratio of two polynomials can also be considered.

## 7.6 Minimum Variance Tracking and Regulation

This strategy concerns the design of a controller ensuring a minimum variance of the controlled variable (plant output) around the reference, in the presence of random disturbances. In a stochastic environment it is assumed that the plant and disturbance are described by the following ARMAX model:

$$A(q^{-1})y(t + d + 1) = B^\star(q^{-1})u(t) + C(q^{-1})e(t + d + 1) \tag{7.149}$$

where $e(t)$ is a sequence of normally distributed independent random variables of zero mean and finite variance $\sigma^2$ (white noise) and the zeros of $C(z^{-1})$ are in $|z| < 1$. The hypotheses made upon the plant model are those made for tracking and regulation with independent objectives. The main restrictive assumption is the requirement that $B^\star(z^{-1})$ has all its zeros inside the unit circle (hypothesis (H3)). The objective is to compute for the plant and disturbance model (7.149) the control $u(t)$:

$$u(t) = f_u(y(t), y(t - 1), \ldots, u(t - 1), u(t - 2), \ldots)$$

which minimizes the following criterion

$$\min_{u(t)} J(t+d+1) = \mathbf{E}\{[y(t+d+1) - y^\star(t+d+1)]^2\}$$
$$\approx \lim_{N\to\infty} \frac{1}{N} \sum_{t=1}^{N} \{[y(t+d+1) - y^\star(t+d+1)]^2\} \quad (7.150)$$

For the case $y^\star(t+d+1) = 0$, the criterion (7.150) becomes:

$$\min_{u(t)} J(t+d+1) = \mathbf{E}\{y^2(t+d+1)\}$$

and corresponds to the original formulation of the minimum variance control (Åström and Wittenmark 1973). To solve this problem, in addition to the plant model, the disturbance model should be known.

## 7.6.1 Design of Minimum Variance Control

Two equivalent approaches can be used for solving the problem

1. direct minimization of the criterion of (7.150);
2. use of the separation theorem.

The interest of examining the two approaches is related to the development of the adaptive version of the control law and this will be discussed in Chap. 11.

### Direct Design

As usual, one should replace $y(t+d+1)$ in (7.150) by an equivalent expression featuring $u(t)$ and then minimize the criterion with respect to $u(t)$. Using the polynomial equation

$$C(q^{-1}) = A(q^{-1})E(q^{-1}) + q^{-d-1}F(q^{-1}) \quad (7.151)$$

where:

$$E(q^{-1}) = 1 + e_1 q^{-1} + \cdots + e_d q^{-d} \quad (7.152)$$
$$F(q^{-1}) = f_0 + f_1 q^{-1} + \cdots + f_{n_A-1} q^{-n_A+1} \quad (7.153)$$

one obtains

$$C(q^{-1})y(t+d+1) = F(q^{-1})y(t) + B^\star(q^{-1})E(q^{-1})u(t)$$
$$+ C(q^{-1})E(q^{-1})e(t+d+1) \quad (7.154)$$

from which the following expression for $y(t+d+1)$ is obtained:

$$y(t+d+1) = \frac{F(q^{-1})}{C(q^{-1})}y(t) + \frac{B^*(q^{-1})E(q^{-1})}{C(q^{-1})}u(t) + E(q^{-1})e(t+d+1) \quad (7.155)$$

This corresponds to an "innovation" representation of $y(t+d+1)$ (see Sect. 2.2). Using (7.155) the variance of the tracking error can be expressed as:

$$\begin{aligned}
&\mathbf{E}\{[y(t+d+1) - y^\star(t+d+1)]^2\} \\
&= \mathbf{E}\left\{\left[\frac{F}{C}y(t) + \frac{B^*E}{C}u(t) - y^*(t+d+1)\right]^2\right\} + \mathbf{E}\{[Ee(t+d+1)]^2\} \\
&\quad + 2\mathbf{E}\left\{Ee(t+d+1)\left[\frac{F}{C}y(t) + \frac{B^*E}{C}u(t) - y^*(t+d+1)\right]\right\} \quad (7.156)
\end{aligned}$$

The third term of the right hand side of (7.156) is zero since the white noise at $t+d+1$ is independent with respect to all the signals appearing at instants up to and including $t$. The second term in the right hand side of (7.156) does not depend upon $u(t)$. Therefore, in order to minimize the variance of the tracking error expressed in (7.156), one should choose $u(t)$ such that:

$$F(q^{-1})y(t) + B^*(q^{-1})E(q^{-1})u(t) - C(q^{-1})y^*(t+d+1) = 0 \quad (7.157)$$

from which one obtains the controller equation:

$$S(q^{-1})u(t) = -R(q^{-1})y(t) + T(q^{-1})y^*(t+d+1) \quad (7.158)$$

with:

$$S(q^{-1}) = B^*(q^{-1})E(q^{-1}) \quad (7.159)$$

$$R(q^{-1}) = F(q^{-1}) \quad (7.160)$$

$$T(q^{-1}) = C(q^{-1}) \quad (7.161)$$

where $E(q^{-1})$ and $F(q^{-1})$ are solutions of (7.151) given in (7.152) and (7.153).

### Use of the Separation Theorem

Using the separation theorem, the control design is done in two stages:

1. Computation of the optimal predictor $\hat{y}(t+d+1/t)$ which minimizes:

$$\mathbf{E}\{[y(t+d+i) - \hat{y}(t+d+1)]^2\} \quad (7.162)$$

Taking into account the results given in Sect. 2.2.2 and (7.155), this is given by:

$$\begin{aligned}
\hat{y}(t+d+1) &= \hat{y}(t+d+1/t) \\
&= \frac{F(q^{-1})}{C(q^{-1})}y(t) + \frac{B^*(q^{-1})E(q^{-1})}{C(q^{-1})}u(t) \quad (7.163)
\end{aligned}$$

2. Computation of $u(t)$ such that $\hat{y}(t+d+1)$ minimizes the associated deterministic criterion:

$$\min_{u(t)} J(t+d+1) = [\hat{y}(t+d+1) - y^*(t+d+1)]^2 \tag{7.164}$$

or equivalently:

$$\hat{y}(t+d+1) - y^*(t+d+1) = 0 \tag{7.165}$$

This leads immediately to (7.157) and respectively to (7.158).

*Remark* The parameters of the controller (coefficients of $R(q^{-1})$ and $S(q^{-1})$) correspond to the coefficients of the predictor for $C(q^{-1})\hat{y}(t+d+1)$.

The *regressor form* of the controller can be expressed as:

$$\theta_C^T \phi_C(t) = C(q^{-1}) y^*(t+d+1) \tag{7.166}$$

where:

$$\theta_C^T = [s_0, \ldots, s_{n_S}, r_0, \ldots, r_{n_R}] \tag{7.167}$$

$$\phi_C^T = [u(t), \ldots, u(t-n_S), y(t), \ldots, y(t-n_R)] \tag{7.168}$$

### Rapprochement with Tracking and Regulation with Independent Objective

Comparing the expression of the controller given in (7.158) through (7.161) with the one for tracking and regulation with independent objective, (7.120), (7.121) and (7.118), one sees that the two controllers have exactly the same structure and the values of the controller parameter are the same if $P(q^{-1}) = C(q^{-1})$. In other terms, choosing the desired closed-loop poles $P(q^{-1}) = C(q^{-1})$ (the model of the disturbance) the tracking and regulation with independent objectives achieves a minimum variance tracking and regulation for the disturbance model in (7.149). Conversely, choosing $P(q^{-1})$ in the deterministic case, leads to a minimum variance controller for a disturbance model $C(q^{-1}) = P(q^{-1})$.

The stochastic approach allows a better understanding of why it is useful to assure an independent behavior in tracking and regulation. The choice of the closed-loop poles should be made with respect to the characteristics of the disturbance and not with respect to the desired behavior in tracking.

### Properties of the Tracking (Regulation) Error

Introducing the control law of (7.158) in (7.149), one obtains:

$$y(t+d+1) = y^*(t+d+1) + E(q^{-1})e(t+d+1) \tag{7.169}$$

and the tracking error becomes:

$$\varepsilon^0(t+d+1) = y(t+d+1) - y^*(t+d+1) = E(q^{-1})e(t+d+1) \tag{7.170}$$

One observes that the tracking (regulation error) in the case of minimum variance tracking and regulation is a moving average of order $d$. It satisfies the property:

$$\mathbf{E}\{\varepsilon^0(t+i)\varepsilon^0(t)\} = 0; \quad \text{for } i > d \tag{7.171}$$

In particular for $d = 0$, the tracking error is a white noise. Note that property of (7.171) suggests a practical test for the optimal tuning of a digital controller if the delay $d$ is known.

### Rejection of Disturbances

This can be formally done by considering instead of the model of (7.1), the following model:

$$A(q^{-1})y(t+d+1) = B^*(q^{-1})u(t) + v(t+d+1) \tag{7.172}$$

$$D(q^{-1})v(t) = C(q^{-1})e(t) \tag{7.173}$$

where $D(q^{-1})$ corresponds to the model of the deterministic disturbance (known also as an ARIMAX model). Multiplying both terms of (7.172) by $D(q^{-1})$ one gets:

$$A(q^{-1})D(q^{-1})y(t+d+1) = B^*(q^{-1})\bar{u}(t) + C(q^{-1})e(t+d+1) \tag{7.174}$$

where:

$$\bar{u}(t) = D(q^{-1})u(t) \tag{7.175}$$

and one returns to the previous case. Solving now the polynomial equation:

$$C = ADE + q^{-d^{-1}}F \tag{7.176}$$

one will get:

$$R = F \quad \text{and} \quad S = DEB^* \tag{7.177}$$

## 7.6.2 Generalized Minimum Variance Tracking and Regulation

As in the deterministic case, one can consider an extension of the minimum variance tracking and regulation in order to weigh the control effort and to be able to apply it to some types of plant models with unstable zeros.

Instead of minimizing the criterion of (7.150), one considers the following one:

$$\min_{u(t)} J(t+d+1) = \mathbf{E}\left\{\left[y(t+d+1) - y^*(t+d+1) + \lambda\frac{Q(q^{-1})}{C(q^{-1})}u(t)\right]^2\right\}$$
$$\lambda \geq 0 \tag{7.178}$$

The quantity $y(t+d+1) + \frac{\lambda Q(q^{-1})}{C(q^{-1})}u(t)$ is interpreted, similar to the deterministic case, as a generalized output and its variations around $y^*(t+d+1)$ will be minimized.

The minimum variance tracking and regulation has been introduced by Clarke and Gawthrop (1975, 1979). The hypothesis upon the plant zeros are the same as for *tracking and regulation with weighted input*, i.e.

(H3) $B^*(q^{-1})$ may have stable or unstable zeros but there is a polynomial $Q$ and $\lambda > 0$ such that $\lambda QA + B^*C$ has all its zeros inside the unit circle.

Using the expression of $y(t+d+1)$ given by (7.155) with the notations of (7.159) and (7.160), the criterion of (7.178) takes the form:

$$\begin{aligned} &J(t+d+1) \\ &\quad = \mathbf{E}\left\{\left[\frac{R(q^{-1})}{C(q^{-1})}y(t) + \frac{S(q^{-1})}{C(q^{-1})}u(t) + \frac{\lambda Q(q^{-1})}{C(q^{-1})}u(t) - y^*(t+d+1)\right]^2\right\} \\ &\qquad + \mathbf{E}\{[E(q^{-1})e(t+d+1)]^2\} \end{aligned} \tag{7.179}$$

Since the second term does not depend upon $u(t)$, the value of $u(t)$ which will force the first term to zero is:

$$[S(q^{-1}) + \lambda Q(q^{-1})]u(t) = -R(q^{-1})y(t) + C(q^{-1})y^*(t+d+1) \tag{7.180}$$

The control law corresponds to the one for *tracking and regulation with weighted input* by choosing $P(q^{-1}) = C(q^{-1})$ and all the considerations for the choice of $\lambda$ and $Q(q^{-1})$ hold.

## 7.7 Generalized Predictive Control

The control strategies presented in the previous sections belong to the class of "one step ahead" predictive control, in the sense that the control objective concerns only the outputs of the plant $d+1$ steps ahead where $d$ is the integer delay.

The *generalized predictive control* belongs to the class of long range predictive control in the sense that the control objective concerns the outputs and the inputs of the plant over a certain horizon in the future beyond $d+1$. These strategies are related to the minimization of a quadratic criterion involving future inputs and outputs in a receding horizon sense, i.e., one computes a sequence of control signals in the future, but only the first one is applied and the optimization procedure is restarted at the next step. Indeed the resulting control law is stationary and the controller takes the form of an RST digital controller. The generalized predictive control has been introduced by Clarke et al. (1987).

The generalized predictive control with an appropriate formulation of the quadratic criterion can be interpreted as generalization of the various "one step ahead" control strategies. Consider the plant model:

$$A(q^{-1})y(t+d+1) = B^*(q^{-1})u(t) + v(t+d+1) \tag{7.181}$$

where $v(t+1)$ is the disturbance which can be deterministic or stochastic (or both). We make similar hypotheses upon the plant model as for *pole placement* except hypothesis (H4) which is replaced by:

(H4) The eventual common factors of $A(q^{-1})$ and $B(q^{-1})$ are inside the unit circle. Consider the quadratic criterion:

$$J(t,h_p,h_c,h_i)=\mathbf{E}\left\{\sum_{j=h_i}^{h_p}[P(q^{-1})y(t+j)-P(q^{-1})y^*(t+j)]^2 + \lambda[Q(q^{-1})u(t+j-h_i)]^2\right\} \quad (7.182)$$

where $h_i$ is the initial horizon, $h_p$ is the prediction horizon, $\lambda \geq 0$ and $P(q^{-1})$ and $Q(q^{-1})$ are transfer operators.

The control objective is to minimize the criterion (7.182) in a receding horizon sense with respect to:

$$U^T(t+h_c-1)=[u(t),u(t+1),\dots,u(t+h_c-1)] \quad (7.183)$$

subject to the constraints:

$$Q(q^{-1})u(t+i)=0; \quad h_c \leq i \leq h_p-d-1 \quad (7.184)$$

The transfer operators $P(q^{-1})$, $Q(q^{-1})$ allow to accommodate a variety of design aspects related to the disturbance characteristics, tracking capabilities, robustness, and also allow to establish close relationship with the various one step ahead predictive control strategies.

The criterion depends on future controls and outputs of the plant and the expectation is conditioned on data available up to and including $t$. Therefore in the effective minimization procedure $y(t+j)$ for $j \geq d+1$ will be replaced by their predictions. The initial horizon $h_i$ is usually taken larger than the delay i.e. $h_i \geq d+1$. The prediction horizon $h_p$ is usually taken close to the time response of the system. The control horizon $h_c$ will define the complexity of the optimization procedure and the objective is to take it as small as possible. However, this may lead to instability for too shorter horizons (see the discussion at the end of this section).

Note that the constraint on the input specified in (7.184) can be simply interpreted in terms of classical linear quadratic control as putting an infinite weight $\lambda(t+i)=\infty$ on the input for $i \geq h_c$. If $h_p, h_c \to \infty$, one obtains the infinite horizon quadratic criterion (which will be discussed in Sect. 7.8). Assuming for example that $Q(q^{-1})=1-q^{-1}$, the constraint (7.184) can be interpreted as imposing that the variations of the control input be zero after $h_c$ steps.

As we will see, $P(q^{-1})$ allows to assign part of the poles of the closed loop. $Q(q^{-1})$ allows to obtain offset-free performances for given disturbance models (it represents the internal model of the disturbance which has to be included in the controller). It also allows the shaping of the controller frequency characteristics for robustness considerations.

The form of the criterion (7.182) allows to establish relationship with several one step ahead control strategies.

1. For $h_p = h_i = d + 1$, $h_c = 1$, $\lambda = 0$ one obtains the *tracking and regulation with independent objectives*.
2. For $v(t) = C(q^{-1})e(t)$, $h_p = h_i = d + 1$, $h_c = 1$, $P(q^{-1}) = 1$, $\lambda = 0$ one obtains the *minimum variance tracking and regulation*.
3. For $h_p = h_i = d + 1$, $h_c = 1$, $Q(q^{-1}) \neq 0$, $\lambda > 0$ one obtains the *tracking and regulation with weighted input*.
4. For $v(t) = C(q^{-1})e(t)$, $h_p = h_i = d + 1$, $h_c = 1$, $P(q^{-1}) = 1$, $Q(q^{-1}) = \frac{1-q^{-1}}{C(q^{-1})}$, $\lambda > 0$ one obtains the *generalized minimum variance tracking and regulation*.

A more general form of the criterion (7.182) is:

$$J(t, h_p, h_c, h_i) = \mathbf{E}\left\{ \sum_{j=h_i}^{h_p} \{P(q^{-1})[y(t+j) - T_1(q^{-1})y^*(t+j)]\}^2 + \lambda\{Q(q^{-1})[u(t+j-h_i) - T_2(q^{-1})y^*(t+j-h_i+d)]\}^2 \right\} \quad (7.185)$$

subject to the constraint:

$$Q(q^{-1})[u(t+j-h_i) - T_2(q^{-1})y^*(t+j-h_i+d)] = 0$$
$$h_c \leq j \leq h_p - d - 1 \quad (7.186)$$

where $T_2(q^{-1})y^*(t+j+d)$ defines a desired trajectory for the control input in a similar way as $T_1(q^{-1})y^*(t+j)$ defines a desired trajectory for the output. Taking for example in (7.185):

$$T_1(q^{-1}) = \beta B^*(q^{-1}); \quad \beta = \frac{1}{B^*(1)} \quad (7.187)$$

$$T_2(q^{-1}) = \beta A(q^{-1}); \quad \beta = \frac{1}{B^*(1)} \quad (7.188)$$

$$Q(q^{-1}) = H_S(q^{-1})/H_R(q^{-1}) \quad (7.189)$$

$$P(q^{-1}) = 1 \quad (7.190)$$

one gets a generalization of the *pole placement* strategy. In this case one has:

$$y(t+j) - \beta B^*(q^{-1})y^*(t+j) = e_y(t+j) \quad (7.191)$$

$$u(t+j) - \beta A(q^{-1})y^*(t+j+d+1) = e_u(t+j) \quad (7.192)$$

(see also Sect. 7.3) and the criterion can be rewritten as:

$$J(t, h_p, h_c, h_i) = \mathbf{E}\left\{ \sum_{j=h_i}^{h_p} [e_y(t+j)]^2 + \lambda[Q(q^{-1})e_u(t+j-h_i)]^2 \right\} \quad (7.193)$$

This criterion should be minimized in a receding horizon sense with respect to:

$$E_u(t+h_c-1) = [e_u(t), e_u(t+1), \ldots, e_u(t+h_c-1)]^T \tag{7.194}$$

subject to the constraints:

$$Q(q^{-1})e_u(t+i) = 0; \quad h_c \le i \le h_p - d - 1 \tag{7.195}$$

Taking $Q(q^{-1}) = H_S(q^{-1})/H_R(q^{-1})$ the resulting controller will contain $H_R(q^{-1})$ and $H_S(q^{-1})$ and they can be interpreted as the internal model of the disturbance which has to be incorporated into the controller in order to obtain offset free performance on $e_y(t)$ for disturbances of the form:

$$H_S(q^{-1})v(t) = H_R(q^{-1})\delta(t) \tag{7.196}$$

where $\delta(t)$ is the Dirac impulse. Using the quantities $e_u(t)$ and $e_y(t)$, the control scheme will take the same form as the one shown in Fig. 7.7 for pole placement.

Remember that in the pole placement the objective is to compute $R'(q^{-1})$ and $S'(q^{-1})$ such that $P(q^{-1})e_y(t) = P(q^{-1})e_u(t) = 0$, while in the generalized predictive control approach, the final objective will be to find $R'$ and $S'$ which minimize criterion (7.193) for given $\lambda, h_p, h_i, h_c, Q(q^{-1})$.[2] The major interest of this formulation (M'Saad et al. 1986; Irving et al. 1986) is that tracking behavior is independent of the resulting closed-loop poles (as in the poles placement when $T(q^{-1}) = \beta P(q^{-1})$). This scheme is known as *partial state reference model* (PSMR) generalized predictive control.

We will focus next on the minimization of the criterion (7.193). Similar methodology is used for the minimization of (7.182) or (7.185) and this is left as exercises. To carry the minimization of (7.193), we will need to express the predictions of $e_y(t+j)$ as a function of $e_u(t+j)$ and therefore we need a relationship between these two quantities. We will assume that no disturbance affects the plant for $j > t$ (the case in which $v$ is a stochastic disturbance will be discussed later on).

Operating (7.191) by $AQ$ and taking in account (7.181) for $v = 0$, one finds:

$$\begin{aligned} AQe_y(t+d+1) &= B^*Q[u(t) - \beta Ay^*(t+d+1)] \\ &= B^*\bar{e}_u(t) \end{aligned} \tag{7.197}$$

where:

$$\bar{e}_u(t) = Qe_u(t) \tag{7.198}$$

A more convenient form is obtained by taking into account the structure of $Q$ given in (7.280) which leads to:

$$AH_Se_y(t+d+1) = B^*H_R\bar{e}_u(t) \tag{7.199}$$

[2]Another formulation allowing an even stronger analogy with the pole placement involves the choices: $P = P_D$ and $Q = P_DH_S/H_R$ instead of (7.189) and (7.190) which will force some of the poles of the closed loop to be equal to $P_D$. However, a similar result is obtained with (7.193) using a predictor for $e_y(t+j)$ with a dynamics defined by $P_D$.

Denoting:

$$\bar{A} = AH_S; \qquad \bar{B}^* = B^* H_R \tag{7.200}$$

Equation (7.199) takes the form:

$$\bar{A} e_y(t+d+1) = \bar{B}^* \bar{e}_u(t) \tag{7.201}$$

The minimization of (7.193) involves two steps:

1. Computation of the future values of $e_y(t+j)$ for $j \in [h_i, h_p]$ as:

   $$e_y(t+j) = f\,[E_u(t+h_c-1)] + \hat{e}_y^0(t+j/t)$$

   where $\hat{e}_y^0(t+j/t)$ will depend on the available information up to the instant $t$:

   $$\hat{e}_y^0(t+j/t) = f[e_y(t), e_y(t-1), \ldots, \bar{e}_u(t-1), \bar{e}_u(t-2)]$$

   and the other term is a function of the future values of $e_u$ up to $h_i + h_c - 1$.
2. Minimization of the criterion (7.193) with respect to $E_u(t+h_c-1)$.

Step 1: *Computation of the future values of* $e_y(t+j)$
Using the polynomial identity:

$$P_D = \bar{A} E_j + q^{-j} F_j \tag{7.202}$$

with $\deg E_j = j-1$ and $\deg F_j = \max(n_{P_D} - j, n_{\bar{A}} - 1)$, one can express the filtered predicted values of $e_y(t+j)$ which are used in the criterion (7.193) as follows (by taking in account (7.201)):

$$P_D e_y(t+j) = E_j \bar{B}^* \bar{e}_u(t+j-d-1) + F_j e_y(t) \tag{7.203}$$

The effect of using (7.203) as prediction equation will be the presence in the resulting closed-loop polynomial of a factor equal to the $P_D$ polynomial. To get the desired form of the prediction, one will use a second polynomial identity:

$$\bar{B}^* E_j = P_D G_{j-d} + q^{-j+d} H_{j-d} \tag{7.204}$$

where:

$$\deg G_{j-d} = j-d-1$$
$$\deg \bar{B}^* E_j = n_{\bar{B}} + j - 2 = n_{\bar{B}^*} + j - 1$$
$$\deg H_{j-d} \le \max(n_{P_D}, n_{\bar{B}^*} + d) - 1$$

The polynomials $G_{j-d}$ and $H_{j-d}$ will have the form:

$$G_{j-d} = g_0 + g_1 q^{-1} + \cdots + g_{j-d-1} q^{-j+d+1} \tag{7.205}$$
$$H_{j-d} = h_0^{j-d} + h_1^{j-d} q^{-1} + \cdots + h_{n-d-2}^{j-d} q^{-n_B-d-2} \tag{7.206}$$

Using the polynomial identity (7.7.24), the filtered prediction $e_y(t+j)$ can be written:

$$P_D e_y(t+j) = P_D G_{j-d} \bar{e}_u(t+j-d-1) + H_{j-d} \bar{e}_u(t-1) + F_j e_y(t) \tag{7.207}$$

Passing this expression through the transfer operator $1/P_D$ and denoting:

$$H_{j-d}\bar{e}_u(t-1) + F_j e_y(t) = P_D \hat{e}_y^0(t+j/t) \tag{7.208}$$

(i.e., $\hat{e}_y^0(t+j/t)$ is the prediction based on the available information up to the instant $t$) one gets:

$$e_y(t+j) = G_{j-d}\bar{e}_u(t+j-d-1) + \hat{e}_y^0(t+j/t) \tag{7.209}$$

with the observation that the term $G_{j-d}\bar{e}_u(t+j-d-1)$ will contain $\bar{e}_u(t)$, $\bar{e}_u(t+1), \ldots, \bar{e}(t+j-d-1)$, i.e., only future values of $\bar{e}_u$.

Step 2. *Minimization of the criterion*

Introducing the notations:

$$E_y^T = [e_y(t+h_i), \ldots, e_y(t+h_p)] \tag{7.210}$$

$$\bar{E}_u^T = [\bar{e}_u(t), \ldots, \bar{e}_u(t+h_c-1)] \tag{7.211}$$

The criterion (7.193) takes the form:

$$J = E_y^T E_y + \lambda \bar{E}_u^T \bar{E}_u \tag{7.212}$$

Now taking into account the expression of $e_y(t+j)$ given by (7.209) and introducing the notations:

$$E_y^0 = [\hat{e}_y^0(t+h_i/t), \ldots, \hat{e}_y^0(t+h_p/t)] \tag{7.213}$$

one has from (7.207) and (7.208):

$$E_y = G\bar{E}_u + E_y^0 \tag{7.214}$$

where:

$$G = \begin{bmatrix} g_{hi-d-1} & \cdots & g_o & 0 & \cdots & 0 \\ g_{hi-d} & \cdots & g_1 & g_o & \cdots & 0 \\ \vdots & \vdots & \vdots & \vdots & \ddots & \vdots \\ g_{hc-1} & \cdots & \cdots & \cdots & \cdots & g_o \\ \vdots & \vdots & \vdots & \vdots & \vdots & \vdots \\ g_{hp-d-1} & \cdots & \cdots & \cdots & \cdots & g_{hp-d-hc} \end{bmatrix} \tag{7.215}$$

for $h_c > h_i - d$ or

$$G = \begin{bmatrix} g_{hi-d-1} & \cdots & g_o & \cdots & \cdots & g_{hi-d-hc} \\ \vdots & \vdots & \vdots & \vdots & \vdots & \vdots \\ \vdots & \vdots & \vdots & \vdots & \vdots & \vdots \\ g_{hp-d-1} & \cdots & \cdots & \cdots & \cdots & g_{hp-d-hc} \end{bmatrix} \tag{7.216}$$

for $h_c \leq h_i - d$ and the criterion (7.193) takes the form:

$$J(t, h_p, h_c, h_i) = [G\bar{E}_u + E_y^0]^T [G\bar{E}_u + E_y^0] + \lambda \bar{E}_u^T \bar{E}_u \tag{7.217}$$

Minimization of this criterion is obtained by searching for $\bar{E}_u$ assuring $\frac{\Delta J}{\Delta \bar{E}_u} = 0$ which yields:

$$\bar{E}_u|_{opt} = -[G^T G + \lambda I_{h_c}]^{-1} G^T E_y^0 \tag{7.218}$$

This formula indicates that the computation of $\bar{E}_u$ is always possible provided that $\lambda > 0$.

Since for the implementation of the control law we need only the first component of $\bar{E}_u$, namely $\bar{e}_u(t)$, only the first row of the matrix $[G^T G + \lambda I_{h_c}]^{-1} G^T$ is of interest, i.e.,

$$\bar{e}_u(t) = -[\gamma_{h_i}, \ldots, \gamma_{h_p}] E_y^0 \tag{7.219}$$

where $\gamma_i$ are the coefficients of the first row of $[G^T G + \lambda I_{h_c}]^{-1} G^T$.

### 7.7.1 Controller Equation

The next step will be to effectively compute $u(t)$. Using (7.208), (7.219) can be further rewritten as:

$$P_D \bar{e}_u(t) = -\left( \sum_{j=h_i}^{j=h_p} \gamma_j H_{j-d} \right) \bar{e}_u(t-1) - \left( \sum_{j=h_i}^{j=h_p} \gamma_j F_j \right) e_y(t) \tag{7.220}$$

or as:

$$\begin{aligned} &\left[ P_D(q^{-1}) + q^{-1} \sum_{j=h_i}^{j=h_p} \gamma_j H_{j-d} \right] \bar{e}_u(t) + \left[ \sum_{j=h_i}^{j=h_p} \gamma_j F_j \right] e_y(t) \\ &\quad = S'(q^{-1}) \bar{e}_u(t) + R'(q^{-1}) e_y(t) = 0 \end{aligned} \tag{7.221}$$

where we identify the $S'$ and $R'$ polynomials by:

$$S'(q^{-1}) = P_D(q^{-1}) + q^{-1} \sum_{j=h_i}^{h_p} \gamma_j H_{j-d}(q^{-1}) \tag{7.222}$$

$$R'(q^{-1}) = \sum_{j=h_i}^{h_p} \gamma_j F_j \tag{7.223}$$

Also taking into account the definition of $\bar{e}_u(t)$ given in (7.198), (7.221) becomes:

$$\begin{aligned} &S'(q^{-1}) H_S(q^{-1}) e_u(t) + R'(q^{-1}) H_R(q^{-1}) e_y(t) \\ &\quad = S(q^{-1}) e_u(t) + R(q^{-1}) e_y(t) = 0 \end{aligned} \tag{7.224}$$

and using now the expression of $e_u(t)$ and $e_y(t)$ given by (7.191) and (7.192), one obtains:

$$S(q^{-1}) u(t) = -R(q^{-1}) y(t) + T(q^{-1}) y^*(t+d+1) \tag{7.225}$$

where:

$$T(q^{-1}) = \beta P = \beta[AS + q^{-d-1}B^*R] \tag{7.226}$$

and $P$ defines the resulting closed-loop poles associated to an RST controller. One immediately remarks that similar tracking behavior with the pole placement will be obtained, since $T$ will compensate from the tracking trajectory, the resulting closed-loop poles. It remains to examine in more detail the resulting poles of the closed loop.

### 7.7.2 Closed-Loop Poles

One has the following result.

**Theorem 7.5** (M'Saad et al. 1993a) *The closed-loop poles for the control system minimizing the criterion* (7.193) *in receding horizon sense subject to the constraints* (7.195) *are*:

$$P(q^{-1}) = P_D(q^{-1})P_{GPC}(q^{-1}) \tag{7.227}$$

*where*:

$$P_{GPC}(q^{-1}) = AH_S + q^{-d-1}\sum_{j=h_i}^{h_p} \gamma_j q^j [B^*H_R - AH_S G_{j-d}] \tag{7.228}$$

*and* $P_D(q^{-1})$ *are the poles of the predictor* (7.203).

Therefore, the poles of the predictor appear in the poles of the closed loop and the additional closed-loop poles are introduced by the criterion minimization. Furthermore, if the stochastic disturbance acting on the system can be described as:

$$\bar{A}e_y(t+d+1) = \bar{B}^*\bar{e}_u(t) + Ce(t+d+1) \tag{7.229}$$

where $e(t)$ is a white noise, then taking $P_D = C$ in (7.202) and (7.203), the optimal predictions of $e_y(t+j)$ will be given by (7.203) where $P_D$ is replaced by $C$ and $E_j$ and $F_j$ are solutions of (7.202) in which $P_D$ is replaced by $C$. Consequently, the resulting closed-loop polynomial will include $C$ (instead of $P_D$).

One can also give immediately two interpretations concerning the choice of $P_D$ (or $C$) in relation with the *pole placement*.

1. If one takes $P_D$ as defining the desired dominant poles, the expected effect of the *GPC* will be to introduce auxiliary poles. Therefore, the choices of $\lambda, h_i, h_p, h_c$ will influence the location of the auxiliary poles.
2. If one replaces in (7.202) and (7.203) $P_D$ by $P_F$ (the desired high-frequency auxiliary poles), the high-frequency behavior (and in particular robustness) will be defined essentially by $P_F$ and the various design parameters of $GPC(\lambda, h_i, h_p, h_c)$ will be used to match the desired (low-frequency) performances. While this procedure is feasible it may require a significative number of trials due to the complicate relationship between $\lambda, h_i, h_p, h_c$ and the performances in regulation. However, note that by the intrinsic structure of the controller associated to (7.193) this will not affect the tracking performances (which are those of the pole placement).

*Proof* The closed-loop poles are given by:

$$P = \bar{A}S' + q^{-d-1}\bar{B}^* R' \tag{7.230}$$

The expression of $q^{-d-1}\bar{B}^* R'$ can be further developed taking into account (7.223) and (7.202):

$$\begin{aligned} & q^{-d-1}\bar{B}^* \left( \sum_{j=h_i}^{h_p} \gamma_j F_j \right) \\ & \quad = q^{-d-1}\bar{B}^* \left( \sum_{j=h_i}^{h_p} \gamma_j q^j (q^{-j} F_j + \bar{A}E_j) \right) - q^{-d-1}\bar{B}^* \sum_{j=h_i}^{h_p} \gamma_j q^j \bar{A} E_j \\ & \quad = -q^{-d-1}\bar{B}^* P_D \sum_{j=h_i}^{h_p} \gamma_j q^j - q^{-d-1}\bar{A} \sum_{j=h_i}^{h_p} \gamma_j q^j \bar{B}^* E_j \end{aligned} \tag{7.231}$$

Considering in addition (7.204), one gets:

$$\begin{aligned} & q^{-d-1}\bar{B}^* R' = q^{-d-1}\bar{B}^* \left( \sum_{j=h_i}^{h_p} \gamma_j F_j \right) \\ & \quad = q^{-d-1}\bar{B}^* P_D \sum_{j=h_i}^{h_p} \gamma_j q^j - q^{-d-1}\bar{A} \sum_{j=h_i}^{h_p} \gamma_j q^j (P_D G_{j-d} + q^{-j+d} H_{j-d}) \\ & \quad = q^{-d-1} P_D \left( \sum_{j=h_i}^{h_p} \gamma_j q^j (\bar{B}^* - \bar{A} G_{j-d}) \right) - q^{-1}\bar{A} \sum_{j=h_i}^{h_p} \gamma_j H_{j-d} \end{aligned} \tag{7.232}$$

On the other hand, using (7.222), one has:

$$\bar{A}S' = \bar{A}P_D + q^{-1}\bar{A} \sum_{j=h_i}^{h_p} \gamma_j H_{j-d} \tag{7.233}$$

Introducing (7.232) and (7.233) in (7.230) yields:

$$P = P_D . P_{GPC} \tag{7.234}$$

where $P_{GPC}$ is given by (7.228). □

The stability of the generalized predictive control will require that the polynomial $P_{GPC}$ given by (7.229) be asymptotically stable. This, however, will depend upon the choice of the design parameters $\lambda, h_i, h_p, h_c$ for a given plant model. A number of results are available (Clarke and Mohtadi 1989; M'Saad et al. 1993a) which allows to give rules for choosing the various design parameters. These rules are summarized below:

1. $h_i = d+1$, $h_c < h_p = h_i + h_c - 1$, $\lambda = 0$, $P_{GPC} = \beta_0 B^*(q^{-1})$, $\beta_0 = \frac{1}{b_1}$.
2. $h_i = n_B + d$, $h_c = n_A + n_D$, $h_p \geq n_A + n_B + d - 1$, $\lambda = 0$, $A$ and $B^*$ do not have common factors $P_{GPC} = 1$ and $P = P_D$ (one gets the pole placement).

   This result extends to the case of using $H_R, H_S \neq 1$, provided that $AH_S = \bar{A}$ and $B^* H_R = \bar{B}^*$ do not have common factors and $n_A, n_B$ are replaced by $n_A + n_{H_S}$ and $n_B + n_{H_R}$ respectively.
3. $h_i = d+1$, $h_c = 1$, $h_p \to \infty$, $\lambda = 0$, $H_S = 1 - q^{-1}$, $H_R = 1$, $P_{GPC} = A(q^{-1})$.

However, in most of the cases the critical design parameter for stability as well as for the complexity of the computations is $h_C$.

The basic rules are as follows:

(a) well damped stable plants

$$h_P T_s \geq t_R \text{ (rise time)}$$
$$h_c = 1$$
$$d + 1 \leq h_i \leq n_B + d$$

(b) other cases:

$$h_p \geq n_A + n_B + d + n_{H_R} + n_{H_S}$$
$$1 \leq h_c \leq n_A + n_{H_S}$$
$$d + 1 \leq h_i \leq n_B + d + n_{H_R}$$

Note that in the case of using $H_R$ and $H_S$, the orders of these polynomials should be taken into account.

Unfortunately even using these rules a formal proof from the stability of the resulting closed-loop system is not available. For this reason the criterion of the form of (7.182), (7.185) or (7.193) are extended in order to include a number of terminal constraints on the input and output variables allowing to establish a stability proof. The resulting control which is called Constrained Receding Horizon Predictive Control (CRHPC) requires the output variables to match the reference value over a specified interval at the end of the prediction horizon, while the control variable is allowed only a limited number of projected control variations (Clarke and Scatollini 1991; De Nicolao and Scatollini 1994; Mosca and Zhang 1992).

### 7.7.3 Recursive Solutions of the Euclidian Divisions

The computation of $E_y^0$ given in (7.213) and of the matrix $G$ given in (7.215) requires a recursive form for the minimal order solutions of (7.202) and (7.204) for various values of $j$. The details are given next.

**Recursive Solution of the Euclidian Division (7.202)** Consider the polynomial equation

$$P_d(q^{-1}) = \bar{A}(q^{-1})E_j(q^{-1}) + q^{-j}F_j(q^{-1}) \tag{7.235}$$

where

$$\bar{A}(q^{-1}) = 1 + \bar{a}_1 q^{-1} + \cdots + \bar{a}_{n_{\bar{A}}} q^{-n_{\bar{A}}} \tag{7.236}$$

$$P_D(q^{-1}) = 1 + p_1^D q^{-1} + \cdots + p_{n_{P_D}}^D q^{-n_{P_D}} \tag{7.237}$$

The involved recursive solution consists in determining the solution $(E_j(q^{-1}), F_j(q^{-1}))$ from the solution $(E_{j-1}(q^{-1}), F_{j-1}(q^{-1}))$. These polynomial pairs respectively represent the minimal order solutions of the polynomial equations

$$P_D(q^{-1}) = \bar{A}(q^{-1})E_j(q^{-1}) + q^{-j}F_f(q^{-1}) \tag{7.238}$$

$$P_D(q^{-1}) = \bar{A}(q^{-1})E_{j-1}(q^{-1}) + q^{-j+1}F_{j-1}(q^{-1}) \tag{7.239}$$

The minimal order solutions of the polynomial equations (7.237)–(7.238) can be given the following forms

$$E_j(q^{-1}) = e_0^j + e_1^j q^{-1} + \cdots + e_{j-1}^j q^{-j+1} \tag{7.240}$$

$$E_{j-1}(q^{-1}) = e_0^{j-1} + e_1^{j-1} q^{-1} + \cdots + e_{j-2}^{j-1} q^{-j+2} \tag{7.241}$$

$$F_j(q^{-1}) = f_0^j + f_1^j q^{-1} + \cdots + f_{n_{F_j}}^j q^{n_{F_j}} \tag{7.242}$$

$$F_{j-1}(q^{-1}) = f_0^{j-1} + f_1^{j-1} q^{-1} + \cdots + f_{n_{F_{j-1}}}^{j-1} q^{n_{F_{j-1}}} \tag{7.243}$$

where $n_{F_j} \leq \max(n_{P_D} - j, n_{\bar{A}} - 1)$. Subtracting term by term (7.238) from (7.237) yields

$$\bar{A}(q^{-1})[E_j(q^{-1}) - E_{j-1}(q^{-1})] + q^{-j+1}[q^{-1}F_j(q^{-1}) - F_{j-1}(q^{-1})] = 0 \tag{7.244}$$

As the polynomial $\bar{A}(q^{-1})$ and $q^{-j+1}$ are coprime, the polynomial equation (7.243) may be rewritten as follows

$$E_j(q^{-1}) = E_{j-1}(q^{-1}) + e_{j-1}^j q^{-j+1} \tag{7.245}$$

and

$$q^{-1}F_j(q^{-1}) = F_{j-1}(q^{-1}) - e_{j-1}^j \bar{A}(q^{-1}) \tag{7.246}$$

Bearing in mind the forms of the polynomials $E_j(q^{-1})$ and $E_{j-1}(q^{-1})$ given by (7.239) and (7.240), and taking in account (7.244) the polynomial $E_j(q^{-1})$ may be written as:

$$E_j(q^{-1}) = e_0 + e_1 q^{-1} + \cdots + e_{j-1} q^{-j+1} \tag{7.247}$$

The required recursive solution is then provided by simple term identification from the polynomial equations (7.244) and (7.245). The minimal order solution for $j = 1$ is used to initialize the algorithm

$$e_0 = 1 \tag{7.248}$$

$$f_i^1 = p_{i+1}^D - e_0 \bar{a}_{i+1}, \quad 0 \le i \le n_{F_1} \tag{7.249}$$

For $j \ge 2$, one has:

$$e_{j-1} = f_0^{j-1} \tag{7.250}$$

$$f_i^j = f_{i+1}^{j-1} - e_{j-1} \bar{a}_{i+1}, \quad 0 \le i \le n_{F_j} \tag{7.251}$$

**Recursive Solution of the Euclidian Division (7.204)** Consider the polynomial equation:

$$\bar{B}^*(q^{-1}) E_j(q^{-1}) = P_D(q^{-1}) G_{j-d}(q^{-1}) + q^{-j+d} H_{j-d}(q^{-1}) \tag{7.252}$$

where:

$$\bar{B}^*(q^{-1}) = \bar{b}_0^* + \bar{b}_1^* q^{-1} + \cdots + \bar{b}_{n_{\bar{B}^*}}^* q^{-n_{\bar{B}^*}}; \quad \bar{b}_i^* = \bar{b}_{i+1} \tag{7.253}$$

$$P_D(q^{-1}) = 1 + p_1^D q^{-1} + \cdots + p_{n_{P_D}}^D q^{-n_{P_D}} \tag{7.254}$$

The involved recursive solution consists in determining the solution ($G_{j-d-1}(q^{-1})$, $H_{j-d}(q^{-1})$) from the solution ($G_{j-d-1}(q^{-1})$, $H_{j-d-1}(q^{-1})$). These polynomial pairs respectively represent the minimal order solutions of the following polynomial equations:

$$\bar{B}^*(q^{-1}) E_j(q^{-1}) = P_D(q^{-1}) G_{j-d}(q^{-1}) + q^{-j+d} H_{j-d}(q^{-1}) \tag{7.255}$$

$$\bar{B}^*(q^{-1}) E_{j-1}(q^{-1}) = P_D(q^{-1}) G_{j-d-1}(q^{-1}) + q^{-j+d+1} H_{j-d-1}(q^{-1}) \tag{7.256}$$

Taking into account the degrees of polynomials $P_D(q^{-1})$ and $\bar{B}^*(q^{-1}) E_j(q^{-1})$, the minimal order solutions of the polynomial equations (7.254) and (7.255) may be given in the following forms:

$$G_{j-d}(q^{-1}) = g_0^{j-d} + g_1^{j-d} q^{-1} + \cdots + g_{j-d-1}^{j-d} q^{-j+d+1} \tag{7.257}$$

$$G_{j-d-1}(q^{-1}) = g_0^{j-d-1} + g_1^{j-d-1} q^{-1} + \cdots + g_{j-d-2}^{j-d-1} q^{-j+d+2} \tag{7.258}$$

$$H_{j-d}(q^{-1}) = h_0^{j-d} + h_1^{j-d} q^{-1} + \cdots + h_{n_H}^{j-d} q^{-n_H} \tag{7.259}$$

$$H_{j-d-1}(q^{-1}) = h_0^{j-d-1} + h_1^{j-d-1} q^{-1} + \cdots + h_{n_H}^{j-d-1} q^{-n_H} \tag{7.260}$$

where $n_H \le \max(n_{p_D} - 1, n_{\bar{B}^*} + d)$. Subtracting term by term (7.255) from (7.254) yields:

$$\begin{aligned} P_D(q^{-1})[G_{j-d}(q^{-1}) - G_{j-d-1}(q^{-1})] + q^{-j+d+1}[q^{-1} H_{j-d}(q^{-1}) \\ - H_{j-d-1}(q^{-1})] - q^{j-d-1} \bar{B}^*(q^{-1})[E_j(q^{-1}) - E_{j-1}(q^{-1})] = 0 \end{aligned} \tag{7.261}$$

As the polynomial $P_D(q^{-1})$ and $q^{-j+1}$ are coprime, the polynomial equation (7.260) may be rewritten as follows:

$$G_{j-d}(q^{-1}) = G_{j-d-1}(q^{-1}) + g_{j-d-1}^{j-d} q^{-j+d+1} \tag{7.262}$$

and:

$$q^{-1}H_{j-d}(q^{-1}) = H_{j-d-1}(q^{-1}) - g_{j-d-1}^{j-d} P_D(q^{-1}) + e_{j-1}q^{-d}\bar{B}^*(q^{-1}) \tag{7.263}$$

Bearing in mind the forms of the polynomials $G_{j-d}(q^{-1})$ and $G_{j-d-1}(q^{-1})$ given by (7.256) and (7.257), the polynomial $G_{j-d}(q^{-1})$ may be given in the form:

$$G_{j-d}(q^{-1}) = g_0 + g_1 q^{-1} + \cdots + g_{j-d-1}q^{-j+d+1} \tag{7.264}$$

The required recursive solution is then provided by simple term by term identification from the polynomial equations (7.261) and (7.262). To do so, let:

$$\bar{B}^*(q^{-1})E_d(q^{-1}) = \beta_0 + \beta_1 q^{-1} + \cdots + \beta_{n_{\bar{B}^*}+d-1}q^{-n_{\bar{B}^*}-d+1} \tag{7.265}$$

and notice that the obvious minimal order solution for $j = d + 1$ allows to initialize the algorithm as follows:

$$g_0 = \beta_0 \tag{7.266}$$

$$h_i^1 = \beta_{i+1} - g_o p_{i+1}^D + e_d \bar{b}_{i+2-d}, \quad 0 \le i \le n_H \tag{7.267}$$

For $j > d + 1$, one has:

$$g_{j-d-1} = h_0^{j-d-1} \quad \text{if } d > 0 \tag{7.268}$$

$$g_{j-d-1} = h_0^{j-d-1} + e_{j-1}\bar{b}_0^* \quad \text{if } d = 0 \tag{7.269}$$

$$h_i^{j-d} = h_{i+1}^{j-d-1} - g_{j-d-1}p_{i+1}^D, \quad 0 \le i \le d-2 \tag{7.270}$$

$$h_i^{j-d} = h_{i+1}^{j-d-1} - g_{j-d-1}p_{i+1}^D + e_{j-1}\bar{b}_{i+2-d}, \quad d-1 \le i \le n_H \tag{7.271}$$

## 7.8 Linear Quadratic Control

The plant model is described by:

$$A(q^{-1})y(t+d+1) = B^*(q^{-1})u(t) + v(t+d+1) \tag{7.272}$$

where $v(t + d + 1)$ is a disturbance which can be deterministic or stochastic. We make similar hypothesis upon the plant model as for the *pole placement* expect hypothesis (H4) which is replaced by:

(H4) The eventual common factors of $A(q^{-1})$ and $B^*(q^{-1})$ are inside the unit circle.

The objective will be to find an admissible control $u(t)$ which minimizes in a receding horizon sense the quadratic criterion:

$$J(t,T) = \lim_{T \longrightarrow \infty} \frac{1}{T} \sum_{j=1}^{T} [P(q^{-1})(y(t+j) - y^*(t+j))]^2 + \lambda[Q(q^{-1})u(t+j)]^2 \tag{7.273}$$

or a criterion of the form (PSMR):

$$J(t,T) = \lim_{T \longrightarrow \infty} \frac{1}{T} \sum_{j=1}^{T} [e_y(t+j)]^2 + \lambda [Q(q^{-1}) e_u(t+j)]^2 \tag{7.274}$$

where $\lambda$ is a positive scalar and $\{e_y(t)\}$ and $\{e_u(t)\}$ respectively denote the output-error and input-error given by

$$e_y(t) = y(t) - \beta B^*(q^{-1}) y^*(t); \quad \beta = 1/B^*(1) \tag{7.275}$$

$$e_u(t) = u(t) - \beta A(q^{-1}) y^*(t+d+1) \tag{7.276}$$

Criterion (7.274) allows a close relationship with the *pole placement* to be established and to obtain a decoupling between tracking and regulation performances. As indicated in Sect. 7.7, criterion (7.274) can be viewed as an extension of the generalized predictive control where the prediction and control horizon goes to $\infty$.

The solution of the linear quadratic control for the model of (7.272) is obtained via a state space reformulation of the problem and the use of standard results in linear quadratic control (Anderson and Moore 1971).

We will consider the quadratic criterion (7.274) and similarly to the development in Sect. 7.7, in order to minimize (7.274) we will use a model expressing the relation between $e_u(t)$ and $e_y(t)$. Using the results of Sect. 7.7 ((7.197) through (7.201)) we will get for $Q(q^{-1}) = H_S(q^{-1})/H_R(q^{-1})$:

$$\bar{A}(q^{-1}) e_y(t) = \bar{B}(q^{-1}) \bar{e}_u(t) \tag{7.277}$$

where:

$$\bar{e}_u(t) = Q(q^{-1}) e_u(t) \tag{7.278}$$

$$\bar{A}(q^{-1}) = A(q^{-1}) H_S(q^{-1}), \qquad \bar{B}(q^{-1}) = q^{-d} B(q^{-1}) H_R(q^{-1}) \tag{7.279}$$

To the input-output representation of (7.277) one associates a state space realization in observable canonical form:

$$x(t+1) = Ax(t) + b\bar{e}_u(t) \tag{7.280}$$

$$e_y(t) = c^T x(t) \tag{7.281}$$

where:

$$A = \begin{bmatrix} -\bar{a}_1 & & & \\ \vdots & & & \\ \vdots & & I_{n-1} & \\ \vdots & & & \\ -\bar{a}_n & 0 & \cdots & 0 \end{bmatrix}, \qquad b = \begin{bmatrix} \bar{b}_1 \\ \vdots \\ \vdots \\ \vdots \\ \bar{b}_n \end{bmatrix}, \qquad c^T = [1, 0, \ldots, 0]$$

where $n$ denotes the design model order, i.e. $n = \max(na + n_{H_S}, nb + d + n_{H_R})$. Notice that the $d$ first elements of the input matrix $b$ are zero and that the last elements of the first column of the state matrix $A$ (resp. the input matrix $b$) are zero if $na + n_{H_S} < n$ (resp. $nb + d + n_{H_R} < n$). With this representation of the model (7.277), the criterion (7.274) takes the form:

$$J(t,T) = \lim_{T\to\infty} \frac{1}{T}\sum_{j=1}^{T} x^T(t+j)cc^T x(t+j) + \lambda\,(\bar{e}_u(t+j))^2 \tag{7.282}$$

Assuming in a first stage that the state $x(t)$ of the state space model (7.280) and (7.281) is measurable, the optimal control law $\bar{e}_u(t)$ is given by:

$$\bar{e}_u(t) = -\frac{b^T\Gamma A}{b^T\Gamma b + \lambda}x(t) \tag{7.283}$$

where $\Gamma$ is the positive definite solution of the Algebraic Riccati Equation (ARE)

$$A^T\Gamma A - \Gamma - A^T\Gamma b(b^T\Gamma b + \lambda)^{-1}b^T\Gamma A + cc^T = 0 \tag{7.284}$$

The existence and the unicity of such positive definite solution for (7.284) is guaranteed by the following conditions:

- The pair $(A, b)$ is stabilizable (this is assured by the hypothesis (H4) made upon the system).
- The pair $(c^T, A)$ is detectable (in the present case, the pair $(c^T, A)$ is observable by construction which implies detectability).

In a second stage, one will use the certainty equivalence principle and the non-accessible state vector $x(t)$ will be replaced by an observed (estimated) state $\hat{x}(t)$ generated by an observer of the form:

$$\hat{x}(t+1) = A\hat{x}(t) + b\bar{e}_u(t) + k[e_y(t) - \hat{e}_y(t)] \tag{7.285}$$

$$\hat{e}_y(t) = c^T\hat{x}(t) \tag{7.286}$$

Given the desired dynamics of the observer specified by a polynomial $P_D(q^{-1}) = \det[I - z^{-1}(A - kc^T)]$, one has:

$$k^T = [k_1, \ldots, k_n] \tag{7.287}$$

with:

$$k_i = p_i - a_i \tag{7.288}$$

(where $p_i$ are the coefficients of the monic polynomials $P_D$). This observer corresponds also to a steady state optimal Kalman filter for a disturbance $\bar{v}(t)$ added to (7.277) and characterized by:

$$\bar{v}(t) = C(q^{-1})w(t); \qquad C(q^{-1}) = P_D(q^{-1}) \tag{7.289}$$

where $w(t)$ is a white noise sequence. Therefore the control law expressed in terms of $\bar{e}_u(t)$ takes the form:

$$\bar{e}_u(t) = -\frac{b^T\Gamma A}{b^T\Gamma b + \lambda}\hat{x}(t) = -l^T\hat{x}(t) \tag{7.290}$$

where $\hat{x}(t)$ is given by (7.285) and (7.286) and $\Gamma$ is the solution of (7.284). Taking into account (7.290), the observer equation in closed loop becomes:

$$\hat{x}(t+1) = A\hat{x}(t) - bl^T\hat{x}(t) - kc^T\hat{x}(t) + ke_y(t) \tag{7.291}$$

and the transfer function between $e_y(t)$ and $e_u(t)$ will be given by:

$$H_c(z) = \frac{R(z)}{S(z)} = l^T(zI - A + kc^T + bl^T)^{-1}k \tag{7.292}$$

where:

$$S'(z) = \det(zI - A + kc^T + bl^T) \tag{7.293}$$

$$R'(z) = l^T adj(zI - A + kc^T + bl^T)k \tag{7.294}$$

This allows to write the controller under the form:

$$\begin{aligned} &S'(q^{-1})H_S(q^{-1})e_u(t) + R'(q^{-1})H_R(q^{-1})e_y(t) \\ &\quad = S(q^{-1})e_u(t) + R(q^{-1})e_y(t) \end{aligned} \tag{7.295}$$

from which one obtains the final RST controller form by taking in account the expressions for $e_u(t)$ and $e_y(t)$:

$$S(q^{-1})u(t) = -R(q^{-1})y(t) + T(q^{-1})y^*(t+d+1) \tag{7.296}$$

where:

$$T(q^{-1}) = \beta P = \beta[AS + q^{-d-1}B^*R] \tag{7.297}$$

This controller assures certain tracking performances (defined by $y^*$ and $B^*$) independently of the resulting closed-loop poles (which will include the poles of the observer defined by $P_D$).

For further details on this control strategy see Samson (1982), Lam (1982), M'Saad et al. (1990), M'Saad and Sanchez (1992).

## 7.9 Concluding Remarks

This chapter has covered a number of digital control strategies which are currently used in practice either with constant controller parameters computed on the basis of identified models or as the underlying control algorithm for adaptive control. Some of their basic features are summarized next:

1. All the resulting controllers belong to the class of controllers with two-degree of freedom which handle both tracking and regulation specification. They have a tri-branched structure and the basic equation is:

   $$S(q^{-1})u(t) + R(q^{-1})y(t) = T(q^{-1})y^*(t+d+1)$$

   where $u(t)$ is the control input, $y(t)$ is the measured plant output, $y^*(t+d+1)$ is the desired tracking trajectory and $R$, $S$, $T$ are polynomials in $q^{-1}$.

2. The complexity of the controller depends primarily upon the complexity of the plant model and is independent of the type of control strategy used.
3. The design of the controller involves two stages:
   (i) Computation of the polynomials $S(q^{-1})$ and $R(q^{-1})$ in order to match the desired regulation performances.
   (ii) Computation of the polynomials $T(q^{-1})$ in order to approach (or to match) the desired tracking performances.
4. Among the digital control strategies presented, one can distinguish:

   - one step ahead predictive control;
   - long range predictive control.

   In one step ahead predictive control, the objective can be expressed in terms of the predicted output at $t + d + 1$. In long range predictive control, the control objective is expressed in terms of future values of the input and the output over a certain horizon.
5. Control strategies for deterministic environments and stochastic environments are similar. The only difference comes from the choice of some of the closed-loop poles (the predictor or observer poles). In the deterministic case, these poles are chosen by the designer. In the stochastic case, their values are determined by the disturbance model. Therefore:

   - every deterministic control design is optimal in a stochastic sense for a specific disturbance, model;
   - every stochastic control design lead to the assignment of some of the closed-loop poles at values defined by the disturbance model.

6. The digital control strategies can be classified with respect to the hypotheses made upon the plant model. A first type of methods makes the assumption that: *either the $B^*$ polynomial is asymptotically stable (stable zeros) or there are polynomials $Q(q^{-1})$ and $P(q^{-1})$ and a scalar $\lambda$ such that $\lambda QA + PB^*$ is asymptotically stable.* The following control strategies belongs to this type:

   - tracking and regulation with independent objectives;
   - tracking and regulation with weighted input;
   - minimum variance tracking and regulation;
   - generalized minimum variance and regulation.

   A second type of methods does not require that $B^*$ be asymptotically stable but requires: *either that $B^*$ and $A$ do not have common factors or if they do have common factors, they are inside the unit circle*. The following control strategies belong to this class:

   - pole placement;
   - generalized predictive control;
   - linear quadratic control.

## 7.10 Problems

**7.1** Consider an analog PI controller in Laplace from:

$$U(s) = K\left[1 + \frac{1}{sT_i}\right][R(s) - Y(s)]$$

where $U(s)$, $R(s)$ and $Y(s)$ are respectively the Laplace transforms of the plant input $u(t)$, of the reference $r(t)$ and of the plant output $y(t)$.

1. Give a discretized version of the analog PI controller using the approximation ($T_s$ = sampling period):

$$\int x(t)dt \approx \frac{T_s}{1 - q^{-1}} x(t)$$

2. Show that the resulting controller has an RST form with $R(q^{-1}) = T(q^{-1})$.
3. Establish the relationship between continuous-time parameters and discrete-time parameters.

**7.2** Consider an analog PID operating on the regulation error $[r(t) - y(t)]$ with the transfer function:

$$H_{PID_1}(s) = K\left[1 + \frac{1}{T_i s} + \frac{T_d s}{1 + \frac{T_d}{N} s}\right]$$

1. Give the discretized version of the analog PID using the approximations:

$$\frac{dx}{dt} \approx \frac{1 - q^{-1}}{T_s} x(t)$$

$$\int x(t)dt \approx \frac{T_s}{1 - q^{-1}} x(t)$$

2. Show that the resulting discrete-time controller has an RST form with $R(q^{-1}) = T(q^{-1})$.
3. Establish the relationship between continuous-time and discrete-time parameters.
4. Assuming that the equivalent $R$-$S$-$T$ digital controller has been designed using a discrete-time control strategy, find the conditions for the existence of an equivalent continuous-time PID having a transfer function of the type $H_{PID_1}(s)$.

**7.3** Consider an analog PID with (filtered) integral action on the error and proportional and (filtered) derivative action on the output only:

$$U(s) = \frac{1}{sT_i(1 + s\frac{T_d}{N})}[R(s) - Y(s)] - KY(s) - \frac{KT_d s}{1 + s\frac{T_d}{N}} Y(s)$$

1. Give the discretized version using the same approximations as in Problem 7.2.
2. Show that the values of the coefficients of $R(q^{-1})$ and $S(q^{-1})$ polynomials are the same as in Problem 7.2 and that $T(q^{-1}) = R(1)$.

**7.4** Consider a continuous-time model:

$$G(s) = \frac{Ge^{-s\tau}}{1+sT}$$

with: $G = 1$, $T = 10$, $\tau = 3$. Choose a sampling period $T_s = 5s$ (the corresponding discrete-time model is: $d = 0$, $B(q^{-1}) = 0.1813q^{-1} + 0.2122q^{-2}$; $A(q^{-1}) = 1 - 0.6065q^{-1}$).

The digital PID has the structure:

$$R(q^{-1}) = r_0 + r_1q^{-1}$$
$$S(q^{-1}) = (1 - q^{-1})(1 + s_1'q^{-1})$$
$$T(q^{-1}) = R(q^{-1}) \quad \text{or} \quad T(q^{-1}) = R(1)$$

1. Compute a digital PID such that the desired closed-loop poles be defined by the poles of the discretized second order continuous-time model $\frac{\omega_0^2}{\omega_0^2+2\zeta\omega_0 s+s}$ with $\omega_0 = 0.15$, $\zeta = 0.8$.
2. Compare the results to a step command for $T(q^{-1}) = R(q^{-1})$ and $T(q^{-1}) = R(1)$. Explain the differences.

**7.5** Design an $R$-$S$-$T$ controller with integral action using tracking and regulation with independent objectives for the plant model:

$$B(q^{-1}) = b_1q^{-1} + b_2q^{-2}$$
$$A(q^{-1}) = 1 - 0.6q^{-1}$$

where $B(q^{-1})$ is given by:

(1) $b_1 = 1$; $b_2 = 0$

(2) $b_1 = 0.7$; $b_2 = 0.3$

(3) $b_1 = 0.55$; $b_2 = 0.45$

Define the same dynamics for tracking and regulation:

$$P(q^{-1}) = A_m(q^{-1}) = (1 - 0.3q^{-1})^2$$

Examine the behavior of the control input and output in the presence of output step disturbances and for reference step changes.

**7.6** For the same data use pole placement. Compare the results with those obtained in Problem 7.5.

**7.7** The discrete-time version of a PI + Smith predictor is shown in Fig. 7.10. The plant model is given by:

$$\frac{q^{-d}B(q^{-1})}{A(q^{-1})} = \frac{q^{-d}(b_1q^{-1})}{1 + a_1q^{-1}}$$

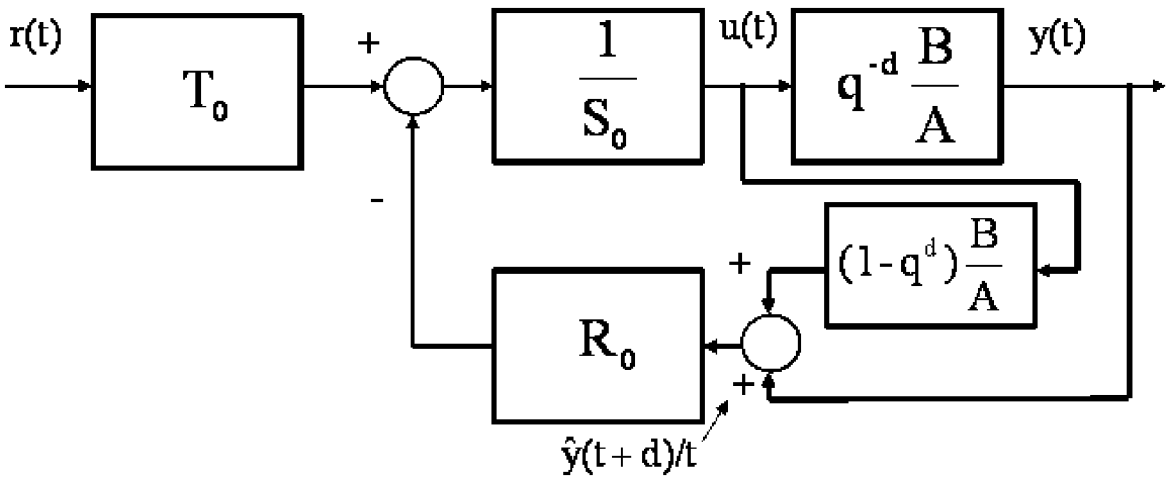

**Fig. 7.10** Discrete-time version of the PI + Smith predictor

The polynomials $R_0, S_0, T_0$ corresponding to the discrete-time PI controller are given by:

$$S_0(q^{-1}) = 1 - q^{-1}; \qquad R_0(q^{-1}) = T_0(q^{-1}) = r_0 + r_1 q^{-1}$$

where $R_0(q^{-1})$ and $S_0(q^{-1})$ are solutions of:

$$A(q^{-1})(1 - q^{-1}) + B(q^{-1})(r_0 + r_1 q^{-1}) = P_0(q^{-1})$$

where $P_0(q^{-1})$ corresponds to some desired closed-loop poles for the case without delay ($d = 0$ and $(\deg P_0)_{\max} = 2$).

1. Give an $R$-$S$-$T$ form for the PI + Smith predictor
2. Show that the closed-loop poles are given by:

$$P(q^{-1}) = A(q^{-1})P_0(q^{-1})$$

3. Show that the transfer operator from the reference to the output is given by:

$$H_{yr}(q^{-1}) = \frac{q^{-d}B(q^{-1})T_0(q^{-1})}{P_0(q^{-1})}$$

4. Show that the control law for the case $d > 0$ can be equivalently expressed by:

$$S(q^{-1})u(t) = T_0(q^{-1})r(t) - R_0(q^{-1})\hat{y}(t + d/t)$$

   where:

$$\hat{y}(t + d/t) = \frac{(1 - q^{-d})B(q^{-1})}{A(q^{-1})}u(t) + y(t)$$

5. Show that the above predictor satisfies the $d$-step ahead predictor equation and that the dynamics of the predictor is defined by $A(q^{-1})$.

**7.8** For the ARMAX model in (7.149) compute the control law $u(t) = f_u[y(t), y(t-1), \ldots, u(t), u(t-1), \ldots]$ which minimizes the following criterion:

$$\min_{u(t)} J(t + d + 1) = \mathbf{E}\{[P_F(q^{-1})y(t + d + 1) - P_F(q^{-1})y^*(t + d + 1)]^2\}$$

where $P_F(q^{-1})$ is an asymptotically stable polynomial. What are the resulting closed-loop poles?

**7.9** For the model in (7.181) assuming $v(t+i) = 0, i > 0$ compute the control which minimizes in a receding horizon sense the criterion:

$$J(t, h_p, h_c, h_i) = E\left\{\sum_{j=h_i}^{h_p} [y(t+j) - y^*(t+j)]^2 + \lambda[Q(q^{-1})u(t+j-hi)]^2\right\}$$

subject to the constraint:

$$Q(q^{-1})u(t+i) = 0; \quad h_c \leq i \leq h_p - d - 1$$

**7.10** Give the expression of the solutions for the minimization of the criterion in (7.193) for the case $h_i = 1$.

**7.11** For the tracking and regulation with weighted input show that choosing $Q(q^{-1}) = 1 - q^{-1}$, preserve the integral action of the controller i.e., $S(q^{-1}) = (1 - q^{-1})S'(q^{-1})$.

**7.12** Derive the expression of the controller for the tracking and regulation with weighted input when $Q(q^{-1}) = H_R(q^{-1})/H_S(q^{-1})$.

**7.13** Consider the system model:

$$A(q^{-1})y(t+1) = B^*(q^{-1})u(t) + v(t+1)$$
$$D(q^{-1})v(t) = C(q^{-1})e(t)$$

where $e(t)$ is a white noise sequence and:

$$A(q^{-1}) = 1 + a_1q^{-1} + a_2q^{-2}$$
$$B(q^{-1}) = b_1q^{-1} + b_2q^{-2}$$
$$C(q^{-1}) = 1 + c_1q^{-1}$$
$$D(q^{-1}) = 1 - q^{-1}$$

Let

$$e_u(t) = u(t) - \beta A(q^{-1})y^*(t+1)$$
$$e_y(t) = y(t) - \beta B^*(q^{-1})y^*(t)$$

where:

$$\beta = [b_1 + b_2]^{-1}$$

and $y^*(t)$ is the reference sequence.

1. Determine the sequence $\{\bar{e}_u(t)\}$ which minimizes the following criterion:

$$J(t, h_p) = \sum_{j=1}^{h_p} \hat{e}_y^2(t+j/t) + \lambda \bar{e}_u^2(t+j-1)$$

subject to the constraint:

$$\bar{e}_u(t+i) = 0; \quad i \geq 1$$

where $\bar{e}_u(t) = D(q^{-1})e_u(t)$, $\hat{e}_y(t+j/t)$ is the $j$-step ahead predictor of $e_y(t+j)$, $h_p$ is the prediction horizon and $\lambda > 0$.

Hint: Compute first the optimal $j$-step ahead predictor minimizing:

$$\mathbf{E}\{[e_y(t+j) - \hat{e}_y(t+j/t)]^2\}$$

2. Show that the resulting control law has the form:

$$S'(q^{-1})\bar{e}_u(t) + R(q^{-1})e_y(t) = 0$$

with

$$S'(q^{-1}) = 1 + s'_1 q^{-1}$$
$$R(q^{-1}) = r_0 + r_1 q^{-1} + r_2 q^{-2}$$

3. Give the expression of the controller generating $u(t)$. Show that it may correspond to a digital PID

**7.14** Consider the plant model:

$$A(q^{-1})y(t) = q^{-d-1}B^*(q^{-1})u(t) + C(q^{-1})e(t)$$

where $e(t)$ is a white noise.

Assume that the polynomials $R_0(q^{-1})$, $S_0(q^{-1})$ are given and that they stabilize the fictitious plant model: $A(q^{-1})y(t) = B^*(q^{-1})u(t)$, i.e.,

$$A(z^{-1})S_0(z^{-1}) + B^*(z^{-1})R_0(z^{-1}) = P_0(z^{-1})$$

with: $P_0(z^{-1}) = 0 \Longrightarrow |z| < 1$.

The $d+1$ step ahead predictor for $y(t)$ is given by:

$$C(q^{-1})\hat{y}(t+d+1/t) = F(q^{-1})y(t) + B^*(q^{-1})E(q^{-1})u(t)$$

where $E$ and $F$ are solutions of:

$$A(q^{-1})E(q^{-1}) + q^{-d-1}F(q^{-1}) = C(q^{-1})$$

The control law applied to the plant is:

$$S_0(q^{-1})u(t) = R_0(q^{-1})[y^*(t+d+1) - \hat{y}(t+d+1/t)]$$

Show that this control law minimizes:

$$J(t+d+1) = \mathbf{E}\{[y(t+d+1) - y^*(t+d+1) + \lambda u_f(t)]^2\}$$

where:

$$u_f(t) = \frac{S_0(q^{-1})}{R_0(q^{-1})}u(t); \qquad \lambda = \frac{r_0}{s_0}$$

# Chapter 8
# Robust Digital Control Design

## 8.1 The Robustness Problem

One of the basic questions in control design is how to account for the discrepancies between the model of the plant used for design and the true model. A number of factors may be responsible for modeling errors. A generic term for this modeling error is *model uncertainty* which can itself be represented mathematically in different ways. Two major classes of uncertainty are: structured uncertainty (more commonly termed parametric uncertainty) and unstructured uncertainty (usually specified in the frequency domain). The use of an unstructured uncertainty description was one of the starting points for the impressive development of robust control design in the 1980s (Zames 1981).

Robust control design is one way of dealing with uncertainty upon the parameters of the plant model in order to guarantee the stability of the closed-loop system and a certain level of performance for a given domain of parameter variations. However, for large variations of the plant model characteristics it may become difficult to assure a satisfactory level of performance and even to guarantee stability. On the other hand, adaptive control is an appropriate solution for dealing with parametric uncertainties.

In the past, the development of adaptation mechanisms has been emphasized in the field of adaptive control without taking into account the high sensitivity of some basic underlying linear control design with respect to plant model uncertainties. This may cause serious problems in practice. Therefore, the robustness of the basic linear design is an important issue in adaptive control.

An adaptive control system responds directly to structured parametric uncertainty by the adjustment of model parameters. Therefore, the unstructured modeling errors should be considered by the linear control law associated with the adaptive scheme. Since the unstructured modeling errors are in general large at high frequencies, the underlying linear controller should be designed such that good robustness is assured with respect to additive unstructured uncertainties located in the high-frequency region.

The aim of this chapter is to present the necessary understanding and some methods for designing robust digital controllers starting from the control strategies presented in Chap. 7.[1] What is perhaps more important, is the fact that this chapter will provide guidelines for the design of the underlying linear controller used in adaptive control. Since identification is a basic step in implementing a high-performance control system, the designer will have at his disposal one or several models which can be used for the design.

In the context of robust control design, the model used for design will be termed the *nominal model*. The design for the nominal model will guarantee the *nominal stability* (i.e., the closed-loop system will be asymptotically stable for the nominal model) and the *nominal performance*. However, in the presence of uncertainties the various true plant models belong to a family of plant models denoted by $\mathscr{P}$ which will be characterized by the nominal model and the uncertainty model (parametric or non parametric). The objective will be to design a control system which will guarantee the stability of the closed-loop system for all the plant models belonging to $\mathscr{P}$. The system will be said to be *robustly stable* if it is asymptotically stable for all the models belonging to $\mathscr{P}$.

The problem of *robust performance* is more difficult to solve and adaptive control is a solution for improving the performance of a robust control system. The approach taken here is to consider unstructured uncertainty in the frequency domain (a disc in the complex plane).[2] There are several reasons for making this choice.

1. One generally uses lower-order models which describe the behavior of the plant in a certain frequency region. Therefore unmodeled unstructured dynamics are present and they are very conveniently described by unstructured uncertainties in the frequency domain.
2. The quality of the estimated models will depend upon the frequency content of the excitation signal. Often, one uses low-frequency excitation signals. Therefore in some frequency ranges (generally above a certain frequency) the precision of the model is low. This model error can also be represented as an unstructured uncertainty in the frequency domain.
3. With a certain degree of conservativeness, unstructured uncertainty can also cover parametric uncertainty.
4. A complete and simple theory for robust stability is available for the case of unstructured uncertainty in the frequency domain.
5. Robustness indicators and design rules can be defined that are appealing and easy to understand.

The assessment of the robustness of a certain design passes through the examination of the frequency characteristics of the modulus of the various sensitivity functions. The sensitivity functions are introduced in Sect. 8.2. The concepts of robustness margins and their relevance in conjunction with model uncertainties and robust stability are discussed in Sect. 8.3. This leads to the definition of *templates* for the sensitivity functions which should be considered for obtaining a robust control system (Sect. 8.4). The properties of the sensitivity functions are examined in Sect. 8.5 in order to identify the *knobs* which allow the sensitivity functions to be shaped. Section 8.6 will present several methods for shaping the sensitivity functions. Other design methods are briefly recalled in Sect. 8.7. An example of robust control design is given in Sect. 8.8.

[1]Robustness with respect to plant model uncertainties is not a *god given* property for some control strategies. It is the wise choice of some design parameters which can assure the robustness of a given control strategy in a specific context.

[2]At a given frequency the point belonging to the Nyquist plot of the true plant model lies in a disc of a given radius centered on the corresponding point belonging to the Nyquist plot of the nominal model.

Robustness analysis and design are presented in the context of pole placement; however, the results are applicable "mutatis-mutandis" to the other control strategies presented in Chap. 7.

## 8.2 The Sensitivity Functions

The closed-loop system operates in presence of disturbances and it is important to assess the effect of these disturbances upon the plant output and input. Their effects can be analyzed through the sensitivity functions. Furthermore these sensitivity functions play a crucial role in the robustness analysis of the closed-loop system with respect to modeling errors. These functions (or some of them) will be *shaped* in order to assure *nominal performances* for the rejection of the disturbances and the stability (and performances) of the closed-loop system in presence of model mismatch (these properties are called *robust stability* and *robust performance*).

Three types of disturbance will be considered: output disturbance, input disturbance and measurement noise (see Fig. 8.1).

The transfer function between the output disturbance $p(t)$ and the plant output $y(t)$ (*output sensitivity function*) is given by:

$$S_{yp}(z^{-1}) = \frac{A(z^{-1})S(z^{-1})}{A(z^{-1})S(z^{-1}) + z^{-d}B(z^{-1})R(z^{-1})} = \frac{A(z^{-1})S(z^{-1})}{P(z^{-1})} \tag{8.1}$$

(This function is often simply called *the sensitivity function* in the robust control literature and is denoted by $\mathcal{S}$.)

The transfer function between the disturbance $p(t)$ and the plant input $u(t)$ (*input sensitivity function*) is given by:

$$S_{up}(z^{-1}) = \frac{-A(z^{-1})R(z^{-1})}{A(z^{-1})S(z^{-1}) + z^{-d}B(z^{-1})R(z^{-1})} = \frac{-A(z^{-1})R(z^{-1})}{P(z^{-1})} \tag{8.2}$$

(This function is often denoted by $\mathcal{U}$ in the literature.)

The transfer function between the measurement noise $b(t)$ and the plant output $y(t)$ (*noise sensitivity function*) is given by:

$$S_{yb}(z^{-1}) = \frac{-z^{-d}B(z^{-1})R(z^{-1})}{A(z^{-1})S(z^{-1}) + z^{-d}B(z^{-1})R(z^{-1})} = \frac{-z^{-d}B(z^{-1})R(z^{-1})}{P(z^{-1})} \tag{8.3}$$

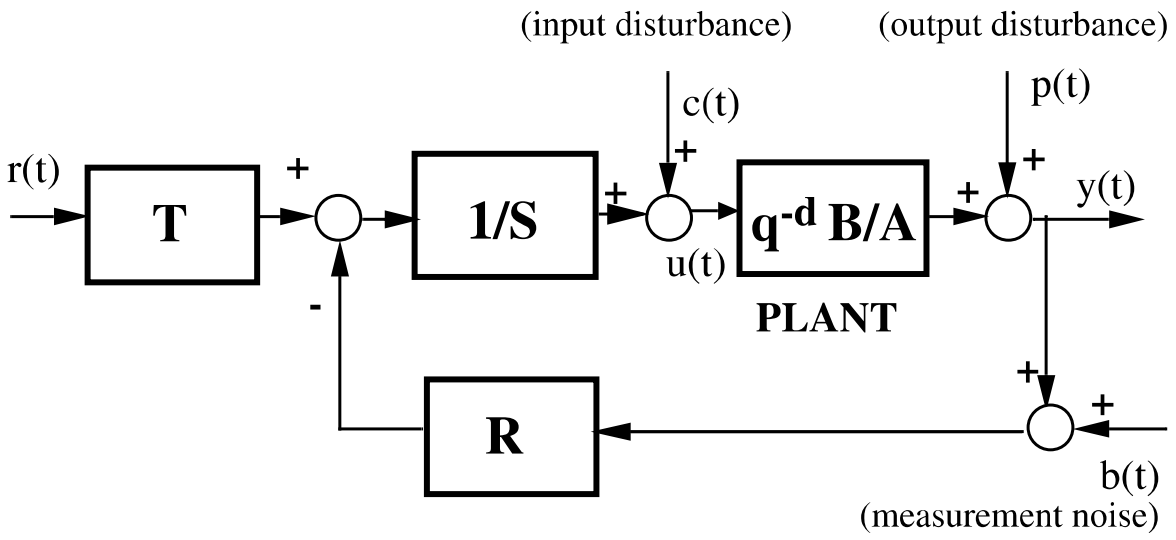

**Fig. 8.1** Closed-loop system with an RST controller in the presence of output disturbances, input disturbances and measurement noise

(This function with a positive sign is called *the complementary sensitivity function* denoted by $\mathscr{T}$ in the robust control literature.) From (8.1) and (8.3) it is obvious that:

$$S_{yp}(z^{-1}) - S_{yb}(z^{-1}) = 1 \tag{8.4}$$

The transfer function between the input disturbance $c(t)$ and the plant output $y(t)$ is given by:

$$S_{yc}(z^{-1}) = \frac{z^{-d}B(z^{-1})S(z^{-1})}{P(z^{-1})} \tag{8.5}$$

In order that the system be internally stable, all these sensitivity functions should be asymptotically stable.

For example, using *tracking and regulation with independent objectives* where $P(z^{-1})$ and $S(z^{-1})$ contains the plant model zeros, one can see that if these zeros(or some of them) are unstable the input sensitivity function $S_{up}(z^{-1})$ will be unstable while the other sensitivity functions will be stable because of poles/zeros cancellations. Using now pole placement with the $A$ polynomial as part of the closed loop poles which corresponds to internal model control, instability of $A$, will show up in $S_{yc}(z^{-1})$ (remember that in this case $R(z^{-1}) = A(z^{-1})R'(z^{-1})$). From these observations, it is clear that, in general, all the four sensitivity functions should be examined once a design is done.

## 8.3 Robust Stability

### *8.3.1 Robustness Margins*

The Nyquist plot of the open-loop transfer function allows one to assess the influence of the modeling errors and to derive appropriate specifications for the controller design in order to assure the *robust stability* of the closed-loop system for certain classes of plant model uncertainties.

The open-loop transfer function corresponding to the use of an RST controller is:

$$H_{OL}(z^{-1}) = \frac{z^{-d}B(z^{-1})R(z^{-1})}{A(z^{-1})S(z^{-1})} \tag{8.6}$$

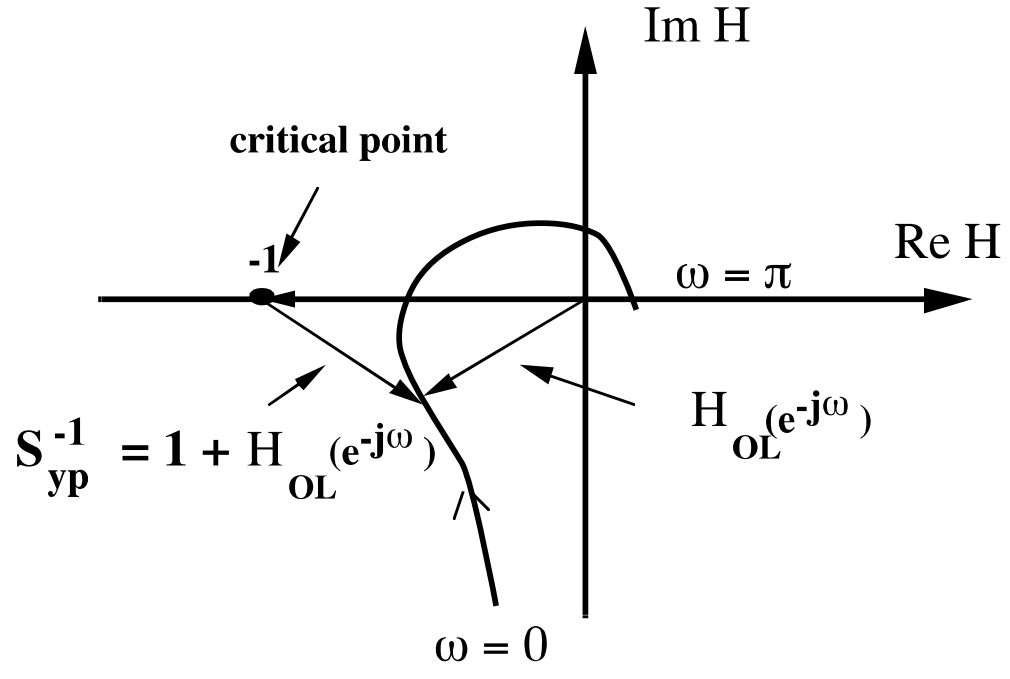

**Fig. 8.2** The Nyquist plot of a discrete-time transfer function and the critical point

By making $z = e^{j\omega}$ where $\omega$ is the normalized frequency ($\omega = \omega T_s = 2\pi f/f_s$; $f_s$ sampling frequency, $T_s$ sampling period) one traces the Nyquist plot of the open-loop transfer function $H_{OL}(e^{-j\omega})$. In general, one considers for the normalized frequency $\omega$ the domain between 0 and $\pi$ (i.e. between 0 and $0.5 f_s$). Note that the Nyquist plot between $\pi$ and $2\pi$ is symmetric with respect to the real axis of the Nyquist plot between 0 and $\pi$. An example of a Nyquist plot is given in Fig. 8.2.

The vector connecting a point of the Nyquist plot with the origin corresponds to $H_{OL}(e^{-j\omega})$ for a certain normalized frequency. The point $[-1, j0]$ on the diagram of Fig. 8.2 corresponds to the *critical point*.

From Fig. 8.2, it results that the vector connecting the *critical point* with the Nyquist plot of $H_{OL}(e^{-j\omega})$ has the expression:

$$1 + H_{OL}(z^{-1}) = \frac{A(z^{-1})S(z^{-1}) + z^{-d}B(z^{-1})R(z^{-1})}{A(z^{-1})S(z^{-1})} = S_{yp}^{-1}(z^{-1}) \qquad (8.7)$$

This vector corresponds to the inverse of the output sensitivity function $S_{yp}(z^{-1})$ given by (8.1) and the zeros of $S_{yp}^{-1}$ are the poles of the closed-loop system. In order that the closed-loop system be asymptotically stable, it is necessary that all the zeros of $S_{yp}^{-1}$ lie inside the unit circle.

The necessary and sufficient conditions in the frequency domain for the asymptotic stability of the closed-loop system are given by the Nyquist criterion. For the case of open-loop stable systems (in our case this corresponds to $A(z^{-1}) = 0$ and $S(z^{-1}) = 0 \Longrightarrow |z| < 1$), the Nyquist criterion is expressed as:

**Stability criterion** (Open-Loop Stable Systems) *The Nyquist plot of $H_{OL}(z^{-1})$ traversed in the sense of growing frequencies (from $\omega = 0$ to $\omega = \pi$) leaves the critical point $[-1, j0]$ on the left.*

Using *pole placement*, the Nyquist criterion will be satisfied for the *nominal* plant model because $R(z^{-1})$ and $S(z^{-1})$ are computed using (7.21) for an asymptotically stable polynomial $P(z^{-1})$ defining the desired closed-loop poles ($P(z^{-1}) = 0 \Longrightarrow |z| < 1$). Of course, we are assuming at this stage that the resulting $S(z^{-1})$ is also stable.

If the open-loop system is unstable either because $A(z^{-1})$ has zeros outside the unit circle (unstable plant model), or the computed controller is unstable in open loop ($S(z^{-1})$ has zeros outside the unit circle), the Nyquist stability criterion takes the form:

**Stability criterion** (Open-Loop Unstable Systems) *The Nyquist plot of* $H_{OL}(z^{-1})$ *traversed in the sense of growing frequencies* (*from* $\omega = 0$ *to* $\omega = 2\pi$) *leaves the critical point* $[-1, j0]$ *on the left and the number of encirclements of the critical point counter clockwise should be equal to the number of unstable poles of the open-loop system.*[3]

The number of encirclements of the critical point is given by:

$$N = P^i_{CL} - P^i_{OL}$$

where $P^i_{CL}$ is the number of unstable poles of the closed loop and $P^i_{OL}$ is the number of unstable poles of the open loop. Positive values of $N$ correspond to clockwise encirclements. In order that the closed-loop system be asymptotically stable one should have:

$$N = -P^i_{OL}$$

If the plant model is open-loop stable and if the controller is computed such that the closed-loop poles are asymptotically stable (for the nominal values of the plant model parameters), then if a Nyquist plot of the form of the curve (a) of the Fig. 8.3 is obtained, one concludes that the controller is unstable in open loop. In general one should avoid such a situation,[4] since the minimal distance with respect to the critical point is generally small, and small changes in parameters will cause instability. The main remedy to this situation is to reduce the desired dynamic performance of the closed loop (change of the closed-loop poles $P(z^{-1})$ either by slowing down the dominant poles or adding additional poles).

The minimal distance between the Nyquist plot of $H_{OL}(z^{-1})$ and the critical point will define a *stability margin*. This minimal distance according to (8.7) will depend upon the maximum of the modulus of the output sensitivity function. This stability margin which we will call subsequently the *modulus margin* could be linked to the uncertainties upon the plant model.

The following indicators serve for characterizing the distance between the Nyquist plot of $H_{OL}(z^{-1})$ and the critical point $[-1, j0]$ (see Fig. 8.4):

- modulus margin ($\Delta M$)
- delay margin ($\Delta\tau$)
- phase margin ($\Delta\phi$)
- gain margin ($\Delta G$)

---

[3]The criterion remains valid in the case of poles zeros cancellations. The number of encirclements should be equal to the number of unstable open-loop poles without taking into account the possible cancellations.

[4]There are some "pathological" systems $\frac{B(z^{-1})}{A(z^{-1})}$, with unstable poles and zeros which can be stabilized only with open-loop unstable controllers.

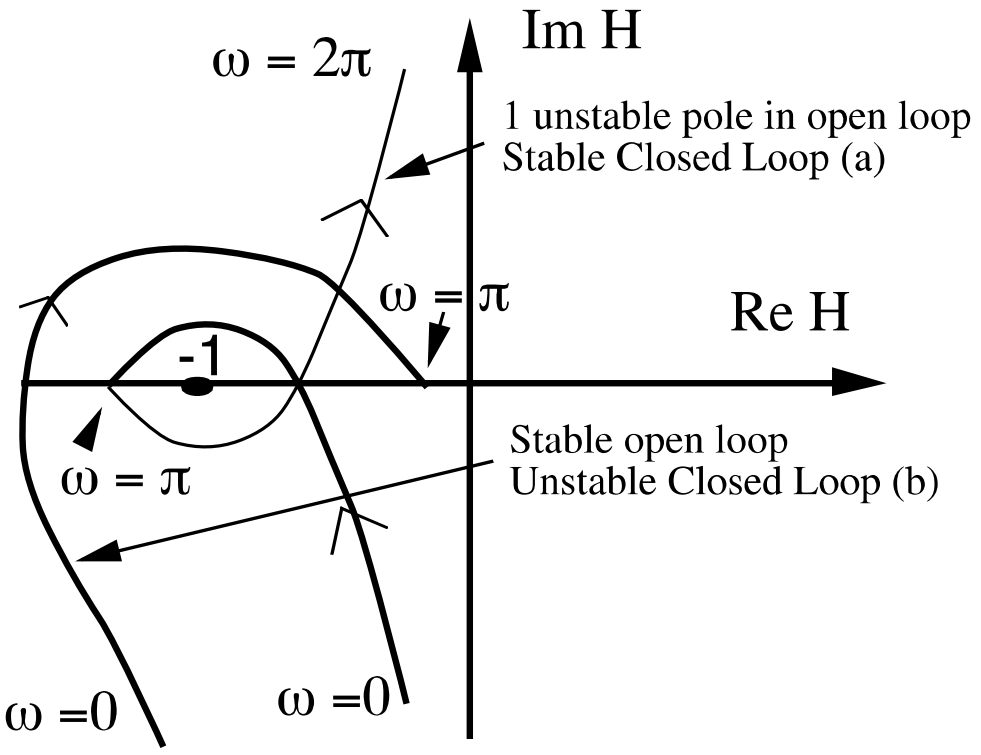

**Fig. 8.3** Two interesting Nyquist plots

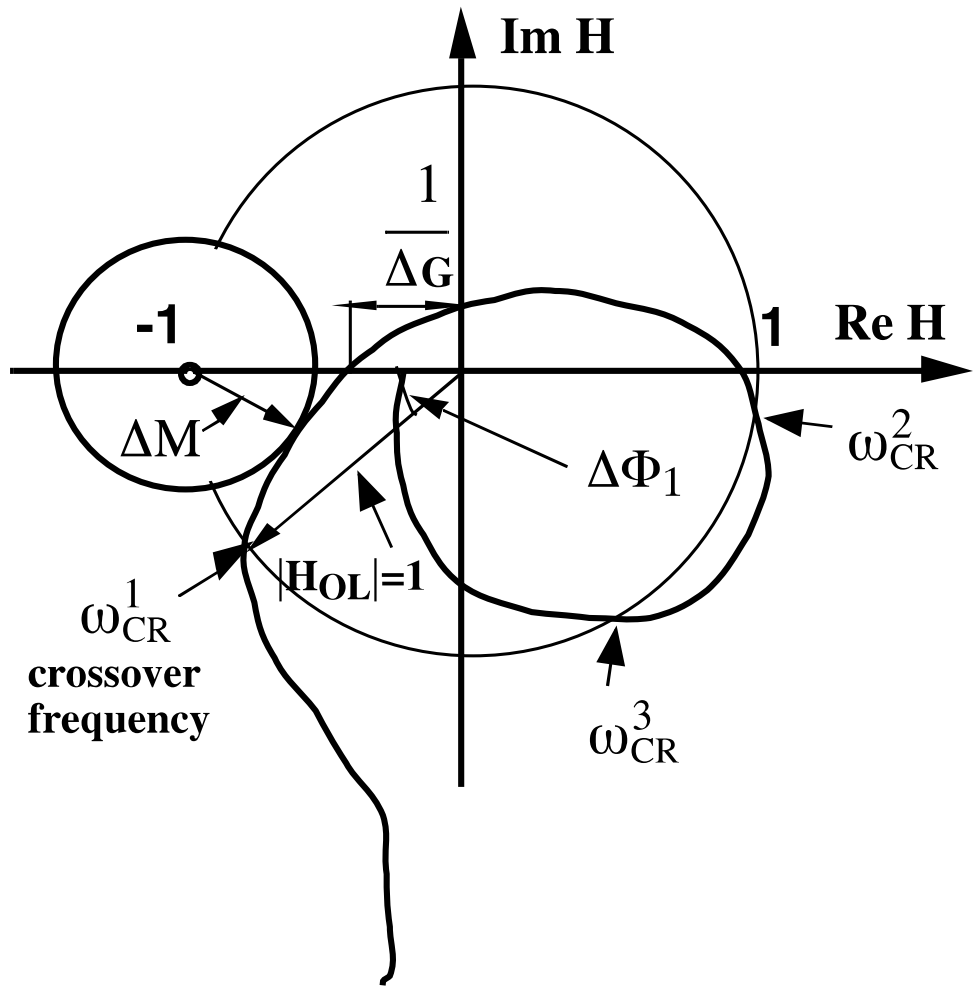

**Fig. 8.4** Modulus, gain and phase margins

Below are the definitions of the modulus margin and delay margin which will be used in the robust control design (for the definition of the gain and phase margin, see any classical control text):

## Modulus Margin ($\Delta M$)

The *modulus margin* ($\Delta M$) is defined as the radius of the circle centered in $[-1, j0]$ and tangent to the Nyquist plot of $H_{OL}(z^{-1})$. From the definition of the vector connecting the critical point $[-1, j0]$ with the Nyquist plot of $H_{OL}(z^{-1})$ (see (8.7)), it results that:

$$\Delta M = |1 + H_{OL}(e^{-j\omega})|_{\min} = (|S_{yp}(e^{-j\omega})|_{\max})^{-1} = (\|S_{yp}\|_{\infty})^{-1} \tag{8.8}$$

As a consequence, the reduction (or minimization) of $|S_{yp}(e^{-j\omega})|_{\max}$ will imply the increase (or maximization) of the modulus margin $\Delta M$.

In other terms the *modulus margin* $\Delta M$ is equal to the inverse of the maximum modulus of the output sensitivity function $S_{yp}(z^{-1})$ (i.e. the inverse of the $H_\infty$ norm of $S_{yp}(z^{-1})$). If the modulus of $S_{yp}(z^{-1})$ is expressed in dB, one has the following relationship:

$$|S_{yp}(e^{-j\omega})|_{\max} \text{ dB} = (\Delta M)^{-1} \text{ dB} = -\Delta M \text{ dB} \tag{8.9}$$

The *modulus margin* is very important because:

- It defines the maximum admissible value for the modulus of the output sensitivity function.
- It gives a bound for the characteristics of the nonlinear and time-varying elements tolerated in the closed-loop system (it corresponds to the circle criterion for the stability of nonlinear systems).

### Delay Margin ($\Delta\tau$)

For a certain frequency the phase lag introduced by a pure time delay $\tau$ is:

$$\angle\phi(\omega) = \omega\tau$$

One can therefore convert the phase margin in a delay margin i.e. to compute the additional delay which will lead to instability. It results that:

$$\Delta\tau = \frac{\Delta\phi}{\omega_{cr}} \tag{8.10}$$

where $\omega_{cr}$ is the crossover frequency (where the Nyquist plot intersects the unit circle; see Fig. 8.4). If the Nyquist plot intersects the unit circle at several frequencies $\omega_{cr}^i$, characterized by the corresponding phase margin $\Delta\phi_i$, the delay margin is defined by:

$$\Delta\tau = \min_i \frac{\Delta\phi_i}{\omega_{cr}^i} \tag{8.11}$$

*Remark* This situation appears systematically for systems with pure time delays or with multiple vibration modes.

The concept of delay margin was first introduced by Anderson and Moore (1971).

The typical values of the robustness margins for a *robust* controller design are:

- modulus margin: $\Delta M \geq 0.5$ ($-6$ dB)[min: 0.4 ($-8$ dB)];
- delay margin: $\Delta\tau \geq T_s$[min: $0.75T_s$].

*Important remarks*

1. A modulus margin $\Delta M \geq 0.5$ implies that $\Delta G \geq 2$ (6 dB) and $\Delta\phi > 29°$. The converse is not generally true. System with satisfactory gain and phase margins may have a very small modulus margin.
2. Phase margin can be misleading according to (8.10). A good phase margin may lead to a very small tolerated additional delay if $\omega_{cr}$ is high.

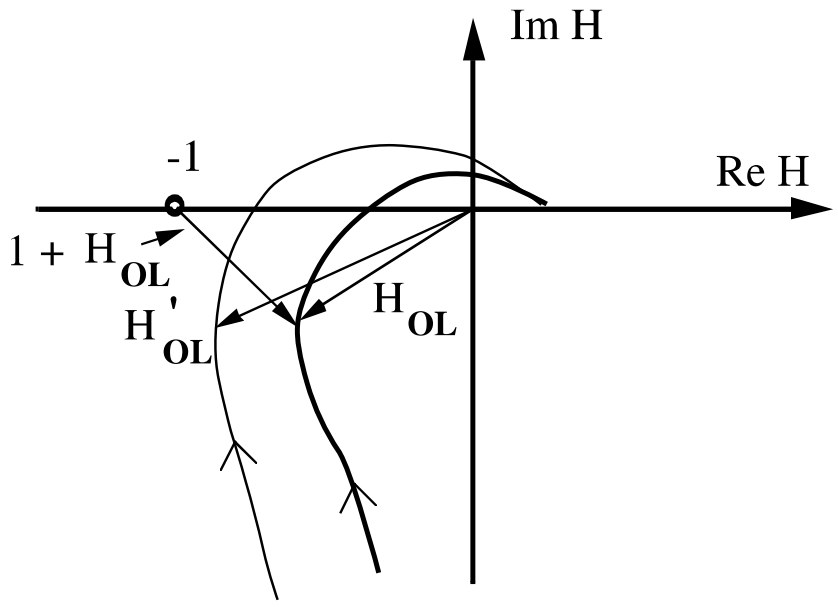

**Fig. 8.5** Nyquist plot of the nominal model and perturbed model

The modulus margin is an intrinsic measure of the stability margin and will be subsequently used together with the delay margin (instead of the phase and gain margin) for the design of robust controllers.

## *8.3.2 Model Uncertainties and Robust Stability*

Figure 8.5 illustrates the effect of uncertainties or of the variations of the parameters of the nominal model on the Nyquist plots of the open-loop transfer function. In general the Nyquist plot corresponding to the nominal model lies inside a *tube* corresponding to the possible (or accepted) tolerances of parameter variations (or uncertainties) of the plant model.

We will consider an open-loop transfer function $H'_{OL}(z^{-1})$ which differs from the nominal one. For simplicity one assures that the nominal transfer function $H_{OL}(z^{-1})$ as well as $H'_{OL}(z^{-1})$ are both stable (the general assumption is that both have the same number of unstable poles, see Åström and Wittenmark 1984; Doyle et al. 1992).

In order to assure the stability of the closed-loop system for an open loop transfer function $H'_{OL}(z^{-1})$ which differs from the nominal one $H_{OL}(z^{-1})$, the Nyquist plot of $H'_{OL}(z^{-1})$ should leaves the critical point $[-1, j0]$ on the left when traversed in the sense of the growing frequencies. Looking at Fig. 8.5 one can see that a sufficient condition for this, is that at each frequency the distance between $H'_{OL}(z^{-1})$ and $H_{OL}(z^{-1})$ be less than the distance between the nominal open-loop transfer function and the critical point. This is expressed by:

$$
\begin{aligned}
|H'_{OL}(z^{-1}) - H_{OL}(z^{-1})| < |1 + H_{OL}(z^{-1})| &= |S_{yp}^{-1}(z^{-1})| \\
&= \left| \frac{P(z^{-1})}{A(z^{-1})S(z^{-1})} \right|
\end{aligned}
\tag{8.12}
$$

In other terms the curve $|S_{yp}(e^{-j\omega})|^{-1}$ in dB (which is obtained by symmetry from $|S_{yp}(e^{-j\omega})|$) will give at each frequency a sufficient condition for the modulus of the tolerated discrepancy between the real open-loop transfer function and the nominal open-loop transfer function in order to guarantee stability.

In general, this tolerance is high in low frequencies where the open-loop gain is high (and $|S_{yp}(e^{-j\omega})|$ is small) and is low at the frequency (or frequencies) where $|S_{yp}(e^{-j\omega})|$ reaches its maximum ($= \Delta M^{-1}$). Therefore low modulus margin will imply small tolerance to parameter uncertainties in a specified frequency region.

The relationship (8.12) expresses a robustness condition in terms of the variations of the open-loop transfer function (controller + plant). It is interesting to express this in terms of the variations of the plant model. One way to do this, is to observe that (8.12) can be rewritten as:

$$\left|\frac{B'(z^{-1})R(z^{-1})}{A'(z^{-1})S(z^{-1})} - \frac{B(z^{-1})R(z^{-1})}{A(z^{-1})S(z^{-1})}\right| = \left|\frac{R(z^{-1})}{S(z^{-1})}\right| \cdot \left|\frac{B'(z^{-1})}{A'(z^{-1})} - \frac{B(z^{-1})}{A(z^{-1})}\right| < \left|\frac{P(z^{-1})}{A(z^{-1})S(z^{-1})}\right| \tag{8.13}$$

Multiplying both sides of (8.13) by $|\frac{S(z^{-1})}{R(z^{-1})}|$ one gets:

$$\left|\frac{B'(z^{-1})}{A'(z^{-1})} - \frac{B(z^{-1})}{A(z^{-1})}\right| \le \left|\frac{P(z^{-1})}{A(z^{-1})R(z^{-1})}\right| = |S_{up}^{-1}(z^{-1})| \tag{8.14}$$

The left hand side of (8.14) expresses in fact an *additive* uncertainty for the nominal plant model. The inverse of the modulus of the input sensitivity function will give a sufficient condition for the tolerated *additive* variations (or uncertainties) of the nominal plant model in order to guarantee stability. Large values of the input sensitivity function in certain frequency range will imply low tolerance to uncertainties in this frequency range. It will also mean that at these frequencies high activity of the input will result under the effect of disturbances.

One can also consider the relative size of the variations of the nominal plant model which are tolerated. One then obtains from (8.14):

$$\frac{|\frac{B'(z^{-1})}{A'(z^{-1})} - \frac{B(z^{-1})}{A(z^{-1})}|}{|\frac{B(z^{-1})}{A(z^{-1})}|} \le \left|\frac{P(z^{-1})}{B(z^{-1})R(z^{-1})}\right| = |S_{yb}^{-1}(z^{-1})| \tag{8.15}$$

The inverse of the modulus of the noise sensitivity function (the complementary sensitivity function) will give a sufficient condition for the tolerated *relative size* variations (or uncertainties) of the nominal plant model in order to guarantee stability.

The above development allows to introduce various models for uncertainties under the form of *disk* uncertainties with a radius depending upon the frequency.

1. Additive uncertainties (Fig. 8.6)

$$G'(z^{-1}) = G(z^{-1}) + \delta(z^{-1})W_a(z^{-1}) \tag{8.16}$$

where $\delta(z^{-1})$ is any stable transfer function having the property $\|\delta(z^{-1})\|_\infty \le 1$ and $W_a(z^{-1})$ is a stable transfer function.

This implies that:

$$|G'(z^{-1}) - G(z^{-1})|_{\max} = \|G'(z^{-1}) - G(z^{-1})\|_\infty = \|W_a(z^{-1})\|_\infty \tag{8.17}$$

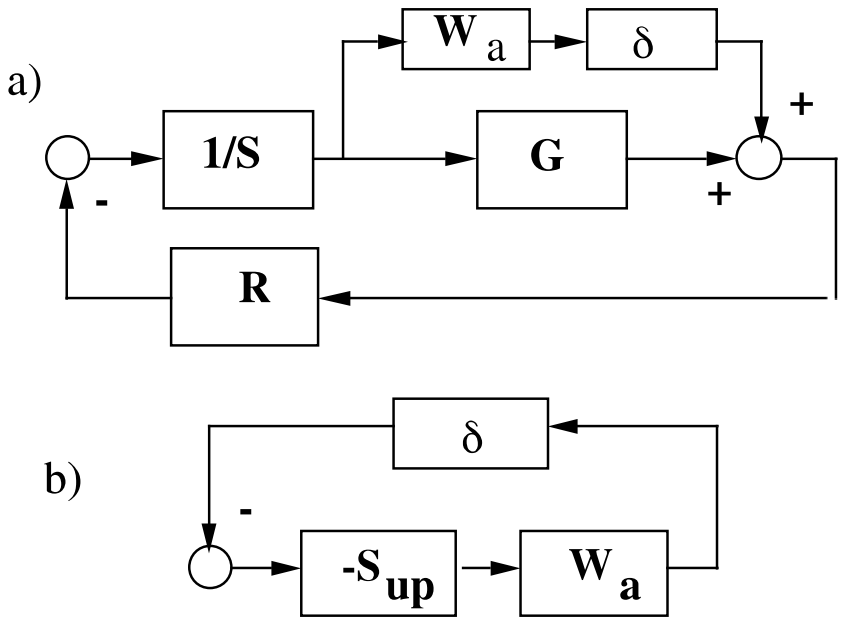

**Fig. 8.6** Additive model uncertainty, (**a**) uncertainty representation, (**b**) equivalent representation of the closed-loop system

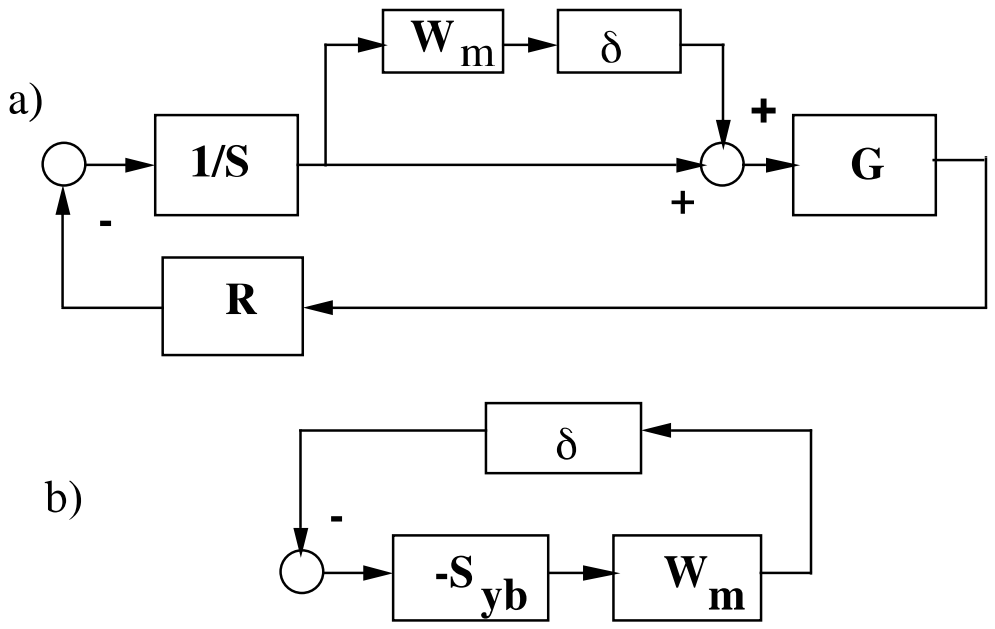

**Fig. 8.7** Multiplicative model uncertainty, (**a**) uncertainty representation, (**b**) equivalent representation of the closed-loop system

In other words $|W_a(z^{-1})|$ characterizes the size of the additive uncertainties in the frequency domain (maximum value given by $\|W_a(z^{-1})\|_\infty$). At a certain frequency, the effect of $\delta(z^{-1})$ will be to allow uncertainties of size $|W_a(z^{-1})|$ and with any direction.

2. Multiplicative uncertainties (Fig. 8.7)

$$G'(z^{-1}) = G(z^{-1})[1 + \delta(z^{-1})W_m(z^{-1})] \tag{8.18}$$

where $W_m(z^{-1})$ is a stable transfer function.

Observes that the additive uncertainties and the multiplicative uncertainties are related by:

$$W_a(z^{-1}) = G(z^{-1})W_m(z^{-1})$$

(i.e., $W_m(z^{-1})$ corresponds to a *relative size* uncertainty).

3. Feedback uncertainties on the input (or output) (Fig. 8.8)

$$G'(z^{-1}) = \frac{G(z^{-1})}{1 + \delta(z^{-1})W_r(z^{-1})} \tag{8.19}$$

($W_r(z^{-1})$ can be interpreted as an *additive* uncertainty on the open-loop transfer function).

We will use one or another of the uncertainty descriptions depending upon the context.

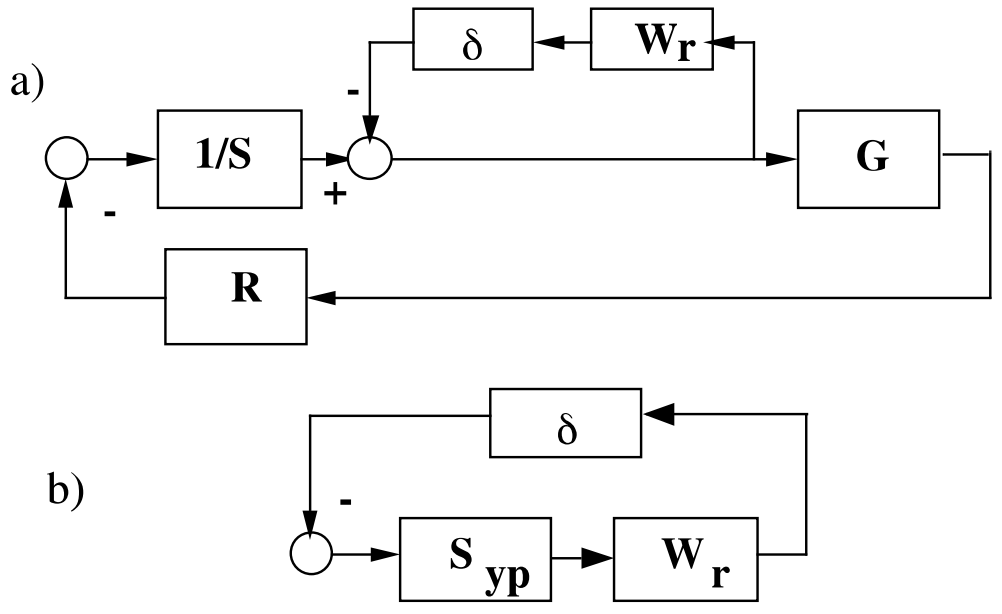

**Fig. 8.8** Feedback uncertainty on the input, (**a**) uncertainty representation, (**b**) equivalent representation of the closed-loop system

Each of the (8.16), (8.18) and (8.19) defines a family of plant models $\mathscr{P}(W,\delta)$. The system will be said to be *robustly stable* if it will be asymptotically stable for all the models $G'(z^{-1})$ belonging to the family $\mathscr{P}(W,\delta)$.

Figures 8.6, 8.7 and 8.8 give equivalent feedback configurations allowing the effects of the uncertainties upon the closed-loop system to be represented. These equivalent feedback structures feature both the transfer function $\delta(z^{-1})$ characterized by the property $\|\delta(z^{-1})\|_\infty \le 1$ and the product between a sensitivity function and the $W$ transfer function characterizing the uncertainty.

Using the *small gain* theorem (Appendix C) and depending upon the type of uncertainty considered one has the following robust stability conditions (Doyle et al. 1992; Kwakernaak 1993; Morari and Zafiriou 1989):

1. Additive uncertainties

$$\|S_{up}(z^{-1})W_a(z^{-1})\|_\infty \le 1 \tag{8.20}$$

   or equivalently

$$|S_{up}(e^{-j\omega})| < |W_a(e^{-j\omega})|^{-1}; \quad 0 \le \omega \le \pi \tag{8.21}$$

2. Multiplicative uncertainties

$$\|S_{yb}(z^{-1})W_m(z^{-1})\|_\infty \le 1 \tag{8.22}$$

   or equivalently

$$|S_{yb}(e^{-j\omega})| < |W_m(e^{-j\omega})|^{-1}; \quad 0 \le \omega \le \pi \tag{8.23}$$

3. Feedback uncertainties on the input (or output)

$$\|S_{yp}(z^{-1})W_r(z^{-1})\|_\infty \le 1 \tag{8.24}$$

   or equivalently

$$|S_{yp}(e^{-j\omega})| < |W_r(e^{-j\omega})|^{-1}; \quad 0 \le \omega \le \pi \tag{8.25}$$

The functions $|W(e^{-j\omega})|^{-1}$ define in fact an upper *template* for the modulus of the various sensitivity functions. Conversely, the frequency profile of the modulus of a sensitivity function can be interpreted in terms of model uncertainties tolerated for the robust stability of the closed-loop system.

### 8.3.3 Robustness Margins and Robust Stability

Next, we will establish relationships between the robustness margins (modulus margin and delay margin) and the *robust stability* conditions. From (8.8) and (8.25) it results that the *modulus margin* corresponds to a *robust stability* condition for a family of feedback uncertainties on the input (it is not the only one) of the form:

$$W_r^{-1}(z^{-1}) = \varDelta M \tag{8.26}$$

$$\delta(z^{-1}) = \lambda f(z^{-1}); \quad -1 \leq \lambda \leq 1 \tag{8.27}$$

where

$$f(z^{-1}) = 1, z^{-1}, z^{-2}, \ldots, \frac{z^{-1}+z^{-2}}{2} \tag{8.28}$$

Two families of models for which the robust stability conditions are satisfied can be immediately derived from (8.26), (8.27) and (8.28):

$$G'(z^{-1}) = G(z^{-1})\frac{1}{1-\lambda \varDelta M} \tag{8.29}$$

$$G'(z^{-1}) = G(z^{-1})\frac{1}{1-\lambda \varDelta M z^{-1}} \tag{8.30}$$

These two representations emphasize the tolerance with respect to gain variation and to the presence of unmodeled dynamics (other forms of unmodeled dynamics as well as variations of the $a_i$ and $b_i$ parameters can be considered, see Morari and Zafiriou 1989; Landau 1995). Conversely, model uncertainties can be converted into a required modulus margin for assuring robust stability.

The *delay margin* can also be converted into a *robust stability* condition which define a template for the sensitivity functions. As an example, consider the case $\varDelta\tau = T_s$, i.e.:

$$G(z^{-1}) = \frac{z^{-d}B(z^{-1})}{A(z^{-1})} \quad \text{and} \quad G'(z^{-1}) = \frac{z^{-d-1}B(z^{-1})}{A(z^{-1})} = G(z^{-1})z^{-1} \tag{8.31}$$

One can interpret the additional delay which should be tolerated in terms of multiplicative uncertainties. From (8.18) one obtains:

$$\begin{aligned} G'(z^{-1}) &= G(z^{-1})z^{-1} = G(z^{-1})(1+\delta(z^{-1})W_m(z^{-1})) \\ &= G(z^{-1})[1+(z^{-1}-1)] \end{aligned} \tag{8.32}$$

which gives for example:

$$\delta(z^{-1}) = 1; \qquad W_m(z^{-1}) = (z^{-1}-1) \tag{8.33}$$

and the corresponding robust stability condition is expressed as:

$$\|S_{yb}(z^{-1}) \cdot (1-z^{-1})\|_\infty \leq 1; \quad z = e^{-j\omega},\ 0 \leq \omega \leq \pi \tag{8.34}$$

or respectively

$$|S_{yb}(z^{-1})| < |(1-z^{-1})|^{-1}; \quad z = e^{-j\omega},\ 0 \leq \omega \leq \pi \tag{8.35}$$

The interpretation of this result is that in order to assure a delay margin of one sampling period, in the medium to high frequencies up to $0.5 f_s$ the modulus of the noise sensitivity function should be below (or equal) the frequency characteristics of a pure integrator.

From (8.4) one has:

$$S_{yp}(z^{-1}) - S_{yb}(z^{-1}) = 1 \tag{8.36}$$

and therefore

$$1 - |S_{yb}(z^{-1})| < |S_{yp}(z^{-1})| < 1 + |S_{yb}(z^{-1})| \tag{8.37}$$

If $S_{yb}(z^{-1})$ satisfies the condition (8.35) then $S_{yp}(z^{-1})$ will satisfy the following condition:

$$1 - |1 - z^{-1}|^{-1} < |S_{yp}(z^{-1})| < 1 + |1 - z^{-1}|^{-1}; \quad z = e^{-j\omega},\ 0 \leq \omega \leq \pi \tag{8.38}$$

Therefore, in order to assure the delay margin $\Delta\tau = T_s$, it is required that the modulus of $S_{yp}(z^{-1})$ lies inside a *tube* defined by a lower template $|W^{-1}|_{\text{inf}} = 1 - |1 - z^{-1}|^{-1}$ and an upper template defined by $|W^{-1}|_{\text{sup}} = 1 + |1 - z^{-1}|^{-1}$. Figure 8.9 gives images of these templates. Similar templates can be defined for other values of the delay margin see Landau (1995).

It is important to note that the template on $S_{yp}$ will not always guarantee the desired delay margin. If the condition on $S_{yb}$ is satisfied, the condition on $S_{yp}$ will be also satisfied. However, if the condition on $S_{yb}$ is violated this will not imply necessarily that the condition on $S_{yp}$ will be also violated. While in general the results using the template on $S_{yp}$ are very reliable, examples can be constructed for which the condition on $S_{yb}$ is slightly violated (i.e. we do not have exactly the desired delay margin) and the condition on $S_{yp}$ is satisfied.

## 8.4 Definition of "Templates" for the Sensitivity Functions

The *nominal performance* requirements and the *robust stability* conditions lead to the definition of desired templates for the sensitivity functions. We will consider first the definition of such a template for the output sensitivity function.

Concerning the *nominal performance*, we are interested in:

1. Rejecting low-frequency disturbances which will define an attenuation band for the output sensitivity function.
2. Having a low-amplification of the disturbances in the high-frequency region which will define an upper value for the modulus of the sensitivity function in this region.

Regarding the *robust stability*, the chosen *modulus margin* will define the maximum value of the modulus of the output sensitivity function (upper template) and the chosen *delay margin* will define an upper and a lower template starting for example around $0.15 f_s$ (for $\Delta\tau = T_s$).

In general, the union of the various templates $W_i^{-1}(e^{-j\omega})$ will allow to define an upper and a lower template. The upper template is defined by:

$$|W^{-1}(e^{-j\omega})|_{\text{sup}} = \min_i [|W_{S_1}^{-1}(e^{-j\omega})|, \ldots, |W_{S_n}^{-1}(e^{-j\omega})|]; \quad i = 1, \ldots, n \tag{8.39}$$

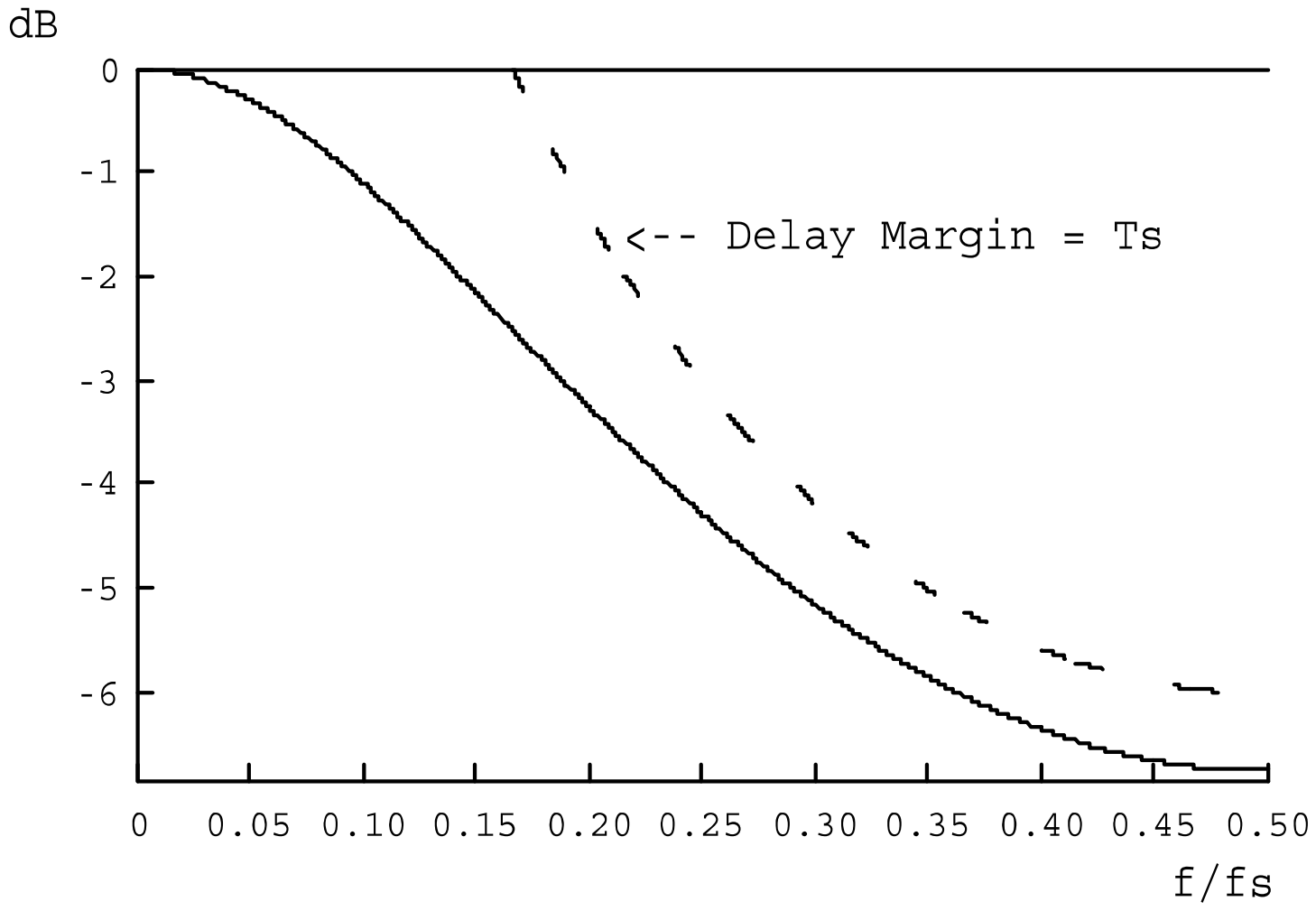

a)

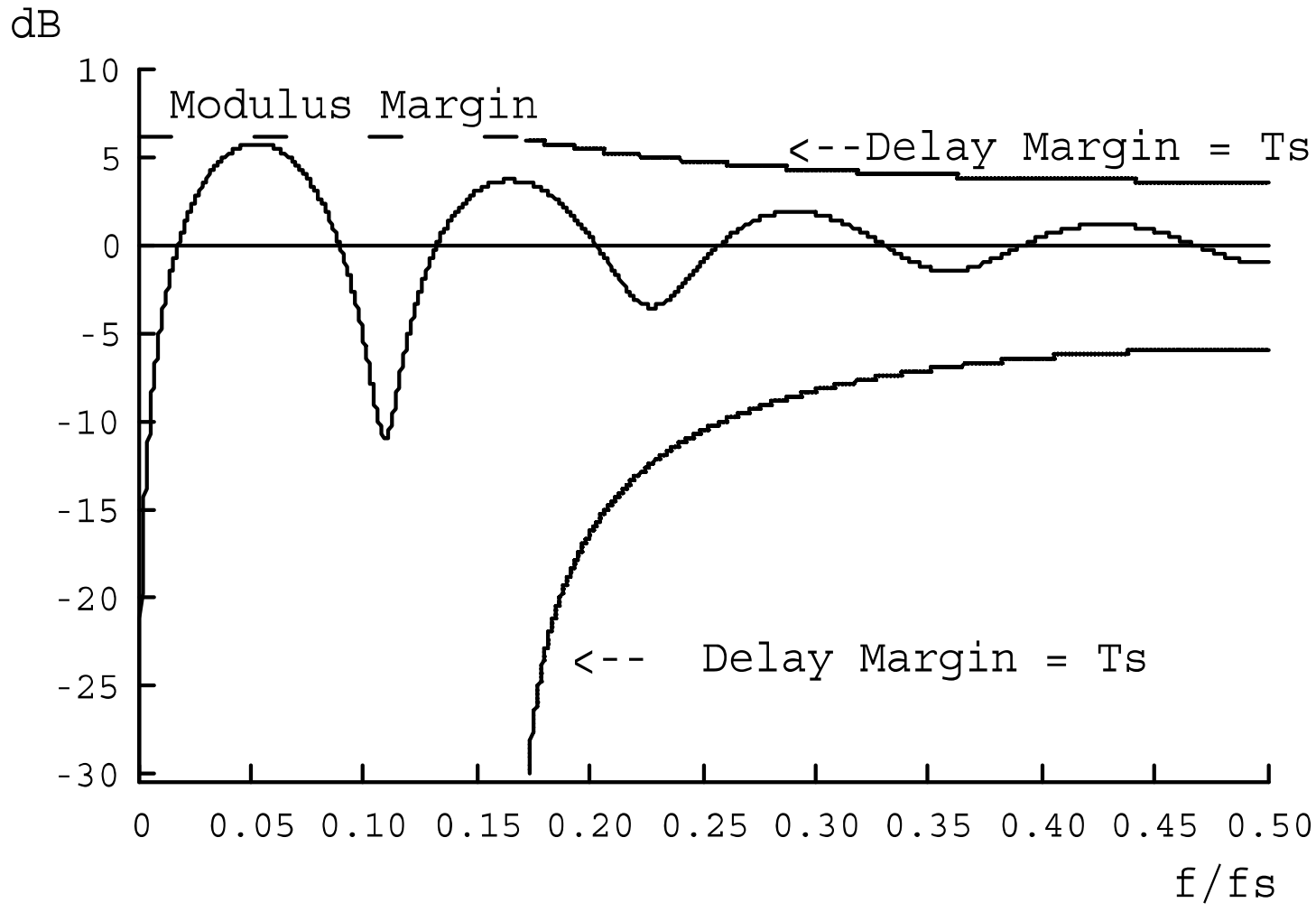

b)

**Fig. 8.9** Templates for the delay margin $\Delta\tau = T_s$, (**a**) on the noise sensitivity function, (**b**) the output sensitivity function (the template corresponding to the modulus margin is also represented)

and the lower template is defined by:

$$|W^{-1}(e^{-j\omega})|_{\inf} = \max_i[|W_{I_1}^{-1}(e^{-j\omega})|, \ldots, |W_{I_n}^{-1}(e^{-j\omega})|]; \quad i = 1, \ldots, n \quad (8.40)$$

The desired template takes in general the form shown in Fig. 8.10.

In the case where disturbance attenuation should occur in several frequency regions and in addition some disturbances occurring at certain frequencies should neither be attenuated nor amplified, one gets a *nominal* template of the form shown in Fig. 8.11. Note that in order to leave unchanged the disturbance effect at a certain frequency, $|S_{yp}|$ should be equal to one at this frequency and this is equivalent to the opening of the loop at this frequency.

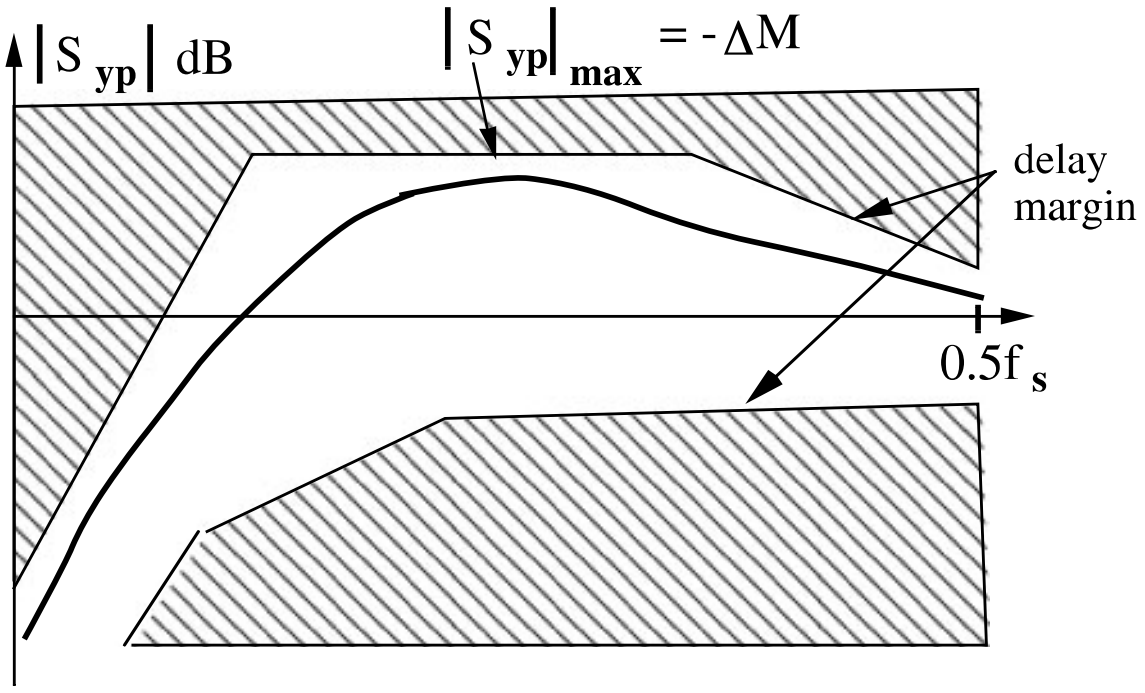

**Fig. 8.10** Desired template for the output sensitivity function (the case of disturbance rejection in low frequencies)

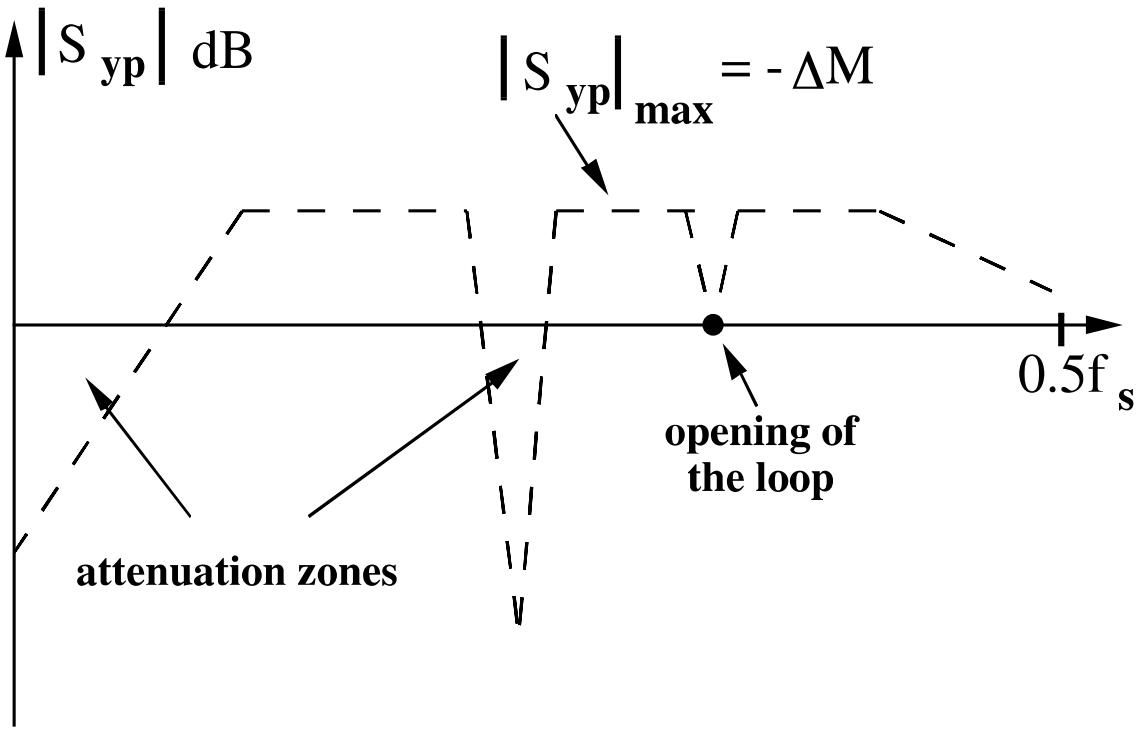

**Fig. 8.11** Desired template for the modulus of the sensitivity function in the case of two disturbance attenuation bands and opening of the loop at a certain frequency

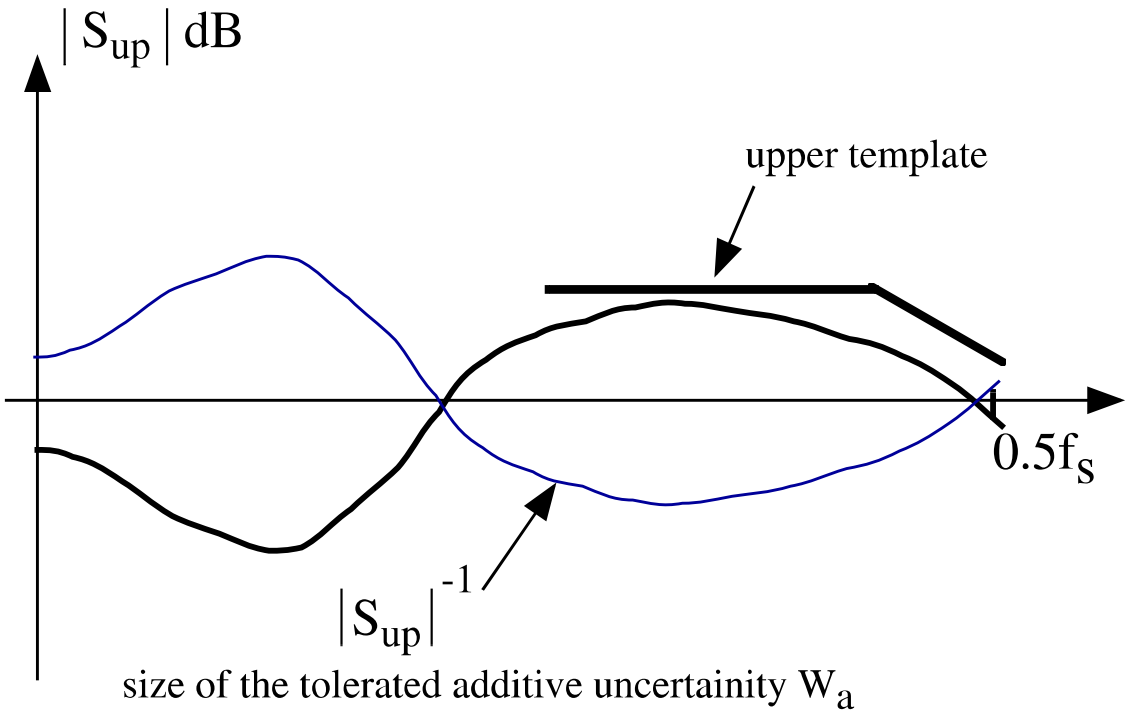

**Fig. 8.12** An example of desired template for the input sensitivity function

A typical form for the *input sensitivity function* is shown in Fig. 8.12.

In the frequency regions where $|S_{up}|$ is low we will have a good tolerance for *additive* uncertainties while in the regions where $|S_{yp}|$ is large one will have on the one hand little tolerance for *additive* uncertainty and on the other hand a strong activity of the input in the presence of disturbances. Therefore the upper value of $|S_{up}|$ should be limited. In particular this value will be high where the gain of the system is low. This occurs in general at high frequencies and in the case of complex zeros (stable or unstable) close to the unit circle (see next sections for details).

The *noise sensitivity function* should normally decrease in the high frequencies faster than an integrator in order to assure a satisfactory delay margin as it was shown in Sect. 8.3.

Therefore, in addition to placing the dominant poles of the closed loop or optimizing a quadratic-type time-domain criterion, it is mandatory to *shape* the sensitivity functions in order to be within the templates for assuring both nominal performances and corresponding robustness margins (which implies robust stability for a certain size of plant model uncertainty). In order to understand how the various sensitivity functions can be shaped it is useful to examine their frequency properties in more detail. This will help to select the closed-loop poles and the fixed parts of the controller.

## 8.5 Properties of the Sensitivity Functions

### 8.5.1 Output Sensitivity Function

Using an RST controller, the output sensitivity function is given by:

$$S_{yp}(z^{-1}) = \frac{A(z^{-1})S(z^{-1})}{A(z^{-1})S(z^{-1}) + z^{-d}B(z^{-1})R(z^{-1})} \tag{8.41}$$

where

$$R(z^{-1}) = H_R(z^{-1})R'(z^{-1}) \tag{8.42}$$

$$S(z^{-1}) = H_S(z^{-1})S'(z^{-1}) \tag{8.43}$$

and

$$A(z^{-1})S(z^{-1}) + z^{-d}B(z^{-1})R(z^{-1}) = P_D(z^{-1})P_F(z^{-1}) = P(z^{-1}) \tag{8.44}$$

In (8.42) and (8.43), $H_R(z^{-1})$ and $H_S(z^{-1})$ correspond to the pre-specified parts of $R(z^{-1})$ and $S(z^{-1})$ respectively. $S(z^{-1})$ and $R(z^{-1})$ are the solutions of (8.44) where $P(z^{-1})$ represents the desired closed-loop poles in pole placement control strategy. The polynomial $P(z^{-1})$ is factorized in order to emphasize the dominant poles defined by $P_D(z^{-1})$ and the auxiliary poles defined by $P_F(z^{-1})$.

**Property 8.1** *The modulus of the output sensitivity function at a certain frequency gives the amplification or the attenuation of the disturbance.*

At the frequencies where $|S_{yp}(\omega)| = 1$ (0 dB), there is no attenuation nor amplification of the disturbance (operation in open loop). At the frequencies where $|S_{yp}(\omega)| < 1$ (0 dB), the disturbance is attenuated. At the frequencies where $|S_{yp}(\omega)| > 1$ (0 dB), the disturbance is amplified.

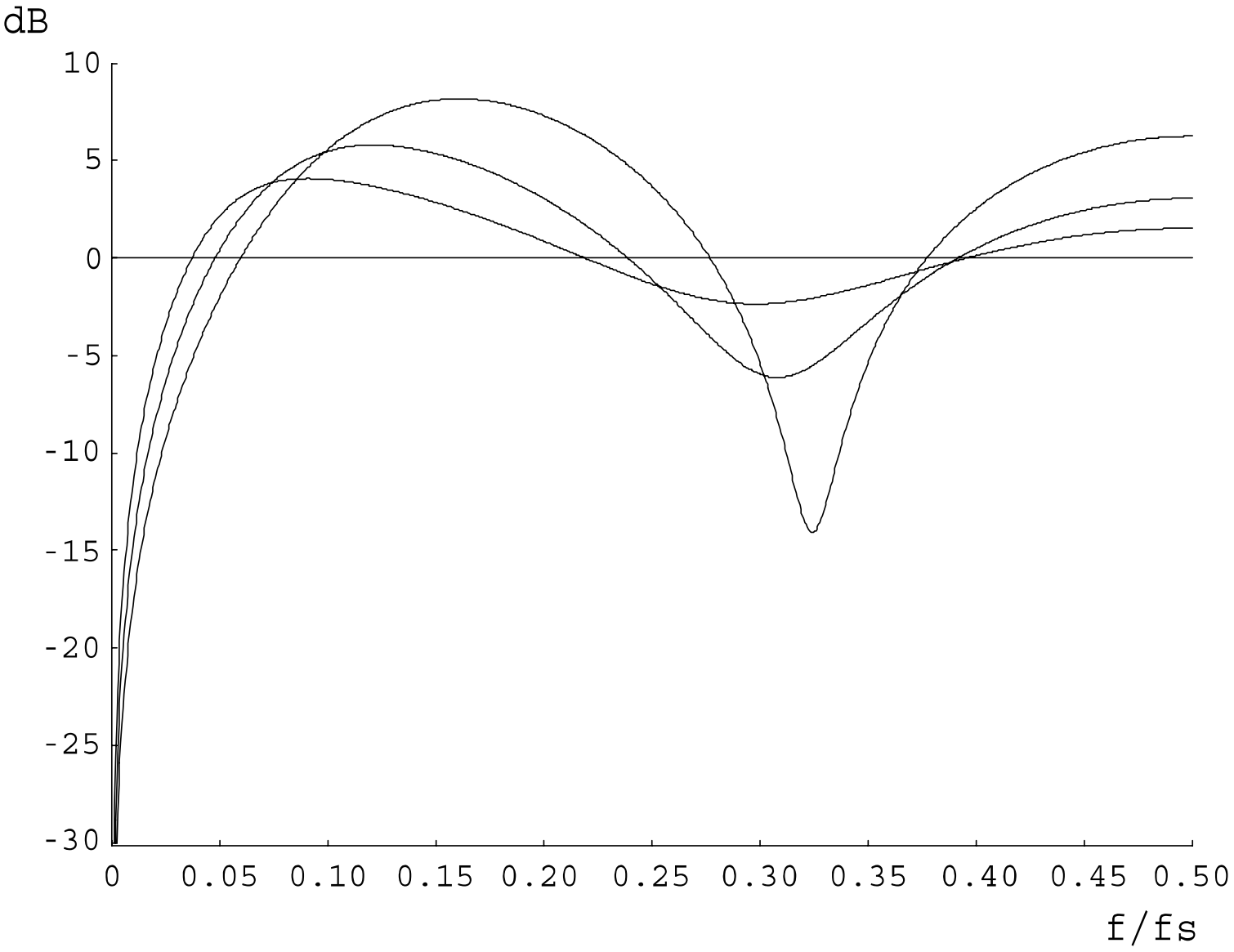

**Fig. 8.13** Modulus of the output sensitivity functions for various attenuation bands

**Property 8.2** *The closed loop being asymptotically stable, the integral of the logarithm of the modulus of the output sensitivity function from* 0 *to* $0.5f_s$ *is equal to* 0 *for the case of stable open-loop systems*:[5]

$$\int_0^{0.5f_s} \log|S_{yp}(e^{-j2\pi f/f_s})|df = 0$$

In other terms, the sum of the areas between the curve of the modulus of the output sensitivity function and the 0 dB axis taken with their sign is null.

**Property 8.3** *The inverse of the maximum of the modulus of the sensitivity function corresponds to the modulus margin* $\Delta M$

$$\Delta M = (|S_{yp}(e^{-j\omega})|_{\max})^{-1} \tag{8.45}$$

From the Properties 8.2 and 8.3, it results that the increase of the attenuation band or of the attenuation in a certain frequency band will in general imply an increase of $|S_{yp}(e^{-j\omega})|_{\max}$ and therefore a decrease of the modulus margin (and therefore less robustness).

Figure 8.13 shows the output sensitivity function for a closed-loop system, corresponding to a plant model $A(z^{-1}) = 1 - 0.7z^{-1}$, $B(z^{-1}) = 0.3z^{-1}$, $d = 2$ and several controllers with integral action. The controllers have been designed using the pole placement for various values of the closed-loop poles obtained from the discretization (with $T_S = 1$) of a second-order system with various natural frequencies $\omega_0(0.4; 0.6; 1\ \text{rad/s})$ but $\zeta = \text{const}$.

[5]See Sung and Hara (1988) for a proof. In the case of unstable open-loop systems but stable in closed loop, this integral is positive.

One can clearly see that the increase of attenuation in a certain frequency region or the increase of the attenuation band implies necessarily a stronger amplification of the disturbances outside the attenuation band. This is a direct consequence of Property 8.2.

**Property 8.4** *The effect of disturbances upon the output is canceled at the frequencies where*

$$A(e^{-j\omega})S(e^{-j\omega}) = A(e^{-j\omega})H_S(e^{-j\omega})S'(e^{-j\omega}) = 0 \tag{8.46}$$

This results immediately from (8.41). Equation (8.46) defines the zeros of the output sensitivity function. The pre-specified part of $S(z^{-1})$, namely $H_S(z^{-1})$, allows zeros at the desired frequencies to be introduced (see Fig. 8.14). For example:

$$H_S(z^{-1}) = 1 - z^{-1}$$

introduces a zero at the zero frequency and assures the perfect rejection of steady state disturbances.

$$H_S(z^{-1}) = 1 + \alpha z^{-1} + z^{-2}$$

with $\alpha = -2\cos\omega T_S = -2\cos(2\pi f/f_s)$ corresponds to the introduction of a pair of undamped complex zeros at the frequency $f$ (or at the normalized frequency $f/f_s$). This will cancel the disturbance effects at this frequency.

$$H_S(z^{-1}) = 1 + \alpha_1 z^{-1} + \alpha_2 z^{-2}$$

corresponds to the introduction of a pair of damped complex zeros at a certain frequency. The chosen damping will depend upon the desired attenuation at this frequency. Figure 8.14 shows the output sensitivity functions for the cases: $H_S(z^{-1}) = 1 - z^{-1}$ and $H_S(z^{-1}) = (1 - z^{-1})(1 + z^{-2})$ (the same plant model as in Fig. 8.13).

**Property 8.5** *The modulus of the output sensitivity function is equal to* 1, *i.e.*:

$$|S_{yp}(e^{-j\omega})| = 1\ (0\ \text{dB})$$

*at the frequencies where*:

$$B(e^{-j\omega})R(e^{-j\omega}) = B(e^{-j\omega})H_R(e^{-j\omega})R'(e^{-j\omega}) = 0 \tag{8.47}$$

This property results immediately from (8.41) since under the condition (8.47) one gets $S_{yp}(j\omega) = 1$.

The pre-specified part $H_R(z^{-1})$ of $R(z^{-1})$ allows a null gain for $R(z^{-1})$ at certain frequencies and therefore at theses frequencies $|S_{yp}(e^{-j\omega})| = 1$. For example:

$$H_R(z^{-1}) = 1 + z^{-1}$$

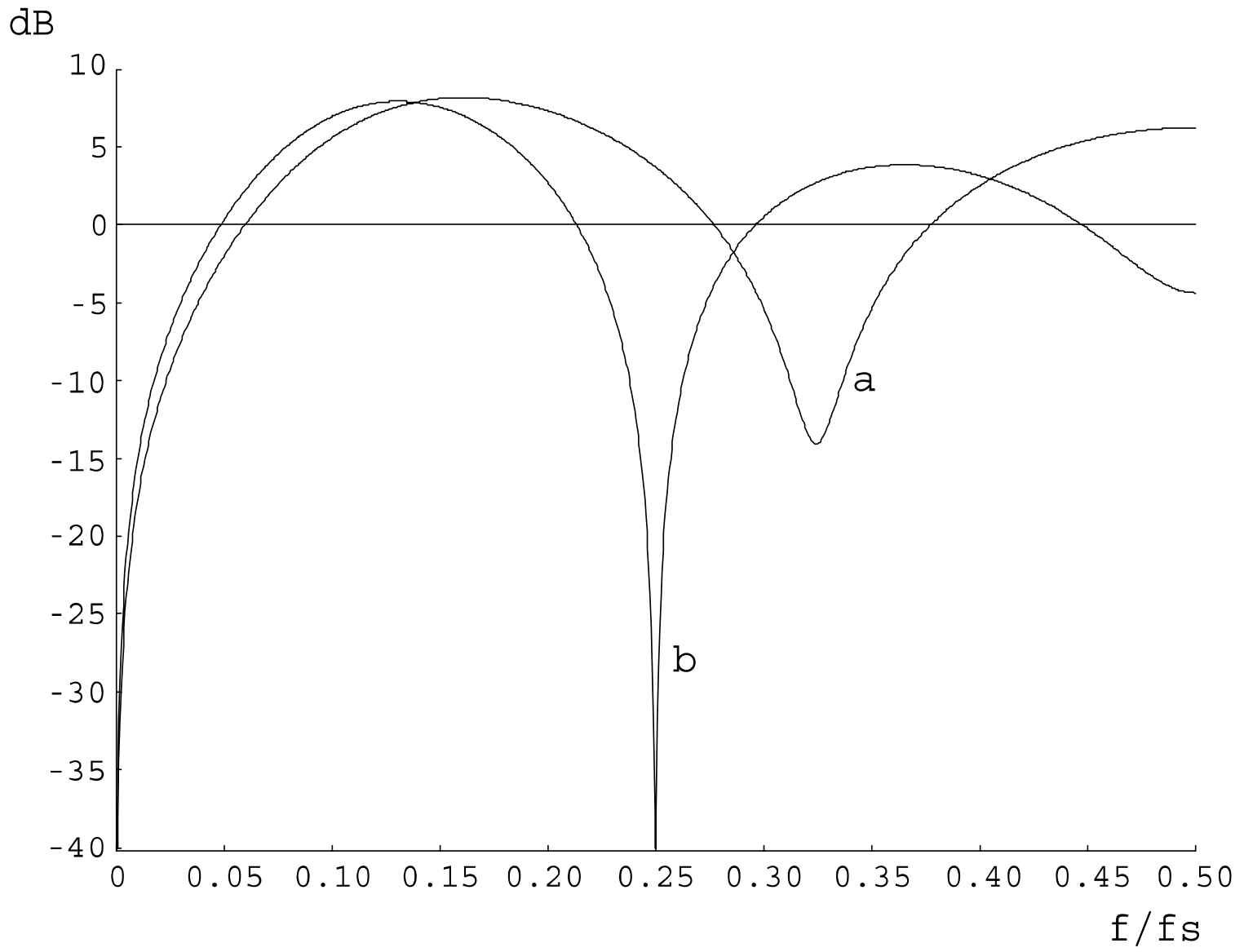

**Fig. 8.14** Output sensitivity function for the cases: (**a**) $H_S(z^{-1}) = 1 - z^{-1}$ and (**b**) $H_S(z^{-1}) = (1 - z^{-1})(1 + z^{-2})$

will introduce a zero at $= 0.5 f_s$ which implies that $|S_{yp}(e^{-j\pi f/fe})| = 1$.

$$H_R(z^{-1}) = 1 + \beta z^{-1} + z^{-2}$$

with $\beta = -2\cos(\omega T_S) = -2\cos(2\pi f/f_s)$ introduces a pair of undamped complex zeros at the frequency $f$. At this frequency $|S_{yp}(e^{-j\pi f/f_e})| = 1$.

$$H_R(z^{-1}) = 1 + \beta_1 z^{-1} + \beta_2 z^{-2}$$

corresponds to the introduction of a pair of damped complex zeros which will influence the attenuation or the amplification of the disturbance at a certain frequency (it will "squeeze" the modulus of the sensitivity function around 0 dB axis).

Figure 8.15 illustrates the effect of $H_R(z^{-1}) = 1 + z^{-2}$, which introduces a pair of undamped complex zeros at $f = 0.25 f_s$. One observes that for this frequency $|S_{yp}(e^{-j\omega})| = 1$, while in the absence of $H_R(z^{-1})$ one has at $f = 0.25 f_s$, $|S_{yp}(e^{-j\omega})| \approx 3$ dB.

Note also that $R(z^{-1})$ defines a number of zeros of the input sensitivity function $S_{up}(z^{-1})$ and of the noise sensitivity function $S_{yb}(z^{-1})$. Therefore at frequencies where $R(z^{-1}) = 0$, these sensitivity functions will be zero (the measurement signal is blocked at certain frequencies).

**Property 8.6** *The introduction of auxiliary asymptotically stable real poles* $P_F(z^{-1})$ *will cause in general a decrease of the modulus of the sensitivity function in the domain of attenuation of* $1/P_F(z^{-1})$.

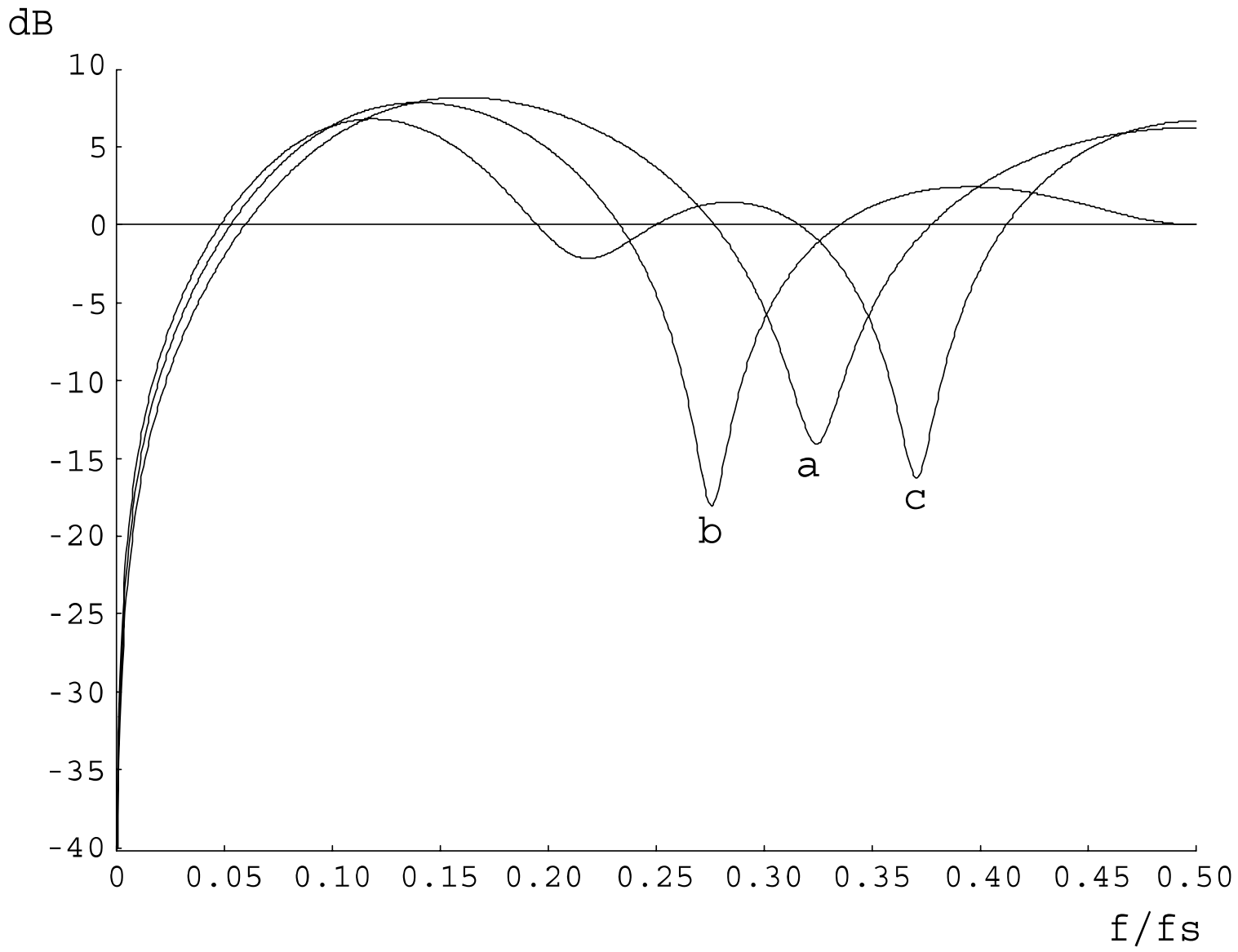

**Fig. 8.15** Output sensitivity function for the cases: (**a**) $H_R(z^{-1}) = 1$, (**b**) $H_R(z^{-1}) = 1 + z^{-1}$, (**c**) $H_R(z^{-1}) = 1 + z^{-2}$

From (8.41), one can see that the term $1/P_D(z^{-1})P_F(z^{-1})$ will introduce a stronger attenuation in the frequency domain than the term $1/P_D(z^{-1})$ if the auxiliary poles $P_F(z^{-1})$ are real (aperiodic) and asymptotically stable. However, since $S'(z^{-1})$ depends upon the poles through (8.44), one cannot guarantee this property for all the values of $P_F(z^{-1})$.

The auxiliary poles are generally chosen as high-frequency real poles under the form:

$$P_F(z^{-1}) = (1 - p_1 z^{-1})^{n_F}; \quad 0.05 \le p_1 \le 0.5$$

where:

$$n_F \le n_p - n_D; \quad n_p = (\deg P)_{\max}; \quad n_D = \deg P_D$$

The effect of the introduction of the auxiliary poles is illustrated in Fig. 8.16 one observes that the auxiliary poles "squeezes" the modulus of the output sensitivity function around 0 dB axis in the high-frequency range.

Note that in many applications the introduction of the high-frequency auxiliary poles allows the requirements for robustness margins to be met.

**Property 8.7** *Simultaneous introduction of a fixed part $H_{S_i}$ and of a pair of auxiliary poles $P_{F_i}$ in the form*

$$\frac{H_{S_i}(z^{-1})}{P_{F_i}(z^{-1})} = \frac{1 + \beta_1 z^{-1} + \beta_2 z^{-2}}{1 + \alpha_1 z^{-1} + \alpha_2 z^{-2}} \tag{8.48}$$

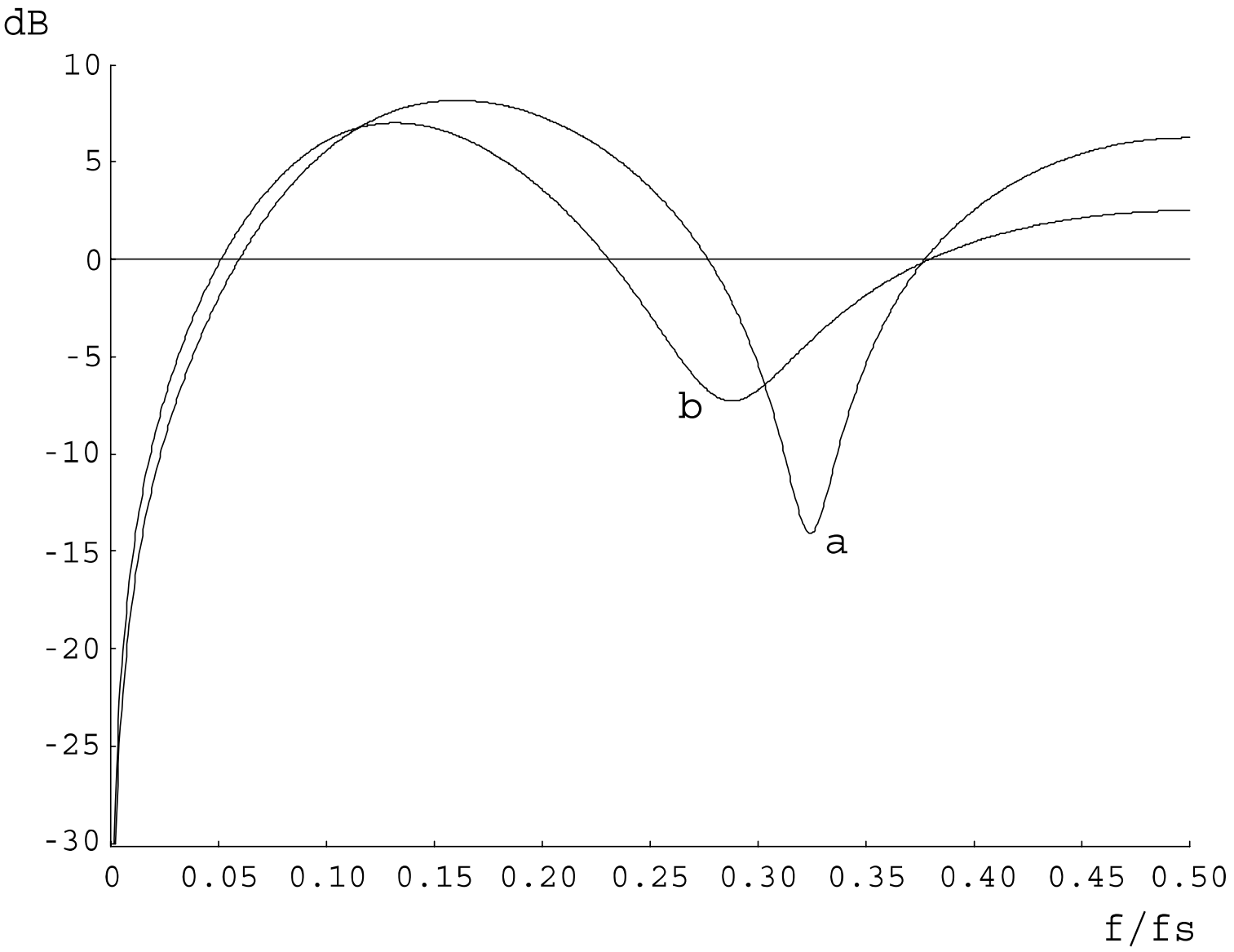

**Fig. 8.16** Effect of auxiliary poles on the output sensitivity function (**a**) without auxiliary poles, (**b**) with auxiliary poles

*resulting from the discretization of the continuous-time filter*:

$$F(s) = \frac{s^2 + 2\zeta_{num}\omega_0 s + \omega_0^2}{s^2 + 2\zeta_{den}\omega_0 s + \omega_0^2} \tag{8.49}$$

*using the bilinear transformation*[6]

$$s = \frac{2}{T_s}\frac{1 - z^{-1}}{1 + z^{-1}} \tag{8.50}$$

*introduces an attenuation* (*a "hole"*) *at the normalized discretized frequency*

$$\omega_{disc} = 2\arctan\left(\frac{\omega_0 T_s}{2}\right) \tag{8.51}$$

*as a function of the ration* $\zeta_{num}/\zeta_{den} < 1$. *The attenuation at* $\omega_{disc}$ *is given by*

$$M_t = 20\log\left(\frac{\zeta_{num}}{\zeta_{den}}\right); \quad (\zeta_{num} < \zeta_{den}) \tag{8.52}$$

*The effect upon the frequency characteristics of* $S_{yp}$ *at frequencies* $f \ll f_{disc}$ *and* $f \gg f_{disc}$ *is negligible.*

[6] The bilinear transformation assures a better approximation of a continuous-time model by a discrete-time model in the frequency domain than the replacement of differentiation by a difference, i.e. $s = (1 - z^{-1})T_s$.

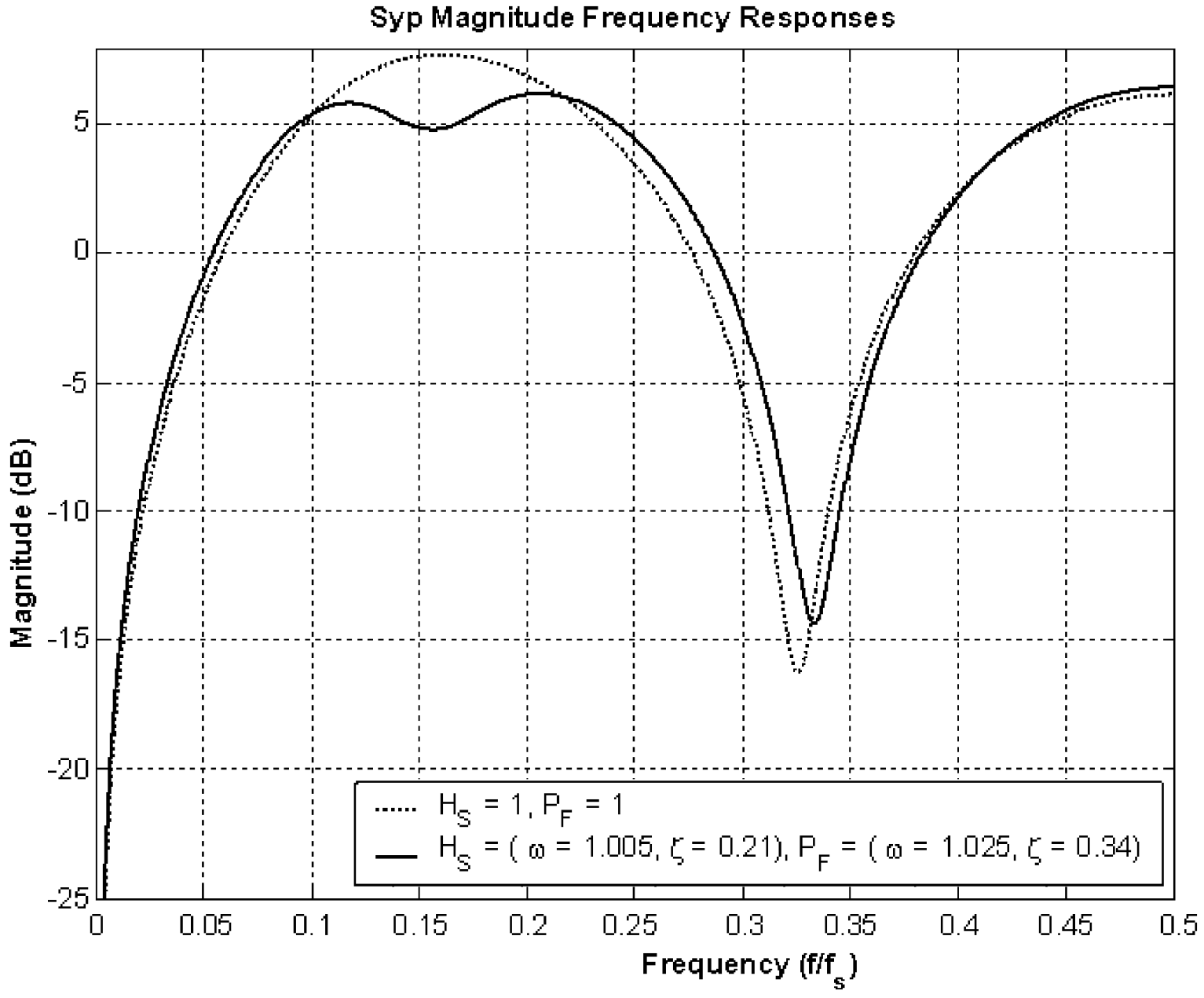

**Fig. 8.17** Effects of a resonant filter $H_{S_i}/P_{F_i}$ on the output sensitivity functions

Figure 8.17 illustrates the effect of the simultaneous introduction of a fixed part $H_S$ and a pair of poles in $P$, corresponding to the discretization of a resonant filter of the form of (8.49). One observes its weak effect on the frequency characteristics of $S_{yp}$, far from the resonance frequency of the filter.

This pole-zero filter is essential for an accurate shaping of the modulus of the sensitivity functions in the various frequency regions in order to satisfy the constraints. It allows one to reduce the interaction between the tuning in different regions.

**Design of the Resonant Pole-Zero Filter $H_{S_i}/P_{F_i}$** The computation of the coefficients of $H_{S_i}$ and $P_{F_i}$ is done in the following way:

Specifications:

- central normalized frequency $f_{disc}$ ($\omega_{disc} = 2\pi f_{disc}$);
- desired attenuation at frequency $f_{disc}$ : $M_t$ dB;
- minimum accepted damping for auxiliary poles $P_{F_i}$ : $(\zeta_{den})_{min}(\geq 0.3)$.

**Step I**: Design of the continuous-time filter

$$\omega_0 = \frac{2}{T_s}\tan\left(\frac{\omega_{disc}}{2}\right); \quad 0 \leq \omega_{disc} \leq \pi; \qquad \zeta_{num} = 10^{M_t/20}\zeta_{den} \tag{8.53}$$

**Step II**: Design of the discrete-time filter using the bilinear transformation of (8.50).

Using (8.50) one gets:

$$F(z^{-1}) = \frac{a_{z0} + a_{z1}z^{-1} + a_{z2}z^{-2}}{a_{z0} + a_{z1}z^{-1} + a_{z2}z^{-2}} = \gamma \frac{1 + \beta_1 z^{-1} + \beta_2 z^{-2}}{1 + \alpha_1 z^{-1} + \alpha_2 z^{-2}} \tag{8.54}$$

which will be effectively implemented as[7]

$$F(z^{-1}) = \frac{H_S(z^{-1})}{P_i(z^{-1})} = \frac{1 + \beta_1 z^{-1} + \beta_2 z^{-2}}{1 + \alpha_1 z^{-1} + \alpha_2 z^{-2}} \tag{8.55}$$

where the coefficients are given by

$$\begin{aligned}
b_{z0} &= \frac{4}{T_s^2} + 4\frac{\zeta_{num}\omega_0}{T_s} + \omega_0^2; \qquad b_{z1} = 2\omega_0^2 - \frac{8}{T_s^2} \\
b_{z2} &= \frac{4}{T_s^2} - 4\frac{\zeta_{num}\omega_0}{T_s} + \omega_0^2 \\
a_{z0} &= \frac{4}{T_s^2} + 4\frac{\zeta_{den}\omega_0}{T_s} + \omega_0^2; \qquad a_{z1} = 2\omega_0^2 - \frac{8}{T_s^2} \\
a_{z2} &= \frac{4}{T_s^2} - 4\frac{\zeta_{den}\omega_0}{T_s} + \omega_0^2
\end{aligned} \tag{8.56}$$

$$\begin{aligned}
\gamma &= \frac{b_{z0}}{a_{z0}} \\
\beta_1 &= \frac{b_{z1}}{b_{z0}}; \qquad \beta_2 = \frac{b_{z2}}{b_{z0}} \\
\alpha_1 &= \frac{a_{z1}}{a_{z0}}; \qquad \alpha_2 = \frac{a_{z2}}{a_{z0}}
\end{aligned} \tag{8.57}$$

The resulting filters $H_{S_i}$ and $P_{F_i}$ can be characterized by the undamped resonance frequency $\omega_0$ and the damping $\zeta$. Therefore, first we will compute the roots of numerator and denominator of $F(z^{-1})$. One gets

$$\begin{aligned}
z_{n1,2} &= \frac{-\beta_1 \pm j\sqrt{4\beta_2 - \beta_1^2}}{2} = A_n e^{j\varphi_n} \\
z_{d1,2} &= \frac{-\alpha_1 \pm j\sqrt{4\alpha_2 - \alpha_1^2}}{2} = A_d e^{j\varphi_d}
\end{aligned} \tag{8.58}$$

One can establish the relation between the filter and the undamped resonance frequency and damping of an equivalent continuous-time filter (discretized with a ZOH). The roots of the second-order monic polynomial in $z^{-1}$ have the expression

$$z_{1,2} = e^{-\zeta_{disc}\omega_{0disc}T_s} e^{\pm j\omega_{0disc}T_s\sqrt{1-\zeta_{disc}^2}} \tag{8.59}$$

[7] The factor $\gamma$ has no effect on the final result (coefficients of $R$ and $S$). It is possible, however, to implement the filter without normalizing the numerator coefficients.

One gets therefore for the numerator and denominator of $F(z^{-1})$

$$\begin{aligned}\omega_{0num} &= \sqrt{\frac{\varphi_n^2 + \ln^2 A_n}{T_s^2}}; & \zeta_{numd} &= -\frac{\ln A_n}{\omega_{0num} T_s} \\ \omega_{0den} &= \sqrt{\frac{\varphi_d^2 + \ln^2 A_d}{T_s^2}}; & \zeta_{dend} &= -\frac{\ln A_d}{\omega_{0den} T_s}\end{aligned} \tag{8.60}$$

where the indices "num" and "den" correspond to $H_S$ and $P_F$, respectively. These filters can be computed using the functions *filter22.sci* (Scilab) *filter22.m* (MATLAB®) and also with *ppmaster* (MATLAB®).[8]

*Remark* For frequencies below 0.17 $f_s$ the design can be done with a very good precision directly in discrete time. In this case, $\omega_0 = \omega_{0den} = \omega_{0num}$ and the damping of the discrete-time filters $H_{S_i}$ and $P_{F_i}$ is computed as a function of the attenuation directly using (8.52).

*Remark* While $H_S$ is effectively implemented in the controller, $P_F$ is only used indirectly. $P_F$ will be introduced in (8.44) and its effect will be reflected in the coefficients of $R$ and $S$ obtained as solutions of (8.42).

### *8.5.2 Input Sensitivity Function*

The input sensitivity function is extremely important in the design of the underlying linear controller used in adaptive control. The modulus of the input sensitivity function should be low at high frequencies in order to assure a good robustness of the system with respect to additive unstructured uncertainties located in the high-frequency region and which are not handled by the adaptation of controller parameters.

The expression of the input sensitivity function using an RST controller with $R$ and $S$ given by (8.42) and (8.43) is:

$$S_{up}(z^{-1}) = -\frac{A(z^{-1})H_R(z^{-1})R'(z^{-1})}{A(z^{-1})H_S(z^{-1})S'(z^{-1}) + q^{-d}B(z^{-1})H_R(z^{-1})R'(z^{-1})} \tag{8.61}$$

**Property 8.8** *The effect of the output disturbances upon the input is canceled* (*i.e.*, $S_{up} = 0$) *at the frequencies where*:

$$A(e^{-j\omega})H_R(e^{-j\omega})R'(e^{-j\omega}) = 0 \tag{8.62}$$

At these frequencies $S_{yp} = 1$ (see Property 8.5 of the output sensitivity functions).

The pre-specified values assuring $S_{up} = 0$ at certain frequencies are the same as those used to make $S_{yp} = 1$ (open-loop operation—see Property 8.5 of the output

[8] To be download from the web site (http://landau-bookic.lag.ensieg.inpg.fr).

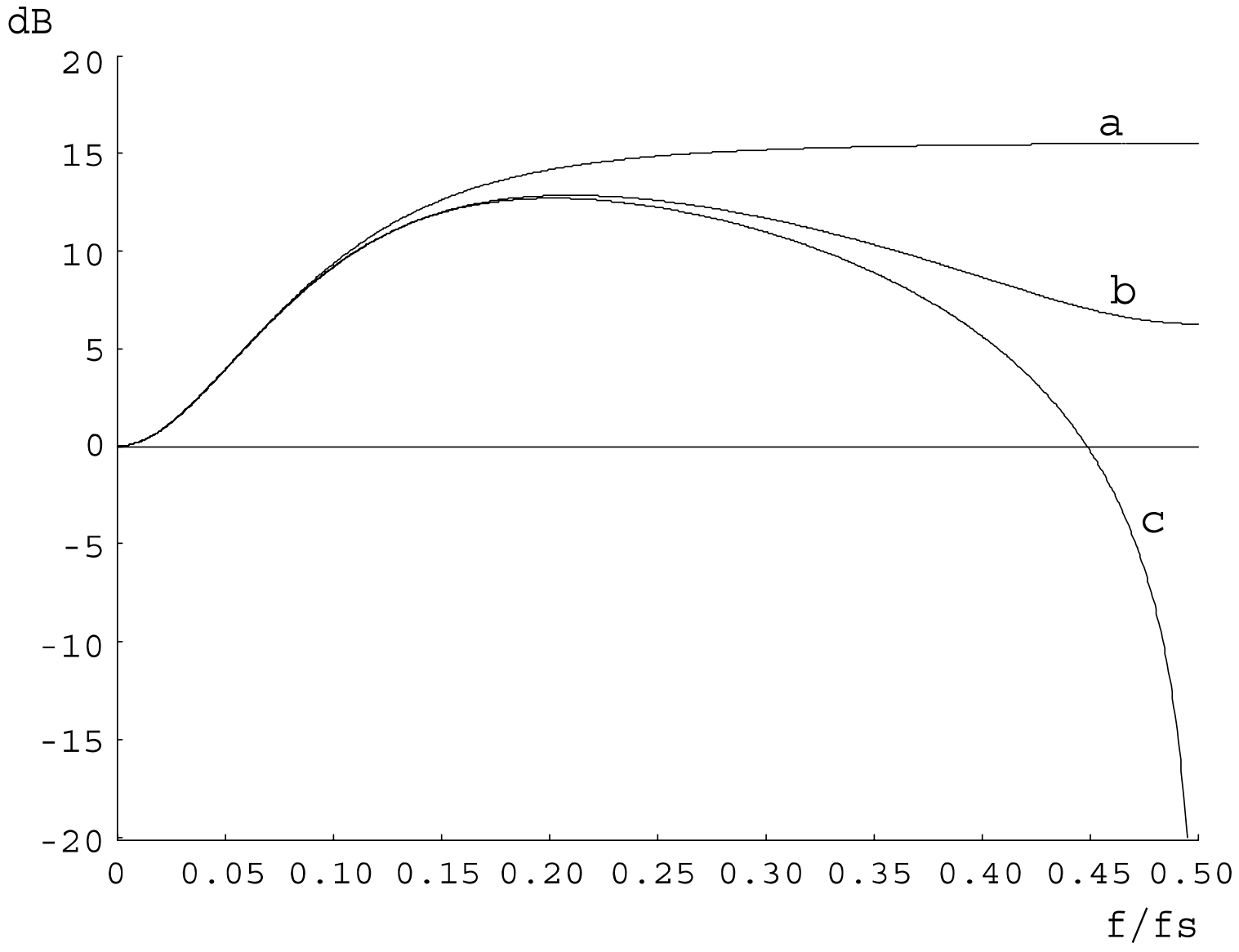

**Fig. 8.18** Effect of $H_R(z^{-1}) = 1 + \alpha z^{-1}$, $0 < \alpha \leq 1$ upon the input sensitivity function: (**a**) $\alpha = 0$, (**b**) $0 < \alpha < 1$, (**c**) $\alpha = 1$

sensitivity functions). Figure 8.18 illustrates the effect upon $S_{up}$ of a pre-specified $H_R(z^{-1})$ of the form:

$$H_R(z^{-1}) = 1 + \alpha z^{-1}; \quad 0 < \alpha \leq 1$$

For $\alpha = 1$, one has $S_{up} = 0$ at $0.5 f_s$. Using $\alpha < 1$ allows to reduce more or less the input sensitivity function around $0.5 f_s$. This structure of $H_R(z^{-1})$ is systematically used for reducing the magnitude of the input sensitivity function in the high-frequency region.

**Property 8.9** *At the frequencies where*:

$$A(e^{-j\omega}) H_S(e^{-j\omega}) S'(e^{-j\omega}) = 0$$

*which corresponds to perfect rejection of the output disturbances* ($S_{yp} = 0$ *at these frequencies*), *one has*:

$$|S_{up}(e^{-j\omega})| = \left| \frac{A(e^{-j\omega})}{B(e^{-j\omega})} \right| \tag{8.63}$$

This implies that perfect rejection of disturbances (or more generally attenuation of disturbances) should be done only in the frequency regions where the gain of the system is large enough. If the gain is too low, $|S_{yp}|$ will be very large. This also indicates that problems will occur if $B$ have complex zeros close to the unit circle (stable or unstable). At these frequencies, rejection of disturbances should be avoided.

**Property 8.10** *Simultaneous introduction of a fixed part $H_{R_i}$ and of a pair of auxiliary poles $P_{F_i}$ having the form*

$$\frac{H_{R_i}(z^{-1})}{P_{F_i}(z^{-1})} = \frac{1+\beta_1 z^{-1}+\beta_2 z^{-2}}{1+\alpha_1 z^{-1}+\alpha_2 z^{-2}} \tag{8.64}$$

*resulting from the discretization of the continuous-time filter*

$$F(s) = \frac{s^2+2\zeta_{num}\omega_0 s+\omega_0^2}{s^2+2\zeta_{den}\omega_0 s+\omega_0^2} \tag{8.65}$$

*using the bilinear transformation*

$$s = \frac{2}{T_s}\frac{1-z^{-1}}{1+z^{-1}} \tag{8.66}$$

*introduces an attenuation (a "hole") at the normalized discretized frequency*

$$\omega_{disc} = 2\arctan\left(\frac{\omega_0 T_s}{2}\right) \tag{8.67}$$

*as a function of the ratio $\zeta_{num}/\zeta_{den} < 1$. The attenuation at $\omega_{disc}$ is given by*

$$M_t = 20\log\left(\frac{\zeta_{num}}{\zeta_{den}}\right); \quad (\zeta_{num} < \zeta_{den}) \tag{8.68}$$

*The effect of the frequency characteristics of $S_{yp}$ at frequencies $f \ll f_{disc}$ and $f \gg f_{disc}$ is negligible.*

The design of these filters is done with the method described in Sect. 8.5.1.

### 8.5.3 Noise Sensitivity Function

The expression of the noise sensitivity function using an RST controller with $R$ and $S$ given by (8.42) and (8.43) is:

$$S_{yb}(z^{-1}) = \frac{z^{-d}B(z^{-1})H_R(z^{-1})R'(z^{-1})}{A(z^{-1})H_S(z^{-1})S'(z^{-1})+z^{-d}B(z^{-1})H_R(z^{-1})R'(z^{-1})} \tag{8.69}$$

**Property 8.11** *The effect of the measurement noise upon the output is canceled (i.e., $S_{yb}=0$) at the frequencies where:*

$$B(e^{-j\omega})H_R(e^{-j\omega})R'(e^{-j\omega}) = 0 \tag{8.70}$$

At these frequencies $S_{yp}=1$ (see Property 8.5 of the output sensitivity function and the related comments concerning the choice of $H_R$).

**Property 8.12** *At the frequencies where:*

$$A(e^{-j\omega})H_S(e^{-j\omega})S'(e^{-j\omega}) = 0$$

(*perfect rejection of the disturbance*) *one has*:

$$S_{yb}(e^{-j\omega}) = 1$$

These various properties give a number of hints for the choice of $H_S$, $H_R$ and $P_F$.

1. $H_S$ should contain the internal model of the disturbance which should be perfectly rejected in steady state.
2. Simultaneous introduction of a fixed part $H_{S_i}$ and of a pair of auxiliary poles $P_{F_i}$ allows to lower the magnitude of the output sensitivity function in a given frequency range
3. $H_R$ allows the loop to be opened at various frequencies and, in particular, at the frequencies where the gain of the system is low or where there are additive unstructured uncertainties.
4. Simultaneous introduction of a fixed part $H_{R_i}$ and of a pair of auxiliary poles $P_{F_i}$ allows to lower the magnitude of the input sensitivity function in a given frequency range.
5. Aperiodic stable high-frequency auxiliary poles will generally force the modulus of the output sensitivity function toward one in the high-frequency region (and therefore it will also reduce the noise sensitivity function).

It is also important to point out that this analysis gives an understanding for the effect of the frequency weighting used in control strategies based on the minimization of a quadratic criterion (see Chap. 7).

## 8.6 Shaping the Sensitivity Functions

In order to achieve the shaping of the modulus of the sensitivity functions such that the constraints defined by the template are satisfied, one has the following possibilities:

1. selection of the desired dominant and auxiliary poles of the closed loop;
2. selection of the fixed parts of the controller ($H_R(q^{-1})$ and $H_S(q^{-1})$);
3. simultaneous selection of the auxiliary poles and of the fixed parts of the controller.

Despite the coupling between the shaping of the output sensitivity function and the input sensitivity function, it is possible, in practice, to deal with these two problems almost independently. In particular, it is useful to remember that at frequencies where $|S_{yp}|$ is close to 1 (0 dB), $|S_{up}|$ is close to 0 ($< -60$ dB).

Automatic methods using convex optimization techniques (Langer and Landau 1999, Langer and Constantinescu 1999; Adaptech 1988) are available in order to solve the shaping problem. However, in most of practical situations, it is relatively easy to calibrate the sensitivity functions using the tools previously presented in this chapter, and taking into account the properties of the sensitivity functions.

An iterative procedure is described in detail in Landau and Zito (2005). The use of a CACSD tool like *ppmaster*[9] (MATLAB®) (Procházka and Landau 2003) will considerably accelerate the procedure.

## 8.7 Other Design Methods

If the final objective is to design a robust controller (without parameter adaptation), other methods can also be considered.

A first approach consists in using $H_\infty$ optimization (Doyle et al. 1992; Green and Limebeer 1995) in order to find a controller such that:

$$\|WS\|_\infty < 1 \tag{8.71}$$

where $S$ is a sensitivity function and $W$ is the size of the type of uncertainty related to the sensitivity function under consideration.

The definition of the upper templates for the various sensitivity functions are also related to this problem since if $W^{-1}$ defines the template, the requirement that the sensitivity function be below the template, leads again to the verification of (8.71).

Nevertheless, in order to apply $H_\infty$ optimization technique, the key problem to be solved first is how to define the appropriate frequency weighting function $W$ in order to obtain the desired performances (placement of the dominant poles and shaping of the sensitivity function).

From the (8.71) one can see that $W$ can be interpreted as the inverse of the desired sensitivity function. Therefore, a method for defining the "desired sensitivity function" has been developed which is then used in an $H_\infty$ optimization in order to obtain the desired controller (Landau and Karimi 1996).

A different approach for the design of a robust controller is based on a convex parameterization of the controller and use of convex optimization procedure (Langer and Landau 1999). This method presents a number of advantages in terms of performance specifications and gives excellent results.

For a comparison of various approaches for the design of a robust controller for a flexible transmission see Landau et al. (1995a, 1995b), Langer and Landau (1999), Kwakernaak (1995).

## 8.8 A Design Example: Robust Digital Control of a Flexible Transmission

The flexible transmission has been presented in Sect. 1.4. Three models have been identified (see Sect. 5.9) for the 0% load (L00), 50% load (L50) and 100% load (L100).

[9] Available on the website (http://landau-bookic.lag.ensieg.inpg.fr).

An RST controller has been designed. Since the adaptive control system responds only to the structured parametric uncertainty by the adjustment of model parameters, the unstructured modeling errors should be taken in account by the linear control law associated to the adaptive scheme. Since the unstructured modeling errors are in general large at high frequencies, the input sensitivity function $S_{up}$ (the transfer function between output disturbance and plant input) which is related to additive uncertainties, should be low in these frequencies. For the flexible transmission system, this can be achieved using the pole placement with sensitivity function shaping (Sect. 8.6 and Landau and Karimi 1998; Landau and Zito 2005), in the following way:

1. choice of a fixed term in the numerator of the controller in the form of $(1 + \alpha z^{-1})^n, 0.5 \leq \alpha \leq 1, n \geq 1$;
2. choice of the auxiliary closed-loop poles near to the high-frequency poles of the plant.

The first one has an effect only in the very high frequencies (between 0.4 to 0.5 $f/f_s$), whereas the second one decrease significantly the input sensitivity function in middle and high frequencies (between 0.2 to 0.5 $f/f_s$). The effects of these techniques can be easily shown in the following example.

Consider the discrete-time model of the flexible transmission system in the no-load case. A first pole placement controller is designed with the following specifications:

1. a pair of complex dominant poles with the frequency of the first vibration mode of the plant model but with a damping factor of 0.8;
2. four auxiliary poles at 0.1 i.e. $(1 - 0.1z^{-1})^4$;
3. an integrator in the controller.

The curve a in Fig. 8.19 shows the magnitude of the input sensitivity function $S_{up}$ with a maximum of 12 dB at high frequencies which signifies a bad robustness with respect to the additive uncertainties. The curve $b$ shows the effect of a fixed term $(1 + z^{-1})$ in the controller numerator. One observes that $S_{up}$ decreases in very high frequencies. But a pair of complex poles in the place corresponding to the second vibration mode of the plant will decrease significantly $S_{up}$ in a very large band (curve $c$). Thus, in order to obtain a robust controller with respect to the modeling errors beyond the closed-loop band pass, auxiliary poles should be chosen near to the plant high-frequency poles. It is advisable to open the loop at $0.5 f/f_s$ and to remove from zero the remaining closed-loop poles which can be assigned.

## 8.9 Concluding Remarks

1. Before considering an adaptive control scheme it is mandatory to design a robust controller.
2. Robustness is not an intrinsic property of a control strategy. It results from an appropriate choice of some design objectives.

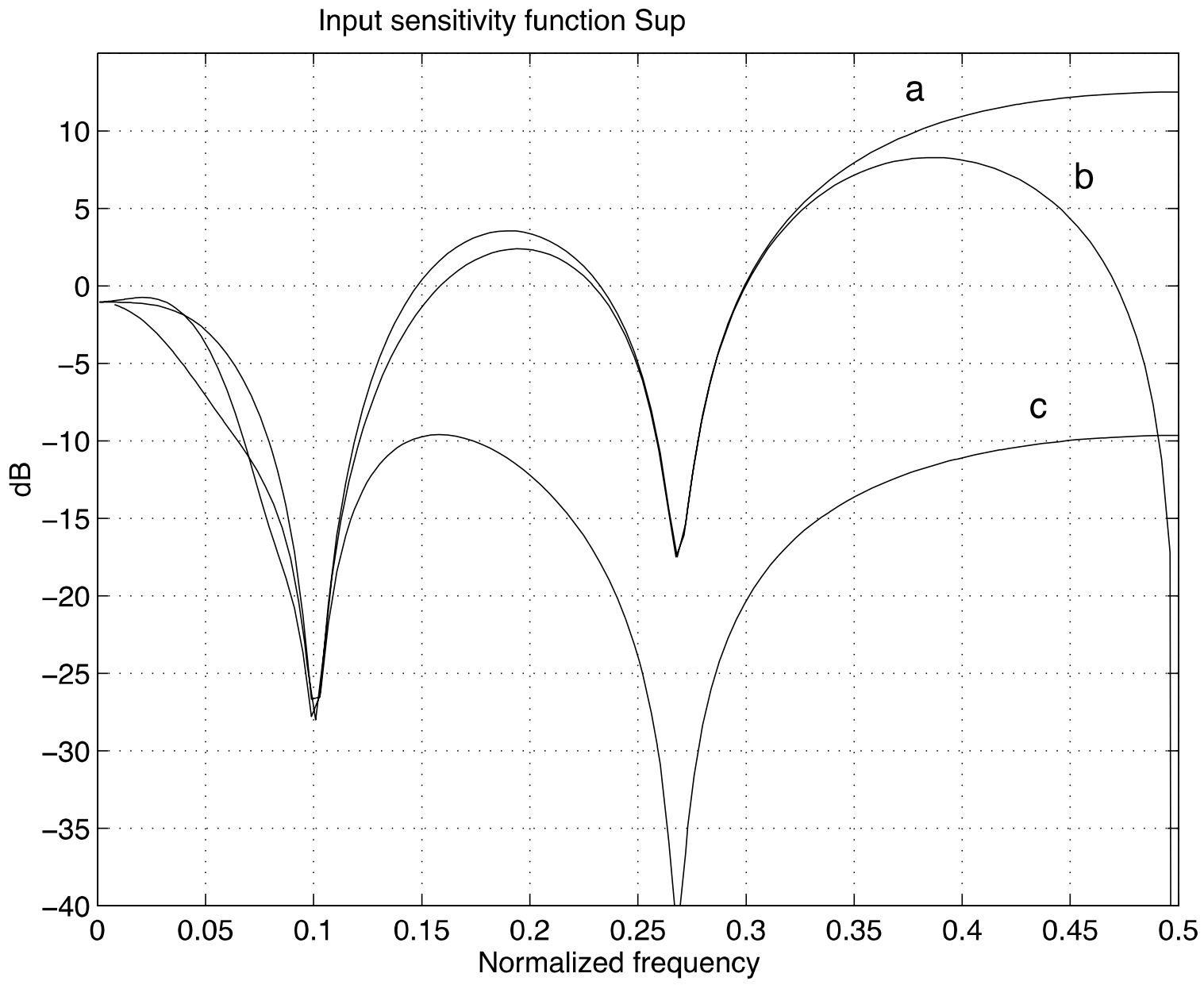

**Fig. 8.19** Input sensitivity function shaping for the flexible transmission

3. The sensitivity functions play a fundamental role in assessing the robustness of a control scheme.
4. Two key robustness indicators are:
   - modulus margin;
   - delay margin.

   The *modulus margin* ($\Delta M$) is the minimum distance between the Nyquist plot of the open loop and the critical point $[-1; j_0]$ and corresponds to the inverse of the maximum of the modulus of the output sensitivity function. A typical value is $\Delta M \geq 0.5$ ($-6$ dB).

   The *delay margin* ($\Delta\tau$) is the additional delay which leads to instability and it is defined as:

$$\Delta\tau = \min_i \frac{\Delta\phi_i}{w_{CR}^i}$$

   where $\Delta\phi_i$ is the phase margin associated with the corresponding crossover frequency $w_{CR}^i$. A typical value is $\Delta\tau \geq T_S$.
5. Since the adaptive control system only responds to the structured parametric uncertainty by the adjustment of model parameters, the unstructured modeling errors should be taken in account by the linear control law associated to the adaptive scheme.
6. The nominal performance specifications together with the robustness requirements lead to the definition of upper and lower templates for the various sensitivity functions.

7. The modulus of the input sensitivity function should be low at high frequencies. Opening of the loop at. $0.5 f_s$ is often recommended.
8. For one step ahead control strategies (pole placement, tracking and regulation with independent objectives...) the shaping of the sensitivity functions in order to match the desired templates can be done by an appropriate choice of the fixed parts of the controller and of the auxiliary poles.
9. For long range predictive control (generalized predictive control, linear quadratic control) the shaping of the sensitivity functions can be done by an appropriate choice of the fixed parts of the controller and of the various design parameters characterizing the quadratic criterion to be minimized.
10. Several methods are available for the shaping of the sensitivity functions.

## 8.10 Problems

**8.1** The discrete-time identified model of a very flexible arm without load for a sampling period of 70 ms is Landau (1993b):

$$\begin{aligned} A(q^{-1}) &= 1 - 2.19358q^{-1} + 1.85893q^{-2} - 1.00016q^{-3} + 1.12598q^{-4} \\ &\quad - 1.43643q^{-5} + 0.726861q^{-6} \\ B(q^{-1}) &= 0.05723q^{-1} - 0.164075q^{-2} + 0.256q^{-3} - 0.13623q^{-4}, \quad d = 1 \end{aligned}$$

The system is characterized by three very low-damped vibration modes and unstable zeros. Use the pole placement combined with shaping of the output sensitivity function in order to design an RST controller based on the following specifications:

- tracking dynamics specified by the discretization of a 2nd-order continuous-time system with $\omega_0 = 3$ rad/s, $\zeta = 0.9$;
- null steady state error;
- dominant poles for the closed loop specified by the discretization of a 2nd-order continuous-time system with $\omega = 3$ rad/s, $\zeta = 0.8$;
- modulus margin $\Delta M \geq 0.5$ (i.e., $|S_{yp}|_{\max} \leq 6$ dB);
- delay margin $\Delta\tau \geq T_S$ ($= 70$ ms).

Test the time response for tracking and disturbance rejection of the resulting design.

**8.2** Follow up the design given in Sect. 8.8 for the control of the flexible transmission in order to improve the performances for 50% load and 100% load by shaping $S/P$.

**8.3** Consider the pole placement design for the case where (internal model control—Sect. 7.3.4):

$$P(q^{-1}) = A(q^{-1})P_0(q^{-1})$$

Give the expressions for $S_{yp}$, and $S_{yb}$ and particularize them for the cases $H_S(q^{-1}) = 1 - q^{-1}$, $H_R(q^{-1}) = 1$ and $H_S(q^{-1}) = 1 - q^{-1}$, $H_R(q^{-1}) \neq 1$.

**8.4** For the case discussed in Problem 8.3, further assume that $B(q^{-1}) = b_1 q^{-1}$

1. Give the expressions of $S_{yp}$ and $S_{up}$ for $P_0(q^{-1}) = 1$, $H_R(q^{-1}) = 1$. Discuss the frequency behavior of $|S_{yp}|$ and $|S_{yb}|$ in this case. What happens with the delay margin for large $d$?
2. Choosing $P_0(q^{-1}) = 1 - \alpha q^{-1}$, $0 < \alpha < 1$, $H_R(q^{-1}) = 1$ find analytically the value of $\alpha$ such that a delay margin of one sampling period is guaranteed (hint: take into account the expression of the template for $\Delta\tau = T_S$ on $|S_{yb}|$).
3. Check the result for the case $A(q^{-1}) = 1 - 0.2q^{-1}$, $d = 7$. Represent $S_{yb}$ and $S_{yp}$ as well as the corresponding templates.
4. Choose $P_0(q^{-1}) = 1$ and $H_R(q^{-1}) = 1 + \beta q^{-1}$. Find analytically the value of $\beta$ such that the delay margin of one sampling period is guaranteed.
5. Check the result for the case $A(q^{-1}) = 1 - 0.2q^{-1}, d = 7$. Represent $S_{yb}$ and $S_{yp}$ as well as the corresponding templates.

**8.5** Consider the following discrete-time plant model (hot dip galvanizing—see Fenot et al. 1993)

$$G(q^{-1}) = \frac{q^{-7}(0.8q^{-1})}{1 - 0.2q^{-1}}$$

The poles of the closed-loop system are defined as:

$$\text{(a)} \quad P(q^{-1}) = (1 - 0.2q^{-1})(1 - 0.3q^{-1})$$
$$\text{(b)} \quad P(q^{-1}) = (1 - 0.2q^{-1})(1 - 0.3q^{-1})(1 - 0.1q^{-1})^7$$

Design an RST controller with $H_S(q^{-1}) = 1 - q^{-1}$. Represent the output sensitivity function and the Nyquist plot for the two cases. Compute the delay margins. Discuss the results (what is the effect of the auxiliary poles?).

**8.6** For the unloaded model of the flexible transmission do a PSMR generalized predictive control design with $H_S(q^{-1}) = 1 - q^{-1}$; $H_R(q^{-1}) = 1$ and $P_d(q^{-1})$ corresponding to the discretization of a second-order continuous-time system with $\omega_0 = 6.46$ rad/s, $\zeta = 0.9$. Study the characteristics of the various sensitivity functions for different values of $\lambda$ between 0.1 and 0.5.

# Chapter 9
# Recursive Plant Model Identification in Closed Loop

## 9.1 The Problem

In practice, many plants cannot be easily operated in open loop with the purpose of carrying an experimentation protocol for identification (ex: integrator behavior, drift of the output, open-loop unstable system). In some situations, a controller may already exist and there is no reason to open the loop when the object of the identification will be to get a better model for either designing a new controller or re-tuning the existing controller.

In adaptive control, one also has to estimate the parameters of the plant model in closed-loop operation (note that, in this case, the controller will be with time-varying parameters depending on the estimated plant model parameters). In the context of adaptive control, the study of recursive plant model identification in closed-loop will emphasize on the one hand the importance of the filters to be used with open-loop recursive identification algorithms and, on the other hand, the use of specific algorithms dedicated to recursive identification in closed loop.

The problem of plant model identification in closed-loop is also of crucial importance in the context of the iterative combination of identification in closed-loop and robust control redesign. This can be viewed as an adaptive control scheme in which data acquisition and parameter identification is done at each sample, but the up-dating of the controller is done from time to time.

Figure 9.1a illustrates the basis of this iterative procedure (for the case of an RST digital controller). The upper part represents the true closed-loop system and the lower part represents the design system. The objective is to minimize the error between the two systems by using new data acquired in closed-loop operation. Since the problem is not directly tractable, the idea is to improve plant model estimation first and then re-design the controller based on the new model. This sequence of operations is carried out one or several times. However, a key point is that the new plant model estimation should be done in order to reduce the error between the two systems. In fact, the objective is to get a better predictor for the closed loop via a better estimation of the plant model. In addition, in most of the situations, iterative identification in closed loop and controller redesign offer, possible improvements in

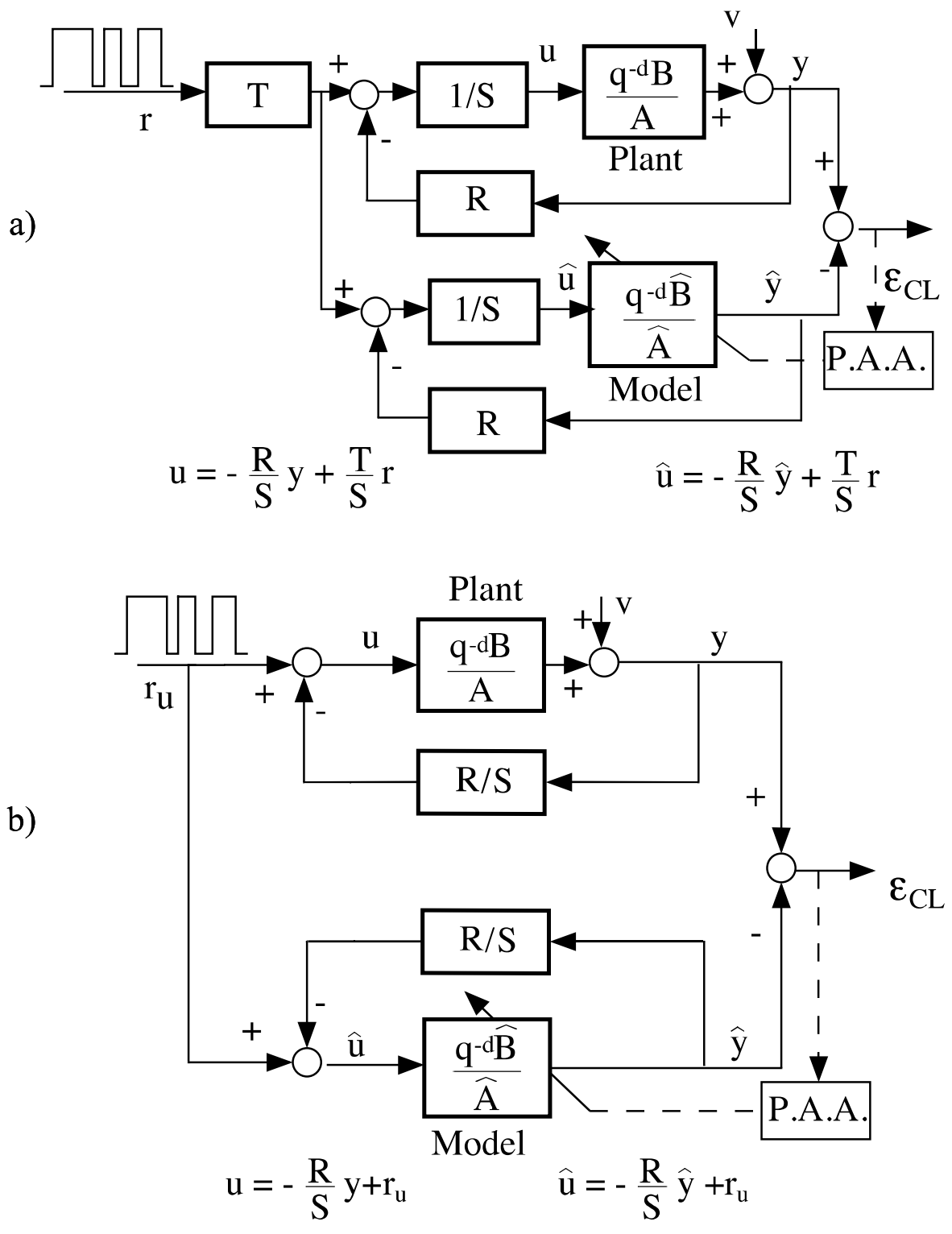

**Fig. 9.1** Identification in closed loop, (**a**) excitation added on the reference, (**b**) excitation added to the output of the controller

the performance of a controller previously designed on the basis of a plant model identified in open-loop.

In the context of recursive algorithms, the problem of plant model identification in closed loop can be formulated as follows: under the assumption that the controller is constant, identify a plant model such that:

1. Global asymptotic stability is assured for any initial parameter estimates and initial error between the output of the true system and that of the closed-loop predictor (in the absence of noise).
2. An asymptotically optimal predictor of the true closed-loop system is obtained.
3. Unbiased estimates of the plant model parameters are obtained asymptotically under appropriate richness conditions (in the presence of noise).

It is assumed that the input-output part of the plant to be identified belongs to the model set.

The problem of identification of a plant model in closed loop can be viewed in two different ways leading however, to similar types of algorithms.

*MRAS point of view* The true closed-loop system corresponds to a reference model and a parallel adjustable system having a feedback configuration is built up. This adjustable feedback system contains a fixed controller and an adjustable model of the plant. The problem is to design a parameter adaptation algorithm assuring the global asymptotic stability of the closed-loop prediction error (or driving the closed-loop prediction error to zero asymptotically in the absence of noise). This is a dual problem with respect to the classical model reference adaptive control problem.

*Identification point of view* Construct a reparameterized adjustable predictor for the closed-loop system in terms of a known fixed controller and of an adjustable plant model.

The external excitation can be applied either on the reference (Fig. 9.1a) or added to the output of the controller (Fig. 9.1b). This does not change the structure of the algorithm, it just affects the external quantity added to the input of the plant (or of the estimated plant model). However, the properties of the estimated model will depend on where the external excitation is applied.

Two families of recursive algorithms for plant model identification in closed loop can be derived from this basic scheme . The first approach starts from the re-writing of the closed-loop system output equation in term of the regressor of the predictor (i.e., in term of $\hat{u}(t)$ and $\hat{y}(t)$) and the previous values of the closed-loop predictor error (see Fig. 9.1). This approach generates the family of *closed-loop output error recursive identification algorithms* (Landau and Karimi 1997a). This family of algorithms is presented in Sect. 9.2.[1]

The second approach starts from the rewriting of the predictor output equation in terms of the measurement vector (i.e., in terms of $u(t)$ and $y(t)$) and the previous values of the closed-loop prediction error. This approach leads to the expression of the closed-loop prediction error as a filtered least square type prediction error for plant model identification from input-output measurements (Åström 1993). This allows one to use many of the open-loop recursive identification algorithms with an appropriate filtering of the data. This family will be called *filtered open-loop recursive identification algorithms* (FOL). Note that these are the types of methods which have been used in indirect adaptive control for plant model parameter estimation. These methods are discussed in Sect. 9.3.

A comparison of the two approaches in terms of the frequency distribution of the asymptotic bias is presented in Sect. 9.4.

Once an identification procedure has been carried out using one or several algorithms, the question arises: what is the best model? The answer may be given by the comparative validation of the identified models in closed-loop operation. The best plant model will be the one which allows the best prediction of the closed-loop behavior. As will be shown, coherent statistical validation tests can be defined as in the case of identification in open loop and further enhanced by a direct comparison of the computed closed-loop poles and the achieved ones (which are obtained by identifying the closed-loop system). The validation aspects are discussed in Sect. 9.5.

A comparative evaluation of the various algorithms based on simulations and real-time time experiments is presented in Sect. 9.6.

---

[1] By duality arguments these algorithms can be used for estimation of reduced order controllers—see Landau et al. (2001b).

It has been observed that identification in closed loop generally improves the results obtained in open-loop identification in terms of resulting control performance. The main reason for this is that in closed-loop operation the external excitation signal is filtered by the sensitivity function. In particular, if the external excitation is added to the output of the controller, it will be filtered by the output sensitivity function which will enhance the plant input signal in the frequency region where the output sensitivity function has its maximum. As a consequence, a more precise model will be obtained in the frequency regions which are critical for control design. For a detailed discussion and experimental results see Langer and Landau (1996), Landau and Karimi (2002).

### 9.1.1 The Basic Equations

The objective is to estimate the parameters of the plant model defined by the transfer operator:

$$G(q^{-1}) = \frac{q^{-d}B(q^{-1})}{A(q^{-1})} \tag{9.1}$$

where:

$$B(q^{-1}) = b_1 q^{-1} + \cdots + b_{n_B} q^{-n_B} = q^{-1}B^*(q^{-1}) \tag{9.2}$$

$$A(q^{-1}) = 1 + a_1 q^{-1} + \cdots + a_{n_A} q^{-n_A} = 1 + q^{-1}A^*(q^{-1}) \tag{9.3}$$

The plant is operated in closed loop with an RST digital controller (without lack of generality). The output of the plant operating in closed loop is given by (see Fig. 9.1b):

$$y(t+1) = -A^*y(t) + B^*u(t-d) + Av(t+1) = \theta^T\varphi(t) + Av(t+1) \tag{9.4}$$

where $u(t)$ is the plant input, $y(t)$ is the plant output, $v(t)$ is the output disturbance noise and:

$$\theta^T = [a_1, \ldots, a_{n_A}, b_1, \ldots, b_{n_B}] \tag{9.5}$$

$$\varphi^T(t) = [-y(t), \ldots, -y(t-n_A+1), u(t-d), \ldots, u(t-n_B+1-d)] \tag{9.6}$$

$$u(t) = -\frac{R}{S}y(t) + r_u \tag{9.7}$$

where $r_u$ is the external excitation added to the output of the controller ($r_u$ is equal to $r\frac{T}{S}$ if the external excitation is applied on the reference).

For a fixed value of the estimated parameters, the predictor of the closed loop (the design system) can be expressed as:

$$\hat{y}(t+1) = -\hat{A}^*\hat{y}(t) + \hat{B}^*\hat{u}(t-d) = \hat{\theta}^T\phi(t) \tag{9.8}$$

where

$$\hat{\theta}^T = [\hat{a}_1, \ldots, \hat{a}_{n_A}, \hat{b}_1 \ldots, \hat{b}_{n_B}] \tag{9.9}$$

$$\phi^T(t) = [-\hat{y}(t), \ldots, -\hat{y}(t - n_A + 1), \hat{u}(t-d), \ldots, \hat{u}(t - n_B + 1 - d)] \tag{9.10}$$

$$\hat{u}(t) = -\frac{R}{S}\hat{y}(t) + r_u \tag{9.11}$$

The closed-loop prediction (output) error is defined as:

$$\varepsilon_{CL}(t+1) = y(t+1) - \hat{y}(t+1) \tag{9.12}$$

Note that as it clearly results from Fig. 9.1, for constant values of the estimated parameters the predictor regressor vector $\phi(t)$ depends only upon the external excitation. Therefore under the assumption that the external excitation ($r$ or $r_u$) and the stochastic disturbance $v$ are independent, $\phi(t)$ and $v(t)$ are not correlated (as well as $\phi(t)$ and $\varepsilon_{CL}(t+1)$). The scheme has the structure of an output error prediction. If furthermore a model for the noise $v(t)$ is assumed, then a predictor can be built such that under the exact knowledge of the input-output plant model and of the noise model, the closed-loop prediction error tends asymptotically towards a white noise.

If known fixed parts should be included in the estimated plant model, the equations of the predictor for the closed loop has to be modified in order to preserve the input/output behavior. See for details (Landau and Zito 2005) and Problem 9.9.

For all the methods the parameter adaptation algorithm (PAA) has the general form:

$$\hat{\Theta}(t+1) = \hat{\Theta}(t) + F(t)\Phi(t)\nu(t+1) \tag{9.13}$$

$$F(t+1)^{-1} = \lambda_1(t)F(t)^{-1} + \lambda_2(t)\Phi(t)\Phi^T(t)$$
$$0 < \lambda_1(t) \le 1;\ 0 \le \lambda_2(t) < 2;\ F(0) > 0;\ F(t)^{-1} > \alpha F^{-1}(0);\ 0 < \alpha < \infty \tag{9.14}$$

$$F(t+1) = \frac{1}{\lambda_1(t)}\left[F(t) - \frac{F(t)\Phi(t)\Phi^T(t)F(t)}{\frac{\lambda_1(t)}{\lambda_2(t)} + \Phi^T(t)F(t)\Phi(t)}\right] \tag{9.15}$$

$$\nu(t+1) = \frac{\nu^\circ(t+1)}{1 + \Phi^T(t)F(t)\Phi(t)} \tag{9.16}$$

where $\nu^\circ(t+1) = f_1(\hat{\Theta}(t), \hat{\Theta}(t-1), \ldots, y(t+1), \nu(t), \nu(t-1), \ldots)$ is the a priori adaptation error, $\nu(t+1) = f_2(\hat{\Theta}(t+1), \hat{\Theta}(t), \ldots, y(t+1), \nu(t), \nu(t-1), \ldots)$ is the a posteriori adaptation error and $\Phi(t)$ is the observation vector.

For each recursive identification algorithm $\Theta, \Phi, \nu^0(t+1)$ will have specific expressions. Note that the sequences $\lambda_1(t)$ and $\lambda_2(t)$ give different laws for the time evolution of the adaptation gain $F(t)$. For convergence analysis in the stochastic environment, it is assumed that a PAA with decreasing adaptation gain is used (i.e., $\lambda_1(t) \equiv 1, \lambda_2(t) > 0$ or $\lambda_2(t) = \lambda_2 > 0$).

## 9.2 Closed-Loop Output Error Algorithms (CLOE)

The key observation is that the output of the closed-loop system given by (9.4) can be expressed as:

$$y(t+1) = \theta^T \varphi(t) + Av(t+1) = \theta^T \phi(t) - \left[A^* + \frac{q^{-d}B^*R}{S}\right]\varepsilon_{CL}(t) + Av(t+1) \tag{9.17}$$

and that:

$$1 + q^{-1}\left[A^* + \frac{q^{-d}B^*R}{S}\right] = \frac{AS + q^{-d}BR}{S} = \frac{P}{S} \tag{9.18}$$

where $P = AS + q^{-d}BR$ defines the poles of the true closed-loop system. Subtracting (9.8) from (9.17) one obtains in the deterministic case $(v(t+1) \equiv 0)$ :

$$\varepsilon_{CL}(t+1) = [\theta - \hat{\theta}]^T \phi(t) - \left[A^* + \frac{q^{-d}B^*R}{S}\right]\varepsilon_{CL}(t) \tag{9.19}$$

Taking into account the relationship (9.18), (9.19) can be re-written as:

$$\varepsilon_{CL}(t+1) = \frac{S}{P}[\theta - \hat{\theta}]^T \phi(t) \tag{9.20}$$

Note that in the linear case with known parameters since $\phi(t)$ and $v(t)$ are uncorrelated, an optimal predictor minimizing $\mathbf{E}\{\varepsilon^2_{CL}(t+1)\}$ is obtained for $\hat{\theta} = \theta$.

### 9.2.1 *The Closed-Loop Output Error Algorithm (CLOE)*

Replacing now the fixed predictor of the closed loop given in (9.8) by an adjustable predictor, i.e.,

- *a priori predicted output*:

$$\hat{y}^\circ(t+1) = \hat{y}[t+1|\hat{\theta}(t)] = \hat{\theta}^T(t)\phi(t) \tag{9.21}$$

- *a posteriori predicted output*:

$$\hat{y}(t+1) = \hat{y}[t+1|\hat{\theta}(t+1)] = \hat{\theta}^T(t+1)\phi(t) \tag{9.22}$$

and defining the a priori prediction error as:

$$\varepsilon^\circ_{CL}(t+1) = y(t+1) - \hat{y}^\circ(t+1) \tag{9.23}$$

and the a posteriori prediction error as:

$$\varepsilon_{CL}(t+1) = y(t+1) - \hat{y}(t+1) \tag{9.24}$$

the equation for the a posteriori prediction error becomes in the deterministic environment $(v(t+1) = 0)$:

$$\varepsilon_{CL}(t+1) = \frac{S}{P}[\theta - \hat{\theta}(t+1)]^T \phi(t) \tag{9.25}$$

Theorem 3.2 (Chap. 3) suggests a PAA with:

$$\hat{\Theta}(t) = \hat{\theta}(t)$$
$$\Phi(t) = \phi(t)$$
$$\nu^\circ(t+1) = \varepsilon^\circ_{CL}(t+1)$$

This is termed the CLOE *closed-loop output error* algorithm (Landau and Karimi 1997a). Using the results of Theorem 3.2, it can be shown that in a deterministic environment the sufficient condition for $\lim_{t\to\infty} \varepsilon_{CL}(t+1) = 0$ together with the boundedness of $\varepsilon_{CL}(t+1)$ for any initial condition is that:

$$H'(z^{-1}) = \frac{S(z^{-1})}{P(z^{-1})} - \frac{\lambda_2}{2} \tag{9.26}$$

is strictly positive real (where $\max_t \lambda_2(t) \le \lambda_2 < 2$). Since $\varepsilon_{CL}(t)$ is bounded this allows to conclude that $\hat{y}(t)$ and $\hat{u}(t)$ are bounded (under the hypothesis that S is stable and that the closed loop is stable) and therefore also $\lim_{t\to\infty} \varepsilon^\circ_{CL}(t+1) = 0$. In a stochastic environment using the averaging method (Theorem 4.1), the equation of the closed-loop prediction error for a fixed value of estimated parameter $\hat{\theta}$, takes the form:

$$\varepsilon_{CL}(t+1,\hat{\theta}) = \frac{S}{P}[\theta - \hat{\theta}]^T \phi(t,\hat{\theta}) + \frac{AS}{P} v(t+1) \tag{9.27}$$

Under the hypothesis that $v(t+1)$ is independent with respect to the external excitation $r_u$, $\phi(t,\hat{\theta})$ and $v(t+1)$ are independent, and Theorem 4.1 can be used (for $\lambda_1(t) \equiv 1, \lambda_2(t) > 0$). The same positive real condition of (9.26) assures asymptotic unbiased estimates under richness condition (Landau and Karimi 1997a).

Note that using the results of Sect. 3.6, the positive real condition (9.26) can be relaxed by using an integral + proportional adaptation instead of the integral adaptation provided by the (9.13).

### 9.2.2 Filtered Closed-Loop Output Error Algorithm (F-CLOE)

Equation (9.20) can also be rewritten as:

$$\varepsilon_{CL}(t+1) = \frac{S}{P}\frac{\hat{P}}{S}[\theta - \hat{\theta}]\frac{S}{\hat{P}}\phi(t) = \frac{\hat{P}}{P}[\theta - \hat{\theta}]\phi_f(t) \tag{9.28}$$

where:

$$\phi_f(t) = \frac{S}{\hat{P}}\phi(t) \tag{9.29}$$
$$\hat{P} = \hat{A}S + q^{-d}\hat{B}R \tag{9.30}$$

In (9.30) $\hat{P}$ is an estimation of the true closed-loop poles based on an initial estimation of the plant model (for example using an open loop experiment). This formulation leads to the *filtered closed-loop output error algorithm* (F-CLOE) (Landau and Karimi 1997a) which uses the same adjustable predictor as CLOE (see (9.21) and (9.22)) and the PAA with:

$$\hat{\Theta}(t) = \hat{\theta}(t)$$
$$\Phi(t) = \phi_f(t)$$
$$\nu^\circ(t+1) = \varepsilon^\circ_{CL}(t+1)$$

It can be shown that neglecting the non-commutativity of time-varying operators (an exact algorithm can, however, be derived), under the sufficient condition that:

$$H'(z^{-1}) = \frac{\hat{P}(z^{-1})}{P(z^{-1})} - \frac{\lambda_2}{2} \tag{9.31}$$

is strictly positive real, both asymptotic stability in deterministic environment and asymptotic unbiasedness in a stochastic environment is assured (Landau and Karimi 1997a).

One can further relax the condition of (9.31) by filtering $\phi(t)$ through a time-varying filter $S/\hat{P}(t)$ where $\hat{P}(t)$ corresponds to the current estimate of the closed loop given by: $\hat{P}(t) = \hat{A}(t)S + q^{-d}\hat{B}(t)R$ where $\hat{A}(t)$ and $\hat{B}(t)$ are the current estimates of the $A$ and $B$ polynomials (the AF-CLOE algorithm).

### *9.2.3 Extended Closed-Loop Output Error Algorithm (X-CLOE)*

For the case $v(t+1) = \frac{C}{A}e(t+1)$, where $e(t+1)$ is a zero mean Gaussian white noise and $C(q^{-1}) = 1 + q^{-1}C^*(q^{-1})$ is an asymptotically stable polynomial, an extended output error prediction model can be defined:

$$\begin{aligned}\hat{y}(t+1) &= -\hat{A}^*\hat{y}(t) + \hat{B}^*\hat{u}(t-d) + \hat{H}^*\frac{\varepsilon_{CL}(t)}{S}\\ &= \hat{\theta}^T\phi(t) + \hat{H}^*\frac{\varepsilon_{CL}(t)}{S} = \hat{\theta}_e^T\phi_e(t)\end{aligned} \tag{9.32}$$

Equation (9.17) for the plant output becomes in this case:

$$y(t+1) = \theta^T\phi(t) + H^*\frac{\varepsilon_{CL}(t)}{S} - C^*\varepsilon_{CL}(t) + Ce(t+1) \tag{9.33}$$

where:

$$H^* = h_1 + h_2 q^{-1} + \cdots + h_{n_H} q^{-n_H+1} = C^*S - A^*S - q^{-d}B^*R \tag{9.34}$$
$$H = 1 + q^{-1}H^* = 1 + CS - P \tag{9.35}$$

and subtracting (9.32) from (9.33), one obtains the following expression for the closed-loop prediction error (for details see Landau and Karimi 1999):

$$\varepsilon_{CL}(t+1) = \frac{1}{C}[\theta_e - \hat{\theta}_e]^T\phi_e(t) + e(t+1) \tag{9.36}$$

where:

$$\theta_e^T = [\theta^T, h_1, \ldots, h_{n_H}] \tag{9.37}$$

$$\hat{\theta}_e^T = [\hat{\theta}^T, \hat{h}_1, \ldots, \hat{h}_{n_H}] \tag{9.38}$$

$$\phi_e^T(t) = [\phi^T(t), \varepsilon_{CLf}(t), \ldots, \varepsilon_{CLf}(t - n_H + 1)] \tag{9.39}$$

$$\varepsilon_{CLf}(t) = \frac{1}{S}\varepsilon_{CL}(t) \tag{9.40}$$

Equation (9.36) clearly shows that for $\hat{\theta}_e = \theta_e$ the closed loop prediction error tends asymptotically towards $e(t+1)$.

Replacing the fixed predictor (9.32) with an adjustable one, a recursive identification algorithm (X-CLOE) can be obtained by using a PAA with:

$$\hat{\Theta}(t) = \hat{\theta}_e(t)$$

$$\Phi(t) = \phi_e(t)$$

$$\nu^\circ(t+1) = \varepsilon_{CL}^\circ(t+1) = y(t+1) - \hat{\theta}_e^T(t)\phi_e(t)$$

The analysis in the deterministic case using Theorem 3.2 shows that global asymptotic stability is assured without any positive real condition (since the a posteriori closed-loop prediction error equation in this case is $\varepsilon_{CL} = [\theta_e - \hat{\theta}_e(t+1)]^T\phi_e(t)$). In the stochastic environment the equation of the a posteriori closed-loop prediction error takes the form:

$$\varepsilon_{CL}(t+1) = \frac{1}{C}[\theta_e - \hat{\theta}_e]^T\phi_e(t) + e(t+1) \tag{9.41}$$

Asymptotic unbiased estimates in a stochastic environment can be obtained under the sufficient condition (Landau and Karimi 1999) that:

$$H'(z^{-1}) = \frac{1}{C(z^{-1})} - \frac{\lambda_2}{2} \tag{9.42}$$

is strictly positive real (where $\max_t \lambda_2(t) \le \lambda_2 < 2$) by using either the averaging method (Theorem 4.1) or the martingale approach (Theorem 4.2).

A similar development can be done for the case of a disturbance model $v(t+1) = \frac{C}{DA}e(t+1)$ (Landau and Karimi 1997b). The resulting algorithm called *generalized closed-loop output error* (G-CLOE) is given in Table 9.1 where closed-loop output error algorithms are summarized.[2]

[2]Routines corresponding to these algorithms in MATLAB® can be downloaded from the web sites: http://www.landau-adaptivecontrol.org and http://landau-bookic.lag.ensieg.inpg.fr.

**Table 9.1** Summary of closed-loop output error algorithms

| | | CLOE | F-CLOE | X-CLOE | G-CLOE |
|---|---|---|---|---|---|
| Plant + noise model | | $y = \frac{z^{-d}B}{A}u + v$ | | $y = \frac{z^{-d}B}{A}u + \frac{C}{A}e$ | $y = \frac{z^{-d}B}{A}u + \frac{C}{AD}e$ |
| Adjustable parameter vector | | $\hat{\Theta}^T(t) = [\hat{a}^T(t), \hat{b}^T(t)]$<br>$\hat{a}^T(t) = [\hat{a}_1, \ldots, \hat{a}_{n_A}]$<br>$\hat{b}^T(t) = [\hat{b}_1, \ldots, \hat{b}_{n_B}]$ | | $\hat{\Theta}^T(t) = [\hat{a}^T(t), \hat{b}^T(t), \hat{h}^T(t)]$<br>$\hat{a}^T(t) = [\hat{a}_1, \ldots, \hat{a}_{n_A}]$<br>$\hat{b}^T(t) = [\hat{b}_1, \ldots, \hat{b}_{n_B}]$<br>$\hat{h}^T(t) = [\hat{h}_1, \ldots, \hat{h}_{n_H}]$ | $\hat{\Theta}^T(t) = [\hat{a}^T(t), \hat{b}^T(t), \hat{h}^T(t), \hat{d}^T(t)]$<br>$\hat{a}^T(t) = [\hat{a}_1, \ldots, \hat{a}_{n_A}]$<br>$\hat{b}^T(t) = [\hat{b}_1, \ldots, \hat{b}_{n_B}]$<br>$\hat{h}^T(t) = [\hat{h}_1, \ldots, \hat{h}_{n_H}]$<br>$\hat{d}^T(t) = [\hat{d}_1, \ldots, \hat{d}_{n_D}]$ |
| Predictor regressor vector | | $\psi^T(t) = [-\hat{y}(t), \ldots,$<br>$-\hat{y}(t-n_A+1),$<br>$\hat{u}(t-d), \ldots,$<br>$\hat{u}(t-d-n_B+1)]$<br>$\hat{u}(t) = -\frac{R}{S}\hat{y}(t) + r_u$ | | $\psi^T(t) = [-\hat{y}(t), \ldots, -\hat{y}(t-n_A+1),$<br>$\hat{u}(t-d), \ldots, \hat{u}(t-d-n_B+1),$<br>$\varepsilon_{CLf}(t), \ldots, \varepsilon_{CLf}(t-n_H+1)]$<br>$\hat{u}(t) = -\frac{R}{S}\hat{y}(t) + r_u$<br>$\varepsilon_{CLf} = \frac{1}{S}\varepsilon_{CL}(t)$ | $\psi^T(t) = [-\hat{y}(t), \ldots, -\hat{y}(t-n_A+1),$<br>$\hat{u}(t-d), \ldots, \hat{u}(t-d-n_B+1),$<br>$\varepsilon_{CLf}(t), \ldots, \varepsilon_{CLf}(t-n_H+1),$<br>$-\alpha(t), \ldots, -\alpha(t-n_D+1)]$<br>$\hat{u}(t) = -\frac{R}{S}\hat{y}(t) + r_u$<br>$\varepsilon_{CLf} = \frac{1}{S}\varepsilon_{CL}(t)$<br>$\alpha(t) = \hat{A}(t)\hat{y}(t) - \hat{B}(t)\hat{u}(t-d)$ |
| Predictor output | a priori<br>a posteriori | $\hat{y}^\circ(t+1) = \hat{\Theta}^T(t)\psi(t)$<br>$\hat{y}(t+1) = \hat{\Theta}^T(t+1)\psi(t)$ | | | |
| Prediction error | a priori<br>a posteriori | $\varepsilon^\circ_{CL}(t+1) = y(t+1) - \hat{y}^\circ(t+1)$<br>$\varepsilon_{CL}(t+1) = y(t+1) - \hat{y}(t+1)$ | | | |
| Adaptation error | | $\nu^\circ(t+1) = \varepsilon^\circ_{CL}(t+1)$ | | | |
| Observation vector | | $\Phi(t) = \psi(t)$ | $\Phi(t) = \frac{S}{\hat{P}}\psi(t)$<br>$\hat{P} = \hat{A}S + z^{-d}\hat{B}R$ | $\Phi(t) = \psi(t)$ | $\Phi(t) = \psi(t)$ |
| Stability condition deterministic env. | | $\frac{S}{P} - \frac{\lambda_2}{2} : SPR$ | $\frac{\hat{P}}{P} - \frac{\lambda_2}{2} : SPR$ | None | None |
| Convergence condition stochastic env. | | $\frac{S}{P} - \frac{\lambda_2}{2} : SPR$ | $\frac{\hat{P}}{P} - \frac{\lambda_2}{2} : SPR$ | $\frac{1}{C} - \frac{\lambda_2}{2} : SPR$ | Unknown |

## 9.3 Filtered Open-Loop Recursive Identification Algorithms (FOL)

The key observation leading to the use of open-loop recursive identification algorithms for identification in closed loop is that the output of the predictor given by (9.8) can be expressed as:

$$\hat{y}(t+1) = \hat{\theta}^T \phi(t) = \hat{\theta}^T \varphi(t) + \left[\hat{A}^* + \frac{q^{-d}\hat{B}^* R}{S}\right]\varepsilon_{CL}(t) \tag{9.43}$$

and that:

$$1 + q^{-1}\left[\hat{A}^* + \frac{q^{-d}\hat{B}^* R}{S}\right] = \frac{\hat{A}S + \hat{B}R}{S} = \frac{\hat{P}}{S} \tag{9.44}$$

where $\hat{P}$ (see also (9.30)) is an estimation of the true closed-loop poles based on the available plant model parameter estimates. Subtracting (9.43) from (9.4) one gets:

$$\varepsilon_{CL}(t+1) = \frac{S}{\hat{P}}[\theta - \hat{\theta}]^T \varphi(t) + \frac{AS}{\hat{P}} v(t+1) = \frac{S}{\hat{P}} \varepsilon_{LS}(t+1) \tag{9.45}$$

where:

$$\varepsilon_{LS}(t+1) = [\theta - \hat{\theta}]^T \varphi(t) + Av(t+1) = y(t+1) - \bar{y}(t+1) \tag{9.46}$$

represents the predictor error when using a standard least square predictor for the plant defined as:

$$\bar{y}(t+1) = -\hat{A}^* y(t) + \hat{B}^* u(t-d) = \hat{\theta}^T \varphi(t) \tag{9.47}$$

Therefore, for fixed parameter estimates and in the absence of noise, *the closed-loop prediction error is equal to the filtered least squares prediction error for the plant output* (ignoring the feedback). The filter to be used is $S/\hat{P}$. This observation was made in Åström (1993).

### *9.3.1 Filtered Recursive Least Squares*

Equation (9.45) can be rewritten as:

$$\varepsilon_{CL}(t+1) = [\theta - \hat{\theta}]^T \varphi_f(t) + \frac{AS}{\hat{P}} v(t+1) \tag{9.48}$$

where

$$\begin{aligned}\varphi_f^T(t) = \frac{S}{\hat{P}} \varphi^T(t) = [&-y_f(t), \ldots, -y_f(t-n_A+1), u_f(t-d), \ldots, \\ &u_f(t-d-n_B+1)]\end{aligned} \tag{9.49}$$

which leads immediately to the use of the standard recursive least squares algorithm but on filtered inputs and outputs.

Note that the filter to be used has a very interesting characteristic, i.e., it will filter out high-frequency signals beyond the estimated closed-loop bandwidth and if the controller contains an integrator i.e., ($S = (1 - q^{-1})S'(q^{-1})$), the d.c. components will be also filtered out. However, asymptotic unbiased estimates can be obtained only for particular noise model structures and closed-loop poles (i.e., for the cases where $\frac{AS}{\hat{P}}v(t+1) = e(t+1)$).

To account for the effect of noise upon the parameter estimates one can use other open-loop recursive identification algorithms belonging to the category of algorithms based on the asymptotic whitening of the prediction error like:

- extended recursive least square (ELS);
- output error with extended prediction model (OEEPM);
- recursive maximum likelihood (RML).

However, from a theoretical point of view (using for example the averaging method, Theorem 4.1) even when $v(t+1) = \frac{C}{A}e(t+1)$, the filtering of the data modifies the conditions for the asymptotic unbiasedness of the estimates. The equation of the filtered output for $v(t+1) = \frac{C}{A}e(t+1)$ is:

$$\begin{aligned} y_f(t+1) &= -A^* y_f(t) + B^* u_f(t) + \frac{CS}{\hat{P}} e(t+1) \\ &\approx -A^* y_f(t) + B^* u_f(t) + C' e(t+1) \end{aligned} \tag{9.50}$$

where $C'(q^{-1})$ is the result of the polynomial division of $CS$ by $\hat{P}$ (typically of higher order than C). The corresponding extended least square predictor will have the form:

$$\bar{y}(t+1) = -\hat{A}^* y_f(t) + \hat{B}^* u_f(t-d) + \hat{C}^* \varepsilon_{LSf}(t) \tag{9.51}$$

but with an order for $\hat{C}^*$ higher than in the open-loop case. The resulting predictor error equation will take the form:

$$\varepsilon_{LSf}(t+1) \approx \frac{1}{C'}[\theta_e - \hat{\theta}_e]^T \varphi_{fe}(t) + e(t+1) \tag{9.52}$$

where

$$\varphi_{fe}^T(t) = [\varphi_f^T(t), \varepsilon_{LSf}(t), \varepsilon_{LSf}(t-1), \ldots] \tag{9.53}$$

and the sufficient condition for asymptotic unbiased estimates becomes:

$$H'(z^{-1}) = \frac{1}{C'} - \frac{\lambda_2}{2} \tag{9.54}$$

be strictly positive real. (where $C'$ is an approximation of $\frac{CS}{\hat{P}}$).

In practice all of the recursive identification algorithms based on the asymptotic whitening of the prediction error can be used on the filtered input-output data through $S/\hat{P}$. Therefore an initial estimate of the plant model is necessary in order to implement the data filter which is not the case for the methods belonging to the "closed-loop output error" identification methods.

### 9.3.2 Filtered Output Error

Output error recursive algorithms for open-loop identification can also be used for plant model identification in closed loop by using an appropriate data filter (Voda-Besançon and Landau 1995b). The open-loop output error predictor is:

$$\bar{y}(t+1) = -\hat{A}^*\bar{y}(t) + \hat{B}^*u(t-d) = \hat{\theta}^T\varphi_{OE}(t) \tag{9.55}$$

where:

$$\varphi_{OE}^T(t) = [\bar{y}(t), \bar{y}(t-1), \ldots, u(t-d), \ldots, u(t-d-n_B+1)] \tag{9.56}$$

Neglecting the effect of noise, the output error prediction obtained by subtracting (9.55) from (9.4) is given by:

$$\varepsilon_{OE}(t+1) = \frac{1}{A}[\theta-\hat{\theta}]^T\varphi_{OE}(t) = \frac{1}{\hat{A}}[\theta-\hat{\theta}]^T\varphi(t) = \frac{1}{\hat{A}}\varepsilon_{LS}(t+1) \tag{9.57}$$

Therefore:

$$\begin{aligned}\varepsilon_{CL}(t+1) &= \frac{S\hat{A}}{\hat{P}}\varepsilon_{OE}(t+1) = \frac{S\hat{A}}{\hat{P}}\frac{1}{A}[\theta-\hat{\theta}]^T\varphi_{OE}(t) \\ &= \frac{1}{A}[\theta-\hat{\theta}]^T\varphi_{OEf}(t) = \varepsilon_{OEf}(t+1)\end{aligned} \tag{9.58}$$

where:

$$\varphi_{OEf}^T(t) = \frac{S\hat{A}}{\hat{P}}\varphi_{OE}(t) \tag{9.59}$$

$$\varepsilon_{OEf}(t+1) = y_f(t+1) - \hat{\theta}^T\varphi_{OEf}(t) \tag{9.60}$$

In other words, the open-loop output error recursive algorithm can be used for plant model identification in closed loop provided that the input-output data are filtered through the estimated output sensitivity function ($\frac{S\hat{A}}{\hat{P}}$). However, since in closed-loop operation $u$ and $v$ are correlated, the parameter estimates will be asymptotically unbiased only for particular noise model structures and closed-loop poles ($\frac{\hat{A}S}{\hat{P}}v = e$ or for $v = \frac{C}{A}e$, $\frac{CS}{\hat{P}} \cdot \frac{\hat{A}}{A} = 1$).

## 9.4 Frequency Distribution of the Asymptotic Bias in Closed-Loop Identification

The method for the analysis of the frequency distribution of the asymptotic bias presented in Sect. 4.4, is very useful for assessing and comparing the behavior of the various algorithms in the presence of noise, or when the plant model and the estimated model do not have the same structure.

For a plant model of the form:

$$y(t) = G(q^{-1})u(t) + W(q^{-1})e(t) \tag{9.61}$$

one gets in open-loop operation a prediction error equation:

$$\varepsilon(t)=\hat{W}^{-1}(q^{-1})\{[G(q^{-1})-\hat{G}(q^{-1})]u(t)+[W(q^{-1})-\hat{W}(q^{-1})]e(t)\}+e(t) \tag{9.62}$$

Minimizing $\mathbf{E}\{\varepsilon^2(t)\}$ leads to the basic formula of asymptotic bias distribution for the open-loop identification methods:

$$\begin{aligned}\hat{\theta}^* = \arg\min_{\hat{\theta}\in\mathscr{D}} \int_{-\pi}^{\pi} &|\hat{W}^{-1}(e^{-j\omega})|^2[|G(e^{-j\omega})-\hat{G}(e^{-j\omega})|^2\phi_u(\omega)\\ &+|W(e^{-j\omega})-\hat{W}(e^{-j\omega})|^2\phi_e(\omega)]d\omega\end{aligned} \tag{9.63}$$

In closed-loop operation, the input $u(t)$ is given by:

$$u(t)=S_{yp}(q^{-1})r_u(t)+S_{up}(q^{-1})W(q^{-1})e(t) \tag{9.64}$$

where $r_u(t)$ is the external excitation added to the output of the controller ($r_u$ is equal to $r\frac{T}{S}$ if the external excitation is applied on the reference) and $S_{yp}$ and $S_{up}$ are the output and input sensitivity functions respectively. Substituting $u(t)$ in (9.62) and (9.63), one gets:

$$\varepsilon(t)=\hat{W}^{-1}\{(G-\hat{G})S_{yp}r_u(t)+[(G-\hat{G})S_{up}W+(W-\hat{W})]e(t)\}+e(t) \tag{9.65}$$

where the argument $q^{-1}$ has been dropped out to simplify the notation and (9.63) will take the form:

$$\begin{aligned}\hat{\theta}^* = \arg\min_{\hat{\theta}\in\mathscr{D}} \int_{-\pi}^{\pi} &|\hat{W}^{-1}|^2[|G-\hat{G}|^2|S_{yp}|^2\phi_{r_u}(\omega)\\ &+|(G-\hat{G})S_{up}W+(W-\hat{W})|^2\phi_e(\omega)]d\omega\end{aligned} \tag{9.66}$$

where the argument $e^{-j\omega}$ has been dropped out and $\phi_{r_u}(\omega)$ corresponds to the spectral density of the external excitation signal. Comparing the above equation with (9.63), one observes that the bias on the estimated model is related to the bias on the estimated noise model, i.e., incorrect estimation of the noise model leads to a biased plant model estimation even when the plant model and the estimated model have the same structure. For a plant model of the form:

$$y(t)=G(q^{-1})u(t)+v(t) \tag{9.67}$$

under the assumptions that $v(t)$ is independent with respect to $r_u$ and has a spectrum $\phi_v(\omega)$ using an output error predictor (taking $\hat{W}=1$ in (9.66) which corresponds to the fact that the noise model is not estimated), one gets in closed-loop operation:

$$\hat{\theta}^* = \arg\min_{\hat{\theta}\in\mathscr{D}} \int_{-\pi}^{\pi} \big[|G-\hat{G}|^2|S_{yp}|^2\phi_{r_u}(\omega)+|(G-\hat{G})S_{up}+1|^2\phi_v(\omega)\big]d\omega \tag{9.68}$$

and this leads to a biased estimation. Both equations (9.66) and (9.68) indicates that for large ratios noise/signal, the estimated plant model approaches the inverse of the controller with the negative sign ($\hat{G}\approx-\frac{S}{R}$). Effectively, for $\phi_{r_u}(\omega)=0$, $G=q^{-d}B/A$, $\hat{G}=q^{-d}\hat{B}/\hat{A}$ and the optimum of (9.66) or (9.68) corresponds to:

$$\frac{\hat{B}}{\hat{A}}+\frac{S}{R}=0$$

### 9.4.1 Filtered Open-Loop Identification Algorithms

In these algorithms an open-loop type identification is performed on the filtered input/output data which leads to the following filtered prediction error:

$$\varepsilon_f(t) = \hat{W}^{-1}L\left\{(G-\hat{G})S_{yp}r_u(t) + \left[(G-\hat{G})S_{up}W + \left(W - \frac{\hat{W}}{L}\right)\right]e(t)\right\} + e(t) \tag{9.69}$$

where $L = L(e^{-j\omega})$ is the input/output data filter. The bias distribution is given by:

$$\begin{aligned}\hat{\theta}^* = \arg\min_{\hat{\theta}\in\mathscr{D}} \int_{-\pi}^{\pi} |\hat{W}^{-1}L|^2 \bigg[ &|G-\hat{G}|^2|S_{yp}|^2\phi_{r_u}(\omega) \\ &+ \left|(G-\hat{G})S_{up}W + \left(W-\frac{\hat{W}}{L}\right)\right|^2 \phi_e(\omega)\bigg]d\omega \end{aligned} \tag{9.70}$$

The following comments upon (9.70) can be made:

- The estimation of the noise model is biased. In fact $\hat{W}$ tends asymptotically to $LW$ instead of $W$. This incorrect estimation leads to a biased estimation of $G$ even when $G$ and $\hat{G}$ have the same structure.
- In the ideal case when $\hat{W} = LW$ the parameter estimates are given by:

$$\hat{\theta}^* = \arg\min_{\hat{\theta}\in\mathscr{D}} \int_{-\pi}^{\pi} |W^{-1}|^2[|G-\hat{G}|^2(|S_{yp}|^2\phi_{r_u}(\omega) + |S_{up}W|^2\phi_e(\omega))]d\omega \tag{9.71}$$

  which means that the data filtering by $L$ will be compensated by the estimated noise model $\hat{W}$ and the bias distribution is almost identical to the non filtered case.
- In the particular case when $\hat{W} = \frac{1}{\hat{A}}$ (recursive least squares) and $L = S/\hat{P}$ we have:

$$\begin{aligned}\hat{\theta}^* = \arg\min_{\hat{\theta}\in\mathscr{D}} \int_{-\pi}^{\pi} |\hat{S}^*_{yp}|^2[&|G-\hat{G}|^2|S_{yp}|^2\phi_{r_u}(\omega) \\ &+ |(G-\hat{G})S_{up}W + (W-(\hat{S}^*_{yp})^{-1})|^2\phi_e(\omega)]d\omega\end{aligned} \tag{9.72}$$

  where $\hat{S}^*_{yp} = \hat{A}S/\hat{P}$ is an estimation of the true output sensitivity function $S_{yp}$. It can be observed that $G-\hat{G}$ is weighted by both the real output sensitivity function and the estimation of the output sensitivity function which should improve the accuracy of the identified model in critical frequency regions where the Nyquist plot is closest to $(-1, j0)$. Unfortunately this positive effect is counteracted by the noise.

For the case of output error structure using an estimation of the output sensitivity function $\hat{S}^*_{yp}$ as a data filter (Voda-Besançon and Landau 1995a) the bias distribution is given by:

$$\hat{\theta}^* = \arg\min_{\hat{\theta}\in\mathscr{D}} \int_{-\pi}^{\pi} |\hat{S}^*_{yp}|^2[|G-\hat{G}|^2|S_{yp}|^2\phi_{r_u}(\omega) + |(G-\hat{G})S_{up}+1|^2\phi_v(\omega)]d\omega \tag{9.73}$$

where $\phi_v(\omega)$ is the spectrum of the disturbance. Of course the estimation of the plant model using the above criterion like in the previous case is biased.

### 9.4.2 Closed-Loop Output Error Identification Algorithms

Consider the plant model (9.67). The closed-loop output error algorithms use the following predictor for the closed-loop system:

$$\hat{y}(t) = \hat{G}\hat{u}(t) \tag{9.74}$$

where $\hat{u}(t) = \hat{S}_{yp} r_u(t)$ and $\hat{S}_{yp}$ is the output sensitivity function of the closed loop predictor. Then the prediction error is given by:

$$\varepsilon_{CL}(t) = Gu(t) + v(t) - \hat{G}\hat{u}(t) \tag{9.75}$$

Adding and subtracting $G\hat{u}(t)$ to the RHS of the above equation, one gets:

$$\varepsilon_{CL}(t) = (G - \hat{G})\hat{u}(t) + G[u(t) - \hat{u}(t)] + v(t) \tag{9.76}$$

and using (9.11) and (9.7), one gets:

$$\varepsilon_{CL}(t) = S_{yp}[(G - \hat{G})\hat{u}(t) + v(t)] \tag{9.77}$$

Substituting now $\hat{u}(t)$ by $\hat{S}_{yp} r_u(t)$ and minimizing $\mathbf{E}\{\varepsilon_{CL}^2(t)\}$, the resulting bias distribution will be:

$$\hat{\theta}^* = \arg\min_{\hat{\theta}\in\mathscr{D}} \int_{-\pi}^{\pi} |S_{yp}|^2[|G - \hat{G}|^2|\hat{S}_{yp}|^2\phi_{r_u}(\omega) + \phi_v(\omega)]d\omega \tag{9.78}$$

This expression shows that:

- The estimation of the plant model parameters is unbiased when $G$ is in the model set.
- The bias distribution is not affected by the spectrum of the noise (which is the case for the filtered open-loop methods).
- The bias distribution is not only weighted by the sensitivity function but is further weighted by the estimated sensitivity function.

The closed-loop output error can be minimized using the classical optimization methods, e.g. Newton-Raphson algorithm, or by the AF-CLOE algorithm. One of the properties of AF-CLOE is that $\phi_f(t) = S/\hat{P}(t)\phi(t)$ is the gradient of $\hat{y}(t)$ (see Problem 9.10) and therefore $-\phi_f(t)\varepsilon_{CL}(t+1)$ is the gradient of $\varepsilon_{CL}^2(t+1)$. Therefore, the AF-CLOE algorithm can be interpreted as the gradient algorithm in Sect. 3.2.1. In a stochastic environment, the asymptotic convergence of the gradient algorithm to the minimum of $\mathbf{E}\{\varepsilon_{CL}^2(t+1)\}$ can be proved using the stochastic approximation method (Robbins and Monro 1951) under some mild conditions on a scalar adaptation gain.

For CLOE algorithm the bias distribution cannot be computed. The other algorithm of this family named F-CLOE has approximately the bias distribution given in (9.78) (better approximation is obtained if $\hat{P}$ is close to $P$).

When the external excitation signal is applied on the reference, the asymptotic bias distribution is given by:

$$\hat{\theta}^* = \arg\min_{\hat{\theta}\in\mathscr{D}} \int_{-\pi}^{\pi} |S_{yp}|^2[|G - \hat{G}|^2|\hat{T}_{ur}^2|\phi_r(\omega) + \phi_v(\omega)]d\omega \tag{9.79}$$

where $\hat{T}_{ur} = T\hat{A}/\hat{P}$ is the transfer operator between $r(t)$ and $\hat{u}(t)$.

Clearly the weighting terms in (9.78) and in (9.79) are different and therefore the estimated models will be different (for same characteristics of the external excitation signal). For a detailed discussion see Landau and Karimi (2002).

For the algorithms in which the noise model is also identified (X-CLOE and G-CLOE) the optimal predictor for the closed-loop system is:

$$\hat{y}(t) = \hat{G}\hat{u}(t) + \hat{\bar{W}}\varepsilon_{CL}(t) \tag{9.80}$$

where:

$$\hat{\bar{W}}(\hat{\theta}) = \hat{W}(\hat{\theta}) - \hat{S}_{yp}^{-1}$$

and the resulting bias distribution is given by Landau and Karimi (1997a), Karimi and Landau (1998):

$$\hat{\theta}^*_{opt} = \arg\min_{\hat{\theta}\in\mathscr{D}} \int_{-\pi}^{\pi} |(1 + \hat{\bar{W}}\hat{S}_{yp})^{-1}|^2[|S_{yp}|^2|G - \hat{G}|^2|\hat{S}_{yp}|^2\phi_{r_u}(\omega)$$
$$+ |\bar{W}S_{yp} - \hat{\bar{W}}\hat{S}_{yp}|^2\phi_e(\omega))]d\omega \tag{9.81}$$

It can be observed that:

- These methods give an unbiased parameter estimation when the plant and noise models and the estimated plant models have the same structure.
- They have the same asymptotic properties as the direct identification method in terms of the bias distribution. In fact if in (9.79) we replace $\bar{W}$ and $\hat{\bar{W}}$ respectively by $W - S_{yp}^{-1}$ and $\hat{W} - \hat{S}_{yp}^{-1}$, (9.66) is obtained.

The conclusion of this analysis is that with respect to the bias distribution, *if the objective is to enhance the accuracy of the model in the critical frequency regions for design*, F-CLOE and AF-CLOE are the most suitable methods since they are not affected by the noise characteristics and they heavily weights the difference between the true plant model and estimated model in the desired frequency region.

## 9.5 Validation of Models Identified in Closed-Loop

The first remark to be made is that independently of the criterion used, the model validation will be *controller dependent*. The objective of the validation will be to find the plant model which gives the best prediction for the output of the closed-loop system with the controller used for identification. Several validation procedures could be defined: statistical validation, pole closeness validation and time-domain validation.

### 9.5.1 Statistical Validation

Statistical validation tests can be developed in a similar way as in the open-loop identification—see Sects. 5.4 and 5.6, with the difference that a closed-loop predictor incorporating the plant model will be used to obtain the residual closed-loop prediction error. Two types of tests can be carried out:

#### Uncorrelation Test

Using the "output error" configuration shown in Fig. 9.1. where the predictor is given by (9.8) through (9.11), the uncorrelation between the closed-loop output error $\varepsilon_{CL}(t+1)$ and the components of the observation vector of $\phi(t)$ ($\hat{y}(t), \hat{u}(t-d)$ and their delayed values) will be tested together with the covariance of the residual closed-loop output prediction error. This type of test is motivated by the fact that uncorrelation between the observations (components of $\phi(t)$) and the closed-loop output prediction error (i) leads to unbiased parameter estimates and (ii) implies the uncorrelation between the prediction error and the external excitation. This means that the residual prediction error does not contain any information depending upon the external excitation and therefore all correlations between the external excitation and the output of the closed loop system are represented by the predictor.

One computes:

$$R_\varepsilon(0) = \frac{1}{N}\sum_{t=1}^{N} \varepsilon_{CL}^2(t) \tag{9.82}$$

$$R_{\hat{u}}(0) = \frac{1}{N}\sum_{t=1}^{N} \hat{u}^2(t) \tag{9.83}$$

$$R_{\hat{y}}(0) = \frac{1}{N}\sum_{t=1}^{N} \hat{y}^2(t) \tag{9.84}$$

$$R_{\varepsilon\hat{y}}(i) = \frac{1}{N}\sum_{t=1}^{N} \varepsilon_{CL}(t)\hat{y}(t-i); \quad i = 1, 2, \ldots, n_A \tag{9.85}$$

$$R_{\varepsilon\hat{u}}(i) = \frac{1}{N}\sum_{t=1}^{N} \varepsilon_{CL}(t)\hat{u}(t-d-i); \quad i = 1, 2, \ldots, n_B \tag{9.86}$$

$$RN_{\varepsilon x}(i) = \frac{R_{\varepsilon x}(i)}{[R_x(0)\cdot R_\varepsilon(0)]^{1/2}}; \quad x = \hat{y}, \hat{u} \tag{9.87}$$

As a confidence test one can use the criterion:

$$|RN(i)| \leq \frac{\alpha}{\sqrt{N}} \tag{9.88}$$

where $\alpha$ is the confidence level (a typical value is 2.17) and $N$ is the number of data (see Sect. 5.4).

In many practical situations, one either has a previous plant model identified in open loop or several identification algorithms are used on the data collected in closed loop. Then a comparative validation has to be done and useful comparison indicators are provided by $R_\varepsilon(0)$ and $\max|RN_{\varepsilon x}|$ for each model (however, other comparison criteria can be considered).

### Whiteness Test

If information upon the noise model affecting the output of the plant is available, through the use of appropriate identification algorithms, a whiteness test on the residual closed-loop prediction error can be done. A predictor of the form given by (9.32) has to be used. If the input-output plant model and the noise model are known, the closed-loop prediction error is a white noise. The whiteness of the residual prediction error implies that:

- unbiased parameter estimates can be obtained;
- the identified model allows the optimal predictor for the closed-loop system to be obtained;
- there is not any correlation between the external excitation and the residual closed-loop prediction error.

One computes:

$$R_\varepsilon(0) = \frac{1}{N}\sum_{t=1}^{N} \varepsilon_{CL}^2(t) \tag{9.89}$$

$$R_\varepsilon(i) = \frac{1}{N}\sum_{t=1}^{N} \varepsilon_{CL}(t)\varepsilon_{CL}(t-i); \quad i = 1, 2, \ldots, \max(n_A, n_B + d) \tag{9.90}$$

$$RN_\varepsilon(i) = \frac{R_\varepsilon(i)}{R_\varepsilon(0)} \tag{9.91}$$

As a confidence test one can use the criterion (9.88) for $i > 0$. For comparative validation of several models $R_\varepsilon(0)$ and $\max|RN_\varepsilon(i)|$ are good indicators.

## *9.5.2 Pole Closeness Validation*

The idea behind this test is the following: if the model identified allows the construction of a good predictor of the closed-loop system for the given controller used during identification, this also implies that the poles of the true closed-loop system and those of the closed-loop predictor are close (assuming the richness of the excitation). Therefore the closeness of the closed-loop predictor poles (which can be computed) and the poles of the true closed-loop system (which can be identified) will indicate the quality of the identified model.

Closeness criteria in the z-plane can be associated with this validation procedure (Landau and Zito 2005), but visual pole chart examination allows the comparison of various models.

Extensive simulation and experimental results have shown that statistical validation and pole closeness validation are two coherent procedures for model validation in closed loop and allow a clear comparison of the identified plant models.

The identification of the closed-loop poles is simply done by identifying the closed loop (between the external excitation and the plant output) using standard open-loop identification and validation techniques. Note that this identification can be carried out using the data which are also used for plant model identification.

### *9.5.3 Time Domain Validation*

While this approach seems very natural (comparing the time response of the true closed-loop system with the one of the closed-loop predictor), experience shows that it does not always enable one to quantify the comparative quality of various models. However, it has been observed that good pole closeness implies excellent superposition of simulated and real-time responses.

## 9.6 Iterative Identification in Closed-Loop and Controller Redesign

In indirect adaptive control, the parameters of the controller are generally updated at each sampling instant based on the current estimates of the plant model parameters. However, nothing forbid us to update the estimates of the plant model parameters at each sampling instant, but to update the controller parameters only every $N$ sampling instants. Arguments for choosing this procedure are related to:

- the possibility of getting better parameter estimates for control design;
- the eventual reinitialization of the plant parameters estimation algorithm after each controller updating;
- the possibility of using more sophisticated control design procedures (in particular robust control design) requiring a large amount of computation.

If the horizon $N$ is sufficiently large and an external excitation is added to the system, one can consider that an identification of the plant model in closed loop using a linear fixed controller is done. In such cases, one obtains an *iterative identification in closed loop and controller redesign* scheme. One of the interests of using this approach is that a link between how the identification in closed loop should be carried on in relation with the objective for control design can be established. Figure 9.2 illustrates the basis of this iterative procedure. The upper part represents the true closed-loop system and the lower part represents the design system. The objective is to minimize the error between the two systems by using new data acquired in closed-loop operation.

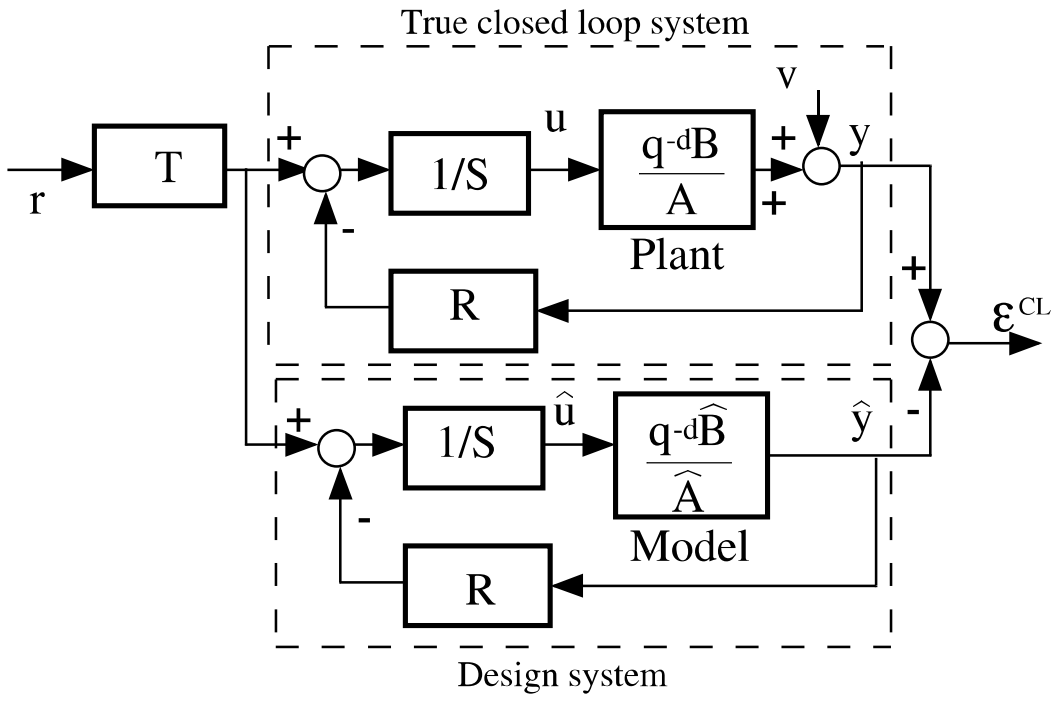

**Fig. 9.2** The true closed-loop system and the design system

Let us denote the true plant model by $G$, the estimated plant model by $\hat{G}$ and the controller computed on the basis of $\hat{G}$, by $C_{\hat{G}}$. We denote the *designed performance* as $J(\hat{G}, C_{\hat{G}})$ which corresponds in fact to both the optimal performance and the achieved one on the *design system.* $J(G, C_{\hat{G}})$ corresponds to the *achieved performance* on the true system. Then one can establish the following triangle inequality (Van den Hof and Schrama 1995):

$$\begin{aligned} &\|J(\hat{G}, C_{\hat{G}})\| - \|J(G, C_{\hat{G}}) - J(\hat{G}, C_{\hat{G}})\| \\ &\quad \le \|J(G, C_{\hat{G}})\| \\ &\quad \le \|J(\hat{G}, C_{\hat{G}})\| + \|J(G, C_{\hat{G}}) - J(\hat{G}, C_{\hat{G}})\| \end{aligned} \tag{9.92}$$

where $J(\cdot)$ and $\|\cdot\|$ are problem dependent. From (9.92), one can conclude that in order to minimize $\|J(G, C_{\hat{G}})\|$ one should have:

1. $\|J(\hat{G}, C_{\hat{G}})\|$ small;
2. $\|J(G, C_{\hat{G}}) - J(\hat{G}, C_{\hat{G}})\|$ small.

Since one cannot simultaneously optimize the model and the controller, this will be done sequentially by an iterative procedure as follows:

1. One searches for an identification procedure such that:

$$\hat{G}_{i+1} = \arg\min_{\hat{G}} \|J(G, C_i) - J(\hat{G}_i, C_i)\| \tag{9.93}$$

   where $i$ corresponds to the step number in the iteration, and $\hat{G}_i$ corresponds to the estimated model in step $i$, used for the computation of the controller $C_i$.
2. Then, one computes a new controller based on $\hat{G}_{i+1}$:

$$C_{i+1} = \arg\min_{C} \|J(\hat{G}_{i+1}, C)\| \tag{9.94}$$

Equation (9.93) is fundamental for designing the closed-loop identification procedure which will depend upon the controller strategy used.

As an example, one can ask what is the identification method which has to be used in conjunction with the pole placement (combined eventually with the shaping of the sensitivity functions). In this case, the closed-loop error (see Fig. 9.2) is a measure in the time domain of the discrepancy between the desired and achieved performances in tracking and has the expression:

$$\varepsilon_{CL}(t+1) = S_{yp}(q^{-1})[G(q^{-1}) - \hat{G}(q^{-1})]\hat{T}_{ur}(q^{-1})r(t) + S_{yp}(q^{-1})v(t+1) \tag{9.95}$$

where $S_{yp}$ is the achieved sensitivity function and $\hat{T}_{ur}$ is the transfer function between $r$ and $\hat{u}$. Over an infinite time horizon and when $r(t)$ has a Fourrier transform, the discrepancy in performance takes the following expression in the frequency domain:

$$J = \int_{-\pi}^{\pi} |S_{yp}(e^{-j\omega})|^2 [G(e^{-j\omega}) - \hat{G}(e^{-j\omega})|^2 |\hat{T}_{ur}(e^{-j\omega})|^2 \phi_v(\omega) + \phi_v(\omega)] d\omega \tag{9.96}$$

This is exactly the identification criterion for AF-CLOE and F-CLOE identification methods given in Sect. 9.4 (for $r_u = 0$).

For other examples of combining control objectives and identification in closed loop see Van den Hof and Schrama (1995), Zang et al. (1995).

## 9.7 Comparative Evaluation of the Various Algorithms

This section deals with the evaluation of the various algorithms in simulation as well as on real-time application via different validation tests.

### *9.7.1 Simulation Results*

In order to compare different closed-loop identification methods and study the behavior of these algorithms, two simulations examples are considered. In the first example, the parameters of the system and controller are chosen such that $S/P$ be a positive real transfer function and the influence of positive realness of $1/C$ on the convergence of various algorithm is studied. Several model validation tests are also examined on the identified models. An unstable system in open loop is considered in the second example and in addition $S/P$ is a non-positive real transfer function. Detailed results can be found in Landau and Karimi (1997b).

*Example 9.1* Figure 9.3 shows the block diagram of the simulation system, where $B/A$ is the plant model, $R/S$ is the controller, $C/A$ is the noise model and $e(t)$ is a zero mean value equally distributed white noise. The parameters of the system are chosen as follows:

$$A(z^{-1}) = 1 - 1.5z^{-1} + 0.7z^{-2} \tag{9.97}$$

$$B(z^{-1}) = z^{-1}(1 + 0.5z^{-1}) \tag{9.98}$$

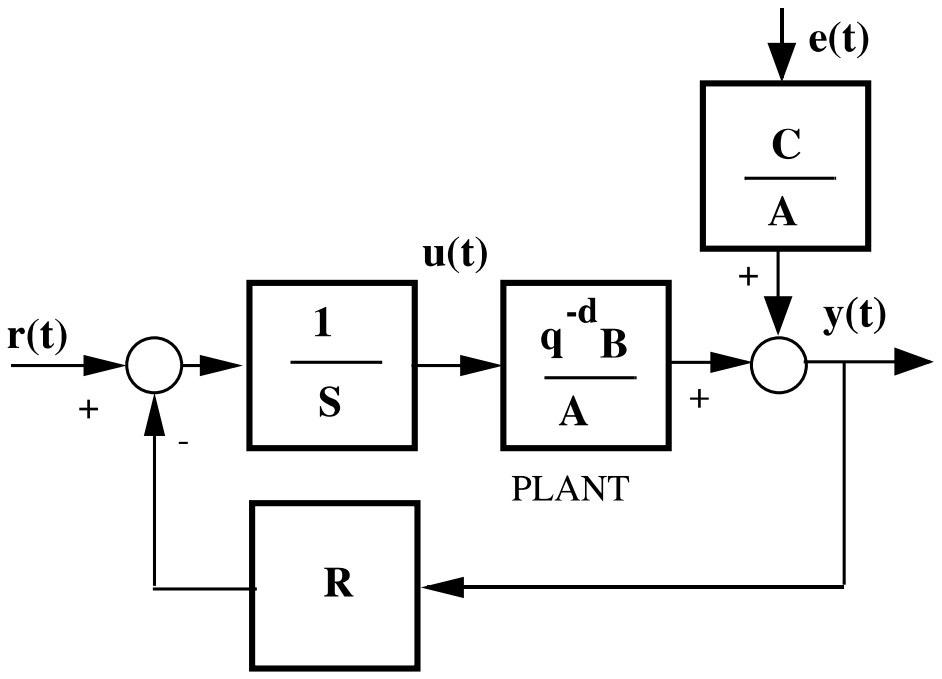

**Fig. 9.3** Block diagram of the simulation system

A controller using the pole placement technique is computed such that $S/P$ be positive real (the controller contains an integrator, i.e., $S$ contains a term $(1-z^{-1})$ which vanishes at the zero frequency). The parameters of the controller and the closed-loop characteristic polynomial are:

$$R(z^{-1}) = 0.8659 - 1.2763z^{-1} + 0.5204z^{-2} \tag{9.99}$$

$$S(z^{-1}) = 1 - 0.6283z^{-1} - 0.3717z^{-2} \tag{9.100}$$

$$P(z^{-1}) = 1 - 1.2624z^{-1} - 0.4274z^{-2} \tag{9.101}$$

For identification of the plant model in closed loop a pseudo-random binary sequence (PRBS) generated by a 7-bit shift register and a clock frequency of $\frac{1}{2}f_s$ ($f_s$ sampling frequency =1) is considered as reference signal. The noise signal ratio at the output of the closed-loop system is about 10% in terms of variance. Two different cases are considered for the noise model:

Case (a): $1/C$ is strictly positive real i.e. $C(z^{-1}) = 1 + 0.5z^{-1} + 0.5z^{-2}$
Case (b): $1/C$ is not positive real i.e. $C(z^{-1}) = 1 + 1.6z^{-1} + 0.9z^{-2}$

The extended recursive least square algorithm with filtered data has been used as FOL identification method.

The parametric distance defined by:

$$D(t) = \left[\sum_{i=1}^{n_A}(a_i - \hat{a}_i(t))^2 + \sum_{i=0}^{n_B}(b_i - \hat{b}_i(t))^2\right]^{1/2} \tag{9.102}$$

may be used as a criterion for comparing different methods in one run. Since we are only interested in the identification of the process, the criterion involves only the parameters of the plant and their estimates. Evolutions of this criterion for Case (a) (i.e., $1/C$ is strictly positive real) using CLOE, F-CLOE, AF-CLOE,X-CLOE and a filtered open-loop identification method (FOL) are illustrated in Fig. 9.4. In this simulation data have been filtered by the exact value of $S/P$ (instead of an estimated value of $P$) for the FOL method as well as for the F-CLOE method.

Figure 9.4 shows that an unbiased estimation of the parameters of the plant is obtained using various recursive closed-loop identification methods while the parameter estimation for the FOL method is biased. This is because a finite order polynomial $\hat{C}$ is used to estimate the parameters of the noise model $CS/P$ while theoretically an infinite order polynomial for $\hat{C}$ has to be used. Increasing the order of $\hat{C}$ in this case might improve the results but it will also augment the number of parameters to estimate which consequently reduces the convergence speed.

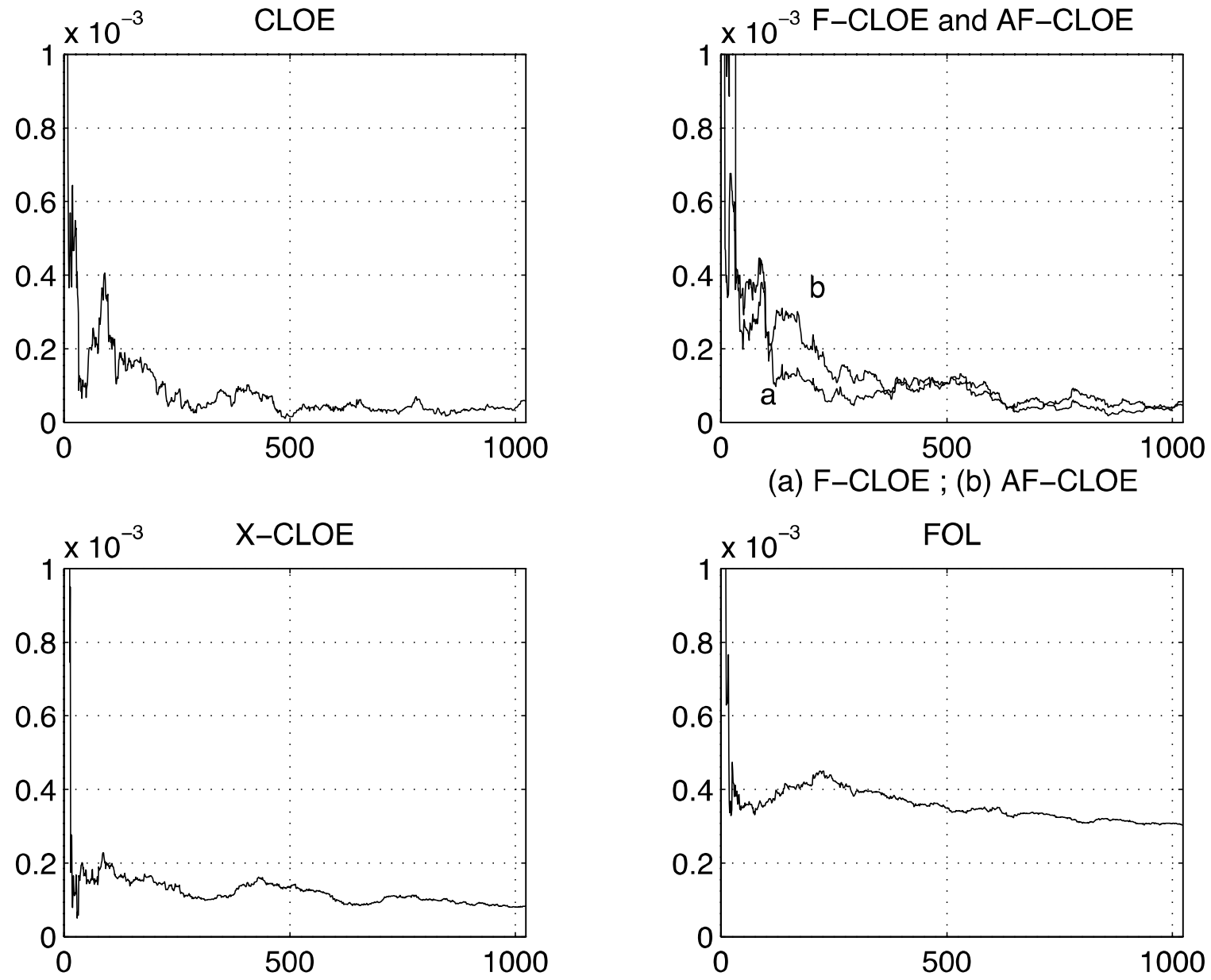

**Fig. 9.4** Evolution of the parametric distance (Example 9.1, Case (a), one run)

It can be observed that the convergence speed is slightly slower in the X-CLOE method which has got more parameters to estimate than that of CLOE or F-CLOE method.

Figure 9.5 shows the closed-loop poles computed on the basis of the plant model identified by the various methods and the true closed-loop poles. The computed closed-loop poles are closer to the real ones when using models estimated with CLOE type algorithms.

Evolutions of the parametric distance for the Case (b) (i.e., $1/C$ is not positive real) are illustrated in Fig. 9.6 (for one run). This figure shows that the positive realness of $1/C$ does not affect the convergence of the CLOE and F-CLOE method but the parameter estimation is biased for the X-CLOE method.

*Example 9.2* The parameters of the system are chosen as follows:

$$A = 1 - 2.5q^{-1} + 0.2q^{-2} \tag{9.103}$$

$$B = q^{-1}(1 + 0.5q^{-1}) \tag{9.104}$$

$$C = 1 + 0.5q^{-1} + 0.5q^{-2} \tag{9.105}$$

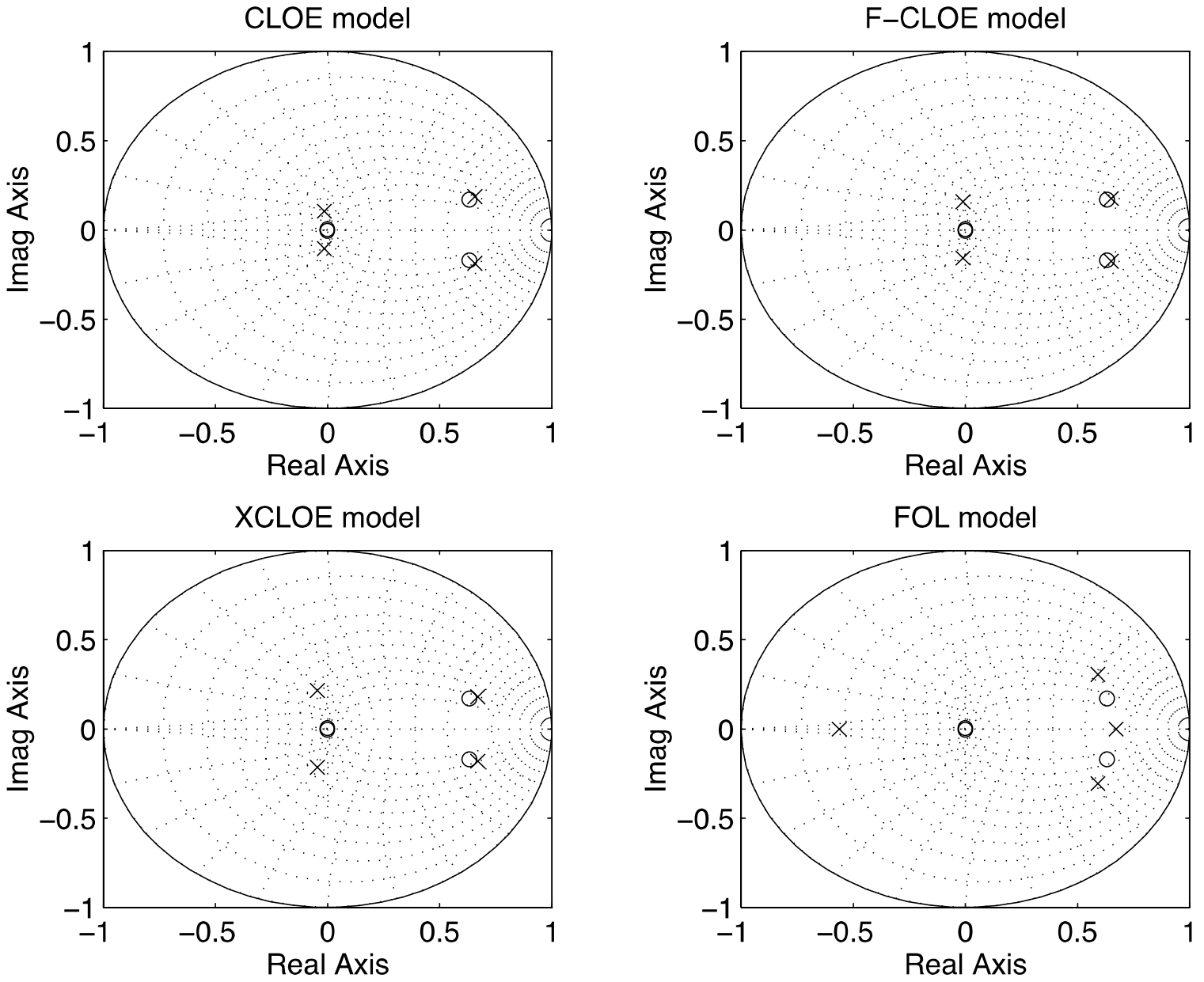

**Fig. 9.5** Pole closeness validation test (simulation): [○] true closed-loop poles, [×] computed closed-loop poles using the closed-loop identified plant model and the controller

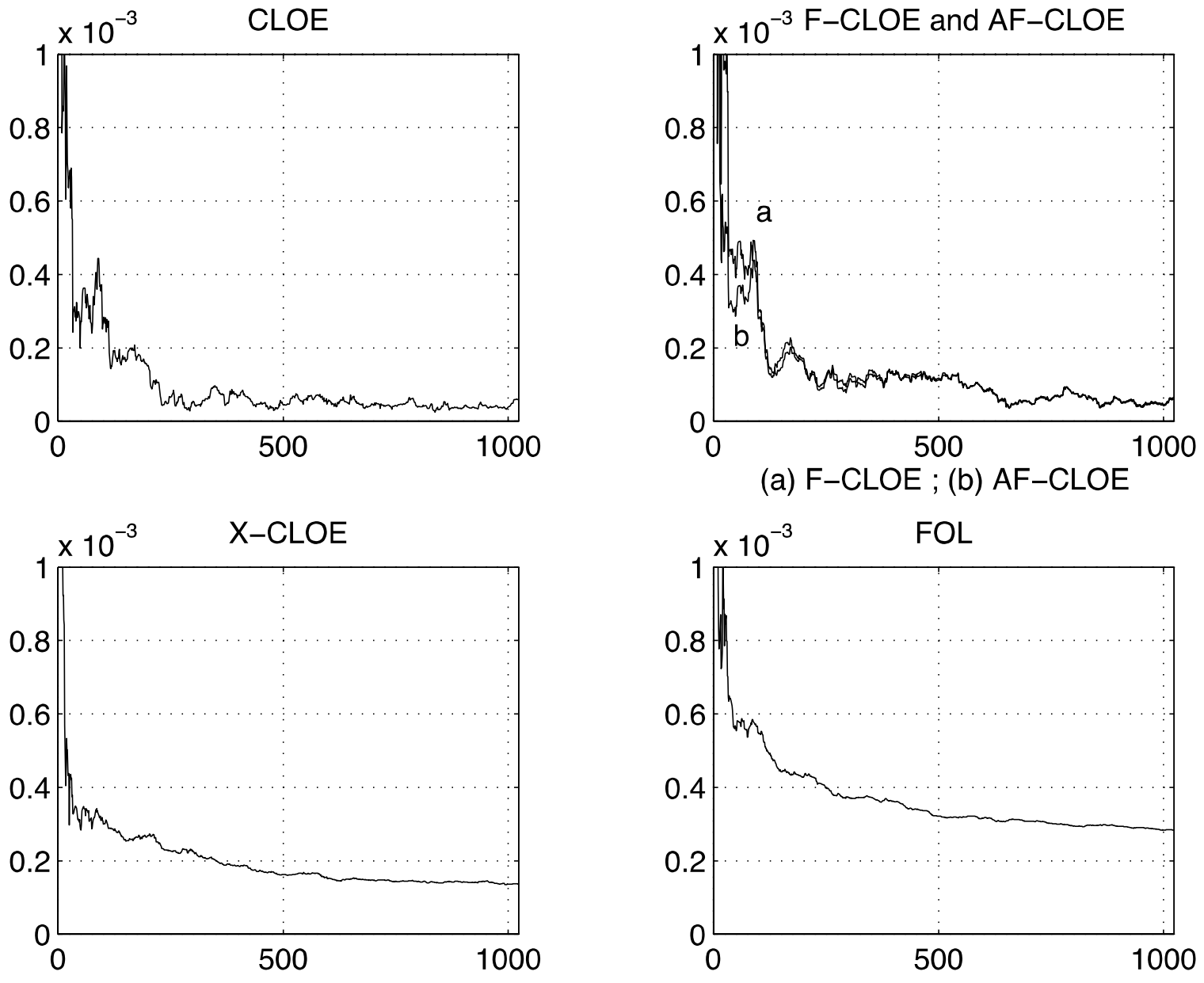

**Fig. 9.6** Evolution of the parametric distance for one run (Example 9.1, Case (b))

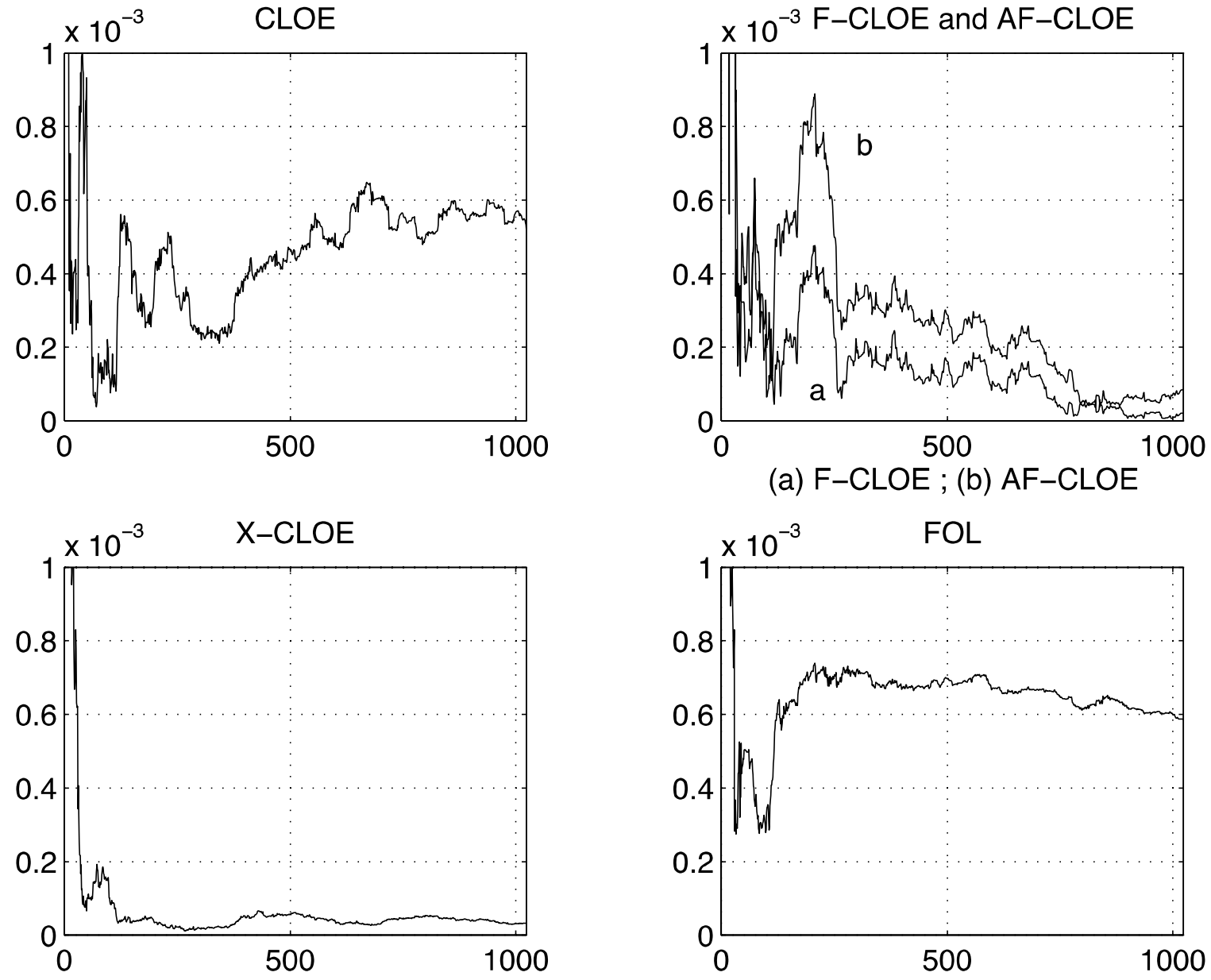

**Fig. 9.7** Evolution of the parametric distance for one run (Example 9.2)

The open-loop system which is in fact unstable can be stabilized using a unit feedback ($R = S = 1$). The closed-loop characteristic polynomial is given by:

$$P = 1 - 1.5q^{-1} + 0.7q^{-2} \tag{9.106}$$

which leads to a non positive real discrete-time transfer function for $S/P$ while $\frac{1}{C} - \frac{1}{2}$ is strictly positive real. The same PRBS which was used in Example 9.1 is used in order to identify the plant model in closed loop. The variance of noise/signal at the output of the system is about 10%.

Figure 9.7 shows the evolutions of the parametric distance for various methods and for one particular realization of the noise. One can observe that although the CLOE method does not converge with a non positive real transfer function for $S/P$ for this particular noise realization, the results of F-CLOE and X-CLOE in terms of parametric distance are satisfactory. This example clearly illustrates the fact that the positive realness condition on $S/P$ can be relaxed using F-CLOE, AF-CLOE or X-CLOE method. One sees also that FOL method cannot be a reliable tool for plant model identification in closed loop in this case.

## *9.7.2 Experimental Results: Identification of a Flexible Transmission in Closed-Loop*

The experimental device is depicted in Fig. 9.8 and has been described in Sect. 1.4.3. It consists of a flexible transmission which is formed by a set of three pulleys cou-pled by two very elastic belts. The system is controlled by a PC via an I/O board. The sampling frequency is 20 Hz.

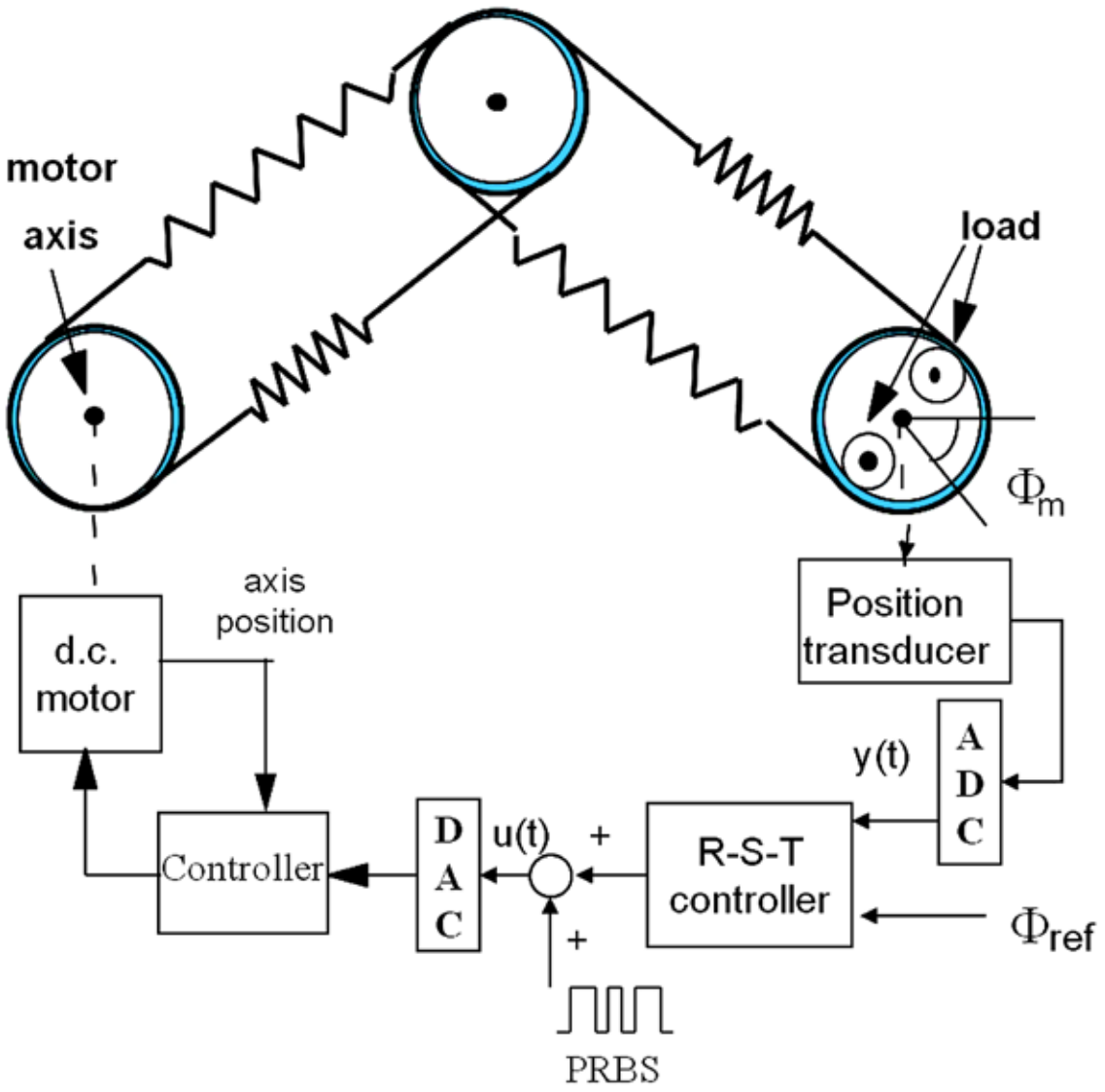

**Fig. 9.8** Block diagram of the flexible transmission

The system identification is carried out in open loop with a PC using WinPIM identification software (Adaptech 1988). The output error with extended prediction model algorithm provided the best results in terms of statistical validation (for details see Sect. 5.9). The model obtained in open loop for the unloaded case which passes the validation test is:

$$A = 1 - 1.3528q^{-1} + 1.5502q^{-2} - 1.2798q^{-3} + 0.9115q^{-4}$$
$$B = 0.4116q^{-1} + 0.524q^{-2}; \quad d = 2$$

The main characteristics of the system are: two very oscillatory modes, an unstable zero and a time delay of two sampling periods. A controller for this system is computed by the pole placement method with WinREG software (Adaptech 1988). The controller is designed in order to obtain two dominant poles with the same frequency of the first mode of the open-loop model but with a damping factor of 0.8. The precompensator $T(q^{-1})$ is chosen to obtain unit closed-loop gain. The parameters of the RST controller are as follows:

$$R(q^{-1}) = 0.4526 - 0.4564q^{-1} - 0.6857q^{-2} + 1.0955q^{-3} - 0.1449q^{-4}$$
$$S(q^{-1}) = 1 + 0.2345q^{-1} - 0.8704q^{-2} - 0.4474q^{-3} + 0.0833q^{-4}$$
$$T(q^{-1}) = 0.2612$$

Implementing the above controller on the real platform using WinTRAC software (Adaptech 1988), the identification of the plant in closed loop is carried out using

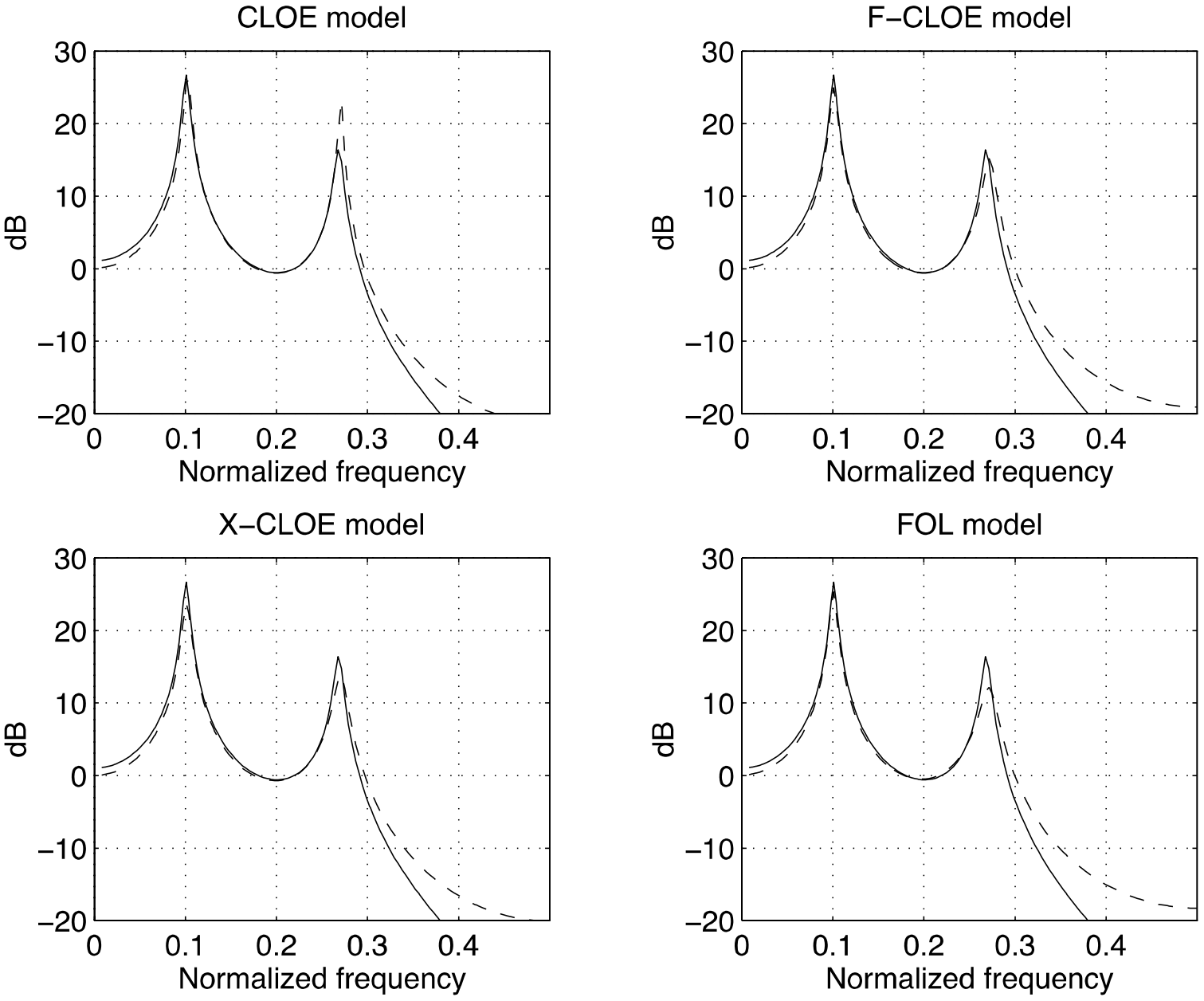

**Fig. 9.9** Frequency response of flexible transmission system models: [—] plant model identified in open loop, [- -] plant model identified in closed loop using different methods

various methods. A *PRBS* generated by a 7-bit shift register and a clock frequency of $\frac{1}{2} f_s$ is considered as reference signal.

The frequency response of the models identified in closed loop using various methods are compared with that of the open-loop model in Fig. 9.9. The first oscillatory identified mode is almost the same as that of the open loop model while the second identified mode varies slightly. A small difference on the steady state gain is also observed.

In order to validate the identified models, the real achieved closed-loop poles (which can be obtained by identification of the whole closed-loop system with standard open-loop identification methods) and the computed ones (using the plant model identified by various methods and the RST controller) are given in Fig. 9.10. It is observed that the computed closed-loop poles using the different models identified in closed loop and the real ones are almost superimposed particularly in low frequencies whereas the computed poles using the open loop identified model are far from the real closed-loop poles. The results of the statistical validation test are also given in Table 9.2 for the comparison purpose. It is clearly seen that the plant models identified in closed loop give better results than the open-loop identified model. It is also observed that the F-CLOE method gives the minimum value of the uncorrelation test while the other methods give more or less the good results. The output error with extended prediction model (OEEPM) is used to identify the plant model in FOL method. An estimation of $\hat{P}$ based on the open-loop identified model is used in the filters for the FOL and F-CLOE method.

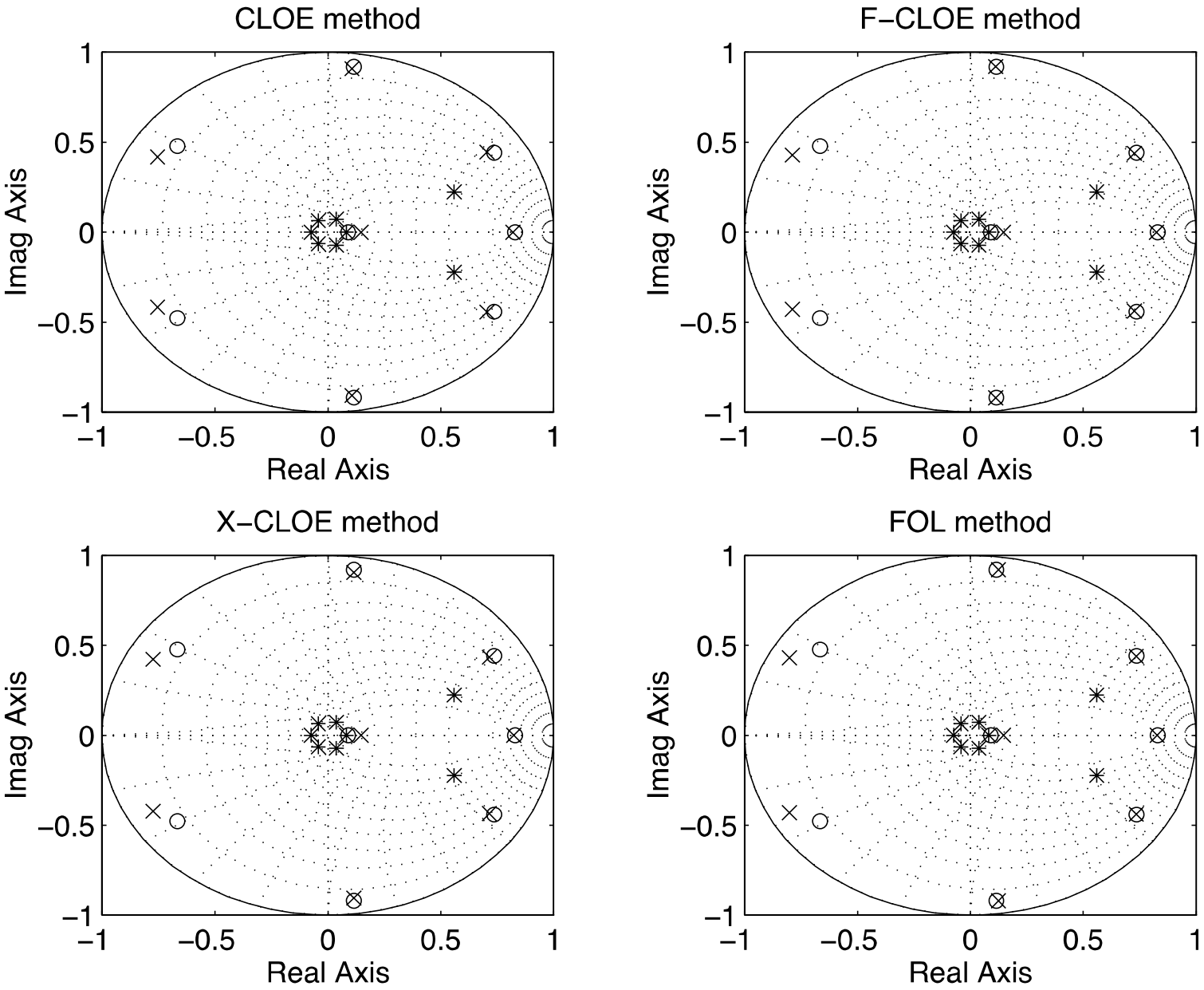

**Fig. 9.10** Closed-loop pole charts for flexible transmission system: [∘] real achieved closed-loop poles, [×] computed closed-loop poles using the closed-loop identified plant model and the controller, [∗] computed closed-loop poles using the open-loop model and the controller

**Table 9.2** Statistical model validation test for the flexible transmission system

| | $R_\varepsilon(0)$ | $\max_i \lvert R_{\varepsilon\hat{y}}(i)\rvert$ | $\max_i \lvert RN_{\varepsilon\hat{y}}(i)\rvert$ |
|---|---|---|---|
| Open-Loop | 0.0367 | 0.0652 | 0.5758 |
| CLOE | 0.0019 | 0.0047 | 0.1978 |
| F-CLOE | 0.0009 | 0.0022 | 0.1353 |
| X-CLOE | 0.0017 | 0.0041 | 0.1840 |
| FOL | 0.0012 | 0.0039 | 0.2055 |

## 9.8 Iterative Identification in Closed Loop and Controller Redesign Applied to the Flexible Transmission

The benefits of identification in closed loop followed by redesign of the controller will be illustrated for the case of the very flexible transmission shown in Fig. 9.8 and which has been described in Sect. 1.4.3.

The system identification is carried out in open loop with a PC using WinPIM identification software for the case without load (Adaptech 1988). The input sequence is a PRBS generated by a 7-bit shift register and with a clock frequency of $f_s/2$ (the data length is $L = 255$ samples). The output error with extended predic-tion model algorithm provided the best results in terms of statistical validation (see Sect. 5.9). The model obtained in open loop for the no load case which passes the validation test is:

$$A(q^{-1}) = 1 - 1.3528q^{-1} + 1.5502q^{-2} - 1.2798q^{-3} + 0.9115q^{-4}$$
$$B(q^{-1}) = 0.4116q^{-1} + 0.524q^{-2}; \quad d = 2$$

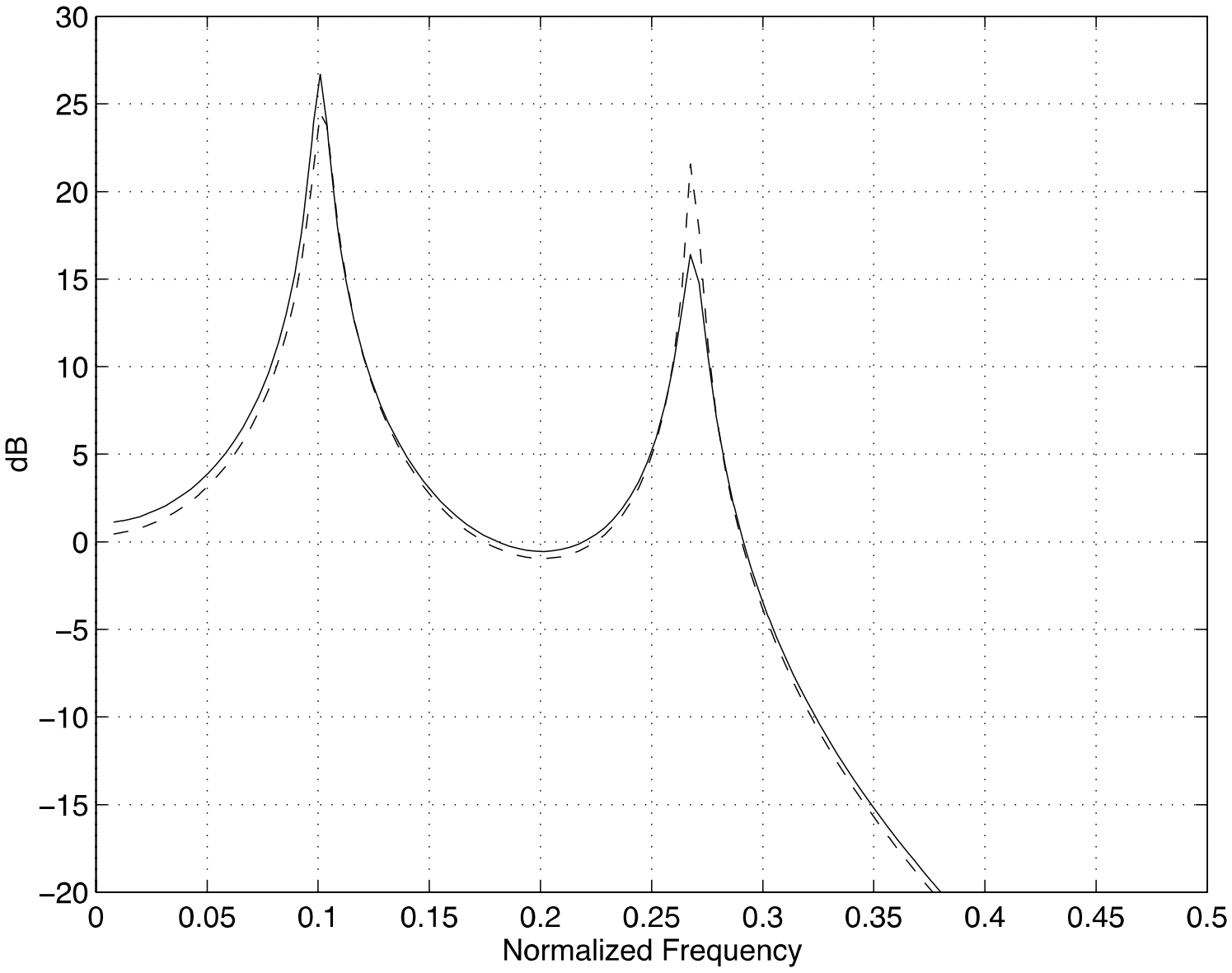

**Fig. 9.11** Magnitude of the frequency response of the plant model: [- -] identified in closed loop, [—] identified in open loop

The frequency characteristics of the model are shown in Fig. 9.11 (with continuous line).

A controller for this system is computed by the combined pole placement/ sensitivity function shaping with WinREG software (Adaptech 1988) (see Chap. 8). The controller is designed in order to obtain two dominant poles with the frequency of the first mode of the open-loop model but with a damping factor of 0.8. The constraints are: a modulus margin $\Delta M \geq -6$ dB, a delay margin $\Delta\tau > 0.75T_s$ and a maximum for $|S_{up}| < 10$ dB beyond $0.4f_s$.

The parameters of the $R$-$S$-$T$ controller are as follows :

$$R(q^{-1}) = 0.3999 - 0.6295q^{-1} + 0.0198q^{-2} + 0.0423q^{-3} + 0.4934q^{-4}$$
$$S(q^{-1}) = 1 + 0.3482q^{-1} - 0.3622q^{-2} - 0.7306q^{-3} - 0.2553q^{-4}$$
$$T(q^{-1}) = 0.3259$$

The controller (called subsequently: *open-loop based controller*) was implemented in real time using WinTRAC software (Adaptech 1988) which is also used

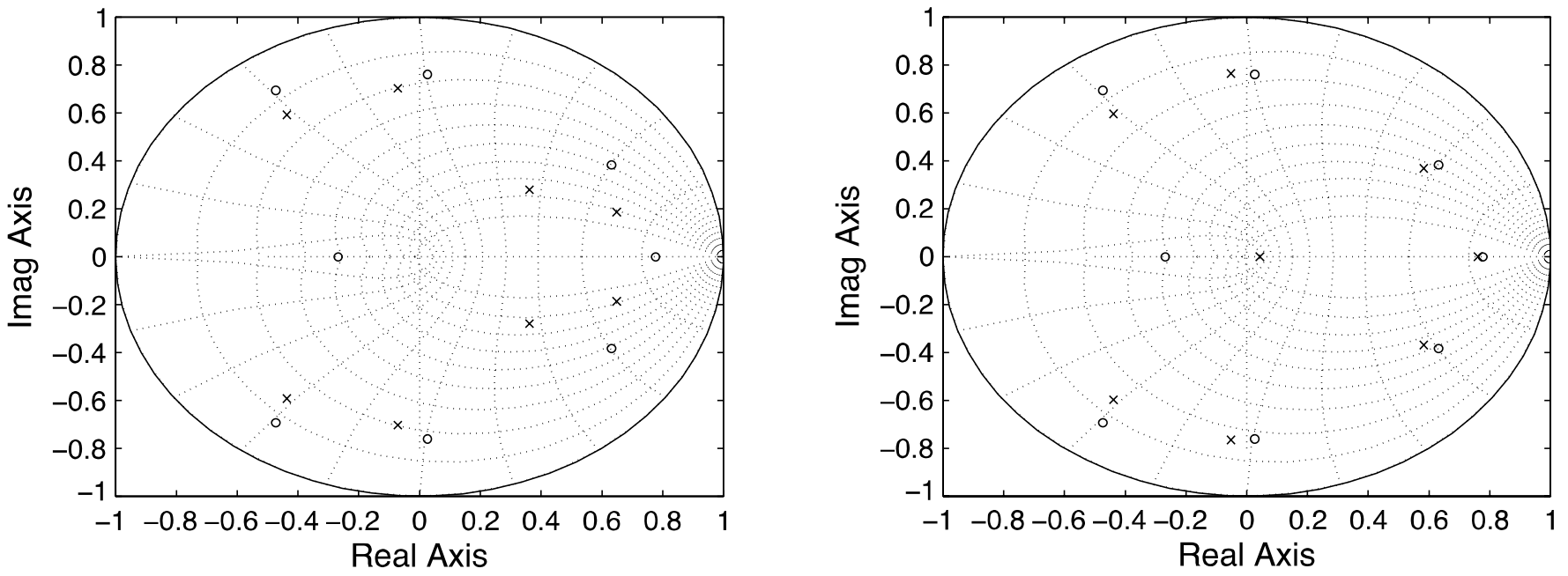

a) b)

**Fig. 9.12** Closed-loop pole chart: (**a**) [∘] achieved closed-loop poles, [×] computed poles using the open-loop identified model, (**b**) [∘] achieved closed-loop poles, [×] computed poles using the closed-loop identified model

for data acquisition in closed loop for system identification. The same PRBS as in open-loop identification was used as the external excitation at the reference input. The frequency response of the closed-loop identified model (using CLOE method, see Chap. 9) is compared with that of the open-loop model in Fig. 9.11.

In order to validate the identified model, the achieved closed-loop poles (which can be obtained by identification of the whole closed-loop system with standard open-loop identification methods) and the computed ones (using the plant model parameters and the $R$-$S$-$T$ controller) are given in Fig. 9.12. It is observed that the computed closed-loop poles using the plant model identified in the closed-loop operation are very near to the achieved ones (Fig 9.12b) whereas this is not the case for the open-loop identified model (Fig 9.12a).

The step response of the real closed-loop system can also be compared with the simulated ones using the open-loop identified model (Fig. 9.13a) and using the closed-loop identified model (Fig. 9.13b). It can be clearly seen that the step response of the real closed-loop system is better approximated with the closed-loop identified model than with the model identified in open loop. It should be mentioned that the same controller is always used in real time and simulation experience.

It will shown that a controller design based on the model identified in the closed-loop operation (which will be called: *closed-loop based controller*) can significantly improve the performance of the system. Figure 9.14 illustrates the step response of the real closed-loop system and of the design system using the new controller designed on the basis of the plant model identified in the closed loop. The same specifications and the same design method as for the case of open-loop identified model have been used. An almost perfect superposition of the achieved and designed step response is observed (an improvement of the achieved performances is also obtained). The response of the system to a step disturbance at the output of the system for two different controllers is shown in Fig. 9.15. The closed-loop based controller is faster than the open-loop based controller in terms of the disturbance rejection.

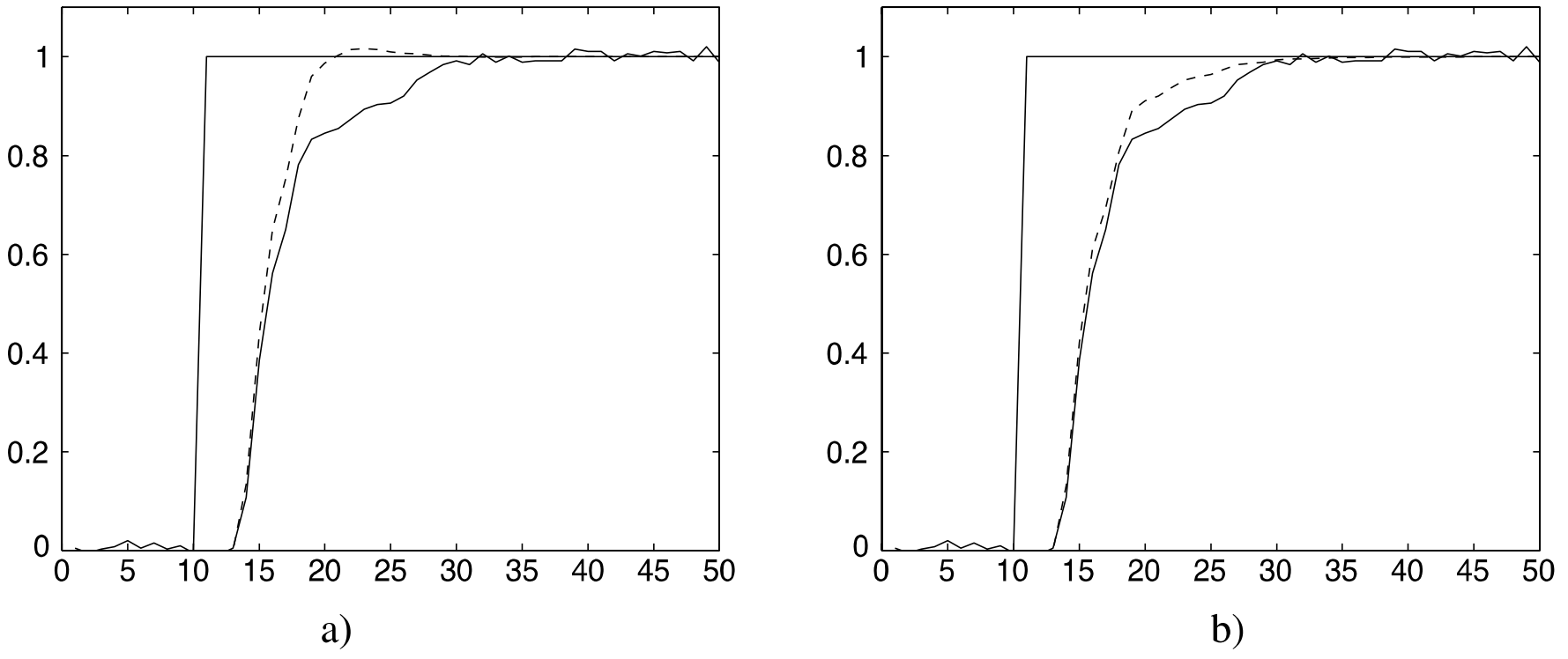

**Fig. 9.13** Step response of the closed-loop system: (**a**) [—] real-time step response, [- -] simulated step response with the open-loop identified model, (**b**) [—] real-time step response, [- -] simulated step response with the closed-loop identified model

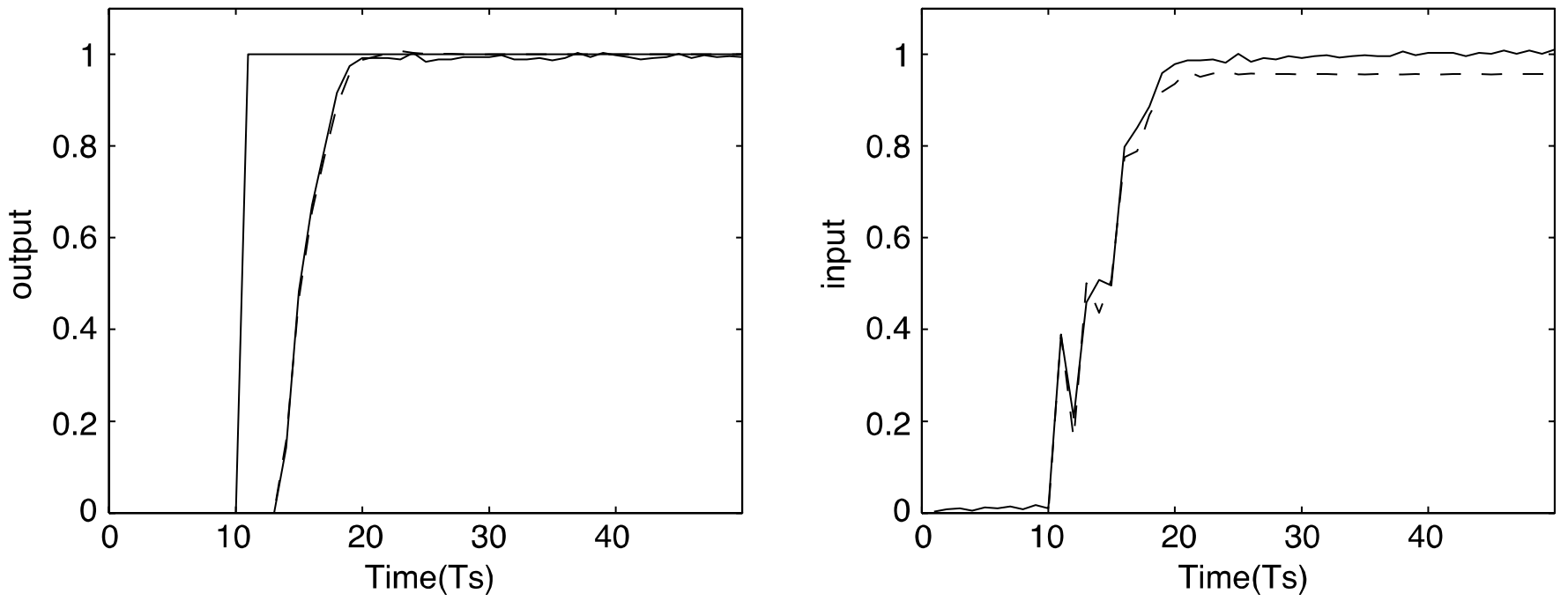

**Fig. 9.14** [—] Achieved and [- -] designed step response of the closed-loop system with the closed-loop based controller

## 9.9 Concluding Remarks

1. Recursive algorithms for plant model identification in closed-loop operation are efficient tools either for improving open-loop identified models or for redesign and re-tuning of existing controllers.
2. Identification in closed loop allows the identification of models for plants which can hardly be operated in open loop (presence of integrators, open-loop unstable plants, drift of the output, etc.).
3. Identification in closed loop is based on the use of an adaptive predictor for the closed loop which is re-parameterized in terms of the plant model to be identified. The estimated parameters minimize asymptotically a criterion in terms of the closed-loop prediction error.
4. Two families of recursive algorithms for identification in closed loop can be distinguished:

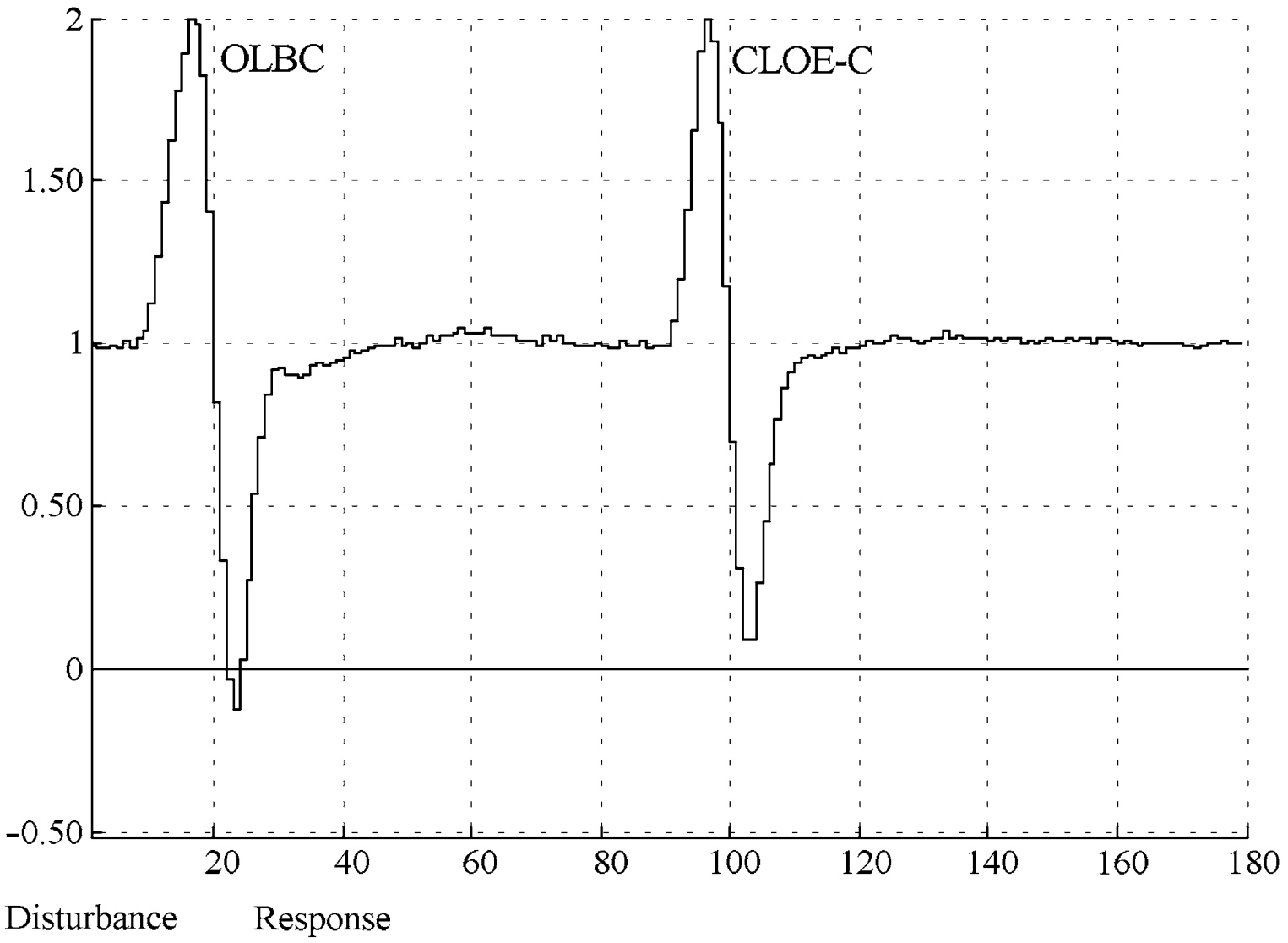

**Fig. 9.15** Output disturbance rejection for the open-loop based controller (OLBC) and the closed-loop based controller (CLOE-C)

- closed-loop output error recursive algorithms;
- filtered open-loop recursive algorithms.

The first family of algorithms uses an adjustable predictor for the closed loop, while the second family of algorithm re-construct the closed-loop prediction error by using open-loop identification algorithms operating on filtered data.

5. As for the case of identification in open loop, there is no single algorithm which gives the best results in all the situations.
6. Comparative validation of the identified models is crucial for the selection of the best identified model.
7. In addition to the statistical validation test, the pole closeness between the true closed-loop poles (obtained through identification of the closed loop) and the computed ones, based on the identified model is a very useful validation tool.
8. A unified framework for recursive plant model identification in closed loop has been provided.

## 9.10 Problems

**9.1** Develop an exact adaptation algorithm for the filtered closed-loop output error (F-CLOE) and provide the corresponding stability analysis in deterministic environment (Hint: use a similar approach as for the filtered output error method given in Sect. 5.5.3).

**9.2** Show that in the case of an ARMAX type plant model, the optimal predictor for the closed loop takes the form (9.32) and the prediction error equation takes the form (9.36).

**9.3** Along the lines for the derivation of the extended closed-loop output error algorithm (X-CLOE), develop an algorithm for the case of a noise model

$$v(t+1) = \frac{C(q^{-1})}{D(q^{-1})A(q^{-1})} e(t+1)$$

where $e(t+1)$ is a white noise (for the result, see Table 9.1, the G-CLOE algorithm).

**9.4** Work out the details of the stability analysis in deterministic environment and convergence analysis in stochastic environment for the algorithms given in Table 9.1.

**9.5** From the basic equations (9.5) through (9.11), define a closed-loop input error as:

$$\varepsilon_{CLU}(t+1) = u(t+1) - \hat{u}(t+1)$$

Develop recursive algorithms for identification in closed loop using this type of error (closed-loop input error methods).

**9.6** Give the details of the derivation of the expression for the asymptotic frequency distribution of the bias for filtered open-loop identification algorithms ((9.70) through (9.73)).

**9.7** Give the details of the derivation of the expression for the asymptotic frequency distribution of the bias for the X-CLOE algorithm (9.79).

**9.8** Consider a true plant model described by:

$$G(q^{-1}) = \frac{(q^{-1} + 0.5q^{-2})}{(1 - 1.5q^{-1} + 0.7q^{-2})(1 - 0.5q^{-1})}$$

whose output is disturbed by noise in closed loop with an RS Controller

$$\frac{R(q^{-1})}{S(q^{-1})} = \frac{0.8659 - 1.2763q^{-1} + 0.5204q^{-2}}{(1 - q^{-1})(1 + 0.3717q^{-1})}$$

Do identification in open loop and closed loop using a second order estimated model with: $d = 0$, $n_B = 2$, $n_A = 2$. Use a PRBS as external excitation.

1. Compare the frequency characteristics of the true plant model and of the estimated model obtained by various closed-loop identification methods.
2. Represent on the same graphic the frequency characteristics of the filter weighting the excitation signal (according to the expression for the asymptotic frequency distribution of the bias) and the sensitivity function weighting the model error.

3. Interpret the results.

**9.9** Modify the CLOE algorithms for the case where the estimated plant model should incorporate a known fixed part, i.e.,

$$\hat{G}(q^{-1}) = K(q^{-1})\frac{q^{-d}\hat{B}'(q^{-1})}{\hat{A}'(q^{-1})} \tag{9.107}$$

where $K(q^{-1})$ is a known transfer operator (only $\hat{B}'$ and $\hat{A}'$ have to be estimated).

**9.10** Show that the regressor vector in the AF-CLOE algorithm is the gradient of $\hat{y}(t)$ with respect to $\hat{\theta}$.

(Hint: use

$$\hat{y}(t) = \frac{T(q^{-1})q^{-d}\hat{B}(q^{-1})}{\hat{P}(q^{-1})}r(t)$$

and take the derivative with respect to $\hat{a}_n$ and $\hat{b}_n$.)

# Chapter 10
# Robust Parameter Estimation

## 10.1 The Problem

In the previous chapters, we assumed that:

1. The true plant model and the estimated plant model have the same structure (the true plant model is described by a discrete-time model with known upper bounds for the degrees $n_A, n_B + d$).
2. The disturbances are zero mean and of stochastic nature (with various assumptions).
3. For parameter estimation in closed-loop operation, the controller
   (a) has constant parameters and stabilizes the closed loop;
   (b) contains the internal model of the deterministic disturbance for which perfect state disturbance rejection is assured.
4. The parameters are constant or piece-wise constant.
5. The domain of possible parameters values is in general not constrained (exception: recursive maximum likelihood and adaptive filtered closed-loop output error).

In this chapter, we will examine the effects of the violation of these hypotheses and see how they can be overcome. Appropriate modification of the parameter adaptation algorithms in order to obtain a robust parameter estimation will be introduced. This will allow us to guarantee certain boundedness properties for the parameter estimates and adaptation error, which will be needed for establishing the boundedness of plant input-output signals in adaptive control.

Probably the major difficulty encountered when analyzing adaptive control schemes is caused by the fact that we cannot assume that the plant regressor vector (containing the plant inputs and outputs) is bounded, i.e., that the adaptive controller stabilizes the plant. The boundedness of the regressor vector is a result of the analysis and not an hypothesis (as in the plant model identification in open or closed loop). The robustification of the parameter adaptation algorithms is crucial both for practical and theoretical reasons.

The basic factors in the robustification of the parameter estimation algorithms are:

- filtering of input/output data;
- PAA with dead zone;
- PAA with projection;
- PAA without integrator effect;
- data normalization.

Next, we will try to give a very brief account of the rationale for these modifications before making a detailed presentation in the subsequent sections.

- Filtering of input/output data
  Filtering of input-output data results naturally if we consider the problem of plant model identification in closed loop using open-loop identification algorithms (see Chap. 9). Filtering of input-output data also represents a solution for enhancing an input-output spectrum in the "positive real" region of the transfer functions which occurs in convergence conditions of some algorithms (see Chaps. 3 and 4). When we would like to estimate a model characterizing the low frequency behavior of a plant, we have to filter the high-frequency content of input-output, in order to reduce the effect of the unmodeled dynamics.
- PAA with dead zone
  In a number of applications, it is difficult to assume a stochastic or a deterministic model for the disturbance, but a hard bound for the level of the disturbance can be defined. In such situations, trying to make the (a posteriori) adaptation error smaller than the bound of the disturbance is irrelevant. Therefore, a dead zone is introduced on the adaptation error such that the PAA stops when the adaptation error is smaller or equal to the upper magnitude of the disturbance.
- PAA with projection
  In many applications, the possible domain of variation of the parameter vector $\theta$ (or of some of its components) is known (for example, the model is stable, or the sign of a component of $\theta$ is known). Similarly, in a number of parameter estimation schemes, $\hat{\theta}$ or part of the components of $\hat{\theta}$, should be restricted to a stability domain. In such cases, the estimated parameters should be restricted to a given convex domain by projecting the estimates on the admissible domain.
- PAA without integrator effect
  As indicated in Chap. 3, the integrator property of the PAA results from the assumptions that the model for the variations of the parameters is a constant and the PAA will find the level of this constant. If the model of variations of the parameters is known, this can be inferred in the algorithm (see Chap. 3). However, in many applications, the parameters of the plant model are slowly time-varying and the model of these variations is unknown. In such cases, one can question the usefulness of the integrator behavior of the PAA since a "final" constant parameter value does not exist. Therefore, one can consider replacing the integrator by a first-order filter. This will also enhance the robustness of the scheme since a passive operator will be replaced by a strictly passive operator. This modification is discussed in Chap. 3, and will not be repeated here.
- Data normalization
  When the estimated model is of lower dimension than the true plant model, the unmodeled response of the plant (if the unmodeled part is stable) can be bounded by a norm or a filtered norm of the reduced order regressor $\phi$ containing the inputs and outputs over a certain horizon. This unmodeled response can also be considered as a disturbance added to a reduced order model.

In the context of adaptive control, since we cannot assume that the regressor $\phi$ containing plant inputs and outputs does not grow unbounded, we have to assure that the parameter estimates and the adaptation error remain bounded. This is obtained by defining "normalized" input-output data. This data corresponds to the input-output data divided by a norm of $\phi$. This new data cannot become unbounded, and the resulting normalized unmodeled response is also bounded. Therefore, the parameter estimator built from this data, will lead to a bounded adaptation error and using a PAA with dead zone or projection the boundedness of the parameter estimates will be assured.

## 10.2 Input/Output Data Filtering

The need for input-output data filtering results if one considers the problem of plant model identification in closed loop, even under the standard hypothesis 1 to 4 as it was shown in Chap. 9.

To estimate the plant model parameters in closed loop, one can either use the family of "closed-loop output error algorithms" or the family of "filtered open-loop parameter estimation algorithms". While in the former the filtering is automatically done through the structure of the adjustable predictor, in the latter the plant input-output signals should be explicitly filtered.

The analysis presented in Sect. 9.4 allows to assess qualitatively the asymptotic frequency distribution of the bias when the true plant model and the estimated model do not have the same structure. The quality of the estimated model can be enhanced in the critical frequency region for design (for example, near the critical point $[-1, 0]$ where the modulus of the output sensitivity function has its maximum). For further details see Karimi and Landau (1998).

The type of data filter used with open-loop parameter estimation algorithms for plant model identification in closed loop is of the form $S/\hat{P}$ or $AS/\hat{P}$, where $\hat{P}$ corresponds to the estimated closed loop poles (or their desired values) and $S$ which is part of the controller contains generally the inverse of the disturbance model. This type of filter will have the following influence upon the data:

1. It will remove the deterministic disturbance from the data.
2. It will filter high-frequency components in the signals beyond the band pass defined by the desired closed-loop poles $\hat{P}$.

As an example, assume that the output is disturbed by a step ($\frac{\delta(t)}{1-q^{-1}}$ where $\delta(t)$ is a Dirac pulse) and the $S$ polynomial contains $(1-q^{-1})$ as a factor (integrator effect). Then the data filter will remove the d.c. disturbance (which other way will cause a drift in the parameters).

Assume that the true plant model has low frequency dynamics which we would like to be captured by the identified model and high-frequency dynamics which will not be modeled. The effect of the filter will be to attenuate the high frequencies beyond a certain frequency mainly defined by the closed-loop poles. In other words, the filter will enhance the data in the frequency range of interest for control and, as a consequence, a lower order model which will try to capture the behavior of the plant in this frequency range can be better identified.

The need for data filtering results also from the fact that a number of algorithms are subject to stability or convergence conditions which rely on the positive realness of a certain transfer function (see Chaps. 3, 4 and 9). Using the "averaging" analysis, it was shown in Chap. 4 that these conditions come from:

$$\mathbf{E}\{l^T \phi(t, \hat{\theta}) H(q^{-1}) \phi^T(t, \hat{\theta}) l\} > 0 \tag{10.1}$$

which using the Parseval theorem can be converted in:

$$\frac{1}{2\pi}\int_{-\pi}^{\pi} \phi_l(\omega) H(e^{-j\omega}) \phi_l(\omega) d\omega = \frac{1}{2\pi}\int_{0}^{\pi} \phi_l(\omega) [\mathrm{Re}\, H(e^{-j\omega})] \phi_l(\omega) d\omega > 0 \tag{10.2}$$

where $\phi_l(\omega)$ is the Fourrier transform of $l^T \phi(t, \hat{\theta})$. If $H(z^{-1})$ is strictly positive real the above condition is satisfied. However, if $H$ is not SPR overall the frequencies, but $\phi(t, \hat{\theta})$ is periodic and contains only frequencies in the range of frequencies where $H$ is SPR, then the condition (10.2) is satisfied. Therefore, filtering the input-output data in the regions where $H$ is not SPR will enhance the contribution of the components of $\phi$ in the "good" frequency region (positive real dominant spectrum) in order to satisfy (10.2).

## 10.3 Effect of Disturbances

In order to introduce the various modifications of the PAA, it is useful to examine first the effect of disturbances upon the parameter adaptation algorithms when these disturbances are not characterized by nice stochastic assumptions like: zero mean, finite variance, uncorrelation with the input, ARMA model and so on.

We will start with a qualitative analysis. Assume that the plant output can be described by:

$$y(t+1) = \theta^T \phi(t) + w(t+1) \tag{10.3}$$

where $w(t+1)$ is the disturbance, $\theta$ is the vector of the model parameters and in this case:

$$\phi^T(t) = [-y(t), -y(t-1), \ldots, u(t), u(t-1), \ldots] \tag{10.4}$$

Consider the parameter adaptation equation:

$$\hat{\theta}(t+1) = \hat{\theta}(t) + F(t)\phi(t)\varepsilon(t+1) \tag{10.5}$$

where:

$$\varepsilon(t+1) = y(t+1) - \hat{\theta}^T(t+1)\phi(t) \tag{10.6}$$

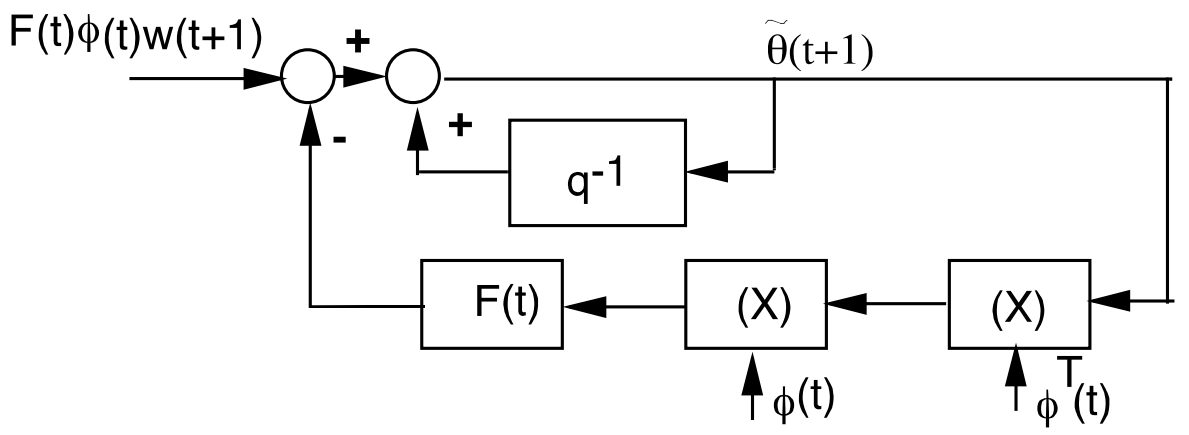

**Fig. 10.1** Equivalent feedback system representation of the PAA in the presence of disturbances

Replacing $y(t+1)$ with its expression given by (10.3) and using the notation:

$$\tilde{\theta}(t+1) = \hat{\theta}(t+1) - \theta \tag{10.7}$$

one obtains from (10.4):

$$\hat{\theta}(t+1) = \hat{\theta}(t) + F(t)\phi(t)[-\phi^T(t)\tilde{\theta}(t+1) + w(t+1)] \tag{10.8}$$

Subtracting $\theta$ in both sides of (10.8), one gets the evolution equation for the parameter error in the presence of disturbances:

$$[I_n + F(t)\phi(t)\phi(t)^T - I_n q^{-1}]\tilde{\theta}(t+1) = F(t)\phi(t)w(t+1) \tag{10.9}$$

where the eigenvalues of $[I_n + F(t)\phi(t)\phi^T(t)]$ satisfies:

$$\lambda_i[I_n + F(t)\phi(t)\phi^T(t)]^{-1} \leq 1; \quad i = 1, \ldots, n \tag{10.10}$$

$$n = \dim\theta \tag{10.11}$$

The equivalent feedback system associated with (10.9) is represented in Fig. 10.1.

The term $F(t)\phi(t)w(t+1)$ represents the exogenous input to the system of (10.9). Therefore, in order to assure that $\tilde{\theta}(t+1)$ (and $\hat{\theta}(t+1)$) is bounded, a necessary condition is that this quantity be bounded. However, even if $\phi(t)$ and $w(t+1)$ are bounded, one cannot guarantee that $\tilde{\theta}(t+1)$ remains bounded because if the sequence $\{\phi(t)\phi(t)^T\}$ is null in average (i.e., it does not satisfy the persistency of excitation condition), the systems of Fig. 10.1 will behave as an integrator and a drift of the parameters will occur if $\{F(t)\phi(t)w(t+1)\}$ is not zero in the average. It will therefore be important to assure that:

1. $\phi(t)$ and $w(t+1)$ are bounded;
2. $\hat{\theta}(t+1)$ remains bounded in the absence of persistency of excitation (when using integral adaptation).

One can complete this qualitative analysis with a quantitative analysis summarized in the following theorem:

**Theorem 10.1** *Assume that the plant output can be described by*:

$$y(t+1) = \theta^T \phi(t) + w(t+1) \tag{10.12}$$

*where* $w(t+1)$ *represents the disturbance*, $\theta$ *corresponds to the vector of model parameters and* $\phi(t)$ *is given by* (10.4). *Assume that the adjustable model is of the form*:

$$\hat{y}^0(t+1) = \hat{\theta}^T(t)\phi(t) \tag{10.13}$$

$$\hat{y}(t+1) = \hat{\theta}^T(t+1)\phi(t) \tag{10.14}$$

*Assume that the PAA used is*:

$$\hat{\theta}(t+1) = \hat{\theta}(t) + F(t)\phi(t)\nu(t+1) \tag{10.15}$$

$$F(t+1)^{-1} = \lambda_1(t)F(t)^{-1} + \lambda_2(t)\phi(t)\phi^T(t)$$

$$\lambda_1(t) \equiv 1;\ 0 \le \lambda_2(t) < 2 \tag{10.16}$$

$$\nu(t+1) = \frac{y(t+1) - \hat{y}^0(t+1)}{1 + \phi^T(t)F(t)\phi(t)} \tag{10.17}$$

*Then*, *one has*:

(i)

$$\tilde{\theta}^T(N+1)F(t+1)^{-1}\tilde{\theta}(N+1) + \sum_{t=0}^{N}[1 + \phi^T(t)F(t)\phi(t)]\nu^2(t+1)$$

$$\le \tilde{\theta}^T(0)F(0)^{-1}\tilde{\theta}(0) + \sum_{t=0}^{N} w^2(t+1) \tag{10.18}$$

*where*:

$$\tilde{\theta}(t) = \tilde{\theta}(t) - \theta \tag{10.19}$$

(ii)

$$\sum_{t=0}^{N}\nu^2(t+1) \le \sum_{t=0}^{N}[1 + \phi^T(t)F(t)\phi(t)]\nu^2(t+1)$$

$$\le c_1 + c_2\sum_{t=0}^{N} w^2(t+1) \quad 0 \le c_1 < \infty;\ 0 < c_2 < \infty \tag{10.20}$$

*If*, *in addition*:

$$|w(t+1)| < \Delta; \quad \forall t \ge 0;\ 0 \le \Delta < \infty \tag{10.21}$$

*then*:

$$\frac{1}{N}\sum_{t=0}^{N}\nu^2(t+1) \le \frac{1}{N}\sum_{0}^{N}[1+\phi^T(t)F(t)\phi(t)]\nu^2(t+1)$$
$$\le c_2\Delta^2 + c_1/N \tag{10.22}$$

*Remark* The above results indicate that the mean square of the a posteriori adaptation error is bounded by the disturbance upper bound. However, nothing can be said in general upon the boundedness of $\tilde{\theta}(t)$.

*Proof* To make the proof clearer, we will first discuss the case $\lambda_2(t) = 0$ which corresponds to a constant adaptation gain ($F(t) = F(0) = F$). From Lemma 3.2, one has:

$$\sum_{t=0}^{N}\tilde{\theta}^T(t+1)\phi(t)\nu(t+1) = \frac{1}{2}\tilde{\theta}^T(N+1)F^{-1}\tilde{\theta}(N+1) - \frac{1}{2}\tilde{\theta}^T(0)F^{-1}\tilde{\theta}(0)$$
$$+\frac{1}{2}\sum_{t=0}^{N}\phi^T(t)F\phi(t)\nu^2(t+1) \tag{10.23}$$

On the other hand, one gets from (10.12) and (10.20):

$$\nu(t+1) = y(t+1) - \hat{y}(t+1) = -\tilde{\theta}^T(t+1)\phi(t) + w(t+1) \tag{10.24}$$

Therefore:

$$\sum_{t=0}^{N}\tilde{\theta}^T(t+1)\phi(t)\nu(t+1) = \sum_{t=0}^{N}[w(t+1)-\nu(t+1)]\,\nu(t+1)$$
$$= \sum_{t=0}^{N}w(t+1)\nu(t+1) - \sum_{t=0}^{N}\nu^2(t+1) \tag{10.25}$$

But for any $\beta > 0$, one has:

$$w(t+1)\nu(t+1) \le \frac{\beta}{2}\nu^2(t+1) + \frac{1}{2\beta}w(t+1)^2 \tag{10.26}$$

(this is obtained by developing $[\beta\nu(t+1) - w(t+1)]^2 \ge 0$) and it results that:

$$\sum_{t=0}^{N}\tilde{\theta}^T(t+1)\phi(t)\nu(t+1) \le \left(\frac{\beta}{2}-1\right)\sum_{t=0}^{N}\nu^2(t+1) + \frac{1}{2\beta}\sum_{t=0}^{N}w^2(t+1) \tag{10.27}$$

Combining (10.27) with (10.23), one gets:

$$\left(\frac{\beta}{2}-1\right)\sum_{t=0}^{N}\nu^2(t+1)+\frac{1}{2\beta}\sum_{t=0}^{N}w^2(t+1)$$
$$\geq \frac{1}{2}\tilde{\theta}^T(t+1)F^{-1}\tilde{\theta}(t+1)$$
$$-\frac{1}{2}\tilde{\theta}^T(0)F^{-1}\tilde{\theta}(0)+\frac{1}{2}\sum_{t=0}^{N}\phi^T(t)F\phi(t)\nu^2(t+1) \tag{10.28}$$

and respectively:

$$\frac{1}{2}\tilde{\theta}^T(N+1)F^{-1}\tilde{\theta}(N+1)$$
$$+\sum_{t=0}^{N}\left\{\left[1-\frac{\beta}{2}+\frac{1}{2}\phi^T(t)F\phi(t)\right]\nu^2(t+1)-\frac{1}{2\beta}w^2(t+1)\right\}$$
$$\leq \frac{1}{2}\tilde{\theta}^T(0)F^{-1}\tilde{\theta}(0) \tag{10.29}$$

Taking $\beta = 1$, one gets:

$$\frac{1}{2}\tilde{\theta}^T(t+1)F^{-1}\tilde{\theta}(t+1)$$
$$+\frac{1}{2}\sum_{t=0}^{N}\{[(1+\phi^T(t)F\phi(t)]\nu^2(t+1)-w^2(t+1)\}$$
$$\leq \frac{1}{2}\tilde{\theta}^T(0)F^{-1}\tilde{\theta}(0) \tag{10.30}$$

which after rearranging the various terms leads to (10.18). Equation (10.20) is a consequence of (10.18) bearing in mind that $\hat{\theta}^T(t)F^{-1}\hat{\theta}(t) \geq 0$.

Consider now the case $\lambda_2(t) \neq 0$, in this case, (10.23) becomes:

$$\sum_{t=0}^{N}\tilde{\theta}^T(t+1)\phi(t)\nu(t+1)$$
$$=\frac{1}{2}\tilde{\theta}^T(N+1)F(N+1)^{-1}\tilde{\theta}(N+1)$$
$$-\frac{1}{2}\tilde{\theta}^T(0)F(0)^{-1}\tilde{\theta}(0)+\frac{1}{2}\sum_{t=0}^{N}\phi^T(t)F(t)\phi(t)\nu^2(t+1)$$
$$-\frac{1}{2}\sum_{t=0}^{N}\lambda_2(t)[\tilde{\theta}^T(t+1)\phi(t)]^2 \tag{10.31}$$

Using (10.24) and (10.25), one gets from (10.31):

$$\begin{aligned}\sum_{t=0}^{N} w(t+1)\nu(t+1) - \sum_{t=0}^{N} \nu^2(t+1) \\ = \frac{1}{2}\tilde{\theta}^T(N+1)F(N+1)^{-1}\tilde{\theta}(N+1) \\ - \frac{1}{2}\tilde{\theta}^T(0)F(0)^{-1}\tilde{\theta}(0) + \frac{1}{2}\sum_{t=0}^{N}\phi^T(t)F(t)\phi(t)\nu^2(t+1) \\ - \frac{1}{2}\sum_{t=0}^{N}\lambda_2(t)[w(t+1)-\nu(t+1)]^2\end{aligned} \tag{10.32}$$

which, after rearranging the various terms, becomes:

$$\begin{aligned}\sum_{t=0}^{N}(1-\lambda_2(t))\,w(t+1)\nu(t+1) - \sum_{t=0}^{N}\left(1-\frac{\lambda_2(t)}{2}\right)\nu^2(t+1) \\ + \sum_{t=0}^{N}\frac{\lambda_2(t)}{2}w^2(t+1) \\ = \frac{1}{2}\tilde{\theta}^T(N+1)F(N+1)^{-1}\tilde{\theta}(N+1) \\ - \frac{1}{2}\tilde{\theta}^T(0)F(0)^{-1}\tilde{\theta}(0) + \frac{1}{2}\sum_{t=0}^{N}\phi^T(t)F(t)\phi(t)\nu^2(t+1)\end{aligned} \tag{10.33}$$

Using now (10.26), one gets:

$$\begin{aligned}\frac{1}{2}\tilde{\theta}^T(N+1)F(N+1)^{-1}\tilde{\theta}(N+1) \\ + \sum_{t=0}^{N}\left[\left(1-\frac{\lambda_2(t)}{2}\right) - (1-\lambda_2(t))\frac{\beta}{2} + \frac{1}{2}\phi^T(t)F(t)\phi(t)\right]\nu^2(t+1) \\ \le \frac{1}{2}\tilde{\theta}^T(0)F(0)^{-1}\tilde{\theta}(0) + \sum_{t=0}^{N}\left[(1-\lambda_2(t))\frac{1}{2\beta} + \frac{\lambda_2(t)}{2}\right]w^2(t+1)\end{aligned} \tag{10.34}$$

which for $\beta = 1$ becomes:

$$\begin{aligned}\frac{1}{2}\tilde{\theta}^T(N+1)F(N+1)^{-1}\tilde{\theta}(N+1) + \frac{1}{2}\sum_{t=0}^{N}[(1+\phi^T(t)F(t)\phi(t)]\nu^2(t+1) \\ \le \frac{1}{2}\tilde{\theta}^T(0)F(0)^{-1}\tilde{\theta}(0) + \frac{1}{2}\sum_{t=0}^{N}w^2(t+1)\end{aligned} \tag{10.35}$$

from which (10.18) results. □

## 10.4 PAA with Dead Zone

The PAA with dead zone is used:

1. When the disturbances on the plant output do not have a stochastic or a deterministic model but a hard bound for the level of the disturbance can be assumed.
2. In order to avoid the drift of the parameters in the presence of disturbances.

A dead zone is introduced on the adaptation error such that the PAA stops when the adaptation error is equal or smaller than a certain level. A prototype for such a PAA takes the following form:

$$\hat{\theta}(t+1) = \hat{\theta}(t) + \alpha(t)F(t)\phi(t)\nu(t+1) \tag{10.36}$$

$$F(t+1)^{-1} = F(t)^{-1} + \alpha(t)\lambda_2(t)\phi(t)\phi(t)^T; \quad 0 \le \lambda_2(t) < 2 \tag{10.37}$$

$$\nu(t+1) = \frac{\nu^0(t+1)}{1+\phi^T(t)F(t)\phi(t)} \tag{10.38}$$

$$\alpha(t) = \begin{cases} 1 & |\nu(t+1)| > \Delta \\ 0 & |\nu(t+1)| \le \Delta \end{cases} \tag{10.39}$$

where the level $\Delta$ will depend upon the disturbance.

One has the following results:

**Theorem 10.2** (PAA with Dead Zone) *Assume that the plant model is described by*:

$$y(t+1) = \hat{\theta}^T\phi(t) + w(t+1) \tag{10.40}$$

*where the disturbance* $w(t+1)$ *satisfies*:

$$|w(t+1)| < \Delta; \quad \forall t \ge 0 \tag{10.41}$$

*Define the adjustable model as*:

$$\hat{y}^0(t+1) = \hat{\theta}^T(t)\phi(t) \tag{10.42}$$

$$\hat{y}(t+1) = \hat{\theta}^T(t+1)\phi(t) \tag{10.43}$$

*Use the PAA of* (10.36) *through* (10.39), *with*:

$$\nu^0(t+1) = y(t+1) - \hat{y}^0(t+1) \tag{10.44}$$

*Then, one has the following properties*:

$$\sum_{t=0}^{\infty} \alpha(t)[\nu^2(t+1) - \Delta^2] < \infty \tag{10.45}$$

$$\lim_{t\to\infty} \alpha(t)[\nu^2(t+1) - \Delta^2] = 0 \tag{10.46}$$

$$\lim_{t\to\infty} \sup |\nu(t+1)| \le \Delta \tag{10.47}$$

$$\begin{aligned}[\hat{\theta}(t)-\theta]^T F(t)^{-1}[\hat{\theta}(t)-\theta] &\le [\hat{\theta}(0)-\theta]F(0)^{-1}[\hat{\theta}(0)-\theta] \\ &\le M < \infty \end{aligned} \tag{10.48}$$

*Proof* Taking the same route as for the proof of Theorem 10.1, one obtains similar to (10.35) the following equation:

$$\begin{aligned} &\frac{1}{2}\tilde{\theta}^T(N+1)F^{-1}(N+1)\tilde{\theta}(N+1) \\ &\quad + \frac{1}{2}\sum_{t=0}^{N}\alpha(t)\{[1+\alpha(t)\phi^T(t)F(t)\phi(t)]\nu^2(t+1) - w^2(t+1)\} \\ &\le \frac{1}{2}\tilde{\theta}^T(0)F(0)^{-1}\tilde{\theta}(0) \end{aligned} \tag{10.49}$$

Neglecting the effect of the term $\alpha(t)\phi^T(t)F(t)\phi(t)\nu^2(t+1)$, in order to guarantee the boundedness of $\tilde{\theta}^T(t+1)F^{-1}\tilde{\theta}(t+1)$, one should avoid terms of the form:

$$\nu^2(t+1) - w^2(t+1) < 0 \tag{10.50}$$

However, taking into account the definition of $\alpha(t)$ given in (10.39), this situation is avoided. One therefore concludes from (10.49) that:

$$\lim_{N\to\infty}\sum_{t=0}^{N}\alpha(t)[\nu^2(t+1) - \Delta^2] < \infty \tag{10.51}$$

and therefore:

$$\lim_{t\to\infty}\alpha(t)[\nu^2(t+1) - \Delta^2] = 0 \tag{10.52}$$

since:

$$\alpha(t)[\nu^2(t+1) - \Delta^2] \ge 0 \tag{10.53}$$

To obtain (10.47), one has to show that $\alpha(t) \to 0$. A contradiction argument will be used. Suppose that $\alpha(t)$ does not converge to zero. It follows that there exists a sequence $t_i$ along which $\alpha(t_i) = 1$ (starting at $t_{i_0}$). But this implies taking into account the stopping rule that along this sequence $\nu^2(t_i) - w^2(t_i) > 0$ and $\lim_{t_i\to\infty}\sum_{t_{i_0}}^{t_i}\alpha(t)[\nu^2(t+1) - w^2(t+1)] = \infty$. This contradicts (10.49), therefore one has $\lim_{t\to\infty}\alpha(t) = 0$ and this implies also:

$$\lim_{t\to\infty}\sup|\nu(t+1)| \le \Delta \tag{10.54}$$

□

*Remark*

1. Slightly stronger results can in fact be obtained from (10.49) by taking advantage of the term $\alpha(t)\phi(t)F(t)\phi(t)\nu^2(t+1)$. In this case, one considers instead of (10.39) the following stopping rule:

$$\alpha(t) = \begin{cases} 1 & [1+\phi^T(t)F(t)\phi(t)]^{1/2}|\nu(t+1)| > \Delta \\ 0 & [1+\phi^T(t)F(t)\phi(t)]^{1/2}|\nu(t+1)| \le \Delta \end{cases}$$

   and the results of Theorem 10.2 will hold for $[1+\phi(t)F(t)\phi(t)]\nu^2(t+1)$ instead of $\nu^2(t+1)$.
2. The fixed disturbance bound $\Delta$, can be replaced by a time-varying bound $\Delta(t)$ ($|w(t+1)| < \Delta(t) < \infty$).

## 10.5 PAA with Projection

As indicated in the introduction of this chapter, in a number of situations it is known that the parameter vector $\theta$ which should be estimated, belongs to a certain closed domain. Such situations include for example the case where the model to be estimated is stable. Therefore, in such situations, one can consider projecting the estimates in the known domain if they are outside. The projection of the estimates within a certain domain is even mandatory in certain cases like:

1. Parameter estimation algorithms using filtered regressors or filtered prediction errors where the filter parameters depends on the estimates (projection into the stability domain).
2. Parameter estimation algorithms in the presence of disturbances where in order to avoid the possible drift of the estimated parameters, one has to project the estimates on a certain restricted domain (note: alternatively one can use the PAA with dead zone).

To introduce the technique, we will consider first the particular case of a unit constant adaptation gain:

$$F(t) = F = I \tag{10.55}$$

Assume that the plant model output is given by:

$$y(t+1) = \theta^T\phi(t) \tag{10.56}$$

and assume that $\theta$ belongs to a known domain $\mathscr{D}$: $\theta \in \mathscr{D}$. Consider the following PAA:

$$\hat{\theta}(t+1) = \hat{\theta}_p(t) + \phi(t)\nu(t+1) \tag{10.57}$$

$$\nu(t+1) = \frac{y(t+1) - \hat{\theta}_p^T(t)\phi(t)}{1+\phi(t)^T\phi(t)} \tag{10.58}$$

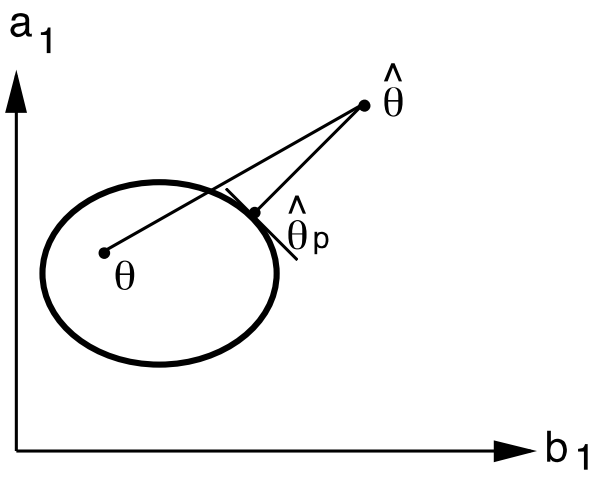

**Fig. 10.2** Orthogonal projection of parameter estimates

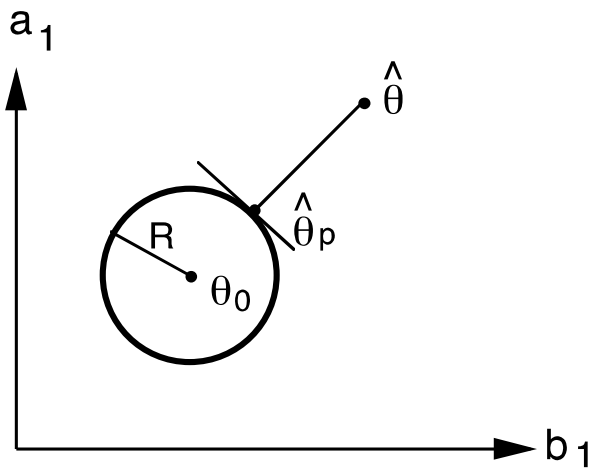

**Fig. 10.3** Projection on a sphere

$$\hat{\theta}_p(t+1) = \begin{cases} \hat{\theta}(t+1) & \text{if } \hat{\theta}(t+1) \in \mathscr{D} \\ \perp \text{ proj of } \hat{\theta}(t+1) \text{ on } \mathscr{D} & \text{if } \hat{\theta}(t+1) \notin \mathscr{D} \end{cases} \tag{10.59}$$

where "$\perp$ proj $\hat{\theta}(t+1)$ on $\mathscr{D}$" corresponds to the orthogonal projection of $\hat{\theta}(t+1)$ on the domain $\mathscr{D}$ (see Fig. 10.2).

*Remark* In the particular case when the domain $\mathscr{D}$ is a n-dimensional sphere defined by a center $\theta_0$ and a radius $R$ (see Fig. 10.3), the orthogonal projection (10.59) becomes:

$$\hat{\theta}_p(t+1) = \begin{cases} \hat{\theta}(t+1) & \text{if } \|\hat{\theta}(t+1) - \theta_0\| \leq R \\ \theta_0 + R\frac{\hat{\theta}(t+1)-\theta_0}{\|\hat{\theta}(t+1)-\theta_0\|} & \text{if } \|\hat{\theta}(t+1) - \theta_0\| > R \end{cases}$$

The main property which results from the orthogonal projection is:

$$\tilde{\theta}_p^T(t+1)\tilde{\theta}_p(t+1) \leq \tilde{\theta}^T(t+1)\hat{\theta}(t+1) \tag{10.60}$$

where:

$$\tilde{\theta}_p(t+1) = \hat{\theta}_p(t+1) - \theta \tag{10.61}$$

$$\tilde{\theta}(t+1) = \hat{\theta}(t+1) - \theta \tag{10.62}$$

This means that the projected vector $\hat{\theta}_p(t)$ is "closer" to $\theta$ than $\hat{\theta}(t)$, or at least at the same distance (in the sense of the Euclidian norm).

Using the intermediate results of Lemma 3.2, one has for $F(t) \equiv F = I$:

$$\begin{aligned} \tilde{\theta}^T(t+1)\tilde{\theta}(t+1) &= \tilde{\theta}_p^T(t)\tilde{\theta}_p(t) + 2\tilde{\theta}(t+1)\phi(t)\nu(t+1) \\ &\quad - \phi^T(t)\phi(t)\nu^2(t+1) \end{aligned} \tag{10.63}$$

Using now (10.60), one obtains:

$$\tilde{\theta}_p^T(t+1)\tilde{\theta}_p(t+1) \leq \tilde{\theta}_p^T(t)\tilde{\theta}_p(t) + 2\tilde{\theta}(t+1)\phi(t)\nu(t+1) - \phi^T(t)\phi(t)\nu^2(t+1) \quad (10.64)$$

Summing up from $t = 0$ to $t = N$, one gets:

$$\sum_{t=0}^{N} \tilde{\theta}^T(t+1)\phi(t)\nu(t+1) \geq \frac{1}{2}\tilde{\theta}_p^T(N+1)\tilde{\theta}_p(N+1) - \frac{1}{2}\tilde{\theta}_p^T(0)\tilde{\theta}_p(0) \quad (10.65)$$

taking into account that:

$$-\tilde{\theta}(t+1)\phi(t) = y(t+1) - \hat{\theta}^T(t+1)\phi(t) = \nu(t+1) \quad (10.66)$$

one gets:

$$\lim_{t\to\infty}\left\{\frac{1}{2}\tilde{\theta}_p^T(N+1)\tilde{\theta}_p(N+1) + \sum_{t=0}^{N}\nu^2(t+1)\right\} \leq \frac{1}{2}\tilde{\theta}_p^T(0)\tilde{\theta}_p(0) \quad (10.67)$$

from which one concludes that:

$$\|\hat{\theta}_p(t) - \theta\| < M < \infty; \quad \forall t \quad (10.68)$$

and:

$$\lim_{t\to\infty} \nu(t+1) = 0 \quad (10.69)$$

In other terms, we recover the same type of properties as for the case without projection. The key technical fact which allows this result to be obtained is (10.60).

Therefore, in the case of a general type adaptation gain, we should have the property that:

$$\tilde{\theta}_p^T(t+1)F(t+1)^{-1}\tilde{\theta}_p(t+1) \leq \tilde{\theta}^T(t+1)F(t+1)^{-1}\tilde{\theta}(t+1) \quad (10.70)$$

The interpretation of this condition in the case of an adaptation gain $F(t+1)$ is that a transformation of coordinates has to be considered by defining:

$$\hat{\theta}'(t+1) = F(t+1)^{-1/2}\hat{\theta}(t+1) \quad (10.71)$$

where:

$$F(t+1)^{-1} = [F(t+1)^{-1/2}]^T F(t+1)^{-1/2} \quad (10.72)$$

and therefore, the domain $\mathscr{D}$ should also be transformed in $\mathscr{D}'$ such that if:

$$x \in \mathscr{D}; \quad x' = F(t+1)^{-1/2}x \in \mathscr{D}'$$

Therefore, the projection algorithm is, in many cases, as follows:

$$\hat{\theta}(t+1) = \hat{\theta}_p(t) + F(t)\phi(t)\nu(t+1) \tag{10.73}$$

$$F(t+1)^{-1} = \lambda_1(t)F(t)^{-1} + \lambda_2(t)\phi(t)\phi(t)^T$$
$$0 < \lambda_1(t) < 1;\ 0 \le \lambda_2(t) < 2 \tag{10.74}$$

$$\nu(t+1) = y(t+1) - \hat{\theta}^T(t+1)\phi(t) = \frac{y(t+1) - \hat{\theta}_p^T(t)\phi(t)}{1 + \phi^T(t)F(t)\phi(t)} \tag{10.75}$$

$$\hat{\theta}'(t+1) = F(t+1)^{-1/2}\hat{\theta}(t+1) \tag{10.76}$$

$$\hat{\theta}'_p(t+1) = \begin{cases} \hat{\theta}'(t+1) & \text{if } \hat{\theta}'(t+1) \in \mathscr{D}' \\ \perp \text{ proj of } \hat{\theta}'(t+1) \text{ on } \mathscr{D}' & \text{if } \hat{\theta}'(t+1) \notin \mathscr{D}' \end{cases} \tag{10.77}$$

$$\hat{\theta}_p(t+1) = F(t+1)^{1/2}\hat{\theta}'_p(t+1) \tag{10.78}$$

The properties of the PAA with projection are summarized in the following theorem:

**Theorem 10.3** (Projection Algorithm) *For the plant model described by* (10.56) *using the algorithm given by* (10.73) *through* (10.78), *with* $\lambda_1(t) \equiv 1$, *one has*:

$$\lim_{N\to\infty} \sum_{t=0}^{N} \nu^2(t+1) < M_1 < \infty \tag{10.79}$$

$$\lim_{t\to\infty} \nu(t+1) = 0 \tag{10.80}$$

$$[\hat{\theta}_p(t+1) - \theta]^T F(t+1)^{-1}[\theta_p(t+1) - \theta] < M_2 < \infty; \quad \forall t \tag{10.81}$$

*Proof* The proof follows the same logical path as the example considered at the beginning of the section, bearing in mind the intermediate results of Lemma 3.2. One gets:

$$\begin{aligned}
&\tilde{\theta}^T(t+1)F(t+1)^{-1}\tilde{\theta}(t+1) \\
&\quad = \tilde{\theta}_p(t)F(t)\tilde{\theta}_p(t) + 2\tilde{\theta}(t+1)\phi(t)\nu(t+1) \\
&\qquad + \lambda_2(t)[\tilde{\theta}(t+1)\phi(t)]^2 - \phi^T(t)F(t)\phi(t)\nu^2(t+1)
\end{aligned} \tag{10.82}$$

Using the projection defined by (10.75) through (10.77), one gets:

$$\begin{aligned}
&\tilde{\theta}_p^T(t+1)F(t+1)\tilde{\theta}_p(t+1) \\
&\quad \le \tilde{\theta}_p(t)F(t)\tilde{\theta}_p(t) + 2\tilde{\theta}(t+1)\phi(t)\nu(t+1) \\
&\qquad + \lambda_2(t)[\tilde{\theta}(t+1)\phi(t)]^2 - \phi^T(t)F(t)\phi(t)\nu^2(t+1)
\end{aligned} \tag{10.83}$$

Defining $\lambda_2$ as:

$$2 > \lambda_2 > \sup_t \lambda_2(t) \tag{10.84}$$

one gets after summing up from 0 to $N$:

$$\frac{1}{2}\tilde{\theta}_p^T(N+1)F(t+1)^{-1}\tilde{\theta}_p(N+1)$$

$$-\sum_{t=0}^{N}\tilde{\theta}(t+1)\phi(t)\left[\nu(t+1)+\frac{\lambda_2}{2}\tilde{\theta}(t+1)\phi(t)\right]$$

$$\leq \frac{1}{2}\tilde{\theta}_p^T(0)F(0)^{-1}\tilde{\theta}_p(0) \tag{10.85}$$

But, taking into account (10.66), one finally gets:

$$\frac{1}{2}\tilde{\theta}_p^T(N+1)F(t+1)^{-1}\tilde{\theta}_p(t+1)+\sum_{t=0}^{N}\left(1-\frac{\lambda}{2}\right)\nu^2(t+1)$$

$$\leq \frac{1}{2}\tilde{\theta}_p^T(0)F(0)^{-1}\tilde{\theta}_p(0) \tag{10.86}$$

from which (10.79), (10.80) and (10.81) result. □

## 10.6 Data Normalization

In Sect. 10.2, it was indicated that data filtering can reduce the effect of unmodeled dynamics by attenuating the high-frequency components of the input and the output. This will allow a better estimation of a low order model, but will not be enough to guarantee the stability of an adaptive control system in the presence of unmodeled dynamics. An additional modification of the data called "data normalization" has to be considered in such cases. The objective of this modification is to obtain (under certain hypotheses) bounded adaptation error in the presence of possibly unbounded plant input-output data.

Some hypotheses have to be made upon the unmodeled dynamics often considered to be "high-frequency". It is fundamental to assume that the unmodeled dynamics are stable. As a consequence, the magnitude of the unmodeled system response can be bounded in general by the norm of the input-output regressor vector used to describe the modeled part of the system. To be specific, let us consider the following example where the "true" plant model is described by:

$$(1-a_1q^{-1}+a_2q^{-2})y(t+1)=(b_1+b_2q^{-2})u(t) \tag{10.87}$$

Let us assume that we would like to estimate this plant model using a lower order model with $n_A=1$ and $n_B=1$, i.e.:

$$\bar{y}(t+1)=-\bar{a}_1y(t)+\bar{b}_1u(t)=\theta^T\phi(t) \tag{10.88}$$

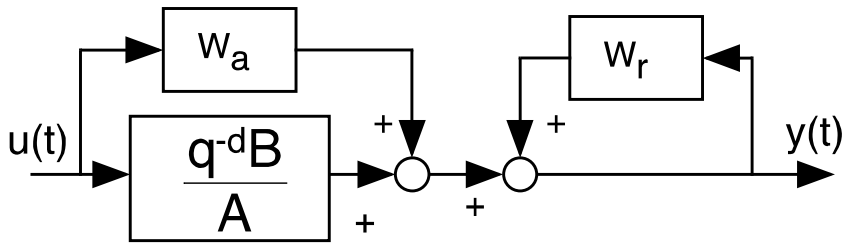

**Fig. 10.4** The reduced order model and the unmodeled dynamics

where:

$$\theta^T = [\bar{a}_1, \bar{b}_1]; \qquad \phi^T(t) = [-y(t), u(t)] \tag{10.89}$$

Therefore, the "true" output of the plant model (10.87) can be written as:

$$y(t+1) = \theta^T \phi(t) + w(t+1) \tag{10.90}$$

where $w(t+1)$ the "unmodelled" response (which can be interpreted also as a disturbance) is given by:

$$\begin{aligned} w(t+1) &= (\bar{a}_1 - a_1)y(t) - a_2 y(t-1) + (b_1 - \bar{b}_1)u(t) + b_2 u(t-1) \\ &= \alpha_1^T \phi(t) + \alpha_2^T \phi(t-1) \\ &= H^T(q^{-1})\phi(t) = q^{-1} H^T(q^{-1})\phi(t+1) \end{aligned} \tag{10.91}$$

where $H(q^{-1})$ is a vector transfer function. One observes that $w(t+1)$ depends upon the input-output signals via $\phi(t)$ and $\phi(t-1)$. In general, one can consider that in the presence of unmodeled dynamics the plant output is given by (10.90) with:

$$\begin{aligned} w(t+1) &= H_u(q^{-1})u(t) + H_y(q^{-1})y(t) = H^T(q^{-1})\phi(t) \\ &= q^{-1} H^T(q^{-1})\phi(t+1) \end{aligned} \tag{10.92}$$

where:

$$H^T(q^{-1}) = [H_u(q^{-1}), H_y(q^{-1})] \tag{10.93}$$

Using a transfer operator form for the reduced order model, $W_a(q^{-1}) = H_u/A$ corresponds to "additive" type uncertainty and $W_r(q^{-1}) = H_y/A$ corresponds to "feedback" type uncertainty on the output as shown in Fig. 10.4 (for more details see Chap. 8).

Similar results are obtained for the various representations of the unmodeled dynamics considered in Chap. 8 (called "uncertainty" in robust control literature and "parasitics" in the adaptive control literature).

Most of the parameter estimation algorithms in the presence of unmodeled dynamics will guarantee that the prediction error will be bounded by $\|\phi(t)\|$ (multiplied by a constant) which means that it will be of the same order of magnitude as the unmodeled response. However, the parameter estimates can become unbounded if $\phi(t)$ grows unbounded (see the discussion in Sect. 10.3). The consequence is that it will not be possible to guarantee the stability of the adaptive control scheme using a lower order estimated plant model. Therefore, one should first build a parameter estimator such that the "equivalent" disturbance caused by the unmodeled dynamics remains bounded, even if $\phi(t)$ tends to become unbounded, and then, use one of the techniques discussed in Sect. 10.4 or 10.5 to assure that the parameter estimates will remain bounded.

In order to obtain a parameter estimator which will have a bounded adaptation error in the presence of unmodeled dynamics, one will use *normalization* of the input-output data. The *normalization* of input-output data is obtained by dividing the data by a quantity $m(t)$. The normalized variables are:

$$\bar{y}(t+1) = \frac{y(t+1)}{m(t)}; \qquad \bar{u}(t) = \frac{u(t)}{m(t)}; \qquad \bar{\phi}(t) = \frac{\phi(t)}{m(t)} \tag{10.94}$$

The corresponding *normalized* output equation will take the form:

$$\bar{y}(t+1) = \theta^T \bar{\phi}(t) + \bar{w}(t+1) \tag{10.95}$$

where:

$$\bar{w}(t) = \frac{w(t+1)}{m(t)} \tag{10.96}$$

The problem is to find a normalizing signal $m(t)$ such that $\bar{w}(t+1)$ be bounded for all $t$. The choice of this signal will depend upon the hypotheses made upon the representation of the unmodeled part of the system output $y(t)$.

Let us return to the representation of $w(t+1)$ given in (10.92). In order to define an appropriate normalizing signal $m(t)$, one should evaluate $|w(t+1)|$. In addition to the transfer operator $H^T(q^{-1})$, one will consider the impulse response of the unmodeled dynamics which allows to write:

$$H^T(z^{-1}) = \sum_0^{\infty} h^T(i) z^{-i} \tag{10.97}$$

where $h(i)$ are the terms of the impulse response. Therefore, one can express $w(t+1)$ as follows:

$$\begin{aligned} w(t+1) &= \sum_{\tau=0}^{t+1} h'^T(t+1-\tau)\phi(\tau) = \sum_{\tau=0}^{t} h^T(t-\tau)\phi(\tau) \\ h'(0) &= 0;\ h'(i) = h(i-1) \end{aligned} \tag{10.98}$$

From (10.98), one gets:

$$\begin{aligned} |w(t+1)| &= \left| \sum_{\tau=0}^{t} h^T(t-\tau)\phi(\tau) \right| \leq \sum_{\tau=0}^{t} |h^T(t-\tau)\phi(\tau)| \\ &= \sum_{\tau=0}^{t} |\mu^{-(t-\tau)} h^T(t-\tau) \mu^{(t-\tau)} \phi(\tau)| \end{aligned}$$

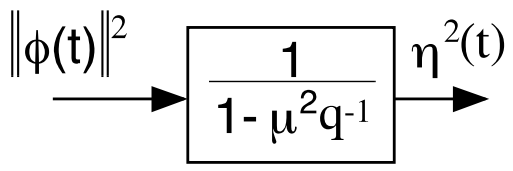

**Fig. 10.5** Interpretation of the signal $\eta(t)$ in (10.100)

$$\leq \left[\sum_{\tau=0}^{t} \|\mu^{-(t-\tau)} h^T(t-\tau)\|^2\right]^{1/2} \left[\sum_{\tau=0}^{t} \|\mu^{(t-\tau)} \phi(\tau)\|^2\right]^{1/2};$$
$$0 < \mu < 1 \tag{10.99}$$

(The Schwarz inequality has been used: $\sum_{\tau=0}^{t} |f(\tau)g(\tau)| \leq [\sum_{\tau=0}^{t} |f(\tau)|^2]^{1/2} [\sum_{\tau=0}^{t} |g(\tau)|^2]^{1/2}$.) Let us define the signal $\eta(t)$ as follows:

$$\eta^2(t) = \sum_{\tau=0}^{t} \|\mu^{(t-\tau)} \phi(\tau)\|^2 = \sum_{\tau=0}^{t} \mu^{2(t-\tau)} \|\phi(\tau)\|^2 \tag{10.100}$$

One sees immediately that $\eta^2(t)$ can be interpreted as the output of a first order discrete-time system whose input is $\|\phi(t)\|^2$ (see Fig. 10.5), i.e.:

$$\eta^2(t) = \mu^2 \eta^2(t-1) + \|\phi(t)\|^2 \tag{10.101}$$

Equation (10.99) can be rewritten as:

$$|w(t+1)| \leq \left(\sum_{\tau=0}^{t} \|\mu^{-(t-\tau)} h^T(t-\tau)\|^2\right)^{1/2} \cdot \eta(t); \quad \eta(t) \geq 0 \tag{10.102}$$

Therefore $\eta(t)$ could be a good candidate for a normalizing signal provided that the multiplying term in the RHS of (10.102) has a finite bound. Using the scaling property of the $z$-transform ($\mathscr{Z}\{\mu^{-i} f(i)\} = F(\mu z)^{-1}$—see Franklin et al. 1990), one has from (10.97):

$$\sum_{0}^{\infty} \mu^{-i} h(i) z^{-i} = H[(\mu z)^{-1}] \tag{10.103}$$

Making the assumption:

- all the poles of $H(z^{-1})$ are inside a circle of radius $\mu(\mu < 1)$, i.e., $H(z^{-1})$ is analytic in $|z| \geq \mu$ and using the Parseval's theorem, one gets:

$$\begin{aligned}
&\sum_{0}^{t+1} \|\mu^{-(t-\tau)} h^T(t-\tau)\|^2 \\
&\quad = \frac{1}{2\pi} \int_{-\pi}^{\pi} H^T(\mu^{-1} e^{-j\omega}) H(\mu e^{j\omega}) d\omega \\
&\quad \leq \max_{-\pi \leq \omega \leq \pi} \|H(\mu^{-1} e^{-j\omega})\|^2 \leq \|H(\mu^{-1} e^{-j\omega})\|_\infty^2
\end{aligned} \tag{10.104}$$

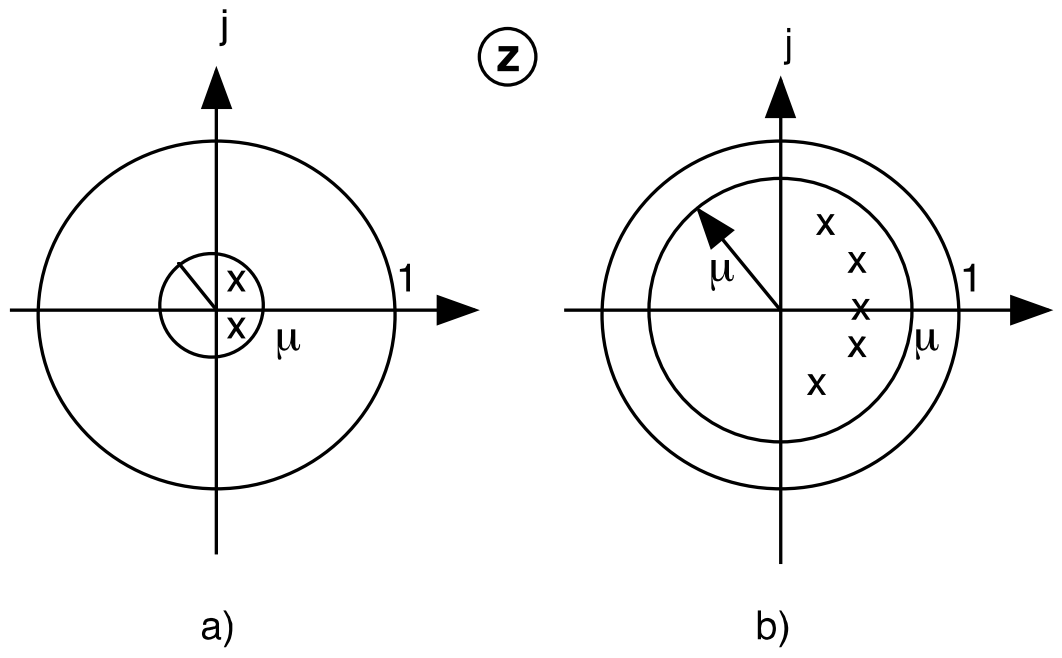

**Fig. 10.6** Fast and slow parasitics

where $\|H(\mu^{-1}e^{-j\omega})\|_\infty$ will be termed the "scaled $\infty$ norm of $H(z^{-1})$". Under the assumption made above this norm exists and is bounded. It follows that $\eta(t)$ given by (10.101) can be used as a normalizing signal (dynamic normalization). The choice of $\mu$ will depend upon the properties of $H(z^{-1})$. For a chosen $\mu$, it is required that the poles of $H(z^{-1})$ lie inside a circle of radius $\mu$. Therefore for "fast" parasitics, $\mu$ can be chosen small but for "slow" parasitics, $\mu$ should have a larger value (see Fig. 10.6). Note also that the value of $\|H(\mu^{-1}z^{-1})\|_\infty$ will depend on how close the poles of $H(z)$ are with respect to the boundary defined by $\mu$. One concludes that in the presence of unmodeled dynamics and under certain conditions, it is possible to bound by *normalization*, the unmodeled response.

The standard assumptions which are made upon $w(t+1)$ are as follows:

- Assumption A:

$$|w(t+1)| \le c\eta(t); \quad 0 < c < \infty \tag{10.105}$$

where:

$$\eta^2(t) = \mu^2\eta^2(t-1) + \|\phi(t)\|^2; \quad 0 < \mu < 1 \tag{10.106}$$

This is exactly the case considered above.

- Assumption B:

$$|w(t+1)| \le c_1 + c_2\|\phi(t)\|; \quad 0 \le c_1,\ c_2 < \infty \tag{10.107}$$

The first term ($c_1$) accounts for bounded disturbances and the second term accounts for the unmodeled response. This term can be considered as an approximation of Assumption A for the case of a fast parasitics ($\mu \approx 0$ in (10.106)).

- Assumption C:

$$\lim_{l\to\infty}\lim_{N\to\infty}\frac{1}{N}\sum_{t=l+1}^{t=l+N}\frac{|w(t+1)|}{\eta(t)} \le c_1 + c_2/N; \quad 0 < c_1,\ c_2 < \infty \tag{10.108}$$

Assumption C is just an "averaging" of Assumption A.

The *normalization* of input-output data is obtained by dividing the data with $m(t)$ where $m(t)$ is given by:

- Assumptions A and C

$$m^2(t) = \mu^2 m^2(t-1) + \max[\|\phi(t)\|^2, 1]; \quad m(0) = 1 \tag{10.109}$$

  The $\max[\|\phi(t)\|, 1]$ has been introduced in order to avoid division by 0 (of course, 1 can be replaced by another constant). Note also that $m^2(t) \geq \eta^2(t)$.
- Assumption B

$$m(t) = 1 + \|\phi(t)\| \tag{10.110}$$

Using the appropriate normalization $\bar{\phi}(t)$ and $\bar{w}(t+1)$ in (10.94) will remain bounded even if $\phi(t)$ tends to become unbounded.

The consequences of the normalization are:

1. the normalized adaptation (prediction) error will be bounded;
2. the unnormalized prediction error will be bounded by $\|\phi(t)\|$;
3. the normalization will slow down the speed of adaptation.

The idea of *normalization* in the context of robustness of direct adaptive control has been mentioned in Egardt (1979) in relation with the stochastic approximation algorithm but has not been used in a proof of robustness with respect to unmodeled dynamics. It is in Praly (1983a) that the concept of *dynamic normalization* is used for the first time as a mean to counteract the effect of unmodeled dynamics. This idea was further developed in Praly (1983b) and the importance of the shifted $H_\infty$ norm of the unmodeled dynamics in defining the dynamic normalization has been emphasized in Ortega et al. (1985). In indirect adaptive control, the normalization can be viewed as a way of extrapolating the robustness results from the case with bounded disturbances (PAA with dead zone for example) (Ortega and Tang 1989; Ioannou and Sun 1996).

### *10.6.1 The Effect of Data Filtering*

When data filtering is used (see Sect. 10.2), the input-output plant model (10.90) becomes:

$$y_f(t+1) = \theta^T \phi_f(t) + w_f(t+1) \tag{10.111}$$

where:

$$y_f(t) = \frac{1}{L(q^{-1})} y(t); \qquad u_f(t) = \frac{1}{L(q^{-1})} u(t) \tag{10.112}$$

$$\phi_f(t) = \frac{1}{L(q^{-1})} \phi(t) \tag{10.113}$$

$$w_f(t+1) = \frac{1}{L(q^{-1})} w(t+1) = [H_u/L, H_y/L]\phi(t) \tag{10.114}$$

As a consequence, an appropriate choice of $L(q^{-1})$ will allow to reduce the value of $\|H(\mu^{-1}z^{-1})\|_\infty$ and respectively of the constant $c$, in (10.105), which will characterize the "size" of uncertainty. This is beneficial for the stability of the adaptive control systems in the presence of unmodeled dynamics.

One has the following results:

**Theorem 10.4** (PAA with Data Normalization)

- *Assume that the plant output can be described by*:

$$y(t+1) = \theta^T \phi(t) + w(t+1) \tag{10.115}$$

*where $w(t+1)$ represents the effects of the unmodeled dynamics or of a disturbance (or both).*

- *Assume that*:

$$|w(t+1)| \le d_1 + d_2\eta(t); \quad 0 < d_1,\ d_2 < \infty \tag{10.116}$$

*where*:

$$\eta^2(t) = \mu^2\eta(t-1) + \|\phi(t)\|^2 \tag{10.117}$$

- *Define the normalized input-output variables*

$$\bar{y}(t+1) = \frac{y(t+1)}{m(t)}; \qquad \bar{u}(t) = \frac{u(t)}{m(t)}; \qquad \bar{\phi}(t) = \frac{\phi(t)}{m(t)} \tag{10.118}$$

*where*:

$$m^2(t) = \mu^2 m^2(t-1) + \max[\|\phi(t)\|^2, 1]; \quad m(0) = 1 \tag{10.119}$$

- *Consider an adjustable model of the form*:

$$\hat{y}^0(t+1) = \hat{\theta}^T(t)\bar{\phi}(t) \tag{10.120}$$

$$\hat{y}(t+1) = \hat{\theta}^T(t+1)\bar{\phi}(t) \tag{10.121}$$

- *Consider the PAA of the form*:

$$\hat{\theta}(t+1) = \hat{\theta}(t) + F(t)\bar{\phi}(t)\bar{\nu}(t+1) \tag{10.122}$$

$$F(t+1)^{-1} = \lambda_1(t)F(t)^{-1} + \lambda_2(t)\bar{\phi}(t)\bar{\phi}(t)^T$$

$$\lambda_1(t) \equiv 1;\ 2 > \lambda_2(t) \ge 0 \tag{10.123}$$

$$\bar{\nu}(t+1) = \frac{\bar{y}(t+1) - \hat{y}^0(t+1)}{1 + \bar{\phi}^T(t)F(t)\bar{\phi}(t)} \tag{10.124}$$

*Then*:

$$\sum_{t=0}^{N} \bar{\nu}^2(t+1) \leq c_1 + c_2 \sum_{t=0}^{N} \bar{w}^2(t+1); \quad 0 < c_1, c_2 < \infty \tag{10.125}$$

*where*:

$$\bar{w}(t+1) = \frac{w(t+1)}{m(t)} \tag{10.126}$$

*and*

$$\frac{1}{N} \sum_{t=0}^{N} \bar{\nu}^2(t+1) \leq c_2 \delta^2 + c_1/N \tag{10.127}$$

*where*:

$$\delta = d_2 + d_1 \tag{10.128}$$

*Proof* The proof of Theorem 10.4 is a straightforward application of Theorem 10.1. By using normalized variables, the normalized output equation takes the form:

$$\bar{y}(t+1) = \theta^T \bar{\phi}(t) + \bar{w}(t+1) \tag{10.129}$$

where now $\bar{w}(t+1)$ is strictly bounded. Applying now the results of Theorem 10.1 to this normalized plant model, one gets (10.125) from (10.20). On the other hand, (10.127) is a direct consequence of (10.125) and (10.116). □

**Corollary 10.1** *Under the same hypotheses as Theorem* 10.4, *if one defines the unnormalized a posteriori prediction error as*:

$$\varepsilon(t+1) = y(t+1) - \hat{\theta}^T(t+1)\phi(t) \tag{10.130}$$

*one has*:

$$\varepsilon^2(t+1) \leq \bar{\nu}^2(t+1) m^2(t) \tag{10.131}$$

*Proof of Corollary 10.1* From the definition of $\bar{\varepsilon}, \bar{y}, \bar{\phi}(t)$ one gets:

$$\varepsilon(t+1) = \bar{y}(t+1)m(t) - \hat{\theta}^T(t+1)\bar{\phi}(t)m(t) = \bar{\nu}(t+1)m(t) \tag{10.132}$$

and (10.131) results. □

### 10.6.2 Alternative Implementation of Data Normalization

Consider the parameter adaptation given in (10.122) through (10.124). Taking into account the definition of the "normalized variables", one has:

$$\begin{aligned}\hat{\theta}(t+1) &= \hat{\theta}(t) + \frac{F(t)\bar{\phi}(t)[\bar{y}(t+1) - \hat{\theta}^T(t)\bar{\phi}(t)]}{1+\bar{\phi}^T(t)F(t)\bar{\phi}(t)} \\ &= \hat{\theta}(t) + \frac{1}{m^2(t)}\frac{F(t)\phi(t)[y(t+1) - \hat{\theta}^T(t)\phi(t)]}{1+\frac{1}{m^2(t)}\phi^T(t)F(t)\phi(t)} \\ &= \hat{\theta}(t) + \frac{F(t)\phi(t)\varepsilon^0(t+1)}{m^2(t)+\phi^T(t)F(t)\phi(t)} \qquad (10.133)\end{aligned}$$

On the other hand:

$$\begin{aligned}F(t+1)^{-1} &= \lambda_1(t)F(t)^{-1} + \lambda_2(t)\bar{\phi}(t)\bar{\phi}^T(t) \\ &= \lambda_1(t)F(t)^{-1} + \frac{\lambda_2(t)}{m^2(t)}\phi(t)\phi^T(t) \qquad (10.134)\end{aligned}$$

In other terms, there is an alternative way to implement *data normalization*. Instead of using (10.94), one can slightly modify the unnormalized parameter adaptation algorithm given by:

$$\hat{\theta}(t+1) = \hat{\theta}(t) + \frac{F(t)\phi(t)[y(t+1) - \hat{\theta}^T(t)\phi(t)]}{1+\phi^T(t)F(t)\phi(t)} \qquad (10.135)$$

$$F(t+1)^{-1} = \lambda_1(t)F(t)^{-1} + \lambda_2(t)\phi(t)\phi^T(t) \qquad (10.136)$$

To get the *data normalization*, one replaces in (10.135) 1 by $m^2(t)$ and in (10.136), one replaces $\lambda_2(t)$ by $\lambda_2(t)/m^2(t)$.

### 10.6.3 Combining Data Normalization with Dead Zone

Replacing the PAA of (10.121) through (10.123) by the PAA given in (10.36) through (10.39), with $\phi(t)$ and $\nu(t+1)$ replaced by their normalized counterpart $\bar{\phi}(t)$ and $\bar{\nu}(t+1)$, and with $\Delta$ replaced by a time-varying bound, one has the following result.

**Theorem 10.5** *Under the same hypotheses as in Theorem 10.4, but using the PAA with dead zone*:

$$\hat{\theta}(t+1) = \hat{\theta}(t) + \alpha(t)F(t)\bar{\phi}(t)\bar{\nu}(t+1) \qquad (10.137)$$

$$F(t+1)^{-1} = F(t)^{-1} + \alpha(t)\bar{\phi}(t)\bar{\phi}^T(t); \quad F(0) > 0 \qquad (10.138)$$

$$\bar{\nu}(t+1) = \frac{\bar{y}(t+1) - \hat{y}^0(t+1)}{1 + \bar{\phi}^T(t)F(t)\bar{\phi}(t)} \tag{10.139}$$

$$\alpha(t) = \begin{cases} 1 & |\bar{\nu}(t+1)| > \bar{\delta}(t+1) \\ 0 & |\bar{\nu}(t+1)| \leq \bar{\delta}(t+1) \end{cases} \tag{10.140}$$

*where* $\bar{\delta}(t)$ *is given by*:

$$\bar{\delta}^2(t+1) = \delta^2(t+1) + \sigma\delta(t+1); \quad \delta(t) > 0; \ \sigma > 0 \tag{10.141}$$

*and*

$$\delta(t+1) = d_2 + \frac{d_1}{m(t)} \geq |\bar{w}(t+1)| \tag{10.142}$$

*one has*:

$$\text{(i)} \quad \sum_{t=0}^{\infty} \alpha(t)[\bar{\nu}^2(t+1) - \bar{\delta}^2(t+1)] < \infty \tag{10.143}$$

$$\text{(ii)} \quad \lim_{t\to\infty} \alpha(t)[\bar{\nu}^2(t+1) - \bar{\delta}^2(t+1)] = 0 \tag{10.144}$$

$$\text{(iii)} \quad \lim_{t\to\infty} \sup |\bar{\nu}(t+1)| \leq \bar{\delta}(t+1) \tag{10.145}$$

$$\text{(iv)} \quad \tilde{\theta}^T(t)F(t)^{-1}\tilde{\theta}(t) \leq \tilde{\theta}^T(0)F(0)^{-1}\tilde{\theta}(0) < \infty \tag{10.146}$$

$$\text{(v)} \quad \lim_{t\to\infty} \sup |\varepsilon(t+1)| \leq \bar{\delta}(t+1)m(t) \tag{10.147}$$

*where* $\varepsilon(t+1)$ *is given by* (10.128)

$$\text{(vi)} \quad F(t), \tilde{\theta}(t) \textit{ converge} \tag{10.148}$$

The interest of this result is that it assures that even if $\|\phi(t)\|$ becomes unbounded, $\|\hat{\theta}(t)\|$ will be bounded and the unnormalized a posteriori prediction error is bounded by $\|\phi(t)\|$ (i.e., does not grow faster than $\phi(t)$) and that $\tilde{\theta}(t)$ (respectively $\hat{\theta}(t)$) converges. The convergence of $\tilde{\theta}(t)$ is a direct consequence of using $\sigma > 0$ in the definition of $\bar{\delta}(t)$.

*Proof* The proof is similar to that of Theorem 10.2 for the results (10.143) to (10.146). Equation (10.147) is a direct consequence of the normalization (see Corollary 10.1). The specific proof concerns the convergence of $F(t)$ and $\tilde{\theta}(t)$ toward fixed values. Convergence of $F(t)$ is obtained as follows. From (10.138) one has for any constant vector $v$:

$$v^T F(t+1)v \leq v^T F(t)v \tag{10.149}$$

since:

$$F(t+1) = F(t) - \frac{\alpha(t)\lambda_2(t)F(t)\phi(t)\phi^T(t)F(t)}{1 + \alpha(t)\lambda_2(t)\phi^T(t)F(t)\phi(t)} \tag{10.150}$$

Note also that $F(t)$ and $F(t)^{-1}$ are positive definite matrices. From (10.149), it follows that $v^T F(t)v \geq 0$ and it is non-increasing. We conclude that $v^T F(t)v$ converges for any constant vector $v$. Consider the $f_{i,j}(t)$ element of $F(t)$ which can be written as: $e_i^T F(t)e_j$ where $e_i$ is a vector having all its elements equal to 0 except for the $i$th component which is equal to 1. We now express $e_i^T F(t)e_j$ as a combination of quadratic terms as follows:

$$e_i^T F(t)e_j = [(e_i + e_j)^T F(t)(e_i + e_j) - e_i^T F(t)e_i - e_j^T F(t)e_j]/2 \tag{10.151}$$

All the terms in the RHS of (10.151) are quadratic terms of the form $v^T F(t)v$ and therefore converge. It follows that $F_{ij}(t)$ converge for all $i$ and $j$. From (10.49), one has (taking into account the use of normalized quantities):

$$\begin{aligned}\tilde{\theta}^T(N+1)F(N+1)^{-1}\tilde{\theta}(N+1) &\leq \tilde{\theta}^T(0)F(0)^{-1}\tilde{\theta}(0)\\ &\quad + \sum_{t=0}^{N}\alpha(t)[\bar{w}^2(t+1) - \bar{\nu}^2(t+1)]\end{aligned} \tag{10.152}$$

But from (10.140) and (10.141) it results:

$$\begin{aligned}\alpha(t)[\bar{w}^2(t+1) - \bar{\nu}^2(t+1)] &\leq \alpha(t)[\bar{\delta}^2(t+1) - \sigma\delta(t+1) - \bar{\nu}^2(t+1)]\\ &\leq -\alpha(t)\sigma\delta(t+1)\end{aligned} \tag{10.153}$$

Introducing this result in (10.152), one gets:

$$\begin{aligned}0 &\leq \tilde{\theta}^T(N+1)F(N+1)^{-1}\tilde{\theta}(N+1)\\ &\leq \tilde{\theta}^T(0)F(0)^{-1}\tilde{\theta}(0) - \sigma\sum_{t=0}^{N}\alpha(t)\delta(t+1)\end{aligned} \tag{10.154}$$

and therefore:

$$\sum_{t=0}^{N}\alpha(t)\delta(t+1) \leq \frac{1}{\sigma}\tilde{\theta}^T(0)F(0)^{-1}\tilde{\theta}(0) \tag{10.155}$$

This result will be necessary for establishing the convergence of $\tilde{\theta}(t)$ toward a constant value. To do this, one considers:

$$\begin{aligned}F(t+1)^{-1}\hat{\theta}(t+1) &= [F(t)^{-1} + \alpha(t)\bar{\phi}(t)\bar{\phi}^T(t)]\\ &\quad \times [\tilde{\theta}(t) + \alpha(t)F(t+1)\bar{\phi}(t)\bar{\nu}^0(t+1)]\end{aligned} \tag{10.156}$$

where:

$$\bar{\nu}^0(t+1) = \bar{y}(t+1) - \hat{\theta}^T(t)\bar{\phi}(t) = -\tilde{\theta}^T(t)\phi(t) + \bar{w}(t+1) \tag{10.157}$$

Introducing (10.157) in (10.156), one gets:

$$\begin{aligned} F^{-1}(t+1)\tilde{\theta}(t+1) &= F^{-1}(t)\tilde{\theta}(t) + \alpha(t)\bar{\phi}(t)\bar{w}(t+1) \\ &= F^{-1}(0)\tilde{\theta}(0) + \sum_{t=0}^{N} \alpha(t)\bar{\phi}(t)\bar{w}(t+1) \end{aligned} \tag{10.158}$$

Since $F(t)$ converges in order to conclude that $\tilde{\theta}(t)$ converges toward a constant value, one has to show that the second term in the RHS of (10.158) converge. To do this, we will use the following result: If $\| \sum_{t=0}^{\infty} a(i) \|$ is bounded for any sequence $a(i)$, then $\sum_{t=0}^{\infty} a(i)$ has a limit (Knopp 1956). In our case:

$$\begin{aligned} \left\| \sum_{t=0}^{N} \alpha(t)\bar{\phi}(t)\bar{w}(t+1) \right\| &\le \sum_{t=0}^{N} \alpha(t) \|\bar{\phi}(t)\| \, \|\bar{w}(t+1)\| \\ &\le \sum_{t=0}^{N} \alpha(t) |\bar{w}(t+1)| \le \sum_{t=0}^{N} \alpha(t)\delta(t+1) \end{aligned} \tag{10.159}$$

since $\|\bar{\phi}(t)\| \le 1$ ($m^2(t) \ge \|\phi(t)\|^2$). But from (10.155), one has:

$$\lim_{N\to\infty} \sum_{t=0}^{N} \alpha(t)\delta(t+1) \le \frac{1}{\sigma} \tilde{\theta}(0) F^{-1}(0) \tilde{\theta}(0) \tag{10.160}$$

and therefore $\lim_{N\to\infty} \sum_{t=0}^{N} \alpha(t)\bar{\phi}(t)\bar{w}(t+1)$ exists. Therefore, $\tilde{\theta}(t)$ (and $\hat{\theta}(t)$) converges. □

## 10.7 A Robust Parameter Estimation Scheme

Figure 10.7 illustrates how the various modifications of the basic parameter adaptation algorithm discussed in this chapter can be integrated in the open-loop type parameter estimation schemes used for adaptive control. When using closed-loop output error parameter estimation schemes, the filtering effect is automatically present as a consequence of the structure of the adjustable predictor. Therefore, in general, the input/output data filtering stage indicated in Fig. 10.7 is no more necessary.

## 10.8 Concluding Remarks

1. Robustification of parameter adaptation algorithms is necessary when:
   - the true plant model and the estimated plant model do not have the same structure;
   - the disturbances are characterized only by an upper bound;
   - enhancement of the input-output data spectrum in a certain frequency region is required.

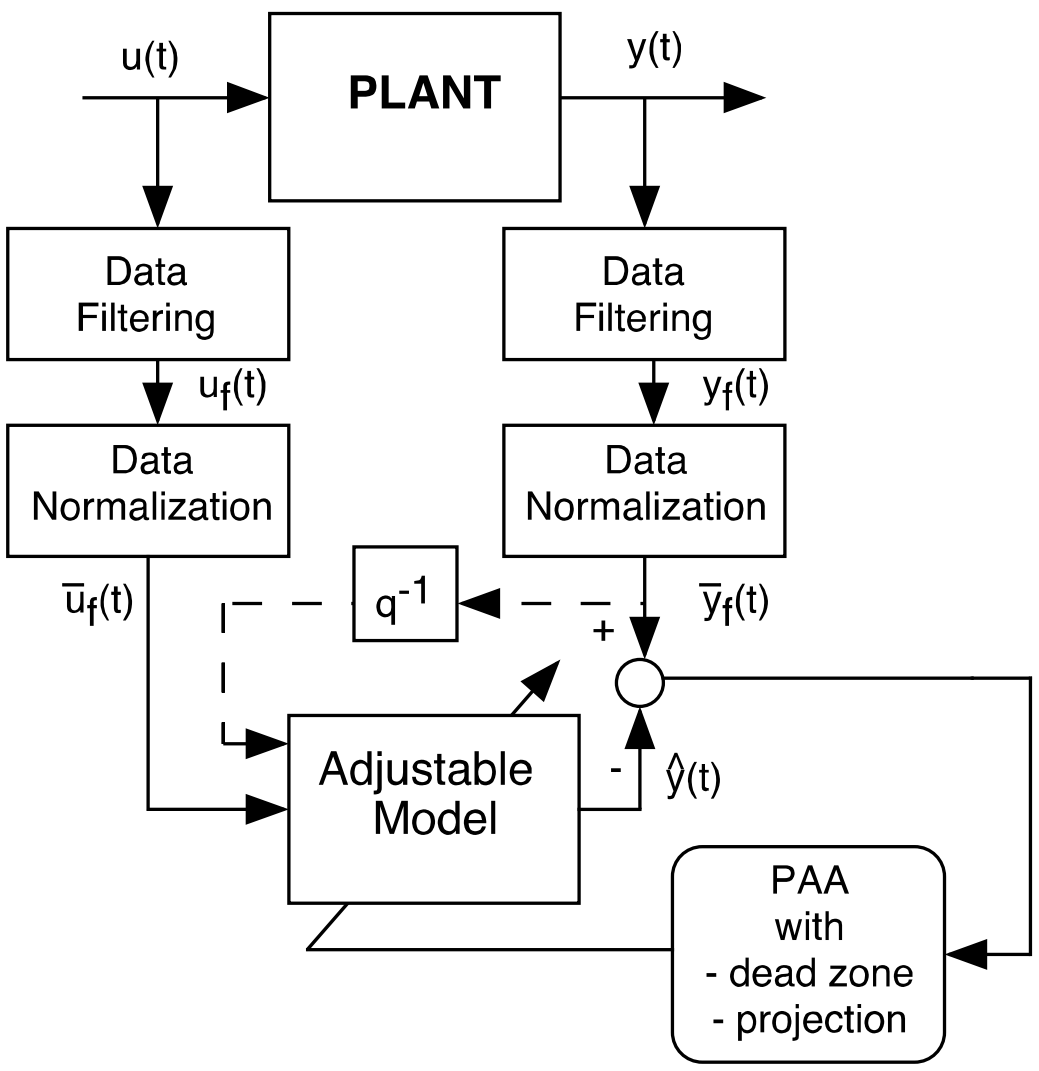

**Fig. 10.7** A robust parameter estimation scheme

2. One of the major tasks of robustification is to assure bounded parameter estimates in the presence of disturbances and possible unbounded input-output data.
3. The main modifications used for robustification of parameter adaptation algorithms are:

   - filtering of input-output data;
   - data normalization;
   - PAA with dead zone;
   - PAA with projection;
   - PAA without integrator effect.

## 10.9 Problems

**10.1** (Dugard and Goodwin 1985) Using a PAA with projection, show that the output error with adjustable compensator (Sect. 5.5) assures:

$$\lim_{t\to\infty} \varepsilon^0(t+1) = \lim_{t\to\infty} \varepsilon(t+1) = 0$$

**10.2** Consider the identification of the following plant + disturbance model:

$$y(t) = \frac{q^{-d}B(q^{-1})}{A(q^{-1})}u(t) + v(t)$$

where $u(t)$ and $v(t)$ are mean square bounded, i.e.:

$$\lim_{N\to\infty}\sum_{t=1}^{N}u^2(t) \le \alpha^2 N + k_u; \quad 0 < \alpha^2;\ k_u < \infty$$

$$\lim_{N\to\infty}\sum_{t=1}^{N}v(t) \le \beta^2 N + k_v; \quad 0 < \beta^2;\ k_v < \infty$$

1. Show that for the output error with fixed compensator algorithm (Sect. 5.5) the a posteriori output error $\varepsilon(t+1)$, the a posteriori adaptation error $\nu(t+1)$, the predicted output $\hat{y}(t+1)$ and the observation vector $\phi(t)$ are mean square bounded provided that:

$$H(z^{-1}) = \frac{D(z^{-1})}{A(z^{-1})} - \frac{\lambda_2}{2}; \qquad \sup_t \lambda_2(t) \le \lambda_2 < 2$$

   is strictly positive real.
2. Do a similar analysis for the other algorithms given in Table 5.1.

**10.3** Consider the identification in closed loop of the plant model used in Problem 10.2. The closed loop uses a RST controller which is stable and stabilizes the closed loop. In this case, it is assumed that the disturbance $v(t)$ and the external excitation $r(t)$ are mean square bounded.

1. Show that using the closed-loop output error algorithm (Sect. 9.2), the a posteriori adaptation error $\hat{y}(t+1)$ and the observation vector $\phi(t)$ are bounded provided that:

$$H(z^{-1}) = \frac{S(z^{-1})}{P(z^{-1})} - \frac{\lambda_2}{2}; \qquad \sup_t \lambda_2(t) \le \lambda_2 < 2$$

   is strictly positive real.
2. Do a similar analysis for the other algorithms given in Table 9.1.

# Chapter 11
# Direct Adaptive Control

## 11.1 Introduction

*Direct adaptive control* covers those schemes in which the parameters of the controller are directly updated from a signal error (adaptation error) reflecting the difference between attained and desired performance. Direct adaptive control schemes are generally obtained in two ways.

1. Define an equation for a signal error (adaptation error) which is a function of the difference between the tuned controller parameters and the current controller parameters. Use this adaptation error to generate a PAA for the controller parameters.
2. Use an indirect adaptive control approach with an adaptive predictor of the plant output reparameterized in terms of the controller parameters and force the output of the adaptive predictor to follow exactly the desired trajectory.

The second approach allows the direct adaptation of the parameters of the controller without solving an intermediate "design equation". As it will be shown the prediction error used in the PAA is in fact an image of the difference between the nominal and the attained performance because of closed-loop operation. Furthermore, the resulting schemes are governed by the same equations as those obtained by the first approach.

Although direct adaptive control is very appealing, it cannot be used for all types of plant model and control strategies. In fact, the situations where direct adaptive control can be used are limited. The typical use of this approach covers:

1. adaptive tracking and regulation with independent objectives (Sect. 11.2);
2. adaptive tracking and regulation with weighted input (Sect. 11.3);
3. adaptive minimum variance tracking and regulation (Sect. 11.4).

As will be shown, the basic hypothesis on the plant model is that, for any possible values of the parameters either

- *the finite zeros of the plant model are inside the unit circle* (this hypothesis is used for schemes 1 and 3),

or

- *there exists a linear combination of the denominator and numerator of the plant transfer function which is an asymptotically stable polynomial* (this hypothesis is used by scheme 2).

For other direct adaptive control schemes dedicated to plant models with non-necessarily stable zeros see M'Saad et al. (1985).

It also has to be mentioned that even if the zeros of the plant model are asymptotically stable, it is not possible (or it becomes very complicated) to develop direct adaptive control schemes for pole placement, linear quadratic control or generalized predictive control. The reason is that it is not possible to obtain an adaptation error equation which is linear in the difference between the nominal and estimated controller parameters. For various attempts to develop direct adaptive control schemes for pole placement see Eliott (1980), Lozano and Landau (1982), Åström (1980).

At first glance, the derivation of PAA for direct adaptive control schemes looks quite close to the one used in recursive identification and adaptive prediction. The PAA structures considered in Chaps. 3 and 4 will be used as well as the synthesis and analysis tools presented in Chap. 3 (for deterministic environment) and Chap. 4 (for stochastic environment). The objective will be that the resulting adaptive control schemes achieve asymptotically the desired performance for the case of known parameters. However, the analysis of adaptive control schemes is much more involved than the analysis of recursive identification and adaptive predictor schemes since in this case the plant input (the control) will depend upon the parameters' estimates. Showing that the plant input and output remain bounded during the adaptation transient is one of the key issues of the analysis. In the analysis of the various schemes, it will be assumed that the unknown parameters are either constant or piece-wise constant and that their variations from one value to another is either slow or is subject to sparse step changes. In the stochastic case, in order to assure the convergence of the adaptive controller toward the optimal one, it is necessary to assume that the parameters are unknown but constant and therefore a decreasing adaptation gain will be used. The robustness aspects related to the violation of the basic hypotheses will be examined in detail in Sect. 11.5.

## 11.2 Adaptive Tracking and Regulation with Independent Objectives

### *11.2.1 Basic Design*

The tracking and regulation with independent objectives for the case of known plant model parameters has been discussed in detail in Sect. 7.4. For the development of a direct adaptive control scheme, the time-domain design discussed in Sect. 7.4.2 is useful.

The plant model (with unknown parameters) is assumed to be described by:

$$\begin{aligned} A(q^{-1})y(t) &= q^{-d}B(q^{-1})u(t) \\ &= q^{-d-1}B^*(q^{-1})u(t) \end{aligned} \tag{11.1}$$

where $u(t)$ and $y(t)$ are the input and the output of the plant respectively.

In the case of known parameters, the objective is to find a control

$$u(t) = f_u[y(t), y(t-1), \ldots, u(t-1), u(t-2), \ldots]$$

such that

$$\varepsilon^0(t+d+1) = P(q^{-1})[y(t+d+1) - y^*(t+d+1)] = 0 \tag{11.2}$$

where $P(q^{-1})$ is an asymptotically stable polynomial defined by the designer. For the case of unknown plant model parameters the objective will be to find a control

$$u(t) = f_u[\hat{\theta}_c(t), y(t), y(t-1), \ldots, u(t-1), u(t-2), \ldots]$$

with $\{u(t)\}$ and $\{y(t)\}$ bounded, such that:

$$\lim_{t\to\infty} \varepsilon^0(t+d+1) = 0 \tag{11.3}$$

where $\hat{\theta}_c(t)$ denotes the current controller parameters estimates. Observe that $\varepsilon^0(t+d+1)$ is a measure of the discrepancy between the desired and achieved performance, and as such is a potential candidate to be the adaptation error. This quantity can be generated at each sample ($y^*(t+d+1)$ is known $d+1$ steps ahead).

The next step toward the design of an adaptive control scheme is to replace the fixed controller given in (7.107) or (7.111) by an adjustable controller. Since the performance error for a certain value of plant model parameters will be caused by the misalignment of the controller parameters values, it is natural to consider an adjustable controller which has the same structure as in the linear case with known plant parameters and where the fixed parameters will be replaced by adjustable ones. It will be the task of the adaptation algorithm to drive the controller parameters towards the values assuring the satisfaction of (11.3). Therefore, the control law in the adaptive case will be chosen as:

$$\hat{S}(t, q^{-1})u(t) + \hat{R}(t, q^{-1})y(t) = P(q^{-1})y^*(t+d+1) \tag{11.4}$$

where:

$$\hat{S}(t, q^{-1}) = \hat{s}_0(t) + \hat{s}_1(t)q^{-1} + \cdots + \hat{s}_{nS}(t)q^{-nS} = 1 + q^{-1}\hat{S}^*(t, q^{-1}) \tag{11.5}$$

$$\hat{R}(t, q^{-1}) = \hat{r}_0(t) + \hat{r}_1(t)q^{-1} + \cdots + \hat{r}_{n_R}(t)q^{-n_R} \tag{11.6}$$

which can be written alternatively as (see also (7.111)):

$$\hat{\theta}_C^T(t)\phi_C(t) = P(q^{-1})y^*(t+d+1) \tag{11.7}$$

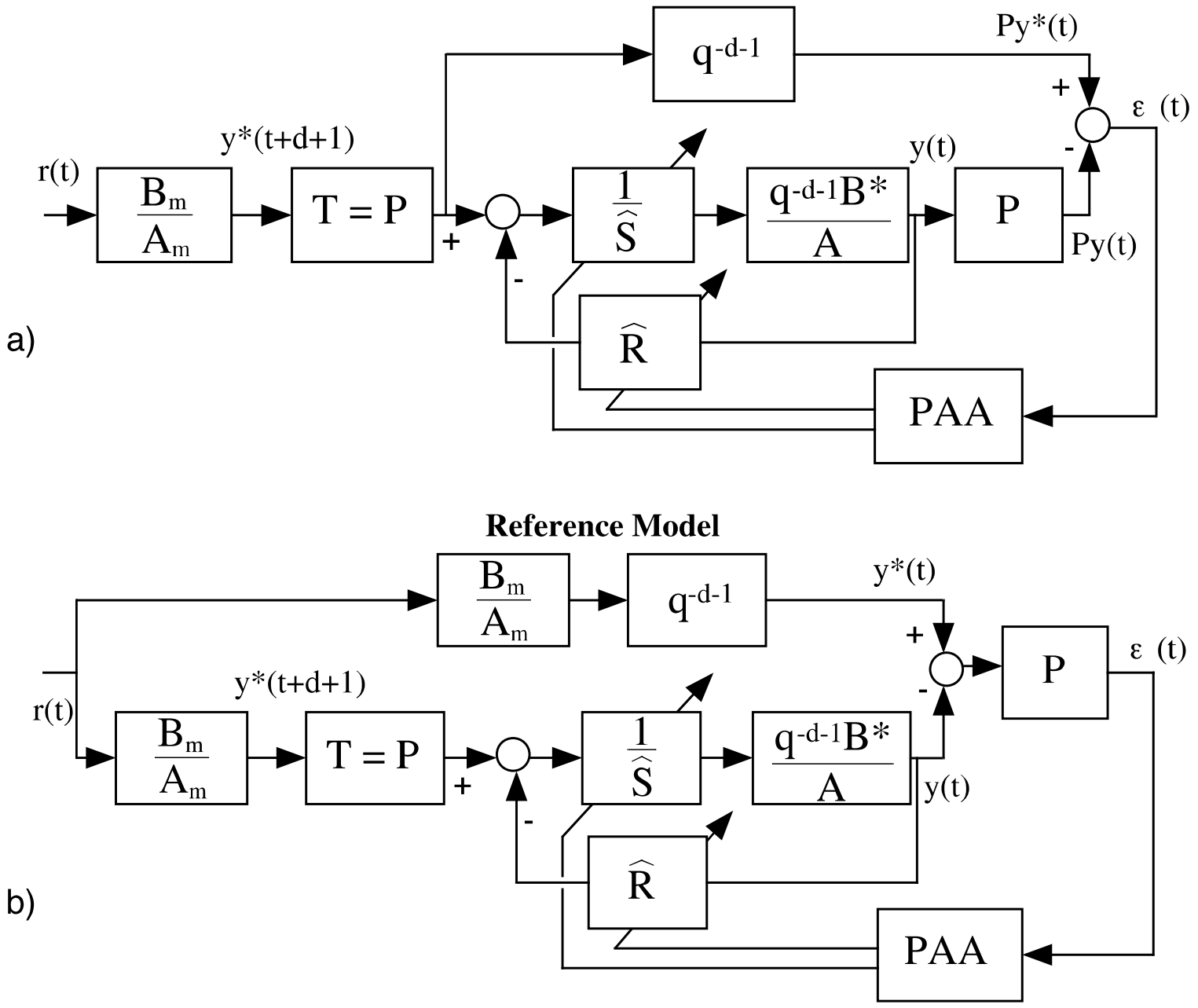

**Fig. 11.1** Adaptive tracking and regulation with independent objective; (**a**) basic configuration, (**b**) a model reference adaptive control interpretation

where:

$$\hat{\theta}_C^T(t) = [\hat{s}_0(t), \ldots, \hat{s}_{n_S}(t), \hat{r}_0(t), \ldots, \hat{r}_{n_R}(t)]; \quad \hat{s}_0(t) = \hat{b}_1(t) \tag{11.8}$$

$$\phi_C^T(t) = [u(t), \ldots, u(t-n_S), y(t), \ldots, y(t-n_R)] \tag{11.9}$$

and the effective control input will be computed as:

$$u(t) = \frac{1}{\hat{s}_0(t)}[P(q^{-1})y^*(t+d+1) - \hat{S}^*(t,q^{-1})u(t-1) - \hat{R}(t,q^{-1})y(t)] \tag{11.10}$$

The choice made for the adaptation error (11.2) and the structure of the adjustable controller leads to the block diagram of the adaptive control system shown in Fig. 11.1a. Figure 11.1a allows a rapprochement with model reference adaptive control (MRAC) to be made, where the output of the reference model generates the quantity $P(q^{-1})y^*(t)$ which is compared to the filtered plant output $P(q^{-1})y(t)$. Alternatively, one can consider (as shown in Fig. 11.1b) that the adaptation error is defined by the difference between the desired trajectory $y^*(t)$ generated by the tracking reference model and the plant output $y(t)$ filtered by $P(q^{-1})$. As explained in Landau and Lozano (1981), $P(q^{-1})$ can be interpreted as a "reference model" for regulation.

Since we have selected a candidate for the adaptation error and a structure for the adjustable controller, the next step will be to give an expression for $\varepsilon^0(t+d+1)$ as a function of the controller parameters misalignment. This will allow one to see to what extent $\varepsilon^0(t+d+1)$ can be considered as an adaptation error to be used in a PAA of the forms discussed in Chap. 3. We note first that for the case of known parameters where the controller parameters are computed by solving the polynomial equation (7.91), one has $\varepsilon^0(t+d+1) \equiv 0$ and using (7.111), one can conclude that the filtered predicted plant output is given by:

$$\theta_C^T \phi_C(t) = P(q^{-1}) y(t+d+1) \tag{11.11}$$

since in the case of known parameters:

$$P(q^{-1}) y^*(t+d+1) = P(q^{-1}) y(t+d+1); \quad \forall t > 0 \tag{11.12}$$

where:

$$\theta_C^T = [s_0, \ldots, s_{n_S}, r_0, \ldots, r_{n_R}] \tag{11.13}$$

defines the parameter vector of the tuned controller (unknown). Subtracting (11.7) which contains the parameters of the adjustable controller from (11.11), one obtains:

$$\begin{aligned} \varepsilon^0(t+d+1) &= P(q^{-1}) y(t+d+1) - P(q^{-1}) y^*(t+d+1) \\ &= [\theta_C - \hat{\theta}_C(t)]^T \phi_C(t) \end{aligned} \tag{11.14}$$

which has the desired form (linear in the parameter error).

For the PAA synthesis, we will use Theorem 3.2.

*The case* $d = 0$: One sees immediately that for $d = 0$, $\varepsilon^0(t+d+1) = \varepsilon^0(t+1)$ can be interpreted as an a priori adaptation error and one can associate an a posteriori adaptation error governed by:

$$\nu(t+1) = \varepsilon(t+1) = [\theta_C - \hat{\theta}_C(t+1)]^T \phi_C(t) \tag{11.15}$$

and Theorem 3.2 can simply be used to derive the PAA assuring

$$\lim_{t \to \infty} \varepsilon(t+1) = 0 \tag{11.16}$$

*The case* $d > 0$: As in the case of $j$-steps ahead adaptive predictors (see Sect. 6.2.1), one will associate to (11.14) an a posteriori adaptation error equation:

$$\varepsilon(t+d+1) = [\theta_C - \hat{\theta}_C(t+d+1)]^T \phi_C(t) \tag{11.17}$$

Equation (11.17) has the form of the a posteriori adaptation error equation considered in Theorem 3.2. Applying Theorem 3.2, the PAA to be used for assuring:

$$\lim_{t \to \infty} \varepsilon(t+d+1) = 0 \tag{11.18}$$

is:

$$\hat{\theta}_C(t+d+1) = \hat{\theta}_C(t+d) + F(t) \phi_C(t) \varepsilon(t+d+1) \tag{11.19}$$

$$F(t+1)^{-1} = \lambda_1(t)F(t)^{-1} + \lambda_2(t)\phi_C(t)\phi_C^T(t)$$
$$0 < \lambda_1(t) \le 1;\ \ 0 \le \lambda_2(t) < 2;\ \ F(0) > 0 \tag{11.20}$$

Notice that because of the form of (11.15) and (11.17), the positive real condition of Theorem 3.2 is automatically satisfied. To make the above PAA implementable, we have to now give an expression of $\varepsilon(t+d+1)$ in terms of the a priori adaptation error $\varepsilon^0(t+d+1)$ and the parameters $\hat{\theta}(t+i)$ up to and including $i=d$. As with the direct adaptive prediction, (6.18), one can rewrite (11.17) as:[1]

$$\begin{aligned}\varepsilon(t+d+1) &= \varepsilon^0(t+d+1) + [\hat{\theta}_C(t) - \hat{\theta}_C(t+d+1)]^T\phi_C(t)\\ &= \varepsilon^0(t+d+1) - \phi_C^T(t)F(t)\phi_C(t)\varepsilon(t+d+1)\\ &\quad + [\hat{\theta}_C(t) - \hat{\theta}_C(t+d)]^T\phi_C(t)\end{aligned} \tag{11.21}$$

from which one obtains:

$$\varepsilon(t+d+1) = \frac{\varepsilon^0(t+d+1) - [\hat{\theta}_C(t+d) - \hat{\theta}_C(t)]^T\phi_C(t)}{1+\phi_C^T(t)F(t)\phi_C(t)} \tag{11.22}$$

which can also be expressed as:

$$\varepsilon(t+d+1) = \frac{Py(t+d+1) - \hat{\theta}_C^T(t+d)\phi_C(t)}{1+\phi_C^T(t)F(t)\phi_C(t)} \tag{11.23}$$

**Analysis**

The fact that the PAA of (11.19), (11.20) and (11.22) assures that (11.18) holds, does not guarantee that the objective of the adaptive control scheme defined by (11.3) will be achieved (i.e., the a priori adaptation error should goes to zero) and that $\{u(t)\}$ and $\{y(t)\}$ will be bounded. This is related to the boundedness of $\phi_C(t)$. Effectively from (11.22), one can see that $\varepsilon(t+d+1)$ can go to zero without $\varepsilon^0(t+d+1)$ going to zero if $\phi_C(t)$ becomes unbounded. Since $\phi_C(t)$ contains the input and the output of the plant, it is necessary to show that the PAA achieves the objective of (11.3) while assuring that $\phi_C(t)$ remains bounded for all $t$.

An important preliminary remark is that the plant model (11.1) is a difference equation with constant coefficients and therefore $y(t+1)$ cannot become unbounded for finite $t$ if $u(t)$ is bounded. Therefore if $u(t)$ is bounded, $y(t)$ and respectively $\phi_C(t)$ can become unbounded only asymptotically.

Bearing in mind the form of (11.10), it is also clear that in order to avoid $u(t)$ becoming unbounded the adjustable parameters must be bounded and $\hat{s}_0(t) =$

[1]The term $[\hat{\theta}_C(t) - \hat{\theta}_C(t+d+1)]^T\phi_C(t)$ is sometimes termed "auxiliary error" and $\varepsilon(t+d+1)$ is sometimes termed "augmented error" in MRAC literature.

$\hat{b}_1(t) \neq 0$ for all $t$. This will assure $\phi_C(t)$ can become eventually unbounded only asymptotically.

In order to assure that $\hat{b}_1(t) \neq 0$ for any $t$, knowing that $b_1$ cannot be zero and its sign is assumed to be known, one fixes a minimum value $|\hat{b}_1(t)| \geq \delta > 0$. If $|\hat{b}_1(t)| < \delta$ for some $t$, one either simply uses the value $\delta$ (with the appropriate sign), or one takes advantage of the weighting sequences $\lambda_1(t), \lambda_2(t)$ and re-computes $\hat{\theta}_C(t)$ for different values of $\lambda_1(t-d-1), \lambda_2(t-d-1)$ (for example: $\lambda_1' = \lambda_1 + \Delta\lambda_1, \lambda_2' = \lambda_2 + \Delta\lambda_2$) such that $|\hat{b}_1(t)| > \delta$.

After these preliminary considerations, the next step is to show that (11.3) is true and $\phi_C(t)$ is bounded. Two approaches are possible:

1. Use a "bounded growth" lemma which will allow to straightforwardly conclude.
2. Derive a dynamic equation for $\phi_C(t)$ and use it for boundedness analysis (Landau and Lozano 1981).

We will use the first approach by outlining the "bounded growth" lemma (Goodwin and Sin 1984).

**Lemma 11.1**

(1) *Assume that*:

$$\|\phi(t)\|_{F(t)} = [\phi^T(t)F(t)\phi(t)]^{1/2} \leq C_1 + C_2 \max_{0\leq k\leq t+d+1} |\varepsilon^0(k)|$$

$$0 < C_1,\ C_2 < \infty,\ F(t) > 0 \tag{11.24}$$

(2) *Assume that*:

$$\lim_{t\to\infty} \frac{[\varepsilon^0(t+d+1)]^2}{1+\|\phi(t)\|^2_{F(t)}} = 0 \tag{11.25}$$

*then*:

$$\|\phi(t)\| \textit{ is bounded} \tag{11.26}$$

$$\lim_{t\to\infty} \varepsilon^0(t+d+1) = 0 \tag{11.27}$$

*Proof* The proof is trivial if $\varepsilon^0(t)$ is bounded for all $t$. Assume now that $\varepsilon^0(t+d+1)$ is asymptotically unbounded. Then there is a particular subsequence such that: $\lim_{t_n\to\infty} \varepsilon^0(t_n) = \infty$ and $|\varepsilon^0(t+d+1)| < |\varepsilon^0(t_n)|;\ t+d+1 \leq t_n$. For this sequence one has:

$$\frac{|\varepsilon^0(t+d+1)|}{(1+\|\phi(t)\|^2_F)^{1/2}} \geq \frac{\varepsilon^0(t+d+1)}{1+\|\phi(t)\|_{F(t)}} \geq \frac{|\varepsilon^0(t+d+1)|}{1+C_1+C_2|\varepsilon^0(t+d+1)|}$$

but:

$$\lim_{t\to\infty} \frac{|\varepsilon^0(t+d+1)|}{1+C_1+C_2|\varepsilon^0(t+d+1)|} = \frac{1}{C_2} > 0$$

which contradicts Assumption 2 and proves that neither $\varepsilon^0(t+d+1)$ nor $\|\phi(t)\|$ can become unbounded and that (11.27) is true. □

The application of this lemma is straightforward in our case. From (11.12) one has:

$$P(q^{-1})y(t+d+1) = P(q^{-1})y^*(t+d+1) + \varepsilon^0(t+d+1) \tag{11.28}$$

Since $P(q^{-1})$ has bounded coefficients and $y^*$ is bounded, one gets immediately from (11.28) that:

$$|y(t)| \leq C_1' + C_2' \max_{0\leq k\leq t+d+1} |\varepsilon^0(k)| \tag{11.29}$$

Using the assumption that $B^*(z^{-1})$ has all its zeros inside the unit circle, the inverse of the system is asymptotically stable and one has:

$$|u(t)| \leq C_1'' + C_2'' \max_{0\leq k\leq t+d+1} |y(t+d+1)| \tag{11.30}$$

From (11.29) and (11.30), one concludes that:

$$\|\phi_C(t)\|_{F(t)} \leq C_1 + C_2 \max_{0\leq k\leq t+d+1} |\varepsilon^0(k)| \tag{11.31}$$

and, on the other hand, from Theorem 3.2 one has (11.25). Therefore the assumptions of Lemma 11.1 are satisfied allowing to conclude that (11.3) is satisfied and that $\{u(t)\}$ and $\{y(t)\}$ are bounded. The results of the previous analysis can be summarized under the following form:

**Theorem 11.1** *Consider a plant model of the form* (11.1) *controlled by the adjustable controller* (11.4) *whose parameters are updated by the PAA of* (11.19), (11.20) *and* (11.23) *where*:

$$\varepsilon^0(t+d+1) = Py(t+d+1) - Py^*(t+d+1)$$

*Assume that*:

(1) *The integer delay $d$ is known.*
(2) *Upper bounds on the degrees of the polynomials $A$ and $B^*$ are known.*
(3) *For all possible values of the plant parameters, the polynomial $B^*$ has all its zeros inside the unit circle.*
(4) *The sign of $b_1$ is known.*

*Then*:

- $\lim_{t\to\infty} \varepsilon^0(t+d+1) = 0$;
- *The sequences $\{u(t)\}$ and $\{y(t)\}$ are bounded.*

*Remarks*

- Integral + Proportional adaptation can also be used instead of the integral PAA considered in (11.19) and (11.20) (see Sect. 3.3.4, (3.270) through (3.274)).
- Various choices can be made for $\lambda_1(t)$ and $\lambda_2(t)$ as indicated in Sect. 3.2.3. This will influence the properties and the performances of the scheme. In particular for $\lambda_1(t) \equiv 1$ and $\lambda_2(t) > 0$, one will obtain a PAA with decreasing adaptation gain. The *constant trace algorithm* is probably the most used algorithm for obtaining an adaptive control scheme which is continuously active.
- Monitoring the eigenvalues of $F(t)$ is recommended in order to assure $0 < \delta < \|F(t)\| < \infty$ for all $t$.
- The choice of $P(q^{-1})$ which influences the robustness of the linear design also influences the adaptation transients. Taking $P(q^{-1}) = 1$, one gets oscillatory adaptation transients (and the linear design will be very sensitive to parameters change) while $P(q^{-1})$ in the range of the band pass of the plant will both improve the robustness of the linear design and the smoothness of the adaptation transient (Landau and Lozano 1981). See Sect. 11.6 for details.

### Alternative Direct Adaptive Control Design via Adaptive Prediction

The basic idea is to use an indirect adaptive control approach as follows:

Step 1: Design an adaptive predictor for the filtered plant output.
Step 2: Force the output of the adaptive predictor to be equal to the desired filtered trajectory.

Remember that using the time-domain design discussed in Sect. 7.4.2, one has the $d+1$ steps ahead predictor given by:

$$P(q^{-1})\hat{y}(t+d+1) = F(q^{-1})y(t) + G(q^{-1})u(t) = \theta_C^T \phi_C(t) \tag{11.32}$$

which has the property that:

$$[P(q^{-1})y(t+d+1) - P(q^{-1})\hat{y}(t+d+1)] = 0 \tag{11.33}$$

Forcing the output of the prediction to be:

$$P(q^{-1})\hat{y}(t+d+1) = P(q^{-1})y^*(t+d+1) = \theta_C^T \phi_C(t) \tag{11.34}$$

one gets the desired result. In the case of the unknown plant parameters, we will use a similar procedure but replacing the linear predictor with fixed coefficient by an adaptive predictor.

Step 1: Design of an adaptive predictor. Using the results of Sect. 6.2.1, one has:

$$P(q^{-1})\hat{y}^0(t+d+1) = \hat{F}(t)y(t) + \hat{G}(t)u(t) = \hat{\theta}_C^T(t)\phi_C(t) \tag{11.35}$$

which has the property that:

$$\lim_{t\to\infty} \varepsilon^0(t+d+1) = \lim_{t\to\infty} [P(q^{-1})y(t+d+1) - P(q^{-1})\hat{y}^0(t+d+1)] = 0$$

when using the PAA of (6.16), (6.17) and (6.19) (assuming that $u(t)$ and $y(t)$ are bounded).

Step 2: Use the ad-hoc separation theorem and compute a control such that the output of the adaptive predictor follows the desired filtered trajectory, i.e.:

$$P(q^{-1})\hat{y}^0(t+d+1) = \hat{\theta}_C^T(t)\phi_C(t) = P(q^{-1})y^*(t+d+1) \tag{11.36}$$

Notice that the control resulting from (11.36) has exactly the same structure as the one given by (11.7).

Furthermore, using (11.36), the equation of the prediction error becomes:

$$\begin{aligned}\varepsilon^0(t+d+1) &= [P(q^{-1})y(t+d+1) - P(q^{-1})y^*(t+d+1)] \\ &= [\theta_C - \hat{\theta}_C(t)]^T \phi_C(t)\end{aligned} \tag{11.37}$$

In other terms, by the choice of the control law, the prediction error becomes the measure of the control performance error and the equation of the prediction error in terms of controller parameter difference is exactly the same as (11.14). One concludes that both schemes are identical and that in the adaptive predictor, one estimates the controller parameters.

Note also that with this particular strategy, the output of the adjustable predictor at each $t$ is equal to the filtered desired trajectory, i.e., it behaves exactly as the filtered value of the reference model used in Fig. 11.1. For this reason, this type of direct adaptive control design was called in the past "implicit model reference adaptive control" (Landau 1981; Landau and Lozano 1981).

Note that this scheme, despite a formal similarity with adaptive predictors, does not operate in open loop and as a consequence $\phi_C(t)$ is not independent of $\hat{\theta}(t), \hat{\theta}(t-1), \ldots$ as for the scheme discussed in Sect. 6.1. This explains why a complementary specific analysis has to be considered. However, since this scheme is equivalent to the previous one, Theorem 11.1 holds.

## *11.2.2 Extensions of the Design*

### Filtering of the Measurement Vector and of the Adaptation Error

In order to guarantee certain convergence conditions (which occur for example when using this scheme in a stochastic environment) or to filter the measurement vector $\phi_C(t)$ outside the closed-loop bandpass (for robustness improvement as well as for finding various other designs considered in the literature), it is useful to consider a more general PAA which uses a filtered measurement vector as observation vector:

$$L(q^{-1})\phi_{Cf}(t) = \phi_C(t) \tag{11.38}$$

where:

$$L(q^{-1}) = 1 + l_1 q^{-1} + \cdots + l_{n_L} q^{-n_L} \tag{11.39}$$

is an asymptotically stable monic polynomial and:

$$\begin{aligned}\phi_{Cf}^T(t) &= \frac{1}{L(q^{-1})}[u(t), \ldots, u(t-n_B-d+1), y(t), \ldots, y(t-n+1)] \\ &= [u_f(t), \ldots, u_f(t-n_B-d+1), y_f(t), \ldots, y_f(t-n+1)]\end{aligned} \tag{11.40}$$

Notice first that in the linear case with known parameters, one can alternatively write the controller equation as:

$$u(t) = L(q^{-1})u_f(t) \tag{11.41}$$

where $u_f(t)$ satisfies the equation:

$$P(q^{-1})y_f^*(t+d+1) = Ry_f(t) + Su_f(t) = \theta_C^T \phi_{Cf}(t) \tag{11.42}$$

in which:

$$L(q^{-1})y_f^*(t+d+1) = y^*(t+d+1) \tag{11.43}$$

The natural extension for the adaptive case is to replace (11.42) by:

$$P(q^{-1})y_f^*(t+d+1) = \hat{\theta}_C^T(t)\phi_{Cf}(t) \tag{11.44}$$

We will now give an expression for the adaptation error. From (11.11) one has:

$$P(q^{-1})y(t+d+1) = \theta_C^T \phi_C(t) \tag{11.45}$$

Remembering that $\theta_C$ is a constant vector one can write (11.45) as:

$$P(q^{-1})y(t+d+1) = L(q^{-1})[\theta_C^T \phi_{Cf}(t)] \tag{11.46}$$

On the other hand, from (11.44), one obtains:

$$P(q^{-1})y^*(t+d+1) = L(q^{-1})[\hat{\theta}_C^T(t)\phi_{Cf}(t)] \tag{11.47}$$

Therefore, the error between the desired filtered trajectory and the achieved filtered output is given by:

$$\begin{aligned}\varepsilon^0(t+d+1) &= Py(t+d+1) - Py^*(t+d+1) \\ &= L(q^{-1})[\theta_C - \hat{\theta}_C(t)]^T \phi_{Cf}(t)\end{aligned} \tag{11.48}$$

The associated a posteriori error equation is defined as:

$$\varepsilon(t+d+1) = L(q^{-1})[\theta_C - \hat{\theta}_C(t+d+1)]^T \phi_{Cf}(t) \tag{11.49}$$

Defining the a posteriori adaptation error as:

$$\nu(t+d+1) = \frac{H_1(q^{-1})}{H_2(q^{-1})}\varepsilon(t+d+1) \tag{11.50}$$

where $H_1(q^{-1})$ and $H_2(q^{-1})$ are asymptotically stable monic polynomials defined as:

$$H_j(q^{-1}) = 1 + (q^{-1})H_j^*(q^{-1}); \quad j = 1, 2 \tag{11.51}$$

one gets:

$$\nu(t+d+1) = H(q^{-1})[\theta_C - \hat{\theta}_C(t+d+1)]^T \phi_{Cf}(t) \tag{11.52}$$

where:

$$H(q^{-1}) = \frac{H_1(q^{-1})L(q^{-1})}{H_2(q^{-1})} \tag{11.53}$$

Applying Theorem 3.2,

$$\lim_{t\to\infty} \nu(t+d+1) = 0 \tag{11.54}$$

will be assured using the PAA

$$\hat{\theta}(t+d+1) = \hat{\theta}(t+d) + F(t)\phi_{Cf}(t)\nu(t+d+1) \tag{11.55}$$

$$F^{-1}(t+1) = \lambda_1(t)F^{-1}(t) + \lambda_2(t)\phi_{Cf}(t)\phi_{Cf}^T(t)$$

$$0 < \lambda_1(t) \le 1; \ 0 \le \lambda_2(t) < 2; \ F(0) > 0 \tag{11.56}$$

provided that there is $2 > \lambda \ge \max_t \lambda_2(t)$ such that:

$$H'(z^{-1}) = H(z^{-1}) - \frac{\lambda}{2} \tag{11.57}$$

is strictly positive real.

It remains to give an expression for $\nu(t+d+1)$ in terms of $\varepsilon^0(t+d+1)$ and $\hat{\theta}(t+i)$ for $i$ up to and including $d$. Using similar developments as in Sect. 3.3, one has from (11.50):

$$\nu(t+d+1) = \varepsilon(t+d+1) + H_1^*\varepsilon(t+d) - H_2^*\nu(t+d)$$

but:

$$\begin{aligned}\varepsilon(t+d+1) = \varepsilon^0(t+d+1) &- L^*(q^{-1})[\hat{\theta}_C(t+d) - \hat{\theta}_C(t+d-1)]^T \phi_{Cf}(t-1)\\ &- L(q^{-1})[\hat{\theta}_C(t+d) - \hat{\theta}_C(t)]^T \phi_{Cf}(t)\\ &- \phi_{Cf}^T(t)F(t)\phi_{Cf}^T \nu(t+d+1)\end{aligned}$$

from which one obtains:

$$\nu(t+d+1) = \frac{\nu^0(t+d+1)}{1+\phi_{Cf}^T(t)F(t)\phi_{Cf}(t)} \tag{11.58}$$

where:

$$\begin{aligned}
&\nu^0(t+d+1)\\
&\quad = \varepsilon^0(t+d+1) - L^*(q^{-1})[\hat{\theta}_C(t+d) - \hat{\theta}_C(t+d-1)]^T\phi_{Cf}(t-1)\\
&\qquad - L(q^{-1})[\hat{\theta}_C(t+d) - \hat{\theta}_C(t)]^T\phi_{Cf}(t) + H_1^*\varepsilon(t+d) - H_2^*\nu(t+d)\\
&\quad = \varepsilon'(t+d+1) - L^*(q^{-1})[\hat{\theta}_C(t+d) - \hat{\theta}_C(t+d-1)]^T\phi_{Cf}(t-1)\\
&\qquad + H_1^*\varepsilon(t+d) - H_2^*\nu(t+d)
\end{aligned} \tag{11.59}$$

with:

$$\varepsilon'(t+d+1) = P(q^{-1})y(t+d+1) - L(q^{-1})\hat{\theta}_C^T(t+d)\phi_{Cf}(t) \tag{11.60}$$

and:

$$\begin{aligned}
\varepsilon(t+d) &= L(q^{-1})[\theta - \hat{\theta}(t+d)]^T\phi_{Cf}(t-1)\\
&= Py^*(t+d) - L(q^{-1})\hat{\theta}_C(t+d)\phi_{Cf}(t-1)
\end{aligned} \tag{11.61}$$

### Taking into Account Measurable Disturbances

In a number of applications, measurable disturbances act upon the output of the process through unknown dynamics. The knowledge of these dynamics would be useful for compensating the effect of the disturbance by using an appropriate controller. This problem has been discussed for the case of known parameters in Sect. 7.4.2.

The plant output is described by (7.130):

$$A(q^{-1})y(t+d+1) = B^*(q^{-1})u(t) + C^*(q^{-1})v(t) \tag{11.62}$$

where $v(t)$ is the measurable disturbance and $C^*(q^{-1})$ is a polynomial of order $n_c - 1$.

In the case of unknown parameters, one uses the same controller structure as in (7.134) but with adjustable parameters, i.e.:

$$P(q^{-1})y^*(t+d+1) = \hat{\theta}_C^T(t)\phi_C(t) \tag{11.63}$$

where:

$$\hat{\theta}_C^T(t) = [\hat{s}_0(t), \ldots, \hat{s}_{n_S}(t), \hat{r}_0(t), \ldots, \hat{r}_{n_R}(t), \hat{w}_0(t), \ldots, \hat{w}_{n_W}(t)] \tag{11.64}$$

$$\phi_C^T(t) = [u(t), \ldots, u(t-n_S), y(t), \ldots, y(t-n_R), v(t), \ldots, v(t-n_W)] \tag{11.65}$$

Then one proceeds exactly as for the case without measurable disturbances. The PAA will be given by (11.19), (11.20) and (11.23) with the remark that in this case $\hat{\theta}_C(t)$ and $\phi_C^T(t)$ will be of higher dimension.

## 11.3 Adaptive Tracking and Regulation with Weighted Input

To develop an adaptive version of this control strategy presented in Sect. 7.5, we will proceed in a similar way as for adaptive tracking and regulation with independent objectives. In Sect. 7.5, it was pointed out that the objective in the case of known parameters is that:

$$\begin{aligned}\varepsilon^0(t+d+1) &= P(q^{-1})y(t+d+1) + \lambda Q(q^{-1})u(t) - P(q^{-1})y^*(t+d+1) \\ &= 0; \quad \forall t > 0 \end{aligned} \tag{11.66}$$

In the case of unknown parameters, the objective will be:

$$\lim_{t\to\infty} \varepsilon^0(t+d+1) = \lim_{t\to\infty} [\bar{y}(t+d+1) - P(q^{-1})y^*(t+d+1)] = 0 \tag{11.67}$$

where:

$$\bar{y}(t+d+1) = Py(t+d+1) + \lambda Qu(t) \tag{11.68}$$

defines a so called "augmented output" or "generalized output".

The adjustable controller will have the same structure as the one given for the case of known parameters, but with adjustable parameters. Therefore from (7.143), one chooses the following adjustable controller:

$$Py^*(t+d+1) - \lambda Qu(t) = \hat{S}(t)u(t) + \hat{R}(t)y(t) = \hat{\theta}_C^T(t)\phi_C(t) \tag{11.69}$$

where:

$$\hat{\theta}_C^T(t) = [s_0, \ldots, s_{n_S}, r_0, \ldots, r_{n_R}] \tag{11.70}$$

$$\phi_C(t) = [u(t), \ldots, u(t-n_S), y(t), \ldots, y(t-n_R)] \tag{11.71}$$

On the other hand (see Sect. 11.2):

$$P(q^{-1})y(t+d+1) = \theta_C^T\phi_C(t) \tag{11.72}$$

Combining (11.66), (11.69) and (11.72), one gets:

$$\varepsilon^0(t+d+1) = [\theta_C - \hat{\theta}_C(t)]^T\phi_C(t) \tag{11.73}$$

which allows to use the same PAA as for adaptive tracking and regulation ((11.19), (11.20) and (11.22)) where $\varepsilon^0(t+d+1) = \bar{y}(t+d+1) - Py^*(t+d+1)$. This will assure that the a posteriori adaptation error:

$$\varepsilon(t+d+1) = [\theta_C - \hat{\theta}_C(t+d+1)]^T\phi_C(t) \tag{11.74}$$

goes asymptotically to zero. It remains to check that using the PAA of (11.19), (11.20) and (11.23), (11.67) will also be satisfied and that $\{u(t)\}$ and $\{y(t)\}$ will be bounded. From (11.67), one has that:

$$|\bar{y}(t+d+1)| \leq C_1' + C_2' \max_{0\leq k\leq t+d+1} |\varepsilon^0(k)| \tag{11.75}$$

On the other hand, $\bar{y}(t+d+1)$ can be expressed as the output of an "augmented plant":

$$\bar{y}(t+d+1) = \frac{\lambda QA + B^*P}{A} u(t) \tag{11.76}$$

From this expression, one concludes that if $\lambda QA + B^*P$ is asymptotically stable, for $\bar{y}(t+d+1)$ bounded we will have a bounded $u(t)$ as well as $y(t)$ (as a consequence of (11.67)). Using now the "bounded growth" lemma (Lemma 11.1), one concludes that (11.67) is true and that $\{y(t)\}$ and $\{u(t)\}$ are bounded. This analysis can be summarized as:

**Theorem 11.2** *Consider a plant model of the form* (11.1) *with not necessarily stable zeros, controlled by the adjustable controller* (11.69) *whose parameters are updated by the PAA of* (11.19), (11.20) *and* (11.23) *where*:

$$\varepsilon^0(t+d+1) = P(q^{-1})y(t+d+1) + \lambda Q(q^{-1})u(t) - P(q^{-1})y^*(t+d+1)$$

*Assume that*:

(1) *The integer delay $d$ is known.*
(2) *Upper bounds on the degrees of the polynomial $A$ and $B^*$ are known.*
(3) *For all possible values of the plant parameters, the polynomial $(\lambda QA + B^*P)$ has all its zeros inside the unit circle where $\lambda$, $Q$ and $P$ are fixed and chosen by the designer.*
(4) *The sign of $b_1$ is known.*

*Then*:

- $\lim_{t\to\infty} \varepsilon^0(t+d+1) = 0$;
- *the sequences $\{u(t)\}$ and $\{y(t)\}$ are bounded.*

*Remarks*

- The resulting closed-loop poles are defined by the polynomial $\lambda AQ + B^*P$ and they will depend upon $A$ and $B^*$. Therefore the location of the closed loop poles cannot be guaranteed.
- The hypothesis 3 is crucial. If it is not satisfied, $\varepsilon^0(t+d+1)$ can go to zero but $u(t)$ and $y(t)$ become unbounded.

## 11.4 Adaptive Minimum Variance Tracking and Regulation

The minimum variance tracking and regulation control algorithm for the case of known plant and disturbance model parameters is presented in Sect. 7.6. The objective of this section is to present the direct adaptive version of this control strategy for the case of unknown plant and disturbance model parameters, to analyze it and to examine several important special cases and in particular the case of adaptive minimum variance regulation (Åström and Wittenmark 1973).

As in the case of parameter identification and adaptive prediction in stochastic environment, due to the presence of noise, one typically uses decreasing adaptation gain in order to have convergence of the estimated parameters toward constant values. The resulting schemes are currently called *self-tuning minimum variance tracking and regulation*.

In the case of known parameters, the strong relationship between tracking and regulation with independent objectives and minimum variance tracking and regulation was pointed out. This strong resemblance carry over in the adaptive case, where the adaptive minimum variance tracking and regulation can be viewed as the stochastic version of tracking and regulation with independent objectives (called also stochastic model reference adaptive control) (Landau 1981, 1982a; Dugard et al. 1982).

As with the development of the adaptive tracking and regulation with independent objectives scheme, one can consider two approaches for deriving the direct adaptive version of the minimum variance tracking and regulation

(1) Use the difference between the plant output $y(t)$ and the reference trajectory $y^*(t)$ as an adaptation error and find an adaptation mechanism for the controller parameters which drive this quantity asymptotically (in stochastic sense) towards the optimal value obtained in the case of known parameters. For the case $d = 0$, this can be expressed for example as:

$$\text{Prob}\left\{\lim_{t\to\infty} \varepsilon^0(t+1) = \lim_{t\to\infty}[y(t+1) - y^*(t+1)] = e(t+1)\right\} = 1$$

where $e(t+1)$ is a white noise sequence, or:

$$\lim_{N\to\infty} \frac{1}{N}\sum_{t=1}^{N}[\varepsilon^0(t+1) - e(t+1)]^2 = 0 \quad \text{a.s.}$$

(2) Use an indirect adaptive approach by building an adaptive predictor reparameterized in terms of the controller parameters and force the predictor output to follow exactly the reference trajectory by an appropriate choice of the control. (This corresponds to the use of the ad-hoc separation theorem.)

Both approaches lead to the same scheme. The corresponding block diagram of adaptive minimum variance tracking and regulation is shown in Fig. 11.2. The major difference with regard to adaptive tracking and regulation with independent objectives is the fact that the closed-loop poles and the parameters of the precompensator $T$, depend on the noise model (polynomial C) which is unknown and should be estimated. Therefore with respect to the deterministic case, there are more parameters to be adapted.

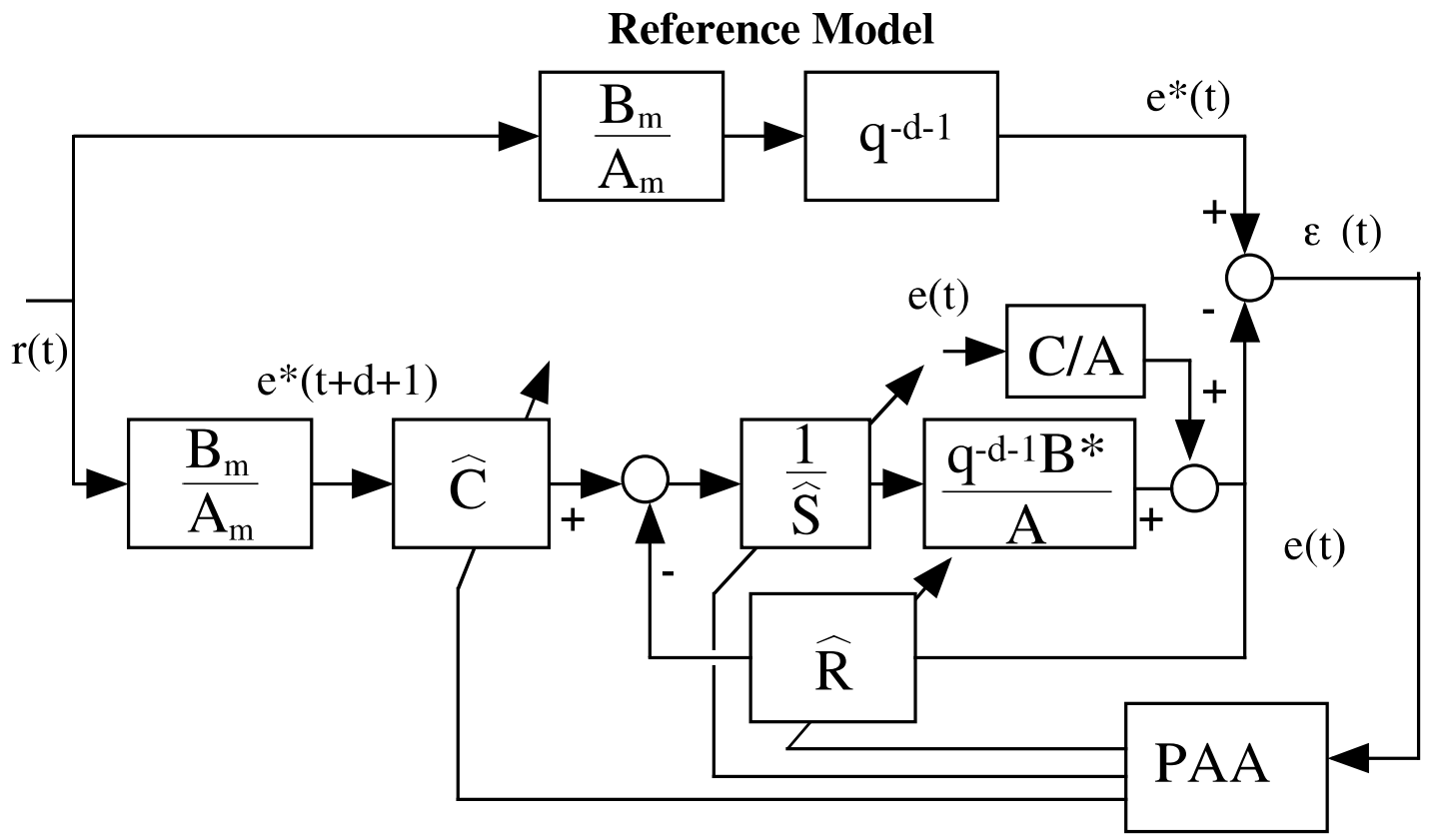

**Fig. 11.2** Adaptive minimum variance tracking and regulation

We will use the second approach for explicitly obtaining the algorithms. We will start with the case $d = 0$ and then extend the algorithms to the case $d \geq 1$. (For the use of the first approach see Landau 1981.) We will proceed then to the analysis of the asymptotic properties of the algorithms using the averaging method presented in Chap. 4, under the hypothesis that the parameter estimates give at each time $t$ a stabilizing controller. The convergence toward the optimal error will depend on a positive real condition on the noise model (like in recursive identification). In connection with this analysis, an adaptive minimum variance tracking and regulation algorithm with filtering will be presented which allows to relax the positive real condition for convergence. We will then present a more complete analysis of self-tuning minimum variance tracking and regulation using martingales.

The plant and the disturbance are described by the following ARMAX model:

$$y(t + d + 1) = -A^* y(t + d) + B^* u(t) + Ce(t + d + 1) \tag{11.77}$$

where $u(t)$ and $y(t)$ are the input and output of the plant and $e(t + 1)$ is a zero mean white noise sequence. For the case $d = 1$, this model takes the form:

$$y(t + 1) = -A^* y(t) + B^* u(t) + Ce(t + 1) \tag{11.78}$$

### *11.4.1 The Basic Algorithms*

#### Exact Adaptive Minimum Variance Tracking and Regulation

**The case** $d = 0$ We will start by building an adaptive predictor.

Step 1: Adaptive prediction

In the case of known parameters, using the polynomial equation:

$$C = AE + q^{-1}F \tag{11.79}$$

which gives for $d = 0 : E = 1, F = C^* - A^*$, the one step ahead predictor associated with the model (11.78) takes the form:

$$\begin{aligned}\hat{y}(t+1) &= -C^*\hat{y}(t) + Fy(t) + B^*u(t)\\ &= -C^*\hat{y}(t) + Ry(t) + Su(t) = \theta_C^T\phi(t)\end{aligned} \tag{11.80}$$

where $R = F$, $S = B^*$ and $C^*$ correspond to the parameters of the minimum variance controller and:

$$\theta_C^T = [s_0, \dots, s_{n_S}, r_0, \dots, r_{n_R}, c_1, \dots, c_{n_C}] \tag{11.81}$$

$$\begin{aligned}\phi^T(t) = [&u(t), \dots, u(t-n_S), y(t), \dots, y(t-n_R),\\ &-\hat{y}(t), \dots, -\hat{y}(t-n_C)]\end{aligned} \tag{11.82}$$

The corresponding adaptive predictor will have the form:

$$\hat{y}^0(t+1) = -\hat{C}^*(t)\hat{y}(t) + \hat{R}(t)y(t) + \hat{S}(t)u(t) = \hat{\theta}_C^T(t)\phi(t) \tag{11.83}$$

$$\hat{y}(t+1) = -\hat{\theta}_C^T(t+1)\phi(t) \tag{11.84}$$

where $\hat{y}^0(t+1)$ and $\hat{y}(t+1)$ define the a priori and a posteriori outputs of the adaptive predictor and:

$$\hat{\theta}_C^T(t) = [\hat{s}_0(t), \dots, \hat{s}_{n_S}(t), \hat{r}_1(t), \dots, \hat{r}_{n_R}(t), \hat{c}_1(t), \dots, \hat{c}_{n_C}(t)] \tag{11.85}$$

Defining the a priori and a posteriori prediction errors as:

$$\varepsilon^0(t+1) = y(t+1) - \hat{y}^0(t+1) \tag{11.86}$$

$$\varepsilon(t+1) = y(t+1) - \hat{y}(t+1) \tag{11.87}$$

the appropriate PAA with decreasing adaptation gain will take the form:

$$\hat{\theta}_C(t+1) = \hat{\theta}_C(t) + F(t)\phi(t)\varepsilon(t+1) \tag{11.88}$$

$$\begin{aligned}&F(t+1)^{-1} = F(t)^{-1} + \lambda_2(t)\phi(t)\phi^T(t)\\ &0 < \lambda_2(t) < 2;\ \ F(0) > 0\end{aligned} \tag{11.89}$$

$$\varepsilon(t+1) = \frac{\varepsilon^0(t+1)}{1 + \phi^T(t)F(t)\phi(t)} \tag{11.90}$$

Step 2: Use of the ad-hoc separation Theorem

Compute $u(t)$ such that $\hat{y}^0(t+1) = y^*(t+1)$. This yields:

$$y^*(t+1) = \hat{y}^0(t+1) = \hat{\theta}_C^T(t)\phi(t) \tag{11.91}$$

from which one obtains the controller expression:

$$\hat{\theta}_C^T(t)\phi(t) = \hat{S}(t)u(t) + \hat{R}(t)y(t) - \hat{C}^*(t)\hat{y}(t) = y^*(t+1) \tag{11.92}$$

or:

$$u(t) = \frac{1}{\hat{b}_1(t)}[y^*(t+1) + \hat{C}^*(t)\hat{y}(t) - \hat{S}^*(t)u(t-1) - \hat{R}(t)y(t)] \tag{11.93}$$

This adjustable controller has almost the same structure as the controller used in the case of known parameters. The only difference is the replacement of $y^*(t) = \hat{y}^0(t)$; $y^*(t-1) = \hat{y}^0(t-1), \ldots$ by the a posteriori predicted outputs $\hat{y}(t), \hat{y}(t-1), \ldots$. Taking into account (11.86), (11.87) and (11.91), $\hat{y}(t)$ can be expressed as:

$$\hat{y}(t) = y^*(t) + [\hat{y}(t) - \hat{y}^0(t)] = y^*(t) + [\hat{\theta}_C(t) - \hat{\theta}_C(t-1)]^T\phi(t) \tag{11.94}$$

This allows (11.92) and (11.93) to be rewritten as follows:

$$\begin{aligned}&\hat{S}(t)u(t) + \hat{R}(t)y(t) - \hat{C}^*(t)y^*(t) - \hat{C}^*(t)[\hat{\theta}_C(t) - \hat{\theta}_C(t-1)]^T\phi(t)\\&\quad = y^*(t+1)\end{aligned} \tag{11.95}$$

or:

$$\begin{aligned}u(t) = \frac{1}{\hat{b}_1(t)}&[y^*(t+1) + \hat{C}^*(t)y^*(t) - \hat{S}^*u(t) - R(t)y(t)\\&+ C^*(t)[\hat{\theta}_C(t) - \hat{\theta}_C(t-1)]^T\phi(t)]\end{aligned} \tag{11.96}$$

where the term depending upon $[\hat{\theta}_C(t) - \hat{\theta}_C(t-1)]$ accounts in a certain sense for the speed of adaptation and will become null asymptotically if $\phi(t)$ is bounded (one uses a decreasing adaptation gain algorithm). Therefore asymptotically $\hat{y}(t)$ tends towards $y^*(t)$ and the controller equation (11.96) will have the structure of the fixed controller for the known parameter case.

To summarize, the *exact adaptive minimum variance tracking and regulation* for $d = 0$ is obtained using the adjustable controller defined by (11.92) whose parameters are updated by the PAA of (11.88) through (11.90) where:

$$\varepsilon^0(t+1) = y(t+1) - y^*(t+1) \tag{11.97}$$

As it can be seen the resulting direct adaptive control uses as primary source of information the error between the measured plant output and the desired output (similar to the deterministic case). For this reason this scheme is also called *stochastic model reference adaptive control* (Landau 1981, 1982a; Dugard et al. 1982).

Because of the division by $\hat{b}_1(t)$ (see (11.96)), the sign of $b_1$ should be assumed known, and the estimate of $\hat{b}_1(t)$ used is defined as:

$$\hat{b}_1'(t) = \begin{cases}\hat{b}_1(t) & |\hat{b}_1(t)| > \delta > 0\\ |\hat{b}_1'(t)| = \delta & |\hat{b}_1(t)| < \delta\end{cases} \tag{11.98}$$

in order to avoid division by zero.

### Approximate Adaptive Minimum Variance Tracking and Regulation

**The case** $d = 0$ Taking into account (11.94), one can consider approximating $\hat{y}(t), \hat{y}(t-1), \ldots$ by $y^*(t), y^*(t-1), \ldots$ both in the controller equation and in the observation vector used in the PAA of (11.88) through (11.90). This corresponds to the use of a priori predictions instead of a posteriori predictions in the observation vector. In this case the adjustable controller is given by:

$$\hat{\theta}_C^T(t)\phi_C(t) = \hat{S}(t)u(t) + \hat{R}(t)y(t) - \hat{C}^*(t)y^*(t) = y^*(t+1) \tag{11.99}$$

where $\hat{\theta}_c(t)$ is given by (11.85) and

$$\phi_C^T(t) = [u(t), \ldots, u(t-n_S), y(t), \ldots, y(t-n_R), -y^*(t), \ldots, -y^*(t-n_C+1)] \tag{11.100}$$

and the PAA is given by:

$$\hat{\theta}_C(t+1) = \hat{\theta}_C(t) + F(t)\phi_C(t)\varepsilon(t+1) \tag{11.101}$$

$$F(t+1)^{-1} = F(t)^{-1} + \lambda_2(t)\phi_C(t)\phi_C^T(t)$$
$$0 < \lambda_2(t) < 2; \;\; F(0) > 0 \tag{11.102}$$

$$\varepsilon(t+1) = \frac{y(t+1) - y^*(t+1)}{1 + \phi_C^T(t)F(t)\phi_C(t)} \tag{11.103}$$

**The case** $d > 0$ One considers a $d+1$ steps ahead adaptive predictor of the form (see Sect. 6.3):

$$\hat{y}^0(t+d+1) = -\hat{C}^*(t)\hat{y}^0(t+d) + \hat{F}(t)y(t) + \hat{G}(t)u(t) \tag{11.104}$$

The control $u(t)$ is obtained by forcing:

$$\hat{y}^0(t+d+1) = y^*(t+d+1) \tag{11.105}$$

which yields:

$$\theta_C^T(t)\phi_C(t) = \hat{R}(t)y(t) + \hat{S}(t)u(t) - C^*(t)y^*(t+d) = y^*(t+d+1) \tag{11.106}$$

with:

$$\theta_C^T(t) = [\hat{s}_0(t), \ldots, \hat{s}_{n_S}(t), \hat{r}_0(t), \ldots, \hat{r}_{n_R}(t), \hat{c}_1(t), \ldots, \hat{c}_{n_C}(t)] \tag{11.107}$$

$$\phi_C^T(t) = [u(t), \ldots, u(t-n_S), y(t), \ldots, y(t-n_F),$$
$$- y^*(t+d), \ldots, -y^*(t+d-n_C+1)] \tag{11.108}$$

In the above equations, $\hat{F}(t)$ and $\hat{G}(t)$ have been replaced by $\hat{R}(t)$ and $\hat{S}(t)$ similar to the known parameter case and $\hat{y}(t+d), \hat{y}^0(t+d-1), \ldots$ by $y^*(t+d)$, $y^*(t+d-1), \ldots$ taking into account (11.105). The used PAA is given by:

$$\hat{\theta}_C(t+d+1) = \hat{\theta}_C(t+d) + F(t)\phi_C(t)\varepsilon(t+d+1) \tag{11.109}$$

$$F^{-1}(t+1) = F^{-1}(t) + \lambda_2(t)\phi_C(t)\phi_C^T(t)$$
$$0 < \lambda_2(t) < 2; \; F(0) > 0 \tag{11.110}$$

$$\varepsilon(t+d+1) = \frac{y(t+d+1) - \hat{\theta}_C^T(t+d)\phi_C(t)}{1 + \phi_C^T(t)F(t)\phi(t)} \tag{11.111}$$

## Adaptive Minimum Variance Regulation

In the case of regulation only $y^*(t) \equiv 0$ and both the controller and the PAA become simpler. Therefore the approximate (or the exact) algorithm for $d = 0$ will take the form:

Adjustable controller

$$\hat{S}(t)u(t) + \hat{R}(t)y(t) = \hat{\theta}_R^T(t)\phi_R(t) \tag{11.112}$$

or

$$u(t) = -\frac{\hat{R}(t)y(t) + \hat{S}^*(t)u(t-1)}{\hat{b}_1(t)} \tag{11.113}$$

where

$$\hat{\theta}_R^T(t) = [\hat{s}_0(t), \ldots, \hat{s}_{n_S}(t), \hat{r}_0(t), \ldots, \hat{r}_{n_R}(t)] \tag{11.114}$$

$$\phi_R^T(t) = [u(t), \ldots, u(t-n_S), y(t), \ldots, y(t-n_R)] \tag{11.115}$$

and the PAA takes the form

$$\hat{\theta}_R(t+1) = \hat{\theta}_R(t) + F(t)\phi_R(t)\varepsilon(t+1) \tag{11.116}$$

$$F^{-1}(t+1) = F^{-1}(t) + \lambda_2(t)\phi_R(t)\phi_R^T(t)$$
$$0 < \lambda_2(t) < 2; \; F(0) > 0 \tag{11.117}$$

$$\varepsilon(t+1) = \frac{y(t+1)}{1 + \phi_R^T(t)F(t)\phi_R(t)} \tag{11.118}$$

From (11.113) one can see that in the particular case of regulation, the control law has one redundant parameter and there is an infinite number of solutions leading to a good result. In practice, one often replaces $\hat{b}_1(t)$ by a “fixed” a priori estimate $\hat{b}_1$ satisfying $\hat{b}_1 > b_1/2$ (Egardt 1979).

*Remark* The algorithm is close to the original Åström and Wittenmark self-tuning minimum variance controller (Åström and Wittenmark 1973) and it can be shown that in this case the resulting biased plant parameters estimates converge toward a controller assuring the minimum variance regulation asymptotically. This algorithm can be obtained straightforwardly by

(1) making a least squares one step ahead predictor (i.e., neglecting the term $Ce(t+1)$) which estimates only $\hat{A}$, $\hat{B}$. The estimates will be biased;
(2) compute the input by forcing $\hat{y}^0(t+1)=0$.

## *11.4.2 Asymptotic Convergence Analysis*

In this section, we will examine the asymptotic properties of the various algorithms using the averaging method presented in Sect. 4.2 (and in particular Theorem 4.1) and we will present an extension of the basic algorithms through filtering the observation vector. While this analysis will give conditions for asymptotic convergence toward minimum variance control, it assumes that during adaptation transient, one has at each $t$ a stabilizing controller (a more complete analysis will be discussed in Sect. 11.4.3).

The analysis begins by deriving an equation for the a priori or a posteriori adaptation error (since the averaging method does not make a distinction between them) and setting the adjustable parameters to a constant value. Then application of Theorem 4.1 will give us the convergence conditions.

### Exact Adaptive Minimum Variance Tracking and Regulation

Using the polynomial equation (11.79), the plant output at $t+1$ can be expressed as:

$$\begin{aligned} y(t+1) &= -C^* y(t) + Ry(t) + Su(t) + Ce(t+1) \\ &= \theta_C^T \phi(t) - C^*[y(t) - \hat{y}(t)] + Ce(t+1) \end{aligned} \tag{11.119}$$

where $\theta_C$ defines the vector of the nominal parameters of the minimum variance controller given in (11.70) and $\hat{y}(t)$ is the a posteriori prediction given by (11.84). Subtracting (11.91) from (11.119) one gets:

$$\varepsilon^0(t+1) = [\theta_C - \hat{\theta}_C(t)]^T \phi(t) - C^* \varepsilon(t) + Ce(t+1) \tag{11.120}$$

and subtracting (11.84) from (11.119) one gets:

$$\varepsilon(t+1) = [\theta_C - \hat{\theta}_C(t+1)]^T \phi(t) - C^* \varepsilon(t) + Ce(t+1) \tag{11.121}$$

from which one obtains:

$$\varepsilon(t+1) = \frac{1}{C(q^{-1})}[\theta_C - \hat{\theta}_C(t+1)]^T \phi(t) + e(t+1) \tag{11.122}$$

Making $\hat{\theta}_C(t) = \hat{\theta}_C$, both (11.120) and (11.121) lead to:

$$\varepsilon^0(t+1, \hat{\theta}_C) = \varepsilon(t+1, \hat{\theta}_C) = \frac{1}{C(q^{-1})}[\theta_C - \hat{\theta}_C]^T \phi(t, \hat{\theta}_C) + e(t+1) \tag{11.123}$$

and since the image of the disturbance in (11.123) is a white noise, Theorem 4.1 can be applied. Assuming that $\hat{\theta}_C(t)$ does not leave the domain $D_S$ of controller parameters for which the closed loop is asymptotically stable and that $\max_t \lambda_2(t) \leq \lambda_2 < 2$, it results immediately from Theorem 4.1 that if:

$$H'(z^{-1}) = \frac{1}{C(z^{-1})} - \frac{\lambda_2}{2} \tag{11.124}$$

is a strictly positive real transfer function one has:

$$\text{Prob}\left\{\lim_{t\to\infty} \hat{\theta}_C(t) \in D_C\right\} = 1 \tag{11.125}$$

where $D_C$ is a set of all possible convergence points characterized by:

$$D_C\{\hat{\theta}_C | [\theta_C - \hat{\theta}_C]^T \phi(t, \hat{\theta}) = 0\} \tag{11.126}$$

Since $\hat{\theta}_c(\infty) \in D_C$ w.p.1, it results from (11.123) that asymptotically:

$$\text{Prob}\left\{\lim_{t\to\infty} \varepsilon^0(t+1) = \lim_{t\to\infty} \varepsilon(t+1) = e(t+1)\right\} = 1 \tag{11.127}$$

which is the desired control objective in the adaptive case. Note that (11.125) does not imply the convergence of $\hat{\theta}_C$ towards $\theta_C$ ($\theta_C$ is just one possible convergence point).

## Approximate Adaptive Minimum Variance Tracking and Regulation

This algorithm can be analyzed in a similar way and same positive real conditions will result. We will subsequently examine the case $d > 0$. Using the results of Sect. 2.2, the output of the plant at $t + d + 1$ is given by:

$$y(t+d+1) = -C^*(t)y(t+d) + Fy(t) + Gu(t) + CEe(t+d+1) \tag{11.128}$$

where $E$ and $F$ are solutions of the polynomial equation:

$$C = AE + q^{-d-1}F \tag{11.129}$$

and:

$$G = B^*E \tag{11.130}$$

However, from Sect. 7.6, one has in the case of minimum variance control

$$S = G \quad \text{and} \quad R = F \tag{11.131}$$

which allows one to rewrite (11.128) as:

$$y(t+d+1) = \theta_C^T \phi_C(t) - C^*[y(t+d) - \hat{y}^0(t+d)] + CEe(t+d+1) \tag{11.132}$$

Taking into account (11.105) and subtracting (11.106) from (11.132), one gets:

$$\begin{aligned}\varepsilon^0(t+d+1) &= y(t+d+1) - \hat{y}^0(t+d+1)\\ &= [\theta_C - \hat{\theta}_C(t)]^T \phi_C(t) - C^* \varepsilon^0(t+d) + CEe(t+d+1)\end{aligned} \quad (11.133)$$

which, after passing the term $C^*\varepsilon^0(t+d)$ on the left hand side, can be put under the form:

$$\varepsilon^0(t+d+1) = \frac{1}{C(q^{-1})}[\theta_C - \hat{\theta}_C(t)]^T \phi_C(t) + Ee(t+d+1) \quad (11.134)$$

Making $\hat{\theta}_C(t) = \hat{\theta}_C$ one gets:

$$\varepsilon^0(t+d+1, \hat{\theta}_C) = \frac{1}{C(q^{-1})}[\theta_C - \hat{\theta}_C]^T \phi_C(t, \hat{\theta}_C) + Ee(t+d+1) \quad (11.135)$$

The term $Ee(t+d+1)$ contains $e(t+d+1)e(t+d)\cdots e(t+1)$ which all are uncorrelated with the components of $\phi_c(t, \hat{\theta}_c)$ ($e(t+i)$ is a white noise) and therefore:

$$\mathbf{E}\{\phi_C(t, \hat{\theta}_C), Ee(t+d+1)\} = 0 \quad (11.136)$$

The form of (11.135) and (11.136) allows to directly apply the results of Theorem 4.1 yielding the same positive real condition given in (11.124). Same result holds for the case of adaptive minimum variance regulation. The result of the above analysis can be summarized as follows:

**Theorem 11.3** *For the adaptive minimum variance tracking and regulation algorithms given either by* (11.106) *through* (11.111), *or by* (11.92), (11.88) *through* (11.90) *and* (11.97) (*for* $d = 0$) *and for adaptive minimum variance regulation algorithm given by* (11.112) *through* (11.118) *provided that*:

(1) *An upper bound for the orders of the polynomials* $A, B, C$ *is known.*
(2) *The integer time delay* $d$ *is known.*
(3) *For all possible values of the plant model parameters,* $B(z^{-1})$ *has its zeros inside the unit circle.*
(4) *The sign of* $b_1$ *is known.*
(5) *The estimated controllers parameters belong infinitely often to the domain for which the closed-loop system is asymptotically stable.*

*One has*:

$$\left.\begin{aligned}&\text{Prob}\left\{\lim_{t\to\infty}[y(t+1) - y^*(t+1)] = e(t+1)\right\} = 1;\ d = 0\\ &\text{Prob}\left\{\lim_{t\to\infty}[y(t+d+1) - y^*(t+d+1)] = Ee(t+d+1)\right\} = 1\\ &\deg E = d\end{aligned}\right\} \quad (11.137)$$

*if there is* $\sup_t \lambda_2(t) \le \lambda_2 < 2$ *such that*:

$$H'(z^{-1}) = \frac{1}{C(z^{-1})} - \frac{\lambda_2}{2} \tag{11.138}$$

*is strictly positive real.*

Assumptions 1, 2, 3 are just assuring that a minimum variance control can be computed. Assumption 4 is necessary in order to use a mechanism for avoiding eventual division by zero during adaptation transient (see (11.98)). Assumption 5 is the typical limitation of the averaging method.

The result of Theorem 11.3 indicates that when $C(q^{-1}) \neq 1$, the adaptive control may not converge towards a value assuring asymptotic optimality of the controller if $\frac{1}{C(z^{-1})} - \frac{\lambda_2}{2}$ is not strictly positive real. Such an example of non convergence can be found in Ljung (1977b). One way to remove this restriction is by over parameterization (Shah and Franklin 1982) but it has the disadvantage of augmenting the number of parameters to be adapted. If one has some a priori knowledge of the domain of variation of $C(q^{-1})$ one can use a filtered observation vector (similar to the method described in the deterministic case—Sect. 11.2) in order to relax the positive real condition (11.138).

### 11.4.3 Martingale Convergence Analysis

The use of the averaging method allows a straightforward analysis of the asymptotic properties of the *adaptive minimum variance tracking and regulation* schemes. However, this analysis does not provide any information on the boundedness of the input and output of the plant during the adaptation transient. The objective of the martingale convergence analysis is to show that not only the asymptotic objectives are achieved but also that all the signals (including input and output of the plant) remain bounded (in a mean square sense). Of course the same positive real condition which occurred in the analysis through the averaging method will be encountered. The martingale convergence analysis of self-tuning minimum variance tracking and regulation has been the subject of a large number of papers (Goodwin et al. 1980b, 1980c; Guo 1993, 1996; Johansson 1995; Fuchs 1982; Chen and Guo 1991; Kumar and Moore 1982; Ren and Kumar 2002; Goodwin and Sin 1984). In particular the convergence in the case of matrix adaptation gain is still a subject of research.

In this section, we will prove the following results for the exact adaptive minimum variance tracking and regulation with $d = 0$.

**Theorem 11.4** (Goodwin and Sin 1984) *For the adaptive minimum variance tracking and regulation algorithm given by* (11.92), (11.97) *and* (11.88) *through* (11.90) *provided that*:

(1) *An upper bound for the orders of the polynomials $A$, $B$ and $C$ is known.*
(2) *The integer delay $d = 0$.*
(3) *For all possible values of the plant model parameters $B^*(z^{-1})$ has its zeros inside the unit circle.*
(4) *The sign of $b_1$ is known.*
(5)

$$\lim_{N\to\infty} \sup \frac{\lambda_{\max} F(N)}{\lambda_{\min} F(N)} \leq K < \infty \tag{11.139}$$

*where $\lambda_{\max} F(N)$ and $\lambda_{\min} F(N)$ are the maximum and minimum values of the eigenvalues of the matrix adaptation gain (bounded condition number assumption).*
(6) *The sequence $\{e(t)\}$ is a martingale difference sequence defined on a probability space $(\Omega, \mathscr{F}, \mathscr{P})$ and adapted to the sequence of increasing algebras $\mathscr{F}_t$ generated by the observations up to and including time $t$. The sequence $\{e(t+1)\}$ is assumed to satisfy the following:*

$$\mathbf{E}\{e(t+1)|\mathscr{F}_t\} = 0 \tag{11.140}$$

$$\mathbf{E}\{e^2(t+1)|\mathscr{F}_t\} = \sigma^2 \tag{11.141}$$

$$\lim_{N\to\infty} \sup \frac{1}{N} \sum_{t=1}^{N} e^2(t) < \infty \tag{11.142}$$

*one has:*

$$\lim_{N\to\infty} \frac{1}{N} \sum_{t=1}^{N} [\varepsilon^0(t+1) - e(t+1)]^2 = 0 \quad a.s. \tag{11.143}$$

$$\lim_{N\to\infty} \frac{1}{N} \sum_{t=1}^{N} y^2(t) < \infty \quad a.s. \tag{11.144}$$

$$\lim_{N\to\infty} \frac{1}{N} \sum_{t=1}^{N} u^2(t) < \infty \quad a.s. \tag{11.145}$$

*if there is:* $\sup_t \lambda_2(t) \leq \lambda_2 < 2$ *such that:*

$$H'(z^{-1}) = \frac{1}{C(z^{-1})} - \frac{\lambda_2}{2} \tag{11.146}$$

*is strictly positive real.*

Equation (11.143) tells us that the difference between the tracking (regulation) error $\varepsilon^0(t) = y(t) - y^*(t)$ and the optimal minimum variance error goes asymptotically to zero in mean square sense with probability 1. While the hypothesis (11.139) is of technical nature for the convergence proof with matrix adaptation gain, it is in any way taken into account in practical implementation of the adaptation algorithm. This may lead to a slight modification of the $F(t)$ given by (11.89) in order to satisfy the condition number requirement (see Chap. 16 for details).

*Proof* The proof of Theorem 11.4 is based on the use of Theorem 4.3 (Chap. 4) plus some additional steps involving the analysis of

- the properties of $[\varepsilon^0(t+1) - e(t+1)]$;
- the relationship between, $u(t)$, $y(t)$, $\hat{y}(t)$ and $[\varepsilon^0(t) - e(t)]$.

From (11.122) one has:

$$\varepsilon(t+1) = \frac{1}{C(q^{-1})}[\theta_C - \hat{\theta}_C(t+1)]^T \phi(t) + e(t+1) \tag{11.147}$$

which combined with the PAA of (11.88) through (11.90) allows the straightforward application of the results of Theorem 4.3 where:

$$r(t) = r(t-1) + \lambda_2(t)\phi^T(t-1)\phi(t-1); \quad 0 < \lambda_2(t) < 2 \tag{11.148}$$

with the observation that $r(0) = \text{tr}\, F^{-1}(0)$ which implies $\text{tr}\, F^{-1}(t) = r(t)$. Provided that (11.146) is strictly positive real one has:

$$\lim_{N\to\infty} \sum_{t=1}^{N} \frac{[\varepsilon(t+1) - e(t+1)]^2}{r(t)} < \infty \tag{11.149}$$

$$\lim_{N\to\infty} \sum_{t=1}^{N} \frac{\phi^T(t)F(t)\phi(t)}{r(t)} \varepsilon^2(t+1) < \infty \tag{11.150}$$

Then one can prove the following result:

**Lemma 11.2** *Under the hypotheses of Theorem* 11.4 *provided that* (11.146) *is strictly positive real, one has*:

$$\lim_{N\to\infty} \sum_{t=1}^{N} \frac{[\varepsilon^0(t+1) - e(t+1)]^2}{r(t)} < \infty \tag{11.151}$$

*This is a property of the measured tracking and regulation error* $\varepsilon^0(t+1)$ ( *a priori adaptation error*).

*Proof* Multiplying both terms of (11.88) from the left by $-\phi^T(t)$ one has:

$$\begin{aligned} -\phi^T(t)\hat{\theta}_C(t+1) &= -\phi^T(t)\hat{\theta}_C(t) - \phi^T(t)F(t)\phi(t)\varepsilon(t+1) \\ &= -\hat{y}(t+1) = -\hat{y}^0(t+1) - \phi^T(t)F(t)\phi(t)\varepsilon(t+1) \end{aligned} \tag{11.152}$$

Adding in both sides $[y(t+1)-e(t+1)]$ one gets:

$$\varepsilon(t+1)-e(t+1)=[\varepsilon^0(t+1)-e(t+1)]-\phi^T(t)F(t)\phi(t)\varepsilon(t+1) \quad (11.153)$$

Using the Schwarz inequality one has:

$$\begin{aligned}[\varepsilon^0(t+1)-e(t+1)]^2 &\le 2[\varepsilon(t+1)-e(t+1)]^2 \\ &\quad + 2[\phi^T(t)F(t)\phi(t)]^2\varepsilon^2(t+1)\end{aligned} \quad (11.154)$$

but taking into account the condition number hypothesis (11.139), one can write:

$$\begin{aligned}[\phi^T(t)F(t)\phi(t)]^2 &\le [\phi^T(t)\phi(t)][\phi(t)F^2(t)\phi(t)] \\ &\le \phi^T(t)\phi(t)\frac{\phi^T(t)F(t)\phi(t)}{r(t)}\end{aligned} \quad (11.155)$$

and therefore from (11.154) one finally obtains:

$$\begin{aligned}&\sum_{t=1}^{N}\frac{[\varepsilon^0(t+1)-e(t+1)]^2}{r(t)} \\ &\quad \le 2\sum_{t=1}^{N}\frac{[\varepsilon(t+1)-e(t+1)]^2}{r(t)} \\ &\qquad + 2\sum_{t=1}^{N}\frac{\phi^T(t)\phi(t)}{r(t)}\frac{\phi^T(t)F(t)\phi(t)}{r(t)}\varepsilon^2(t+1)\end{aligned} \quad (11.156)$$

Now taking into account that $[\phi^T(t)\phi(t)/r(t)] \le 1$ together with (11.149) and (11.150), one concludes that (11.151) is true. □

The next step is to use a "bounded growth" lemma similar to a certain extend with the one used in the deterministic case (Lemma 11.1, Sect. 11.2).

**Lemma 11.3** (Stochastic Bounded Growth Lemma Goodwin and Sin 1984) *If* (11.151) *holds, then if there are constants* $K_1, K_2$ *and* $\bar{N}$ : $0 \le K_1 < \infty$; $0 < K_2 < \infty$; $0 < \bar{N} < \infty$ *such that*:

$$\frac{1}{N}r(N) \le K_1 + \frac{K_2}{N}\sum_{t=1}^{N}[\varepsilon^0(t+1)-e(t+1)]^2; \quad N \ge \bar{N} \quad (11.157)$$

*one has*:

$$\lim_{N\to\infty}\frac{1}{N}\sum_{t=1}^{N}[\varepsilon^0(t+1)-e(t+1)]^2 = 0 \quad a.s. \quad (11.158)$$

$$\lim_{N\to\infty}\sup\frac{1}{N}r(N) < \infty \quad a.s. \quad (11.159)$$

Note that (11.158) and (11.159) correspond to the desired results (11.143) through (11.145) and it will remain to show that indeed for the algorithm considered, (11.157) is verified.

*Proof of Lemma 11.3* For the case $r(t) < K < \infty$ for all $t$, (11.158) results immediately from (11.157) and (11.151). Suppose now that $r(t)$ may become unbounded, then one can apply the Kronecker Lemma (see Appendix D) to (11.151) and conclude that:

$$\lim_{N\to\infty} \frac{N}{r(N)} \frac{1}{N} \sum_{t=1}^{N} [\varepsilon^0(t+1) - e(t+1)]^2 = 0 \tag{11.160}$$

Taking into account (11.157), one gets:

$$\lim_{N\to\infty} \frac{\frac{1}{N}\sum_{t=1}^{N}[\varepsilon^0(t+1) - e(t+1)]^2}{K_1 + \frac{K_2}{N}\sum_{t=1}^{N}[\varepsilon^0(t+1) - e(t+1)]^2} = 0 \tag{11.161}$$

and (11.158) follows by contradiction arguments. Since (11.158) is true it results from (11.160) that also (11.159) is true. □

To complete the proof of Theorem 11.4, it remains to show now that for the considered algorithm $r(t)$ given by (11.148) satisfies the condition (11.157) of the Lemma 11.3. Notice that:

$$\begin{aligned}\frac{1}{N} r(N) = \frac{1}{N} \operatorname{tr} F^{-1}(0) + \sum_{i=0}^{n_B-1} \left[\frac{1}{N}\sum_{t=1}^{N} u^2(t-i)\right] \\ + \sum_{i=0}^{n_A-1}\left[\frac{1}{N}\sum_{t=1}^{N} y^2(t-i)\right] + \sum_{i=0}^{n_C-1}\left[\frac{1}{N}\sum_{t=1}^{N} \hat{y}^2(t-i)\right]\end{aligned} \tag{11.162}$$

Therefore, it will be sufficient to show that $\frac{1}{N}\sum_{t=1}^{N} u^2(t)$, $\frac{1}{N}\sum_{t=1}^{N} y^2(t)$, $\frac{1}{N}\sum_{t=1}^{N} \hat{y}^2(t)$ satisfy a condition of the form (11.157). Notice that $y(t+1)$ can be expressed as:

$$\begin{aligned}y(t+1) &= \varepsilon^0(t+1) + y^*(t+1) \\ &= [\varepsilon^0(t+1) - e(t+1)] + e(t+1) + y^*(t+1)\end{aligned} \tag{11.163}$$

Using again the Schwarz inequality one has:

$$y^2(t+1) \le 3[\varepsilon^0(t+1) - e(t+1)]^2 + 3e^2(t+1) + 3{y^*}^2(t+1) \tag{11.164}$$

Bearing in mind (11.142) and the boundedness of $y^*$ one gets:

$$\frac{1}{N}\sum_{t=1}^{N} y^2(t+1) \le \frac{C_1}{N}\sum_{t=1}^{N}[\varepsilon^0(t+1) - e(t+1)]^2 + C_2$$
$$0 < C_1 < \infty;\ 0 \le C_2 < \infty \tag{11.165}$$

On the other hand, since in (11.78) $B^*(z^{-1})$ is asymptotically stable (i.e., the inverse of the system is stable) and (11.142) holds, one can express $u(t)$ as a function of $y(t)$ and $e(t)$ and one gets:

$$\frac{1}{N}\sum_{t=1}^{N} u^2(t+1) \le \frac{C_3}{N}\sum_{t=1}^{N} y^2(t) + C_4;\quad 0 < C_3 < \infty;\ 0 \le C_4 < \infty \tag{11.166}$$

which taking into account (11.165) allows one to conclude that $u(t)$ will satisfy a relationship of the form (11.157). Consider now the expression of $\hat{y}(t+1)$:

$$\begin{aligned}\hat{y}(t+1) &= y(t+1) - \varepsilon(t+1) = y(t+1) - \frac{\varepsilon^0(t+1)}{1+\phi^T(t)F(t)\phi(t)}\\ &= y^*(t+1) + \varepsilon^0(t+1) - \frac{\varepsilon^0(t+1)}{1+\phi^T(t)F(t)\phi(t)}\\ &= y^*(t+1) + \frac{\phi^T(t)F(t)\phi(t)}{1+\phi^T(t)F(t)\phi(t)}[\varepsilon^0(t+1) - e(t+1)]\\ &\quad + \frac{\phi^T(t)F(t)\phi(t)}{1+\phi^T(t)F(t)\phi(t)}e(t+1)\end{aligned} \tag{11.167}$$

and using now the Schwarz inequality one has:

$$\hat{y}^2(t+1) \le 3y^{*^2}(t+1) + 3[\varepsilon^0(t+1) - e(t+1)]^2 + 3e^2(t+1) \tag{11.168}$$

from which one obtains:

$$\frac{1}{N}\sum_{t=1}^{N} \hat{y}^2(t+1) \le \frac{C_5}{N}\sum_{t=1}^{N}[\varepsilon^0(t+1) - e(t+1)]^2 + C_6 \tag{11.169}$$

Therefore (11.165), (11.166) and (11.169) allow one to conclude that $r(t)$ satisfies the condition (11.157) and therefore that (11.143), (11.144) and (11.145) are true if (11.146) is strictly positive real. This ends the proof of Theorem 11.4. □

## 11.5 Robust Direct Adaptive Control

### 11.5.1 The Problem

In the previous sections, the design of the direct adaptive controllers has been discussed under the assumptions that (called also the *ideal case*):

- the integer delay $d$ is known;
- upper bounds on the degrees of the polynomials $A$ and $B^*$ are known;
- the sign of $b_1$ is known.

When dealing with disturbances, it was assumed that they have a stochastic model of known structure (in particular an ARMAX model has been considered).

The fact that:

1. the true plant model order is assumed to be less or equal to the model order used for design (i.e., the plant model is in the *model set*),
2. the unmeasurable disturbances have a clear (nice) stochastic model,

constitute two major drawbacks of the analysis discussed in the previous section from a practical point of view. Examples are available in the literature (Egardt 1979; Rohrs et al. 1981; Ioannou and Kokotovic 1983) showing that either particular form of unmodeled dynamics or special types of disturbances can cause instability of direct adaptive control schemes. For a detailed review see Ortega and Tang (1989). Therefore, it is extremely important from a practical point of view, to assess the robustness of direct adaptive controller in the presence of:

- unmodeled dynamics (the plant model is not in the model set);
- bounded disturbances (without a specific stochastic model).

Two points of view can be considered:

1. analysis of the robustness of the "ideal case" designs,
2. introduction of modifications in the parameter adaptation algorithms in order to improve the robustness of the "ideal case" designs (robust adaptation).

The first approach is extremely interesting for a deep understanding of the mechanisms leading to the instability in direct adaptive control in the presence of unmodeled dynamics and bounded disturbances (see Ortega and Tang 1989). We will focus on the use of robust parameter adaptation algorithms discussed in Chap. 10, in order to enhance the robustness of direct adaptive control schemes. While one can argue that these modifications are not always necessary (in theory), they definitely enlarge the set of situations which can be handled. In particular, two techniques will be used: PAA with dead zone and PAA with normalization. The presentation is mainly based on Ortega and Lozano-Leal (1987), Lozano and Ortega (1987), Ortega et al. (1985), Ortega (1993), Kreisselmeier and Anderson (1986). Use of PAA with projection can also be considered as shown in Ydstie (1989).

### 11.5.2 Direct Adaptive Control with Bounded Disturbances

Consider the plant model (11.1) in the presence of a disturbance $w'(t+1)$:

$$A(q^{-1})y(t) = q^{-d-1}B^*(q^{-1})u(t) + w'(t+1) \tag{11.170}$$

To simplify the presentation, we will consider the case $d = 0$ and we will use a fixed adaptation gain $F$. Using the polynomial equation:

$$A(q^{-1})S'(q^{-1}) + q^{-1}R(q^{-1}) = P(q^{-1}) \tag{11.171}$$

The filtered predicted output in the presence of disturbances can be written (see also (11.11) through (11.13)) as:

$$\begin{aligned} P(q^{-1})y(t+1) &= \theta_C^T\phi_C(t) + S'(q^{-1})w'(t+1) \\ &= \theta_C^T\phi_C(t) + w(t+1) \end{aligned} \tag{11.172}$$

where:

$$\theta_C^T = [s_0, \dots, s_{n_S}, r_0, \dots, r_{n_R}] \tag{11.173}$$

$$\phi_C^T = [u(t), \dots, u(t-n_S), y(t), \dots, y(t-n_R)] \tag{11.174}$$

$\theta_C$ defines the parameter vector of the tuned controller. Its components are the coefficients of the polynomials $R(q^{-1})$ and $S(q^{-1}) = S'(q^{-1})B^*(q^{-1})$. We will assume that a bound for the disturbance $w(t+1)$ is known, i.e.:

$$|w(t+1)| < \Delta \tag{11.175}$$

The controller parameters according to (11.7) will be given by:

$$\hat{\theta}_C^T(t)\phi_C(t) = P(q^{-1})y^*(t+1) \tag{11.176}$$

The tracking error is defined as in (11.14):

$$\varepsilon^0(t+1) = Py(t+1) - Py^*(t+1) = \tilde{\theta}_C^T(t)\phi_C(t) + w(t+1) \tag{11.177}$$

where:

$$\tilde{\theta}_C(t) = \theta_C - \hat{\theta}_c(t) \tag{11.178}$$

Consider the following PAA (Ortega and Lozano-Leal 1987):

$$\hat{\theta}_C(t+1) = \hat{\theta}_C(t) + \frac{F\phi_C(t)}{1+\phi_C^T(t)F\phi_C(t)} f[\varepsilon^0(t+1)]; \quad F > 0 \tag{11.179}$$

$$f[\varepsilon^0(t+1)] = \begin{cases} \varepsilon^0(t+1) - \Delta & \text{if } \varepsilon^0(t+1) > \Delta \\ 0 & \text{if } |\varepsilon^0(t+1)| \le \Delta \\ \varepsilon^0(t+1) + \Delta & \text{if } \varepsilon^0(t+1) < -\Delta \end{cases} \tag{11.180}$$

where $\varepsilon^0(t+1)$ is given by (11.177). This is a slight modification of the algorithm considered in Sect. 10.4. Note also that for $\Delta = 0$ it is exactly the one used in the ideal case (Sect. 11.2). The basic property of this algorithm which will be used for the analysis of the scheme is that:

$$[\varepsilon^0(t+1) - w(t+1)]f[\varepsilon^0(t+1)] \geq f^2[\varepsilon^0(t+1)]; \quad \forall t \tag{11.181}$$

Using this algorithm one has the following results summarized in Theorem 11.5.

**Theorem 11.5** *Consider the plant model* (11.170) *and its predictor form* (11.172), *the adjustable controller* (11.176) *and the PAA given by* (11.179) *and* (11.180). *Assume that*:

1. *The integer delay $d$ is known.*
2. *Upper bounds on the degrees of the polynomials $A$ and $B^*$ are known.*
3. *For all possible values of the plant parameters, the polynomial $B^*$ has all its zeros inside the unit circle.*
4. *The sign of $b_1$ is known.*
5. *An upper bound for the disturbance $w(t+1)$ in* (11.172) *is known.*

*Then*:

$$\text{(i)} \qquad \lim_{t\to\infty} \sup \|\varepsilon^0(t)| \leq \Delta \tag{11.182}$$

$$\text{(ii)} \qquad \|\tilde{\theta}_C(t)\|_{F^{-1}} < M < \infty; \quad \forall t \tag{11.183}$$

$$\text{(iii)} \qquad \lim_{t\to\infty} \|\hat{\theta}_C(t+1) - \hat{\theta}_C(t)\| = 0 \tag{11.184}$$

$$\text{(iv)} \qquad \|\phi_C(t)\| < M < \infty \tag{11.185}$$

*Proof* In the proof we will need an equation for the evolution of $\phi_C(t)$. This was already established in Sect. 11.2 taking into account that the system has all its zeros inside the unit circle (stable inverse). From (11.31) one has:

$$\|\phi_C(t)\|_F \leq c_1 + c_2 \max_{0<k<t+1} |\varepsilon^0(k+1)|; \quad 0 < c_1,\ c_2 < \infty \tag{11.186}$$

Let us now consider the evolution of the parameter error $\tilde{\theta}_C(t)$. From (11.178) and (11.179) one has:

$$\tilde{\theta}_C(t+1) = \tilde{\theta}_C(t) - \frac{F\phi_C(t)}{1 + \phi_C^T(t)F\phi_C(t)} f[\varepsilon^0(t+1)] \tag{11.187}$$

which allows to write:

$$\tilde{\theta}_C^T(t+1)F^{-1}\tilde{\theta}_C(t+1) = \tilde{\theta}_C^T(t)F^{-1}\tilde{\theta}_C(t) - 2\frac{\tilde{\theta}_C^T(t)\phi_C(t)f[\varepsilon^0(t+1)]}{1+\phi_C^T(t)F\phi_C(t)} + \frac{\phi_C^T(t)F\phi_C(t)}{[1+\phi_C^T(t)F\phi_C(t)]^2}f^2[\varepsilon^0(t+1)] \quad (11.188)$$

But from (11.177):

$$\tilde{\theta}_C^T(t)\phi_C(t) = \varepsilon^0(t+1) - w(t+1) \quad (11.189)$$

and using (11.181), one gets:

$$\tilde{\theta}_C^T(t)\phi_C(t)f[\varepsilon^0(t+1)] \geq f^2[\varepsilon^0(t+1)] \quad (11.190)$$

Therefore:

$$\|\tilde{\theta}_C^T(t+1)\|^2_{F^{-1}} \leq \|\tilde{\theta}_C^T(t)\|^2_{F^{-1}} - \frac{f^2[\varepsilon^0(t+1)]}{1+\phi_C^T(t)F\phi_C(t)} \quad (11.191)$$

As a consequence, the sequence $\{\|\hat{\theta}_C(t+1)\|^2_{F^{-1}}\}$ is a nonincreasing positive sequence bounded below by zero and thus converges. Since $F$ is positive definite, (11.183) results. Equation (11.191) also implies that:

$$\lim_{t\to\infty}(\|\tilde{\theta}_C(t+1)\|^2_{F^{-1}} - \|\tilde{\theta}_C(t)\|^2_{F^{-1}}) = 0 \quad (11.192)$$

and therefore (11.184) holds. From (11.191) one also has that:

$$\lim_{t\to\infty}\frac{f^2[\varepsilon^0(t+1)]}{1+\phi_C^T(t)F\phi_C(t)} = 0 \quad (11.193)$$

Taking into account (11.180), (11.186) can be expressed as:

$$\|\phi_C^T(t)\|_F \leq c_1 + c_2 \max_{0<k<t+1}[|f[\varepsilon^0(t+1)]| + \Delta] \leq c_1' + c_2 \max_{0<k<t+1}|f[\varepsilon^0(k+1)]| \quad (11.194)$$

Using now (11.193), (11.194) and the "bounded growth" lemma (Lemma 11.1) one concludes that (11.182) and (11.185) are true, which ends the proof. □

*Remark* The result can be extended for the case $d > 1$ (Ortega and Lozano-Leal 1987).

### 11.5.3 Direct Adaptive Control with Unmodeled Dynamics

Consider again the equation of the filtered predicted output given in (11.172). In this case, it will be assumed that the disturbance is defined by:

$$|w(t+1)| \leq d_1 + d_2\eta(t), \quad 0 < d_1,\ d_2 < \infty \tag{11.195}$$

$$\eta^2(t) = \mu^2\eta^2(t-1) + \|\phi_C(t)\|^2 \tag{11.196}$$

where $d_1$ accounts for a bounded external disturbance and $d_2\eta(t)$ accounts for the equivalent representation of the effect of unmodeled dynamics. More details about the description of unmodeled dynamics can be found in Sect. 10.6 (the chosen representation for the unmodeled dynamics corresponds to Assumption B in Sect. 10.6). In this case, due to the effect of feedback the unmodeled dynamics in (11.172) are characterized by $H'^T(q^{-1})\phi(t)$ where $H'^T(q^{-1}) = S'(q^{-1})H^T(q^{-1})$ and it is assumed that $H'^T(z^{-1})$ is analytic in $|z| < \mu < 1$ and $d_2 = \|H'(\mu^{-1}z^{-1})\|_\infty$.

It should be noted that even if the parameters of the reduced order plant model are perfectly known, there is no fixed linear controller that could stabilize the plant for all possible value of $d_2$ (which corresponds to the $\mu$-scaled infinity norm of a transfer function depending on the system parameters and design objectives). Nevertheless one can design a controller that stabilizes the plant provided that $d_2 \leq d_2^*$ where $d_2^*$ is a threshold that again depends on the system parameters.

The assumptions made upon the system are:

1. The delay $d$ is known.
2. Upper bounds on the degrees of the polynomials $A$ and $B^*$ are known.
3. For all possible values of the plant parameters, the polynomial $B^*$ has all its zeros inside the unit circle.
4. The sign of $b_1$ is known.
5. The disturbances upper bounds $d_1$ and $d_2$ are known.

Define the normalized input-output variables as:

$$\bar{y}(t) = \frac{y(t+1)}{m(t)}; \qquad \bar{u}(t) = \frac{u(t)}{m(t)}; \qquad \bar{\phi}_C(t) = \frac{\phi_C(t)}{m(t)} \tag{11.197}$$

where:

$$m^2(t) = \mu^2 m^2(t-1) + \max(\|\phi_C(t)\|^2, 1), \qquad m(0) = 1, \quad 0 < \mu < 1 \tag{11.198}$$

The normalized predicted plant output is given by:

$$P(q^{-1})\bar{y}(t+1) = \theta_C\bar{\phi}_C(t) + \bar{w}(t+1) \tag{11.199}$$

where:

$$\bar{w}(t+1) = \frac{w(t+1)}{m(t)} \tag{11.200}$$

The normalized a priori filtered tracking error is given by:

$$\begin{aligned}\bar{\varepsilon}^0(t+1) &= P(q^{-1})\bar{y}(t+1) - \hat{\theta}_C^T(t)\bar{\phi}_C(t) + \bar{w}(t+1)\\ &= \tilde{\theta}_C^T(t)\bar{\phi}_C(t) + \bar{w}(t+1)\end{aligned} \tag{11.201}$$

Consider the following PAA:

$$\hat{\theta}_C(t+1) = \hat{\theta}_C(t) + \frac{F\bar{\phi}_C(t)}{1+\bar{\phi}_C^T(t)F\bar{\phi}_C(t)} f[\bar{\varepsilon}^0(t+1)], \quad F > 0 \tag{11.202}$$

$$f[\bar{\varepsilon}^0(t+1)] = \begin{cases} \bar{\varepsilon}^0(t+1) - \bar{\delta}(t+1) & \text{if } \bar{\varepsilon}^0(t+1) > \bar{\delta}(t+1)\\ 0 & \text{if } |\bar{\varepsilon}^0(t+1)| \le \bar{\delta}(t+1)\\ \varepsilon^0(t+1) + \bar{\delta}(t+1) & \text{if } \bar{\varepsilon}^0(t+1) < -\bar{\delta}(t+1)\end{cases} \tag{11.203}$$

where $\bar{\delta}(t+1)$ is given by:

$$\bar{\delta}(t+1) = d_2 + \frac{d_1}{m(t)} \ge |\bar{w}(t+1)| \tag{11.204}$$

and $\varepsilon^0(t+1)$ is given by (11.201).

**Theorem 11.6** *For the system* (11.170), (11.172), (11.176) *and* (11.177) *where the disturbances and the unmodeled dynamics are defined by* (11.195) *and* (11.196), *using the PAA given by* (11.202), (11.203) *and* (11.204) *with the normalized variables defined in* (11.197) *and* (11.198), *one has*:

$$\text{(i)} \quad \lim_{t\to\infty} \sup |\bar{\varepsilon}^0(t)| \le \bar{\delta}(t) \tag{11.205}$$

$$\text{(ii)} \quad \|\tilde{\theta}_c(t)\| < M < \infty; \quad \forall t \tag{11.206}$$

$$\text{(iii)} \quad \lim_{t\to\infty} \|\hat{\theta}_C(t+1) - \hat{\theta}_C(t)\| = 0 \tag{11.207}$$

$$\text{(iv)} \quad \varepsilon^0(t+1) = \bar{\varepsilon}^0(t+1)m(t) \tag{11.208}$$

(v) *there is a value* $d_2 \le d_2^*$ *such that*:

$$\|\phi_C(t)\| < M_1 \quad \text{and} \quad |\varepsilon^0(t+1)| < M_2, \quad M_1, M_2 < \infty$$

*Proof* The proof of properties (i), (ii) and (iii) is similar to that of Theorem 11.5, except that one operates on normalized variables. One has:

$$\|\tilde{\theta}_C(t+1)\|^2_{F^{-1}} \le \|\tilde{\theta}_C(t)\|^2_{F^{-1}} - \frac{f^2[\bar{\varepsilon}^0(t+1)]}{1+\bar{\phi}_C^T(t)F\bar{\phi}_C(t)} \tag{11.209}$$

Similar to Theorem 11.5, one concludes that (ii) and (iii) hold. From (11.209) one also has:

$$\lim_{t\to\infty} \frac{f^2[\bar{\varepsilon}^0(t+1)]}{1+\bar{\phi}_C^T(t)F\bar{\phi}_C(t)} = 0 \tag{11.210}$$

and since $\|\bar{\phi}_C(t)\| \leq 1$, one concludes that:

$$\lim_{t\to\infty} f[\bar{\varepsilon}^0(t+1)] = 0 \tag{11.211}$$

which implies (11.205). From the definition of the normalized variables, one gets (11.208). Equation (11.186) can also be rewritten as:

$$\begin{aligned}\|\phi_C(t+1)\|^2 &\leq c_1' + c_2' \max_{0<k<t+1} |\varepsilon^0(k+1)|^2 \\ &= c_1' + c_2' \max_{0<k<t+1} |\bar{\varepsilon}^0(k+1)|^2 m^2(k)\end{aligned} \tag{11.212}$$

Assume now that $\phi_C(t+1)$ diverges. It follows that there exists a subsequence such that for this subsequence $t_1, t_2, \ldots, t_n$, one has:

$$\|\phi_C(t_1)\| \leq \|\phi_C(t_2)\| \leq \cdots \leq \|\phi_C(t_n)\| \tag{11.213}$$

or, equivalently:

$$\|\phi_C(t)\| \leq \|\phi_C(t_n)\|; \quad \forall t \leq t_n \tag{11.214}$$

In the meantime as $\phi_C(t)$ diverges, it follows that $\bar{\delta}^2(t) \to d_2^2$. Therefore:

$$|\varepsilon^0(t+1)|^2 \leq d_2^2 m^2(t) \leq d_2^2 m^2(t+1) \tag{11.215}$$

Introducing this result in (11.212), one gets:

$$\|\phi_C(t+1)\|^2 \leq c_1' + c_2' d_2^2 m^2(t+1) \tag{11.216}$$

and, respectively:

$$\frac{\|\phi_C(t+1)\|^2 - c_1'}{m^2(t+1)} \leq c_2' d_2^2 \tag{11.217}$$

The LHS of (11.217) converges toward a positive number $\rho \leq 1$ as $\phi_C(t+1)$ diverges $(m^2(t+1) \leq \gamma + \|\phi(t+1)\|^2, \gamma > 0)$. Therefore, there are values of $d_2 \leq d_2^*$ such that $c_2' d_2^2 < 1$ which leads to a contradiction. Therefore, for $d_2 \leq d_2^*$, $\|\phi_C(t+1)\|$ is bounded and from (11.208), it results also that $\varepsilon^0(t+1)$ is bounded. □

Since $d_2$ corresponds to the "$\mu$-scaled infinity norm" of the unmodeled dynamics $H(z^{-1})$ filtered by $S'(z^{-1})$, one gets the condition that boundedness of $\phi_C(t+1)$ is assured for:

$$\|S'(\mu^{-1}z^{-1}) \cdot H(\mu^{-1}z^{-1})\|_\infty < d_2^* \tag{11.218}$$

In the above approach, the unmodeled response of the plant to be controlled has been represented as an equivalent disturbance. Using normalization it was possible to convert the problem to the bounded disturbance case and using a PAA with dead zone, boundedness of the various signals is assured under a certain condition upon the unmodeled dynamics.

One may question if the PAA with dead zone is really necessary in order to handle unmodeled dynamics in direct adaptive control (even if a small dead zone is always useful in practice). The answer is that it is not always necessary and this can be shown using an equivalent feedback representation of the adaptive systems in the presence of unmodeled dynamics (Ortega et al. 1985; Ortega 1993).

In what follows we will concentrate on the main steps of the analysis and on the interpretation of the results for the design of robust direct adaptive controllers. The proofs of some intermediate steps are omitted. The reader is invited to consult (Ortega et al. 1985; Ortega 1993) for further details. To simplify the presentation we will consider:

- delay $d = 0$;
- no filters on the regressor vector or on the tracking error;
- PAA with constant adaptation gain;
- the unmodeled dynamics is represented by a multiplicative uncertainty on the input.

The plant model will be represented by:

$$A(q^{-1})y(t+1) = B^*(q^{-1})[1 + G(q^{-1})]u(t) \tag{11.219}$$

where $G(q^{-1})$ accounts for the unmodeled dynamics (multiplicative uncertainty). The filtered predicted output is given by:

$$Py(t+1) = (AS' + q^{-1}R)y(t+1) \tag{11.220}$$

where $S'$ and $R$ are solutions of the polynomial equation (11.171). Taking into account (11.219), one has from (11.220):

$$Py(t+1) = Su(t) + SGu(t) + Ry(t) \tag{11.221}$$

where $S = S'B^*$. From the definition of $\varepsilon^0(t+1)$ given in (11.177) and using (11.176), one gets:

$$\varepsilon^0(t+1) = (1+G)Su(t) + Ry(t) - \hat{\theta}_C^T(t)\phi_C(t) \tag{11.222}$$

Bearing in mind that:

$$Su(t) + Ry(t) = \theta_C^T \phi_C(t) \tag{11.223}$$

and adding and subtracting the terms: $\pm GRy(t), \pm G\hat{\theta}_C^T(t)\phi_C(t)$, (11.222) becomes:

$$\varepsilon^0(t+1) = -[1+G]\tilde{\theta}_C^T(t)\phi(t) - GRy(t) + GPy^*(t+1) \tag{11.224}$$

where:

$$\tilde{\theta}_C(t) = \hat{\theta}_C(t) - \theta_C \tag{11.225}$$

From (11.177), one has that:

$$Py(t+1) = \varepsilon^0(t+1) + Py^*(t+1) \tag{11.226}$$

and, therefore:

$$y(t+1) = \frac{1}{P}\varepsilon^0(t+1) + y^*(t+1) \tag{11.227}$$

Using this in (11.224), one gets:

$$\begin{aligned}\varepsilon^0(t+1) = &-[1+G]\tilde{\theta}_C^T(t)\phi(t) - GR_P\varepsilon^0(t+1)\\ &- GR_P Py^*(t+1) + GPy^*(t+1)\end{aligned} \tag{11.228}$$

where:

$$R_P(q^{-1}) = \frac{q^{-1}R}{P} \tag{11.229}$$

Rearranging the various terms in (11.228), it results that:

$$\begin{aligned}\varepsilon^0(t+1) &= -\left(\frac{1+G}{1+GR_P}\right)\tilde{\theta}_C^T(t)\phi(t) + G\frac{[1-R_P]}{1+GR_P}Py^*(t+1)\\ &= -H_1(q^{-1})\tilde{\theta}_C^T(t)\phi(t) + e^*(t+1)\end{aligned} \tag{11.230}$$

where:

$$H_1(q^{-1}) = \frac{1+G}{1+GR_P}; \qquad e^*(t+1) = (H_1-1)Py^*(t+1) \tag{11.231}$$

Equation (11.230) defines a linear block for an equivalent feedback representation of the system (since $\tilde{\theta}_C(t)$ is a function of $\varepsilon^0(t)$) where $H_1(q^{-1})$ represents the transfer operator of this linear block and $e^*(t+1)$ is an exogenous input which is bounded if $(H_1-1)$ is asymptotically stable. One considers the following normalized parameter adaptation algorithm:

$$\hat{\theta}(t+1) = \hat{\theta}(t) + f\bar{\phi}_C(t)\bar{\varepsilon}^0(t+1); \quad 0 < f < 2 \tag{11.232}$$

where:

$$\bar{\phi}_C(t) = \frac{\phi_C(t)}{m(t)}; \qquad \bar{\varepsilon}^0(t+1) = \frac{\varepsilon^0(t+1)}{m(t)} \tag{11.233}$$

$$\begin{aligned}m^2(t) &= \mu^2 m^2(t-1) + \max(\|\phi_C(t)\|^2, 1)\\ m(0) &= 1; \quad 0 < \mu < 1\end{aligned} \tag{11.234}$$

Associated with (11.232) one can define a system with input $\bar{\varepsilon}^0(t+1)$ and output $\tilde{\theta}_C^T(t)\bar{\phi}_C(t)$:

$$\tilde{\theta}_C^T(t)\bar{\phi}_C(t) = H_2\varepsilon^0(t+1) \tag{11.235}$$

where $H_2$ defines the input-output operator. Using the normalized adaptation algorithm one can establish the following result:

**Lemma 11.4** *For the PAA of* (11.232), *one has*:

$$\sum_{0}^{t_1} \tilde{\theta}_C^T(t)\bar{\phi}_C(t)\bar{\varepsilon}^0(t+1) \geq -\frac{f}{2}\sum_{0}^{t_1} \bar{\varepsilon}^0(t+1)^2 - \gamma_2^2; \quad \gamma_2^2 < \infty \tag{11.236}$$

*Remark* The system with the input $\bar{\varepsilon}^0(t+1)$ and the output $\tilde{\theta}_C^T(t)\bar{\phi}_C(t)$ has a lack of input passivity. However, the system with output $\tilde{\theta}_C^T(t)\bar{\phi}_C(t) + (f/2)\varepsilon^0(t+1)$ is passive.

*Proof* From (11.225) and (11.232) one has:

$$\begin{aligned} \|\tilde{\theta}_C(t+1)\|^2 &= \|\tilde{\theta}_C(t)\|^2 + 2f\tilde{\theta}_C^T(t)\bar{\phi}_C(t)\bar{\varepsilon}^0(t+1) \\ &\quad + f^2\|\bar{\phi}_C(t)\|^2\bar{\varepsilon}^0(t+1)^2 \end{aligned} \tag{11.237}$$

$$\begin{aligned} \tilde{\theta}_C^T(t)\phi_C(t)\varepsilon^0(t+1) &= -\frac{f}{2}\|\bar{\phi}_C(t)\|^2\bar{\varepsilon}^0(t+1)^2 + \frac{1}{2f}\|\tilde{\theta}_C(t+1)\|^2 \\ &\quad - \frac{1}{2f}\|\tilde{\theta}_C(t)\|^2 \end{aligned} \tag{11.238}$$

Summing up from 0 to $t_1$ and taking into account that $\|\bar{\phi}_C(t)\| \leq 1$ one gets (11.236). □

One can now consider a normalized version of the system (11.230) with input $-\tilde{\theta}_C^T(t)\bar{\phi}_C(t)$ and output $\bar{\varepsilon}^0(t+1)$, i.e.:

$$\bar{\varepsilon}^0(t+1) = -\bar{H}_1\tilde{\theta}_C^T(t)\bar{\phi}_C(t) + \bar{e}^*(t+1) \tag{11.239}$$

where $\bar{H}_1$ is an input-output operator whose properties are related to those of $H_1(q^{-1})$. Systems (11.239) and (11.235) form a feedback interconnection. This is shown in Fig. 11.3 where the detailed structure of (11.239) is emphasized. One observes the presence of multipliers in the feedforward path, as well as the presence of a delay in the feedback path (this is why the PAA is not passive). The use of multipliers in analysis of stability of feedback systems goes back to the Popov criterion (Popov 1960) and is part of the "loop transformations" used in the analysis of feedback systems (Desoer and Vidyasagar 1975). Provided that the multipliers and their inverse are bounded causal operators, the stability of the equivalent system will imply the stability of the system without multipliers, from which it will be possible to conclude upon the boundedness of $\phi_C(t)$ and $\varepsilon^0(t+1)$.

The proof is divided in two major steps:

1. Analysis of the stability of the normalized equivalent feedback system (Fig. 11.3).
2. Proof of boundedness of $\phi_C(t)$ and of the multipliers and their inverse which implies in fact the stability of the feedback system without multipliers.

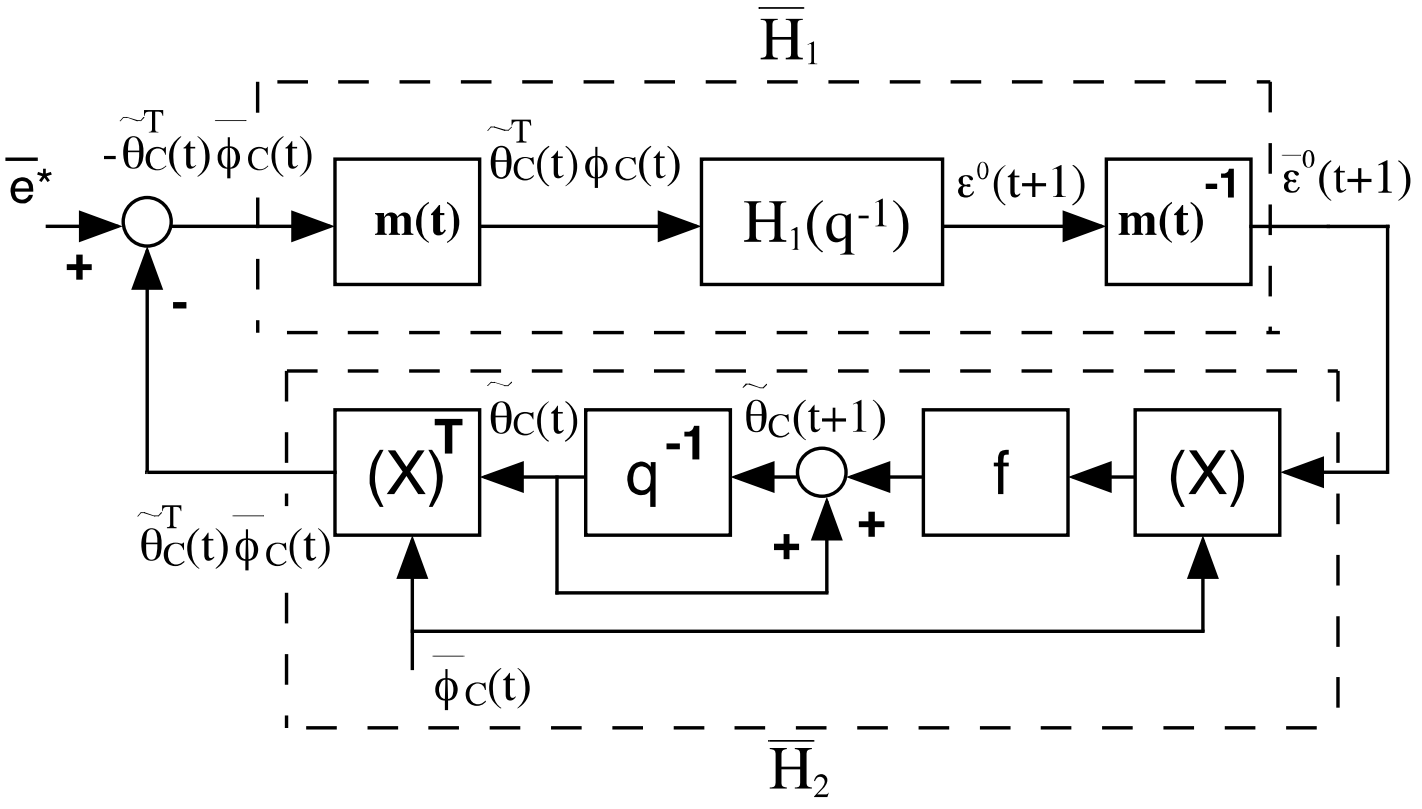

**Fig. 11.3** Equivalent feedback representations of the direct adaptive control scheme in the presence of unmodeled dynamics when using data normalization

We will limit ourselves to the point 1 of the proof. To do this we need an intermediate result upon the passivity of the system (11.239).

**Lemma 11.5** *For the system* (11.239) *with input* $-\tilde{\theta}^T(t)\bar{\phi}_C(t)$ *and output* $\bar{\varepsilon}^0(t+1)$, *one has for* $\bar{e}^*(t) = 0$:

$$-\sum_0^t \tilde{\theta}_C^T(t)\bar{\phi}_C(t)\bar{\varepsilon}^0(t+1) \geq \delta \sum_0^t \bar{\varepsilon}^0(t+1)^2 + \kappa \sum_0^t \hat{\theta}_C^T(t)\bar{\phi}_C(t) - \gamma_1^2$$

$$\delta > 0,\ \kappa > 0,\ \gamma_1^2 < \infty \tag{11.240}$$

*if* $H_1(\mu^{-1}z^{-1})$ *defines a very strictly passive system, i.e.*:

$$\mathrm{Re}\, H_1(\mu^{-1}e^{-j\omega}) \geq \delta |H_1(\mu^{-1}e^{-j\omega})|^2 + \kappa; \quad 0 < \delta,\ \kappa < \infty \tag{11.241}$$

Adding (11.236) and (11.240) one gets:

$$\left(\delta - \frac{f}{2}\right)\sum_0^t \bar{\varepsilon}^0(t+1)^2 + \kappa \sum_0^t [\hat{\theta}_C^T(t)\bar{\phi}_C(t)]^2 \leq \gamma_1^2 + \gamma_2^2 \tag{11.242}$$

and one concludes that if $\delta > f/2$:

$$\lim_{t\to\infty} \bar{\varepsilon}^0(t+1) = \lim_{t\to\infty} \tilde{\theta}_C^T(t)\bar{\phi}_C^T(t) = 0 \tag{11.243}$$

This analysis can be summarized as follows.

**Theorem 11.7** *Assume that there is a* $\delta > f/2$ *for which* (11.241) *holds, then*:

$$\|\bar{\varepsilon}^0(t+1)\|_2,\ \|\tilde{\theta}_C^T\bar{\phi}_C(t)\|_2 \leq c\|\bar{e}^*(t)\|_2 + \beta \tag{11.244}$$

*If, in addition*:

$$\|\bar{e}^*(t)\|_2 \leq \alpha < \infty \tag{11.245}$$

*then*:

$$\lim_{t\to\infty} \bar{\varepsilon}^0(t+1) = \lim_{t\to\infty} \tilde{\theta}_C^T(t)\tilde{\phi}_C^T(t) = 0 \tag{11.246}$$

In Ortega et al. (1985), Ortega (1993), it is shown that $\phi_C(t)$ is bounded and this allows to conclude that the multiplier $m(t)^{-1}$ and its inverse $m(t)$ are bounded, which leads to conclusion that:

$$\|\varepsilon^0(t+1)\|_2, \|\tilde{\theta}_C^T(t)\phi_C(t)\|_2 \leq c\|e^*(t)\|_2 + \beta \tag{11.247}$$

The major implications of this result are:

1. The PAA with dead zone may not be necessary in direct adaptive control in the presence of unmodeled dynamics.
2. It gives a direct relationship between on the one hand, the characteristics of the unmodeled dynamics and of the linear design (see the expression of $H_1(q^{-1})$) and, on the other hand, the value of the adaptation gain. This relationship has been deeply investigated in Cluett et al. (1987) for analysis and design of a robust direct adaptive control scheme.

The condition (11.241) combined with the condition $\delta > f/2$ has a very nice graphical interpretation which is very useful for analysis and design of robust direct adaptive controllers. To do this, we will make first the observation that based on Lemma 11.5 the feedforward block in Fig. 11.3 is very strictly passive provided that the transfer function $H_1(\mu^{-1}z^{-1})$ is strictly positive real and satisfies (11.241). Therefore, Fig. 11.3 takes the form shown in Fig. 11.4a where the feedback pass has a lack of input passivity which can be compensated by adding in parallel a block with a gain $f/2$. The resulting equivalent scheme is shown in Fig. 11.4b. It can be easily checked (see Lemma 11.4) that the new system with input $\bar{\varepsilon}_0(t+1)$ and output $\tilde{\theta}_C^T(t)\bar{\phi}_C(t) + \frac{f}{2}\bar{\varepsilon}^0(t+1)$ is passive.

Using the asymptotic hyperstability theorem (see Appendix C), asymptotic stability (for $\|\bar{e}^*\|_2 < \infty$) is assured if the equivalent new feedforward path is characterized by a strictly positive real transfer function, i.e.:

$$H_1'(\mu^{-1}z^{-1}) = \frac{H_1(\mu^{-1}z^{-1})}{1 - \frac{f}{2}H_1(\mu^{-1}z^{-1})} = SPR \tag{11.248}$$

But using the circle criterion (Zames 1966), this is equivalent to: $H_1(\mu^{-1}z^{-1})$ *should lie inside a circle centered on the real axis and passing through* $(0, \frac{2}{f})$ *with center* $c = 1/f$ *and radius* $r = 1/f$ (the original feedback path lies in a cone $(0, -\frac{f}{2})$). This result is illustrated in Fig. 11.5. Therefore, for assumed nominal reduced order model, desired closed-loop poles and assumed unmodeled dynamics one can:

- compute $H_1'(\mu^{-1}z^{-1})$ given by (11.231);
- find the value of $\mu$ for which $H_1(\mu^{-1}z^{-1})$ has its poles inside $|z| < \mu < 1$ (this value is used also for data normalization);
- trade between $\mu$ and $f$ in order that $H_1(\mu^{-1}z^{-1})$ lies inside the circle with $c = \frac{1}{f}, j0$ and $r = \frac{1}{f}$.

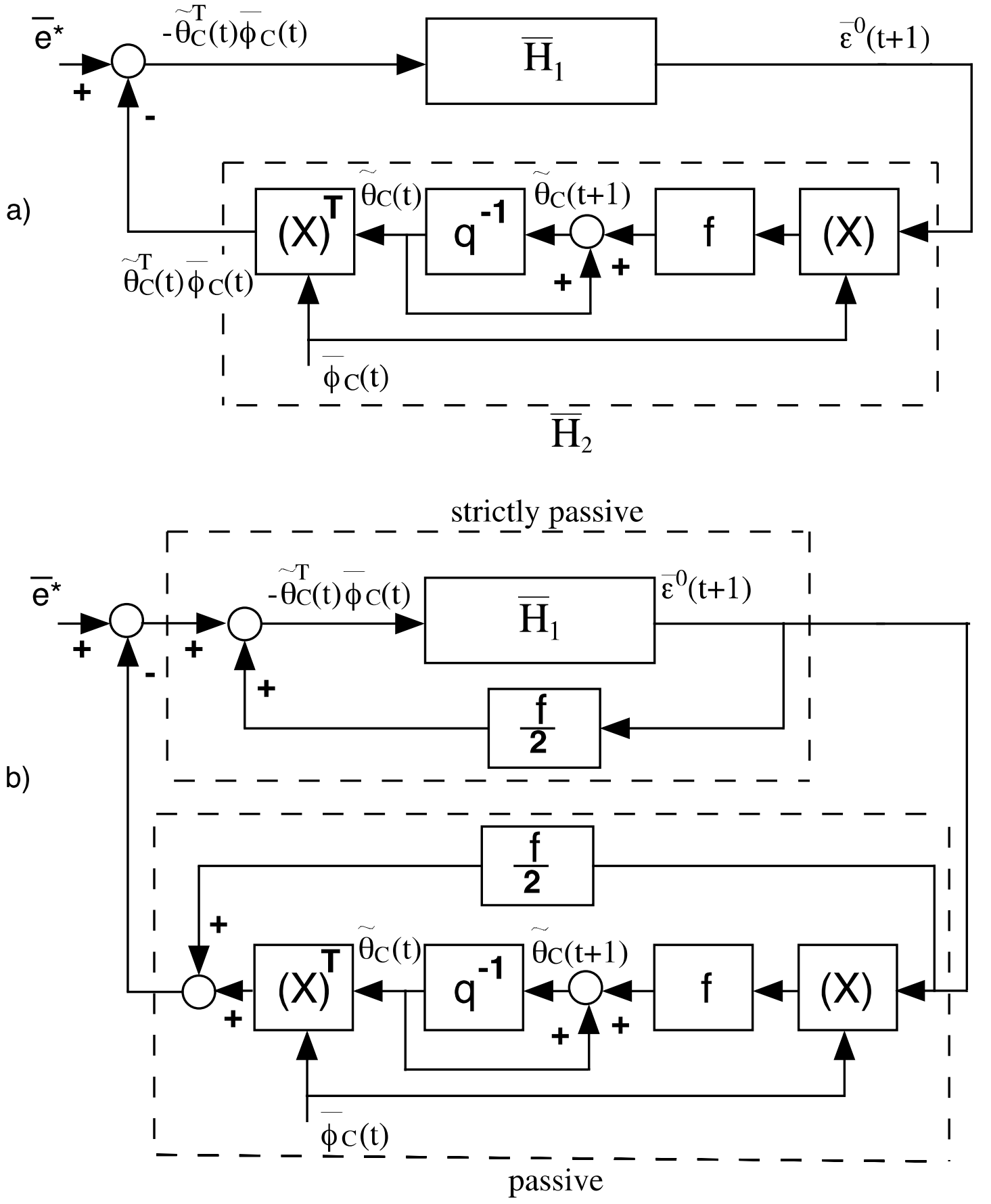

**Fig. 11.4** Equivalent representations for the feedback system of Fig. 11.6

**Fig. 11.5** Graphical interpretation of the condition (11.248)

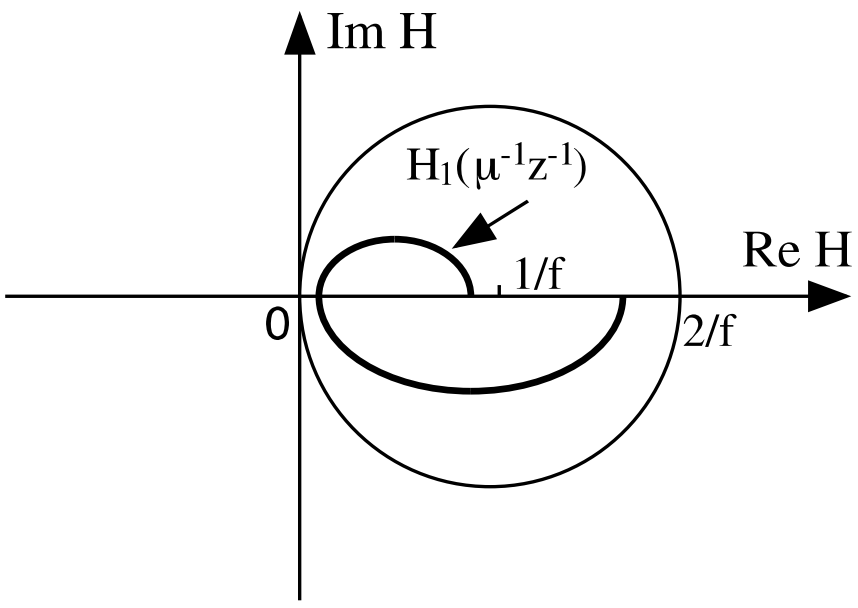

Note that as $f$ increases, the circle becomes smaller. Note also that in the absence of unmodeled dynamics $H_1(z^{-1}) = 1$ and the stability condition is always satisfied. From Fig. 11.5, one also concludes that either condition (11.248) or (11.241) plus the condition $\delta > f/2$ is equivalent to:

$$\left\| H_1(\mu^{-1}z^{-1}) - \frac{1}{f} \right\|_\infty < \frac{1}{f}$$

which can be rewritten as:

$$\| f H_1(\mu^{-1}z^{-1}) - 1 \|_\infty < 1$$

Since 1 inside the norm sign can be interpreted as the value of $H_1(z^{-1})$ for the "tuned" case without unmodeled dynamics, one concludes that the image of the unmodeled dynamics multiplied by $f$ should be relatively close to one at all frequencies.

## 11.6 An Example

In this example, we will illustrate the influence of the regulation dynamics (the polynomial $P(q^{-1})$) on the performance of the adaptive tracking and regulation with independent objectives (Sect. 11.2).

Two different plant models are considered. The plant model before a parameter change occurs is characterized by the discrete transfer operator:

$$G_1(q^{-1}) = \frac{q^{-2}(1 + 0.4q^{-1})}{(1 - 0.5q^{-1})[1 - (0.8 + 0.3j)q^{-1}][1 - (0.8 - 0.3j)q^{-1}]}$$

At time $t = k$, a change of the plant model parameters is made. The new plant model is characterized by the transfer operator:

$$G_2(q^{-1}) = \frac{q^{-2}(0.9 + 0.5q^{-1})}{(1 - 0.5q^{-1})[1 - (0.9 + 0.42j)q^{-1}][1 - (0.9 - 0.42j)q^{-1}]}$$

The simulations have been carried out for two different values of the regulation polynomial:

1. $P_1(q^{-1}) = 1$ (deadbeat control);
2. $P_2(q^{-1}) = 1 - 1.262q^{-1} + 0.4274q^{-2}$.

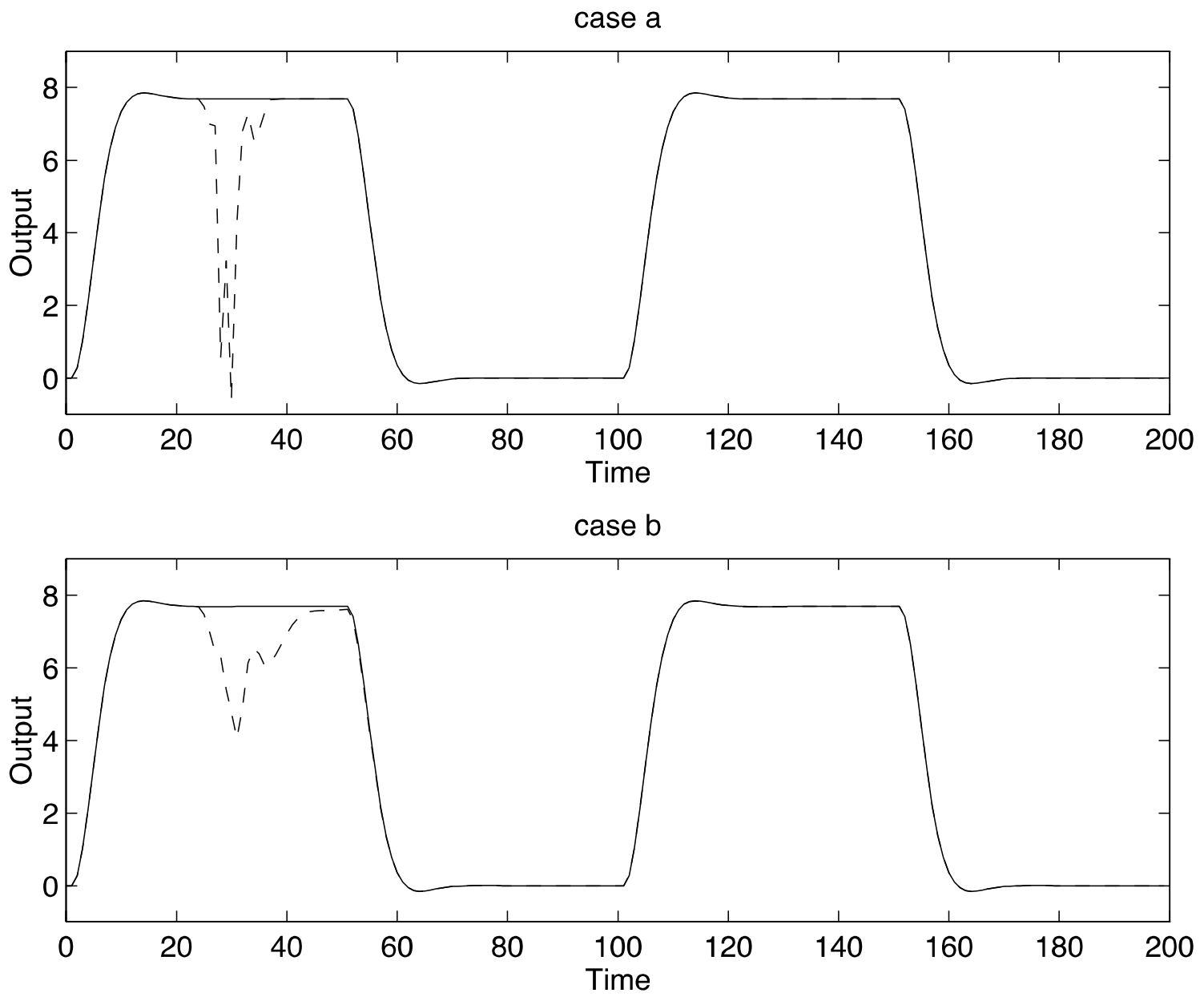

**Fig. 11.6** Tracking behavior in the presence of plant model parameters changes at $t = 25$: [–] desired output, [- -] achieved output. (**a**) Regulation dynamics $P(q^{-1}) = 1$, (**b**) regulation dynamics $P(q^{-1}) = 1 - 1.262q^{-1} + 0.4274q^{-2}$

$P_2(q^{-1})$ corresponds to the discretization ($T_S = 1$ s) of a continuous-time second-order system with $\omega_0 = 0.5$ rad/s and $\zeta = 0.85$.

The tracking reference model is characterized by:

$$\frac{B_m(q^{-1})}{A_m(q^{-1})} = \frac{(0.28 + 0.22q^{-1})}{(1 - 0.5q^{-1})[1 - (0.7 + 0.2j)q^{-1}][1 - (0.7 - 0.2j)q^{-1}]}$$

In all simulations a PAA with constant trace adaptation gain has been used with $\mathrm{tr}\, F(t) = \mathrm{tr}\, F(0)$, $F_0 = \mathrm{diag}[10]$ and $[\lambda_1(t)/\lambda_2(t)] = 1$.

Figure 11.6 shows the desired trajectory and the achieved trajectory during a sequence of step changes on the reference and in the presence of the parameter variations occurring at $t = 25$. One can observe that the adaptation transient is much smoother using the $P_2(q^{-1})$ regulation polynomial than in the case $P_1(q^{-1})$.

Figure 11.7 shows the behavior in regulation, i.e., the evolution of the output from an initial condition at $t = 0$ in the presence of a change in the parameters at $t = 0$. Same conclusion can be drawn: the poles defined by the regulation polynomial strongly influence the adaptation transient. Two fast dynamics in regulation with respect to the natural response of the system will induce undesirable adaptation transients.

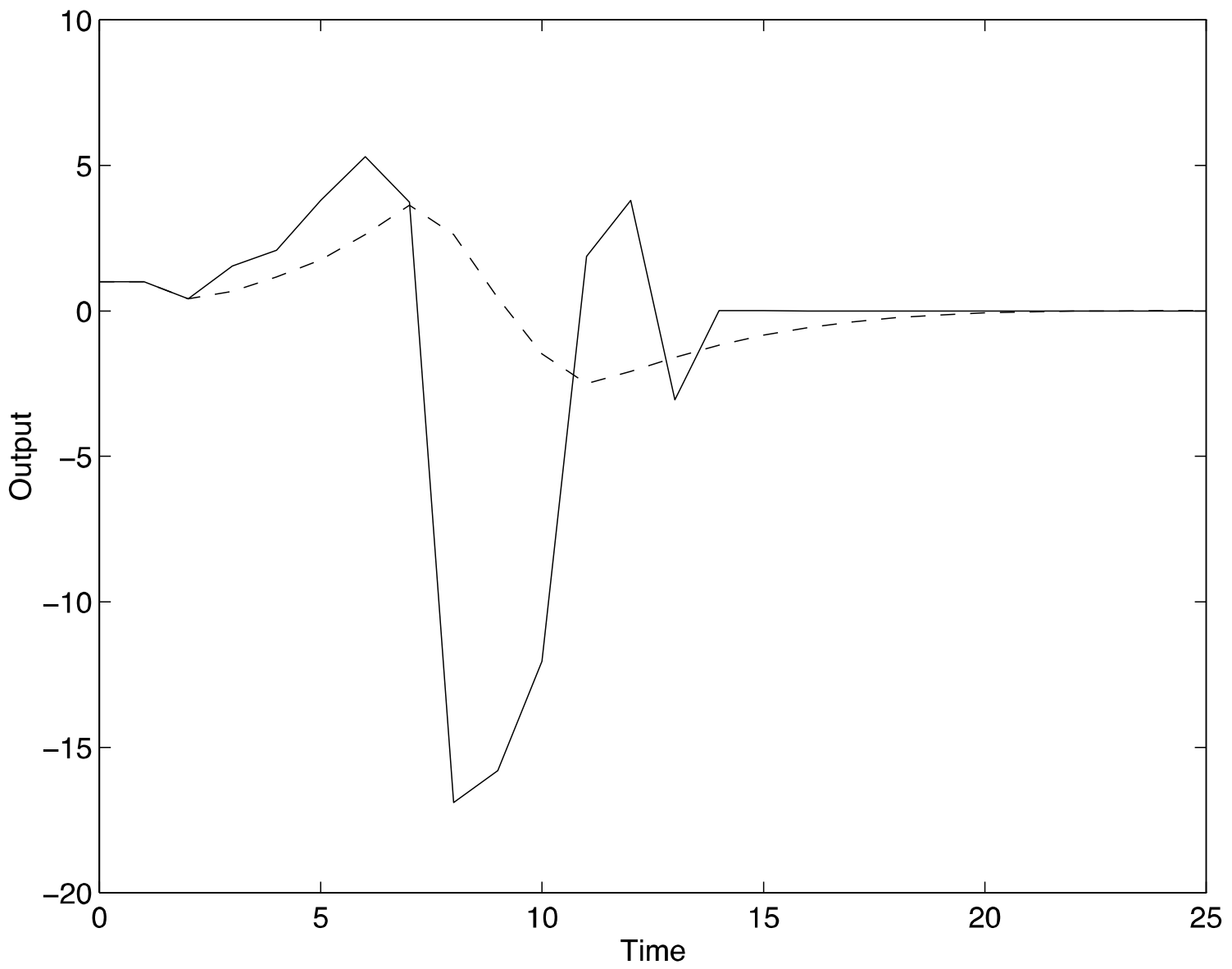

**Fig. 11.7** Regulation behavior in the presence of plant model parameters changes at $t = 0$. [–] Regulation dynamics $P(q^{-1}) = 1$, [- -] regulation dynamics $P(q^{-1}) = 1 - 1.262q^{-1} + 0.4274q^{-2}$

## 11.7 Concluding Remarks

The direct adaptive control algorithms presented in this chapter share a number of features which are summarized below.

1. Direct adaptive control schemes lead to simple implementations since the parameters of the controller are directly updated (however, for large integer delay $d$ it may be more convenient to use an indirect adaptive control scheme).
2. The specification of the desired performance is simple for adaptive tracking and regulation and adaptive minimum-variance tracking and regulation.
3. For adaptive tracking and regulation with weighted input and the adaptive generalized minimum variance, one cannot guarantee the resulting closed-loop performances (the resulting closed-loop poles will depend also upon the current values of $A(q^{-1})$ and $B(q^{-1})$).
4. The weakness of this approach is the fact that it can be used only for a restricted class of plants characterized by stable zeros, for the whole domain of possible variations of the plant model parameters.
5. In general, in the deterministic environment, one uses non-vanishing adaptation gains and in the stochastic environment one uses time-decreasing adaptation gains with resetting from time to time.

6. Robustification of the parameter adaptation algorithms may be necessary in the presence of unmodeled dynamics (i.e., when one uses for design a nominal model of lower order than the true one).
7. Most industrial applications up to 1986 have been accomplished using this approach and several industrial products incorporating these techniques have been developed, see Seborg et al. (1989), Dumont (1992), Åström and Wittenmark (1995).

## 11.8 Problems

**11.1** Give the details of the proof for the Theorem 11.5 using the averaging method (convergence analysis of the adaptive generalized minimum variance and regulation).

**11.2** Show that for a recursive least square one step ahead predictor estimating an ARMAX model:

$$A(q^{-1})y(t+1) = B^*(q^{-1})u(t) + C(q^{-1})e(t+1);\ n_A = n_C$$

(a) ${\theta^*}^T = [(c_1 - a_1), \dots, (c_{n_A} - a_{n_A}), b_1, \dots, b_{n_B}]$ is a possible convergence point.
(b) the sufficient condition for w.p.1 convergence towards

$$D_C = \{\hat{\theta} | [\theta^* - \hat{\theta}]^T \phi(t, \hat{\theta}) = 0\}$$

is the strict positive realness of $\frac{1}{C(z^{-1})} - \frac{\lambda_2}{2}$ for $2 > \lambda_2 \geq \sup_t \lambda_2(t)$.
(c) assuming that $[\theta^* - \hat{\theta}]^T \phi(t, \hat{\theta}) = 0$ has only one solution $\hat{\theta} = \theta^*$, the estimated parameters correspond to the coefficients of the minimum variance regulator.

Interpret the results and compare with the analysis given in Sect. 11.4.2.

**11.3** In the stochastic environment, for an ARMAX plant model with $d = 0$, define an adaptation error as:

$$\varepsilon^0(t+1) = y(t+1) - y^*(t+1)$$

Using the averaging method (Theorem 4.1) derive directly the adaptive minimum variance tracking and regulation scheme assuring:

$$\text{Prob}\left\{ \lim_{t\to\infty} \varepsilon^0(t+1) = \lim_{t\to\infty} [y(t+1) - y^*(t+1)] = e(t+1) \right\} = 1$$

**11.4** The adaptive tracking and regulation with independent objectives (Sect. 11.2) uses in the observation vector entering in the adaptation algorithm $y(t), y(t-1), \dots, u(t), u(t-1)$, which are directly corrupted by noise. In order to remove asymptotically the effect of the measurement noise, try to develop a direct adaptive control scheme which uses the ideas of closed-loop output error recursive identification algorithms (Sect. 9.2).

**11.5** Extend the results of Theorem 11.5 for robust direct adaptive control in the presence of bounded disturbances for the case of a plant with pure time delay $d \geq 1$.

**11.6** Consider the following plant model:

$$(1 + a_1 q^{-1}) y(t+1) = b_1 u(t) + b_2 u(t-1) = b_1 [1 + \beta q^{-1}] u(t)$$

For the case $\beta = \frac{b_2}{b_1} > 1$ ($b_1, b_2 > 0$), the system has an unstable zero. A robust direct adaptive control for the reduced order system with $B^*(q^{-1}) = b_1$ using data normalization is considered. Assuming that the nominal model is characterized by $a_1 = -0.5$, $b_1 = 1$, that the neglected dynamics is characterized by $\beta = 1.5$ or 2 and the desired closed-loop pole is characterized by $P(q^{-1}) = 1 + p_1 q^{-1}$ with $p_1 = -0.5$ or $-0.3$ find:

1. The value of $\mu$ to be used in the dynamic data normalization for the various values of $p_1$ and $\beta$ when using the PAA given in (11.234).
2. Discuss the influence of the values of the unmodeled dynamics and of the desired closed-loop values upon $\mu$.

Hint: Use Lemma 11.5.

**11.7** Consider the system:

$$A(q^{-1}) y(t+1) = q^{-d} B^*(q^{-1}) u(t)$$

Assume that $A(z^{-1})$ is asymptotically stable and well damped but $B^*(z^{-1})$ does not need to be stable.

Consider the pole placement strategy (internal model control—Sect. 7.3.4) where the polynomials $S(q^{-1})$ and $R(q^{-1})$ of the controller are solutions of:

$$A(q^{-1}) S(q^{-1}) + q^{-d-1} B^*(q^{-1}) R(q^{-1}) = A(q^{-1}) P_0(q^{-1})$$

Show that it is possible to build a direct adaptive control scheme in which one directly estimates the parameters of the controller when $A(q^{-1})$ and $B^*(q^{-1})$ are unknown.

**11.8** (Adaptive generalized minimum variance tracking and regulation.) Consider the control objective for $d = 0$:

$$\text{Prob}\left\{ \lim_{t\to\infty} \varepsilon^0(t+1) = \lim_{t\to\infty} \left[ y(t+1) + \lambda \frac{Q}{C} u(t) - y^*(t+1) \right] = e(t+1) \right\} = 1$$

1. Develop an adaptive control for the system given in (11.78).
   Hint: define the generalized output:

$$\bar{y}(t+1) = y(t+1) + \lambda \frac{Q}{C} u(t)$$

   and use the methodology presented in Sect. 11.4 (see also Sect. 11.3).
2. Analyse the asymptotic behavior of the scheme using the averaging method given in Sect. 4.2 (see also Sect. 11.4.2).

# Chapter 12
# Indirect Adaptive Control

## 12.1 Introduction

As indicated in Chap. 1, *indirect adaptive control* is a very general approach to adaptive control since, in principle, one can combine any parameter estimation scheme with any control strategy. The adaptation of the controller parameters is carried out in two steps:

Step 1: On-line estimation of the plant model parameters.
Step 2: On-line computation of the controller parameters based on the estimated plant model.

The resulting control scheme should guarantee that the input and the output of the plant remain bounded and that some indices of performance are achieved asymptotically. This requires a deep analysis of the indirect adaptive control system which is nonlinear and time-varying.

Fortunately, good properties of the indirect adaptive control schemes will be guaranteed if, separately, the parameter estimation algorithm and the control strategy satisfy a number of properties (largely similar to those required for indirect adaptive prediction) which are summarized next:

1. The a posteriori adaptation error in the parameter estimation algorithm goes to zero asymptotically.
2. The a priori adaptation error does not grow faster than the observation vector (containing the input and output of the plant).
3. The estimated plant model parameters are bounded for all $t$.
4. The variations of the estimated plant model parameters go to zero asymptotically.
5. The *design equation*(*s*) provide(s) bounded controller parameters for bounded plant parameter estimates.
6. The estimated plant model is *admissible* with respect to the *design equation* which means that assuming that it corresponds to the exact plant model, the resulting controller has bounded parameters, stabilizes the system and achieves the desired performances.

If these conditions are satisfied, it is then possible to show that the input and the output of the plant remain bounded and that some indices of performances are achieved asymptotically.

Probably, the most difficult problem (at least theoretically) is to guarantee that any estimated model is *admissible* with respect to the control design strategy. For every control strategy, even if the estimated plant parameters are bounded at each time $t$, the current estimated model may not be admissible in the sense that there is no solution for the controller. For example, if pole placement is used as the control strategy, an estimated model at time $t$, which features a pole-zero cancellation, will not allow to compute the controller. Similarly, using a LQ control, if the estimated model at time $t$ is not stabilizable the control cannot be computed. Furthermore, getting close to non-admissibility situations will lead to numerical problems resulting in very large and undesirable control actions. The consequence of this fact is twofold:

1. One has to be aware of the admissibility conditions and take appropriate ad hoc action in practice, in order to deal with the singularities which may occur during adaptation.
2. One can make a theoretical analysis and develop *modifications* of the parameter estimates (or of the algorithms) in order to avoid the singularities corresponding to the non-admissibility of an estimated plant model.

As indicated earlier (Property 6), to remove the singularities one has to assume that the unknown plant model is admissible with respect to the control strategy.

While the various *modifications* of the estimated parameters or of the algorithms resulting from a theoretical analysis are in general complex to implement, they have the merit of showing that there are solutions that avoid the *non-admissibility* of the estimated model and give hints for simple ad hoc modifications to be implemented.

The analysis of the indirect adaptive control schemes is carried out in two steps:

1. One assumes that the estimated models are always in the admissibility set.
2. One modifies the parameter estimation algorithm in order to satisfy the admissibility condition.

The objective of the analysis is to guarantee that a stabilizing controller is obtained without any assumption about the presence or the richness of an external excitation. However, the use of an external or internal excitation signal, even for short periods of time, is useful for speeding up the convergence of the adaptive control scheme since it can be shown that exponential stability is achieved under richness conditions (Anderson and Johnstone 1985).

To implement an indirect adaptive control strategy effectively, we have two major options. The choice is related to a certain extent to the ratio between the computation time and the sampling period.

**Strategy 1**

1. Sample the plant output;
2. Update the plant model parameters;

3. Compute the controller parameters based on the new plant model parameter estimates;
4. Compute the control signal;
5. Send the control signal;
6. Wait for the next sample.

Using this strategy, there is a delay between $u(t)$ and $y(t)$ which will essentially depend upon the time required to achieve (2) and (3). This delay should be small with respect to the sampling period and, of course, smaller than the delay between $u(t)$ and $y(t)$ scheduled in the $I/O$ system (see Chap. 16 for details). In this strategy, a posteriori parameter estimates are used and the a posteriori adaptation error will occur in the stability analysis.

**Strategy 2**

1. Sample the plant output;
2. Compute the control signal based on the controller parameters computed during the previous sampling period;
3. Send the control signal;
4. Update the plant model parameters;
5. Compute the controller parameters based on the new plant model parameter estimates;
6. Wait for the next sample.

Using this strategy, the delay between $u(t)$ and $y(t)$ is smaller than in the previous case. In fact, this is the strategy used with constant parameters controllers (Steps 4 and 5 are deleted). In this strategy, one uses a priori parameter estimates and the a priori adaptation error will appear in the stability analysis.

The analysis of the resulting schemes is very similar except that as indicated earlier, in Strategy 1 the a posteriori adaptation error will play an important role, while in Strategy 2, the properties of the a priori adaptation error will be used.

One uses a non-vanishing adaptation gain to get an indirect adaptive control scheme which can react to changes in plant model parameters. One uses a time-decreasing adaptation gain to implement indirect adaptive control schemes for the case of plant models with unknown but constant parameters (over a large time horizon). In the latter case, adaptation can be restarted either on demand or automatically, based on the analysis of an index of performance.

As indicated in Chap. 1, nothing stops us updating the estimates of the plant model parameters at each sampling instant, but updating the controller parameters only every $N$ sampling instants. The analysis remains the same as long as $N$ is finite. The use of this approach is related to:

- the possibility of getting better parameter estimates for control design,
- the eventual reinitialization of the plant parameters estimation algorithm after each controller updating,

- the use of more sophisticated control designs requiring complex computations (in particular robust control design),
- the reduction of the risk for getting non-admissible estimated plant models.

If the plant to be controlled has constant parameters over a large time horizon and if one considers a large horizon $N$ for plant model parameters estimation, followed by the updating of the controller based on the results of the plant model identification in closed loop, one gets what is called the *iterative identification in closed loop and controller redesign*. This can be considered as a limit case of indirect adaptive control. The main difference is that the plant model parameters estimation is done in closed loop in the presence of a linear controller with fixed parameters. This suggests that *among all the parameter estimation algorithms which allow us to obtain a stable indirect adaptive control scheme, it may be useful to use those dedicated to identification in closed loop* (see Chap. 9).

For the case of plant models with unknown but constant parameters over a large horizon, one may ask what is the best choice among:

- adaptive control with controller updating each sampling instant;
- adaptive control with controller updating each $N$ sampling instants;
- plant model identification in closed loop followed by the redesign of the controller.

The answer is an "engineering" type answer, since essentially a trade-off between computation power available and performances will, to a large extent, dictate the choice of one or another strategy.

Adaptive pole placement and adaptive generalized predictive control are probably the most popular indirect adaptive control strategies used in applications. We have chosen adaptive pole placement as the prototype for the analysis of the indirect adaptive control schemes (Sect. 12.2). The results can be applied *mutatis mutandis* to the other indirect adaptive control strategies. Robustness aspects related to the violation of the basic hypotheses will be examined in detail in Sect. 12.3. Adaptive generalized predictive control and adaptive linear quadratic control will be presented in Sects. 12.4 and 12.5, respectively.

A non-exhaustive list of references detailing various indirect adaptive control strategies is given below.

Adaptive pole placement (Goodwin and Sin 1984; Anderson and Johnstone 1985; de Larminat 1980, 1984, 1986; Guo 1996; Lozano and Goodwin 1985; Lozano 1992; Giri et al. 1990, 1989; de Larminat and Raynaud 1988; Middleton et al. 1988; Mareels and Polderman 1996)

Adaptive generalized predictive control (Clarke and Mohtadi 1989; Bitmead et al. 1990; Giri et al. 1991; M'Saad et al. 1993a)

Adaptive linear quadratic control (Samson 1982; Lam 1982; Ossman and Kamen 1987; Giri et al. 1991; M'Saad and Sanchez 1992)

## 12.2 Adaptive Pole Placement

### 12.2.1 The Basic Algorithm

We will assume that the system operates in a deterministic environment. The basic algorithm combines any of the parameter estimation algorithms presented in Chaps. 3, 5 and 9 with the pole placement control strategy presented in Sect. 7.3. The plant model (with unknown parameters) is assumed to be described by:

$$A(q^{-1})y(t) = q^{-d}B(q^{-1})u(t) = q^{-d-1}B^*(q^{-1})u(t) \tag{12.1}$$

where $u(t)$ and $y(t)$ are the input and the output of the plant and:

$$A(q^{-1}) = 1 + a_1 q^{-1} + \cdots + a_{n_A} q^{-n_A} = 1 + q^{-1}A^*(q^{-1})$$

$$B(q^{-1}) = b_1 q^{-1} + \cdots + b_{n_B} q^{-n_B} = q^{-1}B^*(q^{-1})$$

One assumes that:

- the orders of the polynomials $A(q^{-1})$, $B(q^{-1})$ and of the delay $d$ are known ($n_A, n_B, d$-known);[1]
- $A(q^{-1})$ and $B(q^{-1})$ do not have common factors (admissibility condition).

#### Step I: Estimation of the Plant Model Parameters

As indicated earlier, one can use any parameter estimation algorithm presented in Chaps. 3, 5 and 9 since we operate in a deterministic environment. However, one has to take into account the following:

1. Convergence conditions (positive real type), which are required for some algorithms, will become also convergence conditions for the adaptive control scheme.
2. It is reasonable to use parameter estimation algorithms inspired from those dedicated to the identification in closed loop.

In a deterministic environment, the parameter estimates and the adaptation errors will satisfy the properties resulting from Theorem 3.2 and they will fulfill the points 1 to 5 mentioned in Sect. 12.1. To simplify the analysis, we will assume that a *recursive least squares* type parameter estimation algorithm will be used. The cases of using recursive least squares on filtered data or appropriate algorithms for closed-loop identification are illustrated in the example presented in Sect. 12.4 and in the applications presented in Sect. 12.7.

The plant output can be expressed as:

$$y(t+1) = \theta^T \phi(t) \tag{12.2}$$

[1] What is in fact needed is the knowledge of $n_B + d$. However, the knowledge of $d$ reduces the number of estimated parameters.

where:

$$\theta^T = [a_1, \ldots, a_{n_A}, b_1, \ldots, b_{n_B}] \tag{12.3}$$

$$\phi^T(t) = [-y(t), \ldots, -y(t-n_A+1), u(t-d), \ldots, u(t-d-n_B+1)] \tag{12.4}$$

The a priori output of the adjustable predictor is given by:

$$\hat{y}^0(t+1) = \hat{\theta}^T(t)\phi(t) \tag{12.5}$$

The a posteriori output of the adjustable predictor is given by:

$$\hat{y}(t+1) = \hat{\theta}^T(t+1)\phi(t) \tag{12.6}$$

where:

$$\hat{\theta}^T(t) = [\hat{a}_1(t), \ldots, \hat{a}_{n_A}(t), \hat{b}_1(t), \ldots, \hat{b}_{n_B}(t)] \tag{12.7}$$

Accordingly, the a priori and the a posteriori prediction (adaptation) errors are given by:

$$\varepsilon^0(t+1) = y(t+1) - \hat{y}^0(t+1) \tag{12.8}$$

$$\varepsilon(t+1) = y(t+1) - \hat{y}(t+1) \tag{12.9}$$

The parameter adaptation algorithm is:

$$\hat{\theta}(t+1) = \hat{\theta}(t) + F(t)\phi(t)\varepsilon(t+1) \tag{12.10}$$

$$F(t+1)^{-1} = \lambda_1(t)F(t)^{-1} + \lambda_2(t)\phi(t)\phi^T(t);$$

$$0 < \lambda_1(t) \le 1;\ 0 \le \lambda_2(t) < 2;\ F(0) > 0 \tag{12.11}$$

$$\varepsilon(t+1) = \frac{\varepsilon^0(t+1)}{1 + \phi^T(t)F(t)\phi(t)} \tag{12.12}$$

It is also assumed that:

$$F(t)^{-1} \ge \alpha F(0)^{-1}; \quad F(0) > 0;\ \alpha > 0,\ \forall t \in [0, \infty] \tag{12.13}$$

Selection of $\lambda_1(t)$ and $\lambda_2(t)$ allows various forgetting profiles to be obtained (see Sect. 3.2.3) and this leads to:

- a time-decreasing adaptation gain;
- a time-decreasing adaptation gain with reinitialization;
- a non vanishing adaptation gain.

It also allows the introduction of the various modifications discussed in Chap. 10 in order to have a robust adaptation algorithm (this will be discussed in Sect. 12.3). Using this algorithm, one will have, according to Theorem 3.2, the following properties:

- The a posteriori adaptation error is bounded and goes to zero asymptotically, i.e.:

$$\lim_{t_1\to\infty}\sum_{t=1}^{t_1}\varepsilon^2(t+1)\le C<\infty \tag{12.14}$$

$$\lim_{t\to\infty}\varepsilon(t+1)=0 \tag{12.15}$$

- The a priori adaptation error satisfies:

$$\lim_{t\to\infty}\frac{[\varepsilon^0(t+1)]^2}{1+\phi^T(t)F(t)\phi(t)}=0 \tag{12.16}$$

- The parameter estimates are bounded for all $t$

$$\|\hat{\theta}(t)\|\le C<\infty;\quad \forall t\ge 0 \tag{12.17}$$

- The variations of the parameter estimates go to zero asymptotically:

$$\lim_{t\to\infty}\|\hat{\theta}(t+k)-\hat{\theta}(t)\|=0;\quad \forall k<\infty \tag{12.18}$$

### Step II: Computation of the Controller Parameters and of the Control Law

We will use the Strategy 1 (see Sect. 12.1) for updating the controller parameters. The controller equation generating $u(t)$ is:

$$\boxed{\hat{S}(t,q^{-1})u(t)+\hat{R}(t,q^{-1})y(t)=\hat{\beta}(t)P(q^{-1})y^*(t+d+1)} \tag{12.19}$$

or alternatively:

$$u(t)=-\hat{S}^*(t,q^{-1})u(t-1)-\hat{R}(t,q^{-1})y(t)+\hat{\beta}(t)P(q^{-1})y^*(t+d+1) \tag{12.20}$$

where:

$$\hat{\beta}(t)=1/\hat{B}(t,1) \tag{12.21}$$

$$\hat{S}(t,q^{-1})=1+\hat{s}_1(t)q^{-1}+\cdots+\hat{s}_{n_S}(t)q^{-n_S}=1+q^{-1}\hat{S}^*(t,q^{-1}) \tag{12.22}$$

$$\begin{aligned}\hat{R}(t,q^{-1})&=\hat{r}_0(t)+\hat{r}_1(t)q^{-1}+\cdots+\hat{r}_{n_R}(t)q^{-n_R}\\&=\hat{r}_0(t)+q^{-1}\hat{R}^*(t,q^{-1})\end{aligned} \tag{12.23}$$

and $\hat{S}(t,q^{-1})$, $\hat{R}(t,q^{-1})$ are solutions of:[2]

$$\boxed{\hat{A}(t,q^{-1})\hat{S}(t,q^{-1})+q^{-d-1}\hat{B}^*(t,q^{-1})\hat{R}(t,q^{-1})=P(q^{-1})} \tag{12.24}$$

[2]In fact $\hat{S}(t,q^{-1})=\hat{S}'(t,q^{-1})H_S(q^{-1})$, $\hat{R}(t,q^{-1})=\hat{R}'(t,q^{-1})H_R(q^{-1})$ where $H_S(q^{-1})$ and $H_R(q^{-1})$ are the fixed parts of the controller.

$P(q^{-1})$ in (12.19), (12.20) and (12.24) are the desired closed-loop poles and $y^*$ is the desired tracking trajectory. Equation (12.24) can be reformulated in matrix form:

$$M[\hat{\theta}(t)]\hat{x} = p \tag{12.25}$$

where:

$$\hat{x}^T = [1, \hat{s}_1, \ldots, \hat{s}_{n_S}, \hat{r}_0, \ldots, \hat{r}_{n_R}] \tag{12.26}$$

$$p^T = [1, p_1, \ldots, p_{n_P}, 0, \ldots, 0] \tag{12.27}$$

and $M$ is the Sylvester matrix which has the form (see Sect. 7.3 for details):

$$\left.\underbrace{\begin{bmatrix} \overbrace{\begin{matrix} 1 & 0 & \ldots & 0 \\ \hat{a}_1 & 1 & \ddots & \vdots \\ \vdots & & \ddots & 0 \\ & & & 1 \\ \vdots & & & \hat{a}_1 \\ \hat{a}_{n_{\hat{A}}} & & & \\ 0 & \ddots & & \vdots \\ \vdots & \ddots & & \\ 0 & \ldots & 0 & \hat{a}_{n_{\hat{A}}} \end{matrix}}^{n_{B'}} & \overbrace{\begin{matrix} 0 & \ldots & \ldots & 0 \\ \hat{b}'_1 & 0 & \ddots & \vdots \\ \vdots & & \ddots & \\ & & & 0 \\ \vdots & & & \hat{b}'_1 \\ \hat{b}'_{n_{\hat{B}}} & & & \\ 0 & \ddots & & \vdots \\ \vdots & \ddots & & \\ 0 & \ldots & 0 & \hat{b}'_{n_{\hat{B}}} \end{matrix}}^{n_{\hat{A}}} \end{bmatrix}}_{n_{\hat{A}}+n_{\hat{B}}}\right\} n_{\hat{A}}+n_{\hat{B}} \tag{12.28}$$

where $\hat{a}_i$ are estimated coefficients of the polynomial $A(q^{-1})$ and $\hat{b}'_i$ are the estimated coefficients of $B'(q^{-1}) = q^{-d}B(q^{-1})$ ($b'_i = 0$ for $i = 0, 1, \ldots, d$). Therefore, $\hat{S}(t)$ and $\hat{R}(t)$ are given by:

$$\hat{x} = M^{-1}[\hat{\theta}(t)]p \tag{12.29}$$

The admissibility condition for the estimated model is:

$$|\det M[\hat{\theta}(t)]| \geq \delta > 0 \tag{12.30}$$

which can alternatively be evaluated by the condition number:

$$\frac{\lambda_{\min} M[\hat{\theta}(t)]}{\lambda_{\max} M[\hat{\theta}(t)]} > \delta_0 > 0 \tag{12.31}$$

*Remark* If Strategy 2 is used, (12.19) through (12.31) remain the same except that the index $t$ becomes $(t-1)$. The analysis of the resulting schemes follows the one for Strategy 1.

### 12.2.2 Analysis of the Indirect Adaptive Pole Placement

The objective of the analysis will be to show that:

- $y(t)$ and $u(t)$ are bounded;
- some indices of performance go to zero asymptotically.

This implies that the resulting controller stabilizes the system.

The analysis will be carried on without requiring the persistence of excitation. Before embarking in the analysis of the behavior of $y(t), u(t)$ and of certain error signals, it is useful to write first the equations describing the time evolution of the observation vector $\phi(t)$ which contains $y(t)$ and $u(t)$. To simplify the writing, we will furthermore assume that $d = 0$. We start by making the observation that using (12.2), (12.6) and (12.9), the plant output at time $t$ can be expressed as:

$$\begin{aligned} y(t) &= \hat{y}(t) + \varepsilon(t) = \hat{\theta}^T(t)\phi(t-1) + \varepsilon(t) \\ &= -\sum_{i=1}^{n_A} \hat{a}_i(t)y(t-i) + \sum_{i=1}^{n_B} \hat{b}_i(t)u(t-i) + \varepsilon(t) \end{aligned} \tag{12.32}$$

On the other hand, introducing the expression of $y(t)$ given by (12.2) in the controller equation (12.20), one gets:

$$\begin{aligned} u(t) &= -\hat{S}^*(t, q^{-1})u(t-1) - \hat{r}_0(t)[\theta^T\phi(t-1)] \\ &\quad - \hat{R}^*(t, q^{-1})y(t-1) + \hat{\beta}(t)P(q^{-1})y^*(t+1) \\ &= f[y(t-1), \ldots, u(t-1), \ldots] + \bar{y}^*(t) \\ &= f[\phi(t-1)] + \bar{y}^*(t) \end{aligned} \tag{12.33}$$

where:

$$\bar{y}^*(t) = \hat{\beta}(t)P(q^{-1})y^*(t+1) \tag{12.34}$$

is a quantity perfectly known at time $t$ and which is bounded. Defining now:

$$\hat{r}'_i \stackrel{\Delta}{=} \hat{r}_i(t) - \hat{r}_0(t)a_i, \quad i = 1, 2, \ldots, n_A \tag{12.35}$$

$$\hat{s}'_i \stackrel{\Delta}{=} \hat{s}_i(t) + \hat{r}_0(t)b_i, \quad i = 1, 2, \ldots, n_B \tag{12.36}$$

Equation (12.33) becomes:

$$u(t) = -\sum_{i=1}^{n_A} \hat{r}'_i(t)y(t-i) - \sum_{i=1}^{n_B} \hat{s}'_i(t)u(t-i) + \bar{y}^*(t) \tag{12.37}$$

Combining now (12.32) with (12.37), one obtains a state space equation for $\phi(t)$:

$$\underbrace{\begin{bmatrix} y(t) \\ y(t-1) \\ \vdots \\ y(t-n_A+1) \\ u(t) \\ u(t-1) \\ \vdots \\ u(t-n_B+1) \end{bmatrix}}_{\phi(t)} = \underbrace{\begin{bmatrix} -\hat{a}_1(t) & \dots & & -\hat{a}_n(t) & \hat{b}_1(t) & \dots & \hat{b}_{n_B}(t) \\ 1 & & & 0 & 0 & 0 & 0 \\ & \ddots & & 0 & 0 & 0 & 0 \\ & & 1 & 0 & 0 & 0 & 0 \\ -\hat{r}'_1(t) & \dots & \dots & -\hat{r}'_{n_A}(t) & -\hat{s}'_1(t) & \dots & -\hat{s}'_{n_B}(t) \\ 0 & 0 & 0 & 1 & 0 & 0 & 0 \\ 0 & 0 & 0 & & \ddots & & 0 \\ 0 & 0 & 0 & 0 & 0 & 1 & 0 \end{bmatrix}}_{L(t)}$$

$$\times \underbrace{\begin{bmatrix} y(t-1) \\ y(t-2) \\ \vdots \\ y(t-n_A) \\ u(t-1) \\ u(t-2) \\ \vdots \\ u(t-n_B) \end{bmatrix}}_{\phi(t-1)} + \underbrace{\begin{bmatrix} \varepsilon(t) \\ 0 \\ 0 \\ \bar{y}^*(t) \\ 0 \\ 0 \\ 0 \end{bmatrix}}_{z(t)} \tag{12.38}$$

Moving one step further in time, this equation can be written:

$$\phi(t+1) = L(t+1)\phi(t) + z(t+1) \tag{12.39}$$

where $L(t)$ is a time-varying matrix containing the estimated parameters of the plant model and of the controller and $z(t+1)$ is a vector which contains the a posteriori prediction error $\varepsilon(t+1)$ and the term $\bar{y}^*(t+1)$ which depend on the reference signal. Both $\varepsilon(t+1)$ and $\bar{y}^*(t+1)$ depend on the estimated parameters at time $t+1$. For bounded reference signals and bearing in mind (12.17), it results that $\bar{y}^*(t+1)$ is bounded. If one uses at each time $t$ (12.24) for the computation of the controller parameters, the eigenvalues of the matrix $L(t)$ are the roots of the polynomial equation: $\lambda^{n_A+n_B-1}P(\lambda) = 0$. This can be easily seen if one takes a first order example with $n_A = 1, n_B = 1$.

*Remarks*

1. In general, the eigenvalues of the matrix $L(t)$ for given $\hat{R}(t,q^{-1})$, $\hat{S}(t,q^{-1})$, $\hat{A}(t,q^{-1})$, $\hat{B}(t,q^{-1})$ are the solution of:

$$\lambda^{n_A+n_B+d-1}[\hat{A}(\lambda^{-1})\hat{S}(\lambda^{-1}) + \lambda^{-d}\hat{B}(\lambda^{-1})\hat{R}(\lambda^{-1})] = 0$$

   This equation is used for computing the eigenvalues of $L(t)$ when $\hat{R}(t,q^{-1})$ and $\hat{S}(t,q^{-1})$ are not obtained as solutions of an equation of the type (12.24). Such a situation occurs in the case of adaptive generalized predictive control or adaptive linear quadratic control where $\hat{R}(t,q^{-1})$ and $\hat{S}(t,q^{-1})$ result as solutions of a quadratic criterion minimization.

2. An interesting interpretation of (12.38) is obtained by rewriting it as:

$$\begin{bmatrix} \hat{y}(t) \\ y(t-1) \\ \vdots \\ y(t-n_A+1) \\ u(t) \\ u(t-1) \\ \vdots \\ u(t-n_B+1) \end{bmatrix} = L(t) \begin{bmatrix} y(t-1) \\ y(t-2) \\ \vdots \\ y(t-n_A) \\ u(t-1) \\ u(t-2) \\ \vdots \\ u(t-n_B) \end{bmatrix} + \begin{bmatrix} 0 \\ 0 \\ \\ 0 \\ \bar{y}^*(t) \\ 0 \\ \\ 0 \end{bmatrix}$$

which is the closed-loop equation of the output of the predictor. It corresponds to a pole placement control of a perfectly known time-varying system (the adjustable predictor) (see also Fig. 1.13, Sect. 1.3).

Returning now to (12.39), one sees that it corresponds to a time-varying system with a driving signal. Roughly speaking, in order that $\phi(t)$ be bounded, the system should be asymptotically stable in the absence of the external excitation $z$ and the external excitation either should be bounded, or should not grow faster than $\phi(t)$. While these properties of $z(t)$ can be easily checked, the difficult part is that even if the eigenvalues of $L(t)$ are inside the unit circle at each time $t$, this does not necessarily guarantee the stability of the system.

Let us look now to the performance indices characterizing the adaptive pole placement. To do so, let us first observe that the plant output can be written:

$$y(t+1) = \hat{y}(t+1) + \varepsilon(t+1)$$

which leads to (see (12.32)):

$$\hat{A}(t)y(t) = q^{-d}\hat{B}(t)u(t) + \varepsilon(t) \tag{12.40}$$

where:[3]

$$\begin{aligned} \hat{A}(t) &= 1 + \hat{a}_1(t)q^{-1} + \cdots + \hat{a}_{n_A}(t)q^{-n_A} \\ \hat{B}(t) &= \hat{b}_1(t)q^{-1} + \cdots + \hat{b}_{n_B}(t)q^{-n_B} \end{aligned} \tag{12.41}$$

Together with (12.40), one considers the design equation (12.24) at time $t$:

$$\hat{A}(t)\hat{S}(t) + q^{-d}\hat{B}(t)\hat{R}(t) = P \tag{12.42}$$

[3]The index $q^{-1}$ has been dropped to simplify the notation.

and the controller equation at $t$:

$$\hat{S}(t)u(t) + \hat{R}(t)y(t) = \hat{\beta}(t)Py^*(t+d+1) \tag{12.43}$$

Combining (12.40) and (12.43) by appropriately passing the generated signals through the operators $\hat{A}(t)$, $\hat{B}(t)$, $\hat{R}(t)$ and $\hat{S}(t)$, and taking into account the non-commutativity of the time-varying operators, i.e.:

$$\hat{A}(t)\hat{S}(t)x(t) \neq \hat{S}(t)\hat{A}(t)x(t)$$

and (12.42), one gets:

$$\begin{aligned} P[y(t) - \hat{B}^*(t)\hat{\beta}(t)y^*(t)] = \hat{S}(t)\varepsilon(t) + \Delta_{11}(t)y(t) \\ + \Delta_{12}(t)u(t) + \Delta_{13}(t)y^*(t+1) \end{aligned} \tag{12.44}$$

and:

$$\begin{aligned} P[u(t) - \hat{A}(t)\hat{\beta}(t)y^*(t+1)] = -\hat{R}(t)\varepsilon(t) + \Delta_{21}(t)y(t) \\ + \Delta_{22}(t)u(t) + \Delta_{23}(t)y^*(t+1) \end{aligned} \tag{12.45}$$

where:

$$\begin{aligned} \Delta_{11}(t) &= \hat{A}(t)\hat{S}(t) - \hat{S}(t)\hat{A}(t) \\ \Delta_{12}(t) &= \hat{S}(t)\hat{B}(t) - \hat{B}(t)\hat{S}(t) \\ \Delta_{13}(t) &= \hat{B}(t)P\hat{\beta}(t) - P\hat{B}^*(t)\hat{\beta}(t)q^{-1} \\ \Delta_{21}(t) &= \hat{R}(t)\hat{A}(t) - \hat{A}(t)\hat{R}(t) \\ \Delta_{22}(t) &= \hat{B}(t)\hat{R}(t) - \hat{R}(t)\hat{B}(t) \\ \Delta_{23}(t) &= \hat{A}(t)P\hat{\beta}(t) - P\hat{A}(t)\hat{\beta}(t) \end{aligned}$$

If $y(t)$ and $u(t)$ are bounded and $\varepsilon(t)$ goes to zero, taking into account that $\Delta_{ij}(t)$ are bounded and go to zero, the left hand of (12.44) and (12.45) will go to zero. The left hand sides of (12.44) and (12.45) correspond in fact to the time domain objectives of the pole placement in the known parameter case. These objectives will be achieved asymptotically except that the filtered trajectories which will be followed by $u(t)$ and $y(t)$ may be different if $\hat{B}(\infty) \neq B$ and $\hat{A}(\infty) \neq A$. One has the following result:

**Theorem 12.1** (Strategy 1) *Consider the indirect adaptive pole placement for the plant model* (12.1) *where the plant parameter estimates are given by the algorithm of* (12.5) *through* (12.13) *and the controller is given by* (12.20) *through* (12.24). *Assume that*:

(i) *The plant model is admissible for the pole placement control* (*i.e., the polynomials* $A(q^{-1})$ *and* $B(q^{-1})$ *do not have common factors*).

(ii) *The orders of the polynomials* $A(q^{-1})$, $B(q^{-1})$ *and of the delay* $d$ *are known.*
(iii) *The reference signal is bounded.*
(iv) *The estimated models are admissible for each time* $t$ (*i.e., the estimated polynomials* $\hat{A}(t,q^{-1})$ *and* $\hat{B}(t,q^{-1})$ *do not have common factors*).

*Then*:

1. *The sequences* $\{u(t)\}$ *and* $\{y(t)\}$ *are bounded.*
2. *The a priori prediction error* $\varepsilon^0(t+1)$ *converges to zero, i.e.*:

$$\lim_{t\to\infty} \varepsilon^0(t+1) = 0$$

3. $\lim_{t\to\infty} P[y(t+d) - \hat{B}^*(t,q^{-1})\hat{\beta}(t)y^*(t+d)] = 0.$
4. $\lim_{t\to\infty} P[u(t) - \hat{A}(t,q^{-1})\hat{\beta}(t)y^*(t+d+1)] = 0.$

*Proof* The proof is divided in two stages. In the first stage, one proves that $\{y(t)\}$ and $\{u(t)\}$ are bounded. Then the other properties of the scheme result straightforwardly. The proof of boundedness of $\{y(t)\}$ and $\{u(t)\}$ is based on the following technical lemma. □

**Lemma 12.1** *Consider the system*:

$$\phi(t+1) = L(t)\phi(t) + x(t) \tag{12.46}$$

1. $L(t)$ *has finite coefficients for all* $t$.
2. *The eigenvalues of* $L(t)$ *are inside the unit circle for all* $t$.
3. $\|L(t) - L(t-1)\| \to 0$ *for* $t \to \infty$.

*Then there exists a time* $t$ *such that for* $t \geq t^*$:

$$\|\phi(t+1)\|^2 \leq C_1 + C_2 \max_{0\leq\tau\leq t} \|x(\tau)\|^2; \quad 0 \leq C_1, C_2 < \infty \tag{12.47}$$

*Proof* Since the eigenvalues of $L(t)$ are inside the unit circle, for a given matrix $Q > 0$ there is a matrix $P(t) > 0$ such that:

$$P(t) - L^T(t)P(t)L(t) = Q \tag{12.48}$$

Note also that since $\|L(t) - L(t-1)\| \to 0$ and $\|L(t)\| < \infty$ we have that $\|P(t) - P(t-1)\| \to 0$ and $\|P(t)\| < \infty$. From (12.46) and (12.48) we have:

$$\begin{aligned}
V(t+1) &= \phi^T(t+1)P(t)\phi(t+1) \\
&= [\phi^T(t)L^T(t) + x^T(t)]P(t)[L(t)\phi(t) + x(t)] \\
&= \phi^T(t)L^T(t)P(t)L(t)\phi(t) + 2\phi^T(t)P(t)x(t) + x^T(t)P(t)x(t)
\end{aligned}$$

$$= V(t) - \phi^T(t)Q\phi(t) + 2\phi^T(t)P(t)x(t)$$
$$+ x^T(t)P(t)x(t) + \phi^T(t)[P(t) - P(t-1)]\phi(t) \tag{12.49}$$

Defining:

$$\begin{aligned} \lambda_1 &= \lambda_{\min} Q; \qquad \lambda_2 = \lambda_{\min} P(t); \\ \lambda_3 &= \lambda_{\max} P(t); \qquad \lambda_4 = \max_t \|P(t)\| \end{aligned} \tag{12.50}$$

and:

$$\Delta P(t) = \|P(t) - P(t-1)\| \tag{12.51}$$

it results:

$$V(t+1) \le V(t) - \lambda_1 \|\phi(t)\|^2 + 2\|\phi(t)\|\|x(t)\|\lambda_4$$
$$+ \lambda_3 \|x(t)\|^2 + \Delta P(t)\|\phi(t)\|^2 \tag{12.52}$$

Note that since $2ab \le a^2 + b^2 \ \forall a, b \in R$:

$$2\lambda_4 \|\phi(t)\|\|x(t)\| = 2\lambda_4 \left(\frac{2}{\lambda_1}\right)^{1/2} \|x(t)\| \left(\frac{\lambda_1}{2}\right)^{1/2} \|\phi(t)\|$$
$$\le \lambda_4^2 \frac{2}{\lambda_1} \|x(t)\|^2 + \frac{\lambda_1}{2} \|\phi(t)\|^2 \tag{12.53}$$

From (12.52) and (12.53) and for a time $t \ge t^*$ such that $\Delta P(t) \le \frac{\lambda_1}{4}$ we have:

$$V(t+1) \le V(t) - \frac{\lambda_1}{4} \|\phi(t)\|^2 + \lambda_5 \|x(t)\|^2 \tag{12.54}$$

with:

$$\lambda_5 = 2\frac{\lambda_4^2}{\lambda_1} + \lambda_3$$

Since $V(t) \le \lambda_3 \|\phi(t)\|^2$, one obtains from (12.54):

$$V(t+1) \le \alpha V(t) + \lambda_5 \|x(t)\|^2$$

with:

$$0 < \alpha = 1 - \frac{\lambda_1}{4\lambda_3} < 1$$

Note from (12.48) and (12.50) that $\lambda_3 > \lambda_1$ which implies $\alpha > 0$. Therefore, one immediately gets:

$$V(t+1) = \alpha^{t+1} V(0) + \frac{(1 - \alpha^{t+1})}{1 - \alpha} \lambda_5 \max_{0 \le \tau \le t} \|x(\tau)\|^2 \tag{12.55}$$

which implies (12.47) and ends the proof of Lemma 12.1. □

We will use Lemma 12.1 to analyze the state equation for $\phi(t+1)$ given by (12.39). Taking into account the properties of the parameter estimation algorithm and the *admissibility* hypothesis for the estimated model with respect to the control law, all the assumptions for applying Lemma 12.1 are satisfied. Since $L(t+1)$ has bounded coefficients, $\phi(t)$ can become unbounded only asymptotically, but for $t \geq t^*$ one has:

$$\begin{aligned}\|\phi(t+1)\|^2 &\leq C_1' + C_2 \max_{0\leq\tau\leq t} [\varepsilon(t+1)^2 + \bar{y}^*(t)^2] \\ &\leq C_1 + C_2 \max_{0\leq\tau\leq t} \varepsilon(t+1)^2; \quad C_1, C_2 < \infty \end{aligned} \tag{12.56}$$

From (12.14), it results that $\varepsilon(t+1)^2 < \infty$ for all $t$ and one concludes that $\phi(t)$ is bounded. Recalling that $u(t)$ and $y(t)$ are bounded, that the various $\Delta_{ij}(t)$ in (12.44) and (12.45) go to zero asymptotically and that $\varepsilon(t+1)$ goes to zero, one concludes that properties (1) through (4) of Theorem 12.1 are true.

For Strategy 2, one has a similar result summarized in Theorem 12.2.

**Theorem 12.2** (Strategy 2) *Consider the indirect adaptive pole placement for the plant model* (12.1) *where the plant parameters estimates are given by the algorithm of* (12.5) *through* (12.13) *and the controller is given by*:

$$\hat{S}(t-1, q^{-1})u(t) + \hat{R}(t-1, q^{-1})y(t) = \hat{\beta}(t-1)P(q^{-1})y^*(t+d+1) \tag{12.57}$$

*or alternatively*:

$$\begin{aligned}u(t) = &-\hat{S}^*(t-1, q^{-1})u(t-1) - \hat{R}(t-1, q^{-1})y(t) \\ &+ \hat{\beta}(t-1)P(q^{-1})y^*(t+d+1)\end{aligned} \tag{12.58}$$

*where*:

$$\hat{\beta}(t-1) = 1/\hat{B}(t-1, 1) \tag{12.59}$$

$$\hat{S}(t, q^{-1}) = 1 + \hat{s}_1(t)q^{-1} + \cdots + \hat{s}_{n_S}(t)q^{-nS} = 1 + q^{-1}\hat{S}^*(t, q^{-1}) \tag{12.60}$$

$$\begin{aligned}\hat{R}(t, q^{-1}) &= \hat{r}_0(t) + \hat{r}_1(t)q^{-1} + \cdots + \hat{r}_{n_R}(t)q^{-n_R} \\ &= \hat{r}_0(t) + q^{-1}\hat{R}^*(t, q^{-1})\end{aligned} \tag{12.61}$$

*and* $\hat{R}(t, q^{-1})$ *and* $\hat{S}(t, q^{-1})$ *are solutions of* (12.24). *Assume that the hypotheses* (i) *through* (iv) *from Theorem* 12.1 *hold. Then*:

1. *The sequences* $\{u(t)\}$ *and* $\{y(t)\}$ *are bounded.*
2. *The a priori prediction error converges to zero*, *i.e.*: $\lim_{t\to\infty} \varepsilon^0(t+1) = 0$.
3. $\lim_{t\to\infty} P[y(t+d) - \hat{B}^*(t-1, q^{-1})\hat{\beta}(t-1)y^*(t+d)] = 0$.
4. $\lim_{t\to\infty} P[u(t) - \hat{A}(t-1, q^{-1})\hat{\beta}(t-1)y^*(t+d+1)] = 0$.

*Proof* The proof is similar to that of Theorem 12.1, except that one replaces in the various equations: $\varepsilon(t+1)$ by $\varepsilon^0(t+1)$, $\hat{y}(t+1)$ by $\hat{y}^0(t+1)$ and the time-varying parameters at $t$ by their values at time $t-1$. The major difference occurs in (12.56) which becomes:

$$\begin{aligned}\|\phi(t+1)\|^2 &\leq C_1' + C_2 \max_{0\leq\tau\leq t} [\varepsilon^0(t+1)^2 + \bar{y}^*(t)^2] \\ &\leq C_1 + C_2 \max_{0\leq\tau\leq t} \varepsilon^0(t+1)^2; \quad c_1, c_2 < \infty \qquad (12.62)\end{aligned}$$

since $\bar{y}^*(t)$ is a bounded signal. On the other hand, from the properties of the parameter estimation algorithm one has (see (12.16)):

$$\lim_{t\to\infty} \frac{[\varepsilon^0(t+1)]^2}{1+\phi^T(t)F(t)\phi(t)} = 0$$

and applying the "bounded growth" lemma (Lemma 11.1) one concludes that $\|\phi(t)\|$ is bounded, which implies that $y(t)$ and $u(t)$ are bounded, as well as $\lim_{t\to\infty} \varepsilon^0(t+1) = 0$. □

### *12.2.3 The "Singularity" Problem*

The question of avoiding singularity points in the parameter space during adaptation, i.e., points which correspond to non-admissible models, has received a lot of attention. While most of the techniques proposed are related to the pole placement where the problem is to avoid plant parameter estimates for which the plant model is not controllable (the determinant of the Sylvester matrix is zero), they can be used also for other types of admissibility conditions.

The various techniques can be classified as follows:

- Techniques based on excitation signals internally generated or externally applied, where the singularities are avoided by securing the convergence of the parameter estimates toward the true plant parameters (Anderson and Johnstone 1985; Eliott 1985; Goodwin and Teoh 1985; Kreisselmeier and Smith 1986; Polderman 1989; Giri et al. 1991; M'Saad et al. 1993b).
- Techniques based on the projection of the parameter estimates in a convex region (or a set of convex regions) (Kreisselmeier 1986; Ossman and Kamen 1987; Middleton et al. 1988; Barmish and Ortega 1991).
- Techniques based on the correction of the parameter estimates before using them to compute the controller (de Larminat 1984; Lozano and Goodwin 1985; Lozano 1989, 1992; Lozano et al. 1993; Lozano and Zhao 1994).
- Search for another type of plant model parameterization allowing to define a convex set of admissible models (Barmish and Ortega 1991).

A detailed review of the various techniques can be found in Lozano and Zhao (1994), M'Saad et al. (1993b). We will focus our attention on a technique of correcting the *singular* estimated parameter vector without modifying the asymptotic properties of the adaptation algorithm. When modifying the current parameter estimates, one should provide an algorithm for which:

- The modification is easily constructed.
- The number of possible modifications is finite, and within this finite set, there is at least one modification which gives a nonsingular parameter estimation.
- The modification algorithm stops (it converges) once a satisfactory result is obtained.

The algorithm which will be presented is based on the technique presented in Lozano and Zhao (1994). This technique starts from the observation that for the parameter estimation algorithms covered by Theorem 3.2, one has (see (3.248))

$$\|F(t)^{-1/2}[\hat{\theta}(t)-\theta]\| \leq h_0 \leq \infty; \quad \forall t \tag{12.63}$$

where $\theta$ is the true parameter vector characterizing the unknown plant model.

If one now defines a vector:

$$\beta^*(t) = F(t)^{-1/2}[\theta - \hat{\theta}(t)] \tag{12.64}$$

one can write the true parameter vector as:

$$\theta = \hat{\theta}(t) + F(t)^{1/2}\beta^*(t) \tag{12.65}$$

This means that there is a modification of the vector $\hat{\theta}(t)$ such that the new parameter estimation is equal to the true one for which the determinant of the Sylvester matrix is non-null. This of course implies that there is a modification of $\hat{\theta}(t)$ having the structure:

$$\bar{\theta}(t) = \hat{\theta}(t) + F(t)^{1/2}\beta(t) \tag{12.66}$$

such that the determinant of the Sylvester matrix is non-null. The vector $\beta(t)$ should satisfy in addition the following conditions:

1. $\beta(t)$ should converge in finite time;
2. $\beta(t)$ should be such that the absolute value of the determinant of the Sylvester matrix associated with $\bar{\theta}(t)$ is uniformly bounded from below.

### The Modification Algorithm

1. Define $\beta(t)$ as follows:

$$\beta^T(t) = [\sigma(t), \sigma(t)^m, \sigma(t)^{m^2}, \ldots, \sigma(t)^{m^{(m-1)}}] \tag{12.67}$$

   where:

$$m = n_A + n_B + d; \qquad m^{(m-1)} = l' \tag{12.68}$$

2. $\sigma(t)$ takes values in the set $D$

$$D = [\sigma_1, \sigma_2, \ldots, \sigma_l]; \qquad l = m^m \tag{12.69}$$

with

$$\sigma_i \in R; \quad \sigma_i \geq \sigma_{i-1} + 1; \quad i = 1, 2, \ldots, l \tag{12.70}$$

The values of $\sigma(t)$ will switch from a value to another in $D$ such that an appropriate $\beta(t)$ can be found. In order to stop the searching procedure, a switching function with a constant hysteresis width $\mu > 0$ ($\mu \ll 1$) is introduced. Denote:

$$x(t, \sigma) = |\det M[\hat{\theta}(t) + F(t)^{-1/2}(\sigma, \sigma^m, \ldots, \sigma^{l'})^T]| \tag{12.71}$$

3. The hysteresis switching function defining $\sigma(t)$ is defined as follows:

$$\sigma(t) = \begin{cases} \sigma(t-1) & \text{if } x(t, \sigma_j) < (1+\gamma)x[t, \sigma(t-1)] \\ & \text{for all } \sigma_j \in D \\ \sigma_j & \text{if } j \text{ is the smallest integer such that} \\ & x(t, \sigma_j) \geq (1+\gamma)x[t, \sigma(t-1)] \\ & \text{and } x(t, \sigma_j) \geq x(t, \sigma_i) \quad \forall \sigma_i \in D \end{cases} \tag{12.72}$$

It is shown in Lozano and Zhao (1994) that using this type of algorithm, one has the following result:

**Theorem 12.3** *Subject to the assumptions that the plant model polynomials $A(q^{-1})$ and $B(q^{-1})$ do not have common factors and their orders as well the delay $d$ are known, the parameter estimation algorithm (given by (12.5) through (12.13)) combined with the modification procedure defined by (12.67) through (12.72) assures:*

(i) *a lower bound for the absolute value of the determinant of the Sylvester matrix, i.e.,*

$$|\det M[\bar{\theta}(t)]| \geq \alpha\delta_0; \quad \alpha > 0 \tag{12.73}$$

*where $\delta_0$ is a measure of the controllability of the unknown system (related to the condition number)*

$$0 < \delta_0 \leq |\det M(\theta)| \tag{12.74}$$

(ii) *convergence in finite time toward the modified value $\bar{\theta}(t)$ assuring the property above.*

*Analysis* A full proof of this result is beyond our scope. In what follows, we would like to outline the basic steps and the rationale behind the choice proposed for $\beta(t)$. The key ideas are the following:

1. Using the proposed modification it is possible to write:

$$\det M[\bar{\theta}(t)] = g^T(t)v[\beta(t)] \tag{12.75}$$

where $g(t)$ contains combinations of the coefficient of the unmodified parameter vector $\hat{\theta}(t)$ and $v(\beta(t))$ contain combinations of $\sigma_i$ at various powers.

2. Instead of trying to show that there exists a value $\sigma_i$ such that (12.75):

$$\|g^T(t)v[\beta(t)]\| > 0$$

one constructs a vector collecting the values of $\det M[\bar{\theta}(t)]$ for all $\sigma_i$ in $D$, and one shows that the norm of this vector is bounded from below. This implies that at least one of the determinants is different from zero. An important role in this analysis is played by a Vandermonde matrix (Kailath 1980) and this explains the rationale behind the structure of the modifications defined by (12.67) and (12.69).

To be more specific, let us consider the case of a constant adaptation gain matrix $F(t) = F(0) = I$. In this case, one has:

$$\bar{\theta}(t) = \hat{\theta}(t) + \beta(t)$$

where $\beta(t)$ is given by (12.67). Consider now the case of a plant model with $n_A = 1$ and $n_B = 2$ and furthermore let us assume that there is a common factor for the estimated model at time $t$. One has therefore:

$$\hat{A}(t, q^{-1}) = 1 + \hat{a}_1(t)q^{-1}$$
$$\hat{B}(t, q^{-1}) = \hat{b}_1(t)q^{-1} + \hat{b}_2(t)q^{-1} \quad \text{with } \hat{a}_1(t) = \frac{\hat{b}_2(t)}{\hat{b}_1(t)}$$

The Sylvester matrix is:

$$M[\hat{\theta}(t)] = \begin{bmatrix} 1 & 0 & 0 \\ \hat{a}_1(t) & 1 & \hat{b}_1(t) \\ 0 & \hat{a}_1(t) & \hat{b}_2(t) \end{bmatrix}$$

and:

$$\det M[\hat{\theta}(t)] = \hat{b}_2(t) - \hat{a}_1(t)\hat{b}_1(t) = 0$$

The modified vector will have the form:

$$\begin{bmatrix} \bar{a}_1(t) \\ \bar{b}_1(t) \\ \bar{b}_2(t) \end{bmatrix} = \begin{bmatrix} \hat{a}_1(t) \\ \hat{b}_1(t) \\ \hat{b}_2(t) \end{bmatrix} + \begin{bmatrix} \sigma \\ \sigma^3 \\ \sigma^9 \end{bmatrix}$$

where $\sigma \in D$

$$D = [\sigma_1, \sigma_2, \ldots, \sigma_{27}]$$

which for this example can be chosen as:

$$[1, 2, 3, \ldots, 27]$$

The determinant of the modified Sylverster matrix takes the form:

$$
\begin{aligned}
\det M[\bar{\theta}(t)] &= \det \begin{bmatrix} 1 & \hat{b}_1(t) + \sigma^3 \\ \hat{a}_1(t) + \sigma & \hat{b}_2(t) + \sigma^9 \end{bmatrix} \\
&= \sigma^9 + \sigma^4 - \hat{b}_1(t)\sigma - \hat{a}_1(t)\sigma^3 = g^T(t)v[\beta(t)]
\end{aligned}
$$

where:

$$
\begin{aligned}
v[\beta(t)]^T &= [1, \sigma, \sigma^2, \sigma^3, \sigma^4, \sigma^5, \sigma^6, \sigma^7, \sigma^8, \sigma^9] \\
g^T(t) &= [0 - \hat{b}_1, 0, -\hat{a}_1, 1, 0, 0, 0, 0, 1]
\end{aligned}
$$

For this particular case, the highest power of $\sigma$ is 9 and therefore we can limit the domain $D$ for $i_{\max} = 10$ (but this is a particular case). Defining now a vector whose components are $\det M[\hat{\theta}(t)]$ obtained for various $\sigma_i$, one gets:

$$
p^T(t) = [g^T(t)v(\sigma_1), \ldots, g^T(t)v(\sigma_{10})] = g^T(t)N \tag{12.76}
$$

where $N$ is a $10 \times 10$ Vandermonde matrix:

$$
N = \begin{bmatrix} 1 & 1 & \ldots & \ldots & 1 \\ \sigma_1 & \sigma_2 & & & \sigma_{10} \\ \sigma_1^2 & \sigma_2^2 & & & \sigma_{10}^2 \\ \vdots & \vdots & & & \vdots \\ \vdots & \vdots & & & \vdots \\ \sigma_1^9 & \sigma_2^9 & & & \sigma_{10}^9 \end{bmatrix} \tag{12.77}
$$

Therefore, the objective is to show that indeed $\|p(t)\| > 0$ which will imply that:

$$
\exists \sigma_i \in D \quad \text{such that } |\det M[\bar{\theta}(t)]| > 0
$$

From (12.76) one has:

$$
\|p(t)\|^2 = p^T(t)p(t) = g^T(t)NN^Tg(t) \geq \|g(t)\|^2 \lambda_{\min}[NN^T] \tag{12.78}
$$

From matrix calculus, one has (Lancaster and Tismenetsky 1985):

$$
\begin{aligned}
(\det N)^2 &= \det NN^T = \lambda_{\min}[NN^T] \cdots \lambda_{\max}[NN^T] \\
&\leq \lambda_{\min} NN^T (\operatorname{tr}[NN^T])^{l-1}
\end{aligned} \tag{12.79}
$$

taking into account that:

$$
\operatorname{tr}[NN^T] = \sum_{i=1}^{l} \lambda_i [NN^T]
$$

Combining (12.78) and (12.79), one gets:

$$\|p(t)\| \geq \|g(t)\| \frac{|\det N|}{(\mathrm{tr}[NN^T])^{(l-1)/2}} \tag{12.80}$$

A lower bound for the vector $\|g(t)\|$ can be obtained taking into account the hypothesis that the true plant model polynomials do not have common factors.

$$\begin{aligned} 0 < \delta_0 \leq |\det M(\theta)| &= |\det M[\hat{\theta}(t) + F(t)^{1/2}\beta^*(t)]| \\ &= |g^T(t)v(\beta^*(t))| \leq \|g(t)\|\|v(\beta^*(t))\| \end{aligned} \tag{12.81}$$

From the above relationship, one gets:

$$\|g(t)\| \geq \frac{\delta_0}{\max \|v(\beta^*(t))\|} \tag{12.82}$$

and from (12.80), it results that:

$$\|p(t)\| \geq \frac{\delta_0 |\det N|}{\max \|v(\beta^*(t))\| (\mathrm{tr}\, NN^T)^{(l-1)/2}} \tag{12.83}$$

which is the desired result. The solution to the "singularity" problem presented above can be extended to cover the case when the system is subject to disturbances (see Sect. 12.3). The extension of this solution to the stochastic case is discussed in Guo (1996).

### *12.2.4 Adding External Excitation*

For constant controllers, it was shown in Chap. 9 in the context of plant model identification in closed loop, that richness conditions on the external excitation can be imposed in order to obtain richness conditions on the observation vector. This in turn will lead to a correct estimation of the parameter vector $\theta$ characterizing the plant model.

In the indirect adaptive pole placement control, an additional problem comes from the fact that the controller is time-varying. However, since the variations of the plant parameter estimates tend to zero, the time-varying coefficients of the controller will also exhibit slower and slower coefficients variations.

Let us $\{\bar{\phi}(t)\}$ denote the sequence of $\{\phi(t)\}$ generated by a fixed controller starting at time $t^*$. Since $\lim_{t\to\infty} \|\hat{\theta}(t+k) - \hat{\theta}(t)\| = 0$ and $\|\hat{\theta}(t)\|$ is bounded for all $t$, there is a time $t^*$ such that for $t \geq t^*$, one has:

$$\|\hat{\theta}_c(t^*) - \hat{\theta}_c(t)\| < h_0$$

where:

$$\hat{\theta}_c^T(t) = [\hat{r}_0(t), \hat{r}_1(t), \ldots, \hat{r}_{n_R}(t), \hat{s}_1(t), \ldots, \hat{s}_{n_S}(t)]$$

Since $\phi(t)$ is bounded (we are assuming $\hat{\theta}(t)$ to be in a convex region around the correct plant model) it results that:

$$\|\bar{\phi}(t) - \phi(t)\| < \delta$$

from which one concludes that:

$$\|\bar{\phi}(t)\bar{\phi}^T(t) - \phi(t)\phi^T(t)\| < 0(\delta); \quad \forall t > t^*$$

Therefore, if $\bar{\phi}(t)$, corresponding to a fixed controller for $t \geq t^*$, has the desired richness properties, i.e.:

$$\beta_2 I > \sum_{t=t^*}^{n_A+n_B+d-1+L} \bar{\phi}(t)\bar{\phi}^T(t) > \beta_1 I; \quad \beta_1, \beta_2 > 0$$

then, $\phi(t)$ will also have the desired richness property and this implies that:

$$\lim_{t\to\infty} \hat{A}(t, q^{-1}) = A(q^{-1}); \qquad \lim_{t\to\infty} \hat{B}(t, q^{-1}) = B(q^{-1})$$

Therefore, the external excitation (if it is sufficiently rich) will allow, not only to obtain a stabilizing controller, but to converge toward the controller corresponding to the exact knowledge of the plant model. Furthermore, since convergence under rich excitation is related to asymptotic exponential stability, the adaptation transients will be improved (Anderson and Johnstone 1985; Anderson and Johnson 1982).

It should, however, be noted that the technique based on the introduction of external excitation may have several drawbacks. Introducing an external excitation signal all the time is not always feasible or desirable in practice. Furthermore, due to the presence of the external signals, the plant output cannot reach the exact desired value. Often, the external excitation is used for the initialization of the scheme. To avoid the problems caused by the external excitation, it is possible to use the so-called self-excitation. In this case, the external excitation is introduced only when the plant output is far enough from its desired value (Giri et al. 1991; M'Saad et al. 1993b).

## 12.3 Robust Indirect Adaptive Control

In this section, we will present a design for indirect adaptive control (adaptive pole placement) which is robust with respect to:

- bounded disturbances;
- unmodeled dynamics.

As discussed in Sect. 10.6, the effect of unmodeled dynamics will be represented as a disturbance bounded by a function of $\|\phi(t)\|$, which is not necessarily bounded a priori. Several representations of the unmodeled dynamics are possible. Consider again the plant in (12.1), but with a disturbance $w(t)$ acting as follows:

$$A(q^{-1})y(t) = q^{-d}B(q^{-1})u(t) + w(t) \tag{12.84}$$

The plant (12.84) can be rewritten as (see also (12.2)):

$$y(t+1) = \theta^T\phi(t) + w(t+1) \tag{12.85}$$

In this section, we will focus on disturbances defined by:

$$|w(t+1)| \le d_1 + d_2\eta(t); \quad 0 < d_1, d_2 < \infty \tag{12.86}$$

$$\eta^2(t) = \mu^2\eta^2(t-1) + \|\phi(t)\|^2 \tag{12.87}$$

where $d_1$ accounts for a bounded external disturbance and $d_2\eta(t)$ accounts for the equivalent representation of the effect of unmodeled dynamics. More details can be found in Sect. 10.6 (the chosen representation for the unmodeled dynamics corresponds to Assumption B in Sect. 10.6).

It should be noted that even if the plant parameters are perfectly known, there is no fixed linear controller that could stabilize the plant for all possible values of $d_2$. Nevertheless, we can design a controller that stabilizes the plant provided that $d_2 \le d_2^*$ where $d_2^*$ is a threshold that depends on the system parameters.

The assumptions made upon the system of (12.84) through (12.87) are:

- the orders of the polynomials $A(q^{-1})$, $B(q^{-1})$ and of the delay $d$ are known ($n_A, n_B, d$-known),
- $A(q^{-1})$ and $B(q^{-1})$ do not have common factors,
- the disturbance upper bounds $d_1$ and $d_2$ are known.

We will discuss the robust adaptive pole placement in two situations.

(a) without taking into account the possible singularities in the parameter space during adaptation (standard robust adaptive pole placement),
(b) taking into account the modification of parameter estimates in the singularity points (modified robust adaptive pole placement).

### *12.3.1 Standard Robust Adaptive Pole Placement*

Bearing in mind the type of equivalent disturbance considered in (12.86), we will use a parameter estimation algorithm with data normalization and dead zone. This algorithm and its properties are described in Sect. 10.6, Theorem 10.5.

We will discuss next the direct extension of the design given in Sect. 12.2 for the ideal case using Strategy 1. How to incorporate the modification of the parameter estimates in order to avoid singularities will be discussed afterwards in Sect. 12.3.2. For the analysis of the resulting scheme, one proceeds exactly as in Sect. 12.2.2.

**Step I: Estimation of the Plant Model Parameters**

Define the normalized input-output variables:

$$\bar{y}(t+1) = \frac{y(t+1)}{m(t)}; \qquad \bar{u}(t) = \frac{u(t)}{m(t)}; \qquad \bar{\phi}(t) = \frac{\phi(t)}{m(t)} \tag{12.88}$$

where:

$$m^2(t) = \mu^2 m^2(t-1) + \max(\|\phi(t)\|^2, 1); \quad m(0) = 1; \ 0 < \mu < 1 \tag{12.89}$$

The a priori output of the adjustable predictor is given by:

$$\hat{y}^0(t+1) = \hat{\theta}^T(t)\bar{\phi}(t) \tag{12.90}$$

The a posteriori output of the adjustable predictor is given by:

$$\hat{y}(t+1) = \hat{\theta}^T(t+1)\bar{\phi}(t) \tag{12.91}$$

where:

$$\hat{\theta}^T(t) = [\hat{a}_1(t), \ldots, \hat{a}_{n_A}(t), \hat{b}_1(t), \ldots, \hat{b}_{n_A}(t)] \tag{12.92}$$

The a priori and the a posteriori prediction (adaptation) errors are given by:

$$\bar{\varepsilon}^0(t+1) = \bar{y}(t+1) - \hat{y}^0(t+1) \tag{12.93}$$

$$\bar{\varepsilon}(t+1) = \bar{y}(t+1) - \hat{y}(t+1) \tag{12.94}$$

The parameter adaptation algorithm is:

$$\hat{\theta}(t+1) = \hat{\theta}(t) + \alpha(t)F(t)\bar{\phi}(t)\bar{\varepsilon}(t+1) \tag{12.95}$$

$$F(t+1)^{-1} = F(t)^{-1} + \alpha(t)\bar{\phi}(t)\bar{\phi}(t)^T; \quad F(0) > 0 \tag{12.96}$$

$$\bar{\varepsilon}(t+1) = \frac{\bar{\varepsilon}^0(t+1)}{1 + \bar{\phi}^T(t)F(t)\bar{\phi}(t)} \tag{12.97}$$

$$\alpha(t) = \begin{cases} 1 & |\bar{\varepsilon}(t+1)| > \bar{\delta}(t+1) \\ 0 & |\bar{\varepsilon}(t+1)| \le \bar{\delta}(t+1) \end{cases} \tag{12.98}$$

where $\bar{\delta}(t)$ is given by:

$$\bar{\delta}^2(t+1) = \delta^2(t+1) + \sigma\delta(t+1); \quad \delta(t) > 0; \ \sigma > 0 \tag{12.99}$$

and (see (12.86) and (12.87)):

$$\delta(t+1) = d_2 + \frac{d_1}{m(t)} \ge |w(t+1)|/m(t) \tag{12.100}$$

Using this algorithm, we will have, according to Theorem 10.5, the following properties:

- The parameter vector $\hat{\theta}(t)$ is bounded and converges.
- The normalized a posteriori adaptation error is bounded:

$$\lim_{t\to\infty} \sup |\bar{\varepsilon}(t+1)| \leq \bar{\delta}(t+1) \tag{12.101}$$

- The a posteriori prediction error defined as:

$$\varepsilon(t+1) = y(t+1) - \hat{\theta}^T(t+1)\phi(t) \tag{12.102}$$

satisfies:

$$\lim_{t\to\infty} \sup \varepsilon(t+1)^2 \leq \bar{\delta}^2(t+1)m^2(t) \tag{12.103}$$

### Step II: Computation of the Control Law

We will use Strategy 1, as described in Sect. 12.2, using the parameter estimates provided by the robust parameter adaptation algorithm above. One finally obtains (see (12.2.56)):

$$\|\phi(t+1)\|^2 \leq C_1 + C_2 \max_{0\leq\tau<t} \varepsilon(t+1)^2; \quad 0 \leq C_1, C_2 < \infty \tag{12.104}$$

In the presence of the disturbance $w(t+1)$, $\varepsilon(t+1)$ is no more necessarily bounded as in the ideal case. On the other hand, from Theorem 10.5, one has:

$$\lim_{t\to\infty} \sup \varepsilon^2(t+1) \leq \bar{\delta}^2(t+1)m^2(t) \tag{12.105}$$

where:

$$\bar{\delta}^2(t+1) = \left(d_2 + \frac{d_1}{m(t)}\right)^2 + \sigma\left(d_2 + \frac{d_1}{m(t)}\right) \tag{12.106}$$

Assume now that $\phi(t+1)$ diverges. It follows that there is a subsequence such that for this subsequence $t_1, t_2, \ldots, t_n$, one has:

$$\|\phi(t_1)\| \leq \|\phi(t_2)\| \leq \cdots \leq \|\phi(t_n)\| \tag{12.107}$$

or, equivalently:

$$\|\phi(t)\| \leq \|\phi(t_n)\|; \quad \forall t \leq t_n \tag{12.108}$$

In the meantime, as $\phi(t)$ diverges, it follows that $\bar{\delta}(t)^2 \to d_2^2 + \sigma d_2$, therefore in the limit as $t \to \infty$:

$$\begin{aligned}\lim_{t\to\infty} \sup \varepsilon^2(t+1) &\leq (d_2^2 + \sigma d_2)m^2(t) \\ &\leq (d_2^2 + \sigma d_2)m^2(t+1)\end{aligned} \tag{12.109}$$

Using these results, one gets from (12.104) that:

$$\|\phi(t+1)\|^2 \le C_1 + C_2(d_2^2 + \sigma d_2)m^2(t+1) \tag{12.110}$$

and thus:

$$\frac{\|\phi(t+1)\|^2 - C_1}{m^2(t+1)} \le C_2(d_2^2 + \sigma d_2) \tag{12.111}$$

The LHS of (12.111) converges toward a positive number $\rho$ smaller than 1 as $\phi(t+1)$ diverges ($m^2(t+1) \le \gamma + \|\phi(t+1)\|^2, \gamma > 0$). Therefore, there are values of $d_2 \le d_2^*$ such that $C_2(d_2^* + \sigma d_2^*) < \rho$ which leads to a contradiction. Therefore, $d_2 \le d_2^*$, $\|\phi(t+1)\|$ is bounded and the a posteriori and a priori prediction errors are bounded. Since $d_2$ corresponds to the "$\mu$-scaled infinity norm" of the unmodeled dynamics $H(z^{-1})$, one gets the condition that boundedness of $\phi(t+1)$ is assured for $\|H(\mu^{-1}z^{-1})\|_\infty < d_2^*$.

We will also have:

$$\lim_{t\to\infty} P[y(t+d) - B^*(t,q^{-1})\hat{\beta}(t)y^*(t+d)] = S(t)\varepsilon(t) \tag{12.112}$$

$$\lim_{t\to\infty} P[u(t) - A(t,q^{-1})\hat{\beta}(t)y^*(t+d+1)] = -R(t)\varepsilon(t) \tag{12.113}$$

Taking into account (12.105) and the boundedness of $\phi(t)$, it results that the two indices of performance will be bounded ($\phi(t)$ bounded implies that $m(t)$ is bounded).

### 12.3.2 Modified Robust Adaptive Pole Placement

In order to use the technique for the modification of the parameter estimates in the eventual singularity points, a slight modification of the parameter estimation algorithm given by Theorem 10.5 should be considered. This algorithm and its properties are summarized in the next theorem (Lozano and Zhao 1994).

**Theorem 12.4** *Under the same hypothesis as in Theorem* 10.4 ((10.100) *through* (10.102)) *and Theorem* 10.5 *except that* $w(t+1) = d_1 + d_2\|\phi(t)\|$ *and* $m(t) = 1 + \|\phi(t)\|$ *using a PAA of the form*:

$$\hat{\theta}(t+1) = \hat{\theta}(t) + \alpha(t)F(t+1)\bar{\phi}(t)\bar{\varepsilon}^0(t+1) \tag{12.114}$$

$$F(t+1)^{-1} = F(t)^{-1} + \alpha(t)\bar{\phi}(t)\bar{\phi}^T(t) \tag{12.115}$$

$$\bar{\varepsilon}^0(t+1) = \bar{y}(t+1) - \hat{y}^0(t+1) = \bar{y}(t+1) - \hat{\theta}^T(t)\bar{\phi}(t) \tag{12.116}$$

$$\alpha(t) = \begin{cases} 1 & \varepsilon_a^2(t+1) > \bar{\delta}^2(t+1) \\ 0 & \varepsilon_a^2(t+1) \le \bar{\delta}^2(t+1) \end{cases} \tag{12.117}$$

*where* $\varepsilon_a(t+1)$ *is given by*:

$$\varepsilon_a(t+1) = [\bar{\varepsilon}^0(t+1)^2 + \bar{\phi}^T(t)F(t)\bar{\phi}(t)]^{1/2} \tag{12.118}$$

*and*:

$$\bar{\delta}^2(t+1) = [\delta^2(t+1) + \sigma\delta(t)][1 + \operatorname{tr} F(0)]; \quad \sigma > 0 \tag{12.119}$$

*where*:

$$\delta(t+1) = d_2 + \frac{d_1}{1 + \|\phi(t)\|} \tag{12.120}$$

*one has*:

$$(1) \quad \tilde{\theta}^T(t)F(t)^{-1}\tilde{\theta}(t) \le \tilde{\theta}^T(0)F(0)^{-1}\tilde{\theta}(0) < \infty \tag{12.121}$$

$$\textit{where}: \tilde{\theta}(t) = \hat{\theta}(t) - \theta \tag{12.122}$$

$$(2) \quad \lim_{t\to\infty} \sup[\varepsilon_a^2(t+1) - \bar{\delta}^2(t)] = 0 \tag{12.123}$$

(3) $F(t)$ *and* $\tilde{\theta}(t)$ *converge.*

*Remark* It is worth noting that the properties of the algorithm stated above are independent of the values of $d_1$ and $d_2$ which characterize the equivalent disturbance.

*Proof* From the definition of the normalized variables (10.102) and of the disturbance $w(t)$, it results that:

$$\bar{y}(t+1) = \theta^T\bar{\phi}(t) + \bar{w}(t+1) \tag{12.124}$$

where:

$$\bar{w}(t+1) = w(t+1)/(1 + \|\phi(t)\|) \tag{12.125}$$

and, therefore:

$$\begin{aligned} |\bar{w}(t+1)| &\le \frac{|d_1 + d_2\|\phi(t)\||}{1 + \|\phi(t)\|} \\ &\le d_2 + \frac{d_1}{1 + \|\phi(t)\|} = \delta(t+1) \end{aligned} \tag{12.126}$$

Introducing (12.124) in (12.93), one gets:

$$\bar{\varepsilon}^0(t+1) = -\tilde{\theta}^T(t)\bar{\phi}(t) + \bar{w}(t+1) \tag{12.127}$$

From (12.114) and (12.122), it results that the parameter error is described by:

$$\tilde{\theta}(t+1) = \tilde{\theta}(t) + \alpha(t)F(t+1)\bar{\phi}(t)\bar{\varepsilon}^0(t+1) \tag{12.128}$$

Combining (12.115) and (12.128), one has:

$$\begin{aligned} &\tilde{\theta}^T(t+1)F(t+1)^{-1}\tilde{\theta}(t+1) \\ &\quad = [\tilde{\theta}(t) + \alpha(t)F(t+1)\bar{\phi}(t)\bar{\varepsilon}^0(t+1)]F(t+1)^{-1} \end{aligned}$$

$$\begin{aligned}
&\times [\tilde{\theta}(t) + \alpha(t)F(t+1)\bar{\phi}(t)\bar{\varepsilon}^0(t+1)] \\
&= \tilde{\theta}^T(t)F(t)^{-1}\tilde{\theta}(t) + \alpha(t)[\tilde{\theta}^T(t)\bar{\phi}(t)]^2 \\
&\quad + 2\alpha(t)\tilde{\theta}(t)\bar{\phi}(t)\bar{\varepsilon}^0(t+1) \\
&\quad + \alpha^2(t)\bar{\phi}^T(t)F(t+1)\bar{\phi}(t)[\bar{\varepsilon}^0(t+1)]^2 \\
&= \tilde{\theta}(t)F(t)^{-1}\tilde{\theta}(t) + \alpha(t)[\tilde{\theta}^T(t)\bar{\phi}(t) + \bar{\varepsilon}^0(t+1)]^2 \\
&\quad + \alpha(t)[\varepsilon^0(t+1)]^2[\alpha(t)\bar{\phi}(t)F(t+1)\bar{\phi}(t) - 1]
\end{aligned} \tag{12.129}$$

On the other hand, from (12.115) and the matrix inversion lemma it follows:

$$\bar{\phi}(t)F(t+1)\bar{\phi}(t) = \frac{\bar{\phi}^T(t)F(t)\bar{\phi}(t)}{1 + \alpha(t)\bar{\phi}(t)F(t)\bar{\phi}(t)} \tag{12.130}$$

Therefore:

$$\alpha(t)\bar{\phi}^T(t)F(t+1)\bar{\phi}(t) - 1 = \frac{-1}{1 + \alpha(t)\bar{\phi}(t)F(t)\bar{\phi}(t)} \tag{12.131}$$

Introducing (12.127) and (12.131) in (12.129), one obtains:

$$V'(t+1) = V'(t) + \alpha(t)\bar{w}^2(t+1) - \frac{\alpha(t)[\bar{\varepsilon}^0(t+1)]^2}{1 + \alpha(t)\bar{\phi}(t)F(t)\bar{\phi}(t)} \tag{12.132}$$

where:

$$V'(t) = \tilde{\theta}^T(t)F(t)\tilde{\theta}(t) \tag{12.133}$$

From (12.115) using the matrix inversion lemma (see Chap. 3) one also has:

$$\operatorname{tr} F(t+1) = \operatorname{tr} F(t) - \frac{\alpha(t)\bar{\phi}(t)F^2(t)\bar{\phi}(t)}{1 + \alpha(t)\bar{\phi}(t)F(t)\bar{\phi}(t)} \tag{12.134}$$

Adding (12.132) to (12.134) and using (12.118) one gets:

$$V(t+1) = V(t) + \alpha(t)\bar{w}^2(t+1) - \frac{\alpha(t)\varepsilon_a^2(t+1)}{1 + \alpha(t)\bar{\phi}(t)F(t)\bar{\phi}(t)} \tag{12.135}$$

with:

$$V(t) = V'(t) + \operatorname{tr} F(t) = \tilde{\theta}^T(t)F(t)^{-1}\tilde{\theta}(t) + \operatorname{tr} F(t) \tag{12.136}$$

In view of (10.102), (12.115) and (12.117) one has:

$$1 + \alpha(t)\bar{\phi}(t)F(t)\bar{\phi}(t) \le 1 + \alpha(t)\operatorname{tr} F(t)\|\bar{\phi}(t)\|^2 \le 1 + \operatorname{tr} F(0) \stackrel{\Delta}{=} \gamma \tag{12.137}$$

($\|\bar{\phi}(t)\| \le 1$ from (10.7)). Introducing (12.126) and (12.137) in (12.135) one obtains:

$$V(t+1) \le V(t) + \frac{\alpha(t)}{\gamma}[\gamma\delta^2(t+1) - \varepsilon_a^2(t+1)] \tag{12.138}$$

Using (12.117), (12.118) and (12.119) it follows:

$$\frac{\alpha(t)}{\gamma}[\gamma\delta^2(t+1) - \varepsilon_a^2(t+1)] = \frac{\alpha(t)}{\gamma}[\gamma\bar{\delta}^2(t+1) - \gamma\sigma\delta(t+1) - \varepsilon_a^2(t+1)]$$
$$\leq -\alpha(t)\sigma\delta(t+1) \tag{12.139}$$

Introducing (12.139) into (12.138) one finally gets:

$$0 \leq V(t+1) \leq V(t) - \alpha(t)\sigma\delta(t+1) \leq V(0) - \sigma\sum_{i=0}^{t+1}\alpha(i)\delta(i+1) \tag{12.140}$$

$V(t)$ is therefore a nonincreasing positive sequence (see (12.140) and (12.136)) and thus $V(t)$ converges to a constant value. This proves that $\tilde{\theta}^T(t)F(t)^{-1}\tilde{\theta}(t)$ is bounded. The proofs of (12.123) as well as the convergence of $F(t)$ and $\tilde{\theta}(t)$ are similar to those of Theorem 10.5 and are omitted. □

The algorithm given in Theorem 12.4 will be used together with the modification of parameter estimates discussed in Sect. 12.2.3 in order to avoid the singularities in the computation of the controller parameters. The parameter estimates will be modified according to:

$$\bar{\theta}(t) = \hat{\theta}(t) + F(t)^{1/2}\beta(t) \tag{12.141}$$

where:

$$\bar{\theta}(t) = [\bar{a}_1(t), \ldots, \bar{a}_{n_A}(t), \bar{b}_1(t), \ldots, \bar{b}_{n_B}(t)] \tag{12.142}$$

Since $\|F(t)^{-1/2}\tilde{\theta}(t)\|$ is bounded and $\hat{\theta}(t)$ converges, it is possible to compute $\beta(t)$ as in Sect. 12.2.3 in such a way that $\beta(t)$ converges and that the Sylvester matrix associated to $\bar{\theta}(t)$ is nonsingular.

Let us define the modified a posteriori prediction error:

$$\begin{aligned}\varepsilon(t) &= y(t) - \bar{\theta}(t)\phi(t-1)\\ &= y(t) + \sum_{i=1}^{n_A}\bar{a}_i(t)y(t-i) - \sum_{i=1}^{n_B}\bar{b}_i(t)u(t-d-i)\end{aligned} \tag{12.143}$$

Using similar notations as in (12.1) through (12.4), (12.143) can also be written as:

$$\bar{A}(t)y(t) = q^{-d}\bar{B}(t)u(t) + \varepsilon(t) \tag{12.144}$$

where:

$$\bar{A}(t) = 1 + \bar{a}_1(t)q^{-1} + \cdots + \bar{a}_{n_A}(t)q^{-n_A} \tag{12.145}$$

$$\bar{B}(t) = \bar{b}_1(t)q^{-1} + \cdots + \bar{b}_{n_B}(t)q^{-n_B} \tag{12.146}$$

are relatively prime polynomials having coefficients that converge. A bound for $\varepsilon(t)$ in (12.143), (12.144) which will be required later is given in the following lemma (Lozano 1992).

**Lemma 12.2** *$\varepsilon(t)$ in (12.143), (12.90) satisfies the following inequality*:

$$\lim_{t\to\infty} \sup \left\{ \frac{\varepsilon^2(t)}{(1+\|\phi(t-1)\|)^2} - 3\bar{\delta}^2(t)\beta_{\max}^2 \right\} \leq 0 \tag{12.147}$$

*where $\beta_{\max}$ is such that*:

$$\beta_{\max} = \max(1, \|\beta(t)\|) \tag{12.148}$$

*Proof* Introducing (12.141) into (12.143), one gets:

$$\begin{aligned}\varepsilon(t) &= y(t) - \hat{\theta}^T(t)\phi(t-1) - \beta^T(t)F^{1/2}(t)\phi(t-1) \\ &= [\bar{y}(t) - \hat{\theta}^T(t)\bar{\phi}(t-1) - \beta^T(t)F^{1/2}(t)\bar{\phi}(t-1)] \\ &\quad \times [1 + \|\phi(t-1)\|]\end{aligned} \tag{12.149}$$

Taking into account that:

$$\bar{\phi}(t) = \frac{\phi(t)}{1+\|\phi(t)\|} \tag{12.150}$$

and using (12.86) and (12.93) one obtains:

$$\frac{\varepsilon(t)}{1+\|\phi(t-1)\|} = [\bar{\varepsilon}^0(t) - \beta^T(t)F^{1/2}(t-1)\bar{\phi}(t-1) + z(t)] \tag{12.151}$$

where:

$$\begin{aligned}z(t) &= [\hat{\theta}(t-1) - \hat{\theta}(t)]^T\bar{\phi}(t-1) \\ &\quad + \beta^T(t)[F^{1/2}(t-1) - F^{1/2}(t)]\bar{\phi}(t-1)\end{aligned} \tag{12.152}$$

From (12.118), (12.148), (12.151) and the fact that $(a+b+c)^2 \leq 3(a^2+b^2+c^2)$ for any $a, b$ and $c$, one has:

$$\begin{aligned}\frac{\varepsilon^2(t)}{(1+\|\phi(t-1)\|)^2} &\leq 3[\bar{\varepsilon}^0(t)^2 + \beta_{\max}^2\bar{\phi}^T(t-1)F(t-1)\bar{\phi}(t-1) + z^2(t)] \\ &\leq 3\beta_{\max}^2\varepsilon_a^2(t) + 3z^2(t)\end{aligned} \tag{12.153}$$

Note that $z(t)$ in (12.152) converges to zero because $F(t)$ and $\hat{\theta}(t)$ converge and $\beta(t)$ and $\bar{\phi}(t-1)$ are bounded. Equation (12.153) can also be written as:

$$\frac{\varepsilon^2(t)}{(1+\|\phi(t-1)\|)^2} - 3\beta_{\max}^2\bar{\delta}^2(t) \leq 3\beta_{\max}^2[\varepsilon_a^2(t) - \bar{\delta}^2(t)] + 3z^2(t) \tag{12.154}$$

Since $z(t) \to 0$ as $t \to \infty$ and in view of (12.3.40), the rest of the proof follows. □

From this point, the robust adaptive pole placement is obtained by combining Strategy 1 with the modified parameter estimates provided by (12.141), the parameters being estimated using the algorithm given in Theorem 12.4. The analysis of the resulting scheme is similar to the one discussed in Sect. 12.3.1 since $\varepsilon^2(t+1)$ in the limit is bounded by:

$$\lim_{t\to\infty} \sup \varepsilon^2(t+1) \leq 3\bar{\delta}^2(t+1)\beta^2_{\max}(1+\|\phi(t)\|)^2 \tag{12.155}$$

Note that this expression is very similar to (12.106). Using similar arguments we can obtain an equation of the type of (12.111). Therefore, there is threshold $d_2^*$ such that if $d_2 \leq d_2^*$, one can prove by contradiction that $\|\phi(t)\|$ is bounded. The rest of the analysis follows.

### *12.3.3 Robust Adaptive Pole Placement: An Example*

The example which will be considered here comes from Rohrs et al. (1985). The continuous time plant to be controlled is characterized by the transfer function:

$$G(s) = \frac{2}{s+1} \cdot \frac{229}{(s^2+30s+229)}$$

where the first order part is considered as the dominant dynamics and the second order part as the unmodeled dynamics. The system will be controlled in discrete time with a sampling period $T_S = 0.04$ s. For this sampling period the true discrete time plant model is given by:

$$G(q^{-1}) = \frac{b_1q^{-1} + b_2q^{-2} + b_3q^{-3}}{1 - a_1q^{-1} + a_2q^{-2} + a_3q^{-3}}$$

with:

| $a_1$ | $a_2$ | $a_3$ |
|---|---|---|
| −1.8912 | 1.1173 | −0.21225 |

| $b_1$ | $b_2$ | $b_3$ |
|---|---|---|
| 0.0065593 | 0.018035 | 0.0030215 |

Since this model has unstable zeros, it is reasonable to use an indirect adaptive control strategy which can handle unstable zeros. Adaptive pole placement will be used. The plant model will be estimated using a filtered recursive least squares with dynamic data normalization. Adaptation freezing will be enforced in the absence of enough rich information (i.e., the scheduling variable for the dead zone will depend on the signal richness measured by $\bar{\phi}_f^T(t)F(t)\bar{\phi}_f(t)$—see Chap. 16 for details). Estimated models of different orders will be used ($n = 1, 2, 3$).

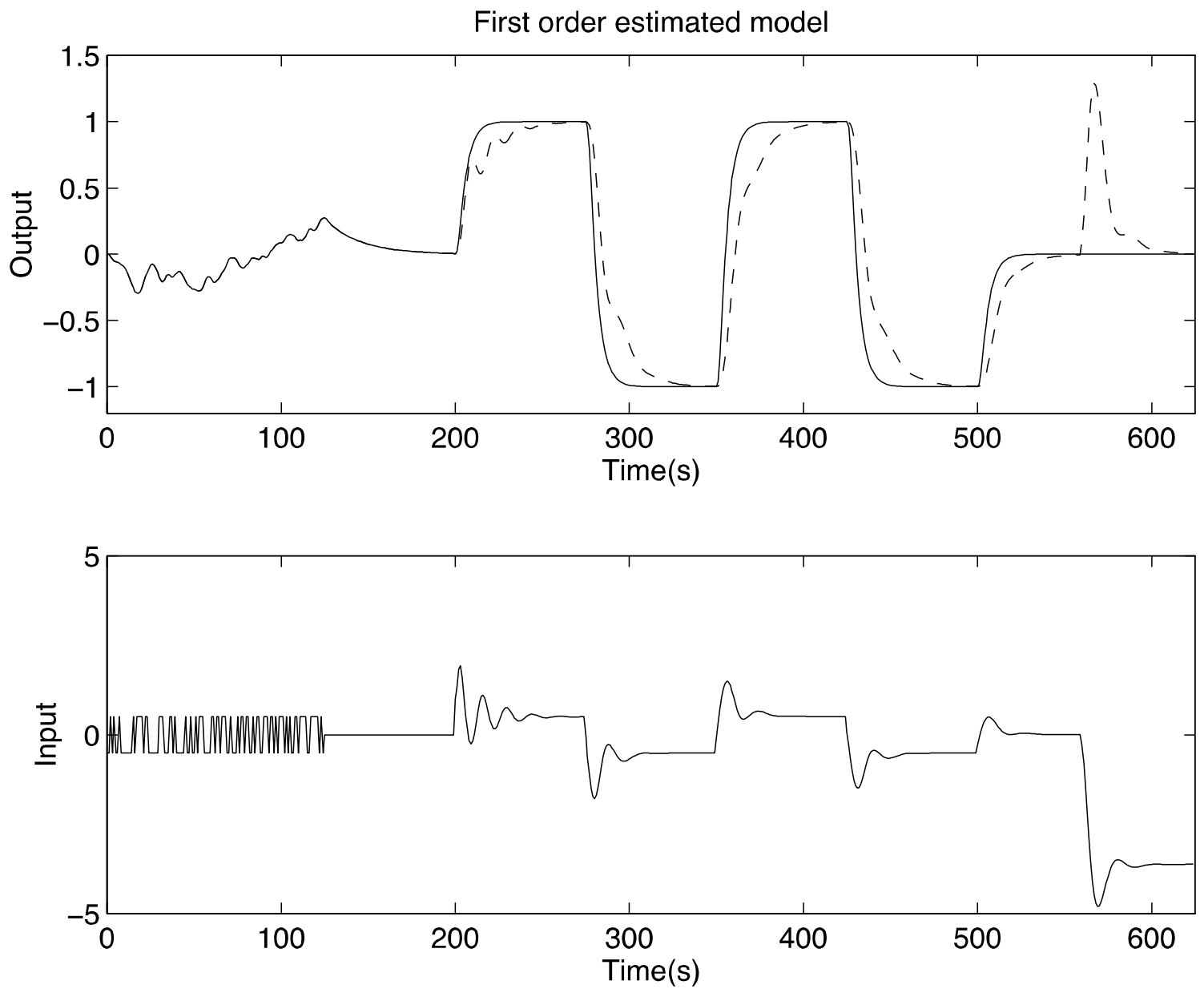

**Fig. 12.1** First order estimated model (— reference trajectory, - - - plant output)

Figures 12.1, 12.2 and 12.3 summarize the results obtained with various orders for the estimated model. The controller is first initialized using an open-loop recursive identification, then the loop is closed and a series of step reference changes is applied followed by the application of an output disturbance.

For a first order estimated model, while the system is stable and the signals have acceptable form, one can see that the tracking performances are not very good. Second order and third order estimated models give almost the same results and they are very good.

Figure 12.4 shows the frequency characteristics of the estimated models for $n = 1, 2, 3$. The identified third order model corresponds exactly to the true model. From this figure, one can see that the second and third order models have similar frequency characteristics in the interesting frequency range for control (related to the desired closed-loop poles) while the first order model cannot cope with the frequency characteristics of the third order model in a sufficiently large frequency region. The desired closed-loop poles used are as follows:

$$\text{for } n = 1: \quad P(q^{-1}) = (1 - 0.8q^{-1})(1 - 0.9q^{-1})$$

$$\text{for } n = 2, 3: \quad P(q^{-1}) = (1 - 0.8q^{-1})(1 - 0.4q^{-1})(1 - 0.2q^{-1})(1 - 0.1q^{-1})$$

The controller has an integrator. For the case $n = 2$, a filter $H_R(q^{-1}) = 1 + q^{-1}$ has been introduced in the controller in order to reduce the "activity" of the control action by reducing the magnitude of the input sensitivity function in the high frequencies. Same dynamics has been used in tracking and regulation.

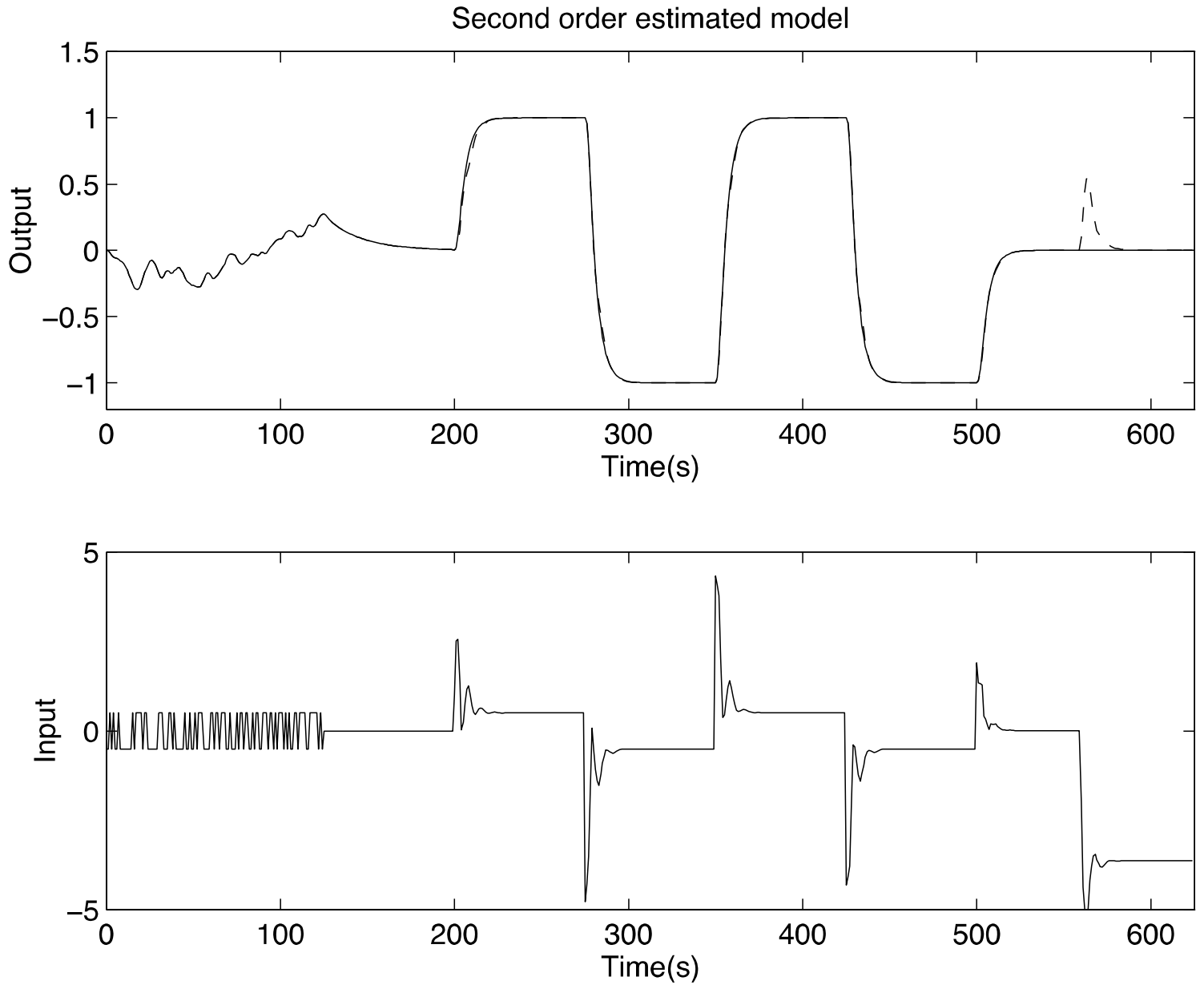

**Fig. 12.2** Second order estimated model (— reference trajectory, - - - plant output)

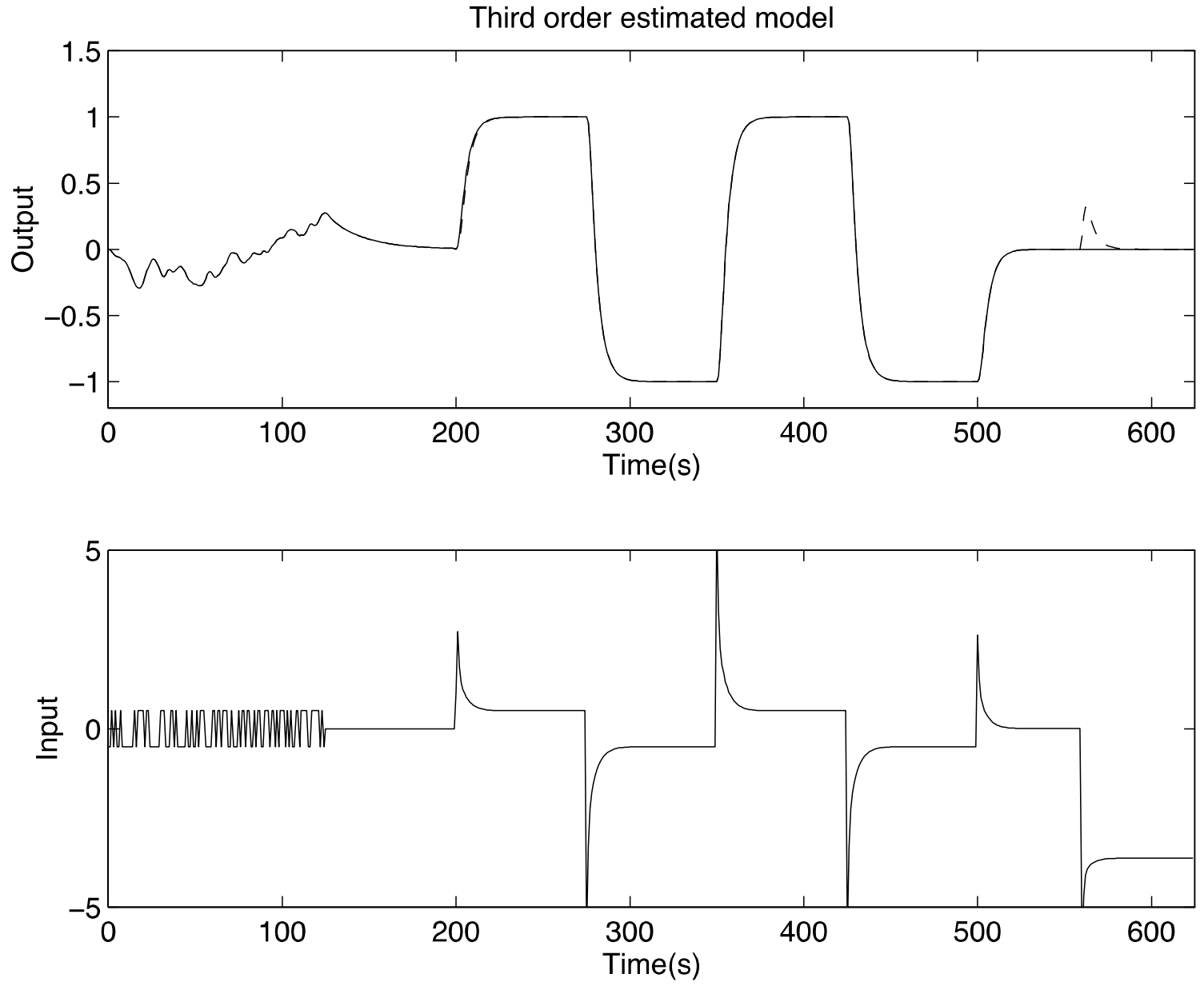

**Fig. 12.3** Third order estimated model (— reference trajectory, - - - plant output)

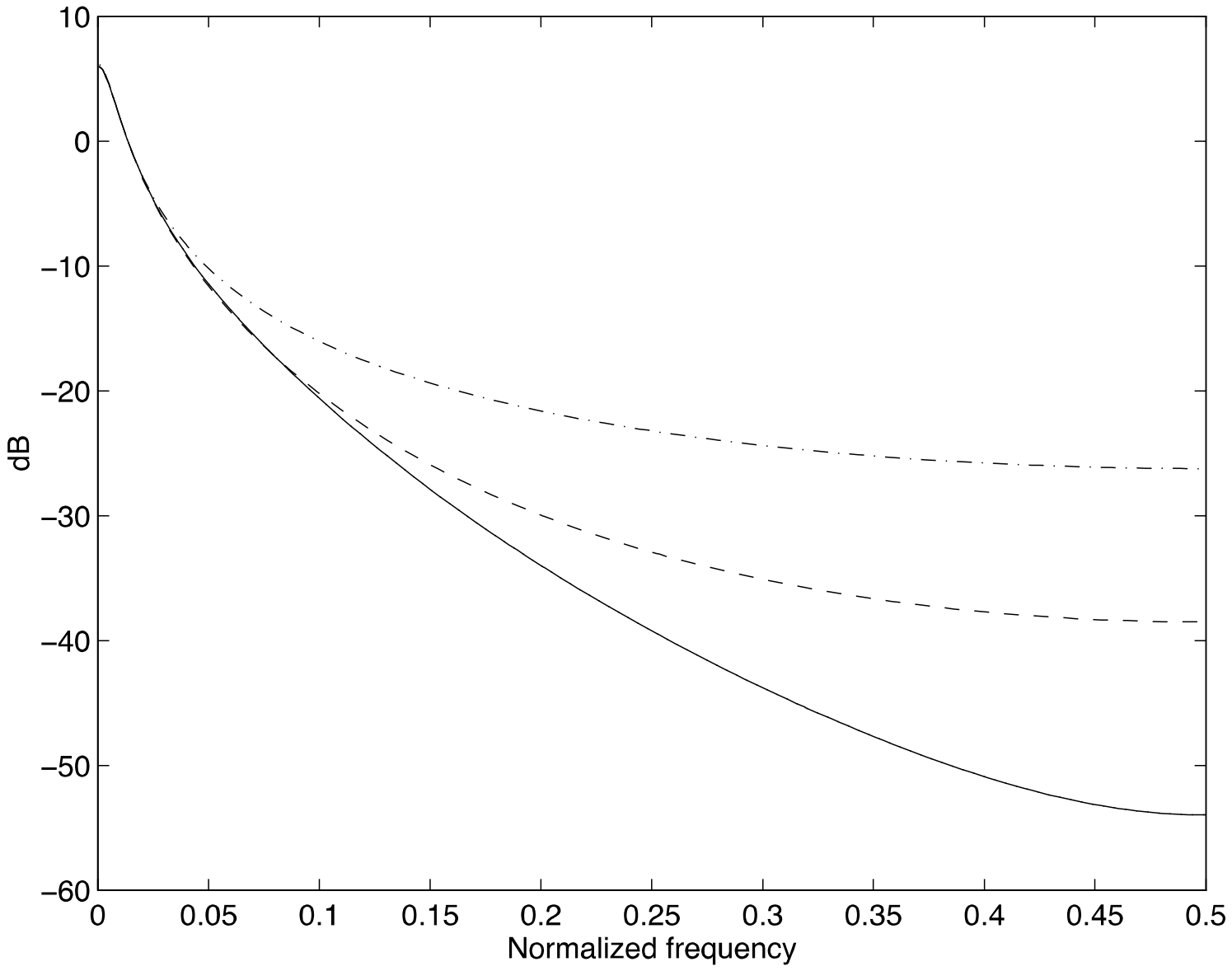

**Fig. 12.4** Frequency characteristics for $n = 1(-.-), 2(--), 3(-)$

An adaptation gain with variable forgetting factor combined with a constant trace adaptation gain has been used ($F(0) = \alpha I$; $\alpha = 1000$ desired trace: $\text{tr}\, F(t) = 6$). The filter used on input/output data is: $L(q^{-1}) = 1/P(q^{-1})$. The normalizing signal $m(t)$ has been generated by $m^2(t) = \mu^2 m^2(t-1) + \max(\|\phi_f(t)\|, 1)$ with $\mu = 0.9$. For $n = 1$, taking into account the unmodeled dynamics, the theoretical value for $\mu$ is $0.6 < \mu < 1$ (the results are not very sensitive with respect to the choice of the desired trace and $\mu$).

The conclusion is that for this example, despite the fact that a stable adaptive controller can be obtained with a first order estimated model corresponding to the dominant dynamics of the true plant model, one should use a second order estimated model in order to obtain a good performance.

The rule which can be established is that low-order modeling can be used in adaptive control but good performance requires that this model be able to copy the frequency characteristics of the true plant model in the frequency region relevant for control design.

## 12.4 Adaptive Generalized Predictive Control

The basic algorithm for implementing Adaptive (PSMR) Generalized Predictive Control is similar to that for adaptive pole placement except for the details of Step II: Computation of the controller parameters and of the control law. In this case, using

Strategy 1, for the updating of the controller parameters (see Sect. 12.1) the controller equation generating $u(t)$ is:

$$\hat{S}(t,q^{-1})u(t)+\hat{R}(t,q^{-1})y(t)=\hat{\beta}(t)\hat{T}(t,q^{-1})y^{*}(t+d+1) \tag{12.156}$$

where:

$$\hat{\beta}(t)=1/\hat{B}(t,1) \tag{12.157}$$

$$\hat{S}(t,q^{-1})=\hat{S}'(t,q^{-1})H_S(q^{-1}) \tag{12.158}$$

$$\hat{R}(t,q^{-1})=\hat{R}'(t,q^{-1})H_R(q^{-1}) \tag{12.159}$$

$$\hat{T}(t,q^{-1})=\hat{A}(t,q^{-1})\hat{S}(t,q^{-1})+q^{-d-1}\hat{B}^{*}(t,q^{-1})\hat{R}(t,q^{-1}) \tag{12.160}$$

From (7.222) and (7.223), one gets for a certain value of the estimated plant model parameter vector $\hat{\theta}(t)$:

$$\hat{S}'(t,q^{-1})=P_D(q^{-1})+q^{-1}\sum_{j=h_i}^{h_p}\hat{\gamma}_j(t)\cdot\hat{H}_{j-d}(t,q^{-1}) \tag{12.161}$$

$$\hat{R}'(t,q^{-1})=\sum_{j=h_i}^{h_p}\hat{\gamma}_j(t)\hat{F}_j(t,q^{-1}) \tag{12.162}$$

where $\hat{\gamma}_j(t)$ are the elements of the first row of the matrix $[\hat{G}^T(t)\hat{G}(t)+\lambda I_{h_c}]^{-1}\hat{G}^T(t)$ and $\hat{G}(t)$ is an estimation of the matrix (7.215). To effectively compute $\hat{\gamma}_j(t)$, $\hat{H}_{j-d}(t,q^{-1})$ and $\hat{F}_j(t,q^{-1})$, one has to solve at each sampling instant $t$ the two polynomial divisions:

$$P_D(q^{-1})=\hat{\hat{A}}(t,q^{-1})\hat{E}_j(t,q^{-1})+q^{-j}\hat{F}_j(t,q^{-1}) \tag{12.163}$$

$$\begin{aligned}\hat{\hat{B}}^{*}(t,q^{-1})\hat{E}_j(t,q^{-1})&=P_D(q^{-1})\hat{G}_{j-d}(t,q^{-1})\\&\quad+q^{-j+d}\hat{H}_{j-d}(t,q^{-1})\end{aligned} \tag{12.164}$$

for the estimated value of the plant model parameter vector $\hat{\theta}(t)$ given in (12.7).

As in the case of the pole placement, one has to check the admissibility of the estimated plant model. For the case of generalized predictive control, one checks if the estimated characteristic polynomial

$$\hat{P}(t,q^{-1})=\hat{A}(t,q^{-1})\hat{S}(t,q^{-1})+q^{-d-1}\hat{B}^{*}(t,q^{-1})\hat{R}(t,q^{-1}) \tag{12.165}$$

is asymptotically stable and that $\hat{\beta}(t)>0$.

## 12.5 Adaptive Linear Quadratic Control

The basic algorithm for implementing (PSMR) Linear Quadratic Control is similar to that for adaptive pole placement except for the details of Step II: Computation of the controller parameters and of the control law. In this case, using strategy I for the updating of the controller parameters one has:

$$\hat{x}(t+1) = \hat{A}(t)\hat{x}(t) + \hat{b}(t)\bar{e}_u(t) + \hat{k}(t)[e_y(t) - \hat{e}_y(t)] \tag{12.166}$$

$$\hat{e}_y(t) = c^T\hat{x}(t) \tag{12.167}$$

$$\hat{k}^T(t) = [\hat{k}_1(t), \ldots, \hat{k}_n(t)]; \quad \hat{k}_i(t) = p_i - \hat{a}_i(t) \tag{12.168}$$

$$\bar{e}_u(t) = -\frac{\hat{b}^T(t)\hat{\Gamma}(t)A(t)}{\hat{b}^T(t)\hat{\Gamma}(t)\hat{b}(t) + \lambda}\hat{x}(t) \tag{12.169}$$

where $p_i$ are the coefficients of the polynomial $P_D(q^{-1})$ defining desired observer dynamics and $\Gamma(t)$ is the positive definite solution of the Algebraic Riccati Equation computed for the estimated values $\hat{A}(t)$ and $\hat{b}(t)$:

$$\begin{aligned} &\hat{A}^T(t)\hat{\Gamma}(t)\hat{A}(t) - \hat{\Gamma}(t) \\ &\quad - \hat{A}^T(t)\hat{\Gamma}(t)\hat{b}(t)[\hat{b}^T(t)\hat{\Gamma}(t)\hat{b}(t) + \lambda]^{-1}\hat{b}^T(t)\hat{\Gamma}(t)\hat{A}(t) \\ &\quad + cc^T = 0 \end{aligned} \tag{12.170}$$

As indicated in Samson (1982), instead of solving an ARE at each sampling instant one can use a Time Varying Riccati Equation with an iteration at each sampling instant, i.e.:

$$\begin{aligned} \hat{\Gamma}(t) = {} & \hat{A}^T(t)\hat{\Gamma}(t-1)\hat{A}(t) - \hat{A}^T(t)\hat{\Gamma}(t-1)\hat{b}(t)[\hat{b}^T(t)\Gamma(t-1)\hat{b}(t) + \lambda]^{-1} \\ & \times \hat{b}^T(t)\hat{\Gamma}(t-1)\hat{A}(t) + cc^T = 0 \end{aligned} \tag{12.171}$$

since if asymptotically $\hat{A}(t)$ and $\hat{b}(t)$ tend toward constant values satisfying the existence condition for a positive definite solution of the ARE, $\hat{\Gamma}(t)$ given by (12.171) will tend toward the solution of the ARE. From (12.166) through (12.169), using (7.293) through (7.297), one gets at each instant $t$ the RST controller form.

As in the case of the pole placement, one has to check the admissibility of the estimated plant model at each instant $t$. For the case of linear quadratic control, one checks if the eventual common factors of $\hat{A}(t, q^{-1})$ and $\hat{B}(t, q^{-1})$ are inside the unit circle.

## 12.6 Adaptive Tracking and Robust Regulation

The idea in this approach is to:

1. use of a robust linear controller with fixed parameters for regulation ($R(q^{-1})$ and $S(q^{-1})$);

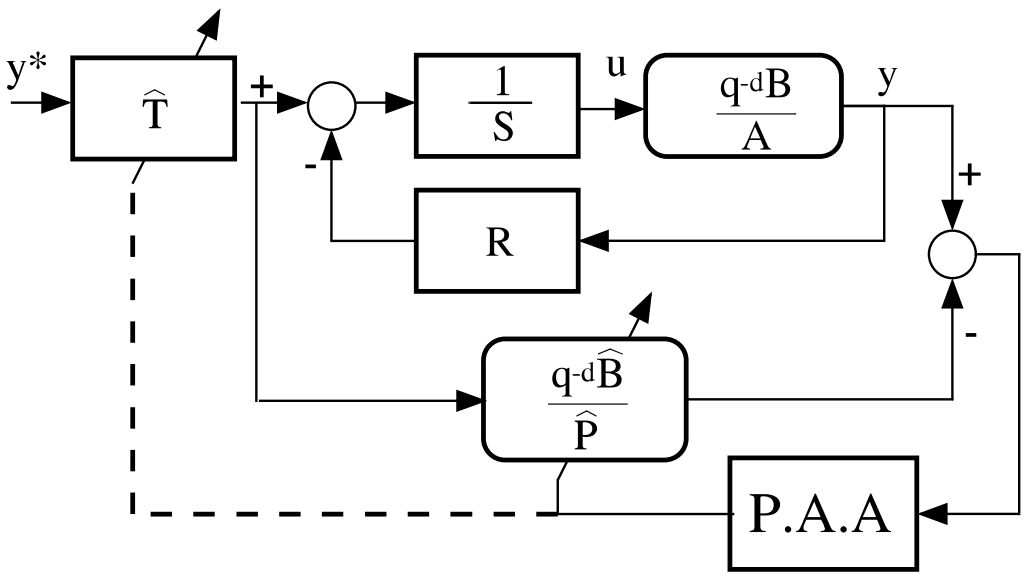

**Fig. 12.5** Adaptive tracking and robust regulation

2. identify the closed-loop poles (either directly or indirectly by identifying the plant model in closed-loop operation and computing the closed-loop poles);
3. adapt the parameters of the precompensator $\hat{T}(t, q^{-1})$ based on the current poles and zeros of the closed loop.

This is illustrated in Fig. 12.5.

This technique is compared with a robust controller and with an indirect adaptive controller in Sect. 12.7 (Fig. 12.14) for the case of the control of the flexible transmission.

## 12.7 Indirect Adaptive Control Applied to the Flexible Transmission

The performance of several indirect adaptive control schemes will be illustrated by their applications to the control of the flexible transmission shown in Fig. 1.19.

### *12.7.1 Adaptive Pole Placement*

In this case, the pole placement will be used as control strategy but two types of parameter estimators will be used:

1. filtered recursive least squares;
2. filtered closed-loop output error (F-CLOE).

As indicated in Chap. 1 in indirect adaptive control the objective of the plant parameter estimation is to provide the best prediction for the behavior of the closed loop system, for given values of the controller parameters. This can be achieved by either using appropriate data filters on plant input-output data or by using adaptive predictors for the closed-loop system parameterized in terms of the controller parameters and plant parameters (Landau and Karimi 1997b). The corresponding parameter estimators are illustrated in Figs. 12.6a and 12.6b and details can be found in Chap. 9.

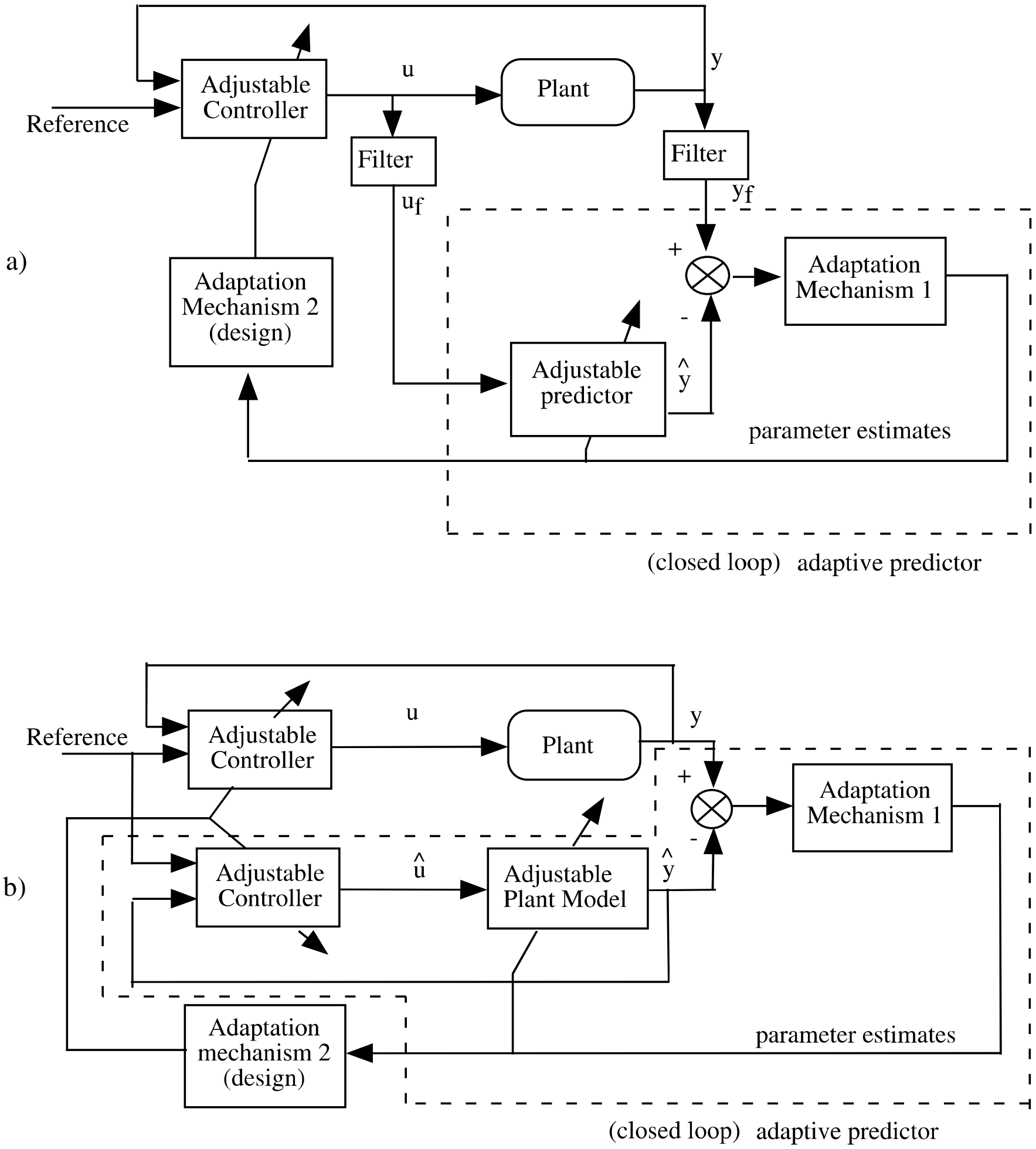

**Fig. 12.6** Indirect adaptive control with closed-loop adjustable predictors, (**a**) using input-output data filters, (**b**) using closed-loop predictors

These experiments have been carried out using the real time simulation package Vissim/RT (Visual Solutions 1995). Figs. 12.7, 12.8 and 12.9 show the behavior of the adaptive pole placement using filtered recursive least squares without adaptation freezing for various loads. In each experiment, an open-loop identification for the initialization of the adaptive scheme is carried out during 128 samples, then one closes the loop and one sends a sequence of step reference changes followed by the application of a position disturbance. The upper curves show the reference trajectory and the output of the system, the curves in the middle show the evolution of the input and the lower curves show the evolution of the estimated parameters. One can observe that the system is almost tuned at the end of the initialization period.

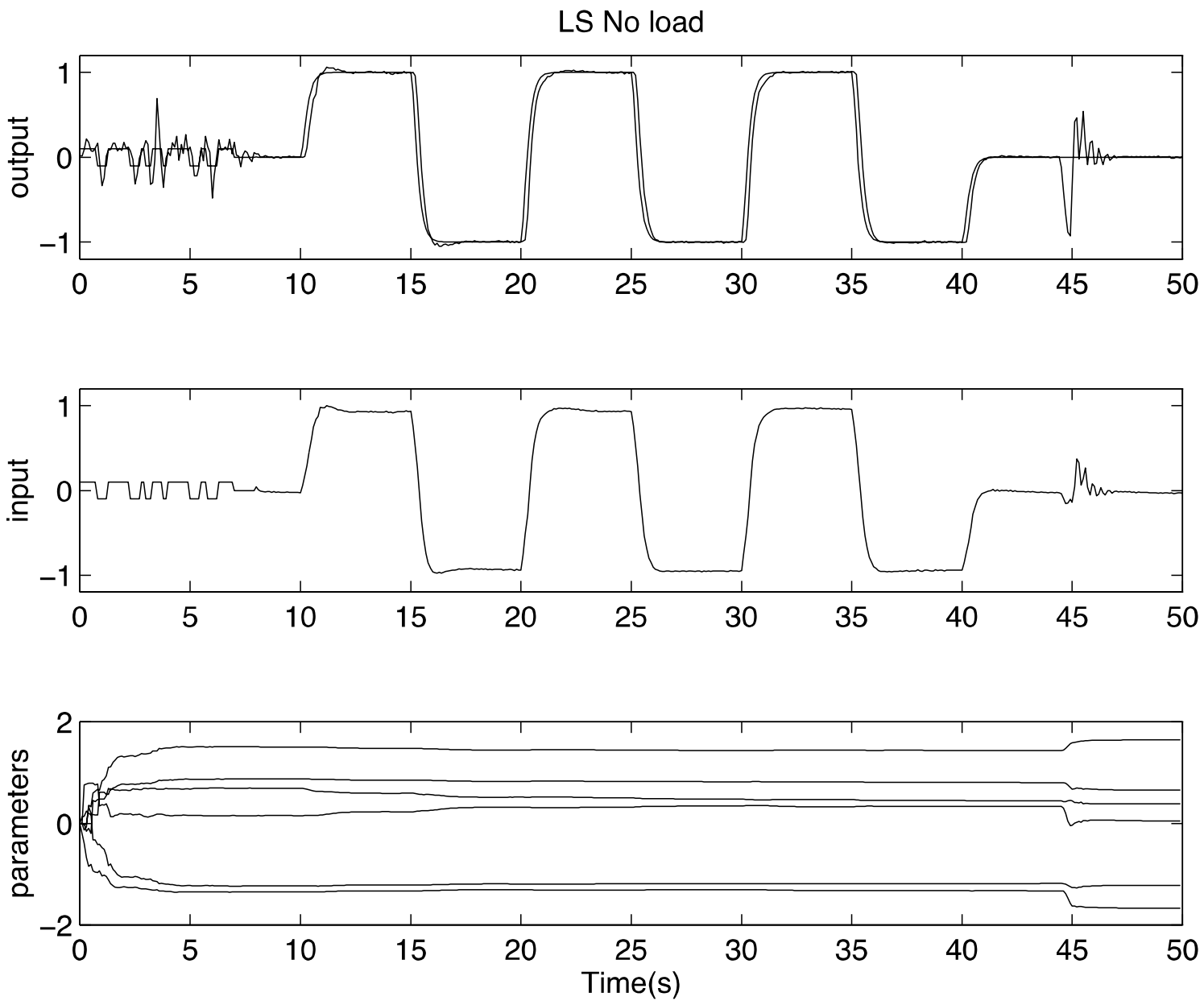

**Fig. 12.7** Flexible transmission—adaptive pole placement using filtered RLS. The no load case, (**a**) reference trajectory and output, (**b**) input, (**c**) estimated parameters

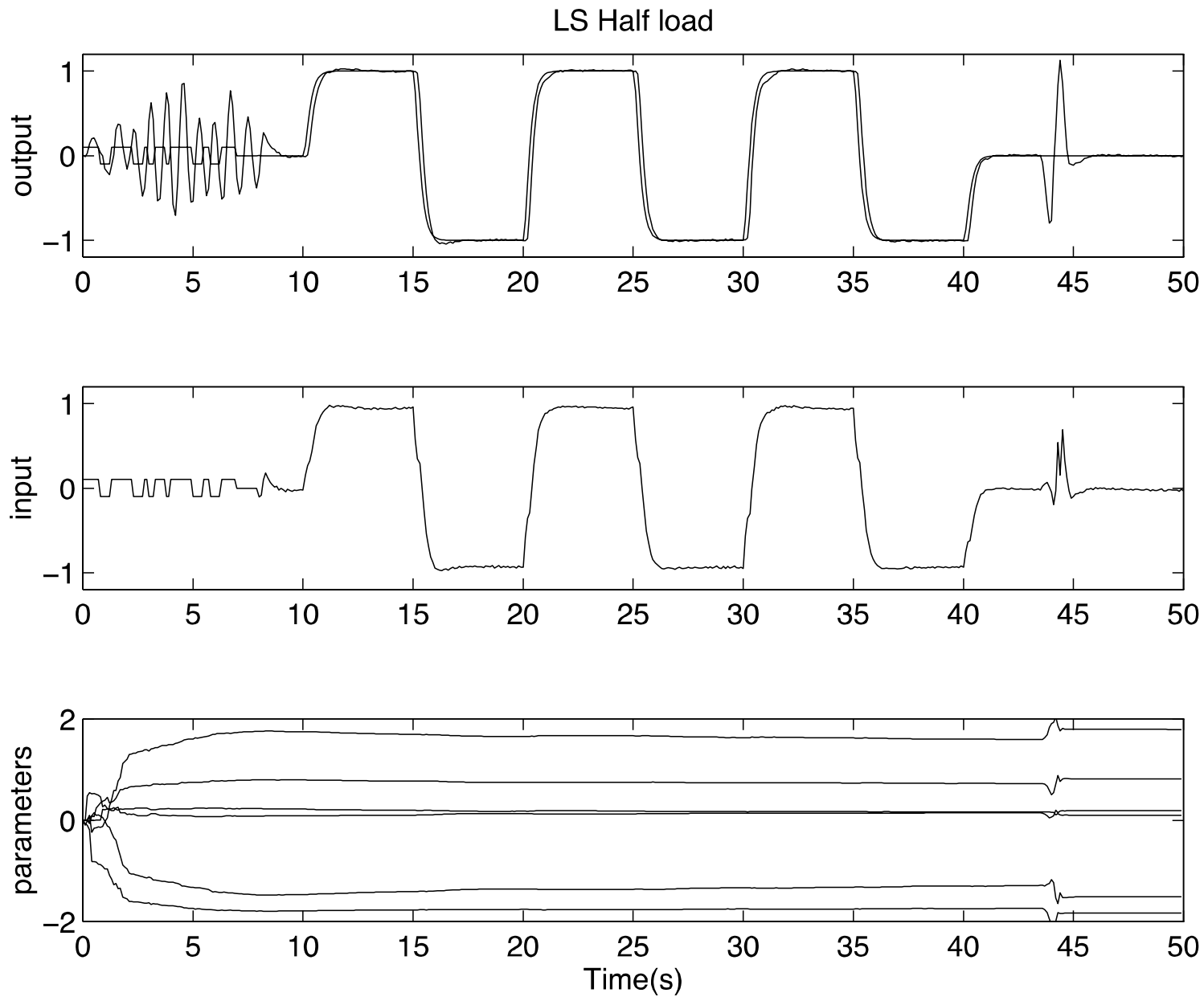

**Fig. 12.8** Flexible transmission—adaptive pole placement using filtered RLS. The half load case, (**a**) reference trajectory and output, (**b**) input, (**c**) estimated parameters

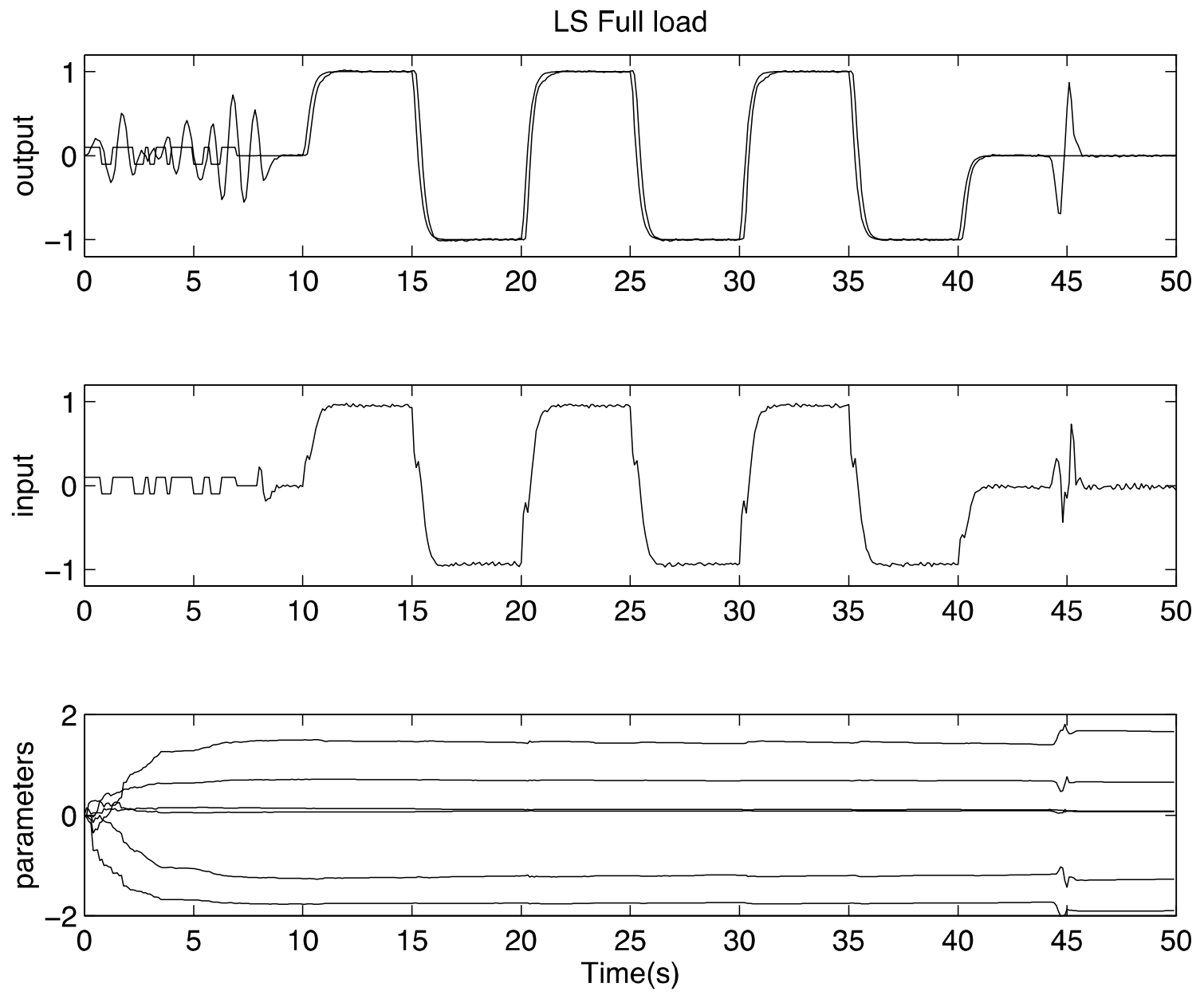

**Fig. 12.9** Flexible transmission—adaptive pole placement using filtered RLS. The full load case, (**a**) reference trajectory and output, (**b**) input, (**c**) estimated parameters

Since adaptation freezing is not used, a drift in the estimated parameters occurs in the presence of the disturbance. Successive applications of such disturbances will destabilize the system. Therefore, in the case of recursive least squares (or any other open-loop type identification algorithm) the use of adaptation freezing is mandatory (see Sect. 12.7.2).

For this application the pole placement design is characterized by the following desired closed-loop poles:

1. A pair of dominant complex poles ($P_D(q^{-1})$) with the frequency of the loaded first vibration mode ($\omega_0 = 6$ rad/s) and with a damping $\zeta = 0.9$.
2. Auxiliary poles as follows: a pair of complex poles ($\omega_0 = 33$ rad/s, $\zeta = 0.3$) and the real poles $(1 - 0.5q^{-1})^3(1 - 0.1q^{-1})^3$.

The controller has an integrator ($H_S(q^{-1}) = 1 - q^{-1}$) and contains a filter $H_R(q^{-1}) = (1 + q^{-1})^2$ which reduces the modulus of the input sensitivity function in the high frequencies. The plant model estimator uses a decreasing adaptation gain combined with constant trace adaptation gain with $\operatorname{tr} F(t) = \operatorname{tr}[I_6]$. The input/output data are filtered by a band pass Butterworth filter with two cells (low frequency 0.05 Hz, high frequency 7 Hz). Figures 12.10, 12.11 and 12.12 show the results for the same type of experiment when using F-CLOE identification algorithm instead of the filtered least squares. Almost similar results are obtained in tracking but there is a significant difference in regulation. Despite the absence of adaptation freezing, the parameter are not drifting in the presence of disturbances and therefore

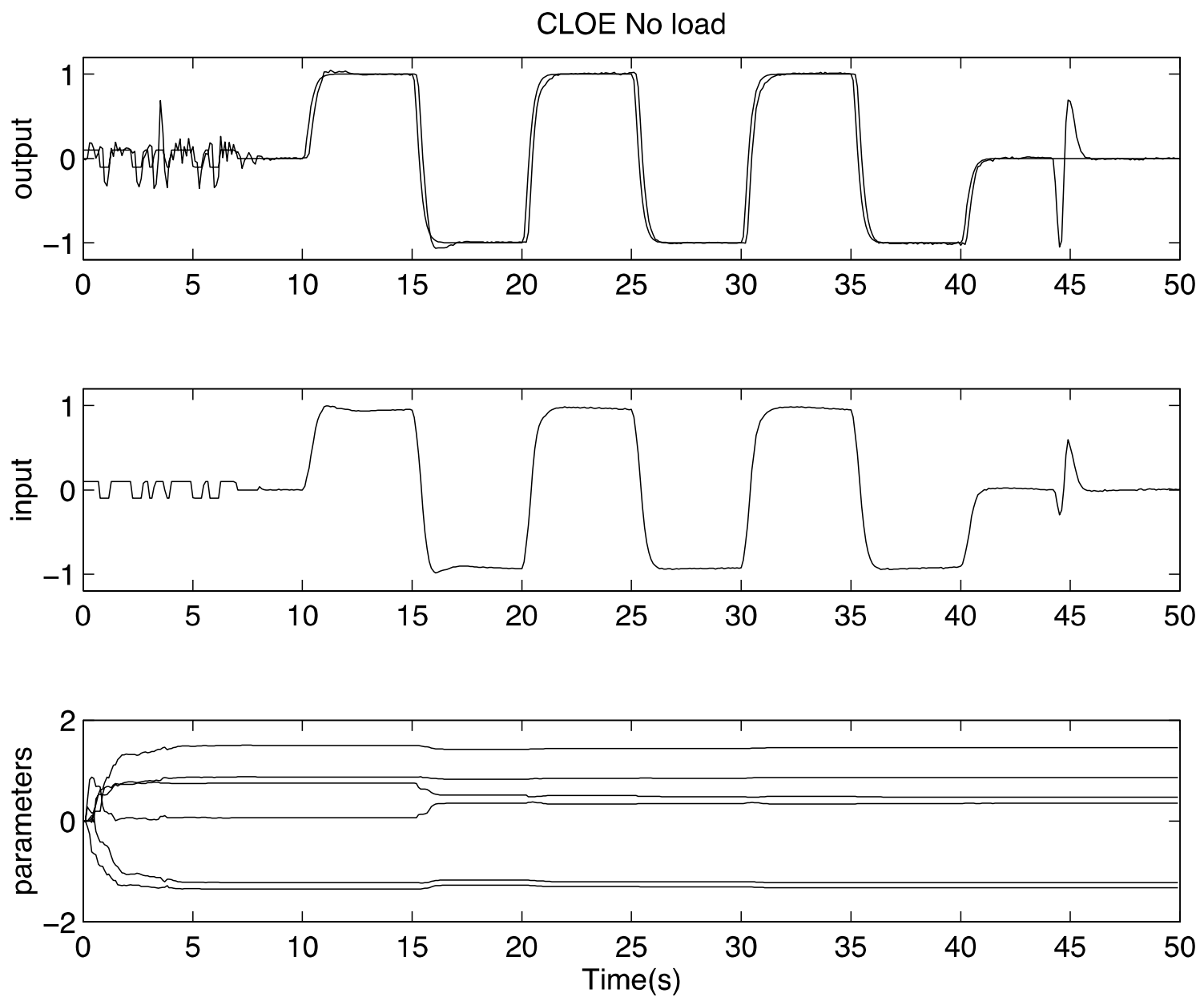

**Fig. 12.10** Flexible transmission—adaptive pole placement using F-CLOE. The no load case, (**a**) reference trajectory and output, (**b**) input, (**c**) estimated parameters

there is no risk of instability. The observation vector used in F-CLOE algorithm is filtered through $(1 - q^{-1})/P_D(q^{-1})$ where $P_D(q^{-1})$ corresponds to the desired dominant closed-loop pole. Figure 12.13 shows the influence of the various linear design choices (the F-CLOE algorithm is used). For this case (full load) the auxiliary poles are all set to zero and $H_R(q^{-1}) = 1 + 0.5q^{-1}$. Comparing these results with those shown in Figs. 12.9 and 12.12, one sees an increase of the variance of the input. This is caused by higher values of the modulus of the input sensitivity function in the high frequencies and justify:[4]

1. introduction of auxiliary poles;
2. use of $H_R(q^{-1}) = (1 + \beta q^{-1})$ or $H_R(q^{-1}) = (1 + \beta q^{-1})^2$ with $0.7 \leq \beta \leq 1$ ($\beta = 1$ corresponds to the opening of the loop at $0.5 f_s$).

Figure 12.14 shows a comparison between an adaptive pole placement controller using F-CLOE parameter estimation algorithm (upper curves), a robust controller designed using the method given in Langer and Landau (1999) which meets all the specifications of the robust digital control benchmark for the flexible transmission (Landau et al. 1995a) (middle curves) and an adaptive tracking with robust regulation using the above robust regulator (lower curves).

[4] See Chap. 8, Sect. 8.7 for details.

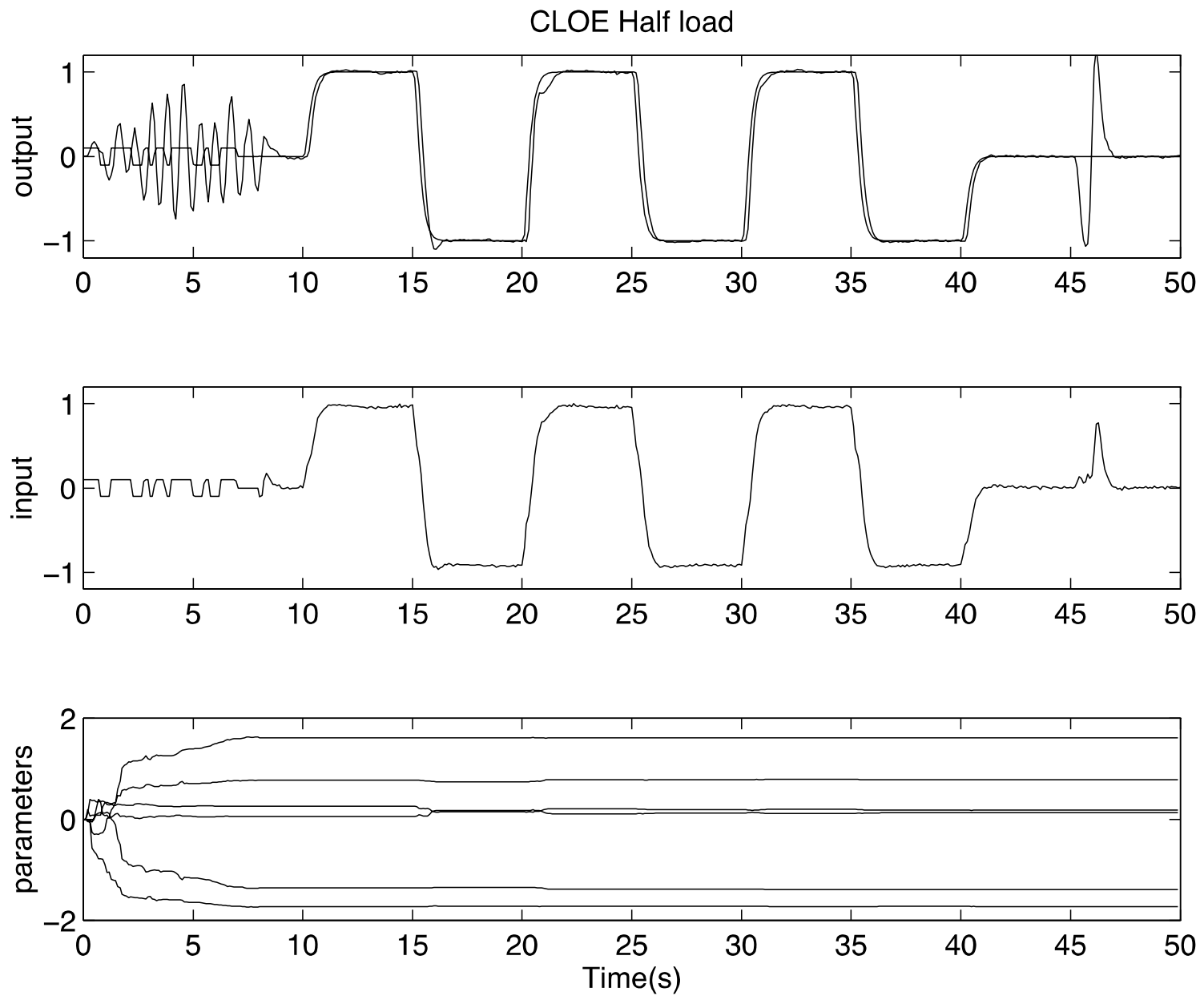

**Fig. 12.11** Flexible transmission—adaptive pole placement using F-CLOE. The half load case, (**a**) reference trajectory and output, (**b**) input, (**c**) estimated parameters

The experiment is carried out from "no load" case at $t = 0$, toward the "full load" case and the load is increased after each step reference change. At the end (full load) a position disturbance is applied. The adaptive scheme is tuned for the no load case at $t = 0$. One clearly sees the improvement of performance when using adaptive control. One also can see that adaptive tracking improves the performance in tracking with respect to the robust controller with a fixed polynomial $T$. The performance can be further improved after several more step reference changes.

## *12.7.2 Adaptive PSMR Generalized Predictive Control*

In this case, the *partial state model reference* generalized predictive control combined with a filtered recursive least squares parameter estimation algorithm will be used. Adaptation freezing in the absence of significant information is incorporated in the algorithm.

The experiments have been carried out using the SIMART package (M'Saad and Chebassier 1997).[5] Figures 12.15, 12.16 and 12.17 show the performance of the adaptive PSMR/GPC for various loads. In each experiment, an open-loop identification for the initialization of the adaptive scheme is carried out during 128 samples, then one closes the loop and one sends a sequence of step reference changes followed by the application of a position disturbance. The upper curve shows the output of the system. The curve below shows the input of the system. The next curves show the evolution of two estimated parameters. The lower curve shows the evolution of the scheduling variable for adaptation freezing (value 1) and of $\phi_f^T(t)F(t)\phi_f(t)$.

[5]These experiments have been carried out by J. Chebassier (Lab. d'Automatique de Grenoble).

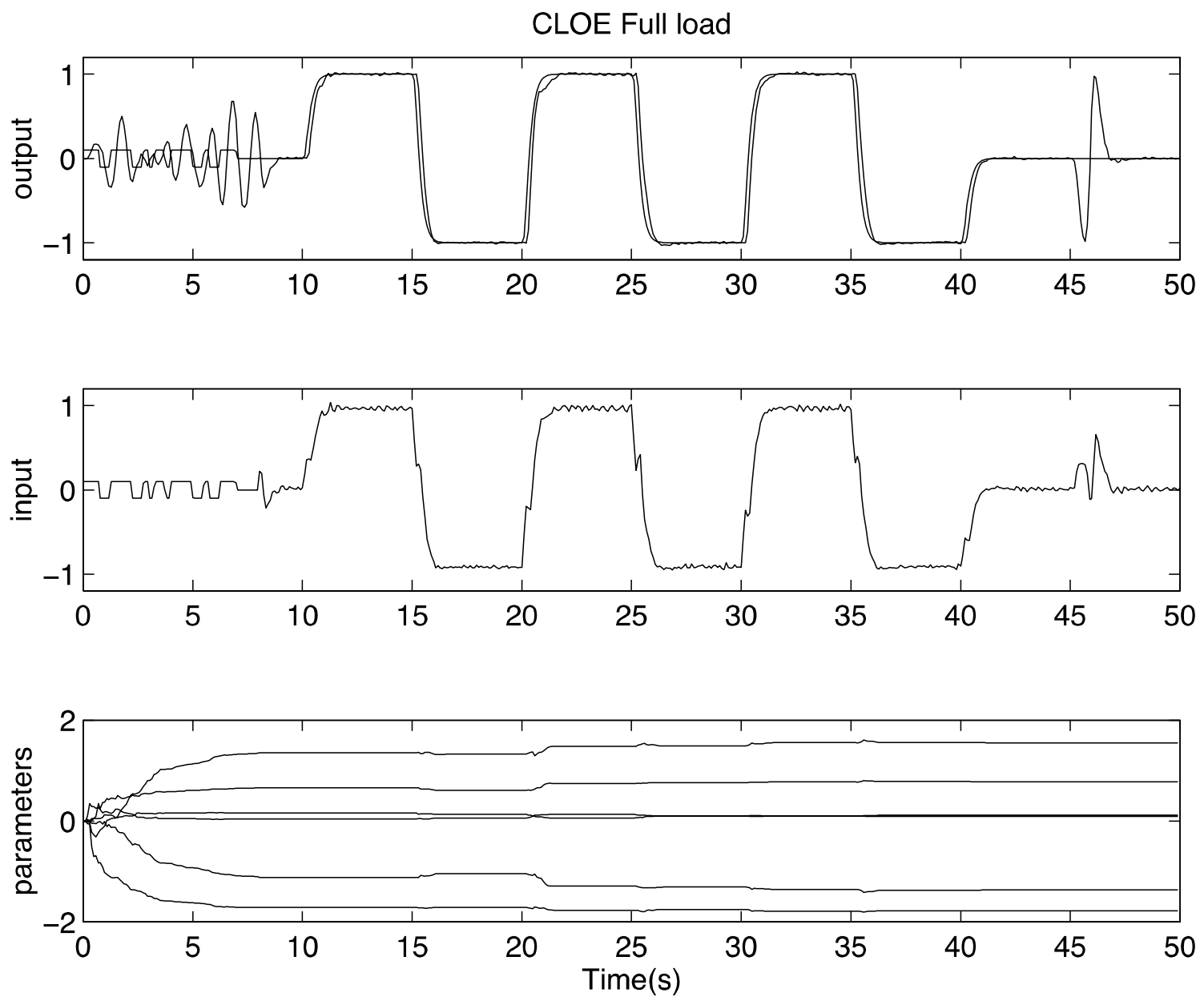

**Fig. 12.12** Flexible transmission—adaptive pole placement using F-CLOE. The full load case, (**a**) reference trajectory and output, (**b**) input, (**c**) estimated parameters

For this application the PSMR/GPC is characterized by:

1. Predictor poles: Two pairs of complex poles corresponding to ($\omega_0 = 10.7$ rad/s, $\zeta = 1$) and ($\omega_0 = 30$ rad/s, $\zeta = 1$) and three real poles $z_1 = z_2 = 0.585$ and $z_3 = 0.7$.
2. Initial horizon $h_i = 3$, prediction horizon $h_p = 20$, control horizon $h_c = 3$, input weighting $\lambda = 0.5$, $H_S(q^{-1}) = 1 - q^{-1}$ (integrator behavior).

The plant model estimator uses an adaptation gain with variable forgetting factor combined with constant trace adaptation gain with $\operatorname{tr} F(t) = \operatorname{tr}[I_6]$. The input/output data are filtered through a band pass Butterworth filter with two cells (low frequency: 0.005 Hz, high frequency: 7 Hz). The adaptation freezing is based on the evaluation of the $\phi_f^T(t)F(t)\phi_f(t)$ over a sliding horizon of 10 samples (i.e., if this quantity is below the threshold during 10 samples, the adaptation is switched off). Figure 12.18 shows the performance of the adaptive PSMR/GPC in the presence of a step load change from the full load case to the no load case (the most difficult situation). The reference signal is a sinusoid. As a consequence of the drastic change in the parameters of the model, an instability phenomena is ignited, but the adaptation succeeds to quickly stabilize the system.

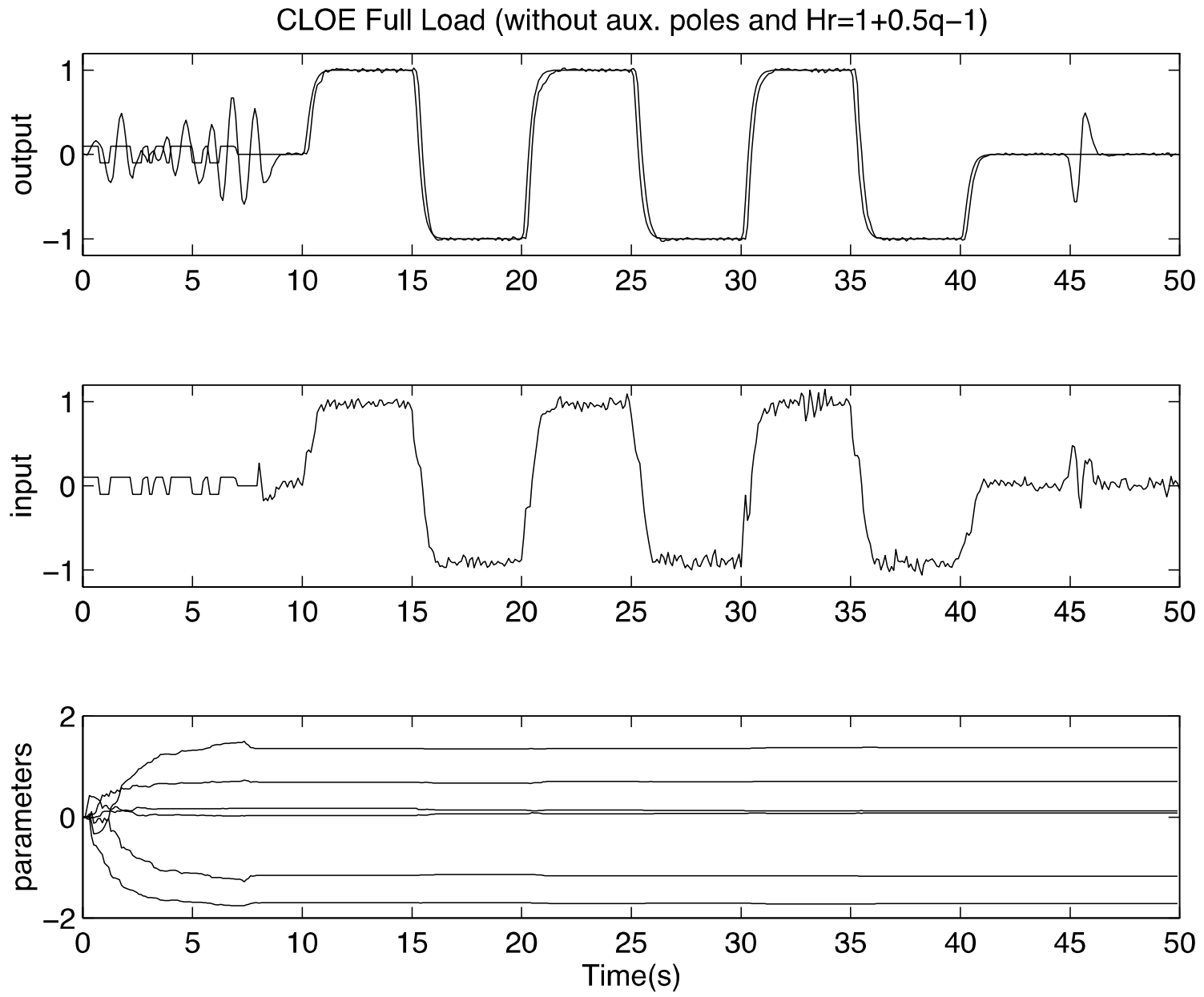

**Fig. 12.13** Adaptive pole placement without auxiliary poles and $H_R(q^{-1}) = 1 + 0.5q^{-1}$ (full load), (**a**) reference trajectory and output, (**b**) input, (**c**) estimated parameters

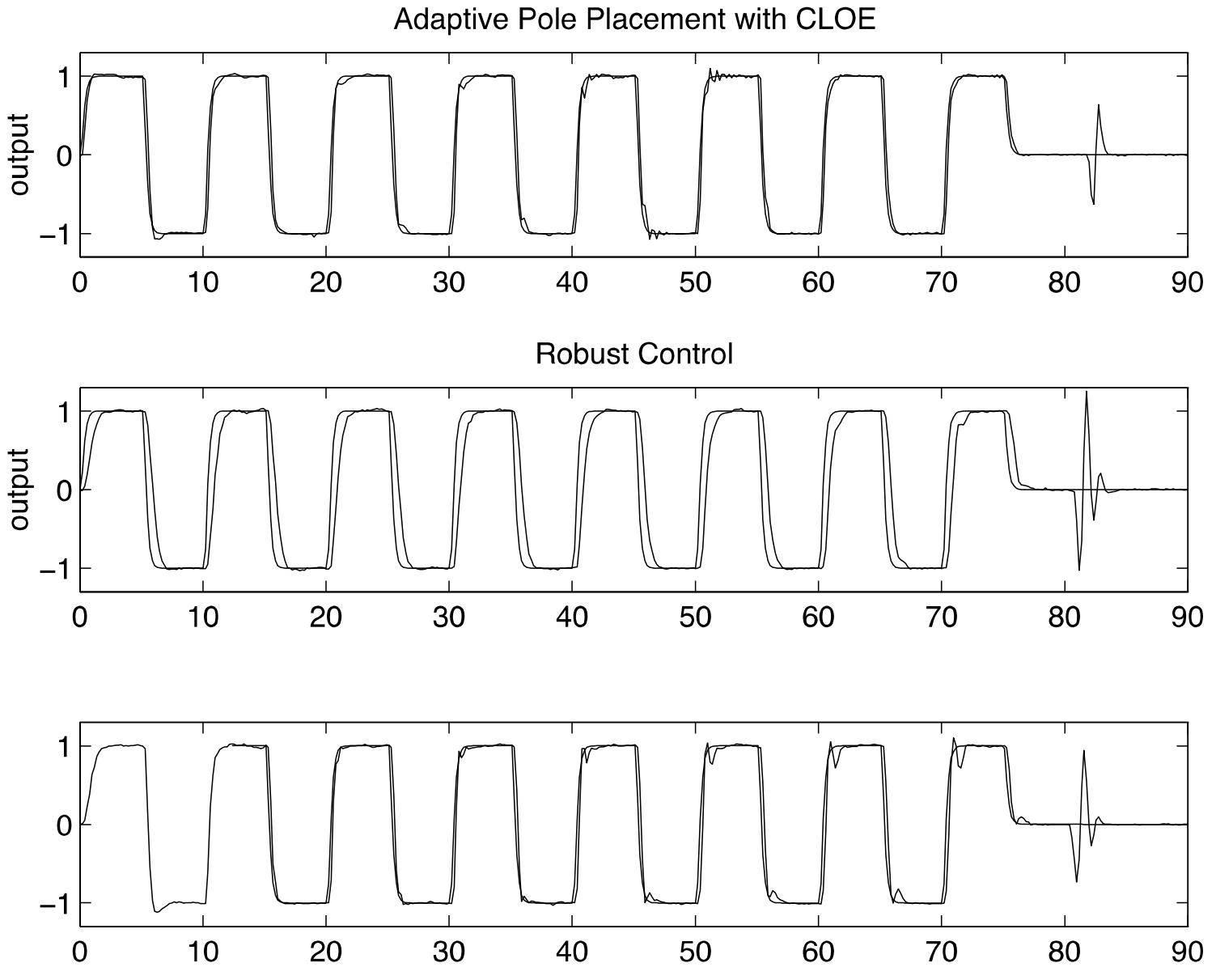

**Fig. 12.14** Flexible transmission—comparison of three control strategies, (**a**) adaptive pole placement with CLOE, (**b**) robust control, (**c**) adaptive tracking with robust regulation

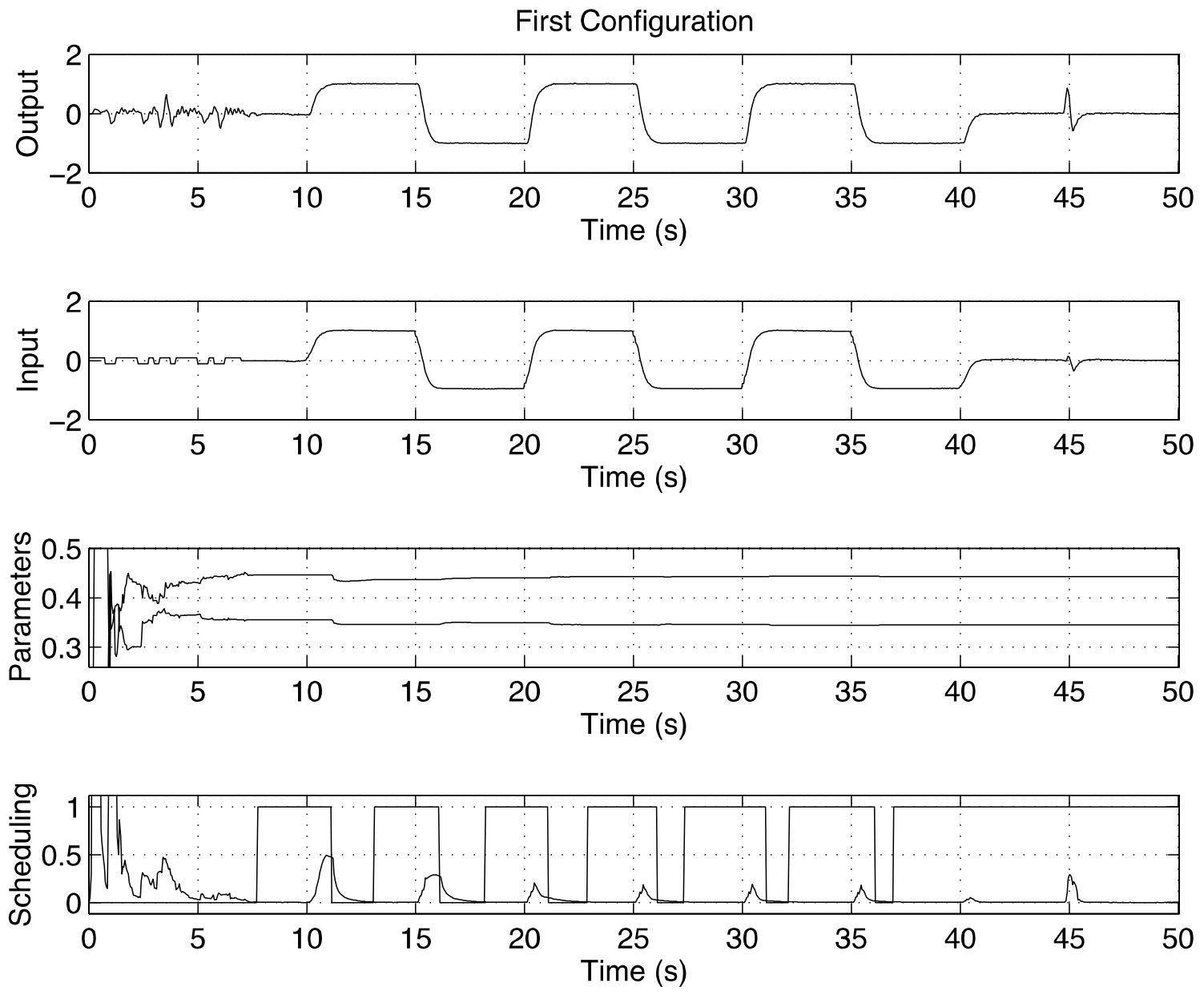

**Fig. 12.15** Flexible transmission—adaptive PSMR/GPC. The no load case, (**a**) output, (**b**) input, (**c**) estimated parameters, (**d**) scheduling variable for adaptation freezing

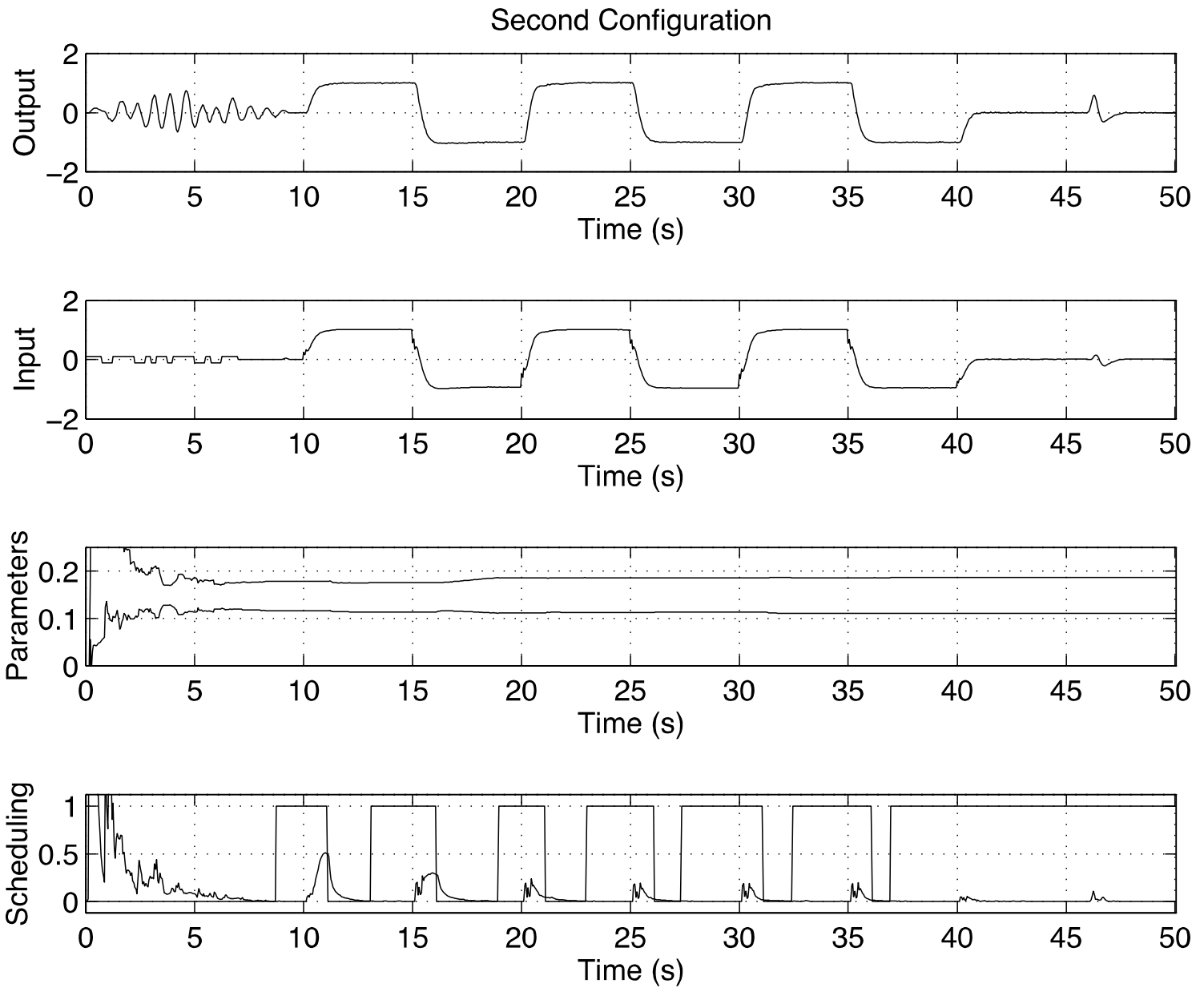

**Fig. 12.16** Flexible transmission—adaptive PSMR/GPC. The half load case, (**a**) output, (**b**) input, (**c**) estimated parameters, (**d**) scheduling variable for adaptation freezing

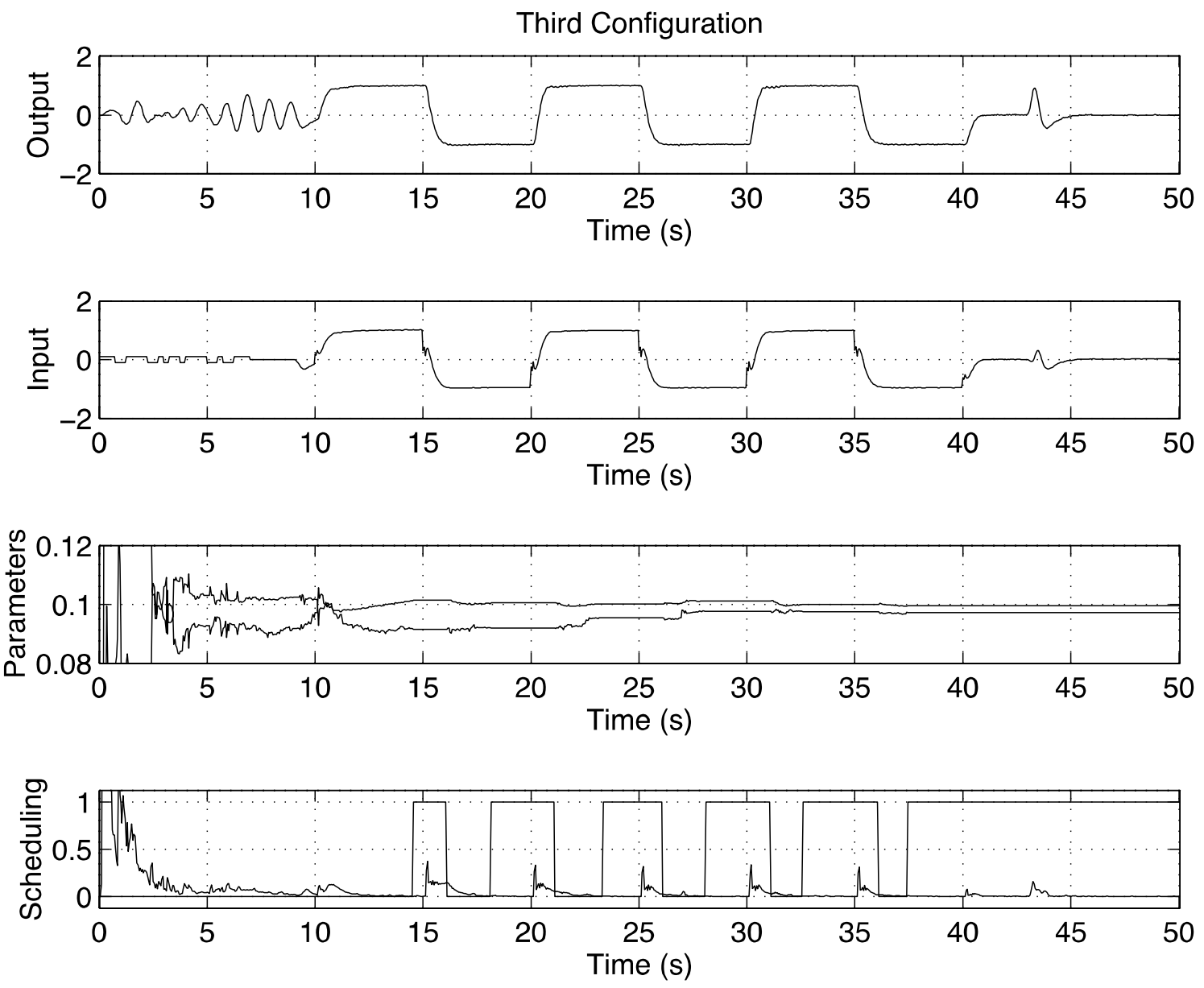

**Fig. 12.17** Flexible transmission—adaptive PSMR/GPC. The full load case, (**a**) output, (**b**) input, (**c**) estimated parameters, (**d**) scheduling variable for adaptation freezing

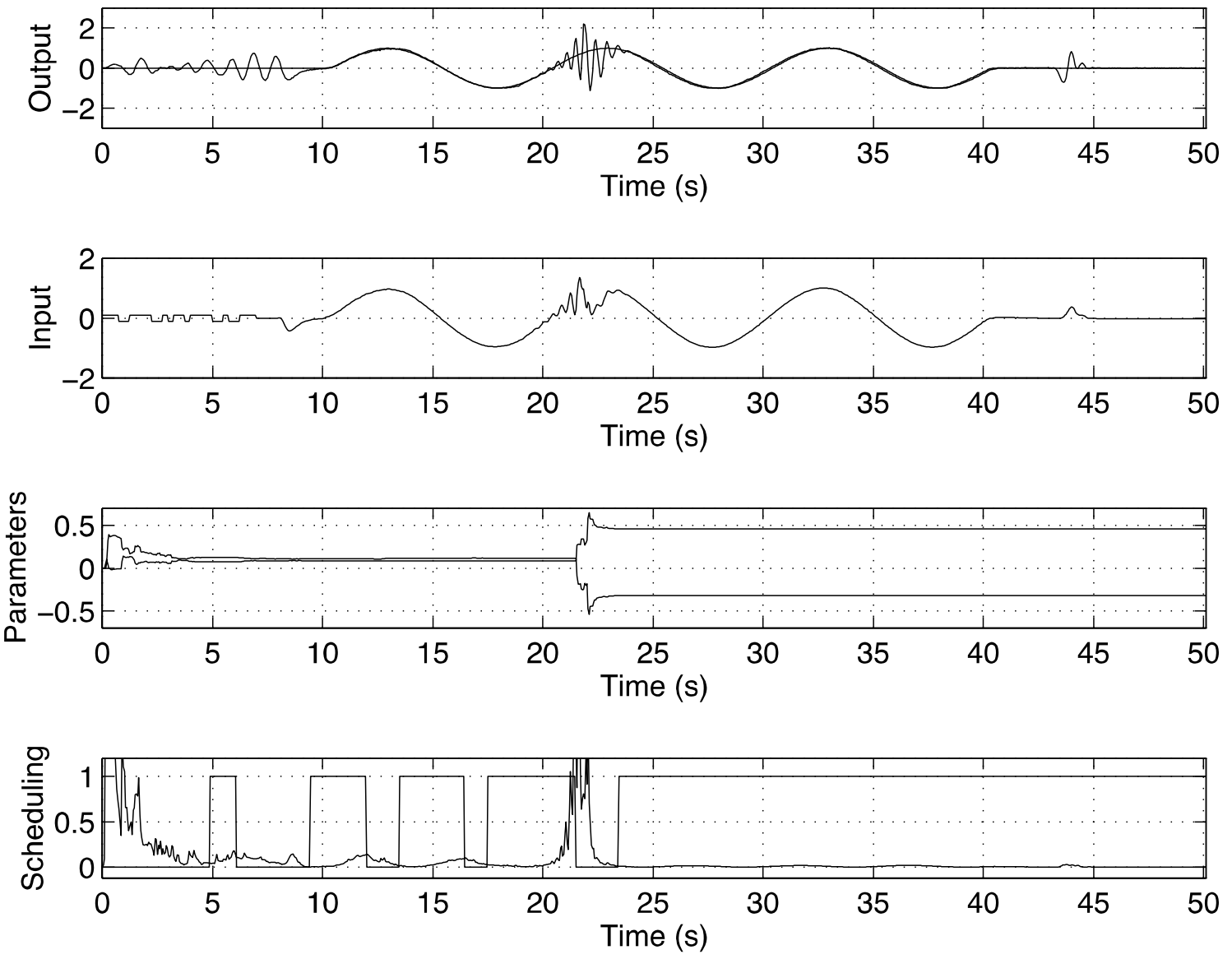

**Fig. 12.18** Flexible transmission—adaptive PSMR/GPC. Effect of step load change (100% → 0%), (**a**) reference trajectory and output, (**b**) input, (**c**) estimated parameters, (**d**) scheduling variable for adaptation freezing

## 12.8 Concluding Remarks

1. Indirect adaptive control algorithms emerged as a solution for adaptive control of systems featuring discrete time models with unstable zeros.
2. Indirect adaptive control offers the possibility of combining (in principle) any linear control strategy with a parameter estimation scheme.
3. The plant model parameter estimator should allow a good prediction of the behavior of the closed loop.
4. The design of the controller based on the estimated plant models should be done such that some robustness constraints on the sensitivity functions be satisfied (in particular on the output and on the input sensitivity functions). This can be achieved by a suitable choice of the desired performances and introduction of some fixed filters in the controller.
5. For each type of underlying linear control strategy used in indirect adaptive control a specific admissibility test has to be done on the estimated model prior to the computation of the control.
6. Robustification of the parameter adaptation algorithms used for plant model parameter estimation may be necessary in the presence of unmodeled dynamics and bounded disturbances.
7. Adaptive pole placement and adaptive generalized predictive control are the most used indirect adaptive control strategies.

## 12.9 Problems

**12.1** Under what conditions, (12.16) will imply $\lim_{t\to\infty} \varepsilon^0(t+1) = 0$?

**12.2** Show that (12.18) does not imply that $\hat{\theta}(t)$ converges towards a constant $\theta$.

**12.3** Using first order polynomials, show that in general

$$\hat{A}(t)\hat{S}(t)y(t) \neq \hat{S}(t)\hat{A}(t)y(t)$$

**12.4** Theorem 12.1 shows that

$$\lim_{t\to\infty} P[y(t+d) - \hat{B}^*(t, q^{-1})\hat{\beta}(t)y^*(t+d)] = 0$$

Explain the meaning of this result (hint: see Sect. 7.3.3).

**12.5** Motivate the input-output variables normalization (12.88) (hint: See Sect. 10.6).

**12.6** Explain what happens if the parameter adaptation is not frozen in the absence of the persistent excitation condition.

# Chapter 13
# Multimodel Adaptive Control with Switching

## 13.1 Introduction

The plants subject to abrupt and large parameter variations are generally very difficult to control. A classical adaptive controller or a fixed robust controller encounter difficulties to solve this problem. An adaptive controller is not fast enough to follow the parameter variations and unacceptable transients occur. Whereas a fixed robust controller normally leads to poor performance because of large uncertainties.

Multimodel adaptive control can be considered as a solution to this problem. In this approach a set of Kalman filters are used to estimate the states of the plant model. A set of state feedback controllers are also designed such that for each Kalman filter there is a state feedback controller in the set. The final control input is the weighted sum of the control inputs from different controllers. The weightings are computed using the covariance matrix of state estimates such that smaller variances lead to larger weights for the corresponding control inputs. This approach was proposed by Athans and Chang (1976), Chang and Athans (1978). Recently a robust version of this scheme has been propose by Fekri et al. (2006). The state feedback controllers are replaced by robust output feedback controllers designed using the $H_\infty$ control approach. Although this method has been successfully applied in a few simulation examples, the stability of the scheme has not yet been proved.

An alternative solution is to choose one of the control inputs that corresponds to the best output estimator and apply it to the real system. This approach which is based on switching among a set of controllers has the advantage that a stability analysis for the closed-loop system can be carried out. The stability of switching systems has been probably analyzed for the first time in Martensson (1986). Some authors consider the case of predetermined switching among a set of controllers (Fu and Barmish 1986; Miller and Davison 1989; Miller 1994). However, switching based on an error signal measured in real-time that represents the difference between the measured output and an estimated one seems to be more interesting. In this approach, it is supposed that a set of models for different operating points is a priori known. Then at every instant a controller corresponding to the model yielding the minimum of a performance index is used to compute the control input. The precision of the control can be further improved using an adaptive model (a model whose parameters are updated with a parameter adaptation algorithm) in the set of models. This method together with a stability analysis was proposed in Narendra and Balakrishnan (1997).

It is well known that switching between stable systems may lead to instability (Liberzon 2003). Therefore, it is essential to assure the stability of the closed-loop system when the switching is controlled by a noisy error signal. The stability can be guaranteed if a hysteresis function is introduced in the switching rule. Another way is to consider a time delay between two consecutive switchings. It can be shown that if this delay is sufficiently large the switching among stable systems is not destabilizing.

This chapter is organized as follows. Section 13.2 describes the principles of adaptive control with switching. A stability analysis based on minimum dwell-time is given in Sect. 13.3. Finally, an adaptive control strategy based on switching and tuning is applied to the flexible transmission system in Sect. 13.4. The CLOE adaptation algorithm is used and the effect on the design parameters on the performance of the closed-loop system are studied.

## 13.2 Principles of Multimodel Adaptive Control with Switching

In this section, the main structure of adaptive control with switching is presented. This structure, shown in Fig. 13.1, contains four blocks: plant model, multi-estimator, multi-controller and supervisor (Anderson et al. 2000; Hespanha et al. 2001).

### 13.2.1 Plant with Uncertainty

Plant model is supposed to be a member of a very large set of admissible models given by:

$$\mathscr{G} = \bigcup_{\theta \in \Theta} G(\theta) \tag{13.1}$$

where $\Theta$ is a compact set and each $G(\theta)$ is a family of models centered around a nominal model $G_0(\theta)$. For example, if we consider multiplicative uncertainty, the plant model can be represented by

$$G(\theta) = G_0(\theta)[1 + W\Delta]$$

where $W$ is the uncertainty filter and $\Delta$ is a stable transfer function with $\|\Delta\|_\infty < 1$. As a result, $\mathscr{G}$ represents the parametric uncertainty (or variation) and for each fixed $\theta$, the subfamily of $G(\theta)$ represents the unmodeled dynamics.

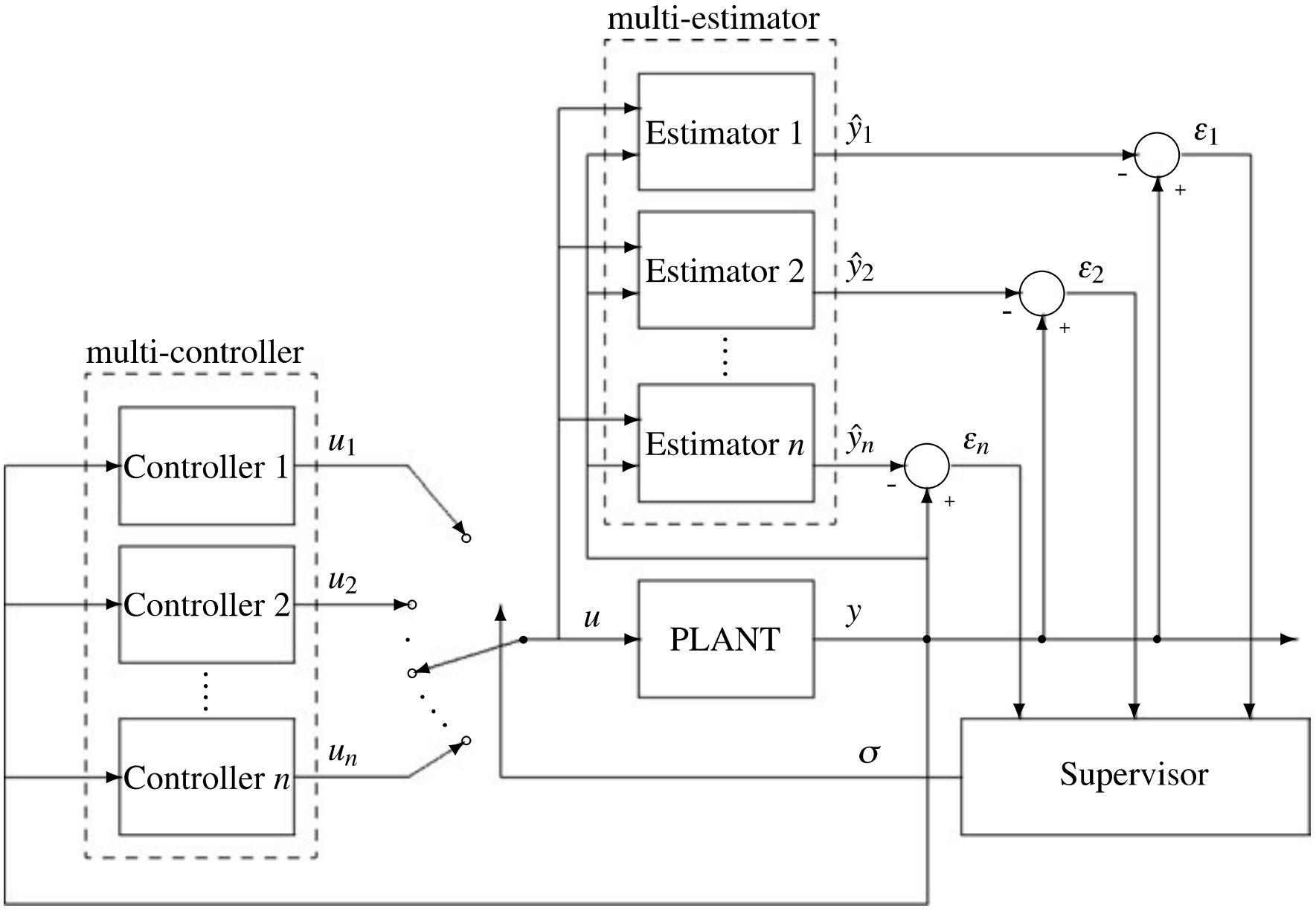

**Fig. 13.1** Block diagram of adaptive control with switching

### *13.2.2 Multi-Estimator*

The multi-estimator is a dynamical system whose inputs are the input and output of the plant and whose outputs are the estimates of the plant output. Suppose that $\Theta$ is a finite set. In this case, multi-estimator can be chosen as a set of fixed models corresponding to each $\theta$. The fixed models can be considered as output error estimators. However, any other estimators like Kalman filter or ARMAX estimator can be used as well. If $\Theta$ is an infinite set, some fixed models and one adaptive model can be considered. Since the number of available models is finite but the number of possible models is generally infinite, the estimation is performed in two steps:

- The model with smallest error with respect to a criterion is rapidly chosen (switching).
- The parameters of the model are adjusted using a parameter adaptation algorithm (tuning).

### *13.2.3 Multi-Controller*

For each estimator (or fixed model) in the multi-estimator block, a controller should be designed that stabilizes and satisfies the desired performance for the estimator (or fixed model). If $\Theta$ is not a finite set but the set of multi-estimators is finite, the controllers should be robust with respect to the parameter uncertainty as well as unmodeled dynamics. If an adaptive model is considered in the multi-estimator set, like the classical adaptive control, the controller should be a function of the model parameters.

### 13.2.4 Supervisor

The supervisor decides the output of which controller should be applied to the real plant at each instant. The decision is based on a monitoring signal and a switching logic. The monitoring signal is a function of the estimation error to indicate the best estimator at each time. It may be defined as follows:

$$J_i(t) = \alpha \varepsilon_i^2(t) + \beta \sum_{j=0}^{t} e^{-\lambda(t-j)} \varepsilon_i^2(j); \quad \alpha \geq 0;\ \beta, \lambda > 0 \tag{13.2}$$

where $j$ is the time index and $\alpha$ and $\beta$ are the weighting factors on the instantaneous measures and the long term accuracy. $\lambda$ is a forgetting factor which also assures the boundedness of the criterion for bounded $\varepsilon_i(t)$. Therefore, the design parameters for the switching part of the control system are $\alpha$, $\beta$ and $\lambda$. If we choose a large value for $\alpha/\beta$ and $\lambda$, we will obtain a very quick response to the abrupt parameter changing but a bad response with respect to disturbances. It means that, an output disturbance will generate an unwanted switching to another controller which results in a poor control. Contrary, a small value for $\alpha/\beta$ and $\lambda$ makes the criterion a good indicator of steady-state identifier accuracy, which reduces the number of unwanted switching but leads to a slow response with respect to the parameter variations.

The switching logic is based on the monitoring signal. A switching signal $\sigma(t)$ indicates which control input should be applied to the plant. In order to avoid chattering, a minimum dwell-time between two consecutive switchings or a hysteresis is usually considered. The switching logic plays an important role on the stability of the switching system which will be discussed in the next section.

Figure 13.2 shows the flowcharts of two common switching logics. A combination of two logics (dwell-time and hysteresis) may also be considered. A small value for $T_d$, dwell-time, gives too frequent switchings and may lead to instability, while a large $T_d$ leads to a slow response system.

A hysteresis cycle with a design parameter $h$ is considered between two switchings. It means that a switching to another controller will occur if the performance index concerning a model is improved by $hJ_i$. With hysteresis, large errors are rapidly detected and a better controller is chosen. However, the algorithm does not switch to a better controller if the performance improvement is not significant.

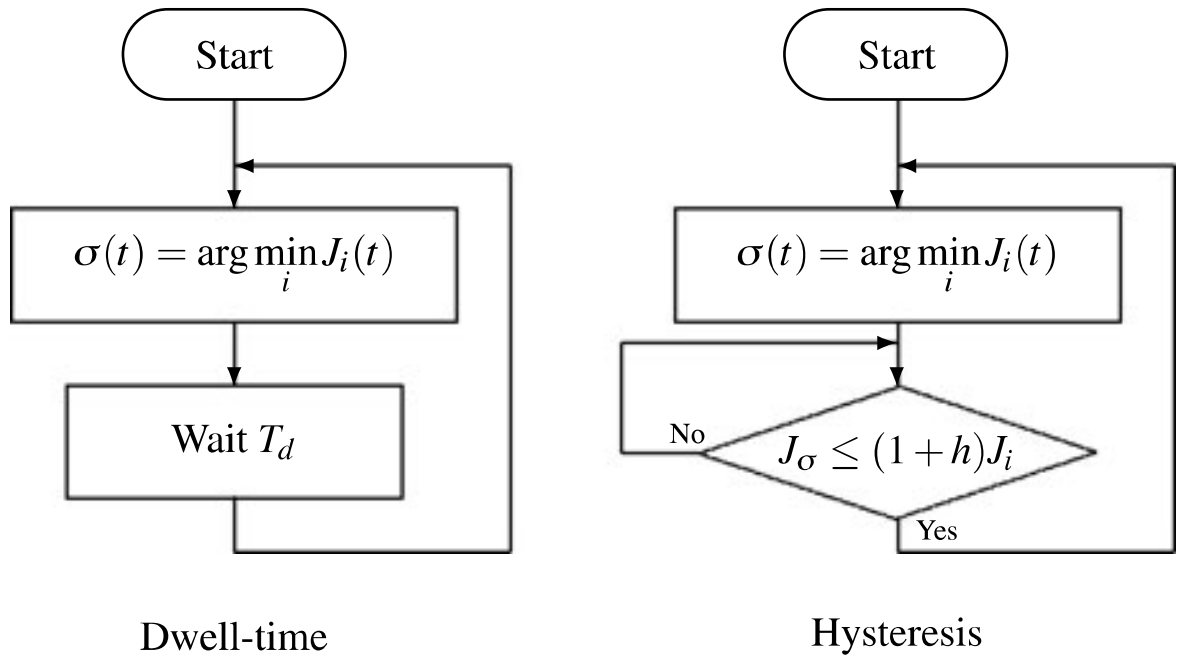

**Fig. 13.2** Flowchart of switching logics

## 13.3 Stability Issues

The objective of this section is not to give a formal proof for the stability of multi-model adaptive control, but to present the basic idea of the proof and a practical way to compute the dwell-time to ensure the closed-loop stability. For a more detailed analysis see Anderson et al. (2000), Hespanha et al. (2001). It should be mentioned that in this section we suppose that the real plant is fixed but unknown. However, the results can be used for sparse large and abrupt parameter changes but not for frequent parameter variations.

The stability is analyzed in two steps. In the first step we show that under some mild assumptions the stability of the system can be reduced to the stability of a subsystem called the "injected system". This subsystem is a switching system that contains only stable models. In the second step the stability of this particular switching system is discussed. It is shown that this system can always be stabilized by choosing an appropriate dwell-time.

### *13.3.1 Stability of Adaptive Control with Switching*

Consider a trivial case in which one of the models in the multi-estimator block matches perfectly with the plant model (no unmodeled dynamics). Suppose that there is no measurement noise and that the plant is detectable.[1] In this case, the estimation error for one of the models, say $\varepsilon_k(t)$ goes to zero. Consequently $\varepsilon_\sigma(t)$ goes to zero and the switching signal $\sigma(t)$ goes to $k$ (switching stops after a finite time). It is evident that $C_k$ which stabilizes the $k$th model will stabilize the plant based on the certainty equivalence stabilization theorem.

However, in practice, because of unmodeled dynamics and measurement noise, the switching will not necessarily stop after a finite time and the analysis becomes more involved.

To proceed, we consider the following assumptions:

[1] A system is detectable if and only if all of its unobservable modes are stable (Zhou 1998).

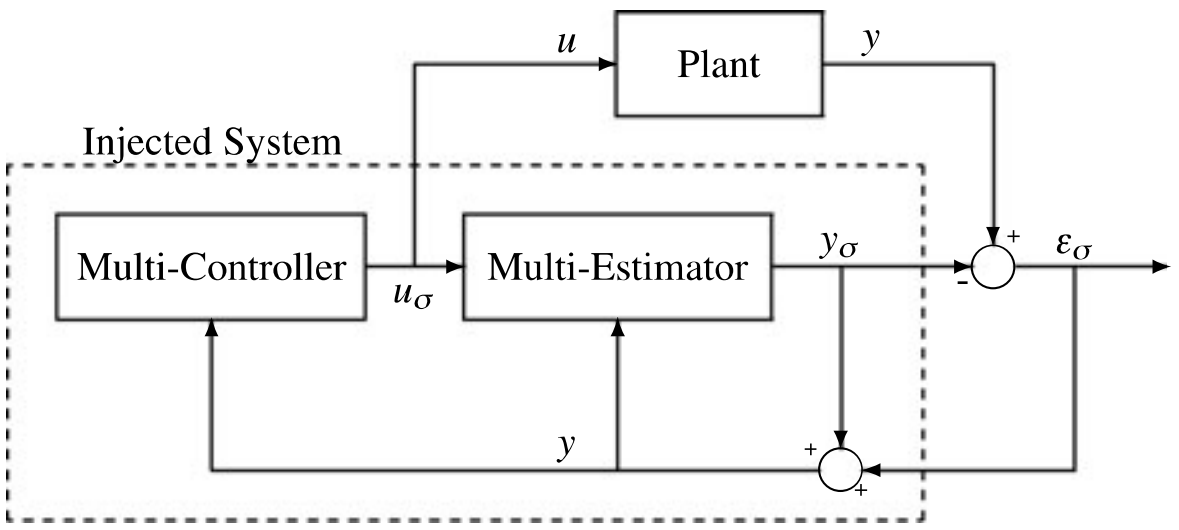

**Fig. 13.3** Representation of the closed-loop system including the injected system

A1: In the multi-estimator block there exists at least one "good" estimator. It means that at least for one estimator, say estimator $k$, the estimation error is "small". The smallness of the estimation error can be defined by an upper bound that depends on the modeling error and noise variance.

A2: $\varepsilon_\sigma(t) = y(t) - y_\sigma(t)$ is small. It means that the monitoring signal, switching criterion $J(t)$, and the switching logic are properly designed ($\alpha$, $\beta$ and $\lambda$ are well tuned and dwell-time or hysteresis are not too large).

It can be shown that under the above assumptions all closed-loop signals and states are bounded if the injected system is stable. The injected system is a dynamical system which has $\varepsilon_\sigma$ as input and $u_\sigma$ (or $u$) and $y_\sigma$ as outputs. Figure 13.3 is the same as Fig. 13.1 but is drawn differently such that the injected system appears as a separate block from the rest of the system. The supervisor and the switch are replaced by index $\sigma$ in the output of multi-estimator and multi-controller. It can be observed that if the injected system is stable, the input of the plant $u$ and $y_\sigma$ are bounded for a bounded $\varepsilon_\sigma$. It is clear that boundedness of $y_\sigma$ implies the boundedness of $y$. So the input and output of the plant are bounded and all internal states of the plant are also bounded if the plant is detectable.

If the switching stops after a finite time, the injected system becomes an LTI system and is naturally stable (each controller stabilizes the corresponding model in the multi-estimator block). However, in presence of noise and unmodeled dynamics, the switching may not stop and the injected system becomes a switching system with arbitrary switching among stable systems. This type of switching systems are not necessarily stable and their stability has been analyzed in the literature (see Liberzon 2003). In the next subsection some recent results on the stability of such systems are given.

### 13.3.2 *Stability of the Injected System*

It is shown that the injected system is a switching system containing only the stable systems constructed by the members of the multi-estimator block and their corresponding controllers in the multi-controller block. In this subsection we study the stability of a switching system containing $n$ stable models in the state space. Consider the following models:

$$x(t+1) = A_1 x(t), \qquad x(t+1) = A_2 x(t), \qquad \ldots, \qquad x(t+1) = A_n x(t)$$

where $A_1, A_2, \ldots, A_n$ are stable matrices. Consider also a switching signal $\sigma(t) \in \{1, 2, \ldots, n\}$. Then $x(t+1) = A_\sigma x(t)$ is stable if $A_1$ to $A_n$ have a common Lyapunov matrix or the minimum time between two switchings is greater than $T_d$.

### Stability with Common Lyapunov Matrix

The existence of a common Lyapunov matrix $P$ for $A_1, A_2, \ldots, A_n$ guarantees the stability of $A_\sigma$. It is easy to show that if we take a Lyapunov function $V(t) = x(t)^T P x(t)$ its finite difference $\Delta V(t) = V(t+1) - V(t) = x(t)^T (A_\sigma P A_\sigma - P) x(t) < 0$ if

$$\begin{aligned} A_1^T P A_1 - P &< 0 \\ A_1^T P A_1 - P &< 0 \\ &\vdots \\ A_n^T P A_n - P &< 0 \end{aligned}$$

The above inequalities are linear with respect to $P$ and represent a set of Linear Matrix Inequalities (LMIs) (Boyd et al. 1994). Therefore, the existence of $P$ can be easily verified using Semi Definite Programming (SDP) solvers (Vandenberghe and Boyd 1996). The main interest of this stability test is that the stability is guaranteed for arbitrary fast switching. However, it is well known that this stability test is too conservative.

### Stability by Minimum Dwell-Time

It can be shown that if $A_1, A_2, \ldots, A_n$ are stable, there exists always some dwell-time $T_d \geq 1$ such that the switching system $x(t+1) = A_\sigma x(t)$ is stable. This result is expressed in the following theorem:

**Theorem 13.1** (Geromel and Colaneri 2006b) *Assume that for some $T_d \geq 1$, there exists a set of positive definite matrices $\{P_1, \ldots, P_n\}$ such that:*

$$A_i^T P_i A_i - P_i < 0, \quad \textit{for } i = 1, \ldots, n \tag{13.3}$$

*and*

$$[A_i^T]^{T_d} P_j [A_i]^{T_d} - P_i < 0, \quad \forall i \neq j = 1, \ldots, n \tag{13.4}$$

*Then the switching signal $\sigma(t) = i \in \{1, \ldots, n\}$ with a dwell-time greater than or equal to $T_d$ makes the equilibrium solution $x = 0$ of the switching system $x(t+1) = A_\sigma x(t)$ globally asymptotically stable.*

*Proof* See Geromel and Colaneri (2006b). □

Very similar result can be given for the continuous-time systems (Geromel and Colaneri 2006a). The inequalities in (13.3) and (13.4) are LMIs with respect to Lyapunov matrices and therefore a feasible value for $T_d$ can be readily computed.

## 13.4 Application to the Flexible Transmission System

In this section several experiments are carried out in order to show the performance of adaptive control with switching in comparison with the classical adaptive control (Karimi and Landau 2000). The flexible transmission system presented in Sect. 1.4.3 with three identified models for 0% load, 50% load and 100% load (see Sect. 5.9) is considered. Different functions have been developed in order to implement the multiple models adaptive control on the VisSim (Visual Solutions 1995) software environment.

### *13.4.1 Multi-Estimator*

Three discrete-time identified models for 0% load, 50% load and 100% load together with an adaptive model using the CLOE algorithm are considered in the multi-estimator set. The adaptive model is always initialized with the parameters of the last fixed model that was chosen as the best model. As a result, the adaptation gain is also initialized with a small value.

It should be noted that a preliminary study in Karimi (1997) shows that the existence of an excitation signal is necessary for the stability of an adaptive control system using the CLOE algorithm in a certain situation. The main reason is that the CLOE algorithm is not sensitive to disturbance signal because its regressor vector contains only the reference signal filtered by closed-loop transfer functions. Therefore, in the absence of reference signal, i.e. $r(t) \equiv 0$, there will be no parameter adaptation even when the true plant parameters change. This may lead to instability of the closed-loop system for large parameter variations. However, this algorithm can be used in the adaptive scheme with switching even if the stability condition for the algorithm is not satisfied. The reason is that in the absence of the excitation signal, the parameters of the CLOE predictor remain unchanged and the adaptive model becomes a fixed model among the other fixed models, for which a stability analysis has already been presented.

### *13.4.2 Multi-Controller*

The multi-controller is replaced by an adaptive controller. At each instant the parameters of the best model $G_\sigma$ are used to compute the parameters of a robust controller designed using the pole placement technique explained in Chap. 8. The adaptive controller is also used to estimate the control input $\hat{u}(t)$, which will be used to determine $\hat{y}(t)$ and $\varepsilon_{CL}(t)$ in the CLOE algorithm.

The adaptive controller has an RST structure i.e.

$$S(q^{-1},\hat{\theta})u(t) = -R(q^{-1},\hat{\theta})y(t) + T(q^{-1},\hat{\theta})y^*(t) \tag{13.5}$$

where $R(q^{-1},\hat{\theta})$, $S(q^{-1},\hat{\theta})$ and $T(q^{-1},\hat{\theta})$ are the controller's polynomials computed using the pole placement method based on the model estimated parameters $\hat{\theta}$. The reference trajectory, $y^*(t)$, is stored in the computer or generated via the reference model. The following specifications are considered for the pole placement control design:

Dominant poles: A pair of complex poles with a frequency of 6 rad/s (0.048 $f/f_s$) and a damping factor of 0.9. This frequency corresponds to the natural frequency of the first vibration mode of the fully loaded model.

Auxiliary poles: A pair of auxiliary poles with a frequency of 33 rad/s (0.263 $f/f_s$) and a damping factor of 0.3 which is close to the high-frequency poles of open-loop models and the real poles $(1-0.5z^{-1})^3(1-0.1z^{-1})^4$. This choice of auxiliary poles helps reducing the input sensitivity function in middle and high frequencies as explained in Sect. 8.8. This makes the controller robust with respect to unmodeled dynamics.

Fixed terms: The controller contains an integrator and a term $(1 + z^{-1})^2$ in the numerator to reduce the input sensitivity function in very high frequencies.

Reference model (tracking): A second order system with $\omega_0 = 6$ rad/s and a damping factor of 1 is chosen as the reference model for tracking.

The main advantage of the pole placement technique for adaptive control with switching is that the closed-loop poles remain unchanged for any value of switching signal $\sigma(t)$. It means that the closed-loop poles of the injected system are invariant with respect to the switching signal. This leads to the existence of a common Lyapunov matrix for the injected system that guarantees the stability of the adaptive system for arbitrary fast switching.

### 13.4.3 Experimental Results

The bloc diagram of the control architecture is illustrated in Fig. 13.4. $G_1$, $G_2$ and $G_3$ represent unloaded, half loaded and fully loaded models, respectively. $\hat{G}$ is the adaptive model using the CLOE algorithm, see Sect. 12.7.1. At every instant, the switching signal $\sigma(t) \in \{0, 1, 2, 3\}$ indicates the best model (0 for adaptive model and 1, 2 and 3 for $G_1$, $G_2$ and $G_3$, respectively) according to the performance index of (13.2) and the control input $u(t)$ is determined based on this model and using the pole placement method.

Three experiments are performed on the real system. The aim and objective of these experiments are summarized as follows:

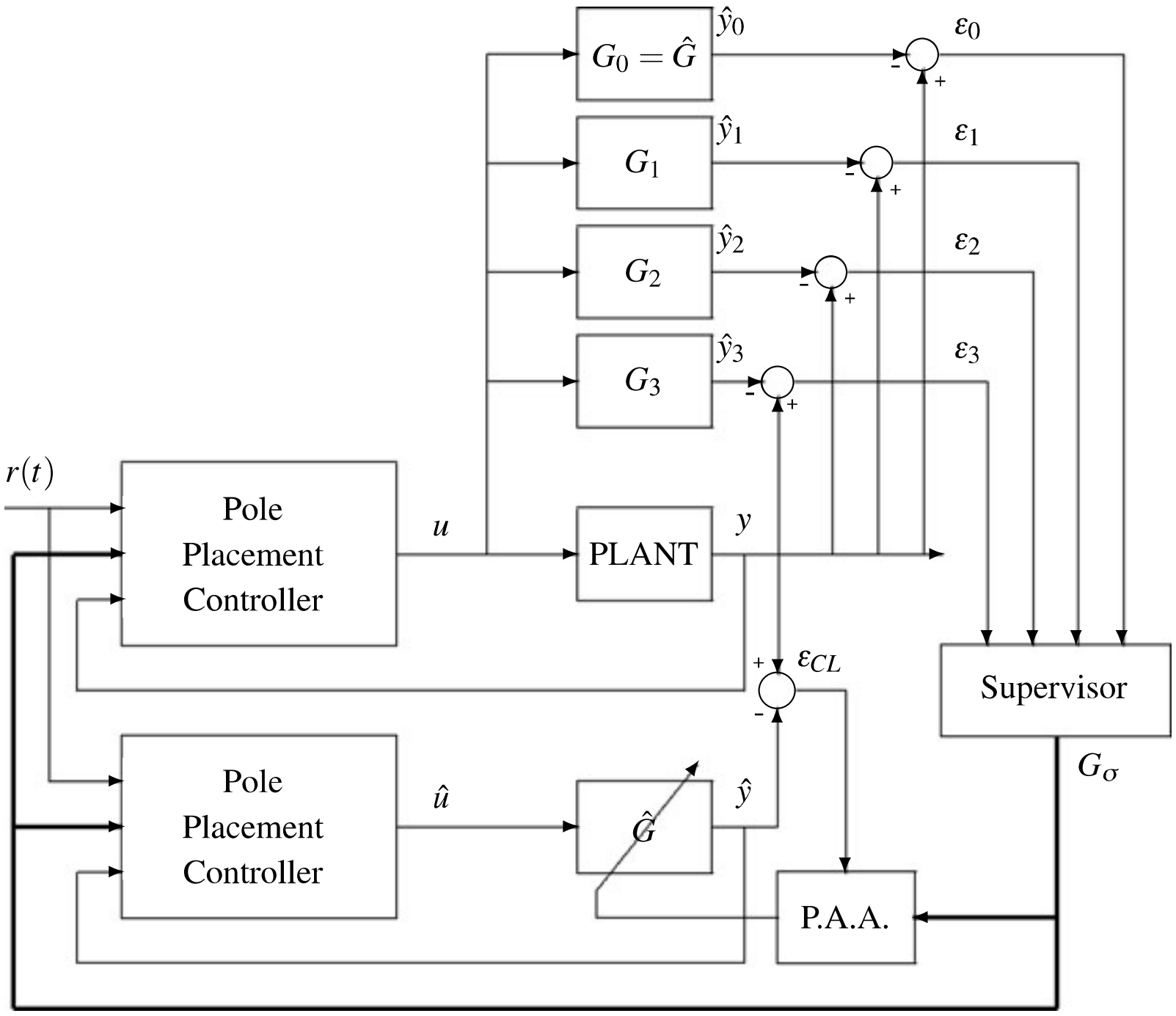

**Fig. 13.4** Bloc diagram of the real-time experiments

Experiment 1: Comparison of fixed robust control versus adaptive control with switching which shows a reduction of tracking error for the adaptive system.

Experiment 2: Comparison of classical adaptive control versus adaptive control with switching which results in a faster parameter adaptation and better reliability for multimodel adaptive system.

Experiment 3: Studying the behavior of the adaptive control system when the fixed models (0%, 50% and 100% load) do not correspond to reality (25%, 75% load) which illustrates fast parameter adaptation.

## Experiment 1

A square wave signal with an amplitude of 1 V and a period of 10 s is applied on the input of the reference model and a disturbance is applied at the output of the plant (the angular position of the third pulley is changed slowly by hand and is released rapidly). The experiment is started without load on the third pulley and at the instants 9 s, 19 s, 29 s and 39 s 25% of the total load is added on the third pulley. Therefore the system without load becomes fully loaded in four stages. The parameters of the adaptive model are initialized to zero. The results of this experiment are compared with a fixed robust controller designed by Kidron and Yaniv (1995), which satisfies the specifications of the flexible transmission benchmark (Landau et al. 1995a) in Fig. 13.5. This figure shows on one hand that very good performance in tracking and disturbance rejection are achieved using adaptive control with switching and on the other hand it shows that adaptive schemes proposed provides better performance than a good robust controller.

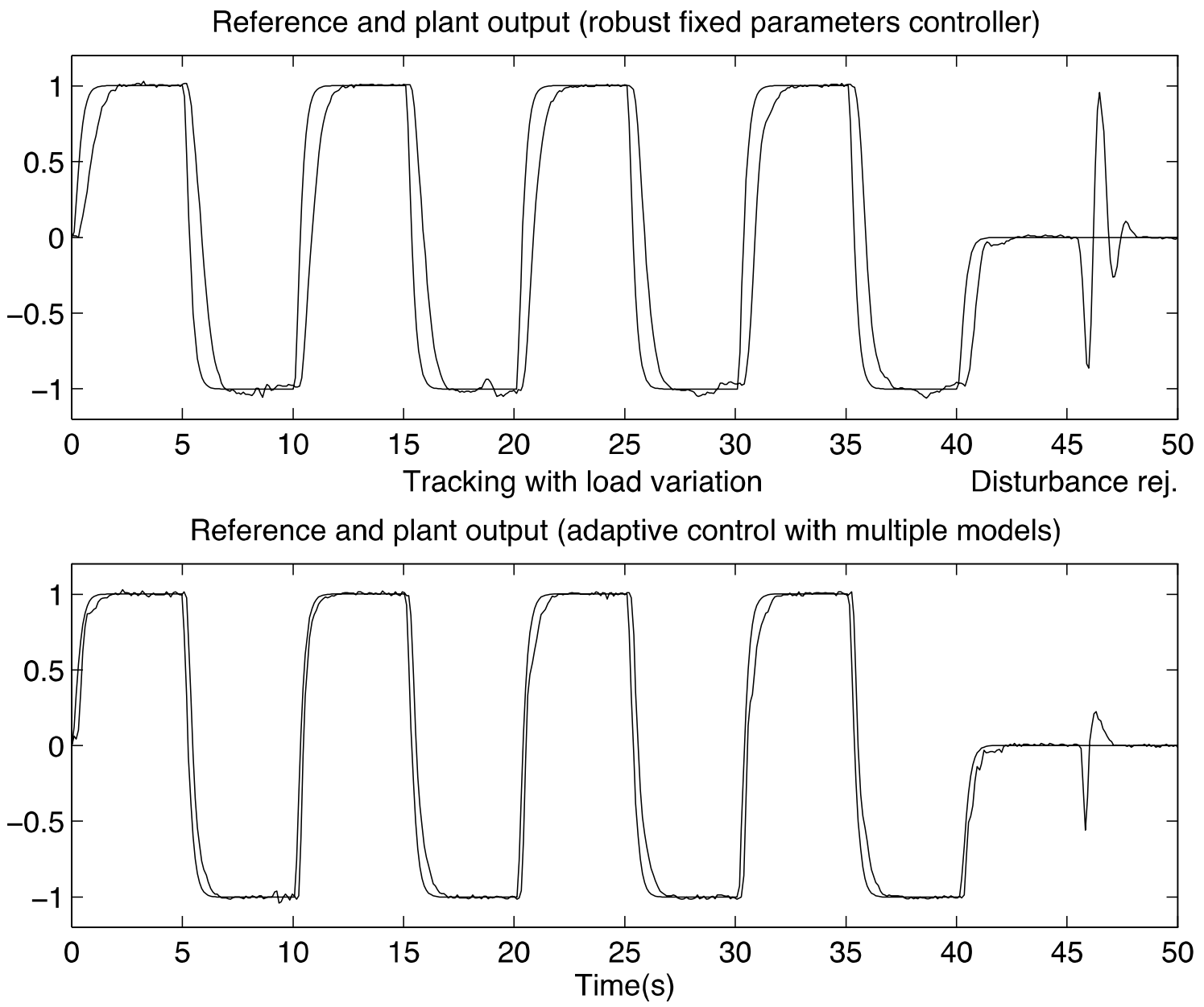

**Fig. 13.5** Comparison of a fixed robust controller with an adaptive multimodel controller (0% → 100% in four stages at 9 s, 19 s, 29 s and 39 s)

## Experiment 2

The same control system with the same design parameters as well as the same reference signal is again considered (without disturbance). The plant is initially fully loaded and it passes to unloaded case in two stages (at 19 s and 29 s). The results illustrated in Fig. 13.6 show the good tracking performance of the control system. The switching diagram in Fig. 13.6d shows the switching signal $\sigma$ or the best model chosen at every instant. The design parameters for the switching rule are: $\alpha = 1$, $\beta = 1$ and $\lambda = 0.1$ and the performance index is computed for 100 samples (from $t - 100T_s$ up to $t$, where $T_s$ is the sampling period). These parameters lead to a rather fast control system with respect to the variation of the plant parameters.

It should be noticed that this experiment cannot be carried out using the classical adaptive controllers, because the control system generates signals greater than the value that can be tolerated by the real system. The maximum variation of angular position of the first and third pulleys is 90° (corresponds to 2 V) from the equilibrium point. A proximity switch is utilized to shut down the system when the third pulley rotates more than 90°. Figure 13.7 shows the simulation results using a classical adaptive controller. The parameters of the model are initialized in open loop using a PRBS (pseudo random binary sequence) and the loop is closed at 9 s. The loads on the third pulley are taken off in two stages (at 19 s and 29 s). In the first stage (at 19 s) only a small deterioration in the response occurs because the full-loaded and half-loaded models are not very different and the controller designed for the first one also stabilizes the second one. The parameter adaptation is not fast at this stage because the adaptation error is small. In the second stage, the controller for half-loaded model does not stabilize the system for no-loaded case. This generates large values of the adaptation error leading quickly to a stabilizing controller. Unfortunately the simulated output of the plant passes above 2 V during the adaptation transient and this is not acceptable on the real system.

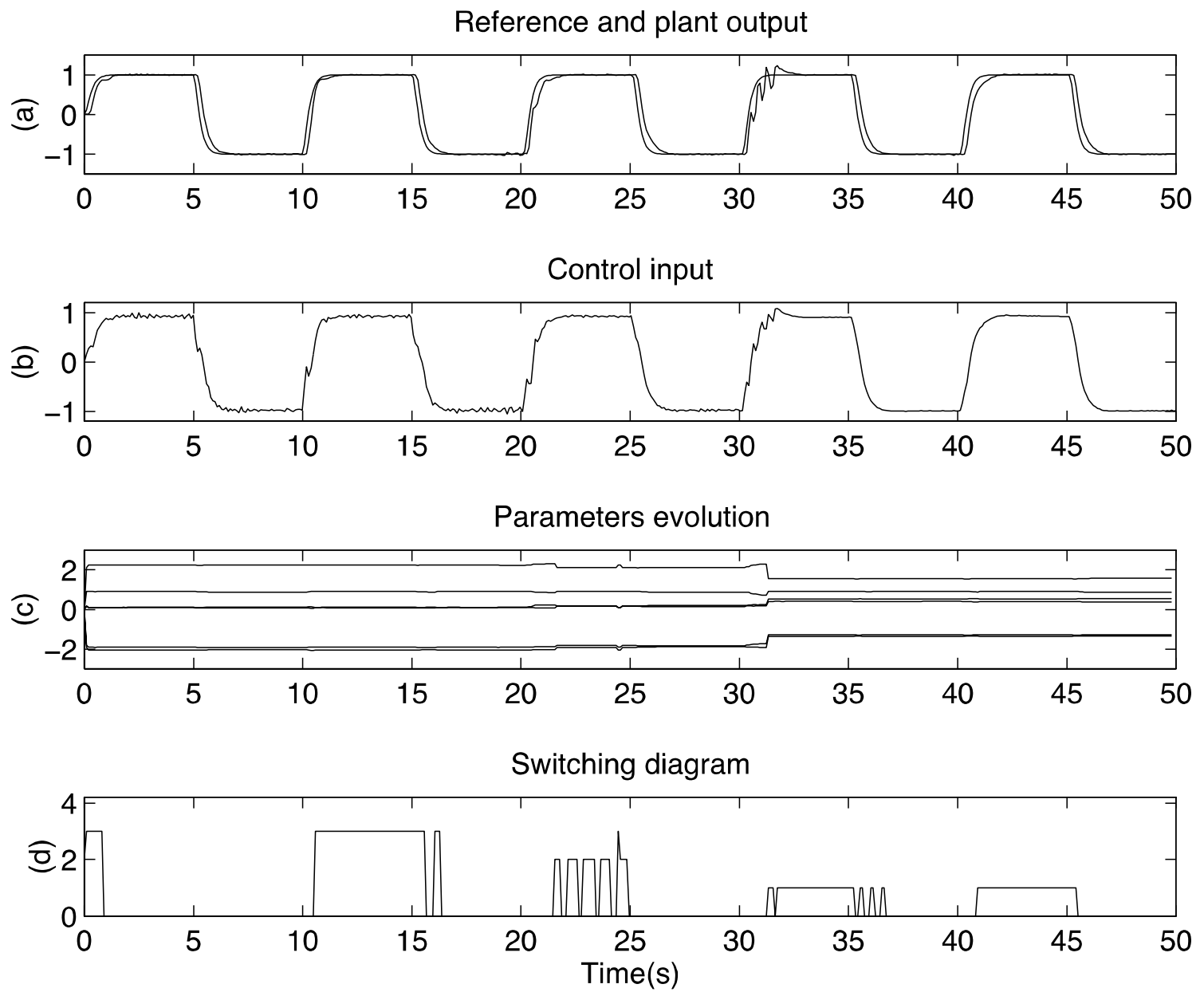

**Fig. 13.6** Results of Experiment 2 (100% → 0% in two stages at 19 s and 29 s)

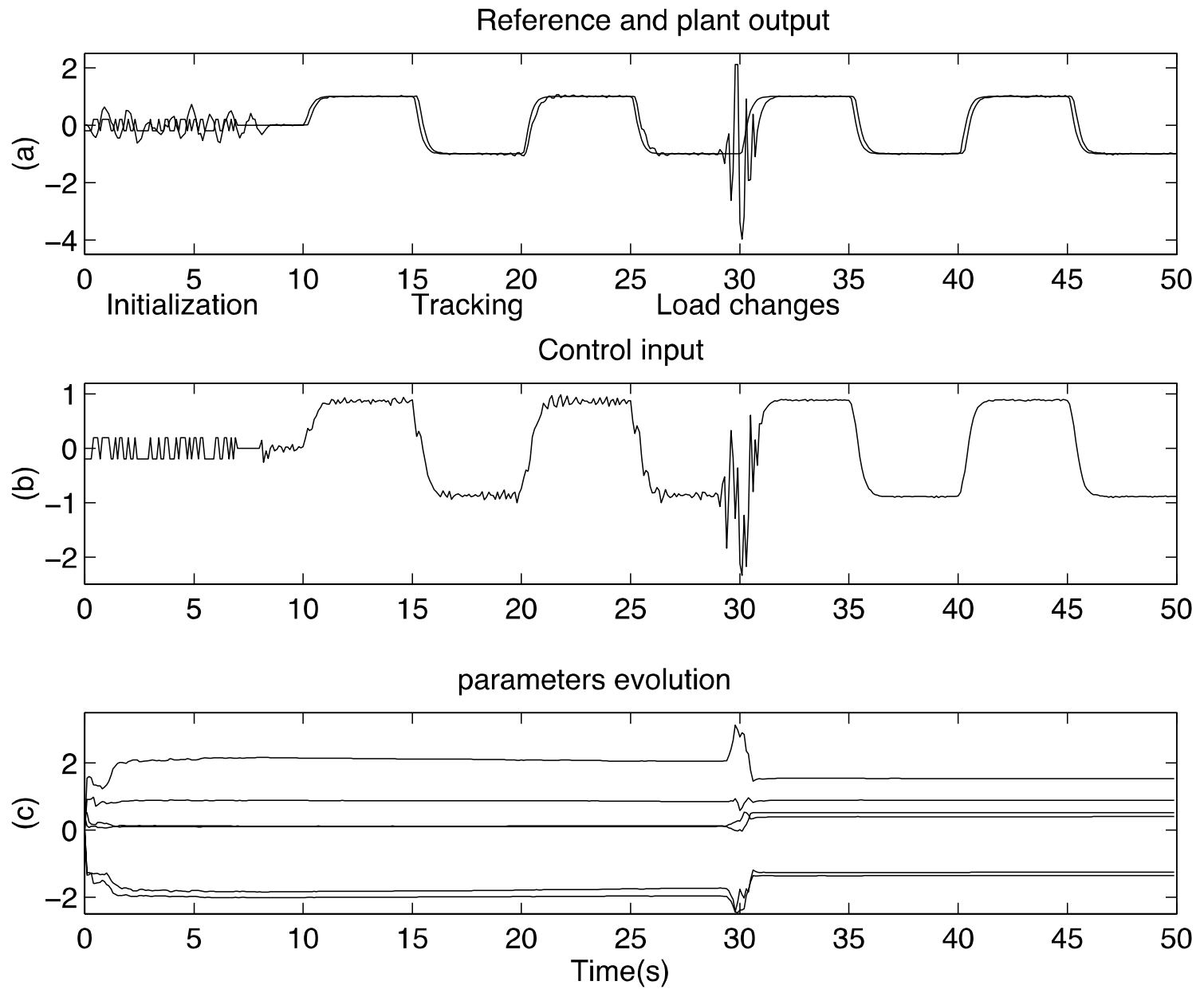

**Fig. 13.7** Simulation results using classical adaptive control (100% → 0% in two stages at 19 s and 29 s)

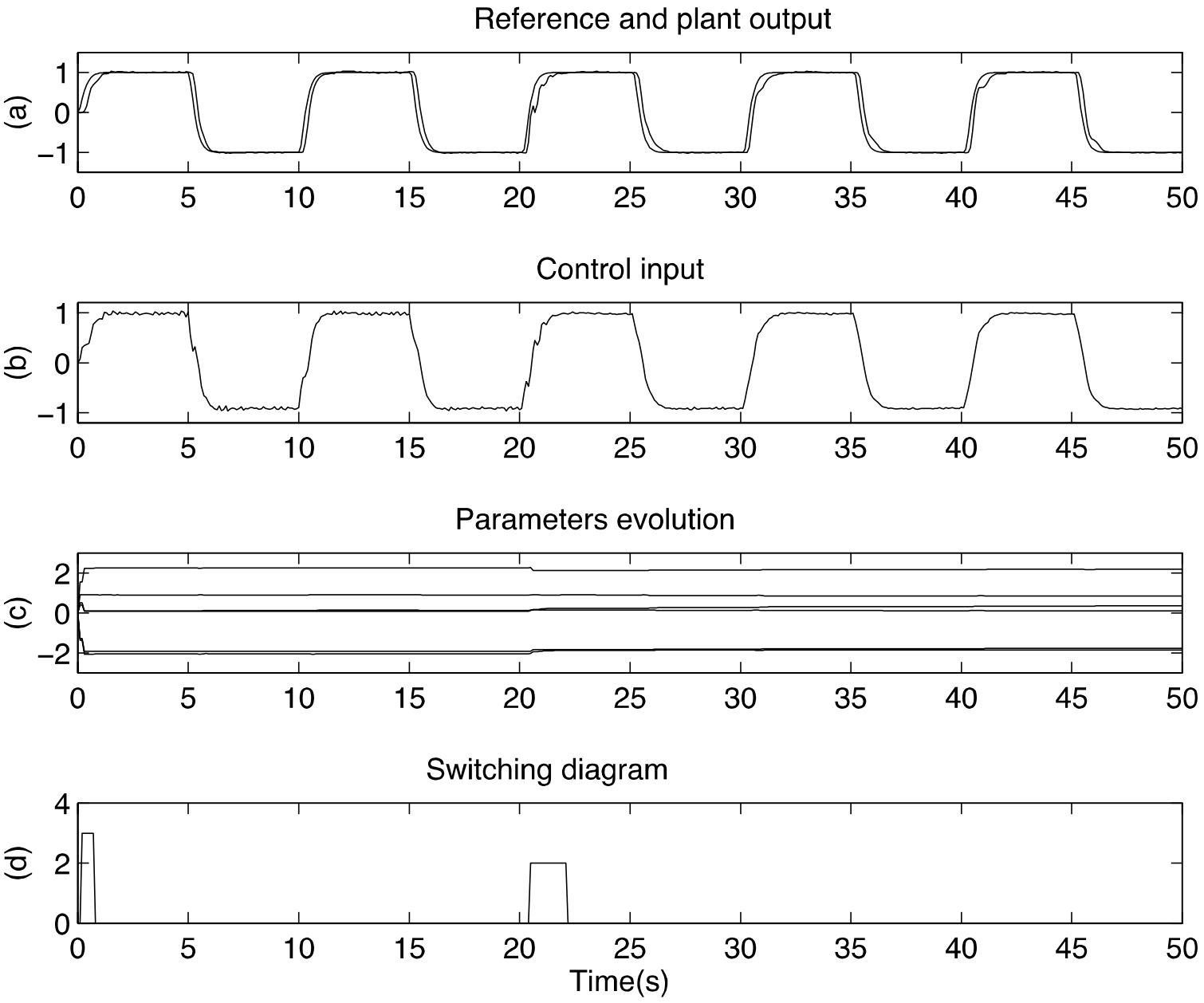

**Fig. 13.8** Results of Experiment 3 (75% → 25% in one stage at 19 s)

### Experiment 3

In order to show how the control system works, specially to emphasize the switching and tuning aspects, the second experiment is carried out again passing from 75% load to 25% load in one stage. This experiment is particular because neither the initial model nor the final model belong to the set of fixed models used in the control scheme. The switching diagram of Fig. 13.8 shows that at the beginning of the experiment the fully loaded model is chosen by the supervisor as the best model for the plant with 75% load (switching). Then the adaptive model initialized by the parameters of the fully loaded model is chosen by the supervisor (from 1 s) and remains as the best model for a period of about 20 s (tuning). Next, just after a load change on the third pulley (19 s), the half loaded model is chosen rapidly as the best model for the plant with 25% load (switching) and then the parameters of the adaptive model are tuned by the adaptation algorithm (tuning). This experiment clearly shows the role of switching to the fixed models in augmenting the adaptation speed of the adaptive control system even when the true plant models do not correspond to the fixed models used in the control scheme.

### 13.4.4 Effects of Design Parameters

In this section, the effect of design parameters in the performance of adaptive control with switching is investigated via some simulations on the flexible transmission system (Karimi et al. 2001). The design parameters for adaptive control with switching are:

- number of fixed and adaptive models,
- type of parameter adaptation algorithm for adaptive models,
- forgetting factor in the monitoring signal,
- minimum delay time between two switchings.

Selection of appropriate values for the design parameters depends upon some information about the plant, like:

- plant model in different operating points,
- existence and type of the reference signal,
- existence and type of the output disturbances,
- variance of the output noise.

For the flexible transmission system, in order to simulate the above mentioned characteristics, thirteen discrete-time identified models of plant relating to different loadings are considered. Then we suppose that the plant is initially unloaded and the small disks are placed on the third pulley one by one until the system becomes fully loaded (with 12 disks). Next, the disks are taken off one by one and system again becomes unloaded. It is supposed that this load changing is repeated cyclically with period $T_c$ and we refer to $f_c = 1/T_c$ as the parameter changing rate.

The reference signal is either null or a filtered square wave signal (filtered by a reference model) with an amplitude of 1 and a period of 10 s. The output disturbance signal is also either null or a pulse train with an amplitude of 0.5 and a period of 20 s. A zero-mean normally distributed white noise is added to the plant output. The noise variance is varied in some simulations to study the noise effect.

The objective of the control system is to follow the reference input and to reject the output disturbances as fast as possible. Thus, in order to compare different design parameters a performance index is defined as follows:

$$J_c = \left( \frac{1}{T_f} \int_0^{T_f} \varepsilon_c^2(t) dt \right)^{1/2} \tag{13.6}$$

where

$$\varepsilon_c(t) = r(t) - y(t) \tag{13.7}$$

and $T_f$ is the simulation time. The performance index is in fact the root-mean-square of the tracking error and, since all the controllers contain an integrator and disturbances are applied while $r(t) = 0$, it represents the root-mean-square of the regulation error as well.

Design of a multimodel adaptive control system consists of the following steps:

1. Determine the number of fixed and adaptive models.
2. Choose the adaptation algorithm (RLS or CLOE).
3. Determine the forgetting factor $\lambda$.
4. Determine the minimum time between switchings $T_d$.
5. Choose a controller for each model or determine a control law based on the model parameters.

The effect of forgetting factor, the choice of minimum dwell-time and design of robust pole placement controller have already been discussed. In the sequel, we study the design of multi-estimator. Particularly, the number of fixed and adaptive model and the role of adaptation algorithm are investigated.

### Number of Fixed and Adaptive Models

The first step in the design of multi-estimator is to determine the number of fixed and adaptive models. The number of fixed models may be chosen equal to the number of operating points. Mainly, better performance will be achieved with more fixed models. However, the price is a more complex control system which leads to large computation time and less reliability. It should be mentioned that if a robust control design is considered, one robust controller may give a good performance for different operating points and can reduce the number of fixed models. An adaptive model can also reduce the number of fixed models under the following conditions:

1. There exists an excitation signal on the reference input.
2. The abrupt changes of parameters are sufficiently spaced in the time (i.e. there is enough time between two changes for parameter adaptation).

Therefore for the systems in regulation (with a fixed reference signal) adaptive models should not be used in the models set.

The following simulations show that one adaptive model can reduce the number of fixed models without changing in the overall performance. In this simulation example, it is supposed that the period of the cyclic changes in the model parameters is $T_c \geq 480$ s (there is at least 20 s between the parameter variations). For the multi-estimator set, three combinations are considered as follows:

1. Two fixed models (No. of disks: 0, 9);
2. Seven fixed models (No. of disks: 0, 2, 4, 6, 8, 10, 12);
3. Two fixed and one adaptive models (No. of disks: 0, 9).

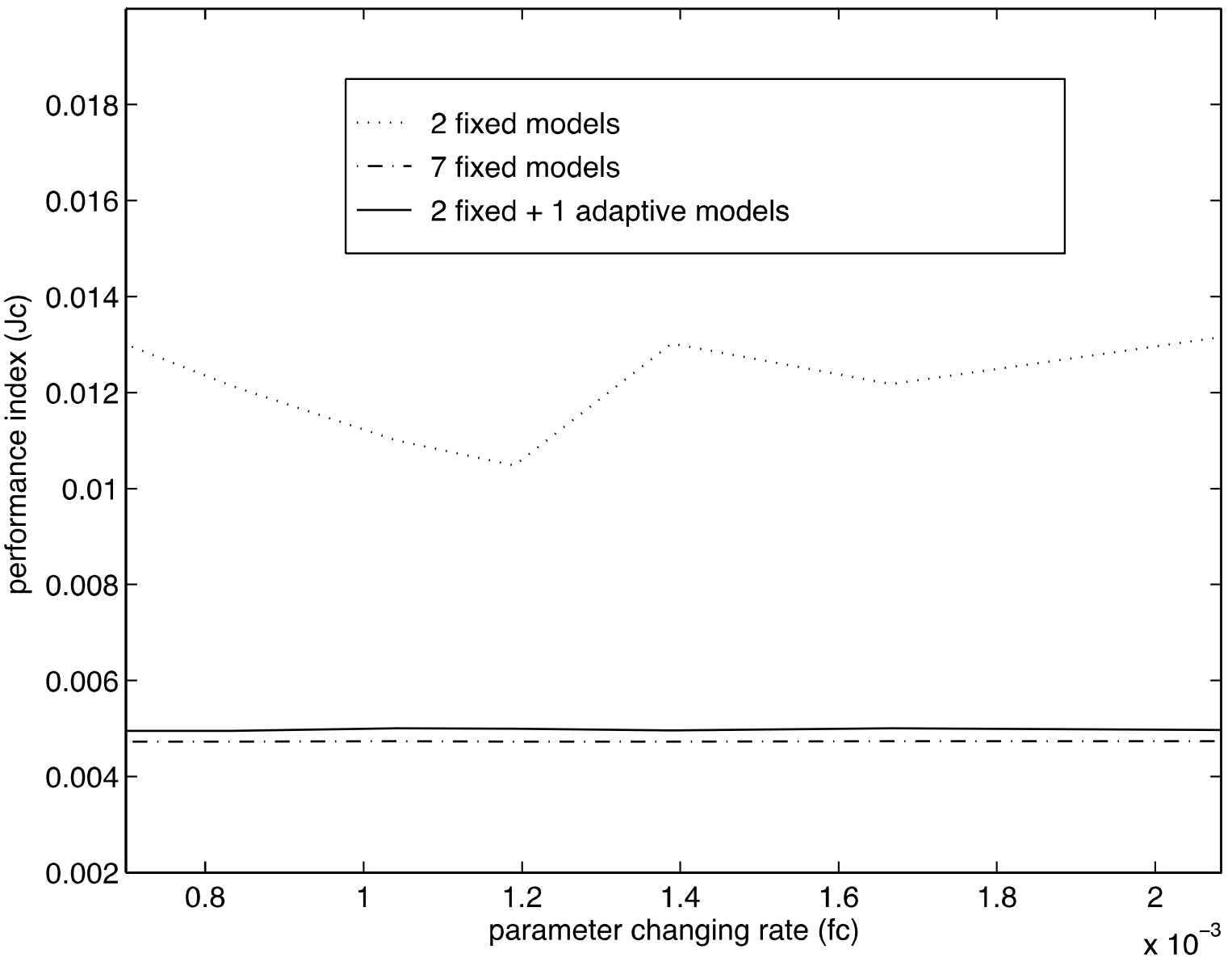

**Fig. 13.9** Performance index versus parameter changing rate

The parameter of the switching rule are: $\lambda = 0.05$ and $T_d = 1$ sampling period (50 ms). The simulation time $T_f$ is 1440 s. The performance index $J_c$ versus the parameter changing rate ($f_c = 1/T_c$) is plotted for the three cases in Fig. 13.9. One can observe that the performance is improved when the number of fixed models is increased. It should be mentioned that the performance cannot be improved significantly using more than 7 fixed models, because the controllers are robust and give also a good performance even when the plant model is not among the fixed models of the control system. It is also observed that in the third case (one adaptive model and 2 fixed models) we have almost the same performance as the second case (7 fixed models) which shows that one adaptive model can replace five fixed models.

### Parameter Adaptation Algorithm

The second step of the multi-estimator design is to choose the type of adaptation algorithm for the adaptive model. In this section we will show that the CLOE adaptation algorithm gives better performance than the classical RLS algorithm in the presence of noise. We consider 3 fixed (0%, 50% and 100% load) and one adaptive model in the multi-estimator set. Two distinct simulations are carried out, one using CLOE algorithm in the adaptive model and the other with RLS adaptation algorithm. The plant model is supposed to be fixed on 25% load which does not belong to the fixed models of the control system. Therefore, the switching will be stopped after a time on the adaptive model and the parameters of the adaptive model will be tuned by the adaptation algorithm. The parameters of the switching part are chosen as follows: $\lambda = 0.05$ and $T_d = 1$ sampling period. The simulation time $T_f$ is 120 s. The variance of the output noise is increased from 0 up to 0.07 and the performance index of the system is plotted versus the noise variance in Fig. 13.10. As depicted in this figure, increasing the noise variance will deteriorate the system performance in both cases, but CLOE algorithm gives better performance specially when the noise variance is high.

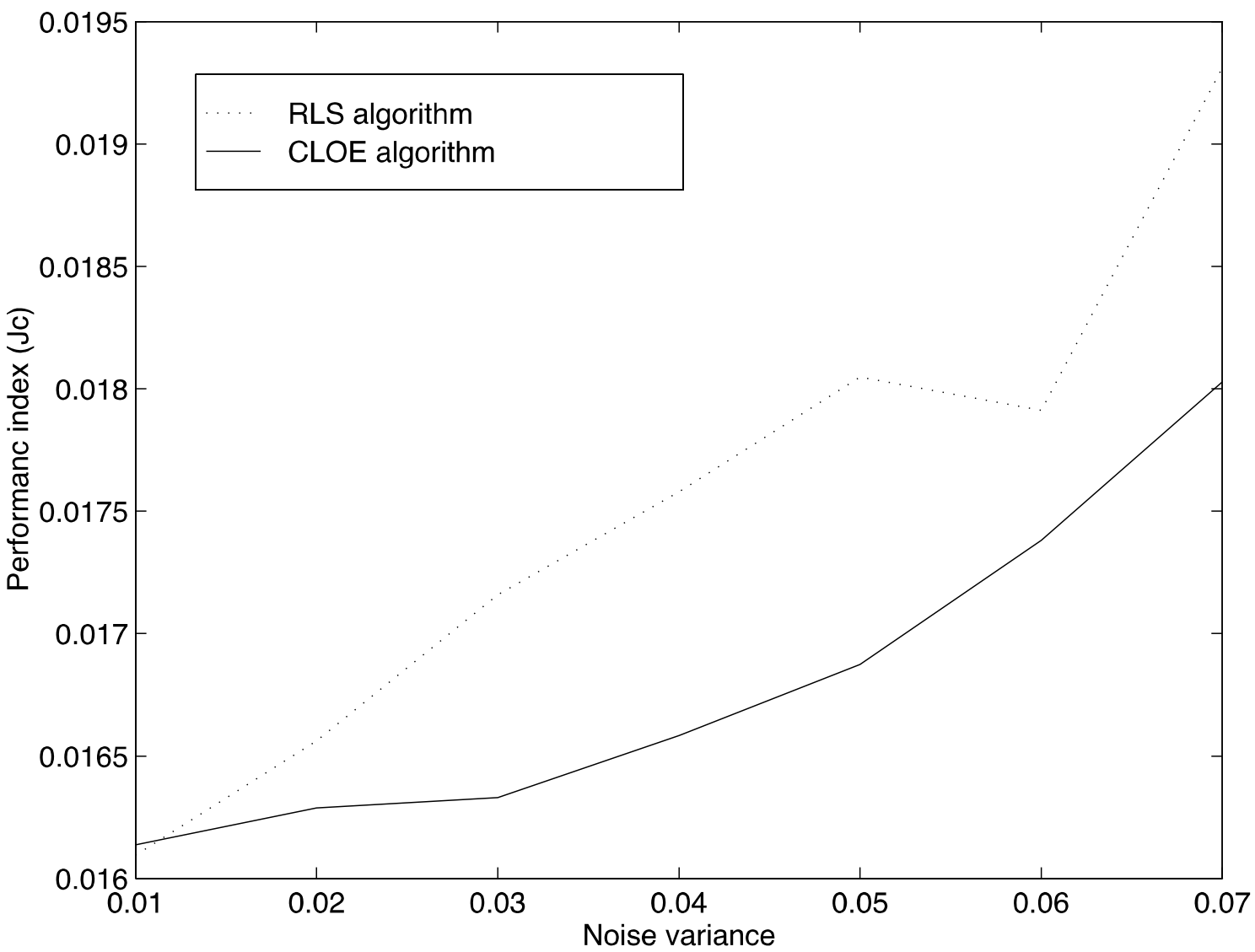

**Fig. 13.10** Performance index versus noise variance (comparison of the adaptation algorithms)

Figure 13.11 shows the output, the reference and the switching diagram for this simulation using the CLOE algorithm with a noise variance of 0.07. The switching diagram shows the best model chosen by the supervisor at each instant. This figure can be compared with Fig. 13.12 corresponding to the RLS algorithm. It can be observed that the larger variations of the output (using RLS algorithm) lead to the unwanted switchings which consequently deteriorate the performance.

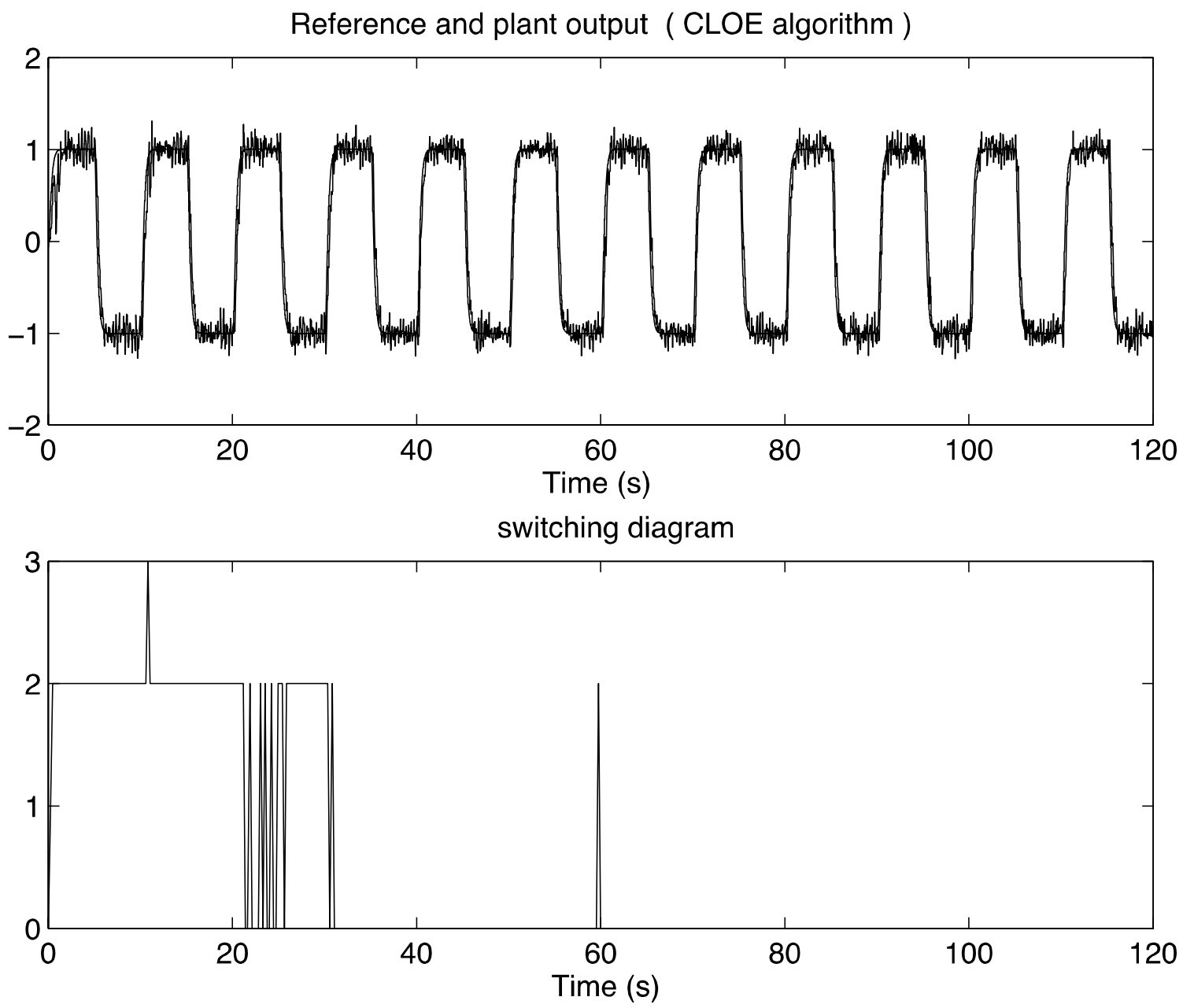

**Fig. 13.11** Simulation results using the CLOE adaptation algorithm

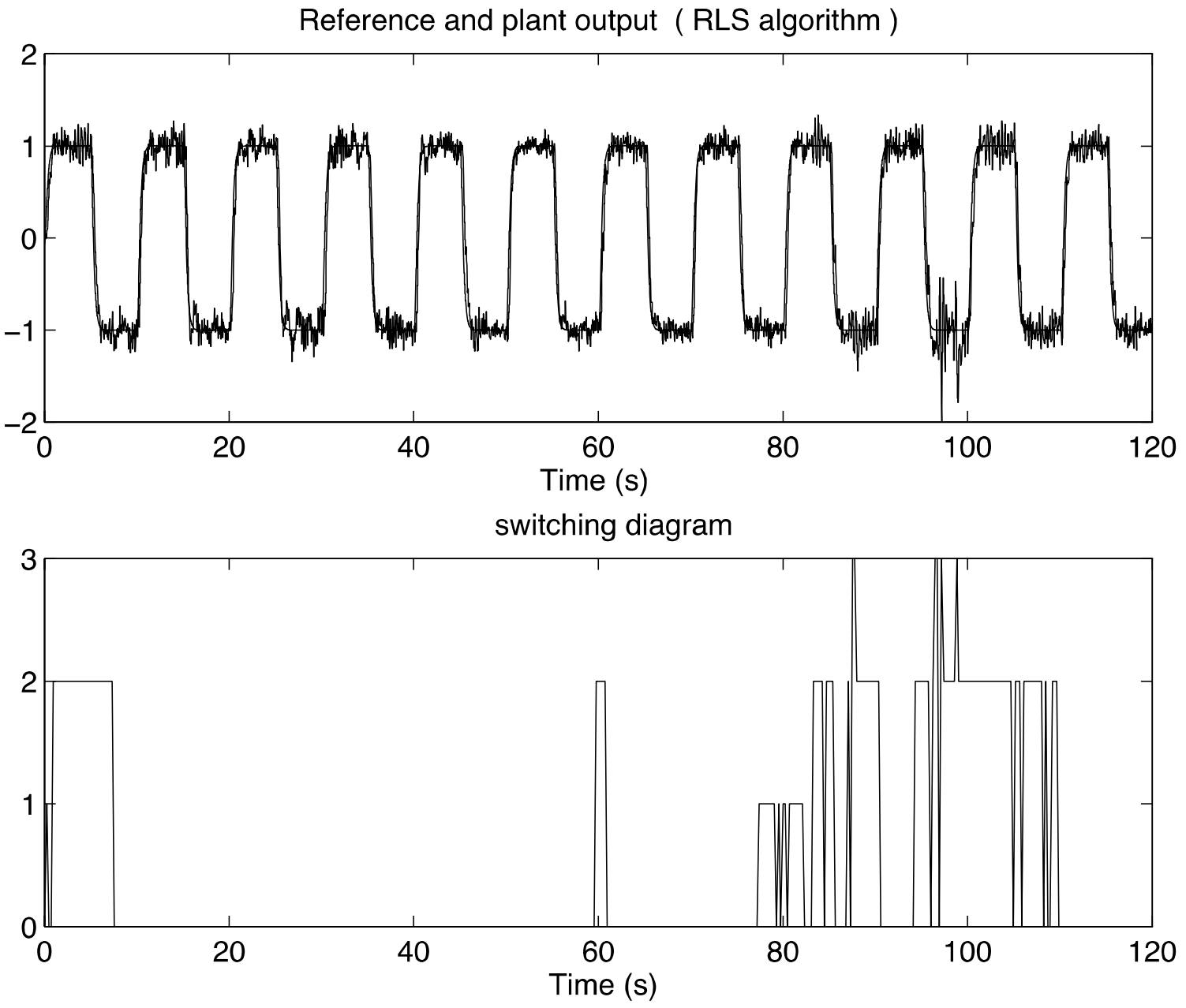

**Fig. 13.12** Simulation results using the RLS adaptation algorithm

## 13.5 Concluding Remarks

1. Large and fast parameter variations may lead to poor transient performance or even instability in classical adaptive control systems.
2. Adaptive control with switching can significantly improve the transient behavior of adaptive systems.
3. The basic idea is to use a multi-estimator instead of a unique estimator in the adaptive control scheme. During the transients, one of the estimator can provide rapidly a good estimate of the plant output and an appropriate controller can be chosen.
4. The main issue in adaptive control with switching is the stability of the closed-loop system. It has been shown that a dwell-time can be computed that guarantees closed-loop stability.
5. An application of adaptive control with switching and tuning to a flexible transmission system shows the good performance of this adaptive system with respect to the classical robust controller design and classical adaptive control.
6. The use of robust pole placement technique in adaptive control with switching guarantees quadratic stability of the injected system and consequently stability of the adaptive system with arbitrary fast switching.
7. The use of the CLOE algorithm for parameter estimation in the framework of adaptive control with switching has the following advantages:
   - it reduces unwanted switching caused by output disturbances;
   - it gives a more precise model in the critical zone for robust control;
   - the estimated parameters are not influenced by noise (asymptotically).

## 13.6 Problems

**13.1** Give a destabilizing switching scenario between two stable systems.

**13.2** Give a flowchart for the switching logic when a combination of dwell-time and hysteresis is considered.

**13.3** Explain why an excitation signal is necessary for the stability of an adaptive control scheme using the CLOE algorithm.

# Chapter 14
# Adaptive Regulation—Rejection of Unknown Disturbances

## 14.1 Introduction

One of the basic problems in control is the attenuation (rejection) of unknown disturbances without measuring them. The common framework is the assumption that the disturbance is the result of white noise or a Dirac impulse passed through the "model of the disturbance". While in general one can assume a certain structure for such a "model of disturbance", its parameters are unknown and may be time varying. This will require to use an adaptive feedback approach.

The classical adaptive control paradigm deals essentially with the construction of a control law when the parameters of the plant dynamic model are unknown and time varying (Åström and Wittenmark 1995; Ioannou and Sun 1996; S Fekri and Pasqual 2006). However, in many situations the plant dynamic model is known (can be obtained by system identification) and almost invariant and the objective is the rejection of disturbances characterized by unknown and time-varying disturbance models. It seems reasonable to call this paradigm "adaptive regulation". In what follows we will try to characterize "adaptive control" and "adaptive regulation" with the help of Fig. 14.1.

In classical "adaptive control" the objective is tracking/disturbance attenuation in the presence of unknown and time-varying plant model parameters. Adaptive control focuses on adaptation with respect to plant model parameters variations. The model of the disturbance is assumed to be known and invariant. Only a level of disturbance attenuation in a frequency band is imposed (with the exception of DC disturbances where the controller may include an integrator). The important remarks to be made are:

- No effort is made to estimate the model of the disturbance in real time.
- In general, the disturbances have an undesirable effect upon the adaptation loop (parameters drift, bursting, etc.) and a class of "robust adaptation algorithms" has been developed in order to reduce this negative impact. See Chap. 10.

In "adaptive regulation" the objective is to asymptotically suppress the effect of unknown and time-varying disturbances. Therefore adaptive regulation focuses on adaptation of the controller parameters with respect to variations of parameters of the disturbance model. The plant model is assumed to be known. It is also assumed that the possible small variations of the plant model can be handled by a robust control design. The important remarks to be made are:

- No effort is made to estimate the model of the plant in real time.
- Indirect (correlated) measurement of the disturbance is not available.

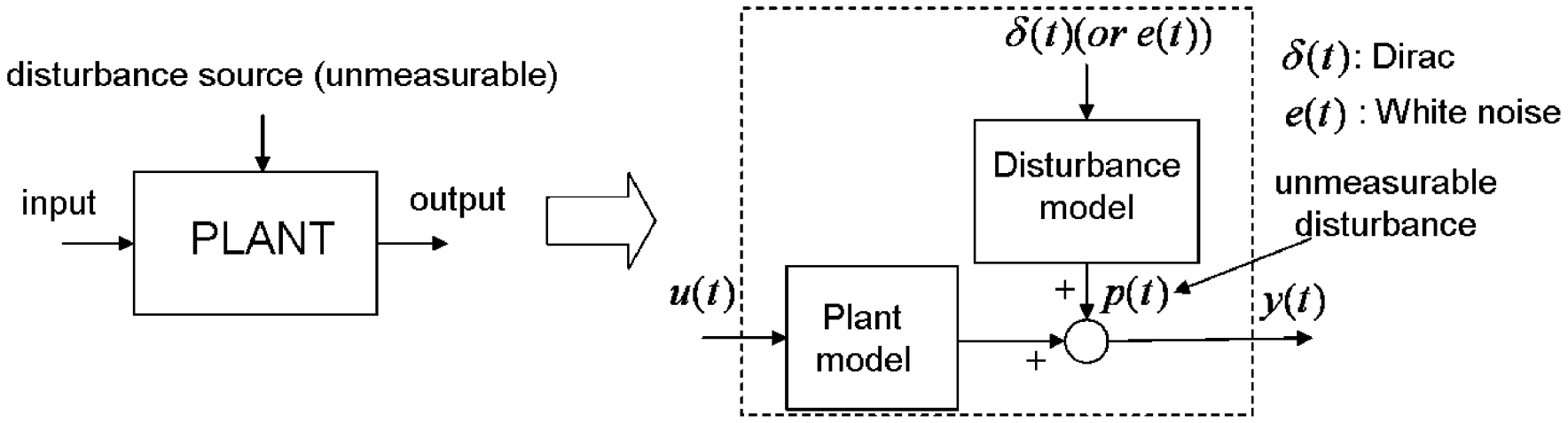

**Fig. 14.1** Plant model and disturbance model

To be more specific, in adaptive regulation the disturbances considered can be defined as "finite band disturbances". This includes single or multiple narrow-band disturbances or sinusoidal disturbances. Furthermore for robustness reasons the disturbances should be located in the frequency domain within the regions where the plant has enough gain (see explanation in Sect. 14.3).

If an image of the disturbance (i.e., a correlated measurement) can be obtained using a well located additional transducer, a feedforward compensation scheme can be considered. This will be discussed in Chap. 15. Feedforward compensation can be used in addition to a feedback approach.

To achieve the rejection of the disturbance (at least asymptotically) without measuring it, a *feedback solution* has to be considered. The common framework is the assumption that the disturbance is the result of white noise or a Dirac impulse passed through the "model of the disturbance".[1] The problem of rejection of the disturbance in the case of known plant and unknown disturbance model has been addressed in Bodson and Douglas (1997), Amara et al. (1999a, 1999b), Valentinotti (2001), Marino et al. (2003), Ding (2003), Gouraud et al. (1997), Hillerstrom and Sternby (1994), Landau et al. (2005) among other references.

The following approaches considered for solving this problem may be mentioned:

1. Use of the internal model principle (Francis and Wonham 1976; Johnson 1976; Bengtsson 1977; Tsypkin 1997; Valentinotti 2001; Amara et al. 1999a, 1999b; Gouraud et al. 1997; Hillerstrom and Sternby 1994; Landau et al. 2005).
2. Use of an observer for the disturbance (Marino et al. 2003; Ding 2003; Serrani 2006).

[1]Throughout this chapter it is assumed that the order of the disturbance model is known but the parameters of the model are unknown (the order can be estimated from data if necessary).

3. Use of the "phase-locked" loop structure considered in communication systems (Bodson and Douglas 1997; Bodson 2005).

Of course, since the parameters of the disturbance model are unknown, all these approaches lead to an adaptive implementation which can be of *direct* or *indirect* type.

From the user point of view and taking into account the type of operation of adaptive disturbance compensation systems, one has to consider two modes of operation of the adaptive schemes:

- *Adaptive* operation. The adaptation is performed continuously with a non vanishing adaptation gain and the controller is updated at each sampling.
- *Self-tuning* operation. The adaptation procedure starts either on demand or when the performance is unsatisfactory. A vanishing adaptation gain is used. The current controller is either updated at each sampling instant once adaptation starts or is frozen during the estimation/computation of the new controller parameters.

Using the internal model principle, the controller should incorporate the model of the disturbance (Francis and Wonham 1976; Johnson 1976; Bengtsson 1977; Tsypkin 1997). Therefore the rejection of unknown disturbances raises the problem of adapting the internal model of the controller and its redesign in real time.

One way of solving this problem is to try to estimate the model of the disturbance in real time and recompute the controller. The estimated model of the disturbance will be included in the controller (as a pre-specified element of the controller). While the disturbance is unknown and its model needs to be estimated, one assumes that the model of the plant is known (obtained for example by identification). The estimation of the disturbance model can be done by using standard parameter estimation algorithms (see for example—Landau et al. 1986; Ljung 1999) provided that a relevant "disturbance observer" can be built. This will lead to an indirect adaptive control scheme. The principle of such a scheme is illustrated in Fig. 14.2. The time consuming part of this approach is the redesign of the controller at each sampling time. The reason is that in many applications the plant model can be of very high dimension and despite that this model is time invariant, one has to recompute the controller because a new internal model should be considered.

This approach has been investigated in Bodson and Douglas (1997), Gouraud et al. (1997), Hillerstrom and Sternby (1994), Landau et al. (2005).

However, by considering the Youla-Kucera parametrization of the controller (known also as the Q-parametrization—see Sect. 7.3.3), it is possible to insert and adjust the internal model in the controller by adjusting the parameters of the $Q$ polynomial (see Fig. 14.3). In the presence of unknown disturbances it is possible to build a direct adaptive control scheme where the parameters of the $Q$ polynomial are directly adapted in order to have the desired internal model without recomputing the controller (polynomials $R_0$ and $S_0$ in Fig. 14.3 remain unchanged). The number of the controller parameters to be directly adapted is roughly equal to the number of parameters of the denominator of the disturbance model. In other words, the size of the adaptation algorithm will depend upon the complexity of the disturbance model.

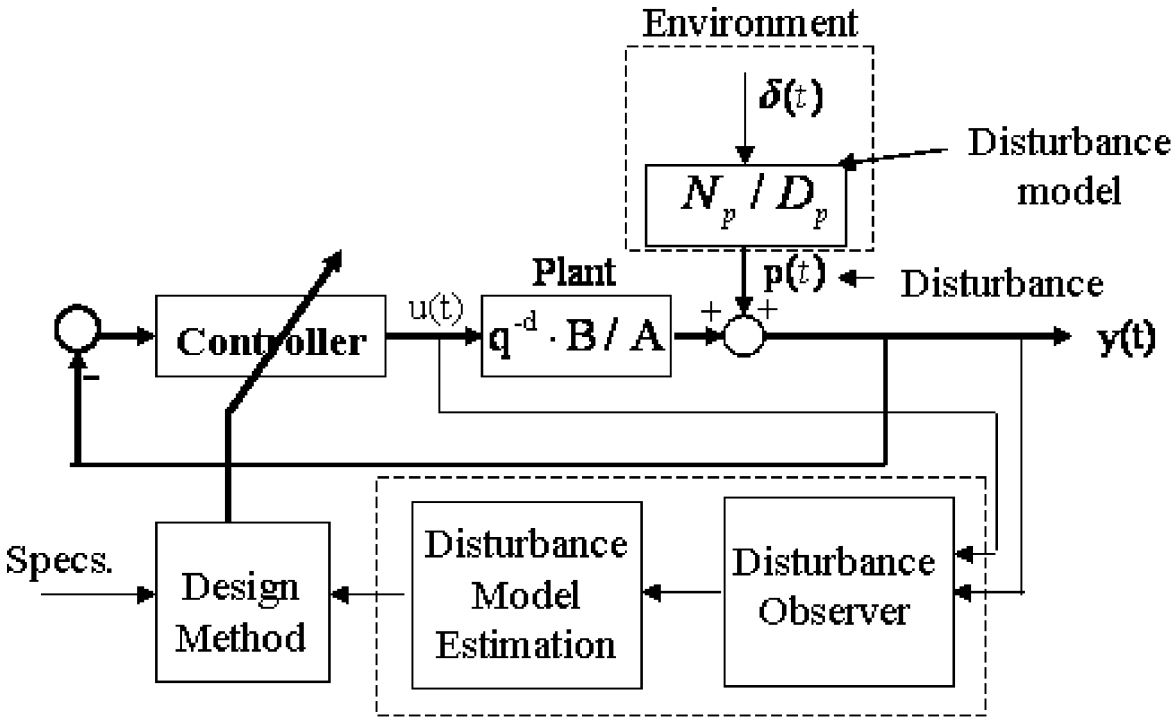

**Fig. 14.2** Indirect adaptive control scheme for rejection of unknown disturbances

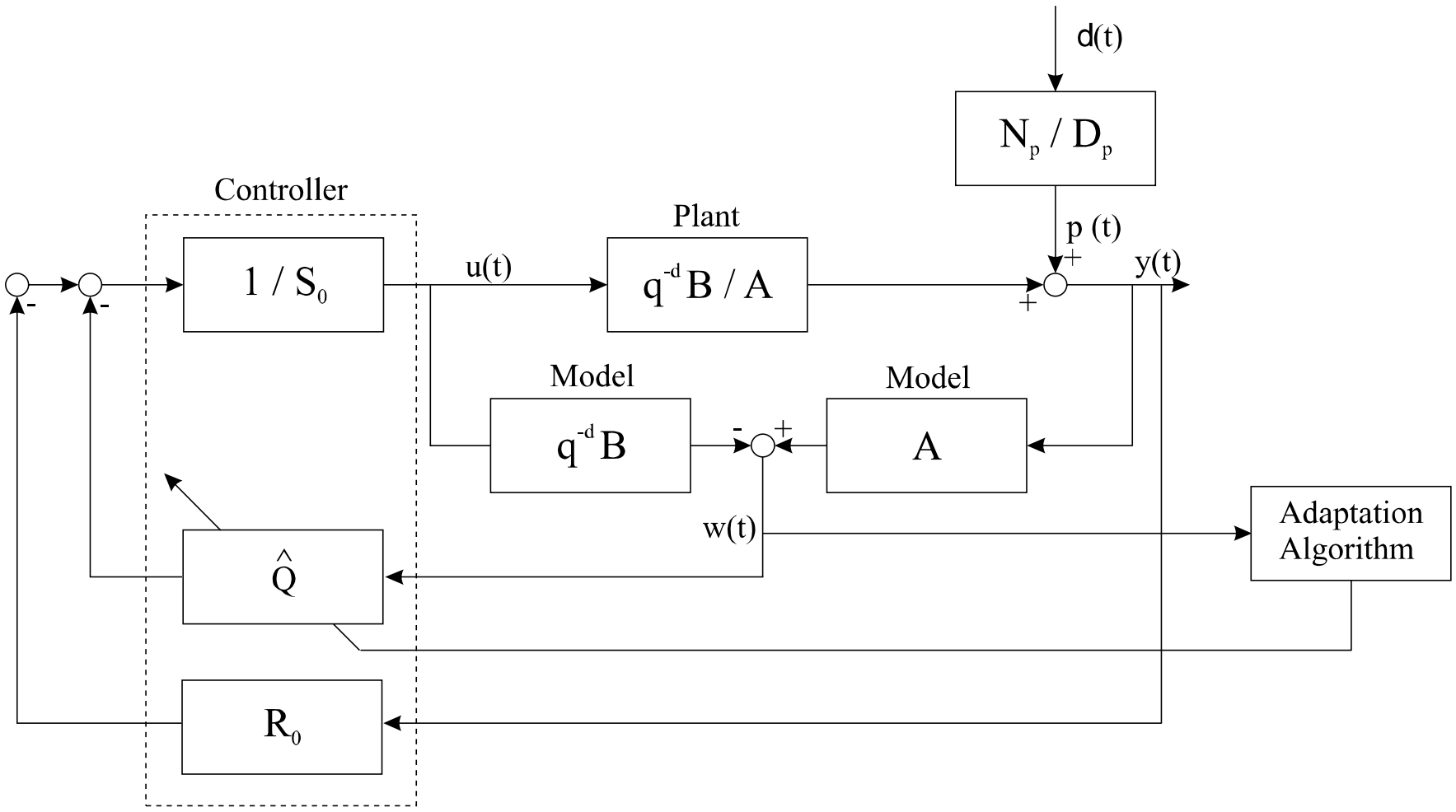

**Fig. 14.3** Direct adaptive control scheme for rejection of unknown disturbances

The chapter focuses on the direct and indirect feedback adaptive regulation methodology for the case of multiple unknown and time-varying narrow band disturbances. The methodology is illustrated by results obtained on an active vibration control system but the methodology is also applicable for active noise control systems (Amara et al. 1999b).

Section 14.2 briefly reviews the plant, disturbance and controller representation as well as the internal model principle. Some specific robustness issues are addressed in Sect. 14.3. A direct adaptive control schemes for disturbance rejection is presented in Sect. 14.4 and a stability analysis is presented in Sect. 14.5. The indirect adaptive control scheme is presented in Sect. 14.6. In Sect. 14.7 real-time results obtained on an active vibration control (AVC) system using an inertial actuator (presented in Sect. 1.4.4) are presented.

## 14.2 Plant Representation and Controller Design

The structure of the linear time invariant discrete-time model of the plant (used for controller design) is:

$$G(z^{-1}) = \frac{z^{-d}B(z^{-1})}{A(z^{-1})} = \frac{z^{-d-1}B^*(z^{-1})}{A(z^{-1})} \tag{14.1}$$

with $d$ equal to the plant pure time delay in number of sampling periods and

$$A(z^{-1}) = 1 + a_1 z^{-1} + \cdots + a_{n_A} z^{-n_A}$$
$$B(z^{-1}) = b_1 z^{-1} + \cdots + b_{n_B} z^{-n_B} = q^{-1}B^*(z^{-1})$$
$$B^*(z^{-1}) = b_1 + \cdots + b_{n_B} z^{-n_B+1}$$

where $A(z^{-1})$, $B(z^{-1})$, $B^*(z^{-1})$ are polynomials in the complex variable $z^{-1}$ and $n_A$, $n_B$ and $n_B - 1$ represent their orders.[2] The model of the plant may be obtained by system identification using the techniques and algorithms presented in Chap. 5. Details on system identification of the models considered in this chapter can be found in Landau and Zito (2005), Constantinescu (2001), Landau et al. (2001a, 2001b), Karimi (2002), Constantinescu and Landau (2003).

Since in this chapter we are focused on regulation, the controller to be designed is a RS-type polynomial controller described in Chap. 7—see also Fig. 14.5. The output of the plant $y(t)$ and the input $u(t)$ may be expressed as:

$$y(t) = \frac{q^{-d}B(q^{-1})}{A(q^{-1})}u(t) + p(t) \tag{14.2}$$

$$S(q^{-1})u(t) = -R(q^{-1})y(t) \tag{14.3}$$

where $q^{-1}$ is the delay (shift) operator ($x(t) = q^{-1}x(t+1)$) and $p(t)$ is the resulting additive disturbance on the output of the system. $R(z^{-1})$ and $S(z^{-1})$ are polynomials in $z^{-1}$ having the orders $n_R$ and $n_S$, respectively, with the following expressions:

$$R(z^{-1}) = r_0 + r_1 z^{-1} + \cdots + r_{n_R} z^{-n_R} = R'(z^{-1})H_R(z^{-1}) \tag{14.4}$$

$$S(z^{-1}) = 1 + s_1 z^{-1} + \cdots + s_{n_S} z^{-n_S} = S'(z^{-1})H_S(z^{-1}) \tag{14.5}$$

where $H_R$ and $H_S$ are pre-specified parts of the controller (used for example to incorporate the internal model of a disturbance or to open the loop at certain frequencies).

We define the following sensitivity functions (see also Chap. 8):

- Output sensitivity function (the transfer function between the disturbance $p(t)$ and the output of the system $y(t)$):

$$S_{yp}(z^{-1}) = \frac{A(z^{-1})S(z^{-1})}{P(z^{-1})} \tag{14.6}$$

[2]The complex variable $z^{-1}$ will be used for characterizing the system's behavior in the frequency domain and the delay operator $q^{-1}$ will be used for describing the system's behavior in the time domain.

- Input sensitivity function (the transfer function between the disturbance $p(t)$ and the input of the system $u(t)$):

$$S_{up}(z^{-1}) = -\frac{A(z^{-1})R(z^{-1})}{P(z^{-1})} \tag{14.7}$$

Here

$$\begin{aligned} P(z^{-1}) &= A(z^{-1})S(z^{-1}) + z^{-d}B(z^{-1})R(z^{-1}) \\ &= A(z^{-1})S'(z^{-1})H_S(z^{-1}) + z^{-d}B(z^{-1})R'(z^{-1})H_R(z^{-1}) \end{aligned} \tag{14.8}$$

defines the poles of the closed loop (roots of $P(z^{-1})$). In pole placement design, $P(z^{-1})$ is the polynomial specifying the desired closed-loop poles and the controller polynomials $R(z^{-1})$ and $S(z^{-1})$ are minimal degree solutions of (14.8) where the degrees of $P$, $R$ and $S$ are given by: $n_P \leq n_A + n_B + d - 1$, $n_S = n_B + d - 1$ and $n_R = n_A - 1$. Using (14.2) and (14.3), one can write the output of the system as:

$$y(t) = \frac{A(q^{-1})S(q^{-1})}{P(q^{-1})} p(t) = S_{yp}(q^{-1})p(t) \tag{14.9}$$

For more details on RS-type controllers and sensitivity functions see Landau and Zito (2005). Suppose that $p(t)$ is a deterministic disturbance, so it can be written as

$$p(t) = \frac{N_p(q^{-1})}{D_p(q^{-1})}\delta(t) \tag{14.10}$$

where $\delta(t)$ is a Dirac impulse and $N_p(z^{-1})$, $D_p(z^{-1})$ are coprime polynomials in $z^{-1}$, of degrees $n_{N_p}$ and $n_{D_p}$, respectively. In the case of stationary disturbances the roots of $D_p(z^{-1})$ are on the unit circle. The energy of the disturbance is essentially represented by $D_p$. The contribution of the terms of $N_p$ is weak compared to the effect of $D_p$, so one can neglect the effect of $N_p$.

**Internal Model Principle** *The effect of the disturbance given in* (14.10) *upon the output*:

$$y(t) = \frac{A(q^{-1})S(q^{-1})}{P(q^{-1})} \frac{N_p(q^{-1})}{D_p(q^{-1})}\delta(t) \tag{14.11}$$

*where $D_p(z^{-1})$ is a polynomial with roots on the unit circle and $P(z^{-1})$ is an asymptotically stable polynomial, converges asymptotically towards zero if and only if*[3] *the polynomial $S(z^{-1})$ in the RS controller has the form*:

$$S(z^{-1}) = D_p(z^{-1})S'(z^{-1}) \tag{14.12}$$

In other terms, the pre-specified part of $S(z^{-1})$ should be chosen as $H_S(z^{-1}) = D_p(z^{-1})$ and the controller is computed using (14.8), where $P$, $D_p$, $A$, $B$, $H_R$ and $d$ are given.[4]

[3] The "only if" does not apply if $D_p$ divides $A$.

[4] Of course it is assumed that $D_p$ and $B$ do not have common factors.

Using the Youla-Kucera parametrization (Q-parametrization) of all stable controllers (Anderson 1998; Tsypkin 1997 and Sect. 7.3.3), the controller polynomials $R(z^{-1})$ and $S(z^{-1})$ get the form:

$$R(z^{-1}) = R_0(z^{-1}) + A(z^{-1})Q(z^{-1}) \tag{14.13}$$

$$S(z^{-1}) = S_0(z^{-1}) - z^{-d}B(z^{-1})Q(z^{-1}) \tag{14.14}$$

The (central) controller $(R_0, S_0)$ can be computed by poles placement (but any other design technique can be used). Given the plant model $(A, B, d)$ and the desired closed-loop poles specified by the roots of $P$ one has to solve:

$$P(z^{-1}) = A(z^{-1})S_0(z^{-1}) + z^{-d}B(z^{-1})R_0(z^{-1}) \tag{14.15}$$

Equations (14.13) and (14.14) characterize the set of all stabilizable controllers assigning the closed-loop poles as defined by $P(z^{-1})$ (it can be verified by simple calculations that the poles of the closed-loop remain unchanged). For the purpose of this chapter, $Q(z^{-1})$ is considered to be a polynomial of the form:

$$Q(z^{-1}) = q_0 + q_1 z^{-1} + \cdots + q_{n_Q} z^{-n_Q} \tag{14.16}$$

To compute $Q(z^{-1})$ in order that the polynomial $S(z^{-1})$ given by (14.14), incorporates the internal model of the disturbance, one has to solve the diophantine equation:

$$S'(z^{-1})D_p(z^{-1}) + z^{-d}B(z^{-1})Q(z^{-1}) = S_0(z^{-1}) \tag{14.17}$$

where $D_p(z^{-1})$, $d$, $B(z^{-1})$ and $S_0(z^{-1})$ are known and $S'(z^{-1})$ and $Q(z^{-1})$ are unknown. Equation (14.17) has a unique solution for $S'(z^{-1})$ and $Q(z^{-1})$ with: $n_{S_0} \le n_{D_p} + n_B + d - 1$, $n_{S'} = n_B + d - 1$, $n_Q = n_{D_p} - 1$. One sees that the order $n_Q$ of the polynomial Q depends upon the structure of the disturbance model.

The central controller may contain some fixed parts ($Hr$ and $Hs_1$) for robustness reasons (opening the loop in low and high frequencies or shaping sensitivity functions) such as:

$$R_0(z^{-1}) = Hr(z^{-1})R_0'(z^{-1}) \tag{14.18}$$

$$S_0(z^{-1}) = Hs_1(z^{-1})S_0'(z^{-1}) \tag{14.19}$$

To preserve these parts, the Q-parametrization of the controller has to be modified as follows:

$$R(z^{-1}) = R_0(z^{-1}) + A(z^{-1})Hr(z^{-1})Hs_1(z^{-1})Q(z^{-1}) \tag{14.20}$$

$$S(z^{-1}) = S_0(z^{-1}) - z^{-d}B(z^{-1})Hr(z^{-1})Hs_1(z^{-1})Q(z^{-1}) \tag{14.21}$$

One can verify that the desired closed-loop poles specified by the roots of $P$ in (14.15) remain unchanged. To compute $Q(z^{-1})$ in order that the controller contains the internal model of the disturbance, one has to solve now:

$$S'(z^{-1})D_p(z^{-1})Hs_1(z^{-1}) + z^{-d}B(z^{-1})Hr(z^{-1})Hs_1(z^{-1})Q(z^{-1}) = S_0(z^{-1}) \tag{14.22}$$

## 14.3 Robustness Considerations

As it is well known, the introduction of the internal model for the perfect rejection of the disturbance (asymptotically) will have as effect to raise the maximum value of the modulus of the output sensitivity function $S_{yp}$. This may lead to unacceptable values for the modulus and delay margins if the controller design is not appropriately done (see Landau and Zito 2005). As a consequence, a robust control design should be considered assuming that the model of the disturbance and its domain of variations are known. The objective is that for all the situations (which means for all possible values of the disturbance model parameters) an acceptable modulus margin and delay margin are obtained. It is important to mention that the "tuned" linear controller will give the maximum achievable performance for the adaptive scheme.

Furthermore at the frequencies where perfect rejection of the disturbance is achieved one has $S_{yp}(e^{-j\omega}) = 0$ and

$$|S_{up}(e^{-j\omega})| = \left| \frac{A(e^{-j\omega})}{B(e^{-j\omega})} \right| \tag{14.23}$$

Equation (14.23) corresponds to the inverse of the gain of the system to be controlled. The implication of (14.23) is that cancellation (or in general an important attenuation) of disturbances on the output should be done only in frequency regions where the system gain is large enough. If the gain of the controlled system is too low, $|S_{up}|$ will be large at these frequencies. Therefore, the robustness vs additive plant model uncertainties will be reduced and the stress on the actuator will become important. Equation (14.23) also implies that serious problems will occur if $B(z^{-1})$ has complex zeros close to the unit circle (stable or unstable zeros) at frequencies where an important attenuation of disturbances is required. It is mandatory to avoid attenuation of disturbances at these frequencies.

Since on one hand we would not like to react to very high-frequency disturbances and on the other hand we would like to have a good robustness, it is often wise to open the loop at $0.5 f_s$ ($f_s$ is the sampling frequency) by introducing a fixed part in the controller $H_R(q^{-1}) = 1 + q^{-1}$ (for details see Sect. 7.3.1 and Landau and Zito 2005).

## 14.4 Direct Adaptive Regulation

The objective is to find an estimation algorithm which will directly estimate the parameters of the internal model in the controller in the presence of an unknown disturbance (but of known structure) without modifying the closed-loop poles. Clearly, the Q-parametrization is a potential option since modifications of the $Q$ polynomial will not affect the closed-loop poles. In order to build an estimation algorithm it is necessary to define an *error equation* which will reflect the difference between the optimal $Q$ polynomial and its current value.[5] Note also that in the time domain, the internal model principle can be interpreted as finding $Q$ such that asymptotically $y(t)$ becomes zero. One has the following result:

**Lemma 14.1** *Define the solution of (14.17) as the optimal polynomial $Q(q^{-1})$ and denote by $\hat{Q}(q^{-1})$ an estimation of the polynomial $Q(q^{-1})$, then*:

$$y(t+1) = [Q(q^{-1}) - \hat{Q}(q^{-1})]\frac{q^{-d}B^*(q^{-1})}{P(q^{-1})}w(t) + x(t+1) \tag{14.24}$$

*where*:

$$w(t) = \frac{A(q^{-1})N_p(q^{-1})}{D_p(q^{-1})}\delta(t) = A(q^{-1})y(t) - q^{-d}B(q^{-1})u(t) \tag{14.25}$$

*and*

$$x(t) = \frac{S'(q^{-1})D_p(q^{-1})}{P(q^{-1})}w(t) = \frac{S'(q^{-1})A(q^{-1})N_p(q^{-1})}{P(q^{-1})}\delta(t) \tag{14.26}$$

*is a signal which tends asymptotically towards zero.*

*Proof* Using the Q-parametrization, the output of the system in the presence of a disturbance of the form (14.10) can be expressed as:

$$\begin{aligned} y(t) &= \frac{A(q^{-1})[S_0(q^{-1}) - q^{-d}B(q^{-1})Q(q^{-1})]}{P(q^{-1})}\frac{N_p(q^{-1})}{D_p(q^{-1})}\delta(t) \\ &= \frac{S_0(q^{-1}) - q^{-d}B(q^{-1})Q(q^{-1})}{P(q^{-1})}w(t) \end{aligned} \tag{14.27}$$

Replacing $S_0(q^{-1})$ from the last equation by (14.17) one obtains (14.24). The signal $x(t)$ given in (14.26) tends asymptotically to zero since it is the output of an asymptotically stable filter whose input is a Dirac. □

Assume that one has an estimation of $Q(q^{-1})$ at instant $t$, denoted $\hat{Q}(t, q^{-1})$. Define $\varepsilon^0(t+1)$ as the value of $y(t+1)$ obtained with $\hat{Q}(t, q^{-1})$. Using (14.24) one gets:

$$\varepsilon^0(t+1) = [Q(q^{-1}) - \hat{Q}(t, q^{-1})]\frac{q^{-d}B^*(q^{-1})}{P(q^{-1})}w(t) + x(t+1) \tag{14.28}$$

One can define now the a posteriori error (using $\hat{Q}(t+1, q^{-1})$) as:

$$\varepsilon(t+1) = [Q(q^{-1}) - \hat{Q}(t+1, q^{-1})]\frac{q^{-d}B^*(q^{-1})}{P(q^{-1})}w(t) + x(t+1) \tag{14.29}$$

[5] In Tsypkin (1997), such an error equation is provided and it can be used for developing a direct adaptive control scheme. This idea has been used in Valentinotti (2001), Amara et al. (1999a, 1999b), Landau et al. (2005).

Define the estimated polynomial $\hat{Q}(t, q^{-1})$ as:

$$\hat{Q}(t, q^{-1}) = \hat{q}_0(t) + \hat{q}_1(t)q^{-1} + \cdots + \hat{q}_{n_Q}(t)q^{-n_Q}$$

and the associated estimated parameter vector:

$$\hat{\theta}(t) = [\hat{q}_0(t)\ \hat{q}_1(t)\ \ldots\ \hat{q}_{n_Q}(t)]^T$$

Define the fixed parameter vector corresponding to the optimal value of the polynomial $Q$ as:

$$\theta = [q_0\ q_1\ \ldots\ q_{n_Q}]^T$$

Denote:

$$w_2(t) = \frac{q^{-d}B^*(q^{-1})}{P(q^{-1})}w(t) \tag{14.30}$$

and define the following observation vector:

$$\phi^T(t) = [w_2(t)\ w_2(t-1)\ \ldots\ w_2(t-n_Q)] \tag{14.31}$$

Equation (14.29) becomes

$$\varepsilon(t+1) = [\theta^T - \hat{\theta}^T(t+1)]\phi(t) + x(t+1) \tag{14.32}$$

One can remark that $\varepsilon(t)$ corresponds to an adaptation error (see Chap. 3) and therefore adaptation algorithms given in Chap. 3 can be used. From (14.28) one obtains the a priori adaptation error:

$$\varepsilon^0(t+1) = w_1(t+1) - \hat{\theta}^T(t)\phi(t) \tag{14.33}$$

with

$$w_1(t+1) = \frac{S_0(q^{-1})}{P(q^{-1})}w(t+1) \tag{14.34}$$

$$w(t+1) = A(q^{-1})y(t+1) - q^{-d}B^*(q^{-1})u(t) \tag{14.35}$$

where $B(q^{-1})u(t+1) = B^*(q^{-1})u(t)$. The a posteriori adaptation error is obtained from (14.29):

$$\varepsilon(t+1) = w_1(t+1) - \hat{\theta}^T(t+1)\phi(t) \tag{14.36}$$

For the estimation of the parameters of $\hat{Q}(t, q^{-1})$ the following parameter adaptation algorithm is used (see Chap. 3):

$$\hat{\theta}(t+1) = \hat{\theta}(t) + F(t)\phi(t)\varepsilon(t+1) \tag{14.37}$$

$$\varepsilon(t+1) = \frac{\varepsilon^0(t+1)}{1 + \phi^T(t)F(t)\phi(t)} \tag{14.38}$$

$$\varepsilon^0(t+1) = w_1(t+1) - \hat{\theta}^T(t)\phi(t) \tag{14.39}$$

$$F(t+1) = \frac{1}{\lambda_1(t)}\left[F(t) - \frac{F(t)\phi(t)\phi^T(t)F(t)}{\frac{\lambda_1(t)}{\lambda_2(t)} + \phi^T(t)F(t)\phi(t)}\right] \tag{14.40}$$

$$1 \geq \lambda_1(t) > 0;\ 0 \leq \lambda_2(t) < 2 \tag{14.41}$$

where $\lambda_1(t), \lambda_2(t)$ allow to obtain various profiles for the evolution of the adaption gain $F(t)$ (for details see Chap. 3).

In order to implement this methodology for disturbance rejection (see Fig. 14.3), it is supposed that the plant model $\frac{z^{-d}B(z^{-1})}{A(z^{-1})}$ is known (identified) and that it exists a controller $[R_0(z^{-1}), S_0(z^{-1})]$ which verifies the desired specifications in the absence of the disturbance. One also supposes that the degree $n_Q$ of the polynomial $Q(z^{-1})$ is fixed, $n_Q = n_{D_p} - 1$, i.e., the structure of the disturbance is known.

The following procedure is applied at each sampling time for *adaptive* operation:

1. Get the measured output $y(t+1)$ and the applied control $u(t)$ to compute $w(t+1)$ using (14.35).
2. Compute $w_1(t+1)$ and $w_2(t)$ using (14.34) and (14.30) with $P$ given by (14.15).
3. Estimate the $Q$-polynomial using the parametric adaptation algorithm (14.37)–(14.41).
4. Compute and apply the control (see Fig. 14.3):

$$S_0(q^{-1})u(t+1) = -R_0(q^{-1})y(t+1) - \hat{Q}(t+1, q^{-1})w(t+1) \tag{14.42}$$

In the *adaptive* operation one uses in general an adaptation gain updating with variable forgetting factor $\lambda(t)$ (the forgetting factor tends towards 1), combined with a *constant trace* adaption gain. Once the trace of the adaptation gain is below a given value, one switches to the constant trace gain updating. See Sect. 3.2.3 for details. The advantage of the *constant trace* gain updating is that the adaptation moves in an optimal direction (least squares) but the size of the step does not goes to zero.

For the *self tuning* operation, the estimation of the $Q$-polynomial starts once the level of the output is over a defined threshold. A parameter adaptation algorithm (14.37)–(14.41) with *decreasing adaption gain* ($\lambda_1(t) = 1$ and $\lambda_2(t) = 1$) or with *variable forgetting factor and decreasing gain* is used.

The estimation can be stopped when the adaption gain is below a pre-specified level.[6] The controller is either updated at each sampling or is updated only when the estimation of the new Q-parameters is finished.

When the fixed parts of the central controller are preserved under the Q parametrization (see (14.20) and (14.21)), one uses the same adaptation algorithm, except that $w_2$ given by (14.30) is replaced by:

$$w_2(t) = \frac{q^{-d}B^*(q^{-1})Hr(z^{-1})Hs_1(z^{-1})}{P(q^{-1})}w(t) \tag{14.43}$$

For more details see Landau et al. (2010, 2011).

## 14.5 Stability Analysis

We will make the following assumptions:

[6]The magnitude of the adaptation gain gives an indication upon the variance of the parameter estimation error—see Chap. 3.

(H1) The available plant model $(\hat{A}, \hat{B})$ is identical to the true plant model (i.e., $\hat{A} = A$, $\hat{B} = B$) and the plant delay is known.
(H2) The model of the disturbance has poles on the unit circle.
(H3) The order of the denominator of the disturbance model $n_{D_P}$ is known.

**Lemma 14.2** *Under hypotheses* (H1), (H2) *and* (H3), *for the adaptive regulation system described by* (14.32) *and using the parametric adaptation algorithm given by* (14.37) *through* (14.41) *one has*:

$$\lim_{t\to\infty} \varepsilon(t+1) = 0 \tag{14.44}$$

$$\lim_{t\to\infty} [\hat{\theta}(t+1) - \theta]^T \phi(t) = 0 \tag{14.45}$$

$$\lim_{t\to\infty} \frac{[\varepsilon^0(t+1)]^2}{1 + \phi^T(t)F(t)\phi(t)} = 0 \tag{14.46}$$

$$\|\phi(t)\| \textit{ is bounded} \tag{14.47}$$

$$\lim_{t\to\infty} \varepsilon^0(t+1) = 0 \tag{14.48}$$

*for any initial conditions* $\hat{\theta}(0)$, $\varepsilon^0(0)$, $F(0)$.

*Proof* Equation (14.32) combined with (14.37) through (14.40) lead to an equivalent feedback system (by defining $\tilde{\theta}(t+1) = \hat{\theta}(t+1) - \theta$) which is formed by a strictly passive linear block (a unit gain) in feedback with a time-varying passive block and with $x(t+1)$ as an external input. However, $x(t+1)$ is a vanishing signal (see Lemma 14.1) and does not influence the asymptotic behavior of $\varepsilon(t+1)$. Equation (14.32) combined with equations. Equations (14.37) through (14.40) (with $x(t+1)$ a vanishing signal) have the form considered in Theorem 3.2 and therefore this theorem can be used to analyze the system. Applying the above mentioned theorem one concludes that (14.44), (14.45) and (14.46) hold.

Under Assumption (H1) the signal $w(t)$ can be written:

$$w(t) = A(q^{-1})p(t) = \frac{A(q^{-1})N_p(q^{-1})}{D_p(q^{-1})}\delta(t) \tag{14.49}$$

Since $D_P$ has the roots on the unit circle and the coefficients of $A$ and $N_P$ are bounded one concludes that $p(t)$ and respectively, $w(t)$ are bounded.

The components of the regressor vector $\phi(t)$ have the form:

$$w_2(t-i) = \frac{q^{-d}B^*(q^{-1})}{P(q^{-1})}w(t-i) \tag{14.50}$$

Since the polynomial $P$ is asymptotically stable by design, one concludes that $w_2(t-i)$ and $\phi(t)$ are bounded. Therefore from (14.46) one concludes that (14.48) holds which ends the proof.[7] □

[7]The convergence towards zero of $\varepsilon^0(t)$ can be proven also with the "bounded growth" lemma of Goodwin and Sin (Chap. 11, Lemma 11.1).

For the convergence of the parameters one has the following result:

**Lemma 14.3** *In the presence of a disturbance characterized by a model of the form* (14.10) *with the order of the polynomial $D_P$ equal to $n_{D_P}$ and an estimated parameter vector $\hat{\theta}$ with $n_{D_P}$ parameters $\hat{q}_i$, $i = 0, 1, \dots, n_{D_P} - 1$,* (14.44) *implies*:

$$\lim_{t\to\infty} [q_i - \hat{q}_i(t+1)] = 0, \quad i = 0, 1, \dots, n_{D_P} - 1 \tag{14.51}$$

*Proof* The proof is similar to that of Theorem 3.5. Since $\varepsilon(t+1)$ goes to zero asymptotically, when $t \to \infty$ one can write from (14.45):

$$\lim_{t\to\infty} \varepsilon(t+1) = \lim_{t\to\infty} \left[ \sum_{i=0}^{n_{D_P}-1} (q_i - \hat{q}_i(t+1)) q^{-i} \right] w_2(t) = 0 \tag{14.52}$$

Since asymptotically $\hat{q}_i$, $i = 0, 1, \dots, n_{D_P} - 1$ will be a constant, (14.52) can hold either if $q_i - \hat{q}_i = 0$, $i = 0, 1, \dots, n_{D_P} - 1$ or if $w_2(t)$ is a solution of the difference equation $[\sum_{i=0}^{n_{D_P}-1} (q_i - \hat{q}_i) q^{-i}] w_2(t) = 0$. In the presence of the external disturbance, $w_2(t)$ which is a filtered version of the disturbance will be characterized by a difference equation of order $n_{D_P}$ and it can not be a solution of a difference equation of order $n_{D_P} - 1$. Therefore $\lim_{t\to\infty} \varepsilon(t+1) = 0$ implies also the parametric convergence in the presence of the disturbance. □

## 14.6 Indirect Adaptive Regulation

The methodology proposed in this section concerns the indirect adaptive control for the attenuation of narrow band disturbances and consists in two steps: (1) Identification of the disturbance model. (2) Computation of a digital controller using the identified disturbance model

For the identification of the disturbance model one has to set first a reliable observer of the disturbance. Provided that the model of the plant is known (which is the case in our context) an estimation $\bar{y}(t)$ of the disturbance $p(t)$ can be obtained by (Bodson and Douglas 1997; Gouraud et al. 1997):

$$\bar{y}(t) = y(t) - \frac{q^{-d} B q^{-1}}{A q^{-1}} u(t). \tag{14.53}$$

However, as we can see from (14.25):

$$\begin{aligned} w(t) &= \frac{A(q^{-1}) N_p(q^{-1})}{D_p(q^{-1})} \delta(t) \\ &= A(q^{-1}) y(t) - q^{-d} B(q^{-1}) u(t) = A(q^{-1}) p(t) \end{aligned} \tag{14.54}$$

The signal $w(t)$ (see also Fig. 14.3) can be viewed as an estimation of the disturbance filtered though the polynomial $A(q^{-1})$. Both solutions for getting an estimation of the disturbances can be used. For numerical reasons it is preferable to use $w$ (since the polynomial $A(q^{-1})$ may have roots close to the unit circle).

The disturbance is considered as a stationary signal having a rational spectrum. As such it may be considered as the output of a filter with the transfer function $N_p(z^{-1})/D_p(z^{-1})$ and a white noise as input:

$$D_p(q^{-1})p(t) = N_p(q^{-1})e(t) \quad \text{or} \quad p(t) = \frac{N_p(q^{-1})}{D_p(q^{-1})}e(t) \tag{14.55}$$

where $e(t)$ represents a Gaussian white noise, and

$$N_p(z^{-1}) = 1 + n_{p_1}z^{-1} + \cdots + n_{p_{n_{N_p}}} z^{-n_{N_p}} = 1 + z^{-1}N_p^*(z^{-1})$$
$$D_p(z^{-1}) = 1 + d_{p_1}z^{-1} + \cdots + d_{p_{n_{D_p}}} z^{-n_{D_p}} = 1 + z^{-1}D_p^*(z^{-1})$$

Therefore the disturbance model can be represented by an ARMA model. As we deal with narrow band disturbances, the filtering effect of the primary path in cascade with the output sensitivity function (when operating in closed loop) around the central frequency of the disturbance can be approximated by a gain and a phase lag which will be captured by the $\frac{N_p(z^{-1})}{D_p(z^{-1})}$ model.

From (14.55) one obtains:

$$p(t+1) = -\sum_{i=1}^{n_{D_p}} d_{p_i} p(t-i+1) + \sum_{i=1}^{n_{N_p}} n_{p_i} e(t-i+1) + e(t+1). \tag{14.56}$$

The problem is, in fact, an on-line adaptive estimation of parameters in presence of noise (see Chap. 5 and Landau et al. 1986). Equation (14.56) is a particular case of identification of an ARMAX model. It is possible to use the *Recursive Extended Least Squares* method given in Sect. 5.3.2, which is dedicated to the identification of this type of model. The parameter adaptation algorithm given in (14.37)–(14.41) is used.

In order to apply this methodology it is supposed that the plant model is known (can be obtained by identification). It is also supposed that the degrees $n_{N_p}$ and $n_{D_p}$ of $N_p(z^{-1})$ respectively $D_p(z^{-1})$ are fixed.

The controller containing the disturbance dynamics is computed by solving the diophantine equation (14.8) and using (14.5)

$$A(z^{-1})H_S(z^{-1})S'(z^{-1}) + z^{-d}B(z^{-1})R(z^{-1}) = P(z^{-1}) \tag{14.57}$$

with $H_S(z^{-1}) = \hat{D}_p(z^{-1})$ (the current estimated model of the disturbance).

In the case of single or multiple narrow band disturbances, if only a certain level of attenuation of the disturbance is necessary, one can either modify the damping of the estimated disturbance model before solving (14.57) or use the stop band filters presented in Sect. 8.5.1.

In *adaptive* operation, the parameters of the controller have to be recomputed at each sampling instant based on the current estimation of the disturbance model (the disturbance estimation algorithm will use a non vanishing adaptation gain). For the *self tuning* operation, the algorithm for disturbance model estimation uses a parameter adaptation algorithm with decreasing adaptation gain. One can either update the controller parameters at each sample, or one can keep constant the parameters of the controller during disturbance estimation and up-date the parameters of the controller once the estimation has converged (a measure of the convergence is the trace of the adaptation gain).

The major drawback of this approach is the complexity of the computation which has to be done at each sampling since one has to solve a Bezout equation of a dimension which depends on the order of the model of the plant and the order of the disturbance model (often the order of the model of the secondary path is very high). A slight reduction of the complexity is obtained if one uses a Youla-Kucera parameterization of the controller. In this case one has to solve (14.17) instead of (14.57).

## 14.7 Adaptive Rejection of Multiple Narrow Band Disturbances on an Active Vibration Control System

### *14.7.1 The Active Vibration Control System*

The principle and the structure of the active vibration control system using an inertial actuator has been presented in Sect. 1.4.4. The scheme of the system is shown in Fig. 14.4. The inertial actuator will create vibrational forces to counteract the effect of vibrational disturbances. It is fixed to the chassis where the vibrations should be attenuated. The controller will act (through a power amplifier) on the position of the mobile part. The sampling frequency is 800 Hz. The system has to be considered as a "black box".

The equivalent control scheme is shown in Fig. 14.5. The system input, $u(t)$ is the position of the mobile part (see Figs. 14.4, 14.5), the output $y(t)$ is the residual force measured by a force sensor. The transfer function $(q^{-d_1}\frac{C}{D})$, between the disturbance force, $u_p$, and the residual force $y(t)$ is called *primary path*. In our case (for testing purposes), the primary force is generated by a shaker controlled by a signal given by the computer. The plant transfer function $(q^{-d}\frac{B}{A})$ between the input of the system, $u(t)$, and the residual force is called *secondary path*. The input of the system being a position and the output a force, the secondary path transfer function has a double differentiator behavior.

### *14.7.2 Experimental Results*

The performance of the system for rejecting multiple unknown time-varying narrow band disturbances will be illustrated using the direct adaptive regulation scheme presented in Sect. 14.4. A comparison between direct and indirect regulation strategies will also be provided.[8]

[8]These experiments have been carried out by M. Alma (GIPSA-LAB).

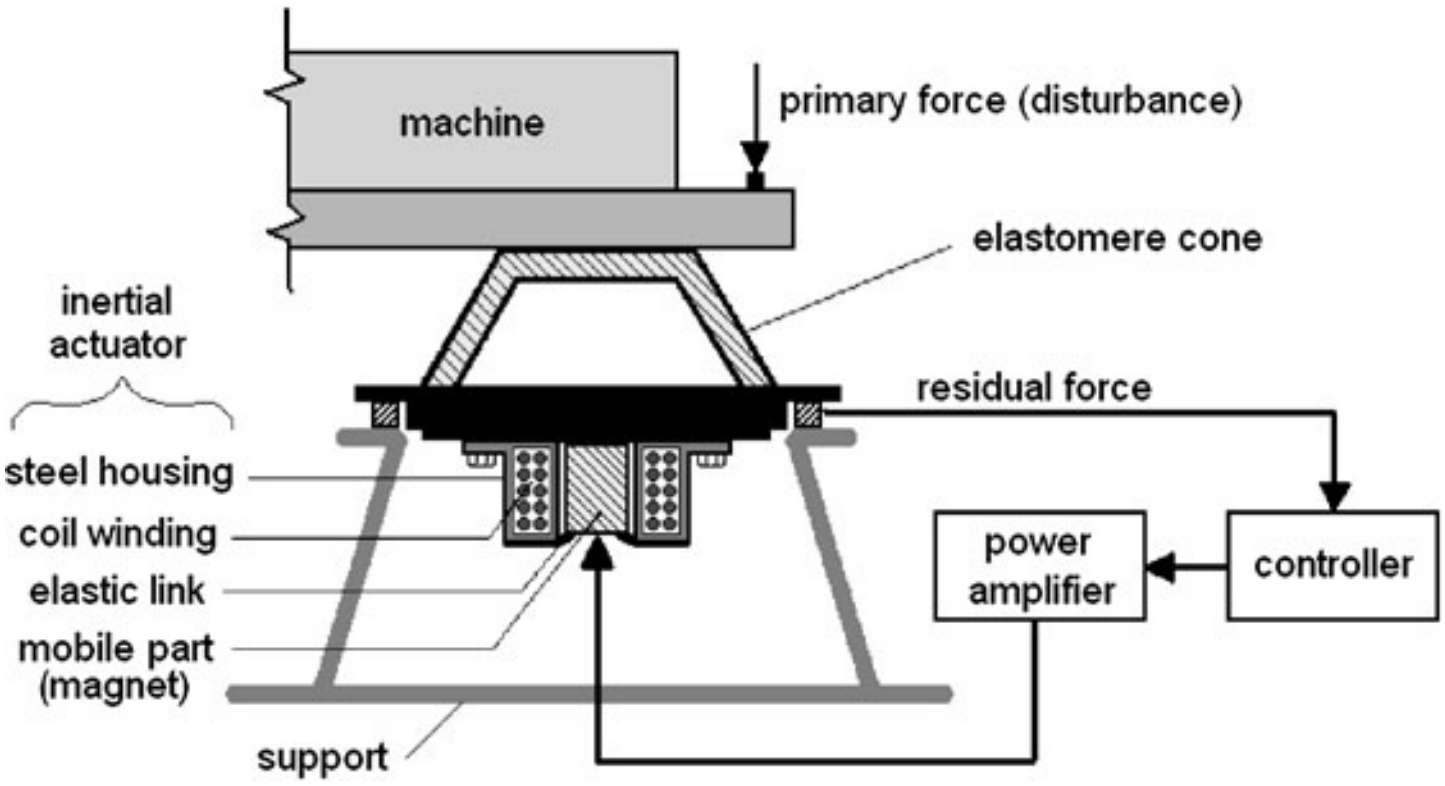

**Fig. 14.4** Active vibration control using an inertial actuator (scheme)

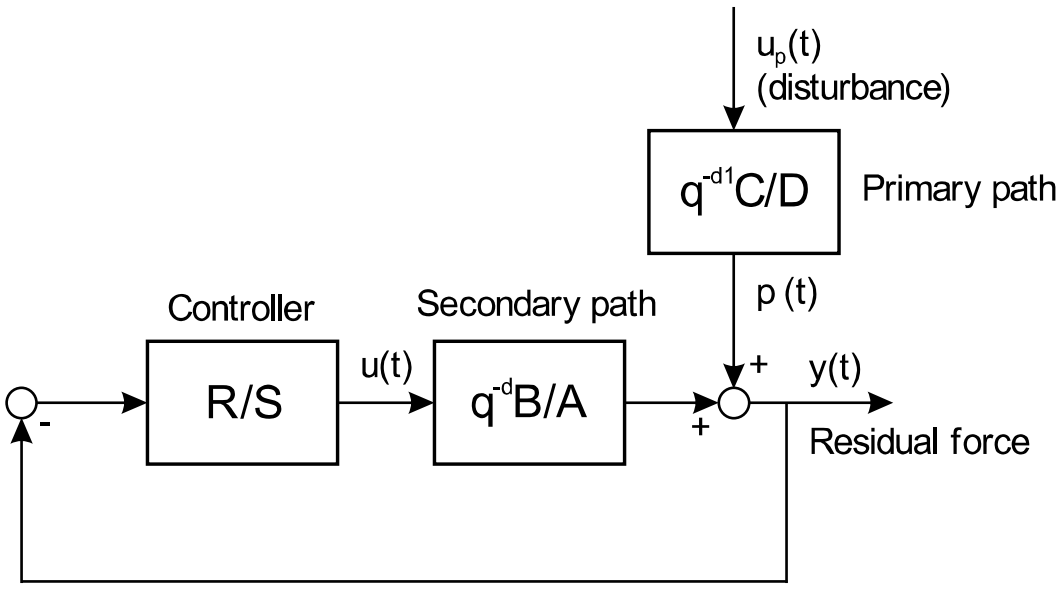

**Fig. 14.5** Block diagram of the active vibration control system

## Plant Identification and Central Controller Design

Procedures for system identification in open and closed loop as described in Chaps. 5 and 9 have been used. The frequency characteristics of the primary path (identification in open loop) and of the secondary path (identification in closed loop) are shown in Fig. 14.6. The secondary path has the following complexity: $n_B = 12$, $n_A = 10$, $d = 0$. The identification has been done using as excitation a PRBS (with frequency divider $p = 2$ and $N = 9$). There exist several low-damped vibration modes in the secondary path, the first vibration mode is at 51.58 Hz with a damping of 0.023 and the second at 100.27 Hz with a damping of 0.057.

The central controller (without the internal model) has been designed using pole placement with sensitivity shaping (Chap. 8 and Landau and Zito 2005).

## Direct Adaptive Regulation Under the Effect of Two Sinusoidal Disturbances

In this case one should take $n_{D_p} = 4$ and $n_Q = n_{D_p} - 1 = 3$. Only the "adaptive" operation regime has been considered for the subsequent experiments.

Figure 14.7 shows the spectral densities of the residual force obtained in open loop and in closed loop using the direct adaptation scheme (after the adaptation algorithm has converged). The results are given for the simultaneous applications of two sinusoidal disturbances (65 Hz and 95 Hz). One can remark a strong attenuation of the disturbances (larger than 45 dB). Note that there exists a significant measurement noise at 50 Hz (power network) which is not amplified in closed loop. Harmonics of the disturbances are present in open-loop operation but they disappear in closed-loop operation since the fundamental frequencies are strongly attenuated.

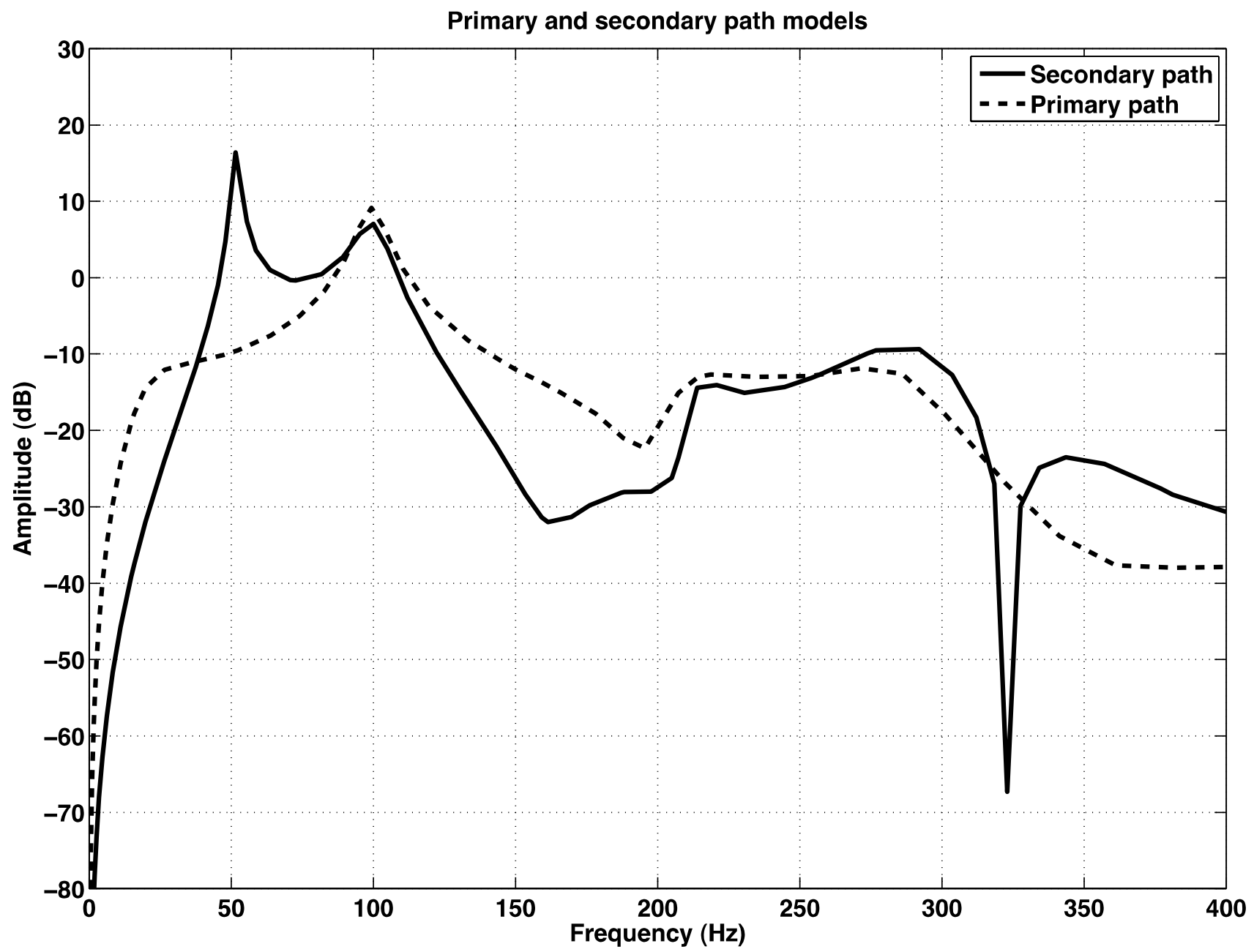

**Fig. 14.6** Frequency characteristics of the primary and secondary paths (inertial actuator)

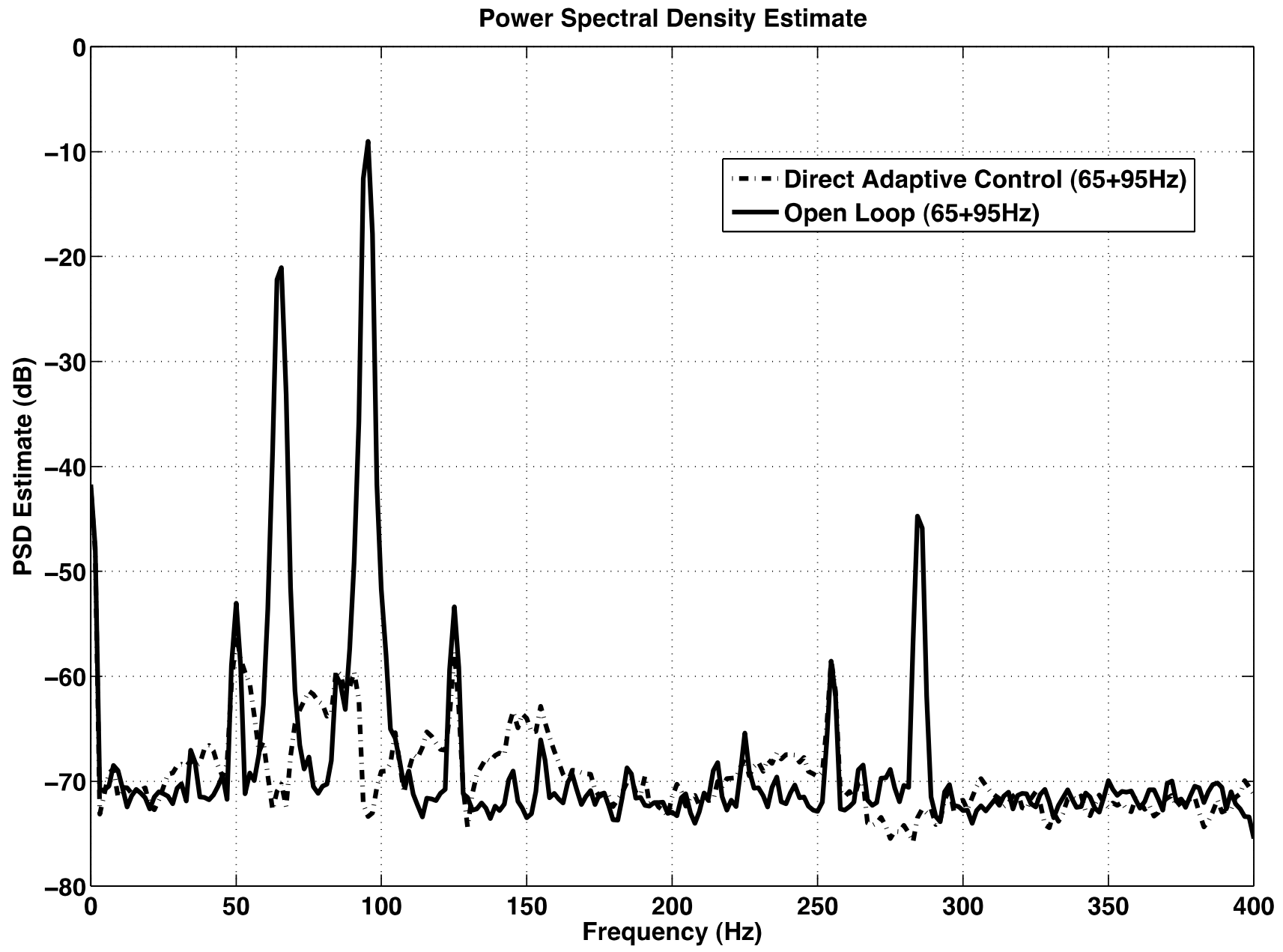

**Fig. 14.7** Spectral densities of the residual force in open and in closed loop, with the direct adaptation method in adaptive operation

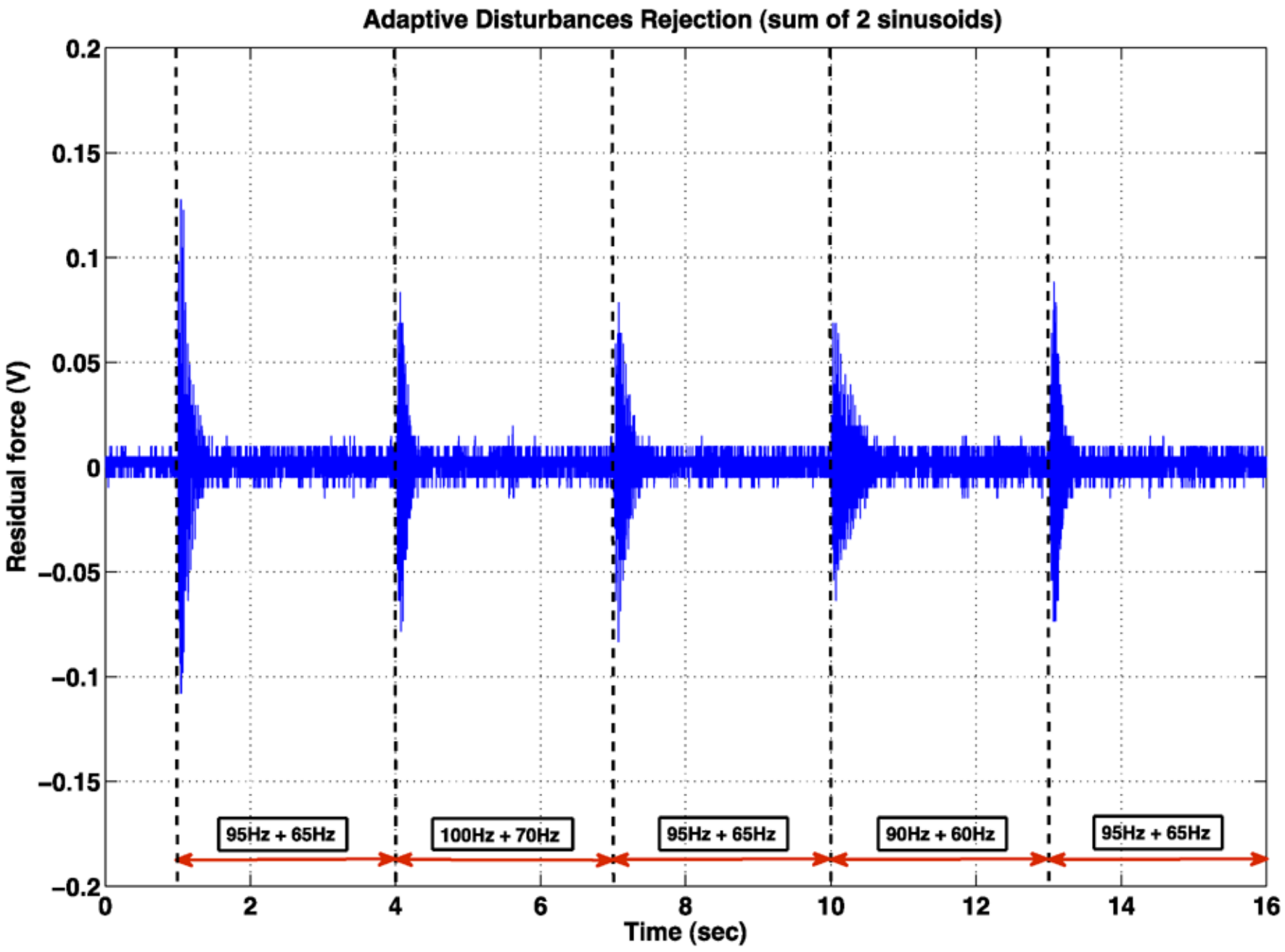

**Fig. 14.8** Time domain results with direct adaptation method for simultaneous step frequency changes of two sinusoidal disturbances

Time-domain results obtained with direct adaptation scheme in "adaptive" operation regime for step changes of the frequency of the disturbances are shown in Fig. 14.8. The disturbances are applied at 1 s (the loop has already been closed) and step changes of their frequencies occur every 3 s. Figure 14.9 shows the corresponding evolution of the parameters of the polynomial $Q$. The convergence of the output requires less than 0.4 s in the worst case.

#### Comparison Between Direct and Indirect Adaptive Regulation

In Fig. 14.10 the results for direct adaptive regulation for the case of a single sinusoidal disturbance in presence of step changes in frequency are shown. In Fig. 14.11 the results obtained in the same context with an indirect adaptive regulation scheme are shown. The performances for indirect and direct scheme are quite similar. Of course the complexity of the indirect adaptive regulation scheme is much higher. For other experimental results in adaptive regulation see Landau et al. (2005, 2010, 2011).

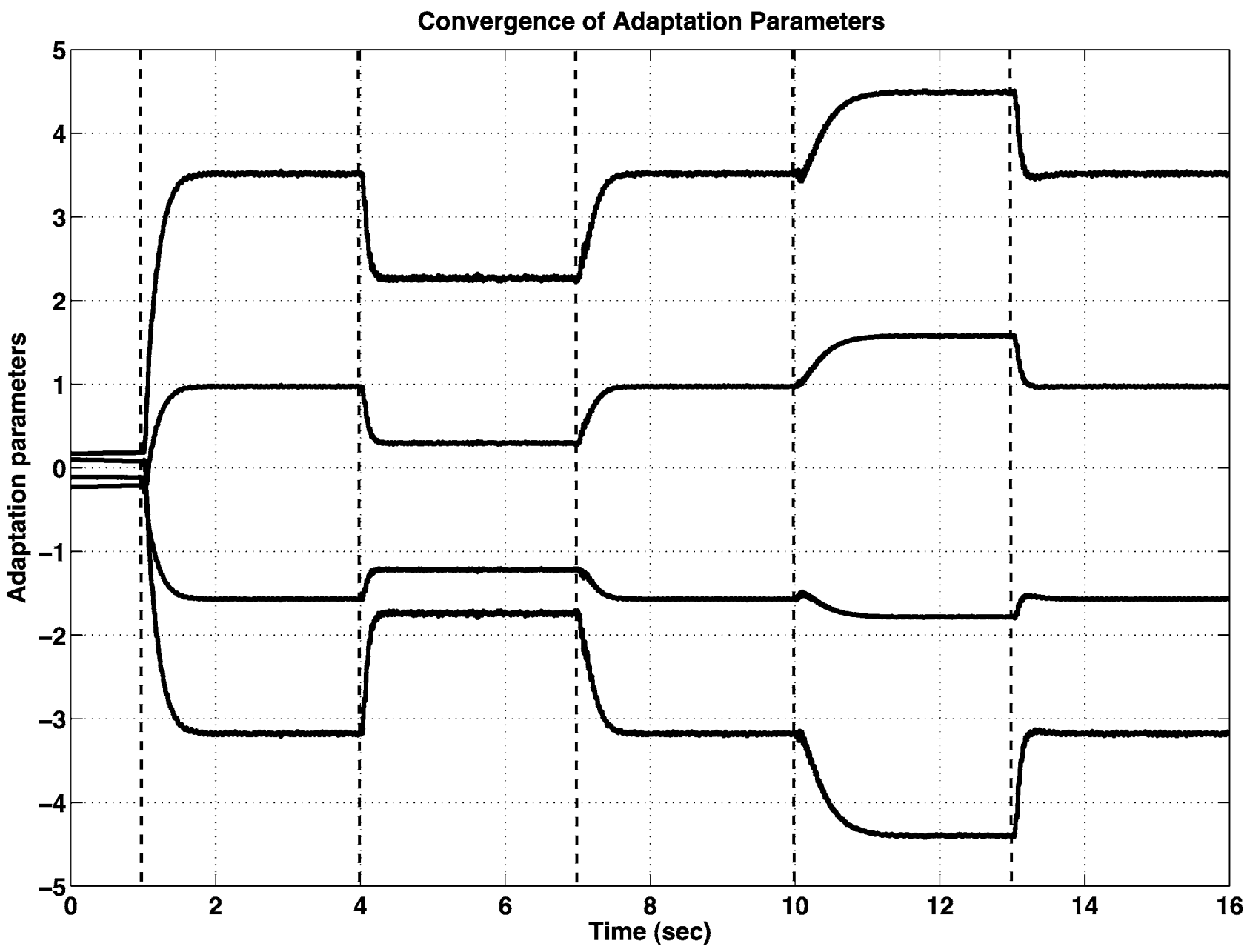

**Fig. 14.9** Evolution of the parameters of the polynomial $Q$ during adaptation

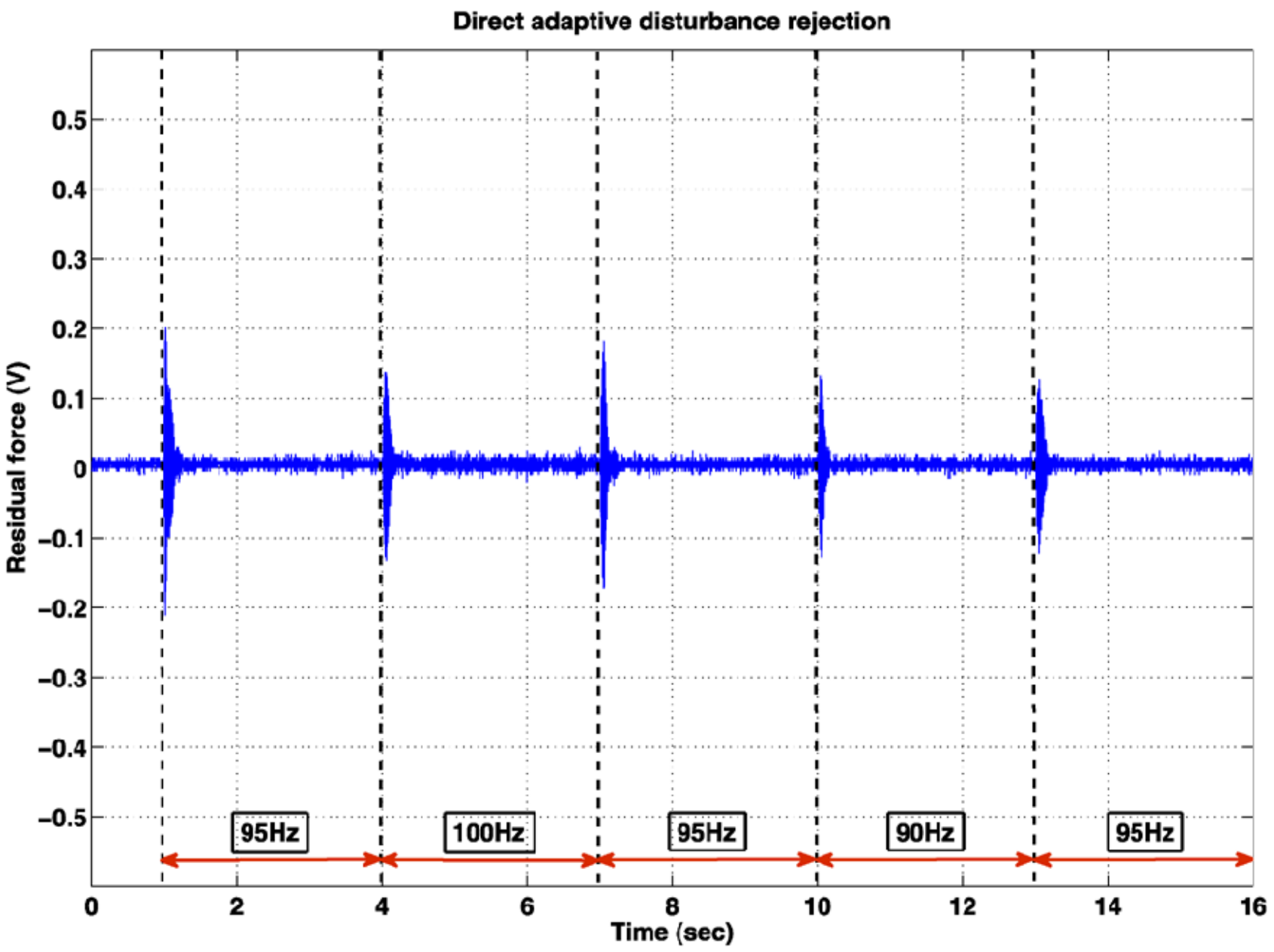

**Fig. 14.10** Time domain results with direct adaptation for a single sinusoidal disturbance with step changes in frequency (adaptive regime)

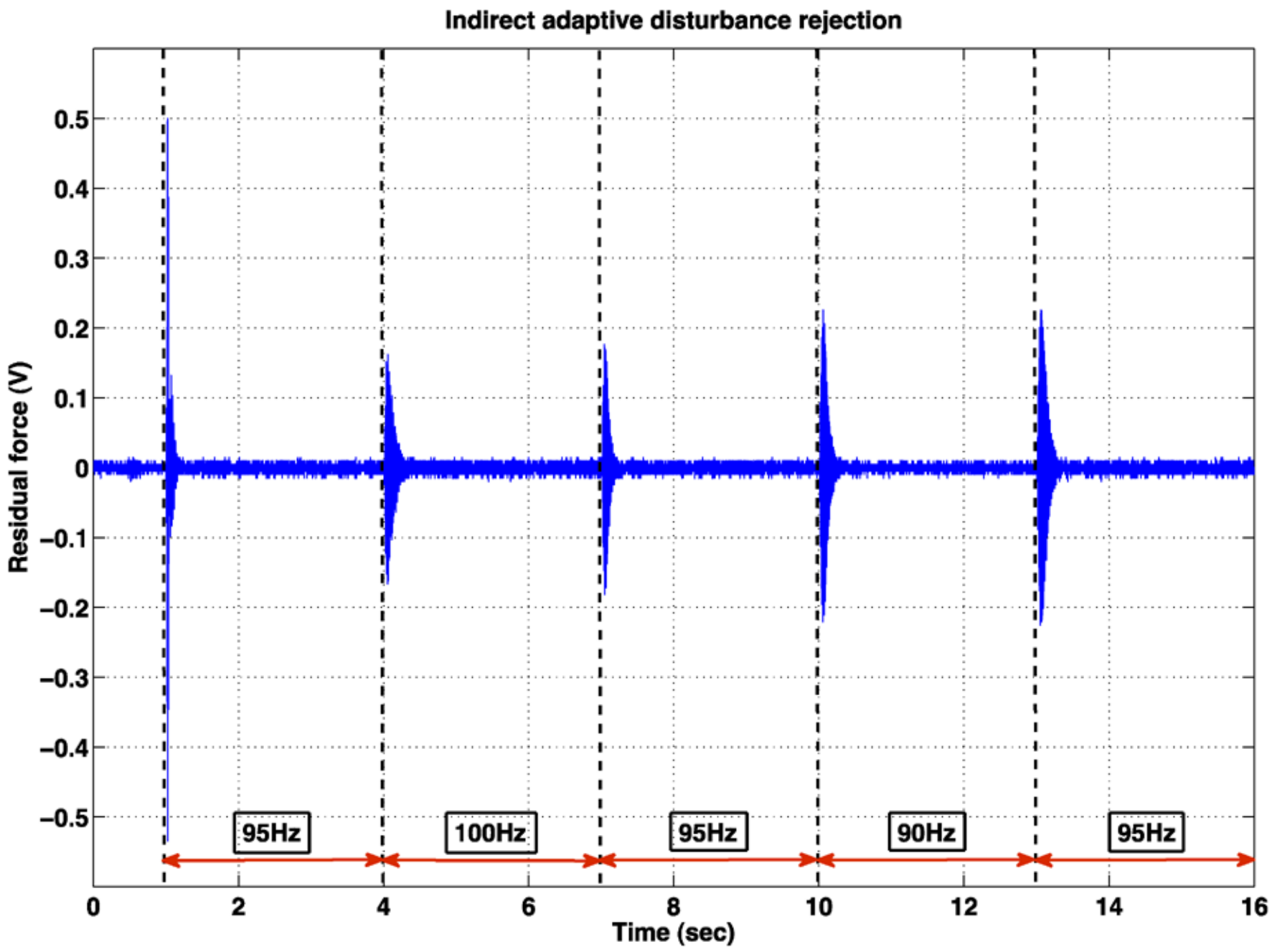

**Fig. 14.11** Time domain results with indirect adaptation for a single sinusoidal disturbance with step changes in frequency (adaptive regime)

## 14.8 Concluding Remarks

1. Adaptive regulation is an important practical issue in several classes of applications where one can assume that the plant model is almost invariant, while the model of the disturbance to be attenuated or suppressed is unknown and/or time-varying.
2. The use of the internal model principle allows us to provide solutions for suppressing the effect of the disturbances upon the output provided that, the model of the disturbance can be estimated.
3. Indirect and direct adaptive regulation schemes can be built.
4. The use of the Youla-Kucera controller parametrization allows us to built direct adaptive regulation schemes which are simpler that indirect adaptive regulation schemes.
5. Direct adaptive regulation schemes using the Youla-Kucera controller parametrization can be extended to the multivariable case (Ficocelli and Amara 2009).

## 14.9 Problems

**14.1** Work in detail the algorithm for direct adaptive regulation when preserving fixed parts in the controller (i.e., one uses (14.20) and (14.21) instead of (14.13) and (14.14)). Show that the algorithm given by (14.30) through (14.41) remains unchanged except for (14.30) which is replaced by (14.43).

**14.2** Develop the details for the computation of the control law in indirect adaptive control for the case of a single sinusoidal disturbance to be attenuated by 30 dB using either the internal model principle or the stop band filters described in Sect. 8.5.1.

# Chapter 15
# Adaptive Feedforward Compensation of Disturbances

## 15.1 Introduction

In a number of applications areas including active noise control (ANC) and active vibration control (AVC), an image (a correlated measurement) of the disturbances acting upon the system can be made available. This information is very useful for attenuating the disturbances using a feedforward compensation scheme. However, the feedforward compensator will depend not only upon the dynamics of the plant but also upon the characteristics of the disturbances. Since the characteristics (the model) of the disturbances are generally unknown and may be time varying, adaptive feedforward compensation was proposed many years ago. Probably one of the first references is Widrow and Stearns (1985).

The basic idea for adaptive rejection of unknown disturbances by feedforward compensation is that a "well located" transducer can provide a measurement, highly correlated with the unknown disturbance (a good image of the disturbance). This information is applied to the control input of the plant through an adaptive filter whose parameters are adapted such that the effect of the disturbance upon the output is minimized. This is illustrated in Fig. 15.1.

Adaptive feedforward broadband vibration (or noise) compensation is currently used in ANC and AVC when an image of the disturbance is available (Elliott and Nelson 1994; Elliott and Sutton 1996; Jacobson et al. 2001; Zeng and de Callafon 2006; Kuo and Morgan 1996, 1999). However, at the end of the nineties it was pointed out that in most of these systems there is a physical "positive" feedback coupling between the compensator system and the measurement of the image of the disturbance (vibration or noise) (Kuo and Morgan 1999; Hu and Linn 2000; Jacobson et al. 2001; Zeng and de Callafon 2006). This is clearly illustrated in Fig. 15.2 which represents an AVC system using a measurement of the image of the disturbance and an inertial actuator for reducing the residual acceleration. For more details about this system and a photo see Sect. 1.4.5 and Landau and Alma (2010).

The system consists of three mobile metallic plates (M1, M2, M3) connected by springs. The first and the third plates are also connected by springs to the rigid part of the system formed by two other metallic plates, themselves connected rigidly. The upper and lower mobile plates (M1 and M3) are equipped with inertial actuators. The one on the top serves as disturbance generator (see Fig. 15.2), the one at the bottom serves for disturbance compensation. The system is equipped with a measure of the residual acceleration (on plate M3) and a measure of the image of the disturbance made by an accelerometer posed on plate M1. The path between the disturbance (in this case, generated by the inertial actuator on top of the structure), and the residual acceleration is called the *global primary path*, the path between the measure of the image of the disturbance and the residual acceleration (in open loop) is called the *primary path* and the path between the inertial actuator and the residual acceleration is called the *secondary path*. When the compensator system is active, the actuator acts upon the residual acceleration, but also upon the measurement of the image of the disturbance (a positive feedback). The measured quantity $\hat{u}(t)$ will be the sum of the correlated disturbance measurement and of the effect of the actuator used for compensation.

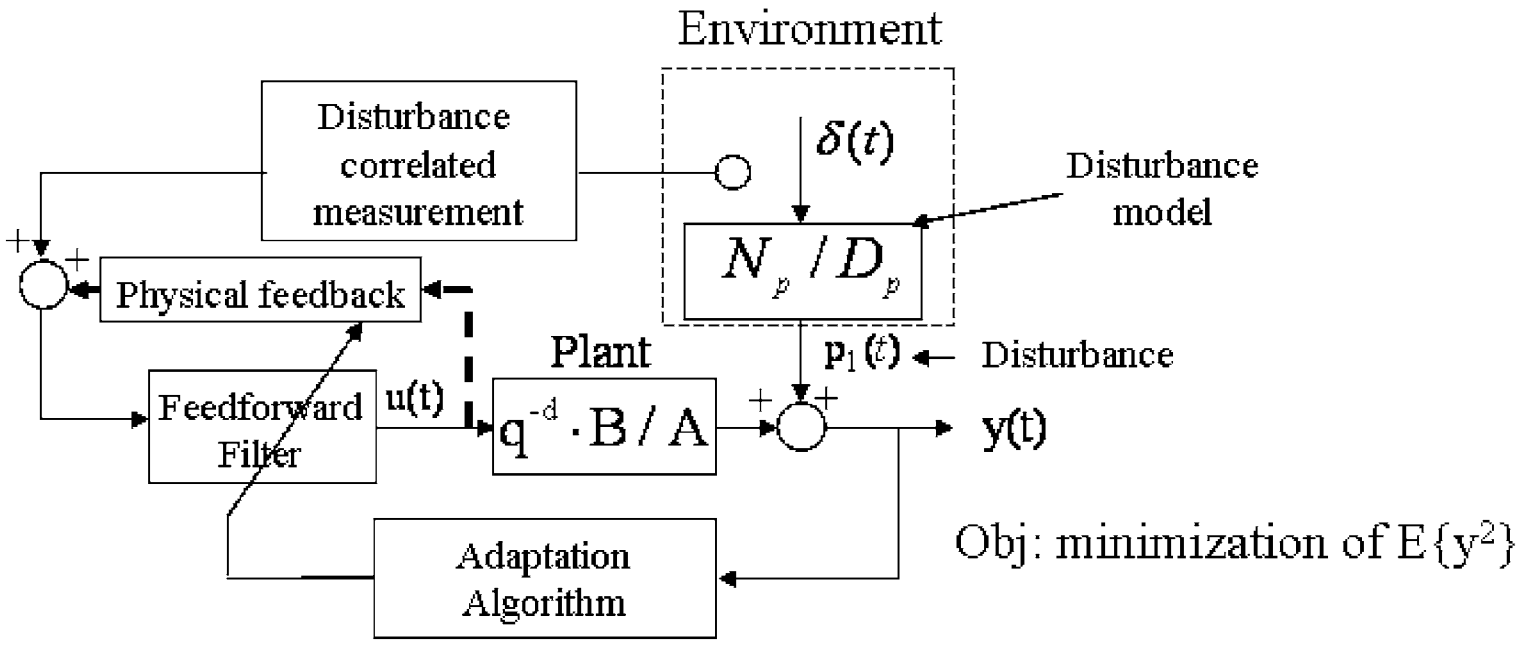

**Fig. 15.1** Adaptive feedforward compensation of unknown disturbances

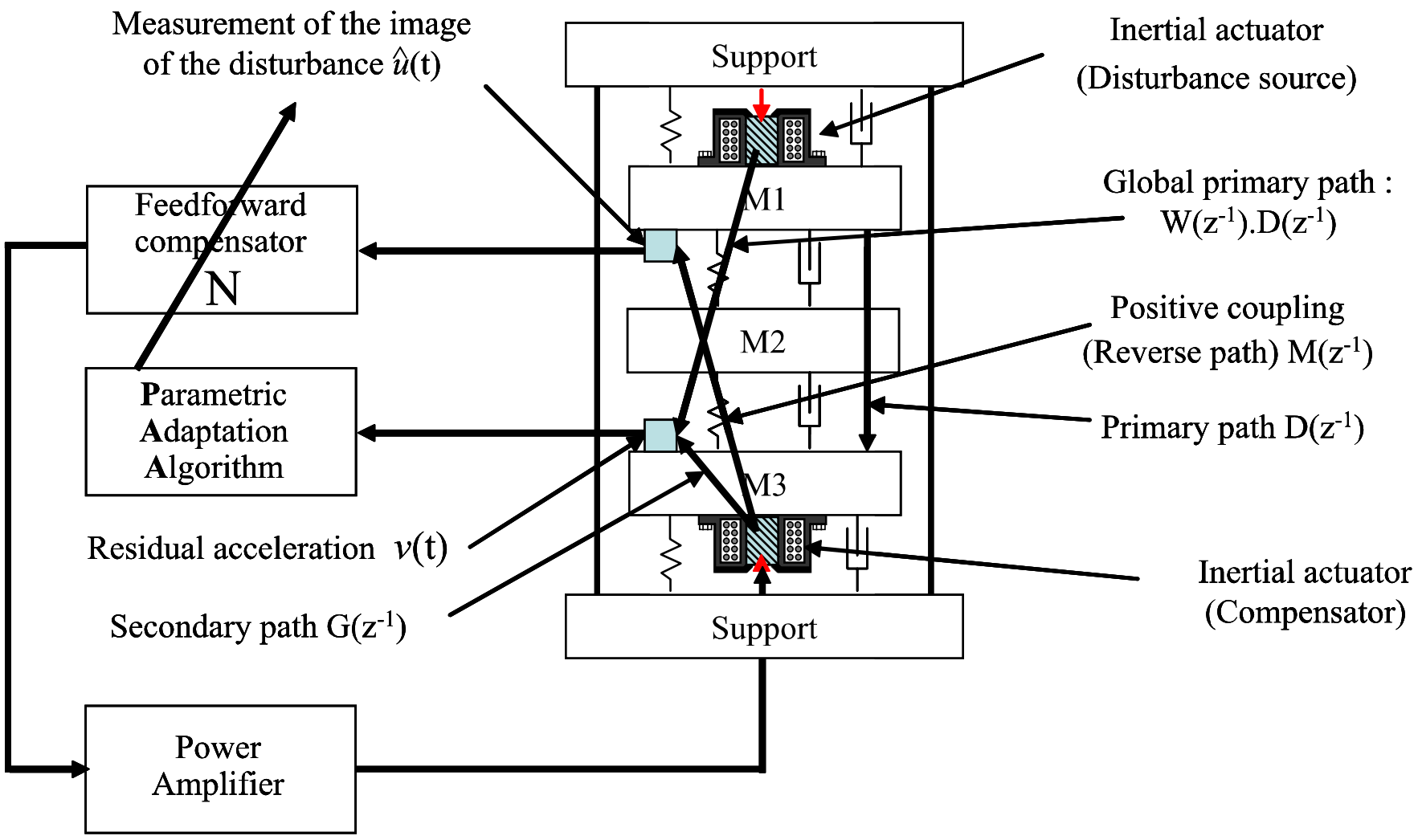

**Fig. 15.2** An AVC system using a feedforward compensation-scheme

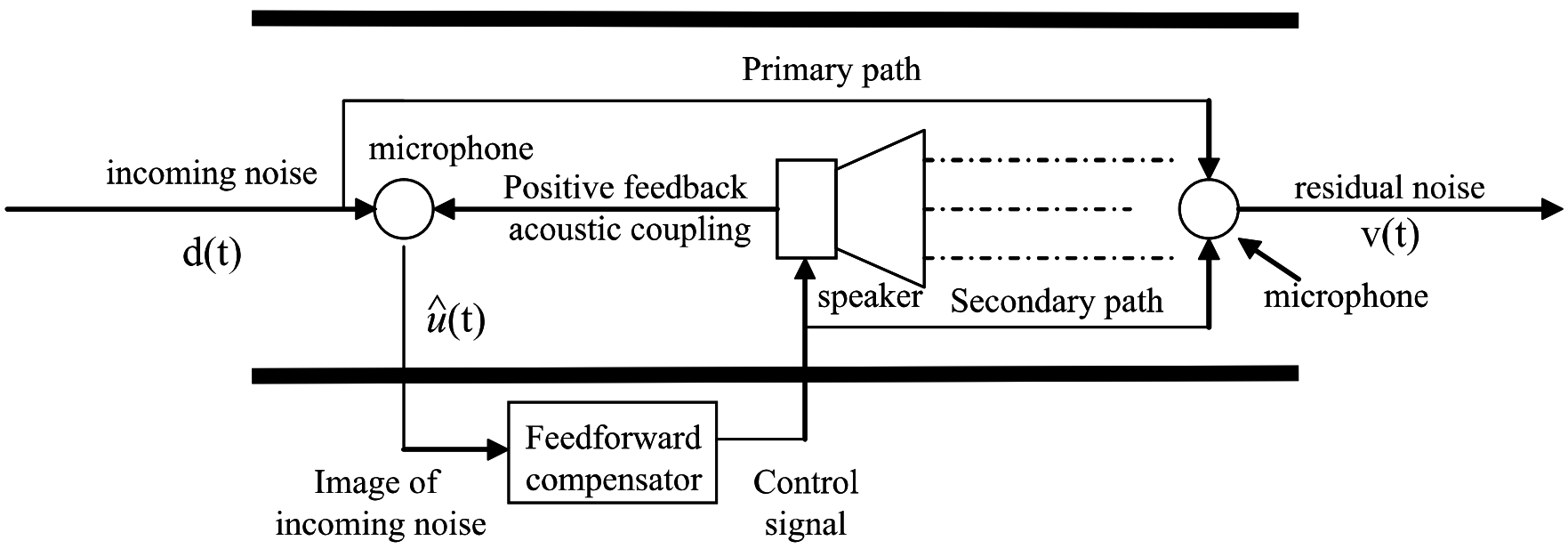

**Fig. 15.3** An ANC system using a feedforward compensation-scheme

Similar structures also occur in feedforward ANC (Jacobson et al. 2001; Zeng and de Callafon 2006). This is illustrated in Fig. 15.3 which represents an ANC system used to reduce noise in an airduct. It uses a loudspeaker as an actuator for noise reduction. However, since the loudspeaker will generate waves both down stream and upstream, this will influence the measurement of the image of the disturbance creating an acoustic positive feedback coupling.

The corresponding block diagrams in open-loop operation and with the compensator system are shown in Fig. 15.4. The signal $d(t)$ is the image of the disturbance measured when the compensator system is not used (open loop). The signal $\hat{u}(t)$ denotes the effective output provided by the measurement device when the compensator system is active and which will serve as input to the adaptive feedforward filter $\hat{N}$. The output of this filter denoted by $-\hat{y}(t)$ is applied to the actuator through an amplifier. The transfer function $G$ (the secondary path) characterizes the dynamics from the output of the filter $\hat{N}$ to the residual acceleration measurement (amplifier + actuator + dynamics of the mechanical system). Subsequently we will call the transfer function between $d(t)$ and the measurement of the residual acceleration the "primary path".

The coupling between the output of the feedforward filter compensator and the measurement $\hat{u}(t)$ through the compensator actuator is denoted by $M$. As indicated in Fig. 15.4, this coupling is a "positive" feedback. The positive feedback may destabilize the system.[1] The system is no longer a pure feedforward compensator. In many cases, this unwanted coupling raises problems in practice and makes the analysis of adaptive (estimation) algorithms more difficult. The problem is to estimate and adapt the parameters of the feedforward filter in the presence of this internal positive feedback.

It is important to make the following remarks, when the feedforward filter is absent (open-loop operation):

[1]Different solutions for reducing the effect of this internal positive feedback are reviewed in Kuo and Morgan (1996, 1999).

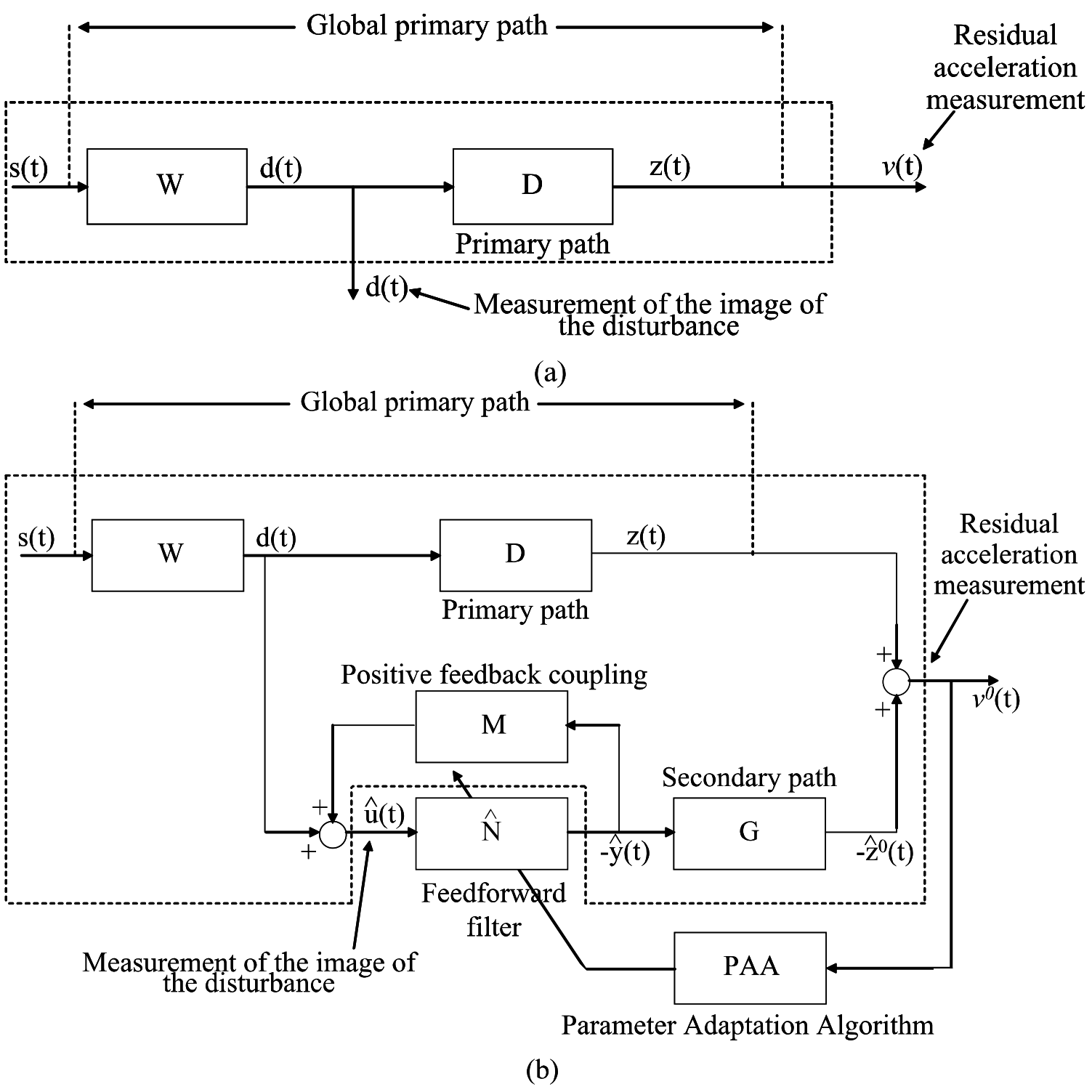

**Fig. 15.4** Feedforward AVC: in open loop (**a**) and with adaptive feedforward compensator (**b**)

1. One can identify very reliable models for the secondary path and the "positive" feedback path by applying appropriate excitation on the actuator (for example PRBS).
2. One can get an estimation of the primary path transfer function from the spectral densities of $d(t)$ and $\nu(t)$ (the actuator being at rest).

It is also important to note that the estimation of the feedforward filter in Fig. 15.4 can be interpreted as an identification-in-closed-loop operation (see Chap. 9 and Landau and Karimi 1997b) or as estimation in closed loop of a (reduced order) controller as in Landau et al. (2001b). Therefore, to a certain extent, the results given in Chap. 9 and Landau et al. (2001b) can be used to this problem.

The objective is to develop recursive algorithms for on-line estimation and adaptation of the parameters of the feedforward filter compensator $N$ (which will be denoted $\hat{N}$) for broadband disturbances $d(t)$ (or $s(t)$) with unknown and variable spectral characteristics and in the presence of possible variations of the primary path model ($D$). The resulting algorithms, while minimizing the residual error (acceleration or force in AVC, noise in ANC), should assure the stability of the internal positive feedback loop created by the mechanical or acoustical coupling. Like for *adaptive regulation* (see Chap. 14) the *adaptive* operation and the *self-tuning* operation of the system should be considered.

In Sect. 15.2 system representation and feedforward filter structure will be given. The algorithms for adaptive feedforward compensation will be developed in Sect. 15.3 and analyzed in Sect. 15.4. Section 15.5 will present simulation results and real-time results obtained on an AVC system. The results of this chapter are also applicable to ANC systems.

## 15.2 Basic Equations and Notations

The objective is to estimate (and to adapt) the parameters of the feedforward filter $N(q^{-1})$ such that the measured residual error (acceleration or force in AVC, noise in ANC) be minimized in the sense of a certain criterion. The description of the various blocks will be made with respect to Fig. 15.4.

The primary path is characterized by the asymptotically stable transfer operator:[2]

$$D(q^{-1}) = \frac{B_D(q^{-1})}{A_D(q^{-1})} \tag{15.1}$$

where

$$B_D(q^{-1}) = b_1^D q^{-1} + \cdots + b_{n_{B_D}}^D q^{-n_{B_D}} \tag{15.2}$$

$$A_D(q^{-1}) = 1 + a_1^D q^{-1} + \cdots + a_{n_{A_D}}^D q^{-n_{A_D}} \tag{15.3}$$

The unmeasurable value of the output of the primary path (when the compensation is active) is denoted $z(t)$. The secondary path is characterized by the asymptotically stable transfer operator:

$$G(q^{-1}) = \frac{B_G(q^{-1})}{A_G(q^{-1})} \tag{15.4}$$

where:

$$B_G(q^{-1}) = b_1^G q^{-1} + \cdots + b_{n_{B_G}}^G q^{-n_{B_G}} = q^{-1} B_G^*(q^{-1}) \tag{15.5}$$

$$A_G(q^{-1}) = 1 + a_1^G q^{-1} + \cdots + a_{n_{A_G}}^G q^{-n_{A_G}} \tag{15.6}$$

The positive feedback coupling is characterized by the asymptotically stable transfer operator:

$$M(q^{-1}) = \frac{B_M(q^{-1})}{A_M(q^{-1})} \tag{15.7}$$

[2] The complex variable $z^{-1}$ will be used for characterizing the system's behavior in the frequency domain and the delay operator $q^{-1}$ will be used for describing the system's behavior in the time domain.

where:

$$B_M(q^{-1}) = b_1^M q^{-1} + \cdots + b_{n_{B_M}}^M q^{-n_{B_M}} = q^{-1} B_M^*(q^{-1}) \tag{15.8}$$

$$A_M(q^{-1}) = 1 + a_1^M q^{-1} + \cdots + a_{n_{A_M}}^M q^{-n_{A_M}} \tag{15.9}$$

Both $B_G$ and $B_M$ have a one step discretization delay. The identified models of the secondary path and of the positive feedback coupling will be denoted $\hat{G}$ and $\hat{M}$, respectively.

The optimal feedforward filter (unknown) is defined by:

$$N(q^{-1}) = \frac{R(q^{-1})}{S(q^{-1})} \tag{15.10}$$

where:

$$R(q^{-1}) = r_0 + r_1 q^{-1} + \cdots + r_{n_R} q^{-n_R} \tag{15.11}$$

$$S(q^{-1}) = 1 + S_1 q^{-1} + \cdots + S_{n_S} q^{-n_S} = 1 + q^{-1} S^*(q^{-1}) \tag{15.12}$$

The estimated filter is denoted by $\hat{N}(q^{-1})$ or $\hat{N}(\hat{\theta}, q^{-1})$ when it is a linear filter with constant coefficients or $\hat{N}(t, q^{-1})$ during estimation (adaptation) of its parameters.

The input of the feedforward filter is denoted by $\hat{u}(t)$ and it corresponds to the measurement provided by the primary transducer (force or acceleration transducer in AVC or a microphone in ANC). In the absence of the compensation loop (open-loop operation) $\hat{u}(t) = d(t)$. The "a posteriori" output of the feedforward filter (which is the control signal applied to the secondary path) is denoted by $-\hat{y}(t+1) = -\hat{y}(t+1|\hat{\theta}(t+1))$. The input-output relationships for the estimated feedforward filter are given by the "a priori" output:

$$\begin{aligned} -\hat{y}^0(t+1) &= -\hat{y}(t+1|\hat{\theta}(t)) = -[-\hat{S}^*(t, q^{-1})\hat{y}(t) + \hat{R}(t, q^{-1})\hat{u}(t+1)] \\ &= -\hat{\theta}^T(t)\phi(t) = -[\hat{\theta}_S^T(t), \hat{\theta}_R^T(t)] \begin{bmatrix} \phi_{\hat{y}}(t) \\ \phi_{\hat{u}}(t) \end{bmatrix} \end{aligned} \tag{15.13}$$

where

$$\hat{\theta}^T(t) = [\hat{s}_1(t), \ldots, \hat{s}_{n_S}(t), \hat{r}_0(t), \ldots, \hat{r}_{n_R}(t)] = [\hat{\theta}_S^T(t), \hat{\theta}_R^T(t)] \tag{15.14}$$

$$\begin{aligned} \phi^T(t) &= [-\hat{y}(t), \ldots, -\hat{y}(t - n_S + 1, \hat{u}(t+1), \hat{u}(t), \ldots, \hat{u}(t - n_R + 1)] \\ &= [\phi_{\hat{y}}^T(t), \phi_{\hat{u}}^T(t)] \end{aligned} \tag{15.15}$$

and $-\hat{y}(t)$, $-\hat{y}(t-1), \ldots$ are the "a posteriori" outputs of the feedforward filter generated by:

$$-\hat{y}(t+1) = -\hat{y}(t+1|\hat{\theta}(t+1)) = -\hat{\theta}^T(t+1)\phi(t) \tag{15.16}$$

while $\hat{u}(t+1), \hat{u}(t), \ldots$ are the measurements provided by the primary transducer.[3] The "a priori" output of the secondary path will be denoted $-\hat{z}^0(t+1)$

$$-\hat{z}^0(t+1) = -\hat{z}(t+1|\hat{\theta}(t)) = \frac{B_G^*(q^{-1})}{A_G(q^{-1})}[-\hat{y}(t)] \tag{15.17}$$

[3] $\hat{u}(t+1)$ is available before adaptation of parameters starts at $t+1$.

The "a posteriori" unmeasurable value of the output of the secondary path is denoted by:

$$-\hat{z}(t+1) = -\hat{z}(t+1|\hat{\theta}(t+1)) \tag{15.18}$$

The measured primary signal (called also reference) satisfies the following equation:

$$\hat{u}(t+1) = d(t+1) + \frac{B_M^*(q^{-1})}{A_M(q^{-1})}\hat{y}(t) \tag{15.19}$$

The measured residual error satisfies the following equation:

$$\nu^0(t) = z(t) - \hat{z}^0(t) \tag{15.20}$$

It will be termed the "a priori" adaptation error. The "a posteriori" adaptation (residual) error (which is computed) will be given by:

$$\nu(t) = z(t) - \hat{z}(t) \tag{15.21}$$

When using an estimated filter $\hat{N}$ with constant parameters: $\hat{y}^0(t) = \hat{y}(t), \hat{z}^0(t) = \hat{z}(t)$ and $\nu^0(t) = \nu(t)$.

## 15.3 Development of the Algorithms

The algorithms for adaptive feedforward compensation will be developed under the following hypotheses:

(H1) The signal $d(t)$ is bounded, i.e.,

$$|d(t)| \le \alpha; \quad \forall t \ (0 \le \alpha \le \infty) \tag{15.22}$$

or

$$\lim_{N\to\infty} \sum_{t=1}^{N} d^2(t) \le N\varepsilon^2 + K_r$$
$$0 \le \varepsilon^2 < \infty, \ 0 < K_r < \infty \tag{15.23}$$

(which is equivalently to say that $s(t)$ is bounded and $W(q^{-1})$ in Fig. 15.4 is asymptotically stable).

(H2) Perfect matching condition. There exists a filter $N(q^{-1})$ of finite dimension such that:[4]

$$\frac{N}{(1-NM)}G = -D \tag{15.24}$$

and the characteristic polynomial of the "internal" feedback loop:

$$P(z^{-1}) = A_M(z^{-1})S(z^{-1}) - B_M(z^{-1})R(z^{-1}) \tag{15.25}$$

is a Hurwitz polynomial.

[4]In many cases, the argument $q^{-1}$ or $z^{-1}$ will be dropped out.

(H3) The effect of the measurement noise upon the measured residual error is neglected (deterministic context).

Once the algorithms will be developed under these hypotheses, hypotheses (H2) and (H3) will be removed and the algorithms will be analyzed in this modified context.

A first step in the development of the algorithms is to establish a relation between the errors on the estimation of the parameters of the feedforward filter and the measured residual acceleration. This is summarized in the following lemma.

**Lemma 15.1** *Under hypotheses* (H1), (H2) *and* (H3), *for the system described by* (15.1) *through* (15.21) *using a feedforward compensator* $\hat{N}$ *with constant parameters*, *one has*:

$$\nu(t+1) = \frac{A_M(q^{-1})G(q(^{-1})}{P(q^{-1})}[\theta - \hat{\theta}]^T \phi(t) \tag{15.26}$$

*where*

$$\theta^T = [s_1, \ldots, s_{n_S}, r_0, r_1, \ldots, r_{n_R}] = [\theta_S^T, \theta_R^T] \tag{15.27}$$

*is the vector of parameters of the optimal filter* $N$ *assuring perfect matching*,

$$\hat{\theta}^T = [\hat{s}_1, \ldots, \hat{s}_{n_S}, \hat{r}_0, \ldots, \hat{r}_{n_R}] = [\hat{\theta}_S^T, \hat{\theta}_R^T] \tag{15.28}$$

*is the vector of constant estimated parameters of* $\hat{N}$,

$$\begin{aligned}\phi^T(t) &= [-\hat{y}(t), \ldots, -\hat{y}(t-n_S+1, \hat{u}(t+1), \hat{u}(t), \ldots, \hat{u}(t-n_R+1)] \\ &= [\phi_{\hat{y}}^T(t), \phi_{\hat{u}}^T(t)]\end{aligned} \tag{15.29}$$

*and* $\hat{u}(t+1)$ *is given by*

$$\hat{u}(t+1) = d(t+1) + \frac{B_M^*(q^{-1})}{A_M(q^{-1})}\hat{y}(t) \tag{15.30}$$

*Proof* Under Assumption (H2) (perfect matching condition) the output of the primary path can be expressed as:

$$z(t) = G(q^{-1})y(t) \tag{15.31}$$

where $y(t)$ is a dummy variable given by:

$$\begin{aligned}y(t+1) &= -S^*(q^{-1})y(t) + R(q^{-1})u(t+1) \\ &= \theta^T \varphi(t) = [\theta_S^T, \theta_R^T]\begin{bmatrix}\varphi_y(t) \\ \varphi_u(t)\end{bmatrix}\end{aligned} \tag{15.32}$$

where

$$\begin{aligned}\varphi^T(t) &= [-y(t), \ldots, -y(t-n_S+1), u(t+1), \ldots, u(t-n_R+1)] \\ &= [\varphi_y^T(t), \varphi_u^T(t)]\end{aligned} \tag{15.33}$$

and $u(t)$ is given by

$$u(t+1) = d(t+1) + \frac{B_M^*(q^{-1})}{A_M(q^{-1})} y(t) \tag{15.34}$$

For a fixed value of the parameter vector $\hat{\theta}$ characterizing the estimated filter $\hat{N}(q^{-1})$ of same dimension as the optimal filter $N(q^{-1})$, the output of the secondary path can be expressed by (in this case $\hat{z}(t) = \hat{z}^0(t)$ and $\hat{y}(t) = \hat{y}^0(t)$):

$$-\hat{z}(t) = G(q^{-1})[-\hat{y}(t)] \tag{15.35}$$

where

$$-\hat{y}(t+1) = -\hat{\theta}^T \phi(t) \tag{15.36}$$

The key observation is that the dummy variable $y(t+1)$ can be expressed as:

$$\begin{aligned} y(t+1) &= \theta^T \phi(t) + \theta^T [\varphi(t) - \phi(t)] \\ &= \theta^T \phi(t) + \theta_S^T [\varphi_y - \phi_{\hat{y}}] + \theta_R^T [\varphi_u - \phi_{\hat{u}}] \end{aligned} \tag{15.37}$$

Define the dummy error (for a fixed vector $\hat{\theta}$)

$$\varepsilon(t+1) = y(t+1) - \hat{y}(t+1) \tag{15.38}$$

and the residual error

$$\nu(t+1) = z(t) - \hat{z}(t) = G(q^{-1})\varepsilon(t+1) \tag{15.39}$$

It results from (15.37) that:

$$y(t+1) = \theta^T \phi(t) - S^*(q^{-1})\varepsilon(t) + R(q^{-1})[u(t+1) - \hat{u}(t+1)] \tag{15.40}$$

But taking into account the expressions of $u(t)$ and $\hat{u}(t)$ given by (15.34) and (15.19) (or (15.30)) respectively one gets:

$$y(t+1) = \theta^T \phi(t) - (S^*(q^{-1}) - \frac{R(q^{-1})B_M^*(q^{-1})}{A_M(q^{-1})})\varepsilon(t) \tag{15.41}$$

and therefore:

$$\varepsilon(t+1) = [\theta - \hat{\theta}]^T \phi(t) - \left( S^*(q^{-1}) - \frac{R(q^{-1})B_M^*(q^{-1})}{A_M(q^{-1})} \right) \varepsilon(t) \tag{15.42}$$

This gives:

$$\frac{A_M S - R B_M}{A_M} \varepsilon(t+1) = [\theta - \hat{\theta}]^T \phi(t) \tag{15.43}$$

which can be rewritten as:

$$\varepsilon(t+1) = \frac{A_M(q^{-1})}{P(q^{-1})} [\theta - \hat{\theta}]^T \phi(t) \tag{15.44}$$

Taking now into account (15.39) one gets (15.26). □

Filtering the vector $\phi(t)$ through an asymptotically stable filter $L(q^{-1}) = \frac{B_L}{A_L}$, (15.26) for $\hat{\theta} =$ constant becomes:

$$\nu(t+1) = \frac{A_M(q^{-1})G(q^{-1})}{P(q^{-1})L(q^{-1})}[\theta - \hat{\theta}]^T \phi_f(t) \tag{15.45}$$

with:

$$\phi_f(t) = L(q^{-1})\phi(t) \tag{15.46}$$

Equation (15.45) will be used to develop the adaptation algorithms neglecting for the moment the non-commutativity of the operators when $\hat{\theta}$ is time varying (however, an exact algorithm can be derived in such cases—see Sect. 5.5.3).

Replacing the fixed estimated parameters by the current estimated parameters, (15.45) becomes the equation or the a posteriori residual error $\nu(t+1)$ (which is computed):

$$\nu(t+1) = \frac{A_M(q^{-1})G(q^{-1})}{P(q^{-1})L(q^{-1})}[\theta - \hat{\theta}(t+1)]^T \phi_f(t) \tag{15.47}$$

Equation (15.47) has the standard form for an a posteriori adaptation error given in Chap. 3, which immediately suggests to use the following parameter adaptation algorithm:

$$\hat{\theta}(t+1) = \hat{\theta}(t) + F(t)\Phi(t)\nu(t+1) \tag{15.48}$$

$$\nu(t+1) = \frac{\nu^0(t+1)}{1 + \Phi^T(t)F(t)\Phi(t)} \tag{15.49}$$

$$F(t+1) = \frac{1}{\lambda_1(t)}\left[F(t) - \frac{F(t)\Phi(t)\Phi^T(t)F(t)}{\frac{\lambda_1(t)}{\lambda_2(t)} + \Phi^T(t)F(t)\Phi(t)}\right] \tag{15.50}$$

$$1 \geq \lambda_1(t) > 0;\ \ 0 \leq \lambda_2(t) < 2;\ \ F(0) > 0 \tag{15.51}$$

$$\Phi(t) = \phi_f(t) \tag{15.52}$$

where $\lambda_1(t)$ and $\lambda_2(t)$ allow to obtain various profiles for the adaptation gain $F(t)$ (see Sect. 3.2.3 and Sect. 15.5) in order to operate in *adaptive* regime or *self-tuning* regime.

Three choices for the filter $L$ will be considered, leading to three different algorithms:

Algorithm I: $L = G$.
Algorithm II: $L = \hat{G}$.
Algorithm III:

$$L = \frac{\hat{A}_M}{\hat{P}}\hat{G} \tag{15.53}$$

where:

$$\hat{P} = \hat{A}_M\hat{S} - \hat{B}_M\hat{R} \tag{15.54}$$

is an estimation of the characteristic polynomial of the internal feedback loop computed on the basis of available estimates of the parameters of the filter $\hat{N}$.

Taking into account (15.42) and (15.39) the "a posteriori" adaptation error for Algorithm I can be expressed also as follows (neglecting the non commutativity of time-varying operators):

$$\nu(t+1) = [\theta - \hat{\theta}(t+1)]^T \phi_f(t) - \left( S^*(q^{-1}) - \frac{R(q^{-1})B_M^*(q^{-1})}{A_M(q^{-1})} \right) \nu(t) \quad (15.55)$$

Defining the measured quantity of the residual error at instant $t+1$ (see (15.20)), as the a priori adaptation error one has:

$$\nu^0(t+1) = [\theta - \hat{\theta}(t)]^T \phi_f(t) - \left( S^*(q^{-1}) - \frac{R(q^{-1})B_M^*(q^{-1})}{A_M(q^{-1})} \right) \nu(t) \quad (15.56)$$

Relationship between $\nu^0(t+1)$ and $\nu(t+1)$ given by (15.49) clearly results from (15.56) and (15.55).

Using the methodology presented in Sect. 3.3.4, (3.231) through (3.241), same relationship between a priori and a posteriori adaptation error is obtained for Algorithms II and III.

The following procedure is applied at each sampling time for *adaptive* operation:

1. Get the measured image of the disturbance $\hat{u}(t+1)$ and the measured residual error $\nu^0(t+1)$.
2. Compute $\phi(t)$ and $\phi_f(t)$ using (15.29) and (15.46).
3. Estimate the parameter vector $\hat{\theta}(t+1)$ using the parametric adaptation algorithm (15.48) through (15.52).
4. Compute the control (using (15.16)) and apply it.

## 15.4 Analysis of the Algorithms

### *15.4.1 The Deterministic Case—Perfect Matching*

For Algorithms I, II and III the equation for the a posteriori adaptation error has the form:

$$\nu(t+1) = H(q^{-1})[\theta - \hat{\theta}(t+1)]^T \Phi(t) \quad (15.57)$$

where:

$$H(q^{-1}) = \frac{A_M(q^{-1})G(q^{-1})}{P(q^{-1})L(q^{-1})}, \qquad \Phi = \phi_f \quad (15.58)$$

Neglecting the non-commutativity of time-varying operators, one can straightforwardly use Theorem 3.2, Chap. 3, for the analysis of the algorithms. One has the following result:

**Lemma 15.2** *Assuming that* (15.57) *represents the evolution of the a posteriori adaptation error and that the parameter adaption algorithm* (15.48) *through* (15.52) *is used*, *one has*:

$$\lim_{t\to\infty} \nu(t+1) = 0 \tag{15.59}$$

$$\lim_{t\to\infty} \frac{[\nu^0(t+1)^2]}{1+\Phi(t)^T F(t)\Phi(t)} = 0 \tag{15.60}$$

$$\|\Phi(t)\| \text{ is bounded} \tag{15.61}$$

$$\lim_{t\to\infty} \nu^0(t+1) = 0 \tag{15.62}$$

*for any initial conditions* $\hat{\theta}(0), \nu^0(0), F(0)$, *provided that*:

$$H'(z^{-1}) = H(z^{-1}) - \frac{\lambda_2}{2}, \quad \max_t [\lambda_2(t)] \le \lambda_2 < 2 \tag{15.63}$$

*is a strictly positive real transfer function.*

*Proof* Using Theorem 3.2, under the condition (15.63), (15.59) and (15.60) hold. However, in order to show that $\nu^0(t+1)$ goes to zero one has to show first that the components of the observation vector are bounded. The result (15.60) suggests to use the Goodwin's "bounded growth" lemma (Lemma 11.1). Provided that one has:

$$|\Phi^T(t)F(t)\Phi(t)|^{\frac{1}{2}} \le C_1 + C_2 \max_{0\le k\le t+1} |\nu^0(k)| \tag{15.64}$$

$$0 < C_1 < \infty,\ 0 < C_2 < \infty,\ F(t) > 0 \tag{15.65}$$

$\|\Phi(t)\|$ will be bounded. So it will be shown that (15.64) holds for Algorithm I (for Algorithms II and III the proof is similar). From (15.39) one has:

$$-\hat{z}(t) = \nu(t) - z(t) \tag{15.66}$$

Since $z(t)$ is bounded (output of an asymptotically stable system with bounded input), one has:

$$\begin{aligned} |-\hat{y}_f(t)| = |-G\hat{y}(t)| = |-\hat{z}(t)| &\le C_3 + C_4 \max_{0\le k\le t+1} |\nu(k)| \\ &\le C_3' + C_4' \max_{0\le k\le t+1} |\nu^0(k)| \end{aligned} \tag{15.67}$$

$$0 < C_3, C_4, C_3', C'4 < \infty \tag{15.68}$$

since $|\nu(t)| \le |\nu^0(t)|$ for all $t$. Filtering both sides of (15.19) by $G(q^{-1})$ one gets in the adaptive case:

$$\hat{u}_f(t) = \frac{B_G}{A_G} d(t) + \frac{B_M}{A_M} \hat{y}_f(t) \tag{15.69}$$

Since $A_G$ and $A_M$ are Hurwitz polynomials and that $d(t)$ is bounded, it results that:

$$|\hat{u}_f(t)| \le C_5 + C_6 \max_{0\le k\le t+1} |\nu^0(k)|; \quad 0 < C_5, C_6 < \infty \tag{15.70}$$

Therefore (15.64) holds, which implies that $\Phi(t)$ is bounded and one can conclude that (15.62) also holds. □

It is interesting to remark that for Algorithm III taking into account (15.53), the stability condition is that:

$$\frac{A_M}{\hat{A}_M}\frac{\hat{P}}{P}\frac{G}{\hat{G}} - \frac{\lambda_2}{2} \tag{15.71}$$

should be a strictly positive real transfer function. However, this condition can be rewritten for $\lambda_2 = 1$ as (Ljung and Söderström 1983):

$$\left|\left(\frac{A_M}{\hat{A}_M}\cdot\frac{\hat{P}}{P}\cdot\frac{G}{\hat{G}}\right)^{-1} - 1\right| < 1 \tag{15.72}$$

for all $\omega$. This roughly means that it always holds provided that the estimates of $A_M$,$P$, and $G$ are close to the true values (i.e., $H(e^{-j\omega})$ in this case is close to a unit transfer function).

### 15.4.2 The Stochastic Case—Perfect Matching

There are two sources of measurement noise, one acting on the primary transducer which gives an image of the disturbance and the second acting on the measurement of the residual error (force, acceleration).

For the primary transducer the effect of the measurement noise is negligible since the signal to noise ratio is very high. The situation is different for the residual error where the effect of the noise can not be neglected. In the presence of the measurement noise, the equation of the a posteriori residual error becomes:

$$\nu(t+1) = H(q^{-1})[\theta - \hat{\theta}(t+1)]^T \Phi(t) + w(t+1) \tag{15.73}$$

The averaging method presented in Chap. 4 can be used to analyse the asymptotic behavior of the algorithm in the presence of noise. Taking into account the form of (15.73), one can directly use Theorem 4.1.

The following assumptions will be made:

1. $\lambda_1(t) = 1$ and $\lambda_2(t) = \lambda_2 > 0$.
2. $\hat{\theta}(t)$ generated by the algorithm belongs infinitely often to the domain $D_S$:

$$D_S \overset{\Delta}{=} \{\hat{\theta} : P(z^{-1}) = 0 \Rightarrow |z| < 1\}$$

   for which stationary processes:

$$\Phi(t,\hat{\theta}) \overset{\Delta}{=} \Phi(t)|_{\hat{\theta}(t)=\hat{\theta}=\text{const}}$$
$$\nu(t,\hat{\theta}) = \nu(t)|_{\hat{\theta}(t)=\hat{\theta}=\text{const}}$$

   can be defined.
3. $w(t)$ is a zero mean stochastic process with finite moments and independent of the sequence $d(t)$.

From (15.73) one gets:

$$\nu(t+1,\hat{\theta}) = H(q^{-1})[\theta - \hat{\theta}]^T \Phi(t,\hat{\theta}) + w(t+1,\hat{\theta}) \tag{15.74}$$

Since $\Phi(t,\hat{\theta})$ depends upon $d(t)$ one concludes that $\Phi(t,\hat{\theta})$ and $w(t+1,\hat{\theta})$ are independent. Therefore using Theorem 4.1 it results that if:

$$H'(z^{-1}) = \frac{A_M(z^{-1})G(z(^{-1})}{P(z^{-1})L(z(^{-1})} - \frac{\lambda_2}{2} \tag{15.75}$$

is a strictly positive real transfer function, one has:

$$\text{Prob}\left\{\lim_{t\to\infty} \hat{\theta}(t) \in D_C\right\} = 1$$

where, $D_C = \{\hat{\theta} : \Phi^T(t,\hat{\theta})(\theta - \hat{\theta}) = 0\}$. If furthermore $\Phi^T(t,\hat{\theta})(\theta - \hat{\theta}) = 0$ has a unique solution (richness condition), the condition that $H'(z^{-1})$ be strictly positive real implies that: $\text{Prob}\{\lim_{t\to\infty} \hat{\theta}(t) = \theta\} = 1$.

## *15.4.3 The Case of Non-Perfect Matching*

If $\hat{N}(t,q^{-1})$ does not have the appropriate dimension there is no any chance to satisfy the perfect matching condition. Two questions are of interest in this case:

1. The boundedness of the residual error.
2. The bias distribution in the frequency domain.

### Boundedness of the Residual Error

For analyzing the boundedness of the residual error, results from Landau and Karimi (1997b), Landau et al. (2001b), can be used. The following assumptions are made:

1. There exists a reduced order filter $\hat{N}$ characterized by the unknown polynomials $\hat{S}$ (of order $n_S$) and $\hat{R}$ (of order $n_R$), for which the closed loop formed by $\hat{N}$ and $M$ is asymptotically stable, i.e., $A_M\hat{S} - B_M\hat{R}$ is a Hurwitz polynomial.
2. The output of the optimal filter satisfying the matching condition can be expressed as:

$$-\hat{y}(t+1) = -[\hat{S}^*(q^{-1})\hat{y}(t) - \hat{R}(q^{-1})\hat{u}(t+1) + \eta(t+1)] \tag{15.76}$$

   where $\eta(t+1)$ is a norm bounded signal

Equation (15.76) can be interpreted as a decomposition of the optimal filter into two parallel blocks, one is the reduced order filter and the other block with output $\eta(t)$ corresponds to the neglected dynamics (input additive uncertainty). The boundedness of $\eta(t)$ requires that the neglected dynamics be stable. Using the results of Landau and Karimi (1997b) (Theorem 4.1, pp. 1505–1506) and assuming that $d(t)$ is norm bounded, it can be shown that all the signals are norm bounded under the passivity condition (15.63), where $P$ is computed now with the reduced order estimated filter.

**Bias Distribution**

Using the Parseval's relation, the asymptotic bias distribution of the estimated parameters in the frequency domain can be obtained starting from the expression of $\nu(t)$, by taking in account that the algorithm minimize (almost) a criterion of the form $\lim_{N\to\infty} \frac{1}{N}\sum_{t=1}^{N} \nu^2(t)$. For details see Sect. 4.4.

The bias distribution (for Algorithm III) will be given by:

$$\hat{\theta}^* = \arg\min_{\hat{\theta}} \int_{-\pi}^{\pi} \left[ \left| D(e^{-j\omega}) - \frac{\hat{N}(e^{-j\omega})G(e^{-j\omega})}{1-\hat{N}(e^{-j\omega})M(e^{-j\omega})} \right|^2 \phi_d(\omega) + \phi_w(\omega) \right] d\omega \tag{15.77}$$

where $\phi_d$ and $\phi_w$ are the spectral densities of the disturbance $d(t)$ and of the measurement noise. Taking in account (15.24), one obtains

$$\hat{\theta}^* = \arg\min_{\hat{\theta}} \int_{-\pi}^{\pi} [|S_{NM}|^2 |N-\hat{N}|^2 |S_{\hat{N}M}|^2 |G|^2 \phi_d(\omega) + \phi_w(\omega)] d\omega \tag{15.78}$$

where $S_{NM}$ and $S_{\hat{N}M}$ are the output sensitivity functions of the internal closed loop for $N$ and respectively $\hat{N}$:

$$S_{NM} = \frac{1}{1-NM}; \qquad S_{\hat{N}M} = \frac{1}{1-\hat{N}M}$$

From (15.77) and (15.78) one concludes that a good approximation of $N$ will be obtained in the frequency region where $\phi_d$ is significant and where $G$ has a high gain (usually $G$ should have high gain in the frequency region where $\phi_d$ is significant in order to counteract the effect of $d(t)$). However, the quality of the estimated $\hat{N}$ will be affected also by the output sensitivity functions of the internal closed loop $N - M$.

## 15.4.4 *Relaxing the Positive Real Condition*

For Algorithms I and II, it is possible to relax the strictly positive real (SPR) conditions taking in account that:

1. The disturbance (input to the system) is a broadband signal.[5]
2. Most of the adaptation algorithms work with a low adaptation gain.

Under these two assumptions, the behavior of the algorithm can be well described by the "averaging theory" developed in Anderson et al. (1986) and Ljung and Söderström (1983) (see also Sect. 4.2). When using the averaging approach, the basic assumption of a slow adaptation holds for small adaptation gains (constant and scalar in Anderson et al. (1986) with $\lambda_2(t) = 0, \lambda_1(t) = 1$; matrix and time decreasing asymptotically in Ljung and Söderström (1983) and Sect. 4.2 with $\lim_{t\to\infty} \lambda_1(t) = 1, \lambda_2(t) = \lambda_2 > 0$).

[5]The fact that the disturbance is a broadband signal will imply that one has persistence of excitation.

In the context of averaging, the basic condition for stability is that:

$$\lim_{N\to\infty} \frac{1}{N} \sum_{t=1}^{N} \Phi(t) H'(q^{-1}) \Phi^T(t) = \frac{1}{2} \int_{-\pi}^{\pi} \Phi(e^{j\omega})[H'(e^{j\omega}) + H'(e^{-j\omega})]\Phi^T(e^{-j\omega}) d\omega > 0 \quad (15.79)$$

be a positive definite matrix ($\Phi(e^{j\omega})$ is the Fourier transform of $\Phi(t)$). One can view (15.79) as weighted energy of the observation vector $\Phi$. Of course the SPR sufficient condition upon $H'(z^{-1})$ (see (15.63)) allows to satisfy this condition. However, in the averaging context it is only needed that (15.79) is true which allows that $H'$ be non positive real in a limited frequency band. Expression (15.79) can be rewritten as follows:

$$\int_{-\pi}^{\pi} \Phi(e^{j\omega})[H' + H'^*]\Phi^T(e^{-j\omega}) d\omega = \sum_{i=1}^{r} \int_{\alpha_i}^{\alpha_i + \Delta_i} \Phi(e^{j\omega})[H' + H'^*]\Phi^T(e^{-j\omega}) d\omega - \sum_{j=1}^{p} \int_{\beta_j}^{\beta_j + \Delta_j} \Phi(e^{j\omega})[\bar{H}' + \bar{H}'^*]\Phi^T(e^{-j\omega}) d\omega > 0 \quad (15.80)$$

where $H'$ is strictly positive real in the frequency intervals $[\alpha_i, \alpha_i + \Delta_i]$ and $\bar{H}' = -H'$ is positive real in the frequencies intervals $[\beta_j, \beta_j + \Delta_j]$ ($H'^*$ denotes the complex conjugate of $H'$). The conclusion is that $H'$ does not need to be SPR. It is enough that the "positive" weighted energy exceeds the "negative" weighted energy. This explains why Algorithms I and II will work in practice in most of the cases. It is, however, important to remark that if the disturbance is a single sinusoid (which violates the hypothesis of broadband disturbance) located in the frequency region where $H'$ is not SPR, the algorithm may diverge (see Anderson et al. 1986; Ljung and Söderström 1983).

Without doubt, the best approach for relaxing the SPR conditions, is to use Algorithm III (given in (15.53)) instead of Algorithm II. This is motivated by (15.71) and (15.72). As it will be shown in the next section this algorithm gives the best results both in simulations and on real-time experiments.

## 15.5 Adaptive Attenuation of Broad Band Disturbances on an Active Vibration Control System

The mechanical system equipped with an adaptive vibration control system using an inertial actuator which will be considered, has been presented in Sect. 1.4.5.

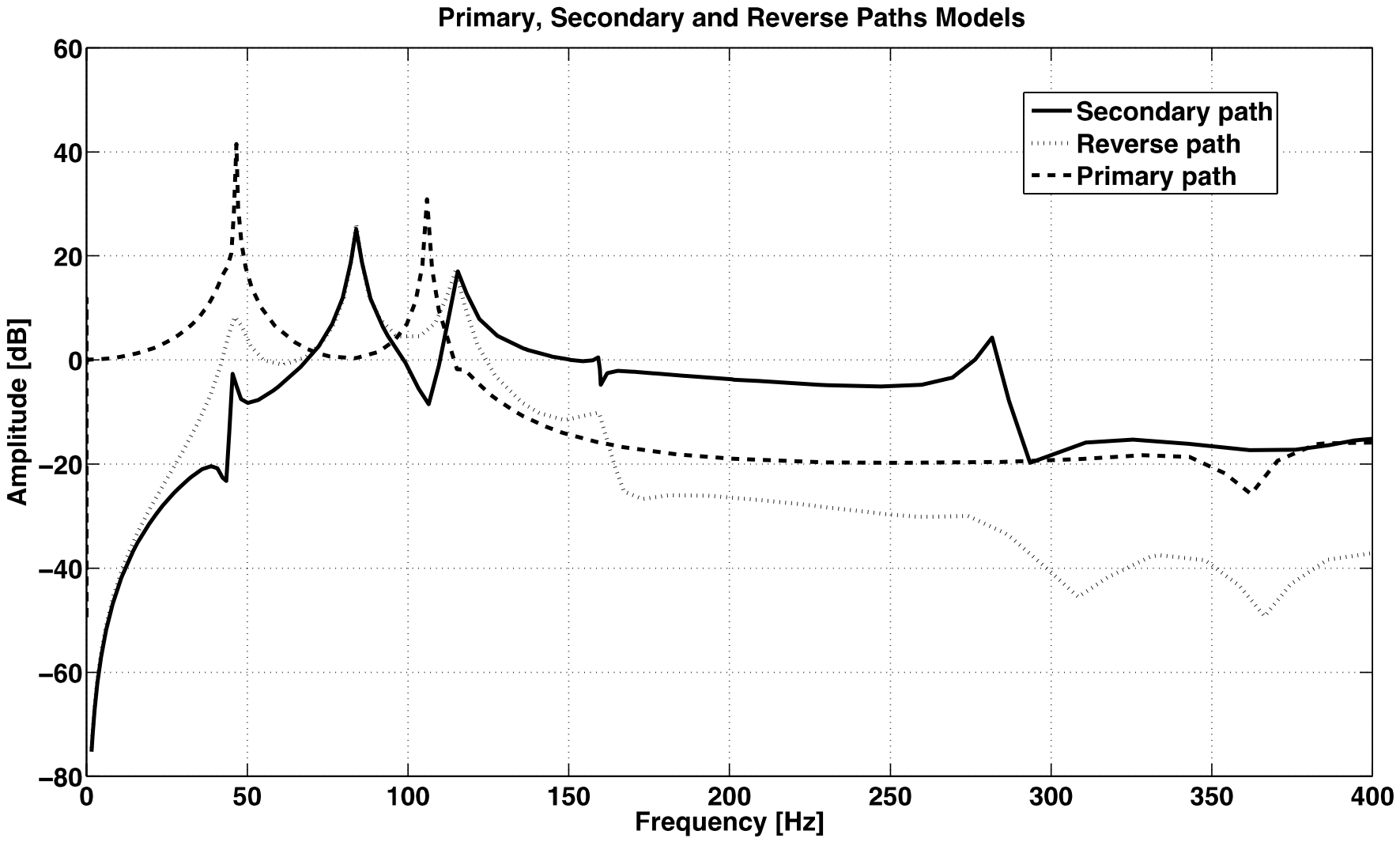

**Fig. 15.5** Frequency characteristics of the primary, secondary and reverse paths

The structure of the system is described in Fig. 15.2. For a view of the system see Sect. 1.4.5, Fig. 1.23.

The incoming disturbance is the position of the mobile part of the inertial actuator on top of the structure (see Fig. 15.2).[6] The residual acceleration $\nu(t)$ and the input to the feedforward compensator $\hat{u}(t)$ are measured by accelerometers. The control input is the position of the mobile part of the inertial actuator located on the bottom of the structure. The input of $G(q^{-1})$, $M(q^{-1})$ (as well as for $W(q^{-1})$) being a position and the output a force, these transfer functions have a double differentiator behavior.

### *15.5.1 System Identification*

The models of the plant may be obtained by system identification using the methods provided in Chaps. 5 and 9. The secondary path, $G(q^{-1})$, between control signal $-\hat{y}(t)$ and the output $\nu(t)$ has been identified in open loop. The excitation signal was a PRBS generated with a shift register with $N = 10$ and a frequency divider of $p = 4$. The estimated orders of the model are: $n_{B_G} = 15$, $n_{A_G} = 13$. The best results in terms of model validation were obtained with *Recursive Extended Least Square* method. The frequency characteristic of the secondary path is shown in Fig. 15.5 (solid). There exist several very low-damped vibration modes in the secondary path. The first vibration mode is at 46.56 Hz with a damping of 0.013 and the second at 83.9 Hz with a damping of 0.011, the third one at 116 Hz has a damping of 0.014. There are two zeros on the unit circle corresponding to a double differentiator behavior and also a pair of low-damped complex zeros at 106 Hz with a damping of 0.021.

[6]The inertial actuator is driven by an external source.

The reverse path, $M(q^{-1})$, has been identified in open loop with the same PRBS excitation ($N = 10$ and a frequency divider of $p = 4$) applied at $-\hat{y}(t)$ and measuring the output signal of the primary transducer $\hat{u}(t)$. The orders of the obtained model are: $n_{B_M} = 15$, $n_{A_M} = 13$. The best results in terms of model validation were obtained with *Recursive Extended Least Square* method. The frequency characteristic of the reverse path is presented in Fig. 15.5 (dotted). There exist several very low-damped vibration modes at 46.20 Hz with a damping of 0.045, at 83.9 Hz with a damping of 0.01, at 115 Hz with a damping of 0.014 and some modes in high frequencies. There are two zeros on the unit circle corresponding to a double differentiator.

The primary path has been identified from the spectral densities of $d(t)$ and $v(t)$ in open-loop operation. The frequency characteristic is presented in Fig. 15.5 (dashed) and may serve for simulations and detailed performance evaluation. Note that the primary path features a resonance at 106 Hz exactly where the secondary path has a pair of low-damped complex zeros (almost no gain). Therefore one can not expect good attenuation around this frequency.

### 15.5.2 Experimental Results

The performance of the system for rejecting broadband disturbances will be illustrated using the adaptive feedforward scheme.[7] The adaptive filter structure is $n_{B_N} = 19$; $n_{A_N} = 20$ (total of 40 parameters) and this complexity does not allow to verify the "perfect matching condition" (not enough parameters).

For the *adaptive* operation the Algorithms II and III have been used with adaptation gain updating with variable forgetting factor $\lambda(t)$ (the forgetting factor tends towards 1), combined with a *constant trace* adaptation gain. For details, see Sect. 3.2.3.

A PRBS excitation on the global primary path will be considered as the disturbance ($s(t)$). The corresponding spectral densities of $d(t)$ in open loop and of $\hat{u}(t)$ when feedforward compensation is active are shown in Fig. 15.6. One can clearly see the effect of the internal positive feedback when the compensation system is active.

For comparison purposes, Fig. 15.7 shows simulation results obtained with adaptive feedforward compensation using Algorithms II and III in terms of the spectral densities of the residual acceleration. The spectral density of the residual acceleration in open loop is also represented. Clearly the performance of Algorithm III[8]

[7]These experiments have been carried out by M. Alma (GIPSA-LAB).

[8]The filter used in Algorithm III has been computed using the estimated values of $N$ obtained with Algorithm II after 40 s.

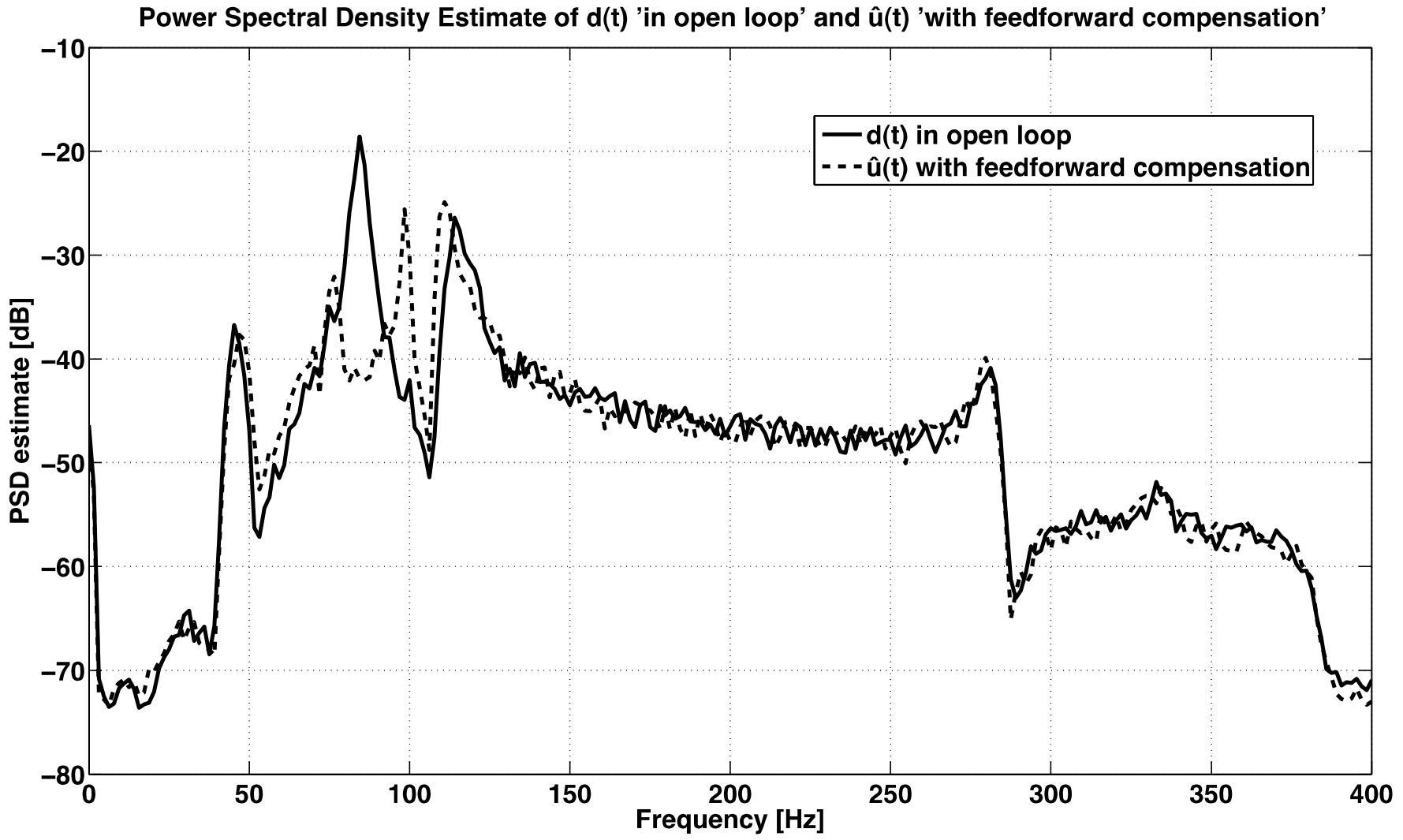

**Fig. 15.6** Spectral densities of the image of the disturbance in open loop $d(t)$ and with the feedforward compensation scheme $\hat{u}(t)$ (experimental)

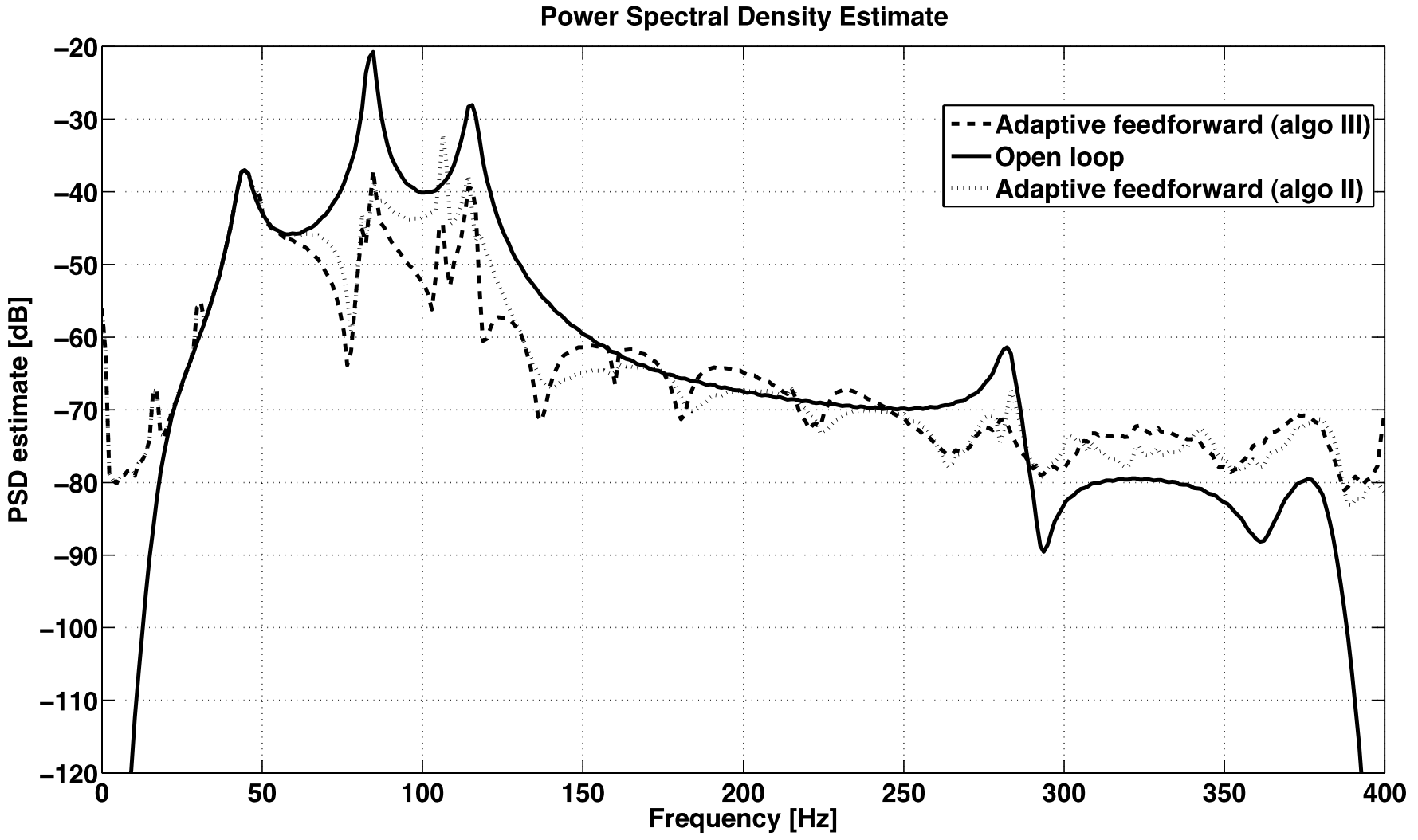

**Fig. 15.7** Spectral densities of the residual acceleration in open loop and with adaptive feedforward compensation (simulation)

is better than that of Algorithm II. This is also confirmed by the measurements of the variance of the residual acceleration in open loop and with the adaptive feedforward compensation. Algorithm III gives a global attenuation of 23.52 dB while Algorithm II gives an global attenuation of 18.80 dB. It is interesting to remark on

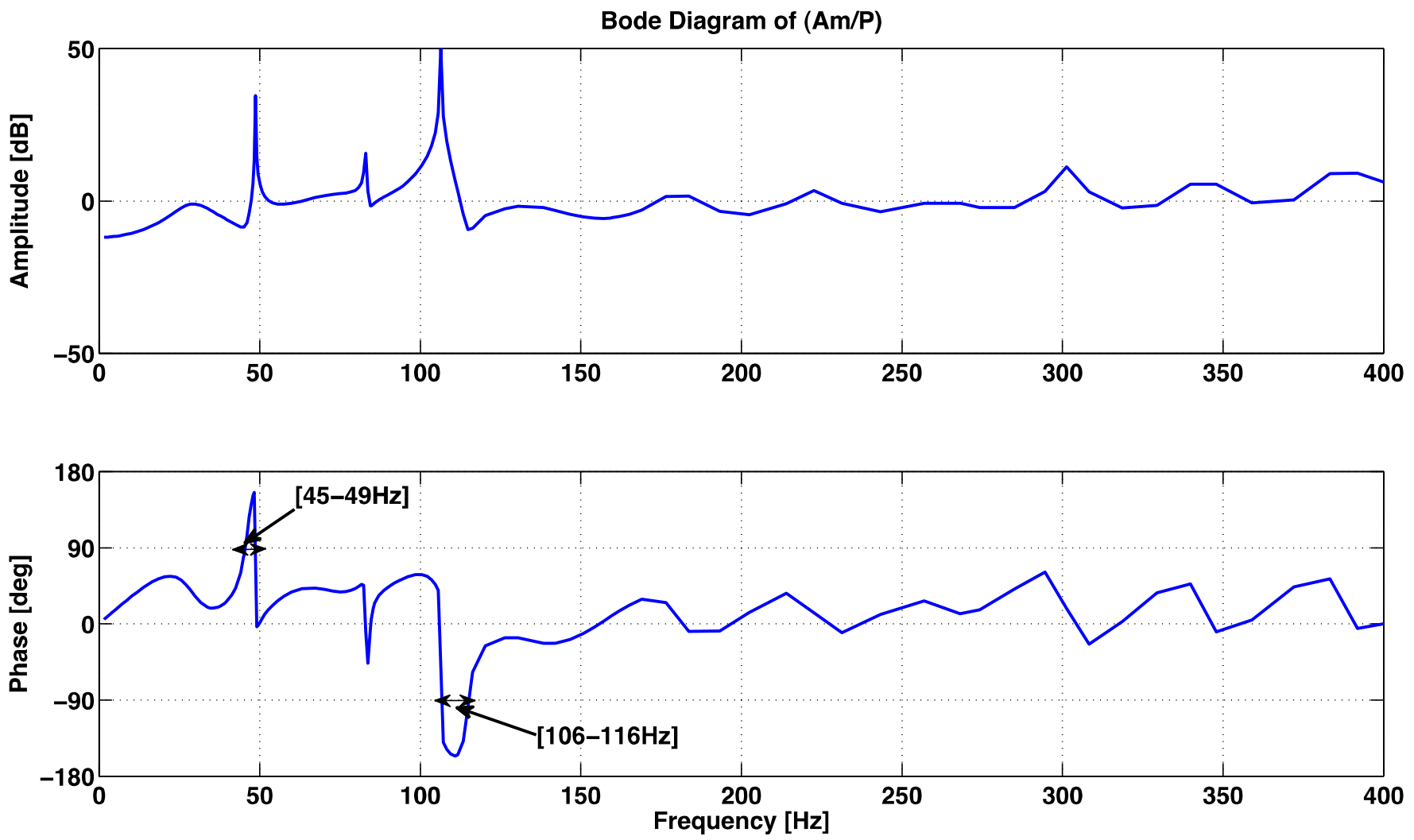

**Fig. 15.8** Bode diagram of the estimated transfer function $\frac{A_M}{P}$ (simulation)

Fig. 15.7 that Algorithm II performs poorly around 110 Hz. Looking now to the Bode diagram of the estimated transfer function $\frac{\hat{A}_M}{\hat{P}}$ shown in Fig. 15.8 one can see that around this frequency, the transfer function is not positive real (the phase lag is over $-90$ degrees) and has a significant gain (in simulation $\hat{G} = G$). A similar but less significant effect occurs also around 48 Hz.

Experimental time-domain results obtained in open loop and with adaptive feedforward compensation Algorithms II and III on the AVC system shown in Fig. 15.2 are presented in Fig. 15.9. The variance of the residual force in open loop is: $\mathrm{var}(\nu(t) = z(t)) = 0.0325$. With adaptive feedforward compensation Algorithm II (after the adaptation algorithm has converged), the variance is: $\mathrm{var}(\nu(t)) = 0.0057$. This corresponds to a global attenuation of 15.12 dB. Using Algorithm III[9] the variance of the residual acceleration is: $\mathrm{var}(\nu(t)) = 0.0047$. The corresponding global attenuation is 16.82 dB, which is an improvement with respect to Algorithm II.

Figure 15.10 shows the spectral densities of the residual acceleration measured on the AVC in open loop and using adaptive feedforward compensation (after the adaptation algorithm has converged). Algorithm III performs slightly better than Algorithm II. Same phenomena observed in simulations occur for Algorithm II in the regions where $\frac{A_M}{P}$ is not positive real. The improvement of performance using Algorithm III is less significant than in simulation. Probably a better estimation of $\frac{\hat{A}_M}{\hat{P}}$ will improve the results. For other experimental results see Landau and Alma (2010).

[9]The filter used in simulation has been also used in real time.

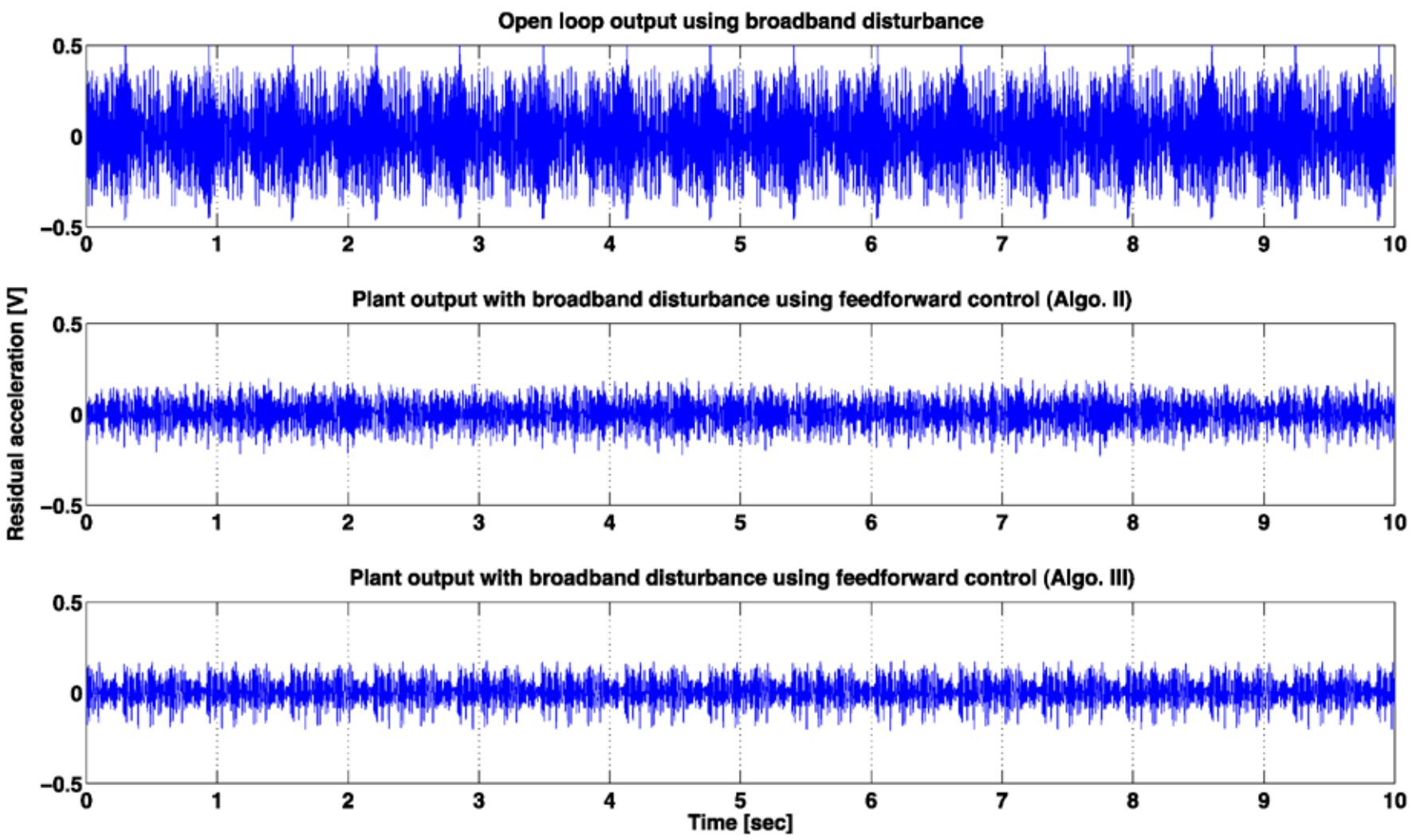

**Fig. 15.9** Real-time results obtained with a disturbance $s(t) = \text{PRBS}$ (experimental)

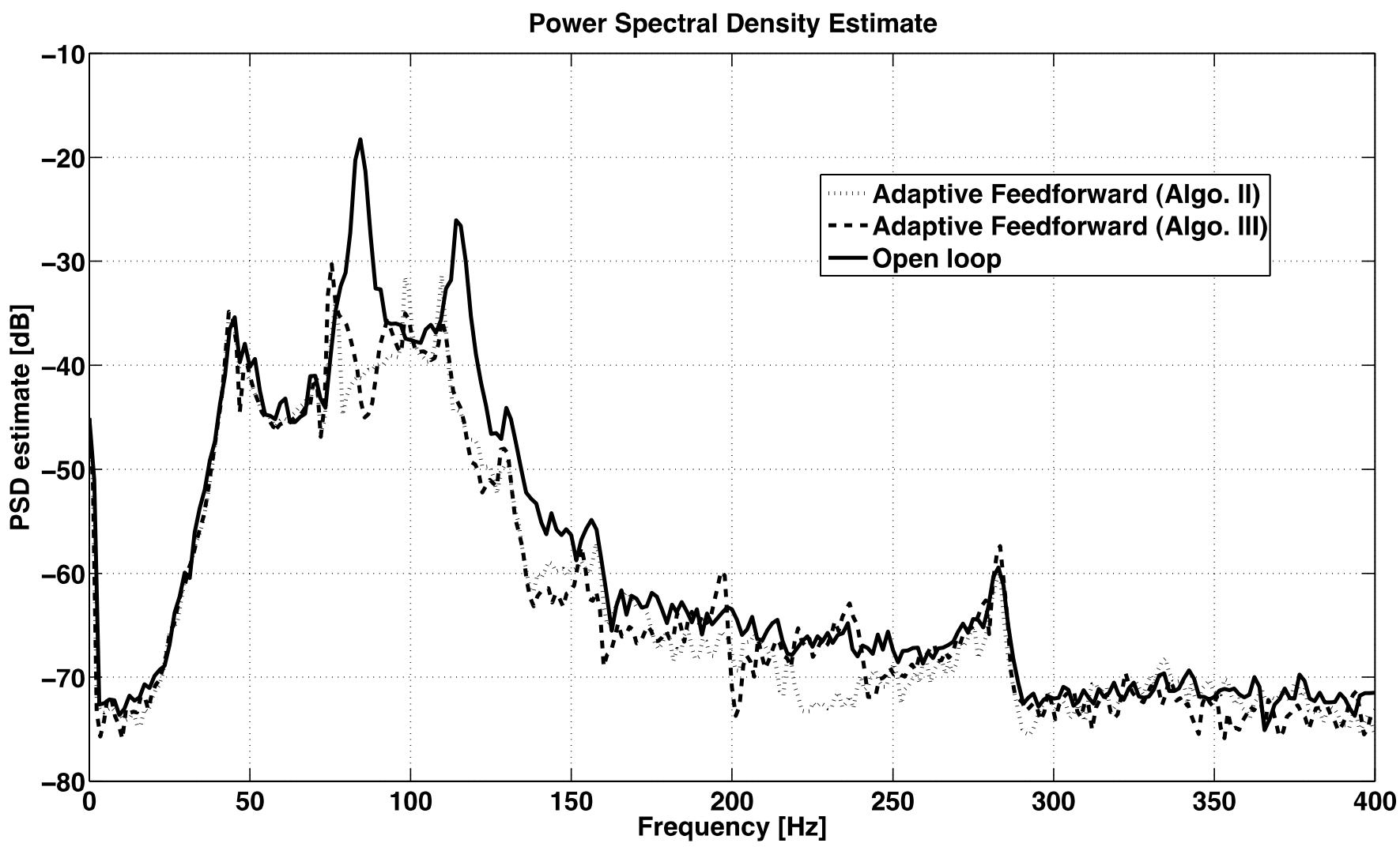

**Fig. 15.10** Spectral densities of the residual acceleration in open loop and with adaptive feedforward compensation (experimental)

## 15.6 Concluding Remarks

1. If a measurement correlated with the disturbance is available an adaptive feedforward compensation scheme can be built.

2. This approach is currently used for active noise control and active vibration control.
3. It is important to emphasize the existence of an inherent positive feedback coupling between the actuator and the measure of the image of the disturbance (correlated measurement).
4. The problem of estimating or adapting the feedforward compensator is linked to the problem of plant identification in closed loop operation.
5. Stable algorithms preserving the stability of the inner positive feedback loop can be developed using the approach used for identification in closed loop.
6. Adaptive feedforward compensation can be used on top of a feedback loop used to minimize the residual error.

## 15.7 Problems

**15.1** Instead of the optimal feedforward filter given in (15.11) and (15.12) consider a Youla-Kucera parameterized optimal feedforward filter given by:

$$R(q^{-1}) = R_0(q^{-1}) - A_M(q^{-1})Q(q^{-1}) \tag{15.81}$$

$$S(q^{-1}) = S_0(q^{-1}) - B_M(q^{-1})Q(q^{-1}) \tag{15.82}$$

where $A_M(q^{-1})$ and $B_M(q^{-1})$ denote respectively the denominator and numerator of the positive feedback coupling path and $R_0(q^{-1})$ and $S_0(q^{-1})$ are the polynomials of the central (stabilizing) filter. The optimal polynomial $Q(q^{-1})$ is considered to be of the form

$$Q(q^{-1}) = q_0 + q_1 q^{-1} + \cdots + q_{n_Q} q^{-n_Q} \tag{15.83}$$

The central filter ($R_0(q^{-1})$ and $S_0(q^{-1})$) stabilizes the internal feedback loop i.e., the characteristic polynomial:

$$P(z^{-1}) = A_M(z^{-1})S_0(z^{-1}) - B_M(z^{-1})R_0(z^{-1}) \tag{15.84}$$

is a Hurwitz polynomial.

One assumes that there exists a value of the $Q$ polynomial such that (perfect matching condition):

$$\frac{GA_M(R_0 - A_M Q)}{A_M S_0 - B_M R_0} = -D \tag{15.85}$$

Estimates of $G$ and $M(\hat{G}, \hat{M})$ are available and the effect of the measurement noise on the residual error can be neglected.

1. Show that for a constant estimated vector:

$$\hat{Q}(q^{-1}) = \hat{q}_0 + \hat{q}_1 q^{-1} + \cdots + \hat{q}_{n_Q} q^{-n_Q} \tag{15.86}$$

the residual error can be expressed as:

$$\nu(t+1) = \frac{A_M(q^{-1})G(q^{-1})}{P(q^{-1})}(Q - \hat{Q})\alpha(t+1) \tag{15.87}$$

where:

$$\alpha(t+1) = B_M \hat{y}(t+1) - A_M \hat{u}(t+1) = B_M^* \hat{y}(t) - A_M \hat{u}(t+1) \tag{15.88}$$

2. Show that the residual error can be also expressed as:

$$\nu(t+1) = \frac{A_M(q^{-1})G(q^{-1})}{P(q^{-1})}[\theta - \hat{\theta}]^T \phi(t) \tag{15.89}$$

where:

$$\theta^T = [q_0, q_1, q_2, \ldots, q_{n_Q}] \tag{15.90}$$

$$\hat{\theta}^T = [\hat{q}_0, \hat{q}_1, \hat{q}_2, \ldots, \hat{q}_{n_Q}] \tag{15.91}$$

$$\phi^T(t) = [\alpha(t+1), \alpha(t), \ldots, \alpha(t - n_Q + 1)] \tag{15.92}$$

3. Develop adaptation algorithms for estimating the Q-parameters
4. Make a stability analysis of the resulting adaptive system (hint: use the methodology developed in Sect. 15.4.1).

**15.2** Develop an exact adaption algorithm for the case of Algorithm I ($L = G$) given in Sect. 15.3 without neglecting the non-commutativity of time-varying operators (hint: use the methodology developed in Sect. 5.5.3).

# Chapter 16
# Practical Aspects

## 16.1 Introduction

The objective of this chapter is to examine in detail the practical aspects related to the implementation of the various adaptive control schemes presented in the previous chapters.

As indicated in Chap. 1, *adaptive control* can be viewed as a hierarchical system:

Level 1: Conventional feedback control (in our case: linear digital control),
Level 2: Adaptation loop.

The efficient use of adaptation requires that, except when plant model parameters vary or knowledge of them is lacking, a correct implementation of a digital controller is available. The only effect of adaptation will be the adjustment of the controller parameters in real time. Section 16.2 will review a number of basic facts relating to the implementation of digital controllers.

The basic component of the adaptation loop is the *parameter adaptation algorithm* (PAA). Various forms have been discussed throughout the book and more specifically in Chaps. 3 and 10. In general, PAAs with time-varying adaptation gain are used in practice because they offer better performances than PAAs with constant adaptation gain. Various aspects related to their implementation will be discussed in Sect. 16.3. In particular, the key practical issue for assuring the positive definiteness of the matrix adaptation gain despite possible round-off errors will be examined.

Another important aspect is the programming of the design equation for indirect adaptive control. Standard books like (Press et al. 1988) offer solutions for numerical robust programming of the various design equations considered.

Once the adjustable digital controller and the parameter adaptation algorithm have been implemented, one considers the implementation of the various adaptive control strategies. This is discussed in Sect. 16.4. The initialization of adaptive control schemes is discussed in Sect. 16.5. In a number of situations, the safe operation of an adaptive control scheme requires the monitoring of the information contained in the signals sent to the parameter adaptation algorithm, as well as of the results delivered by the parameter adaptation algorithm. This leads to the introduction of a supervision loop and this is discussed in Sect. 16.6.

## 16.2 The Digital Control System

In this section, we review the basic requirements for the implementation of a digital control system. For a more detailed discussion of the various aspects see Landau et al. (1990b, 1993b), Landau and Zito (2005), Åström and Wittenmark (1984), Franklin et al. (1990).

### *16.2.1 Selection of the Sampling Frequency*

A good rule for the selection of the sampling frequency is:

$$f_s = (6 \to 25) f_B^{CL} \tag{16.1}$$

where:

- $f_s$ = sampling frequency (in Hz);
- $f_B^{CL}$ = desired bandwidth of the closed-loop system (in Hz).

Of course, the desired closed-loop bandwidth is related to the bandwidth of the system to be controlled. The formula (16.1) gives enough freedom for the selection of the sampling frequency.

It is important to recall that many systems have a pure time delay or their behavior can be approximated by a time delay plus dynamics. One has the following relationship between continuous-time delay and discrete-time delay:

$$\tau = dT_s + L; \quad 0 < L < T_s \tag{16.2}$$

where:

- $\tau$ = time delay of the continuous-time system;
- $T_s$ = sampling period ($T_s = 1/f_s$);
- $d$ = integer discrete-time delay;
- $L$ = fractional delay.

Except in very particular cases, all the discrete-time models will feature a fractional delay. Fractional delays are reflected as zeros in the transfer function of the discrete-time models. For simple continuous-time models (like first order plus delay), the zeros induced by the fractional delay become unstable for $L \geq 0.5T_s$ (Landau 1990b; Landau and Zito 2005; Franklin et al. 1990). Therefore, the selection of the sampling frequency becomes crucial if we would like to use control laws which cancel the system zeros (tracking and regulation with independent objectives, minimum variance tracking and regulation). For continuous-time systems with a relative degree higher or equal to 2, high-frequency sampling will induce unstable zeros (Åström et al. 1984). As a general rule, one tries to select the lower sampling frequency compatible with the desired performances.

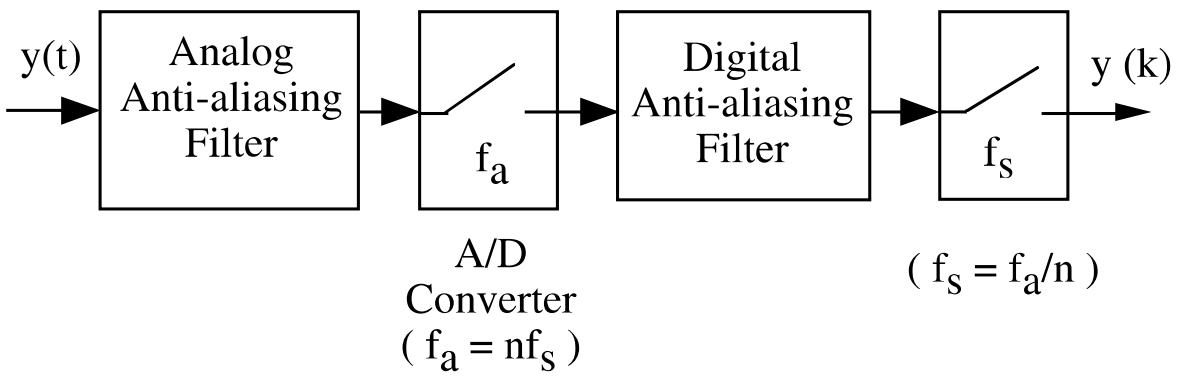

**Fig. 16.1** Data acquisition with oversampling

### *16.2.2 Anti-Aliasing Filters*

The theory of discrete-time systems indicates that the maximum frequency ($f_{\max}$) of a signal sent to the analog to digital converter should satisfy:

$$f_{\max} < f_s/2 \tag{16.3}$$

where $f_s/2$ is called the Nyquist or Shannon frequency. Sending frequencies over $f_s/2$ produces distortion of the recovered discrete-time spectrum called *aliasing*. Therefore, *anti-aliasing* filters should always be introduced in order to remove the undesirable components of the signal. Anti-aliasing filters are constituted in general as one or several second order filters in cascade (Bessel, ITAE, Butterworth type). They should introduce a consequent attenuation of the signal beyond $f_s/2$ but their bandwidth should be larger than the desired closed-loop bandwidth. Their design will also depend on the level of undesirable signal at frequencies beyond $f_s/2$. The anti-aliasing filters introduce a high-frequency dynamics which can in general be approximated by an additional small time delay. Since one directly estimates a discrete-time model, their effect is captured by the estimated model.

In a number of practical situations (i.e., very low sampling frequency for process control), it is more convenient to do the anti-aliasing filtering in two steps by using over sampling for data acquisition ($f_a = nf_s$). This is illustrated in Fig. 16.1. A high-sampling frequency (integer multiple of the desired one) is used for data acquisition, with an appropriate analog anti-aliasing filter. Then the sampled signal is passed through an anti-aliasing digital filter followed by a frequency divider which delivers a sampled signal with the required frequency.

### *16.2.3 Digital Controller*

The basic equation for the two-degree of freedom RST digital controller is:

$$S(q^{-1})u(t) = T(q^{-1})y^*(t+d+1) - R(q^{-1})y(t) \tag{16.4}$$

where:

$$S(q^{-1}) = s_0 + s_1 q^{-1} + \cdots + s_{n_S} q^{-n_S} = s_0 + q^{-1} S^*(q^{-1}) \tag{16.5}$$

$$R(q^{-1}) = r_0 + r_1 q^{-1} + \cdots + r_{n_R} q^{-n_R} \tag{16.6}$$

$$T(q^{-1}) = t_0 + t_1 q^{-1} + \cdots + t_{n_T} q^{-n_T} \tag{16.7}$$

$u(t)$ is the plant input, $y(t)$ is the plant output and $y^*(t+d+1)$ is the desired tracking trajectory either stored in the computer or generated by a tracking reference model:

$$y^*(t+d+1) = -A_m^*(q^{-1}) y^*(t+d) + B_m(q^{-1}) r(t) \tag{16.8}$$

where:

$$A_m(q^{-1}) = 1 + a_1^m q^{-1} + \cdots + a_{n_A}^m q^{-n_A} = 1 + q^{-1} A_m^*(q^{-1}) \tag{16.9}$$

$$B_m(q^{-1}) = b_0^m + b_1^m q^{-1} + \cdots + b_{n_B}^m q^{-n_B} \tag{16.10}$$

and $r(t)$ is the reference signal. Equation (16.4) can also be written as:

$$u(t) = \frac{1}{s_0}[T(q^{-1}) y^*(t+d+1) - S^*(q^{-1}) u(t-1) - R(q^{-1}) y(t)] \tag{16.11}$$

Note that for a number of control algorithms (like pole placement) $s_0 = 1$ in (16.11).

### *16.2.4 Effects of the Digital to Analog Converter*

The values of the control signal computed by the control algorithm in fixed or floating arithmetics with 16, 32 or 64 bits have in general a number of distinct values higher than the distinct values of the digital-to-analog converter often limited to 12 bits (4096 distinct values). The characteristics of a digital-to-analog ($D/A$) converter are illustrated in Fig. 16.2. In the equation of a digital controller, the control generated at time $t$ depends upon the previous control values *effectively applied* to the system. It is therefore necessary to round off the control signal $u(t)$ inside the controller and to use these values for the computation of the future values of the control signal. Equation (16.11) becomes:

$$u(t) = \frac{1}{s_0}[T(q^{-1}) y^*(t+d+i) - S^*(q^{-1}) u_r(t-1) - R(q^{-1}) y(t)] \tag{16.12}$$

where:

$$|u_r(t) - u(t)| \le \frac{1}{2} Q \tag{16.13}$$

and $u_r(t)$ is the rounded control effectively sent to the $D/A$ converter and $Q$ is the quantization step. If this rounding operation is not implemented, an equivalent noise is induced in the system which may cause undesirable small oscillations of the output in some situations.

**Fig. 16.2** Input-output characteristics of a digital-to-analog converter

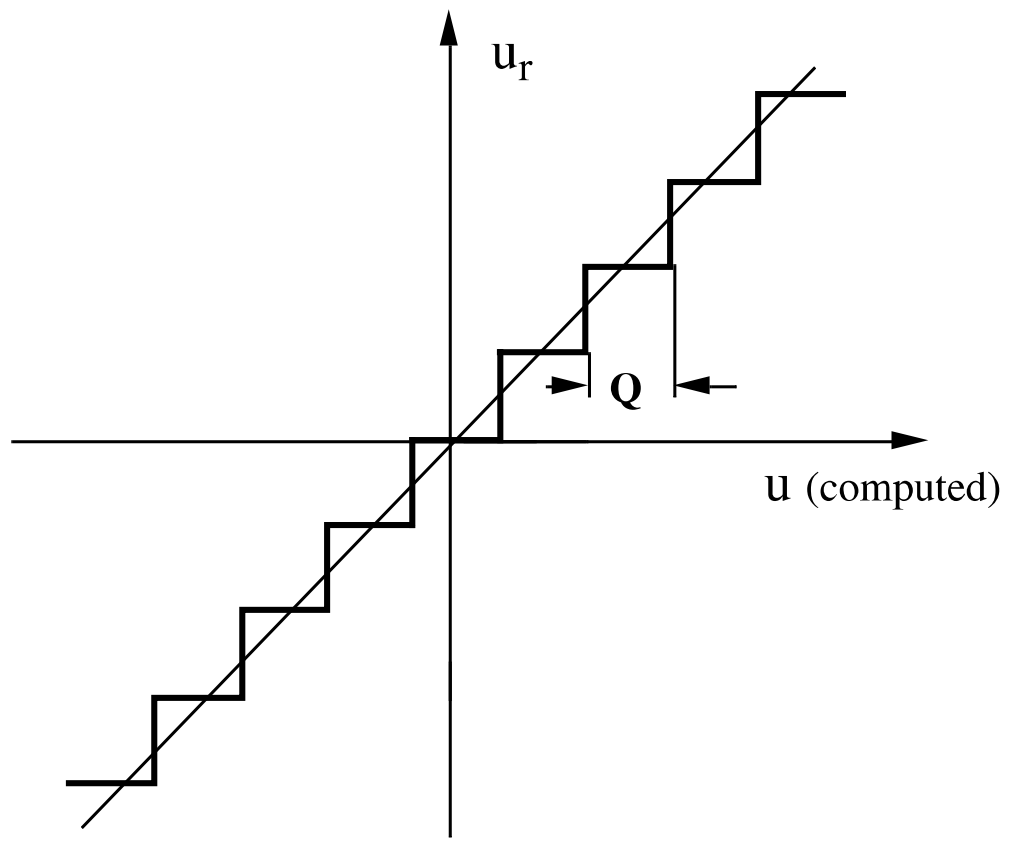

**Fig. 16.3** Digital controller with anti-windup device

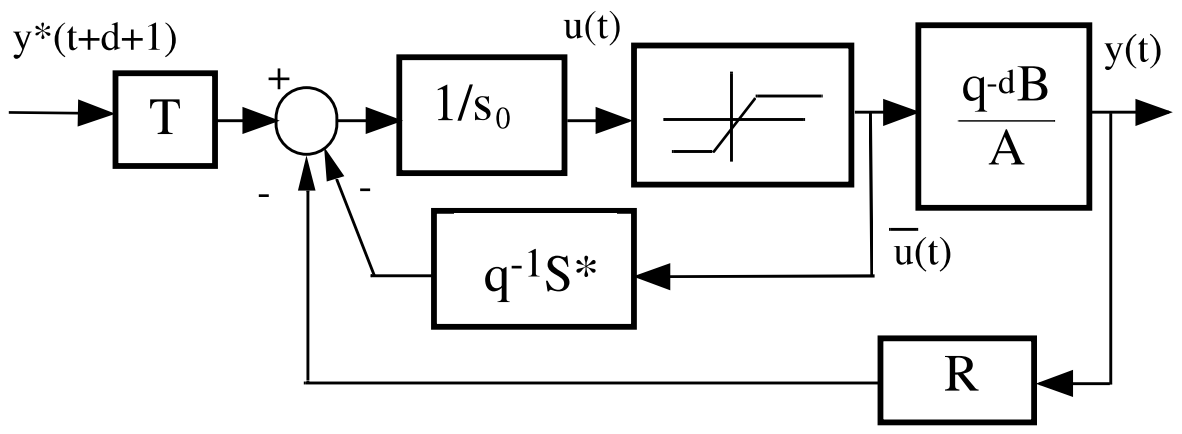

### *16.2.5 Handling Actuator Saturations (Anti-Windup)*

The effect of actuator saturation can have an adverse effect upon the behavior of the control system and, in particular, when the controller contains an integrator. To avoid such troubles, one has to take into account the fact that the control at instant $t$ depends upon the previous inputs effectively applied to the system. However, in order to avoid the use of an additional $A/D$ converter, a copy of the nonlinear characteristics of the actuator should be incorporated in the controller. This is illustrated in Fig. 16.3. The control signal will be given by:

$$u(t) = \frac{1}{s_0}[T(q^{-1})y^*(t+d+i) - S^*(q^{-1})\bar{u}(t-1) - R(q^{-1})y(t)] \qquad (16.14)$$

In (16.14), $\bar{u}(t-1), \bar{u}(t-2), \dots, \bar{u}(t-n_s)$ correspond to the values of $u(t-1)$, $\dots, u(t-n_s)$ passed through the nonlinear characteristics, i.e.:

$$\bar{u}(t) = \begin{cases} u(t) & \text{if } |u(t)| < u_{sat} \\ u_{sat} & \text{if } u(t) \geq u_{sat} \\ -u_{sat} & \text{if } u(t) \leq -u_{sat} \end{cases} \qquad (16.15)$$

It is also possible to impose certain dynamic when the system leaves the saturation. This is illustrated in Fig. 16.4. The desired dynamic is defined by a polynomial

**Fig. 16.4** Digital controller with pre-specified anti-windup dynamics

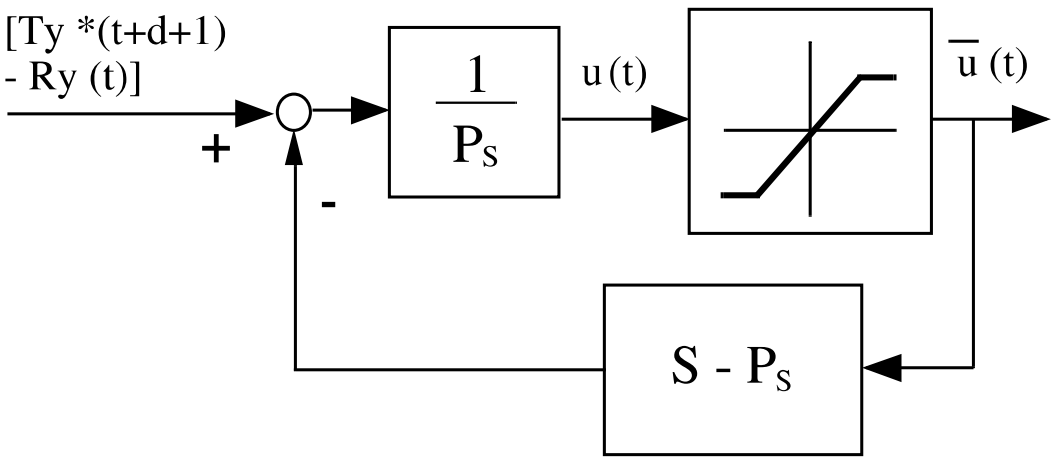

$P_S(q^{-1})$:

$$P_S(q^{-1}) = s_0 + q^{-1} P_S^*(q^{-1}) \tag{16.16}$$

This configuration assures, however, that within the linear operation domain ($|u(t)| < u_{sat} \rightarrow \bar{u}(t) = u(t)$), the transfer operator from $[T(q^{-1})y^*(t+d+1) - R(q^{-1})y(t)]$ to $u(t)$ is still $1/S(q^{-1})$. Effectively, in the linear region one has:

$$\frac{\frac{1}{P_S(q^{-1})}}{1 - \frac{S(q^{-1}) - P_S(q^{-1})}{P_S(q^{-1})}} = \frac{1}{S(q^{-1})}$$

Therefore, in the presence of a nonlinear characteristics the controller equation takes the form:

$$\begin{aligned} P_S(q^{-1})u(t) = T(q^{-1})y^*(t+d+i) - R(q^{-1})y(t) \\ - [S^*(q^{-1}) - P_S^*(q^{-1})]\bar{u}(t-1) \end{aligned} \tag{16.17}$$

or, equivalently:

$$\begin{aligned} u(t) = \frac{1}{s_0}\{T(q^{-1})y^*(t+d+i) - R^*(q^{-1})y(t) - [S^*(q^{-1}) - P_S^*(q^{-1})]\bar{u}(t-1) \\ - P_S^*(q^{-1})u(t-1)\} \end{aligned} \tag{16.18}$$

For more details see Landau and Zito (2005).

### 16.2.6 Manual to Automatic Bumpless Transfer

To avoid undesirable effects of the transfer from open-loop operation to closed-loop operation, it is necessary to initialize the controller "memory", i.e., to provide $y(t-1), y(t-2), \ldots, u(t-1), u(t-2), \ldots, y^*(t+d), y^*(t+d-1), \ldots$. One method for initializing the controller is as follows (Landau 1993b):

1. replace the reference and the desired output by the current output measurement ($y^*(t+d+1) = y(t), r(t) = y(t)$);
2. store the open-loop control $u(t)$ in the controller memory;
3. repeat steps 1 and 2 for $n = \max(n_A + n_S, d + n_B + n_R)$;
4. close the loop.

It can be shown that if $y(t) =$ constant during the initialization phase and the controller has an integrator, one gets after $n$ samples $u(t) = u(t-1)$ (i.e., the system input generated by the controller will be equal to the one applied in open loop).

### *16.2.7 Effect of the Computational Delay*

The effect of the computation time has to be taken into account in standard digital control systems but it becomes even more important in adaptive control since the number of computations is larger. Two basic situations can be distinguished:

1. Computation time equal to or greater than $0.5T_s$: In this case, the measured values at instant $t$ are used to compute the control $u$ which will be sent at $t+1$ (see Fig. 16.5a). The computer introduces an additional time delay of 1 sampling period and the new delay to be considered for the design of the controller will be:

$$d' = d + 1$$

   This delay has to be introduced in the estimated plant model.
2. Computation time less than $0.5T_s$: In this case, the control is sent at the end of the computation (see Fig. 16.5b). More precisely, it is sent with a larger delay than the maximum possible computation time. The computer introduces a fractional time delay which has the effect of either introducing a zero or of modifying the existing zeros in the pulse transfer function of the system to be controlled. This effect is captured by the discrete-time estimated plant model. The fractional delay introduced by the computation time is negligible if the computation time is considerably smaller than the sampling period.

   In computing the current values, one manipulates a number of variables already available (previous inputs and outputs) before the data acquisition of the last measurement at time $t$. In order to reduce the computation time, it is advisable to organize the program such that these computations be done between $t-1$ and $t$.

### *16.2.8 Choice of the Desired Performance*

The choice of desired performance in terms of time response is linked to:

- the dynamic of the open-loop system;
- the power availability of the actuator during transients;
- the desired robustness with respect to model uncertainties (for the regulation dynamics).

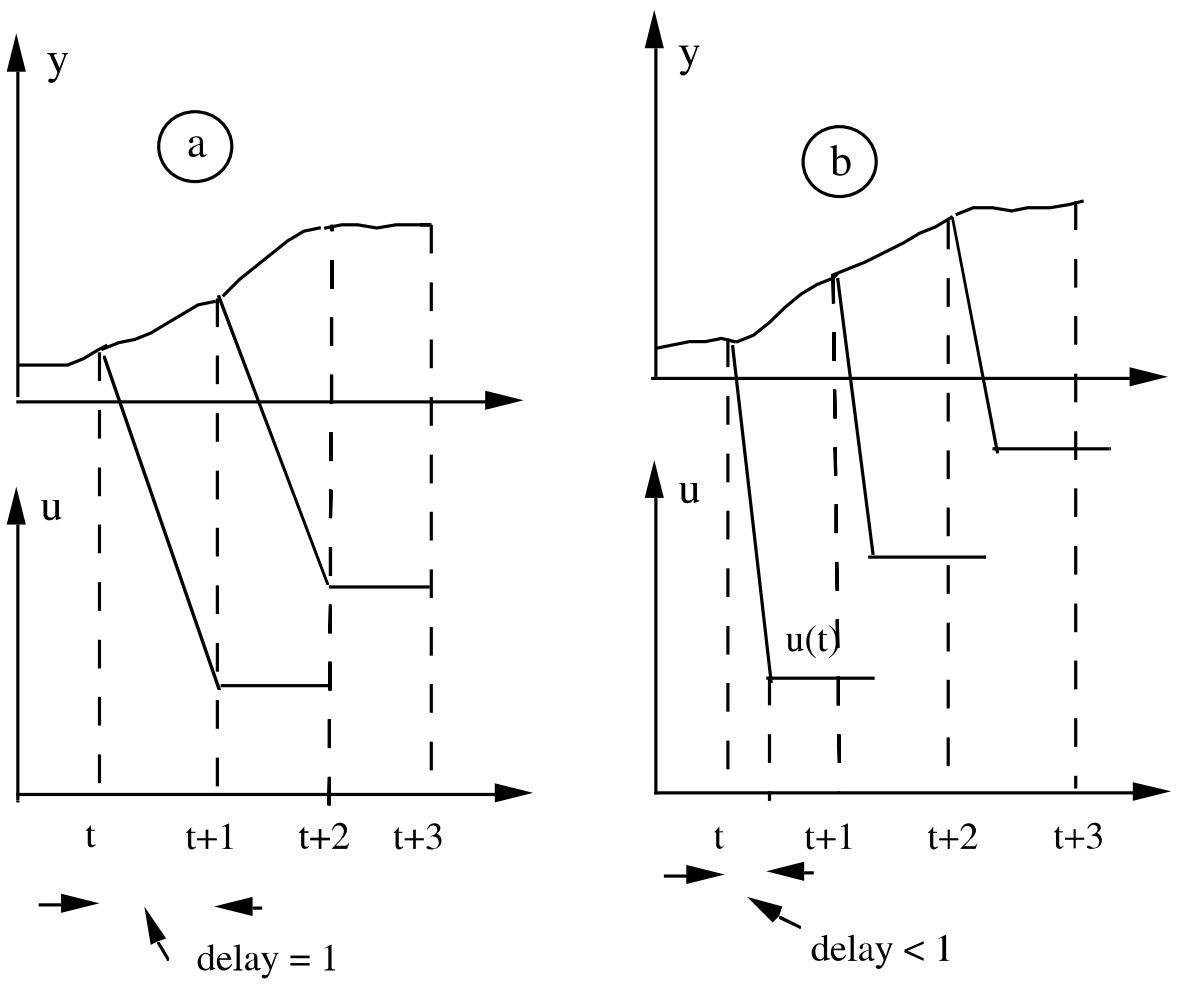

**Fig. 16.5** (**a**) Computation time greater than $0.5T_s$, (**b**) computation time less than $0.5T_s$

We will recall here the fact that if we want to accelerate by a factor $n$ the time response of the closed loop system with respect to the open-loop time response, it is necessary to apply over a certain period of the transient response a maximum input level equal to $n$ times the final steady state input value. The following relation may be then considered:

$$\frac{|u_{\max}|}{|u_{stat}|} \approx \frac{\text{open-loop time response}}{\text{desired-time response}} \approx \frac{\text{desired bandwidth}}{\text{open-loop bandwidth}}$$

where $|u_{\max}|$ is the maximum level of the input during the transient and $|u_{stat}|$ is the steady state input to be applied to the system in steady state. Therefore, the achievable closed-loop performances will depend upon the actuator power availability during transients. Conversely, the actuators have to be chosen as a function of the desired performances and the open-loop time response of the system.

For the structure of the tracking reference model used to generate the desired tracking trajectory, 2nd order or 4th order models are preferred over 1st order models for the same time response. This choice is dictated by the fact that for the same time response, the transient stress in the actuators will be weaker with 2nd or 4th order models as a result of smoother response slope at the beginning of the transient.

The effect of demanding performance in regulation upon the sensitivity functions and as a consequence upon the robustness of the closed-loop systems has been discussed in detail in Chap. 8. The specific choice of the closed-loop poles and of the fixed parts of the controller have also been discussed in Chap. 8. We will recall here just some of the basic facts.

The closed-loop performance will be mainly defined by the dominant closed-loop poles. However, it is advisable to assign all the poles and to avoid poles at $z = 0$ (which corresponds to the default location of non-assigned poles). Concerning the choice of the fixed part of the controller (the filter $H_R(q^{-1})$ and $H_S(q^{-1})$), we recall that they allow:

- to perfectly reject disturbances with known models in steady state ($H_S(q^{-1})$),
- to open the loop or to reduce the gain of the controller at certain frequencies ($H_R(q^{-1})$),
- to shape the sensitivity functions.

In order to avoid an undesirable stress on the actuator, a low value of the modulus of the input sensitivity function is needed (i.e., low gain of the controller) at the frequencies where the system to be controlled has a very low gain. Typical situations concern:

(a) low system gain at high frequencies near $0.5 f_s$;
(b) low system gain at the frequencies corresponding to oscillatory zeros (deep zeros).

In order to reduce the gain of the controller at these frequencies a filter $H_R(q^{-1})$ should be introduced. For case (a), one generally uses $H_R(q^{-1}) = 1 + \beta q^{-1}$ or $(1+\beta q^{-1})^2$, $0.7 < \beta \leq 1$ and for case (b), one uses $H_S(q^{-1}) = 1 + \beta_1 q^{-1} + \beta_2 q^{-2}$ with a corresponding natural frequency ($\omega_0$) and damping factor ($\zeta$) (see Chap. 8 for details).

## 16.3 The Parameter Adaptation Algorithm

In Chap. 10, we have examined in detail the various forms of parameter adaptation which include: data filtering, data normalization, parameter projection, dead-zone. All of these options may not be necessary for a specific application. It is, however, possible to give a general formula (for integral adaptation) which includes the various options. To do this, we need to recall the definition of filtered and normalized variables.

1. Filtered variables:

$$\begin{aligned} y_f(t) &= L(q^{-1})y(t), \qquad u_f(t) = L(q^{-1})u(t) \\ \phi_f(t) &= L(q^{-1})\phi(t) \end{aligned} \tag{16.19}$$

   Filtering is used in order to remove steady state components (very low frequencies) and high-frequency components beyond he desired bandwidth from the input-output data. The filters can also enhance the signal in the critical frequency region for design (in order to improve the quality of the estimated model at these frequencies). As indicated in Chap. 9, using dedicated algorithms for identification in closed loop, filtering may not be necessary or can be simplified.

2. Normalized filtered variables

$$\bar{y}_f(t) = \frac{y_f(t)}{m(t)}, \qquad \bar{u}_f(t) = \frac{u_f(t)}{m(t)}, \qquad \bar{\phi}_f(t) = \frac{\phi_f(t)}{m(t)} \tag{16.20}$$

where:[1]

$$m^2(t) = \mu^2 m^2(t-1) + \max(\|\phi_f(t)\|^2, 1)$$
$$m(0) = 1,\ 0 < \mu < 1 \tag{16.21}$$

or:

$$m(t) = 1 + \|\phi_f(t)\| \tag{16.22}$$

Normalization is introduced in the presence of unmodeled dynamics. $\|\bar{\phi}(t)\| \leq 1$ for all $t$ and has as effect the reduction of the adaptation speed.

3. Parameter adaptation algorithm (integral adaptation)

$$\hat{\theta}(t+1) = \hat{\theta}(t) + \alpha_1(t)\frac{F(t)\bar{\phi}_f(t)\bar{\varepsilon}^0(t+1)}{1+\bar{\phi}_f^T(t)F(t)\bar{\phi}_f(t)}$$
$$= \hat{\theta}(t) + \alpha_1(t)\frac{F(t)\phi_f(t)\varepsilon^0(t+1)}{m^2(t)+\phi_f^T(t)F(t)\phi_f(t)} \tag{16.23}$$

$$\varepsilon^0(t+1) = y_f(t+1) - \hat{\theta}^T(t)\phi_f(t) \tag{16.24}$$

$$F(t+1) = \frac{1}{\lambda_1(t)}\left[F(t) - \alpha_1(t)\frac{F(t)\bar{\phi}_f(t)\bar{\phi}_f^T(t)F(t)}{\frac{\lambda_1(t)}{\lambda_2(t)} + \bar{\phi}_f^T(t)F(t)\bar{\phi}_f(t)}\right] + \alpha_2(t)Q$$
$$= \frac{1}{\lambda_1(t)}\left[F(t) - \alpha_1(t)\frac{F(t)\phi_f(t)\phi_f^T(t)F(t)}{\frac{\lambda_1(t)}{\lambda_2(t)}m^2(t) + \phi_f^T(t)F(t)\phi_f(t)}\right] + \alpha_2(t)Q$$
$$Q > 0,\ F(0) > 0,\ 0 < \lambda_1(t) \leq 1,\ 0 \leq \lambda_2(t) < 2 \tag{16.25}$$

Taking $\lambda_1(t) = \lambda_2(t)$, (16.25) becomes:

$$F(t+1) = \frac{1}{\lambda_1(t)}\left[F(t) - \alpha_1(t)\frac{F(t)\phi_f(t)\phi_f^T(t)F(t)}{m^2(t)+\phi_f^T(t)F(t)\phi_f(t)}\right] + \alpha_2(t)Q \tag{16.26}$$

In the above equations, $\alpha_1(t)$ and $\alpha_2(t)$ are scheduling variables (0 or 1) which will be defined subsequently.

For vanishing adaptation, one uses a decreasing adaptation gain (see Sect. 3.2.3), i.e., one takes $\lambda_1(t) \equiv 1$; $\lambda_2(t) > 0$. In order to maintain enough adaptation gain at the beginning of the adaptation, to forget the eventual bad initial conditions and to speed up adaptation, one uses a variable forgetting factor, i.e.:

$$\lambda_1(t) = \lambda_0\lambda_1(t-1) + 1 - \lambda_0, \quad 0 < \lambda_0 < 1,\ \lambda_1(0) < 1 \tag{16.27}$$

[1] One can also use a normalization filter with a unitary steady state gain. In this case, (16.21) becomes $m^2(t) = \mu^2 m^2(t-1) + (1-\mu^2)\max(\|\phi_f(t)\|^2, 1)$.

Typical values for $\lambda_0$ and $\lambda_1(0)$ are: $\lambda_0 = 0.5$ to 0.99, $\lambda_1(0) = 0.95$ to 0.99. Using (16.27), one has $\lim_{t\to\infty} \lambda_1(t) = 1$.

For non-vanishing adaptation, one has to assure the adaptation alertness to eventual parameter changes. This can be done using appropriate forms for the updating of the adaptation gain like:

- regularized constant trace algorithm;
- reinitialization of the adaptation gain (adding a positive definite matrix $Q$ at certain instants).

Taking $\lambda_1(t) = \lambda_2(t)$ the trace of the adaptation gain is kept constant (or at a desired value) by selecting $\lambda_1(t)$ as follows:

$$\lambda_1(t) = 1 - \frac{\operatorname{tr}[F(t)\phi_f(t)\phi_f^T(t)F(t)]}{\operatorname{tr} F(t)[m^2(t) + \phi_f^T(t)F(t)\phi_f(t)]} \tag{16.28}$$

To start the adaptation algorithm, one generally uses the variable forgetting factor (or decreasing adaptation gain) and one switches to the constant trace algorithm when $\operatorname{tr} F(t) \leq$ desired trace (see also Sect. 3.2.3). One should also assure that (regularization):

$$f_{\min} I_{n_p} \leq F(t) \leq f_{\max} I_{n_p}, \qquad 0 < f_{\min} \leq f_{\max} < \infty \tag{16.29}$$

where $n_p$ is the number of parameters to be adapted.

The scheduling variable $\alpha_2(t)$ is essentially used for the reinitialization of the adaptation gain. When parameter changes occur or are detected ($\alpha_2(t) = 1$).

### 16.3.1 Scheduling Variable $\alpha_1(t)$

For dead zone type adaptation algorithms, one has:

$$\alpha_1(t) = \begin{cases} 1 & |f[\varepsilon^0(t+1)]| > \Delta(t) \\ 0 & |f[\varepsilon^0(t+1)]| \leq \Delta(t) \end{cases} \tag{16.30}$$

where $\Delta(t)$ is a time-varying threshold and $f[\varepsilon^0(t+1)]$ is a function of the a priori adaptation error which definitions depend upon the various used control strategies (see Sects. 10.4, 11.6 and 12.3).

The scheduling variable $\alpha_1(t)$ is also used in order "to adapt only when the signals are relevant". In particular, one should avoid adapting when all the signals are constant. In such situations, the incoming information cannot improve the parameter adaptation process. A good test quantity for the relevance of the signals for adaptation is Clary and Franklin (1985):

$$\sigma(t) = \bar{\phi}_f^T(t)F(t)\bar{\phi}_f(t) \geq \sigma_0 \tag{16.31}$$

to which threshold $\sigma_0$ can be assigned. This threshold will depend on the residual noise in the system as well as on $\lambda_1$ and $\lambda_2$. The dependence on $\lambda_1$ and $\lambda_2$ will be illustrated next.

If the signals are constant, i.e., $\phi_f(t) = \phi_f(t-1) = \cdots = \phi_f(0)$ and $\lambda_1(t) = \lambda_2(t) = \lambda$, one gets:

$$F(t)^{-1} = \lambda^t F(0)^{-1} + \lambda \frac{1-\lambda^t}{1-\lambda} \bar{\phi}_f(0) \bar{\phi}_f^T(0) \tag{16.32}$$

from which one obtains (using the matrix inversion lemma):

$$\bar{\phi}_f^T(t) F(t) \bar{\phi}_f(t) = \frac{(1-\lambda)\bar{\phi}_f^T(0) F(0) \bar{\phi}_f(0)}{\lambda^t (1-\lambda) + \lambda \bar{\phi}_f^T(0) F(0) \bar{\phi}_f(0)} \tag{16.33}$$

and therefore for constant signals:

$$\lim_{t \to \infty} \bar{\phi}_f^T(t) F(t) \bar{\phi}_f(t) = \frac{1-\lambda}{\lambda} \tag{16.34}$$

Therefore, the threshold $\sigma_0$ will be defined as:

$$\sigma_0 = \beta \frac{1-\lambda}{\lambda}; \quad \beta > 1 \tag{16.35}$$

When using $\sigma(t)$ given by (16.31), the scheduling variable $\alpha_1(t)$ is defined as:

$$\alpha_1(t) = \begin{cases} 1 & \text{if } \sigma(t) > \sigma_0 \text{ over a finite interval} \\ 0 & \text{if } \sigma(t) \le \sigma_0 \text{ over a finite interval} \end{cases} \tag{16.36}$$

One can also use an average value of $\sigma(t)$ over a sliding horizon. Other indicators can also be used for defining the scheduling variable as a function of data. However, in all the cases, the selection of the threshold $\sigma_0$ is critical for the good operation of the adaptive control system. Unfortunately, the choice of the threshold is *problem dependent*. If the threshold is too low, one may get a drift in estimated parameters (particularly under the effect of constant disturbances applied from time to time), and if the threshold is too high, the parameter estimates will be incorrect because the adaptation will not operate.

A specific situation occurs when *filtered closed-loop output error algorithm* (F-CLOE) is used for parameter estimation. In this case, the filtering of the data is not necessary (d.c. components are removed by the filtering of the observation vector) and furthermore, the filtered observation vector will be null if excitation is not applied to the system. Therefore, when using this algorithm, the need for finding a threshold for starting or stopping the adaptation algorithm as a function of the richness of data is removed. This represents a significative advantage over the filtered open-loop identification algorithms currently used in adaptive control.

### 16.3.2 Implementation of the Adaptation Gain Updating—The U-D Factorization

The adaptation gain equation is sensitive to round-off errors. This problem is comprehensively discussed in Bierman (1977) where a U-D factorization has been developed to ensure the numerical robustness of the PAA. To this end, the adaptation gain matrix is rewritten as follows

$$F(t) = U(t)D(t)U^T(t) \tag{16.37}$$

where $U(t)$ is an upper triangular matrix with all diagonal elements equal to 1 and $D(t)$ is a diagonal matrix. This allows the adaptation gain matrix to remain positive definite so that the rounding errors do not affect the solution significantly.

As the reinitializing matrix sequence $\{\alpha_2(t)Q\}$ is not explicitly used in the proposed parameter adaptation algorithm, it can be omitted. Let

$$G(t) = D(t)V(t) \tag{16.38}$$

$$V(t) = U^T(t)\phi_f(t) \tag{16.39}$$

$$\beta(t) = m^2(t) + V^T(t)G(t) \tag{16.40}$$

$$\delta(t) = \frac{\lambda_1(t)}{\lambda_2(t)}m^2(t) + V^T(t)G(t) \tag{16.41}$$

Define:

$$\Gamma(t) = \frac{U(t)G(t)}{\beta(t)} = \frac{F(t)\phi_f(t)}{m^2(t) + \phi_f^T(t)F(t)\phi_f(t)} \tag{16.42}$$

The U-D factorization algorithm of the parameter adaptation gain is given below.

Initialize U(0) and D(0) at time t = 0, this provides the initial value of the adaptation gain matrix $F(0) = U(0)D(0)U^T(0)$. At time $t+1$, determine the adaptation gain $\Gamma(t)$ while updating $D(t+1)$ and $U(t+1)$ by performing the steps 1 to 6.

1. Compute $V(t) = U^T(t)\phi_f(t)$, $G(t) := D(t)V(t)$, $\beta_0 = m^2(t)$ and $\delta_0 = \frac{\lambda_1(t)}{\lambda_2(t)} \times m^2(t)$
2. For $j = 1$ to $n_p$ (number of parameters) go through the steps 3 to 5
3. Compute

$$\beta_j(t) := \beta_{j-1}(t) + V_j(t)G_j(t)$$

$$\delta_j(t) := \delta_{j-1}(t) + V_j(t)G_j(t)$$

$$D_{jj}(t) := \frac{\delta_{j-1}(t)}{\delta_j(t)\lambda_1(t)}D_{jj}(t)$$

$$\Gamma_j(t) := G_j(t)$$

$$M_j(t) := -\frac{V_j(t)}{\delta_{j-1}(t)}$$

4. If $j = 1$ then go to step 6 else for $i = 1$ to $j - 1$ go through step 5
5. Compute

$$U_{ij}(t+1) = U_{ij}(t) + \Gamma_i(t)M_j(t)$$
$$\Gamma_i(t) = \Gamma_i(t) + U_{ij}(t-1)G_j(t)\Gamma_j(t)$$

6. For $i = 1$ to $np$ do

$$\Gamma_i(t) = \frac{1}{\beta_{np}(t)}\Gamma_i(t) \tag{16.43}$$

A lower bound on the adaptation gain is simply obtained by maintaining the values of the elements of the diagonal matrix $D(t)$ above some specified threshold $d_0$ as follows:

$$d_i(t) = \begin{Bmatrix} d_0 \text{ or } d_i(t-1) & \text{if } d_i(t) \le d_0 \\ d_i(t) & \text{otherwise} \end{Bmatrix} \tag{16.44}$$

Notice that the implementation of such an algorithm is indeed simple to legitimate its use.[2]

## 16.4 Adaptive Control Algorithms

### *16.4.1 Control Strategies*

Various control strategies have been discussed in Chap. 7 for the case of known plant model parameters. Their use in adaptive control schemes has been studied in Chaps. 11, 12 and 13. We will briefly recall here the specificity of their use.

*Tracking and regulation with independent objectives* and *minimum variance tracking and regulation* require the knowledge of the plant delay and the stability of the plant model zeros. Often the later condition is not satisfied in practice. *Tracking and regulation with weighted input* and *generalized minimum variance tracking and regulation* allow to release the stability condition upon the plant model zeros. However, in the adaptive context the performances will not be guaranteed.

*Pole placement*, *generalized predictive control* and *linear quadratic control* can be used for the adaptive control of systems featuring unstable zeros.

The *partial state reference model* versions of the generalized predictive control and linear quadratic control have been introduced in order to incorporate in an easy way desired tracking capabilities. Partial state reference model control has been tested on a number of pilots plants involving thermal processes (M'Saad et al. 1987), chemical processes (Najim et al. 1982, 1994), flexible mechanical structures (M'Saad et al. 1993a; M'Saad 1994), fermentation processes (Queinnec et al. 1992).

[2]The function *udrls.m* (MATLAB) available from the websites: www.landau-adaptivecontrol.org and landau-bookIC.lag.ensieg.inpg.fr, implements this algorithm.

It has also been successfully used for several adaptive control benchmarks (M'Saad et al. 1989, 1990; M'Saad and Hejda 1994).

When large and rapid variations of the plant parameters occur or when a number of plant models for various regimes of operation are available, the *Multimodel Adaptive Control with Switching* is strongly recommended. It has been tested on several types of plants including a flexible transmission (Karimi and Landau 2000; Karimi et al. 2001), a simulated PH neutralization process (Böling et al. 2007), air traffic control tracking (Li and Bar-Shalom 2002) and control of F-8C aircraft (Athans et al. 2002).

## 16.4.2 Adaptive Control Algorithms

Adaptive control algorithms are obtained by combining robust parameter adaptation with controller parameter updating. In direct adaptive control, the implementation of the algorithm is done as follows:

Strategy 1:

1. sample the output;
2. update the controller parameters;
3. compute the control signal;
4. send the control signal;
5. wait for the next sample.

As indicated in Sect. 16.2, if the computation time for step 2 and 3 is too significative with respect to the sampling period, one uses the following strategy:

Strategy 2:

1. sample the output;
2. send the control signal (computed in the previous sampling period);
3. update the controller parameters;
4. compute the control signal;
5. wait for the next sample.

Using the Strategy 2, a delay of one sample is introduced in the system and this has to be taken into account for controller design.

For indirect adaptive control, the implementation of the algorithm is done as follows:

Strategy 1:

1. sample the plant output;
2. update the plant model parameters;
3. compute the controllers parameters based on the new plant model parameter estimates;
4. compute the control signal;

5. send the control signal;
6. wait for the next sample.

If the computation time for steps 2 and 3 is too long, one can use the following strategy:

Strategy 2:

1. sample the plant output;
2. compute the control signal based on the controller parameters computed during the previous sampling period;
3. send the control signal;
4. update the plant model parameters;
5. compute the controller parameters based on the new plant model parameter estimates;
6. wait for the next sample.

As indicated in Chap. 1, one can also estimate the parameters at each sampling, but update the controller parameters only every $N$ sampling instants.

## 16.5 Initialization of Adaptive Control Schemes

For a certain type of system to be controlled, the design of an adaptive controller will in general require an a priori system identification in one or several operating situations. The system identification will allow to set up:

- the structure of the controller,
- the performance specifications,
- the underlying linear design method.

Information about the operating conditions of the system and the type of parameter variations will allow the type of adaptive control scheme to be selected and the conditional updating of the adaptation gain to be set up. The basic choices are:

- identification in closed loop followed by controller redesign;
- adaptive control with vanishing adaptation gain and resetting;
- adaptive control with non-vanishing adaptation gain.

The first two schemes are used when the plant model parameters are constant over large time horizons and change significantly from time to time. When the plant model parameters vary almost continuously, adaptive control with non-vanishing gain should be used. Constant trace adaptation gain is suitable in these situations. The choice of the desired constant trace is a trade off between the speed of variation of the parameters and the level of noise. The test of the signal richness in these cases is necessary in general (an exception: use of F-CLOE algorithm). To start the adaptation algorithm, one combines variable forgetting factor algorithm with constant trace algorithm (see Sects. 16.3 and 3.2.3).

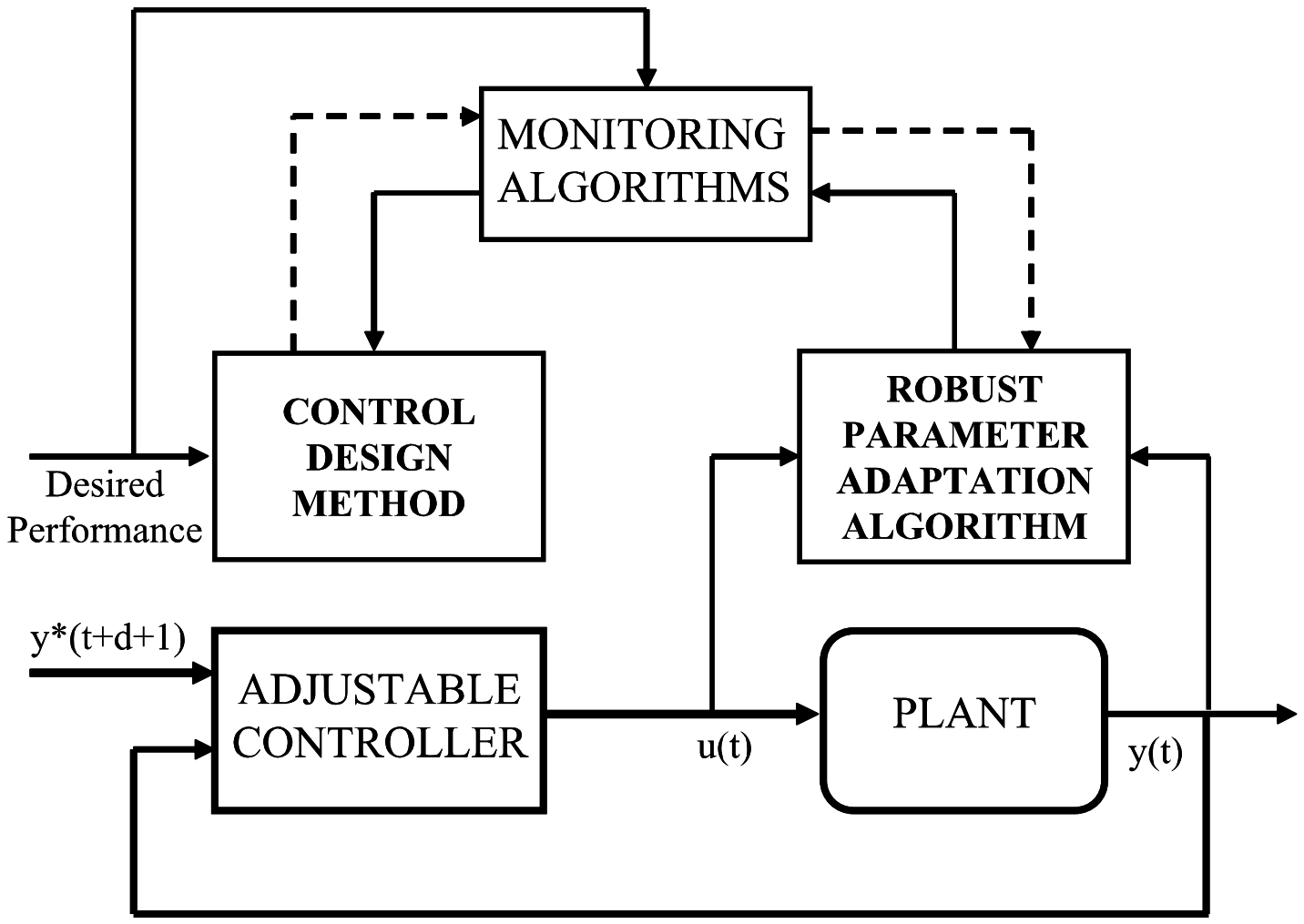

**Fig. 16.6** Adaptive control system with monitoring

To effectively start up an adaptive control scheme, one uses a parameter estimation in open loop or in closed loop (with a fixed parameters controller) using an external excitation signal over a short time horizon $\geq 10n_p$ (if the excitation is sufficiently rich), where $n_p$ is the number of parameters to be estimated and then one switches to the adaptive control scheme (with appropriate initialization of the controller as discussed in Sect. 16.2).

## 16.6 Monitoring of Adaptive Control Systems

From the examination of the various adaptive control schemes and parameter adaptation algorithms, one can conclude that in addition to a number of design choices which have to be made when adaptive control is set up, one has to monitor in real time a number of aspects in order to assure a correct operation of the adaptive control system. The basic situations which have to be monitored are:

1. the admissibility of the estimated plant model with respect to controller design,
2. the parameter adaptation.

The admissibility problem is discussed in Chap. 12 and solutions for removing the eventual singularities are proposed. Parameter projection is another way to remove singularities if an admissible parameter domain can be defined. Other ad-hoc solutions can be considered (like: do not modify the controller and eventually add an excitation signal to improve the parameter estimation).

Monitoring of the parameter adaptation is much more sensitive. Two basic problems have to be considered:

- Reinitialization of the adaptation gain when large parameter variations occur (scheduling variable $\alpha_2(t)$ in (16.25)). This requires a detection procedure.
- Freezing the adaptation when the signals are not likely to improve the parameter estimation (scheduling variable $\alpha_1(t)$ in (16.23) and (16.25)).

While the monitoring is implemented in the algorithms, one has conceptually a third hierarchical level in an adaptive control system, i.e.,

Level 1: Conventional feedback control.
Level 2: Adaptation loop.
Level 3: Monitoring of the adaption loop.

This is illustrated in Fig. 16.6.

## 16.7 Concluding Remarks

This chapter has summarized the techniques and design issues to be considered for a successful implementation of adaptive control schemes.

The techniques provided throughout the book allow us to build robust adaptive control schemes with respect to unmodeled dynamics and disturbances. However, their performance will depend on the choice of some parameters and in particular on the design of the underlying linear control. The influence of some of the key design parameters has been illustrated.

# Appendix A
# Stochastic Processes

**Probability Space** The outcome of a phenomenon will be denoted by $\xi$. $\xi$ belongs to a set $\Omega$ which contains all the possible outcomes of a phenomenon. Therefore, $\Omega$ is the *sure event*. One denotes the *impossible event* by $\phi$. The set of events is denoted by $\mathcal{F}$. A probability measure $P$ is associated to $\Omega$, i.e., for subsets of $\Omega$ probability of occurrence is assigned. The triplet $(\Omega, \mathcal{F}, P)$ is called a *probability space*.

A $\sigma$-algebra $\mathcal{F}$ is a set of subsets of $\Omega$ which contains the empty set $\phi$ and is closed under complements and countable unions.

A random variable generates a $\sigma$-algebra $\mathcal{F}$ since all possible outcomes define subsets in the event space $\Omega$.

**Axioms of Probability** Consider two events $a$ and $b$ belonging to $\Omega$. To each event one assigns a probability $P(\cdot)$ satisfying the following axioms:

1. $P(a) \geq 0;\ \forall a \in \Omega$
2. $P(\Omega) = 1$
3. $P(a \cup b) = P(a) + P(b)$ if $a \cap b = \emptyset$

**Random Variable** A *random variable* $X$ is a function from the *event space* $\Omega$ to $R^1$ and will be denoted by $X(\xi)$ with $\xi \in \Omega$. (The argument will be omitted in general.)

**Induced Probability** Since to each subset (event) of $\Omega$, a probability has been assigned and since $x$ is a function of $\xi$, there is a probability that $X$ takes a value in a subset of $R^1$, i.e.:

$$P\{X(\xi) \leq x_1\} \quad \text{or} \quad P\{x_1 \leq X(\xi) \leq x_2\}$$

**Distribution Function**

$$F_X(x) = P\{X \leq x\}; \quad -\infty < x < \infty$$

is called the *distribution function*.

**Probability Density Function**

$$F_X(x) = P\{X \leq x\} = \int_{-\infty}^{x} f_X(x)dx$$

where $f_X(x)$ is the *probability density function* (i.e., $F_X(x)$ is continuous and the derivatives with respect to $x$ exist).

**Gaussian (Normal) Distribution**

$$f_X(x) = \frac{1}{\sigma\sqrt{2\pi}} e^{-x^2/2\sigma^2}$$

**Expectation (Mean Value) of a Random Variable**

$$m(X) = \mathbf{E}\{X(\xi)\} = \int_{-\infty}^{\infty} f_X(x)dx$$

**Covariance of a Random Variable**

$$\operatorname{cov} X = \mathbf{E}[X - m(x)]^2 = \mathbf{E}\{[X(\xi) - m(x)]^2\}$$

**Independent Random Variables** Two random variables $X$ and $Y$ are characterized by the *joint distribution function*

$$F_{XY}(xy) = P\{X \le x; Y \le y\}$$

The random variables $X$ and $Y$ are called *independent* if:

$$F_{XY}(x, y) = F_X(x)F_X(y) = P\{X \le x\}P\{Y \le y\}$$

**Uncorrelated Random Variables** Two random variables $X$ and $Y$ are uncorrelated if:

$$\mathbf{E}\{XY\} = \mathbf{E}\{X\}\mathbf{E}\{Y\}$$

Note: independence implies uncorrelation

**Orthogonal Random Variables** Two random variables $X$ and $Y$ are orthogonal if:

$$\mathbf{E}\{XY\} = 0$$

**Sequence of Independent Random Variables** Given $n$ real random variables $X_1, X_2, \ldots, X_n$ they are independent if:

$$P\{X_1 \le x_1, \ldots, X_n \le x_n\} = P\{X_1 \le x_1\}, \ldots, P\{X_n \le x_n\}$$

**Discrete-Time Stochastic Process** A *discrete-time stochastic process* is defined as a sequence of random variables defined on a common probability space and indexed by the discrete time $t = 0, 1, 2 \ldots$. A stochastic process $x$ is a function of $t$ and $\xi$: $X(t, \xi)$. For a fixed $t$, one obtains a random variable. For each $\xi$, one obtains a *realization* of the stochastic process.

**Mean Value of a Stochastic Process**

$$m_X(t) = \mathbf{E}\{X(t, \xi)\}$$

**Covariance Function**

$$R_{XY}(t, l) = \mathbf{E}\{[X(t) - m_X(t)][Y(l) - m_Y(l)]\}$$

**Discrete-Time (Gaussian) White Noise $\{e(t)\}$** It is a sequence of independent equally distributed (normal) random variables of zero mean ($m(x) = 0$) and variance $\sigma^2$. Often the sequence will be denoted $\{e(t)\}$ and will be characterized by the parameters $(0, \sigma)$ where 0 corresponds to zero mean and $\sigma$ is the standard deviation (square root of the variance). This sequence has the following properties:

1. $\mathbf{E}\{e(t)\} = 0$
2. $R(t,l) = \mathbf{E}\{e(t)e(l)\} = \begin{cases} \sigma^2 & l = t \\ 0 & l \neq t \end{cases}$

**Weakly Stationary Stochastic Processes** They are characterized by:

1. $m_X(t) = m_X$
2. $R_{XX}(t,l) = R_{XX}(t-l)$

and the Fourrier transform of $R_{XX}$ can be defined:

$$\phi_X(\omega) = \frac{1}{2\pi} \sum_{t=-\infty}^{t=\infty} R_{XX}(t) e^{-j\omega t}$$

which is called the *spectral density function*. For the discrete-time white noise:

$$\phi_e(\omega) = \frac{\sigma^2}{2\pi}$$

**Conditional Probability** Given an event $b$ with non zero probability ($P(b) > 0$), the conditional probability of the event $a$ assuming that $b$ has occurred is defined by:

$$P(a|b) = \frac{P(a \cap b)}{P(b)}$$

This definition is extended to events $x_1, x_2, \ldots, x_n \in \mathcal{F}$ and more generally to $\mathcal{F}$.

**Conditional Expectation**

$$\mathbf{E}\{a|b\} = \frac{\mathbf{E}\{a \cap b\}}{\mathbf{E}\{b\}}$$

This can be extended to the case where $b$ is the $\sigma$-algebra $\mathcal{F}_t$ generated by the sequence of random variables up to and including $t$. The conditional expectation of $X$ with respect to $\mathcal{F}_t$ is written: $\mathbf{E}\{X|\mathcal{F}_t\}$. An increasing $\sigma$-algebra $\mathcal{F}_t$ has the property: $\mathcal{F}_{t-1} \subset \mathcal{F}_t$.

**Martingale Sequence** Let $(\Omega, \mathcal{F}, P)$ be a probability space and suppose that for each $t$ there is a sub-$\sigma$-algebra $\mathcal{F}$ of $\mathcal{F}$ such that $\mathcal{F}_{t-i} \subset \mathcal{F}_t$ for $i \geq 0$. Then the sequence $X(t)$ is said to be adapted to the sequence of increasing $\sigma$-algebras $\mathcal{F}_t$ if $X(t)$ is $\mathcal{F}_t$ measurable for every $t$ (every stochastic process is adapted to its own past).

Under the above conditions $X(t)$ is a *martingale* sequence provided that:

1. $\mathbf{E}\{|X(t)|\} < \infty$ almost sure
2. $\mathbf{E}\{X(t+1)|\mathcal{F}_t\} = 0$

**Convergence w.p.1 (a.s.)** If $X(t,\xi)$ is a sequence of random variables defined on a probability space $(\Omega, \mathcal{F}, P)$ such that:

$$\text{Prob}\left\{\lim_{t\to\infty} X(t) = x^*\right\} = 1$$

we say that: "$X(t)$ converges to $x^*$ with probability 1 (w.p.1)" or "$X(t)$ converges to $x^*$ almost sure (a.s.)".

**Ergodic Stochastic Process (in the Mean)** For any outcome $\xi$ (with the exception of a subset with zero probability) one has:

$$P\left\{\lim_{N\to\infty}\frac{1}{N}\sum_{t=1}^{N}X(t,\xi)=\mathbf{E}\{X(t)\}\right\}=1$$

For ergodic stochastic process time average replaces set average.

**Spectral Factorization Theorem** (Åström 1970) Consider a stationary stochastic process with a spectral density $\phi(e^{j\omega})$ which is rational in $\cos\omega$.

1. There exists a linear system with the pulse transfer function

$$H(z^{-1})=\frac{C(z^{-1})}{D(z^{-1})}$$

with poles in $|z|<1$ and zeros in $|z|\leq 1$ such that:

$$\phi(e^{j\omega})=H(e^{-j\omega})H(e^{j\omega})$$

2. The spectral density of the output of $H(z^{-1})$ when the input is a discrete-time Gaussian white noise is a stationary stochastic process with spectral density $\phi(e^{j\omega})$.

*Remarks*

1. Without the restriction on poles and zeros the factorization of $\phi(e^{j\omega})$ is not unique.
2. Spectral factorization theorem allows to consider all the stationary stochastic processes with a rational spectral density as being generated by an asymptotically stable linear system driven by a white noise.
3. Positive real lemma (see Appendix D) plays an important role in solving the factorization problem (see Faurre et al. 1979).

**Innovation Process** A consequence of the factorization theorem is that a stochastic process can be expressed as:

$$y(t+1)=f[y(t),y(t-1),\ldots]+h(0)e(t+1)$$

where $e(t+1)$ is a white noise and $h(0)$ is the coefficient of the impulse response for $l=t+1-i=0$ ($y(t+1)=\sum_{i=-\infty}^{t+1}h(t+1-i)e(i)$). Therefore $y(t+1)$ can be expressed as a sum of two terms, a term which depends only on the past (measurable) quantities and a pure stochastic term which is the "true" new information unpredictable at $t$. This formulation can be further extended for $y(t+d)$, i.e.

$$y(t+d)=f_y[y(t),y(t-1),\ldots]+f_e[e(t+d),e(t+d-1),\ldots,e(t+1)]$$

The term $f_e[e(t+d),e(t+d-1)\ldots]$ is called the *innovation*. As a consequence:

$$\mathbf{E}\{y(t+d)|y(t),y(t-1),\ldots\}=f_y[y(t),y(t-1),\ldots]$$
$$\mathbf{E}\{f_y\cdot f_e\}=0$$

The innovation representation plays an important role in prediction problems.

# Appendix B
# Stability

We will consider free (unforced) discrete-time systems of the form:

$$x(t+1) = f[x(t), t] \tag{B.1}$$

where $x(t)$ denotes the $n$-dimensional state vector of the system. In the linear time invariant case (B.1) becomes:

$$x(t+1) = Ax(t) \tag{B.2}$$

The solutions of the system (B.1) will be denoted by $x(t, x_0, t_0)$ where $x_0$ is the initial condition at time $t_0$.

**Definition B.1** A state $x_e$ of the system (B.1) is an *equilibrium state* if:

$$f(x_e, t) = 0; \quad \forall t \geq t_0$$

**Definition B.2** (Stability) An equilibrium state $x_e$ of the system (B.1) is *stable* if for arbitrary $t_0$ and $\varepsilon > 0$, there exists a real number $\delta(\varepsilon, t_0)$ such that:

$$\|x_0 - x_e\| \leq \delta(\varepsilon, t_0) \quad \Longrightarrow \quad \|x(t, x_0, t_0) - x_e\| \leq \varepsilon; \quad \forall t \geq t_0$$

**Definition B.3** (Uniform Stability) An equilibrium state $x_e$ of the system (B.1) is *uniformly stable* if it is stable and if $\delta(\varepsilon, t_0)$ does not depend on $t_0$.

**Definition B.4** (Asymptotic Stability) An equilibrium state $x_e$ of the system (B.1) is *asymptotically stable* if:

1. $x_e$ is stable
2. there exists a real number $\delta(t_0) > 0$ such that:

$$\|x_0 - x_e\| \leq \delta(t_0) \quad \Longrightarrow \quad \lim_{t \to \infty} \|x(t, x_0, t_0) - x_e\| = 0$$

**Definition B.5** (Uniformly Asymptotically Stable) An equilibrium state $x_e$ of the system (B.1) is *uniformly asymptotically stable* if:

1. $x_e$ is uniformly stable
2. there exists a real number $\delta > 0$ independent of $t_o$ such that:

$$\|x_0 - x_e\| \leq \delta \quad \Longrightarrow \quad \lim_{t\to\infty} \|x(t, x_0, t_0) - x_e\| = 0$$

**Definition B.6** (Global Asymptotic Stability) An equilibrium state $x_e$ of the system (B.1) is *globally asymptotically stable* if for all $x_0 \in R_n$:

1. $x_e$ is stable
2. $\lim_{t\to\infty} \|x(t, x_0, t_0) - x_e\| = 0$

If, in addition, $x_e$ is uniformly stable then $x_e$ is *uniformly globally asymptotically stable*.

**Theorem B.1** (Lyapunov) *Consider the free dynamic system* (B.1) *with*:

$$f(0,t) = 0; \quad -\infty < t < +\infty \ (x_e = 0 \textit{ is an equilibrium state})$$

*If there exists a real scalar function* $V(x,t)$ *such that*:

1. $V(0,t) = 0 \ -\infty < t < \infty.$
2. $V(x,t) \geq \alpha(\|x\|) > 0, \ \forall x \neq 0, x \in R_n$ *and* $\forall t$ *where* $\alpha(\cdot)$ *is a continuous non-decreasing scalar function such that* $\alpha(0) = 0$ ($V(x,t)$ *is positive definite*).
3. $V(x,t) \to \infty$ *with* $\|x\| \to \infty$ ($V(x,t)$ *is radially unbounded*).
4. $\Delta V(x,t) = V(x,t+1) - V(x,t) \leq -\mu(\|x\|) < 0, \ \forall x \neq 0, \ x \in R_n, \ \forall t$ *along the trajectories of the system* (B.1) *where* $\mu(\cdot)$ *is a continuous non-decreasing scalar function such that* $\mu(0) = 0$. ($\Delta V(x,t)$ *is negative definite.*) *Then the equilibrium state* $x_e = 0$ *is* globally asymptotically stable *and* $V(x,t)$ *is a* Lyapunov function *for the system* (B.1).
   *If, in addition*:
5. $V(x,t) \leq \beta(\|x\|)$ *where* $\beta(\cdot)$ *is a continuous, non-decreasing scalar function such that* $\beta(0) = 0$, *then* $x_e = 0$ *is* uniformly globally asymptotically stable.

**Corollary B.1** *If in Theorem* B.1, $V(x,t)$ *is replaced by* $V(x)$, *the equilibrium state* $x_e = 0$ *of the autonomous system*:

$$x(t+1) = f[x(t)] \tag{B.3}$$

*is* globally uniformly asymptotically stable *if*:

1. $V(0) = 0$;
2. $V(x) > 0, \ \forall x \neq 0, \ x \in R_n$;
3. $V(x) \to \infty$ *with* $\|x\| \to \infty$;
4. $\Delta V[x(t)] = V[x(t+1)] - V[x(t)] < 0$ *along the trajectories of* (B.3) *for all* $x \neq 0, \ x \in R_n$.

**Corollary B.2** *In Corollary* B.1, *condition* 4 *can be replaced by*:

1. $\Delta V(x) \leq 0, \ \forall x \neq 0, \ x \in R_n$;
2. $\Delta V(x)$ *is not identically zero along any trajectory of* (B.3).

**Theorem B.2** (Lyapunov) *The equilibrium state $x_e = 0$ of a linear time invariant free system*:

$$x(t+1) = Ax(t) \tag{B.4}$$

*is globally uniformly asymptotically stable if and only if, given a positive definite matrix $Q$, there exists a positive definite matrix $P$ which is the unique solution of the matrix equation*:

$$A^T PA - P = -Q \tag{B.5}$$

Consider the system:

$$x(t+1) = f[x(t), u(t)] \tag{B.6}$$

$$y(t) = h[x(t), t] \tag{B.7}$$

**Definition B.7** The system is said to be weakly finite gain stable if:

$$\|y(t)\|_{2T} \leq \kappa \|u(t)\|_{2T} + \beta[x(t_0)] \tag{B.8}$$

where $x(t_0) \in \Omega$, $0 < \kappa < \infty$, $|\beta[x(t_0)]| < \infty$ and $\|y(t)\|_{2T} = \sum_{t=t_0}^{T} y^2(t)$.

Consider the system:

$$x(t+1) = Ax(t) + Bu(t) \tag{B.9}$$

$$y(t) = Cx(t) + Du(t) \tag{B.10}$$

**Lemma B.1** (Goodwin and Sin 1984) *Provided that the following conditions are satisfied*:

(i) $|\lambda_i(A)| \leq 1;\ i = 1, \dots, n$.
(ii) *All controllable modes of* $(A, B)$ *are inside the unit circle.*
(iii) *Any eigenvalues of $A$ on the unit circle have a Jordan block of size* 1.

*Then*:

(a) *There exist constants $K_1$ and $K_2$ with $0 \leq K_1 < \infty$ and $0 \leq K_2 < \infty$, which are independent of $N$ such that*:

$$\sum_{t=1}^{N} \|y(t)\|^2 \leq K_1 \sum_{t=1}^{N} \|u(t)\|^2 + K_2; \quad \forall N \geq 1 \tag{B.11}$$

(*the system is finite gain stable*).

(b) *There exist constants $0 \leq m_3 < \infty$ and $0 \leq m_4 < \infty$ which are independent of $t$ such that*:

$$|y(t)| \leq m_3 + m_4 \max_{1 \leq \tau \leq N} \|u(\tau)\|; \quad \forall 1 \leq t \leq N \tag{B.12}$$

**Lemma B.2** *If the system* (B.9), (B.10) *is a minimal state space realization and* $|\lambda_i(A)| < 1;\ i = 1, \ldots, n$, *then*:

$$\lim_{N\to\infty} \frac{1}{N} \sum_{t=1}^{N} \|y(t)\|^2 \leq K_1 \lim_{N\to\infty} \frac{1}{N} \sum_{t=1}^{N} \|u(t)\|^2 \tag{B.13}$$

For definitions and proofs concerning the stability of discrete time systems see Kalman and Bertram (1960), Mendel (1973).

# Appendix C
# Passive (Hyperstable) Systems

## C.1 Passive (Hyperstable) Systems

The concept of passive (hyperstable) systems emerged in control through the seminal work of Popov, Kalman and Yakubovitch (the Popov-Kalman-Yakubovitch lemma) in connection with the problem of stability of nonlinear feedback systems (Popov 1963, 1964). The concept of *passive* (hyperstable) *system* can be viewed also as a particularization of the concept of *positive dynamic systems*. The term *positive dynamic system* is an extension of the mathematical concept of *positivity*. This is a necessary and sufficient condition for a mathematical object to be factorizable as a product (for example a *positive definite matrix* $Q$ can be factored as $Q = LL^T$).

The study of passive (hyperstable) systems and their relations on the one hand with positive dynamic systems and on the other hand with the properties of passive network have been initiated in Popov (1964). The terms "hyperstable" has been coined to describe some input-output properties of a system which appears to be a generalization of the property of passive systems, which requires that at every moment the stored energy be equal or less than the initial energy stored in the system, plus the energy supplied to the system's terminal over the considered time horizon. (This generalization has been done more or less in the same way as for Lyapunov functions which are positive definite functions, and for a particular state representation they can be related with the energy of the system.) The main use of this property was to treat stability problems related to the interconnection of hyperstable systems. A deep treatment of the subject can be found in Popov (1966, 1973) as well as in Anderson and Vongpanitlerd (1972). Through the work of Willems (1971), Desoer and Vidyasagar (1975), the term *passive* systems (instead of hyperstability) became widely accepted, even if in this context, *passive* is not directly related to the energy supplied to a system. References for this subject include Zames (1966), Anderson (1967), Hitz and Anderson (1969), Hill and Moylan (1980), Faurre et al. (1979).

## C.2 Passivity—Some Definitions

The norm of a vector is defined as:

$$\|x(t)\| = (x^T x)^{1/2}; \quad x \in R_n$$

The norm $L_2$ is defined as:

$$\|x(t)\|_2 = \left( \sum_0^\infty x^T(t)x(t) \right)^{1/2}$$

where $x(t) \in \mathbb{R}_n$ and $t$ is an integer (it is assumed that all signals are 0 for $t < 0$). To avoid the assumption that all signals go to zero as $t \to \infty$, one uses the extended $L_2$ space denoted $L_{2e}$ which contains the *truncated* sequences:

$$x_T(t) = \begin{cases} x(t) & 0 \le t \le T \\ 0 & t > T \end{cases}$$

Consider a system $S$ with input $u$ and output $y$ (of same dimension). Let us define the input-output product:

$$\eta(0, t_1) = \sum_{t=0}^{t_1} y^T(t)u(t)$$

**Definition C.1** A system $S$ is termed *passive* if:

$$\eta(0, t_1) \ge -\gamma^2; \quad \gamma^2 < \infty; \ \forall t_1 \ge 0$$

**Definition C.2** A system $S$ is termed *input strictly passive* if:

$$\eta(0, t_1) \ge -\gamma^2 + \kappa\|u\|_{2T}^2; \quad \gamma^2 < \infty; \ \kappa > 0; \ \forall t_1 \ge 0$$

**Definition C.3** A system $S$ is termed *output strictly passive* if:

$$\eta(0, t_1) \ge -\gamma^2 + \delta\|y\|_{2T}^2; \quad \gamma^2 < \infty; \ \delta > 0; \ \forall t_1 \ge 0$$

**Definition C.4** A system $S$ is termed *very strictly passive* if:

$$\eta(0, t_1) \ge -\gamma^2 + \kappa\|u\|_{2T}^2 + \delta\|y\|_{2T}^2; \quad \gamma^2 < \infty; \ \delta > 0; \ \kappa > 0; \ \forall t_1 \ge 0$$

In all the above definitions, the term $\gamma^2$ will depend upon the initial conditions. If a state space representation (state vector: $x$) can be associated to the system $S$, one can introduce also the following lemmas:

**Lemma C.1** *The system resulting from the parallel connection of two passive blocks is passive.*

**Lemma C.2** *The system resulting from the negative feedback connection of two passive blocks is passive.*

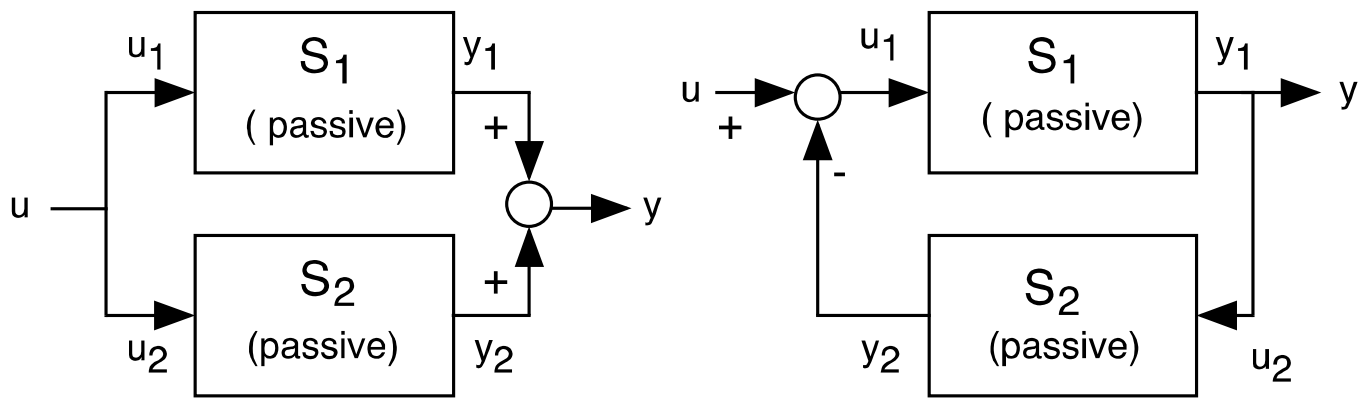

**Fig. C.1** Parallel and feedback connections of two passive blocks

*Proof* The resulting systems are illustrated in Fig. C.1. The proof is based on the observation that in the *parallel* connection, one has for the resulting system (input $u$, output $y$):

$$u = u_1 = u_2; \qquad y = y_1 + y_2$$

and in the case of *feedback* connection, one has:

$$u = u_1 + y_2; \qquad y = y_1 = u_2$$

where $u_j, y_j, j = 1, 2$ are the input and output of the elementary blocks. The above observations allow to write for the resulting system:

$$\begin{aligned}\eta(0, t_1) &= \sum_{t=0}^{t_1} y^T(t)u(t) \\ &= \sum_{t=0}^{t_1} y_1^T(t)u_1(t) + \sum_{t=0}^{t_1} y_2^T(t)u_2(t) \geq -\gamma_1^2 - \gamma_2^2; \quad \gamma_1, \gamma_2 < \infty\end{aligned}$$ □

**Definition C.5** A symmetric square matrix $P$ is termed positive (semidefinite) definite if:

$$x^T Px(\geq) > 0; \quad \forall x \neq 0; \ x \in R^n$$

A positive definite matrix $P$ can be expressed as: $P = LL^T$ with $L$ a nonsingular matrix.

**Definition C.6** A matrix function $H(z)$ of the complex variable $z$ is a *Hermitian matrix* (or simply *Hermitian*) if:

$$H(z) = H^T(z^*)$$

(where $*$ means *conjugate*).

*Hermitian* matrices feature several properties, including:

1. A *Hermitian* matrix is a square matrix and the diagonal terms are real.
2. The eigenvalues of a *Hermitian* matrix are always real.
3. If $H(z)$ is a *Hermitian* matrix and $x$ is a vector with *complex* components, the quadratic form $x^T Hx^*$ is always real.

**Definition C.7** A Hermitian matrix is positive (semidefinite) definite if:

$$x^T H x^*(\geq) > 0; \quad \forall x \neq 0; \ x \in C^n$$

## C.3 Discrete Linear Time-Invariant Passive Systems

Consider the linear time invariant discrete time system:

$$x(t+1) = Ax(t) + Bu(t) \tag{C.1}$$

$$y(t) = Cx(t) + Du(t) \tag{C.2}$$

where $x$ is a $n$-dimensional state vector, $u$ and $y$ are m-dimensional vectors representing the input and the output respectively and $A, B, C, D$ are matrices of appropriate dimension. One assumes that $[A, B, C, D]$ is a minimal realization of the system. The system of (C.1) and (C.2) is also characterized by the square transfer matrix:

$$H(z) = D + C(zI - A)^{-1}B \tag{C.3}$$

**Definition C.8** A $m \times m$ discrete matrix $H(z)$ of real rational functions is *positive real* if and only if:

1. All elements of $H(z)$ are analytic on $|z| > 1$ (i.e., they do not have poles in $|z| > 1$).
2. The eventual poles of any element of $H(z)$ on the unit circle $|z| = 1$ are simple and the associated residue matrix is a positive semidefinite Hermitian.
3. The matrix $H(e^{j\omega}) + H^T(e^{-j\omega})$ is a positive semidefinite Hermitian for all real values of $\omega$ which are not a pole of any element of $H(e^{j\omega})$.

In the case of a scalar transfer function $H(z)$ condition 3 is replaced by:

$$\mathrm{Re}\, H(z) \geq 0; \quad \forall |z| = 1$$

**Definition C.9** The system (C.1)–(C.2) is termed *hyperstable* if:

$$\eta(0, t_1) \geq \beta_1 \|x(t_1 + 1)\|^2 - \beta_0 \|x(0)\|^2; \quad \beta_1, \beta_0 > 0; \ \forall t_1 \geq 0$$

and for all $u(t)$ and $x(t)$ that satisfy (C.1).

The inequality considered in Definition C.9 is a generalization of the passivity condition from physics. With an appropriate choice of the state vector, the stored energy at $t_1 + 1$ may be expressed as $\beta \|x(t_1 + 1)\|^2$ with $\beta > 0$ and $\eta(0, t_1)$ is the energy supplied to the system's terminals over $(0, t_1)$. Passivity requires that at every moment the stored energy be equal to or less than the initial energy plus the energy supplied to the system.

Hyperstability implies passivity in the sense of Definition C.1 ($\gamma = \beta_0 \|x(0)\|^2$). The converse is also true under the minimality assumption on the system (C.1)–(C.2).

**Lemma C.3** (Positive real lemma) *The following propositions concerning the system of* (C.1) *and* (C.2) *are equivalent to each other*:

1. $H(z)$ *given by* (C.3) *is a positive real discrete time transfer matrix.*
2. *The system* (C.1)–(C.2) *is* passive.
3. *There is a positive definite matrix* $P$, *a positive semidefinite matrix* $Q$ *and matrices* $S$ *and* $R$ *such that*:

$$A^T PA - P = -Q \tag{C.4}$$

$$C - B^T PA = S^T \tag{C.5}$$

$$D + D^T - B^T PB = R \tag{C.6}$$

$$M = \begin{bmatrix} Q & S \\ S^T & R \end{bmatrix} \geq 0 \tag{C.7}$$

4. *There is a positive definite matrix* $P$ *and matrices* $K$ *and* $L$ *such that*:

$$A^T PA - P = -LL^T \tag{C.8}$$

$$C - B^T PA = K^T L^T \tag{C.9}$$

$$D + D^T - B^T PB = K^T K \tag{C.10}$$

5. *Every solution* $x(t)$, $(x(0), u(t))$ *of* (C.1), (C.2) *satisfies the following equality*:

$$\sum_{t=0}^{t_1} y^T(t)u(t) = \frac{1}{2}x^T(t_1+1)Px(t_1+1) - \frac{1}{2}x^T(0)Px(0) + \frac{1}{2}\sum_{t=0}^{t_1}[x^T(t), u^T(t)]M\begin{bmatrix} x(t) \\ u(t) \end{bmatrix} \tag{C.11}$$

6. *The system* (C.1)–(C.2) *is* hyperstable.

*Proof* Detailed proof can be found in Hitz and Anderson (1969), Popov (1973). Property (2) results immediately from property (5) ($\gamma^2 = -\frac{1}{2}x^T(0)Px(0)$). Equivalence between (3) and (4) results immediately from (C.7), which taking into account the properties of positive semidefinite matrices can be expressed as:

$$\begin{bmatrix} Q & S \\ S^T & R \end{bmatrix} = NN^T = \begin{bmatrix} L \\ K^T \end{bmatrix}[L^T \; K] = \begin{bmatrix} LL^T & LK \\ K^T L^T & K^T K \end{bmatrix} \geq 0$$

where $L$ is a $(n \times q)$-dimensional matrix and $K^T$ is a $(m \times q)$-dimensional matrix. Replacing $Q$ by $LL^T$, $S^T$ by $K^T L^T$ and $R$ by $K^T K$ in (C.4) through (C.6), one gets (C.8) through (C.10). We will limit ourselves to show that (3) → (5), since this will be used subsequently for other proofs.

Using (C.4) and (C.1), one has:

$$\begin{aligned} x^T(t)Qx(t) &= -x^T(t+1)Px(t+1) + x^T(t)Px(t) \\ &\quad + u^T(t)B^T Px(t+1) + x^T(t+1)PBu(t) - u^T(t)B^T PBu(t) \\ &= -x^T(t+1)Px(t+1) + x^T(t)Px(t) + u^T(t)B^T P[Ax(t) + Bu(t)] \\ &\quad + [x^T(t)A^T + u^T(t)B^T]PBu(t) - u^T(t)B^T PBu(t) \end{aligned} \tag{C.12}$$

Adding the term $2u^T(t)S^T x(t) + u^T(t)Ru(t)$ in both sides of (C.12) and taking into account (C.5) and (C.6), one gets:

$$\begin{aligned} &x^T(t)Qx(t) + 2u^T(t)S^T x(t) + u^T(t)Ru(t) \\ &\quad = -x^T(t+1)Px(t+1) + x^T(t)Px(t) \\ &\qquad + 2u^T(t)Cx(t) + u^T(t)[D + D^T]u(t) \\ &\quad = -x^T(t+1)Px(t+1) \\ &\qquad + x^T(t)Px(t) + 2y^T(t)u(t) \end{aligned} \tag{C.13}$$

Summing up from 0 to $t_1$, one gets (C.11). □

**Definition C.10** A $m \times m$ discrete transfer matrix $H(z)$ of real rational functions is *strictly positive real* if and only if:

1. All the elements $H(z)$ are analytic in $|z| \geq 1$.
2. The matrix

   $$H(e^{j\omega}) + H^T(e^{-j\omega})$$

   is a positive definite Hermitian for all real $\omega$.

**Definition C.11** The system (C.1)–(C.2) is termed *state strictly passive* if:

$$\eta(0, t_1) \geq \beta_1 \|x(t_1 + 1)\|^2 - \beta_0 \|x(0)\|^2 + \mu \|x\|_{2T}^2; \quad \beta_1, \beta_0, \mu > 0; \ \forall t_1 \geq 0$$

and for all $u(t)$ and $x(t)$ that satisfy (C.1).

**Lemma C.4** (Generalized Positive Real Lemma) *Consider the system* (C.1)–(C.2) *characterized by the transfer matrix* $H(z)$ *given in* (C.3). *Assume that there is a symmetric positive matrix* $P$, *a semi-positive definite matrix* $Q$, *symmetric matrices* $R$, $\Delta$, $K$, $M_o$ *and a matrix* $S$ *of appropriate dimensions such that*:

$$A^T PA - P = -Q - C^T \Delta C - M_0 \tag{C.14}$$

$$C^T - B^T PA = S^T + D^T \Delta C \tag{C.15}$$

$$D + D^T - B^T PB = R + D^T \Delta D + K \tag{C.16}$$

$$M = \begin{bmatrix} Q & S \\ S^T & R \end{bmatrix} \geq 0 \tag{C.17}$$

*Then*:

1. *Every solution* $x(t, x(0), u(t))$ *of* (C.1)–(C.2) *satisfies the following equality*:

$$\sum_{t=0}^{t_1} y^T(t)u(t) = \frac{1}{2}x^T(t_1+1)Px(t_1+1) - \frac{1}{2}x^T(0)Px(0)$$
$$+\frac{1}{2}\sum_{t=0}^{t_1}[x^T(t)u^T(t)]M\begin{bmatrix} x(t) \\ u(t)\end{bmatrix} + \frac{1}{2}\sum_{t=0}^{t_1}x^T(t)M_0x(t)$$
$$+\frac{1}{2}\sum_{t=0}^{t_1}u^T(t)Ku(t) + \frac{1}{2}\sum_{t=0}^{t_1}y^T(t)\Delta y(t) \tag{C.18}$$

2. *For* $M_0 = 0$, $\Delta = 0$, $K = 0$
   - *The system* (C.1)–(C.2) *is passive*
   - $H(z)$ *is positive real*
3. *For* $M_0 \geq 0$, $\Delta \geq 0$, $K > 0$
   - *The system* (C.1)–(C.2) *is input strictly passive*
   - $H(z)$ *is positive real*
4. *For* $M_0 \geq 0$, $\Delta > 0$, $K \geq 0$
   - *The system* (C.1)–(C.2) *is output strictly passive*
   - $H(z)$ *is positive real*
5. *For* $M_0 > 0$, $\Delta \geq 0$, $K > 0$
   - *The system* (C.1)–(C.2) *is input strictly passive*
   - *The system* (C.1)–(C.2) *is state strictly passive*
   - $H(z)$ *is strictly positive real*
6. *For* $M_0 > 0$, $\Delta > 0$, $K \geq 0$
   - *The system* (C.1)–(C.2) *is output strictly passive*
   - *The system* (C.1)–(C.2) *is state strictly passive*
   - $H(z)$ *is strictly positive real*
7. *For* $M_0 > 0$, $\Delta > 0$, $K > 0$
   - *The system* (C.1)–(C.2) *is very strictly passive*
   - *The system* (C.1)–(C.2) *is state strictly passive*
   - $H(z)$ *is strictly positive real*

*Proof* The properties (2) through (7) result from the property (1) bearing in mind the various definitions of passivity and the properties of positive and strictly positive real transfer matrices. We will show next that (C.14) through (C.17) implies property (C.18). Denoting:

$$\bar{Q} = Q + C^T\Delta C + M_0 \tag{C.19}$$
$$\bar{S} = S + C^T\Delta D \tag{C.20}$$
$$\bar{R} = R + D^T\Delta D + K \tag{C.21}$$
$$\bar{M} = \begin{bmatrix} \bar{Q} & \bar{S} \\ \bar{S}^T & \bar{R}\end{bmatrix} \tag{C.22}$$

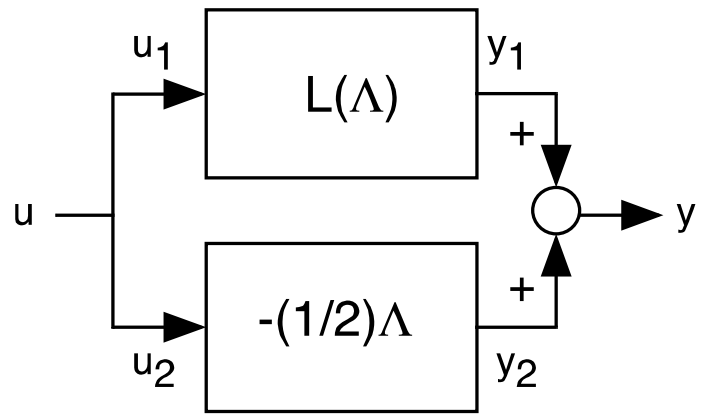

**Fig. C.2** The class $L(\Lambda)$

one can use the result (C.11) of Lemma C.3, which yields:

$$\sum_{t=0}^{t_1} y^T(t)u(t) = \frac{1}{2}x^T(t_1+1)Px(t_1+1) - \frac{1}{2}x^T(0)Px(0) + \frac{1}{2}\sum_{t=0}^{t_1}[x^T(t), u^T(t)]\bar{M}\begin{bmatrix} x(t) \\ u(t) \end{bmatrix} \quad \text{(C.23)}$$

Taking into account the particular form of $\bar{M}$, one obtains:

$$\begin{aligned}\sum_{t=0}^{t_1} y^T(t)u(t) = {} & \frac{1}{2}x^T(t_1+1)Px(t_1+1) - \frac{1}{2}x^T(0)Px(0) \\ & + \frac{1}{2}\sum_{t=0}^{t_1}[x^T(t)u^T(t)]M\begin{bmatrix} x(t) \\ u(t) \end{bmatrix} \\ & + \frac{1}{2}\sum_{t=0}^{t_1}x^T(t)M_0x(t) + \frac{1}{2}\sum_{t=0}^{t_1}u^T(t)Ku(t) \\ & + \frac{1}{2}\sum_{t=0}^{t_1}[Cx(t)+Du(t)]^T\Delta[Cx(t)+Du(t)] \end{aligned} \quad \text{(C.24)}$$

and using (C.2), one gets (C.18). □

**Definition C.12** (Landau and Silveira 1979) The linear time invariant system described by (C.1) and (C.2) is said to belong to the class $L(\Lambda)$ if and only if for a given positive definite matrix $\Lambda$ of appropriate dimension:

$$H'(z) = C(zI-A)^{-1}B + D - \frac{1}{2}\Lambda = H(z) - \frac{1}{2}\Lambda \quad \text{(C.25)}$$

is a strictly positive real transfer matrix.

The interpretation of this definition is given in Fig. C.2.

The linear time invariant system (C.1)–(C.2) characterized by the discrete time transfer function $H(z)$ belongs to the class $L(\Lambda)$ if the parallel combination of this block with a block having the gain $-\frac{1}{2}\Lambda$ is a system characterized by a strictly positive real transfer matrix.

**Lemma C.5** *If the system* (C.1)–(C.2) *belongs to the class* $L(\Lambda)$, *one has*:

$$\begin{aligned}\sum_{t=0}^{t_1} y^T(t)u(t) &= \frac{1}{2}x^T(t_1+1)Px(t_1+1) - \frac{1}{2}x^T(0)Px(0)\\ &\quad + \frac{1}{2}\sum_{t=0}^{t_1}[x^T(t)u^T(t)]M\begin{bmatrix} x(t)\\ u(t)\end{bmatrix} + \frac{1}{2}\sum_{t=0}^{t_1} x^T(t)M_0x(t)\\ &\quad + \frac{1}{2}\sum_{t=0}^{t_1} u^T(t)\Lambda u(t) + \frac{1}{2}\sum_{t=0}^{t_1} u^T(t)Ku(t)\\ \Lambda > 0;\ M_0 > 0;\ K > 0 \end{aligned} \tag{C.26}$$

In fact, as it results from (C.26) a system belonging to $L(\Lambda)$ is input strictly passive and the excess of passivity is defined by $\Lambda$ (see Lemma C.4).

## C.4 Discrete Linear Time-Varying Passive Systems

Consider the discrete linear time-varying system:

$$\bar{x}(t+1) = A(t)\bar{x}(t) + B(t)\bar{u}(t) \tag{C.27}$$

$$\bar{y}(t) = C(t)\bar{x}(t) + D(t)\bar{u}(t) \tag{C.28}$$

where $\bar{x}, \bar{u}, \bar{y}$ are the state, the input and the output vectors respectively ($\bar{u}$ and $\bar{y}$ are of the same dimension), and $A(t)$, $B(t)$, $C(t)$, $D(t)$ are sequences of time-varying matrices of appropriate dimensions defined for all $t \geq 0$.

**Lemma C.6** *The system* (C.27)–(C.28) *is* passive *if one of the following equivalent propositions holds*:

1. *There are sequences of time-varying positive semidefinite matrices* $\bar{P}(t)$, *of positive semidefinite matrices* $Q(t)$ *and* $R(t)$ *and a matrix sequence* $S(t)$ *such that*:

$$A^T(t)\bar{P}(t+1)A(t) - \bar{P}(t) = -Q(t) \tag{C.29}$$

$$C(t) - B^T(t)\bar{P}(t+1)A(t) = S(t) \tag{C.30}$$

$$D(t) + D^T(t) - B^T(t)P(t+1)B(t) = R(t) \tag{C.31}$$

$$\bar{M}(t) = \begin{bmatrix} Q(t) & S(t)\\ S^T(t) & R(t)\end{bmatrix} \geq 0 \tag{C.32}$$

   *with* $\bar{P}(0)$ *bounded.*
2. *Every solution* $x(t)(x(0), u(t), t)$ *of* (C.27)–(C.28) *satisfies the following equality*:

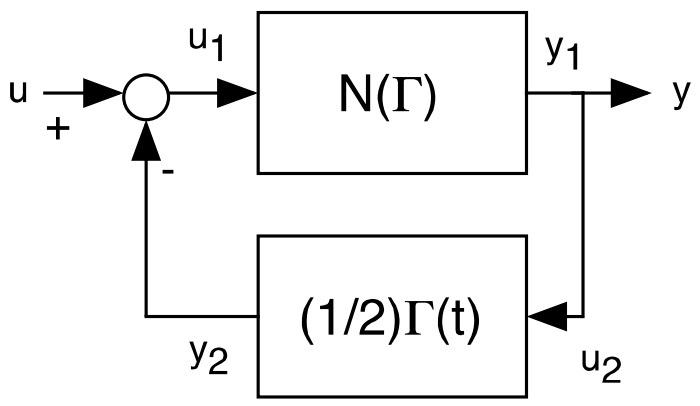

**Fig. C.3** The class $N(\Gamma)$

$$\sum_{t=0}^{t_1} \bar{y}^T(t)\bar{u}(t) = \frac{1}{2}x^T(t_1+1)\bar{P}(t+1)x(t_1+1) - \frac{1}{2}x^T(0)\bar{P}(0)x(0)$$

$$+\frac{1}{2}\sum_{t=0}^{t_1}[\bar{x}^T(t), \bar{u}^T(t)]\bar{M}(t)\begin{bmatrix}\bar{x}(t)\\ \bar{u}(t)\end{bmatrix} \quad \text{(C.33)}$$

*Proof* The passivity property results from (C.33). To obtain (C.33) from (C.32), one follows exactly the same steps as for the proof of Lemma C.3. □

For the study of the stability of interconnected systems, it is useful to consider the class of linear time-varying discrete time systems defined next.

**Definition C.13** (Landau and Silveira 1979) The linear time-varying system (C.27)–(C.28) is said to belong to the class $N(\Gamma)$ if for a given sequence of symmetric matrices $\Gamma(t) \geq 0$ of appropriate dimension, one has:

$$\sum_{t=0}^{t_1} \bar{y}^T(t)\bar{u}(t) = \frac{1}{2}\bar{x}^T(t_1+1)\bar{P}(t_1+1)\bar{x}(t_1+1) - \frac{1}{2}\bar{x}^T(0)\bar{P}(0)\bar{x}(0)$$

$$+\frac{1}{2}\sum_{t=0}^{t_1}[\bar{x}^T(t), \bar{u}^T(t)]\bar{M}(t)\begin{bmatrix}\bar{x}(t)\\ \bar{u}(t)\end{bmatrix}$$

$$-\frac{1}{2}\sum_{t=0}^{t_1}\bar{y}^T(t)\Gamma(t)\bar{y}(t) \quad \text{(C.34)}$$

with $\bar{P}(t) \geq 0, \bar{M}(t) \geq 0, \forall t \geq 0$ and $\bar{P}(0)$ bounded.

The interpretation of this definition is given in Fig. C.3.

Equation (C.34) indicates that the block resulting from the negative feedback connection of the linear time-varying system of (C.27) and (C.28) belonging to the class $N(\Gamma)$ with a block having a gain $\frac{1}{2}\Gamma(t)$ is passive. The resulting block has the input $\bar{u}^R(t) = \bar{u}(t) + \frac{1}{2}\Gamma(t)\bar{y}(t)$ and the output $\bar{y}^R(t) = y(t)$. In fact, the system belonging to the class $N(\Gamma)$ has a lack of passivity which is expressed by the last term of (C.34).

Algebraic conditions upon $A(t), B(t), C(t), D(t)$ in order to satisfy (C.34) are given next:

**Lemma C.7** *The system of* (C.27) *and* (C.28) *belongs to the class* $N(\Gamma)$ *if for a given sequence of symmetric matrices* $\Gamma(t) \geq 0$ *there exist three sequences of non-negative definite matrices* $\bar{P}(t)$, $R(t)$, $Q(t)$ *and a matrix sequence* $S(t)$ *such that*:

$$A^T(t)\bar{P}(t+1)A(t) - \bar{P}(t) = -Q(t) + C^T(t)\Gamma(t)C(t) \tag{C.35}$$

$$C^T(t) - B^T(t)\bar{P}(t+1)A(t) = S^T(t) - D^T(t)\Gamma(t)C(t) \tag{C.36}$$

$$D(t) + D^T(t) - B^T(t)\bar{P}(t+1)B(t) = R(t) - D^T(t)\Gamma(t)D(t) \tag{C.37}$$

*and*:

$$\bar{M}(t) = \begin{bmatrix} Q(t) & S(t) \\ S^T(t) & R(t) \end{bmatrix} \geq 0 \tag{C.38}$$

*with* $\bar{P}(0)$ *bounded.*

*Proof* A detailed proof can be found in Landau and Silveira (1979). The proof is similar to the one for Lemma C.4 by replacing constant matrices $Q, S, R, P$ by the corresponding time-varying matrices $Q(t), R(t), S(t), \bar{P}(t)$, replacing $\Delta$ by $-\Gamma(t)$ and taking $K = 0$, $M_0 = 0$. □

## C.5 Stability of Feedback Interconnected Systems

**Theorem C.1** (Landau 1980) *Consider a linear time invariant block described by* (C.1) *and* (C.2) *belonging to the class* $L(\Lambda)$ *in feedback connection with a discrete linear time-varying block described by* (C.27) *and* (C.28) *belonging to the class* $N(\Gamma)$. *If*:

$$\Lambda - \Gamma(t) \geq 0 \tag{C.39}$$

*the following properties hold*:

(P1)

$$\lim_{t\to\infty} x(t) = 0; \qquad \lim_{t\to\infty} y(t) = 0; \qquad \lim_{t\to\infty} u(t) = -\lim_{t\to\infty} \bar{y}(t) = 0 \tag{C.40}$$

*with* $\bar{P}(t), Q(t), S(t), R(t)$ *bounded or unbounded for* $t > 0$.

(P2)

$$\lim_{t\to\infty} [\bar{x}(t)^T, \bar{u}(t)^T] \begin{bmatrix} Q(t) & S(t) \\ S(t)^T & R(t) \end{bmatrix} \begin{bmatrix} \bar{x}(t) \\ \bar{u}(t) \end{bmatrix} = 0 \tag{C.41}$$

*with* $Q(t), S(t), R(t)$ *bounded or unbounded for* $t > 0$.

(P3)

$$\bar{x}(t)^T P(t)\bar{x}(t) \leq C_1 < \infty; \quad \forall t > 0 \tag{C.42}$$

*with* $P(t)$ *bounded or unbounded for* $t > 0$.

(P4)

$$\lim_{t\to\infty} \bar{x}(t)^T P(t)\bar{x}(t) = \text{const} \tag{C.43}$$

*with* $P(t)$ *bounded or unbounded for* $t > 0$.

*Proof* The feedback connection corresponds to:

$$u(t) = -\bar{y}(t) \tag{C.44}$$

$$y(t) = \bar{u}(t) \tag{C.45}$$

which implies:

$$\sum_{t=0}^{t_1} y(t)^T u(t) = -\sum_{t=0}^{t_1} \bar{y}(t)^T \bar{u}(t) \tag{C.46}$$

1. Evaluating the left hand side term of (C.46) from (C.26), and the right hand side term of (C.46) from (C.34), one obtains after rearranging the various terms (and taking into account (C.44) and (C.45)):

$$\begin{aligned} & x^T(t_1+1)Px(t_1+1) + \sum_{t=0}^{t_1} x^T(t)M_0x(t) + \sum_{t=0}^{t_1} u^T(t)Ku(t) \\ & \quad + \bar{x}^T(t_1+1)\bar{P}(t_1+1)\bar{x}(t_1+1) + \sum_{t=0}^{t_1}[\bar{x}^T(t)y^T(t)]\bar{M}(t)\begin{bmatrix} c\bar{x}(t) \\ y(t) \end{bmatrix} \\ & \quad + \sum_{t=0}^{t_1} \bar{y}^T(t)[\Lambda - \Gamma(t)]\bar{y}(t) + \sum_{t=0}^{t_1}[x^T(t), -\bar{y}^T(t)]M\begin{bmatrix} x(t) \\ -\bar{y}(t) \end{bmatrix} \\ & = \bar{x}^T(0)\bar{P}(0)\bar{x}(0) + x^T(0)Px(0) \end{aligned} \tag{C.47}$$

Since $\Lambda - \Gamma(t) \geq 0$, one obtains:

$$\begin{aligned} & \lim_{t_1\to\infty}\left[x^T(t_1+1)Px(t_1+1) + \sum_{t=0}^{t_1} x^T(t)M_0x(t) + \sum_{t=0}^{t_1} u^T(t)Ku(t)\right] \\ & \leq x(0)Px(0) + \bar{x}^T(0)\bar{P}(0)\bar{x}(0) \end{aligned} \tag{C.48}$$

with $\bar{P}(t)$, $Q(t)$, $S(t)$, $R(t)$ bounded or unbounded for $t > 0$. Since $P > 0$, $M_0 > 0$, $K > 0$, this implies that:

$$\lim_{t\to\infty} x(t) = 0; \qquad \lim_{t\to\infty} u(t) = 0; \qquad \forall x(0),\ \forall \bar{x}(0) < \infty \tag{C.49}$$

and from (C.2), it also results that:

$$\lim_{t\to\infty} y(t) = 0 \tag{C.50}$$

Bearing in mind that $u(t) = -\bar{y}(t)$, one has:

$$\lim_{t\to\infty} \bar{y}(t) = 0 \tag{C.51}$$

2. From (C.47), taking into account (C.45), one obtains:

$$\lim_{t\to\infty}\left\{\bar{x}^T(t_1+1)\bar{P}(t_1+1)\bar{x}(t_1+1) + \sum_{t\to\infty}^{t_1}[\bar{x}^T(t)\bar{u}^T(t)]\begin{bmatrix} Q(t) & S(t) \\ S^T(t) & R(t)\end{bmatrix}\begin{bmatrix}\bar{x}(t)\\ \bar{u}(t)\end{bmatrix}\right\} \le x^T(0)Px(0)+\bar{x}^T(0)\bar{P}(0)\bar{x}(0) \quad \text{(C.52)}$$

with $\bar{P}(t)$, $S(t)$, $Q(t)$, $R(t)$ bounded or unbounded for $t>0$. This implies that:

$$\lim_{t\to\infty}[\bar{x}^T(t), \bar{u}^T(t)]\begin{bmatrix} Q(t) & S(t) \\ S^T(t) & R(t)\end{bmatrix}\begin{bmatrix}\bar{x}(t)\\ \bar{u}(t)\end{bmatrix} = 0 \quad \text{(C.53)}$$

3. From (C.52) one also has:

$$\bar{x}^T(t)\bar{P}(t)\bar{x}(t) \le C_1 < \infty; \quad \forall t > 0 \quad \text{(C.54)}$$

4. Denote

$$V(t) = x^T(t)Px(t) + \bar{x}^T(t)\bar{P}(t)\bar{x}(t) \quad \text{(C.55)}$$

Again from (C.47), one obtains:

$$\begin{aligned} V(t+1)-V(t) &= -\bar{y}^T(t)[\Lambda - \Gamma(t)]\bar{y}(t) - x^T(t)M_0(t)x(t) \\ &\quad - u^T(t)Ku(t) - [x^T(t), u^T(t)]M\begin{bmatrix}x(t)\\ u(t)\end{bmatrix} \\ &\quad - [\bar{x}^T(t), y^T(t)]\bar{M}(t)\begin{bmatrix}\bar{x}(t)\\ y(t)\end{bmatrix} \le 0 \end{aligned} \quad \text{(C.56)}$$

Since $V(t)>0$; $\forall x(t) \neq 0$; $\forall \bar{x}(t)\neq 0$ and $V(t+1)-V(t)\le 0$, $V(t)$ will converge to a fixed value. One has proven already that $x(t)\to 0$. One concludes that:

$$\lim_{t\to\infty}\bar{x}^T(t)\bar{P}(t)\bar{x}(t) = \text{const} \quad \text{(C.57)}$$

□

Theorem C.1 can be viewed as an extension of the asymptotic hyperstability theorem (see next paragraph) which corresponds to $\Lambda = 0$; $\Gamma(t)\equiv 0$. In fact, in the asymptotic hyperstability theorem, there is no specific structure attached to the feedback block (which can be linear, nonlinear, time-varying). The only condition is that the feedback block is *passive*. By considering a state space structure for the feedback block, it is possible to get certain properties for the state of the feedback block. This is needed in adaptive control.

## C.6 Hyperstability and Small Gain

**Theorem C.2** (Asymptotic Hyperstability) *Consider the feedback connection (Fig. C.4) of a linear time-invariant block $H_1$ (state: $x(t)$), characterized by a*

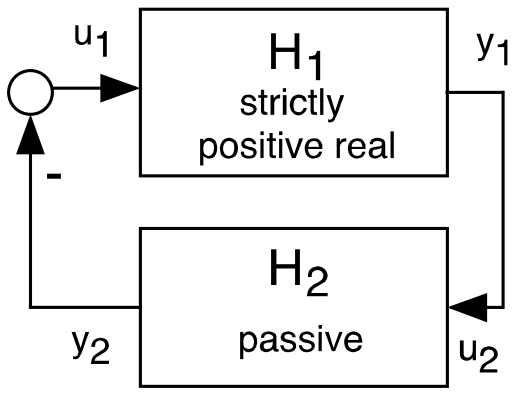

**Fig. C.4** Feedback connection (hyperstability)

*strictly positive real transfer function* (*which implies that* $H_1$ *is input strictly passive*) *with a block* $H_2$ (*linear or nonlinear, time invariant or time-varying*) *characterized by*:

$$\eta_2(0, t_1) = \sum_{t=0}^{t_1} y_2^T(t)u_2(t) \geq -\gamma_2^2; \quad \gamma_2^2 < \infty; \ \forall t_1 \geq 0 \tag{C.58}$$

*Then*:

$$\lim_{t\to\infty} x(t) = 0; \qquad \lim_{t\to\infty} u_1(t) = \lim_{t\to\infty} y_1(t) = 0; \quad \forall x(0) \tag{C.59}$$

The proof is similar to that of Theorem C.1 and it is omitted.

**Definition C.14** Consider a system $S$ with input $u$ and output $y$, the infinity norm of the system $S$ denoted $\|S\|_\infty$ is such that:

$$\|y\|_2^2 \leq \|S\|_\infty^2 \|u\|_2^2$$

**Definition C.15** Given a transfer function $H(z)$, the infinity norm $\|H\|_\infty$ is:

$$\|H\|_\infty = \max_\omega |H(e^{j\omega})|; \quad 0 \leq \omega \leq 2\pi$$

**Lemma C.8** (Small Gain Lemma) *The following propositions concerning the system of* (C.1)–(C.2) *are equivalent to each other*:

1. $H(z)$ *given by* (C.3) *satisfies*:

$$\|H(z^{-1})\|_\infty \leq \gamma; \quad 0 < \gamma < \infty \tag{C.60}$$

2. *There exist a positive definite matrix* $P$, *a positive semidefinite matrix* $Q$ *and matrices* $S$ *and* $R$ *such that*:

$$A^T PA - P = -Q - C^T C \tag{C.61}$$

$$-B^T PA = S^T + D^T C \tag{C.62}$$

$$B^T PB = R + D^T D - \gamma^2 I \tag{C.63}$$

$$M = \begin{bmatrix} Q & S \\ S^T & R \end{bmatrix} \geq 0 \tag{C.64}$$

3. *There is a positive definite matrix* $P$ *and matrices* $K$ *and* $L$ *such that*:

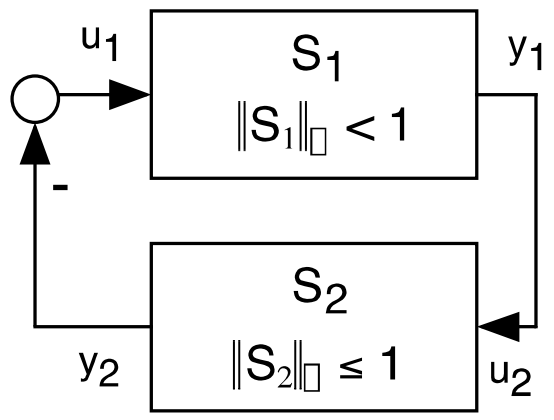

**Fig. C.5** Feedback connection (small gain)

$$A^T PA - P + C^T C = -LL^T \tag{C.65}$$
$$-B^T PA - D^T C = K^T L^T \tag{C.66}$$
$$B^T PB - D^T D + \gamma^2 I = K^T K \tag{C.67}$$

4. *Every solution* $x(t)(x(0), u(t))$ *of* (C.1)–(C.2) *satisfies the following equality*:

$$\begin{aligned}\sum_{t=0}^{t_1} y^T(t)y(t) &= \gamma^2 \sum_{t=0}^{t_1} u^T(t)u(t) + x^T(0)Px(0) - x^T(t_1+1)Px(t_1+1)\\ &\quad - \sum_{t=0}^{t_1}[x^T(t), u^T(t)]M\begin{bmatrix} x(t)\\ u(t)\end{bmatrix}\end{aligned} \tag{C.68}$$

*Remark* From (C.68) one gets for $x(0) = 0$:

$$\|y(t)\|_2^2 \le \gamma^2 \|u\|_2^2 \le \|H\|_\infty^2 \|u\|^2 \tag{C.69}$$

*Proof* We will only show that (C.61) through (C.64) implies (C.68). From (C.61), one gets:

$$\begin{aligned}x^T(t)Qx(t) &= -x^T(t)C^T Cx(t) + x^T(t)Px(t) - x^T(t)A^T PAx(t)\\ &= -x^T(t)C^T Cx(t) + x^T(t)Px(t) - x^T(t+1)Px(t+1)\\ &\quad + 2u^T(t)B^T PAx(t) - u^T(t)B^T PBu(t)\end{aligned} \tag{C.70}$$

Using (C.62) and (C.63) and adding the terms $2u^T(t)S^T x(t) + u^T(t)Ru(t)$ in the both sides of (C.58), one obtains:

$$\begin{aligned}[x^T(t), u^T(t)]M\begin{bmatrix} x(t)\\ u(t)\end{bmatrix} &= -y^T(t)y(t) + \gamma^2 u^T(t)u(t) + x^T(t)Px(t)\\ &\quad - x^T(t+1)Px(t+1)\end{aligned} \tag{C.71}$$

Summing up now from 0 to $t_1$, one gets (C.65). □

**Theorem C.3** (Small Gain) *Consider the feedback connection* (*Fig.* C.5) *between a linear time invariant block* $S_1$ (*state*: $x$) *characterized by* $\|S_1\|_\infty < 1$ *and a block* $S_2$ *characterized by* $\|S_2\|_\infty \le 1$. *Then*:

$$\lim_{t\to\infty} x(t) = 0; \qquad \lim_{t\to\infty} u_1(t) = \lim_{t\to\infty} y_1(t) = 0 \tag{C.72}$$

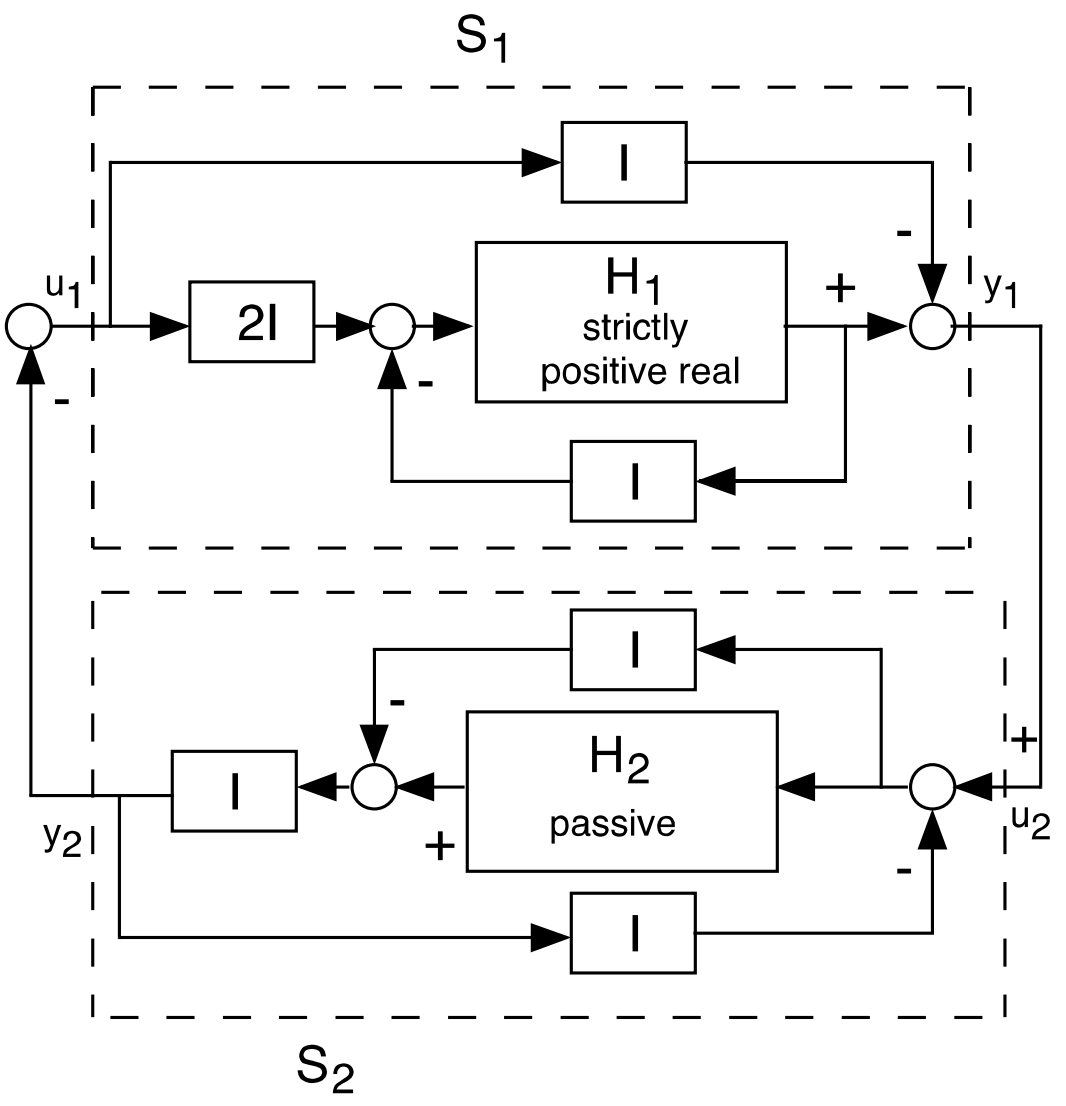

**Fig. C.6** Equivalent representation of the system shown in Fig. C.4

*Proof* The small gain theorem (Theorem C.3) can be obtained from Theorem C.2, using a series of equivalent loop transformation on the scheme of Fig. C.4 leading to the scheme shown in Fig. C.6, where the input-output operator $S_1$ ($y_1 = S_1 u_1$) is given by:

$$S_1 = (H_1 - I)(H_1 + I)^{-1} \tag{C.73}$$

and the operator $S_2(y_2 = S_2 u_2)$ is given by:

$$S_2 = (H_2 - I)(H_2 + I)^{-1} \tag{C.74}$$

It can be shown (Desoer and Vidyasagar 1975) that under the hypotheses of Theorem C.2, $\|S_1\|_\infty < 1$ and $\|S_2\|_\infty \leq 1$. □

# Appendix D
# Martingales

This appendix gives a number of results related to the properties of feedback systems in the presence of stochastic disturbances represented as a martingale sequence (for the definition of a martingale sequence see Appendix A). The properties of passive (hyperstable) systems are extensively used in relation to some basic results concerning the convergence of non-negative random variables. The results of this appendix are useful for the proofs of Theorems 4.2 and 4.3, as well as for the convergence analysis of various recursive identification and adaptive control schemes.

**Theorem D.1** (Neveu 1975) *If $T(t)$ and $\alpha(t+1)$ are non-negative random variables measurable with respect to an increasing sequence of $\sigma$-algebras $\mathcal{F}_t$ and satisfy*:

$$(1)\quad \mathbf{E}\{T(t+1)|\mathcal{F}_t\} \le T(t) + \alpha(t+1) \tag{D.1}$$

$$(2)\quad \sum_{t=1}^{\infty} \alpha(t) < \infty; \quad a.s. \tag{D.2}$$

*then*:

$$\lim_{t\to\infty} T(t) = T \quad a.s. \tag{D.3}$$

*where $T$ is a finite non-negative random variable*.

Consider the stochastic feedback system shown in Fig. D.1 and described by:

$$x(t+1) = Ax(t) + Bu(t+1) \tag{D.4}$$

$$y(t+1) = Cx(t) + Du(t+1) \tag{D.5}$$

$$\bar{u}(t+1) = y(t+1) + \omega(t+1) \tag{D.6}$$

$$\bar{x}(t+1) = A(t)\bar{x}(t) + B(t)\bar{u}(t+1) \tag{D.7}$$

$$\bar{y}(t+1) = -u(t+1) = C(t)\bar{x}(t) + D(t)\bar{u}(t+1) \tag{D.8}$$

Note that the output of the linear time-invariant feedforward block described by (D.4) and (D.5) is disturbed by the sequence $\{\omega(t+1)\}$. The following assumptions are made upon the system of (D.4)–(D.8).

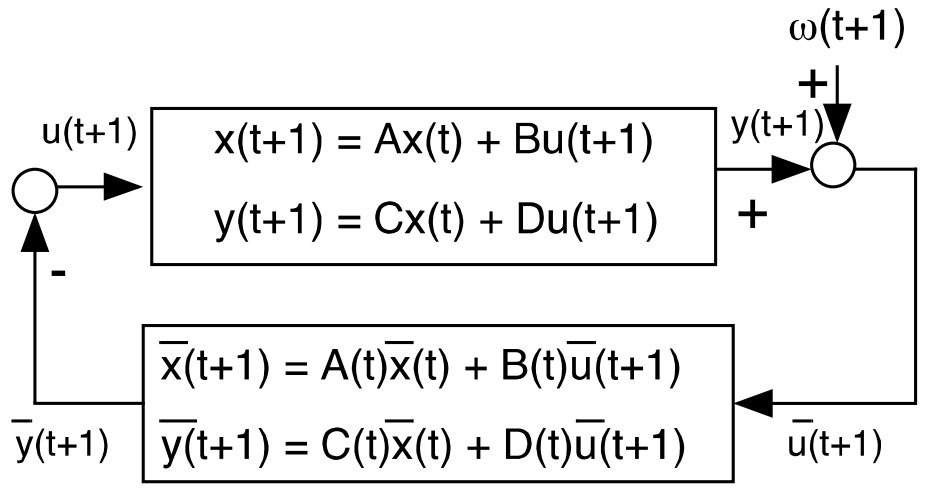

**Fig. D.1** Stochastic feedback system associated with (D.4) through (D.8)

D.1 There exists a symmetric matrix $\Lambda$ such that the linear time-invariant block described by (D.4) and (D.5) belongs to the class $L(\Lambda)$ (see Definition C.3).

D.2 There exists a sequence of matrices $\Gamma(t)$ such that the feedback linear time-varying block described by (D.7) and (D.8) belongs to the class $N(\Gamma)$ (see Definition C.8).

D.3 $\{\omega(t)\}$ is a vectorial martingale difference sequence defined on a probability space $(\Omega, \mathcal{A}, \mathcal{P})$ adapted to the sequence of increasing $\sigma$-algebras $\mathcal{F}_t$ generated by the observations up to and including time $t$. The sequence $\{\omega(t+1)\}$ satisfies:

$$\mathbf{E}\{\omega(t+1)|\mathcal{F}_t\} = 0 \quad \text{a.s.} \tag{D.9}$$

$$\mathbf{E}\{\omega(t+1)\omega^T(t+1)|\mathcal{F}_t\} = \text{diag}[\sigma_i^2] \quad \text{a.s.} \tag{D.10}$$

$$\lim_{N\to\infty} \sup \frac{1}{N}\sum_{t=1}^{N} \omega^T(t+1)\omega(t+1) < \infty \quad \text{a.s.} \tag{D.11}$$

**Theorem D.2** (Landau 1982b) *Let Assumptions* D.1, D.2 *and* D.3 *hold for the system of* (D.4)–(D.8). *If*:

$$(1) \quad \alpha(t+1) = \frac{1}{t+1}\mathbf{E}\{\bar{y}^T(t+1)\omega(t+1)|\mathcal{F}_t\} \geq 0; \quad \forall t \geq 0 \tag{D.12}$$

$$(2) \quad \sum_{t=1}^{\infty} \alpha(t) < \infty \quad a.s. \tag{D.13}$$

$$(3) \quad \Lambda - \Gamma(t) \geq 0; \quad \forall t \geq 0 \tag{D.14}$$

*then*:

$$(a) \quad \lim_{t\to\infty} \frac{1}{t}\sum_{i=1}^{t} \|x(i)\|^2 = \lim_{t\to\infty} \frac{1}{t}\sum_{i=1}^{t} \|u(i)\|^2 = \lim_{t\to\infty} \frac{1}{t}\sum_{i=1}^{t} \|y(i)\|^2 = 0 \quad a.s. \tag{D.15}$$

$$(b) \quad \lim_{t\to\infty} \bar{x}^T(t)\left(\frac{\bar{P}(t)}{t}\right)\bar{x}(t) = 0 \quad a.s. \tag{D.16}$$

$$(c) \quad \textit{If, in addition, } \lim_{t\to\infty} \frac{1}{t}\bar{P}(t) > 0 \; a.s., \textit{ then: } \lim_{t\to\infty} \bar{x}(t) = 0 \; a.s. \tag{D.17}$$

*Proof* The proof is based on the use of the near supermartingale convergence theorem (Theorem D.1). The appropriate near supermartingale is:

$$T(t) = \frac{V(t)}{t} + \sum_{i=1}^{t-1} \frac{1}{i+1} \frac{V(i)}{i} \tag{D.18}$$

where:

$$\begin{aligned} V(t) &= x^T(t)Px(t) + \bar{x}^T(t)\bar{P}(t)\bar{x}(t) + \sum_{i=0}^{t-1} x^T(i)M_0x(i) \\ &\quad + \sum_{i=0}^{t-1} u^T(i+1)Ku(i+1) \end{aligned} \tag{D.19}$$

Taking into account (D.6) and (D.8), one has:

$$\begin{aligned} \sum_{i=0}^{t} y^T(i+1)u(i+1) &= -\sum_{i=0}^{t} \bar{y}^T(i+1)\bar{u}(i+1) \\ &\quad + \sum_{i=0}^{t} \bar{y}^T(i+1)\omega(i+1) \end{aligned} \tag{D.20}$$

The expressions for $\sum_{i=0}^{t} y^T(i+1)u(i+1)$ and $\sum_{i=0}^{t} \bar{y}^T(i+1)\bar{u}(i+1)$ can be obtained from Lemmas C.5 and C.7, since the two blocks involved belong to the classes $L(\Lambda)$ and $N(\Gamma)$ respectively. Combining (C.26) and (C.34) through the use of (D.20) and rearranging the various terms taking into account (D.19), one obtains:

$$\begin{aligned} V(t+1) &= x^T(t+1)Px(t+1) + \bar{x}^T(t+1)\bar{P}(t+1)\bar{x}(t+1) \\ &\quad + \sum_{i=0}^{t} x^T(i)M_0x(i) + \sum_{i=0}^{t} u^T(i+1)Ku(i+1) \\ &= x^T(0)Px(0) + \bar{x}^T(0)\bar{P}(0)\bar{x}(0) - \sum_{i=0}^{t} [x^T(i), u^T(i+1)]M \begin{bmatrix} x(i) \\ u(i+1) \end{bmatrix} \\ &\quad - \sum_{i=0}^{t} [\bar{x}^T(i), \bar{u}^T(i+1)]\bar{M}(t) \begin{bmatrix} \bar{x}(i) \\ \bar{u}(i+1) \end{bmatrix} \\ &\quad - \sum_{i=0}^{t} \bar{y}^T(i+1)[\Lambda - \Gamma(i)]\bar{y}(i+1) \\ &\quad + 2\sum_{i=0}^{t} \bar{y}^T(i+1)\omega(i+1) \end{aligned} \tag{D.21}$$

Since $\Lambda - \Gamma(t) \geq 0$ (from (D.14)) and $\bar{M}(t) \geq 0$ (from (C.34)), one concludes from (D.21) that:

$$V(t+1) \leq V(t) + 2\bar{y}(t+1)\omega(t+1) \tag{D.22}$$

and:

$$\frac{V(t+1)}{t+1} \le \frac{V(t)}{t} - \frac{1}{t+1}\frac{V(t)}{t} + \frac{2}{t+1}\bar{y}(t+1)\omega(t+1) \tag{D.23}$$

Adding in both sides of (D.23) the term $\sum_{i=1}^{t}(\frac{1}{i+1})\frac{V(i)}{i}$, one concludes that $T(t)$ given by (D.18) satisfies:

$$T(t+1) \le T(t) + \frac{2}{t+1}\bar{y}^T(t+1)\omega(t+1) \tag{D.24}$$

and:

$$\mathbf{E}\{T(t+1)|\mathcal{F}_t\} \le T(t) + \frac{2}{t+1}\mathbf{E}\{\bar{y}^T(t+1)\omega(t+1)\} \tag{D.25}$$

Conditions (D.12) and (D.13) together with (D.25) lead to the satisfaction of Theorem D.1 and therefore:

$$\lim_{t\to\infty}\left[\frac{V(t)}{t} + \sum_{i=1}^{t-1}\frac{1}{i+1}\frac{V(i)}{i}\right] < \infty \quad \text{a.s.} \tag{D.26}$$

But since $V(t) \ge 0$, it results in order to avoid contradiction that:

$$\text{Prob}\left\{\lim_{t\to\infty}\frac{V(t)}{t} = 0\right\} = 1 \tag{D.27}$$

Bearing in mind that the matrix $M_0$ in (C.26) is positive definite and $K > 0$, it results from (D.27) that:

$$\lim_{t\to\infty}\frac{1}{t}\sum_{i=1}^{t}x^T(i)M_0x(i) = \lim_{t\to\infty}\frac{1}{t}\sum_{i=1}^{t}\|x(i)\|^2 = 0 \quad \text{a.s.} \tag{D.28}$$

and:

$$\lim_{t\to\infty}\frac{1}{t}\sum_{i=1}^{t}\|u(i+1)\|^2 = 0 \quad \text{a.s.} \tag{D.29}$$

From (D.5), one has:

$$\|y(i+1)\|^2 \le \alpha\|x(i)\|^2 + \beta\|u(i+1)\|^2; \quad 0 < \alpha,\ \beta < \infty \tag{D.30}$$

and therefore:

$$\lim_{t\to\infty}\frac{1}{t}\sum_{i=1}^{t}\|y(i+1)\|^2 = 0 \quad \text{a.s.} \tag{D.31}$$

The results (D.16) and (D.17) are directly obtained from (D.27) and (D.19). □

**Theorem D.3** *Let Assumptions* D.1, D.2 *and* D.3 *hold for the system* (D.4)–(D.8). *If*:

$$(1)\quad r(t+1) \geq r(t), \quad r(t) > 0,\ \forall t \geq 0 \tag{D.32}$$

$$(2)\quad \alpha(t+1) = E\left\{ \frac{\bar{y}^T(t+1)\omega(t+1)}{r(t+1)} | \mathcal{F}_t \right\} \geq 0, \quad \forall t \geq 0 \tag{D.33}$$

$$(3)\quad \sum_{t=1}^{\infty} \alpha(t) < \infty \quad a.s. \tag{D.34}$$

$$(4)\quad \Lambda - \Gamma(t) \geq 0; \quad \forall t \geq 0 \tag{D.35}$$

*then*:

$$\text{(a)}\quad \lim_{t\to\infty} \sum_{i=0}^{t} \frac{\|x(i)\|^2}{r(i+1)} < \infty \quad a.s. \tag{D.36}$$

$$\text{(b)}\quad \lim_{t\to\infty} \sum_{i=0}^{t} \frac{\|u(i+1)\|^2}{r(i+1)} < \infty \quad a.s. \tag{D.37}$$

$$\text{(c)}\quad \lim_{t\to\infty} \sum_{i=0}^{t} \frac{\|y(i+1)\|^2}{r(i+1)} < \infty \quad a.s. \tag{D.38}$$

$$\text{(d)}\quad \lim_{t\to\infty} \sum_{i=0}^{t} \frac{[\bar{x}^T(i), \bar{u}^T(i+1)]\bar{M}(i)\left[\begin{smallmatrix} \bar{x}(i) \\ \bar{u}(i+1) \end{smallmatrix}\right]}{r(i+1)} < \infty \quad a.s. \tag{D.39}$$

*Remark* This theorem is useful when one cannot conclude that, for a given scheme, that (D.13) is satisfied. One can then choose eventually a sequence $r(t)$ to satisfy the conditions of (D.35) and (D.34). The results of Theorem D.3 are weaker than those of Theorem D.2, but from the specific properties of the sequence $r(t)$ is then possible in certain cases to conclude upon the convergence of certain schemes (Goodwin and Sin 1984; Goodwin et al. 1980c). Note that if in addition of (D.34) one assumes $\lim_{t\to\infty} r(t) = \infty$, using Kronecker's lemma one obtains from (D.36), (D.37), (D.38) and (D.40) results of the form of Theorem D.2.

*Proof* One considers the following near supermartingale:

$$\begin{aligned} T(t) = {} & \frac{1}{r(t)}[x^T(t)Px(t) + \bar{x}^T(t)\bar{P}(t)\bar{x}(t)] \\ & + \sum_{i=0}^{t-1} \frac{[x^T(i), u^T(i+1)]M\left[\begin{smallmatrix} x(i) \\ u(i+1) \end{smallmatrix}\right]}{r(i+1)} + \sum_{i=0}^{t-1} \frac{x^T(i)M_0x(i)}{r(i+1)} \\ & + \sum_{i=0}^{t-1} \frac{[\bar{x}(i), \bar{u}(i+1)]\bar{M}(t)\left[\begin{smallmatrix} \bar{x}(i) \\ \bar{u}(i+1) \end{smallmatrix}\right]}{r(i+1)} \\ & + \sum_{i=0}^{t-1} \frac{u^T(i+1)Ku(i+1)}{r(i+1)} \end{aligned} \tag{D.40}$$

where $P$, $\bar{P}(t)$, $M$, $M_0$, $\bar{M}(t)$, $K$ result from Lemmas C.5 and C.7 respectively. Denote:

$$V(t) = x^T(t)Px(t) + \bar{x}^T(t)\bar{P}(t)\bar{x}(t) \tag{D.41}$$

Using Lemmas C.5 and C.7 and taking also into account (D.6) and (D.8), one has:

$$\begin{aligned} V(t+1) = V(t) - [x^T(t), u^T(t+1)]M\begin{bmatrix} x(t) \\ u(t+1) \end{bmatrix} - x^T(t)M_0x(t) \\ - [\bar{x}^T(t), \bar{u}^T(t+1)]\bar{M}(t)\begin{bmatrix} \bar{x}(t) \\ \bar{u}(t+1) \end{bmatrix} - \bar{y}(t+1)[\Lambda - \Gamma(t)]\bar{y}(t+1) \\ - u^T(t+1)Ku(t+1) + 2\bar{y}^T(t+1)\omega(t+1) \end{aligned} \tag{D.42}$$

From (D.42), taking into account (D.32), one gets:

$$\begin{aligned} \frac{V(t+1)}{r(t+1)} \le \frac{V(t)}{r(t)} - \frac{1}{r(t+1)}\Bigg\{[x^T(t), u^T(t+1)]M\begin{bmatrix} x(t) \\ u(t+1) \end{bmatrix} + x^T(t)M_0x(t) \\ + [\bar{x}^T(t), \bar{u}^T(t+1)]\bar{M}(t)\begin{bmatrix} \bar{x}(t) \\ \bar{u}(t+1) \end{bmatrix} + \kappa\|u(t+1)\|^2\Bigg\} \\ + 2\frac{\bar{y}(t+1)^T\omega(t+1)}{r(t+1)} \end{aligned} \tag{D.43}$$

Adding in both sides of (D.43):

$$\begin{aligned} \sum_{i=0}^{t}\frac{[x^T(i), u^T(i+1)]M\left[\begin{smallmatrix} x(i) \\ u(i+1) \end{smallmatrix}\right]}{r(i+1)} + \sum_{i=0}^{t}\frac{x^T(t)M_0x(t)}{r(i+1)} \\ + \sum_{i=0}^{t}\frac{[\bar{x}^T(i), \bar{u}^T(i+1)]\bar{M}^T(t)\left[\begin{smallmatrix} \bar{x}(i) \\ \bar{u}(i+1) \end{smallmatrix}\right]}{r(i+1)} + \kappa\sum_{i=0}^{t}\frac{\|u(i+1)\|^2}{r(i+1)} \end{aligned} \tag{D.44}$$

and taking the conditional expectation one gets using (D.40):

$$\mathbf{E}\{T(t+1)|\mathcal{F}_t\} \le T(t) + 2\mathbf{E}\left\{\frac{\bar{y}^T(t+1)\omega(t+1)}{r(t+1)}|\mathcal{F}_t\right\} \tag{D.45}$$

Under the hypotheses (D.33) and (D.34), it results using Theorem D.1:

$$\text{Prob}\left\{\lim_{t\to\infty} T(t) < \infty\right\} = 1 \tag{D.46}$$

Taking into account the structure of $T(t)$ given in (D.40), it results immediately:

$$\lim_{t\to\infty}\sum_{i=0}^{t}\frac{x^T(i)M_0x(i)}{r(i+1)} < \infty; \quad M_0 > 0 \text{ a.s.} \tag{D.47}$$

and:

$$\lim_{t\to\infty}\kappa\sum_{i=0}^{t}\frac{\|u(i+1)\|^2}{r(i+1)} < \infty; \quad \kappa > 0 \text{ a.s.} \tag{D.48}$$

which implies (D.36) and (D.37). From (D.5), one has:

$$\|y(i+1)\|^2 \leq \alpha \|x(i)\|^2 + \beta \|u(i+1)\|^2; \quad \alpha, \beta > 0 \tag{D.49}$$

and therefore (D.38) holds. (D.39) results also from the structure of $T(t)$. □

**Lemma D.1** (Kronecker Lemma): (Goodwin and Sin 1984) *Assume that* $\{x(t)\}$ *and* $\{b_t\}$ *are sequences of reals such that*:

(i) $\lim_{N\to\infty} \sum_{t=1}^{N} x(t) < \infty$ (ii) $\{b_N\} : b_N \geq b_{N-1}$ (iii) $\lim_{N\to\infty} b_N = \infty$

*Then*:

$$\lim_{N\to\infty} \frac{1}{b_N} \sum_{t=1}^{N} b_t x(t) = 0 \tag{D.50}$$

# Index